\documentclass[11pt]{article}
 \usepackage{titlesec}
\usepackage{hyperref}
\usepackage{graphicx}
\usepackage{stackrel}

\usepackage{blindtext}
\usepackage{subfiles} % Best loaded last in the preamble

\titleclass{\subsubsubsection}{straight}[\subsection]

\usepackage[utf8]{inputenc}
\usepackage{amsmath,amsfonts,amssymb,stackengine,graphicx}

\usepackage[utf8]{inputenc}
\newif\iffigs\figstrue

\usepackage{cite}
\usepackage[title]{appendix}
\usepackage{epsfig,latexsym}
\usepackage{hyperref}
\usepackage{amsmath}
\usepackage{verbatim}
\usepackage{color}
\usepackage{mathrsfs}
\usepackage{slashed}
\usepackage{amssymb}

\def\sla{\raise.15ex\hbox{$/$}\kern-.57em}

\DeclareMathAlphabet{\mathpzc}{OT1}{pzc}{m}{it}

 \csname
@addtoreset\endcsname{equation}{section}

\def\gz0{\gamma^{0}}

\def\sign{\rm sign}

\def\q{\quad}

\def\nn{\nonumber}

 \def\det{{\rm det\,}}

\def\a{\alpha}
\def\b{\beta}
\def\g{\gamma}

\def\e{\epsilon}

\def\l{\lambda}

\def\m{\mu}
\def\n{\nu}

\def\r{\rho}

\def\s{\sigma}

\def\t{\tau}

\def\vf{\varphi}

\def\cL{{\cal L}}

\def\beq{\begin{equation}}
\def\eeq{\end{equation}}
\def\be{\begin{equation}}
\def\ee{\end{equation}}
\def\bea{\begin{eqnarray}}
\def\eea{\end{eqnarray}}
\def\bec{\begin{center}}
\def\ec{\end{center}}
\def\beal{\begin{align}}
\def\enal{\end{align}}

\def\12{\frac{1}{2}}

\def\pr{\partial}

\newcounter{subsubsubsection}[subsubsection]
\renewcommand\thesubsubsubsection{\thesubsubsection.\arabic{subsubsubsection}}
\titleformat{\subsubsubsection}
  {\normalfont\normalsize\bfseries}{\thesubsubsubsection}{1em}{}
\titlespacing*{\subsubsubsection}
{0pt}{3.25ex plus 1ex minus .2ex}{1.5ex plus .2ex}

\makeatletter
\renewcommand\paragraph{\@startsection{paragraph}{5}{\z@}%
  {3.25ex \@plus1ex \@minus.2ex}%
  {-1em}%
  {\normalfont\normalsize\bfseries}}
\renewcommand\subparagraph{\@startsection{subparagraph}{6}{\parindent}%
  {3.25ex \@plus1ex \@minus .2ex}%
  {-1em}%
  {\normalfont\normalsize\bfseries}}
\def\toclevel@subsubsubsection{4}
\def\toclevel@paragraph{5}
\def\toclevel@paragraph{6}
\def\l@subsubsubsection{\@dottedtocline{4}{7em}{4em}}
\def\l@paragraph{\@dottedtocline{5}{10em}{5em}}
\def\l@subparagraph{\@dottedtocline{6}{14em}{6em}}
\makeatother

\usepackage{makeidx}
\makeindex

\begin{document}

\begin{flushright}
{\today}
\end{flushright}

\vspace{10pt}

\begin{center}

%%%%%%%%%%%%%%%%%%%%%%%%%%%%%%%%%%%%%%%%%%%%%%%%%%%%%%%%%%%%%%%%%%%%

{\Large\sc Supersymmetry Breaking with Fields, Strings and Branes}\vskip 12pt

%%%%%%%%%%%%%%%%%%%%%%%%%%%%%%%%%%%%%%%%%%%%%%%%%%%%%%%%%%%%%%%%%%%%

\vspace{15pt}
{\sc E.~Dudas${}^{\; a}$, J.~Mourad${}^{\; b}$  \ and \ A.~Sagnotti${}^{\; c}$\\[15pt]

${}^a$\sl\small CPHT, CNRS, École polytechnique, Institut Polytechnique de Paris\\ 91120 Palaiseau, FRANCE
\\ e-mail: {\small \it
emilian.dudas@polytechnique.edu}
\vspace{10 pt}

${}^b$\sl\small APC, UMR 7164-CNRS, Universit\'e   Paris Cit\'e  \\
10 rue Alice Domon et L\'eonie Duquet \\75205 Paris Cedex 13 \ FRANCE
\\ e-mail: {\small \it
mourad@apc.univ-paris7.fr}\vspace{10
pt}

{${}^c$\sl\small
Scuola Normale Superiore and INFN\\
Piazza dei Cavalieri, 7\\ 56126 Pisa \ ITALY \\
e-mail: {\small \it sagnotti@sns.it}}\vspace{10pt}
}

%%%%%%%%%%%%%%%%%%%%%%%%%%%%%%%%%%%%%%%%%%%%%%%
\vspace{15pt} {\sc\large Abstract}\end{center}
%%%%%%%%%%%%%%%%%%%%%%%%%%%%%%%%%%%%%%%%%%%%%%%
\vskip 8pt
The first part of this review tries to provide a self--contained view of supersymmetry breaking from the bottom--up perspective. We thus describe $N=1$ supersymmetry in four dimensions, the Standard Model and the MSSM, with emphasis on the ``soft terms'' that can link it to supergravity. The second part deals with the top--down perspective. It addresses, insofar as possible in a self--contained way, the basic setup provided by ten--dimensional strings and their links with supergravity, toroidal orbifolds, Scherk--Schwarz deformations and Calabi--Yau reductions, before focusing on a line of developments that is closely linked to our own research. Its key input is drawn from ten--dimensional non--tachyonic string models where supersymmetry is absent or non--linearly realized, and runaway ``tadpole potentials'' deform the ten--dimensional Minkowski vacua. We illustrate the perturbative stability of the resulting most symmetrical setups, which are the counterparts of circle reduction but involve internal intervals. We then turn to a discussion of fluxes in Calabi-Yau vacua and the KKLT setup, and conclude with some aspects of Cosmology, emphasizing some intriguing clues that the tadpole potentials can provide for the onset of inflation. The appendices collect some useful material on global and local $N=1$ supersymmetry, in components and in superspace, on string vacuum amplitudes, and on convenient tools used to examine the fluctuations of non--supersymmetric string vacua.

\setcounter{page}{1}

\pagebreak

\tableofcontents

\pagebreak

\section{\sc Introduction}

Symmetry principles have long served as a guidance in the search for the laws of Nature, before their implementation in detailed dynamical setups, as was the case for General Relativity and gauge theories. Supersymmetry is the maximal extension of the Poincar\'e algebra~\cite{coleman-mandula,{Haag:1974qh}}, and enforces a correspondence between Bose and Fermi degrees of freedom and their interactions, with softening effects on the ultraviolet behavior of Quantum Field Theory. The correspondence is not apparent in Nature, since it is violated by the known types of elementary particles, so that supersymmetry must be broken, if present at all. This review is devoted to the mechanisms leading to broken versions of supersymmetry that find some motivations within String Theory~\cite{stringtheory}, a prominent high--energy extrapolation including gravity.

Supersymmetry made its first appearance in 1971, as an extension of ordinary Lie algebras also including fermionic generators, in the work of Golfand and Likhtman \cite{Golfand:1971iw}, and also emerged as a two--dimensional fermionic symmetry of the Neveu--Schwarz--Ramond (NSR) dual models~\cite{NSR1,NSR2} that eventually became String Theory, in the work of Gervais and Sakita \cite{Gervais:1971ji}.
These results were followed in 1973 by the Volkov--Akulov nonlinear supersymmetry for the corresponding Goldstone spinor~\cite{Volkov:1973ix}, which they tentatively
identified with a neutrino. The underlying idea was to link massless neutrinos to Goldstone's theorem~\cite{goldstone} for a new, spontaneously broken, fermionic symmetry. In retrospect, neutrinos are not exactly massless and cannot be regarded as Goldstone fermions for other reasons, related to their interactions and quantum numbers, but this approach still plays an important role in the theory. Wess and Zumino~\cite{Wess:1974tw1,Wess:1974tw2} then constructed the first detailed four-dimensional models where linearly realized supersymmetry relates scalar and spinor fields of equal masses. Shortly thereafter, Ferrara and Zumino~\cite{Ferrara:1974pu} and, independently, Salam and Strathdee~\cite{Salam:1974yz}, obtained
their Yang-Mills counterparts, and with these ingredients Fayet~\cite{fayet1,fayet2,fayet3} began to explore supersymmetric extensions of the Standard Model (SM), also coining the familiar terms ``photino” and ``gaugino” (for reviews on supersymmetry, see~\cite{susy-books}).

Combining supersymmetry with gravity turns it into a local symmetry within an elegant class of theories that are generically called supergravity \cite{sugra1,sugra2} (for reviews, see~\cite{sugrarev}). The corresponding gauge fields, one or more spin--$3/2$ gravitini, accompany in them the spin-2 field of General Relativity.  This remarkable fact continues to be fascinating and resonates with the clash between gravity and global symmetries, which has been increasingly appreciated in recent years~\cite{noglobalgrav}. Supersymmetry also plays a key role, in several respects, in the formulation of String Theory, which is the most serious candidate to unify gravity with the other interactions in the quantum realm and makes its presence inevitable. It is naturally present in the NSR string world-sheet to describe space-time fermions, and in spacetime in the best understood ten--dimensional models~\cite{GSO,gs82,IIB,gs,heterotic1,heterotic2,heterotic3}. Moreover, it reveals deep links among them via generalized duality transformations~\cite{witten}.

At low energies, where gravitational interactions between fundamental particles become very weak, one can decouple gravity altogether. In this fashion, the global (or rigid) limit of supersymmetry acquires an important role on its own in applications to Particle Physics (for reviews on Quantum Field Theory and the Standard Model, see~\cite{recent_books_SM}). For this reason, we open our discussion, in Section~\ref{sec:susy_algebras}, with a review of supersymmetry algebras and their particle multiplets, and then address, in Section~\ref{sec:broken_global}, models with global $N=1$ supersymmetry, the only option that allows chiral fermions in four dimensions, the mechanisms for its spontaneous breaking~\cite{O'R,fayetilio} and the difficulties that are encountered when trying to generate realistic spectra in renormalizable models. The discussion touches upon mass sum rules~\cite{fgp}, the effect of non--renormalizable interactions and the ``soft terms''~\cite{girardello-grisaru}, which are essential to obtain realistic models and whose emergence can find a rationale in supergravity~\cite{gravity-mediation1,gravity-mediation2,hlw}.

In Section~\ref{sec:sm}~, after recalling some facts about the Standard Model \cite{SM1,SM2,SM3}, we elaborate on some inherent difficulties and puzzles that can be solved, at least in part, by the introduction of spontaneously broken supersymmetry. In
Section~\ref{sec:mssm} we turn to the Minimal Supersymmetric Standard Model (MSSM)~\cite{fayet1,fayet2,fayet3,dg}, where supersymmetry is broken also by soft terms, highlighting its main features and some additional difficulties that emerge when following this route. 

Moving on toward higher energies, in Section~\ref{sec:SUGRA} we review some basic facts about $N=1$ four--dimensional supergravity, with emphasis on its important role in the spontaneous breaking of supersymmetry at low energies. In Section~\ref{sec:VAnonlinear} we address some basic facts about
non--linear supersymmetry, and in Section~\ref{sec:sugra1110} we review the salient properties of supergravity in ten and eleven dimensions, in view of their close links with String Theory.
This chapter concludes the first part of the review.

The second part of the review begins, in Section~\ref{sec:critical_strings}, with a detailed discussion of string quantization and of the resulting spectra, with emphasis on the GSO projection and the supersymmetric models. 
We then discuss T-duality~\cite{Tduality1,Tduality2,Tduality3,Tduality4} (for a review, see~\cite{Tduality_rev}), D-branes~\cite{Dbranes}, the orientifold projection~\cite{orientifolds1,orientifolds2,orientifolds3,orientifolds4,orientifolds5,orientifolds6,orientifolds7,orientifolds8} (for reviews, see~\cite{orientifolds_rev1,orientifolds_rev2,orientifolds_rev3,orientifolds_rev4}) and the world--sheet consistency conditions underlying the different options, before addressing perturbative and non--perturbative links among the different supersymmetric ten--dimensional strings and the overall M-theory picture~\cite{witten} that connects them to the eleven--dimensional supergravity of Cremmer, Julia and Scherk~\cite{CJS}. We conclude the section by turning to the additional ten--dimensional strings that satisfy all world--sheet consistency conditions but whose spectra are not supersymmetric, with emphasis on the three tachyon--free options where supersymmetry is absent~\cite{agmv1,agmv2,as95,as97} or non--linearly realized~\cite{sugimoto,dmnonlinear1,dmnonlinear2,dmnonlinear3}.

Toroidal compactifications of closed strings and their orientifolds, together with their continuous and discrete deformations, are the subject of Section~\ref{sec:highertori}, while the basic string realizations of the Scherk--Schwarz mechanism \cite{ss1,ss2}~\cite{ss_closed1,ss_closed2,ss_closed3,ss_closed4,ss_closed5,ss_closed6}, where supersymmetry breaking is induced in circle compactifications, and some orientifolds thereof~\cite{ads1,ads2,fh,dm-tachyonfree1,bcd}, are the subject of Section~\ref{sec:toroidal_ss}.

Section~\ref{sec:6dstrings} is devoted to supersymmetric six--dimensional orbifold compactifications for closed strings, corresponding orientifolds, and variants that include brane supersymmetry breaking and magnetic deformations~\cite{ftse,acny,bachas,aads,bsb1,bsb2,bsb3,bsb4,berlinmadrid1,berlinmadrid2,berlinmadrid3} or, equivalently, brane rotations~\cite{berkoozdl,berlinmadrid4,berlinmadrid5,berlinmadrid6,berlinmadrid7,berlinmadrid8,berlinmadrid9,berlinmadrid10,berlinmadrid11} (for reviews, see~\cite{bsb_rev,berlinmadrid_rev}). 
Brane supersymmetry breaking occurs in vacua hosting combinations of branes and orientifolds that preserve incompatible fractions of supersymmetry, and brings non--linear realizations to the forefront~\cite{dmnonlinear1,dmnonlinear2} without giving rise to tachyonic modes. Its simplest ten--dimensional incarnation~\cite{sugimoto}, discussed in Section~\ref{sec:critical_strings}, is only a possible option, but brane supersymmetry breaking can be an inevitable feat in special lower--dimensional settings, the simplest of which occurs in six dimensions~\cite{bsb1,bsb2,bsb3,bsb4}. In this review, which tries to address a mixed audience, the detailed analysis of these six--dimensional examples should suffice to give a clear view of the underlying formalism. Therefore, we have left out a similar analysis of four--dimensional models, which are more involved but whose construction proceeds along similar lines. The reader can find more details in the original works~\cite{4d1,4d2,4d3,4d4,4d5,4d6,4d7,abpss,kakudual,4d8,4d9,4d10,ibanez1,ibanez2,ibanez3,ibanez4,bert1,bert2,bert3,bert4}, in the reviews~\cite{orientifolds_rev1,orientifolds_rev2,berlinmadrid_rev,bert_rev,orientifolds_rev3,orientifolds_rev4}, and in the references therein.

In the remaining portion of the review, we leave aside the tools of String Theory proper and return to the low--energy supergravity.
In Section~\ref{sec:calabi-yau} we review some basic facts about Calabi--Yau compactifications~\cite{calabiyau} (for a recent review, see~\cite{tomasiello}). These yield solutions of the low--energy equations (with some higher derivative additions) that connect minimal ten--dimensional supergravity to $N=1$ four--dimensional supergravity and the ten--dimensional Type II theories to $N=2$ supergravity. There is a huge variety of these manifolds, and the orbifold compactifications discussed in Section~\ref{sec:6dstrings} are singular limits related to a four--dimensional counterpart of them, the K3 surface (for reviews, see~\cite{aspinwall}). 

The following two sections are devoted to a line of developments stimulated by~\cite{dm_vacuum} that was central to our activity in recent years. The non--tachyonic ten--dimensional strings with broken supersymmetry~\cite{agmv1,agmv2,as95,as97} do not admit Minkowski vacua, due to the emergence of exponential ``tadpole potentials'' 
that are precisely determined by the vacuum amplitudes. The compactifications on an internal circle leave way, in this case, to a class of vacua~\cite{dm_vacuum} that involve an internal interval, at whose ends the low--energy theory develops a singular behavior. Their properties are reviewed in detail in Section~\ref{sec:SUSY_breaking_com}, where we show that these vacua are, surprisingly, perturbatively stable, in sharp contrast to $AdS \times S$ solutions, which also exist in this case and where the preceding singular behavior is absent~\cite{gm,ms_16,bms1,bms2,bms3,bms4,ms23_1}. 
In Section~\ref{sec:directsusybcom} we turn to a class of solutions of the type IIB theory where supersymmetry is broken by compactifications on internal intervals, but where some supersymmetry is recovered at one end. In these vacua, the string coupling is bounded everywhere, while curvature singularities are still present, but one can again provide arguments that point to the perturbative stability. In Section~\ref{sec:KKLT} we address the role of internal fluxes in the stabilization of massless scalar modes (moduli) emerging from Calabi--Yau compactifications and their role in the overall KKLT setup~\cite{KKLT} aimed at connecting ten--dimensional supersymmetric strings to four--dimensional de Sitter vacua.
In the last section, which is devoted to some aspects of Cosmology, we also elaborate on spatially flat solutions driven by a ``tadpole potential''. These have a striking feature: their early dynamics changes drastically when the logarithmic slope of the potential grows up to that of the non--supersymmetric tachyon--free ten--dimensional orientifolds. The resulting scenarios can provide a picture of the onset of inflation~\cite{inflation1,inflation2,inflation3,inflation4,inflation5,inflation6,inflation7,inflation_rev,WMAP, PLANCK1,PLANCK2,PLANCK3} within a weak string--coupling regime~\cite{bsb_cosmology1,bsb_cosmology2,bsb_cosmology3,bsb_cosmology4,bsb_cosmology5,bsb_cosmology6,bsb_cosmology7}, with some interesting indications.

All these results appear encouraging, but vacuum stability in the presence of broken supersymmetry remains, in general, a vexing open problem.

The main body of the review is followed by some Appendices, which are devoted to complements of various types. Appendix~\ref{app:conventions} summarizes our conventions, while Appendices~\ref{app:superfields_global} and \ref{app:superfields_local} collect some basic results on the global and local $N=1$ superspace formalisms in four dimensions. Appendix~\ref{app:bosonic_orientifold} reviews basic properties of genus--one amplitudes for the bosonic string, while Appendix~\ref{app:so2n} collects some properties of the $SO(2N)$ level--one characters that play an important role in the formulation of ten--dimensional strings. Finally, Appendix~\ref{app:hypergeometric} collects useful material on some exactly solvable potentials that can closely capture the modes present in the vacua of Sections~\ref{sec:SUSY_breaking_com} and~\ref{sec:directsusybcom}.

The presentation is mostly self--contained, but a number of very valuable textbooks and reviews on supersymmetry~\cite{susy-books}, supergravity~\cite{sugrarev} and String Theory~\cite{stringtheory} can conveniently complement the material collected here. We work to a large extent in components, in a ``mostly plus'' signature, and we often elaborate on four--dimensional $N=1$ supersymmetry and supergravity, also resorting to $N=1$ superfields, but we also address some higher--dimensional settings that are relevant for String Theory. Even if it will not be the ultimate step in the quest for a quantum theory of gravity, String Theory has arguably been a major source of inspiration over the years. For example, the remoteness from experiments of its characteristic scale, which is often a source of criticism, has stimulated extensive bottom--up searches for general principles and constraints that any theory coupled to quantum gravity should satisfy. In writing this review, we focused on a selection of topics that we deem important but above all reflect our interests and our competencies. Several other topics were inevitably left out, including what can be found in the vast recent literature on this fascinating research line, which aims to identify low--energies theories in and out of the ``swampland'' . Some reviews on it can be found in~\cite{swampland}.

The attempts to incorporate supersymmetry in the theory of the Fundamental Interactions that we are summarizing here are clearly in strong need of some guidance from new experiments, and, as often happened in the past, Nature may end up challenging our current views. It will be the duty and privilege of the community to uncover possible new clues, continuing the fascinating journey toward the microscopic laws of the Universe that has long accompanied the development of Physics.

\newpage
{}
\vskip 5cm 
\begin{center}
    {\Large \sc Part I} 
    \vskip 2cm
    {\Large \sc Bottom--Up Approach to Supersymmetry Breaking}
\end{center}
\emph{The first portion of the review is devoted to setting the stage for supersymmetry and supersymmetry breaking, drawing some important motivations from Particle Physics. A key message is that the MSSM, the minimal supersymmetric extension of the Standard Model, appears deeply linked to supergravity. This fact resonates with general arguments against global symmetries in the presence of gravity, and naturally leads to String Theory, which regulates the ultraviolet behavior of gravity and supergravity. }
\newpage

\section{\sc Supersymmetry Algebras and their Representations} \label{sec:susy_algebras}

The four--dimensional Minkowski spacetime possesses isometries corresponding to the Poincar\'e group, which include translations, with generators $P_\mu$, and Lorentz transformations, with generators $J_{\mu \nu}$. One can also distinguish, among them, the rotation generators $J_i = \frac{1}{2}\, \epsilon_{ijk }J_{jk}$ and the boost generators $K_i = J_{i0}$. In the absence of gravity, all models of Particle Physics must possess these symmetries.
Additional generators $T_a$ of internal symmetries, if present, form a Lie algebra and commute with the generators of spacetime symmetries.

In detail, the  algebra  of Poincar\'e and internal symmetries takes  the form
\bea
&& [J^{\mu \nu},J^{\rho \sigma}] \ = \ i \left( \eta^{\mu \rho}  J^{\nu \sigma} - \eta^{\mu \sigma}  J^{\nu \rho} - \eta^{\nu \rho}  J^{\mu \sigma} + \eta^{\nu \sigma}  J^{\mu \rho}  \right)  \ , \nonumber \\
 && [P^{\mu},J^{\nu \rho}] \ = \ -\ i \left(  \eta^{\mu \nu} P^\rho  -   \eta^{\mu \rho} P^\nu  \right)  \ , \quad  [P^{\mu},P^\nu] \ = \ 0   \ , \nonumber \\
 && [T_a ,T_b] \ = \ i \,C_{ab}^c T_c \ , \quad [T_a,P_\mu ] \ = \ 0 \ , \quad [T_a ,J_{\mu \nu }] \ = \ 0  \ ,  \label{repsusy1}
  \eea
where the $C_{ab}^c$ are structure constants of the internal Lie algebra. The Jacobi identities imply the quadratic relations
\beq
C_{ab}^d\,  C_{dc}^e \ + \  C_{bc}^d\,  C_{da}^e \ + \ C_{ca}^d\,  C_{db}^e \ = \ 0    \label{repsusy2}
\eeq
among them.

Let us briefly recall the key properties that characterize the representations of the four--dimensional Lorentz algebra. The commutation relations can be recast in the convenient form
\beq
[J_i, J_j] \ = \ i \,\epsilon_{ijk} J_k \ , \quad [J_i, K_j] \ = \ i \,\epsilon_{ijl} K_l  \ , \quad [K_i, K_j] \ = \ - \ i \,\epsilon_{ijk} J_k   \ ,  \label{lorentz1}
\eeq
and consequently  the Lorentz algebra can be split into a pair of mutually commuting angular momentum algebras,
\beq
[J_{1i}, J_{1j}] \ = \ i\, \epsilon_{ijk} J_{1k} \ , \quad [J_{2i}, J_{2j}] \ = \ i \,\epsilon_{ijk} J_{2k}  \ , \quad   [J_{1i}, J_{2j}] \ =  \ 0  \ .  \label{lorentz3}
\eeq
where
\beq
{\bf J}_1 \ = \ \frac{1}{2} \left({\bf J} \,+\, i\, {\bf K}\right) \ , \qquad {\bf J}_2 \ = \ \frac{1}{2} \left({\bf J} \ - \ i \,{\bf K}\right)   \ . \label{lorentz2}
\eeq
Denoting by $j_1,j_2$ the angular momentum quantum numbers corresponding to ${\bf J}_1, {\bf J}_2$, a generic finite--dimensional Lorentz representation associated to a quantum field can thus be labeled by a pair $(j_1,j_2)$, and has dimension
$(2j_1+1)(2j_2+1)$. The angular momentum generator is then ${\bf J} = {\bf J}_1+ {\bf J}_2$, and contains the representations characterized by
\beq
j = |j_1-j_2| \ , \ |j_1-j_2|+1, \  \cdots \ ,\  j_1+j_2-1 \ , \  j_1+j_2 \ .  \label{lorentz4}
\eeq
Note that $\mathbf{J}_{1}$ and $\mathbf{J}_{2}$ are Hermitian operators, but $\mathbf{K}$ is anti--hermitian, so that these finite--dimensional representations are not unitary.

A famous theorem by Coleman and Mandula \cite{coleman-mandula} proves that the most general Lie algebra of symmetries of the S matrix can only contain the generators $J_{\mu \nu}, P_\mu$ and $T_a$, so that space-time and internal symmetries cannot mix in a non--trivial way. The theorem is valid under the following assumptions: \\
- there is an S matrix corresponding to a relativistic Quantum Field Theory in four dimensions; \\
- for any $M > 0$, there is a finite number of types of particles with masses less than or equal to $M$; \\
- any two-particle state undergoes reactions at all energies, except perhaps for an isolated set; \\
- the amplitudes for elastic two-body scattering are analytic functions of the scattering angle for almost all energies and angles; \\
- there is a gap between the vacuum and the one-particle states, which strictly speaking excludes massless particles, thus making the S matrix well defined.

 The Coleman-Mandula theorem played an important role in the 1960's, since it provided a rationale for the failure of several attempts to combine, in a relativistic theory, internal symmetries with space-time ones. These attempts included a natural $SU(6)$ generalization~\cite{gurseyradicati} of Gell-Mann's $SU(3)$ symmetry~\cite{eightfold}.

 After the first supersymmetric models were proposed \cite{Golfand:1971iw,Volkov:1973ix,Wess:1974tw1,Wess:1974tw2}, an important paper of Haag, Lopuszanski and
 Sohnius~\cite{Haag:1974qh} showed that the Coleman--Mandula theorem could be bypassed in the presence of graded Lie algebras including fermionic generators. Graded Lie algebras contain generators that we shall generically denote by $G_A$, which can be bosonic or fermionic. The basic (anti)commutation relations take the form
\beq
G_A G_B  \ - \ (-1)^{\eta_A \eta_B} G_B G_A \ = \ i\, C_{AB}^C G_C  \ ,  \label{repsusy3}
\eeq
where $\eta_{A,B} = 0$ for bosonic generators and  $\eta_{A,B} = 1$ for fermionic ones, and
the corresponding structure constants satisfy the relations
\beq
C_{BA}^C \ = \ - \,(-1)^{\eta_A \eta_B} C_{AB}^C \quad , \quad C_{AB}^C \ = \ 0 \quad {\rm unless} \quad \eta_C \ =  \ \eta_A \ + \ \eta_B \ ({\rm mod} 2)  \ . \label{repsusy4}
\eeq
The super--Jacobi identities
\beq
(-1)^{\eta_A \eta_C} [[G_A,G_B\},G_C \}  +  (-1)^{\eta_A \eta_B} [[G_B,G_C\},G_A \}  + (-1)^{\eta_B \eta_C} [[G_C,G_A\},G_B \} = 0 \ , \label{repsusy5}
\eeq
 where
\beq
 [O,O' \} \ = \ O O' \ -\  (-1)^{\eta (O) \eta (O')} O' O \ ,
\eeq
translate into the following quadratic constraints on the structure constants:
 \beq
(-1)^{\eta_A \eta_C} C_{AB}^D \, C_{DC}^E  + (-1)^{\eta_A \eta_B} C_{BC}^D \, C_{DA}^E + (-1)^{\eta_B \eta_C} C_{CA}^D \, C_{DB}^E \ = \ 0  \ .  \label{repsusy6}
 \eeq

One can describe the set of generators valued in a $(j_1,j_2)$ Lorentz representation as $Q_{m_1 m_2}^{j_1j_2}$, and their conjugates $(Q_{m_1 m_2}^{j_1j_2})^*$ are then valued in the
$(j_2,j_1)$ representation. A given generator is bosonic (fermionic) if $j_1+j_2$ is integer (half-odd-integer). Given a pair of fermionic generators,
their anticommutator $\{Q_{m_1 m_2}^{j_1j_2} , (Q_{m_1 m_2}^{j_1j_2})^*\}$ will contain bosonic generators up to the largest representation $(j_1+j_2,j_1+j_2)$.  According to the
Coleman-Mandula theorem, the bosonic generators are Poincar\'e and internal generators, which are only valued in the representations $(\frac{1}{2},\frac{1}{2}),(1,0)$, $(0,1)$ and $(0,0)$. This is only possible if $j_1=\frac{1}{2}$, $j_2=0$, or vice versa.
The new spin--$\frac{1}{2}$ generators and their complex conjugates will be denoted by $Q_{\alpha}^A$ and ${\bar Q}_{\dot \alpha ,A}$, where $A= 1, \ldots, N$, and Lorentz invariance restricts the anticommutators to the form
\beq
\{ Q_{\alpha}^A , {\bar Q}_{\dot \beta, B}  \} \ = \ 2 N_B^A \, \sigma^\mu_{\alpha \dot \beta} P_\mu \ ,  \label{repsusy7}
\eeq
where $N_B^A$ is a Hermitian positive definite matrix and
\beq
\sigma^{\mu} \ = \ \left(- 1, \sigma^i \right) \ , \qquad {\bar \sigma}^{\mu} = (-1, - \sigma^i ) \ ,\label{sigmasigmabar_intro}
\eeq
with $\sigma^i$ are the Pauli matrices.
A unitary redefinition of the $Q$'s and ${\bar Q}$'s leads to the standard form for the anticommutator,
\beq
\{ Q_{\alpha}^A , {\bar Q}_{\dot \beta, B}  \} \ =\ 2 \,\delta_B^A \, \sigma^{\mu}_{\alpha \dot \beta} P_\mu \ .  \label{repsusy8}
\eeq

Let us now consider the commutators $[Q_\alpha^A,P_\mu]$. According to the Lorentz representation decomposition $(\frac{1}{2},0) \otimes (\frac{1}{2},\frac{1}{2}) = (0,\frac{1}{2}) \oplus (1,\frac{1}{2})$, but the
Coleman--Mandula theorem tells us that the representation $(1,\frac{1}{2})$ cannot be a symmetry of the S matrix. The commutator must therefore be of the form
\beq
[Q_\alpha^A,P^\mu] \ = \ {\cal Y}^{AB} \sigma_{\alpha \dot \beta}^\mu \, {\bar Q}_B^{\dot \beta} \ ,  \label{repsusy9}
\eeq
and the Jacobi identity
\beq
[P^\nu, [P^\mu,Q_\alpha^A]] +  [Q_\alpha^A,  [P^\nu,P^\mu]]  + [P^\mu, [Q_\alpha^A,P^\nu]]  =  4 ({\cal Y} {\cal Y}^*)_C^A  \left( \sigma^{\mu \nu} Q^C \right)_{\alpha} = 0 \ ,  \label{repsusy10}
\eeq
implies that ${\cal Y}^{AB}$ vanishes\footnote{Contracting (\ref{repsusy12}) with $\epsilon^{\alpha \beta}$ one can show that ${\cal Y}$ is a symmetric matrix, therefore
${\cal Y} {\cal Y}^*={\cal Y} {\cal Y}^{\dagger}$, so that the vanishing of the preceding expression implies that ${\cal Y}=0$.}.

Let us now consider the anticommutators $\{  Q_\alpha^A , Q_\beta^B \}$.  From the Lorentz product decomposition $(\frac{1}{2},0) \otimes (\frac{1}{2},0) = (0,0) \oplus (1,0)$, one can write
\beq
\{  Q_\alpha^A , Q_\beta^B \} \ = \ \epsilon_{\alpha \beta} {\cal Z}^{AB} \ + \ \left(\sigma^{\mu \nu} \right) _{\alpha \beta} J_{\mu \nu} {\cal W}^{AB} \ ,  \label{repsusy11}
\eeq
where $\sigma_{\mu\nu}$ is defined in Appendix~\ref{app:conventions} starting from $\sigma^\mu$ and ${\bar \sigma}^{\mu}$,
and clearly ${\cal Z}^{AB}= -\,{\cal Z}^{BA}$. From the Jacobi identity
\beq
[P^\mu, \{Q_\alpha^A  ,Q_\beta^B\}] - \{Q_\beta^B , [P^\mu, Q_\alpha^A]  \} +  \{Q_\alpha^A , [Q_\beta^B, P^\mu]  \}  = 0 \ ,  \label{repsusy12}
 \eeq
one then finds that ${\cal W}^{AB}=0$, while the other Jacobi identities force the ${\cal Z}^{AB}$ to commute among themselves and with the other generators. For this reason, the ${\cal Z}^{AB}$
are usually called {\it central charges}.

Let us finally consider the commutators $[Q_{\alpha}^A, J^{\mu \nu}]$.  In the product of Lorentz representations $(\frac{1}{2},0) \otimes [ (0,1)+(1,0)]$, only $(\frac{1}{2},0)$ is allowed. Therefore, the general form of this commutator is
\beq
[Q_{\alpha}^A, J^{\mu \nu}] \ = \ c (\sigma^{\mu \nu})_{\alpha}{}^{\beta} \, Q_{\beta}^A  \ , \label{repsusy012}
\eeq
where $c$ is a constant.  From the Jacobi identity
\beq
[ [Q_{\alpha}^A, J^{\mu \nu}],J^{\rho \sigma}] \ + \ [ [J^{\rho \sigma} ,Q_{\alpha}^A], J^{\mu \nu}] \ + \ [[J^{\mu \nu}, J^{\rho \sigma}] , Q_{\alpha}^A ] \ = \ 0 \ ,
\eeq
one finds $c=i$, and
the final form of supersymmetry algebra is therefore
\bea
&& \{ Q_{\alpha}^A , {\bar Q}_{\dot \beta, B}  \} \ = \ 2 \delta_B^A \sigma^\mu_{\alpha \dot \beta} P_\mu  \quad , \quad \{  Q_\alpha^A , Q_\beta^B \} \ = \ \epsilon_{\alpha \beta} Z^{AB} \ , \nonumber \\
&& [Q_{\alpha}^A , P^\mu] \ = \ 0   \quad , \quad [Q_{\alpha}^A , J^{\mu \nu}] \ = \ i  (\sigma^{\mu \nu})_{\alpha}^{\ \beta} Q_{\beta}^A   \  ,  \label{repsusy13}
\eea
where $Z$ is an $N \times N$ antisymmetric matrix.

One direct consequence of the supersymmetry algebra is that
\beq
[Q_{\alpha}^A , P_\mu P^\mu] \ = \ 0 \ ,
\eeq
so that \emph{all particles in a given supermultiplet have identical masses}.
If $| j m \rangle $ are angular momentum eigenvectors, so that
\beq
J^2 | j m_j \rangle  \ = \ j(j+1) | j m_j \rangle  \ ,  \qquad J_3 | j m_j \rangle  \ = \ m_j | j m_j \rangle  \ ,
\eeq
using the commutation relation of the supercharges with
the angular momentum generators, one can show that
\bea
&& J_3 \ Q_{1}^A | j m_j \rangle \ = \ \left(m_j \,- \,\frac{1}{2}\right) Q_{1}^A | j m_j \rangle \ , \qquad J_3 ({\bar Q}_{2}^A | j m_j \rangle) \ = \ \left (m_j \,+\,\frac{1}{2} \right) {Q}_{2}^A | j m_j \rangle \  ,  \nonumber \\
&& J_3 \ {\bar Q}_{\dot{1}}^A | j m_j \rangle = \left(m_j \,+ \,\frac{1}{2}\right) Q_{\dot{1}}^A | j m_j \rangle \ , \qquad J_3 ({\bar Q}_{\dot{2}}^A | j m_j \rangle) = \left (m_j \,-\,\frac{1}{2} \right) {\bar Q}_{\dot{2}}^A | j m_j \rangle \  .
\label{repsusy013}
\eea
Eqs.~(\ref{repsusy013}) indicate that the different components of the supercharges raise or lower the ``magnetic quantum number'' $m_j$, and thus the spin, by $\frac{1}{2}$.

In the absence of central charges, the supersymmetry algebra has a $U(N)$  symmetry, which is an R-symmetry~\footnote{The R-symmetry is actually $SU(4)$ for $N=4$ in the global case, and indeed the maximal Yang--Mills theory contains six real scalars described via an $SU(4)$--valued tensor $\phi^{ij}$ subject to the self--duality condition $\phi^{ij} \ = \ \frac{1}{2}\ \epsilon^{ijkl}\ \overline{\phi}_{kl}$. In a similar fashion, the R--symmetry is $SU(8)$ for $N=8$, but there is a subtlety for the $N=4$ supergravity multiplet, where it is $U(4)$, due to the axion--dilaton pair.} This qualification is meant to stress that the R-charges of bosons and fermions within a given multiplet
are different. The $N=1$ case corresponds to minimal supersymmetry, in which case the central charges vanish by antisymmetry. For $N > 1$, one talks about $N$-extended supersymmetry, and central charges are allowed.

%%%%%%%%%%%%%%%%%%%%%%%%%%%%%%%%%%%%%%%%%%%%%%%%%%%%%%%%%%%%%%%%%%
 \subsection{\sc Massless Multiplets}

The supersymmetry multiplets of massless particles are of special importance, since masses are mostly introduced via spontaneous symmetry breaking.  As we have seen, the supercharges commute with the momentum operator, and consequently all particles belonging to a given supermultiplet can have identical momenta, which can be chosen to be $P_{\mu} = (-E,0,0,E)$ in the massless case. The only non--trivial anticommutation relation is then
 \beq
 \{ Q_{\alpha}^A , {\bar Q}_{\dot \beta, B}  \} \  =  \  4 \,E \, \delta_{\alpha\,1}\, \delta_{\dot{\beta}\,1} \   \delta_B^A \ , \label{repsusy14}
  \eeq
  while all others vanish. In particular, setting $A=B$, $\alpha=2$, $\dot{\beta}=2$, \emph{l.h.s.} reduces to the sum of two positive terms that must vanish separately, so that $Q_{2}^A$ and ${\bar Q}_{\dot{2},A}$ can be set to zero. One is thus led to identify the $N$ creation and annihilation operators \cite{FSZ1,FSZ2,FSZ3}
\beq
b^A \ = \ \frac{1}{2 \sqrt{E}} \,Q_{1}^A  , \qquad b^{\dagger}_A \ = \ \frac{1}{2 \sqrt{E}}\, {\bar Q}_{\dot{1},A} = (b^A)^{\dagger} \qquad \left(A=1,\dots, N \right) \  \label{repsusy15}
\eeq
which satisfy the algebra
\beq
\{  b^A , b_B^{\dagger}\} \ = \ \delta^A_B \ , \qquad \{  b^A , b^B\} \ =  \ \{  b_A^{\dagger} , b_B^{\dagger}\} \ = \ 0 \ . \label{repsusy16}
\eeq
These relations are typical of fermionic oscillators. As a result, massless representations require vanishing central charges. In this case conformal symmetry, and even its supersymmetry extensions, are possible.

With our choice for the momentum $P_\mu$, the helicity $\lambda$ can be identified with the $J_3$ eigenvalue, 
and the operators $b_A^{\dagger}$ ($b^A$) increase (decrease) it by $\frac{1}{2}$.
Within a given multiplet, there is therefore a state of lowest helicity $|  \lambda_{\rm min} \rangle $, which satisfies $b^A | \lambda_{\rm min} \rangle= 0$, and all other states are built acting on it with the $b_{A} ^{\dagger}$:
\beq
| A_1 \cdots A_n ;  \lambda_{\rm min} + \frac{n}{2} \rangle \ = \ b_{A_n} ^{\dagger} \cdots b_{A_1} ^{\dagger} | \lambda_{\rm min} \rangle  \ . \label{repsusy17}
\eeq
Due to their anticommuting nature, $N$ is the maximum number of fermionic creation operators that can be applied to the right-hand side of  (\ref{repsusy17}), and consequently the total number of states that are built in this fashion is
 \beq
 \sum_{n=0}^N \begin{pmatrix} N \\ n  \end{pmatrix} \ = \ 2^N \ , \label{repsusy18}
  \eeq
with identical numbers of bosonic and fermionic degrees of freedom.
However, there is a subtlety. Any relativistic Quantum Field Theory should respect the CPT theorem,
and CPT transformations map a state of helicity $\lambda$ into another of helicity $-\lambda$. The preceding construction generally does not guarantee that vector spaces built in this fashion
automatically satisfy this requirement. When this is not the case, they must be completed by adding CPT conjugate helicities to the original ones.

In order to see this in detail, let us begin by considering the vector space built in $N=1$ supersymmetry starting from $\lambda_{\rm min} = -\,\frac{1}{2}$. Acting with $b^{\dagger}$ on this Fermi vacuum,
one finds a bosonic state with $\lambda = 0$. The resulting multiplet has $2^N=2$ degrees of freedom, as expected from the counting (\ref{repsusy18}), but clearly violates the CPT theorem,
since there is no state with $\lambda = \frac{1}{2}$~\footnote{Alternatively, there is no fermion in four dimensions with only one on-shell degree of freedom.}. Therefore, two more states obtained from a bosonic vacuum with $\lambda_{\rm min} = 0$, with $\lambda= 0$ and $\lambda=\frac{1}{2}$, must be added. Equivalently, two more states of opposite helicities could be added to the original set, ending up with
two bosonic states with $\lambda=0$, which can be obtained from a complex scalar $z$, and two fermionic states with $\lambda = \pm \,\frac{1}{2}$, which can be obtained from a Weyl fermion $\psi$. The latter field is a chiral fermion, and the result, which is usually
called a {\it chiral multiplet}, will play a crucial role in supersymmetric extensions of the Standard Model.

The second example of great interest for minimal $N=1$ supersymmetry
is obtained starting with $\lambda_{\rm min} = -1$. Acting with $b^{\dagger}$,
one thus finds a fermionic state with $\lambda = -\,\frac{1}{2}$, but the CPT theorem demands the addition of another pair of states with $\lambda = \frac{1}{2},1$.  These states originate from a massless gauge boson $A_{\mu}$, with helicities $\lambda= \pm 1$, and from a fermion (called {\it gaugino}), with helicities $\lambda = \pm \,\frac{1}{2}$, and build the so--called vector multiplet.  
One can also conclude that non-CPT invariant
multiplets lead, after adding their CPT mirrors, to $2^{N+1}$ degrees of freedom, with identical numbers of Bose and Fermi ones.

The third example obtains starting with $\lambda_{\rm min} = -2$. Acting with $b^{\dagger}$,
one thus finds a fermionic state with $\lambda = -\,\frac{3}{2}$, but the CPT theorem demands the addition of another pair of states with $\lambda = \frac{3}{2},2$.  These states originate from a massless vielbein $e_\mu{}^a$, with helicities $\lambda= \pm 2$, and from a fermion (called {\it gravitino}), with helicities $\lambda = \pm \,\frac{3}{2}$, and build the so--called supergravity multiplet. 

Another interesting example obtains considering $N=2$ extended supersymmetry and starting from $\lambda_{\rm min} = -1 $.  Acting with $b_A^{\dagger}$
one thus finds two fermionic states with $\lambda = -\,\frac{1}{2}$, and one more action leads to one scalar state  $b_{2} ^{\dagger}  b_{1} ^{\dagger} | \lambda_{-1} \rangle$ of
helicity $\lambda =0$. CPT invariance requires the addition of another set of states with helicities $\lambda=0,\frac{1}{2},1$, and the end result is the $N=2$ vector multiplet containing one gauge boson, two fermions
(gaugini) and a complex scalar. Note that the two Weyl gaugini can be assembled into a Dirac fermion. An explicit construction of the Lagrangian shows that the fermion has inevitably non-chiral interactions. This is actually a general result:  only the minimal $N=1$ supersymmetry can accommodate chiral fermions, and chirality is a key
ingredient for constructing realistic extensions of the Standard Model.  Therefore, if extended supersymmetry does exist in nature, it is presumably broken at high energy scales (at or above
the typical unification scale of minimal supersymmetric theories $M_{\rm GUT} \sim 2 \times 10^{16}$ GeV) into an $N=1$ supersymmetric theory, or maybe directly to the Standard Model. 

As a last example, let us consider $N=4$ supersymmetry starting again from $\lambda=-1$. In this case one finds four states with $\lambda=-\,\frac{1}{2}$, six states with $\lambda=0$, four states with $\lambda=\frac{1}{2}$ and finally one state with $\lambda=1$. This set is automatically consistent with the CPT theorem, and the maximal N=4 supersymmetric Yang-Mills theory contains indeed one vector, four Weyl spinors and six real scalars. Similar considerations hold for the CPT self--conjugate multiplet of $N=8$ supergravity, whose helicity content corresponds to the vielbein, 8 gravitini, 28 vectors, 56 spinors and 70 scalars~\cite{N8SUGRA}.

In general, the state of maximal helicity in a given supermultiplet has $\lambda_{\rm max } = \lambda_{\rm min} + \frac{N}{2}$. Renormalizable theories are only possible for a maximum spin $S_{\rm max}=1$, and consequently
renormalizable supersymmetric theories demand that $N \leq 4$. On the other hand, if one allows at most helicities $\pm 2$, thus including gravity, the upper bound on $N$ becomes $N \leq 8$.
 All these theories contain one or more spin--$\frac{3}{2}$ partners of the graviton, the gravitini, which are the gauge particles of supersymmetry, and thus play a role which is similar to that played by Yang-Mills gauge bosons in gauge theories.  In these models, which are called supergravities, supersymmetry is necessarily a local symmetry.

Theories with $N=1$ supersymmetry are often said to have four supercharges, in view of the dimension of the corresponding $Q_\alpha$, while those with $N=2$ are said to have eight supercharges, and therefore the maximal supergravity in four dimensions, corresponding to $N=8$, is said to have 32 supercharges.

 %%%%%%%%%%%%%%%%%%%%%%%%%%%%%%%%%%%%%%%%%%%%%%%
 \subsection{\sc Massive Multiplets}

For massive fields \cite{FSZ1,FSZ2,FSZ3} one can choose to work in the rest frame, where $P_{\mu} = (-M,0,0,0)$, a condition that is not affected by arbitrary $SO(3)$ rotations. As a result, in building supersymmetry multiplets one is to start from complete SO(3) multiplets. 

If there are no central charges, the supersymmetry algebra becomes
\beq
 \{ Q_{\alpha}^A , {\bar Q}_{\dot \beta B}  \} \ = \ 2 M \delta_B^A \delta_{\alpha \dot \beta}   \quad , \quad \{  Q_\alpha^A , Q_\beta^B \} \ = \
\{ {\bar Q}_{\dot \alpha A} , {\bar Q}_{\dot \beta B}  \}  =0  \ , \label{repsusy19}
\eeq
and in this case one can define the $2N$ creation and annihilation operators
\beq
b_{\alpha}^A \ = \ \frac{1}{\sqrt{2M}} \ Q_{\alpha}^A \quad , \quad  (b_{\alpha}^A)^{\dagger} \ = \ \frac{1}{\sqrt{2M}} \ {\bar Q}_{\dot \alpha A}  \ , \label{repsusy20}
\eeq
 satisfying the fermionic oscillator algebra
 \beq
 \{  b_{\alpha}^A , (b_{\beta}^B)^{\dagger}\} \ = \ \delta_A^B  \delta_{\alpha \dot \beta} \quad , \quad \{  b_{\alpha}^A , b_{\beta}^B\} \ = \  \{  (b_{\alpha}^A)^{\dagger} , (b_{\beta}^B)^{\dagger}\} = 0
 \ . \label{repsusy21}
 \eeq
The vacuum, annihilated by $b_{\alpha}^A$, must now be a complete multiplet of spin eigenstates $| j,j_3 \rangle$, with $j_3=-j,\ldots,j$, and the states in the supersymmetry multiplet are then
\beq
| \alpha_1 A_1, \ldots  ,\alpha_n A_n \rangle \ = \  (b_{\alpha_1}^{A_1})^{\dagger} \cdots (b_{\alpha_n}^{A_n})^{\dagger} | j,j_3 \rangle  \ . \label{repsusy22}
\eeq
 The total number of such states (the number of on-shell degrees of freedom in the multiplet) is consequently
\beq
 \sum_{n=0}^{2N} \ (2j+1)  \begin{pmatrix} 2N \\ n  \end{pmatrix} \ = \ \left(2j+1\right)  2^{2N} \ , \label{repsusy23}
\eeq
and the state with the highest values of $j$ and $j_3$ is obtained applying only the creation operators $\left(b_1^A\right)^\dagger$, which raise the $j_3$ eigenvalue. Due to their fermionic nature, only $N$ such operators can be applied, so that the maximum spin $j_{\rm max} = j + N/2$. Therefore, for $N > 1$ and in the absence of central charges, all massive multiplets contain states of spin $1$ or higher.

 Consider for simplicity the case of minimal $N=1$ supersymmetry, denoting the vacuum of spin $j_0$ by $| \rm vac \rangle$.
 One can then construct the multiplet
 \bea
 &&  | \rm vac \rangle \quad , \quad {\rm spin} \ j_0 \ , \nonumber \\
 && b_{\dot \beta}^{\dagger}   | \rm vac \rangle \quad , \quad {\rm spin} \ j_0 \ \pm \ 1/2 \qquad  \left({\rm if  } \ j_0 \geq \frac{1}{2}\right)   \ , \nonumber \\
 &&  b_{\dot \alpha}^{\dagger}  b_{\dot \beta}^{\dagger}    | \rm vac \rangle \ = \ - \ \frac{1}{2} \, \epsilon_{\dot \alpha \dot \beta}\, b_{\dot \gamma}^{\dagger} \, b^{\dagger \dot \gamma}    | \rm vac \rangle \quad ,
 \quad {\rm spin} \ j_0   \ . \label{repsusy24}
\eea
For $j_0=0$ one obtains one complex scalar and one Weyl fermion, of equal mass. This is the massive counterpart of the {\it chiral multiplet} that we already encountered in the massless case.
For $j_0=1/2$ one obtains two Weyl fermions of spin $1/2$, one gauge boson of spin projections $\pm 1$ and $0$ and one real scalar, all with identical nonzero masses. This  {\it massive $N=1$ vector multiplet} can be regarded as arising from the combination of a massless vector multiplet and a massless chiral multiplet, where the vector in the former eats one scalar of the latter.

In the presence of central charges $Z^{AB}$ there are special representations with fewer degrees of freedom, which are usually called "short multiplets". One can
write the central charge matrix $Z$ as
\beq
Z \ = \ H \, V \ ,
\eeq
where $H$ is a positive definite hermitian matrix and $V$ is a unitary matrix, relying on the polar decomposition theorem. This can be justified considering the positive--definite operator\footnote{Not using the van der Waerden notation reviewed in Appendix~\ref{app:conventions}, $S_\alpha$ is usually written in the more compact form $Q_{\alpha}^A - \epsilon_{\alpha \beta} V^{AB} {\bar Q}_{\beta B}$ in the rest frame.} 
\beq
\sum_{\alpha,A} \left\{ S_\alpha^A \,,\,\left(S_\alpha^A\right)^\dagger\right\} \ \geq  \ 0 \ ,  
\eeq
built from the combination
\beq
S_\alpha^A \ = \ Q_{\alpha}^A \ - \ \epsilon_{\alpha \gamma} \frac{{\bar \sigma}^{\dot{\beta} \gamma} \cdot P}{\sqrt{ - P \cdot P}} \,V^{AB} {\bar Q}_{\dot{\beta} B} \ , \qquad 
\eeq
so that the explicit form of the sum and the supersymmetry algebra~\eqref{repsusy13} imply the inequality
\beq
M \ \geq \ \frac{1}{2N} \ \mathrm{Tr} (Z Z^{\dagger} )^{1/2} \ . \label{repsusy26}
\eeq

For example, for $N=2$ supersymmetry $Z^{AB} = \epsilon^{AB} Z$ and then $M \geq \frac{1}{2} |Z|$. When the equality   holds in eq.~(\ref{repsusy26}), one talks about ``BPS saturation''~\cite{BPS1,BPS2}, and the corresponding
``short" multiplets satisfy the condition
\beq
S_{\alpha}^A \, | {\rm short \ multiplet} \rangle \ = \ 0 \ . \label{repsusy27}
\eeq
The BPS conditions (\ref{repsusy27}) reduce the number of fields present in short multiplets, compared to the standard massive multiplets, for which the inequality (\ref{repsusy26}) is not saturated.
In particular, for $N=2$ the short multiplets have the same content as the corresponding massless ones.

Note, finally, that the numbers of bosonic and fermionic degrees of freedom are identical in all the preceding multiplets. This is a generic property: if supersymmetry
is linearly realized, any fermionic mode of a given mass is accompanied by a corresponding bosonic mode with the same mass.

\section{\sc Spontaneously Broken \texorpdfstring{$N=1$} \ \ Global  Supersymmetry} \label{sec:broken_global}

We can now begin to explore supersymmetry breaking, with reference to the widely studied four--dimensional $N=1$ global case. This setting can accommodate the parity and time reversal violations present in the weak interactions, while also affording a convenient formulation in $N=1$ superspace~\cite{Salam:1974jj} (for a detailed review, see~\cite{superspace_review}, while a short account can also be found in Appendix~\ref{app:superfields_global}). The construction is based on two main ingredients:
\begin{itemize}
    \item The chiral multiplet $\Phi$
\beq
\Phi \ : \  \left(z, \psi, F\right) \ ,
\eeq
which combines a \emph{complex} scalar $z$, a Weyl spinor $\psi$ and a non-dynamical \emph{complex} auxiliary field $F$. The two dynamical fields have identical masses if supersymmetry is unbroken.
\item The vector multiplet $V$
\beq
V \ : \  \left(A_\mu, \lambda, D\right) \ ,
\eeq
which combines a \emph{real} gauge vector $A_{\mu}$, a Weyl spinor $\lambda$ and a non--dynamical \emph{real} auxiliary field $D$. These fields are massless, to begin with, as demanded by the gauge symmetry, and are valued in the adjoint representation of a compact semi-simple Lie group and/or a product of $U(1)$ factors.
\end{itemize}

In Section~\ref{sec:susy_algebras} we have seen that an exact Poincar\'e supersymmetry would require equal masses for the Bose and Fermi fields belonging to any given multiplet. In particular, massless fermionic ``superpartners" of photons and gluons, usually called ``photinos'' and ``gluinos'', should exist, together with light scalar partners of quarks and leptons, usually called ``squarks'' and ``sleptons''. Experimentally, this is clearly not the case. Therefore, if supersymmetry were to play a role in Nature, it ought to be spontaneously broken. However, as we shall see, a spontaneously broken global supersymmetry would imply the existence of a massless Fermi field, usually called ``goldstino'', along the lines of what happens for ordinary continuous global symmetries and their Goldstone bosons.
There is apparently no place in Nature for a goldstino, and this is one reason to expect that supersymmetry be a local gauge symmetry. If supersymmetry were a local symmetry, the goldstino would be absorbed to grant the additional degrees of freedom needed to make its gauge field, a spin-$\frac{3}{2}$ gravitino, massive. Moreover, making supersymmetry local brings along Einstein gravity, and supergravity models naturally connect to String Theory. We shall return to these issues in the following sections.

A characteristic signature of a Goldstone boson $\theta$ is its nonlinear shift
\beq
\delta \theta \ = \  v \, \alpha
\eeq
under a continuous symmetry transformation of parameter $\alpha$. Here, $v$ is
typically the vacuum value (\emph{v.e.v.}) of a scalar field, which plays the role of
order parameter for symmetry breaking. 
The counterpart of Goldstone's theorem in supersymmetry leads one to distinguish two cases, according to whether or not the vacuum is invariant under supersymmetry:
\bea
&& Q_{\alpha} | 0 \rangle = 0 \quad \to \quad {\rm unbroken \ supersymmetry} \ , \nonumber \\
&& Q_{\alpha} | 0 \rangle \not= 0 \quad \to \quad {\rm broken \ supersymmetry} \   \  .  \label{susybreak3}
\eea
The supersymmetry algebra then \emph{defines} the Hamiltonian in terms of the supercharges, as
\be
H \ = \ \frac{1}{4} \sum_{\alpha = 1,2} \left( {\overline Q}_{\alpha} {Q}_{\alpha} \ + \  {Q}_{\alpha}  {\overline Q}_{\alpha} \right)   \  ,  \label{susybreak4}
\ee
so that an additive constant cannot be added to it at will.
This contraction combines the dotted and undotted indices of Appendix~\ref{app:conventions}, while respecting the invariance of $H$ under spatial rotations. With this prescription for $H$ dictated by the supersymmetry algebra, unbroken (broken) supersymmetry implies a vanishing (positive) vacuum energy. 

In components, the most general Lagrangian describing renormalizable interactions of matter chiral multiplets $(z^i, \psi^i, F^i)$ with Yang-Mills supermultiplets $(A_{\mu}^a, \lambda^a, D^a )$ reads (for reviews, see~\cite{susy-books})
\bea
 {\cal L} &=& -\, | {\cal D}_\mu  z^i|^2  - i \psi^i  \sigma^\mu  {\cal D}_\mu {\bar \psi}_i   + |F^i|^2 - \frac{1}{4} (F_{\mu\nu}^{a})^2  - i \lambda^a \sigma^\mu  {\cal D}_\mu {\bar \lambda}^a   + \frac{1}{2} D^a D^a  \nonumber \\
&-&  i \sqrt{2} g   {\bar \lambda^a} {\bar \psi}_i (T^a)^i{}_j\, z^j   + i \sqrt{2} g \bar{z}{}_i (T^a)^i{}_j  {\psi}^j   {\lambda^a}    -
\frac{1}{2} \frac{\partial^2 W}{\partial z^i \partial z^j} \psi^i \psi^j - \frac{1}{2} \frac{\partial^2 \bar W}{\partial \bar{z}_i \partial \bar{z}_j} {\bar \psi}_i {\bar \psi}_j \nonumber \\
&+&   F^i \frac{\partial W}{\partial z^i} +    {\bar F}_i  \frac{\partial \overline W}{\partial \bar{z}_i}  + g \sum_a  \bar{z}_i (T^a)^i{}_j  z^j D^a +   \xi_a D^a   \  ,  \label{superym10off}
\eea
where summations over repeated indices are implicit and the covariant derivatives are
\beq
{\cal D}_{\mu}  z^i \ = \ \partial_{\mu} z^i \ +\  i g A_{\mu}^a (T^a)^i{}_j  z^j \ , \qquad  {\cal D}_{\mu}  \psi^i \ = \ \partial_{\mu} \psi^i \ +\  i g A_{\mu}^a (T^a)^i{}_j  \psi^j  \  .  \label{superym11}
\eeq

The constant parameters $\xi^a$ are an important option that is available for Abelian vector multiplets: $D$-type auxiliary fields of Abelian multiplets transform into total derivatives under global supersymmetry, and 
only in these cases one can add to the Lagrangian the supersymmetric and gauge invariant Fayet-Iliopoulos terms~\cite{fayetilio}
\beq
\Delta\,{\cal L} \ = \ \xi^a \, D^a \ .
\label{FI}
\eeq
The $ (T^a)_j^i$ are the matrices for general group generators in the representations in which matter fields are valued, and fermions are here in the two--component notation of Wess and Bagger in~\cite{susy-books}, which is reviewed in Appendix~\ref{app:conventions}. Moreover, $W(z^i)$, usually called superpotential, is a holomorphic gauge--invariant function of the scalar fields that determines both the scalar potential and the Yukawa couplings. A classic result of supersymmetry is that the holomorphic superspace integrals involving the superpotential are not renormalized to all orders of perturbation theory (for a review, see~\cite{superspace_review}).

The Lagrangian~\eqref{superym10off} is invariant, up to total derivatives, under the supersymmetry transformations
\bea
&& \delta z^i \ = \ \sqrt{2}\, \epsilon\, \psi^i  \ , \nonumber \\
 && \delta \psi^i  \ = \  i \sqrt{2} \sigma^{\mu} {\bar \epsilon} \ {\cal D}_{\mu} z^i \ + \ \sqrt{2}  {\epsilon} \  F^i \ , \nonumber \\
 && \delta F^i  \ = \  i \sqrt{2}\, {\bar \epsilon}\, {\bar \sigma}^{\mu} {\cal D}_{\mu} \psi^i \ + \ 2 i g    (T^a)^i{}_j  z^j {\bar \epsilon} \, {\bar \lambda^a} \ , \nonumber \\
 && \delta A_{\mu}^a  \ = \  i ({\bar \epsilon} {\bar \sigma}_{\mu} \lambda^a \ - \ {\bar \lambda^a} {\bar \sigma}_{\mu} \epsilon ) \ , \nonumber \\
&& \delta \lambda^a  \ = \  i \epsilon D^a \  + \ \sigma^{\mu \nu} \epsilon F_{\mu \nu}^a \ , \nonumber \\
&& \delta D^a  \ = \  - \ \epsilon \,\sigma^{\mu} {\cal D}_{\mu} {\bar \lambda^a} \ - \  {\cal D}_{\mu} {\lambda^a} \sigma^{\mu} {\bar \epsilon} \ . \label{superym12}
\eea

If one eliminates the auxiliary fields $F^i$ of the matter multiplets and the auxiliary fields $D^a$ of the Yang--Mills multiplets via their equations of motion,
 \beq
 F^i \ = \ - \ \frac{\partial \overline W}{\partial \overline{z}_i} \ , \qquad D^a\ =\  - \ g  \bar{z}_i (T^a)^i{}_j  z^j \ - \ \xi^a \ , \label{supersym13}
\eeq
the Lagrangian~\eqref{superym10off} becomes
\bea
{\cal L} &=& - \ | {\cal D}_{\mu}  z^i|^2  \ - \ i\, \psi^i\, \sigma^{\mu}  {\cal D}_{\mu} {\bar \psi}_i  \ - \ \frac{1}{4} \,(F_{\mu \nu}^{a})^2  \ - \ i\, \lambda^a \sigma^{\mu}  {\cal D}_{\mu} {\bar \lambda}^a     \nonumber \\
&-& i \,\sqrt{2} \,g \,  {\bar \lambda^a} {\bar \psi}_i (T^a)^i{}_j   \ z^j   \ + \ i\, \sqrt{2}\, g\, {\bar z}_i (T^a)^i{}_j   {\psi}^j   {\lambda^a}   \ - \
\frac{1}{2} \,\frac{\partial^2 W}{\partial z^i \partial z^j} \psi^i \psi^j \nonumber \\ &-& \frac{1}{2}\, \frac{\partial^2 \overline W}{\partial {\bar z}_i \partial {\bar z}_j} \, {\bar \psi}_i \,{\bar \psi}_j \ - \ \left|\frac{\partial W}{\partial z^i}\right|^2
- \frac{1}{2} \sum_a \Big[ g \ {\bar z}_i \ (T^a)^i{}_j \ z^j + \xi ^a \Big]^2     \  .   \label{superym10}
\eea
In terms of the auxiliary fields, the scalar potential reads
\beq
V (z^i, \bar{z}_j)\ = \ \sum_i |F^i|^2 \ + \ \frac{1}{2} \,\sum_a D^a D^a \ ,  \label{superym14}
\eeq
and consequently the most general renormalizable superpotential is a cubic polynomial of the form
\beq
W (z) \ = \ \lambda_i \,z^i \ + \ \frac{1}{2}\, m_{ij} \,z^i \, z^j \ + \ \frac{1}{3} \,\lambda_{ijk} \, z^i \,z^j \, z^k   \ .\label{chirals9cub}
\eeq

In a Lorentz invariant vacuum Fermi fields vanish, and yet their supersymmetry variations 
\be
\langle \delta \psi^i \rangle\ = \  \sqrt{2} \, {\epsilon} \langle F^i \rangle \  , \qquad  \langle \delta \lambda^a \rangle \ = \ i \,\epsilon \,\langle D^a \rangle  \  ,  \label{susybreak2}
\ee
are nontrivial in the presence of vacuum values of $F^i$ and $D^a$. The auxiliary fields of the chiral and vector multiplets are thus order parameters for supersymmetry breaking.
When searching for supersymmetric vacua, there is no need to solve the classical field equations for the scalar fields. In fact, if there is a solution to the equations
\beq
F^i \ = \ D^a \ = \ 0 \ ,  \label{superym15}
\eeq
which depends on the scalar fields, eq.~\eqref{superym14} implies that it is automatically a minimum of the scalar potential. The corresponding vacuum energy vanishes and supersymmetry is then unbroken.
The vacuum energy determined by $H$ in eq.~\eqref{susybreak4} is positive if supersymmetry is broken, so that supersymmetric solutions are stable minima. Relying on eq.~\eqref{supersym13}, it is thus easier to find the supersymmetric vacua of eq.~\eqref{superym15} than those of an arbitrary non-supersymmetric theory, if they exist, or to prove that supersymmetry is broken if eqs.~\eqref{superym15} have no solution.

Extremizing the scalar potential yields the conditions
 \be
F^j \frac{\partial^2 W}{\partial z^i \partial z^j} \ + \ g\, \langle D^a\rangle\,{\bar z}_j (T^a)^j{}_i \  = \ 0 \  ,  \label{susybreak5}
 \ee
while the gauge invariance of the superpotential yields a second set of conditions,
\be
\delta W (z^i) \ = \ \frac{\partial W}{\partial z^i} \ i \alpha^a (T^a)^i{}_j \,z^j\  = \ 0 \quad \to \quad \langle {\bar F}_i \rangle\,  (T^a)^i{}_j\, z^j \ = \ 0  \ , \label{susybreak6}
\ee
taking eqs.~\eqref{supersym13} into account,
together with their complex conjugates. One can conveniently group these two sets of equations in matrix form as
\be
\begin{pmatrix} \langle F^i \rangle & \frac{i}{\sqrt{2}} \, \langle D^b \rangle \end{pmatrix}
\begin{pmatrix}  \frac{\partial^2 W}{\partial z^i \partial z^j} &  - \,i \sqrt{2} g {\bar z}_j   (T^a)^j{}_i  \\
- \,i \sqrt{2} g {\bar z}_i   (T^b)^i{}_j & 0
 \end{pmatrix}
 \ = \ 0 \ . \label{susybreak7}
\ee
The fermionic mass matrix can be extracted from the quadratic terms in the Lagrangian, which can be cast in the form
\bea
&& {\cal L}_m \ = \  - \ \frac{1}{2} \frac{\partial^2 W}{\partial z^i \partial z^j} \,\psi^i \,\psi^j  \ + \ i \sqrt{2}\, g \,{\bar z}_i (T^a)^i{}_j \, {\psi}^j \,  {\lambda^a} \ + \ {\rm h.c.}     \nonumber \\
&& = \
-\,\frac{1}{2} \begin{pmatrix} \psi^i  & \lambda^b \end{pmatrix}
\begin{pmatrix}  \frac{\partial^2 W}{\partial z^i \partial z^j} &  - \,i \sqrt{2} \,g \,{\bar z}_j (T^a)^j{}_i  \\
-\, i \sqrt{2} \,g \,{\bar z}_i  (T^b)^i{}_j & 0
 \end{pmatrix}  \begin{pmatrix} \psi^j  \\ \lambda^a \end{pmatrix}   \ + \ {\rm h.c.}  \  .  \label{susybreak8}
\eea
Therefore, if  $\langle F^i \rangle = \langle D^a \rangle = 0$, eq.~\eqref{susybreak7} is empty, but if  $\langle F^i \rangle \not=0$ and/or
$ \langle D^a \rangle \not= 0$ the mass matrix has a normalized eigenvector with vanishing eigenvalue, 
\be
\eta = \frac{1}{\cal F} \left(\langle F^i \rangle \,\psi^i \ + \ \frac{i}{\sqrt{2}} \langle D^a \rangle \,\lambda^a\right)  \  , \label{susybreak9}
\ee
which is in fact the {\it goldstino}.

The scale of supersymmetry breaking ${\cal F}$ can thus be defined as
\be
{\cal F}^2 = \langle \left|F^i\right|^2 \rangle \ + \ \frac{1}{2} \langle D^a D^a \rangle  \ ,
\ee
and is directly linked to the vacuum energy $V_0$
at the minimum,
\beq
\Lambda_\mathrm{SUSY}{}^4 \ = \ V_0 \ = \  {\cal F}^2 \ .
\eeq
Taking the Goldstone and Higgs theorems into account, one can thus distinguish four different cases for the breaking of a gauge symmetry and/or supersymmetry, whose manifestations on a generic potential are sketched
in Fig.~\ref{fig:susybreaking}.

The existence of a massless fermion, the goldstino, is a general feature of the spontaneous breaking of global $N=1$ supersymmetry. All available indications for Particle physics and Cosmology exclude goldstini that are charged under Standard--Model symmetries. Moreover, even if the goldstino were a gauge singlet, its couplings to the Standard Model fields should be highly suppressed, and massless fermions of this type are generally excluded by cosmological considerations.
Gauging supersymmetry converts the goldstino into the longitudinal component of the gravitino, and due to the equivalence theorem~\cite{fayet-goldstino} the longitudinal polarizations of the gravitino interact more strongly than the transverse ones. We shall return to the equivalence theorem in the following. 

Summarizing, the correspondence between standard continuous global symmetries and supersymmetry runs as follows:
\bea
& \mathrm{\underline{bosonic \ global \ symmetry}}  \qquad \qquad & \underline{N=1 \ \mathrm {global \ supersymmetry}}  \  \nonumber \\
& \mathrm{symm. \ transf. \ parameter} \ \Lambda \qquad\qquad  & \mathrm{SUSY \ transf. \ parameter} \ \epsilon^{\alpha} \ (\mathrm {spinor})  \  \nonumber \\
& \mathrm{Goldstone \ boson} \  \theta  \qquad\qquad & \mathrm {Goldstone \ fermion \ ( goldstino ) }\ \eta^{\alpha}  \  \nonumber \\
&  \mathrm{scalar \ field\ } v.e.v.  \ \langle v \rangle  \qquad\qquad & \mathrm{ auxiliary \ field\ } v.e.v. \ \langle F^i \rangle ,   \ \langle D^a \rangle \  .  \label{susybreak1}
\eea

%%%%%%%%%%%%%%%%%%%%%%%%%%%%%%%%%%%%%%%%%%%%%%%%%%%
\subsection{\sc \texorpdfstring{$D$} \ -term Breaking: the Fayet-Iliopoulos Model}

The  Fayet-Iliopoulos model \cite{fayetilio} was the first example of a Lagrangian with spontaneously broken supersymmetry. It contains a massive charged Dirac fermion and its
scalar superpartners, together with an Abelian vector multiplet, and a Fayet-Iliopoulos term. After eliminating the auxiliary fields, the Lagrangian takes the form
\bea
{\cal L}_{\rm FI} &=& - \ |(\partial_\mu + i e A_\mu) z_+|^2 \ - \ |(\partial_\mu - i e A_\mu) z_-|^2 \ - \ \psi_+ \sigma^\mu (i \partial_\mu + e A_\mu) {\bar \psi}_+  \nonumber \\
&-& \psi_- \sigma^\mu (i \partial_\mu - e A_\mu) {\bar \psi}_-  \ - \ \frac{1}{4} F_{\mu\nu}^2 \ - \ i  \lambda \sigma^\mu \partial_\mu  {\bar \lambda} \ - \ m(\psi_+ \psi_- + {\bar \psi}_+ {\bar \psi}_-)
\nonumber \\ &-&  m^2 (|z_+|^2 + |z_-|^2) \ - \  i \sqrt{2} e (z_+ {\bar \psi}_+ {\bar \lambda} - {\bar z}_+ {\psi}_+ {\lambda} - z_- {\bar \psi}_- {\bar \lambda} + {\bar z}_- {\psi}_- {\lambda}) \nonumber \\ &-& \frac{ 1}{2} \left[ e (|z_+|^2- |z_-|^2) + \xi \right]^2   \  . \label{fi2}
\eea
The scalar potential is
\be
V (z_+,z_-) \ = \ m^2 (|z_+|^2 + |z_-|^2) \ + \ \frac{ 1}{2} \left[ e (|z_+|^2- |z_-|^2) + \xi \right]^2  \   , \label{fi3}
\ee
and the algebraic equations for the auxiliary fields are
\bea
&& D \ = \ - \ e (|z_+|^2\,-\, |z_-|^2) \ - \ \xi \ , \nonumber \\
&& F_+ \ = \ - \ m {\bar z}_-\ , \qquad F_- \ = \ - \ m {\bar z}_+  \  . \label{fi4}
\eea
\begin{figure}[ht]
\centering
%\begin{figure}
\begin{tabular}{ccc}
%\mbox{graphic} & \mbox{table} \\
\includegraphics[width=65mm]{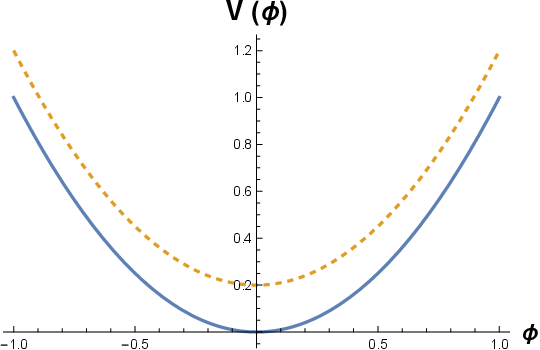} \quad  &
\includegraphics[width=65mm]{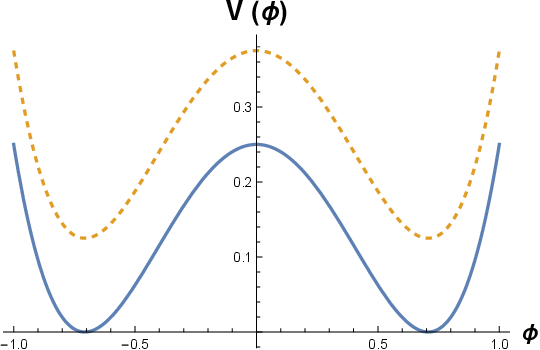}  \\
\end{tabular}
 \caption{\small The four different cases for gauge and (global) supersymmetry breaking. Left panel: gauge symmetry unbroken and supersymmetry unbroken (solid) or broken (dashed). Right panel: gauge symmetry broken and supersymmetry unbroken (solid) or broken (dashed). }
\label{fig:susybreaking}
\end{figure}

There is clearly no solution with vanishing vacuum values for all the auxiliary fields, so that supersymmetry is inevitably broken. Taking by convention $\xi > 0$,
there are two qualitatively different scenarios, depending on the parameters of the model:

\begin{itemize}

\item[$\bullet$] $m^2  > e \,\xi$.  In this case, the global minimum of the scalar potential is reached for vanishing vacuum values of the scalar fields, and therefore  $\langle F_+ \rangle = \langle F_- \rangle = 0$,
$\langle D \rangle = -\, \xi$ . The gauge symmetry remains unbroken, while
supersymmetry is broken and the vacuum energy is
\beq
V_0 \ = \ \frac{1}{2}\ \xi^2 \ .
\eeq
This case corresponds to the dashed curve in the left panel of  Fig.~\ref{fig:susybreaking}. The resulting masses in this region of parameter space are as follows:
\bea
&& m_{z_\pm}^2 \ = \  \ m^2 \,\pm\, e \xi \ , \qquad m_{\psi_\pm} \ = \ m  \ , \nonumber   \\
&& m_{A} \ = \ 0 \ , \qquad m_\lambda \ = \ 0   \ .  \label{fi5}
\eea
Since $\langle \delta \lambda \rangle = i \epsilon \langle D \rangle \not = 0$, the goldstino is the gaugino $\lambda$, which is a massless field in this case.
Note that one of the scalars, $z_-$, is lighter than the fermions $\psi_{\pm}$, while the other  scalar $z_+$ is heavier, and moreover the three masses of these fields  obey the ``sum rule''
\beq
m_{z_+}^2 \ + \ m_{z_-}^2 \ = \  2 \, m_{\psi}^2 \ .
\eeq

\item[$\bullet$] $m^2  < e \xi$. In this case the scalar $z_-$ has a tachyonic mass at the origin, so that supersymmetry and the gauge symmetry are both spontaneously broken.
In detail, minimizing the scalar potential gives
\beq
\langle z_+ \rangle\  = \ 0 \quad , \quad \langle z_- \rangle \ = \ \frac{v}{\sqrt{2}}  \ ,
\eeq
where
\beq
\frac{e^2 v^2}{2} \ + \ m^2 \ - \ e \xi \ = \ 0  \ .  \label{fi6}
\eeq
Note that in this case
\beq
\langle F_- \rangle = 0 \ , \qquad \langle F_+ \rangle \ =\ -\ m\,v /{\sqrt{2}} \ ,  \qquad\langle D \rangle = - m^2/e \ .
\eeq
Performing the redefinition
\beq
z_- \ = \  \frac{ v+ \phi_1 +i \phi_2}{\sqrt{2}}  \ ,
\eeq
one finds that the quadratic part of the Lagrangian becomes
\bea
{\cal L}^{(2)} &=& - \ \frac{e^2 v^2}{2} A_{\mu} A^{\mu} \ - \ 2 m^2  |z_+|^2 \ - \ \frac{e^2 v^2}{2} \phi_1^2 \  \nonumber \\
&-& m (\psi_+ \psi_- \ + \ {\bar \psi}_+ {\bar \psi}_-) \ - \ i e v (\psi_- \lambda \ - \  {\bar \psi}_- {\bar \lambda}) \ .  \label{fi7}
\eea
As a result, the physical masses are
\beq
m_A \ = \ m_{\phi_1} \ = \ e\,v\ , \qquad m_{z_+}\ =\ \sqrt{2} \,m \ ,
\eeq
while the fermions acquire a Dirac mass term $- \ m_{\psi} \psi_1 \psi_2$, with
\be
m_{\psi} \ = \ \sqrt{m^2+e^2v^2} \ , \quad \psi_1 \ = \ \psi_- \quad , \quad \psi_2 \ = \ \frac{m \psi_+ \ + \ i e v \lambda}{\sqrt{m^2\ +\ e^2\,v^2}}   \ .  \label{fi8}
\ee
The scalar $\phi_2$ is absorbed by the gauge field $A_{\mu}$, as pertains to the Higgs mechanism, while the massless goldstino, defined to be orthogonal to
$\psi_2$, is now
\be
 \eta  \ =\  \frac{m \lambda \ -\  i e v \psi_+}{\sqrt{m^2+e^2v^2}}  \ ,  \label{fi9}
  \ee
in agreement with the general formula (\ref{susybreak9}). For phenomenological reasons, the Abelian symmetry of this model cannot be the hypercharge $U(1)_Y$ of the Standard Model,
but it might be associated to an additional gauge symmetry $U(1)_X$. The reason is that charged scalar fields get contributions to their masses from the auxiliary field $D$, given by
\be
\delta m_i^2 \ = \ X_i \langle D \rangle  \ ,  \label{fi09}
\ee
where $X_i$ is the $U(1)_X$ charge of the multiplet containing the scalar, for example a quark or lepton multiplet. Since in the Standard Model there are fields with hypercharges of both signs, some of them
would be inevitably tachyonic, breaking spontaneously the gauge symmetries associated to color or the electric charge, which is clearly not the case in Nature. This case corresponds to the dashed curve in the right panel of Fig.~\ref{fig:susybreaking}.
\end{itemize}

The ``mass super-trace''
\beq
\mathrm{Str} \, {\cal M}^2 \ = \ \sum_j (-1)^{2j} (2j+1) \mathrm{Tr} {\cal M}_j^2 \ = \ \mathrm{Tr} {\cal M}_0^2 \ - \ 2 \,\mathrm{Tr} {\cal M}^{\dagger }{}_{1/2}\, {\cal M}_{1/2} \ + \ 3\, \mathrm{Tr} {\cal M}_1^2 \ ,  \label{fi10}
\eeq
is a useful tool for characterizing supersymmetry breaking, and plays a role in the study of quantum corrections.
Here $j$ is the spin, ${\cal M}_0$ is the scalar mass matrix, ${\cal M}_{1/2}$ is the mass matrix of the Weyl fermions, and ${\cal M}_1$ is the mass matrix
of the spin-$1$ gauge fields. 

Intuitively, $Str {\cal  M}^2$ compares the average masses of Bose and Fermi fields, and vanishes identically in supersymmetric vacua.
Note, however, that in the preceding examples $ Str {\cal M}^2$ still vanishes, although supersymmetry is spontaneously broken. This is often the case, and creates tensions with data when one tries to raise the masses of superpartners compared to those of Standard--Model particles. As we shall see, the situation improves in the presence of non--renormalizable interactions, and in particular in supergravity.

%%%%%%%%%%%%%%%%%%%%%%%%%%%%%%%%%%%%%%%%%%%%%%%%%%%%%%%%%%
\subsection{\sc \texorpdfstring{$F$} \ -Term Breaking: The Fayet--O'Raifeartaigh Models}

In the basic Fayet-Iliopoulos (FI) model, supersymmetry was broken in the presence of gauge interactions. It is actually possible to break supersymmetry spontaneously even in the absence of gauge interactions.
The simplest example can be constructed with just one chiral superfield $W$, with a linear superpotential. For a minimal K\"ahler potential, the model is therefore described in superspace by
\beq 
K = X^{\dagger} X \quad , \quad W = f X \ . \ \label{polony1}
\eeq 
This generates a constant positive scalar potential 
\beq 
V = f^2 \ , \ \label{polony2}
\eeq 
and supersymmetry is therefore broken. However, this is just a free theory, of no interest for Particle Physics. Still, as we shall see, this model can become richer and interesting if one considers a non-minimal K\"ahler potential. This would induce non--renormalizable interactions, and since for the time being we are restricting our attention to renormalizable examples, we postpone a discussion of these cases.  

The simplest nontrivial renormalizable examples of F-term breaking contain at least three chiral fields. These are usually called O'Raifeartaigh models, or F-term breaking models.
There are several types of O'Raifeartaigh (O'R) models \cite{O'R,ISS}, and the simplest renormalizable setting  contains three chiral multiplets, with a superpotential of the type
\be
W = z_1 f_1 (z_2) + z_3 f_2 (z_2) \ , \ \label{or1}
\ee
where $f_1$ and $f_2$ are two holomorphic functions whose zeroes lie at different values of $z_2$. This property grants that the auxiliary fields, which are determined as
\be
- {\bar F}_1 =  f_1 (z_2) \quad , \quad - {\bar F}_3 =  f_2 (z_2)  \ , \label{or2}
\ee
 cannot vanish simultaneously, so that supersymmetry is inevitably broken.  One peculiar feature of these models is the existence of a ``flat direction''. The scalar potential is indeed
\be
V \ = \ |f_1 (z_2)|^2 \ + \ |f_2 (z_2)|^2 \ + \ |z_1 f_1' (z_2) \,+\, z_3 f_1' (z_2)|^2 \ , \label{or3}
\ee
and therefore only a linear combination of $z_1$ and $z_3$ is determined by the minimization. The orthogonal combination does not feel the potential or, as is usually said, is a flat direction. In other words, a scalar field is inevitably massless at tree level. However, quantum corrections will generically lift flat directions, giving masses to the corresponding scalar modes.

Two renormalizable examples of O'R models are based on the superpotentials
\bea
&& W_1 \ = \ h_1 z_1 z_2^2 \ + \ z_3 (h_2 z_2^2 - M^2) \ , \nonumber \\
&&  W_2 \ = \ m z_1 z_2 \ +  \ z_3 (h  z_2^2 - M^2)  \ ,   \label{or4}
\eea
where $W_1$ contains two cubic contributions depending on the constant parameters $h_1$ and $h_2$ and a linear one depending on $M^2$, while $W_2$  involves one cubic contribution depending on $h$, a quadratic one depending on $m$ and a linear one depending on $M^2$~\footnote{A model similar to the first of these was actually constructed slightly earlier by Fayet in~\cite{fayetor}.}. 

Let us discuss the second of these models, whose scalar potential
\be
V_2 \ = \ m^2 |z_2|^2 \ + \ |h z_2^2 \ - \ M^2|^2 \ + \ |m z_1 \ + \ 2 h z_2 z_3|^2 \    \label{or5}
\ee
contains a valley of minima that depends on the parameter range. For ${m^2 > 2 h M^2}$, the minimum lies at the origin and supersymmetry is broken, since $\langle {\overline F}_1 \rangle=0$ but $\langle {\overline F}_3 \rangle = -M^2$.
For  ${m^2 < 2 h M^2}$,  the minimum is displaced away from the origin, and supersymmetry is also broken, but now both vacuum values $\langle F_1 \rangle$ and $\langle F_3 \rangle$ do not vanish.

Fayet-Iliopoulos and O'Raifeartaigh type models have continuous R-symmetries. For the FI model, up to gauge transformations, one can choose $R_{z_+} = 3$, $R_{z_-} = 0$, and the
R-symmetry is unbroken in the ground state. For the O'R models, one can choose $R_{z_1} = R_{z_3}=3$, $R_{z_2} = 0$, and  R-symmetry is also unbroken
in the ground state.

The existence of an unbroken continuous R-symmetry implies the vanishing  of a Majorana gaugino mass $- \ \frac{M}{2}\,  \lambda \lambda $, since $\lambda$ acquires a phase under R-symmetry transformations and a mass term of this type would not be invariant. As a result, gaugino Majorana masses cannot be generated at any order in perturbation theory. On the other hand, a discrete version of R symmetry, usually called R-parity, can be compatible with Majorana gaugino masses, and plays a key role, as we shall see, in extensions of the Standard Model.

\subsection{\sc Tree-level Mass Formulae for Renormalizable Interactions} \label{sec:sum_rules}

We can now highlight some general features of supersymmetry breaking in renormalizable extensions of the Standard Model, without focusing on particular examples.
To this end, it is important to take into account that 
\begin{itemize}
\item  the superpotentials of renormalizable models are at most cubic holomorphic functions of the scalar fields $z^i$;
\item the auxiliary fields are determined, in general, by eqs.~\eqref{supersym13}; 
\item the Fayet-Iliopoulos terms $\xi^a$ can only be included for Abelian generators.
\end{itemize}

In the comparison with Particle Physics, the mass spectrum of superpartners should be compatible with the current lack of evidence for them. In order to characterize the general mass patterns for scalars, fermions and gauge fields in supersymmetric models, let us first introduce the compact notation
 \bea
&& W_i \ = \ \frac{\partial W}{\partial z^i} \ , \quad {\overline W}^i \ = \ \frac{\partial \overline W}{\partial {\bar z}_i}  \ , \quad  W_{ij} \ = \ \frac{\partial^2 W}{\partial z^i \partial z^j} \ , \quad {\overline W}^{ij} \ = \ \frac{\partial^2  \overline W}{\partial {\bar z}_i\, \partial {\bar z}_j}  \ , \label{tree1} \\
 && D_i^a \ \equiv \ \frac{\partial D^a}{\partial z^i} \ = \ - \ g\, {\bar z}_j (T^a)^j{}_i \ , \quad D^{a\,i} \ = \ \frac{\partial D^a}{\partial {\bar z}_i} \ = \ - \ g\,  (T^a)^i{}_j\, z^j \ , \quad D^{a\,i}{}_j \ = \ - \ g\,  (T^a)^i{}_j  \ .   \nonumber
 \eea
 
The scalar potential  (\ref{superym14a}) then takes the form
\beq
V (z^i, {\bar z}_j) \ = \ W_i \,{\overline W}^i \ + \ \frac{1}{2}  \,D^a \,D^a \ ,  \label{tree2}
\eeq
where, as usual, some summations are left implicit.
In this notation, the vector mass matrix is
\beq
({\cal M}_1^2)^{ab} \ = \ 2 \, g^2 \langle z^{\dagger} T^a T^b z \rangle \ = \ 2 \langle D_i^a D^{bi} \rangle \ ,
\eeq
so that
\beq
\quad 3\,  \mathrm{Tr} {\cal M}_1^2 \ = \ 6 \langle D_i^a D^{ai} \rangle  \ ,  \label{tree3}
\ee
while the fermionic mass matrix in (\ref{susybreak8}) reads
\beq
 {\cal M}_{1/2} =
\begin{pmatrix}  \langle W_{ij} \rangle &   i \sqrt{2}  \langle { D_i^a} \rangle    \\
  i \sqrt{2}  \langle { D_j^b}  \rangle   & 0
 \end{pmatrix}
 \eeq
 and consequently
 \beq
  -\, 2 \, \mathrm{Tr} {\cal M}_{1/2}^{\dagger } {\cal M}_{1/2} \,=\, -\, 2   \langle  {\overline W}^{ij} W_{ij} \rangle \,-\, 8 \langle D_i^a D^{ai} \rangle  \  .  \label{tree4}
\eeq
Finally, the scalar mass matrix in a complex basis is
\beq
  {\cal M}_{0}^2 \ =\ 
\begin{pmatrix}  \left\langle  \frac{\partial^2 V}{\partial z^i \partial {\bar z}{}_k} \right\rangle &     \left\langle\frac{\partial^2 V}{\partial z^i \partial z^l}  \right\rangle    \\ \\
  \left\langle  \frac{\partial^2 V}{\partial {\bar z}{}_j \partial {\bar z}{}_k}    \right\rangle   &   \left\langle  \frac{\partial^2 V}{\partial {\bar z}{}_j \partial z^l}    \right\rangle
   \end{pmatrix}  \  ,  \label{tree5}
\eeq
and consequently
\be
\mathrm{Tr} {\cal M}_0^2 \ = \ 2 \left\langle  \frac{\partial^2 V}{\partial z^i \partial {\bar z}{}_i} \right\rangle \ = \ 2 \left(  \langle W_{ij} {\overline W}^{ij} + D_i^a D^{ai} + D^a (D^a){}^i{}_i  \rangle  \right)  \  .  \label{tree6}
\ee
Collecting the three contributions, one thus finds \cite{fgp}
\be
\mathrm{Str} {\cal M}^2 \ =  \ 2\,\sum_a \langle D^a  \rangle \langle (D^a){}^i{}_i  \rangle\ = \ - \ 2 \sum_a g_a  \langle D^a  \rangle \,Tr T^a \  ,  \label{tree7}
\ee
where we used the definitions in eq.~\eqref{tree1}, while also reinstating an explicit sum over the gauge group generators $T^a$. 

A non--vanishing $Str  {\cal M}^2$ implies the presence of quadratic divergences for charged scalar masses at one loop. Their absence would resonate with the stability of the large ratio between the Planck and Weak scales in Nature, and has served, over the years, as an important motivation for low--energy supersymmetry. In the interesting cases then $Str  {\cal M}^2=0$, the average masses of bosons and fermions are bound to coincide, which is not easily reconciled with data.

Supersymmetry-breaking contributions are easily identifiable since they are proportional to the expectation values of auxiliary fields. The preceding results show that Fermi fields and spin--1 fields do not lead to any net contribution to the supertrace.
The only supersymmetry--breaking contributions to the scalar masses originate from
\beq
 ({\cal M}_{0}^2)_{susy-breaking} = \begin{pmatrix}   \langle D^a (D^a){}^k{}_i  \rangle  & \langle W_{ikl} {\overline W}^k \rangle   \\
   \langle {\overline W}^{jki} {W}_i \rangle   &  \langle D^a (D^a){}^j{}_l  \rangle \end{pmatrix} \  ,  \label{tree8}
 \eeq
 and we can now reconsider the preceding examples from this perspective.
  \begin{itemize}
\item[$\bullet$] In the presence of F-term breaking, as in the O'R models, $\langle F_i \rangle \not=0$, $\langle D^a \rangle = 0$, and then $Str  {\cal M}^2=0$.  Moreover, taking  into account the off-diagonal form of the supersymmetry-breaking terms
in eq.~(\ref{tree8}), one can anticipate the presence of negative contributions to the squared masses of scalar fields. Letting
\beq
z^i \ = \ \frac{A_i + i B_i}{\sqrt{2}} \ ,
\eeq
one finds the supersymmetry-breaking contributions to the Lagrangian
\beq
-\, \frac{1}{2}  \langle {\overline W}^{ijk} {\overline F_k} \rangle \left[  A_i A_j - B_i B_j - i (A_i B_j + A_j B_i) \right] + {\rm h.c.}  =
 \langle {\overline W}^{ijk} {\overline F_k} \rangle ( -A_i A_j + B_i B_j) \  ,  \label{tree9}
 \eeq

where in the last equality we focused, for simplicity, on real values for $\langle {\overline W}^{ijk} {\overline F_k} \rangle  $. Contributions concerning the real scalars $A_i$ and $B_i$ are thus of opposite sign. This type of spectrum is not realistic for scalar partners of quarks (the squarks) and leptons (sleptons), since these options would contradict experimental data. Some superpartners of light quarks and leptons, the squarks and sleptons, would become tachyonic for realistic supersymmetry breaking scales, and this would induce patterns of spontaneous breaking of color and electric charges, which is clearly unacceptable.

\item[$\bullet$]  With D-term breaking, focusing for simplicity on the case $\langle F_i \rangle=0$, $\langle D^a \rangle  \neq 0$, one can obtain a non--vanishing $\mathrm{Str}  {\cal M}^2$ provided $\mathrm{Tr} T^a \not=0$, where the $\left(T^a\right)^j{}_i \ = \ \delta_i^j \, X_i ^a$ are Abelian gauge group generators. The supersymmetry-breaking contributions to the scalar masses
are
\beq
2 g_a \langle D^a \rangle z^i (T^a)^j{}_i \,{\bar z}_j \ = \ 2 \sum_i \, g_a \langle D^a \rangle X_i ^a |z^i|^2 \ = \ \sum_i \, g_a \langle D^a \rangle X_i ^a (A_i^2 + B_i^2) \  ,  \label{tree10}
\eeq
where the $X_i^a$ determine the charges of the fields $A_i,B_i$ with respect to the Abelian gauge fields $A_\mu^a$.
This case could be realistic if all charges had the same sign, which would be possible, in principle, if $Tr X^a \not=0$.
However, realistic models of this type are complicated by the need to cancel gauge anomalies. In order for supersymmetry breaking to be transmitted at tree level to the observable scalar superpartners of quarks and leptons, the corresponding
 multiplets must be charged under a new $U(1)_X$ gauge symmetry. Due to D-term mass contributions as in~(\ref{fi09}), one would need observable fields that have charges of the same sign, in order to avoid tachyons. Additional exotic fields of opposite charge would be needed to cancel the gauge anomalies, which would get additional contributions
 to their masses from F-terms. While there is no theorem excluding these types of scenarios, no realistic model was constructed, until now, along these lines, although Fayet made substantial efforts, over the years, along these directions.
 \end{itemize}

Gaugino masses are another source of difficulty when one tries to build realistic supersymmetric extensions of the Standard Model. If O'R models are coupled to gauge theories, the scalar fields $z_i$ are gauge singlets, and therefore no Majorana gaugino masses will be generated, in perturbation theory, in renormalizable models. In the presence of FI terms, if the gauge symmetry is broken, as we saw in (\ref{fi8}), the gaugino pairs with a matter fermion to build a Dirac mass, but due to the reasons discussed above it is difficult to construct realistic models. For example, gluinos, the fermionic partners of gluons, cannot acquire masses in this way, since this would require that a colored scalar acquire a vacuum expectation value, thus breaking the color gauge symmetry.

All attempts to generate realistic supersymmetric extensions of the Standard Model with renormalizable interactions only are thus fraught with a number of difficulties. These include the tree--level mass sum rules that we have discussed, and have led to different attempts based on more sophisticated setups for supersymmetry breaking, which fall generically into two classes.

\begin{itemize}
\item[1) \ ] {\sc Tree--Level Mediation via non--Renormalizable Interactions}

``Gravity mediation'' \cite{gravity-mediation1,gravity-mediation2,hlw} is the most elegant example in this class. This framework includes a ``hidden sector'', responsible for supersymmetry breaking, which couples to the observable sector  only  via non-renormalizable interactions related to gravity. One can then find qualitatively correct patterns of supersymmetry breaking in a non--trivial decoupling limit where the kinetic terms of the hidden sector can be ignored. Its dynamical fields, which are usually called ``spurions'', can then be integrated out, but the corresponding auxiliary fields can still acquire nonzero vacuum values.
 The effects of supersymmetry breaking are thus encoded in supersymmetric couplings, which are typically non renormalizable, linking the spurions to the observable sector. They are thus suppressed by a heavy mass scale,
 called $M$ in what follows, which becomes the Planck mass $M_P$ if supergravity sources the mediation. We shall return to these issues in Section~\ref{sec:SUGRA}.

\item[2) \ ] {\sc Gauge Mediation}

In the most widely explored scenario of this type, the couplings between hidden and observable sectors originate from flavor--blind gauge loops (for a review, see~\cite{gauge-mediation}).  Supersymmetry breaking is transmitted from the hidden sector by messenger fields $\Phi,{\overline \Phi}$, whose mass $M$ can be much lower than $M_{P}$. The typical superpotential of such models is
\beq
W = W_0 (X) + (M + \lambda X) \Phi {\widetilde \Phi} \ , \label{gm1}
\eeq
where $X$ is the spurion breaking supersymmetry.
The messenger fields $\Phi,{\widetilde \Phi}$  have conjugate quantum numbers under the Standard Model gauge group, and fill vector-like representations, typically complete $SU(5)$ representations, in order not to spoil the gauge coupling unification that occurs in the MSSM.
The spin-0 messenger mass matrix is of the form
\beq
M_{\Phi}^2 =
\begin{pmatrix}  M^2 &     \lambda {\overline F}{}_X  \\ \\
\lambda F_X   &   M^2
   \end{pmatrix}  \ ,
 \  \label{gm2}
\eeq
where
\beq
{\overline F}{}_X \ = \ - \ \frac{\partial \,W_0}{\partial\,X} \ ,
\eeq
and in order to avoid tachyonic messenger masses one must demand that
\beq
M^2 \geq \lambda |F_X|
\ .  \  \label{gm3}
\eeq

One can see that in such cases gaugino masses $M_{1/2}$ emerge at one loop, while scalar masses $m_0^2$ emerge at two loops. The leading contributions coming from gauge mediation are thus of the form
\beq
M_{1/2}\  \sim \ \frac{g^2}{16 \pi^2} \ \left|\frac{F_X}{M}\right| \ , \qquad
m_0^2 \ \sim \ \left(\frac{g^2}{16 \pi^2}\right)^2\  \left|\frac{F_X}{M}\right|^2 \ , \label{gm4}
\eeq
where $g$ is a SM gauge coupling and $F_X$ is the auxiliary component of the spurion responsible for the effect. As a result, this loop pattern yields comparable values for gaugino and scalar masses. Moreover, since the Standard Model loops are flavor blind, the scalar masses are flavor independent, and therefore large flavor changing contributions that could jeopardize the GIM mechanism~\cite{gim}, a key property of the Standard model that we shall review shortly, are avoided. Combining (\ref{gm3}) and (\ref{gm4}) and demanding that soft masses lie at least in the TeV range, one finds for the messenger masses $M$ a lower bound of about $100$ TeV.

\end{itemize}
\begin{figure}[ht]
\begin{center}
    \includegraphics[width=2.5in]{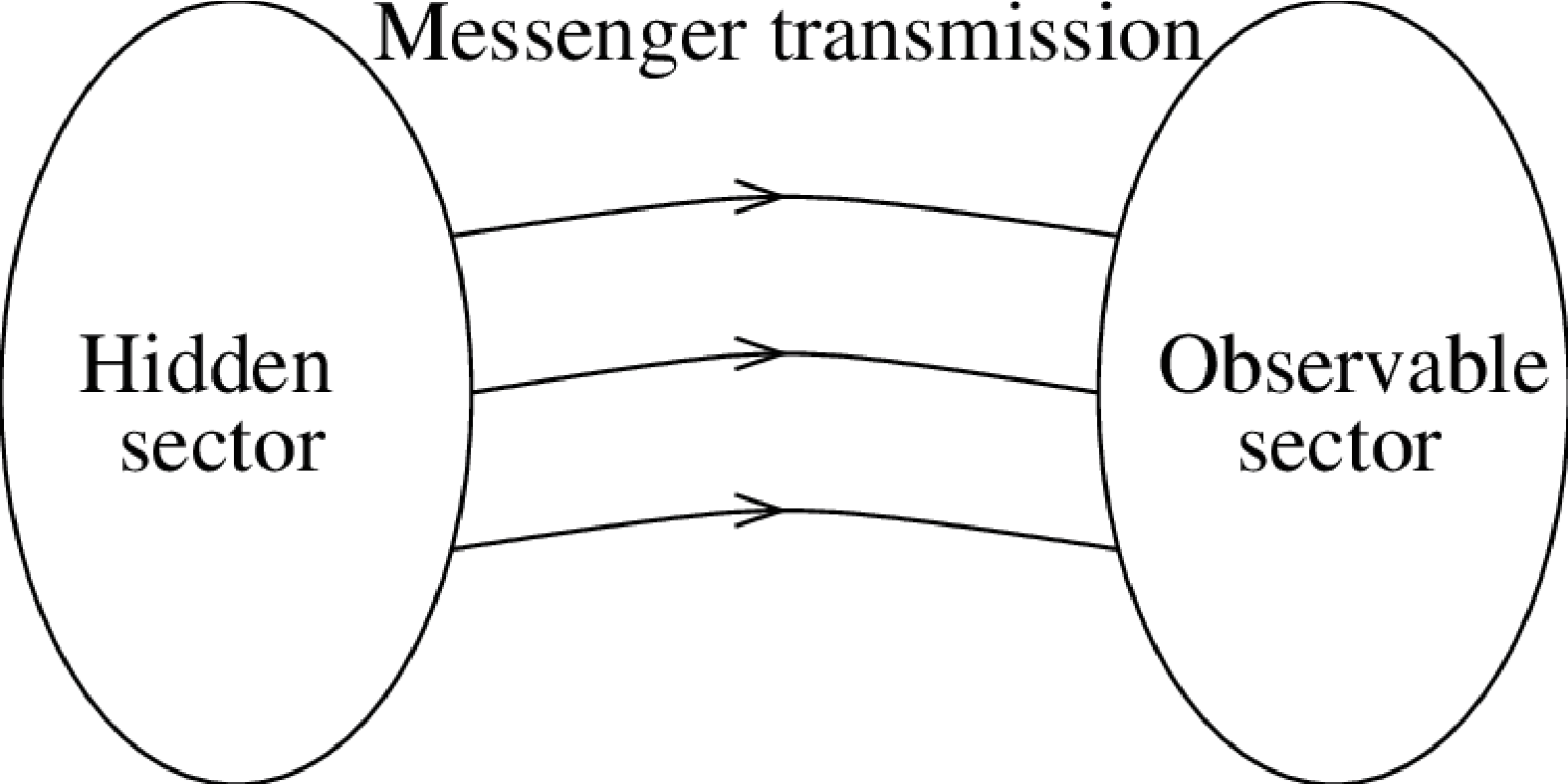}
\end{center}
\caption{The hidden sector breaks supersymmetry, which is mediated to the visible sector by gravity or other non--renormalizable interactions or via quantum loops.}
\label{fig:hidden}
\end{figure}

While in both types of mechanisms supersymmetry breaking in the hidden sector is typically of non--perturbative origin, its effects in the observable sector are captured, to a large extent, by standard perturbative methods.

%%%%%%%%%%%%%%%%%%%%%%%%%%%%%
\subsection{\sc Beyond Renormalizable Interactions } \label{sec:beyond_renormalizable}

Non-renormalizable theories with global $N=1$ supersymmetry exhibit a beautiful mathematical structure called K\"ahler geometry~\cite{Zumino:1979et}. K\"ahler manifolds are complex manifolds, whose local coordinates we shall denote by $(z^i, {\bar z}{}^{j})$. These manifolds are endowed with a real function ${\cal K} (z^i , {\bar z}{}^{j}) $, which is usually called K\"ahler potential. Denoting its derivatives by $\partial_i$ and $\partial_{\bar{j}}$, the metric on a K\"ahler manifold can be expressed in terms of the K\"ahler potential as
 \be
  g_{i \bar j} \ = \ \partial_i\, \partial_{\bar{j}}\,{\cal K}  \ \equiv \ \frac{\partial^2\,{\cal K}}{\partial z^i\, \partial {\bar z}{}^j} \ . \label{bri1}
 \ee
Note that the preceding expression is manifestly invariant under the so-called K\"ahler transformations
\beq
{\cal K} \ \to \ {\cal K} \ + \ h (z^i) \ +\  {\overline h} ({\bar z}{}^{i}) \ , \label{kahlertr}
\eeq
where ($\overline h$) $h$ are (anti)holomorphic functions of the (${\bar z}{}^{j}$) $z^i$.  In these manifolds, holomorphic reparametrizations of the $z^i$ and anti--holomorphic reparametrizations of the $\bar{z}{}^i$ play a central role, as ordinary reparametrizations do in ordinary Riemannian geometry.
 Eq.~\eqref{bri1} implies for the metric integrability conditions of the type
 \beq
\partial_k\, g_{i \bar j} \ =  \ \partial_i\, g_{k \bar j} \ , \label{bri2}
 \eeq
and one can verify that the only non-vanishing Christoffel symbols,
\beq
 \Gamma^k{}_{ij} \ = \ g^{k \bar l} \partial_i g_{j \bar l}  \ , \qquad  \Gamma^{\bar k}{}_{\bar i \bar j} \ = \ g^{{\bar k}  l} \partial_{\bar i} g_{l \bar j}  \ , \label{bri3}
 \eeq
have purely holomorphic or purely anti--holomorphic labels.
In analogy with General Relativity, one can define a K\"ahler covariant derivative, which defines tensors compatibly with the holomorphic structure of the manifold. For example, if $V_i$ is a vector field
 with a holomorphic index $i$, then
\beq
\nabla_i V_j \ = \ \partial_i V_j \ - \ \Gamma^k{}_{ij}\, V_k \ , \qquad \nabla_{\bar i} V_j \ = \ \partial_{\bar i} V_j  \ .  \label{bri4}
\eeq

These local expressions actually have a global meaning, since in complex manifolds the transition functions between different patches are holomorphic.
As in General Relativity, one can deduce the expression for the Riemann tensor of a K\"ahler manifold from the commutator of two covariant derivatives, and this procedure leads to identify two types of curvature components:
 \beq
 [\nabla_i , \nabla_{\bar j}] V_k \ = \ {\cal R}_{{i \bar j} k}{}^l \ V_l \ , \qquad   [\nabla_i , \nabla_{ j}] V_k \ = \ {\cal R}_{{i j} k}{}^l \ V_l  \ .  \label{bri5}
 \eeq
 The K\"ahler metric is used to raise and lower indices, so that for example
 \beq
 V^i \ = \ g^{i \bar j} \,V_{\bar j} \ , \qquad {\cal R}_{i \bar j k \bar l} \ = \ g_{m \bar l} \,{\cal R}_{i \bar j k}{}^m   \ .  \label{bri6}
  \eeq
The structure of the Christoffel symbols implies that the only non-vanishing components of the Riemann tensor arise from $R_{i \bar j k \bar l}$ and the complex conjugate. Moreover, from the usual definition
 of the Riemann tensor, one finds
 \bea
&& {\cal R}_{i \bar j k \bar l} \ = \ \partial_i\, \partial_{\bar{j}}\, g_{k \bar l} \ -\  g^{m \bar n} \partial_{\bar{j}}\, g_{m \bar l}  \ \partial_i\,g_{k\bar{n}} \ = \ g_{m \bar l}\,\partial_{\bar{j}}\, \Gamma^m{}_{ik}  \ , \nonumber \\
  &&  {\cal R}_{i \bar j k \bar l} \ = \ -  \ {\cal R}_{i \bar j \bar l k} \ =\  - \  {\cal R}_{\bar j i k \bar l} \ =\   {\cal R}_{\bar j i  \bar l k}   \ .  \label{bri7}
 \eea

We can now make use of these properties of K\"ahler manifolds, returning to supersymmetric Lagrangians and considering, to begin with, only chiral multiplets $\Phi^i$. In $N=1$ superspace, whose main properties are reviewed in Appendix~\ref{app:superfields_global}, the most general Lagrangian describing their supersymmetric interactions is of the form
\beq
{\cal L} \ = \ \int d^4 \theta \ {\cal K} (\Phi^i , \overline{\Phi}{}^{\,j}) \ + \ \left( \int d^2 \theta \ {\cal W} (\Phi^i) \,+\, {\rm h.c. } \right) \  . \label{bri8}
\eeq
This Lagrangian is manifestly invariant under K\"ahler transformations~\eqref{kahlertr}, since superspace integrals act like derivatives: the real function ${\cal K}$ is the K\"ahler potential, while the holomorphic function ${\cal W}$ is the superpotential, and the reality of ${\cal K}$ implies that
\beq
{\cal K} (\Phi^i , \overline{\Phi}{}^{\,j}) \ = \ {\overline {\cal K}} (\overline{\Phi}{}^{\,i} , {\Phi}^{ j}) \ .
\eeq
In renormalizable theories the K\"ahler potential is simply
\beq
{\cal K} \  = \  \Phi^i \ \overline{\Phi}{}^{\,j} \ \delta_{i\,\bar j } \, 
\eeq
and
${\cal W}$ is at most a cubic polynomial, as we have seen, but in this section their forms are not restricted.

The more general models that we are about to describe contain at least one intrinsic mass scale, but we shall leave all these dimensionful quantities implicit in the ensuing discussion.  Moreover,
we shall resort again to a shorthand notation, so that
\bea
&& {\cal K}_i \ = \ \partial_i\ {\cal K} \ , \qquad   {\cal K}_{i \bar j} \ = \ g_{i \bar j} \ =\ \partial_i\,\partial_{\bar{j}}\,{\cal K} \ , \nonumber \\
&& {\cal W}_i \ = \ \partial_i \,{\cal W}  \ , \qquad  {\cal W}_{ij} \ =\ \partial_i\, \partial_j \, {\cal W} \ ,\qquad
{\overline {\cal W}}{}_{\bar i} \ = \ \partial_{\bar i }\, {\overline {\cal W}} \ , \qquad {\rm etc} \ . \label{bri9}
\eea

The component Lagrangian is recovered expanding superpotential and K\"ahler potential and retaining the highest superfield components of the resulting expressions, so that
\bea
\int d^2 \theta \ {\cal W} (\Phi^i) &=& F^i \,{\cal W}_i \ -\  \frac{1}{2} \,\psi^i \,\psi^j \,{\cal W}_{ij}  \ , \label{bri10} \\
\int d^4 \theta \ {\cal K} (\Phi^i , \overline{\Phi}{}^{\,j}) &=& {\cal K}_{i \bar j} \left( -\  \partial_{\mu} z^i \, \partial^{\mu} {\bar z}^{j} \ -\  i\, {\bar \psi}^{j} {\bar \sigma}^{\mu} D_{\mu} \psi^i \ + \ F^i {\bar F}{}^{j} \right)  \nonumber \\
&-& \frac{1}{2}\, {\cal K}_{l \bar i} \,\Gamma^{ l}{}_{ j  k}\, {\bar F}{}^{i} \,{\psi}^{j} \, {\psi}^{ k} \ - \ \frac{1}{2} \,{\cal K}_{i \bar l} \,\Gamma^{\bar l}{}_{\bar j \bar k} \,F^i \,{\bar \psi}{}^{j}  \, {\bar \psi}{}^{k}
\ + \ \frac{1}{4} \,{\cal K}_{i \bar j k \bar l} \, {\psi}^{i}  {\psi}^{ k} {\bar \psi}{}^{j}   {\bar \psi}{}^{l}   \ . \nonumber
\eea
Here we defined the spacetime K\"ahler covariant derivatives of the Fermi fields,
\beq
D_{\mu} \psi^i \ = \ \partial_{\mu} \psi^i \ + \ \Gamma^i{}_{kl} \,\partial_{\mu} z^k \,\psi^l  \ , \label{bri11}
\eeq
and the off-shell Lagrangian is the sum of the two terms in (\ref{bri10}). The on-shell version can be obtained eliminating the auxiliary fields via their field equations
\beq
{\cal K}_{i \bar j} \, F^i \ - \ \frac{1}{2} {\cal K}_{i \bar j} \, \Gamma^i{}_{ k l }\,  {\psi}^{k}  \,{\psi}^{ l} \ +\  {\overline {\cal W}}_{\bar j} \ =\  0   \ , \label{bri12}
\eeq
where the middle term does not vanish, here and elsewhere, since the Fermi bilinear is symmetric under the interchange of $i$ and $j$ due to the implicit contraction of the spinor indices as in eq.~\eqref{sp14}.

Substituting into  eqs.~(\ref{bri8}) and (\ref{bri10}) and using eqs.~\eqref{bri7}, one finds the final on-shell Lagrangian
\bea
{\cal L} &=&  {\cal K}_{i \bar j} \left( - \ \partial_\mu z^i\, \partial^\mu {\bar z}{}^{j}  \, - \, i {\bar \psi}{}^{j} {\bar \sigma}^\mu D_\mu \psi^i  \right) \ +\
 \frac{1}{4} \,{\cal R}_{i \bar j k \bar l}\,  {\psi}^{i}  {\psi}^{ k} {\bar \psi}{}^{j}   {\bar \psi}{}^{l}  \ \nonumber \\
 &-& \frac{1}{2} \,\nabla_i \nabla_j {\cal W} \ {\psi}^{i}\,  {\psi}^{ j} \ - \   \frac{1}{2} \,\nabla_{\bar i} \nabla_{\bar j} {\overline {\cal W}}\   {\bar \psi}{}^{i} \, {\bar \psi}{}^{j}  \ - \ {\cal K}^{i \bar j} \, \nabla_i {\cal W} \,  \nabla_{\bar j} {\overline {\cal W}}
 \ , \label{bri13}
\eea
where  ${\cal K}^{i \bar j}$ is the inverse K\"ahler metric,
\beq
\nabla_i\,{\cal W} \ = \ {\cal W} _i \quad , \quad \nabla_i \nabla_j {\cal W}  \ = \ {\cal W} _{ij} \ - \ \Gamma^k{}_{ij} \,{\cal W}_k  \ . \label{bri013}
\eeq
and the quartic spinor term can be expressed in terms of the curvature, taking eq.~\eqref{bri2} into account.

One can now add gauge fields in the adjoint of a gauge group $G$, relying on the discussion presented in Appendix~\ref{app:superfields_global}.  If the original K\"ahler potential and superpotential are invariant under the global version of $G$, it suffices to replace
$\overline{\Phi}$ with $\overline{\Phi} \,e^{2V}$, while also adding kinetic terms for the gauge fields, so that the superspace Lagrangian becomes
\beq
{\cal L} \ = \ \int d^4 \theta \ {\cal K} \left(\Phi^i , (\overline{\Phi}\, e^{2 V})^{j}\right) \ +\  \left[ \int d^2 \theta \  \left( \frac{1}{4} f_{ab} (\Phi^i) W^{a\,\alpha} \,W^b{}_{\alpha} \ + \ {\cal W} (\Phi^i) \right) \ + \ {\rm h.c. } \right] \ . \label{bri15}
\eeq
Here $ f_{ab} (\Phi^i)$ is a holomorphic function of the chiral superfields, which is usually called ``gauge kinetic function'', and in components the second term above becomes
\bea
{\cal L}_{\rm gauge} &=&  \int d^2 \theta \   \frac{1}{4} f_{ab} (\Phi^i) W^{a\,\alpha} \,W^b{}_{\alpha} \  + \ {\rm h.c. } \label{bri16}  \\
&=& Re f_{ab} \left( -\,\frac{1}{4}  F_{\mu \nu}^a F^{b \mu \nu}  \,-\, i \lambda^a \sigma^\mu D_\mu {\bar \lambda}^b
 \,+\, \frac{1}{2} D^a D^b  \right) \ - \  \frac{1}{4}  Im f_{ab} \,F_{\mu \nu}^a \,{\tilde F}^{b \mu \nu} \nonumber \\
 &+& \left[ f_{ab i} \left( i \sqrt{2} \, \psi^i \lambda^a D^b - \sqrt{2} \,\lambda^a \sigma^{\mu\nu} \psi^i F_{\mu\nu}^b + F^i \lambda^a  \lambda^b \right) \ +  \ \frac{1}{2} f_{ab ij} \lambda^a  \lambda^b
 \psi^i \psi^j + {\rm h.c.}  \right]  \ , \nonumber
\eea
where, for example,
\beq
{\tilde F}^{a \mu\nu} = \frac{1}{2} \epsilon^{\mu\nu\rho\sigma} F^{a}_{\rho\sigma} \ ,  \qquad f_{ab i} = \partial_i\, f_{ab} \ , \qquad \qquad f_{ab i j} = \partial_i\,\partial_j\, f_{ab} \ .
\eeq
The matter part of the Lagrangian,
\bea
{\cal L}_{\rm matter}&=&
{\cal K}_{i \bar j} \left( -\, D_\mu z^i D^\mu {\bar z}{}^{j}\,- \,\frac{i}{2}\, {\bar \psi}{}^{j} {\bar \sigma}^\mu { D}_\mu \psi^i \,+\, \frac{i}{2} { D}_\mu  {\bar \psi}{}^{j} {\bar \sigma}^\mu \psi^i\,+\, F^i {\bar F}^{j} \right)   \label{bri17} \\
&+& \left( F^i {\cal W}_i \,-\, \frac{1}{2} \,\psi^i \psi^j {\cal W}_{ij} \,-\, \frac{1}{2} \,{\cal K}_{l \bar i} \Gamma_{ j  k}^{ l} {\bar F}^{i} {\psi}^{j}  {\psi}^{ k}  \,+\, {\rm h.c.} \right) \nonumber \\
&+& \frac{1}{4}\, {\cal R}_{i \bar j k \bar l}\,  {\psi}^{i}  {\psi}^{ k} {\bar \psi}{}^{j}   {\bar \psi}{}^{l}  \,+\, i \sqrt{2} {\cal K}_{\bar i j}\,{\bar z}{}^{i} \left(T^a\right)^j{}_k \psi^k \lambda^a
\nonumber \\ &-&  i \sqrt{2} \, {\cal K}_{\bar i j} \,{z}^{j} \left(T^{*\,a}\right)^{\bar i}{}_{\bar k} \,{\bar \psi}^{k} {\bar \lambda}^a  \, +\,    {\bar z}^{k} \left(T^{*\, a}\right)^{\bar i}{}_{\bar k} {\cal K}_{\bar i} \,D^a \ , \nonumber
\eea
where the gauge and K\"ahler covariant derivatives are 
\bea
&& D_\mu\,z^i \ = \ \partial_\mu z^i \ - \ i\, A_\mu^a \,\left( T^a \right)^i{}_j z^j \ , \nonumber \\
&& D_\mu\,{\bar z}{}^i \ = \ \partial_\mu {\bar z}{}^i \ + \ i\, A_\mu^a \,\left(T^{*\,a}\right)^{\bar i}{}_{\bar j}\,  {\bar z}^j \ , \nonumber \\
&& {D}_\mu \psi^i \ =\ \partial_\mu \psi^i \ -\ i  A_\mu^a \,\left( T^a \right)^i{}_j \psi^j \ + \ \Gamma^i{}_{jk} \, \partial_\mu z^j \psi^k \ , \nonumber \\
&& {D}_\mu {\bar \psi}{}^{i} \ = \ \partial_\mu {\bar \psi}{}^{i}{  \ + \ i A_\mu^a \left(T^{*\,a}\right)^{\bar i}{}_{\bar j}\, {\bar \psi}^{j}} \ + \ \Gamma^{\bar i}{}_{\bar j \bar k} \, \partial_\mu {\bar z}{}^{j} {\bar \psi}{}^{k}  \ , \label{bri18}
\eea
is the gauged version of (\ref{bri10}).  This extension requires that the K\"ahler manifold possess the Killing vectors needed to grant a linear realization of the gauge group. 
If this is the case, up to the last three terms, eq.~\eqref{bri17} could be simply obtained introducing gauge and K\"ahler covariant derivatives for fermions and standard gauge covariant derivatives for scalars.

More precisely, the K\"ahler manifold should admit G as an isometry, so that for each generator of G there should be a holomorphic Killing vector $\xi^{i a}(z^i)$ (together with its conjugate), so that
\beq
\delta\,z^i \ = \ \epsilon^a \, \xi^{i}{}_a(z^i) \ , \qquad \delta\,\bar{z}^i \ = \ \epsilon^a \, {\xi}^{\bar{i}}{}_a(\bar{z}^i)
\eeq
and ${\cal K}$, ${\cal W}$ and $f_{ab}$ should be compatible with these symmetries, so that
\beq
{\cal K}_i \, \xi^{i}{}_a \ + \ {\cal K}_{\bar{i}} \, \xi^{\bar{i}}{}_a \ = \ 0 \ , \qquad {\cal W}_i \, \xi^{i}{}_a \ = \ 0 \ , \qquad f_{abi} \, \xi^{i}{}_c \ = \ 0 \ . \label{invariance}
\eeq
In eqs.~\eqref{bri18} we wrote, for brevity
\beq
\xi^{i a}(z^i)  \ = \ \left(T^a\right)^i{}_j \, z^j \ ,
\eeq
and we shall often do it in the following, but strictly speaking we have in mind the general case.

${\cal L}_{\rm gauge} + {\cal L}_{\rm matter}$ is the most general supersymmetric off-shell Lagrangian with
chiral and vector multiplets. The auxiliary fields can be eliminated via their field equations
\bea
F^i &=& - \ g^{i \bar j} \left( {\overline {\cal W}}{}_{\bar j} \ - \ \frac{1}{2}\, {\cal K}_{kl \bar j} \,\psi^k \psi^l \ - \   \frac{1}{4}\,  {\overline f}_{ab \bar j} \,{\overline \lambda}^a{\overline \lambda}^b \right) \ , \nonumber \\
D^a &=& - \ (Re f)_{ab}^{-1} \left( {\bar z}{}^{j} \left(T^{*\,b}\right)^{\bar i}{}_{\bar j}\, {\cal K}_{\bar i} \ + \ \frac{i}{2 \sqrt{2}}\, f_{bc i}\, \psi^i \lambda^c \ - \ \frac{i}{2 \sqrt{2}} \, {\overline f}_{bc \bar i} \,{\bar \psi}{}^{i}\, {\overline \lambda}^c  \right)
 \ , \label{bri19}
 \eea
and the resulting scalar potential is again the sum of two non--negative terms, originating from F-term and D-term contributions:
\beq
V \ \equiv\  V_F \ + \ V_D  \ = \  g^{i \bar j} \, {\cal W}_i \,{\overline {\cal W}}_{\bar j} \ +\ \frac{1}{2} (Re f)_{ab}^{-1} {\bar z}{}^{j} \Big(T^{*\,a}\Big)^{\bar i}{}_{\bar j}\,{\cal K}_{\bar i}\  {\bar z}{}^{l}  \Big(T^{*\,b}{}\Big)^{\bar k}{}_{\bar l}\, {\cal K}_{\bar k}   \ . \label{bri20}
\eeq
The comparison with the preceding section rests on the identification
\beq
\frac{1}{g^2} \ = \ \left\langle \mathrm{Re}\, f\right\rangle \ ,
\eeq
where all indices are left implicit.

The form of the auxiliary fields in  (\ref{bri19}) allows more options for supersymmetry breaking. For one matter, the Fermi bilinears $\psi \psi$  and  ${\lambda} {\lambda}$
can contribute by condensing non-perturbati\-vely. There are various arguments suggesting that a ``dynamical breaking of supersymmetry'' of this type
can be realized under some conditions \cite{dsb}, within a hidden sector, at some non--perturbatively generated mass scale.  Supersymmetry breaking in the hidden sector can then be transmitted to the
observable sector by ``messengers'', along the lines of what we discussed above, as illustrated schematically in fig.~\ref{fig:hidden}. The hidden--sector paradigm has played a central role in supersymmetric
extensions of the Standard Model, and in particular in scenarios inspired by String Theory, during the last decades.

%%%%%%%%%%%%%%%%%%%%%%%%%%%%%%%%%%%%%%%%%%%%%%%%%%%
\subsection{\sc The Witten Index } 
\label{sec:wittenindex}

It would be desirable to have some more general criteria for supersymmetry breaking, especially in view of non--perturbative settings. Indeed, having primarily in mind scenarios in which supersymmetry is dynamically broken in a hidden sector, typically at an intermediate energy scale,  Witten introduced~\cite{windex} a quantity, called later the Witten index, defined as
\beq 
Tr \ (-1)^F \ = \ N_b - N_f \ ,  \label{windex1}
\eeq
where $N_b$ ($N_f$) is the number of bosonic (fermionic) states in the theory. In what follows, let us focus on zero--momentum states. In this case, the Hamiltonian of the system can be cast in the form
\beq 
H = \frac{1}{2} \left( Q \,{Q}^\dagger \ + \ {Q}^\dagger \,Q  \right)  \ ,  \label{windex2}
\eeq 
where $Q$ denotes one of the two supercharges $Q_{\alpha}$ ($\alpha = 1,2$) of the theory. $Q$ and $Q^\dagger$ satisfy the familiar algebra of the fermionic harmonic oscillator when they act on eigenstates of $H$ of nonzero energy, which thus occur in pairs, which correspond to the empty state $|0\rangle$ and the full state $|1\rangle$. In detail
\bea 
&& Q \ | n \rangle \ = \ \sqrt{2 E_n} \ | n' \rangle \ , \qquad  {Q}^\dagger \ | n \rangle \ = \ 0 \ , \nonumber \\
&& {Q} \ | n' \rangle \ = \ 0  \ , \qquad 
{Q}^\dagger \ | n' \rangle \ = \ \sqrt{2 E_n} \ | n \rangle 
\ .  \label{windex3}
\eea
One of the two degenerate states is bosonic and the other is fermionic, so that all states of non--vanishing energy are boson-fermion pairs, while zero-energy states can potentially contribute to the index (\ref{windex1}). 

If the Witten index does not vanish, then there are necessarily zero-energy states. They are automatically the ground states of the system, and therefore supersymmetry is unbroken in this case. On the other hand, if the index vanishes the situation is inconclusive, since it can also vanish if there are equal non--vanishing numbers of bosonic and fermionic states of zero energy, not only if there are no zero-energy states altogether. In the first case, supersymmetry is unbroken, whereas in the second case it is broken. 
One can therefore summarize the criterion as follows:
\bea
&& Tr \ (-1)^F \ \neq \ 0 \qquad \to \qquad {\rm unbroken \ SUSY} \ , \nonumber \\
&& Tr \ (-1)^F \ = \ 0 \qquad \to \qquad {\rm inconclusive}  \ .  \label{windex4}
\eea

An interesting question is the behavior of the Witten index when the values of some parameters in the theory are changed. In this case, one expects states to move up and down in energy, but always in pairs. As long as changing parameters does not create additional zero-energy states, the value of the index is unchanged.  In order to address this point more precisely, let us consider a  Wess-Zumino model with
\beq 
W \ = \ \frac{m}{2} \Phi^2 \ - \ \frac{\lambda}{3} \Phi^3 \quad , \quad 
V \ = \ \left| m z \ - \  \lambda z^2\right|^2 \ .  \label{windex5}
\eeq 
For $\lambda \neq 0$, the scalar potential has two zero-energy states
\beq 
z = 0 \quad , \quad z = \frac{m}{\lambda} \ .  \label{windex52}
\eeq 
The Witten index equals two and supersymmetry is therefore unbroken. However, the limit $\lambda \to 0$ is special, since the second vacuum in (\ref{windex5}) is sent to infinity and therefore disappears. In this limit, the index jumps from $2$ to $1$. One can therefore refine the statement that the index is invariant under continuous deformations of the parameters that do not affect the asymptotic behavior of the scalar potential. In fact, the potential for $\lambda = 0$ is different for large field values compared to the case of a generic $\lambda$. 

Until now, we have implicitly assumed that only massive particles be present.  The case of massless particles is also subtle.  
If the mass parameter is allowed by the symmetries, one can compute $Tr (-1)^F$ for nonzero mass and then take the massless limit at the end of the computation. However, if the mass is forbidden by a symmetry, a separate analysis must be performed. This is in particular the case for gauge theories.  

Very often, in order to have a discrete spectrum, it is easier to consider the system in a finite volume (with boundary conditions for Bose and Fermi fields that are compatible with supersymmetry), and take the limit of infinite volume at the end of the computation. If supersymmetry is unbroken in a finite volume $E(V)=0$, taking the infinite-volume limit will not affect the result, so supersymmetry remains unbroken. On the other hand, if supersymmetry is broken in a finite volume, the energy can become zero in the infinite volume limit, restoring supersymmetry.  This clearly implies that with this method it is easier to prove that a theory has a supersymmetric ground state, but it is much harder to prove that a theory breaks supersymmetry. 

%%%%%%%%%%%%%%%%%%%%%%%%%%%%%%%%%%%%%%%%%%%%%%%%%%%%%%%%%%%%%%%%%%%%
\subsection{\sc Soft Breaking Terms} \label{sec:softerms}

At energies below the scale of supersymmetry breaking, and in the decoupling limit of hidden--sector interactions, the observable sector is described by a renormalizable theory, up to
relics of the transmission of supersymmetry breaking, which are usually called ``soft breaking terms''~\cite{girardello-grisaru}.  The effective Lagrangian can be captured by a ``spurion'' analysis,
to which we now turn.

One can introduce two types of spurions, chiral and vector ones, $\varphi_a$ and $V_I$, whose superfield components
\be
\varphi_a \ = \  v_a \ - \ \theta^2 F_a \ , \qquad V_I \ = \ - \ \frac{1}{2} \,\theta^2 \,{\bar \theta}^2\, D_I \ ,  \label{soft1}
\ee
only contain vacuum values compatible with the Lorentz symmetry, together with auxiliary components that, as usual, have mass dimension two.

In the low-energy Lagrangian, one retains the couplings to the hidden sector while also assuming that its dynamics generates somehow the auxiliary components in (\ref{soft1}).  The vacuum expectation values and the auxiliary fields in (\ref{soft1}) should be regarded as low--energy relics of some unspecified high--energy dynamics.  As we have anticipated, this reflects the idea that the supersymmetry breaking sector involves very massive fields, which can be integrated out at low energies.
Naively, such a massive sector should completely decouple, leaving no signs at low energies. However, the decoupling is not complete, and in a well-defined low-energy limit that we are about to discuss, the breaking of supersymmetry can be perceived in the observable sector.

The couplings between observable and hidden sectors are of the form
\bea
{\cal L}_{\rm eff} &=& \int d^4 \theta \  \overline{\Phi}{}^{{i}} \ e^{2  V + 2 q_i^I V_I} \Phi^i    \ + \ \left\{ \int d^2 \theta \left[ \frac{1}{4} \, f \ \mathrm{Tr} \left(W^{\alpha} W_{\alpha}\right) \right.
\right.
\nonumber \\
&+& \left. \left. {\lambda}_i \! \left(\frac{\varphi_a}{M}\right)  \Phi^i\, +\, \frac{1}{2} \,{\mu}_{ij} \!\left(\frac{\varphi_a}{M}\right)  \Phi^i \Phi^j\,+\, \frac{1}{3} \,{\lambda}_{ijk} \!\left(\frac{\varphi_a}{M}\right) \Phi^i \Phi^j \Phi^k \right] + {\rm h.c.} \right\}
 \label{soft2}
\eea
where, for simplicity, the kinetic functions $f$ are diagonal and the kinetic terms of the scalar fields are canonical. This Lagrangian
includes non-renormalizable interactions with the hidden sector parametrized by the three sets of functions ${\lambda}{}_i\left(\frac{\varphi_a}{M}\right)$, ${\mu}{}_{ij}\left(\frac{\varphi_a}{M}\right)$ and ${\lambda}{}_{ijk}\left(\frac{\varphi_a}{M}\right)$ (not to be confused with derivatives), which depend on the ratio $\frac{\varphi_a}{M}$ between the spurions $\varphi_a$  and a high scale $M$, together with $D$--term contributions associated to the spurions $V_I$, to which the fields couple with charges $q_i{}^I$, in addition to the standard gauge interactions. The non--renormalizable interactions can originate from
supergravity, from tree-level exchanges of massive fields, or from standard field theory loops.
The low--energy limit of interest obtains letting $M \to \infty$,
while maintaining the quantities $\left(\frac{F_a}{M}\right)^2$ and $D_I$ at fixed values of order $m_{soft}^2$.

Here $M$ is a typical mass scale of the hidden sector that breaks supersymmetry. The preceding limit is consistent if $F_a, D_I \ll M^2$, so that $m_{soft}$, which defines the superpartner masses, can be a low-energy scale, possibly lying in the (multi) TeV range for low-energy supersymmetric models.  Strictly speaking,  $\left(\frac{F_a}{M}\right)^2$ and  $D_I$ could be different mass scales,
giving rise to some type of hierarchy among the superpartners~\cite{bd,split1,split2}, but the simplest scenarios rest on a single mass scale governing all soft--breaking terms.

At energies below the heavy mass scale $M$, the effective Lagrangian is thus
\be
{\cal L} \ = \ {\cal L}_{\rm SUSY} \left(\Phi^i, \overline{\Phi}{}^{{j}},V\right) \ + \  {\cal L}_{\rm susy-breaking} \left(z^i,\bar{z}^{i},\lambda,\bar{\lambda}\right)  \ , \label{soft3}
\ee
where in the couplings of the supersymmetric portion,
\bea
{\cal L}_{\rm SUSY} &=&  \int d^4 \theta \  \overline{\Phi}{}^{{i}}\ e^{2 {g} V } \Phi^i   \  + \ \left( \int d^2 \theta \left[ \frac{1}{4} \mathrm{Tr} \left(W^{\alpha} W_{\alpha}\right) \right. \right.
\nonumber \\
&+& \left. \left. {\lambda}_i
\left(\frac{v_a}{M}\right)  \Phi^i \ + \ \frac{1}{2}\, {\mu}_{ij}\!\left(\frac{v_a}{M}\right)  \Phi^i \Phi^j \ +\  \frac{1}{3} {\lambda}_{ijk}\!\left(\frac{v_a}{M}\right) \Phi^i \Phi^j \Phi^k\  +\  {\rm h.c.} \right] \right)
\ , \label{soft4}
\eea
the spurions are replaced with their lowest components, and 
\beq
\frac{1}{g^2} \ = \ \mathrm{Re} \,f\!\left(\frac{v_a}{M}\right)   \ . \label{soft5}
\eeq

The soft--breaking portion of the Lagrangian originates from terms involving the auxiliary spurion components, and reads
\be
 {\cal L}_{\rm soft} \ = \ -\ m_i^2 |z^i|^2 \ - \ \left[ M_{1/2}\, tr (\lambda \lambda) + A_i z^i + \frac{1}{2} b_{ij} z^i z^j +   \frac{1}{3} A_{ijk} z^i z^j z^k  + {\rm h.c.} \right] \ . \label{soft6}
\ee
Here the $z^i$ are the scalar fields belonging to the chiral multiplets, and
\bea
&&m_i^2 =  q_i^I D_I   \ , \ \ M_{1/2} = \frac{{\partial_a f}|_{v_a}}{4 \,\mathrm {Re} (f)}  \frac{F_a}{M} \ , \ \
A_i =   { \partial_a  { \lambda}_i}|_{v_a} \frac{F_a}{M} \ , \  \nonumber \\
&& b_{ij} = 
{\partial_a {\mu}_{ij}}|_{v_a}   \frac{F_a}{M}  \ , \qquad  A_{ijk} = 
{\partial_a {\lambda}_{ijk}}|_{v_a}   \frac{F_a}{M} \ , \label{soft7}
 \eea
where the derivatives of the functions $f,{\lambda}_i, {\mu}_{ij}, {\lambda}_{ijk}$ are taken with respect to their arguments.  The $F^a$ and $D_I$ are free parameters in the final Lagrangian.

The contributions proportional to $m_i^2, A_i, b_{ij}, A_{ijk}, M_{1/2}$ are called ``soft supersymmetry breaking terms'', or briefly ``soft terms''. They are the most general set of explicit
supersymmetry breaking terms that, when added to the supersymmetric Lagrangian (\ref{soft4}),  do not introduce quadratic divergences in quantum corrections \cite{girardello-grisaru}. The heuristic reason for this benign ultraviolet behavior is that they originate from a supersymmetric microscopic Lagrangian, and in general spontaneous breakings do not affect the main ultraviolet properties of the parent theory. Note that $m_i^2$ are non--holomorphic scalar masses, $M_{1/2}$ are Majorana gaugino masses, while
$ A_i, B_{ij}, A_{ijk}$ are holomorphic couplings of the same structure as the corresponding terms in the superpotential.

Interestingly, supersymmetry-breaking masses for the fermions of chiral multiplets are not soft. The reason is that fermion masses in chiral multiplets are determined by parameters
in the superpotential that also appear in Yukawa and trilinear scalar interactions. Changes in fermion masses alone would affect these relations, and thus the cancellation of quadratic divergences.
Similarly, supersymmetry-breaking terms of the type $z^i z^j {\bar z}{}^k$ or $z^i z^j  {\bar z}{}^k {\bar z}{}^l$ are also hard, and generically introduce quadratic divergences.

When added to a supersymmetric Lagrangian, soft terms can lead to realistic spectra. The non-holomorphic scalar mass terms $m_i^2$ can lift the masses of squarks and sleptons, the super-partners of
quarks and leptons, while Majorana gaugino masses can lift the masses of the fermionic superpartners (photino, gluinos and ``electroweakinos'', supersymmetric partners of electroweak fields) above those of the gauge fields.

If the $z^i$ vanish in the vacuum, the contributions to the supertrace originate solely from the soft--breaking terms, and
\beq
\mathrm{Str} {\cal M}^2 \ = \ 2 \, \sum_i \ q_i^I\,D_I \ - \ 2\, \mathrm{dim}(G) \left| \frac{{\partial_a f}|_{v_a}}{4 \,\mathrm {Re} (f)}  \frac{F_a}{M}  \right|^2\ .
\eeq

\section{\sc The Standard Model} \label{sec:sm}

We now leave supersymmetry momentarily aside and turn to a quick overview of the Standard Model, in order to highlight some related puzzles that supersymmetry might help to overcome. The Standard model relies on gauge symmetry and its spontaneous breaking to grant common grounds to the weak and electromagnetic interactions, and also to the strong interactions of quarks and gluons. Let us begin our discussion with an overview of some of the main steps that led to it~\footnote{A recent historical review by one of the founding fathers can found in~\cite{iliopoulos25}.}.

\subsection{\sc A Brief Historical Note}

Before 1961, QED had already experienced a remarkable success~(for a collection of the original papers, see~\cite{QED}) as a reliable computational scheme for subtle corrections to atomic spectra (the Lamb shift) and to intrinsic properties of electrons and positrons (the anomalous magnetic moment). Strong nuclear interactions were provisionally accounted for in terms of a (strongly coupled) quantum field theory \`a la
Yukawa, with baryons (including nucleons) as matter fields and mesons (including pions) as mediators. With the discovery of more and more baryons and mesons, this theory seemed to need more and more fields and couplings and became more and more baroque. On the other hand, weak interactions were described by the
Fermi theory~\cite{fermi}, and the known weak interactions could be accounted for at low energies in terms of a universal coupling $G_F$, up to the introduction of a new parameter, the Cabibbo angle~\cite{cabibbo}. However, it was clearly a non--renormalizable theory whose amplitudes violated unitarity beyond energies of order $E \sim 1/\sqrt{G_F} \simeq 100$ GeV.

In 1961 Glashow~\cite{glashow} proposed that weak and electromagnetic interactions could originate from
a Yang-Mills theory with gauge group $SU(2)_L \times U(1)_Y$, and that gauge--boson masses (which were added by hand at the time) could render the weak interactions properly of short range.
However, the photon remained massless, so that electromagnetism could be a long--range interaction. Glashow's model also predicted the existence of a neutral current, in addition to the charged currents that were already present in the Fermi
theory.  The Standard Model was completed in 1967-68 by Weinberg~\cite{weinberg} and Salam~\cite{salam}, who incorporated the Brout--Englert--Higgs~\cite{beh1,beh2,beh3,gurhagkib} mechanism, showing how three of the four gauge fields (called $W^{\pm}$ and $Z$ later on) could get masses, via the spontaneous breaking around a ground state respecting the electromagnetic gauge symmetry.
They also introduced in the theory lepton fields with appropriate quantum numbers,  showing that spontaneous symmetry breaking could yield proper mass patterns for the charged leptons via Yukawa couplings, although gauge invariance would not allow Dirac mass terms to begin with.  Quarks, introduced by Gell-Mann and Zweig in 1964~\cite{quarks1,quarks2},  were incorporated into the theory in 1970 by Glashow, Iliopoulos and Maiani~\cite{gim}. They also showed that in addition to the up, down and strange quarks, a fourth one, the charm, was needed to this end. In a series of papers in 1970-1972, 't Hooft and Veltman~\cite{TV1,TV2,TV3,TV4} finally showed that Yang-Mills theories, with or without spontaneous symmetry
breaking, are renormalizable (for a detailed discussion of renormalizability, see~\cite{ren_rev}). Bouchiat, Iliopoulos and Meyer~\cite{BIM} and Gross and Jackiw~\cite{GJACK} then showed that the Standard Model has no quantum gauge anomalies~\cite{ABJ1,ABJ2,ABJ3,wzcc}, a key feature granting its consistency.
In 1973, Koyabashi and Maskawa~\cite{ckm} showed that, in the presence of a third generation of quarks and leptons, a Cabibbo--like~\cite{cabibbo} mixing would allow to parametrize the violation of CP in hadronic
charged current interactions of quarks that was revealed in the 1960's~\cite{croninfitch}, while Gross and Wilczek~\cite{gw}, and Politzer~\cite{politzer}, showed that (massless) non--Abelian gauge theories imply that the interactions between quarks become weaker at high-energies, a key property commonly referred to as ``asymptotic freedom''.  An $SU(3)_c$  gauge theory of strong interactions was introduced  by  Gell-Mann, Fritzsch and  Leutwyler~\cite{QCD1,QCD2} shortly thereafter, while the CERN Gargamelle bubble chamber~\cite{gargamelle} presented the first direct evidence for the weak neutral current.  The intermediate vector bosons were discovered at CERN in 1983~\cite{ivb1,ivb2,ivb3,ivb4}, and finally the last building block of the Standard Model, the Higgs scalar,  was discovered at the CERN LHC in 2012 \cite{higgs1,higgs2}. All these contributions, and many others, have granted the Standard Model its present status of a highly successful low--energy description of strong, weak, and electromagnetic interactions.

\subsection{\sc Gauge Group and Matter Content}

\begin{figure}[ht]
\begin{center}
    \includegraphics[width=2in]{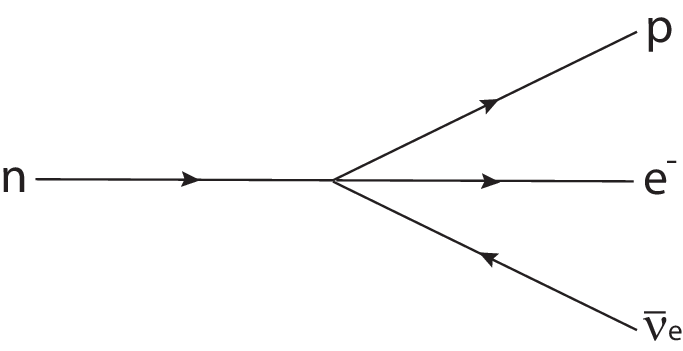}
\end{center}
\caption{Fermi theory of weak interactions: the beta decay $n \rightarrow p e^- \,{\bar \nu}_e$ at low energies $E << M_W$ can
be described via an effective four-fermion interaction, in line with Fermi's initial proposal. Here the arrows are entering (outgoing) for incoming particles (antiparticles) and outgoing (entering) for outgoing particles (antiparticles).}
\label{fig:fermi}
\end{figure}
 The Standard Model (SM) gauge group is
\beq
G \ = \  SU(3)_c  \ \times SU(2)_L \ \times U(1)_Y \ , \nonumber
\eeq
and the corresponding gauge bosons are the eight gluon fields $G_\mu^A$, valued in the adjoint of $SU(3)$, the three weak bosons $A_\mu^a$, valued in the adjoint of $SU(2)$, and $B_\mu$, the $U(1)_Y$ gauge boson.
In addition, there are three generations of quarks and leptons $(i=1,2,3)$,

Quarks : \ $q_i = \ \begin{pmatrix} u_i \\ d_i
\end{pmatrix} _L  \ : \ ({\bf 3,2})_{Y=\frac{1}{3}} \ , \  u_{iR} \ : \ ({\bf 3,1})_{Y=\frac{4}{3}} \ , \qquad d_{iR} \ : \ ({\bf 3,1})_{Y=-\,\frac{2}{3}} $

Leptons : \ $l_i = \ \begin{pmatrix} \nu_i \\ e_i
\end{pmatrix} _L  \ : \ ({\bf 1,2})_{Y=-1} \quad , \quad e_{iR} \ : \ ({\bf 1,1})_{Y=-2}$

and a single scalar doublet

Higgs : \ $ \Phi = \begin{pmatrix} \Phi^{+} \\ \Phi^0
\end{pmatrix}  : \quad ({\bf 1,2})_{Y=1} \ , $

\noindent where the boldface entries summarize the $SU(3)$ and $SU(2)$ representations of the matter fields.

In the Standard Model, a vacuum value of the {\it Higgs doublet} breaks the electroweak gauge symmetry $SU(2)_L \times U(1)_Y$ to the electromagnetic one $U(1)_Q$. With our conventions, the electric charge is related to the third
component of the weak isospin $T_3$ and to the hypercharge $Y$ according to
\beq
Q \ = \ T_3 \ + \ \frac{Y}{2} \ . \label{sm05}
\eeq
Note that only {\it left-handed}\ quarks and leptons interact with the $SU(2)_L$  gauge fields, and so parity is maximally violated.

One can distinguish three portions in the Lagrangian. The first is
\beq
{ \cal L}_{SM} \ = \ {\cal L}_{\rm kin} \ - \ V(\Phi) \ + \ {\cal L}_{\rm Yuk}  \ , \label{sm01}
\eeq
where
\bea
{\cal L}_{\rm kin} &=& - \ \frac{1}{4} (G_{\mu\nu}^A)^2 \ - \ \frac{1}{4} (F_{\mu\nu}^a)^2\ - \ \frac{1}{4} B_{\mu\nu}^2 \ - \ |D_\mu \Phi|^2 \nonumber \\
&+& {\bar \Psi}_L i \gamma^\mu D_\mu \Psi_L \ +  \ {\bar \Psi}_R i \gamma^\mu D_\mu \Psi_R  \ . \label{sm02}
\eea
Here we are using the four--component notation, so that for any Dirac spinor $\Psi$
\beq
\Psi_{L} \ = \ \frac{1 \ - \ \gamma_5}{2} \ \Psi \ , \qquad \Psi_{R} \ = \ \frac{1 \ + \ \gamma_5}{2} \ \Psi \ ,
\eeq
are the corresponding Weyl projections, and
\bea
&& G_{\mu\nu}^A \ = \ \partial_\mu\, A_\nu^A \ - \ \partial_\nu\, A_\mu^A \ + \ g_s \, f^{ABC} \, A_\mu^B\, A_\nu^C \ , \nonumber \\
&& F_{\mu\nu}^a \ = \ \partial_\mu\, A_\nu^a \ - \ \partial_\nu\, A_\mu^a \ + \ g \, \epsilon^{abc} \, A_\mu^b\, A_\nu^c \ , \nonumber \\
&& B_{\mu\nu} \ = \ \partial_\mu\, B_\nu \ - \ \partial_\nu\, B_\mu \ .
\eea
Moreover
\bea
&& D_\mu \Psi_L \ = \ \left(\partial_\mu \ - \ i g_s \frac{\lambda_A}{2} A_\mu^A \ - \ i g \frac{\tau_a}{2} A_\mu^a \ -\  i g' \frac{Y_L}{2} B_\mu \right) \Psi_L \nonumber \\
&& D_\mu \Psi_R \ = \ \left(\partial_\mu \ - \ i g_s \frac{\lambda_A}{2} A_\mu^A \ - \ i g' \frac{Y_R}{2} B_\mu \right) \Psi_R \ , \label{sm03}
\eea
where the $\lambda^A$ are $SU(3)$ Gell--Mann matrices and $(g_s,g,g')$ are the coupling constants associated to there factors $SU(3)_c$, $SU(2)_L$ and $U(1)_Y$. The leptons have no gluon couplings, and
the Higgs potential is
\beq
V (\Phi) \ = \ - \ \mu^2 \Phi^{\dagger} \Phi \ + \ \lambda (\Phi^{\dagger} \Phi)^2 \ . \label{sm04}
\eeq
We shall return to the Yukawa sector ${\cal L}_{\rm Yuk}$ after describing the effect of spontaneous symmetry breaking on the gauge fields.

%%%%%%%%%%%%%%%%%%%%%%%%%%%%%%%%%%%%%%%%%%%%%%%%%%%%%%%%%%%%%%%%%%%%%%%%%%%%%%%%%%%%%%%%%%%%%%%%%%%%%%%%%%%%%%
\subsection{\sc Weak Mixing Angles and Gauge Boson Masses}

After an $SO(4)$ rotation, the Higgs \emph{v.e.v.} can be cast in the form
\beq
\Phi \ = \ \begin{pmatrix} 0 \\ \frac{v}{\sqrt{2}}
\end{pmatrix}  \ , \label{ssb}
\eeq
where the value $v^2 = \frac{\mu^2}{\lambda} \ \simeq$ (246 {GeV})${}^2$ is deduced from experiments.

The gauge boson masses are induced by the kinetic term of the Higgs field, taking the vacuum value~\eqref{ssb} into account. The resulting quadratic terms are
\bea
|D_{\mu} \Phi|^2 &\longrightarrow& \frac{g^2 v^2}{8} |A_{\mu}^{(1)}-i A_{\mu}^{(2)}|^2 \ + \ \frac{v^2}{8} |g A_{\mu} ^{(3)}-g' B_{\mu}|^2 \nonumber \\
&=& \frac{g^2 v^2}{4} W^{+\,\mu} W_{\mu}^- \ + \ \frac{(g^2+{g'}^2) v^2}{8} Z^{\mu} Z_{\mu} \ ,
\eea
and consequently the different electroweak gauge bosons and the corresponding masses are
\bea
&& W_\mu^{\pm} \ = \ \frac{1}{\sqrt{2}} (A_\mu^{(1)} \ \mp \ i A_\mu^{(2)}) \quad , \quad M_W \ = \ \frac{gv}{2} \ , \nonumber \\
&& Z_\mu \ = \ \frac{g A_\mu^{(3)}\ -\ g' B_\mu}{\sqrt{g^2\ + \ {g'}^2}} \quad , \qquad M_Z \ =\ \frac{v}{2} \sqrt{g^2\ + \ {g'}^2} \ , \nonumber \\
&& A_\mu \ = \ \frac{g' A_\mu^{(3)}\ + \ g B_\mu}{\sqrt{g^2\ +\ {g'}^2}} \quad , \qquad M_A \ = \ 0 \ . \label{sm1}
\eea
The preceding results lead to define the {\it electroweak angle} $\theta_w$, such that
\beq
\cos \theta_w \ =  \ \frac{g}{\sqrt{g^2+{g'}^2}} \ = \ \frac{M_W}{M_Z} \ , \qquad  \tan \theta_w \ =\  \frac{g'}{g} \ , \label{sm2}
\eeq
which relates the mass eigenstates to the weak basis according to
\bea
\begin{pmatrix} Z_{\mu} \\ A_{\mu} \end{pmatrix} \ =\ \begin{pmatrix} \cos \theta_w & \ - \ \sin \theta_w \\ \sin \theta_w & \cos \theta_w
\end{pmatrix}   \begin{pmatrix} A_{\mu}^{(3)} \\ B_{\mu} \end{pmatrix}  \ , \label{sm3}
\eea
and allows one to link the electric charge $e$ to $g$ according to
\beq
e \ = \ g \sin \theta_w\ .
\eeq

Note the peculiar value of the ratio
\beq
\rho \ \equiv \ \frac{M_W^2}{M_Z^2 \cos^2 \theta_w} \ =\ 1  \ . \label{sm4}
\eeq
The $\rho$ parameter receives quantum corrections in the SM, which are dominated by the top-quark contribution, and any experimental deviation from its SM value is a possible hint for new physics. Conversely, any model of new physics must yield a $\rho$ parameter that is close to one: this feature underlies one of the {\it precision tests} of the Standard Model, and has excluded many proposals for possible extensions~\footnote{More precisely, the value is around 1.1, due to loop contributions dominated by the top quark, but the precision of measurements is at the level of one part in a thousand.}.

As we have anticipated, the W and Z gauge bosons were discovered in 1983 at CERN~\cite{ivb1,ivb2,ivb3,ivb4}, and their masses are 
\beq
M_W \ \simeq \ 80.4  \ \mathrm{GeV} \ , \qquad M_Z \ \simeq \ 91.2 \  \mathrm{GeV} \ ,  
\eeq
so that 
\beq
\sin^2\theta_w \  \simeq \ 0.22 \ .
\eeq

%%%%%%%%%%%%%%%%%%%%%%%%%%%%%%%%%%%%%%%%%%%%%%%%%%%%%%%%%%%%%%%%%%%%%%%%%%%%%%%%%%%%%%%%%%%%%%%%%
\subsection{\sc Neutral and Charged Currents}

The charged and neutral currents are the fermion bilinears that couple to the charged $W^\pm$ and neutral $Z$ weak gauge bosons.
They can be deduced expressing the covariant derivative in terms of the mass eigenstates, so that
\bea
&&D_\mu \, = \, \partial_\mu \, - \, i \,e\, Q A_\mu \, - \, \frac{i g}{2 \sqrt{2}} ( W_\mu^+ \tau_+ \,+\,  W_\mu^- \tau_- ) \, - \, \frac{i g}{\cos \theta_w} Z_\mu
(T_3 \,-\, \sin^2 \theta_w Q) \,.\label{cc1}
\eea
Starting from the fermionic kinetic terms one can thus obtain
\beq
{\cal L} \ = \  {\bar \Psi}_i i {\gamma^{\mu} \partial_{\mu}} \Psi_i \ + \  \frac{g}{\sqrt{2}} (W_{\mu}^+ J^{\mu}_{W^+} \,+\, W_{\mu}^- J^{\mu}_{W^-})  \ + \
\frac{g}{\cos \theta_w} Z_{\mu} J_Z^{\mu}  \ + \ e\, A_{\mu} J_{em}^{\mu} \label{cc2}
\eeq
from which one can identify the charged weak currents $J^{\mu}_{W^\pm}$
\beq
J^{\mu}_{W^+} \ = \  {\bar \nu}_L^i \gamma^\mu e_L^i + {\bar u}_L^i \gamma^\mu d_L^i  \ , \qquad
J^{\mu}_{W^-} \ = \ {\bar e}_L^i \gamma^\mu \nu_L^i + {\bar d}_L^i \gamma^\mu u_L^i  \ ,
\eeq
the electromagnetic current
\beq
J^{\mu}_{em} \ = \ - \ {\bar e}^i \gamma^\mu e^i \ + \ \frac{2}{3}\, {\bar u}^i \gamma^\mu u^i \ -\
\frac{1}{3} \,{\bar d}^i \gamma^\mu d^i  \ ,
\eeq
and the weak neutral current
\bea
J^{\mu}_Z &=& J^\mu_3 \ - \ \sin^2 \theta_w \,J^{\mu}_{em} \ = \  \frac{1}{2} {\bar \nu}_L^i \gamma^\mu \nu_L^i \ - \
\left(\frac{1}{2} \,-\, \sin^2 \theta_w\right) {\bar e}_L^i \gamma^\mu e_L^i \ + \ \sin^2 \theta_w \,{\bar e}_R^i \gamma^\mu e_R^i  \nonumber \\
&+& \left(\frac{1}{2} \, - \, \frac{2}{3} \sin^2 \theta_w\right) {\bar u}_L^i \gamma^\mu u_L^i \ - \ \frac{2}{3} \sin^2 \theta_w \,{\bar u}_R^i \gamma^\mu u_R^i
 \nonumber \\
&-&  \left(\frac{1}{2} \,-\,\frac{1}{3} \sin^2 \theta_w\right) {\bar d}_L^i \gamma^\mu d_L^i \ + \ \frac{1}{3} \sin^2 \theta_w \,{\bar d}_R^i \gamma^\mu d_R^i \nonumber \\
& =& \frac{1}{2} \sum_i {\bar \Psi}_i \gamma^\mu \left(g_V^i \,-\, g_A^i \gamma_5\right) \Psi_i \ \quad .
\quad  \label{cc3}
\eea
Here the $\Psi_i$ denote collectively the quarks and leptons of the Standard Model, and
\be
g_V^i \ = \ I_3^i \ - \ 2\, Q_i \,\sin^2 \,\theta_W \ , \qquad g_A^i \ = \ I_3^i \  \label{cc03}
\ee
are the vector and axial fermionic couplings to the $Z$ boson.
At low energies $E << M_W$ or $M_Z$, the exchange of $W$ and $Z$ bosons recovers the charged current Fermi interaction, together with a similar neutral current interaction
\be
{\cal L}_F \ = \ - \ 2 \,\sqrt{2} \,G_F \left[ J^{\mu}_{W^+}  J_{\mu \,W^-} \ + \ \rho\,
 J^{\mu}_Z \ J_{\mu \,Z} \right] \ , \label{cc4}
\ee
where $\rho$ was defined in eq.~\eqref{sm4}.

Weak interaction experiments led to a precise determination of the electroweak angle. When quantum corrections are taken into account, it acquires an energy dependence, but at energy scales close to the Z mass the corrected value is
\beq
\sin^2 \theta_w \ \simeq \ 0.223 \ .
\eeq
%%%%%%%%%%%%%%%%%%%%%%%%%%%%%%%%%%%%%%%%%%%%%%%%%%%%%%%%%%%%%%%%%%%%%%%%%%%%%%%%%%%%%%%%%%%%%%%%%%%%%%%%%%%%%%%%%%%%

\subsection{\sc Fermion Masses and the CKM Matrix}

Dirac mass terms are not allowed in the SM, since they would not be gauge invariant, because of the chiral nature of the electroweak interactions. However,  using the Higgs field, one can write the Yukawa--like interactions
\beq
{\cal L}_{\rm Yuk} \ = \  - \ h_{ij}^{u} \,{\bar q}^i_L u_R^j {\Phi}^c \ -  \ h_{ij}^{d} \,{\bar q}^i_L d_R^j \Phi \ -
 \ h_{ij}^{e} \,{\bar l}^i_L e_R^j \Phi   \ + \ \mathrm{h.c.}\ , \label{ckm1}
\eeq
where
\beq
{\Phi} \ = \ \begin{pmatrix} {\Phi^+} \\ \Phi^0
\end{pmatrix} \ , \qquad {\Phi}^c \ = \ \begin{pmatrix} {\overline \Phi}{}^0 \\ -\, {\overline \Phi}{}^+
\end{pmatrix}
\eeq
are the Higgs doublet and its charge-conjugate, and $i,j=1,2,3$ are flavor indices.

The electroweak symmetry breaking induced by the vacuum value~\eqref{ssb} generates quark and lepton masses via the preceding Yukawa couplings, according to
\beq
{\cal L}_{\rm mass} \ = \ - \ m_{ij}^u \,{\bar u}^i_L u_R^j \ - \  m_{ij}^d\, {\bar d}^i_L d_R^j \ - \  m_{ij}^e\, {\bar e}^i_L e_R^j + {\rm c.c.} \ ,
\label{ckm2}
\eeq
in terms of the three mass matrices $m^u$, $m^d$ and $m^l$
\beq
m_{ij}^{u} \ = \ \frac{v}{\sqrt{2}}\ h_{ij}^{u}  \ , \qquad m_{ij}^{d} \ = \ \frac{v}{\sqrt{2}}\ h_{ij}^{d}  \ , \qquad m_{ij}^{e} \ = \ \frac{v}{\sqrt{2}}\ h_{ij}^{e}  \ ,
\eeq
which have no prescribed symmetries. 

In matrix notation
\beq
{\cal L}_{\rm mass} \ = \ - \ {\bar u}_L m^u  u_R \ - \  {\bar d}_L m^d d_R \ - \  {\bar e}_L m^e e_R \ +  \ {\rm c.c.} \ , \label{ckm3}
\eeq 
and one can turn to the {\it mass eigenstate basis}  with the help of pairs of $3 \times 3$ unitary transformations $V_{L,R}^{u,d,e}$~\footnote{These transformations are not innocent: there are quantum anomalies that play a role in connection with the strong--CP problem~\cite{strong_CP} and the solution proposed in~\cite{PQ1,PQ2}. }, according to
\beq
u_{L,R} \ = \ V_{L,R}^u \,u'_{L,R} \ , \qquad d_{L,R} \ = \ V_{L,R}^d \,d'_{L,R} \ , \qquad e_{L,R} \ = \ V_{L,R}^e \,e'_{L,R} \ ,  \label{ckm9}
\eeq
so that
\bea
(V_L^u)^{\dagger} m^u V_R^u &=& \mathrm{diag}\left(m_u,m_c,m_t\right) \ ,  \nonumber \\
(V_L^d)^{\dagger} m^d V_R^d &=& \mathrm{diag}\left(m_d,m_s,m_b\right) \ ,  \nonumber \\
(V_L^e)^{\dagger} m^e V_R^e &=& \mathrm{diag}\left(m_e,m_\mu,m_\tau\right)
\ , \label{ckm10}
\eea
where the need for two different matrices $V_L$ and $V_R$, due to the generic form of the mass matrices, implies that the diagonal elements are generally complex.

When turning to the mass basis, the neutral and e.m. currents remain the same. Neutrinos are essentially massless, so that one is free to perform the same unitary transformation on them and on their $SU(2)_L$ charged lepton partners, defining $\nu_{L} \ = \ V_{L}^e \nu'_{L}$.
In the new basis, the Standard Model Lagrangian preserves the lepton numbers of the individual species $L_e$, $L_{\mu}$ and $L_{\tau}$. These conservation laws are indeed observed experimentally with great accuracy: for example, no transition of type $\mu \to e \gamma$ was observed until now.
However, the story is more intricate for the quark sector. When expressed in terms of mass eigenstates, the hadronic charged current becomes
\beq
(J^{\mu}_{W^+})_{\rm quarks} \ = \ \frac{1}{\sqrt{2}}\, {\bar u}'_L \gamma^{\mu} V d'_L \ \equiv \
\frac{1}{\sqrt{2}} \, {\bar u}'_L \gamma^{\mu} {\hat d}_L \ , \label{ckm11}
\eeq
where 
\beq
V \ = \ (V_L^u)^{\dagger}\,  V_{L}^d
\eeq
is the unitary  Cabibbo--Kobayashi--Maskawa (CKM) matrix \cite{cabibbo,ckm}.
In detail
\bea
{\hat d}_L  \ = \ V d'_L \ , \quad \mathrm{where} \quad
\begin{pmatrix}  {\hat d}_L \\ {\hat s}_L \\ {\hat b}_L \end{pmatrix}
\ = \ \begin{pmatrix}  V_{ud} & V_{us} & V_{ub} \\ V_{cd} & V_{cs} & V_{cb} \\
V_{td} & V_{ts} & V_{tb} \end{pmatrix}
\begin{pmatrix}  d'_L \\ s'_L \\ b'_L \end{pmatrix}  , \label{ckm12}
\eea
so that there is \emph{flavor--changing charged--current transitions} in the standard model, as, for example, $s \rightarrow u W^-$. Experimental measurements suggest for $V$ a {\it hierarchical structure} of the type
\beq
\begin{pmatrix}  1 - \frac{\lambda^2}{2} & \lambda & A \lambda^3 (\rho- i \eta) \\ - \lambda & 1 - \frac{\lambda^2}{2} & A \lambda^2 \\
A \lambda^3 (1-\rho- i \eta) & - A \lambda^2 & 1 \end{pmatrix} \ , \label{ckm13}
\eeq
with $\lambda = \sin \theta_c \simeq 0.22$ the Cabibbo angle, which is usually referred to as Wolfenstein parametrization~\cite{wolfenstein}. $A$, $\rho$ and $\eta$ are parameters of order one, which include a phase. This should be regarded as the leading contribution to an expansion in $\lambda$, which is compatible with unitarity up to the order $\lambda^4$. Cabibbo introduced in 1962 a parameter that, in modern language, determines the upper $2 \times 2$ portion of this matrix~\cite{cabibbo} in the form
\bea
\begin{pmatrix}  \cos \theta_c & \sin \theta_c \\ -\sin \theta_c & \cos \theta_c  \end{pmatrix} \ . \label{ckm14}
\eea

One can verify that, after field redefinitions,   $V$ contains {\it three rotation angles} and a {\it CP violating phase}\footnote{With $N$ generations of quarks and leptons, a simple counting would reveal the presence of $N(N-1)/2$ rotation angles and $(N-1)(N-2)/2$ CP
violating phases. A fourth generation of quarks and leptons is now essentially excluded, but if it existed one would expect new sources of CP violation, together with apparent violations of unitarity for the three--family CKM matrix, which has long been the object of detailed scrutiny.}.
The fact that the CKM matrix includes a CP violating phase is transparent from the terms in the Lagrangian that couple the charged current to the W gauge bosons. In vector/matrix notation, these terms transform under CP according to
\beq
{\bar u}_L \gamma^\mu V d_L W_\mu^+ \ + \ {\bar d}_L \gamma^\mu V^{\dagger} u_L W_\mu^- \ \longrightarrow \ {\bar u}_L \gamma^\mu V^{*} d_L W_\mu^+ \ + \ {\bar d}_L \gamma^m V^t u_L W_m^-  \ , \label{ckm014}
\eeq
and therefore CP is violated if the matrix has complex entries, which can be the case with three generations.
Note also that CP violation in the Standard Model is {\it suppressed} by $\lambda^3$ in the parametrization~\eqref{ckm13} of $V$.

The {\it unitarity} of the CKM matrix
\beq
V_{ik} V_{jk}^* \ = \ \delta_{ij} \quad , \quad V_{ki}^* V_{kj} \ = \ \delta_{ij} \label{ckm15}
\eeq
has important experimental consequences. One of them is {\it the GIM mechanism}~\cite{gim}, to which we shall soon return.

%%%%%%%%%%%%%%%%%%%%%%%%%%%%%%%%%%%%%%%%%%%%%%%%%%%%%%%%%%%%%%%
\subsection{\sc Higgs Couplings}

The minimal option for the Higgs mechanism  relies on a single doublet $\Phi$, as above.
Starting from the Standard Model Lagrangian (\ref{sm01}), one can work out the Higgs boson couplings to fermions and gauge fields. In the unitary gauge, where
\begin{equation}
 \Phi = \begin{pmatrix} 0 \\ \frac{v+h}{\sqrt{2}}
\end{pmatrix} \ , \label{higgs1}
\end{equation}
one thus finds the masses and couplings in
\begin{eqnarray}
{\cal L}_{\rm Higgs} &=& - M_W^2 \left(1 + \frac{h}{v}\right)^2 W_{\mu}^+  W^{\mu\,-} \ - \
 M_Z^2 \left(1 + \frac{h}{v}\right)^2 Z_{\mu}  Z^{\mu} \ - \ \frac{1}{2} \,m_h^2 \left(1 + \frac{h}{2v}\right)^2
 h^2  \nonumber \\
&-&  \left[\left(1 + \frac{h}{v}\right) {\bar u}_L  m^u u_R \ + \ \left(1 + \frac{h}{v}\right) {\bar d}_L  m^d d_R
\ + \  \left(1 + \frac{h}{v}\right) {\bar e}_L m^e e_R \ + \ {\rm h.c.} \right]  \, . \label{higgs02}
\end{eqnarray}
A key test of the Standard Model one--Higgs--doublet setup is therefore the proportionality between the couplings of the Higgs boson and the masses of the particles it interacts with. This proportionality would cease to hold if the Higgs sector involved more doublets and/or other representations.

Note also that the transformations that diagonalize the fermion mass matrices have the same effect on the Higgs couplings to fermions. No flavor transitions
are thus mediated in the presence of the minimal Higgs doublet, which is very welcome in light of the tight
constraints from flavor changing neutral current processes. However, if for example two Higgs doublets
$\Phi_1,\Phi_2$ coupled to the same types of quarks or leptons,  the Yukawa coupling would have the more general form,
\beq
{\cal L'}_{\rm Yuk} \ = \  - \ {\bar q}^i_L u_R^j
\left(h_{ij}^{1,u} {\Phi}_1 + h_{ij}^{2,u} {\Phi}_2\right)  \ .  \label{ckm01}
\eeq
It would then be generically impossible to diagonalize simultaneously the fermion mass matrices
and the fermion couplings to the Higgs scalars~\cite{GlashWei}.
In the simplest multi-Higgs extensions of the Standard Model with no Higgs-induced flavor changing neutral current (FCNC) effects, the three generations of the same type of quarks (or leptons) couple to just one Higgs doublet, which could be enforced by discrete symmetries. For example, two-Higgs doublet models should contain Yukawa couplings of the type
\beq
{\cal L}_{\rm Yuk} \ = \  - \ h_{ij}^{u} {\bar q}^i_L u_R^j {H}_2 \ - \  h_{ij}^{d} {\bar q}^i_L d_R^j H_1 \ -
 \ h_{ij}^{e} {\bar l}^i_L e_R^j H_1 \ , \label{ckm02}
\eeq
where $H_2$ and $H_1$ play the role of ${\Phi}^c$ and $\Phi$, in such a way that the mass of each quark or lepton flavor originates from a single Higgs field.

%%%%%%%%%%%%%%%%%%%%%%%%%%%%%%%%%%%%%%%%%%%%%%%%%%%%%%%%%%%%%%%%%%%%%%%%%%%%%%%%%%%%%%%%%%%%%%%%%%%%%%%%%%%
\subsection{\sc The GIM Mechanism}

If neutrino masses are neglected, FCNC processes in the Standard Model can only be induced by loop diagrams where a quark line of a given charge and flavor turns into another quark line of the same charge but of a different flavor. FCNC effects were widely investigated experimentally in the 1960's and turned out to be highly suppressed. This result was puzzling for a while, in view of the quark spectrum known at the time. Interesting examples of this type involve neutral mesons such as the $K^0-{\overline K}{}^{\,0}$ system.
Taking the pseudoscalar nature of these particles into account, the CP  transformations act as $CP  |  K^0  \rangle  = -  |  \overline{K^0}  \rangle  $ ,   $CP  |  \overline{K^0}  \rangle   = -  |  K^0  \rangle $, so that the CP eigenstates are
\bea
&&  |  K_1^0  \rangle = \frac{1}{\sqrt 2}  \left(   |  K^0  \rangle  \ - \  |  \overline{K^0}  \rangle   \right) \ , \qquad CP    |  K_1^0  \rangle  =  |  K_1^0  \rangle  \ , \nonumber \\
&&   |  K_2^0  \rangle = \frac{1}{\sqrt 2}  \left(   |  K^0  \rangle \  + \  |  \overline{K^0}  \rangle   \right) \ , \qquad CP    |  K_2^0  \rangle  =  - |  K_2^0  \rangle   \  .     \label{K01}
\eea

If CP were an exact symmetry of the weak interactions, the only possible decay modes would be $ K_1^0 \to \pi \pi$,  $ K_2^0 \to \pi \pi \pi$.  Due to phase-space suppression, one would therefore have a relatively long-lived $K_L$ meson decaying into
three pions and a short-lived one $K_S$ decaying into two pions.  In fact, even CP is slightly violated in weak interactions, and $K_S$ can also decay into three pions after a long time. The true 
eigenstates of the Hamiltonian for the neutral kaon system are actually modified by terms involving a small parameter $\epsilon$, according to
\bea
&&  |  K_S  \rangle \ \simeq \   \frac{1}{\sqrt {2} }\left[  (1+\epsilon)  |  K^0  \rangle  \ - \  (1- \epsilon)  |  \overline{K^0}  \rangle  \right]    \ , \nonumber \\
&&    |  K_L  \rangle \ \simeq \  \frac{1}{\sqrt{ 2 }} \left[  (1+\epsilon)  |  K^0  \rangle  \ + \  (1- \epsilon)  |  \overline{K^0}  \rangle  \right]     \  .     \label{K02}
\eea
Note that these states are not orthogonal, since the effective Hamiltonian for them is not Hermitian, but CPT demands that
\beq
\langle K^0| H |  K^0 \rangle \ = \ \langle \overline{K^0}  | H |  \overline{K^0}  \rangle \ ,
\eeq
and this determines the form of eq.~\eqref{K02}.

The experimental data indicate a very small mass difference between the two eigenstates, a large hierarchy between their lifetimes and a suppressed CP violation parameter: 
\bea 
&& M_{ K^0} ({\rm average}) = 497 \ \mathrm{MeV} \quad , \quad \Delta M = 3.5 \times 10^{-12} \ \mathrm{MeV} \ , \nonumber \\
&& \tau_S \simeq 90 \ ps \quad , \quad \tau_L  \simeq 51800 \  ps \quad , \quad {\rm Re}  \ \epsilon  \simeq 1.65 \times 10^{-3}   \  .     \label{K03}
\eea

Within the various decay channels of $K_L$, it was difficult to understand why the decay mode $K_L \to \mu^+ \mu^-$ (which changes strangeness by
one unit, and in modern language is dominated by a loop diagram) has a branching ratio of about $6.8 \times 10^{-8}$, but the decay  $K_L \to \pi^+ e^-  {\bar \nu}_e$, which in modern language is dominated by a tree--level diagram, has a branching ratio of about $0.4$, while a rough estimate would indicate a relative suppression of about $10^{-4}$.
This state of affairs persisted until 1970, when it was explained in the SM by Glashow, Iliopoulos and Maiani (GIM) \cite{gim}.  The GIM mechanism rests on the introduction of the charm quark: the authors realized that the unitarity of the CKM matrix results in a suppression of FCNC processes. Combining this setup with experimental data, in 1974 Gaillard and Lee~\cite{GaiL} could estimate the mass of the charm quark, which is about $1.5$ GeV.

Let us consider in more detail the $K^0-{\overline K}{}^{\,0}$ mixing, which arises at the loop level in the SM.
\begin{figure}[ht]
\begin{center}
    \includegraphics[width=5in]{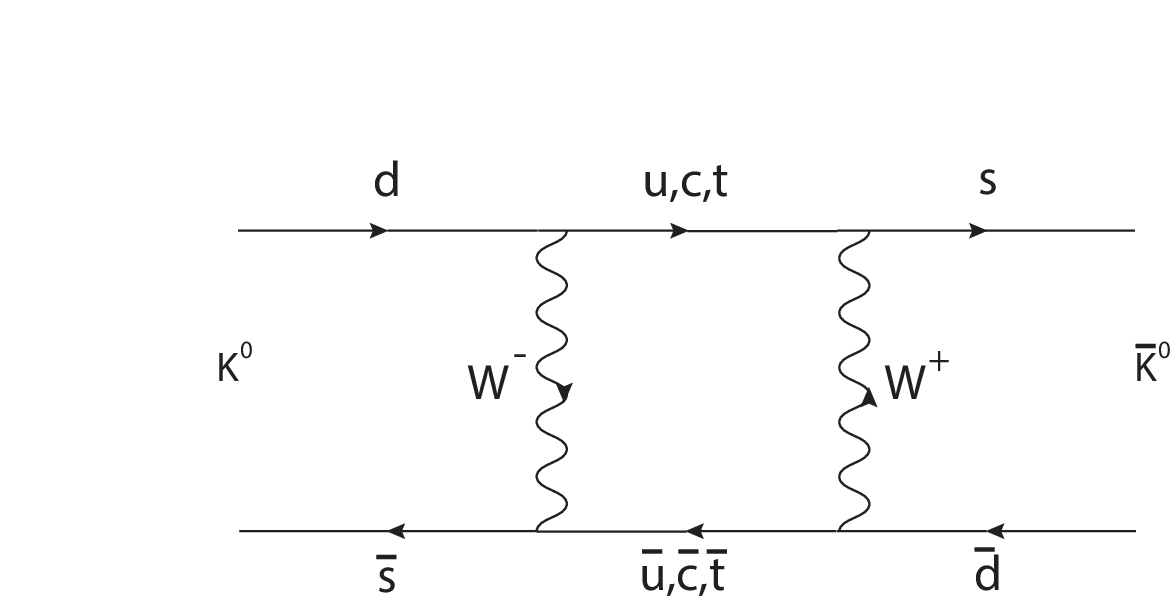}
\end{center}
\caption{$K^0$-${\bar K}^0$ mixing generated at loop level in the Standard Model, with quarks $u_i= u,c,t$ running in the loop. Here the arrows are again entering (outgoing) for incoming particles (antiparticles) and outgoing (entering) for outgoing particles (antiparticles).}
\label{fig:gim}
\end{figure}
The amplitude of this process has the form (see fig.~\ref{fig:gim})
\be
A_{K^0{\bar K}^0} \ \sim \ \frac{g^4}{M_W^2} \left(\sum_i V_{id} V_{is}^* \ \sum_j V_{js}^* V_{jd} \ F (x_i,x_j) \right) \ , \label{gim01}
\ee
where $x_i = \frac{m_i^2}{M_W^2}$ and $F(x_i,x_j)$ is a function arising from the loop that depends on the ratio between up-type quark masses and the $W$ mass. Insofar as the first two generations are concerned, one could well replace $F(x_i,x_j)$ with $F(0,0)$, while the contribution from the top quark is highly suppressed by the CKM matrix,  as we saw in eq.~\eqref{ckm13}. In view of the unitarity of the CKM matrix, the amplitude is thus essentially proportional to $\delta_{ds}$, which vanishes. This result provided a striking evidence for the existence of the charmed quark before its bound state $J/\Psi$ was actually revealed.

%%%%%%%%%%%%%%%%%%%%%%%%%%%%%%%%%%%%%%%%%
\subsection{\sc The Custodial Symmetry}

  The $\rho$ parameter of eq.~\eqref{sm4} determines the relative strength of neutral and charged current interactions. At tree-level, as we have seen, $\rho=1$ in the Standard Model. If the scale of the new physics were in the TeV range, there could be, in principle, large corrections, which strongly constrains model building. The experiments give $\rho \simeq 1.1 \pm 0.001$, where the deviation from unity originates from the top quark, but the small uncertainty implies that there is little room for new physics contributions in the quantum corrections.

The tree-level value $\rho=1$ in the Standard Model reflects an approximate symmetry, called
{\it custodial symmetry}, as was shown in~\cite{custodial}.  

As we have seen, the tree--level gauge boson mass matrix of the Standard Model is of the form
\bea
M_W^2 \begin{pmatrix}  1 & 0 & 0 & 0 \\ 0 & 1 & 0 & 0 \\
0 & 0 & 1 & - \ \tan \theta_W \\ 0 & 0 & - \ \tan \theta_W & \tan^2 \theta_W \end{pmatrix} \ , \label{custodial1}
\eea
which accounts for a massless photon,
\beq
\left(\begin{array}{c} 0 \\ 0 \\ \sin\theta_W \\\cos\theta_W \end{array}\right)
\eeq
and has the special feature of containing three identical entries along the main diagonal. This hints to an $SU(2)_d$ symmetry, under which the $SU(2)$ gauge bosons behave as a triplet. The symmetry would be exact in the limit of zero hypercharge coupling, $g'=0$, where $\tan\theta_W=0$, if in addition the masses of the quarks belonging to any given doublet were to coincide.

The custodial symmetry would naturally act on the combination of the Higgs doublet and its charge conjugate, which can be combined into the $2 \times 2$ matrix
\beq
{\cal H} = \begin{pmatrix}  i \,\sigma_2 \,\Phi^* & \Phi \end{pmatrix}
= \begin{pmatrix}  \Phi_0^* & \Phi_+ \\ - \,\Phi_+^* & \Phi_0
\end{pmatrix} \ ,
\eeq
and moreover the Higgs potential depends on the combination
\beq
\Phi^\dagger\,\Phi \ = \ \frac{1}{2}\ \mathrm{Tr}\left({\cal H}^\dagger\,{\cal H}\right) \ ,
\eeq
so that it is clearly invariant under ${\cal H} \rightarrow U_L \,{\cal H}\, U^{\dagger}{}_R$, with $U_{L,R}$  $2 \times 2$
unitary matrices implementing $SU(2)_L \times SU(2)_R$ transformations.

When the Higgs acquires a vacuum value, ${\cal H}$ becomes proportional to the identity matrix and $SU(2)_L \times SU(2)_R$ is broken to the diagonal $SU(2)_d$ that we already met, defined by $U_L=U_R=U_d$. 
The presence of different Yukawa couplings {\it breaks} the custodial symmetry. However, the particular
coupling
\beq
{\cal L}_{\rm Yuk} \ = \ h \ \begin{pmatrix}  {\bar t}_L & {\bar b}_L \end{pmatrix} {\cal H}
\begin{pmatrix}  {t}_R \\ {b}_R \end{pmatrix} \ , \label{custodial8}
\eeq
which corresponds to the limit of equal masses in the quark doublet $h_t = h_b$,
is invariant under $SU(2)_L \times SU(2)_R$, if one transform the quark doublets according to
\beq
\begin{pmatrix}  {t_L} \\ {b_L} \end{pmatrix}
\ \to \ U_L  \begin{pmatrix}  {t_L} \\ {b_L} \end{pmatrix} \quad , \quad \begin{pmatrix}  {t_R} \\ {b_R} \end{pmatrix}
\ \to \ U_R  \begin{pmatrix}  {t_R} \\ {b_R} \end{pmatrix} \ ,  \label{custodial9}
\eeq
for the left and right-handed quark chiralities. The same transformation holds for the quark doublets of the first two generations, so that in the quark sector one could identify the custodial symmetry with the strong isospin. This guarantees, in particular, that the custodial symmetry is exactly conserved by the strong interactions. With the usual definition of the SM without right-handed neutrinos the custodial symmetry is always broken by the Yukawa couplings of charged leptons. In the presence of right-handed neutrinos, one can define the custodial symmetry for leptons as in eq.~(\ref{custodial9}), which would distinguish it from strong isospin. 

The explicit breaking of custodial symmetry results in calculable quantum corrections that affect the tree--level value of $\rho$ by about 10\%, as we have anticipated. $\rho$ remains one, at tree level, with arbitrary numbers of $SU(2)$ Higgs doublets and singlets, but more general choices can seriously affect it, and are typically incompatible with experimental data.

\subsection{\sc Anomalies in Global and Gauge Currents} 
\label{sec:anomalies}

We can now turn to a phenomenon of utmost importance for the Standard Model and its extensions, including String Theory. This is the violation of classical conservation laws by quantum effects. If the symmetry at stake is a global one, new processes that are naively forbidden become possible, which is how the effect originally emerged. While the role of global symmetries in theories including gravity is presently called into question, global anomalies continue to play an important role when gravitational effects can be neglected and, most importantly, their non--renormalization places important constraints on the connection between ultraviolet and infrared limits for a given theory. However, if the symmetry is local, the lack of conservation impinges on the consistency of the theory and places important constraints on its content. In this section, we focus on the four--dimensional case, but in Section~\ref{sec:sugra1110} we shall supplement this discussion with some novelties concerning ten--dimensional strings.

\subsubsection{\sc General Considerations} \label{sec:anomaly_general}

A famous theorem by Emmy Noether associates to any continuous symmetry a conserved current $J_\mu$, such that $\partial^\mu J_\mu = 0$, and a conserved charge $Q = \int d^3 {\bf x} J_0 (x)$. Detailed derivations can be found in any of the books in~\cite{recent_books_SM} (see also~\cite{thooft_anom, zwyz, msz, abdk})~\footnote{In recent years these ideas were reconsidered, following~\cite{gaisei}, in a way that aims at a unified treatment for discrete and continuous symmetries. Some recent reviews can be found in~\cite{rev_anom_unif}.}. At the quantum level, the correspondence implies that matrix elements of the divergence of the current should vanish
\beq
\langle p_1 \cdots p_n , out | \partial^\mu J_\mu (x) | q_1 \cdots q_l , in \rangle \ = \ 0 \ , \label{genan1}
\eeq
for all choices of initial and final states.
One can actually relate initial and final states of momenta $p_i$ or $q_j$ to the insertion of corresponding operators $O_i (x_i)$ or ${\tilde O}_j (y_j)$ in correlation functions, so that the preceding condition translates into the vanishing of vacuum-to-vacuum correlation functions of the type
\beq
\langle 0 , out | \partial^\mu J_\mu (x) \ O_1 (x_1) \cdots O_n (x_n) \, {\tilde O}_1 (y_1) \cdots {\tilde O}_n (y_l)  | 0 , in \rangle \ = \ 0 \ , \label{genan2}
\eeq
to all orders in perturbation theory. 

Problems can arise in gauge theories when the external particles are gauge vectors, and the main one concerns the matrix element between the vacuum and a pair of gauge fields
\bea
&& \langle 0 , out | \partial^\mu J_\mu (x) | \gamma_b (q_1) \gamma_c (q_2) , in \rangle \ =  \ \nonumber \\
&& \int d^4 y_1  d^4 y_2  \ e^{-i (q_1 \cdot y_1 + q_2 \cdot y_2)} \left[ \partial_\nu \partial_\rho - \left(1-\frac{1}{\xi}\right) \eta_{\nu\rho} \Box \right]_{y_1} \left[ \partial_{\nu'} \partial_{\rho'} - \left(1-\frac{1}{\xi}\right) \eta_{\nu'\rho'} \Box \right]_{y_2} 
\times \nonumber \\
&\times&  \langle 0 , out | T \partial^\mu J_\mu (x)  A^{\rho b} (y_1) A^{\rho' c} (y_2) | 0 , in \rangle \epsilon_b^{\nu} (q_1) \epsilon_c^{\nu'} (q_2)  \ , \label{genan3}
\eea
 where $\xi$ is a gauge--fixing parameter $\epsilon_b^{\nu} (q_1)$ and $\epsilon_c^{\nu'} (q_2)$ are the polarization vectors of the two gauge fields. 
In four dimensions there is a one-loop contribution from a triangle diagram, with fermions running in the loop, which is generated by expanding the interaction Lagrangian to quadratic order
 \bea
&& \langle 0 , out | \partial^\mu J_\mu (x) | \gamma_b (q_1) \gamma_c (q_2) , in \rangle =  \nonumber \\
&& \int d^4 y_1  d^4 y_2  e^{-i (q_1 \cdot  y_1 + q_2 \cdot y_2)} \left[ \partial_\nu \partial_\rho - \left(1-\frac{1}{\xi}\right) \eta_{\nu\rho} \Box \right]_{y_1} \left[ \partial_{\nu'} \partial_{\rho'} - \left(1-\frac{1}{\xi}\right) \eta_{\nu'\rho'} \Box \right]_{y_2}  \nonumber \\
& \times&  \int d^4 z_1  d^4 z_2 \langle 0  | T \partial^\mu J_\mu (x)  J_\sigma^b (z_1) J_{\sigma'}^c (z_2) | 0  \rangle  D^{\rho\sigma} (y_1-z_1)  D^{\rho'\sigma'} (y_2-z_2)  \epsilon_b^{\nu} (q_1) \epsilon_c^{\nu'} (q_2)  = \nonumber \\
&-&  \frac{g_b \,g_c}{2}   \int d^4 y_1  d^4 y_2  e^{-i (q_1 \cdot y_1 + q_2 \cdot y_2)}   \langle 0  | T \partial^\mu J_\mu (x)  J_{\nu}^b (y_1) J_{\rho}^c (y_2) | 0  \rangle 
\epsilon_b^{\nu} (q_1) \epsilon_c^{\rho} (q_2)   \ . \label{genan4}
\eea
There is a quantum anomaly in the conservation of the current $J_\mu$ if
\beq
  \langle 0  | T \partial^\mu J_\mu (x)  J_{n}^b (y_1) J_{r}^c (y_2) | 0  \rangle \ \neq \ 0  \ , \label{genan5}
\eeq 
and the preceding argument applies independently of whether $J_\mu$ is the current of a global or a local symmetry. As we shall see, in the global case anomalies are possible
and even necessary to explain some particle decays. On the other hand, in the local case, the lack of current conservation would imply
that the quantum effective action is not gauge invariant, and thus leads to inconsistencies. 
Gauge invariance is crucial in gauge theories for eliminating negative--norm states, and for the whole quantization procedure.

One can further elaborate on this point by considering the quantum effective action for gauge fields, obtained by integrating out fermions,
\bea
&& e^{i \Gamma_{qu} (A)} \ = \ \langle 0 | T e^{ig \int d^4 z J^{\mu \,a} (\psi) A_\mu^a} | 0 \rangle_{\psi} \nonumber \\
&=& \sum_n \frac{(ig)^n}{n ! }   \int d^4 z_1 \cdots d^4 z_n \langle 0  | T J_{\mu_1}^{a_1} (z_1)  \cdots   J_{\mu_n}^{a_n} (z_n) | 0  \rangle_{\psi} A^{\mu_1}_{a_1} (z_1)  \cdots   A^{\mu_n}_{a_n} (z_n)
\ . \label{genan6}
\eea
 In the Abelian case, the gauge invariance of  $\Gamma_{qu} (A)$ under $\delta A_\mu = \partial_\mu \alpha$ leads to
 \beq
\partial^{\mu_1}_{z_1}  \langle 0  | T J_{\mu_1} (z_1)  \cdots   J_{\mu_n} (z_n) | 0  \rangle \ = \ 0  \ . \label{genan7}
\eeq 
In the non-Abelian case, the invariance of the quantum action $\Gamma_{qu}$ under  
\beq
\delta A_\mu^a \ = \ \partial_\mu \alpha^a \ + \ g f^{abc} A_\mu^b \alpha^c \ \equiv \ D_\mu \alpha^a
\eeq
leads, for the particular case of three currents, to
 \bea
&& \partial_\mu^{x}  \langle 0  | T J^{\mu a} (x)   J^{\nu b} (y) Jc^{\rho c} (z) | 0  \rangle \nonumber \\ &=& i f^{abd} \delta^4 (x-y)  \langle 0  | T J^{\nu d} (y) J^{\rho c} (z) | 0  \rangle
\ + \  i \,f^{acd} \delta^4 (x-z)  \langle 0  | T J^{\nu b} (y) J^{\rho d} (z) | 0  \rangle \ . \label{genan8}
\eea
Eqs.~(\ref{genan7}) and (\ref{genan8}) and called {\it naive} Ward identities. They can be deduced from current algebra equations of the type
\beq
[Q^a , J_\mu^b (z)] = i f^{abc} J_\mu^c (z) \quad \to \quad [J_0^a (x) , J_\mu^b (z)] = i f^{abc} \delta^3 ({\bf x} - {\bf z}) J_\mu^c (z) \ + \ {\rm S.T.}  \ , \label{genan9}
\eeq
where S.T. in (\ref{genan9}) denotes Schwinger terms, which are irrelevant for the present argument. Using (\ref{genan9}),  one can deduce the Ward identity 
 \bea
&& \partial_{\mu}^{x}  \langle 0  | T J_a^{\mu} (x)   J_b^{\nu} (y) J_c^{\rho} (z) | 0  \rangle \ = \    \langle 0  | T  \partial_\mu^{x} J_a^{\mu} (x)   J_b^{\nu} (y) J_c^{\rho} (z) | 0  \rangle \nonumber \\
&+&  i f^{abd} \delta^4 (x-y)  \langle 0  | T J_d^{\nu} (y) J_c^{\rho} (z) | 0  \rangle
\ + \  i f^{acd} \delta^4 (x-z)  \langle 0  | T J_b^{\nu} (y) J_d^{\rho} (z) | 0  \rangle \ , \label{genan10}
\eea
 which is compatible with the naive one (\ref{genan8}) provided  
 \beq
 \langle 0  | T  \partial_\mu^{x} J_a^{\mu} (x)   J_b^{\nu} (y) J_c^{\rho} (z) | 0  \rangle \ = \ 0 \ .
 \eeq

\begin{figure}[ht]
\begin{center}
    \includegraphics[width=3in]{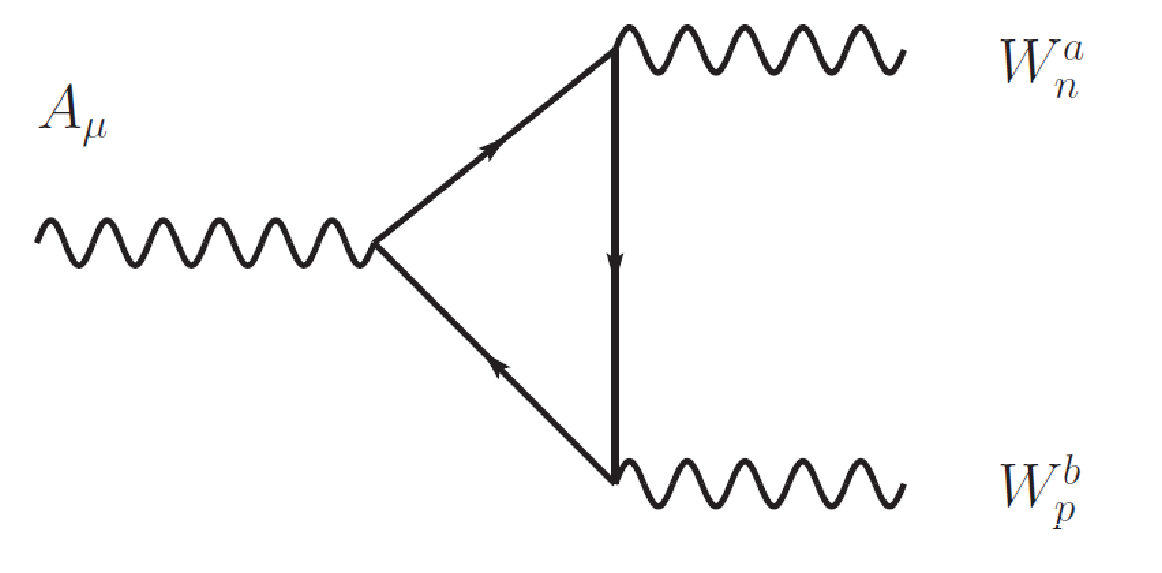}
\end{center}
\caption{Adler-Bell-Jackiw triangle anomalies.}
\label{fig:AGGtriangle2}
\end{figure}

The simplest example of a quantum anomaly is found considering a Dirac fermion coupled to a $U(1)$ gauge field with
\beq
{\cal L} \ = \ i\,{\bar \Psi} \, \gamma^\mu D_\mu \Psi \ - \ M \,{\bar \Psi} \Psi \ . \label{tan1} 
\eeq
In the massless limit $M \rightarrow 0$, the model has a vector and an axial symmetry, with a corresponding $U(1)_V \times U(1)_A$ group.
At the classical level, the corresponding Noether currents
\beq
J_\mu \ = \ {\bar \Psi} \gamma_\mu \Psi \ , \qquad  J_\mu^5 \ = \ {\bar \Psi} \gamma_\mu \gamma_5 \Psi \label{tan2}
\eeq
satisfy
\beq
\partial^\mu J_\mu \ = \ 0 \ , \qquad \partial^\mu J_\mu^5 \ = \ 2 \,i\, M \,{\bar \Psi} \gamma_5 \Psi  \ , \label{tan3}
\eeq
but at the quantum level these conservation laws are modified. It was first shown in \cite{ABJ1,ABJ2,ABJ3} that, even in the massless limit, it is impossible to preserve simultaneously the vector and axial symmetries, due to subtleties of triangle graphs (see fig.~(\ref{fig:AGGtriangle2})). We can now elaborate on this important result.

%%%%%%%%%%%%%%%%%%%%%%%%%%%%%%%%%%%%%%%%%%%%%%%%%%%%%%%%%%%%%%%%%
\subsubsection{\sc Triangle Anomalies: General Case} 

In studying the quantum anomaly for a system of left-handed and right-handed fermions coupled to gauge fields via the current
\beq 
J^{\mu}_a \ = \ \bar\psi_L {t}_a^L \gamma^{\mu} \psi_L \ + \ \bar\psi_R {t}_a^R \gamma^{\mu} \psi_R 
\ , \label{tan01} 
\eeq
we shall work, for definiteness, with only left-handed Fermi fields.
In four dimensions, this is always possible by resorting to charge conjugation. Indeed, if $\chi$ is a Dirac fermion, $\frac{1-\g_5}{2}$ projects it onto the left-handed component, while
$\frac{1+\g_5}{2}$ projects it onto the right-handed one. We shall instead resort to $\chi^c = C {\bar \chi}^T$, the charge-conjugated spinor, and work with $\frac{1-\g_5}{2} \chi$ and $\frac{1-\g_5}{2} \chi^c$. With this choice, a generic charge generator $T_a$ splits according to
\beq
T^a  \ = \
\begin{pmatrix}
t_L^a & 0 \\
0 & - (t_R^a)^{\star} 
\end{pmatrix}  \ , \label{tan6}
\eeq 
and the corresponding current takes the form
\beq
J_\mu^a \ = \ \bar{\chi} \,\gamma_\mu \,t_L^a  \,\frac{1-\g_5}{2} \chi \ - \ \bar{\chi}^c \,\gamma_\mu \,\left(t_R^a\right)^\star   \frac{1-\g_5}{2} \chi^c \ .
\eeq

The three-current correlator of interest is
\beq
\Gamma^{\mu\nu\rho}_{abc}(x,y,z) \ = \ \langle 0 | T (J_a^{\mu}(x)J_b^{\nu}(y)J_c^{\rho}(z))| 0  \rangle
\ , \label{tan8}
\eeq
The leading correction emerges from the fermion loop, and results from the contributions of all Fermi fields that couple to the currents.
There are two diagrams for the correlator, which grant its total symmetry and can be evaluated to yield
\bea
\Gamma^{\mu\nu\rho}_{abc}(x,y,z)&=&{\rm Tr}\left[S_F(x-y)T_b\gamma^{\n}
P_{L} S_F(y-z) T_c \gamma^{\rho}P_L S_F(z-x) T_a \gamma^{\m} P_L \right]
\nn\\
&+& {\rm Tr}\left[S_F(x-z) T_c \gamma^{\r}P_{L} S_F(z-y) T_b \gamma^{\n}P_L S_F(y-x) T_a \gamma^{\m} P_L \right] \ , 
\label{tan07}
\eea
with
\beq
P_{L}\ = \ \frac{1\ - \ \gamma^5}{2} \ , \qquad S_F(x) \ = \ \int \frac{d^4 p}{ (2\pi)^4}\ \frac{i \sla{p}}{ p^2 + i \e} \, e^{ip\cdot x}
\label{tan08} \ . \eeq
Substituting one obtains
\bea
\Gamma^{\mu\nu\rho}_{abc}(x,y,z)
\!\!&=& \!\!\!- \, i \int \frac{d^4 k_1}{(2\pi)^4}
 \frac{d^4 k_2}{ (2\pi)^4}e^{-i(k_1+k_2)\cdot x+ik_1\cdot y+ik_2\cdot z}\int \frac{d^4 p}{ (2\pi)^4}
 \times  \label{tan9}\\
&& \!\!\! \bigg\{ {\rm Tr}\left[{\sla{p}-\sla{k_1}+\sla{a_1}\over (p-k_1+a_1)^2+ i \e }\gamma^{\n}
{\sla{p}+\sla{a_1}\over (p+a_1)^2+ i \e }\gamma^{\r}
{\sla{p}+\sla{k_2}+\sla{a_1}\over (p+k_2+a_1)^2+ i \e }\gamma^{\m}P_L\right]\! {\rm tr}[{T}_b {T}_c {T}_a] \nn\\
&+& \!\!{\rm Tr}\left[{\sla{p}-\sla{k_2}+\sla{a_2}\over (p-k_2+a_2)^2+ i \e }\gamma^{\r}
{\sla{p}+\sla{a_2}\over (p+a_2)^2+ i \e }\gamma^{\n}
{\sla{p}+\sla{k_1}+\sla{a_2}\over (p+k_1+a_2)^2+ i \e }\gamma^{\m}P_L\right]\! {\rm tr}[{T}_c {T}_b {T}_a]  \bigg\}
\,. \nn
\eea
We have shifted the integrated momenta in the two diagrams using two vectors $a_{1,\m}$ and $a_{2,\m}$.
This choice reflects an ambiguity of the triangle graph, which cannot be dealt with in dimensional regularization due to the presence of $\gamma^5$ in $P_L$. The shifts affect the current operators and can move the anomaly from one current to another. The identities
\beq 
\sla{k_1} + \sla{k_2} = (\sla{p} + \sla{k_2} + \sla{a_1}) - (\sla{p} - \sla{k_1} + \sla{a_1})
= (\sla{p} + \sla{k_1} + \sla{a_2}) - (\sla{p} - \sla{k_2} + \sla{a_2})  \label{tan10}
\eeq
allow to remove one of the denominators, and one finds
\bea
\partial_{\m} \Gamma^{\mu\nu\rho}_{abc}(x,y,z)
&=& - \int {d^4 k_1\over (2\pi)^4}
 {d^4 k_2\over (2\pi)^4}e^{-i(k_1+k_2)\cdot x+ik_1\cdot y+ik_2\cdot z}\int {d^4 p\over (2\pi)^4}
 \times\nn\\
&& \bigg\{ {\rm tr}[{T}_b {T}_c {T}_a] {\rm Tr}\left[{\sla{p}-\sla{k_1}+\sla{a_1}\over (p-k_1+a_1)^2 + i \e }\gamma^{\n} {\sla{p}+\sla{a_1}\over (p+a_1)^2 + i \e }\gamma^{\r} P_L\right] \nn\\
&-&  {\rm tr}[{T}_c {T}_b {T}_a] {\rm Tr}\left[{\sla{p}+\sla{a_2}\over (p+a_2)^2 + i \e } \gamma^{\n}
{\sla{p}+ \sla{k_1}+ \sla{a_2}\over (p+k_1+a_2)^2 + i \e }\gamma^{\r} P_L\right]  \nn \\
&+& {\rm tr}[{T}_c {T}_b {T}_a] {\rm Tr}\left[{\sla{p}-\sla{k_2}+\sla{a_2}\over (p-k_2+a_2)^2 + i \e }\gamma^{\r} {\sla{p}+\sla{a_2}\over (p+a_2)^2 + i \e }\gamma^{\n} P_L\right] \nn\\
&-&  {\rm tr}[{T}_b {T}_c {T}_a] {\rm Tr}\left[{\sla{p}+\sla{a_1}\over (p+a_1)^2 + i \e }\gamma^{\r}
{\sla{p}+ \sla{k_2}+ \sla{a_1}\over (p+k_2+a_1)^2 + i \e }\gamma^{\n} P_L\right] 
 \bigg\}
\ . \label{tan11}
\eea
It is now convenient to separate the group theory trace into symmetric and antisymmetric parts, according to
\bea 
&& {\rm tr}[{T}_b {T}_c {T}_a] \ = \ d_{abc} \ +\  \frac{i}{2} N f_{abc} \ , \nonumber \\
&&  {\rm tr}[{T}_c {T}_b {T}_a] \ =\  d_{abc} \ - \ \frac{i}{2} N f_{abc} \ , \label{tan12}
\eea
where $f_{abc}$ are the structure constants, $N$ is the number of left--handed fermions circulating in the loop and the additional tensor
\beq
d_{abc} \ = \ \frac{1}{2} {\rm tr} [ \{T_a, T_b \} T_c]
\eeq
is totally symmetric in its three labels. The contribution of the term proportional to the structure constants $f_{abc}$ reproduces
the naive Ward identity (\ref{tan8}) and has nothing to do with the quantum anomaly. The remaining symmetric part is
\bea
\partial_{\m} \Gamma^{\mu\nu\rho}_{abc}(x,y,z)
&=& - \ d_{abc} \int {d^4 k_1\over (2\pi)^4}
 {d^4 k_2\over (2\pi)^4}e^{-i(k_1+k_2)\cdot x+ik_1\cdot y+ik_2\cdot z}\int {d^4 p\over (2\pi)^4}
 \times\nn\\
&& \bigg\{  {\rm Tr}\left[{\sla{p}-\sla{k_1}+\sla{a_1}\over (p-k_1+a_1)^2 + i \e }\gamma^{\n} {\sla{p}+\sla{a_1}\over (p+a_1)^2 + i \e }\gamma^{\r} P_L\right] \nn\\
&-&  {\rm Tr}\left[{\sla{p}+\sla{a_2}\over (p+a_2)^2 + i \e } \gamma^{\n}
{\sla{p}+ \sla{k_1}+ \sla{a_2}\over (p+k_1+a_2)^2 + i \e }\gamma^{\r} P_L\right]  \nn \\
&+&  {\rm Tr}\left[{\sla{p}-\sla{k_2}+\sla{a_2}\over (p-k_2+a_2)^2 + i \e }\gamma^{\r} {\sla{p}+\sla{a_2}\over (p+a_2)^2 + i \e }\gamma^{\n} P_L\right] \nn\\
&-&  {\rm Tr}\left[{\sla{p}+\sla{a_1}\over (p+a_1)^2 + i \e }\gamma^{\r}
{\sla{p}+ \sla{k_2}+ \sla{a_1}\over (p+k_2+a_1)^2 + i \e }\gamma^{\n} P_L\right] 
 \bigg\}
\ . \label{tan13}
\eea 
Grouping together the first two and the last two trace factors leads to
\bea
&& \partial_{\m} \Gamma^{\mu\nu\rho}_{abc}(x,y,z)
=  \ - \ d_{abc} \int {d^4 k_1\over (2\pi)^4}
 {d^4 k_2\over (2\pi)^4}e^{-i(k_1+k_2)\cdot x+ik_1\cdot y+ik_2\cdot z}\int {d^4 p\over (2\pi)^4}
\times\nn\\
&& \bigg\{  {\rm Tr} \left[ \gamma^{\tau} \gamma^{\n} \gamma^{\l} \gamma^{\r} \frac{1-\gamma_5}{2} \right] 
\ I_{\tau \lambda} (a_1-a_2-k_1,a_2,a_2+k_1)  \nn\\
&+& {\rm Tr} \left[ \gamma^\tau \gamma^{\r} \gamma^{\l} \gamma^{\n} \frac{1-\gamma_5}{2} \right] 
\ I_{\tau \lambda} (a_2-a_1-k_2,a_1,a_1+k_2)  
 \bigg\}
\ , \label{tan14}
\eea
where 
\bea 
 I_{\tau,\l} (k,c,d) &=& \int \frac{d^4 p}{(2 \pi)^4} \ [ f_{\tau,\l} (p+k,c,d) - f_{\tau,\l} (p,c,d) ]
\ , \nonumber \\
f_{\tau,\l} (p,c,d) &=& \frac{(p+c)_\tau (p+d)_{\l}}{[(p+c)^2 + i \e ] 
[(p+d)^2 + i \e ]}  \ . \label{tan15}
\eea
These integrals can be calculated by Taylor expanding in powers of $k$. Only terms linear and quadratic in $k$ in the expansion contribute to the resulting surface integral, and finally
\beq 
I_{\tau,\l} (k,c,d) \ = \ \frac{i}{96 \pi^2} \left[ 2 k_{\l} c_\tau + 2 k_\tau d_{\l} -
k_{\l} d_\tau - k_\tau c_{\l} - \eta_{\tau \l} k (k + c+d) \right]  \ . \label{tan16}
\eeq  
Demanding that there be no anomaly in the vector currents implies that the term without $\gamma_5$
in (\ref{tan14}) should vanish. Due to the symmetry of this term in $\tau,\l$ and $\n,\r$, this term appears in the combination
\bea
&& I_{\tau,\l} (a_1-a_2-k_1,a_2,a_2+k_1) \ + \ I_{\l,\tau} (a_1-a_2-k_1,a_2,a_2+k_1) \nonumber \\
&+&  I_{\tau,\l} (a_2-a_1-k_2,a_1,a_1+k_2) \ + \ I_{\l,\t} (a_2-a_1-k_2,a_1,a_1+k_2)  \ . \label{tan17}
\eea
This combination vanishes for $a_2=-a_1$, and the same is true for vector anomalies in the other currents. We shall abide to this choice in the following.
The vector $a_1 $ therefore parametrizes the residual scheme dependence of the triangle graph.
The trace containing $\gamma_5$ is
\beq  
{\rm Tr} [ \gamma^{\n}  \gamma^{\r} \gamma^{\s} \gamma^{\t} \g_5] \ = \ - \ 4\, i \ \e^{\n \r \s \t}  \ , \label{tan017}
\eeq 
and  using  eq.~(\ref{tan017}) finally yields
\beq
\partial_{\mu}\Gamma^{\mu\nu\rho}_{abc}(x,y,z)\ =\ - \ {d_{abc}\over 8\pi^2}\int {d^4 k_1\over (2\pi)^4}
 {d^4 k_2\over (2\pi)^4}e^{-i(k_1+k_2)\cdot x+ik_1\cdot y+ik_2\cdot z}~\epsilon^{\nu\rho\s\t}~a_{1,\s}(k_1+k_2)_{\t} \ , 
\label{tan18}\eeq
\beq
\partial_{\n}\Gamma^{\mu\nu\rho}_{abc}(x,y,z)\ =\ - \ {d_{abc} \over 8\pi^2}\int {d^4 k_1\over (2\pi)^4}
 {d^4 k_2\over (2\pi)^4}e^{-i(k_1+k_2)\cdot x+ik_1\cdot y+ik_2\cdot z}~\epsilon^{\m\rho\s\t}~(a_1+k_2)_{\s}(k_1)_{\t} \ , 
\label{tana19}\ee
\beq
\partial_{\r}\Gamma^{\mu\nu\rho}_{abc}(x,y,z)\ =\ -\ {d_{abc} \over 8\pi^2}\int {d^4 k_1\over (2\pi)^4}
 {d^4 k_2\over (2\pi)^4}e^{-i(k_1+k_2)\cdot x+ik_1\cdot y+ik_2\cdot z}~\epsilon^{\m\n\s\t}~(k_1-a_1)_{\s}(k_2)_{\t} \ . 
\label{tan20} \eeq

The choice
$a_1 \sim k_1+k_2$ eliminates the anomaly from the current $J_a$, while $a_1 \sim \pm k_1-k_2$  
($a_1 \sim k_1 \pm k_2$) eliminates it from the current $J_b$ ($J_c$), but for generic vectors
$k_1,k_2$ it is impossible to remove it altogether. 
A generic choice of scheme (i.e. $a_{1,\m}$) indicates that the divergence structure is asymmetric among the three vertices of the triangle graph, so that the choice should be dictated by physical requirements. In the configuration investigated in the next subsection, $J_a$ is the current of
a global symmetry, while $J_b$ and $J_c$ are currents of gauge symmetries, coupling to gauge fields.
In this case, one must choose $a_1$ so that only $J_a$ has an anomaly, which leads to the unique solution
\beq 
a_1 = k_1 - k_2 \ , \label{tan21} 
\eeq
and the anomaly in the current $J_a$ becomes
\beq 
\partial_{\mu}\Gamma^{\mu\nu\rho}_{abc}(x,y,z) \ = \ \frac{1}{4 \pi^2}\  
d_{abc} \,\epsilon^{\a \b \n \r} \, \frac{\partial \delta^4 (y-x)}{\partial y^{\a}} \,
 \frac{\partial \delta^4 (z-x)}{\partial z^{\b}} \ .  \label{tan22} 
\eeq
One can interpret this result as a quantum contribution to the current in the presence of gauge fields
\beq 
\langle J_a^{\m} \rangle_q \ = \  \frac{g_b \,g_c}{2} \int d^4 x  d^4 y \ \Gamma^{\mu\nu\rho}_{abc}(x,y,z)\,
A_{\n}^b \,A_{\r}^c \ ,  \label{tan23} 
\eeq 
leading to the anomalous divergence
\beq 
\langle \partial_{\m } J_a^{\m} \rangle_{\rm an} \ = \ \frac{g_b \,g_c}{8 \pi^2} \,d_{abc}\, \epsilon^{\a \n \b \r}
\,\partial_{\a} A_{\n}^b \,\partial_{\b} A_{\r}^c   \ .  \label{tan024} 
\eeq 

In the non-Abelian case, there are in principle additional contributions from square and pentagon
diagrams, but the pentagon diagram is convergent, and one can show that it does not contribute, as in~\cite{zwyz}. The square correction builds nonetheless an expression satisfying the Wess--Zumino consistency conditions~\cite{wzcc}, but the terms that we have computed explicitly, including those in eq.~\eqref{genan8}, suffice to identify the consistent result
\beq 
\langle D_{\m } J_a^{\m} \rangle_{\rm an} \ = \ \frac{g_b \,g_c}{8 \pi^2} \,d_{abc}\, \epsilon^{\a \n \b \r}
\,\partial_{\a} A_{\n}^b \,\partial_{\b} A_{\r}^c \ + \ \ldots   \ ,  \label{tan24} 
\eeq
where the ellipsis indicate the higher--order terms obtained from the square diagram. 
Note that the explicit form of the overall group-theory coefficient determined by the left-handed and right-handed
fermions is
\beq 
d_{abc} \ = \ \frac{1}{2} {\rm Tr} [ \{ t_a^L,t_b^L \} t_c^L ] \ - \ \frac{1}{2} {\rm Tr} [ \{ t_a^R,t_b^R \} t_c^R ] \ .  \label{tan25}
\eeq

Summarizing, we have seen that some symmetries of the classical action can have {\it anomalies} at the quantum level, which first emerge in one-loop {\it triangle diagrams}. The Adler--Bardeen theorem~\cite{ABJ3} shows that no higher--loop anomalous terms emerge, since the corresponding diagrams can be regulated by the addition of higher--derivative terms.

We can now elaborate further on the two classes of anomalies that we have identified.
%%%%%%%%%%%%%%%%%%%%%%%%%%%%%%%%%%%%%%%%%%%%%%%%%%%%%%%%%%%%%%%%%%%%%%%%%%%%%%%%%%%%%%%%%%%%%%%%%%%%%%%%%%%%%%%%%%%%%%
\subsubsection{\sc Anomalies in Global Symmetries} 

For global symmetries, quantum anomalies do not create consistency problems; they actually play an important role in QCD, in connection with the
so-called $\eta'$ problem, and in the electromagnetic decay of the neutral pion $\pi^0 \to \gamma \gamma$.  In general, they also provide important constraints linking the ultraviolet and infrared degrees of freedoms present in theories, precisely due to their non--renormalization properties~\cite{thooft_anom}.

For a global symmetry with Noether current $J_m^a$ of generator $T^a$, the anomaly in operatorial form is
\beq
\partial^\mu J_\mu^a \ = \ - \  \frac{g^2}{16 \pi^2}\ 
\epsilon^{\mu\nu\rho\sigma} \  \mathrm{tr} \left( T^a F_{\mu\nu} \ F_{\rho\sigma} \right)  \ , 
\eeq
where
\beq
\mathrm{tr} ( T^a F_{\mu\nu} \ F_{\rho\sigma}) \ = \ \frac{1}{2} \ \mathrm{tr} \left( T^a \{T^A ,T^B \}\right) \ F_{\mu\nu}^A \ F_{\rho\sigma}^B  \ , 
\label{an01} 
\eeq
with $g$ is the gauge coupling for $A_\mu = A_\mu^A T^A$. 
In (\ref{an01}) the trace is computed over the fermionic spectrum of the theory, considered in this section to be of Dirac type, i.e. with $t^L=-t^R$ in eq.  (\ref{tan25}). 

Let us consider, to begin with, a 
Dirac fermion coupled to a $U(1)$ gauge field, so that
\beq
{\cal L} \ = \ {\bar \Psi} i \gamma^\mu D_\mu \Psi \ - \ M {\bar \Psi} \Psi \ . \label{an1}
\eeq
As we saw in Section~\ref{sec:anomaly_general}, in the massless limit the model has a phase symmetry $U(1)_V$ and an axial symmetry $U(1)_A$, with the currents in eq.~\eqref{tan2}. Taking into account the anomaly, the two currents
satisfy
\beq
\partial^\mu J_\mu \ = \ 0 \ , \qquad \partial^\mu J_\mu^5 \ = \ 2 i M {\bar \Psi} \gamma_5 \Psi \ - \ { \frac{e^2}{16 \pi^2}
\epsilon^{\mu\nu\rho\sigma} \  F_{\mu\nu} \ F_{\rho\sigma}} \ . \label{an3}
\eeq
If $U(1)_V$ is a gauge symmetry, as in QED, the consistency of the theory demands that one choose a regularization preserving its conservation, thus accepting that the quantum anomaly violates the conservation of the axial current.
This setup is regarded as a justification of why the $\eta'$ meson is not a pseudo-Goldstone boson for the dynamical chiral symmetry breaking
$U(2)_L \times U(2)_R = SU(2)_L \times SU(2)_R \times U(1)_B \times U(1)_A \Rightarrow SU(2)_V \times U(1)_B$ in QCD. Indeed, in this case the $U(1)_A$ axial current
\beq
J_\mu^{U(1)_A} \ = \ {\bar u \gamma_\mu \gamma_5 u} \ + \  {\bar d \gamma_\mu \gamma_5 d}
\eeq
has the anomaly
\beq
 \partial^\mu J_\mu^{U(1)_A} \ = \ 2 i (m_u {\bar u} \gamma_5 u + m_d {\bar d} \gamma_5 d )  \ - \ {\frac{g_3^2}{16 \pi^2}\
\epsilon^{\mu\nu\rho\sigma} \  G_{\mu\nu}^A \ G_{\rho\sigma}^A} \ , \label{an4}
\eeq
where $G^A$ is the gluon field strength and $g_3$ is the color $SU(3)_c $ gauge coupling. The explicit breaking of the
axial symmetry by quark masses and non--perturbative instanton effects make the $\eta'$ more massive than the pions $\pi^{\pm},
\pi^0$, which are the pseudo-goldstone bosons of the axial $SU(2)_A$ symmetry. 

The electromagnetic decay $\pi^0 \rightarrow \gamma \gamma$ is another important manifestation of the axial anomaly. In this case there are two relevant
$SU(2)$ currents,
\beq
J_\mu^a \ = \ {\bar q} \gamma_\mu \frac{\tau^a}{2} q \ , \qquad J_\mu^{5,a} \ =\  {\bar q} \gamma_\mu \gamma_5 \frac{\tau^a}{2} q \ , \label{an5}
\eeq
where $q = (u,d)^T$ and $\tau^a$ are the Pauli matrices. The fact that the pions are {\it Goldstone bosons} of the axial
$SU(2)_A$ implies that the corresponding currents have a non vanishing matrix element between the vacuum and a one pion state of the form
\beq
\langle 0 | J_\mu^{5,a} (x)| \pi^b (p) \rangle \ =  \ i \ p_\mu \ f_{\pi} \ \delta^{ab} \ e^{- i px} \ , \label{pi01}
\eeq
where the mass parameter $f_{\pi}$ is called the pion decay constant. 
The axial isospin currents have no QCD anomalies since $\mathrm{tr} \left(\tau^a \{T^A ,T^B \}\right) = \mathrm{tr} \left(\tau^a\right) \, \mathrm{tr}\left(\{T^A ,T^B \}\right) = 0$, consistent with the isospin symmetry of strong
interactions, but $J_\mu^{5,3}$ has an {\it anomaly} from the electromagnetic coupling. Neglecting the mass difference between the up and down quarks,
\beq
\partial^\mu J_\mu^{5,3} \ = \  - \ \frac{1}{16 \pi^2}\,
\epsilon^{\mu\nu\rho\sigma} \  F_{\mu\nu} \ F_{\rho\sigma} \ \mathrm{tr} \left(Q^2 \frac{\tau_3}{2}\right)\ = \ - \ \frac{N_c \,e^2}{96 \,\pi^2}\,
\epsilon^{\mu\nu\rho\sigma} \  F_{\mu\nu} \ F_{\rho\sigma} \ , \label{pi02}
\eeq
where $Q = \mathrm{diag}\left(\frac{2e}{3}, -\,\frac{e}{3}\right)$ is the matrix of the quark electric charges and $N_c=3$ is the number of quark colors. 
\begin{figure}[ht]
\begin{center}
    \includegraphics[width=3in]{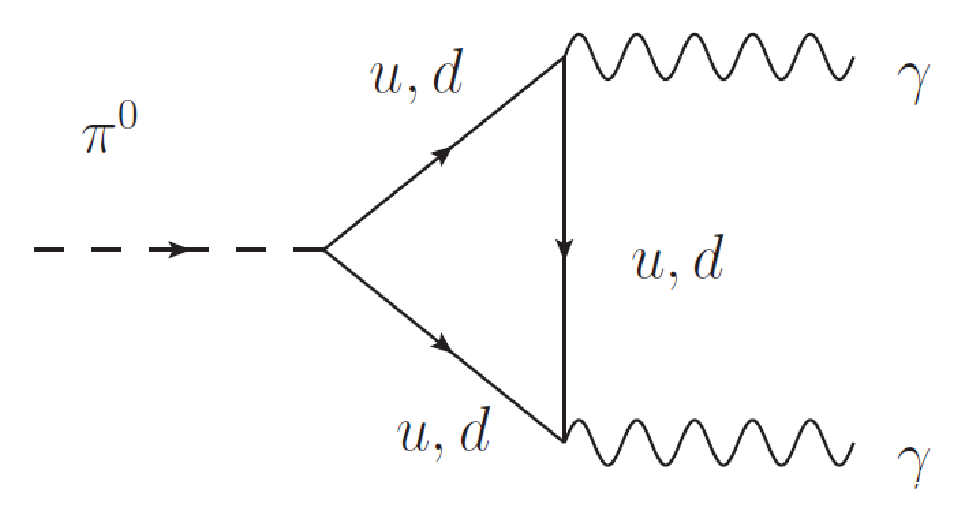}
\end{center}
\caption{The electromagnetic pion decay $\pi^0 \rightarrow \gamma \gamma$ is related to the axial $U(1)_A$ anomaly.}
\label{fig:piondecay}
\end{figure}

Making use of eqs.~(\ref{pi01}) and (\ref{pi02}), and taking into account that under the axial $SU(2)_A$ the up and down quarks transform as
$\delta q = i \alpha \gamma_5 \frac{\tau_3}{2} q$, while the pion transforms like a Goldstone boson, with $\delta \pi^0 = \alpha f_{\pi}$, one can translate the anomaly into an effective pion-photon-photon coupling
\beq
{\cal L}_{\rm eff} \ = \ \frac{\pi^0}{f_{\pi}} \ \partial^\mu J_\mu^{5,3} \ = \ - \ \frac{N_c e^2}{96 \pi^2 f_{\pi}} \, \pi^0 \,
\epsilon^{\mu\nu\rho\sigma} \,  F_{\mu\nu} \, F_{\rho\sigma} \ . \label{pi03}
\eeq
The anomaly matching in eq.~(\ref{pi03}) rests on the use of Noether's theorem: the variation of the Lagrangian under a transformation generated by the
parameter $\alpha$ is equal to $\delta {\cal L} = \alpha \ \partial^\mu J_\mu$, where $J_\mu$ is the corresponding Noether current. The
effective Lagrangian (\ref{pi03}) is its low-energy manifestation, and its variation under the axial transformation reproduces
the anomaly of the microscopic theory.
Using this effective coupling, the pion decay amplitude is
\beq
{\cal M} (\pi^0 \to \gamma \gamma) \ = \ - \ \frac{\alpha}{\pi f_{\pi}} \epsilon^{\mu\nu\rho\sigma} \,\epsilon_\nu^* \epsilon_\sigma^*
p_\mu k_\rho \ , \label{pi04}
\eeq
where $(p,\epsilon_n) $ and $(k, \epsilon_s)$ are momenta and polarizations of the two photons. Summing over photon polarizations gives
\beq
\sum_{\rm pol.} \ | \epsilon^{\mu\nu\rho\sigma} \epsilon_\nu^* \epsilon_\sigma^* p_\mu k_\rho |^2 \ = \ 2 \ (p \cdot k)^2 \ = \ \frac{m_{\pi}^4}{2} \ ,
\label{pi05}
\eeq 
and one can finally obtain the pion decay width
\beq  
\Gamma (\pi^0 \rightarrow \gamma \gamma) \ = \frac{1}{32 \pi m_{\pi}} \sum_{\rm pol.} |{\cal M} (\pi^0 \to \gamma \gamma)|^2  
\ =  \ \frac{\alpha^2}{64 \pi^3} \frac{m_{\pi}^3}{f_{\pi}^2} \ . \label{pi06}
\eeq 
This result is in excellent agreement with the experimental branching ratio of the pion decay into two photons
$\Gamma (\pi^0 \rightarrow \gamma \gamma) = (1.19 \pm 0.08) \times 10^{16} s^{-1}$.

Global anomalies find another important application in the discussion of the {strong CP problem}, which we shall address in Section~\ref{sec:strongcp}.
%%%%%%%%%%%%%%%%%%%%%%%%%%%%%%%%%%%%%%%%%%%%%%%%%%%%%%%%%%%%%%%%%%%%%%%%%%%%%%%%%%%%%%%%%%%%%%%%%%%%%%%%%%%%%%%%%%%%%%%%%%%%%%%%%
\subsubsection{\sc Gauge Anomalies} 

Anomalies in gauge currents, if present, generate inconsistencies \cite{BIM,GJACK}, since they violate gauge invariance. The gauge variation of the Lagrangian is determined by Noether's theorem and reads
\beq
\delta {\cal L} \ \sim \ \alpha_A \ \partial^\mu J_\mu^A \ + \ \ldots \ . \label{an6}
\eeq

\begin{figure}[ht]
\begin{center}
    \includegraphics[width=3in]{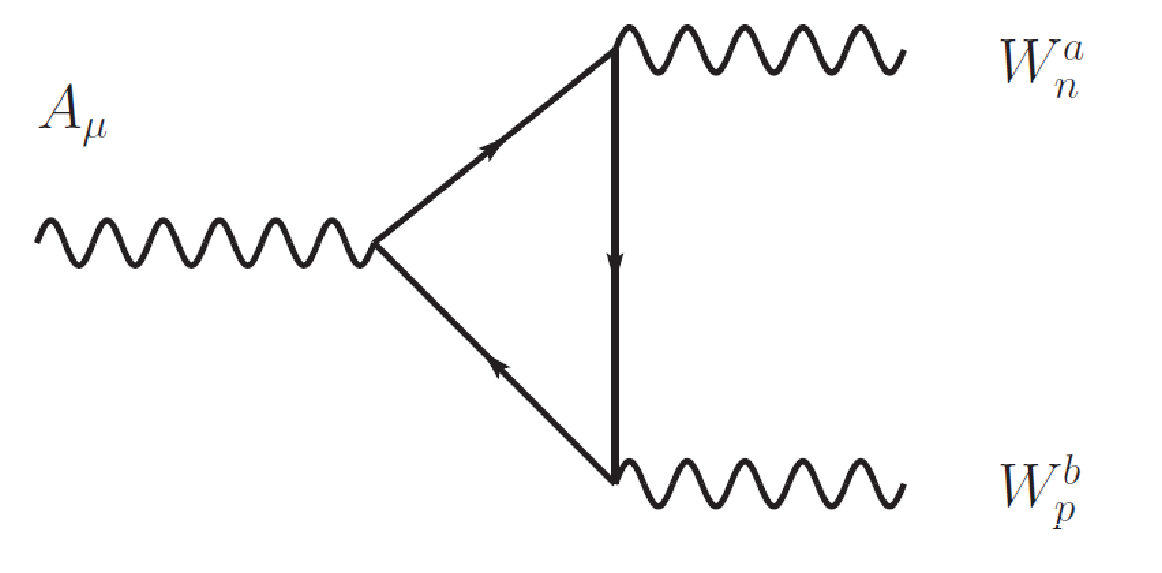}
\end{center}
\caption{Gauge anomalies, if present, render the theory {\it inconsistent} at the quantum level.}
\label{fig:AGGtriangle02}
\end{figure}

Maintaining gauge invariance at the quantum level is crucial for a consistent quantization, since it guarantees the decoupling of unphysical states. The gauge currents contributing to the anomaly are chiral
\beq
J_\mu^A  \ = \ {\bar \Psi}_L \gamma_\mu T^A_L \Psi_L
\ + \ {\bar \Psi}_R \gamma_m  T^A_R  \Psi_R  \ , \label{tan44}
\eeq
with $T^A_R \neq T^A_L$,
and according to our previous considerations their divergence is proportional to
\beq
D^\mu J_\mu^A \ = \ \frac{g_B g_C}{32 \pi^2} \ d^{ABC} \epsilon^{\mu\nu\rho\sigma} \  F_{\mu\nu}^B \ F_{\rho\sigma }^C \ +  \ \ldots \ , \label{tan45}
\eeq
where the ellipsis refers to additional terms needed to satisfy the Wess-Zumino consistency conditions.
The anomaly coefficients that must vanish are then
\beq
2 d^{ABC} \ = \ tr \ ( \{T^A,T^B\}T^C)_L \ - \ tr \ ( \{T^A,T^B\}T^C)_R  \ , \label{tan46}
\eeq 
where the trace is taken over {\it all} fermions present in the theory.  Non-chiral fermions do
not contribute to gauge anomalies, and the same is true for fermions in {\it self--conjugate} representations of the gauge group. On the other hand, fermions in {\it complex} representations of the gauge group, for which no standard mass terms are possible,
do contribute. This is precisely the case for the quarks and leptons in the Standard Model.  The properties of the Pauli matrices, and more generally the pseudo--reality of $SU(2)$ representations, imply that there are no pure cubic $SU(2)^3$ anomalies, so that the only possible gauge anomalies are of the types
\beq 
 SU(2)_L^2 \ U(1)_Y \ , \qquad U(1)_Y^3 \ , \qquad  {\rm and} \ , \qquad SU(3)_c^2\  U(1)_Y \ , \label{an9}
\eeq 
to which one must add the mixed gravitational anomaly
\beq
T \ T \ U(1)_Y \ ,
\eeq
where $T$ denotes the energy--momentum tensor.

The quantum numbers of the known quarks and leptons imply that the gauge anomaly coefficients cancel generation by generation in the Standard Model. For the first generation
\bea
&& \mathrm{tr} \left( \left\{\frac{\tau^a}{2},  \frac{\tau^b}{2} \right\} Y\right)_L = \delta^{ab} \mathrm{tr}\Big( Y\Big)_L = \delta^{ab} \left(Y_\ell + N_c \, Y_q\right) = 
 \delta^{ab}\left(-1 + \frac{N_c}{3}\right)  \ = \ 0 \ , \nonumber \\
&&  \mathrm{tr} \Big( Y^3 \Big)_{L-R}  =  2 Y_\ell^3 - Y_{e_R}^3+ N_c\left(2 Y_q^3 - Y_{u_R}^3 - Y_{d_R}^3\right) = \ 12 (-N_c + 3) \ = \ 0 \ , \nonumber \\
&&  \mathrm{tr} \left( \left\{\frac{\lambda^A}{2},  \frac{\lambda^B}{2} \right\} Y\right)_{L-R} = \delta^{AB} \, N_c\left[Y_q - \frac{1}{2}\left(Y_{u_R}+Y_{d_R}\right)\right] \  = \  0 \ , \nonumber\\
&& N_c\left(2 Y_q - Y_{u_R}-Y_{d_R}\right)  + 2 Y_\ell - Y_{e_R} \ = \ \frac{N_c}{3} - 1 \ ,
\label{an010}
\eea
where in the third equation in~(\ref{an010}) the $\lambda^A$ are the $SU(3)$ Gell-Mann matrices. 
Note that anomaly cancellation occurs {\it precisely} for three quark colors, $N_c=3$. This seems to reflet a deep {\it connection between quarks and leptons} in the Standard Model, and may be regarded as a possible {\it hint} toward Grand Unified Theories.  

Anomaly cancellation generally gives strong constraints on the possible quantum numbers under new symmetries or on the spectrum of {\it new chiral particles}. For example, one can show that $U(1)_{B-L}$, defined according to
$$
\begin{array}{|c|c|c|c|c|c|}
\hline
{\rm \ Field} \quad &  \quad \ q_i \quad & \quad u_{iR} \quad & \quad d_{iR} \quad & \quad l_{i} \quad & \quad e_{iR} \\
\hline
{\rm U(1)_{B-L} {\rm charge}} \quad & \quad \frac{1}{3} \quad & \quad   \frac{1}{3} \quad & \quad  \frac{1}{3} \quad &
\quad
- 1 \quad & \quad  -1 \\
\hline
\end{array}
$$
would be free of anomalies
\emph{only if} a right--handed neutrino, with $Y_{\nu_R}=-1$, were also present.

\subsection{\sc The Strong CP Problem and the QCD Axion} \label{sec:strongcp}

The QCD action allows in principle the presence of the term
\beq
 {\cal S}_\theta \ = \ \frac{\theta\,g_3^2}{16\,\pi^2} \ \int d^4 x \ \mathrm{Tr} \left( G_{\mu\nu}\, {\widetilde G}^{\mu\nu} \right) \ = \  \frac{\theta\,g_3^2}{32\,\pi^2} \ \epsilon_{\mu\nu\rho\sigma} \ \int d^4 x \ \mathrm{Tr} \left( G^{\mu\nu}\, {G}^{\rho\sigma}\right)  \ ,  \label{Stheta}
\eeq
which is here parametrized in terms of the $SU(3)$ coupling constant $g_3$ and of a dimensionless parameter, $\theta$.
This contribution is a total derivative, and therefore does not affect the classical equations of motion, but can nonetheless play a role in the quantum theory, where non--trivial field configurations can yield non--vanishing contributions to the functional integral. However, this term violates two symmetries, C and P: its presence should have manifested itself in experiments, which do not provide any clues in this respect. One should thus explain why the dimensionless parameter $\theta$ vanishes, or is at least negligibly small.

The Euclidean path integral for QCD actually receives contributions that originate from (anti)selfdual \emph{Euclidean} field configurations, for which
\beq
G_{\mu\nu} \ = \  \pm \, {\widetilde G}{}_{\mu\nu} \ \equiv \ \pm \ \frac{1}{2}\ \epsilon_{\mu\nu\rho\sigma}\ G^{\rho\sigma} \ .
\eeq
These configurations give rise to quantized values $N\,\theta$ for ${\cal S}_\theta$, and are local minima of the classical Euclidean action, as can be seen noting that
\beq
\mathrm{Tr}\left(G_{\mu\nu} \pm {\tilde G}_{\mu\nu}\right)^2 \ = \ 2 \, \mathrm{Tr}\left(G_{\mu\nu} \ {G}^{\mu\nu} \  \pm \ G_{\mu\nu}  {\widetilde G}{}^{\mu\nu}\right) \ \geq \ 0 \ , \eeq
so that
\beq
S \ = \ \frac{1}{2} \int d^4 x   \ \mathrm{Tr} \left(G_{\mu\nu} \ {G}^{\mu\nu}\right)  \ \geq   \  \frac{8 \pi^2} {g_3^2}  |N|  \  . \ \label{scp3}
\eeq
Note that ${\cal S}_\theta$ remains accompanied by an imaginary unit in the Euclidean path integral, and consequently $\theta$ is a real (angular) parameter, defined up to multiples of $2 \pi$.
Instanton (anti-instanton) configurations correspond to the simplest selfdual (anti selfdual) solutions with $N=1$ ($N=-1$): these topological configurations give rise to CP violating effects sized by the  $\theta$ parameter. 

Actually, even if for some reason the original $\theta$ parameter vanished in the QCD Lagrangian, a similar contribution would be generated by the unitary redefinitions (\ref{ckm9}) needed to diagonalize the quark mass matrices.
Indeed, diagonalizing fermion masses leads to terms of the type
\beq
{\cal L}_{\rm mass} \ = \  - \ \sum_i \left(  m_i^u e^{i \alpha_i^u} {\bar u}_L^i {u}_R^i \ + \  m_i^d e^{i \alpha_i^d} {\bar d}_L^i {d}_R^i \ + \ \cdots   \right)   \  , \ \label{scp4}
\eeq
where, as we stressed after eq.~\eqref{ckm10}, the mass parameters are in general complex.
The $U(1)$ chiral rotations
\beq
u^i \to e^{-\frac{i}{2} \gamma_5  \alpha_i^u}  u^i \qquad , \qquad d^i \to e^{-\frac{i}{2} \gamma_5  \alpha_i^d}  d^i  \   \ \label{scp5}
\eeq
that make them real have a color anomaly
\beq
\delta {\cal L} \ = \ \frac{g_3^2}{16 \pi^2}  \sum_i  \left( \alpha_i^u +  \alpha_i^d\right) \mathrm{Tr} \left(G_{\mu\nu} \ {\widetilde G}{}^{\mu\nu}\right)   \  , \label{scp6}
\eeq
and so the redefinitions change the $\theta$ parameter into
\beq
{\bar \theta} \ = \  \theta  \ +  \ \sum_i  ( \alpha_i^u +  \alpha_i^d)  \ \equiv \  \theta \ +   \ \mathrm{arg} \left( \det \ m^q \right) \  , \label{scp7}
\eeq
where
\beq
\det \ m^q \ = \  \det \ m^u \ \det \ m^d
\eeq
is the product of the quark mass matrices.

A non--vanishing effective $\theta$ parameter {\it violates} the CP symmetry of strong interactions, and induces a neutron electric dipole moment of order~\cite{baluni,crewther}
\beq
d_n \ \sim  \ \left| {\bar \theta} \right| \ \frac{e m_{\pi}^2}{m_N^3} \ \sim \ 10^{-16} \left| {\bar \theta} \right| \ e \,cm \ .
\eeq
This effect conflicts with experimental data unless $ {\bar \theta} < 10^{-10}$. It is unlikely that the two contributions to ${\bar \theta}$ from QCD and electroweak interactions, coming from the anomalous
chiral transformation (\ref{scp5}) and of completely different origin, cancel to such a high accuracy. This leads to the so-called {\it strong CP problem}. The problem would be absent if the up-quark mass were zero, since in this case the theta parameter could be removed by an up-quark chiral redefinition. However, a massless up-quark appears excluded by recent lattice simulations (see~\cite{flag}).

There are several potential solutions to the strong CP problem (see, for example,~\cite{nelson-barr}), but in what follows we shall concentrate on the possible existence of a new approximate global symmetry, which points to the existence of a new light pseudoscalar particle, the axion.

  %%%%%%%%%%%%%%%%%%%%%%%%%%%%%%

\subsubsection{\sc  The Peccei--Quinn Solution:  \texorpdfstring{$U(1)_{PQ}$} \ \ and the QCD Axion}
\label{sec:pq}

The solution of the strong CP problem that is still regarded as most elegant and intriguing, also in view of its experimental implications, postulates the existence of a new
approximate Abelian global symmetry $U(1)_{PQ}$ \cite{PQ,axion,rubakov} with a corresponding light pseudo--Goldstone boson, a pseudoscalar particle usually called the axion $a$, and a symmetry breaking scale $f_a$. $U(1)_{PQ}$ ought to have mixed triangle anomalies with the QCD gauge group $U(1)_{PQ} SU(3)_c^2$, so that $Tr (X_{\rm PQ} T^A T^A)$ should not vanish.
The anomaly generates {\it new couplings} in the effective Lagrangian that shift the $\theta$ parameter according to
\beq
\frac{g_3^2}{16 \pi^2} \ \xi \ \frac{a (x)}{f_a} \ \mathrm{Tr} \left(G_{\mu\nu} \ \widetilde{G}{}^{\mu\nu}\right) \ \longrightarrow \ \theta_{\rm eff} \ = \ {\bar \theta} \ +\  \xi \frac{a}{f_a}
\ , \label{pq1}
\eeq
where $\xi$ is model dependent and defines the strength of the axion couplings to matter. The $\theta$ parameter thus
becomes a dynamical quantity that is determined by minimizing the scalar potential of the axion. A theorem due to Vafa and Witten \cite{vw} proves than
the minimum energy in QCD (and, more generally, in vector-like theories) obtains for
\beq
\langle \theta_{\rm eff} \rangle \ = \  0 \ ,
\eeq
so that the strong CP problem is solved if  $\theta_{\rm eff}$
becomes dynamical.

An example, the so-called invisible axion or KSVZ model \cite{ksvz}, will make these considerations more concrete.  The model contains superheavy colored fermions $\Psi$ with vector-like strong interactions, whose masses originate
from Yukawa couplings with a complex scalar $\Phi$ that acquires a large vacuum expectation value. The relevant terms in the effective action are
\beq
{\cal L}_{\rm KSVZ} \ = \  {\bar \Psi} i \gamma^\mu D_\mu \Psi - \left| \partial_\mu \Phi\right|^2 - h ({\bar \Psi}_L \Phi  \Psi_R + {\rm h.c.}) - V(\Phi) + \frac{{\bar \theta} \, g_3^2}{16 \pi^2} \ \mathrm{Tr} \left( G_{\mu\nu} \ {\widetilde G}^{\mu\nu}\right) + \cdots
\ , \label{pq2}
\eeq
where $V$ is the scalar potential. At the classical level, the Lagrangian is invariant under the {\it global } chiral symmetry $U(1)_{PQ}$ that acts as
\beq
\Psi \to e^{i \gamma^5 \alpha} \Psi \qquad , \qquad \Phi \to e^{- 2 i  \alpha} \Phi \ . \label{pq3}
\eeq
The symmetry is spontaneously broken well above the electroweak scale, since we are assuming that
\beq
\langle \Phi\rangle \ = \ \frac{f_a}{\sqrt{2}} \ \gg \ v \ .
\eeq
The complex scalar can be parametrized as
\beq
\Phi \ =  \ e^{ \frac{i a}{f_a}} \left(\frac{f_a+ \varphi}{\sqrt{2}}\right) \ , \label{aphi}
\eeq
where $a$ is the Goldstone boson of the chiral symmetry  $U(1)_{PQ}$ while $\varphi$ is a very heavy scalar that can be integrated-out and ignored in the low-energy dynamics of the axion $a$.
The resulting low-energy action takes the form
\beq
{\cal L}_{\rm KSVZ} \ = \ {\bar \Psi} i \gamma^\mu D_\mu \Psi - \frac{1}{2} (\partial_\mu a)^2 - m_{\Psi} ({\bar \Psi}_L  e^{ \frac{i a}{f_a}}   \Psi_R + {\rm h.c.}) + \frac{{\bar \theta} \, g_3^2}{16 \pi^2}\,
\mathrm{Tr} \left(G_{\mu\nu} \ {\widetilde G}{}^{\mu\nu}\right) + \cdots \ . \label{pq4}
\eeq
One can now remove the axion coupling to the heavy fermions by the field redefinition
\beq
\Psi \ \to \ e^{- \,\frac{ia}{2f_a} \,\gamma_5} \ \Psi \ ,
\eeq
but this acts like a chiral transformation and has an anomaly. Assuming for
simplicity that the heavy fermions $\Psi$ are valued in the fundamental representation of $SU(3)_c$, one thus gets the transformed action
\beq
{\cal L}_{KSVZ} \ = \   {\cal L}_{kin} - m_{\Psi} {\bar \Psi} \Psi + \left({\bar \theta} + \frac{a}{f_a}\right) \ \frac{g_3^2}{16 \pi^2} \mathrm{Tr}\left( G_{\mu\nu} \ {\widetilde G}{}^{\mu\nu}\right) \ + \
\frac{1}{2f_a} \, \partial_\mu a  \, {\bar \Psi} \gamma^\mu \gamma^5 \Psi + \cdots \ . \label{pq5}
\eeq
Note that the $\bar \theta$ parameter can now be removed by shifting the axion according to
\beq
a \ \to \ a \ - \ \bar \theta f_a \ .
\eeq

If the heavy fermions $\Psi$ have a nontrivial hypercharge, this redefinition will induce anomalous
axion couplings to the photon and the $Z$ boson, and if they also transform in a nontrivial representation of $SU(2)_L$ this will also generate axion couplings to the $W_\mu^{\pm}$ gauge bosons. The general lesson is therefore that the invisible axion typically couples to gauge fields. For the time being, the axion seems a true Goldstone boson and, therefore, it ought to be massless. However, in the next section we shall see
that it actually acquires a tiny mass from non--perturbative QCD effects.

At low energies one can also ignore the heavy fermions $\Psi$ to concentrate on the axion, the SM gauge fields and the light quarks, turning the axion couplings to gluons into axion couplings to the light $u,d$ quarks. This can be achieved with the
chiral field redefinition
\beq
 \begin{pmatrix} u \\   d   \end{pmatrix}  \to  e^{ \frac{ia}{2f_a} \gamma_5 Q_a}    \begin{pmatrix} u \\   d   \end{pmatrix}   \quad , \quad {\rm with} \quad Tr Q_a = 1 \ ,  \label{pq6}
 \eeq
  which leads to the equivalent low--energy action
\bea
{\cal L} &=& {\cal L}_{\rm kin.} - \left\{ \begin{pmatrix} {\bar u}_L   {\bar d}_L   \end{pmatrix}  e^{ \frac{ia}{2f_a}  Q_a}
 \begin{pmatrix} m_u &  0 \\ 0 & m_d  \end{pmatrix}   e^{ \frac{ia}{2f_a} Q_a}   \begin{pmatrix} u_R \\   d_R   \end{pmatrix} + {\rm h.c.} \right\}  \nonumber \\
&+& \frac{1}{2 f_a}  \,\partial_\mu a   \begin{pmatrix} {\bar u} &   {\bar d}   \end{pmatrix}  \gamma^\mu \gamma^5 Q_a   \begin{pmatrix} u \\   d   \end{pmatrix}   \ . \label{pq7}
\eea
 This last form is more suited for the chiral Lagrangian approach containing axion-pion couplings, to which we now turn.

%%%%%%%%%%%%%%%%%%%%%%%%%%%%%%%%%%%%%%%%
\subsubsection{\sc  The Chiral Lagrangian and the Axion Potential}
\label{sec:axionchiral}

We shall now see that the QCD axion is typically very light. Consequently, the low-energy axion interactions with hadrons are better described resorting to an effective field theory description of its couplings
to mesons, which are the lightest hadrons, ignoring any other heavier fields. The appropriate low-energy action is then the chiral Lagrangian, which emerges as the mass of the $\varphi$ particle of eq.~\eqref{aphi} is sent to
infinity. The mesonic degrees of freedom are encoded in a $2 \times 2$ unitary matrix $\Sigma $, which transforms under $SU(2)_L \times SU(2)_R$ chiral symmetries as
\beq
\Sigma = e^{\frac{i}{f_{\pi}} \tau^a \pi^a} \qquad , \qquad \Sigma \to U_L \Sigma U_R^{\dagger } \ , \label{cl1}
\eeq
where $U_{L,R} \in SU(2)_{L,R}$, and the low-energy Lagrangian is
\beq
{\cal L} \ = \ - \ \frac{ f_{\pi}^2 }{4} Tr (\partial_\mu \Sigma) (\partial^\mu \Sigma^{-1}) \ -  \ \frac{1}{2} (\partial_\mu a)^2\ + \ \frac{ f_{\pi}^2 }{2} m_0 Tr  ( \Sigma^{\dagger} M + M^{\dagger} \Sigma )  \ , \label{cl2}
\eeq
where $m_0$ is a mass parameter that will determine the pion masses.
In (\ref{cl2}), the meson kinetic term is invariant under $SU(2)_L \times SU(2)_R$ chiral symmetries, while the term linear in $\Sigma$ breaks the chiral symmetry and accounts for
the up and down quark mass terms of QCD. After the chiral redefinition (\ref{pq6}), the quark mass matrix contains the axion field, and takes the form
\beq
M =  e^{ \frac{ia}{2f_a}  Q_a}
 \begin{pmatrix} m_u &  0 \\ 0 & m_d  \end{pmatrix}   e^{ \frac{ia}{2f_a} Q_a}   \quad , \quad {\rm with} \quad Tr Q_a = 1 \ .  \label{cl3}
\eeq
The charged pions do not play an important role in what follows, and can be consistently set to zero for our purposes. For simplicity, we shall also restrict our attention to a diagonal $Q_a$ matrix, deducing the axion mass and its potential from
\beq
\Sigma \ = \ e^{\frac{i}{f_{\pi}} \tau^3 \pi^0} \quad , \quad  Q_a \ =\  \begin{pmatrix} Q_a^u &  0 \\ 0 & Q_a^d  \end{pmatrix} \quad , \quad {\rm with} \quad  Q_a^u + Q_a^d \ = \ 1
 \ .  \label{cl4}
\eeq

The scalar potential derived from  (\ref{cl2}) is then easily computed, and reads
\beq
V(a,\pi^0) \ = \ - f_{\pi}^2 m_0 \left\{  m_u \cos \left(\frac{\pi^0}{f_{\pi}} - Q_a^u \frac{a}{f_a}\right) \ + \  m_d \cos \left(\frac{\pi^0}{f_{\pi}} + Q_a^d \frac{a}{f_a}\right) \right\}  \ .  \label{cl5}
\eeq
Its minimum, for generic choices of $Q_a^u$ and $Q_a^d$, lies at $\langle \pi^0 \rangle = \langle a \rangle=0$, and the resulting mass matrix is
\beq
{\cal M}^2 \ = \  m_0 \begin{pmatrix} m_u+m_d &  \frac{f_{\pi}}{f_a} (Q_a^d m_d-Q_a^u m_u)  \\ \frac{f_{\pi}}{f_a} (Q_a^d m_d-Q_a^u m_u) &   \frac{f_{\pi}^2}{f_a^2}
[ (Q_a^u)^2 m_u + (Q_a^d)^2 m_d]  \end{pmatrix}  \ .  \label{cl6}
\eeq
In the physically relevant limit $f_a \gg  f_{\pi}$, one is thus led to the masses
\beq
m_{\pi}^2 \ = \ m_0 (m_u+m_d) \ , \qquad m_a^2 \ = \ \frac{m_um_d}{(m_u+m_d)^2} \frac{f_{\pi}^2 m_{\pi}^2}{f_a^2} \ .  \label{cl7}
\eeq
Note that the axion mass vanishes in the chiral limit $m_u \to 0$, as expected. Since we work in the limit $m_{\pi} \gg m_a$, the pion field can be integrated out in order to find the low-energy
potential of the axion. This can be readily done by ignoring the quantum fluctuations of the pion, and therefore solving  its {\it classical} field equation and substituting the result into the scalar potential.
The solution is
\beq
\tan \frac{\pi_0}{f_{\pi}} \ = \ \frac{ m_u  \sin (Q_a^u \frac{a}{f_a}) \ - \  m_d  \sin (Q_a^d \frac{a}{f_a} )}{ m_u  \cos (Q_a^u \frac{a}{f_a}) \ + \  m_d  \cos (Q_a^d \frac{a}{f_a}) } \ ,  \label{cl8}
\eeq
and the final low-energy Di Vecchia--Veneziano axion potential is~\cite{dvv}
\beq
V(a) \ = \ - \ f_{\pi}^2 m_{\pi}^2 \sqrt{1 \ - \ \frac{4m_um_d}{(m_u+m_d)^2} \sin^2 \frac{a}{2f_a}} \ .  \label{cl9}
\eeq

Some comments are in order. First, this scalar potential differs from a popular one appearing in the literature, proportional to $\Lambda_{QCD}^4 \cos \left(\frac{a}{f_a}\right)$.  It has the same periodicity $a \to a + 2 \pi f_a$,
but a different shape, and in particular the quartic axion self-coupling is significantly different. It is clear that eq.~\eqref{cl9} is more reliable than the simple cosine potential and becomes properly trivial in the massless limit for the
up quark. It also reproduces correctly the axion mass computed earlier in~\eqref{cl7}.

\subsection{\sc  Comments on Axion Phenomenology}
\label{sec:axioncouplings}

The coupling between axions and photons has suggested promising ways for their detection, following to a large extent the original work of Sikivie~\cite{sikivie}. The most general couplings of the QCD axion to gluons, photon and SM fermions can be parametrized as
\bea
{\cal L}  &=&    {\cal L}_{\rm kin.} \ + \   \frac{a}{f_a}  \frac{\alpha_s}{4 \pi} \ \mathrm{Tr} \left(G_{\mu\nu} \ {\widetilde G}^{\mu\nu}\right)
\ + \  \frac{a}{f_a}   \ \frac{\alpha_{em}}{8 \pi}  \frac{{\cal E}}{N} F_{\mu\nu} \ {\tilde F}^{\mu\nu}   \nonumber \\
&-&  \left(  {\bar u} \,m_u \, u \,+\,  {\bar d} \, m_d \, d  \,+\,  {\bar e}\, m_e\,   e  \right) \  + \ \frac{a}{f_a}\ \partial_\mu J_{PQ}^{\mu \,{(0)}}
\ , \  \label{dfsz11}
\eea
after redefining the Fermi fields so as to remove the axion from the Yukawa couplings. Here
$J_{PQ}^{\mu \,{(0)}}$ is the fermionic Peccei--Quinn axial current
\beq
J_{PQ}^{\mu \,{(0)}} \ \sim  \ \sum_i X_i \, \overline{\psi}{}_i \,\gamma_\mu\,\gamma_5\,\psi_i \ ,
\eeq
the $X_i$ are the PQ charges of the different fermions, and
\beq
\alpha_s \ = \ \frac{g_3^2}{4 \pi} \ , \qquad \alpha_{em} \ = \ \frac{e^2}{4 \pi} \ ,
\qquad 
\frac{\cal E}{N} \ = \ 8 \ \frac{{\mathrm Tr}\left(X_{PQ}\, Q^2 \right)}{{\mathrm Tr}\left(X_{PQ}\, T^A \,T^A \right)}\ ,
\eeq
where $Q$ and $T^A$ denote the electric charge and the $SU(3)$ generators.

In order to use the chiral Lagrangian axion-meson formulation presented in Section~\ref{sec:axionchiral}, one can perform
a further chiral field redefinition~\eqref{pq6} on the first generation up and down quarks, which removes the axion coupling to the gluons and shifts it into the light--quark mass matrix. After the chiral field redefinition and restricting ourselves to the axion couplings to the lightest (up and down) quarks and the photon,
the Lagrangian generalizing eq.~\eqref{pq7} becomes
\bea
{\cal L} &=& {\cal L}_{\rm kin.} \ - \ \left\{ \begin{pmatrix} {\bar u}_L &   {\bar d}_L   \end{pmatrix}  e^{ \frac{ia}{2f_a}  Q_a}
 \begin{pmatrix} m_u &  0 \\ 0 & m_d  \end{pmatrix}   e^{ \frac{ia}{2f_a} Q_a}   \begin{pmatrix} u_R \\   d_R   \end{pmatrix}\  +\  {\rm h.c.} \right\}  \nonumber \\
&+& \frac{1}{ f_a}  \  a  \ \partial^\mu\ J_{\mu,PQ}  \  +  \   \frac{1}{4}  \,a  \ g_{a \gamma \gamma} F_{\mu\nu} \ {\widetilde F}^{\mu\nu}  \ ,  \label{dfsz12}
\eea
where
\bea
g_{a \gamma \gamma}  &=&  \frac{\alpha_{em}}{2 \pi f_a} \left[  \frac{{\cal E}}{N}  \ - \ 6 Tr (Q_a Q^2)  \right] \ ,   \nonumber \\
J_{\mu,PQ} &=&   J_{\mu,PQ}^{(0)}  \ - \  \frac{1}{2} \begin{pmatrix} {\bar u} &   {\bar d}   \end{pmatrix}  \gamma_\mu \gamma^5 \,Q   \begin{pmatrix} u \\   d   \end{pmatrix}   \ , \label{dfsz13}
\eea
where $Q^a$ was defined in eq.~\eqref{cl3} and $Q = \mathrm{diag} \left(\frac{2}{3},-\,\frac{1}{3}\right)$ is the charge matrix of the up/down quarks. 
Note that if the fermions are on-shell, after an integration by parts the typical derivative coupling of the axion to fermions can also be written as
\beq
\frac{1}{f_a} \partial^\mu a \  {\bar \psi} \gamma_\mu \gamma_5 \psi \ = \ - \ 2\, i \,\frac{m_{\psi}}{f_a}\  a \  {\bar \psi}  \gamma_5 \psi  \ . \label{dfsz14}
\eeq

In addition to their potential role in connection with the strong CP problem,  axions are also excellent dark  matter candidates, even for typical values of their masses and couplings to matter.
The axions are being intensively searched experimentally since the $1980$'s. They have peculiar couplings to gauge fields and fermions, as we have seen. In particular, in the presence of electromagnetic fields, they can convert into photons. This property has led to attractive proposals~\cite{sikivie} for dark-matter detection experiments of the ``light shining through the wall" type, using cavities with large magnetic fields inside.
Since axions have highly suppressed couplings to matter, they can easily pass through the wall cavities, but the strong magnetic field can convert them into photons, by Primakoff--like processes~\cite{primakoff}.  

Although originally the typical QCD axion parameters were considered to be $ 10^{-5} < m_a(eV) < 10^{-2}$ and $10^{9}< f_a(GeV) < 10^{12}$, significant recent activity,
both on the experimental side and in theoretical model building, considerably enlarged the axion parameter space. Axions as light as $10^{-24}$ eV were studied as ``fuzzy dark matter" candidates~\cite{tremaine},
with wavelengths of the size of galaxies, and light axions with large decay constants $f_a \sim 10^{16}-10^{17}$ GeV were shown to naturally arise in String Theory.  Axions that are (or not)
dark-matter candidates have interesting astrophysical, cosmological, and experimental signatures, even if they do not have the right properties to solve the strong CP problem. For such axions, the mass $m_a$ and the decay constants $f_a$ are not tied as in eq.~(\ref{cl7}), but they are considered as independent free parameters.  Such axions are typically called ALP's (axion-like particles), while the
name QCD axion is reserved to the one that (hopefully) solves the strong CP problem.
On the theoretical side, axions are also present in SUSY and other extensions of the SM, and play a crucial role in supergravity and String Theory, from various points of view, including
gauge and gravitational anomaly cancellation and the stabilization of moduli fields describing the geometry of the internal space \cite{svrcek-witten}.

\subsection{\sc Some Puzzles with the Standard Model}

As we have seen, the Standard Model is a highly successful description of the strong, weak and electromagnetic interactions, up to the typical energies that are currently accessible at LHC. Still, it is widely regarded as an effective low--energy description, rather than a fundamental theory. There are several reasons for this, which we can now elaborate on.

In general terms, when the Standard Model was conceived, in the 1960's and 1970's, it possessed the striking feature of accounting for the strong, weak and electromagnetic interactions within a renormalizable framework, relying on the Higgs mechanism to describe massive gauge bosons. While renormalizable models still retain their interest, due to their highly predictive nature, during the last decades the emphasis has shifted somewhat. Once
an effective cutoff is introduced, and to date we have no compelling reason to exclude that Nature demands it, the special feature of renormalizable interactions is simply that they dominate the dynamics at low energies, where contributions of higher--dimension operators are suppressed. The ultimate need for a high--energy or small--distance cutoff has long lurked on the Fundamental Interactions, due to the peculiar behavior of gravity at high energies~\cite{gs85,gs86}, and was widely reinforced by the success in describing critical phenomena in discrete models via continuum field theories~(for reviews see \cite{wk,statistical}). Even the four--dimensional Minkowski spacetime, which is the arena of the Standard Model, might be a low--energy approximation, because extra dimensions might be present, albeit too small to leave, so far, detectable signs in experiments. Alternatively, an underlying discrete structure of spacetime might play a role at the Planck scale, where gravity interactions are expected to become stronger than the others. All these deep questions have long been debated, together with a number of others of comparable importance, if of a more exquisitely phenomenological nature, to which we can now turn.

\begin{itemize}

\item \emph{There are no  neutrino masses in the original form of the Standard Model}, but there is now evidence for small neutrino mass differences of order at most $10^{-3}$ eV  (for a review, see, \emph{e.g.}, \cite{RamMoh}). Although Dirac neutrino masses can be added by introducing right-handed (gauge singlet) neutrinos with extremely small Yukawa couplings to the Higgs, this is usually regarded as not very economical and elegant. The reason is that there is no experimental hint of the existence of light sterile right-handed neutrinos, while extremely small values (of order $10^{-12}$) of the corresponding Yukawa couplings would beg for an explanation.  On the other hand, assuming that only left-handed neutrinos exist in the Standard Model, it is impossible to write neutrino masses only relying on renormalizable couplings.  However, we now know that neutrinos are massive. Therefore, their masses and mixings could originate from non--renormalizable couplings, and for this reason they are often regarded as a hint of a new high--energy mass scale beyond the Standard Model itself that determines the corresponding operators. 

An elegant setting for neutrino masses is provided by the ``seesaw'' mechanism~\cite{GMRS}, in which the relevant fermions are the three leptonic doublets $l_i$, together with three heavy sterile neutrinos $N_i$. The relevant Lagrangian for neutrino masses,
\begin{equation}
{\cal L}_\nu \ = \ - \ \sqrt{2} \, {\bar N}\,\lambda\,  l \Phi 
\ - \  
\frac{1}{2}\ {\bar N} C \,M\,{\bar N}^T \ + \ \mathrm{h.c.}  \ , \label{neu5}
\end{equation}
where we are using a matrix notation for flavor indices, so that $\lambda$ and 
$M$ are $3\times 3$ matrices. Moreover, $C$ is the charge--conjugation matrix and $M$ is naturally large, since it is not protected by chiral symmetries, so that integrating out the heavy $N^i$ fermions leads to Weinberg's operator
\begin{equation}
{\cal L}_{\rm Majorana \ mass} \ = \ 
\ (l \Phi)^T C \,\lambda^T\,M^{-1}\,\lambda\, (l \Phi) \ , \label{neu9}
\end{equation}
where $l \Phi$ denotes the $SU(2)$ singlet combination $\epsilon_{ab}\,l^a \Phi^b$. After electroweak symmetry breaking this coupling induces the Majorana mass term
\beq
{\cal L}_{\rm eff} \ = \ \frac{v^2}{2}\,  \nu^T\,C\,\lambda^T\,M^{-1}\,\lambda\,\nu  \,+\, \mathrm{h.c.}  \ .
\eeq

The resulting neutrino mass matrix can be diagonalized by a unitary transformation, which is usually called the MNSP matrix, and contains two more complex phases compared to the corresponding CKM matrix for the quarks. 
Neutrino oscillations are possible as for neutral mesons, and they have been observed.
If the masses in the neutrino sector of the Standard Model have a Majorana origin, there are thus potential new sources of CP violation. This motivated a recent ambitious experimental program aimed at further investigations of neutrino physics.

Cosmology and the experimental data on neutrino oscillations favor small neutrino masses, typically of order $10^{-2}$ eV, and for Yukawa couplings $\lambda_{ij}$ of order one this demands a high mass scale  $M \sim 10^{15}$ GeV. Interestingly enough, this value is close to the energy scale where gauge couplings tend to unify, as we shall recall shortly. For this reason, it is often said that neutrino masses might be also a first hint of new physics at a very high scale $M$. Majorana masses violate the individual and also the overall lepton number $L$ by two units
$\Delta L = 2$. This violation would be potentially observable in neutrino-less double beta decay experiments, in which nuclei of atomic number $Z$ and atomic mass $A$ undergo transformations of the type $(Z,A) \to (Z+2,A) + e^- + e^-$. However, no decays of this type have been detected to date, and this sets bounds on a linear combination of neutrino masses and mixings.

\item \emph{The Standard Model possesses another peculiarity: its gauge group, $SU(3) \times SU(2) \times U(1)$ is a product of three factors, and there are thus three independent coupling constants.} This might be merely the result of a spontaneous breaking, at high energies, of a more fundamental gauge group, with a single coupling constant, and there are three canonical examples of this type: $SU(5)$~\cite{su51}, $SO(10)$~\cite{so10} and $E_6$~\cite{e6}. If this were the case, there would be additional interactions of very short range mediated by 12, 33 or 66 ultra--massive gauge vectors. Most interestingly, any of the tree choices would provide a rationale for charge quantization, since in all cases the electric charge operator would be a traceless matrix. The electric charges of quarks and leptons belonging to the same irreducible matter representations would thus be proportional, and actually with proper factors of $\frac{1}{3}$. However, this richer setup would bring along proton decay, which was not detected so far, and in order to suppress it the scale where the full symmetry would be recovered should lie not below $10^{16}-10^{17}$ GeV, and thus intriguingly close to the Planck scale $M_P = 10^{19}$ GeV, where Einstein gravity is expected to become strong. The last two gauge groups would also predict the existence of sterile neutrinos, singlets with respect to all gauge groups of the Standard Model. A key issue was then finding dynamical clues for the convergence of the three independent gauge couplings of the Standard Model into a single value corresponding to the high--energy unification. This is possible, in principle, since the couplings acquire a slow (logarithmic) energy dependence due to quantum effects, and the estimated range for their convergence is indeed around $2 - 3 \times 10^{16}$ GeV~\cite{su52,susyunif}. However, the convergence of the couplings becomes strikingly more precise in the minimal supersymmetric extension of the Standard Model (MSSM) that we shall describe shortly.

\item \emph{The values of quark/lepton masses and mixings are completely mysterious.} Many (discrete and continuous) symmetry principles have been explored over the years to account for them, but there is no general consensus on their origin. Quark and lepton hierarchies might even hide new flavor symmetries, or a geometrical origin related to wave function profiles in a higher-dimensional space~\cite{DDG}. Fermionic masses are protected by approximate chiral symmetries, so much so that the leading quantum corrections to them depend at most logarithmically on the cutoff scale, but bosonic masses are far more sensitive to the cutoff, since their corrections depend quadratically on it. If one takes into account the Planck scale of gravity, or the large mass scales that emerge from neutrino masses or grand unification, there is a clear stability issue~\cite{hierarchy} when quantum correction are taken into account. How can one justify the apparent fine tuning?

\item \emph{The Standard Model has no viable {dark--matter} candidate}, many signs of which have long emerged in the dynamics of galaxies and galaxy clusters, which seem to spin too fast if only luminous matter is present in them \cite{nfw}. The most pressing problem from the Particle Physics vantage point is perhaps understanding the origin and the properties of dark matter candidate(s) that, as was ascertained in the last decade or so, ought to provide about $30 \%$ of the energy density of the Universe \cite{bertone}. 

\item \emph{The Standard Model has a strong CP problem}. The most popular solution, as we have seen, postulates the existence of a new light particle, the axion. Scalar particles of this type, with only derivative couplings, are ubiquitous in String Theory, and often play a central role in its quantum consistency, but none have been detected so far.
\end{itemize}

If one aims at a complete description of all known interactions, gravity should be included into the picture, but this creates additional problems.

\begin{itemize}
\item  \emph{Einstein Gravity cannot be incorporated into a renormalizable field theory framework}. Its singular ultraviolet behavior~\cite{gs85,gs86} is improved in supergravity, within special ``pure'' models that include a single supermultiplet. Although these are usually regarded as toy models, recent explicit computations have excluded the onset of divergences in the maximal $N=8$ four--dimensional supergravity up to and including five loops~\cite{sugra_div_recent}.

\item \emph{The hierarchy problem}. When gravity is included, the Fundamental Interactions involve the two vastly different electroweak and Planck scales, and large quantum corrections can destabilize the former. As we have anticipated, supersymmetry softens the ultraviolet behavior, removing quadratically divergent corrections, which can grant a stable hierarchy~\cite{hierarchy1,hierarchy2,hierarchy3,hierarchy4,hierarchy5}. It also allows for a precise unification of gauge couplings at high energies~\cite{su51,su52,susyunif1,susyunif2,susyunif3} (for a review, see~\cite{keith}).

\item \emph{Dark Matter}. Most of the matter present in the Universe is of unfamiliar type, and is generically dubbed ``dark matter''. A special ingredient accompanying the supersymmetric extension of the Standard Model, a discrete ``R-parity'' \cite{farrar-fayet1,farrar-fayet2}, can grant the stability of ordinary matter and of a realistic dark-matter candidate~\cite{dark_matter}.

\item \emph{Gravity brings along a cosmological constant problem}. Comparing the microscopic estimate from Quantum Field Theory and the macroscopic estimate from the history of the Universe,  one arrives at a stunning result:
\beq
\frac{\Lambda_{\mathrm macro}}{\Lambda_{\mathrm micro}} \  \sim \ 10^{-120} \ .
\eeq
This is perhaps the most embarrassing discrepancy in the history of Physics (for reviews, see~\cite{weinberg_cc}). There is no general consensus, at present, on whether, or how, broken supersymmetry and also String Theory can help with this issue.
\end{itemize}

\section{\sc The Minimal Supersymmetric Standard Model } \label{sec:mssm}

We can now turn to the Minimal Supersymmetric Standard Model (MSSM) \cite{fayet1,fayet2,fayet3,dg}. As we shall see, the simplest renormalizable setting for breaking supersymmetry spontaneously that we discussed in Section~\ref{sec:broken_global} is not viable, since it does not allow for the breaking of electroweak symmetry and the generation of realistic particle spectra\footnote{For notable important early attempts, see \cite{fayet1,fayet2,fayet3}.}. All these features can be accommodated by introducing soft terms, which do not ruin the ultraviolet properties and can be still associated to the spontaneous breaking of supersymmetry, albeit in a hidden sector that communicates only indirectly with the SM. The MSSM associates to the leptons $l^i$ and the quarks $q^i$ complex scalars, which are called {\it sleptons} $(\tilde{l}{}^i)$ and {\it squarks} $(\tilde{q}{}^i)$. In addition, it associates spin-$\frac{1}{2}$ Majorana fermions $\lambda$, which are generically called \emph{gaugini}, to the gauge fields. 

There is an important difference with respect to the SM, the need for \emph{two Higgs doublets}. There are two reasons for this. The first is that their fermionic partners (the Higgsini) are chiral, and consequently with a single Higgs multiplet the MSSM would be inconsistent, due to gauge anomalies. The second is related to the holomorphy of the superpotential, which forbids the use of both $\Phi$ and the charge conjugate field ${\tilde \Phi} = \Phi^c$ to construct quark Yukawa couplings. The two Higgs doublets are accompanied by Weyl fermion doublets, the so--called \emph{Higgsini} $\psi_{h_{1}}$ and $\psi_{h_{2}}$.

The full spectrum of the MSSM thus comprises the following multiplets:
\begin{itemize}
\item Lepton doublet multiplets :
$$L_i \ : \ l^i = \begin{pmatrix} \nu^i \\ e^i
\end{pmatrix} _L  \ , \  {\tilde l}^i = \begin{pmatrix} {\tilde \nu}^i \\ {\tilde e}^i
\end{pmatrix} _L  \qquad \in \qquad ({\bf 1,2})_{Y=-1} \ , $$
\item Lepton singlet multiplets :
$$ E^{ic} : \quad  e^{ic}  \ , \   {\tilde e}^{ic} \qquad \in \qquad ({\bf 1,1})_{Y=2}\ , $$
\item Quark doublet multiplets :
$$ Q^i : \quad q^i = \ \begin{pmatrix} u^i \\ d^i
\end{pmatrix} _L  \ , \ {\tilde q}^i = \ \begin{pmatrix} {\tilde u}^i \\ {\tilde d}^i
\end{pmatrix} _L   \qquad \in \qquad ({\bf 3,2})_{Y=1/3} \ ,$$
\item Up-quark multiplets :
$$  U^{ic} \ : \quad u^{ic} \ , \  {\tilde u}^{ic} \qquad \in \qquad ({\bf {\bar 3},1})_{Y=-4/3} \ , $$
\item Down-quark multiplets :
$$ D^{ic} \ : \quad  d^{ic} \ , \ {\tilde d}^{ic} \qquad \in \qquad ({\bf {\bar 3},1})_{Y=2/3} \ , $$
\item Gauge multiplets :
$$  V_1 \ : \quad (B_\mu, \lambda_1) \  , \qquad  V_2 \ : \ (A_\mu^a, \lambda_2^a)  \ , \qquad   V_3 \ : \ (G_\mu^A, \lambda_3^A) \ , $$
\item Higgs multiplets :
$$ H_1 \ : \quad h_1 = \begin{pmatrix} h_1^{0} \\ h_1^{-}
\end{pmatrix}  \ , \ \psi_{h_1} = \begin{pmatrix} \psi_{h_1}^{0} \\ \psi_{h_1}^{-}
\end{pmatrix}   \qquad \in \qquad ({\bf 1,2})_{Y=-1} \ , $$
$$ H_2 \ : \quad  h_2 = \begin{pmatrix} h_2^{+} \\ h_2^{0}
\end{pmatrix}  \ ,\ \psi_{h_2} = \begin{pmatrix} \psi_{h_2}^{+} \\ \psi_{h_2}^{0}
\end{pmatrix}   \qquad \in \qquad ({\bf 1, 2})_{Y=1} \ . $$
\end{itemize}
In this collection of fields the superscript $c$ indicates charge conjugation, the three generations of matter are accounted for by the label $i=1,2,3$, while the $SU(3)\times SU(2)\times U(1)_Y$ quantum numbers shown on the side complete their definitions. It is also convenient to introduce for vector multiplets the compact matrix notation
\be
V_3 \ = \ V_3^A \ \frac{\lambda^A}{2} \ , \qquad V_2 \ = \ V_2^a \ \frac{\tau^a}{2} \ ,  \label{mssm01}
\ee
where the $\lambda^A$ are $SU(3)$ Gell-Mann matrices and the $\tau^a$ are $SU(2)$ Pauli matrices.
In all cases capital letters, as for instance $L_i$, denote the corresponding superfields.

One can distinguish three type of contributions to the MSSM Lagrangian, so that
\be
{\cal L}_{\rm MSSM}  \ = \ {\cal L}_{\theta^4} \ + \ {\cal L}_{\theta^2} \ +\  {\cal L}_{\rm soft}   \ .   \label{mssm02}
\ee
Let us begin by describing the supersymmetric portion of the Lagrangian, which corresponds to the first two contributions. This will also explain the unavoidable need for the third.
The first contribution to the Lagrangian,
\bea
 {\cal L}_{\theta^4} &=& \int d^4 \theta \left[ Q^{i \dagger} e^{2 g_3 V_3 + 2 g_2 V_2 + \frac{1}{3} g_1 V_1} Q^i \ +  \ U^{i c \dagger} e^{-2 g_3 V_3 - \frac{4}{3} g_1 V_1} U^{ic} \right.
\nonumber \\
&+& \left. D^{i c \dagger} e^{-2 g_3 V_3 + \frac{2}{3} g_1 V_1} D^{ic} \ + \  L^{i \dagger} e^{2 g_2 V_2  -  g_1 V_1} L^i \right. \nonumber \\ &+&  \left. E^{i c \dagger} e^{2 g_1 V_1} E^{ic} \ + \ H_1^{\dagger} e^{2 g_2 V_2 - g_1 V_1} H_1 \ + \ H_2^{\dagger} e^{2 g_2 V_2 + g_1 V_1} H_2 \right] \ ,
 \    \label{mssm03}
\eea
originates from complete, and thus real, $\theta$ integrals, and describes the covariant kinetic terms for the chiral multiplets.
The second contribution,
\beq
{\cal L}_{\theta^2} \,=\, \int d^2 \theta  \left[  \frac{1}{2} \,\mathrm{Tr}\, W_3^{\alpha} W_{3,\alpha}  \,+\,  \frac{1}{2}\, \mathrm{Tr}\, W_2^{\alpha} W_{2,\alpha} \,+\,   \frac{1}{4}\, W_1^{\alpha} W_{1,\alpha}
+ {\cal W}_{\rm MSSM}  \right] \,+\, {\rm h.c.}   \  ,  \label{mssm04}
\eeq
originates from holomorphic, chiral $\theta$ integrals and their conjugates, and includes vector kinetic terms and the superpotential
\beq
{\cal W}_{\rm MSSM} \,=\, h_{ij}^u Q^i  U^{jc} H_2 \,-\,  h_{ij}^d Q^i  D^{jc} H_1 \,-\, h_{ij}^e L^i  E^{jc} H_1 \,+\, \mu H_1 H_2 \ , \label{mssm1}
\eeq
where one can always make $\mu$ real by redefining the overall phase of the product $H_1\,H_2$. Holomorphy is a key property in supersymmetric constructions, and no other options are available for supersymmetric mass terms. The notation used for the Lagrangian is somewhat compact, and all products are meant to build $SU(2)$ singlets. For example, since $H_1,H_2$ are both in the ${\bf 2}$ of $SU(2)_L$, it is implicit that in the product ${\bf 2} \times {\bf 2} = {\bf 3}+ {\bf 1}$ one has to select the singlet component, so that
\beq
H_1 H_2 \ = \   \epsilon^{mn} \, H_{1m}\  H_{2n} \ = \ H_1^0 H_2^0 \ - \ H_1^- H_2^+ \ .
\eeq

\subsection{\sc The Higgs Potential}

The portion of the supersymmetric scalar potential for the Higgs fields can be deduced from the general setting of Section~\ref{sec:broken_global} and reads
\bea
V_{\mathrm{SUSY}}(h_1,h_2) &=& \mu^2\left( h_1^{\dagger}\, h_1  \ + \  h_2^{\dagger}\, h_2 \right)  \nonumber \\
&+& \ \frac{g^2 + g'^2}{8} \left( h_1^{\dagger} h_1 \,-\,  h_2^{\dagger} h_2\right)^2 \ + \  \frac{g^2}{2} \left( h_1^{\dagger} h_2\right) \left(h_2^{\dagger} h_1\right) \ .\label{pot_mssm}
\eea
Its quartic couplings only originate from D-terms, whose
Higgs content can be cast in the form
\be
D_Y \ = \ \frac{g'}{2} \left(|h_1|^2\ -\ |h_2|^2\right) \quad , \qquad D^a \ = \ - \ \frac{g}{2} \left(h_1^\dagger \tau^a h_1 \ + \ h_2^\dagger \tau^a h_2\right)   \ , \label{mssm5}
\ee
using for the Pauli matrices the standard completeness conditions \beq
\tau_{mn}^a \, \tau_{pq}^a \ =\  2 \,\delta_{mq} \,\delta_{np} \ - \ \delta_{mn} \,\delta_{pq} \ .
\eeq

The potential~\eqref{pot_mssm} is the sum of positive terms, and the minimum is obtained when all vacuum values vanish. Consequently, the electroweak symmetry cannot be broken with this supersymmetric Lagrangian, and soft terms are needed. Before addressing them let us note that the MSSM superpotential  (\ref{mssm1}) also generates the Yukawa couplings
\beq
- \ { {\cal L}_{\rm MSSM}^{\rm Yuk.}\ = \  h_{ij}^u q^i  u^{jc} h_2 \ + \  h_{ij}^d q^i  d^{jc} h_1 \ + \  h_{ij}^e l^i  e^{jc} h_1 \ + \ \mathrm{h.c.} } \ . \label{mssm001}
\eeq
Since $h_2$ ($h_1$) only couple to up (down) quarks, no flavor--changing neutral current (FCNC) effects are generated from the Higgs sector of the MSSM.

As we have seen, the soft terms of eq.~\eqref{soft6} are inevitable to obtain the electroweak symmetry breaking, but they are also needed to give large enough masses to superpartners, while also bypassing the sum rules of Section~\ref{sec:sum_rules}. They could be generated by field theory loops, transmitting
effects that originate in a (hidden) sector breaking supersymmetry, or by supergravity, in ways that were already touched upon in Section~\ref{sec:broken_global} and on which we shall elaborate further in Section~\ref{sec:SUGRA}. Focusing for simplicity on non--holomorphic soft scalar masses $m^{2,i} _{\tilde q}$ that are diagonal in the family/generation space, the most general collection of soft terms is 
\bea
  {\cal L}_{\rm soft} &=& - \ m^{2,i} _{\tilde q} |{\tilde q}^i|^2 \ - \ m^{2,i} _{\tilde u} |{\tilde u}^{ic}|^2 \ - \  m^{2,i} _{\tilde d} |{\tilde d}^{ic}|^2  \ - \ m^{2,i} _{\tilde l} |{\tilde l}^i|^2 \ - \
m^{2,i} _{\tilde e} |{\tilde e}^{ic}|^2 \nonumber \\  &-& m_{h_1}^2 |h_1|^2  \ - \ m_{h_2}^2 |h_2|^2 \ - \  \left( b_{12} h_1 h_2 \ + \ A_u^{ij}  {\tilde q}^i  {\tilde u}^{jc}  h_2 \ - \  A_d^{ij}  {\tilde q}^i  {\tilde d}^{jc} h_1 \ - \ A_e^{ij}  {\tilde l}^i  {\tilde e}^{jc}    h_1 \right.  \nonumber \\
&+& \left. \frac{M_1}{2} \lambda_1 \lambda_1 \ + \ \frac{M_2}{2} \lambda_2^a \lambda_2^a
\ + \ \frac{M_3}{2} \lambda_3^A \lambda_3^A \ + \ {\rm h.c.} \right) \ , \label{mssm4}
\eea
where $\lambda_{1,2,3}$ denote the gaugini that are superpartners of the $U(1)$, $SU(2)$ and $SU(3)$ gauge bosons. 

Taking the soft contributions into account, the Higgs potential becomes
\bea
V &=& \left(\mu^2+m_{h_1}^2\right) h_1^{\dagger}\, h_1  \ + \  \left(\mu^2+m_{h_2}^2\right) h_2^{\dagger} \,h_2 \ + \ b_{12}\left(h_1\,h_2 \ + \ \mathrm{h.c.}\right) \nonumber \\
&+& \ \frac{g^2 + g'^2}{8} \left( h_1^{\dagger} h_1 \,-\,  h_2^{\dagger} h_2\right)^2 \ + \  \frac{g^2}{2} \left( h_1^{\dagger} h_2\right) \left(h_2^{\dagger} h_1\right) \ ,
\label{pot_mssm4}
\eea
and in terms of the neutral and charged components it reads
\bea
 V &=& \left(\mu^2\,+\,m_{h_1}^2\right) \left(|h_1^{0}|^2 \,+\,|h_1^{-}|^2 \right)  \ + \ \left(\mu^2\,+\,m_{h_2}^2\right)  \left(|h_2^{+}|^2 \,+\,|h_2^{0}|^2  \right) \nonumber \\ &+& b_{12}  \left(h_1^0 h_2^0 \,-\,  {h}_1^{-} {h}_2^{+} \,+ \, {\rm h.c.}\right)  \ + \ \frac{g^2 \,+\, g'^2}{8} \left( |h_1^0|^2\,+\, |h_1^-|^2 \,-\,  |h_2^+|^2 \,-\,  |h_2^0|^2 \right)^2 \nonumber \\ &+&  \frac{g^2}{2} \left| {\bar h}_1^{0} h_2^+ \ + \ {\bar h}_1^{-} h_2^0 \right|^2
  \ . \label{mssm06}
\eea

If one only allows \emph{v.e.v.}'s for the uncharged scalar components of the two Higgs doublets, 
\beq
\langle h_1^0 \rangle \ = \ v_1 \ , \qquad \langle h_2^0 \rangle \ = \ v_2 \ , \label{mssm2}
\eeq
in order not to break electromagnetism, the actual potential to be minimized reduces to 
{
\bea
V &=& \left(\mu^2\,+\,m_{h_1}^2\right) |h_1^0|^2  \ + \ \left(\mu^2\,+\,m_{h_2}^2\right) |h_2^0|^2 \ + \ b_{12} \left(h_1^0 h_2^0 \,+\, {\bar h}_1^{0} {\bar h}_2^{0}\right) \nonumber \\ &+& \frac{g^2 \,+\, g'^2}{8} \left(|h_1^0|^2 \,-\,  |h_2^0|^2\right)^2  \ , \label{mssm6}
\eea
}
where $\mu$, as we have anticipated, can be considered a real coupling.

Note that the quartic terms are governed solely by the gauge couplings, a fact that will have important consequences for the Higgs mass. The quartic terms have actually the flat directions
\beq
h_2^0 \ = \ e^{i \alpha} \, h_1^{*\,0} \ , 
\eeq
along which the Higgs potential reduces to
\beq
V \left(h_1^0,e^{i \alpha} h_1^{*0} \right)  \ = \ \left(2 \mu^2\,+\,m_{h_1}^2 \,+\, m_{h_2}^2 \,+\, 2 b_{12} \cos \alpha\right) |h_1^0|^2 \ .
\eeq
This residual potential is bounded from below
at infinity if this term is positive for any $\alpha$, which is guaranteed if the inequality
\beq
2 \mu^2\ + \ m_{h_1}^2 \ + \ m_{h_2}^2 \ \geq \ 2 \,\left|b_{12}\right| \label{mssm7}
\eeq
holds.

Electroweak symmetry breaking demands that the quadratic portion of the scalar potential~\eqref{mssm6} have at least one negative eigenvalue. However, the condition (\ref{mssm7}) excludes that the trace of the matrix, which coincides with its left portion, be negative, so that at most one of its eigenvalues can
be negative. This happens if the determinant of the quadratic portion in~\eqref{mssm6} is negative, which translates into the inequality
\be
\left(m_{h_1}^2 \ +\ \mu^2\right) \left(m_{h_2}^2 \ +\ \mu^2\right) \ \leq  \ b_{12}^2 \ . \label{mssm8}
\ee
The two conditions (\ref{mssm7}) and (\ref{mssm8}) are compatible with each other only if $m_{h_1} \not = m_{h_2}$. Summarizing, two different soft Higgs masses are necessary to attain electroweak symmetry breaking, while also avoiding that the Higgs potential be unbounded from below.

Note that $b_{12}$ is the only possibly complex coefficient in eq.~(\ref{mssm6}), but can be made real and negative by a phase redefinition of $h_1^0,h_2^0$. The minimization of the scalar potential leads to
\bea
&& \left[ m_{h_1}^2 \ + \ \mu^2 \ +\  \frac{g^2 \,+\, g'^2}{4} \left(|h_1^0|^2 \,-\, |h_2^0|^2 \right)   \right]  h_1^0 \ + \ b_{12} \, h_2^{0*} \ = \  0 \ , \nonumber \\
&&  \left[ m_{h_2}^2\ + \ \mu^2 \ - \ \frac{g^2 \,+\, g'^2}{4} \left(|h_1^0|^2 \,-\, |h_2^0|^2 \right)   \right]  h_2^0 \ + \  b_{12} \, h_1^{0*} \ = \ 0 \ ,  \label{mssm9}
\eea
and combining these equations one can see that $\langle h_1^0\,h_2^0 \rangle$ is real and positive. One can then conclude that $\langle h_1^0\rangle$ and $\langle h_2^0\rangle$ are both real and positive, after an overall hypercharge rotation, which multiplies them by opposite phases.
As a result, the ground state of the MSSM does not involve any complex parameters, and thus respects CP.

Letting now
\be
\tan \beta \ = \ \frac{v_2}{v_1}   \ , \label{mssm11}
\ee
the minimization conditions become
\bea
&& m_{h_1}^2 \ + \ \mu^2 \ + \ \frac{g^2 \,+\, g'^2}{4} \left(v_1^2 \ - \ v_2^2 \right) \ + \  b_{12}  \tan \beta \ = \ 0 \ , \nonumber \\
&&  m_{h_2}^2 \ + \ \mu^2 \ -\  \frac{g^2 \,+\, g'^2}{4} \left(v_1^2 \,-\, v_2^2 \right)   \ + \  \frac{b_{12}}{ \tan \beta}   \ =  \ 0 \ .  \label{mssm12}
\eea
Adding them gives
\be
\sin 2 \beta \ = \ - \ \frac{2 b_{12}}{m_{h_1}^2+m_{h_2}^2+2 \mu^2}  \   \label{mssm13}
\ee

\subsection{\sc Electroweak Symmetry Breaking and Higgs Spectrum}
From the kinetic terms of the Higgs doublet, one can see that the gauge boson masses are 
\be
M_Z^2 \ = \ \frac{g^2 \,+ \, g'^2}{2} \left(v_1^2+v_2^2\right) \ , \quad M_W^2 \ = \ \frac{g^2}{2} \left(v_1^2\,+\,v_2^2\right)  \ ,  \label{mssm14}
\ee
so that the $\rho$ parameter remains one. The minimization conditions (\ref{mssm12}) become
\bea
&& m_{h_1}^2 \ + \ \mu^2 \ + \ \frac{1}{2} M_Z^2 \cos 2 \beta \ +  \ b_{12}  \tan \beta \ = \ 0 \ , \nonumber \\
&&  m_{h_2}^2 \ +\  \mu^2\  -   \  \frac{1}{2} M_Z^2 \cos 2 \beta  \ + \  \frac{b_{12}}{ \tan \beta} \ = \ 0 \ ,  \label{mssm15}
\eea
and therefore the $Z$--boson  mass,
\beq
M_Z^2 \ = \ -\ m_{h_1}^2 \ - \ m_{h_2}^2 \ - \ 2 \mu^2 \ + \ \frac{m_{h_2}^2 \ - \ m_{h_1}^2}{\cos 2 \beta} \ ,  \label{mssm16}
\eeq
depends on four parameters. These considerations should be supplemented by a global stability analysis aimed at detecting vacua with lower energy, in order to characterize the (meta)stability of these.

The two Higgs doublets contain altogether eight real degrees of freedom. Three of them are absorbed in the Higgs mechanism: one real pseudoscalar is absorbed by $Z$ and one complex pseudoscalar pair is
absorbed by $W^\pm$.  This leaves five physical degrees of freedom: two neutral scalars that we now denote by $h$ and $H$, where $h$ by convention is the lightest of the two, a neutral pseudoscalar $A$ and a charged Higgs scalar pair $H^{\pm}$.

One can diagonalize the mass matrix determined by the potential~\eqref{mssm06}, taking eq.~\eqref{mssm11} into account, independently in these different sectors, and the results are as follows:
\begin{itemize}
\item One of the eigenvalues of the neutral pseudoscalar mass matrix vanishes, and the corresponding field is absorbed by the longitudinal component of $Z$ in the Higgs mechanism, while the other pseudoscalar $A$ is physical and its squared mass is given by
\be
M_A^2 \ =\  - \ \frac{2 b_{12}}{\sin 2 \beta} = m_{h_1}^2+m_{h_2}^2 + 2 \mu^2  \ ,  \label{higgsmass3}
\ee
where the first two contributions arise from soft terms.
\item The two eigenvalues of the neutral scalar mass matrix can be expressed as
\bea
&& M_h^2 \ = \ \frac{1}{2}  \left[ M_A^2 \,+\, M_Z^2 \,-\, \sqrt{(M_A^2 \,-\, M_Z^2)^2 \,+\, 4 \sin^2 2 \beta M_A^2 M_Z^2 } \right] \ , \nonumber \\
&& M_H^2 \ = \ \frac{1}{2}  \left[ M_A^2 \,+\, M_Z^2 \,+\, \sqrt{(M_A^2 \,-\, M_Z^2)^2 \,+\, 4 \sin^2 2 \beta M_A^2 M_Z^2 } \right]  \ ,  \label{higgsmass5}
\eea
where $M_A$ and the angle $\beta$ of eq.~\eqref{mssm13} depend explicitly on soft terms, while $M_Z$ is defined in eq.~\eqref{mssm16}.
Note that, according to these tree-level results, the lightest of these Higgs bosons would be lighter than the $Z$ boson. However, perturbative quantum corrections, which are largely driven by Yukawa couplings, suffice to invert the inequality, consistently with the masses of the Standard Model Higgs particle (125 GeV) and of the $Z$ boson (90 GeV).

\item[$\bullet$] Charged Higgs Masses

One charged Higgs pair is absorbed in the Higgs mechanism, while the other is physical, and its mass is
\be
M^2_{H^{\pm}} \ = \   M_W^2 \ + \ M_A^2 \ .  \label{higgsmass9}
 \ee
\end{itemize}

After electroweak breaking the superpotential (\ref{mssm1}) generates supersymmetric fermionic mass terms via the vacuum value $v_1$ and $v_2$ that we identified, which read
\beq
- \, {\cal L}_{\rm mass} = h_{ij}^u v_2 u_L^i  u^{jc}  + h_{ij}^d v_1 d_L^i  d^{jc}  \ +\  h_{ij}^e v_1 l^i  e^{jc}  + \mu \left(  \psi_{h_1}{}^0\,\psi_{h_2}{}^0 - \psi_{h_1}{}^-\,\psi_{h_2}{}^+\right) \ + \ \mathrm{h.c.} \label{mssm3}
\eeq
in the two--component notation. Alternatively, in the four--component notation these mass terms become
\bea
- \  {\cal L}_{\rm mass} &=&
 h_{ij}^u v_2 \bar{u}_R^{j}\,u_L^i    \ + \  h_{ij}^d v_1 \bar{d}_R^j  d^{i}{}_L  \ +\  h_{ij}^e v_1 \bar{e}_R^j  l^{i}{}_L   \nonumber  \\ &+& \mu \left(  \psi_{h_1\,L}^{0\,T}\,C\,\psi_{h_2\,L}^0 \ - \ \psi_{h_1\,L}^{-\,T}\,C\,\psi_{h_2\,L}^+\right) \ + \ \mathrm{h.c.} \ . \label{mssm32}
\eea
These contributions are complete for chiral multiplet fermions, which include SM particles and partners of the two Higgses. Their masses are not affected by soft terms. 

%%%%%%%%%%%%%%%%%%%%%%%%%%%%%%%%%%%%%%%%%%%%%%%%%%%%%%%%%%%%%%%%%%%%%%%%%%

%%%%%%%%%%%%%%%%%%%%%%%%%%%%%%%%%%%%%%%%%%%%%%%%%%%%%%%%%%%%%%%%%%%%%%%%%%%%%%%%
\subsection{\sc Further Details on the MSSM Spectrum}

The spectrum and the interactions of the MSSM are rather involved, and therefore we shall content ourselves with a brief discussion. The model comprises five different types of particles:

\begin{itemize}

\item[$\bullet$] Standard Model and Higgs particles, which we have already discussed;

\item[$\bullet$]  squarks $\left({\tilde q}^i, {\tilde u}^{ic},  {\tilde d}^{ic}\right)$ and sleptons $\left({\tilde l}^i, {\tilde e}^{ic}\right)$;

\item[$\bullet$] gluinos $\lambda_3^A$;

\item[$\bullet$] neutral fermionic superpartners
$\left( \psi_{h_1}^0, \psi_{h_2}^0, \lambda_1, \lambda_2^3\right)$. The corresponding mass eigenstates are usually called {\it neutralini};

\item[$\bullet$] charged fermionic superpartners
$\left(\psi_{h_1}^-, \psi_{h_2}^+, \lambda_2^+, \lambda_2^-\right)$. The corresponding mass eigenstates are usually called {\it chargini}.
\end{itemize}
Here and in what follows, the index $i=1,2,3$ continues to refer to the three generations of quarks and leptons.  We already discussed the scalar and fermionic spectrum originating from the Higgs multiplets, and there is nothing special to add about gluinos, since the only contribution
to their mass comes from the  Majorana mass parameter called $M_3$ in  eq.~(\ref{mssm4}). Let us therefore concentrate on squarks and sleptons, the neutralino sector and the chargini.

\subsubsection{\sc Squarks and Sleptons}

The main problem concerning matter fields in the MSSM has to do with the typical presence of FCNC processes. In the Standard Model, the CKM mixing of the charged--current interactions is accompanied by the complete diagonalization of fermionic mass terms. In the MSSM one can still follow the same procedure, performing identical redefinitions for quarks (leptons) and squarks (sleptons), according to
\be
u_L \ = \ V_L^u \,u'_L \quad , \quad {\tilde u}_L \ = \ V_L^u \,{\tilde u'}_L \quad , \quad  u^c \ = \ (V_R^u)^{\dagger}\, u^{'c} \quad, \quad  {\tilde u}^c \ =\  (V_R^u)^{\dagger} \,{\tilde u}^{'c}  \quad , \quad {\rm etc} \ ,  \label{squarks3}
\ee
and in this fashion the fermion masses are again diagonalized,
\be
(V_L^u)^{t} \,m^u \, (V_R^u)^{\dagger} \ = \ m_{\rm diag}^u \quad , \quad (V_L^d)^{t} \, m^d  \, (V_R^d)^{\dagger}\  =  \ m_{\rm diag}^d  \quad , \quad (V_L^e)^{t} \, m^e \, (V_R^e)^{\dagger} \ = \  m_{\rm diag}^e
\ .  \label{squarks03}
\ee
However, this property does not extend, in general, to squark (or slepton) masses. Still, this procedure has the advantage of yielding diagonal gluino and gaugino couplings, for instance
\be
-
 i \, \sqrt{2} \   g_3 \ {\tilde q}^{i*}
 \,T^A  \, q^i \, \lambda_3^A \ = \  - \ i \, \sqrt{2}  \  g_3 \ {\tilde q}^{'i*}
 \,T^A q^{'i} \, \lambda_3^A \ .  \label{squarks4}
\ee
This choice, often called super-CKM basis, is preferred in the literature, since the alternative option of diagonalizing all fermionic and scalar mass terms via two different sets of unitary transformations would make the couplings in eq.~\eqref{squarks4} non diagonal. Either way, there are in general violations of the GIM mechanism in processes mediated by superpartners.

\begin{figure}[ht]
\begin{center}
    \includegraphics[width=3in]{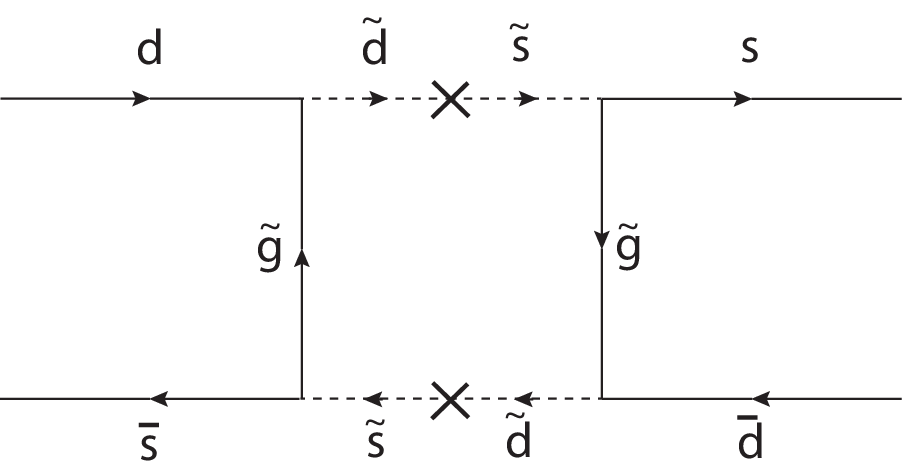}
\end{center}
\caption{A gluino--squark diagram contributing to $K^0- {\overline K^0}$ mixing. The crosses indicate mixed mass terms for the squarks. Here the arrows are again entering (outgoing) for incoming particles (antiparticles) and outgoing (entering) for outgoing particles (antiparticles).}
\label{fig:susy-fcnc}
\end{figure}

The non-holomorphic soft scalar mass matrices are Hermitian and the A-term matrices of eq.~\eqref{mssm4} are in general complex.
Without a microscopic understanding of their origin from the supersymmetry breaking sector, there is no reason to assign them a specific structure. In particular,
in the super-CKM basis there is no reason to believe that they are diagonal or real. If they were not diagonal in generation space, they would provide new sources of FCNC processes at the quantum level, as in fig.~\ref{fig:susy-fcnc},
in addition to the Standard Model contributions encoded in the CKM matrix. Moreover, the possible complex phases generate new sources of CP violation. Both FCNC and CP violation are strongly constrained
by experimental data, by processes like $K^0-{\overline{K^0}}$ mixing, $\mu \to e \gamma$, $b \to s \gamma$.  For example, if there is no particular structure in these matrices, these constraints push the masses of squarks and sleptons to above $10^3$ TeV, even in the absence of CP violation.
If there are complex phases, CP violation constraints raise these masses to about $10^5$ TeV!~\cite{susy-FCNC}
Figure \ref{fig:susy-fcnc} shows a one-loop contribution to the $\Delta F =2$ FCNC process $K^0- {\overline K^0}$ mixing with gluino-quark-squark vertices.
In the leptonic sector, the strongest constraints typically come from the unobserved decay $\mu \to e \gamma$.

In order to comply with the tight experimental  constraints, some simplifying choices are usually made.
The simplest is to assume that all squarks and slepton non-holomorphic scalar mass matrices are equal and proportional
to the identity, and are thus governed by a single mass scale $m_0$. Moreover, if the A-terms of eq.~\eqref{mssm4} were proportional to the corresponding Yukawa couplings, they would be diagonalized by the same unitary matrices as the corresponding fermion mass matrices. In this case, their contribution to FCNC processes would also vanish. An even simpler option would obtain if the A-terms were determined by just one parameter, $A_0$. All these simplifying assumptions restrict the available parameters according to
\beq
m_{\tilde q}^2 \ = \  m_{\tilde u}^2 \ = \ m_{\tilde d}^2 \ = \ m_{\tilde l}^2 \ = \ m_{\tilde e}^2 \ = \ m_0^2 \ \underline{1}  \quad , \qquad A^{u,d,e} \ = \ h^{u,d,e} \,A_0  \ , \label{squarks7}
\eeq
where the $h$'s are SM Yukawa couplings and $\underline{1}$ denotes a $3 \times 3$ identity matrix.

The preceding assumptions on the squark mass matrix originally appear natural in supergravity, where supersymmetry breaking is only mediated by gravitational interactions. These cases can result in identical unitary $V$ matrices for quark(lepton) and squark(slepton) fields, but this type of scenario appears difficult to realize in String Theory, where the needed hierarchy of Yukawa couplings associates different generations to different regions of the internal space. An alternative would rest on supersymmetry breaking by a hidden sector mediated by Standard--Model gauge interactions, which are themselves flavor blind.

\subsubsection{\sc Neutralini and Chargini}

These fields originate from the Higgsini (the spinor partners of the Higgs doublets, which are a charged pair and two neutral ones), from the charged $SU(2)_L$ gaugino partners of the $W^\pm$ bosons, from the neutral partner of the $W_\mu^3$ boson, and from the bino, the superpartner of the hypercharge gauge boson $B_\mu$. The resulting mass eigenstates are a quartet of uncharged fermions, the neutralini, and two charged ones, the chargini.
The neutralini are typically denoted by $\chi_1^0,\chi_2^0, \chi_3^0,\chi_4^0$, with masses by convention in increasing order. These originate from the terms in eq.~\eqref{mssm4}, with additional contributions from gauge interactions mixing Higgsini with gaugini after electroweak symmetry breaking. Since electroweak mass scales are naturally smaller than $M_1$, $M_2$ and $\mu$ in eq.~\eqref{mssm4}, neutralino masses are close to $M_1$, $M_2$ and $\pm |\mu|$, where the last pair is better described by a Dirac mass equal to $|\mu|$.
  If the lightest fermionic superpartner (LSP) were a neutralino, it could be an ideal dark matter candidate of the WIMP (weekly interacting massive particle) type, if R-parity, a discrete symmetry that we are about to discuss, were conserved. Very often, in models of supersymmetry breaking and after renormalization group running to low energy, the LSP turns out to be predominantly bino.

 The chargini are often denoted by $\chi_1^{\pm}$, $\chi_2^{\pm}$, with the masses again in increasing order, which originate again from the terms in eq.~\eqref{mssm4}, with additional contributions from gauge interactions. The resulting mass matrix is simple enough to conclude that, if the supersymmetric mass parameters $M_2$, $\mu$ are well above the electroweak entries, the chargino masses are approximately $\mu$ and $M_2$, with one chargino predominantly Higgsino and the other predominantly wino.

%%%%%%%%%%%%%%%%%%%%%%%%%%%%%%%%%%%%%%%%%%%%%%%%%%%%%%%%%%%%%%%%%%%%%%%%
\subsection{\sc R-Parity}

In the Standard Model, the baryon and lepton numbers are accidentally conserved classically, due to its peculiar gauge symmetry and matter content.
In supersymmetric theories in general, and in the MSSM in particular, these conservation laws are not guaranteed anymore, even if one restricts the attention to renormalizable couplings.
To wit, supersymmetry would allow to add
\be
{\cal W} \ = \ \mu'_i L^i H_2 \ + \ \lambda_{ijk}^{''} U^{ic} D^{cj} D^{ck} \ + \ \lambda_{ijk}^{'}  L^i Q^j D^{ck}\  +  \ \lambda_{ijk}  L^i L^j E^{ck} \ , \label{rparity1}
\ee
to the standard superpotential in eq.~\eqref{mssm1}. Among these additional terms, 
those proportional to $\lambda^{''}$ violate the baryon number $B$, while those proportional to $\mu'_i$, $\lambda$ and $\lambda^{'}$ violate the lepton number $L$.
For brevity, in  eq.~(\ref{rparity1}) the $SU(2)_L$ and $SU(3)_c$ labels are left implicit. All $SU(2)_L$ representations are either doublets (with no anti-doublets) or singlets, so that
for example $L^i H_2 = N^i H_2^0 - E^i H_2^+$ is gauge invariant, where we denoted by $N^i$ the chiral superfields containing the neutrinos $\nu^i$ and their scalar partners (sneutrini) ${\tilde \nu}^i$. A closer look reveals  that the couplings in eq.~\eqref{rparity1} have the symmetry properties
\be
 \lambda_{ijk} =  - \lambda_{jik} \quad , \quad  \lambda_{ijk}^{''} =  - \lambda_{ikj}^{''}  \  . \label{rparity2}
\ee

\begin{figure}[ht]
\begin{center}
    \includegraphics[width=3in]{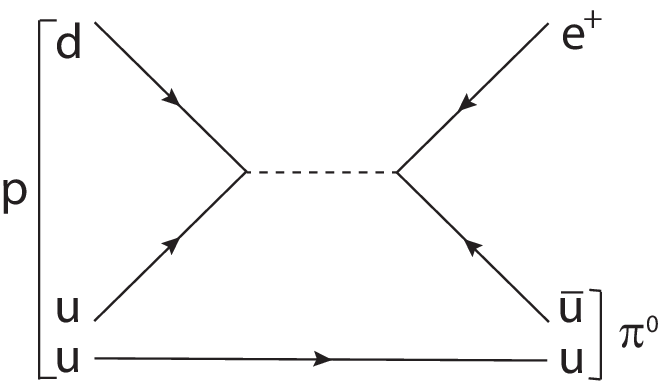}
\end{center}
\caption{Proton decay in MSSM with R-parity violation. Here the arrows are again entering (outgoing) for incoming particles (antiparticles) and outgoing (entering) for outgoing particles (antiparticles). }
\label{fig:proton-decay-mssm}
\end{figure}

 Even in the Standard Model
$B$ and $L$ are violated, but only by non--perturbative effects, which do not induce a sizable proton decay rate. On the other hand, the interactions described by (\ref{rparity1})
 give rise to proton decay via sizable perturbative processes like $p \to e^+ \pi^0$, which depend on the product of the couplings $  \lambda^{'}$ and $\lambda^{''}$. The superpotential  (\ref{rparity1}) yields indeed couplings of the type
 \beq
2 \lambda_{ijk}^{''} u^{ic} d^{cj} {\tilde d}^{ck} \ - \ \lambda_{ijk}^{'}  e_L^i u_L^j {\tilde d}^{ck}   \  ,  \label{rparity3}
 \eeq
where the fermions are presented in two-component notation,
which lead to the diagram displayed in fig.~\ref{fig:proton-decay-mssm}.
An order of magnitude estimate for this process gives a partial width
\be
\Gamma_{p \to e^+ \pi^0} \ \sim \ \sum_{i=2,3} \frac{m_p^5}{m_{\tilde d}^{4,i}} \,| \lambda_{11i}^{'} \lambda_{11i}^{''} |^2 \ , \label{rp1}
\ee
which could translate into a disastrous lifetime of about one second for couplings  $\lambda_{11i}^{'}$ and $\lambda_{11i}^{''}$ of order one and squark masses in the TeV range.

 In the early days of low-energy supersymmetry model building, it was therefore crucial
to eliminate such terms, supplementing the MSSM with an additional symmetry. There is a particularly simple and attractive option to this end,  which is called R-parity \cite{farrar-fayet1,farrar-fayet2} and rests on the discrete symmetry
\beq R_p \ = \ (-1)^{3(B-L)+ 2s} \ ,
\eeq
where $s$ is the spin. All superpartners of quarks, leptons, Higgs and gauge fields have $R_p=-1$, while all Standard Model fields (quarks, leptons, gauge fields and the Higgs scalars)  have $R_p=1$. Demanding  that R-parity be preserved forbids the unwanted couplings  (\ref{rparity1}), while allowing the usual MSSM superpotential. 
This discrete $\mathbb{Z}_2$ symmetry demands that the product of the R-parity eigenvalues for all fields entering an interaction vertex be equal to one. In collisions of Standard Model particles, like at LHC, the initial state has $R=1$, and therefore, if R-parity is conserved, the final state is also bound to have $R=1$, with the key consequence that
superpartners can be only produced in pairs. It is actually conceivable to allow for a small amount of R-parity violation \cite{r-parity-review}, and dedicated works showed that this is indeed possible. The resulting phenomenology would then be vastly different, and superpartners could manifest themselves at lower energies. Alternatively, one could work in terms of the matter parity
\beq 
{R}_m \ = \ (-1)^{3(B-L)} \ ,
\eeq
since $(-1)^{2s}= (-1)^F$, where $F$ is the fermion number, is a clearly symmetry of the Lagrangian. All quark and lepton superfields have then $R_m=-1$, while the Higgs superfields have $R_m=1$. Enforcing this discrete symmetry clearly allows the standard superpotential~\eqref{mssm1} and excludes the additional contributions in eq.~\eqref{rparity1}. 

It is important to realize that R-parity, in contrast with continuous symmetries, is also compatible with the associated soft-breaking terms of eq.~\eqref{mssm4}. This is true, in particular, for the gaugino mass terms $M_{1/2} \lambda \lambda$. Note that the R-parity breaking mass parameters $\mu'_i$ can generate a mass for a neutrino. Since neutrino masses are small, these mass parameters are also constrained to be small.

The most important consequence of imposing R-parity as an exact symmetry in the MSSM or its extensions is that the lightest superpartner (LSP) is stable. Being the lightest superpartner, the LSP could only
decay kinematically into Standard Model particles, but this is forbidden by R-parity. If electrically neutral, such a stable particle could be a good dark matter candidate \cite{lsp}. If its mass were to lie in the TeV range, it would fall within the class of particles generically called WIMP's (Weakly Interacting Massive Particles).

%%%%%%%%%%%%%%%%%%%%%%%%%%%%%%%%%%%%%%%%%%%

\subsection{\sc Concluding Remarks on the MSSM}

In addition to providing a possible solution to the hierarchy problem, by removing the quadratic cutoff dependence of scalar masses, and to suggesting a possible WIMP dark matter candidate, low-energy supersymmetry has the remarkable property
of leading to gauge--coupling unification with high accuracy at energies of order $M_{\rm GUT} \sim 2-3 \times 10^{16}$ GeV~\cite{susyunif1,susyunif2}. It is possibly the only extension of the Standard Model embodying all these features.

There is another aspect of the Standard Model that supersymmetry ameliorates. After the discovery of the Higgs and the measurement of its mass at LHC in 2012, it was shown that quantum corrections to the Higgs self-coupling $\lambda $ make its value decrease as the energy increases, so that it can potentially vanish at intermediate energies.  The interpretation of this result is that a deeper minimum, as compared to our vacuum, appears at these scales with a consequent instability or more likely a metastability of our vacuum~\cite{sm-instability}.  On the other hand, in the MSSM the Higgs self-coupling is not an independent parameter, but at tree level it is determined by gauge couplings as $\lambda = (g^2 + g^{'2})/8$. Moreover, there is no metastability problem in the MSSM, since the running to high energies is governed by its full spectrum and not the SM one.

More generally, supergravity typically emerges as a low--energy approximation to String Theory, and supersymmetry plays a central role in the stability of String Theory and its compactifications.
For all these reasons, the possible low-energy incarnations of supersymmetry were regarded as a main target of the Large Hadron Collider at CERN, after the Higgs discovery. For the time being, there are no
clear experimental signs of supersymmetry, and generally of new physics at LHC. The apparent lack of superpartners sets lower bounds on their masses:
\begin{itemize}
\item for colored particles like gluinos and
squarks, the production cross-sections are large, due to their strong interactions, and the current experimental limits on gluino masses or flavor--independent squark masses lie in the multi-TeV range. There are some caveats, however, since the processes typically involve cascades, and one eventually reveals LSP's tracking missing energy, which can be easily undetectable if the mass differences with respect to other particles are small enough 
\cite{ATLAS-SUSY}; 

\item for non-colored particles, like electro-weakini, neutralini, sleptons (with mass patterns not affecting FCNC constraints), and the additional scalar Higgs bosons, the production cross-sections are smaller and the current limits lie in the $200-500$ GeV range;

\item  for generic flavor--dependent mass matrices of squarks and sleptons, the resulting violations of the GIM mechanism push the corresponding scales to above $10^3$ TeV, even in the absence of CP violation.
If complex phases are also present, CP violation constraints raise these limits even further, to about $10^5$ TeV \cite{susy-FCNC} !
 \end{itemize}

Ongoing experiments aimed at the  direct detection of dark matter \cite{direct-detection-dm} are also carving significant portions of the WIMP parameter space, under the assumption of a naturally large mass difference between the LSP and the next to lightest
supersymmetric partner (NLSP).  The experimental searches become much more challenging in the presence of R-parity violations, which would allow for single superpartner production,
 or for almost degenerate LSP and NLSP, in which case ``missing energy" searches in colliders would be considerably more difficult \cite{R-parity-searches}.

An intriguing feature of the MSSM is that it cannot be formulated without resorting to soft breaking terms. As we are about to see, soft-breaking terms have a natural origin in supergravity, so that, surprisingly, local supersymmetry, and thus gravity itself, emerge somehow as underlying even a gauge theory like the MSSM.

 Surely enough, any theory claiming to describe Nature must be validated by experiments. For the time being, LHC found
no clear signs of new physics beyond the Standard Model, but led to the discovery of a scalar particle with a mass around $125$ GeV, which is naturally identified with the Higgs boson~\cite{atlas,cms}. It is still early to judge the ultimate relevance of low-energy supersymmetry. A close scrutiny of the Higgs couplings to fermions, gauge bosons and to itself is likely to shed new light on the problem. If these couplings were found to deviate from SM expectations, this would indicate the existence of new forces, particles, or resonances at the explored energy scales. LHC or future colliders might be able to discover these types of effects after a few years of running.
There are already some mild deviations from flavor universality in meson decays. If confirmed, due to the high sensitivity of flavor observables to new physics, these results could point to the existence of new flavor violating processes induced by hitherto unknown particles, possibly even in the multi-TeV energy range \cite{B-anomalies}.

 We can now turn to illustrate the novelties that supergravity brings along for the key issue of supersymmetry breaking.

\section{\sc Supersymmetry Breaking in \texorpdfstring{$N=1$} \ \  Supergravity} \label{sec:SUGRA}

When supersymmetry is promoted to a local symmetry, it brings along gravitational interactions, within beautiful generalizations of Einstein gravity that are generically called supergravity~\cite{sugra1,sugra2} (for reviews see~\cite{sugrarev}). There are several versions of supergravity, which are distinguished by the couplings of the gravitational multiplet to others, when these are allowed, and primarily by the number $N$ of local supersymmetries. This number has eight as an upper bound, beyond which the construction becomes impossible in usual terms, since supersymmetry would start to require the inclusion of (infinitely many) higher--spin fields. As anticipated in Section~\ref{sec:susy_algebras}, $N$--extended supergravity models involve $N$ spin--$\frac{3}{2}$ fields, called gravitini, which are the gauge fields of local supersymmetry and afford consistent couplings, thus overcoming long--recognized problems~\cite{velo-zwanziger}. Only $N=1$ supergravity allows the introduction of chiral matter in four dimensions, as needed in extensions of the Standard Model, so that in this section we shall focus on this most widely studied case.

\subsection{\sc Basic Properties of Four--Dimensional \texorpdfstring{$N=1$} \ \ Supergravity}

As we saw in Section~\ref{sec:susy_algebras}, together with chiral and vector multiplets $N=1$ supersymmetry allows the introduction of a gravity multiplet describing the modes of a massless spin--2 particle and the corresponding super--partner, a spin--$\frac{3}{2}$ gravitino. The Lagrangian for ``pure'' $N=1$ supergravity is reviewed in Appendix~\ref{app:superfields_local}: it combines the Einstein--Hilbert term with a Rarita--Schwinger term for the gravitino, possibly supplemented by a gravitino mass term and a negative cosmological constant.

General $N=1$ supergravity models~\cite{cfgvp} rest on combinations of the gravity multiplet with arbitrary numbers of chiral and vector multiplets. Consequently, their Lagrangians depend on the K\"ahler potential ${\cal K}(z^i,{\bar z}{}^i)$, a real function of the scalar fields $z^i$ belonging to chiral multiplets and their conjugates ${\bar z}{}^{i}$, which was already introduced in Section~\ref{sec:susy_algebras}. In addition, they generally depend on two holomorphic functions of the scalar fields $z^i$, the superpotential ${\cal W}(z^i)$ and the gauge kinetic function $f_{ab}(z^i)$ of the vectors, which were also introduced in Section~\ref{sec:susy_algebras}. 

The K\"ahler potential determines the kinetic terms of the scalar fields,
\beq
- {\cal K}_{i \bar{j}}\,  g^{\mu\nu}\, D_\mu \,\bar{z}^{j} \, D_\nu \,z^i \ , 
\eeq
where $D_\mu$ is a gauge covariant derivative, as defined in eq.~\eqref{bri18},
while the gauge kinetic functions of the vector multiplets determine kinetic terms and axion couplings of the gauge fields, according to
\beq
\ - \ \frac{1}{4}\, \mathrm{Re}\left(f_{ab}\right)\, \mathrm{Tr} \left(F^a{}_{\mu\nu}\, F^{b\,\mu\nu}\right) \ - \ \frac{1}{4} \,\mathrm{Im}\left(f_{ab}\right)\, \mathrm{Tr} \left(F^a{}_{\mu\nu}\, \widetilde{F}^{b\,\mu\nu}\right) \ .
\eeq

The scalar potential can be expressed in terms of the dimensionless combination
\beq
{\cal G} \ = \ \kappa^2\,{\cal K} \ + \ \log\left|\kappa^3\,{\cal W}\right|^2 \ ,
\eeq
as
\beq
\kappa^4\,V \ = \ e^{\cal G} \left( \frac{1}{\kappa^2}\, {\cal G}^{i \bar{j}}\ {\cal G}_i \ {\cal G}{}_{\bar j} \ - \ 3 \right) \ + \ \frac{\kappa^4}{2}\,\mathrm{Re} f_{ab}\, D^a\,D^b \ ,  \label{potn1}
\eeq
where
\beq
D^a \ = \ \left({\cal K}_i \ + \ \frac{1}{\kappa^2} \,
\frac{{\cal W}_i}{\cal W}\ \right)\left(T^a\right)^i{}_j \,z^j \ = \ \frac{1}{\kappa^2}\  {\cal G}_i \ \left(T^a\right)^i{}_j \,z^j  \label{Da}
\eeq
and $T^a$ are the generators of the gauge group. When expressed in terms of ${\cal K}$ and ${\cal W}$ the potential becomes
\beq
 V \ = \ {\cal K}_{i \bar{j}}\ F^i  \, \overline{F}^{\bar j} \ - \ 3\,\kappa^2\,e^{\kappa^2\cal K} \left|{\cal W}\right|^2 \ + \ \frac{1}{2}\,\mathrm{Re} f_{ab}\, D^a\,D^b \ , \label{sugra67}
\eeq
where the different contributions can be related to the auxiliary fields of the matter multiplets according to~\footnote{Here we are writing the $D^a$  in a simplified form. In general, one should replace $\left(T^a\right)^i{}_j \, z^j$ with a Killing vector $\xi^{i a}(z^i)$, as described in Section~\ref{sec:beyond_renormalizable}.}
\beq
F^i \ = \  - \ e^\frac{\kappa^2\,\cal K}{2} \ {\cal K}^{i \bar{j}}\, \left( \partial_{\bar{j}}\,\overline{\cal W} \ + \ \kappa^2\,{\cal K}_{\bar{j}}\,\overline{\cal W} \right) \ = \ - \ \frac{1}{\kappa^3}\ e^\frac{\cal G}{2} \ {\cal K}^{i \bar{j}}\, \partial_{\bar{j}}\,{\cal G} \ . \label{Fisugra}
\eeq
For ordinary gauge symmetries, the superpotential ${\cal W}$ is invariant, and therefore the expression for $D^a$ can be simplified and turned into
\beq
D^a \ =\ {\cal K}_i \, \left(T^a\right)^i{}_j \,z^j \ .
\eeq

Note also that
\beq
{\cal K}_{i \bar{j}} \ = \ \frac{1}{\kappa^2}\ {\cal G}_{i \bar{j}} \ ,
\eeq
so that the scalar kinetic terms can be also expressed solely in terms of ${\cal G}$, which is invariant under the K\"ahler transformations
\beq
{\cal K}\left(z^i,\bar{z}^i\right) \ \longrightarrow \ {\cal K}\left(z^i,\bar{z}^i\right) \ + \ \Lambda\left(z^i\right) \ + \ \overline{\Lambda}\left(\bar{z}^i\right) \ , \qquad {\cal W}  \ \longrightarrow \ e^{\,-\,\kappa^2\,\Lambda\left(z^i\right) }\,{\cal W} \ , \label{kahler_shift}
\eeq
under which the K\"ahler covariant derivative of ${\cal W}$ transforms as
\beq
 \partial_i\,{\cal W} \ + \ \kappa^2\,{\cal K}_{i}\,{\cal W} \ \to \ e^{\,-\,\kappa^2\,\Lambda\left(z^i\right) } \,  \left( \partial_i\,{\cal W} \ + \ \kappa^2\,{\cal K}_{i}\,{\cal W}\right)  \ .
\eeq

Summarizing, the bosonic couplings in $N=1$ supergravity coupled to matter can be cast in the form
\bea
\frac{1}{e}\ {\cal L} &=& \frac{1}{2\,\kappa^2}\, R - \ \frac{1}{\kappa^2}\ {\cal G}_{i \bar{j}}\, g^{\mu\nu}\, D_\mu \,\bar{z}^{j} \, D_\nu \,z^i \ - \  \frac{1}{4}\, \mathrm{Re}\left(f_{ab}\right)\, \mathrm{Tr} \left(F^a{}_{\mu\nu}\, F^{b\,\mu\nu}\right) \nonumber \\  &-& \frac{1}{4} \,\mathrm{Im}\left(f_{ab}\right)\, \mathrm{Tr} \left(F^a{}_{\mu\nu}\, \widetilde{F}^{b\,\mu\nu}\right) \ - \ V\left(z^i,\bar{z}^i\right) \ , \label{lag_bose}
\eea
and in terms of ${\cal G}$ the potential $V$ takes the compact form of eq.~\eqref{potn1}.
Moreover, the fermionic terms in the Lagrangian of~\cite{cfgvp}, which we are not displaying, would reveal that the negative contribution to the potential is proportional to the squared absolute value of the gravitino mass term
\beq
m_\frac{3}{2} \ = \ \kappa^2\, \langle e^{\frac{\kappa^2}{2} \, {\cal K}}\ \left|{\cal W}\right| \rangle \ = \ \frac{1}{\kappa}\ \langle e^{\frac{1}{2}\,{\cal G}} \rangle \ . \label{5.14}
\eeq
This result is along the lines of what is discussed in Appendix~\ref{app:superfields_local}. In that case, which can be recovered by removing the matter multiplets while retaining a constant superpotential ${\cal W}$, the gravitino is massless in the AdS vacuum. In this more general context, there are more options, and in particular the gravitino can also be massive in Minkowski space.

Due to the negative contribution related to the gravitino mass in eq.~\eqref{sugra67},  in supergravity \emph{the breaking of local supersymmetry can indeed occur with $V=0$ in the vacuum}, i.e. without the emergence of a non--vanishing vacuum energy~\cite{deser-zumino}. When the auxiliary fields acquire a vacuum value, their positive contributions to the vacuum energy can be compensated for, and then, on account of Eq.~\eqref{5.14}, the gravitino acquires a mass term. In this ``super--Higgs" mechanism, the gravitino absorbs the goldstino, which provides the additional modes of a massive Rarita--Schwinger particle, consistent with the standard interpretation of the mass term in Minkowski space reviewed in Appendix~\ref{app:superfields_local}.

\subsection{\sc \texorpdfstring{$R$} \ \ --Symmetries, Freedman's Model and Fayet--Iliopoulos Terms}

In supergravity, the constant Fayet-Iliopoulos terms of standard Abelian gauge symmetries must
be accompanied by metric determinants, and this makes them generally incompatible with gauge
symmetry and local supersymmetry. However, their consistency can be restored in the case of
gauged Abelian R-symmetries, as was originally shown by Freedman in [282]. In this setting, the gravitino is charged, the Fayet--Iliopoulos terms are proportional to its charge, and the spinor of the single vector multiplet present in Freedman's model is the goldstino. 

In the superconformal approach briefly summarized in Appendix~\ref{app:superfields_local}, the action for Freedman's model reads
\beq
{\cal L} \ = \  \left[ 
\overline{S}_0 \ e^{-\,\frac{2}{3} \xi \kappa^2 V} S_0 \right]_D \ +
 \ \left[ \frac{1}{4} \,W^{\alpha} W_{\alpha} \right]_F \ .  
\label{freed1}
\eeq
Under a gauge transformation with chiral superfield parameter $\Lambda$, the vector multiplet and the compensator $S_0$ transform as
\beq 
V \ \to \ V \ + \ i \left(\Lambda - {\bar \Lambda}\right) \quad , \quad
S_0 \ \to \ e^{\frac{2i\kappa^2}{3} \xi \Lambda} \ S_0 \ . 
\label{freed2} 
\eeq
In the Wess-Zumino gauge the vector multiplet $V$ contains the component fields $(V_{\mu}, \lambda,D)$, and the gauge parameter  of the component formulation is proportional to the real part of the lowest component of $\Lambda$, 
\beq
\alpha \ = \ 2 Re \left.\Lambda\right|_{\theta=\bar{\theta}=0} \ . 
\eeq
The only nontrivial field transformations are thus
\beq 
\delta_{\alpha} S_0 \ = \ \frac{i \kappa^2}{3}\, \xi\, \alpha 
\, S_0 \ , \qquad \delta_{\alpha} \,V_{\mu} \ = \ \partial_{\mu} \,\alpha \ . 
\label{freed3} 
\eeq 
The compensator superfield  $S_0$ contains the compensator field, which we shall also call $S_0$, in its lowest component, and its highest component is the complex auxiliary field $u$ of the old--minimal formulation of supergravity  reviewed in Appendix~\ref{app:superfields_local}.
The conformal $U(1)$ symmetry of parameter $\beta$ acts according to
\bea 
\delta_{\beta} S_0 &=& i \beta S_0 \ , \qquad\qquad 
\delta_{\beta} D \ =\  2\,i \,\beta \,D \ , \nonumber \\
\delta_{\beta} \lambda &=& \frac{3i}{2} \,\gamma_5 \,\beta \,\lambda \ , \qquad \delta_{\beta} \psi_{\mu} \ = \ \frac{3i}{2}\, \gamma_5 \,\beta \,\psi_{\mu} \ , \nonumber \\ 
\delta_{\beta} A_{\mu} &=& \partial_{\mu} \,  \beta 
\ , \label{freed4} 
\eea
where $A_\mu$ is the auxiliary vector of the supergravity multiplet.
After gauge fixing the superconformal symmetry, as needed to eliminate the compensator field $S_0$, $\alpha$ and $\beta$ are related according to 
\beq 
\beta \ = \ - \ \frac{\kappa^2}{3} \xi \alpha \ , \qquad S_0 \ = \ \kappa^{-1}  \label{freed5} 
\eeq 
and consequently one finds the final gauge transformations
\bea 
&& \delta \lambda \ = \ - \ \frac{i \kappa^2}{2}\, \xi \alpha \gamma_5 \lambda \ , \qquad 
\delta \psi_{\mu} \ = \ - \ \frac{i \kappa^2}{2} \xi \alpha \gamma_5 \psi_{\mu} \ , \nonumber \\
&& \delta V_{\mu} \ = \ \partial_{\mu} \alpha \ , \qquad \delta A_{\mu} \ = \ - \ \frac{\kappa^2}{3}\, \xi \partial_{\mu} \alpha 
\ . \label{freed52}
\eea 
The Lagrangian of the model thus includes some minimal couplings, and is given by
\bea
e^{-1}\ {\cal L} &=& \frac{1}{2\,\kappa^2} R \ - \ \frac{i}{2} {\bar \Psi}_{\mu} \gamma^{\mu \nu \rho} D_{\nu} \Psi_{\rho} 
\ +\  \frac{1}{3} \left(A_{\mu}+ \frac{\kappa^2}{3} \xi V_{\mu}\right) 
\left(A^{\mu}+ \frac{\kappa^2}{3} \xi V^{\mu}\right) - \frac{1}{3} |u|^2 
\ \nonumber \\
&-& \frac{1}{4} V_{\mu \nu}  V^{\mu \nu} \ +\ 
 i {\bar \lambda} \gamma^{\mu} D_{\mu} \lambda 
\  + \ \frac{1}{2} D^2 \ - \ \xi D \ \nonumber \\
 &-& \frac{i \kappa}{{2}} {\bar \Psi}_{\rho} \gamma^{\mu \nu} V_{\mu \nu} \gamma^{\rho} \lambda \ +\ 
 \frac{\xi\,\kappa}{{2}} {\bar \Psi}_{\mu} \gamma^{\mu} \gamma_5 \lambda \ - \ \frac{\kappa^2}{4} ({\bar \Psi}_{\rho}
 \gamma^{\mu \nu} \gamma^{\rho} \lambda) 
 ({\bar \Psi}_{\mu} \gamma_{\nu} \lambda)
 \ , \label{freed6}
\eea
where $V_{\mu\nu}$ is the field strength of $V_\mu$, and where the covariant derivatives
\bea
D_{\mu} \Psi_{\nu} &=& \left( \partial_{\mu} \ + \
\frac{1}{4} \omega_{\mu}^{ab} \gamma_{ab} \ + \ \frac{i}{2} \,\xi \,\kappa^2 V_{\mu} \gamma_5 \right) \Psi_{\nu} \ , \nonumber \\
D_{\mu} \lambda &=& \left( \partial_{\mu} \ + \
\frac{1}{4} \omega_{\mu}^{ab} \gamma_{ab} \ + \ \frac{i}{2} \,\xi \, \kappa^2 V_{\mu} \gamma_5 \right) \lambda 
\ , \label{freed7}
\eea
couple the gravitino $\Psi_{\mu}$ and the gaugino
$\lambda$ to the axial gauge field $V_{\mu}$, with charges that are proportional to the FI term.

After eliminating the auxiliary fields, one can recognize that supersymmetry is spontaneously broken \`a la Fayet-Iliopoulos, with
\beq 
\langle D \rangle \ = \  \xi \ , \qquad \left\langle A_{\mu}\ + \ \frac{1}{3} \,\xi V_{\mu} \right\rangle \ = \ 0 \ , \qquad \langle u \rangle \ = \ 0  \ , \label{freed8}
\eeq 
so that $\lambda$ is the goldstino.  The final on-shell Lagrangian 
\bea
e^{-1}\ {\cal L} &=& \frac{1}{2\,\kappa^2} R \ - \ \frac{i}{2} {\bar \Psi}_{\mu} \gamma^{\mu \nu \rho} D_{\nu} \Psi_{\rho} 
\ - \ \frac{1}{4} V_{\mu \nu}  V^{\mu \nu} \ + \
 i {\bar \lambda} \gamma^{\mu} D_{\mu} \lambda 
 \ - \ \frac{1}{2}\, \xi^2 \ \nonumber \\
 &-&  \frac{i \,\kappa}{{2}} {\bar \Psi}_{\rho} \gamma^{\mu \nu} V_{\mu \nu} \gamma^{\rho} \lambda \ +\ 
 \frac{\xi\,\kappa}{{2}} {\bar \Psi}_{\mu} \gamma^{\mu} \gamma_5 \lambda \ -\ \frac{\kappa^2}{4} ({\bar \Psi}_{\rho}
 \gamma^{\mu \nu} \gamma^{\rho} \lambda) 
 ({\bar \Psi}_{\mu} \gamma_{\nu} \lambda)
  \label{freed9}
\eea
includes a positive cosmological constant proportional to the square of the FI term, and in the unitary gauge one can set $\lambda = 0$. 
There have been interesting recent developments in this respect, leading to a construction with supersymmetry broken by $D$-terms that does not rely on $R$--symmetries~\cite{cftvp}, to which we shall return when we shall discuss nonlinear supersymmetry.

More generally, matter with a superpotential can also be included. This is necessarily not invariant under the $U(1)$ R symmetry, but has a charge that must equal $\xi \kappa^2 $. The $D$ term can then be cast in the form
\beq
D \ = \  {\cal G}_i \ r_i \,z^i \ =  \   {\cal K}_i \ r_i \,z^i \ + \ \xi \ ,
\eeq
where ${\cal G}_i$ and ${\cal K}_i$ denote derivatives with respect to $z^i$ and $r_i$ and the corresponding charges. Moreover
\beq
\xi \ = \ \sum_i \ \frac{1}{\cal W }\ {\cal W}_i \ r_i \,z^i \ , 
\eeq
which is a constant in view of Euler's theorem.

Whether or not gauged R-symmetries can be realized in String Theory is still subject to debate, but string--inspired models often include field-dependent Fayet--Iliopoulos--like terms~\cite{Dine:1987xk}, which have important physical consequences. One simple example of this type rests on the K\"ahler potential
\beq
{\cal K} \ = \  - \ \ln \ (S + S^{\dagger} \ - \ \eta \ V)
\ , \label{fdfi1}
\eeq
here written in the superspace notation elaborated upon in Appendices~\ref{app:superfields_global} and \ref{app:superfields_local}, where
$S$ is a chiral superfield, whose scalar component is a closed--string modulus, $\eta$ is a constant and $V$ is an Abelian vector superfield.  In a Taylor expansion, the term linear in $V$ in eq. (\ref{fdfi1}) defines a field-dependent FI term
\beq
\xi \ = \ \frac{\eta}{s \ + \ s^\dagger} \ ,
\eeq
where $s$ denotes the lowest component of $S$. However, the quadratic term in $V$ makes the gauge field massive, unlike the one entering the gauged R-symmetry example of~\cite{freedman}. These additions are related to lower--dimensional generalizations of the Green--Schwarz anomaly cancellation mechanism~\cite{gs} that lies at the heart of ten--dimensional superstrings.
Gauge invariance is guaranteed by the  transformations
\beq
\delta V \ = \ i \left(\Lambda \ -  \ {\bar \Lambda }\right) \ , \qquad \delta S \ = \  i\,\eta\, \Lambda \ , \label{fdfi2}
\eeq
and the Abelian gauge symmetry, which is not an R-symmetry in this case, is realized \emph{\`a la} Stueckelberg.

We can now pause to add some general comments on the nature of $F$ and $D$ terms, which are related in the presence of a superpotential with a non--vanishing vacuum value. To this end, it is convenient to express the former in terms of the ${\cal G}$-function, as in eqs.~\eqref{Da} and \eqref{Fisugra},
and the latter as in eq.~\eqref{Da}, which shows that they are both proportional to derivatives of the ${\cal G}$ function. If ${\cal G}$ is well defined, which is the case if the superpotential $W$ and the gravitino mass do not vanish in the vacuum, it is thus impossible to have only $D$-term contributions. However, the $D$-terms can still vanish as a result of cancellations among different $F$-term contributions, even when not all of these vanish. Freedman's model has only a D-term contribution to supersymmetry breaking but lacks a superpotential, so the ${\cal G}$ function is not well defined for it.

\subsection{\sc Gravity Mediation and Tree--Level Mass Formulas}

We are now ready to describe how $N=1$ supergravity can generate soft terms in the rigid limit. To this end, we shall assume that supersymmetry is spontaneously broken by vacuum values of $F$ or $D$ terms, leaving aside the microscopic origin of the phenomenon.

At low energies gravitational interactions are weak, and the global supersymmetric limit usually provides a good approximation, up to some important effects that we can now highlight.
If the auxiliary fields do not acquire a vacuum value, supersymmetry remains unbroken, and the emergence of a gravitino mass term signals that the vacuum is actually an anti-de Sitter space, as in Appendix~\ref{app:superfields_local}. 
In the absence of $R$-symmetries, the flat limit can be reached decoupling gravity in the Lagrangian~\eqref{lag_bose} and retaining, in the rest, the dominant terms as $\kappa \to 0$. The potential~\eqref{potn1} then reduces to
 \beq
 V_0 \ = \ {\cal K}_{i \bar{j}}\ F^i  \, \overline{F}^{\bar j}  \ + \ \frac{1}{2}\,\mathrm{Re} f_{ab}\, D^a\,D^b \ , \label{sugra67f}
\eeq
with
\beq
F^i \ = \  - \ {\cal K}^{i \bar{j}}\,\partial_{\bar{j}} \,\overline{\cal W}\ , \qquad 
D^a \ = \ {\cal K}_i \, \left(T^a\right)^i{}_j \,z^j \ ,
\eeq
which are the expressions discussed in Section~\ref{sec:broken_global} for ordinary gauge symmetries. In this limit the gravitino becomes massless, as can be seen from eq.~\eqref{5.14}, while $V_0$ becomes the potential of the global case. When supersymmetry is broken, this potential acquires the mean value
\beq
\Lambda_{\mathrm{SUSY}}{}^4 \ \equiv \ \langle V_0 \rangle  \label{mink_constraint}
\eeq
which can be identified with the scale of supersymmetry breaking. 

However, in local supersymmetry demanding that the supergravity vacuum be a Minkowski spacetime leads to the condition
\beq
\langle V \rangle \ = \ 0 \  , \label{zero_cosm_const}
\eeq
and now using eqs.~\eqref{sugra67} and \eqref{5.14} one can conclude that
\beq
m_{3/2}^2 \ = \ \frac{\kappa^2}{3} \ {\Lambda_{\mathrm{SUSY}}{}^4} \ = \  \frac{\Lambda_{\mathrm{SUSY}}{}^4}{3\,M_P{}^2} \ ,
\eeq
so that, as is commonly stated, the scale of supersymmetry breaking is the geometric mean
of the Planck and gravitino mass scales~\cite{deser-zumino}:
\beq
\Lambda_{\mathrm{SUSY}} \ \sim \ \left( m_{3/2}\, M_{\mathrm{P}}\right)^\frac{1}{2} \ . \label{gravimass}
\eeq
There are actually two interesting decoupling limits for gravity:
\begin{enumerate}
    \item $\Lambda_{\mathrm{SUSY}}$ can be kept far below $M_P$ if the gravitino mass becomes essentially zero, as in the preceding discussion.
    \item Alternatively, if $m_{3/2}$ is kept fixed in the decoupling limit, $\Lambda_{\mathrm{SUSY}}$ can remain large and gravity can give rise to interesting modifications of the mass sum rules. 
\end{enumerate}

As we explained in Section~\ref{sec:sum_rules}, renormalizable models of global supersymmetry lead to mass sum rules that hamper attempts to build realistic models of supersymmetry breaking. In Section~\ref{sec:beyond_renormalizable} we already saw that non--renormalizable interactions of non--gravitational origin can improve matters. Since supersymmetry is expected to be a gauge symmetry, for reasons that we already discussed, it is natural to inquire how supergravity interactions can affect the sum rules.

We can now illustrate these crucial effects by referring to an explicit setting in which they were originally revealed~\cite{gravity-mediation1,gravity-mediation2,hlw}.
The model of interest combines an observable sector with a hidden sector responsible for breaking supersymmetry. The two sectors contain one chiral multiplet each, $\Phi$ for the first and $Z$ for the second, whose scalar components $\phi$ and $z$ have canonical kinetic terms and interact only via supergravity. This is a simple instance of the so-called {\it minimal supergravity}, and the Lagrangian is determined by
\beq
{\cal K} \ = \  Z^{\dagger} Z \ + \ \Phi^{\dagger} \Phi \ , \qquad 
{\cal W} \ = \  {\cal W}_h (Z) \ + \ {\cal W}_o(\Phi) \ ,  \label{masugra4} 
\eeq
where the two contributions to the superpotential refer to the hidden and observable sectors. Moreover, the superpotential
\beq 
{\cal W}_o(\Phi) \ = \  \frac{M}{2} \ \Phi^2 \ +\ \frac{\lambda}{3} \ \Phi^3 \  \label{masugra5} 
\eeq 
is chosen so that the interactions in the observable sector become renormalizable in the rigid limit. 

The hidden sector breaks supersymmetry, but its properties need not be further specified beyond providing the \emph{v.e.v.} $z_0$ of the hidden sector scalar field, which is assumed to lie not far below the Planck scale, since this suffices to determine the scalar potential and the gravitino mass
\bea
V \!\!&=&\!\! e^{\frac{|z|^2 + |\phi|^2}{M_P^2}}\biggl\{  
\left|\partial_z {\cal W}_h + \frac{\bar z}{M_P^2} ({\cal W}_h + {\cal W}_o)\right|^2  + \left|\partial_\phi {\cal W}_o + \frac{\bar \phi}{M_P^2} ({\cal W}_h + {\cal W}_o)\right|^2 - \frac{3}{M_P^2} |{\cal W}_h + {\cal W}_o|^2 \biggr\} \, , \nonumber \\
m_{3/2} \!\!&=&\!\! \frac{{\cal W}_h (z_0)}{M_P^2} \ e^{\frac{|z_0|^2}{2 M_P^2}} \ . 
\label{masugra6} 
\eea 
With $z \simeq z_0$ and $\left|z_0\right| \gg \left|\phi\right|$ the potential reduces to
\beq
V \ \simeq \ e^{\frac{|z_0|^2}{M_P^2}}\biggl\{  
\left|\partial_z {\cal W}_h + \frac{{\bar z}{}_0}{M_P^2} ({\cal W}_h + {\cal W}_o)\right|^2  + \left|\partial_\phi {\cal W}_o + \frac{\bar \phi}{M_P^2} ({\cal W}_h + {\cal W}_o)\right|^2 - \frac{3}{M_P^2} |{\cal W}_h + {\cal W}_o|^2 \biggr\} \, , 
\label{masugra61} 
\eeq
and the cancellation of the cosmological constant links the hidden--sector auxiliary field to the gravitino mass according to
\beq 
F^z (z_0) \ \equiv  \ e^{\frac{|z_0|^2}{2 M_P^2}} \ \overline{ \left(\partial_z {\cal W}_h (z_0) + \frac{\bar z_0}{M_P^2} {\cal W}_h (z_0)\right)} \ = \ \sqrt{3} \ m_{3/2} M_P 
\ , \label{masugra8} 
\eeq 
which grants that
\beq 
\left|\partial_z {\cal W}_h (z_0) \ + \ \frac{\bar z_0}{M_P^2} {\cal W}_h (z_0)\right|^2 \ = \ \frac{3}{M_P^2} |{\cal W}_h (z_0)|^2   \ . 
\label{masugra7} 
\eeq
The preceding condition reduces the potential to
\bea
V &\simeq&  
\left|\sqrt{3} \,m_{3/2}\,M_P  + \frac{{\bar z}{}_0}{M_P^2} \, {\widehat{\cal W}}{}_o\right|^2  + \left|\partial_\phi {\widehat{\cal W}}{}_o \ + \ m_{3/2}\,\bar{\phi} \ + \ \frac{\bar \phi}{M_P^2} \ {\widehat{\cal W}}{}_o\right|^2 \nonumber \\ &-& 
3\,\left(m_{3/2}\,{\widehat{\cal W}}{}_o \,+\, h.c.\right) -  3\, \, \frac{\left|{\widehat{\cal W}}{}_o\right|^2}{M_P{}^2} \ - \ 3\,m_{3/2}^2\,M_P{}^2\ , 
\label{masugra613} 
\eea
where
\beq
{\widehat{\cal W}_o}(\phi) \ = \  \frac{\hat M}{2} \phi^2 \ + \ \frac{\hat \lambda}{3} \,\phi^3 \ , \label{masugra10} 
\eeq
with
\beq 
{\hat M} \ = \ e^{\frac{|z_0|^2}{2 M_P^2}} M \ , \qquad 
{\hat \lambda} \ = \ e^{\frac{|z_0|^2}{2 M_P^2}} 
\lambda \ , \label{masugra9}
\eeq 
and in the $M_P \to \infty$ limit one is left with
\bea
V &\simeq& \left| \partial_\Phi\,{\widehat{\cal W}}{}_o \right|^2 \ + \  m_{3/2}^2 \left|\phi\right|^2  \ + \ 
m_{3/2} \left[ \sqrt{3} \, \frac{{\bar z}{}_0}{M_P } \, {\widehat{\cal W}}{}_o \ + \ \left(\phi\,\partial_\phi\,-\,3\right) {\widehat{\cal W}}{}_o   \ + \ h.c.\right] \ ,
\label{masugra612} 
\eea
so that some contributions to the observable sector remain present.

The supersymmetric potential that emerges in the observable sector at low energies takes the standard form 
\beq
V_{SUSY} (\phi) \ = \ \left|\frac{\partial {\widehat{\cal W}}_o}{\partial \phi}\right|^2 \ = \ \left|{\hat M} \phi \ + \  {\hat \lambda} \,\phi^2\right|^2 
\eeq
in terms of ${\hat M}$ and ${\hat \lambda}$,
but there are additions that depend on $m_{3/2}$, so that
\beq 
V \ = \ V_{SUSY} (\phi) \ + \ V_{\rm soft} (\phi)
\ , \label{masugra11}
\eeq
which includes the soft terms encoded in
\beq 
V_{\rm soft} (\phi) \ = \ m_{3/2}^2 \, |\phi|^2 \ + \
\left(  \frac{1}{2} B_{\phi} \phi^2 \ + \
 \frac{1}{3} A_{\lambda} \phi^3 \ + \ {\rm h.c.} \right) \ , \label{masugra12}    
\eeq 
where
\beq
B_{\phi} \ = \ m_{3/2} \, {\hat M} \ \left(  \frac{\sqrt{3}\, \bar{z}{}_0}{M_P} \ - \ 1 \right) \ , \qquad
A_{\lambda}  \ = \  {\hat \lambda} \ m_{3/2}\, 
\frac{\sqrt{3}\, \bar{z}{}_0}{M_P}  \ . \label{masugra13}
\eeq

The decoupling limit of gravitational interactions thus yields precisely the types of soft terms that were previously identified by Girardello and Grisaru~\cite{girardello-grisaru} as special non--supersymmetric contributions that preserve the softer supersymmetric ultraviolet behavior. 

The mass sum rules are modified, due to the diagonal mass term proportional to $m_{3/2}^2$ in eq.~\eqref{masugra12} for the complex scalar $\phi$, and consequently
\beq 
Str \  {\cal M}_{\rm obs}^2 \ = \ 2 \ m_{3/2}^2 
\ . \label{masugra14}
\eeq 
However, the sign of the supertrace is not necessarily positive, since additional contributions from the vector multiplet can alter it. These can be described starting from a superspace action of the type (see Appendix~\ref{app:superfields_global})
\beq
\int d^2 \,\theta \ \frac{1}{4}\ f(Z) \ W^\alpha\, W_\alpha  \ ,
\eeq
with a non--trivial kinetic function $f(Z)$,
where $Z$ denotes again the hidden--sector superfield. In components, this addition gives rise to the terms
\beq
\Delta\,{\cal L} \ = \ - \ \frac{1}{4} \ Re(f) \Big( F^{\mu\nu} \,F_{\mu\nu}  -  i \bar{\lambda} \,\gamma^\mu D_\mu\,\lambda \Big) \ - \
\frac{1}{4} \ Im(f)  F^{\mu\nu} {\tilde F_{\mu\nu}}  \ - \ \frac{1}{4}\ \partial_z f \ F_Z \, \bar{\lambda}\, \lambda \ ,
\eeq
among others,
where $F_Z$ denotes the auxiliary field of the $Z$ multiplet, where only the lowest component is taken in the superfield--dependent terms. After normalizing the Fermi kinetic term, one can read the gaugino masses
\beq
M_\frac{1}{2} \ = \ \frac{1}{2}\ F_Z\ \partial_z \log f(z) \ ,
\eeq
which also affect the sum rules.

The gravity--induced soft terms discussed above do not address a puzzling feature of the MSSM. The Higgsino mass originates from the supersymmetric $\mu$--term that, for phenomenological reasons, should be comparable in scale to the soft terms. However, as a supersymmetric mass term, it would seem unrelated to the scale of supersymmetry breaking. This so-called "$\mu$-problem" affords a simple and elegant solution proposed long ago by Giudice and Masiero~\cite{giudice-masiero}, to which we now turn. The solution can be described in the simple toy model that we used above to illustrate the emergence of soft terms. 

Consider a modification of the K\"ahler potential containing the non--minimal terms 
\beq
\Delta {\cal K} \ = \ \left(c_0 \ + \ c_1\, \frac{Z^\dagger}{M_P}\right) \Phi^2  \ + \ \mathrm{h.c.} \ , \label{gm101} 
\eeq 
with $c_0$ and $c_1$ complex numbers of order unity.
The (anti)holomorphic contributions would have no effects in global SUSY, but in SUGRA they play a role. Although the induced supersymmetric mass term could be computed starting from eq.~\eqref{potn1}, a few simple steps suffice to illustrate the point. One can indeed perform a K\"ahler transformation, as in eq.~\eqref{kahler_shift}, so that
\beq
{\cal K} \ \to \ {\cal K} \ + \ \Lambda \ +\  {\overline \Lambda} \ , \qquad {\cal W} \ \to \  e^{-\,\frac{\Lambda}{M_P^2}} {\cal W} \ ,
\eeq
with
\beq 
\Lambda \ = \ - \ \left(c_0 \ +\  c_1 \frac{z_0^\dagger}{M_P}\right) \Phi^2 \ . \label{gm20} 
\eeq 
where $z_0$, as above, denotes the vacuum value of $z$.
In the decoupling limit of gravitational interactions, taking eq.~\eqref{5.14} into account, one can recognize that this transformation generates the supersymmetric mass term 
\beq
\left(c_0 \ + \ c_1\, \frac{z_0^\dagger}{M_P}\right) m_{3/2} \ ,
\eeq
where, as in the preceding discussion, we have included an additional factor $e^{\frac{\cal K}{2}}$ introduced by the rescaling to the low-energy description. There is another contribution to the low-energy supersymmetric mass, 
\beq
c_1 \frac{F^z}{M_P} \ = \  - \ \sqrt{3} \ c_1 m_{3/2} \ ,
\eeq
which is easily recognized even in a global SUSY setup. One thus finds the contribution
\beq 
\Delta {\hat M} \ =\  \left(c_0 \ + \ c_1 \frac{z_0^\dagger}{M_P} - \sqrt{3} \ c_1\right)  m_{3/2} \ ,  \label{gm30}
\eeq
which is of the order the gravitino mass, like the soft terms. In conclusion, if the original superpotential (\ref{masugra5}) lacked a supersymmetric mass term, the gravity-induced contribution (\ref{gm30}) would be of the right size to generate a correct low-energy phenomenology in the MSSM. 

General expressions can be written for soft terms and $\mu$-like superpotential mass terms, which afford a nice geometrical interpretation in the K\"ahler space spanned by the scalar fields. For example, one can show~\cite{sw,kl,fkz, bim, sudhir} that the supersymmetric mass that we just illustrated, when generalized to several chiral multiplets $\frac{1}{2} \mu_{ij} \Phi^i \Phi^j$, takes the form
\beq
\mu_{ij} \ = \  m_{3/2} \,\nabla_i G_j \ .
\eeq

\subsection{\sc No--Scale Models} \label{sec:no_scale_models}

There is a class of supergravity models whose scalar potentials are naturally positive definite \cite{Cremmer:1983bf}, with a minimum corresponding to a vanishing vacuum energy. In these models there are generically flat directions, which translate into the presence of massless particles, usually called moduli, in the low-energy effective theory. Their K\"ahler potentials describe symmetric spaces, while the superpotentials are independent of the moduli fields $z^\alpha$. The condition defining no--scale models reads
\begin{equation}
g^{\alpha \bar{\beta}}\ {\cal G}_\alpha\, {\cal G}_{\bar{\beta}} \ = \ 3 \ ,
\label{eq:ns1}
\end{equation}
so that the negative term in the scalar potential is identically compensated for by the contribution of F-term moduli fields.

Let us describe a simple example of this type, which finds some motivation in String Theory and includes only one modulus field, $T$, together with its conjugate $\overline{T}$. Denoting by $\Phi^i$ all other chiral fields in the theory (and leaving aside, for brevity, the gauge multiplets), the effective theory is described by
\beq
{\cal K} \ =\  -\  3 \log \left(T \,+\, {\overline T} \,-\, |\Phi^i|^2\right) \ , \quad {\cal W} \ = \ {\cal W} (\Phi^i) \ , \label{eq:ns3}
\eeq
which leads to the scalar potential
\beq
V \ = \ \frac{1}{3 \left(T \,+\, {\overline T} \,-\, |\Phi^i|^2\right)^2} \
\frac{\partial \cal W}{\partial \Phi^i} \
\frac{\partial \cal \overline W}{\partial {\overline \Phi}_i}
\ , \label{eq:ns4}
\eeq
where the sum is left implicit.
Note that if the conditions
\beq
\frac{\partial\, \cal W}{\partial\, \Phi^i} \ = \ 0  
\eeq
can be solved,
the vacuum energy $V$ vanishes and the modulus $\langle T \rangle $ is not determined. Supersymmetry is then generally broken, but the gravitino mass~\eqref{5.14} is typically not determined, as the soft terms. All the effective string models that will emerge in the next sections, with exact or spontaneously broken supersymmetry, will be of no-scale type. There are also interesting scenarios in which these solutions do not exist, so that only $\mathrm{Im}(T)$ remains a modulus.

\subsection{\sc Comments on Gauge Mediation of Supersymmetry Breaking}

As we saw in Section~\ref{sec:broken_global}, in gauge mediation, the soft terms induced by gauge loops are determined by messenger masses $M$ that typically lie well below the Planck mass. The gravitino mass in flat space is determined by the Deser-Zumino relation,
\beq
m_{3/2} \ \sim \ {\cal O}\left(\frac{F_X}{M_P}\right)
\ , \label{sgm10}
\eeq
and in this type of scenario the gravitino is naturally the lightest supersymmetric particle. As a result, it should be produced in colliders, and a lot of activity has been devoted, over the years, to explore its potential signatures at LHC. Furthermore, in these scenarios, the gravitino is also the natural dark-matter candidate.

In Supergravity with gauge mediation, which we touched upon in Section~\ref{sec:sum_rules}, around eq.~\eqref{gm1},  soft terms receive contributions from both Standard Model gauge loops and supergravity tree--level interactions. Qualitatively, the resulting scalar masses are of the form
\beq
(m_0^{i})^2 \ = \ c^i \left(\frac{g^2}{16 \pi^2}\right)^2 \left(\frac{F_X}{M}\right)^2 \ + \ \alpha^i \,m_{3/2}^2 \ ,  \label{sgm20}
\eeq
with $c^i,\alpha^i$ numerical coefficients of order unity. A key feature of these terms is that the gauge mediation coefficients $c^i$ are flavor blind, while the $\alpha^i$, which originate from supergravity, are typically flavor dependent~\cite{hlw,kl,fkz,bim}. In order to avoid significant FCNC effects, one should thus suppress the supergravity contributions, and detailed estimates taking eq.~\eqref{sgm20} into account indicate the need for a relative factor of at least $10^4$. This condition implies that the messenger masses should lie below $10^{14}$ GeV or so. Combining this constraint with the lower bound deduced in Section~\ref{sec:broken_global} from stability considerations, one can conclude that the phenomenologically allowed range for messenger masses is
 \beq
100 \ \mathrm{TeV} \ \lesssim \ M \ \lesssim \ 10^{14} \ \mathrm{GeV}  \ ,   \label{sgm30}
\eeq
and the corresponding range for the gravitino mass is
\beq
10^{-5} \ \ {\rm eV} \ \lesssim  \  m_{3/2} \ \lesssim  \ 1 \ \ {\rm GeV}
\ .   \label{sgm4}
\eeq

\section{\sc Volkov-Akulov Model and Nonlinear Supersymmetry}\label{sec:VAnonlinear}

It is interesting and instructive to address the breaking of supersymmetry in limiting cases when some super-partners become very massive, and possibly disappear altogether from the spectrum. This type of limiting behavior has familiar counterparts in ordinary symmetries. For example, the linear $O(n)$ $\sigma$ model
\beq
{\cal S} \ = \ \int d^D x \left[ \ - \ \frac{1}{2} \ \partial_\mu \phi^T\ \partial^\mu \phi \ - \ \frac{g}{4} \left( \phi^T \, \phi \ - \ \rho^2\right)^2 \ \right]  \ , \label{linear_sigma}
\eeq
where $\phi$ is an $n$-component real vector,
describes $n-1$ massless Goldstone modes parametrizing the coset space $O(n)/O(n-1)$, together with a Higgs--like excitation with
\beq
m^2 \ = \ 2\,g\,\rho^2 \ .
\eeq
In the $g \to \infty$ limit the Higgs--like mode becomes infinitely massive, and the leftover Goldstone modes are described by the non--linear $\sigma$ model
\beq
{\cal S} \ = \ - \ \frac{1}{2}\, \int d^D x \ \partial_\mu \phi^T\ \partial^\mu \phi  \ ,
\eeq
with $\phi$ now subject to the quadratic constraint
\beq
\phi^T \, \phi \ = \ \rho^2 \ .
\eeq
The final Lagrangian and the constraint are both manifestly $O(n)$ invariant. However, only the $O(n-1)$ subgroup of $O(n)$ that stabilizes the vacuum value is linearly realized. 

This type of constraint played an important role, in the 1960's, in the descriptions of pions, which can be related to an $SU(2)_L\times SU(2)_R$--valued field $\Sigma$ subject to the  constraint
\beq
\Sigma\Sigma^\dagger \ =\ 1 \ . \label{nlsigmaconstraint}
\eeq
As we saw in Section~\ref{sec:axionchiral}, letting
\beq
\Sigma \ = \ e^{\frac{i}{f_{\pi}} \tau^a \pi^a} 
\eeq
one can write for these (pseudo--)Goldstone modes the low--energy Lagrangian
\beq
{\cal L} \ = \ - \ \frac{ f_{\pi}^2 }{4} Tr (\partial_\mu \Sigma) (\partial^\mu \Sigma^{-1}) \ + \ \frac{ f_{\pi}^2 }{2} m_0 Tr  ( \Sigma^{\dagger} M + M^{\dagger} \Sigma )  \ , \label{lowpi}
\eeq
where $f_\pi$ is the pion decay constant.

We can now illustrate the counterparts of these ideas that play a role in connection with supersymmetry breaking.

%%%%%%%%%%%%%%%%%%%%%%%%%%%%%%%%%%%%%%%%%%%%%%%%%%%%%%%%%%%
\subsection{\sc Non--Linear Supersymmetry}\label{section:AVlagrangian}

Nonlinear realizations of supersymmetry originally emerged in the Volkov--Akulov model~\cite{Volkov:1973ix}. In order to recall its key features, let us begin by noting that the transformations
\beq
\delta_\epsilon\lambda^\alpha\ = \ f\epsilon^\alpha\ - \ \frac{i}{f}\left(\lambda\sigma^\mu\overline\epsilon \ -\ \epsilon\sigma^\mu\overline\lambda\right)\partial_\mu\lambda^\alpha \ ,
\label{AVtransformation}
\eeq
which include the shift pertaining to a Goldstone fermion,
close surprisingly on the supersymmetry algebra, although they only concern a Weyl spinor, with no bosonic partners. From what we saw already on the issue of supersymmetry breaking and auxiliary fields, one can anticipate that the parameter $f$, of mass dimension two, reflects the scale of supersymmetry breaking in the microscopic theory including the superpartners. This can be appreciated by comparing these transformations with eqs.~\eqref{superym12} when the auxiliary fields acquire vacuum values. 
However, this construction has the striking feature of violating the equality of Fermi and Bose degrees of freedom, which is the basic tenet of linear supersymmetry. The mismatch can be ascribed to the disappearance of the bosonic partners that, as the Higgs--like field of eq.~\eqref{linear_sigma}, have acquired an infinite mass in the singular $g \to \ \infty$ singular limit of eq.~\eqref{linear_sigma}.

The Volkov--Akulov setup affords an elegant geometrical presentation that mimics the vielbein formulation of gravity. In order to illustrate it, it is convenient to introduce the vielbein
\beq
E_\mu{}^\nu \ = \ \delta^\nu_\mu\ + \ \frac{i}{f^2}\left(\lambda\sigma^\nu\partial_\mu\overline\lambda \,-\,\partial_\mu\lambda\sigma^\nu\overline\lambda\right) \ , \label{vielbein_VA}
\eeq
since supersymmetry acts on it as a diffeomorphism of parameter
\beq
\xi^\mu \ = \ -\ \frac{i}{f}\left(\lambda\sigma^\mu\overline\epsilon \ - \ \epsilon\sigma^\mu\overline\lambda\right) \ . \label{VAparameter}
\eeq
As a result, the measure in
\beq
{\cal S} \ = \ - \ \frac{f^2}{2}\, \int d^4 x\ \det\, E 
\eeq
is invariant under the transformation~\eqref{AVtransformation}, so that this action is invariant under non--linear supersymmetry, up to a total derivative. Its quadratic portion yields the properly normalized kinetic term for the goldstino $\lambda$, which is accompanied by a special collection of self-interactions.

Couplings to additional scalar or spinor fields can be introduced insisting on diffeomorphisms with parameters~\eqref{VAparameter}~\cite{Ivanov:1978}, which act for example on a scalar $\phi$ according to
\beq
\delta_\epsilon\phi\ =\ -\ \frac{i}{f}\left(\lambda\sigma^\mu\overline\epsilon \ -
 \ \epsilon\sigma^\mu\overline\lambda\right)\partial_\mu\phi \ .
\label{fieldsNonLinearSUSYTransform}
\eeq
For fields of more general types, this transformation would involve the Lie derivative with parameter~\eqref{VAparameter}. This work stimulated the subsequent developments discussed
below.
These realizations can be reached by dressing any given field $\phi'$ belonging to a linear SUSY multiplet with the Volkov--Akulov vielbein \cite{Pashnev:1974,Ivanov:1978}, according to
\beq
\phi\ \equiv \ e^{\epsilon Q +\overline\epsilon \overline Q}\,\phi'\big\vert_{\epsilon=-\lambda/f} \ .
\label{dressedFieldsNonLinearSUSY}
\eeq

In this fashion, SUSY transformations are realized as local diffeomorphisms with parameter~\eqref{VAparameter} on all fields, aside from the goldstino, which combines this transformation with a constant shift as in eq.~\eqref{AVtransformation}. One can thus
define covariant derivatives \cite{Clark:1996aw,Clark:2000rv}
\be
{\cal D}_\mu\ = \ \left(E^{-1}\right)_\mu{}^\nu\partial_\nu
\ee
 for both spinors $\psi$ and scalars $\phi$,
which transform as total derivatives under SUSY variations. Consequently, any flat--space Lagrangian can be turned into one invariant under non--linear supersymmetry via the usual covariantization to curved space, albeit with the metric corresponding the vielbein of eq.~\eqref{vielbein_VA}, so that the original matter Lagrangian becomes
\beq
{\cal L}(\phi,{\partial}_\mu\phi,\psi,{\partial}_\mu\psi,\lambda,\partial_\mu \lambda) \ = \ \det(E)\left(- \ \frac{f^2}{2} \ +\ \cL_\text{matter}(\phi,{\cal D}_\mu\phi,\psi,{\cal D}_\mu\psi)\right) \ .
\eeq
The first term in the modified Lagrangian, as we have seen, defines kinetic term and self-interactions of the goldstino $\lambda$, and generalizations to gauge theories follow similar principles \cite{Clark:1996aw,Clark:2000rv}. The original Lagrangians are recovered removing the goldstino.

A similar, if technically more involved, procedure can be implemented for supergravity, and yields for the goldstino a transformation that depends on the fields in the supergravity multiplet (here displayed up to three-spinor terms) \cite{Kapustnikov:1981de}
\bea
\delta_\epsilon\lambda &=&  f\,\epsilon \ - \ \frac{i}{f}\left(\lambda\sigma^\mu\overline\epsilon \,-\,\epsilon\sigma^\mu\overline\lambda\right)\left(\hat {\cal D}_\mu\lambda\,-\,i\,\frac{u}{18}\sigma_\mu\overline\lambda\right) \ + \ \frac{\overline u}{3 \, f}\ \epsilon \, \lambda^2\nonumber \\ &+& \frac{A_\mu}{3\,f}\left(\frac{\sigma^\mu\overline\lambda(\lambda\epsilon)}{3}\,+\,\frac{\lambda(\epsilon\sigma^\mu\overline\lambda)}{2}\,-\,\frac{\sigma^\mu\overline\epsilon(\lambda^2)}{12}\right)  \ + \ \ldots \ ,
\label{nonLinearSUGRA}
\eea
where we used the notation of Wess and Bagger in~\cite{susy-books}, so that $u$ and $A_\mu$ are again the complex scalar and axial vector auxiliary fields of the supergravity multiplet discussed in Appendix~\ref{app:superfields_local}. The invariant action for the goldstino will include a mixing with the gravitino, in such a way that the goldstino is eaten, in unitary gauge, in the super--Higgs effect. In supergravity $\epsilon$ is coordinate dependent and can be used to remove the goldstino altogether in the unitary gauge. The complications due to goldstino couplings then disappear, leaving behind a mass term for the gravitino, whenever this is allowed. In fact, a ten--dimensional counterpart of this transformation, together with the dressing of matter/supergravity fields as in \eqref{dressedFieldsNonLinearSUSY}, plays a role in the non--linear realization of local supersymmetry that underlies the phenomenon of ``brane supersymmetry breaking'' in String Theory \cite{dmnonlinear1,dmnonlinear2,dmnonlinear3}, which will be illustrated in the following sections. In this case a mass term for the gravitino cannot arise, which is not contradictory since flat space is not a vacuum due to the emergence of a tadpole potential, as we shall see.

%%%%%%%%%%%%%%%%%%%%%%%%%%%%%%%%%%%%%%%%%%%%%%%%%%%%%%%%%%%
\subsection{\sc Constrained Superfields}\label{section:constrainedSuperfields}

There are alternative presentations of non--linear supersymmetry, and one of them is along the lines of what we saw for ordinary symmetries.
Theories where $N=1$ or $N=2$ SUSY is non--linearly realized can indeed be directly formulated in superspace, relying on superfields subject to constraints that eliminate some components while
not enforcing any equations of motion. One option is to start from the goldstino or matter fields and redefine them according to \cite{Kapustnikov:1981de,Samuel:1982uh}
\beq
\Lambda^\alpha\ = \ e^{\theta Q+\overline \theta \overline Q}\,\lambda^\alpha \ , \quad \Psi\ = \ e^{\theta Q+\overline \theta \overline Q}\,\psi \ ,
\label{nonLinearSuperfields}
\eeq
thus building complete superfields.
In this fashion, their general couplings can be written via the superspace integral
\be
\cL=\int d^4\theta \ \Lambda^2\overline\Lambda^2\left(- \ \frac{1}{2f^2}\ + \ \frac{1}{f^4}\cL_\text{matter}(\Psi,\partial_\mu\Psi)\ +\ ...\right) \ ,
\ee
where the ellipsis indicates that higher--order terms, or terms involving supersymmetric derivatives of the different superfields, can be included.

Note that the goldstino superfield $\Lambda^\alpha$ satisfies the constraints \cite{Samuel:1982uh}
\be
D_\beta\Lambda_\alpha \ = \ f\,\epsilon_{\alpha\beta} \ , \qquad {\overline D}{}^{\dot \beta}\Lambda^\alpha\ =\ \frac{2i}{f}\,(\overline\sigma^\mu\Lambda)^{\dot \beta}\,\partial_\mu\Lambda^\alpha \ ,
\ee
where the spinorial covariant derivatives $D$ and $\overline D$ are defined in eq.~\eqref{chirals1}, and that those constraints entirely characterize the superfield. In particular, they suffice to show that the lowest component $\lambda^\alpha$ of $\Lambda^\alpha$ transforms as the Volkov--Akulov goldstino. They can also be generalized to supergravity, in order to attain a non-linear realization of local supersymmetry equivalent to \eqref{nonLinearSUGRA}.

In four--dimensional models where superspace is a fully developed tool, supersymmetric constraints on superfields can lead to non-linear realizations. A simple constraint was originally proposed in ~\cite{rocek,nonlinear,brignole,ks}: in this case the goldstino is identified with the fermionic component of a chiral superfield $X$ subject to the algebraic constraint
 \be
 X^2\ = \ 0 \ ,
 \label{RocekConstraint}
 \ee
so that its scalar component $x$ is expressed in terms of its spinor $\lambda$ and of its auxiliary field $F$ according to
\beq
X \ = \ \frac{\lambda\,\lambda}{2\,F} \ + \ \sqrt{2}\,\theta^\alpha\,\lambda_\alpha \ + \theta^2\,F \ .
\eeq
The most general action for $X$ with conventional kinetic terms is thus
\be
\int d^4\theta \ XX^\dagger \ + \ \left(\int d^2\theta \ fX\, + \,h.c.\right) \ , \label{eq:vasuperspace}
\ee
where $f$ is the scale of SUSY breaking already introduced in eq.~\eqref{AVtransformation}. One can show that the action thus obtained is equivalent to the Volkov--Akulov action, albeit after a non--trivial field redefinition \cite{kt}.

The difference between the usual linear supersymmetric theories and this one with a superfield constraint is closely reminiscent of the more familiar examples of linear versus nonlinear sigma model discussed at the beginning of this chapter. 
The constraint in eq. (\ref{RocekConstraint}) is indeed similar to the nonlinear constraint (\ref{nlsigmaconstraint}), and can be similarly obtained by decoupling a heavy sgoldstino scalar $x$, which gets a large mass after supersymmetry breaking. 

In order to illustrate this correspondence, let us consider, to begin with, a supersymmetric model based on
\beq 
K \ = \ X^{\dagger} X \ - \ \frac{1}{M^2} (X^{\dagger} X)^2 \ , \qquad W \ =\  f\, X \ , \label{eq:microva1} 
\eeq
where $M$ will play the role of a UV cutoff scale for the resulting low--energy the theory. 
If the components of the chiral superfield $X$ are denoted by
$(\phi,\lambda,F_X)$, the relevant part of the Lagrangian is
\beq 
{\cal L} \ = \ |F_X|^2 \ + \ f \left(F_X \ + \ {\bar F}_X\right) 
\ - \ \frac{1}{M^2} |2 \,\phi\, F_X \ - \ \lambda \lambda|^2 \ + \ \ldots
 \ . \label{eq:microva2} 
\eeq 
The scalar potential is
\beq 
V \ = \  K_{X {\bar X}}^{-1} \,f^2 \ = \ \frac{f^2}{1 \ - \  \frac {4 |\phi|^2}{M^2}}  \ , \label{eq:microva3} 
\eeq 
and the minimum lies at $\phi=0$, while the mass of the sgoldstino is 
\beq 
m_x^2 \ = \ \frac{4 f^2}{M^2} \ . \label{eq:microva4}
\eeq 
At energy scales below its mass one can integrate out the scalar field $\phi$, and the low-energy theory will only contain the goldstino as a physical degree of freedom. Integrating out the sgoldstino via its classical field equation one finds 
\beq 
\phi \ = \ \frac{\lambda \lambda}{2 F_X} \ , \label{eq:microva5}
\eeq 
which recovers the superfield constraint
\eqref{RocekConstraint}. Although we neglected derivatives in  (\ref{eq:microva2}), it turns out that the solution~(\ref{eq:microva5}) is exact, and therefore the Volkov-Akulov superspace Lagrangian 
(\ref{eq:vasuperspace}) is the exact low-energy action. This action is regarded as universal, in the sense that it is the unique low-energy action obtained starting from a supersymmetric two-derivative Lagrangian in the UV, after breaking spontaneously supersymmetry and integrating out the scalar sgoldstino at energies below its mass.  

The constraint \eqref{RocekConstraint} establishes a clear link between a UV model in which SUSY is linearly realized in superspace and the non-linear Lagrangian that arises after SUSY breaking. In particular, the coupling of the goldstino superfield $X$ to matter rests on a standard superspace integral. An explicit D-brane setting for this constraint was presented in~\cite{sorokin1,sorokin2}.

More general constraints can eliminate different components from supersymmetric multiplets.  For example, one can eliminate the lowest component of a generic superfield $Y$~\cite{DallAgata:2016syy} via the constraint
\be
XX^\dagger Y \ = \ 0 \ ,
\label{minimalConstraintSuperfield}
\ee
where $X$ is the goldstino superfield, which verifies eq.~\eqref{RocekConstraint}. In particular, if $Y$ is $\Phi$ or $D_\alpha \Phi$, with $\Phi$  a chiral superfield, the constraint~\eqref{minimalConstraintSuperfield} eliminates its scalar or spinor components, and all these steps have counterparts in supergravity~\cite{DallAgata:2015zxp}.

This approach proved convenient for coupling the goldstino to the MSSM \cite{Antoniadis:2010hs,Antoniadis:2012hs,Bellazzini:2017neg}, for inflation models \cite{adfs,fkl1,fkl2,fkl3}, taking the Starobinsky~\cite{starobinsky} model as a starting point, and for string--motivated effective field theories~\cite{follow_ups1,follow_ups2,follow_ups3,follow_ups4,follow_ups5,follow_ups6,follow_ups7,follow_ups8} with broken SUSY. There is a subtlety with auxiliary fields, since proceeding as above, one can also remove auxiliary fields by making use of the goldstino superfield. For instance, demanding that $XX^\dagger D^2\Phi=0$ for a chiral superfield $\Phi$, or even that $\frac{1}{4}D^2X=f$, where $f$ is the constant entering the Volkov--Akulov model, the auxiliary field can be replaced by combinations of other fields in the theory. For example, the constraint proposed in \cite{ks} to remove fermions from chiral superfields, $\overline D_{\dot\alpha}(X\overline\Phi)=0$, also removes the auxiliary field in $\Phi$. However, while the removal of scalars or fermions can be connected to high--mass limits, with auxiliary fields one runs into a problem. The preceding steps require the introduction of higher--derivative operators, whose presence impinges on causality issues, so that the resulting procedure is not based on similarly solid grounds \cite{Bonnefoy:2022rcw}.

In general, the emergence of constrained superfields reflects the fact that supersymmetry breaking can induce mass splittings, within matter multiplets, which are so large that only some of the original fields remain in the EFT below the scale of SUSY breaking. There is an apparent contradiction with the mass sum rules, but linear models giving rise to constrained superfields also include higher--dimension operators and, often, higher--derivative terms, which can modify the mass sum rules, as we have seen.

Constrained superfields provide instructive playgrounds for applications of supergravity to Cosmology, since for one matter they reduce to a minimum the fields involved in model building. This program started with the example of the Starobinsky model~\cite{starobinsky}, whose realization via constrained superfields~\cite{adfs} obtains coupling supergravity to an ordinary Wess--Zumino multiplet and a constrained one (for reviews, see, for example,~\cite{nonlinear_reviews}). Many interesting models then followed this example and~\cite{fkl1,fkl2,fkl3}, which include different realizations of the minimal coupling of supergravity to the Volkov--Akulov model~\cite{follow_ups1,follow_ups2,follow_ups3,follow_ups4,follow_ups5,follow_ups6,follow_ups7,follow_ups8}. The upshot is that, insofar as the constraints do not affect the auxiliary fields, one can apply the standard formula for the supergravity potential of~\cite{cfgvp},
\beq
{\cal V} \ = \ e^{\cal G} \left[ \left({\cal G}^{-1}\right)^{i \bar{j}} \, {\cal G}_i\,{\cal G}_{\bar{j}}  \ - \ 3\right]
\, \quad {\rm with} \qquad {\cal G} \ = \ {\cal K} \ + \ \log \left|{\cal W}\right|^2 \ , \eeq
while enforcing superfield constraints only at the end.

\subsection{\sc \texorpdfstring{${\cal N}=2 \to {\cal N}=1$} \ \  Breaking}

The simplest ${\cal N}=2$ multiplet is the vector multiplet, which comprises, in ${\cal N}=1$ language, a vector multiplet $V$ and a chiral multiplet $\Phi$. In an ${\cal N}=1$ superspace formulation, the Lagrangian is determined by a holomorphic function ${\cal F} (\Phi)$ called the ``prepotential''. The most general Lagrangian for a vector multiplet is then
\beq
{\cal L} \ = \ \int d^4 \theta \ {\cal K} (\Phi, \bar \Phi)\ + \ \int d^2 \theta \left[\frac{1}{4}\, f(\Phi) \ W^{\alpha}\, W_{\alpha}\ + \ {\cal W} (\Phi)\ + \ {\rm h.c.} \right] \ , \label{pb1}
\eeq
where
\bea
&& {\cal K} \ = \ \frac{i}{2}  \left(   \Phi \ \frac{\partial \bar {\cal F}}{\partial \bar \Phi} \ - \   \bar \Phi \ \frac{\partial  {\cal F}}{\partial  \Phi} \right) \ , \nonumber \\
&& f \ =\ -\ {i} \  \frac{\partial^2  \cal F}{\partial \, \Phi^2}  \ , \qquad {\cal W} \ = \ -\ \frac{i\, e_c}{2} \,\Phi \ -\ \frac{i\, m }{2}\ \frac{\partial \, \cal F}{\partial \, \Phi}    \ . \label{pb2}
\eea
In general, ${\cal N}=2$ Fayet--Iliopoulos terms are given in terms of electric and magnetic fluxes $e_X$ and $m_X$, $(X=1,2,3)$, which are triplets of the $SU(2)$ R-symmetry. In ${\cal N}=1$ language, the complexified electric and magnetic charges 
\beq
e_c\ =\ e_1\ -\ i\,e_2\ , \qquad m \ = \ m_1\ +\ i m_2
\eeq
enter the superpotential, while the third components are ${\cal N}=1$ Fayet--Iliopoulos terms. Using the $SU(2)$ R-symmetry, one can always set $e_3=0$ and $m_2=m_3=0$, so that one is left with a complexified electric charge $e_c$ and a real magnetic charge $m$, as in the preceding expressions, both of mass dimension two.

 One of the supersymmetries is then manifest in superspace, while the other mixes the two superfields according to
\bea
& \delta_2 \, \Phi & = \ i\, \sqrt{2}\,\epsilon_2^\alpha\, W_\alpha \ , \nonumber \\
& \delta_2\, W_\alpha & = \ - \ i\,\sqrt{2} \left(m \ -\ \frac{1}{4} \ \overline{D}^{\,2} \,\overline{\Phi}\right)\epsilon_{2\alpha} \ + \  \sqrt{2}\, \sigma^\mu_{\alpha\dot{\alpha}}\, \partial_\mu \Phi \ {\overline{\epsilon}_2}^{\dot{\alpha}} \ . \label{second_SUSY}
\eea
The magnetic charge $m$ is clearly a crucial ingredient in the simple model of eq.~\eqref{pb1}, which was introduced by Antoniadis, Partouche and Taylor in~\cite{apt}, since it allows a nonlinear realization of the second supersymmetry of parameter $\epsilon_2$, with the gaugino being the corresponding goldstino.

An original development that emerged slightly later, in the 1990's, led to the recovery of the supersymmetric Born--Infeld model, which realizes the ${\cal N}=2 \to {\cal N}=1$ breaking~\cite{bg}, via a \emph{quadratic} constraint linking to one another a chiral and a vector multiplet according to
\beq
{\cal C}_2: \ W^\alpha\, W_\alpha \ - \ 2\, \Phi \left( \frac{1}{4} \ \overline{D}^{\,2}\, \overline{\Phi} \ - \ m \right) \ = \ 0 \ . \label{2_1_constraint}
\eeq
This constraint eliminates \emph{all} the degrees of freedom contained in the chiral multiplet $\Phi$.
The crucial difference between this constraint and the preceding ones is that, starting from ${\cal N}=2$ supersymmetry, in this case one ends up with ${\cal N}=1$.

There was for a while some discussion as to whether or not a partial breaking would be possible, and these developments owe to the original considerations in~\cite{hughes_polchinski1,hughes_polchinski2} and to the explicit linear models of~\cite{apt} and~\cite{cgfp1,cgfp2,cgfp3}. In retrospect, the end result can be regarded as a toy model for a standard $D$ brane, whose insertion in the vacuum halves its original amount of supersymmetry.

The supersymmetric Born--Infeld theory, with a scale determined by the real ``magnetic'' charge $m$, can be obtained starting in eq.~(\ref{pb1}) from a quadratic prepotential ${\cal F} = i \frac{\Phi^2}{2}$. The nonlinear Born-Infeld action then emerges from the Polonyi term, the contribution linear in $\Phi$, since the constraint eliminates all kinetic terms in
\beq
{\cal L} \ = \ Re\left[ \frac{1}{2}\ \int d^2 \theta \, W^\alpha\,W_\alpha + (m \,-\,i\,e_c) \int d^2 \theta \,\Phi \right] \ + \ \int\,d^4 \theta \ {\overline \Phi} \, \Phi \ . \label{pb3}
\eeq
In this simple case $e_c$ is redundant and can be set to zero.
After solving the quadratic constraint, the supersymmetric Born-Infeld Lagrangian takes the form~\cite{cecottiferrara}
\beq
 {\cal L} = \frac{1}{4} \int d^2 \theta \ W^2 \,+\, {\rm h.c.} \ + \  \frac{1}{4 m^2} \int d^4 \theta \frac{W^2 \,{\overline W^2}}{1 - \frac{1}{2} A + \sqrt{1-A + \frac{1}{4} B^2}} \ ,  \label{pb31}
\eeq
where
\beq
A = \frac{1}{2 m^2} \left( D^2 W^2 \ +\   {\overline D^2} {\overline W^2} \right) \quad , \quad
B = \frac{1}{2 m^2} \left( D^2 W^2 \ - \  {\overline D^2} {\overline W^2} \right) \ .  \label{pb4}
\eeq
One can show that the bosonic part of (\ref{pb4}) includes the Born-Infeld action
\beq
{\cal L}_{BI} = - m^2 \sqrt{- \ \det  (\eta_{\mu \nu} + \frac{1}{m} F_{\mu \nu})} \ .  \label{pb5}
\eeq

These results afford a natural generalization to the case of a number of Abelian multiplets, whereby the constraint of eq.~\eqref{2_1_constraint} becomes~\cite{fps1,fps2}
\beq
d_{ABC} \left[ W^{\alpha\,B}\, W_\alpha^C \ - \ 2\, \Phi^B \left( \frac{1}{4} \ \overline{D}^{\,2}\, \overline{\Phi}^C \ - \ m^C \right)\right] \ = 0 \ ,
\eeq
which might provide some clues on the long--sought non--Abelian Born--Infeld theory associated to $D$-brane stacks. For a review, see~\cite{tseytlin_review}.

\subsection{\sc \texorpdfstring{${\cal N}=2 \to {\cal N}=0$} \ \ Breaking}
Let us conclude this discussion with a brief overview of a more recent result, which was presented in~\cite{dfs} and concerns the ${\cal N}=2 \to {\cal N}=0$ breaking in a vector multiplet via constrained superfields. In this example all the original extended supersymmetry is non--linearly realized, and therefore this setting can be regarded as a four--dimensional toy model for ``brane supersymmetry breaking''. In Sections~\ref{sec:critical_strings} and~\ref{sec:toroidal_ss} we shall encounter this peculiar phenomenon in its simplest manifestation, a ten--dimensional string where local supersymmetry is non--linearly realized but no tachyonic instabilities are present. 

The relevant constraint is now cubic in $N=2$ superspace, and in $N=1$ language it translates into the following set of constraints:
\beq
 \Phi\, W^\alpha\, W_\alpha \ - \ \Phi^2
\left( \frac{1}{4} \ \overline{D}^{\,2}\, \overline{\Phi}
\ - \ m \right) \ = \ 0 \ , \qquad \  \Phi^3 \ =
 \ 0 \, , \qquad \Phi^2\, W_\alpha \ = \ 0 \ .
 \eeq
One can show that, aside from a singular corner where the ${N}=2 \to {N}=1$ breaking is recovered, this \emph{cubic} constraint eliminates the complex scalar of the Wess-Zumino multiplet that accompanies the ${N}=1$ vector multiplet in the description of ${N}=2$ Yang--Mills theory via ${ N}=1$ superfields. However, this projection does not affect the gauge field and the two Weyl fermions of the original multiplets, which are the two goldstini of the system. Notice that, in contrast with the ${N}=1 \to {N}=0$ case of~\cite{bg}, here a Born--Infeld structure of the remaining low--energy interactions is not implied. Other recent works dealing with the ${N}=2 \to {N}=0$ case, also from the vantage point of the Volkov--Akulov model, can be found in~\cite{recent201,recent202}.

\subsection{\sc Concluding Remarks on the Partial Breaking of Supersymmetry}

The possibility of spontaneously attaining the partial breaking of supersymmetry has been a controversial subject for a while. Indeed, starting from the standard anticommutation relations for N supercharges,
\beq
\{ Q_{\alpha}^A , {\bar Q}_{\dot \beta, B}  \} = 2 \delta_B^A \sigma^\mu_{\alpha \dot \beta} P_\mu  \quad , \label{witten-partial1}
\eeq
with $A,B = 1 \cdots N$, and taking a trace in the two dimensional spinor space, one can link the charges to the Hamiltonian according to
\beq
 H  \  \delta_B^A = \frac{1}{4} \sum_{\alpha}
\  \{ Q_{\alpha}^A , {\bar Q}_{\dot \alpha, B}  \}
 \quad . \label{witten-partial2}
\eeq
Consequently, if there is an unbroken supersymmetry, say for $A=1$, so that
\beq
Q_{\alpha}^1 \ | 0 \rangle =  {\bar Q}_{\dot \alpha, 1} \ | 0 \rangle = 0  \quad , \label{witten-partial3}
\eeq
the structure of eq.~\eqref{witten-partial2} implies that 
\beq
 H  \  | 0 \rangle = \frac{1}{4} \sum_{\alpha}
\  \{ Q_{\alpha}^1 , {\bar Q}_{\dot \alpha, 1}  \}
| 0 \rangle \ = \ 0
 \quad , \label{witten-partial4}
\eeq
so that $H$ annihilates the vacuum. Then, in view of eq.~\eqref{witten-partial2}, for any fixed value of $A$
\beq
\langle 0 | H  | 0 \rangle = \frac{1}{4} \sum_{\alpha}
\ \left( || Q_{\alpha}^A \ | 0 \rangle ||^2  +
||  {\bar Q}_{\dot \alpha, A} \ | 0 \rangle ||^2 \right)  \ = \ 0
 \quad , \label{witten-partial5}
\eeq
and the positivity of the individual contributions implies that
\beq
Q_{\alpha}^A \ | 0 \rangle =  {\bar Q}_{\dot \alpha, A} \ | 0 \rangle = 0  \quad , \label{witten-partial6}
\eeq
so that if one supersymmetry is unbroken one would be tempted to conclude that they are all unbroken. In other words, this simple argument would seem to imply that the partial breaking of supersymmetry is impossible \cite{witten-partial}, contrary to what we saw already in previous sections.

However, placing extended objects in the vacuum breaks part of the Poincar\'e symmetries and also part of the supersymmetries. In the weak--coupling limit D-branes, which will be discussed in the following chapters related to String Theory, are hyperplanes that break indeed some Poincar\'e symmetries when they are present, preserving only translations and Lorentz rotations within their world volumes. Moreover, when they are BPS, D-branes break precisely half of the supersymmetries. One can thus circumvent the naive argument if the space-time symmetry is reduced, or more generally whenever $\delta^{AB}\ P_\mu$ is replaced by a (field--dependent) matrix of lower rank. This type of modification can be induced by boundary terms~\cite{hughes_polchinski1, hughes_polchinski2} that contribute to the definition of the charges. Starting from type II strings, with the original ten--dimensional supercharges denoted $Q$ and ${\tilde Q}$, in the presence of a BPS Dp brane the boundary conditions preserve only half of the original supersymmetries, the fraction determined by the linear combination
\beq 
\left( Q \ - \ \Gamma_{\perp} {\tilde Q} \right) | Dp \rangle \ = \ 0 \ , 
\eeq
where $\Gamma_{\perp}$ denotes the product of the gamma matrices in directions perpendicular to the Dp brane world volume. The brane solutions that we shall describe in the next chapter will illustrate in detail this fact.
Alternatively, partial breaking may occur when some positivity conditions are somehow violated. The first example of this type, presented in~\cite{apt} rests, as we have seen, on the introduction of Fayet--Iliopoulos terms that are not purely real.

\section{\sc Supergravity in Eleven and Ten Dimensions} \label{sec:sugra1110}

We can now explore some properties of supergravity in ten and eleven dimensions that will play a pivotal role in our discussion of String Theory. As we shall see, they underlie the dualities~\cite{witten} linking supersymmetric strings to one another and to an elusive common principle usually dubbed M-theory. We begin with the eleven--dimensional supergravity of Cremmer, Julia and Scherk (CJS)~\cite{CJS}. We then describe the different forms of ten--dimensional supergravity~\cite{IIB,I,gs} and the coupling of $(1,0)$ supergravity to the ten--dimensional supersymmetric Yang--Mills theory introduced in~\cite{GSO,BSS}.  We also provide some useful complements on Weyl rescalings, Kaluza--Klein reductions, and $R$+$R^2$ gravity, which will bring up Starobinsky's model of inflation~\cite{starobinsky}, which was already mentioned in the previous chapter.

\subsection{\sc Supergravity in Eleven Dimensions}

Supergravity affords a unique eleven--dimensional formulation~\cite{CJS}, whose existence was foreseen in~\cite{nahm}, from which many lower--dimensional models can be derived by Kaluza--Klein reductions and/or truncations. The end results are generally complicated, due to the presence of non--polynomial scalar interactions, and major progress was only attained when scalar couplings were granted proper geometric characterizations. All other Bose fields are gauge fields, and are thus bound to enter the action principle polynomially, via their field strengths. These contributions are related by supersymmetry to other terms where derivatives are traded for pairs of fermions. The constructions terminate, in a few steps, with quartic Fermi couplings, which are, however, tedious to derive.

In eleven dimensions, the multiplet is to contain the graviton, which has in general $\frac{(D-2)(D-1)}{2}-1$ degrees of freedom, 44 in this case, and a gravitino, which has in general $2^{\left[\frac{D}{2}\right]-1}(D-3)$ degrees of freedom, 128 in this case. The Majorana condition is indeed an available option in eleven dimensions, while the Dirac equation halves, on shell, the number of components. The other factor, $(D-3)$, reflects the presence of the gauge condition $\Gamma \cdot \psi=0$, of the constraint $\partial \cdot \psi=0$ that it implies in conjunction with the Rarita--Schwinger equation, and of a residual on--shell gauge transformation that remains available as in Maxwell's theory. The remaining 84 Bose degrees of freedom are carried by a three--form gauge field, and the Lagrangian reads
\begin{eqnarray}
{\cal S} &=& \frac{1}{2\, k_{11}^2}\
\int d^{11} x \, e\, \bigg[ e^M{}_A\, e^N{}_B \, {R_{MN}}^{AB}(\omega) \nonumber  -  i\,\overline{\psi}_M\, \Gamma^{MNP}\, D_N\left(\frac{\omega+\hat{\omega}}{2}\right)\,\psi_P  \nonumber \\
   &-& \frac{i}{192} \ \left(\overline{\psi}_{M_1} \, \Gamma^{M_1 \ldots M_6} \, \psi_{M_2} \ + \ 12\, \overline{\psi}^{M_3} \, \Gamma^{M_4 M_5} \, \psi^{M_6}\right)\left(F_{M_3 \ldots M_6} \ + \ {\widehat{F}}_{M_3 \ldots M_6}\right) \nonumber \\
   &-& \frac{1}{48}\ F_{M_1\ldots M_4}\,F^{M_1\ldots M_4} \ - \ \frac{1}{6 (24)^2} \ \epsilon^{M_1 \ldots M_{11}}\ F_{M_1\ldots M_4}\ F_{M_5\ldots M_8}\, A_{M_9 M_{10} M_{11}} \bigg] \ . \label{eqd11}
\end{eqnarray}
Note that the last term within the brackets could be written more compactly $F \wedge F \wedge A$ in form language.

The spin connection $\omega$ solves its field equation and ``hats'' denote supercovariant quantities, {\it i.e.} completions of the expressions by terms involving the gravitino that eliminate all derivatives of $\epsilon$ arising from supersymmetry transformations. The supersymmetry transformations read
\bea
&& \delta\, e_M{}^A \ = \ \frac{i}{2}\ {\overline \epsilon}\,\gamma^A \, \psi_M \ , \nonumber \\
&& \delta\,\psi_M \ = \ D_M\left(\hat{\omega}\right)\, \epsilon \ + \ \frac{1}{288} \left( {\Gamma^{M_1 \ldots M_4}}_M \ - \ 8\, \delta_M^{M_1}\,\Gamma^{M_2 \ldots M_4} \right)  \hat{F}_{M_1 \ldots M_4}\, \epsilon \ , \nonumber \\
&& \delta\,A_{MNP} = \ - \ \frac{i}{4} \ {\overline \epsilon} \, \Gamma_{[MN}\, \psi_{P]} \ .
\eea
The CJS action includes a Chern--Simons term, which is gauge invariant only up to a total derivative, and whose presence has important consequences for String Theory.

\subsection{\sc Weyl Rescalings} \label{sec:Weyl}

It is often useful to see how the Einstein--Hilbert action behaves under a Weyl rescaling, 
\beq
G_{\mu\nu} \ = \ \Omega\ g_{\mu\nu} \ ,
\eeq
which is an overall redefinition of the metric tensor by a positive coordinate--dependent factor $\Omega$. To this end, let us begin from the effect on the Christoffel connection, 
\beq
\Gamma^\alpha{}_{\mu\nu}(G) \ = \ \Gamma^\alpha{}_{\mu\nu}(g) \ + \ \Delta\, \Gamma^\alpha{}_{\mu\nu} \ ,
\eeq
which follows directly from its definition,
where
\beq
\Delta\, \Gamma^\alpha{}_{\mu\nu} \ = \ \frac{1}{2}\left(\delta_\mu^\alpha \ \partial_\nu\,\log\,\Omega \ + \ \delta_\nu^\alpha \ \partial_\mu\,\log\,\Omega \ - \ g_{\mu\nu}\, g^{\alpha\beta}\partial_\beta\,\log\,\Omega \right) \ .
\eeq
This implies that the Ricci tensor and the Ricci scalar transform as
\bea
R_{\mu\nu}(G)  &=&  R_{\mu\nu}(g) \ -\  \frac{D-2}{2} \ \nabla_\mu\,\nabla_\nu\,\log\,\Omega \ - \ \frac{1}{2} \ g_{\mu\nu}\ g^{\alpha\beta}\,\nabla_\alpha\,\nabla_\beta\,\log\,\Omega \nonumber \\ &+& \frac{D-2}{4}\ \bigl( \partial_\mu\,\log\,\Omega\ \partial_\nu\,\log\,\Omega\ - \ g_{\mu\nu}\, g^{\alpha\beta}\,\partial_\alpha \log\,\Omega \ \partial_\beta \log\,\Omega \bigr)  \ , \\
R(G)  &=&  \frac{1}{\Omega}\,R(g) - \frac{D-1}{\Omega}\, g^{\alpha\beta}\,\nabla_\alpha\,\partial_\beta\,\log\,\Omega  -   \frac{(D-1)(D-2)}{4\,\Omega} \ g^{\alpha\beta}\,\partial_\alpha\,\log\,\Omega \ \partial_\beta\,\log\,\Omega \nonumber \ .
\eea
Therefore, when expressed in terms of $g$ the original Einstein--Hilbert action 
\beq
{\cal S}[G] \ = \ \frac{1}{2\, k^2}\ \int d^D x \ \sqrt{-G}\Bigl( R(G) \ - \ 2\,\lambda\Bigr)
\eeq
becomes
\bea
{\cal S}[G] &=& \frac{1}{2\, k^2}\ \int d^D x \ \sqrt{-g}\ \biggl[  \Omega^{\frac{D}{2}-1}\, R(g) \ - \ 2\, \lambda\ \Omega^\frac{D}{2} \nonumber \\ &+& \frac{(D-1)(D-2)}{4} \ \Omega^{\frac{D}{2}-1}\, g^{\mu\nu}\,\partial_\mu\,\log \Omega \ \partial_\nu\,\log \Omega\,\biggr] \ . \label{weylresceh}
\eea

A special case of this relation will prove particularly useful in the following. It corresponds to the choice
\beq
\log\,\Omega \ = \ - \ \frac{4}{D-2}\ \phi \ ,
\eeq
with $\phi$ the dilaton and $G$ and $g$ the metric tensors in the Einstein and string frames. The preceding transformations, together with
\beq
\int d^D x \ \sqrt{-G}\ G^{\mu\nu}\, \partial_\mu\,\phi\ \partial_\nu\,\phi \ = \ \int d^D x \ \sqrt{-g}\ e^{\,-\,2\,\phi}\ g^{\mu\nu}\, \partial_\mu\,\phi\ \partial_\nu\,\phi \ ,
\eeq
then imply that
\bea
&& \frac{1}{2\,k^2} \int d^D x \ \sqrt{-G}\ \biggl[  R(G) \ - \ \frac{4}{D-2}\  G^{\mu\nu}\, \partial_\mu\,\phi\ \partial_\nu\,\phi \biggr]   \\ &&= \ -\  \ \frac{1}{2\,k^2}  \int  d^D x \ \sqrt{-g}\ e^{\,-\,2\,\phi}\ \biggl[ R(g) \ + \ 4\,  g^{\mu\nu}\, \partial_\mu\,\phi\ \partial_\nu\,\phi\biggr]\nonumber \ .
\eea
This relation connects the contributions of gravity and dilaton in the more conventional Einstein frame and in the string frame, which is tailored to the world--sheet properties of String Theory.

\subsection{\sc Dual Forms for \texorpdfstring{$R+R^2$} \ \  Gravity and Starobinsky's Model}

We can now address the Starobinsky model~\cite{inflation1} and its relation to $R+R^2$ gravity.
Our starting point is the action
\beq
{\cal S} \,=\,\int d^D x \sqrt{-g} \biggl[  \frac{R}{2\,k^2}\  + \ \chi \left( \frac{R}{k}\,-\,m^2\,\phi\right) \ +\ \frac{1}{2}\ m^2\,\phi^2 \biggr] \ , \label{erchiphi}
\eeq
which depends on the metric tensor and on two scalar fields,
$\chi$ and $\phi$. Here $\chi$ and $\phi$ are Lagrange multipliers, and varying $\chi$ leads to
\beq
\phi \ = \ \frac{R}{m^2\,k} \ .
\eeq
Replacing $\phi$ by this expression reduces the content of the action to a combination of two terms, one involving $R$, as in the standard Einstein--Hilbert case, and one involving $R^2$:
\beq
{\cal S} \,=\,\int d^D x \sqrt{-g} \biggl[ \, \frac{R}{2\,k^2}\ \ +\ \frac{R^2 }{2\,m^2\,k^2}\biggr] \ .  \label{err2}
\eeq

Alternatively, one can eliminate $\phi$, which also enters algebraically the action principle of eq.~\eqref{erchiphi}, obtaining
\beq
{\cal S} \,=\,\int d^D x \sqrt{-g} \biggl[  \, \frac{1\,+\, 2\,k\,\chi}{2\,k^2}\ R \ -\ \frac{1}{2}\ m^2\,\chi^2 \biggr] \ , \label{erchi}
\eeq
and the results of the preceding section allow to recast this result in a more familiar form. To this end, one can resort to eq.~\eqref{weylresceh}, with
\beq
\Omega \ = \ \frac{2}{D-2} \ \log\left( 1\,+\, 2\,k\,\chi \right) \ ,
\eeq
for $(1\,+\, 2\,k\,\chi)>0$, so that the Weyl rescaling yields
\beq
\int d^D x \sqrt{-g} \,\left(1\,+\, 2\,k\,\chi\right) \frac{R}{2\,k^2}\ = \ \int d^D x \ \sqrt{-G} \left[  \,\frac{R}{2\,k^2} \ -\  2\,\frac{D-1}{D-2}\, \frac{G^{\mu\nu}\, \partial_\mu\,\chi\ \partial_\nu\,\chi}{(1\,+\, 2\,k\,\chi)^2} \right] \ .
\eeq
To this expression one must add the contribution of the scalar mass term,
and the final Einstein--frame form of the action,
\beq
{\cal S} \ = \ \int d^D x  \ \sqrt{-G} \,\biggl[ \frac{R}{2\,k^2} \ - \  2\,\frac{D-1}{D-2}\, \frac{G^{\mu\nu}\, \partial_\mu\,\chi\ \partial_\nu\,\chi}{(1\,+\, 2\,k\,\chi)^2} \ -\  \frac{m^2}{2}\ \frac{\chi^2}{(1\,+\, 2\,k\,\chi)^\frac{D}{D-2}}\,\biggr] \ ,
\eeq
describes indeed gravity and a massive scalar field with some self--interactions.

One can now perform the change of variables
\beq
1 \ + \ 2\,k\,\chi \ = \ e^{\alpha\,\varphi} \ , \label{staro_red}
\eeq
with
\beq
\alpha \ = \ k \, \sqrt{\frac{D-2}{D-1}} \ ,
\eeq
which covers the whole relevant region $(1+2 k \chi)>0$, 
in order to obtain a canonically normalized kinetic term for $\varphi$. With this choice the action becomes in general
\beq
{\cal S} \, = \, \int d^D x \ \sqrt{-G}\ \biggl[ \frac{R}{2\,k^2} \ - \  \frac{1}{2}\ G^{\mu\nu}\, \partial_\mu\,\varphi\ \partial_\nu\,\varphi \ - \ \frac{m^2\, e^{-\,\frac{\alpha \,D\,\varphi}{D-2}}}{8\, k^2}\, \Bigl( e^{\alpha\varphi} - 1\Bigr)^2\,\biggr] \  ,
\eeq
and, in particular, in four dimensions it reduces to Starobinsky's model of inflation~\cite{starobinsky}, for which
\bea
{\cal S} \, = \, \int d^D x \ \sqrt{-G}\ \biggl[ \frac{R}{2\,k^2} \ - \  \frac{1}{2}\ G^{\mu\nu}\, \partial_\mu\,\varphi\ \partial_\nu\,\varphi \ - \ \frac{m^2}{8\, k^2}\, \left( 1 - e^{-k \varphi\,\sqrt{\frac{2}{3}}}\right)^2\,\biggr] \ . \label{starobinsky}
\eea

More general $R+f(R)$ theories of gravity are equivalent to gravity coupled to a scalar field with proper self--interactions, as we can now see generalizing the preceding construction. The starting point
is now
\beq
{\cal S} \,=\,\int d^D x \sqrt{-g} \ \biggl[  \, \frac{R}{2\,k^2}\  + \  \chi \left( \frac{R}{k}\ -\ m^2\,\phi\right) \ +\ \frac{m^2}{2\,k^2}\ f(k\phi)\, \biggr] \ , \label{erchiphif}
\eeq
and extremizing with respect to $\chi$ leads to
\beq
{\cal S} \,=\,\int d^D x \sqrt{-g} \biggl[  \,\frac{R}{2\,k^2} \ +\ \frac{m^2}{2\,k^2}\ f\left(\frac{R}{m^2}\right)\, \biggr] \ . \label{erchiphif2}
\eeq
On the other hand, solving for $\phi$ leads to the condition
\beq
\chi \ = \ \frac{1}{2\,k}\ f'(k \,\phi) \ ,
\eeq
and to the action principle
\beq
{\cal S} \,=\,\int d^D x \sqrt{-g} \biggl[  \, \frac{1\,+\, 2\,k\,\chi}{2\,k^2}\ R \ +\  \frac{m^2}{2\,k^2 }\ \left( f(k\,\phi) \ - \ k\,\phi\,f'(k\,\phi) \right) \biggr] \ , \label{erchigen}
\eeq
where the kinetic term can be turned again into a canonical form via eq.~\eqref{staro_red}. However, the form of the potential is somewhat implicit, as it is based on the Legendre transform of $f$.

\subsection{\sc Circle Kaluza--Klein Theory} \label{sec:kaluzaklein}

Our next topic concerns the low--energy interpretation of theories in the presence of gravity and small extra dimensions. It is generally referred to as Kaluza--Klein theory, but for the sake of clarity, we shall actually begin from the simpler cases of free scalar and vector fields in flat space before turning to gravity proper. We shall also confine our attention to the simplest example of Kaluza--Klein theory, circle compactification.

\subsubsection{\sc Scalar Field}

Let us consider a $(D+1)$--dimensional space-time that is the direct product of a $D$--dimensional Minkowski space and a circle of radius $R$.
Let us begin from the case of a real massive scalar field in $D+1$ dimensions ($x^M=(x^\mu,y)$), with $0 \leq y \leq 2 \pi \ R$ parametrizing the length along a circle, described by the action
\beq
{\cal S} \ = \ \int d^{D+1}x \left[ - \ \frac{1}{2} \ \partial^M \phi\ \partial_M \phi \ - \ \frac{{\cal M}^2}{2}\, \phi^2 \right] \ .
\eeq
The $D$--dimensional interpretation rests on the Fourier decomposition 
\beq
\phi(x,y) \ = \ \sum_{k \in Z} \varphi_k(x) \ \frac{e^{i \frac{ky}{R}}}{\sqrt{2\,\pi\,R}} \ ,
\eeq
where the reality of $\phi$ implies that $\varphi_{-n}(x) \ = {\varphi}^*{}_n(x)$, and expanding the action leads to
\bea
{\cal S} &=& - \ \int d^{D}x \left\{ \frac{1}{2}\, \partial^\mu {\varphi}_{0}\ \partial_\mu \varphi_0\ + \  \frac{{\cal M}^2}{2}\,\varphi_0{}^2 \right. \nonumber \\ &+& \left. \sum_{k > 0} \Bigl[ \partial^\mu {\varphi}^*{}_{k}\ \partial_\mu \varphi_k \ + \  \Bigl( {\cal M}^2 \ +\  \frac{k^2}{R^2} \Bigr){\varphi}^*{}_{k}\,\varphi_k\Bigr] \right\} \ .
\eea
At low energies, a $D$--dimensional observer would thus perceive a single real scalar field, $\varphi_0$, with mass ${\cal M}$. The additional complex scalar fields with masses
\beq
{\cal M}_k \ = \ \sqrt{{\cal M}^2 \ + \ \frac{k^2}{R^2}}\qquad k=1,2,\ldots \quad \ ,
\eeq
could only be produced if the available energies were at least of order $\frac{1}{R}$ in natural units. Therefore, one cannot exclude the existence of extra dimensions with length scales below $10^{-18}cm$, since they would be so small to have escaped detection up to the energies that have been explored in detail so far. The validity of Newtonian gravity was tested down to distances of about $0.1$ mm, so that one could even imagine a fascinating alternative option for the Universe: extra dimensions of mesoscopic size might exist, where only gravity propagates~\cite{add,dark}. 

\subsubsection{\sc Maxwell Field}

Our second example concerns a Maxwell field, which presents a couple of novelties. To begin with, when reinterpreted in $D$ dimensions, a $D+1$--dimensional vector potential $A_M(x,y)$ comprises a $D$--dimensional vector potential $A_\mu(x,y)$ and a scalar, its internal component $\Pi(x,y)$.
These fields can be expanded as above, so that
\beq
A_M(x,y) \ = \ \sum_{k \in Z} {\cal A}_{M\,k}(x) \ \frac{e^{i \frac{ky}{R}}}{\sqrt{2\,\pi\,R}} \ ,
\eeq
and the original gauge transformations then translate into
\beq
\delta\,{\cal A}_{\mu\,k}(x) \ =\  \partial_\mu\,\Lambda_k \ , \qquad \delta\,\Pi_{k}(x) \ =\  i\,\frac{k}{R}\,\Lambda_k \ .
\eeq
As a result, for $k=0$ there is a massless vector together with a real massless scalar, while for $k\neq 0$ there is a tower of vectors accompanied by corresponding Stueckelberg fields $\Pi_{k}$. They can be removed by gauge fixing, leaving infinitely many massive vectors with masses $\frac{|k|}{R}$. The resulting action principle in $D$ dimensions,
\bea
{\cal S}&=& \int d^D x \left[ - \ \frac{1}{4} \ {\cal F}_{\mu\nu\,0}\,{\cal F}_{0}^{\mu\nu} \ - \ \frac{1}{2} \ \partial^\mu\,\Pi_0 \, \partial_\mu\,\Pi_0 \right] \nonumber \\
&+& \sum_{k > 0} \ \int d^D x \left[ - \ \frac{1}{2} \ {\cal F}^*{}_{\mu\nu\,k}\,{\cal F}_{k}^{\mu\nu} \ - \ \frac{k^2}{R^2}\  {\cal A}^*{}_{\mu\,k} \ {\cal A}^{\mu}{}_{k}\right] \ ,
\eea
where
\beq
{\cal F}_{\mu\nu\,k} \ = \ \partial_\mu\, {\cal A}_{\nu\,k} \ - \ \partial_\nu\, {\cal A}_{\mu\,k}
\eeq
describes indeed a real massless scalar, a real massless vector and an infinite tower of complex massive vectors.

\subsubsection{\sc Einstein Gravity}

We can now turn to describing how to derive the massless modes arising from the circle compactification of gravity, together with their interactions. The vielbein formalism is more convenient to this end, for a reason that will become apparent shortly. We also start from the form of the action that obtains integrating by parts, which reads
\beq
{\cal S} \ = \ \frac{1}{2\,k_{D+1}^2} \ \int d^{D+1}x \ {\hat e} \left(\, {{\hat \omega}_A}^{\ \ A C}\ {{\hat \omega}^B}_{\ \ B C} \ - \ {\hat \omega}_{BAC}\ {\hat \omega}^{ABC}\, \right) \ .
\eeq

The Ansatz
\beq
{{\hat e}_M}^{\ \ A} \ = \ \left( \begin{array}{ll} e^{\beta\varphi}\, e_\mu{}^a \ \ & e^{\varphi}\, A_\mu \\ 0 & e^{\varphi} \end{array} \right) \label{vielbEMA}
\eeq
for the vielbein,
where capital letters refer to the original $D+1$--dimensional space and small ones to the final $D$--dimensional one, is obtained by partially fixing the original local Lorentz symmetry, and the determinant then factorizes according to
\beq
{\hat e} \ = \ e^{\left(\beta D+1\right)\varphi}\ e \ .
\eeq
$\beta$ is a parameter that allows one to end in Einstein frame (or in string frame) in $D$ dimensions, and eq.~\eqref{vielbEMA} determines the inverse vielbein
\beq
{\hat e}^M{}_A \ = \ \left( \begin{array}{ll} e^{-\,\beta\varphi}\, e^\mu{}_a \quad \ & 0  \\ -\, e^{-\,\beta\varphi}\, e^\nu{}_a\,A_\nu \quad & e^{-\,\varphi} \end{array} \right) \ . \label{vielbEAM}
\eeq

Note that the Ansatz~\eqref{vielbEMA} involves precisely the massless modes that one could have expected from the preceding example, namely a tensor, a vector and scalar, all of which at the linearized level would be associated to fluctuations of $g_{\mu\nu}$, $g_{\mu4}$ and $g_{44}$. At the full non--linear level matters are more complicated, and indeed
\beq
{\hat g}_{MN} \ = \ {{\hat e}_M}^{\ \ A} \ \eta_{AB}\ {{\hat e}_N}^{\ \ B} \ = \ \left( \begin{array}{ll} e^{2\beta\varphi}\, g_{\mu\nu}\,+\, e^{2\varphi}\, A_\mu\,A_\nu \ \ & e^{2\varphi}\, A_\mu \\ e^{2\varphi}\,A_\nu & e^{2\varphi} \end{array} \right) \ , \label{kkmetric}
\eeq
so that tensor and vector components of the metric mix. In detail, this implies that
\beq
ds^2 \ = \left( e^{2\beta\varphi}\, g_{\mu\nu}\,+\, e^{2\varphi}\, A_\mu\,A_\nu \right)dx^\mu\,dx^\nu \ + \ 2 \, e^{2\varphi}\, A_\mu \, dx^\mu\, dy \ + \ e^{2\varphi} \, dy^2 \ ,
\eeq
and a main lesson here is that the vacuum value $\langle e^\varphi \rangle$ determines the actual size of the internal dimension.

In order to obtain the reduced action, one must begin by determining $\omega_{ABC}$ from the (antisymmetrized) vielbein postulate
\beq
\partial_M\,{{\hat e}_N}{}^{\ A} \ - \ \partial_N \,{{\hat e}_M}{}^{\ A} \ + \ {{\hat \omega}_M}^{\ \ AB}\ {\hat e}_{NB} \ - \ {{\hat \omega}_N}^{\ \ AB}\ {\hat e}_{MB} \ = \ 0 \ ,
\eeq
which can be solved for the spin connection, and leads to
\bea
{\hat \omega}_{ABC} &=& \frac{1}{2} \ {{\hat e}}^{M}{}_A \ {{\hat e}}^{N}{}_B \left( \partial_M\,{\hat e}_{NC} \ - \ \partial_N \,{\hat e}_{MC} \right) \ +\  \frac{1}{2} \ {{\hat e}}^{M}{}_C \ {{\hat e}}^{N}{}_A \left( \partial_M\,{\hat e}_{NB} \ - \ \partial_N \,{\hat e}_{MB} \right) \nonumber \\
&-& \frac{1}{2} \ {{\hat e}}^{M}{}_B \ {{\hat e}}^{N}{}_C \left( \partial_M\,{\hat e}_{NA} \ - \ \partial_N \,{\hat e}_{MA} \right) \ .
\eea

One can now compute from this expression all the components in $D$--dimensional notation, finding
\beq
{\hat \omega}_{abc} \ = \ e^{-\,\beta\varphi} \, \omega_{abc} \ + \ \beta\,e^{-\,\beta\varphi}\left( \eta_{ab}\,e^\mu{}_c\, \partial_\mu\,\varphi \ - \ \eta_{ac}\,e^\mu{}_b\,\partial_\mu\,\varphi \right)\ , \label{omegabc}
\eeq
and
\bea
{\hat \omega}_{4bc} &=&  - \ \frac{1}{2}\ e^{\left(1-2\beta\right)\varphi}\ F_{bc}\ , \nonumber \\
{\hat \omega}_{ab4}  &=&  \frac{1}{2}\ e^{\left(1-2\beta\right)\varphi}\ F_{ab} \ , \nonumber \\
{\hat \omega}_{44c}  &=&  e^{-\,\beta\varphi}\ e^\mu{}_c \,\partial_\mu\,\varphi \ . \label{omegarest}
\eea

Consequently
\bea
{{\hat \omega}_A}^{\ Ac} &=& e^{-\,\beta\varphi}\, {{\omega}_a}^{\ ac} \ + \ e^{-\,\beta\varphi} \left[ \beta(D-1)+1 \right] e^{\mu c} \, \partial_\mu \, \varphi \ , \nonumber \\
{{\hat \omega}_A}^{\ A4} &=& 0 \ ,
\eea
and therefore
\bea
{{\hat \omega}_A}^{\ AC}\ {{\hat \omega}^B}_{\ BC} &=& e^{-\,2\beta\varphi}\biggl\{ {{\omega}_a}^{\ ac}\ {{ \omega}^b}_{\ bc} \ + \ {2} \left[ \beta(D-1)+1 \right]{{\omega}_a}^{\ ac}\, e^{\mu}{}_c \,\partial_\mu\varphi \nonumber \\ &+& \left[ \beta(D-1)+1 \right]^2 \ g^{\mu\nu}\, {\partial_\mu\, \varphi\ \partial_\nu\, \varphi} \biggr\} \ .
\eea
In a similar fashion, one can show that
\bea
{\hat \omega}_{BAC}\ {\hat \omega}^{ABC} &=&e^{-\,2\beta\varphi}\biggl\{{\omega}_{bac}\ {\omega}^{abc} \ + \ 2\, {\beta} \ {{ \omega}_a}^{\ ac}\,e^\mu{}_c\, \partial_\mu\, \varphi \ + \ \frac{1}{4}\ e^{2\left(1-\beta\right)\varphi}\ F^{ab}\,F_{ab}\nonumber \\
&+& \left[ \beta^2(D+1)+1 \right]  g^{\mu\nu}\, \partial_\mu\, \varphi\ \partial_\nu\, \varphi \biggr\} \ ,
\eea
so that the reduced action takes the form
\bea
{\cal S} &=& \frac{2\pi r}{2\,k_{D+1}^2} \, \int d^D x \ e\ e^{[\beta(D-2)+1]\varphi} \biggl\{ \left( {{\omega}_a}^{\ ac}\ {{ \omega}^b}_{\ bc} \, - \, {\omega}_{bac}\ {\omega}^{abc}\right) \nonumber \\
 &-& \frac{e^{2\left(1-\beta\right)\varphi}}{4}\ F^{ab}\,F_{ab} \ + \ {2} \left[ \beta(D-2)+1 \right]{{\omega}_a}^{\ ac}\, e^\mu{}_c \,\partial_\mu \varphi \nonumber \\&+& \beta(D-1)[\beta (D-2) +2]\  g^{\mu\nu}\, \partial_\mu\, \varphi\ \partial_\nu\, \varphi \biggr\} \ .
\eea

Up to total derivatives, the first two terms reconstruct the Einstein--Hilbert Lagrangian, so that
\bea
{\cal S} &=& \frac{2\pi r}{2\,k_{D+1}^2} \, \int d^D x \ e\ e^{[\beta(D-2)+1]\varphi} \biggl\{ e^\mu{}_a\,e^\nu{}_b\,R_{\mu\nu}{}^{ab}(\omega) \ - \ \frac{e^{2\left(1-\beta\right)\varphi}}{4}\ F^{ab}\,F_{ab}  \nonumber \\&+& \beta(D-1)[\beta (D-2) +2]\ g^{\mu\nu}\, \partial_\mu\, \varphi\ \partial_\nu\, \varphi \biggr\} \ .
\eea
For $D>2$, one can now choose
\beq
\beta \ = \ - \ \frac{1}{D-2}
\eeq
and redefine $\varphi$ according to
\beq
\varphi \ = \ \sigma\,\sqrt{\frac{D-2}{2(D-1)}}  \ .
\eeq
The final result is the action in Einstein frame with a canonically normalized scalar, which takes the form
\bea
{\cal S} &=& \frac{1}{2\,k_{D}^2} \, \int d^D x \ e\ \biggl\{ e^\mu{}_a\,e^\nu{}_b\,R_{\mu\nu}{}^{ab}(\omega)  \, - \, \frac{e^{\sigma\,\sqrt{\frac{2(D-1)}{D-2}}}}{4}\ F^{ab}\,F_{ab} -  \frac{1}{2} \ g^{\mu\nu}\, \partial_\mu\, \sigma\ \partial_\nu\, \sigma \biggr\} \ .
\eea

Some comments are now in order. First, the $D$--dimensional Newton constant is related to its $(D+1)$--dimensional counterpart according to
\beq
\frac{1}{k_{D}^2} \ = \ \frac{2 \pi r}{k_{D+1}^2}  \ .
\eeq
Moreover, the canonically normalized scalar $\sigma$ enters the effective internal radius, as we have seen, for which the theory yields no prediction, since they both rest on the undetermined vacuum value $\langle \sigma \rangle$. These undetermined values are a recurring problem with gravity, and are usually called ``moduli''. In this case
\beq
r_{eff} \ = \ r \ \left\langle e^{\sigma\,\sqrt{\frac{D-2}{2(D-1)}}} \right\rangle \ .
\eeq

These considerations extend directly to compactifications on products of circles, and the global symmetries corresponding to translations along any of them translates into a product of local $U(1)$ symmetries for the corresponding Maxwell fields.
On the other hand, compactifications on internal spaces with non--Abelian isometries, and in particular on spheres, would give rise to non--Abelian gauge fields in the resulting low--energy theory (D.J.~Gross gave in~\cite{DJG} a very nice account of O.~Klein's prescient work in this respect). 
As we shall see, in String Theory a non-Abelian symmetry can emerge, in special circumstances, even from circle compactification.

\subsection{\sc Supergravity in Ten Dimensions}

There are a only few options for supergravity in ten dimensions:
\begin{enumerate}
  \item \textbf{Type--IIA, or (1,1) Supergravity.} This theory is characterized by a pair of supercharges of opposite chiralities. It can be linked to Supergravity in eleven dimensions~\cite{CJS} via a Kaluza--Klein circle reduction. It is not chiral in spacetime, and is thus free of anomalies. It provides the low--energy effective Lagrangian of type-IIA string theory, to which we shall return;
  \item\textbf{ Type--IIB, or (2,0) Supergravity.} This theory~\cite{IIB} is characterized by a pair of supercharges of identical chiralities.  It is a novelty of ten dimensions, and is chiral in spacetime but nonetheless free of anomalies, as first shown in~\cite{agwitt}. It provides the low--energy effective Lagrangian of type-IIB string theory, to which we shall return;
  \item \textbf{Type-I, or (1,0) Supergravity.} This theory is a truncation of the two preceding models~\cite{I}. It is inconsistent by itself due to gravitational anomalies, but thanks to the Green--Schwarz mechanism~\cite{gs} it provides the low--energy description of the two supersymmetric heterotic string models of~\cite{heterotic1,heterotic2,heterotic3} with gauge groups $SO(32)$ and $E_8 \times E_8$, and also of the supersymmetric type-I $SO(32)$ string model, if coupled to ten--dimensional supersymmetric Yang--Mills theory~\cite{GSO,BSS}. Only the $SO(32)$ gauge group can in fact be realized with the Chan--Paton construction for open strings~\cite{cp1,cp2,cp3,cp4,cp5}, starting from the type--IIB string, via the orientifold construction~\cite{orientifolds1,orientifolds2,orientifolds3,orientifolds4,orientifolds5,orientifolds6,orientifolds7,orientifolds8,orientifolds_rev1,orientifolds_rev2,orientifolds_rev3}, as we shall review in Section~\ref{sec:critical_strings}. In addition, ten--dimensional $(1,0)$ supergravity underlies Sugimoto's $USp(32)$ model~\cite{sugimoto}, where supersymmetry is non--linearly realized, as was pointed out in~\cite{dmnonlinear1,dmnonlinear2}.
\end{enumerate}

\subsubsection{\sc Type--IIA Supergravity}

One can give a fairly detailed account of the structure of this model following some of the steps described in our discussion of Kaluza--Klein theory. 

Let us begin with the Fermi fields. Starting from a gravitino in $D=11$, $\psi_M$, one can recognize the emergence of two ten--dimensional gravitini, $\psi_\mu^{1}$ and $\psi_\mu^{2}$, obtained by left and right Weyl projections, and of pair of spinors with opposite chiralities emerging in a similar fashion from $\psi_{10}$. This is a non--chiral Fermi spectrum. The Kaluza--Klein theory of the preceding section tells us that $e_M{}^A$ gives rise to a ten--dimensional vielbein $e_\mu{}^a$, an Abelian vector $A_\mu$ and a scalar $\varphi$ (which will play the role of a dilaton of the corresponding IIA string theory, with $e^{\langle \varphi\rangle}$ the corresponding string coupling), while the three form $A_{MNP}$ gives rise, in a similar fashion, to a three form $A_{\mu\nu\rho}$ and a two form $B_{\mu\nu}$ from $A_{\mu\nu 10}$.

It is instructive to take a closer look at the bosonic terms, starting from the eleven--dimensional Einstein--Hilbert action, which yields the ten--dimensional Einstein--frame contributions
\bea
{\cal S}_{10\,,\,{\cal E}} &=& \frac{1}{2\,k_{10}^2} \, \int d^{10} x \ \sqrt{-G}\ \biggl\{ E_a{}^\mu\,E_b{}^\nu\,{R_{\mu\nu}}^{ab}(\omega) \, - \, \frac{e^{\,\frac{3}{2}\,\varphi}}{4}\ F^{\mu\nu}\,F_{\mu\nu} \nonumber \\
&-&  \frac{1}{2} \ {\ G^{\mu\nu}\, \partial_\mu\,\varphi\ \partial_\nu\,\varphi} \biggr\} \ ,
\eea
according to Section \ref{sec:kaluzaklein}, where $E$ denotes the Einstein--frame vielbein. As we have seen, the transition to the string frame is effected by the Weyl rescaling $G_{\mu\nu} = g_{\mu\nu}\, e^{\,-\,\frac{\varphi}{2}}$, which turns the action into
\bea
{\cal S}_{10\,,\, {\cal S}} &=& \frac{1}{2\,k_{10}^2} \, \int d^{10} x \ \sqrt{-g}\, \biggl\{ e^{\,-\,2\,\varphi} \left[ e_a{}^\mu\,e_b{}^\nu\,{R_{\mu\nu}}^{ab}(\omega)\ +\  4 \ {g^{\mu\nu}\, \partial_\mu\,\varphi\ \partial_\nu\,\varphi} \right] \nonumber \\
&- &  \frac{1}{4}\ F^{\mu\nu}\,F_{\mu\nu} \biggr\} \ ,
\eea
where $e$ denotes the string--frame vielbein.

Let us now recall that in the Einstein frame this reduction rests on the eleven--dimensional metric
\beq
ds^2 \ = \ e^{\,-\,\frac{\varphi}{6}} \ G_{\mu\nu}\, dx^\mu\,dx^\nu \ +\ e^{\,\frac{4\,\varphi}{3}} \left(dy \ + \ A_\mu\,dx^\mu \right)^2 \ ,
\eeq
while in the string frame it rests on
\beq
ds^2 \ = \ e^{\,-\,\frac{2\,\varphi}{3}} \ g_{\mu\nu}\, dx^\mu\,dx^\nu \ +\ e^{\,\frac{4\,\varphi}{3}} \left(dy \ + \ A_\mu\,dx^\mu \right)^2 \ ,
\eeq
since according to Section~\ref{sec:Weyl}
\beq
G_{\mu\nu} \ = \ e^{\,-\, \frac{\varphi}{2}} \ g_{\mu\nu} \ .
\eeq

In both cases, the eleven--dimensional circle decompactifies in the strong--coupling limit $\varphi \to \infty$~\cite{witten},
since
\beq
M_{P,11} \ r_{11} \ \sim \ e^{\,\frac{2\,\varphi}{3}} \ = \ \left(g_s\right)^\frac{2}{3} \ , \label{r11phi}
\eeq
where $M_{P,11}$ denotes the eleven--dimensional Planck mass.

It is instructive to complete the derivation of the bosonic terms in the string frame. To this end, note that the four--form field strength kinetic term, proportional to $F_{\mu\nu\rho\sigma}^2$, is not accompanied by any dilaton factor in the string frame. Indeed, the ten--dimensional portion of the vielbein determinant carries a factor $e^{-10 \varphi/3}$, its eleven--dimensional portion carries a factor $e^{2\varphi/3}$ and the four inverse ten--dimensional metrics involved in the contraction carry an overall factor $e^{8\varphi/3}$. In a similar fashion, one can conclude that the three--form field strength contribution, which we shall denote $H_{\mu\nu\rho}^2$ following common practice, is accompanied by a factor $e^{-2 \varphi}$, since it involves three inverse space--time metrics and one inverse internal metric. 

All in all, the bosonic contributions to the string--frame IIA action add up to
\bea
{\cal S}_{10\,,\, {\cal S}} &=& \frac{1}{2\,k_{10}^2} \, \int d^{10} x \ e\, \biggl\{ e^{\,-\,2\,\varphi} \biggl[ e_a{}^\mu\,e_b{}^\nu\,{R_{\mu\nu}}^{ab}(\omega)\ +\  4 \ {g^{\mu\nu}\, \partial_\mu\,\varphi\ \partial_\nu\,\varphi}  \\ &-& \frac{1}{12} \ H_{\mu\nu\rho}\, H^{\mu\nu\rho}\biggr] \ - \   \frac{1}{4}\ F^{\mu\nu}\,F_{\mu\nu} \ - \ \frac{1}{48} \ F^{\mu\nu\rho\sigma}\, F_{ \mu\nu\rho\sigma}  \ + \ 3 \,F_4\wedge F_4\wedge B_2\biggr\} \ , \nonumber
\eea
where $F_{\mu\nu\rho\sigma}$ denotes the curl of $A_{\mu\nu\rho}$ and $H_{\mu\nu\rho}$ denotes the curl of $A_{\mu\nu 10}$.
There are thus two groups of terms, but only the first is accompanied by a factor $e^{-2\varphi}$~\cite{witten}. 

\subsubsection{\sc Type--IIB Supergravity} This theory originates in ten dimensions~\cite{IIB}, and, as we anticipated, is chiral. This is actually the case for both Fermi and Bose spectra. The former includes a complex Weyl gravitino and a complex spinor of opposite chiralities, while the latter includes a complex scalar 
\beq
\tau \ = \ a \ + \ i e^{- \,\varphi}
\eeq
parametrizing the $SL(2,R)/U(1)$ coset, whose low--energy action
\beq{}{}{}{}{}{}{}{}{}{}{}{}{}{}{}{}{}{}
{\cal S} \ = \ - \ \int d^{10}\,x\  \sqrt{-g}\ \frac{\partial_M \overline{\tau}\,\partial^M \tau}{\left({\rm Im}\ \tau\right)^2} , \label{dilaxion_action}
\eeq
is invariant under $PSL(2,R)$, which transforms $\tau$ according to
\beq{}{}{}{}{}{}{}{}{}{}{}{}{}{}{}{}{}{}
\tau \ \to \ \frac{\alpha \,\tau \ + \ \beta}{\gamma\,\tau \ + \ \delta}  \ , \qquad
\alpha\,\delta \ - \ \beta\,\gamma \ = \ 1 \ .  \label{mobius_sym} \eeq 
The spectrum includes
an $SL(2,R)$ doublet of two-form potentials, and also 
a chiral Bose field, a four--form potential whose equation of motion is a super--covariant completion (with some mixings with the two--forms) of the self--duality condition
\beq
F_{\mu\nu\rho\sigma\tau} \ = \ \widetilde{F}_{\mu\nu\rho\sigma\tau}  \ . \label{sdual5form}
\eeq

We shall often use the form language, where the Hodge dual of an $n$-form $\omega$ is a $D-n$ form given by
\beq
\star \ \omega \ = \ \frac{\omega_{\mu_1 \ldots \mu_n}}{n!} \ \star \left(dx^{\mu_1} \wedge \ldots  dx^{\mu_n}\right) \ = \ \frac{\epsilon^{\mu_1 \ \ldots \mu_n}{}_{ \mu_{n+1} \ldots \mu_{D}}\,\omega_{\mu_1 \ldots \mu_n}}{n! (D-n)!} \ dx^{\mu_{n+1}} \wedge \ldots dx^{\mu_{D}} \ ,
\eeq
with $\epsilon$ the totally antisymmetric Levi--Civita symbol and $\epsilon_{0 1...D-1} = 1$. Note that
\beq
\star\, \star \ = \ - \ (-1)^{n(D-n)} \ ,
\eeq
so that eq.~\eqref{sdual5form} is consistent for a real form in ten dimensions. A similar condition would be inconsistent for a real Maxwell field strength in four dimensions with Minkowski signature, but it is consistent with a Euclidean signature.

To date, there is no overall agreement on a fully satisfactory action principle for the IIB model, due to the peculiar first--order equation of motion for the four--form potential, but the complete covariant field equations are known~\cite{IIB}, together with a corresponding on--shell superspace formulation (hence, without the auxiliary fields) that encodes them~\footnote{Much progress along these lines was recently made by K. Mkrtchyan and collaborators. See, for example, the review~\cite{karapet} and references therein.}. $SU(1,1)$ is a symmetry of this supergravity model that acts in a non--trivial fashion on the dilaton, and an $SL(2,Z)$ subgroup is expected to remain a symmetry of the full IIB string theory~\cite{hull-townsend1,hull-townsend2}. 

\subsubsection{\sc Fermionic Terms and Supersymmetries for the Type--II Theories} \label{sec:susyIIAB}

Following~\cite{bergshoeff}, we can now present the string--frame type--II actions up to quadratic order in the Fermi fields in a compact unified notation, 
\bea
{\cal S}_{10} &=& \frac{1}{2\,k_{10}^2} \, \int d^{10} x \ e\, \biggl\{ e^{\,-\,2\,\varphi} \biggl[ R \ +\  4 \ {g^{\mu\nu}\, \partial_\mu\,\varphi\ \partial_\nu\,\varphi}  \ - \  \frac{1}{12} \ H_{\mu\nu\rho}\, H^{\mu\nu\rho} \nonumber \ + \ 2\,\partial^\mu\varphi\,\chi_\mu^{(1)}  \nonumber \\  &-& \frac{1}{6}\, H^{\mu\nu\rho}\,\chi_{\mu\nu\rho}^{(3)} \ - \ 2\ \bar{\psi}_{\mu}\,\Gamma^{\mu\nu\rho}\,D_\nu\,\psi_\rho \ + \ 2\,\bar{\lambda}\,\Gamma^\mu\,D_\mu\,\lambda \ - \ 4 \bar{\lambda}\,\Gamma^{\mu\nu}\,D_\mu\,\psi_\nu \biggr] \nonumber  \\ 
&-&  \frac{1}{2}\ \sum_{n=0,1/2}^{5,9/2} \frac{1}{(2n)!}\  {\cal H}_{\mu_1\ldots \mu_{2n}} \left[ \frac{1}{2}\ {\cal H}^{\mu_1\ldots \mu_{2n}}\ +\ \Psi^{\mu_1\ldots \mu_{2n}}\right] 
\biggr\} \ , \label{formal_action}
\eea
where the notation for the scalar curvature has been adapted to our own. This formal expression includes the relevant field strengths and their duals for both theories and also the Romans mass, a peculiar deformation introduced in~\cite{romansm}. The formal action principle in eq.~\eqref{formal_action} can encompass the two cases, with any of the following two choices of independent field strengths:
\begin{enumerate}
\item[IIA. ] $H$ and the ${\cal H}$'s for integer values of $n$;
\item[IIB. ] $H$ and the ${\cal H}$'s for half--odd integer values of $n$.
\end{enumerate}
Moreover, the self--duality condition for the five--form field strength in type--IIB should only be enforced in the resulting equations of motion. The field strengths are defined in general as
\beq
H \ = \ d\,B \ , \qquad {\cal H} \ = \ d\,A \ - \ dB \wedge A \ + \ m \ e^B \ ,
\eeq
where the Romans mass is only present in type IIA, and thus together with even--rank field strengths. With these definitions, the Chern--Simons term is included in the kinetic terms involving these field strengths. 

The Fermi fields are doublets (of opposite chiralities in type--IIA and of identical chiralities in type--IIB), and the fermionic couplings read
\bea
\chi_\mu^{(1)} &=& - \ 2\,\bar{\psi}_\nu\,\Gamma^\nu\,\psi_\mu \ - \ 2\,\bar{\lambda}\,\Gamma^\nu\,\Gamma_\mu\,\psi_\nu \ , \nonumber \\
\chi_{\mu\nu\rho}^{(3)} &=& \frac{1}{2}\, \bar{\psi}_\sigma\,\Gamma^{[\sigma}\,\Gamma_{\mu\nu\rho}\,\Gamma^{\tau]}\, {\cal P} \psi_{\tau}\ + \ \bar{\lambda}\,\Gamma_{\mu\nu\rho}{}^{\sigma}\,{\cal P}\,\psi_\sigma \ - \ \frac{1}{2}\,\bar{\lambda}\,{\cal P}\,\Gamma_{\mu\nu\rho}\,\lambda \ , \\
e^\varphi\, \Psi_{\mu_1 \ldots \mu_{2n}}^{(2n)} \!\!\! &=&\frac{1}{2}\, \bar{\psi}_\sigma\,\Gamma^{[\sigma}\,\Gamma_{\mu_1 \ldots \mu_{2n}}\,\Gamma^{\tau]}\, {\cal P}_n \, \psi_{\tau} \, + \, \frac{1}{2} \,\bar{\lambda}\, \Gamma_{\mu_1 \ldots \mu_{2n}}\,\Gamma^\sigma\,{\cal P}_n\,\psi_\sigma \, - \, \frac{1}{4}\, \bar{\lambda}\,\Gamma_{[\mu_1 \ldots \mu_{2n-1}} {\cal P}_n\, \Gamma_{\mu_{2n}]}\,\lambda \ , \nonumber
\eea
where the $\Psi^{(2n)}$ and the field strengths satisfy the self--duality conditions
\bea
\Psi^{(2n)} &=& (-1)^{[n]+1} \ \star \ \Psi^{(2n)}  \ , \nonumber \\
{\cal H}^{(2n)} \ + \ \Psi^{(2n)} &=& \ (-1)^{[n]}\, \star\,{\cal H}^{(10-2n)} \ .
\eea

Finally, the supersymmetry transformations can be deduced truncating, as explained above, the two expressions
\bea
\delta\,e_\mu{}^a &=& \bar{\epsilon} \gamma^a\,\psi_\mu \ , \nonumber \\
\delta\,\psi_\mu &=& \left[\partial_\mu \ +\  \frac{1}{4}\,{\slashed{\omega}_\mu} \ \ \ + \ \frac{1}{8}\, {\cal P}\, \slashed{H}_\mu \ + \ \frac{1}{16}\, e^\varphi \sum_{n=0,1/2}^{5,9/2} \frac{\slashed{\cal H}_{2n}}{(2n)!}\, \Gamma_\mu\,{\cal P}_n  \right] \epsilon \ , \nonumber \\
\delta\,B_{\mu\nu} &=& -\ 2\,\bar{\epsilon}\,\Gamma_{[\mu} \,{\cal P}\,\psi_{\nu]} \ , \nonumber \\
\delta\,A_{\mu_1 \ldots \mu_{2n-1}} &=& - \ e^{-\varphi}\, \bar{\epsilon}\,\Gamma_{[\mu_1 \ldots \mu_{2n-2}}\,{\cal P}_n\,\left[(2n-1) \psi_{\mu_{2n-1}]} \,-\,\frac{1}{2}\,\Gamma_{\mu_{2n-1}]}\,\lambda \right] \nonumber \\
&+& (n-1)(2n-1)\,A_{[\mu_1 \ldots \mu_{2n-3}}\,\delta\,B_{\mu_{2n-2}\mu_{2n-1}]} \ , \nonumber \\ 
\delta\,\lambda &=& \left[\slashed{\partial}\, \varphi \ + \ \frac{1}{12}\, {\cal P}\, \slashed{H} \ +\ \frac{1}{8}\, e^\varphi \sum_{n=0,1/2}^{5,9/2} (-1)^{2n}\left(5-2 n\right) \frac{\slashed{\cal H}_{2n}}{(2n)!}\, {\cal P}_n \right]\epsilon \nonumber \\
\delta\,\varphi &=& \frac{1}{2}\,\bar{\epsilon\,\lambda} \ . \label{fermi_transf_gen}
\eea
These expressions combine contributions involving the field strengths present in type IIA and type IIB with their duals. The integer values of $n$ concern the IIA theory, where the two values $n=0$ and $5$ correspond to the Romans mass and its dual~\cite{romansm}, while half--odd integer values of $n$ concern the IIB theory, where $n=\frac{5}{2}$ corresponds to the self--dual five--form field strength. Moreover,
\begin{enumerate}
    \item for type IIA \  ${\cal P}=\gamma_{11}$, ${\cal P}_{n} = \left(\gamma_{11}\right)^{n}$, with $n$ integer;
    \item for type IIB \  ${\cal P}=-\, \sigma_3$, ${\cal P}_\frac{3}{2} = {\cal P}_\frac{7}{2} =\sigma_1$, and ${\cal P}_{n} = i \sigma_2$ in the remaining cases.
\end{enumerate}
In the preceding expressions, as in~\cite{bergshoeff}, all antisymmetrizations have strength one.

\subsubsection{\sc Type--I Supergravity Coupled to  Super Yang--Mills} \label{sec:(1,0)sugra}
A portion of this theory can be regarded as a truncation of the two preceding models, whose fields include the vielbein $e_\mu{}^a$, the dilaton and a two-form potential, together with a left--handed Majorana--Weyl gravitino $\psi_\mu$ and a right-handed Majorana--Weyl spinor $\lambda$, while the ten--dimen\-sional supersymmetric Yang--Mills theory~\cite{GSO,BSS} combines Yang--Mills fields $A_\mu^a$ and adjoint--valued left--handed Majorana--Weyl spinors $\chi^a$. Up to quadratic order in spinor fields, the resulting action reads
\bea
{\cal S}_{10\,,\, H} &=& \frac{1}{2\,k_{10}^2} \, \int d^{10} x \ e\ e^{\,-\,2\,\varphi} \biggl[ e_a{}^\mu\,e_b{}^\nu\,{R_{\mu\nu}}^{ab}(\omega)\ +\  4 \ {g^{\mu\nu}\, \partial_\mu\,\varphi\ \partial_\nu\,\varphi}  \\ 
&-& 2\,\bar{\psi}_\mu\,\Gamma^{\mu\nu\rho}\,D_\nu\,\psi_\rho \ + \ 2\,\bar{\lambda}\,\Gamma^\mu\,D_\mu\,\lambda \ - \ 4 \bar{\lambda}\,\Gamma^{\mu\nu}\,D_\mu\,\psi_\nu \ - \ 2 \bar{\chi}^a \Gamma^\mu D_\mu \,\chi^a \nonumber \\
&-& 4 \partial_\mu\,\varphi \left(\bar{\psi}_\nu\,\Gamma^{\nu}\,\psi^\mu \ + \bar{\lambda} \,\Gamma^{\nu}\,\Gamma^{\mu}\,\psi_\nu\right) \ - \  \frac{1}{12} \ H_{\mu\nu\rho}\, H^{\mu\nu\rho}  \ - \   \frac{1}{4}\ F^{a \mu\nu}\,F_{\mu\nu}^a \nonumber \\
&+& \frac{1}{12} \ H^{\mu\nu\rho}\left( \bar{\psi}_\sigma\,\Gamma^{[\sigma}\,\Gamma_{\mu\nu\rho}\,\Gamma^{\tau]}\,\psi_\tau \ + \ 2\,\bar{\lambda}\,\Gamma_{\mu\nu\rho}{}^\sigma\,\psi_\sigma \ - \ \bar{\lambda}\,\Gamma_{\mu\nu\rho}\,\lambda \ +\ \bar{\chi}{}^a\,\Gamma_{\mu\nu\rho}\,\chi^a\right) \nonumber \\
&-& \frac{1}{{2}}\, \bar{\chi}{}^a \Gamma^\mu\,\Gamma^{\rho\sigma}\left(\psi_\mu \,-\, \frac{1}{12}\,\Gamma_\mu\,\lambda \right) F_{\rho\sigma}^a\biggr]\ , \nonumber \label{n1het}
\eea
if one refers to the heterotic strings. In this case, up to cubic spinor terms, the supersymmetry transformations read
\bea
\delta\,e_\mu{}^a &=& \bar{\epsilon}\,\gamma^a\,\psi_\mu \ , \qquad \delta\,\psi_\mu \ = \ D_\mu\,\epsilon \ + \ \frac{1}{8}\ \Gamma^{\nu\rho}\,\epsilon\ H_{\mu\nu\rho} \ , \nonumber \\
\delta\,\varphi &=& \frac{1}{2}\ \bar{\epsilon}\,\lambda \ , \qquad \delta\,\lambda \ = \ \Gamma^\mu\,\epsilon \, \partial_\mu\,\varphi \ + \ \frac{1}{12}\, \Gamma^{\mu\nu\rho}\,\epsilon\ H_{\mu\nu\rho} \ , \nonumber \\
\delta\,B_{\mu\nu} &=& - \ \bar{\epsilon}\left( \Gamma_\mu\,\psi_\nu \ - \ \Gamma_\nu\,\psi_\mu\right) \ , \nonumber \\
\delta\,A_\mu^a &=& \bar{\epsilon} \,\Gamma_\mu\,\chi^a \ , \qquad \delta\,\chi^a \ = \ - \ \frac{1}{8} \ \Gamma^{\mu\nu}\,\epsilon \ F_{\mu\nu}^a \ , \label{susy_typeI}
\eea
and the three--form field strength is modified so that
\beq
H \ = \ d\, B \ - \ \Omega_3 \ , \label{mod_H}
\eeq
with
\beq
\Omega_3 \ = \ \mathrm{Tr} \left( A \wedge dA \ + \ \frac{2}{3}\, A \wedge A \wedge A \right) \label{omega3}
\eeq
a Chern--Simons form. 

The counterpart of the results for the type I string can be obtained by first transforming the preceding expressions into the Einstein frame, 
letting
\bea
e_\mu{}^a &=&  e^\frac{\varphi}{4} \, \tilde{e}_\mu{}^a \ , \qquad D_\mu \ = \ \tilde{D}_\mu \,+\, \frac{1}{8}\,\Gamma_{\mu\nu}\, \partial^\nu\,\varphi\ , \nonumber \\
\psi_\mu &=& e^{\frac{1}{8}\,\varphi} \, {\tilde{\psi}}_\mu \ , \qquad \lambda \,=\, e^{-\frac{1}{8}\,\varphi} \, \tilde{\lambda} \ , \qquad  \chi^a \,=\, e^{-\frac{1}{8}\,\varphi} \, \tilde{\chi}{}^a \ ,
\eea
while also redefining the supersymmetry parameter according to
\beq
\epsilon \ = \ e^{\frac{\varphi}{8}}\,\tilde{\epsilon} \ .
\eeq
The dilaton is then to be redefined according to $\varphi = \,-\,\tilde{\varphi}$
before returning to the string frame, and finally the gaugini $\chi^a$ are to be subject to the redefinition
\beq
\tilde{\chi}{}^a \ = \ \hat{\chi}{}^a \ e^\frac{\tilde{\varphi}}{2} 
\eeq
so that their kinetic terms carry the $e^{-\,\tilde{\varphi}}$ that is to accompany disk--level contributions.

As we saw in eq.~\eqref{mod_H}, an interesting feature of these theories is that the field strength of $H$ is not simply the curl of the corresponding two--form $B$, but involves a peculiar construct, the $\Omega_3$ Chern--Simons form of eq.~\eqref{omega3}
built from the gauge potential $A$.  An important consequence of this fact is that the two-form is affected by vector gauge transformations. 

In order to deduce the complete gauge transformation of $B$, let us first note that although $\Omega_3$ is not gauge invariant, its exterior derivative is, since
\beq
d\,\Omega_3 \ = \ {\rm Tr} \left( F \wedge F \right) \ ,
\eeq
where the field strength is
\beq
F \ = \ d\,A \ + A \wedge A \ .
\eeq
This relation holds because the wedge product of four potentials is anti--cyclic and, therefore,
\beq
{\rm Tr} \left( A \wedge A \wedge A \wedge A\right) \ = \ 0 \ ,
\eeq
so that a quartic term in $A$ can be added at no cost.

If one varies $\Omega_3$, the result can be recast in the form
\beq
\delta\,\Omega_3 \,=\,{\rm Tr} \left( 2\,\delta\,A\wedge F \,+\,A \wedge d\,\delta\,A \,-\, \delta\,A \wedge d\,A \right) \ ,
\eeq
and that these terms can be written more conveniently as
\beq
\delta\,\Omega_3 \,=\, -\,d\,{\rm Tr} \left[A \wedge (d\,\Lambda + A\,\Lambda-\Lambda\,A)\right] \,+\, 2\,d\,{\rm Tr} \left[\Lambda (dA + A \wedge A) \right] \,,
\eeq
using the Bianchi identity for $F$. Putting these terms together and recalling that $d^2=0$ gives finally
\beq
\delta\,\Omega_3 \,=\, d\,\Omega_2{}^1 \ ,
\eeq
where
\beq
\Omega_2{}^1 \ = \ {\rm Tr} \left( \Lambda\, d\,A\right) \ .
\eeq
Consequently, the modified field strength
\beq
H \ = \ d\, B \ - \ \Omega_3  
\eeq
is gauge invariant provided $B$ transforms under vector gauge transformations! The complete transformation of $B$ is thus
\beq
\delta\,B \ = \ d\,\Lambda_1 \ + \ {\rm Tr} \left( \Lambda\, d\,A\right) \ ,
\eeq
where $\Lambda_1$ denotes the standard one--form gauge parameter for the two--form potential.

This subtle behavior emerged from the coupling between supergravity and Yang--Mills theory in ten dimensions~\cite{I}, and has a profound meaning. It underlies the Green--Schwarz mechanism~\cite{gs}, which eliminates a portion of the apparent ten--dimensional anomalies via local counterterms that are automatically included in String Theory~\footnote{To this end, the Yang--Mills Chern--Simon term is to be completed by a corresponding higher--derivative gravitational Chern-Simons term.}. We shall see that there are two options for ten--dimensional type-I supergravity coupled to ten--dimensional supersymmetric Yang--Mills that can be realized in String Theory, where the anomaly is canceled by this mechanism, with gauge groups $SO(32)$ and $E_8 \times E_8$. The latter originates from the heterotic $E_8 \times E_8$ theory, while the former has two distinct ``dual'' realizations in the $SO(32)$ heterotic theory and in the type-I model of open and closed strings.

\newpage
{}
\vskip 5cm 
\begin{center}
    {\Large \sc Part II} 
    \vskip 2cm
    {\Large \sc Top--Down Approach to Supersymmetry Breaking}
\end{center}
\emph{The second portion of the review is devoted to some key aspects of String Theory. Our focus is on the basic constraints underlying string vacua and on the challenges that supersymmetry breaking adds to the current picture, which are mostly connected to the vexing issue of vacuum stability. We also discuss some incomplete, and yet encouraging, results in this respect, which concern non--supersymmetric interval compactifications. We conclude reviewing some mechanisms for stabilizing moduli and the KKLT picture, before turning to some key facts of Cosmology, with an eye on indications provided by String Theory. These include an intriguing link between string--inspired mechanisms for supersymmetry breaking at high energy scales and the onset of the inflationary phase.}
\newpage

\section{\sc Critical Strings and their Circle Compactifications} \label{sec:critical_strings}

In the preceding sections, we have reviewed a number of results concerning what can be termed the ``bottom-up'' approach to supersymmetry breaking.  We have also highlighted some indications that point to a deep link between the MSSM and supergravity. 

In Einstein gravity and supergravity quantum corrections cannot be handled by renormalizing a few parameters as in the Standard Model. This ultraviolet behavior improves in supergravity, but motivates nevertheless the recourse to String Theory, which overcomes it altogether by introducing an effective ultraviolet cutoff at the string scale. In this extension, ultraviolet finite amplitudes are deduced from a two--dimensional setup deeply rooted on (super)conformal invariance. The techniques are relatively well developed in flat space, where long--range gravitational and gauge interactions emerge from low--lying string modes. However, only ten--dimensional supersymmetry grants that the background be stable and remain flat even when the string back-reaction is taken into account. In this context, one can also provide enticing arguments to the effect that all ten--dimensional supersymmetric strings are related to one another and also to an elusive eleven--dimensional theory. While there are thus strong reasons to believe that all supersymmetric strings stem from a unique underlying principle, there is no fully satisfactory formulation yet beyond the two--dimensional setup. This section is devoted to a brief introduction to ten--dimensional strings.

However, all attempts to connect these ideas to Nature cannot forego the need for two additional ingredients. The first is some sort of spontaneous compactification hiding the six (or seven) extra dimensions, while the second is the breaking of supersymmetry. Geometric compactifications, and generalizations thereof, where at least some supersymmetry remains unbroken, were widely explored during the last decades. They can lead to four--dimensional Minkowski spaces hosting supersymmetric matter and interactions that are qualitatively along the lines of what one would like to find. Supersymmetry breaking can then be realized in the resulting field theories, along the lines of what we saw in previous chapters. Different aspects of this approach are reviewed in Sections~\ref{sec:6dstrings} and \ref{sec:calabi-yau}.  

The spontaneous breaking of supersymmetry via compactifications is another interesting option. This will be illustrated, at the string level, in Section~\ref{sec:toroidal_ss}, where we shall address the Scherk--Schwarz mechanism in the simplest context of circle reductions, and then, at the field theory level, in a more intricate context, in Section~\ref{sec:directsusybcom}. In this case, the breaking of supersymmetry will be induced in the IIB string compactified to four dimensions combining special internal fluxes and warping.

Alternatively, one can consider string models where supersymmetry is absent or non--linearly realized, at the price of complications that are both technical and conceptual. Typically, these settings are fraught with the emergence of unstable modes, and even in the few cases where this does not occur, there are severe modifications of the vacuum. As a result, two--dimensional techniques continue to provide indications on string spectra, but the low-energy effective field theory is, in general, the only available tool to analyze the resulting dynamics. These string theories are reviewed in this section and the following two, while vacuum modifications and the techniques based on the low--energy effective field theory will be illustrated in Section~\ref{sec:SUSY_breaking_com}. 

\subsection{\sc A Brief Overview of String Spectra} \label{sec:string_spectra}

In order to illustrate the key properties of ten--dimensional string spectra, it is convenient to begin by considering the simpler case of the bosonic string in flat spacetime, whose world--sheet action can be presented in the form
\beq
S\left[g_{\alpha\beta},X^\mu\right] \ =\  - \ \frac{1}{4 \pi \alpha^\prime} \int d^2 \xi \sqrt{-g}\,
g^{\alpha\beta}
\partial_\alpha X^\mu \ \partial_\beta X^\nu \ \eta_{\mu\nu}  \ . \label{bosonicaction}
\eeq
Here $g_{\alpha\beta}$ is a world--sheet metric, 
\beq
T \ = \ \frac{1}{2\,\pi\alpha'}
\eeq
is the string tension,
$\xi^\alpha=(\tau,\sigma)$ are world--sheet coordinates, with $0 \leq \sigma \leq \pi$, and $X^\mu(\xi^\alpha)$ $\mu=(0,\ldots,D-1)$ are the string coordinates. The world--sheet and the ambient spacetime have both ``mostly plus'' Minkowski signatures, to begin with.

The reparametrization invariance of the action guarantees that the world--sheet metric can be brought locally to the diagonal form
\beq
g_{\alpha\beta} \ = \ \eta_{\alpha\beta} \, e^\lambda \label{conf_2dmetric}
\eeq
by a choice of $\xi$-coordinates, while the classical field equation for the world--sheet metric demands that the energy momentum tensor
\beq
T_{\alpha\beta} \ = \ \partial_\alpha X^\mu \partial_\beta X_\mu \ -
\ {\textstyle\frac{1}{2}}\, \eta_{\alpha\beta} \ \partial^\gamma X^\mu \partial_\gamma
X_\mu  \label{emt}
\eeq
vanish. Note that the conformal factor $\lambda$ in eq.~\eqref{conf_2dmetric} has disappeared altogether from $T_{\alpha\beta}$, which is a manifestation of the Weyl invariance of the action~\eqref{bosonicaction}. The equation for the string coordinates
\beq
\partial_\alpha \left( \sqrt{-g} \,g^{\alpha\beta} \partial_\beta \
X^\mu \right) = 0 
\eeq
undergoes a similar simplification once eq.~\eqref{conf_2dmetric} is taken into account, and reduces to the standard free wave equation
\be
\left( \frac{\partial^2}{\partial \tau^2} \ -\
\frac{\partial^2}{\partial \sigma^2}
\right) X^\mu \ = \ 0 \ . \label{wave}
\ee

This equation is solved by
\beq
X^\mu \ =\  x^\mu \ +\  2 \alpha^\prime\, p^\mu \ \tau \ + \ \frac{i
\sqrt{2\alpha^\prime}}{2}
\ \sum_{n \not= 0} \left( \frac{\alpha^\mu_n}{n} \ e^{-2 i n (\tau -
\sigma)} \ + \
\frac{\tilde{\alpha}^\mu_n}{n}\  e^{-2 i n (\tau + \sigma)} \right) \label{closedmodes}
\, ,
\eeq
with the periodic boundary condition that pertains to the closed string, and by
\beq
X^\mu \ = \ x^\mu \ + \ 2 \alpha^\prime\, p^\mu \ \tau \ + \ i
\sqrt{2\alpha^\prime}
\ \sum_{n \not= 0} \frac{\alpha^\mu_n}{n} \ e^{- i n \tau} \cos(n
\sigma)  \ . \label{opennmodes}
\eeq
with the Neumann boundary conditions that pertain to the open string. One can also consider non--periodic modes for closed strings, or open strings with Dirichlet conditions at one or both ends in settings with reduced spacetime symmetries. We shall return to these more general options in the following sections.

Even after the gauge fixing of eq.~\eqref{conf_2dmetric}, an infinite-dimensional symmetry is left, whereby
\beq
\tilde{\tau} \ + \tilde{\sigma} \ = \ f(\tau \ + \ \sigma) \ , \qquad - \ \tilde{\tau} \ + \tilde{\sigma} \ = \ g(- \ \tau \ + \ \sigma)  \ .
\eeq
This is the Minkowski--space counterpart of (anti)analytic reparametrizations in the complex plane. $f$ and $g$ are independent functions in the closed--string case, while
only one can be chosen at will for open strings, due to the boundary conditions at the ends. We shall refer to this symmetry, independently of the signature that we shall often change to the Euclidean one, as two--dimensional conformal invariance.

The two equations implied by the vanishing of the (trace--free) energy--momentum tensor ~\eqref{emt} are quadratic in the $X$ coordinates but can be linearized and solved by a procedure that is reminiscent of the light-cone formulation of Electrodynamics. This procedure is a convenient shortcut to describe string spectra, but other covariant approaches exist and are discussed at length in~\cite{stringtheory}.

To begin with, one can use the residual conformal invariance to set~\footnote{As is well known, the bosonic string originates from the work of Veneziano~\cite{Veneziano68} on what is now recognized as the amplitude for four open--string tachyons. This was followed by the work of Virasoro~\cite{Virasoro} and Shapiro~\cite{Shapiro} on what is now recognized as the amplitude for four closed--string tachyons, and many other important contributions added on during the first years. Details can be found in~\cite{stringtheory}. }
\beq
X^+  \ \equiv \ \frac{X^0\,+\,X^{D-1}}{\sqrt{2}} \ = \ x^+ \ +\ 2 \alpha^\prime\, p^+ \ \tau \ ,
\eeq
thus eliminating all oscillators in the $+$ direction. The condition on the energy--momentum tensor then yields linear equations that determine 
\beq
X^-  \ = \ \frac{-\, X^0\,+\,X^{D-1}}{\sqrt{2}}
\ee
in terms of the $\alpha^i$:
\beq
\pm \ 2 \sqrt{2} \, \alpha'\, p^+ \ \partial_{\pm} X^- \ - \ {(\partial_{\pm}
X^i)}^2 \ = \  0 \, .
\label{constraints}
\eeq
Here
\beq
\partial_{\pm} \ = \ \frac{1}{\sqrt{2}} \left( \pm \ \frac{\partial}{\partial\,\tau} \ + \ \frac{\partial}{\partial\,\sigma}\right) \ , \label{partialpm}
\eeq
and the second term in eq.~\eqref{constraints}, when expanded in terms of the $D-2$ transverse $\alpha^i$'s and $\tilde{\alpha}{}^i$'s, reads
\bea
{(\partial_{+} X^i)}^2 &=& 4\,\alpha'\ \sum_{m,n} \tilde{\alpha}{}_m^i \,\tilde{\alpha}{}_n^i \ e^{-2 i (m+n) (\tau + \sigma)} \ \equiv \ 8 \,\alpha' \ \sum_{m} \widetilde{L}{}_m \ e^{-2 i m (\tau + \sigma)} \ , \nonumber \\
{(\partial_{-} X^i)}^2 &=& 4\,\alpha'\ \sum_{m,n} \alpha_m^i \,\alpha_n^i \ e^{-2 i (m+n) (\tau - \sigma)} \ \equiv \ 8 \,\alpha' \ \sum_{m} L_m \ e^{-2 i m (\tau - \sigma)} \ .
\eea
The Fourier modes of the quadratic combinations define the Virasoro bilinears $\widetilde{L}{}_m$ and $L_m$ for the closed string, and for later convenience we have set
\beq
\alpha_0{}^i \ = \ \frac{1}{2}\, \sqrt{2\alpha'} \ p^i  \ , \qquad \tilde{\alpha}{}_0{}^i \ = \ \frac{1}{2}\, \sqrt{2\alpha'} \ p^i \ ,
\eeq
which are obtained splitting the momentum evenly between left and right modes. On the other hand, for the open string one would obtain
\beq
{(\partial_{+} X^i)}^2  \ = \ \alpha'\ \sum_{m,n} {\alpha}{}_m^i \,{\alpha}{}_n^i \ e^{-i (m+n) (\tau + \sigma)} \ \equiv \ 2 \,\alpha' \ \sum_{m} {L}{}_m \ e^{-i m (\tau + \sigma)} \ ,
\eeq
with
\beq
\alpha_0{}^i \ = \ \sqrt{2\alpha'} \ p^i \ .
\eeq

The quantization of the two--dimensional theory is obtained treating the transverse coordinates $X^i$ as operators. For the closed string, there are thus independent commutation relations for the two families of transverse oscillators $\alpha_m^i$ and  $\tilde{\alpha}{}_m^i$, which read
\beq
\left[\alpha_m^i \,, \alpha_n^j \right]  \ = \ m \,\delta^{ij} \ \delta_{m+n,0} \ , \qquad \left[\tilde{\alpha}_m^i \,, \tilde{\alpha}_n^j \right]  \ = \ m \,\delta^{ij} \ \delta_{m+n,0} \ , \label{oscill}
\eeq
together with the commutation relation
\beq
\left[x^i \,,\, p^j \right]  \ = \ i \ \delta^{ij} \label{xpcom}
\eeq
for the zero modes.

Therefore, the $\alpha_n^i$ with $n>0$ ($n<0$) are annihilation (creation) operators.
The Fourier modes of the conditions~\eqref{constraints} involve the transverse (normal--ordered) Virasoro operators
\beq
L_m \ = \ {\textstyle\frac{1}{2}} \ : \, \sum_n \ {\alpha}_{m-n}^i \,{\alpha}_n^i \,:  \ + \ a\, \delta_{m,0}\ , \label{ell0}
\eeq
and the additional contribution $a$ in eq.~\eqref{ell0} is a normal--ordering constant. It can be deduced noting that
\bea
\sum_{n \neq 0} \alpha_{-n}^i \,\alpha_n^i &=& 2\,\sum_{n=1}^\infty \alpha_{-n}^i \,\alpha_n^i \ + \ \sum_{n=1}^\infty \left[ \alpha_{n}^i \ , \ \alpha_{-n}^i\right] \nonumber\\
&=& 2\,N \ + \ (D-2) \ \sum_{n=1}^\infty\ n \ , \label{9.14}
\eea
where
\beq
N \ = \ \sum_{n=1}^\infty \, \alpha_{-n}^i \,\alpha_{-n}^i
\eeq
is the number operator for the oscillators. In a similar fashion, one can deduce that
\beq
\sum_{n \neq 0} \tilde{\alpha}_{-n}^i \,\tilde{\alpha}_n^i \ = \ 2\,\tilde{N} \ + \ (D-2) \ \sum_{n=1}^\infty\ n \label{9.15} \ .
\eeq
The end results
\beq
L_0 \ = \ 2\,N \ - \ \frac{D-2}{12} \ , \qquad \widetilde{L}{}_0 \ = \ 2\,N \ - \ \frac{D-2}{12} \ , \label{l0lbar0}
\eeq
can be justified considering the regulated sum
\beq
\sum_{n=1}^\infty  n  \ \rightarrow \  \sum_{n=1}^\infty  e^{- \epsilon\,n} \ n \to \ - \ \frac{1}{12} \ ,
\eeq
and retaining the finite part as $\epsilon \to 0$, or alternatively relying on the correspondence with Riemann's $\zeta$ function. In general, as $\epsilon \to 0^+$ the expressions
\beq
\zeta_\alpha(-1,\epsilon) \ = \ \sum_{n=1}^\infty \, (n+\alpha) \, e^{-(n+\alpha)\epsilon} \ \to \
\zeta_\alpha(-1,0^+) \ =\  - \ \frac{6\alpha(\alpha-1)+1}{12} \label{zriem}
\eeq
provide a convenient
way to recover the counterparts of $a$ that would emerge with more general conditions at the ends of the range for $\sigma$ in the $\epsilon \to 0$ limit.

The $L_m$ satisfy the Virasoro algebra
\beq
[ L_m , L_n ] \ = \ (m - n) L_{m+n} \ + \ \frac{c}{12} \ m\left(m^2 - 1\right)
\delta_{m+n,0} \ ,
\eeq
which characterizes the conformal group,
with a ``central charge''
\beq
c \ = \ D \ - \ 2 \ ,
\eeq
so that each transverse coordinate $X^i$ contributes one to its strength. 

The zero modes of eqs.~\eqref{constraints} determine the mass spectrum
\beq
M^2 \equiv\ - \ 2 p^+ p^- \ - \ p^i p^i \ = \ \frac{4}{\alpha'} \left( N \ - \  \frac{D-2}{24}
\right) \ = \ \frac{4}{\alpha'} \left( \bar{N} \ - \  \frac{D-2}{24}
\right) \ , \label{closedmass}
\eeq
for the closed string, and the corresponding mass spectrum
\beq
M^2 \ = \ \frac{1}{\alpha'} \left( N \ - \ \frac{D-2}{24} \right) 
\label{openmass}
\eeq
for the open string. The closed spectrum~\eqref{closedmass} is usually described via the
half-sum of the two conditions,
\beq
M^2  \ = \ \frac{2}{\alpha'} \left( N \ + \ \bar{N} \ - \  \frac{D-2}{12}
\right)  \ , \label{closedmass2}
\eeq
supplemented by the ``level--matching'' condition
\beq
N \ = \ \bar{N} \label{levelmatching}
\eeq
on all physical states. Note the presence of a tachyonic mode in each of the preceding spectra.

The first excited states yield important consistency conditions on the open and closed spectra that determine the ``critical dimension'' $D$, since they indicate that eqs.~\eqref{closedmass} and \eqref{openmass} are compatible with Lorentz invariance only for $D=26$. Indeed, the first excitations only involve \emph{transverse} oscillators acting on vacua carrying a momentum $p^+$, and are
\beq
\epsilon_{i,j}\, \alpha_{-1}^i \, \tilde{\alpha}_{-1}^j\left| 0\,\tilde{0} \,;\,p^+ \right\rangle \ ,
\eeq
for the closed string, where in $\epsilon_{i,j}$ one can distinguish a symmetric traceless portion $h_{ij}$, an antisymmetric portion $B_{ij}$ and a trace $\varphi$, and
\beq
A_{i}\,\alpha_{-1}^i \, \left| 0 \,;\,p \right\rangle 
\eeq
for the open string. The purely transverse oscillators $\alpha_{-1}^i$ and $\tilde{\alpha}{}_1^i$ suffice to describe Lorentz--invariant spectra only if these modes are massless, which is the case in the critical dimension $D=26$, and with $p^i=0$ to exclude longitudinal polarizations. The massless closed--string spectrum then contains gravity--like modes $h_{ij}$ that would emerge linearizing Einstein gravity around flat space, a massless scalar mode $\varphi$ that is usually called dilaton, and the possibly less familiar two--form modes $B_{ij}$. In a similar fashion, the massless modes of the open string describe the transverse polarizations of massless vectors.
While this description in terms of transverse oscillators is effective and clear for the massless modes, matters become more complicated for the massive spectrum, whose polarizations are spread among different combinations of string modes. There is a long--held feeling that a closer look into the systematics of the massive spectrum can help to illuminate the elusive foundations of String Theory. Some progress in this respect was made in~\cite{chrysoula}, working in the covariant formalism, and more recently in~\cite{bucciotti}, working in the light--cone formalism. 

All preceding states can be built acting on the vacuum with vertex operators that combine the exponential $e^{i p \cdot X}$ with derivatives of $X$. The perturbative prescription for $n$--point scattering amplitudes rests on the sum
\beq
{\cal A} \ = \ \sum_\Sigma e^{\left(n- \, \chi\right)\varphi} \ {\cal A}_\Sigma \ , \label{genus_expansion}
\eeq
which involves contributions from two--dimensional Riemann surfaces $\Sigma$ of increasing genera weighted by the exponentials $e^{- \, \chi\,\varphi}$ built from their Euler characters $\chi$ and the dilaton vacuum value. The ${\cal A}_\Sigma$ are determined by integrating (Wick--rotated) correlation functions of the vertices over their positions and over surface moduli, additional parameters that characterize their shapes, as discussed in~\cite{stringtheory}.  For example, for the scattering amplitudes of closed strings
\beq
\chi \ = \ 2 \, - \, 2\, h  \ , \label{chi_closed}
\eeq
where $h$ is the number of handles, while for open strings there are additional surfaces with different numbers of boundaries (and further complications in the unoriented case), to which we shall return. For the purposes of this review, however, it will largely suffice to elaborate on the nature of one--loop vacuum amplitudes, but let us stress that 
\beq
g_s \ = \ e^\varphi
\eeq
plays the role of a coupling constant in the preceding expansion.
Comparing this setup with Einstein's theory, one can see that the $D$--dimensional Planck mass and the string coupling $g_s$ determine the string scale $\frac{1}{\sqrt{\alpha'}}$ according to
\beq
M_P^{D-2} \ \sim \ \frac{1}{{\alpha'}^\frac{D-2}{2}\, g_s^2} \ . \label{MP_string}
\eeq
Consequently, with $g_s <1$ the string scale $\frac{1}{\sqrt{\alpha'}}$ lies below the Planck scale. 

Returning to the two--form modes, let us note that their dynamics is described by the Lagrangian
\beq
{\cal L} \ = \ - \ \frac{1}{12} \, H_{\mu\nu\rho}\,H^{\mu\nu\rho} \ ,
\eeq
where the field strength
\beq
H_{\mu\nu\rho} \ = \ \partial_\mu\,B_{\nu\rho} \ + \ \partial_\nu\,B_{\rho\mu} \ + \ \partial_\rho\,B_{\mu\nu} 
\eeq
is invariant under the gauge transformations
\beq
\delta\,B_{\mu\nu} \ = \ \partial_\mu\,\Lambda_\nu \ - \ \partial_\nu\,\Lambda_\mu \ ,
\eeq
following steps similar to those needed to recover transverse photon polarizations in the light--cone gauge from Maxwell's theory.
A field of this type would actually provide a dual description of a free scalar in four dimensions, but it is independent for $D>4$. In fact, we already met two--form gauge fields in Section~\ref{sec:sugra1110}, in connection with the Green--Schwarz mechanism, but generalizations to the case of generic $p+1$ forms $B_{p+1}$ play an important role in String Theory, as we shall see, so that we can now pause to explain a few facts that will play a role in the following sections.

General $p$-form gauge fields can be conveniently described, in form language, via the Lagrangians
\beq
{\cal L} \ = \ \frac{1}{2} \, F_{p+2}\,\wedge\,\star\ F_{p+2} \ ,
\eeq
with
\beq
F_{p+2} \ = \ d\, A_{p+1} \ , 
\eeq
which are invariant under the gauge transformations
\beq
\delta\,A_{p+1} \ = \ d\,\Lambda_p \ .
\eeq
The generalized Maxwell equations for these form fields
\beq
d \, \star\, F_{p+2} \ = \ \star\, J_{p+1} 
\eeq
then demand that the ``electric'' current $J$ be a conserved $p+1$-form. 

In Maxwell's theory the current is a conserved one-form, and describes particles of charge $q_0$ in motion,
interacting with the Maxwell potential via
\beq
q_0  \int A_1 \ ,
\eeq
where the integral is along the world lines of the particles.
In a similar fashion, ``electric'' objects couple to the two--form gauge fields $B_{\mu\nu}$ via surface integrals
\beq
q_1  \int B_2 \ ,
\eeq
over two--dimensional surfaces swept by strings in their motion, and
$p+1$-forms $J_{p+1}$ describe $p$-dimensional ``electric'' objects in motion, generically called $p$-branes, with $p$ spatial dimensions, which interact with gauge fields $A_{p+1}$ via world--volume integrals of the type
\beq
q_{p}  \int A_{p+1} \ . \label{geometric_ap1}
\eeq

In four dimensions, the Maxwell equations
\beq
d\, \star\, F_{2} \ = \ \star\, J_{e}  \ , \qquad d \, F_{2} \ = \ 0 
\eeq
can be generalized to also allow for magnetic charges and currents, into the more symmetrical form
\beq
d\, \star\, F_{2} \ = \ \star\, J_{e}  \ , \qquad d \, F_{2} \ = \ \star\,J_{m} \ .
\eeq
Electric-magnetic duality turns the Maxwell two-form field strength $F$ into its dual $\star F$, which is also a two-form, and at the same time turns the electric current $J_{e}$ into the magnetic current $J_{m}$. In a generic dimension $D$, starting from a $p+1$-form gauge field, as we have seen, one is led to electric $p$-branes with conserved $p+1$-form currents. The dual magnetic currents are defined via
\beq
d \, F_{p+2} \ = \ \star\,J_{m} \ ,
\eeq
and the sources are thus $D-p-4$-branes with magnetic charges $q_{D-p-4}$. Therefore, even in Maxwell's theory, monopoles become extended objects in $D>4$.

In $D=2,6,10$, where
\beq
\star^2 \ = \ 1 \ ,
\eeq
one can also impose on real field strengths self--duality conditions of the type
\beq
F_\frac{D}{2} \ = \ \star {F}_\frac{D}{2} \ . \label{selfdual}
\eeq
These first--order equations imply the more familiar second--order ones
\beq
d\, F_\frac{D}{2} \ = \ 0 \ , \qquad d\, \star {F}_\frac{D}{2} \ = 0 \ ,
\eeq
but give rise to chiral bosonic fields contributing to gravitational anomalies, as we have stressed in our discussion of type--IIB supergravity in Section~\ref{sec:sugra1110}.

In view of eq.~\eqref{genus_expansion}, the low--energy effective field theory of the closed bosonic string combines the kinetic terms for gravity, two-form, dilaton and tachyon, albeit in the somewhat unusual presentation known as ``string frame'', whereby
\beq
{\cal S}_{closed} \ = \ \frac{1}{2\,\kappa^2} \ \int d^{26}
x \ e^{-2 \varphi} \left[ R \ + \ 4 \,\partial^\mu\,\varphi\,\partial_\mu\,\varphi \ - \ \frac{1}{12}
\, H^{\mu\nu\rho}\,H_{\mu\nu\rho} \ + \ \ldots \right] \ ,
\eeq
where in $D$ dimensions
\beq
\frac{1}{\kappa^2} \ \sim \ M_P^{D-2} \ .
\eeq
The open string first manifests itself at the disk level, and therefore its contribution to the effective action involving gauge vectors reads
\beq
{\cal S}_{open} \ = \ \frac{1}{2\,\kappa^2} \ \int d^{26} x 
\left[  - \ \frac{\alpha'}{4}\, e^{- \varphi} F^{\mu\nu}\,F_{\mu\nu}  \right] \ .
\eeq

The preceding discussion has revealed two major drawbacks of bosonic strings: they lack Fermi modes, and moreover, they contain tachyonic instabilities in open and closed spectra. Both difficulties can be overcome including in the world sheet sets of Majorana--Weyl fermionic coordinates $\psi^\mu(\xi)$ and/or $\tilde{\psi}^\mu(\xi)$ of positive and negative chiralities. The extension from two--dimensional gravity, as in eq.~\eqref{bosonicaction}, to two--dimensional supergravity~\cite{BDVH1,BDVH2} is then instrumental to end up with purely transverse $X^i$ and $\psi^i$.

In the superconformal gauge 
\beq
g_{\alpha\beta} \ = \ e^\lambda\ \eta_{\alpha\beta} \, \qquad \chi_\alpha \ = \ 0 \ ,
\eeq
the metric is diagonal, as before, and the two--dimensional gravitino $\chi_\alpha$ is eliminated, so that one ends up with the free action
\beq
 {\cal S} \ = \ - \ \frac{1}{4 \pi \alpha^\prime} \int d^2 \xi \biggl[
\partial^\alpha X^\mu \ \partial_\alpha X^\nu \ \eta_{\mu\nu} \ +\ i\,
{{\psi}{}^{\mu}} \left(\partial_\tau \,+\,\partial_\sigma\right) {\psi}{}_{\mu}  \ +\ 
i\,\tilde{\psi}{}^{\mu} \left(\partial_\tau \,-\,\partial_\sigma\right) \tilde{\psi}{}_{\mu} \biggr] \ ,
\eeq
from which one can simply deduce the equations of motion
\beq
\partial_+\,\partial_- \, X^\mu \ = \ 0 \ , \qquad \partial_{+}\, {\psi}^\mu \ = \ 0 \ , \qquad \partial_{-}\, \tilde{\psi}^\mu \ = \ 0 \ .
\eeq
The solution for $X^\mu$ is as in eqs.~\eqref{closedmodes} and \eqref{opennmodes}, but there is an interesting novelty for the fermionic equations. To begin with, there are boundary terms for fermions at the two ends of open strings
\beq
\Big[ {\tilde{\psi}{}^{\mu}}\, \delta\,{\tilde{\psi}{}_{\mu}} \ - \ {{\psi}{}^{\mu}}\,\delta\,{{\psi}{}_{\mu}} \Big]_0^\pi \ = \ 0 \ ,
\eeq
which can be eliminated in two different ways by local conditions, demanding that
\beq
\psi^\mu(\tau,0) \ = \ \tilde{\psi}^\mu(\tau,0) \ , \qquad \psi^\mu(\tau,\pi) \ = \ \ \pm \ \tilde{\psi}^\mu(\tau,\pi) \ ,
\eeq
compatibly with the maximal Poincar\'e symmetry, whose generators are quadratic in these fields.
The corresponding Fourier decompositions depend on the sign choice in the second equation. With the ``minus'' sign one obtains the Neveu--Schwarz sector, for which
\beq
\psi^\mu \ = \ \sqrt{\frac{\alpha'}{2}} \sum_{r \,\in \,\mathbb{Z}+\frac{1}{2}} \ b_r{}^\mu \  e^{-\,i r (\tau - \sigma)} \ , \qquad \tilde{\psi}^\mu \ = \ \sqrt{\frac{\alpha'}{2}} \sum_{r \,\in\, \mathbb{Z}+\frac{1}{2}} \ b_r{}^\mu \  e^{-\,i r (\tau + \sigma)}  \ ,
\eeq
while with the ``plus'' signs one obtains the Ramond sector, for which the Fourier decompositions become
\beq
\psi^\mu \ = \ \sqrt{\frac{\alpha'}{2}} \sum_{n \,\in \,\mathbb{Z}} \ d_n{}^\mu \  e^{-\,i n (\tau - \sigma)} \ , \qquad \tilde{\psi}^\mu \ = \ \sqrt{\frac{\alpha'}{2}} \sum_{n \,\in\, \mathbb{Z}} \ d_n{}^\mu \  e^{-\,i n (\tau + \sigma)}  \ .
\eeq
For the closed string $\psi^\mu$ and $\tilde{\psi}{}^\mu$ have independent Fourier coefficients and can be antiperiodic (periodic) for the Neveu--Schwarz (Ramond) sectors. For the two sets of fermions the Fourier decompositions compatible with maximal symmetry thus read
\bea
\psi^\mu &=&  \sqrt{2\,\alpha'} \sum_{r \,\in \,\mathbb{Z}+\frac{1}{2}} \ b_r{}^\mu \  e^{-\,2 i \, r (\tau - \sigma)} \ , \qquad \psi^\mu \ = \ \sqrt{2\,\alpha'} \sum_{n \,\in \,\mathbb{Z}} \ d_n{}^\mu \  e^{-\,2 i n (\tau - \sigma)} \ , \nonumber \\
\tilde{\psi}{}^\mu &=&  \sqrt{2\,\alpha'} \sum_{r \,\in \,\mathbb{Z}+\frac{1}{2}} \ \tilde{b}{}_r{}^\mu \  e^{-\,2 i \, r (\tau + \sigma)} \ , \qquad \tilde{\psi}^\mu \ = \ \sqrt{2\,\alpha'} \sum_{n \,\in \,\mathbb{Z}} \ \tilde{d}{}_n{}^\mu \  e^{-\,2 i n (\tau + \sigma)}
\ .
\eea

There are again non--linear constraints involving the energy--momentum tensor $T_{\alpha\beta}$ and the supersymmetry current $J_\alpha$ that originate from the metric and gravitino equations. The residual super-conformal symmetry can now be used to set
\beq
X^+ \ = \ x^+ \ + \ 2\,\alpha'\,p^+\,\tau \ , \qquad \psi^+ \ = \ 0 \ , \qquad \tilde{\psi}{}^+ \ = \ 0 \ , 
\eeq
and the constraints 
\bea
T_{++} &\equiv& \partial_+ \, X^\mu \, \partial_+ \, X_\mu \ + \ \frac{i}{\sqrt{2}} \,\tilde{\psi}^\mu\,\partial_+\,\tilde{\psi}_\mu \ = \ 0  \ , \nonumber \\
T_{--} &\equiv& \partial_- \, X^\mu \, \partial_- \, X_\mu \ + \ \frac{i}{\sqrt{2}} \, {\psi}{}^\mu\,\partial_-\,{\psi}{}_\mu \ = \ 0 \ , \nonumber \\
J_+ &\equiv& i\, \tilde \psi^\mu\,\partial_+\,X_\mu \ = \ 0  \ , \nonumber \\ J_- &\equiv& i\, {\psi}^\mu\,\partial_-\,X_\mu \ = \ 0
\eea
determine $X^-$, $\psi^-$ and $\tilde{\psi}^-$ in terms of transverse modes. For closed strings, quantization turns the corresponding Fourier coefficients into oscillators. The bosonic commutators are as in eqs.~\eqref{oscill} and\eqref{xpcom}, while for the Fermi fields the anticommutation relations read
\bea
\left\{ b_r{}^i \,,\, b_s{}^j \right\} \ = \ \delta^{ij}\,\delta_{r+s,0} \ , \qquad \left\{ \tilde{b}{}_r{}^i \,,\, \tilde{b}{}_s{}^j \right\} \ = \ \delta^{ij}\,\delta_{r+s,0} 
\eea
for left-moving and right-moving Neveu--Schwarz sectors, and
\bea
\left\{ d_m{}^i \,,\, d_n{}^j \right\} \ = \ \delta^{ij}\,\delta_{m+n,0} \ , \qquad \left\{ \tilde{d}{}_m{}^i \,,\, \tilde{d}{}_n{}^j \right\} \ = \ \delta^{ij}\,\delta_{m+n,0} 
\eea
for left-moving and right-moving Ramond sectors. Similar results hold for the single set of (NS or R) oscillators of the open string.

The mass--shell conditions for the closed string involve modified $L_0$ and $\bar{L}{}_0$ operators that include fermionic contributions.
The normal--ordering constants are determined by eq.~\eqref{zriem}, which gives, for the contributions to the number operators, $- \,\frac{1}{24}$ for each transverse bosonic direction, $- \,\frac{1}{48}$ for each fermionic Neveu--Schwarz direction and $\frac{1}{24}$ for each fermionic Ramond direction. Note, in fact, that an additional ``minus'' sign is needed for anticommuting oscillators, so that the total shift is $-\,\frac{D-2}{16}$ for Neveu--Schwarz sectors and zero for Ramond sectors.
The corresponding mass--shell conditions are then
\beq
M^2 \ = \ \frac{4}{\alpha'} \left( N_X\ + \ N_{NS} \ - \ \frac{D-2}{16} \right) \ , \qquad M^2 \ = \  \frac{4}{\alpha'} \ \left(N_X \ + \ N_R\right) \ ,
\eeq
for left--moving modes in the NS and R sectors, and 
\beq
M^2 \ = \  \frac{4}{\alpha'} \left( \bar{N}{}_X\ + \ \bar{N}{}_{NS} \ - \ \frac{D-2}{16} \right) \ , \qquad
M^2 \ = \  \frac{4}{\alpha'} \ \left( \bar{N}{}_X \ + \ \bar{N}{}_R\right)   \label{left_ms}
\eeq
for right-moving modes in the NS and R sectors, where 
\beq
N_X \ = \  \sum_{n=1}^\infty \alpha_{-n}{}^i\,  \alpha_{n}{}^i \ , \qquad N_{NS} \ = \   \sum_{r \, \in\, \mathbb{N}-\frac{1}{2}} r \ b_{-r}{}^i\,  b_{r}{}^i \ , \qquad N_R \ = \   \sum_{n=1}^\infty n \ d_{-n}{}^i\,  d_{n}{}^i  \ , \label{right_ms}
\eeq
together with similar expressions for the $\bar{N}$'s involving the ``tilde''-oscillators. 

More conveniently, as for the closed bosonic string, one can work with the half-sum of these conditions and their difference, the ``level--matching conditions'', while for the open string the mass--shell conditions are as in eq.~\eqref{left_ms} for the two sectors, but with $\frac{4}{\alpha'}$ replaced with $\frac{1}{\alpha'}$.
Note that the $d_0^i$ do not contribute to the mass formulas but introduce a degeneracy of the ground states, on which they act like $D-2$ $\gamma$ matrices. 

We can now take a closer look at the resulting spectra, starting from the open string. The vacuum of the Neveu--Schwarz sector still contributes a tachyonic scalar, while the low--lying modes are now obtained acting on it with the $b_{-\frac{1}{2}}^i$, which build the transverse polarizations of a vector, for which
\beq
M^2 \ = \ \frac{10-D}{16\,\alpha'} \ ,
\eeq
so that the Lorentz symmetry demands that $D=10$ in this case. On the other hand, the Ramond sector starts at zero mass, but its states are spacetime fermions. In fact, they are built by oscillators acting on a degenerate vacuum acted upon by the $d_0{}^i$. The vacuum is therefore a fermion in this sector, with $2^{\left[\frac{D-2}{2}\right]}$ components, and captures the on--shell degrees of freedom of a ten--dimensional Majorana spinor.

As first noticed by Gliozzi, Scherk and Olive (GSO)~\cite{GSO}, the spectra obtained so far are not satisfactory since, aside from containing tachyonic modes, the NS sectors are built with arbitrary numbers of anticommuting oscillators, thus giving rise to bosonic modes that violate the spin--statistics theorem, unless they are suitably truncated. More general options exist, the original paper identified a special interesting truncation leading to open spectra with equal numbers of Bose and Fermi modes at all mass levels. It retains NS states built with odd numbers of $b$ oscillators, and the Ramond sector is also projected, letting only even (odd) numbers of $d$ oscillators act on the two eight--dimensional Weyl projections of the original vacuum. The tachyon is thus removed, and the massless modes correspond to the $N=1$ supersymmetric Yang-Mills theory for the $U(1)$ group. 

In this fashion, GSO recovered spacetime supersymmetry from the NSR string, which rests on world--sheet symmetry and leads to $D$--dimensional Poincar\'e invariance but had apparently no links with spacetime supersymmetry, also establishing a connection with the ten--dimensional supersymmetric Yang--Mills theory~\cite{GSO,BSS}~\footnote{P.~Fayet was probably the first to identify the $N=4$ supersymmetric Yang--Mills multiplet in four dimensions, early in 1976, but his work was completed later~\cite{fayetn4}, after he clarified the role of $R$-symmetry.}.

The construction can be generalized to any classical group ($U(N)$, $USp(2N)$ or $O(N)$) via the introduction of Chan--Paton factors~\cite{cp1,cp2,cp3,cp4,cp5}, which corresponds to adding charges at the ends of open strings. The only anomaly--free options, as we  saw in Section~\ref{sec:sugra1110}, are $SO(32)$ and $E_8 \times E_8$. Only the first can be realized with open strings, while both are available in the heterotic setup~\cite{heterotic1,heterotic2,heterotic3}, obtained by combining the left--moving NSR modes with the right--moving modes of the 26--dimensional bosonic string compactifying 16 of its dimensions in a peculiar way, as we shall see.

Subsequent work~\cite{gs82} identified similar truncations for closed spectra, where left and right Fermi modes can be treated independently. The different supersymmetric options, along the lines of GSO, lead to left and right NS sectors that start with a massless vector, and to R sectors that can start with a left or right Majorana--Weyl spinor. Combining left and right string modes leads to the type-IIA and type--IIB strings, whose low--lying modes are those of the type--IIA and type--IIB supergravities described in Section~\ref{sec:sugra1110}.

In detail, the NS-NS sector starts with the modes of a symmetric tensor, to be identified with the graviton, together with those of a two-form and the dilaton, all of which are massless. In addition, if the left and right R vacua have the same chirality, the NS-R and R-NS sectors start with the $\gamma$-traceless modes of a Weyl gravitino of that chirality and with a Weyl spinor of opposite chirality, often referred to as a dilatino, whose modes are described by the corresponding $\gamma$ traces. On the other hand, if the left and right R vacua have opposite chiralities, the low--lying NS-R and R-NS sectors combine into the non--chiral modes of a Majorana gravitino and a Majorana spinor. As we saw in Section~\ref{sec:sugra1110}, the low--lying spinor modes are indeed chiral for type--IIB but not for type--IIA supergravity. The RR sector contains additional bosonic modes that can be identified by decomposing Weyl bi-spinors $\Psi_{\alpha\beta}$ of identical chiralities (for type IIB) or opposite chiralities (for type IIA) into forms of different degrees.  In the absence of Weyl projections, one can indeed expand an $SO(8)$ bi-spinor into forms of different degrees, as
\beq
\Psi_{\alpha\beta}  \ = \ \sum_{n=0}^8  \, \frac{1}{n!}\ \gamma^{i_1 \ldots i_n}_{\alpha\beta}\, A_{i_1 \ldots i_n} \ ,
\eeq
where $A_{i_1 \ldots i_n}$ are the physical, transverse modes, of $n$-form gauge fields.
If the left and right vacua have identical (say, left) chiralities, the sum is restricted to forms of even ranks $n \leq 4$, or more explicitly to $n=0,2,4$, and the four-form is chiral. These are the light--cone modes of a scalar, a two-form, and a selfdual four-form gauge field, which emerge from the covariant formulation with field strengths of ranks 1, 3, 5, the last of which is subject to the selfduality condition~\eqref{selfdual}. On the other hand, if the chiralities are opposite, the sum is restricted to forms of odd ranks 1, 3, which are the light-cone modes of a vector and a three-form gauge field. These RR states complete the massless spectra of type--IIB and type--IIA supergravity, as the reader can verify by comparing with the discussion presented in Section~\ref{sec:sugra1110}.

\subsection{\sc Supersymmetric Closed Strings in Ten Dimensions} \label{sec:closed10}

Once the allowed choices for perturbative string spectra are identified, as we explained in the previous section, the corresponding interactions are not arbitrary like in Field Theory, but are determined by inserting on Riemann surfaces local operators creating the corresponding states. We shall see shortly that, for supersymmetric closed strings in ten dimensions, there are three more options, two of which are oriented like the IIA and IIB ones, and will be addressed in this section, while one is unoriented and will be addressed in the next section. The four oriented options can be identified based on bosonic and Neveu--Schwarz--Ramond (NSR) light--cone formulations~\cite{NSR1,NSR2}, supplemented with suitable Gliozzi--Scherk--Olive (GSO) projections~\cite{GSO}. The world--sheet theories provide the two--dimensional oscillators (Bose and Fermi) that build string states, while the GSO projections select consistent mutually subsets of the modes. The allowed projections must satisfy two important consistency conditions: spacetime \emph{spin--statistics} and \emph{modular invariance}, which we can now explain. 
Both conditions are encoded in the torus amplitude: this is the string counterpart of the one--loop vacuum amplitude of Field Theory but contains important information on the string interactions among different sectors of the spectrum. 

If taken at face value the GSO projections identified in the previous section might seem ad hoc, but in fact string spectra are highly constrained by the simplest loop amplitudes. In particular, if only closed strings are present, the relevant contribution, as we just stated, is the torus amplitude swept by the strings. A convenient way to define it starts from the Field Theory expression for the vacuum energy density, 
\beq
{\cal E} \ = \ \sum_i \frac{(-1)^{F_i}}{2}\ \int^{\Lambda} \frac{d^{D-1} p}{(2\pi)^{D-1}} \ \sqrt{\vec{p}^{\, 2} \ + \ m_i{}^2}  \ ,\label{vacuumenergy_particles}
\eeq
where $\Lambda$ is an ultraviolet cutoff needed for this otherwise singular expression.
The sum is over the different particle species, and the factors $(-1)^{F_i}$ associate opposite signs to the zero--point contributions of bosons and fermions. 

From the path-integral vantage point, the vacuum energy density originates from functional determinants, so that, after a Wick rotation,
\beq
{\cal E}  \ = \ \frac{1}{V} \ \sum_i \frac{(-1)^{F_i}}{2}\ \log\det\left( - \, \Box \,+\, m_i^2\right) \ = \ \sum_i \frac{(-1)^{F_i}}{2}\, \int ^\Lambda \frac{d^D p}{(2 \pi)^D}\  \log\left(p^2 \ + \ m_i{}^2\right) \ .
\eeq
The last expression takes the more manageable form 
\beq
{\cal E}  \ \sim  \ - \ \sum_i \frac{(-1)^{F_i}}{2} \ \int_\epsilon^\infty \frac{dt}{\left(4 \pi \right)^\frac{D}{2}\ t^{\frac{D}{2}+1}} \ e^{-\,t\,m_i^2} \ ,
\eeq
where $\epsilon \sim \frac{1}{\Lambda^2}$ is an ultraviolet cutoff, 
up to an additional contribution that we leave aside for reasons that we shall soon try to justify, if one resorts to the ``Pauli--Villars''--like identity
\beq
\log\,\frac{a}{b} \ = \ - \ \int_0^\infty \frac{dt}{t} \left( e^{-t a} \ - \ e^{-t b} \right) \ .
\eeq
Therefore, the general case of interest in String Theory is neatly summarized by the expression
\be
{\cal E}  \ = \ - \ \frac{1}{2 (4 \pi)^\frac{D}{2}} \ \int_\epsilon^\infty \
\frac{dt}{t^{\frac{D}{2}+1}} \ {\rm Str} \left( e^{- t M^2} \right) \ ,
\label{gammatotn}
\ee
where $M$ is an operator whose eigenvalues are the masses $m_i$ of string states and the ``supertrace'' ${\rm Str}$ counts the signed multiplicities of Bose and Fermi modes. 

In general, from what we saw in the previous section,
\beq
M^2 \ = \ \frac{2}{\alpha'} \left( N \ + \ \bar{N} \ - \ \Delta\ - \ \bar{\Delta}\right)\ ,
\eeq
where $\Delta$ and $\bar{\Delta}$ are constant shifts,
and one must take into account the ``level--matching condition'', which is generally of the form
\beq
N \ - \ \bar{N} \ - \ \Delta \ + \ \bar{\Delta} \ = \ 0 \ . \label{levelmatchinggen}
\eeq
One can thus write
\beq
{\cal E}  \ = \  - \ \frac{1}{2 (4 \pi)^{\frac{D}{2}}}  \int_{-\, \frac{1}{2}}^{\frac{1}{2}} ds
\int_\epsilon^\infty
\frac{dt}{t^{\frac{D}{2}+1}} \ \mathrm{Str} \left(
\ e^{- \frac{2}{\alpha'} \left(N + \bar{N} - \Delta - \bar{\Delta}\right)t} \ e^{2\pi i (N -
\widetilde{N} - \Delta + \bar{\Delta} ) s}
\right)
\, ,
\label{gammaclosed2}
\eeq
and defining the ``complex'' Schwinger parameter
\beq
\tau \ = \ \tau_1 \ + \ i \,\tau_2 \ = \ s \ + \ i \, \frac{t}{\alpha' \pi} \, ,
\eeq
and letting
\beq
q = e^{2 \pi i \tau} \, , \qquad  \bar{q} = e^{-2 \pi i
\bar{\tau}} \, ,
\eeq
eq. (\ref{gammaclosed2}) takes the more elegant form
\beq
{\cal E} 
\ = \  - \ \frac{1}{2 (4 \pi^2 \alpha')^{\frac{D}{2}}} \ \int_{- \, \frac{1}{2}}^{\frac{1}{2}}
d \tau_1 \
\int_{\tilde{\epsilon}}^\infty \
\frac{d \tau_2}{\tau_2^{\frac{D}{2}+1}} \ \mathrm{Str}\left( q^{N - \Delta} \ 
\bar{q}^{\bar{N} - \bar{\Delta}}\right)
\ ,
\label{gammaclosed3}
\eeq
with $\tilde{\epsilon} \ = \ \frac{\epsilon}{\alpha'\,\pi}$ a dimensionless ultraviolet cutoff. Eq.~\eqref{gammaclosed3} defines the torus amplitude of String Theory, up to a crucial subtlety to which we now turn.
\begin{figure}[ht]
\centering
%\begin{figure}
%\begin{tabular}{ccc}
%\mbox{graphic} & \mbox{table} \\
\includegraphics[width=90mm]{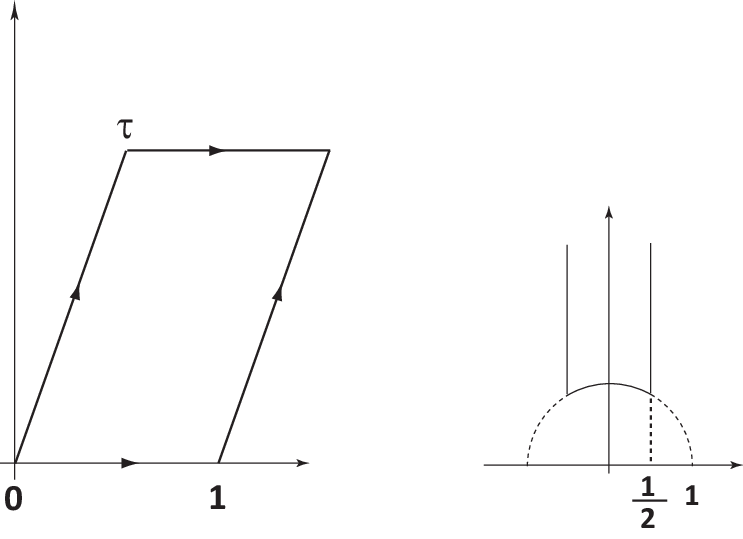}
%\includegraphics[width=60mm]{FundamentalRegion.eps}  \\
%\end{tabular}
 \caption{\small Left panel: the torus as a rectangle in the complex $\tau$ plane. Right panel: the fundamental region for $\tau$.}
\label{fig:torusn}
\end{figure}

At one loop a closed string sweeps indeed a torus, which can be characterized by the complex Schwinger
parameter $\tau$ or, better, subjecting the complex $z$ plane to the identifications 
\beq
z \ \sim \ z+1 \ , \qquad z \ \sim \ z \ + \ \tau \ , 
\eeq
with $Im(\tau)>0$, which defines a fundamental cell (left panel of fig.~\ref{fig:torusn}). The conformal symmetry underlying String Theory then implies that the independent tori are in one-to-one correspondence with two-dimensional lattices, up to overall rescalings, and the choice of fundamental cell is clearly not unique. In fact, the tori with $\tau$ and $\tau+1$ are equivalent, since the two cells have the same areas, and the same is true for those with $\tau$ and $- \, \frac{1}{\tau}$, which interchanges the sides. The two transformations
\beq
T \ : \ \tau \ \to \ \tau \ + \ 1 \ , \qquad S \ : \ \tau \ \to \ - \ \frac{1}{\tau} 
\eeq
generate the modular group $PSL(2,Z)$, which acts on $\tau$ according to
\beq
\tau \ \to \ \frac{a \,\tau \ + \ b}{c\,\tau \ + \ d} \ , \qquad a d - bc \ = \ 1 \ .
\eeq
Letting $\tau$ span the strip $\{-\, \frac{1}{2} < \tau_1 \le
\frac{1}{2}, \epsilon < \tau_2 < \infty\}$ would count infinitely many times the independent contributions, and for this reason a gauge fixing is to be effected. One must restrict
the integration to a fundamental region, so that each inequivalent torus is counted once. The standard choice is
the region ${\cal F}$ of the upper half--plane delimited by the half-circle $\left|\tau\right| = 1$ and the two vertical lines $\mathrm{Re}(\tau) = \pm \,\frac{1}{2}$. This
introduces an effective ultraviolet cutoff, of the order of the string
scale, for all modes (see the right panel of fig.~\ref{fig:torusn}).

Removing the overall factor 
\beq
f \ = \ - \ \frac{1}{2\left(4 \pi^2 \alpha'\right)^\frac{D}{2}} 
\eeq
in eq.~\eqref{gammaclosed3} leads to the torus amplitude
\beq {\cal T} \ = \ \int_{\cal F} \ \frac{d \tau_1\, d\,\tau_2}{\tau_2^2}
\ \frac{1}{\tau_2{}^\frac{D-2}{2}} \ \mathrm{Str}\left( q^{N - \Delta} \ 
\bar{q}^{\bar{N} - \bar{\Delta}}\right)
\ ,
\label{Zclosedn}
\eeq
which has a Diophantine nature since it only involves coefficients that are integer numbers.
Note that we have sorted out a modular--invariant integration measure, and therefore the consistency of the theory demands that the remaining portion of the integrand be modular invariant, which poses strong restrictions on the possible GSO projections. It is important to keep in mind that ${\cal T}$ is related to the vacuum energy ${\cal A}$ according to 
\beq
{\cal E}  \ = \ f\ {\cal T} \ .
\eeq

It is instructive to compute explicitly the torus amplitude
\eqref{Zclosedn} for the
bosonic string. To this end, we recall that $N$ and
$\bar{N}$ are effectively number operators for two infinite sets of
harmonic oscillators. In particular, in terms of conventionally
normalized creation and annihilation operators, for each transverse
space-time dimension
\beq
N \ = \ \sum_n \ n \ a^\dagger_n \ a_n \ ,
\eeq
while for each $n$
\beq
{\rm tr} \ q^{n  \, a^\dagger_n a_n} \ = \  1 + q^n + q^{2 n} + \ldots
\ = \ \frac{1}{1 - q^n} \, ,
\eeq
and putting all these contributions together for the full
spectrum gives
\beq
{\cal T}\  = \ \int_{\cal F} \ \frac{d^2 \tau}{\tau_2^2} \
\frac{1}{\left(\sqrt{\tau_2}\left|\eta(\tau)\right|^2\right)^{D-2}} \label{bosonictorusn}
\ ,
\eeq
where we have defined the Dedekind $\eta$ function
\beq
\eta(\tau) \ = \ q^\frac{1}{24} \, \prod_{n=1}^\infty \ (1 - q^n ) \ .
\eeq

As originally noticed by Shapiro \cite{shapmod}, the integrand is invariant
under the two generators $S$ and $T$, independently of $D$, since the two transformations
\cite{jacobi}
\beq
T: \ \eta(\tau + 1 ) = e^\frac{i \pi}{12} \, \eta( \tau ) \, ,
\qquad S:
\ \eta\left( -\, \frac{1}{\tau}  \right) \ = \ \sqrt{- i \tau } \  \eta( \tau ) \, , \label{STeta}
\eeq
imply that the combination $\sqrt{\tau_2} \left|\eta\right|^2$ is
invariant.
In other words, modular invariance
holds separately for the contribution of each transverse string
coordinate, independently
of the total central
charge $c$. This is a crucial
property of the conformal field theories that {\it define} the torus
amplitudes for all consistent models of oriented closed strings.
 \begin{figure}[ht]
\begin{center}
    \includegraphics[width=1.5in]{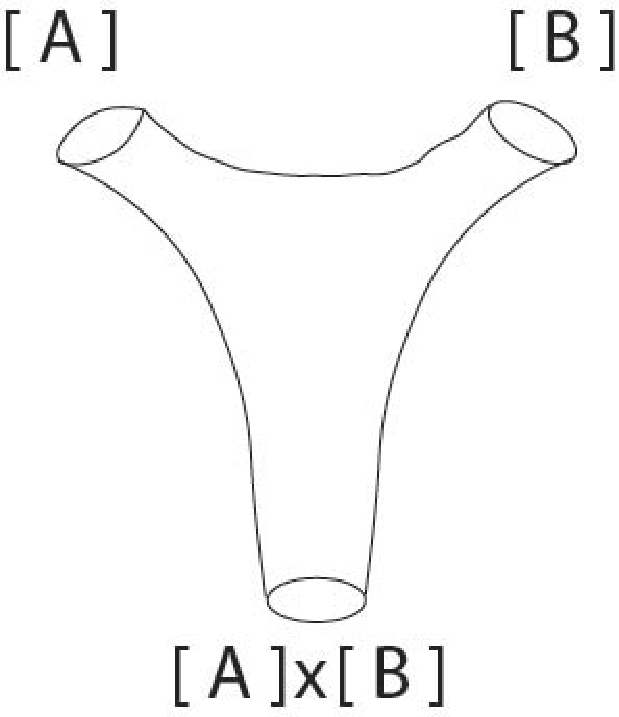}
\end{center}
\caption{\small Fusion rules identify which sectors of the spectrum emerge from interactions involving a pair of others.}
\label{fig:fusion}
\end{figure}

In approaching the GSO projection, it is very convenient to work with $SO(8)$ level--one characters, which provide an orthogonal decomposition of the spectrum of Fermi oscillators into independent sectors, with definite spin--statistics properties both in space time and on the string world sheet. These characters are special cases of the more general level--one $SO(2n)$ characters of Appendix~\ref{app:so2n} that are selected by the manifest transverse $SO(8)$ symmetry that is left, in ten--dimensional Minkowski space, after gauge--fixing the local symmetries of the string world sheet theory. Their main properties are reviewed in Appendix~\ref{app:so2n}, so that here we can limit ourselves to recalling that
\bea
{O_{8}} &=&  \mathrm{Tr}\left[\frac{1 \,+\, (-1)^{F_L}}{2}\ q^{N_{NS} \,-\,\frac{1}{6}}  \right] \ = \ \frac{\theta^4\left[\substack{0\\0}\right]\left(0|\tau\right)\, + \ \theta^4\left[\substack{0\\{1/2}}\right]\left(0|\tau\right)}{2\,\eta^4(\tau)}\ ,   \nonumber\\
V_{8} &=& \mathrm{Tr}\left[\frac{1 \,-\, (-1)^{F_L}}{2}\ q^{N_{NS} \,-\,\frac{1}{6}}  \right] \ = \ \frac{\theta^4\left[\substack{0\\0}\right]\left(0|\tau\right)\, - \ \theta^4\left[\substack{0\\{1/2}}\right]\left(0|\tau\right)}{2\,\eta^4(\tau)}\ , \nonumber \\
\quad S_{8} &=& \mathrm{Tr}\left[\frac{1 \ + \ \gamma_9\,(-1)^{F_L}}{2}\ q^{N_{R} + \frac{1}{3}}  \right] \ = \ 
 \frac{\theta^4\left[\substack{{1/2}\\0}\right]\left(0|\tau\right)\, + \,\theta^4\left[\substack{{1/2}\\{1/2}}\right]\left(0|\tau\right)}{2\,\eta^4(\tau)} \ , \nonumber \\
\quad C_{8} &=& \mathrm{Tr}\left[\frac{1 \ - \  \gamma_9\,(-1)^{F_L}}{2}\ q^{N_{R} + \frac{1}{3}}  \right] \ = \ 
 \frac{\theta^4\left[\substack{{1/2}\\0}\right]\left(0|\tau\right)\,- \,  \theta^4\left[\substack{{1/2}\\{1/2}}\right]\left(0|\tau\right)}{2\,\eta^4(\tau)} \ , \label{fermi10D}
\eea
start with a scalar, a vector and the two types of conjugate spinors. The two characters $S_8$ and $C_8$ are conveniently distinguished, despite being numerically equal, since the Weyl projections are opposite for the states they count, level by level. $S$ and $T$ transformations mix these characters via the matrices
\beq
 S \ = \ \frac{1}{2} \left( \begin{array}{rrrr} 1 & 1 & 1 & 1 \\
1 & 1 & -1 & -1 \\
1 & -1 & 1 & -1 \\
1 & -1 & -1 & 1  \end{array} \right)\, ,  \quad T \ = \ e^{\,-\,\frac{i 2 \pi}{3}} \ \left( \begin{array}{rrrr} 1 & 0 & 0 & 0 \\
0 & -1 & 0 & 0 \\
0 & 0 & -1 & 0 \\
0 & 0 & 0 & -1  \end{array} \right)\,  .  \label{modularn}
\eeq

The operators that create the states of these four different sectors also underlie their interactions through the so--called ``fusion rules'' (see fig.~\ref{fig:fusion}). If the operators belonging to the different sectors are collectively indicated by the symbols used for the characters enclosed within square brackets, these rules are the following :
\bea
&& \left[ V_8\right ] \times \left[ V_8\right ] \, = \, \left[ V_8\right ] ; \ \left[ V_8\right ] \times \left[ O_8\right ] \, = \, \left[ O_8\right ] ; \ \left[ V_8\right ] \times \left[ S_8\right ] \, = \, \left[ S_8\right ] ; \ \left[ V_8\right ] \times \left[ C_8\right ] \, = \, \left[ C_8\right ] \ , \nonumber\\
&& \left[ O_8\right ] \times \left[ V_8\right ] \, = \, \left[ O_8\right ] ; \ \left[ O_8\right ] \times \left[ O_8\right ] \, = \, \left[ V_8\right ] ; \ \left[ O_8\right ] \times \left[ S_8\right ] \, = \, \left[ C_8\right ] ; \ \left[ O_8\right ] \times \left[ C_8\right ] \, = \, \left[ S_8\right ] , \nonumber\\
&& \left[ S_8\right ] \times \left[ V_8\right ] \, = \, \left[ S_8\right ] ; \ \left[ S_8\right ] \times \left[ O_8\right ] \, = \, \left[ C_8\right ] ; \ \left[ S_8\right ] \times \left[ S_8\right ] \, = \, \left[ V_8\right ] ; \ \left[ S_8\right ] \times \left[ C_8\right ] \, = \, \left[ O_8\right ] \ ,  \nonumber\\
&& \left[ C_8\right ] \times \left[ V_8\right ] \, = \, \left[ C_8\right ] ; \ \left[ C_8\right ] \times \left[ O_8\right ] \, = \, \left[ S_8\right ] ; \ \left[ C_8\right ] \times \left[ S_8\right ] \, = \, \left[ O_8\right ] ; \ \left[ C_8\right ] \times \left[ C_8\right ] \, = \, \left[ V_8\right ]  , \label{fusion_rules}
\eea
and apply to both left--moving and right--moving contributions. For example, if the $[S_8]$ sector is present, the $[V_8]$ sector must also be present, since it arises from its self-interaction, in view of $\left[ S_8\right ] \times \left[ S_8\right ] \, = \, \left[ V_8\right ]$. These properties are directly implied by the Verlinde formula~\cite{verlinde}, applied, however, to $(O_8,V_8,-S_8,-C_8)$, the Minkowski spacetime sectors. Recall, in fact, that bosons and fermions must contribute with opposite signs to the vacuum amplitudes, as we saw in eq.~\eqref{vacuumenergy_particles}. The peculiar behavior of these ``spacetime characters'' was first understood in~\cite{schellekens_spacetime}, where it was also related to the two--dimensional gravitino (for a review, see~ \cite{schellekens_spacetime_rev}). 

The partition functions of the two ten--dimensional supersymmetric IIA and IIB superstrings introduced in the previous section can be neatly expressed in terms of the preceding $SO(8)$ characters, and read
\beq
{\cal T}_{IIA} \ = \ \int_{\cal F} \ \frac{d^2 \tau}{\tau_2{}^2} \ \frac{(V_8 \,-\,S_8)(\overline {V}_8 \,-\,\overline{C}_8)}{\tau_2{}^4 \ \eta^8 \, \overline{\eta}^8} \, , \quad {\cal T}_{IIB} \ = \ \int_{\cal F} \ \frac{d^2 \tau}{\tau_{2}{}^2} \ \frac{\left|V_8 \,-\,S_8\right|^2}{\tau_2{}^4 \ \eta^8 \, \bar{\eta}^8} \ \ . \label{IIA_IIB}
\eeq
A remarkable $\theta$--function identity due to Jacobi  (see~\cite{jacobi}) translates into the numerical equivalence of three of the characters,
\beq
V_8 \ = \ S_8 \ = \ C_8 \ , \label{Jacobi_identity}
\eeq
so that the contributions to the vacuum energy vanish in both cases, as pertains to supersymmetric strings. In fact $V_8 - S_8$ and $V_8 - C_8$ are the only two supersymmetric characters, and their modular properties justify the claim that the IIA and IIB strings are the only independent options for string models of this type (with left and right superconformal symmetry), up to an overall parity.

If one writes the integrand of a generic torus amplitude combining the preceding Minkowski characters with the bosonic contribution as
\beq
\sum_{i,j}\ \overline{\chi}_i(\bar{\tau}) \ h^i{}_j \ \chi^j(\tau) \ ,
\eeq
modular invariance requires that the matrix constraints
\beq
S^\dagger\, h\ S \ = \ h \ , \qquad T^\dagger\, h\ T \ = \ h 
\eeq
be satisfied,
and the allowed choices for the matrix $h$ are further restricted by two conditions:
\begin{enumerate}
    \item spin-statistics: the entries $h^i{}_j$ must be non--negative integers;
    \item operator correspondence: the integers should be actually 0 or 1, since the closed sectors are uniquely identified by the four characters and their conjugates. These are in direct correspondence with the available operators, so that no multiplicities are allowed. For open strings, on the other hand, multiplicities will be allowed and will contain information on the possible Chan--Paton groups, as we shall see.
\end{enumerate}
As a result, the two preceding partition functions are the only two consistent options for type II strings.

The other two ten--dimensional supersymmetric strings are more complicated. They are hybrids of the superstring and the bosonic string~\cite{heterotic1,heterotic2,heterotic3}. They were discovered since the Green--Schwarz mechanism~\cite{gs} allowed exceptional gauge groups, which could not be included in the Chan--Paton construction~\cite{cp1,cp2,cp3,cp4,cp5}. They are built combining somehow the left--moving modes of the superstring and the right--moving modes of the bosonic string. Leaving aside the 8 transverse spacetime coordinates, one needs to this end to account for 16 additional right--moving bosonic contributions, which can be done, equivalently, resorting to 32 Majorana-Weyl right--moving fermions. Supersymmetry is then guaranteed by the left--moving modes, while modular invariance leaves two options, which are clearly encoded in the two partition functions 
\beq
 {\cal T}_{HE} \, = \, \int_{\cal F} \ \frac{d^2 \tau}{\tau_{2}{}^2} \ \frac{(V_8 - S_8)(\overline{O}_{16} + \overline{S}_{16})^2}{\tau_{2}{}^4 \ \eta^8 \, \overline{\eta}^8} \ , \ {\cal T}_{HO} \, = \, \int_{\cal F} \ \frac{d^2 \tau}{\tau_{2}{}^2} \ \frac{(V_{8} - S_{8})(\overline{O}_{32} + \overline{S}_{32})}{\tau_{2}{}^4 \ \eta^8 \, \overline{\eta}^8}  \ . \label{heterotics}
 \eeq

The four characters of SO(16) or SO(32) start with a scalar, a vector, a left spinor and a right spinor, which makes the low--lying spectrum essentially manifest, up to level--matching condition. This can be seen tracking, in the partition functions, the powers of $q$ and $\bar{q}$ that accompany the left--moving and right--moving modes.

For the HE theory, the left--moving sector starts with $\Delta=0$, and can be matched with two types of contributions from the right--moving sector. The first is obtained applying the bosonic operators $\tilde{\alpha}_{-1}^j$ to the right--moving vacuum, which had $\Delta =1$, and the result again yields the massless modes of a traceless symmetric tensor, a two-form and the dilaton if combined with $V_8$, and a gravitino and a dilatino if combined with $S_8$. The second type of contribution arises from the internal oscillators, for which the roles of the $O$ and $V$ sectors are interchanged, so that the former plays the role of the fusion identity. In this case the massless states that compensate $\Delta$ and satisfy level-matching are obtained from
\beq
b_{-\,\frac{1}{2}}{}^I \,b_{-\,\frac{1}{2}}{}^J |0_{NS1} \ 0_{NS2}\rangle \ , \qquad |0_{R1} 0_{NS2}\rangle \ , 
\eeq
$(J=1,\ldots,16)$, which build the adjoint of the first $E_8$ from the adjoint of SO(16) and a spinor, of dimension 120 and 128, and similarly for the second SO(16) factor. Similar considerations hold for the factor corresponding to the second $E_8$, so that combining these options with $V_8-S_8$ one recovers the modes of the  ten--dimensional supersymmetric Yang--Mills theory with an $E_8 \times E_8$ gauge group. In the $HO$ theory the $(1,0)$ gravity
is recovered in the same way, while the quantum numbers of the Yang--Mills modes originate fully from $\overline{O}{}_{32}$, the character that describes the internal NS sector, since the Ramond states in $\overline{S}{}_{32}$ are all massive. 

Note that in these heterotic models the ten--dimensional Yang--Mills coupling is related to the string scale and the string coupling $g_s$ according to
\beq
\frac{1}{g_{YM}^2} \ \sim \ \frac{1}{g_s^2 \ {\alpha'}^3} \ .  \label{YM_string}
\eeq
Combining this with eq.~\eqref{MP_string} relates the ten--dimensional Yang--Mills coupling to the Planck and string scales according to
\beq
g_{YM}^2 \ = \ M_P^8\ \alpha'\ .
\eeq

Our compact notation has the virtue of highlighting the crucial difference between the two partition functions in eq.~\eqref{IIA_IIB}: the IIB partition function is \emph{symmetric} under the complex conjugation that interchanges left--moving and right--moving modes, while its low--lying excitations are \emph{chirally asymmetric} in space--time. The opposite is true for the IIA partition function,
which is \emph{not symmetric} under complex conjugation, while its low--lying excitations are \emph{chirally symmetric} in space--time. In principle, the IIB theory could have gravitational anomalies, but they cancel automatically, as was shown in~\cite{agwitt}, since the anomalous contributions arising from the chiral Fermi fields are compensated by others from the self--dual form. The symmetry of the IIB partition function links it to the type-I theory of open and closed strings, as we shall see in the next section.

The type--IIB and type--IIA models can be turned into each other by a few steps that can also illustrate, in a simple context, the orbifold construction~\cite{orbifolds1,orbifolds2}, a key procedure that allows one to build new string models from given ones. To this end, let us begin by projecting the IIB spectrum by the operation $(-1)^{G_R}$, where $G_R$ is the \emph{spacetime} fermion number of right--moving excitations, not to be confused with the world-sheet fermion number that was introduced in the previous section. This step amounts to inserting in the original partition function, before performing the Str operation, the projection
\beq
\frac{1}{2} \left[ 1 \ + \ (-1)^{G_R}\right] \ ,
\eeq
which replaces the original contribution in eq.~\eqref{IIA_IIB} with
\beq
 {{\cal T}_{IIB} \ \rightarrow \ \frac{1}{2}\ {\cal T}_{IIB}\ +\ \frac{1}{2} \ \int_{\cal F} \ \frac{d^2 \tau}{\tau_{2}{}^2}\, \frac{\left(V_8 \,-\,S_8\right)\left({\overline V}_8 \,+\,{\overline S}_8 \right)}{\tau_{2}{}^4 \ \eta^8 \, \bar{\eta}^8} } \ , \label{T_Untw}
 \eeq
but now the second term is not invariant under the modular transformation corresponding to the matrix $S$ in eq.~\eqref{modularn}. The remedy is to add the term thus generated to eq.~\eqref{T_Untw},
\beq
 {\frac{1}{2} \ \int_{\cal F} \ \frac{d^2 \tau}{\tau_{2}{}^2} \ \frac{\left(V_8 \,-\,S_8\right)\left({\overline O}_8 \,-\,{\overline C}_8 \right)}{\tau_{2}{}^4 \eta^8 \, \bar{\eta}^8} } \ , \label{T_tw1}
 \eeq
together with another one obtained letting $\tau \to \tau+1$ in it, and thus determined by the matrix $T$ of eq.~\eqref{modularn},
\beq
{\frac{1}{2} \ \int_{\cal F} \ \frac{d^2 \tau}{\tau_{2}{}^2} \ \frac{\left(V_8 \,-\,S_8\right)\left(\,-\,{\overline O}_8 \,-\,{\overline C}_8 \right)}{\tau_{2}{}^4\ \eta^8 \, \bar{\eta}^8} } \ . \label{T_tw2}
\eeq
Collecting all terms in eqs.~\eqref{T_Untw}, \eqref{T_tw1} and \eqref{T_tw2} builds a modular invariant result, but this is precisely the partition function of type IIA in eq.~\eqref{IIA_IIB}. In other words, as we had anticipated, IIA is an orbifold of IIB by $(-1)^{G_R}$ (and vice versa). In a similar fashion, one can also turn the $HE$ model into the $HO$ one, projecting the first with $(-1)^{G_1+G_2}$, which flips the sign of the two $S_{16}$ contributions. Therefore, two of the four models in Eqs.~\eqref{IIA_IIB} and \eqref{heterotics} are not independent, but can be derived from the others.

The four models that we have just described exhaust the available options for supersymmetric oriented closed strings. However, another supersymmetric model of a different type exists: it is the type--$I$ $SO(32)$ superstring, which is not really independent since it is an \emph{open descendant} or \emph{orientifold} of type IIB~\cite{orientifolds1}.

In view of eq.~\eqref{genus_expansion}, the low--energy effective field theory of the  different closed superstrings combines the kinetic terms for gravity, NS-NS two-form, dilaton and gauge vectors (in the heterotic cases) in the string frame, where they are all accompanied by a factor $e^{-\,2\,\varphi}$:
\beq
{\cal S}_{closed} \, = \, \frac{1}{2\,\kappa^2} \ \int d^{\,10}
x \ e^{-2 \varphi} \left[ R + 4 \,\partial^\mu\,\varphi\,\partial_\mu\,\varphi - \frac{1}{12}
\, H^{\mu\nu\rho}\,H_{\mu\nu\rho} - \frac{\alpha'}{4} \,  F^a_{\mu\nu}\, F^{a \,\mu\nu} + \ldots \right] \ .
\eeq
In addition, when RR fields are present, they have no accompanying powers of $e^\varphi$, as first noted in~\cite{witten}.

\subsection{\sc The Orientifold Construction and the \texorpdfstring{$SO(32)$} \ \ Type-I Theory} \label{sec:10dor}

We can now illustrate how the $SO(32)$ type--I superstring emerges as a descendant (or orientifold) of the left--right symmetric type--IIB model that we have described, following~\cite{orientifolds1,orientifolds2,orientifolds3,orientifolds4,orientifolds5,orientifolds6,orientifolds7,orientifolds8} (for reviews, see~\cite{orientifolds_rev1,orientifolds_rev2,orientifolds_rev3,orientifolds_rev4}). The procedure is reminiscent of the orbifold construction that we have just illustrated, but it is more complicated since it also affects the two--dimensional surfaces swept by the string in the vacuum amplitudes and thus the very nature of strings.  The simpler construction for the bosonic string, where the different surfaces are characterized in detail, is discussed in Appendix~\ref{app:bosonic_orientifold}.
\begin{figure}[ht]
\centering
%\begin{figure}
\begin{tabular}{ccc}
%\mbox{graphic} & \mbox{table} \\
\includegraphics[width=38mm]{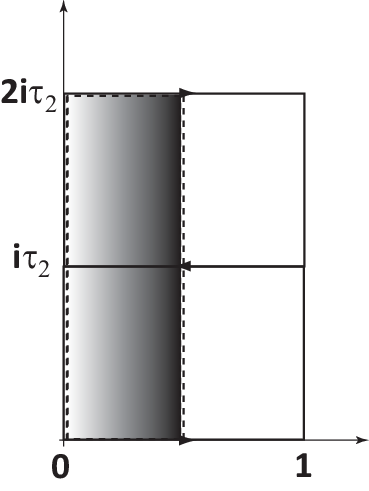} &
\includegraphics[width=38mm]{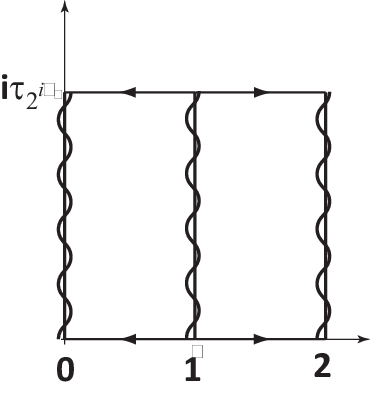} &  
\includegraphics[width=48mm]{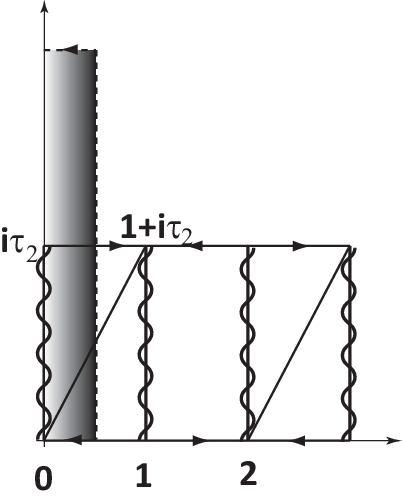} \\
\end{tabular}
 \caption{\small Left panel: the Klein bottle (lower rectangle or shaded region) and the doubly covering torus with $\tau= 2 i \tau_2$. Middle panel: the annulus and its doubly covering torus with $\tau = \frac{i\,\tau_2}{2}$. Right panel: the M\"obius strip (left rectangle or shaded region) and its doubly covering torus with $\tau= \frac{1}{2}+ i \frac{\tau_2}{2}$.}
\label{fig:KAM}
\end{figure}

To begin with, exploiting the symmetry of the type--IIB model under the interchange of its left and right modes, one can project the spectrum so as to retain states that are invariant under world-sheet parity. The underlying operation, often denoted by $\Omega$, acts on the two--dimensional fields as
\beq
\Omega \ X (\tau, \sigma) \ \Omega^{-1} \ =\
X (\tau, - \sigma) \ , \qquad  \Omega \ \psi(\tau, \sigma) \ \Omega^{-1} \ = \ \pm \ 
\tilde{\psi} (\tau, - \sigma) \label{omega1} \ ,
\eeq
thus interchanging the two world-sheet spinors.

The projection is effected by adding to the (halved) torus amplitude a contribution defined on the Klein bottle, whose properties are described in Appendix~\ref{app:bosonic_orientifold}. This step leads to
\beq
\frac{1}{2} \  {\cal T}_{IIB} \ + \  {\cal K} \ ,
\eeq
where
\beq
{\cal K} \ =\ \frac{1}{2} \ \int_0^\infty \ \frac{d\tau_2}{\tau_2^2} \ \frac{V_8\left(2 i \tau_2\right) \,-\,S_8\left(2 i \tau_2\right)}{\tau_2^4 \ \eta^8\left(2 i \tau_2\right)}  \label{Klein_I}
\eeq
is the ``direct--channel'' (or ``loop--channel'') Klein--bottle amplitude. As explained in Appendix~\ref{app:bosonic_orientifold}, the argument of the functions involved is $2 i \tau_2$, which corresponds to working with the product $q \,\overline{q}$. We also stress that, consistently with their dependence on $q \,\overline{q}$, the contributions $V_8$ and $S_8$ present in the Klein bottle amplitude have a bosonic nature. The former completes the symmetrization of the NS-NS sector, while the latter completes the anti-symmetrization of the RR sector. As a result, starting from the $8 \times 8 = 64$ states present in the massless NS-NS or RR sectors of the type--IIB oriented closed string, one ends up with 36 states in the former and 28 states in the latter. These combinations are symmetric and antisymmetric under the interchange of left and right sectors, as determined by the signs of the corresponding contributions to ${\cal K}$. The massless NS-NS sector thus looses the two--form and retains the dilaton and graviton modes, while the massless RR sector looses the scalar and the self--dual four--form and retains the two--form. The Fermi modes are simply halved, since they receive no contributions from the Klein bottle amplitude. As a result, only a type--IIB gravitino and dilatino pair is left. At the massless level, the result is the spectrum of $(1,0)$ supergravity in ten dimensions.

As explained in Appendix~\ref{app:bosonic_orientifold}, the Klein--bottle amplitude is \emph{not invariant} under modular transformations. However, it exhibits an interesting behavior if $\tau_2$ is halved, which amounts to referring the measure to the doubly covering torus in the left panel of Fig.~\ref{fig:KAM}, and then a $S$ transformation is performed. This turns the second contribution into a \emph{tree--level exchange diagram} for the closed--string spectrum between a pair of crosscaps (real projective planes or, if you will, spheres with opposite points identified). The final result reads
\beq
{\widetilde {\cal K}} \ = \ \frac{2^5}{2} \ \int_0^\infty \ {d\ell} \ \frac{V_8 \left(i \ell\right)\,-\,S_8\left(i \ell\right)}{\eta\left(i \ell\right)^8}  \ , \label{Ktilde}
\eeq
where $V_8$ and $S_8$ reflect the propagation of the NS-NS and RR sectors, and the overall link between the integration variables of the tree-level (or ``transverse") and loop (or ``direct") channels is
\beq
\ell = \frac{1}{2\,\tau_2}.
\eeq

The powers of $\ell$ have disappeared, consistent with the absence of momentum flow in the tree-level propagation of the closed string in the transverse channel, as explained in Appendix~\ref{app:bosonic_orientifold}. Note that the two contributions to ${\widetilde {\cal K}}$ involve precisely the two ``diagonal'' sectors $V_8 \ \overline{V}{}_8$ and $S_8 \ \overline{S}{}_8$ that propagate in the tube, and the actual ``spacetime characters'', $V_8$ and $-\,S_8$, enter ${\cal K}$ with a relative ``plus'' sign. The antisymmetrization of the Ramond--Ramond sector could also be justified noting that $\Omega$ interchanges a pair of mutually anticommuting spinors, which introduces the additional sign upon rearrangement, but a Klein-bottle amplitude involving $V_8\,+\,S_8$ would readily lead to an inconsistency. In fact, it would imply the presence of $O_8-C_8$ in the transverse channel, two sectors that are not present in the IIB spectrum. As a result, the only choice allowed for ${\cal K}$ rests on the supersymmetric character $V_8-S_8$, and therefore does not give any contribution to the vacuum energy. However, there are two independent tadpole contributions, and the RR one signals the presence of gauge and gravitational anomalies that cannot be canceled by the Green--Schwarz mechanism. This can be seen by noting that the counterpart of the tadpole potential discussed in the Appendix~\ref{app:bosonic_orientifold} would be, in this case, a geometric coupling to a ten-form potential similar to~\eqref{geometric_ap1}, which would make the low--energy equations inconsistent unless the overall coefficient of this term vanished. From a spacetime viewpoint, the contributions to the gravitational anomaly from gravitino and dilatino modes do not cancel, and the introduction of an open sector, with fermions contributing to anomaly cancellation, is inevitable in this case.

The loop--channel type--I open--string amplitudes rest on the GSO--projected NS and R spectra discussed in Section~\ref{sec:string_spectra}. As for the bosonic string, both contributions depend on $\frac{i\,\tau_2}{2}$, rather than on ${2\,i\,\tau_2}$, as demanded by their reduced Regge slope. The annulus amplitude reads
\beq
{{\cal A}} \ = \  \frac{1}{2} \ {\cal N}^2 \ \int_0^\infty \ \frac{d\tau_2}{\tau_2^2} \ \frac{V_8\left(\frac{i \tau_2}{2}\right) \,-\,S_8\left(\frac{i \tau_2}{2}\right)}{\tau_2^4 \ \eta^8\left(\frac{i \tau_2}{2}\right)} \ , \label{annulus_I}
\eeq
where ${\cal N}$ is a Chan--Paton multiplicity factor~\cite{cp1,cp2,cp3,cp4,cp5} associated with each end of the open strings, which will be shortly shown to be an integer number. This sector is bound to depend on the supersymmetric combination by spin--statistics: the $V_8$ and $S_8$ sectors describe in fact bosonic and fermionic fields that fill, at the massless level, the spectrum of ten--dimensional supersymmetric Yang--Mills theory.

The action of $\Omega$ extends naturally to open strings, and swaps their ends according to
\beq
\Omega \ X^\mu (\tau, \sigma) \ \Omega^{-1} \ = \
X^\mu (\tau, \pi- \sigma) \ , \qquad  \Omega \ \psi^\mu(\tau, \sigma) \ \Omega^{-1} \ = \
\pm \ \tilde{\psi}^\mu (\tau, \pi- \sigma)  \ ,  \label{omega12}
\eeq
thus introducing alternating signs in the contributions of different oscillators. This action is
consistent with the shift by $\frac{1}{2}$ present in the contributions to the M\"obius amplitude implied by the shape of its doubly covering torus, up to phases that can be ascribed to the vacua of the different sectors and are accounted for by the ``hatted'' redefinitions described in Appendix~\ref{app:bosonic_orientifold}. For the type-I string
\beq
{{\cal M}} \ =\   \frac{\epsilon}{2} \ {\cal N} \ \int_0^\infty \ \frac{d\tau_2}{\tau_2^2} \ \frac{{\hat V}_8\left(\frac{i \tau_2}{2}+\frac{1}{2}\right) \,-\,{\hat S}_8\left(\frac{i \tau_2}{2}+\frac{1}{2}\right)}{\tau_2^4 \ {{\hat \eta}^8\left(\frac{i \tau_2}{2}+\frac{1}{2}\right)}}  \ , \label{Moebius_I}
\eeq
where the overall factor ${\cal N}$ is associated to the M\"obius boundary and $\epsilon$ is a sign. In this construction, which defines the type--I superstring, both ${\cal A}$ and ${\cal M}$ involve the supersymmetric combination $V_8 - S_8$, and therefore the introduction of this supersymmetric open sector does not affect the vacuum energy. As for the bosonic string, the overall powers of $\tau_2$ in these amplitudes reflect the presence of loop momenta. At the massless level, the open string spectrum describes $\frac{{\cal N}({\cal N} \ \pm \ 1)}{2}$ massless gauge bosons and an equal number of massless gaugini depending on whether $\epsilon=\pm 1$, which are thus valued in the adjoint representation of a  $USp(N)$ or $SO(N)$ gauge group, in the two cases.
The shifted argument $\frac{1}{2}+i \frac{\tau_2}{2}$ of the M\"obius amplitude then implies the presence of massive states with Chan-Paton multiplicities equal to $\frac{{\cal N}({\cal N} \ \pm \ 1)}{2}$ at alternate levels.

The two amplitudes ${\cal A}$ and ${\cal M}$ can be turned to the transverse channel by the $S$ transformation in eq.~\eqref{modularn} and by the $P$ transformation (see eq. (\ref{P-trans}) of Appendix \ref{app:bosonic_orientifold})
\beq
 P \ = \ \left( \begin{array}{rrrr} -1 & 0 & 0 & 0 \\
0 & 1 & 0 & 0 \\
0 & 0 & 1 & 0 \\
0 & 0 & 0 & 1  \end{array} \right) \label{P8}
\eeq
for the SO(8) ``hatted'' characters, and proceeding as in Appendix~\ref{app:bosonic_orientifold} leads to~\footnote{The absence of powers of $\ell$ is a special case of a property originally noted in~\cite{clasha}: only in the critical dimension loop amplitudes of this type, with one string intermediate state, have correctly pole singularities rather than cuts.}
\bea
{\widetilde {\cal A}} &=& \frac{2^{-5}\,{\cal N}^2}{2} \, \int_0^\infty \ {d\ell} \ \frac{V_8\left(i \ell\right) \,-\,S_8\left(i \ell\right)}{\eta^8\left(i \ell\right)} \ , \nonumber \\
{\widetilde {\cal M}} &=&  \ 2\ \epsilon\, \frac{{\cal N}}{2} \ \int_0^\infty \ {d\ell} \ \frac{\widehat{V}_8\left(i \ell\,+\,\frac{1}{2}\right) \,-\,\widehat{S}_8\left(i \ell\,+\,\frac{1}{2}\right)}{\widehat{\eta}^8\left(i \ell\,+\,\frac{1}{2}\right)}  \ , \label{AMtilde}
\eea
where $V_8$ and $S_8$ now reflect the tree-level propagation of the NS-NS and RR closed--string sectors. The overall coefficients depend on ${\cal N}^2$ and ${\cal N}$, and the multiplicities can be associated to the reflection coefficients at two boundaries present in $\widetilde{\cal A}$ and at the single boundary present in $\widetilde{\cal M}$. 

The transverse--channel amplitudes in eqs.~\eqref{Ktilde} and \eqref{AMtilde} involve the same closed spectrum, and the two tadpole contributions from the NS-NS and RR sectors are proportional to
\beq
\frac{2^5}{2} \left( 1 + {2^{-10}\, \cal N}^2 \ + \ 2\, \epsilon \,\times \, 2^{-5}\, \cal N\right) \ . \label{tadpole}
\eeq
Both contributions cancel for $\epsilon=-1$ and ${\cal N}=32$, and thus for the $SO(32)$ group first identified in~\cite{gs}. The elimination of the RR tadpole associated with $S_8$ is a crucial prerequisite of the Green--Schwarz anomaly cancellation mechanism, since it removes the irreducible contribution to the twelve--dimensional anomaly polynomial, as discussed in Section~\ref{sec:sugra1110}. In this supersymmetric model, the $V_8$ term also cancels due to supersymmetry, and we shall soon elaborate on what happens in more complicated cases, when this does not occur.

Summarizing, the preceding steps have connected the type--IIB string, with (2,0) supersymmetry, to a model with $(1,0)$ supersymmetry combining a projected closed sector that begins, at the massless level, with $(1,0)$ supergravity, with an open sector that is forced to begin with the $(1,0)$ supersymmetric Yang--Mills theory with an SO(32) gauge group, and where symmetric and antisymmetric group representations alternate at different mass levels. 

In this case, the Yang--Mills action emerges at the disk level, and consequently the relation between the Yang--Mils coupling and the string scale,
\beq
\frac{1}{g_{YM}^2} \ \sim \ \frac{1}{g_s \ {\alpha'}^3} \ ,
\eeq
is different from what we saw for the heterotic string. The string--frame low--energy effective field theory 
\beq
{\cal S}_{I}  =  \frac{1}{2\,\kappa^2}  \int d^{\,10}
x \left\{ e^{-2 \varphi} \left[ R + 4 \left(\partial\,\varphi\right)^2\right]  - \frac{\alpha'\,e^{-\,\varphi}}{4} \,  F^a_{\mu\nu}\, F^{a \,\mu\nu}  - \frac{1}{12}
\, H^{\mu\nu\rho}\,H_{\mu\nu\rho} + .. \right\} 
\eeq
contains a vector contribution from the disk level and a RR two-form, which therefore carries no power of $e^\varphi$.

\subsection{\sc Circle Compactification and \texorpdfstring{$T$}\ --Duality} \label{sec:1Dtorus}

In Section~\ref{sec:kaluzaklein} we discussed Kaluza--Klein compactification on a circle of radius $R$ in Field Theory. This revealed the emergence of discrete momenta and a corresponding discrete spectrum of excitations, of masses $M^2 \ = \ \frac{m^2}{R^2}$ with $m=0,1,2,\ldots$. We can now analyze how strings behave in the same type of compactification, starting from the case of closed strings. This will readily reveal that, surprisingly, for the closed bosonic string a circle of radius $R$ is indistinguishable from another of the ``$T$-dual'' radius~\cite{Tduality1,Tduality2}
\beq
R' \ =\ \frac{\alpha'}{R} \label{T_dual} \ .
\eeq
On the other hand, we shall see that, for the open bosonic string, this duality links two descriptions, one with Neumann conditions in the direction of the circle and another with Dirichlet conditions. Elaborating on this fact will reveal the role of solitonic objects usually called $D$-branes, where open strings can end~\cite{Tduality3,Dbranes}. Wilson lines built from constant internal gauge fields along the circle shift string momenta, and in the $T$-dual description displace the $D$-branes. 
Moreover, we shall see that for orientifold vacua the $\Omega$ projection is also modified and turned into $\Omega\,P$, where $P$ denotes the parity operation along the circle, so that in the $T$-dual description the circle is turned into an interval with a pair of orientifolds at the ends.  Moreover, for closed superstrings $T$-duality interchanges the IIA and IIB closed models, and also the HE and HO models, in the presence of internal Wilson lines breaking their gauge groups to $SO(16) \times SO(16)$. Finally, the type-I $SO(32)$ superstring with Wilson lines breaking $SO(32)$ is turned into the so--called type--I model, with D8 branes along the interval, O8 orientifolds at its ends, and the IIA theory in the bulk. 

\subsubsection{\sc \texorpdfstring{$T$}\ --Duality for Closed Strings on a Circle}

Let us begin by considering the closed bosonic string, which can wrap the circle any number of times, so that its spectrum comprises infinitely many topological sectors. Consequently, the internal string coordinate 
\beq
X\  =\  x \ + \ 2 \alpha' \, \frac{m}{R} \ \tau \ + \  2 \, n \, R \,
\sigma \ + \  \frac{i
\sqrt{2\alpha^\prime}}{2}
\ \sum_{n \not= 0} \left( \frac{\alpha_n}{n} \ e^{-2 i n (\tau -
\sigma)} \ + \
\frac{\tilde{\alpha}_n}{n}\  e^{-2 i n (\tau + \sigma)} \right) \label{closedmodes2}
\, .
\eeq
depends not only on the momentum quantum number $m$, but also on the ``winding number'' $n$.
The name reflects the fact that, for any given value of $n$, as $\sigma$ spans the $[0,\pi]$ interval, $X$ increases by $2 n \pi R$.
The structure of the zero modes is better emphasized by
defining the two combinations
\beq
X_{\rm L,R} \ = \ {\frac{1}{2}} \,x \ + \  \alpha' \, p_{L,R} (\tau \pm \sigma)
\  +  \ {\rm (oscillators)_{L,R}} \  , \label{XLR}
\eeq
and the two chiral momenta $p_{\rm L}$ and $p_{\rm R}$
associated to the compact coordinate
\beq
p_{\rm L,R} \ = \ \frac{m}{R} \ \pm \ \frac{n R}{\alpha'}  \ , \label{onedimmom}
\eeq
so that
\beq
X \ = \ X_{\rm L} \ + \ X_{\rm R} \ .
\eeq
Eq.~\eqref{closedmass} is now modified, and becomes
\beq
M^2 \ = \ \left(\frac{m}{R} \ - \ \frac{n\,R}{\alpha'}\right)^2 \ + \ \frac{4}{\alpha'} \left( N \ - \  1
\right) \ = \ \left(\frac{m}{R} \ + \ \frac{n\,R}{\alpha'}\right)^2 \ +\ \frac{4}{\alpha'} \left( \bar{N} \ - \  1
\right)  \label{closedmass_torus}
\eeq
for the mass spectrum perceived in the 25 non--compact directions. Equivalently, one can work with
\beq
M^2 \ = \  \left(\frac{m}{R}\right)^2 \ + \ \left(\frac{n\,R}{\alpha'}\right)^2 \ + \ \frac{2}{\alpha'}\left(N \ + \ \bar{N} \ - \ 2 \right) \ ,
\eeq
together with the level--matching condition
\beq
{N} \ - \ \bar{N} \ = \  m n \ .
\eeq

The spectrum is clearly invariant under the $T$-duality operation
\beq
R \ \to \ R' \ , \qquad m \ \longleftrightarrow \ n \ , 
\eeq
but then $p_L \ \to \ -\ p_L$ and $p_R \ \to \ p_R$. In other words, $T$-duality entails a parity that only affects the left--moving modes. This is actually a two--dimensional analog of electric--magnetic duality, which turns $R$ into $R'$ and $X$ into $X'$, with
\beq
\partial_+\,X \ = \ \partial_+\,X' \ , \qquad \partial_-\,X \ = \ - \ \partial_-\,X' \ , \label{2d-duality}
\eeq
without affecting the zero mode $x$. For the superstring, this flips the sign of $\psi$, the left--moving Fermi field along the circle, while leaving $\tilde{\psi}$ invariant. Therefore, it also flips the relative chirality of the two R sectors of superstrings, and for this reason $T$--duality turns the IIA and IIB models into one another.

For generic values of $R$, one finds two massless Kaluza--Klein vectors corresponding to $m=n=0$, $N=\tilde{N}=1$, and thus to 
\beq
\alpha_{-1}^i \,\tilde{\alpha}_{-1} \left| 0 \tilde{0} \right\rangle \ , \qquad  \alpha_{-1}{} \,\tilde{\alpha}_{-1}^i \left| 0 \tilde{0} \right\rangle \ , 
\eeq
but at the self--dual radius
\beq
R \ = \ \sqrt{\alpha'} 
\eeq
an interesting new phenomenon occurs. There are indeed four additional massless vectors,
\beq
\alpha_{-1}^i \left| m=\pm 1, n=\mp 1\right\rangle \ , \qquad  \tilde{\alpha}_{-1}^i \left| m = \pm 1, n = \pm 1 \right\rangle \ , 
\eeq
and the string interactions among them reveal that the gauge group has enhanced to $SU(2) \times SU(2)$. For generic radii, the circle thus yields a $U(1) \times U(1)$ group, associated to internal components of the metric and the two-form in the standard Kaluza--Klein picture. However, the enhancement taking place at the selfdual radius yields two rank--one gauge groups. A mechanism of this type on 16--dimensional tori for the right--moving internal heterotic coordinates underlies the emergence of the rank--16 gauge groups $E_8 \times E_8$ and $SO(32)$ for the HE and HO models.

\subsubsection{\sc \texorpdfstring{$T$} \ --Duality for Open Strings on a Circle}

For open strings on a circle, the momentum is quantized as for particles, but no windings are allowed by the Neumann conditions at the ends, so that
\beq
X \ = \ x \ + \ 2 \alpha^\prime\, \frac{m}{R} \ \tau \ + \ i
\sqrt{2\alpha^\prime}
\ \sum_{n \not= 0} \frac{\alpha^\mu_n}{n} \ e^{- i n \tau} \cos(n
\sigma)  \ . \label{opennmodes_circle}
\eeq
A $T$-duality turns $X$ into $X'$, momentum excitations into windings and Neumann conditions into Dirichlet ones, so that
\beq
X' \ = \ x'\ + \ 2\,n\,R'\,\sigma \ + \ 
\sqrt{2\alpha^\prime}
\ \sum_{n \not= 0} \frac{\alpha^\mu_n}{n} \ e^{- i n \tau} \sin(n
\sigma)  \ , \label{opennmodesT}
\eeq
which describes open strings with both ends located at a generic point $x'$ on the dual circle, which wrap the circle an arbitrary number of times. One can interpret this expansion as revealing the presence, in the dual spacetime, of a D24-brane, an extended object located at $x'$, whose world volume invades all the other dimensions. In a similar fashion, for the superstrings one would thus identify a D8-brane, and by analogy one talks about D25 (or D9) branes for the standard open strings with Neumann boundary conditions. Performing additional compactifications and corresponding $T$-dualities can similarly lead to branes of lower dimensionalities.
For both open and closed strings, up to $T$-dualities, one can thus restrict the attention to the region $R \geq \sqrt{\alpha'}$. All these D-branes have tensions proportional to $\frac{1}{g_s}$, since their modes start emerging at the disk level, as we saw for the open sector of the type-I $SO(32)$ superstring in ten dimensions.

An important class of continuous deformations of the open spectrum is obtained via internal gauge--field components that are constant along the circle and result in $\left[U(1)\right]^r$ Wilson lines 
\beq
A \ = \ \oplus_{I=1}^r \ \frac{a^I}{R} \ = \  \frac{1}{R} \ \mathrm{diag}\left(a_1,\ldots,a_r\right) \ ,
\eeq
where $r$ denotes the rank of the gauge group. These
modify the string Lagrangian for the internal coordinate by the addition of boundary terms at the string ends, which for a single $U(1)$ carry opposite charges, so that in the $(I,J)$ sector
\beq
{\cal S}_{IJ} \ = \ \frac{1}{4\,\pi\,\alpha'}\ \int d\tau\,d\sigma \left[\left(\partial_\tau\,X\right)^2 \ - \ \left(\partial_\sigma\,X\right)^2 \right] \ - \ \frac{a^I}{R} \int_{\sigma=0} d\tau \ \dot{X} \ + \ \frac{a^J}{R} \int_{\sigma=\pi}  d\tau \ \dot{X} \ , \label{S_Wilson}
\eeq
This deformation
shifts the momentum of the internal open-string coordinate in that sector, turning it into
\beq
X \ = \ x \ + \ 2 \alpha^\prime\, \frac{m\,+\,a^I \, - \, a^J}{R} \ \tau \ + \ i
\sqrt{2\alpha^\prime}
\ \sum_{n \not= 0} \frac{\alpha^\mu_n}{n} \ e^{- i n \tau} \cos(n
\sigma)  \ , \label{opennmodes_circle_W}
\eeq
while in the $T$-dual description the internal coordinate becomes
\beq
X' \ = \ x'\ - \ 2\,\pi\,a^I\,R'\ + \ 2 \Big(n\,+\,a^I\,-\,a^J\Big) \,R'\,\sigma \ + \ 
\sqrt{2\alpha^\prime}
\ \sum_{n \not= 0} \frac{\alpha^\mu_n}{n} \ e^{- i n \tau} \sin(n
\sigma)  \ , \label{opennmodesT_w}
\eeq
and pertains to open strings stretched between branes $I$ and $J$ located at $x'\,-\, 2\,\pi\,a^I\,R'$ and $x' \,-\, 2\,\pi\,a^J\,R'$. 

The ordinary formulation  describes at most $r$ independent Wilson lines deforming the string momenta, and these considerations extend to open superstrings, so that the resulting $(D-1)$--dimensional mass spectrum becomes
\beq
M^2\ = \ \left(\frac{m\,+\,a^I \, - \, a^J}{R}\right)^2 \ + \ \frac{N-1}{\alpha'} \ . \label{mass_omega}
\eeq
On the other hand, the $T$-dual formulation can be associated to at most $r$ non--coincident D-branes with open strings stretching among them, so that the $(D-1)$--dimensional mass spectrum becomes
\beq
M^2\ = \ \left(n\,+\,a^I \, - \, a^J\right)^2\,\frac{\left(R'\right)^2}{\alpha'} \ + \ \frac{N-1}{\alpha'} \ . \label{mass_omegaP}
\eeq

\subsubsection{\sc \texorpdfstring{$T$} \ --Duality for Unoriented Strings on a Circle}

We can now address the $T$-dual description of the type--I superstring, and to this end we shall begin by linking the world--sheet parity $\Omega$ and its $T$-dual $\Omega'$, whose effects on the internal coordinate should be consistent with eq.~\eqref{2d-duality}. In detail, we shall deduce $\Omega'$ by demanding that
\beq
\left(\Omega\,X\,\Omega\right)'\ = \ \Omega'\,X'\,\Omega' \ . \label{omegaomega'}
\eeq
In this fashion, we shall see that 
\beq
\Omega'\ = \ \Omega\,P \ ,  \label{omega'}
\eeq
where $P$ is the parity operation along the circle. 

Let us begin by defining the flipped string coordinate along the circle
\beq
X_\Omega \ = \ \Omega \, X\,\Omega \ .
\eeq
Then in the closed--string case, starting from
\beq
X \ = \ X_+(\tau+\sigma,R) \ + \ X_-(\tau - \sigma,R) \ + \ x \ , \label{xpm}
\eeq
where $x$ is the zero mode, one finds
\beq
X_\Omega \ = \ X_+(\tau-\sigma,R) \ + \ X_-(\tau + \sigma,R) \ + \ x \ ,
\eeq
while, using eq.~\eqref{2d-duality}, its $T$-dual is
\beq
\left(X_\Omega\right)' \ = \ - \ X_+(\tau-\sigma,R') \ + \ X_-(\tau + \sigma,R') \ + \ x' \ .
\eeq
On the other hand, starting from eq.~\eqref{xpm} one can conclude that
\beq
X' \ = \ X_+(\tau+\sigma,R') \ - \ X_-(\tau - \sigma,R') \ + \ x' \ , \label{xpm2}
\eeq
so that eq.~\eqref{omegaomega'} indeed holds, with $\Omega'$ defined as in eq.~\eqref{omega'}, but
$x'$ must be a fixed point of $P$, which can be chosen to be $0$ or $\pi R$.

There are two fixed points of this operation on the dual circle, at $X'=0,\pi R'$. These identify two orientifold O8-planes for the $T$-dual of the $SO(32)$ superstring, which is often referred to as the type-I' theory. 

An open--string state carrying Chan--Paton labels $(I,J)$ is turned by $\Omega$ into a state of $(J,I)$ type, in such a way that
\beq
\Omega
\ \prod_{u=1}^s \alpha_{k_u}^{\mu_u} \ \prod_{v=1}^t \alpha_{k_v} |0,x,m\rangle_{IJ} \ = \ (-1)^{\sum_u k_u+\sum_v k_v}\  \prod_{l=1}^s \alpha_{k_u}^{\mu_u} \ \prod_{v=1}^t \alpha_{k_v}  \, \Omega\,|0,x,m\rangle_{IJ} \ ,
\eeq
where $x$  and $m$ denote the zero-mode of $X$ and the Kaluza--Klein momentum quantum number, while in the $T$-dual picture
\beq
\Omega\,P 
\ \prod_{u=1}^s \alpha_{-\,k_u}^{\mu_u} \ \prod_{v=1}^t \alpha_{-\,k_v}  |0',x',n\rangle_{IJ} \ = \ (-1)^{\sum_u k_u+\sum_v k_v}\  \prod_{l=1}^s \alpha_{-\,k_u}^{\mu_u} \ \prod_{v=1}^t \alpha_{-\,k_v} \ \Omega\,P\,|0',x',n\rangle_{IJ} \ ,
\eeq
where $x'$ $(0 \ \mathrm{or} \ \pi R)$ denotes the zero-mode of $X'$, and thus the common position of the $D$-branes in this example, which is fixed under $\Omega'$, while $n$ denotes the winding number in the internal direction.
The action of $\Omega$ (or $\Omega\,P$) on the vacuum interchanges the two Chan--Paton labels $(I,J)$ and brings along an additional sign $\epsilon=\pm 1$, so that
\beq
\Omega\, |0,x,m\rangle_{IJ} \ = \ \epsilon \, |0,x,m\rangle_{JI} \ , \qquad \Omega\ P\,|0',x', n\rangle_{IJ} \ = \ \epsilon \ |0',-x',n\rangle_{JI} \ .
\eeq
Although we have focused for simplicity on bosonic vacua acted upon by bosonic oscillators only, the preceding properties, together with the additional signs induced by the oscillators, characterize the behavior of all states belonging to a given sector.

We now have all the ingredients to describe the effect of $T$-duality on the type-I $SO(32)$ superstring. The circle becomes an interval of length $\pi R'$, and in the bulk the type IIB string is replaced by type IIA. The open sector now involves a set of 16 $D8$ branes localized at one of the fixed points $x'=(0,\pi\,R')$ along the interval, and the gauge group coincides, as it should, in the two descriptions.

\subsubsection{\sc \texorpdfstring{$T$}\ --Duality and Vacuum Amplitudes on a Circle} \label{sec:wilson}

We can now describe the effects of circle compactification and $T$-duality on the vacuum amplitudes, including the effect of Wilson lines and their $T$-dual description.
To begin with, if a non--compact coordinate is replaced by a circle of radius $R$, the corresponding contribution to the torus amplitude is modified according to
\beq
{\cal T} \ : \quad \frac{R}{\sqrt{\alpha'}}\  \frac{1}{\sqrt{\tau_2}\  \eta(\tau) \,\eta(\bar{\tau})} \ \longrightarrow \ \Lambda_{m,n} \ \equiv \ \sum_{m,n \in \mathbb{Z}}
\frac{ q^{\frac{\alpha'}{4} \left(\frac{m}{R} \ - \ \frac{n R}{\alpha'}\right)^2} \ \bar{q}^{\frac{\alpha'}{4} \left(\frac{m}{R} \ + \ \frac{n R}{\alpha'}\right)^2}}{\eta(\tau)\,
\eta(\bar{\tau})} \ , \label{TRinf}
\eeq
since the momentum integration in that direction is replaced by a sum over discrete momenta and windings. The shorthand $\Lambda_{m,n}$ for the lattice sum will prove useful in the following sections.

Note that in performing this replacement, we are only factoring out the volume of the non--compact spacetime directions, so as to define the effective $(D-1)$--dimensional spectrum resulting from the compactification, and this grants the amplitude a Diophantine nature. The right--hand side is then manifestly invariant under $T$-duality. Moreover, the lattice sum can be recast in a manifestly modular invariant form, 
\beq
\Lambda_{k,n} \ = \ \frac{R}{\sqrt{\alpha'}}\  \frac{\sum_{k,n \in Z} \ e^{\,-\,\frac{\pi\,R^2}{\alpha'\,\tau_2} \, \left|k - n \tau\right|^2}}{\sqrt{\tau_2} \,\eta(\tau)\, \eta({\overline{\tau}})} \ . \label{lag_lattice}
\eeq
by resorting to the Poisson summation formula
\beq
\sum_{k \in Z} \ f(k) \ = \ \sum_{m \in \mathbb{Z}} \ \int_{-\infty}^\infty \ dx \ f(x) \ e^{\,-\,2\,\pi\,i\,m\,x} \ . \label{poisson}
\eeq
In fact, the transformation $\tau \to \tau+1$ can be compensated by replacing $k$ with $k-n$, while $\tau \to -\, \frac{1}{\tau}$ can be compensated by interchanging $k$ and $-\,n$. In this representation, the $k=n=0$ term recovers the noncompact limit.

Similarly, for the sake of comparison, the $(D-1)$--dimensional contribution to the Klein-bottle amplitude with $\Omega$ only involves internal momenta, and reads
\beq
{\cal K}_{\Omega}(R) \ = \ \frac{1}{2}\ \frac{\sum_{m \in \mathbb{Z}} e^{ -\, \pi\,\tau_2\, \frac{\alpha'\,m^2}{R^2}}}{\eta(2 i \tau_2)}\ . 
\eeq
On the other hand, the contribution with $\Omega P$ only involves windings, and reads
\beq
{\cal K}_{\Omega\,P}(R') \ = \ \frac{1}{2}\ \frac{\sum_{n \in \mathbb{Z}} e^{ -\, \pi\,\tau_2\, \frac{n^2\,R'^2}{\alpha'}}}{\eta(2 i \tau_2)} \ .
\eeq
These expressions are turned into one another by eq.~\eqref{T_dual}. 

With the standard Chan--Paton multiplicity, the $(D-1)$--dimensional contribution to the annulus amplitude is
\beq
{\cal A}_{\Omega}(R)  \ = \  \frac{N^2}{2}\  \frac{\sum_{m \in \mathbb{Z}} e^{ -\, \pi\,\tau_2\, \frac{\alpha'\,m^2}{R^2}}}{\eta\left(\frac{i \tau_2}{2}\right)}\ , \label{AomegaR}
\eeq
and in the $T$--dual description it becomes
\beq
{\cal A}_{\Omega\,P}(R') \ = \   \frac{N^2}{2}\ \frac{\sum_{n \in \mathbb{Z}} e^{ -\, \pi\,\tau_2\, \frac{n^2\,R'^2}{\alpha'}}}{\eta\left(\frac{i \tau_2}{2}\right)} \ , \label{AomegaPR}
\eeq
while the corresponding $(D-1)$--dimensional contributions to the M\"obius strip amplitude are
\beq
{\cal M}_{\Omega}(R)  \ = \ - \ \frac{N}{2}\ \frac{\sum_{m \in \mathbb{Z}} e^{ -\, \pi\,\tau_2\, \frac{\alpha'\,m^2}{R^2}}}{\hat{\eta}\left(\frac{1 + i \tau_2}{2}\right)}\ , \label{MomegaR}
\eeq
and
\beq
{\cal M}_{\Omega\,P}(R')  \ = \ - \ \frac{N}{2}\ \frac{\sum_{n \in \mathbb{Z}} e^{ -\, \pi\,\tau_2\, \frac{n^2\,R'^2}{\alpha'}}}{\hat{\eta}\left(\frac{1+ i \tau_2}{2}\right)} \ . \label{MomegaPR}
\eeq
All these expressions involve the standard denominators $\eta(2 i \tau_2)$, $\eta\left(\frac{i\tau_2}{2}\right)$ and $\hat{\eta}\left(\frac{1+i\tau_2}{2}\right)$, but without the factor $\sqrt{\tau_2}$ that is only recovered in the $R \to \infty$ ($R' \to 0$) limit in the two cases of $\Omega$ ($\Omega\,P$).

We can now complete the analysis by illustrating the role of open--string Wilson lines in orientifold vacua.
As we have seen in eq.~\eqref{mass_omega}, internal gauge fields translate the momenta of open-string states according to their charges. Alternatively, in the $T$--dual picture, they displace $D$-branes. 
 \begin{figure}[ht]
\begin{center}
    \includegraphics[width=2in]{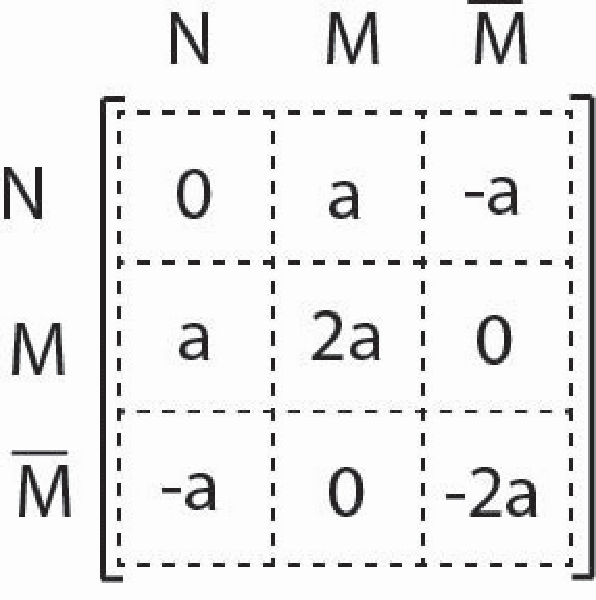}
\end{center}
\caption{\small The momentum shifts in the presence of Wilson lines result from the contributions of three types of ends for the open strings, which combine as in the figure. In this example, $N$ branes have no Wilson lines,  $M$ have Wilson lines $a/R$, and their $M$ orientifold images  have Wilson lines $-a/R$. The result breaks in general $SO(32)$ to $SO(N) \times U(M)$, with $N+2M=32$. Numerically, $M=\bar{M}$, but distinguishing them in direct--channel amplitudes allows one to distinguish different Chan--Paton representations, just like distinguishing $S_8$ and $C_8$ allowed to distinguish different fermionic sectors containing identical number of states.}
\label{fig:CP_torus}
\end{figure}

Referring to the first formulation, a general Wilson line induced by a constant internal gauge--field component $A$ gives a contribution to the direct--channel annulus and M\"obius amplitudes determined by
\beq
\mathrm{Tr}\left({\cal W}\right)  \ = \ \mathrm{Tr} \exp\left[ \oplus_{r=1}^{16} \ i \alpha^r\,\sigma_2 \right] \ , \qquad 
\mathrm{Tr}\left({\cal W}^2 \right) \ = \  \mathrm{Tr} \exp\left[ \oplus_{r=1}^{16} \ 2\,i \alpha^r\,\sigma_2 \right] \ , 
\eeq
since in the geometric regularization that we are adopting (see Section~\ref{sec:10dor}) the M\"obius strip has a boundary of double length with respect to the annulus.
It will suffice to illustrate a simple instance, with
\beq
\alpha^r \ = \ 0 \ \ \left(r=1,\ldots,\frac{N}{2}\right) \ , \qquad \alpha^r \ = \ a \ \left(r=\frac{N}{2}+1,\ldots,\frac{N}{2}+M\right) \ , 
\eeq
where
\beq
N \ + \ 2 \, M \ = \ 32 \ .
\eeq
Then
\beq
\mathrm{Tr}\left({\cal W}\right) \ = \ N \ + \ M \, e^{i \alpha} \ + \ \overline{M} \, e^{-\,i \alpha} \ , \qquad
\mathrm{Tr}\left({\cal W}^2\right) \ = \ N \ + \ M \, e^{2 i \alpha} \ + \ \overline{M} \, e^{-\,2 i \alpha}  \ ,  \label{Wilson_line}
\eeq
where $M = \bar{M}$, but we distinguish them since the corresponding states will be in the fundamental or anti-fundamental representation of $U(M)$. This deformation induces, in general, the breaking of $SO(32)$ to $SO(N) \times U(M)$, with
\beq
N \ + \ 2 \, M \ = \ 32 \ .
\eeq

The deformed annulus amplitude is then determined by eq.~\eqref{mass_omega}, and reads
\bea
{\cal A} &=&  \int_0^\infty \frac{d\tau_2}{\tau_2^2} \ \frac{V_8\left(\frac{i \tau_2}{2}\right)  - S_8\left(\frac{i \tau_2}{2}\right) }{\tau_2^\frac{7}{2} \ \eta^8\left(\frac{i \tau_2}{2}\right)  }\  \sum_{m} \
\biggl\{
\biggl( M \bar{M} \ + \ \frac{1}{2} \ N^2 \biggr)
q^{\frac{\alpha'
m^2}{2 R^2}} \nonumber \\ &+& M N \
q^{\frac{\alpha' (m+a)^2}{2 R^2}} \ + \ \bar{M} N \
q^{\frac{\alpha' (m-a)^2}{2 R^2}}
\ + \ 
\frac{M^2}{2} \ q^{\frac{\alpha' (m+2a)^2}{2
R^2}} \ + \
\frac{\bar{M}^2}{2}
\ q^{\frac{\alpha'
(m-2a)^2}{2 R^2}} \biggr\} \ , \label{annulus_wilson}
\eea
while the corresponding M\"obius amplitude is
\bea
{\cal M} &=& \int_0^\infty \frac{d\tau_2}{\tau_2^2} \ \frac{\hat{V}{}_8\left(\frac{1+i \tau_2}{2}\right)  - \hat{S}{}_8\left(\frac{1+i \tau_2}{2}\right) }{\tau_2^\frac{7}{2} \ \hat{\eta}^8\left(\frac{1+i \tau_2}{2}\right)} \nonumber \\
&\times& \sum_{m} \ \biggl\{ - \ 
 \frac{1}{2}\ N  q^{\frac{\alpha'
m^2}{2 R^2}}
\ - \ 
\frac{1}{2} \  M \ q^{\frac{\alpha' (m+2a)^2}{2
R^2}}
\ - \  \frac{1}{2} \ \bar{M} \
q^{\frac{\alpha'
(m-2a)^2}{2 R^2}}
\biggr\} \ , \label{moebius_wilson}
\eea
where in these expressions
\beq
q \ = \ e^{-\,2\pi \tau_2} \ . \label{q_ell}
\eeq

The first two terms in ${\cal A}$ and the first term in ${\cal M}$ that completes the second indicate that the gauge group is $SO(N)\times U(M)$, consistent with the Wilson lines present in this example. The massless states are valued in the adjoint representations of these gauge groups, while the massive modes of strings with two ends of the $N$-type alternate between symmetric and antisymmetric representations. The remaining contributions present in ${\cal A}$ are of two types: those proportional to $M N$ and $\bar{M} N$ describe states in bi--fundamental representations, while those proportional to $M^2$ or $\bar{M}^2$, together with their completions in the M\"obius strip describe, for generic values of $a$, towers of massive states that alternate between symmetric and antisymmetric representations. However, for half--integer values of $a$ the low--lying states of these towers become massless, and the amplitudes then depend on $M+\bar{M}$, so that the gauge group enhances to $SO(2M) \times SO(32-2M)$. 

The setup of Section~\ref{sec:1Dtorus} implies that $T$-duality turns momenta into windings, so that the Wilson lines present in eqs.~\eqref{annulus_wilson} and \eqref{moebius_wilson} afford an alternative interpretation in terms of brane displacements, as we saw in eq.~\eqref{opennmodesT}.  Moreover, according to eq.~\eqref{S_Wilson}, the displacements are felt oppositely by the two open-string ends.
The multiplicity $N$ thus identifies the branes that remain at the fixed point, while $M$ and $\bar{M}$ branes are displaced by $\pm 2\pi a R'$ on the circle. The $M \bar{M}$ sector does not feel any stretching, while the $M N$ and $\bar{M} N$ sectors are stretched as determined by the distance between a brane in the bulk and one at a fixed point, and finally the $M^2$ and $\bar{M}^2$ sectors are doubly stretched on the dual circle. In particular, integer values of $a$ are equivalent and describe branes all at the same fixed point, while half--odd integer values of $a$, and in particular $a=\frac{1}{2}$, describe the $N$ branes at one fixed point and the $M$ and $\bar{M}$ branes at the other fixed point, and in this case the gauge group enhances to $SO(2M) \times SO(32-2M)$.

The structure of this deformation may also be appreciated considering the corresponding transverse-channel amplitudes
\bea
\widetilde{\cal A} &=& \frac{2^{-5}}{2} \ \frac{R}{\sqrt{\alpha'}} \ \int_0^\infty d\ell \ 
\frac{V_8(i \ell) \ - \ S_8(i \ell)}{\eta^8(i \ell)}
\nonumber \\ &\times&
 \sum_{n} \
q^{\frac{n^2 R^2}{ 4 \alpha'}} \ \left( N \ + \  M
\, e^{2 i \pi a n} \ + \
\bar{M} \, e^{- 2 i \pi a n}    \right)^2  \label{aawil}
\eea
and
\bea
\widetilde{\cal M} &=& - \ \frac{2}{2} \frac{R}{\sqrt{\alpha'}} \ \int_0^\infty d\ell \ 
\frac{\hat{V}_8\left(\frac{1}{2}+i \ell\right) - \hat{S}_8\left(\frac{1}{2}+i \ell\right)}{\hat{\eta}\left(\frac{1}{2}+i \ell\right) } 
\nonumber \\ &\times&
 \sum_{n} \
q^\frac{n^2 R^2}{ \alpha'}
\ \left( N + M\,
e^{4 i \pi a n} +
\bar{M}\, e^{- 4 i \pi a n}    \right) \, , \label{amwil}
\eea
where in these expressions
\beq
q \ = \ e^{-\,2\pi \ell}  \label{q_}
\eeq
and the phases reflect the Wilson line~\eqref{Wilson_line}
of the internal components of the gauge vectors.  The preceding expressions, where the combinations involving the Chan--Paton multiplicities build perfect squares, clearly display the symmetry enhancement that occurs for $a=\frac{1}{2}$: in this
case, the two ``complex" charges $M$ and $\bar{M}$ have
identical reflection coefficients in $\tilde{\cal A}$ and
$\tilde{\cal M}$, while the direct-channel amplitudes only
depend on their sum, as we have seen. For this value of $a$ the ${\rm U}(16-N)$ gauge group thus
enhances to ${\rm SO}(32 - 2 N)$. As in previous sections, the different contributions identify the couplings of the closed sector to branes and orientifolds.

The presence of different
lattice sums in the vacuum channel amplitudes has an important consequence: in the $R \to 0$ limit
all winding modes collapse to zero mass, and if $a=0$
the resulting odd-level tadpoles in $\tilde{\cal A}$ are unmatched.
This problem can be actually cured by introducing Wilson lines, and to
this end let us reconsider eq.~(\ref{aawil}) for the special case
$a=\frac{1}{2}$.
In the T-dual picture, this choice corresponds to placing two sets of
branes on top of the two O8 planes, and then
\beq
\widetilde{\cal A} \ = \  \frac{2^{-4}}{4} \ \frac{R}{\sqrt{\alpha'}} \ \int_0^\infty d\ell \ 
\frac{V_8(i \ell) \ - \ S_8(i \ell)}{\eta^8(i \ell)} \ 
 \sum_{n \in \mathbb{Z}} \
q^{\frac{n^2 R^2}{ 4 \alpha'}} \ \left[ N \ + \  N' (-1)^n \right]^2 \, , \label{aawilc}
\eeq
where we have let $M=\frac{N'}{2}$, so that the gauge group is $SO(N) \times SO(N')$, with $N+N'=32$.
We can now compare this expression with $\widetilde{\cal K}$ and $\widetilde{\cal M}$, which
we cast in the same form introducing projectors on even values of $n$, twice in $\widetilde{\cal K}$ and once in $\widetilde{\cal M}$ so that
\bea
\widetilde{\cal K} &=&\frac{2^5\,R}{2\,\sqrt{\alpha'}} \;
(V_8 - S_8)  \sum_{n \in \mathbb{Z}}  \ \frac{q^{n^2
R^2/ \alpha'}}{\eta} \ = \ \frac{2^5\,R}{2\,\sqrt{\alpha'}} \;
(V_8 - S_8)  \sum_{n \in \mathbb{Z}}  \frac{ \left[1 + (-1)^n\right]^2}{4} \  \frac{q^{n^2
R^2/4 \alpha'}}{\eta} \ , \nonumber \\
\widetilde{\cal M} &=& - \ \frac{2 \times\,R}{2\,\sqrt{\alpha'}} \ 
(\hat{V}_8 -
\hat{S}_8)  \sum_{n \in \mathbb{Z}}
 \left[N + N' (-1)^n \right] \ \frac{1 + (-1)^n}{2}\ \frac{q^{n^2
R^2/4 \alpha'}}{\hat{\eta}} \ , \label{ktildemtilde_int}
\eea
where as above $q$ is defined in eq.~\eqref{q_ell} and the factor $(-1)^n$ accompanying $N'$ in $\widetilde{\cal M}$ is ineffective due to the projection, but is nevertheless convenient in what follows.

The $T$--dual version of the three amplitudes,
\bea
\widetilde{\cal K} &=& \frac{2^5\,\sqrt{\alpha'}}{2\,R'} \;
(V_8 - S_8)  \sum_{m \in \mathbb{Z}}  \frac{ \left[1 + (-1)^m\right]^2}{4} \  \frac{q^{\alpha' m^2/4 \left(R'\right)^2}}{\eta} \ , \nonumber \\
\widetilde{\cal A} &=& \frac{2^{-5}}{2} \ \frac{\sqrt{\alpha'}}{R'} \ \frac{V_8 \ - \ S_8}{\eta^8} \ 
 \sum_{m \in \mathbb{Z}} \
\frac{q^{\alpha' m^2/4 \left(R'\right)^2}}{\eta} \ \left[ N \ + \  N' (-1)^m \right]^2 \ , \nonumber \\
\widetilde{\cal M} &=& - \ 2 \ \frac{\sqrt{\alpha'}}{2\,R'}  \ 
(\hat{V}_8 -
\hat{S}_8)  \sum_{m \in \mathbb{Z}} \frac{q^{\alpha' m^2/4 \left(R'\right)^2}}{\hat{\eta}}
 \left[N + N' (-1)^m \right] \ \frac{1 + (-1)^m}{2} \label{KAM_T}
\eea
affords a simple geometrical interpretation in terms of branes and orientifolds at the ends of an interval of length $\pi R'= \frac{\pi \alpha'}{R}$. Note, in fact, that the zero--mode contributions to an orientifold state located at $x$ admit the decomposition 
\beq
\left|{\cal O} \right\rangle_x \ = \ {\cal O} _0 \ \sum_{m \in \mathbb{Z}} e^{i\, \frac{m x}{R}} T_{\cal O} \left| m \right\rangle \ ,
\eeq
where $\,{\cal O}_0$ is a normalization constant and $T_{\cal O}$ denotes the orientifold tension, and the total crosscap state is then
\beq
\left| {\cal C} \right\rangle  \ = \ \left|{\cal O} \right\rangle_0 \ + \ \left|{\cal O} \right\rangle_{\pi R} \ .
\eeq
The transverse--channel Klein--bottle amplitude $\widetilde{\cal K}$ describes the propagation of the closed spectrum according to
\beq
\widetilde{\cal K} \ = \ \left\langle {\cal C} \right| e^{-\,\frac{1}{2}\, \pi \ell \alpha' H} \left| {\cal C} \right\rangle \ = \ \Big( \left\langle {\cal O} \right|_0 \ + \ \left\langle{\cal O} \right|_{\pi R}\Big) \ e^{-\,\frac{1}{2}\, \pi \ell \alpha' H} \Big( \left|{\cal O} \right\rangle_0 \ + \ \left|{\cal O} \right\rangle_{\pi R}\Big) \ ,
\eeq
and the overlap between zero--mode contributions produces precisely the alternating signs in the first of eqs.~\eqref{ktildemtilde_int}. The same considerations apply to boundaries, and determine $\widetilde{\cal A}$, and finally to $\widetilde{\cal M}$, which is determined by the overlap between boundary and crosscap states. These states first appeared in the work of Cremmer and Gervais~\cite{cremm_gerv}, and were later extended to the superstring in~\cite{ext_superstring,pc1}.

Collecting the contributions to the three
transverse-channel amplitudes, one discovers that the overall factor is proportional to
\bea
&& \sum_{n \in \mathbb{Z}}  q^{n^2 R^2/4 \alpha'} \left\{ \left[ N +  N' (-1)^n \right]^2 + 2^8 \left[1 + (-1)^n\right]^2 - 2 \times 2^4 \left[ N \ + \  N' (-1)^n \right]  \left[1 + (-1)^n\right] \right\} \nonumber \\
&=&
\sum_{n \in \mathbb{Z}}  q^{n^2 R^2/4 \alpha'} \; \left[ (16 - N) +
(16 - N') (-1)^n \right]^2 \, , \label{tadpolecirclelocal}
\eea
which clearly shows that the ${\rm SO}(16) \times {\rm SO}(16)$ gauge group 
is the unique choice that eliminates the tadpole contributions from {\it all} winding modes.
In the T-dual picture, this configuration corresponds to the saturating
tadpoles {\it locally} in the two ${\rm O}_-$ planes, since the
cancellation continues to hold in the $R' \to \infty$ limit,
when branes not coincident with the orientifolds would move an infinite distance
away from them. In other words, when local tadpole cancellation occurs, the sources in the field equations are canceled locally, and the dilaton and RR fields can have constant values. On the other hand, if tadpoles are not canceled locally, the localized sources in the field equations give rise to varying dilaton and RR fields, which in the $R' \to \infty$ limit generate a large back-reaction and strong coupling. The special $SO(16) \times SO(16)$ setup with local tadpole cancellation  played a role in the analysis of~\cite{pw}, and is naturally linked to the Horava-Witten heterotic M-theory \cite{HW1,HW2}.

Leaving local tadpole cancellation aside, there is an interesting special choice for the Klein--bottle projection where $\Omega$ is combined with a shift by half of the circle, so that when it acts on Kaluza--Klein modes with quantum number $m$ it yields a factor $(-1)^m$:
\beq
{\cal K}_{\Omega\,\delta}(R) \ = \ \frac{1}{2}\ \left(V_8 - S_8\right) \frac{\sum_{m \in \mathbb{Z}} \ (-1)^m \ e^{ -\, \pi\,\tau_2\, \frac{\alpha'\,m^2}{R^2}}}{\eta(2 i \tau_2)}\ . 
\eeq
Here the implicit argument is $2 i \tau_2$, and this choice leads to a purely massive $\widetilde{\cal K}$, 
\beq
\widetilde{\cal K}_{\Omega\,\delta}(R) \ = \ \frac{2^5}{2}\ \frac{R}{\sqrt{\alpha'}} \ \left(V_8 - S_8\right) \frac{\sum_{n \in \mathbb{Z}} \ e^{ -\, \frac{\pi\,\ell\, R^2}{2\,\alpha'}\,\left(2n+{1}\right)^2}}{\eta(i \ell)}\ , 
\eeq
so that \emph{no open sector is needed}, since only massive states contribute to $\widetilde{\cal K}$~\cite{dp,gepner}. Alternatively, in the $T$-dual picture 
\beq
{\cal K}_{\Omega\,P\,\delta'}(R') \ = \ \frac{1}{2}\ \frac{\sum_{n \in \mathbb{Z}} (-1)^n\ e^{ -\, 2\,\pi\,\tau_2\, \frac{n^2\,\left(R'\right)^2}{\alpha'}}}{\eta(2 i \tau_2)} \ ,
\eeq
where $\delta'$ acts on windings as $(-1)^n$, and consequently
\bea
\widetilde{\cal K}_{\Omega\,P\,\delta'}(R') &=& \frac{2^5}{2}\ \frac{\sqrt{\alpha'}}{R'} \ \left(V_8 - S_8\right) \frac{\sum_{m \in \mathbb{Z}} \ e^{ -\, \frac{\pi\,\ell\, \alpha'}{2 \,\left(R'\right)^2}\,\left(2 m+{1}\right)^2}}{\eta(i \ell)} \nonumber \\
&=& 
\frac{2^5}{2}\ \frac{\sqrt{\alpha'}}{R'} \ \left(V_8 - S_8\right) \frac{\sum_{m \in \mathbb{Z}} \ e^{ -\, \frac{\pi\,\ell\, \alpha'\,m^2 }{2 \,\left(R'\right)^2}}}{\eta(i \ell)} \left[\frac{1 \ - \ (-1)^m}{2}\right]^2
\ .
\eea
Note the sign difference inside the brackets with respect to $\widetilde{\cal K}$ in eqs.~\eqref{KAM_T}: this configuration describes indeed two O8 planes at the ends of an interval that have opposite values for tension and charge (an O${}_+$ and an O${}_-$, where the former is a variant that we shall meet again in the following sections), so that the total RR charge vanishes.

\subsection{\sc String Dualities and the Link to Eleven Dimensions}

\begin{figure}[ht]
\centering
\includegraphics[width=70mm]{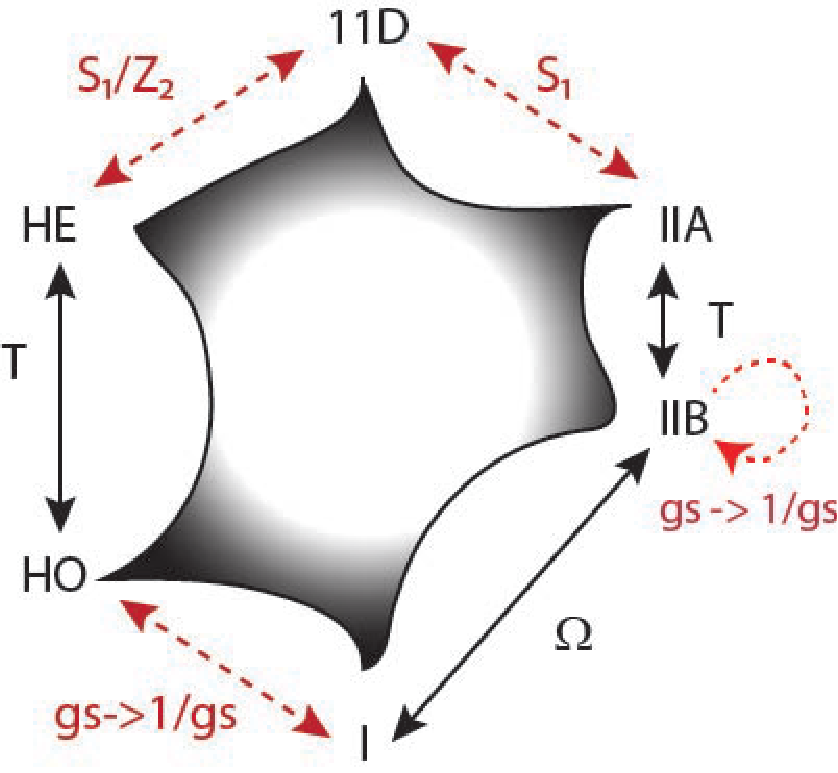}
%\vspace{-5cm}
\caption{\small The duality hexagon for ten--dimensional supersymmetric superstrings.}
\label{fig:susyduality}
\end{figure}

The 10D--11D supersymmetric duality hexagon of fig.~\ref{fig:susyduality} is arguably one of the highest achievements in String Theory~\cite{stringtheory}. The five ten--dimensional superstrings of types IIA, IIB, HE, HO and I, whose low--energy limits are captured by the ten--dimensional supergravities~\cite{I, IIB,gs,cp1,cp2,cp3,cp4,cp5}, of types (1,1), (2,0) and (1,0) that we described in Section~\ref{sec:sugra1110} (see~\cite{sugrarev} for reviews) are linked by generalized dualities to one another and to an eleven--dimensional theory. This grants unprecedented clues that \emph{all} of String Theory, despite its elusive foundations, somehow originates from a unique underlying principle.

The black continuous links in fig.~\ref{fig:susyduality} are proved in perturbation theory, and we already discussed them: $T$ denotes the $T$-dualities linking the IIA and IIB theories and the two heterotic HE and HO theories, while $\Omega$ denotes the orientifold projection reviewed in Section~\ref{sec:10dor}. The other links are conjectured strong--weak dualities that find their ground in the different forms of ten--dimensional supergravity~\cite{witten}, consistently with properties of the extended objects that source their form fields. 

To begin with, the lower dashed link is to be regarded as an actual equivalence, which is strongly suggested by the uniqueness of the (1,0) supergravity. The HO and type-I theories have identical $SO(32)$ gauge groups, and their low--energy presentations,
\bea
{\cal S}_{HO} &=& \frac{1}{2\,\kappa^2} \ \int d^{\,10}
x \ e^{-2 \varphi} \left[ R + 4 \left(\partial\,\varphi\right)^2 - \frac{1}{12}
\, H^{\mu\nu\rho}\,H_{\mu\nu\rho} - \frac{\alpha'}{4} \,  F^a_{\mu\nu}\, F^{a \,\mu\nu} + \ldots \right] ,  \\
{\cal S}_{I}  &=&  \frac{1}{2\,\kappa^2}  \int d^{\,10}
x \left\{ e^{-2 \varphi} \left[ R + 4 \left(\partial\,\varphi\right)^2\right]  - \frac{\alpha'\,e^{-\,\varphi}}{4} \,  F^a_{\mu\nu}\, F^{a \,\mu\nu}  - \frac{1}{12}
\, H^{\mu\nu\rho}\,H_{\mu\nu\rho} + \ldots \right\} , \nonumber 
\eea
can be turned into each other by transforming them into the Einstein frame and letting $\varphi \to - \, \varphi$ in one of them, which inverts the corresponding string coupling. The strong--coupling of one theory is thus captured by the weak--coupling limit of the other, which is under control in string perturbation theory. Moreover, the type--IIB string is conjectured to have a similar self--duality, a discrete remnant of the $SL(2,R)$ symmetry of $(2,0)$ supergravity that acts on its two scalar fields as in eq.~\eqref{mobius_sym}. These include the transformation $
\tau \to - \ \frac{1}{\tau}$, which can connect regimes of strong and weak coupling. 
The need for quantization can be understood noting, for example, that generic $SL(2,R)$ transformations mix the fundamental string with the D1-brane, while the discrete $SL(2,Z)$ interchanges them, consistently with charge quantization.

However, the effective actions of the five ten--dimensional superstrings are also connected, via the remaining sides of the hexagon, to the eleven--dimensional supergravity of Cremmer, Julia and Scherk~\cite{CJS}. 

The link between the IIA theory and the eleven--dimensional supergravity can be justified by referring to the discussion presented in Section~\ref{sec:sugra1110}, and in particular to eq.~\eqref{r11phi}. A circle Kaluza--Klein compactification linking the two theories rests, in fact, on an internal radius that scales as 
\beq
r_{11} \ \sim \ (g_s)^\frac{2}{3} \label{r11PL}
\eeq
in eleven--dimensional Planck units. Remarkably, the strong--coupling limit of the IIA string can thus be argued to open up an additional spacetime dimension! A consistency check of these facts is provided by the Kaluza--Klein mass spectrum, which scales as $\frac{1}{r_{11}}$, as we have seen. However, combining eq.~\eqref{r11phi} with the link between eleven--dimensional and string units,
\beq
r_{11}\ \left(M_{P,11}\right)^9 \ = \ \frac{\left(M_s\right)^8}{g_s^2} \ ,
\eeq
shows that
\beq
\frac{1}{r_{11}} \ = \ \frac{M_s}{g_s} \ ,
\eeq
consistently with the interpretation of Kaluza--Klein modes as D0-branes, whose tension scales proportionally to $\frac{1}{g_s}$, as we have seen.

Moreover, the Horava--Witten construction~\cite{HW1,HW2} links the eleven--dimensional theory to the HE string, in a very appealing picture where the two factors of the $E_8 \times E_8$ gauge group reside at the two ends of an internal interval. The interval originates from a $\mathbb{Z}_2$ orbifold action consisting of a parity in the eleven--dimensional circle, under which the three-form is odd, which acts on Fermi fields as the ten--dimensional chiral rotation $\psi_\mu'(z) = \gamma_{11} \, \psi_\mu(-z)$, $\psi_{10}'(z) = - \, \gamma_{11} \, \psi_{10}(-z)$. This gives rise to anomalies localized at the ends of the interval that are canceled by the addition of $E_8$ super Yang--Mills at each end, with suitable modifications of the Bianchi identities~\cite{HW1,HW2}.
The two ends superpose in the weak--coupling limit, where in view of eq.~\eqref{r11PL} the whole spacetime becomes ten-dimensional. 

Many authors contributed to the hexagon picture~\cite{Tduality1,Tduality2,Tduality3,Tduality4,orientifolds1,orientifolds2,orientifolds3,orientifolds4,orientifolds5,orientifolds6,orientifolds7,orientifolds8,Dbranes,hull-townsend1,hull-townsend2,HW1,HW2}, but Witten~\cite{witten} was arguably the driving force behind this achievement, obtained, as we have tried to explain, by combining input from string theory with conjectures based on lessons drawn from low--energy supergravity. In addition, the dashed links in fig.~\ref{fig:susyduality} rest heavily, one way or another, on the existence of BPS extended objects, D-branes and NS-branes, whose tensions scale proportionally to $\frac{1}{g_s}$ and $\frac{1}{g_s^2}$. As we have already stated, these branes are the counterparts of point charges for the form fields present in low--energy spectra, and emerge as generalizations of the solitons that had long surfaced in Field Theory. The dualities among the different ten--dimensional strings subsumed in fig.~\ref{fig:susyduality} often interchange ordinary string excitations with these types of solitons, or map one type of soliton into another. The dualities can be checked for BPS--protected quantities, whose features are insensitive to strong-coupling limits, and no contradictions have emerged in all these cases.
Still, the eleven--dimensional theory, which couples an eleven--dimensional vielbein $e_M^A$, a Majorana gravitino $\Psi_M$ and a three-form gauge field ${\cal A}_{MNP}$, as we saw in Section~\ref{sec:sugra1110}, lacks the two typical signatures of strings, which are the dilaton $\varphi$ and a two-form gauge field. However, it is surprisingly linked to the ten--dimensional strings by duality maps that involve the reduction to ten dimensions on a circle or an interval. In this wider picture eleven--dimensional membranes~\cite{BST} replace somehow strings, since they couple naturally to a three--form potential, while the latter only emerge effectively when membranes are wrapped around the eleventh dimension. These remarkable findings are usually summarized appealing to an unknown ``M-theory'' that will eventually encompass all cases, as different as they are, as special limits. The non--renormalizable world--volume theory of membranes of~\cite{BST} is precisely tailored to this case, but cannot provide a complete picture of the microscopic dynamics.

\subsection{\sc Boundaries and  Crosscaps \emph{vs} D-Branes and Orientifold Planes}

In Polchinski's spacetime picture~\cite{Dbranes}, the boundaries that emerged in the previous sections and in Appendix~\ref{app:bosonic_orientifold} signal the presence of dynamical extended objects, D-branes, where the endpoints of open strings terminate. 
In a similar fashion, crosscaps are associated to non--dynamical extended objects, end-of-the-world mirrors, which are usually called ``orientifold planes''. This connection is somewhat formal in ten  dimensions, since the extended objects invade the whole spacetime, but it becomes sharper for lower--dimensional extended objects.
From this vantage point, the orientifold construction that we described introduces in the IIB vacuum a ten--dimensional $O9_-$ orientifold~\footnote{Note that this convention is opposite to the one of~\cite{wittenOplus} and of the review~\cite{orientifolds_rev2}, where $O_\mp$ would be called $O_\pm$.}, with negative tension and charge, and of a number of $D9$--branes, whose positive tensions and charges compensate the preceding contributions. 

D-branes are, in general, solitonic objects whose tension scales proportionally to $\frac{1}{g_s}$, where $g_s$ is the string coupling. Therefore, in the perturbative regime that one refers to implicitly, they are rigid and manifest themselves as hyperplanes to which the ends of open strings are attached. 
When describing a generic D$p$-brane, one must distinguish the $p+1$ longitudinal coordinates $x^\mu$ ($\mu=0,\ldots, p$) parallel to the brane and the remaining $9-p$ coordinates $x^i$ orthogonal to it. If both ends of an open string lie on such a brane, its $X^\mu$ coordinates satisfy Neumann conditions there, while the remaining $9-p$ coordinates $X^i$ satisfy Dirichlet conditions. These D-branes and orientifolds are generalized solitons that can be treated as probes or, when combined into the orientifold construction, can give rise to new string models, as we saw for the type--IIB and type-I theories in the previous section. In Section~\ref{sec:1Dtorus} we saw that $T$-dualities interchange Neumann and Dirichlet conditions, thus generating D-branes of different dimensions. Performing successive $T$-dualities, one can conclude that the IIB theory contains BPS D${}_{-1}$, 
 D1, D3, D5, D7 and D9 branes, while the IIA theory contains BPS D0, D2, D4, D6 and D8 branes, consistent with their spectra of form fields. In ten dimensions, one can add these objects to the IIA or IIB theories, as would be the case for solitons in Field Theory, but this step introduces two new ingredients in the vacuum, their tension and charge, which manifest themselves, in the low--energy effective theory, via 
\beq
\Delta\,{\cal S} \ = \ - \ T_p \int d^{p+1} \,x \ \sqrt{-\,\tilde{g}} \ e^{-\,\varphi} \ - \ q_p \ \int \, \tilde{A}_{p+1} \ .
\eeq
Here $\tilde{g}$ denotes the metric induced on the world--volume of the brane, 
\beq
\tilde{g}_{ab} \ = \ g_{\mu\nu} \ \partial_a\,X^\mu\, \partial_b\,X^\nu\ ,
\eeq
and
\beq
\tilde{A}_{a_1 \ldots a_{p+1}} \ = \ A_{\mu_1 \ldots \mu_{p+1}}\ \partial_{a_1}\,X^{\mu_1}\, \ldots \,\partial_{a_{p+1}}\,X^{\mu_{p+1}}  \ ,
\eeq
where the $X^\mu$ are the embedding coordinates for the brane in spacetime. The brane-world-volume can also host gauge fields, but we leave them out for now, for simplicity. 
\begin{table}[h!]
\centering
\begin{tabular}{||c | c c||}
 \hline
Object & Tension & $RR$ Charge  \\ [0.5ex]
 \hline\hline
Dp${}$ & $+ $ &  $+ $ \\ \hline
$\overline{\mathrm Dp}$ & $+$ &  $-$ \\ \hline
Op${}_{-}$ & $- $ &  $- $ \\ \hline
Op${}_{+}$ & $+ $ &  $+$ \\ \hline
${\overline{\mathrm Op}}{}_{-}$ & $- $ &  $+$ \\ \hline
${\overline{\mathrm Op}}{}_{+}$ & $+ $ &  $-$ \\
 \hline
\end{tabular}
\caption{ The signs of the tensions and RR charges for the D-branes and O-planes that appear in perturbative orientifold constructions. The charges of Dp branes are related to the ones of the Op-planes via $Q_{Dp} = \pm 2^{5-p} Q_{Op}$.}
\label{tab:DpOp}
\end{table}

The D-branes give rise to a back-reaction on spacetime, which generally has important effects. We shall often refer to the first contribution as a ``tadpole potential'', due to the presence of the exponential factor that reflects its link to the disk amplitude. Whenever one needs to take it into account, Minkowski space ceases to be a vacuum, but the second contribution has even more dramatic effects if some directions transverse to the brane are compact. Faraday lines have nowhere to end in a compact space, as can be appreciated considering, for example, a sphere. Consequently, the charge can only be compensated by an opposite one, which can originate from anti--branes or orientifolds (see Table~\ref{tab:DpOp}). This is the role in spacetime of the RR tadpole cancellation that determines the SO(32) gauge group for the type-I string in ten dimensions.

In the presence of D$p$-branes there are several open--string sectors. For example, the bosonic coordinates of open superstrings stretched between two parallel D$p$-branes separated by distances $\delta^i$ along the $i$-th directions expand as
\beq
X^i \ = \ x^i \ + \  \frac{\delta^i \, \sigma}{\pi} \ + \ i
\sqrt{2\alpha^\prime}
\ \sum_{n \not= 0} \frac{\alpha^\mu_n}{n} \ e^{- i n \tau} \sin(n
\sigma) \ , \label{annulus_delta}
\eeq
with integer--mode sine functions,
and the $p+1$--dimensional mass-shell condition becomes
\beq
M^2 \ = \ p^\mu \, p_\mu \ = \ \frac{1}{\alpha'}\left(N^{(\mu)} \ + \ N^{(i)}\ - \ \Delta\right) \ + \ \left(\frac{\delta^i}{2\pi\alpha'}\right)^2 \ ,
\eeq
where $N^{(\mu)}$ and $N^{(i)}$ refer to the oscillators along spacetime and internal directions. The stretching gives rise to an additional contribution involving $\delta^i$, which vanishes if both open--string ends lie on the same brane. These modes are localized on the brane, and for a single D$p$ brane the massless open spectrum is associated to a dimensional reduction of the supersymmetric Yang--Mills theory that we identified in the critical dimension. The world-volume of a single D${}p$ brane thus hosts a $U(1)$ vector $A_\mu$ living on it and 9-$p$ scalars associated to its transverse displacements, together with their superpartners in the superstring case. On the other hand, the spectrum of open superstrings stretched between a pair of parallel branes separated by a distance $\delta^i$ is massive, and starts from the massless level only when the separation vanishes, enhancing to $U(2)$ the $U(1)\times U(1)$ gauge group of the two individual branes. At the same time, the transverse coordinates become $2 \times 2$ matrices.
In a similar fashion, $n$ coincident D$p$-branes give rise to a $U(n)$ gauge group. The interaction between two parallel branes separated by a distance {$\delta^i$ can be deduced from the transverse--channel annulus amplitude based on eq.~\eqref{annulus_delta}, where one only retains the massless closed-string exchange. For a pair of D${}p$ branes, starting from the direct--channel amplitude
\beq
{{\cal A}} \ = \  \frac{1}{2} \ \int_0^\infty \ \frac{d\tau_2}{\tau_2^2} \ \frac{V_8\left(\frac{i \tau_2}{2}\right) \,-\,S_8\left(\frac{i \tau_2}{2}\right)}{\tau_2^4 \ \eta^8\left(\frac{i \tau_2}{2}\right)} \ \tau_2^\frac{9-p}{2} \ e^{-\,\tau_2 \frac{\delta^2}{(4 \pi \alpha')}}\ , \label{annulus_In}
\eeq
it becomes
\beq
{\widetilde{\cal A}} \ = \  \frac{2^{-\,\frac{p+1}{2}}}{2} \ \int_0^\infty \ d\ell \ \frac{V_8\left(i \,\ell\right) \,-\,S_8\left(i \,\ell\right)}{ \eta^8\left(i \, \ell\right)} \ \ell^\frac{p-9}{2} \ e^{-\,\frac{\delta^2}{(2 \pi \alpha' \,\ell)}}\  \label{annulus_In_t}
\eeq
in the transverse channel. The Jacobi identity~\ref{Jacobi_identity} reveals that the effective interaction vanishes, as ought to be the case for BPS objects, as a result of a compensation between a NS-NS attraction, encoded in the term containing $V_8$, and a RR repulsion, encoded in the term containing $S_8$. Furthermore, the dominant contributions at large distances, which originate from the low--lying massless modes in either sector, are proportional to
\beq
\int_0^\infty \ d\ell \ \ell^\frac{p-9}{2} \ e^{-\,\frac{\delta^2}{(2 \pi \alpha' \,\ell)}}\ \sim \ \frac{1}{\delta^{7-p}} \ ,
\eeq
as pertains to a static Coulomb or Newton potential in the $9-p$ dimensions orthogonal to the brane world volume. This analysis also reveals that the branes are charged with respect to the RR fields, and the vanishing total exchange reveals that their tensions and charges coincide.

For a $p$-brane, the combination $V_8 - S_8$ describing the GSO--projected oscillator spectrum can be recast (for odd values of $p$) in the form
\beq
V_8 \ - \ S_8 \ = \ V_{p-1} O_{9-p} - S_{p-1} S_{9-p} +  O_{p-1} V_{9-p} - C_{p-1} C_{9-p} \ ,
\eeq
in terms of the characters of $SO(p-1)\times SO(9-p)$, where the factors correspond to the world--volume portion of the $p+1$--dimensional Lorentz group and to the internal symmetry group. This decomposition distinguishes, in open spectra, contributions depending on (even or odd) numbers of world--volume NS oscillators (for $V_{p-1}$ or $O_{p-1}$), which are accompanied by corresponding odd or even numbers of internal NS oscillators (for $O_{9-p}$ or $V_{9-p}$), consistent with the GSO projection. At the massless level, one thus gets vectors from $V_{p-1}O_{9-p}$ and scalars from $O_{p-1}V_{9-p}$, which originate from vector polarizations transverse to the brane. Similarly, in the fermionic spectrum described by $S_8$, which collects states with an even number of positive helicity eigenstates, one distinguishes those containing even numbers of helicities in the world volume and also in the internal space (described by $S_{p-1} S_{9-p}$) from those containing odd numbers of positive helicity components in both 
(described by $C_{p-1} C_{9-p}$).

There is another case of interest. If the string end at $\sigma=\pi$ is free, the expansion becomes
\beq
X^i \ = \ x^i \ + \ i
\sqrt{2\alpha^\prime}
\ \sum_{n \not= 0} \frac{\alpha^\mu_n}{n+\frac{1}{2}} \ e^{- i \left(n+\frac{1}{2}\right) \tau} \sin\left[\left(n+\frac{1}{2}\right)\sigma\right] \ .
\eeq
This type of expansion concerns strings joining two branes such that $X^i$ is orthogonal to the first and parallel to the second.

In general, the tensions $T_p$ and charges $Q_p$ of branes and orientifolds are neatly encoded, up to an overall normalization, in the partition functions. In the ten--dimensional type--I example, up to an overall factor, $\widetilde{K}$ thus reveals the squares of the O9-orientifold tension and charge, $\widetilde{A}$ (leaving aside the overall factor ${\cal N}^2$) the squares of the D9-brane tension and charge, and finally $\widetilde{M}$ (leaving aside the overall factor ${\cal N}$) the products of these tensions and charges, up to an overall combinatoric factor of two.
Given that $D$--branes have by definition a positive tension and charge, the overall signs for the O$9_-$-orientifold are then fully determined (see Table~\ref{tab:DpOp}, where the signs of tensions and charges for the types of branes and orientifolds entering perturbative string vacua are displayed).

Moreover, a minor modification of eq.~\eqref{annulus_In_t} reveals the key features of brane--antibrane interactions, which are encoded in
\beq
{\widetilde{\cal A}} \ = \  \frac{2^{-\,\frac{p+1}{2}}}{2} \ \int_0^\infty \ d\ell \ \frac{V_8\left(i \,\ell\right) \,+\,S_8\left(i \,\ell\right)}{ \eta^8\left(i \, \ell\right)} \ \ell^\frac{p-9}{2} \ e^{-\,\frac{\delta^2}{2 \pi \alpha' \,\ell}}\ . \label{annulus_In_tbab}
\eeq
The opposite sign of the RR contribution reflects the RR attraction between a brane and an antibrane with opposite RR charges, which breaks supersymmetry. The corresponding direct--channel amplitude is now
\beq
{{\cal A}} \ = \  \frac{1}{2} \ \int_0^\infty \ \frac{d\tau_2}{\tau_2^2} \ \frac{O_8\left(\frac{i \tau_2}{2}\right) \,-\,C_8\left(\frac{i \tau_2}{2}\right)}{\tau_2^4 \ \eta^8\left(\frac{i \tau_2}{2}\right)} \ \tau_2^\frac{9-p}{2} \ e^{-\,\tau_2 \frac{\delta^2}{4 \pi \alpha'}}\ , \label{annulus_In_bab}
\eeq
and the spectrum start with a scalar tachyonic mode that can be lifted if $\delta$ is large enough.

A subset of the possible D-branes, the BPS Dp branes, are stable and satisfy $T_p = |Q_p|$. When inserted in the vacuum, these objects preserve half of the original supersymmetries, but non-BPS Dp branes also exist~\cite{sen1,sen2,sen3,sen4,sen5,sen6,sen7}. They are not supersymmetric and can carry no charge altogether, but some of them can also be stable. As we have seen, the type IIB string has a number of BPS brane types
($D1$, $D3$, $D5$, $D7$ and $D9$), all with odd $p$, together with a $D_{-1}$ instanton,   and non-BPS branes with even $p$, while the type IIA string has BPS branes
with even $p$ ($D0$, $D2$, $D4$, $D6$ and $D8$) and non-BPS branes for odd $p$. Dp and $D(6-p)$ BPS branes are electric-magnetic duals to each other. The orientifold projection removes some RR forms and the corresponding branes, so that the type I string has only D1, D5 and D9 BPS branes, while other D-branes with different dimensionalities are non-BPS in this case. A detailed analysis of the D-branes present in all ten--dimensional string theories can be found in~\cite{dms}.

D-branes also play a crucial role in scenarios with large extra dimensions \cite{add,dark} (for earlier models with large extra dimensions, see \emph{e.g.} \cite{fayet-large}) and their string implementation with a low string scale \cite{aadd}.  

\subsection{\sc Non--Supersymmetric Ten--Dimensional Strings} \label{sec:nonsusy_10d}

\begin{figure}[ht]
\centering
\includegraphics[width=95mm]{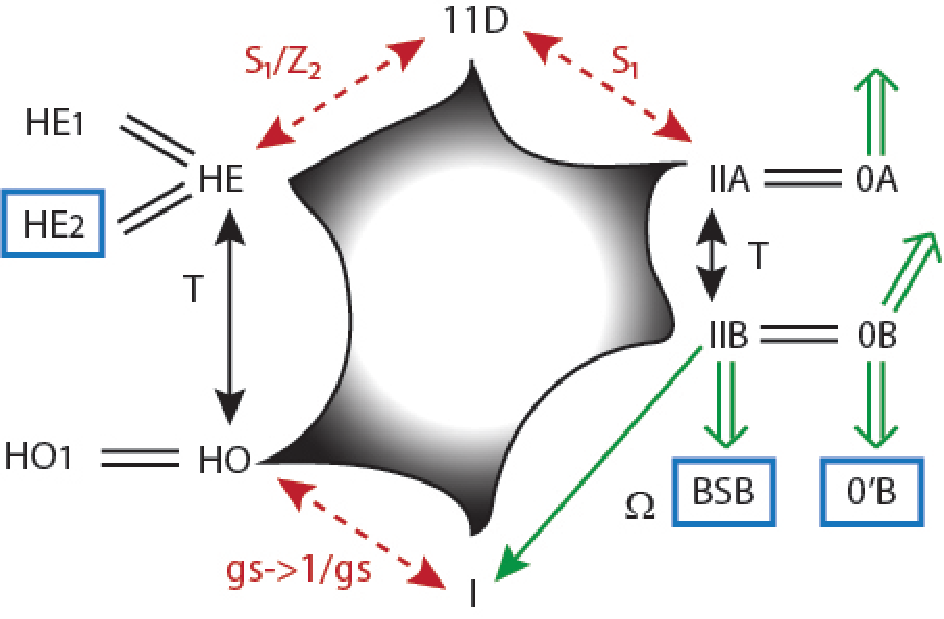}
%\vspace{-5cm}
\caption{\small The larger duality diagram including the ten--dimensional non--supersymmetric superstrings. The green lines identify orientifold projections, most of which were first considered in~\cite{orientifolds5} by Bianchi and one of us. The blue boxes identity the three non--tachyonic models: $HE2$ stands for the $SO(16) \times SO(16)$ heterotic model of~\cite{agmv1,agmv2}, $0'B$ for the $U(32)$ orientifold of~\cite{as95} and BSB for Sugimoto's orientifold in~\cite{sugimoto}.}
\label{fig:nonsusyduality}
\end{figure}

The hexagon diagram~in fig.~\ref{fig:susyduality} collects only a fraction of the available options for ten--dimensional superstring models. There are in fact many more solutions in ten dimensions, whose origin is sketched in fig.~\ref{fig:nonsusyduality}. The additional options~\cite{sw0,agmv1,orientifolds5,as95,sugimoto} lack space--time supersymmetry, and are thus a unique laboratory to gather some information on what String Theory tells us on the key issue of supersymmetry breaking. Surely enough, any attempt to connect String Theory with the Standard Model of Particle Physics is confronted with a bottom--up approach to supersymmetry breaking, which was the subject of the first portion of this review. However, fig.~\ref{fig:nonsusyduality} indicates that String Theory itself is calling for an understanding of this key phenomenon from a top--down perspective.

The vast majority of the new options contain tachyons, and following the fate of their vacua, while possible in principle, at least in the open sectors, appears altogether prohibitively difficult. Important progress was made in fact in the early 2000's, in connection with the tachyon of the open bosonic string~\cite{opentachyon1,opentachyon2}, but the corresponding closed--string tachyon is still fraught with mysteries.
Therefore, it appears reasonable, at this stage, to focus on the three new options, identified by blue boxes, where supersymmetry is broken, or is present but non--linearly realized, and yet the low--lying spectra contain no tachyons (for a review, see~\cite{bsb_rev}). There are two models of the first kind.
The first is the $SO(16)\times SO(16)$ heterotic string of~\cite{agmv1,agmv2}, whose massless spectrum contains states corresponding to $(e_M^A, B_{MN},\phi)$, together with adjoint vectors, left--handed spinors in the $(128,1)+(1,128)$ and right--handed spinors in the $(16,16)$. The second is the $U(32)$ $0'B$ orientifold~\cite{as95,as97}~\footnote{The actual gauge group is $SU(32)$, since a $U(1)$ is anomalous, as shown in~\cite{as97}. However, we shall leave this subtlety aside in the following, for brevity, referring simply to a $U(32)$ model.}, whose massless spectrum contains bosonic states corresponding to $(e_M^A,\phi,a,A_{MN},A_{MNPQ}^+)$, together with adjoint vectors and left--handed fermions in the $496$ +$\overline{496}$.

Finally, the third model is Sugimoto's $USp(32)$ string~\cite{sugimoto}, whose massless spectrum combines the states of (1,0) supergravity with massless vectors in the adjoint of $USp(32)$ and Majorana--Weyl fermions in its (reducible) antisymmetric representation. The singlet spinor contained in the antisymmetric of $USp(32)$ plays a key role in this model: it is the goldstino, to be eaten by the gravitino in a spontaneous breaking of supersymmetry. Local supersymmetry is indeed present in this case, albeit in a nonlinear phase~\cite{dmnonlinear1,dmnonlinear2,dmnonlinear3}. Still, one cannot even write a mass term for the gravitino, which is a Majorana--Weyl spinor--vector in ten dimensions. Ten--dimensional Minkowski space is not a vacuum, however, and this contributes to endow the breaking of supersymmetry with some unusual features.

Supersymmetry is indeed broken, in the $USp(32)$ model, by the simultaneous presence, in the vacuum, of spacetime-filling extended objects, branes and orientifolds, which preserve complementary portions of supersymmetry. Their presence leaves behind a dilaton tadpole potential, and thus deforms the original Minkowski spacetime. The non--dynamical nature of the orientifolds is the very reason behind the absence of tachyonic modes. We have referred to this type of phenomenon as ``brane supersymmetry breaking'', after identifying a first manifestation of it, in six dimensions in~\cite{bsb1}. This resolved a puzzle reviewed in~\cite{erice}, the impossibility of combining certain supersymmetric four--dimensional Klein-bottle projections with supersymmetric open sectors, a more complicated incarnation of the same phenomenon of brane supersymmetry breaking that is inevitable in those cases.
The lack of a gravitino mass term in the $USp(32)$ model does not contradict any known notion: all three non--tachyonic models, and this one in particular, are not defined around ten--dimensional Minkowski space since, as we have stressed, the breaking of supersymmetry induces an important back-reaction. This is signaled by the emergence of runaway (``tadpole'') potentials for the dilaton, which take the form
\beq{}
\Delta {\cal S} \ = \ - \ \frac{T}{2\,\kappa^2} \int d^{10}x \ \sqrt{-g} \ e^{\gamma_S\,\phi}  \label{tadpole_potential_S}
\eeq
in the string frame, with $\gamma_S=-1$ for the two orientifolds and $\gamma_S=0$ for the heterotic model, which reflects the origin of $T$ from the residual tension of D-branes and orientifolds that manifests itself in (projective--) disk amplitudes, and consequently
\beq{}
\Delta {\cal S} \ = \ - \ \frac{T}{2\,\kappa^2}  \int d^{10}x \ \sqrt{-g} \ e^{\gamma\,\phi}  \label{tadpole_potential_E}
\eeq
in the Einstein frame, with $\gamma=\frac{3}{2}$ for the two orientifold models with $U(32)$ and $USp(32)$ gauge groups, and $\gamma = \frac{5}{2}$ for the $SO(16) \times SO(16)$ string. These specific values will play an important role in the following sections.

One should also meditate on another conundrum brought up by fig.~\ref{fig:nonsusyduality}. The very presence of a fundamental string model in ten dimensions where supersymmetry is non--linearly realized should be regarded, in our opinion, as a puzzle on the par with the surprising link to eleven dimensions. One is used to think of non--linear realizations as limiting forms of linear ones that emerge in singular limits, but where is the linear realization in this case? 

\subsection{\sc Non--Tachyonic Ten--Dimensional Strings} \label{sec:10D}

We can now take a closer look at the ten--dimensional string models where supersymmetry is broken, focusing on the three cases where no tachyons are present in the spectrum. Let us begin, as in the preceding sections, by describing the model with closed strings only. This is the $SO(16) \times SO(16)$ string~\cite{agmv1,agmv2}, the unique heterotic model that is not supersymmetric and yet is free of tachyons. It can be obtained as an orbifold of the HE string by the operation $(-1)^{G_L+\overline{F}{}_1+\overline{F}{}_2}$, where $G_L$ is the left spacetime fermion number while $\overline{F}{}_{1,2}$ are the internal fermion numbers associated to the $E_8$ group factors,
and the partition function reads
\bea
{\cal T}_{SO(16)\times SO(16)} \!\!\!&=&\!\!\! \int_{\cal F} \ \frac{d^2 \tau}{\tau_2^2} \ \frac{1}{\tau_2^4 \ \eta^8 \, \bar{\eta}^8} \  \left[O_8 (\bar{V}_{16}\,\bar{C}_{16}+\bar{C}_{16}\,\bar{V}_{16}) +
V_8 (\bar{O}_{16}\,\bar{O}_{16}\,+\,\bar{S}_{16}\,\bar{S}_{16}) \right. \nonumber \\ &-&\left.
S_8 (\bar{O}_{16}\,\bar{S}_{16}\,+\,\bar{S}_{16}\,\bar{O}_{16}) -
C_8 (\bar{V}_{16}\,\bar{V}_{16}\,+\,\bar{C}_{16}\,\bar{C}_{16})\right] \ . \label{so16so16}
\eea

The detailed discussion of supersymmetric string spectra presented in Section~\ref{sec:closed10}  should help the reader identify the massless field content
\begin{eqnarray}&& { ( e_\mu^a, B_{\mu\nu}, \phi) \oplus A_\mu^{(120,1)\oplus(1,120)} \oplus \psi_L^{(128,1)\oplus(1,128)} \oplus \psi_R^{(16,16)} } \ . \label{so16so16_spectrum}
\end{eqnarray}
Note that these massless fields do not include a gravitino, as pertains to spectra not related to supersymmetry, but interestingly no tachyons are present, since the tachyonic ground state of $O_8$ in the first group of contributions to eq.~\eqref{so16so16} does not satisfy level matching with its right--moving partners, whose spectrum begins $\frac{3}{2}$ units above the right--moving ground state, and thus at mass level $\frac{1}{2}$.

Here we encounter a first manifestation of a ubiquitous problem with broken supersymmetry, as we had anticipated: the partition function in eq.~\eqref{so16so16} \emph{does not vanish} after enforcing the Jacobi identity~\eqref{Jacobi_identity}. The resulting vacuum energy, which is positive and was first computed in~\cite{agmv2}, indicates that the system exerts a back-reaction on space time. Its net result at the torus level is the emergence of a string--scale space--time curvature ${\cal O}(1/\alpha^\prime)$, which invalidates the original description of the spectrum around a Minkowski background. As we have stressed, there is no better tool, at present, to investigate the fate of this class of vacua than the low--energy (super)gravity, even in the presence of string-scale vacuum energies. In Sections~\ref{sec:SUSY_breaking_com} and~\ref{sec:SUSY_breaking_cosmo} we shall see how the low--energy theory leads nonetheless to intriguing indications. 

In order to complete our presentation of non--tachyonic models, it is now necessary to divert from our main theme and briefly describe two tachyonic models of oriented closed strings~\cite{sw0}. They are usually called $0A$ and $0B$ strings, and their partition functions read
\bea
{\cal T}_{0A} &=& \int_{\cal F} \ \frac{d^2 \tau}{\tau_2^2} \ \frac{|O_8|^2 + |V_8|^2 + S_8\,\bar{C}_8+ S_8\,\bar{C}_8}{\tau_2^4 \ \eta^8 \, \bar{\eta}^8}\ , \nonumber \\
{\cal T}_{0B} &=& \int_{\cal F} \ \frac{d^2 \tau}{\tau_2^2} \ \frac{|O_8|^2 + |V_8|^2+|S_8|^2 + |C_8|^2}{\tau_2^4 \ \eta^8 \, \bar{\eta}^8} \ , \label{0A_0B}
\eea
so that their spectra are purely bosonic. These partition functions can be obtained as orbifolds of type--IIA and type--IIB strings by $(-1)^{G_L+G_R}$, where $G_{L,R}$ denote the left and right spacetime fermion numbers. The low--lying excitations comprise a tachyon $T$ and the other massless modes below:
\beq { 0A: \ ( e_\mu^a,B_{\mu\nu}, \phi, A_\mu^{1,2}, A_{\mu\nu\rho}^{1,2})}\ , \qquad {0B: \ ( e_\mu^a,B_{\mu\nu}^{1,2,3},\phi^{1,2,3},A_{\mu\nu\rho\sigma})}\ . \label{0AB_spectra}
\eeq
The presence of the tachyon is signaled by the first contribution to either of these partition functions, which involves the $O_8$ character in isolation. However, one of their open descendants is the tachyon--free U(32) model of~\cite{as95,as97}, which we can now illustrate.

In fact, there is a tachyon--free orientifold of the $0B$ string~\cite{as95}, whose applications to the phenomenon of confinement have been pursued in~\cite{armoni1,armoni2,armoni3,armoni4,armoni5,armoni6}. It is obtained starting from the Klein--bottle amplitude,
\beq
{\cal K}_{0'B} \ = \ \frac{1}{2} \ \int_0^\infty \ \frac{d\tau_2}{\tau_2^2} \ \frac{\,-\,O_8\,+\,V_8 \,+\,S_8\,-\,C_8}{\tau_2^4 \ \eta^8}  \ , \label{K0bprime}
\eeq
which is precisely designed to project the closed--string tachyon out of the spectrum. The arguments of the characters and the Dedekind function in this expression are $2 i \tau_2$, as we explained in connection with the type--I superstring.
The peculiar sign choices determine a pattern of (anti)symmetrizations that is consistent with the interactions of the various string sectors, or if you will with the ``spacetime fusion rules'' of the SO(8) level--one Conformal Field Theory that we summarized in eqs.~\eqref{fusion_rules}. It differs from the standard choice,
\beq
{\cal K}_{0B} \ = \ \frac{1}{2} \ \int_0^\infty \ \frac{d\tau_2}{\tau_2^2} \ \frac{O_8\,+\,V_8 \,-\,S_8\,-\,C_8}{\tau_2^4 \ \eta^8}  \ , \label{K0b}
\eeq
by flips of signs for the $O_8$ and $S_8$ sectors, the first of which eliminates the closed--string tachyon. The other sign is needed for consistency with the spacetime fusion rules~\eqref{fusion_rules}: since the $O_8$ originates from the fusion of the $S_8$ and $C_8$ ones, it can be antisymmetric under the interchange of left and right modes only if the fermionic sectors behave oppositely in that respect.
As a result, the massless closed--string spectrum contains, as in all other cases, a graviton and dilaton, but no tachyon, from the NS-NS sectors, together with a two form, an axion and a self--dual four--form from the RR sectors.

There are other tachyonic descendants of
the 0A and 0B models, some of which originate from the Klein--bottle amplitude of eq.~\eqref{K0b}. They were introduced in~\cite{orientifolds5}, and contain fermions in the open sector, while variants of the U(32) model were also introduced in~\cite{as95}~\footnote{In the geometric language of~\cite{Dbranes}, they include additional brane-antibrane pairs, which give rise to tachyons in the open sector.}

The construction of the $U(32)$ model, which is sometimes referred to as 0'B string, was an outgrowth of the previous work in~\cite{pss1,pss2,pss3}, where the two--dimensional consistency conditions for crosscaps of~\cite{crosscap_constraint}, obtained extending the results in~\cite{lewellen} to non--orientable surfaces, were generalized allowing for exotic Klein--bottle projections (these results relied on key inputs by the late Yassen S.~Stanev, and were further extended in~\cite{schellekens1,schellekens2}). A direct investigation of the open descendants of WZW~\cite{WZW} models exhibited indeed an earlier counterpart of the sign flips that underlie the peculiar Klein bottle amplitude of eq.~\eqref{K0bprime}.

The open spectrum accompanying the Klein--bottle projection~\eqref{K0bprime} is determined, as we anticipated in Section~\ref{sec:building_rules}, by letting the allowed sectors, which here are all the four available ones, flow in the annulus vacuum exchange $\widetilde{\cal A}$. However, there is a novelty, since this model is perhaps the simplest instance in which a unitary gauge group, with corresponding ``complex'' charge multiplicities, plays a role. Consequently, its annulus and M\"obius amplitudes read
\bea
{\cal A}_{0'B} &=&  \int_0^\infty \ \frac{d\tau_2}{\tau_2^2} \ \frac{{ {\cal N}} \, { {\cal {\overline N}}} \ V_8 \,-\, \frac{1}{2} \ ( {\cal N}^2 \,+\,{\cal {\overline N}}^2 ) \ C_8}{\tau_2^4 \ \eta^8}  \ , \nonumber \\
{\cal M}_{0'B} &=& \frac{{ {\cal N}\,+\, {\cal {\overline N}}}}{2}\ \int_0^\infty \ \frac{d\tau_2}{\tau_2^2} \ \frac{\hat{C}_8}{\tau_2^4 \ \hat{\eta}^8} \ , \label{open_0'B}
\eea
while the corresponding vacuum exchange amplitudes read
\bea
{\widetilde{\cal K}}_{0'B} &=&  - \ \frac{2^6}{2} \ \int_0^\infty \ {d\ell} \ \frac{C_8}{ \eta^8}  \ , \nonumber \\
{\widetilde{\cal A}}_{0'B} &=&  \frac{2^{-6}}{2} \ \int_0^\infty \ {d\ell} \ \frac{({\cal N}+{\overline{\cal N}})^2 \,(V_8 - C_8) \ - \ ({\cal N}-{\overline{\cal N}})^2 \,(O_8 - S_8)}{ \eta^8}  \ , \nonumber \\
{\widetilde{\cal M}}_{0'B} &=& 2\ \frac{{ {\cal N}\,+\, {\cal {\overline N}}}}{2}\ \int_0^\infty \ {d\ell} \  \frac{\hat{C}_8}{ \hat{\eta}^8}  \ .
\eea
Note that the constraint ${\cal N}= {\overline{\cal N}}$, which reflects the numerical coincidence of the dimensions of the fundamental and conjugate fundamental representations of unitary groups, eliminates the contribution involving $S_8$ from these tree--level vacuum amplitudes, a key property closely linked to the cancellation of gauge and gravitational anomalies. Thus, there are no physical couplings between the background D9 branes and the closed--string states contained in $O_8$ and $S_8$, as needed for consistency, since the tachyon in $O_8$ was removed by the orientifold projection. The same is true for $S_8$: due to its symmetrization in the Klein bottle, the corresponding RR sector contains a zero-form and the dual eight-form, together a selfdual four-form, but no ten-form that could couple to the D9 branes. On the other hand, the antisymmetrization of $C_8$ leads to a two form, its dual six form and a ten form, that can physically couple to the D9 branes. This issue is discussed at length in~\cite{as97}. Note also that the $O_8$ and $S_8$ \emph{unphysical} contributions would have unusual signs if they were not eliminated: this is typical of systems with ``complex'' charge multiplicities.

The presence of $C_8$ in the ${\widetilde{\cal K}}_{0'B}$ amplitude reveals that the orientifold plane in this model carries a RR charge, which is positive, as can be seen from the sign of the transverse M\"obius amplitude, while the absence of $V_8$ reveals that the orientifold plane has a vanishing tension. It is somehow a bound state of an orientifold and an anti--orientifold of type I, in such a way that the tensions cancel while the charges add up~\footnote{This deconstruction involves the $O_+$, with positive tension and charge, and an $\overline{O}_-$, with negative tension and, again, positive charge; The resulting charge is $\sqrt{2}$ times the value for the BPS objects~\cite{branecft1,dms}, consistent with the structure of the 0'B partition functions.}.
The RR tadpole condition demands the cancellation of the terms involving $C_8$ in the vacuum channel, and reads
\beq
- \ \frac{2^6}{2}  \ - \ \frac{2^{-6}}{2} \ ({\cal N}+{\overline{\cal N}})^2 \ + \ 2 \ ({\cal N}+{\overline{\cal N}}) \ = \ 0 \ ,
\eeq
so that, taking the constraint ${\cal N}= {\overline{\cal N}}$ into account, one can conclude that the gauge group is U(32), with massless vectors properly valued in the adjoint and massless fermions in the antisymmetric and its conjugate, as is manifest in eqs.~\eqref{open_0'B}. However, the actual gauge group is $SU(32)$, since the factor $U(1)$ is anomalous, as shown in~\cite{as97}.  Note that, despite the absence of supersymmetry, there are equal numbers of bosonic and fermionic modes in the open sector, up to contributions ${\cal O}\left(\frac{1}{\cal N}\right)$. This peculiar feature lies at the heart of the work of~\cite{armoni1,armoni2,armoni3,armoni4,armoni5,armoni6}, where lessons for large--N QCD were drawn from this setup. In summary, the massless spectrum of this model has the following content:
\beq 
 \mathrm{closed:} \ ( e_\mu^a,\phi, A_{\mu {\nu}}, a, A_{\mu\nu\rho\sigma}^{+}) \ , \qquad \mathrm{open:} \ ( A_\mu^i{}_j,\lambda_{L\,[ij]},\lambda_L^{[ij]}) \ , \label{0Bp_spectra}
\eeq
where the gauge field is valued in the adjoint representations and the two sets of fermions are valued in the antisymmetric and conjugate antisymmetric representations. Moreover, the notation $A_{\mu\nu\rho\sigma}^{+}$ is meant to stress that the corresponding field strength is selfdual.

Compactifications of tachyon--free ten--dimensional strings do not grant the absence of tachyons in lower dimensions. Lower--dimensional counterparts of the $U(32)$ model were identified long ago in~\cite{ang_98}, and the construction of four-dimensional heterotic strings that are tachyon-free for any value of moduli fields is possible. Some of the recent activity in this direction can be found in \cite{4dnonsusyheterotic,admavr,recent_v}.

As was the case for the SO(16)$\times$SO(16) heterotic model, this non--supersymmetric system also exerts a back--reaction on spacetime, which is now due both to the torus amplitude and to the insertion in the vacuum of branes and orientifolds, which are not BPS to begin with. The contribution to the vacuum energy from the torus amplitude can be manifested directly by making use of Jacobi's \emph{aequatio}~\eqref{aequatio}, but it is naively infinite. The infinity can be ascribed, in this case, to the tachyon mode of the unprojected closed string, whose contribution is, however, canceled by the limiting infrared portion of the direct--channel Klein--bottle amplitude. The Ramond--Ramond contributions cancel, in compliance with anomaly cancellation, but there is a residual NS-NS contribution in ${\widetilde{\cal A}}_{0'B}$ involving $V_8$ that does lead to a divergence. This divergence can be associated to a massless exchange at zero momentum, and signals the emergence of a positive ``tadpole potential'' contribution to the effective action, which takes the form
\beq
- \ \frac{T}{2\,\kappa^2} \ \int d^{10}\,x \ \sqrt{-g} \ e^{-\phi}  \label{tadpole2}
\eeq
in the string frame. Once this contribution is sorted out, the overall vacuum energy is finite. The resulting torus contribution is sub--dominant at weak coupling, and therefore in the following chapters we shall focus on the effects of the tadpole potential~\eqref{tadpole2}.

The third non--tachyonic string, Sugimoto's model~\cite{sugimoto}, is apparently a minor variant of the $SO(32)$ superstring. It also is descendant of the IIB theory based on $\Omega$, so that the Klein--bottle and annulus amplitudes are still those of eqs.~\eqref{Klein_I} and \eqref{annulus_I}.  However, the action of $\Omega$ on the open sector is different, and the M\"obius amplitude reads
\beq
{{\cal M}} \ = \   \frac{1}{2} \ {\cal N} \ \int_0^\infty \ \frac{d\tau_2}{\tau_2^2} \ \frac{ \hat{V}{}_8 \,+\,\hat{S}{}_8}{\tau_2^4 \ \hat{\eta}^8} \left(\frac{1 \,+\, i \,\tau_2}{2}\right)  \ ,
\eeq
so that the spectrum undergoes a subtle change: there are now $\frac{{\cal N}({\cal N} + 1)}{2}$
gauge bosons, while the anomaly cancellation, driven by the Ramond--Ramond contribution to the vacuum channel amplitude
\beq
{\widetilde{\cal M}} \ = \   \frac{2}{2} \ {\cal N} \ \int_0^\infty \ {d\ell}\ \frac{ \hat{V}{}_8 \,+\,\hat{S}{}_8}{\hat{\eta}^8} \left(i \ell + \frac{1}{2}\right)  \ ,
\eeq
continues to require the presence of $\frac{{\cal N}({\cal N} - 1)}{2}$ Fermi fields, with ${\cal N}=32$. The gauge group is thus USp(32), but the massless fermions remain in the antisymmetric representation, which is \emph{reducible} in USp(32) and contains a \emph{singlet}. The singlet is most important in the overall picture, since it is the goldstino. Indeed, while the projected closed spectrum is not affected, and still comprises the whole $N=(1,0)$ supergravity described in Section~\ref{sec:(1,0)sugra}, \emph{supersymmetry is broken, and it is actually non--linearly realized in the open sector}~\cite{dmnonlinear1,dmnonlinear2,dmnonlinear3}, \emph{without an order parameter capable of recovering it}. The total vacuum amplitude does not vanish anymore, consistently with the back-reaction that is expected when supersymmetry is broken. The occurrence of this type of phenomenon is startling, precisely because no order parameter is present in ten dimensions to recover a supersymmetric vacuum.

The physical phenomenon that entails all this, which is usually referred to as ``brane supersymmetry breaking'', manifested itself earlier, if indirectly, in the failed attempts summarized in~\cite{erice}. Surprisingly, some tadpole conditions, more complicated counterparts of eq.~\eqref{tadpole} in six and four dimensions, appeared to yield inconsistent results in certain supersymmetric models. The whole story remained a puzzle until its origin was clarified in~\cite{bsb1,bsb2,bsb3,bsb4} in a class of six--dimensional orientifolds. However, in all these cases the novel options resulted from different Klein--bottle projection, which implied the need to combine (anti)branes of different types, thus breaking supersymmetry, while the model of~\cite{sugimoto} is far simpler, since it involves a single type of brane and orientifold.

In the space--time picture, these novelties reflect the replacement of the standard $O_-$ orientifold by an $O_+$, with positive tension and charge. The positive charge requires anti $D$--branes for its cancellation, with a consequent breaking of supersymmetry, and a net tension is thus present in the vacuum. As a result, the M\"obius amplitude appears ultraviolet divergent, but in ${\widetilde{\cal M}}$ the effect reveals its infrared origin, which can be ascribed to a massless NS-NS exchange at zero momentum, as in the 0'B model, and leads again to a ``tadpole potential'', as in eq.~\eqref{tadpole}. Once this contribution is sorted out, the remainder is again finite, and no torus contribution is present in this case, but the vacuum is not Minkowski space anymore.
Rather, one ought to work around another vacuum that solves the equations of motion~\cite{vacuum_redefinitions1,vacuum_redefinitions2,vacuum_redefinitions3,vacuum_redefinitions4,vacuum_redefinitions5}, but as of today this can be done efficiently only at the level of the low--energy field theory. The key question concerns the actual vacuum configurations, but one should also address a related problem, characterizing how the (charged or uncharged~\cite{sen1,sen2,sen3,sen4,sen5,sen6,sen7}) branes available in these systems~\cite{branecft1,dms} adjust themselves to deformed backgrounds. This problem was recently dealt with in~\cite{mrs24_1,mrs24_2}, thus complementing the old results obtained by conformal field theory techniques while ignoring the tadpole in~\cite{dms}. The uncharged branes remain exact solutions around the key vacuum of the system, which was first constructed by two of us in~\cite{dm_vacuum}, while the deformed charged branes were studied so far only at large distances, where the equations linearize.

The link between the emergence of tachyons and the asymptotics of the spectrum in models of oriented closed strings was first considered in~\cite{kut_sei}. The conclusion was that, in the absence of supersymmetry, differences can exist at individual mass levels between the numbers of Bose and Fermi modes, but nonetheless they must compensate for one another and disappear in the full spectrum, lest tachyonic modes emerge. The proposal of “misaligned supersymmetry”~\cite{mis_1,mis_2} refined the picture, stressing that the differences between the cumulative numbers of Bose and Fermi excitations must oscillate, in stable systems, for increasing levels, in a way determined by a reduced central charge $c_{eff}<c$, and conjectured that $c_{eff}=0$. This idea was further examined in~\cite{carlo1}, and led to intriguing links with Riemann’s zeta function. The arguments in~\cite{carlo2} aimed at proving that $c_{eff}=0$ were recently completed in~\cite{carlo3}. Extensions to cases including open strings were recently considered in~\cite{carlo4}. In this case the issue is more subtle, since rather than being individually modular invariant like the torus amplitude, the additional contribution to the vacuum energy form related pairs, as we have seen. There is another interesting issue, also related to the genus--one vacuum energy for oriented closed strings, that we are leaving aside. It is possible to build models of oriented closed strings with broken supersymmetry and a vanishing genus--one vacuum energy~\cite{harvey, kks,carlo5,carlo6}, but there seems to be no consensus on similar results for genus larger than one.

Note that, in both the $0'B$ and USp(32) models, the difference between the numbers of Bose and Fermi modes present in the open spectrum oscillates in the massive levels, in a way that resonates with Dienes's misaligned supersymmetry proposal~\cite{dienes}. For recent developments along these lines, see~\cite{carlogiorgio1,carlogiorgio2,carlogiorgio3}.
\begin{figure}[ht]
\centering
\includegraphics[width=80mm]{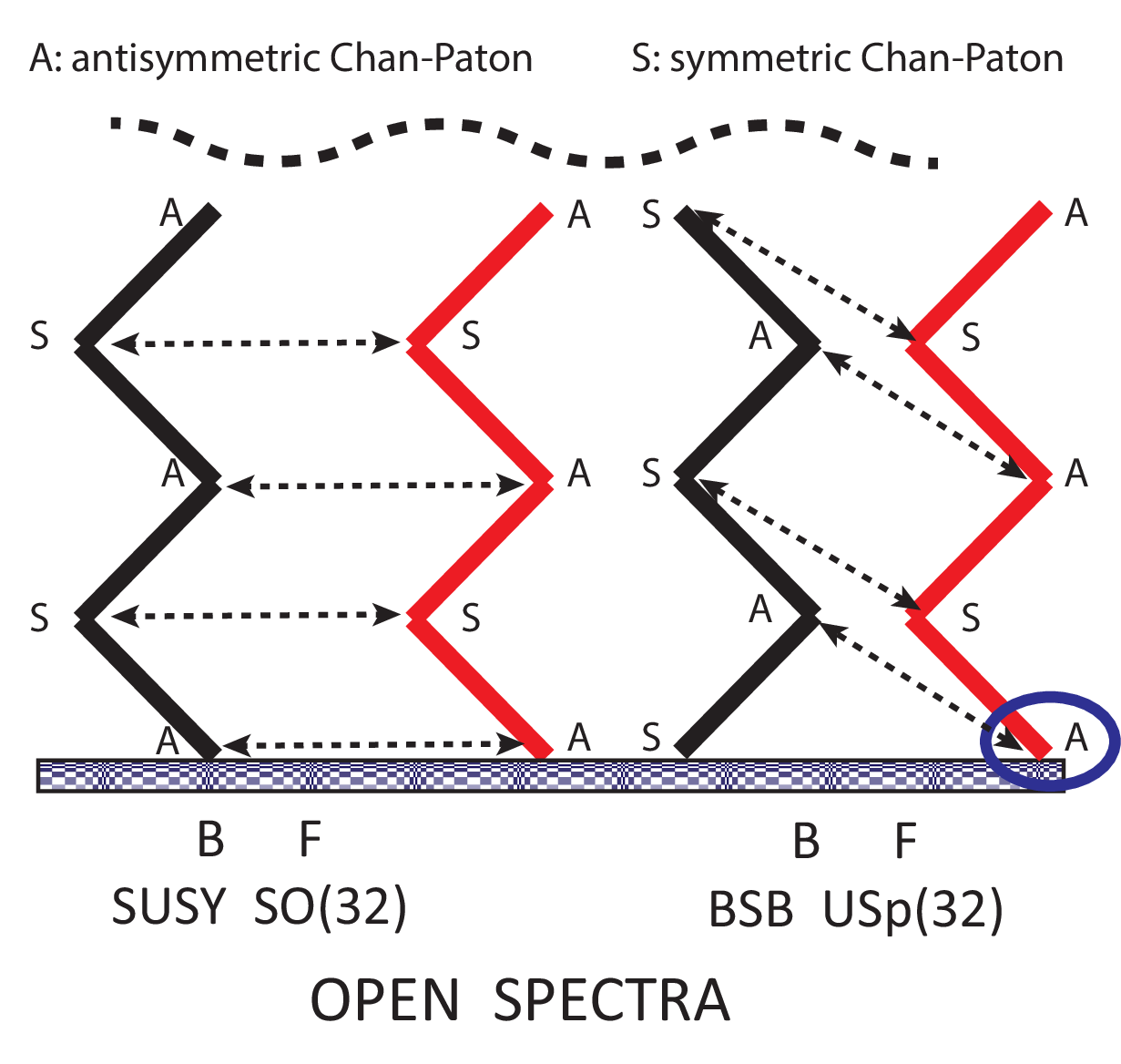}
\caption{\small A cartoon of ``brane supersymmetry breaking'', which induces a shift by one level of Bose Chan--Paton representations with respect to the supersymmetric type--$I$ open spectrum.}
\label{fig:bsb_spectrum}
\end{figure}

Supersymmetry is broken in this $USp(32)$ model due to the simultaneous presence, in the vacuum, of objects, O9${}_+$ and $\overline{D9}$, which are individually BPS but not mutually. As suggested by the cartoon in fig.~\ref{fig:bsb_spectrum}, in this model supersymmetry breaking occurs at the string scale, which makes one wonder how well the effective low--energy supergravity can possibly capture the phenomenon. Notice that, in comparison with the $SO(32)$ superstring, the whole ``belt'' of representations for the bosonic states has merely moved up by one step.
For one matter, one clearly misses its relatively ``soft'' nature, which becomes apparent if one considers the whole spectrum as sketched in fig.~\ref{fig:bsb_spectrum}. 

There is a difference between the effects of supersymmetry breaking in the torus amplitude and the leading effects in the open and unoriented sectors, which is worth stressing. The former is a genuine quantum effect, while the latter reflect the tension present in the vacuum, and are thus, somehow, classical effects. Still, due to the mutual attraction between (anti)branes and orientifolds, the resulting spacetimes would seem prone to collapsing.

When the vacuum is modified, there are no tools at present to deal with the phenomenon within full-fledged String Theory. All one can do, with some degree of generality, is to explore matters relying on the low--energy field theory. Extracting information, in this fashion, about String Theory proper is not an easy task, since the two are connected by a double expansion, in powers of $g_s$ and in powers of the curvature in units of the string scale $\frac{1}{\sqrt{\alpha'}}$. When both couplings are small, the low--energy effective field theory ought to yield reliable indications, but one can at best approach this ideal setting. As we shall see in Sections~\ref{sec:SUSY_breaking_com} and \ref{sec:SUSY_breaking_cosmo}, a closer look at these models is nevertheless quite rewarding. These systems seem to perform beyond expectations, with non--trivial stability properties in the static case and surprising lessons for Cosmology. The analysis based on the low--energy field theory thus yields novel insights and even glimpses of a wider picture. We shall return to all this in the following sections.

\subsection{\sc World--sheet Consistency rules for Orientifold Spectra} \label{sec:building_rules}
The key example of the type--I superstring contains, in the simplest possible setting, the consistency conditions that underlie more complicated orientifold spectra. In compactifications, more options are available for the Klein--bottle projection and the open spectrum changes accordingly, as we already saw for the 0'B theory~\cite{as95,as97} in ten dimensions. Moreover, even for the open sector there are generally multiple choices, as we already saw for Sugimoto's model~\cite{sugimoto}. For the reader's convenience, we can now summarize the main steps of the construction. This should serve as a guidance for the following sections.

\begin{itemize}

\item[1. ] The starting point is a consistent closed--string spectrum that is left--right symmetric, as the IIB, 0A and 0B theories in ten dimensions, or more generally is invariant if $\Omega$ is combined with other transformations into a more general involution $\Omega'$. This occurs, for example, in the IIA theory in nine dimensions, where $\Omega'= \Omega \, P$, with $P$ the internal parity.

\item[2.] The Klein--bottle projection completes the (anti-)symmetrization of the closed spectrum. The fusion rules of the spacetime characters $(O_8,V_8,-S_8,-C_8)$ may allow more than one choice, as was illustrated for the 0B theory in eqs.~\eqref{K0bprime} and \eqref{K0b}. In the following sections, we shall see that the same can be true for the internal characters that will play a role in compactifications. In the vacuum channel, these restrictions translate into the positivity of the different contributions when they are expressed in the proper basis compatible with spin statistics. For the IIB theory the Klein--bottle amplitude is unique, although the USp(32) model~\cite{sugimoto} is based on the different orientifold projection $\Omega' = \Omega (-1)^G$, where $G$ denotes the spacetime fermion number, which introduces an O${}_+$.

\item[3. ]The vacuum--channel annulus amplitude $\widetilde{\cal A}$ contains, in general, a combination of spacetime and internal characters, with arbitrary (non--negative) coefficients, while $\widetilde{\cal M}$ involves the sectors common to $\widetilde{\cal K}$ and $\widetilde{\cal A}$. The corresponding coefficients are, in general, twice the geometric means of those present in the two cases, although we shall see a refinement of this rule in Section~\ref{sec:orientifold_twisted}. D-brane and orientifold tensions and charges can be read from the vacuum--channel amplitudes, up to a common overall factor. 

Starting from the IIB string, one could have considered the more general transverse--channel annulus amplitude
\beq
{\widetilde {\cal A}} \ = \ \frac{2^{-5}}{2} \, \int_0^\infty \ {d\ell} \ \frac{\alpha^2\ V_8\left(i \ell\right) \,-\,\beta^2\ S_8\left(i \ell\right)}{\eta^8\left(i \ell\right)} \ , \label{Atildeb}
\eeq
where $\alpha$ and $\beta$ are independent coefficients for the two sectors associated to $V_8$ and $-S_8$, which enter symmetrically the IIB partition function, and are the only ones allowed in the transverse channel. The relative sign of the two contributions reflects the attractive nature of the force mediated by NS-NS fields and the repulsive nature of that mediated by RR fields. The overall normalization, while inessential since $\alpha$ and $\beta$ are arbitrary at this stage, although they are proportional to the total tension and charge, has the virtue of removing powers of two from 
\beq
{\widetilde {\cal M}} \ = \  - \ 2\ \frac{1}{2} \ \int_0^\infty \ {d\ell} \ \frac{\alpha\ \widehat{V}_8\left(i \ell\,+\,\frac{1}{2}\right) \,-\,\beta\ \widehat{S}_8\left(i \ell\,+\,\frac{1}{2}\right)}{\widehat{\eta}^8\left(i \ell\,+\,\frac{1}{2}\right)}  \ , \label{Mtildeb}
\eeq
whose coefficients are, for both the $\hat{V}_8$ and $- \,\hat{S}_8$ sectors, geometric means of those present in $\widetilde{\cal K}$ of eq.~\eqref{Ktilde} and in $\widetilde{\cal A}$, up to an overall factor of two, as demanded by unitarity. With this overall sign, when the two coefficients coincide, one recovers the type--I SO(32) superstring.

Turning these amplitudes to the direct channel gives
\bea
{\cal A} &=& \frac{1}{2}  \ \int_0^\infty \ \frac{d\tau_2}{\tau_2^2} \frac{\frac{\alpha^2+\beta^2}{2}\left[V_8\left(\frac{i \tau_2}{2}\right)-S_8\left(\frac{i \tau_2}{2}\right)\right] \ + \ \frac{\alpha^2-\beta^2}{2}\left[O_8\left(\frac{i \tau_2}{2}\right)-C_8\left(\frac{i \tau_2}{2}\right)\right]}{\tau_2^4 \, \eta\left(\frac{i \tau_2}{2}\right)^8} \ , \nonumber \\
{\cal M} &=& \frac{1}{2}  \ \int_0^\infty \ \frac{d\tau_2}{\tau_2^2} \frac{\alpha\,\hat{V}{}_8\left(\frac{1 +i \tau_2}{2}\right)\ - \ \beta \ \hat{S}{}_8\left(\frac{1 +i \tau_2}{2}\right)}{\tau_2^4 \, \hat{\eta}\left(\frac{1 +i \tau_2}{2}\right)^8} \ ,
\eea
and the issue is how to link the coefficients to Chan-Paton factors.
 \begin{figure}[ht]
\begin{center}
    \includegraphics[width=1.5in]{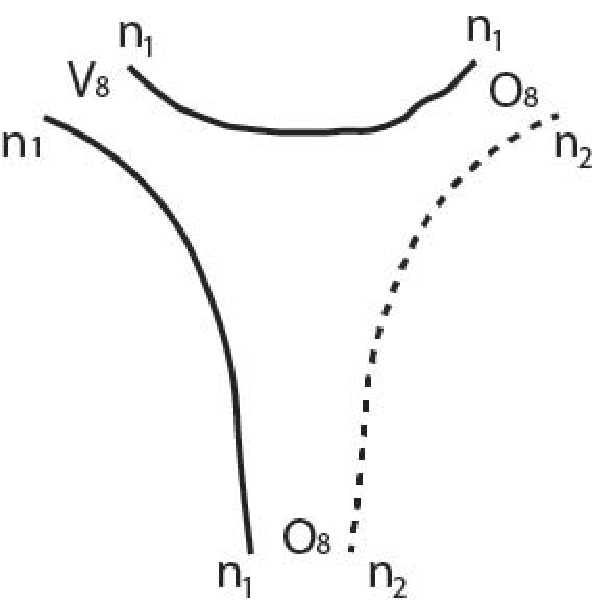}
\end{center}
\caption{\small The interaction between two open string states belonging to the $O_8$ sector, with Chan--Paton factors of multiplicities $n_1$ and $n_2$ at the ends and an open string belonging to the $V_8$ sector with a Chan-Paton of multiplicity $n_1$ at both ends.}
\label{fig:fusion_open}
\end{figure}

\item[4.]  In order to understand the final form of the amplitudes, let us note that, for example, the interaction between two open--string states belonging to the $O_8$ or $C_8$ sectors yields states in the $V_8$ or $S_8$ sectors, owing to the fusion rules~\eqref{fusion_rules} (see fig.~\ref{fig:fusion_open}). One is thus led to the parametrization
\beq
\frac{\alpha^2+\beta^2}{2} \ = \ n_1{}^2 \ + \ n_2{}^2 \qquad \mathrm{and} \qquad \frac{\alpha^2-\beta^2}{2} \ = \ 2\ n_1\, n_2  \ , \label{alphabeta}
\eeq
where $n_1$ and $n_2$ are two integers that characterize the number of rows and columns of the Chan--Paton matrices for the different open--string states. The preceding equations are solved by
\beq
\alpha \ = \ n_1\ + \ n_2 \ , \qquad \beta \ = \ n_1\ -\ n_2 \ ,
\eeq
and consequently the direct--channel amplitudes read
\bea
{\cal A} &=&  \int_0^\infty \ \frac{d\tau_2}{\tau_2^2} \frac{\frac{n_1^2+n_2^2}{2}\left[V_8\left(\frac{i \tau_2}{2}\right)-S_8\left(\frac{i \tau_2}{2}\right)\right] \ + \ n_1 \, n_2 \left[O_8\left(\frac{i \tau_2}{2}\right)-C_8\left(\frac{i \tau_2}{2}\right)\right]}{\tau_2^4 \,\eta\left(\frac{i \tau_2}{2}\right)^8} \ , \nonumber \\
{\cal M} &=& - \ \frac{1}{2}  \ \int_0^\infty \ \frac{d\tau_2}{\tau_2^2} \frac{\left(n_1+n_2\right)\,\hat{V}{}_8\left(\frac{1 +i \tau_2}{2}\right)\ - \ \left(n_1-n_2\right) \ \hat{S}{}_8\left(\frac{1 +i \tau_2}{2}\right)}{\tau_2^4 \, \hat{\eta}\left(\frac{1 +i \tau_2}{2}\right)^8} \ , \label{AMtachyond}
\eea
and consistency with spin-statistics for the open strings demands that 
\beq
n_1 \, n_2 \ \geq \ 0 \ , \label{n1n2}
\eeq
since otherwise a violation would occur in their $O_8$ and $C_8$ sectors. Note that supersymmetry is broken if $n_2 \neq 0$: this can be seen from the M\"obius amplitude, which associates different representations to bosons and fermions of this type. Moreover, tachyonic modes are present whenever $n_1\,n_2 \neq 0$.
 \begin{figure}[ht]
\begin{center}
    \includegraphics[width=2.3in]{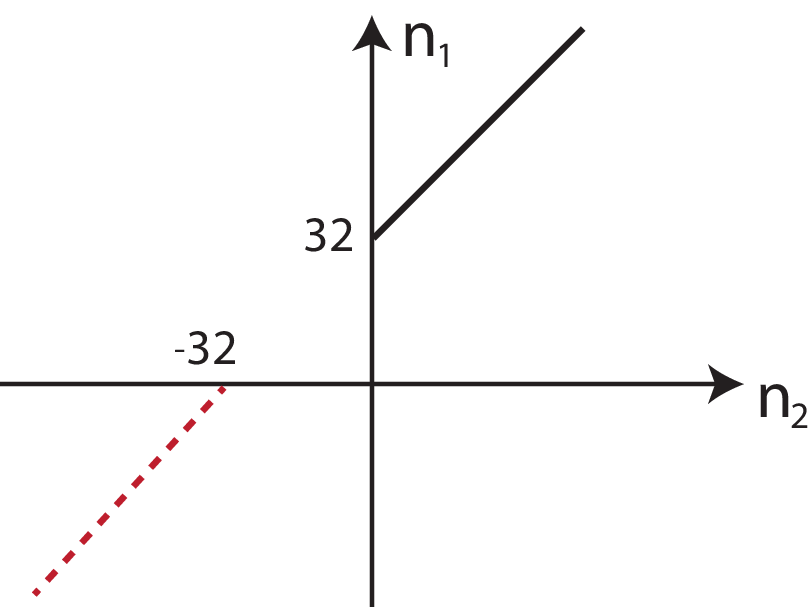}
\end{center}
\caption{\small The two disjoint lines in the $(n_1,n_2)$ plane that terminate at two separate stable points, the $SO(32)$ superstring (black, solid) and the $USp(32)$ model (red, dashed).}
\label{fig:disjoint_lines}
\end{figure}
The $V_8$ and $S_8$ sectors are associated to open strings carrying charges of identical types at their ends, with multiplicities $n_1$ and $n_2$ (or, if you will, to open strings that have square $\left|n_1\right| \times \left|n_1\right|$ and $\left|n_2\right| \times \left|n_2\right|$ Chan--Paton matrices). These matrices are antisymmetric if the $n_i$ are positive and symmetric if they are negative, given our convention for the overall sign in ${\cal M}$. Taking eq.~\eqref{n1n2} into account, the gauge group is in general a semi--simple combination $SO({n_1}) \times SO({n_2})$ if the $n_i>0$ and $USp({\left|n_1\right|}) \times USp({\left|n_2\right|})$ if the $n_i<0$ (and even). The massless modes of the $V_8$ sector are valued in the adjoint, while the massless modes of the $S_8$ sector are valued in the adjoint of the first gauge group and, if $n_2 \neq 0$, in the symmetric (antisymmetric) representation of the second. Note that, whenever $n_2 \neq 0$, this second representation is \emph{reducible} and contains a singlet, which can be identified with the goldstino that accompanies the breaking of supersymmetry.  In addition, the $O_8$ and $C_8$ sectors, if present, are associated to open strings with charges of different types at their ends (or, if you will, they have generally rectangular $\left|n_1\right| \times \left|n_2\right|$ Chan--Paton matrices, so that they are valued in the bi--fundamental of the semi--simple group $G_{n_1} \times G_{n_2}$). The shifted argument of the M\"obius amplitude then implies that symmetric and antisymmetric representations alternate for massive states of the $V_8$ and $S_8$ sectors. 

\item[5. ] When they are expressed in terms of $n_1$ and $n_2$, the transverse--channel amplitudes \eqref{Atildeb} and \eqref{Mtildeb} read 
\bea
{\widetilde {\cal A}} &=& \frac{2^{-5}}{2} \, \int_0^\infty \ {d\ell} \ \frac{\left(n_1+n_2\right)^2\ V_8\left(i \ell\right) \,-\,\left(n_1-n_2\right)^2\ S_8\left(i \ell\right)}{\eta^8\left(i \ell\right)} \ , \nonumber \\
{\widetilde {\cal M}} &=&  - \ 2\ \frac{1}{2} \ \int_0^\infty \ {d\ell} \ \frac{\left(n_1+n_2\right)\ \widehat{V}_8\left(i \ell\,+\,\frac{1}{2}\right) \,-\,\left(n_1-n_2\right) \widehat{S}_8\left(i \ell\,+\,\frac{1}{2}\right)}{\widehat{\eta}^8\left(i \ell\,+\,\frac{1}{2}\right)}  \ .  \label{AMtachyont}
\eea
Combining them with the Klein--bottle amplitude of eq.~\eqref{Ktilde}, one can identify the RR tadpole condition
\beq
n_1 \ - \ n_2 \ = \ 32\ , \label{tadpole10dtach}
\eeq
so that only the difference between $n_1$ and $n_2$ is fixed, although the signs of $n_1$ and $n_2$ must agree, as we saw in eq~\eqref{n1n2}. Therefore, if $n_1$ and $n_2$ are positive, the
gauge group is finally $SO(32+n_2) \times SO(n_2)$, and the signs of the two sectors in $\widetilde{\cal M}$ reveal that the vacuum contains an $O9_-$ orientifold, with negative tension and charge equal to -32 in our units, $32+n_2$ branes and $n_2$ antibranes. On the other hand, if $n_1$ and $n_2$ are both negative, the gauge group is $USp(2m) \times USp(32+2m)$, and the vacuum contains an $O9_+$ orientifold, with positive tension and charge equal to 32 in our units, $32+2m$ antibranes and $2m$ branes.
The allowed range in the $(n_1,n_2)$ plane thus comprises two disconnected half-lines, which terminate at the $SO(32)$ superstring and at Sugimoto's model, the only two stable options, and depart from them due to the addition of brane--antibrane pairs (that are expected to annihilate), compatibly with the tadpole condition.
\end{itemize}

These points are extensively reviewed in~\cite{orientifolds_rev1,orientifolds_rev2,orientifolds_rev3,orientifolds_rev4}, where they are also illustrated in detail in several examples. In the following, we shall focus mainly on lower--dimensional constructions related to supersymmetry breaking, relying on the results reviewed in Appendices~\ref{app:bosonic_orientifold} and~\ref{app:so2n}.

\section{\sc Higher--Dimensional Toroidal Compactifications} \label{sec:highertori}

We can now address higher--dimensional toroidal compactifications. These generalize what we saw for circle compactification in
important ways, and had the virtue of first revealing the existence of a huge space of vacua for String Theory. This became manifest when Narain~\cite{narain1,narain2,T-duality} noted that 
a $d$--dimensional compactification of the heterotic string gives rise to vacua corresponding to points in $O(16+d,d;R)/\left(O(d;R) \times O(16+d;R)\right)$ due to the undetermined vacuum values of the internal metric $G_{ab}$, the internal NS-NS field $B_{ab}$ and the 16 independent Wilson lines in the Cartan subalgebra of $SO(32)$ or $E_8 \times E_8$. This ``moduli space'' needs an important amendment, since $T$--duality is properly extended, and becomes in this case a discrete group $O(16+d,d;Z)$ of identifications in the moduli space, but nonetheless there is a huge space of options.
These vacua leave tangible signs in the low--energy supergravity: $N=4$ supergravity coupled to $N=4$ Yang-Mills multiplets with a rank--16 gauge group possesses a scalar manifold that is precisely as above. The different vacua correspond to different points in the scalar manifold of the low--energy field theory, although the discrete identifications are purely a string effect.

The extension of Narain's construction to $\Omega$--projected orientifold vacua~\cite{orientifolds7} revealed a novelty, a residual role for quantized values for $B_{ab}$, although the field itself cannot give rise to continuous deformations since it is projected out of the spectrum. These quantized values reduce the rank of the Chan--Paton gauge group and, in the $T$-dual formulation, the effect can be ascribed to a second type of BPS orientifold~\cite{wittenOplus}, with positive tension and charge. We already encountered a manifestation of this type of orientifold in our discussion of Sugimoto's $USp(32)$ model~\cite{sugimoto} but, as we shall see, for $D<10$ this opens up a wide number of options, including settings where supersymmetry is inevitably nonlinearly realized.

\subsection{\sc Closed Strings and Narain's Construction} \label{sec:narain}
In generalizing circle compactification to d-dimensional tori, let us begin by considering the closed bosonic string. Let us also denote compact coordinates by $X^a$, with the identifications~\footnote{For more details see, for instance,~\cite{T-duality} or the books in~\cite{stringtheory}.}
\beq 
X^a\ = \ X^a\ + \ 2\pi\ ,\qquad  a=1,\ldots, d \ . \label{Xa_periodic}
\eeq 
We shall mainly focus on the bosonic string, whose different toroidal backgrounds depend on a constant internal metric $G_{ab}$ and a constant antisymmetric two-form $B_{ab}$, occasionally mentioning the novelties that are met when considering closed superstrings. For heterotic strings, one can also consider Wilson lines, and we shall further elaborate on this case. 

The string world-sheet action is the sum of the flat background term~\eqref{bosonicaction} for $D-d$ non--compact coordinates and the torus contribution, as 
\beq 
S\ =\ S_{flat}\ + \ S_{Torus}\ ,
\eeq 
with
\beq
S_{Torus} \left[g_{\alpha\beta},X^\a\right] \ =\  - \ \frac{1}{4 \pi \alpha^\prime} \int d^2 \xi\,  
\partial_\alpha X^a \ \partial_\beta X^b \ \left(\sqrt{-g}\,
g^{\alpha\beta}G_{ab}\ + \ \epsilon^{\alpha\beta}B_{ab}\right)  \ , \label{bosonicactionT}
\eeq
where, as in Section~\ref{sec:critical_strings}, $g_{\alpha\beta}$ denotes the world--sheet metric and $\epsilon^{01}=-1$.
The zero--mode conjugate momenta deduced from the action are thus, in the conformal gauge,
\beq 
p_a\ =\ \frac{1}{2\,\alpha'}\left[\left(\dot{x}\right)^b \,G_{ab}\ +\ \left(x'\right)^b \, B_{ab}\right]\ , \label{momentum}
\eeq
where ``dot'' and ``prime'' denote, as usual, $\tau$ and $\sigma$ derivatives of $x^a$, the zero mode part of $X^a$.
Momenta are quantized, in view of~\eqref{Xa_periodic}, in integer values denoted by $m_a$, and there are also $d$ winding numbers, so that 
\beq 
x^{a}(\tau,\sigma)\ =\ x^a\ +\ 2\left(\alpha' \,m_b\ - \ B_{bc}\,n^c\right)G^{ab}\tau \ +\ 2 \,n^a\,\sigma \ .\label{X0_GB}
\eeq 
In analogy with eqs.~\eqref{XLR} and \eqref{onedimmom}, one can define the left and right momenta
\beq 
p_{LR}^a \ =\ \frac{1}{\alpha'}\ \left[\left(\alpha' m_b\ -\ B_{bc}\,n^c\right)G^{ab} \ \pm \  n^a\right]\ ,
\eeq 
or equivalently
\beq 
p_{LR, a} \ =\ m_a - \frac{1}{\alpha'} \left(B_{ab} \ \mp\  G_{ab} \right) n^b \ , \label{plrd}
\eeq 
so that  in matrix notation the string coordinates are
\bea
X\  & =&\  x \ + \ 2 \alpha' G^{-1}(\alpha'\,m\,-\,Bn)\tau  \ + \  2 \, n \,\sigma 
\nonumber \\ &+& \  
\frac{i\sqrt{2\alpha^\prime}}{2}
\ {\cal E} \sum_{n \not= 0} \left( \frac{\alpha_n}{n} \ e^{-2 i n (\tau -
\sigma)} \ + 
\frac{\tilde{\alpha}_n}{n}\  e^{-2 i n (\tau + \sigma)} \right) \label{Bclosedmodes2}
\, .
\eea
Here ${\cal E}$, the moving basis, satisfies
\beq
{\cal E}\, {\cal E}^T=G^{-1} \ ,
\eeq
and 
  $\alpha_n$ and $\tilde\alpha_n$ obey  the commutation relations
  \beq 
  [\alpha_n^A,\alpha_m^B]\ = \ n\delta^{AB}\delta_{n+m,0}\ ,\qquad [\tilde \alpha_n^A,\tilde \alpha_m^B]\ = \ n\delta^{AB}\delta_{n+m,0} \ ,
  \eeq
  which are independent of $B$ and $G$.
The mass--shell condition in the residual $26-d$ dimensions and the level matching condition can be obtained as before from
\bea
M^2 & =& p_{R a}\, G^{ab}\, p_{R b} \ +\ \frac{4}{\alpha'} \left( N \ - \  1
\right) \nonumber \\ &=& p_{L a}\,G^{ab}\,p_{L b}\ + \ \frac{4}{\alpha'} \left( \bar{N} \ - \  1
\right) \ . \label{closedmassT}
\eea
These results generalize eqs.~\ref{closedmass_torus} to higher--dimensional tori.
In particular, the level matching condition now becomes
\beq 
\frac{\alpha'}{4} \left( p_{L a}\,G^{ab}\,p_{L b} \ -\ p_{R a}\, G^{ab}\, p_{R b} \right) \ = \ {N} \ - \ \bar{N} \ . \label{T-inv}
\eeq
In terms of the moving basis ${\cal E}_a{}^A$ associated to the metric $G$, the previous relation can be recast in the form
\beq 
\Pi^T\ \eta\  \Pi\ \in \ 2\, \mathbb{Z} \ , 
\eeq
where
\beq 
\Pi \ \equiv \ \sqrt{\frac{\alpha'}{2}} \left( p_L^a\,{\cal E}_a{}^A\,,\,p_R^a\,{\cal E}_a{}^A \right)  
\eeq
is a vector with $2d$ components and $\eta$ is a Minkowski--like metric with $(d,d)$ signature. 
The $\Pi$ vectors span a lattice $\Gamma$, usually called Narain lattice, which depends on the background metric $G$ and the background two-form $B$. The toroidal partition function
\beq
{\cal T} \ = \ \frac{\sum_{\{m_a,n^b\} \in \mathbb{Z}^{2d}}\ q^{\frac{\alpha'}{4}\,p_{R a}\, G^{ab}\, p_{R b}}\ \bar{q}^{\frac{\alpha'}{4}\,p_{L a}\, G^{ab}\, p_{L b}}}{\eta^d(\tau)\, \eta^d(\bar{\tau})}  \ \equiv \  \frac{\Lambda_{\mathbf{m},\mathbf{n}}}{\eta^d(\tau)\, \eta^d(\bar{\tau})} \label{Tnarain}
\eeq
is a building block of closed--string amplitudes, and its correspondence with its $R \to \infty$ limit,
\beq
\frac{1}{\tau_2^\frac{d}{2}\ \eta^d(\tau)\, \eta^d(\bar{\tau}) } \ ,
\eeq
generalizes what we discussed in eq.~\eqref{TRinf}.

It
is invariant under $\tau \to \tau+1$, due to the condition~\eqref{T-inv}, and moreover the transformation $\tau \to - \,\frac{1}{\tau}$ gives rise, after Poisson summations in $m$ and $n$, to a similar expression, but where the lattice is replaced by its dual. Therefore, one can conclude that the Narain lattice is \emph{even} and \emph{self--dual}, and we shall make this property manifest shortly.
If $\Gamma_0$ is such a lattice, then $\Lambda\Gamma_0$, where $\Lambda$ is in $O(d,d,R)$, is also an even selfdual lattice, and all Narain lattices can be obtained in this way. The elements that map a lattice to itself are denoted by $O(d,d,Z)$, and represent generalized T-duality groups. $d$ dimensional rotations do not affect the background, so the moduli space is $O(d,d,R)/(O(d,R)\times O(d,R)) \backslash O(d,d,Z)$, a coset space up to the discrete identifications in $O(d,d,Z)$.

Note that the contributions to the mass formula and to the level--matching condition for the zero modes read, in matrix notation,
\bea
\frac{\alpha'}{4}\left( p_{L a}\, G^{ab}\, p_{L b} \, + \,  p_{R a}\, G^{ab}\, p_{R b}\right) &=& \frac{\alpha'}{2} \, m^T G^{-1} m  - m^T G^{-1} B n  + \frac{1}{2\,\alpha'}\,n^T \left( G- B G^{-1} B\right)n \nonumber \\
&=& \frac{\alpha'}{2} \left( m \ - \ \frac{B}{\alpha'}\,n\right)^T G^{-1} \left( m \ - \ \frac{B}{\alpha'}\,n\right) \ + \ \frac{1}{2\alpha'} \, n^T G \,n 
 \ , \nonumber \\
\frac{\alpha'}{4}\left( p_{L a}\, G^{ab}\, p_{L b} \, - \,  p_{R a}\, G^{ab}\, p_{R b}\right) &=& m^T n \label{numerators}
 \ .
\eea
After a Poisson summation in the $m_a$ the partition function takes the form
\beq
{\cal T} \ = \ \sqrt{\det\left(\frac{G}{\alpha'}\right)}\ \frac{\sum_{\{k^a,n^b\} \in \mathbb{Z}^{2d}}\ e^{-\,\frac{\pi}{\alpha'\,\tau_2}\left(k^a - n^a\,\bar{\tau} \right)G_{ab}\left(k^b - n^b\,{\tau} \right)} \ e^{\,-\,\frac{2\pi i}{\alpha'}\, n^a B_{ab} k^b }}{\tau_2^\frac{d}{2}\, \eta^d(\tau)\, \eta^d(\bar{\tau})} \ , \label{narain1}
\eeq
which generalizes the circle expression in eq.~\eqref{lag_lattice} and is manifestly invariant under $\tau \to - \, \frac{1}{\tau}$.  Note that the mass spectrum in eq.~\eqref{numerators}, and actually this whole expression, are invariant under $B_{ab} \to B_{ab} \ + \ \alpha'$, so that $B_{ab}$ is a set of angular variables. 
Alternatively, taking the antisymmetry of $B$ into account, this expression can be cast in the form
\beq
{\cal T} \ = \ \sqrt{\det\left(\frac{G}{\alpha'}\right)}\ \frac{\sum_{\{k^a,n^b\} \in \mathbb{Z}^{2d}}\ e^{-\,\frac{\pi}{\tau_2}\left(k^a - n^a\,\bar{\tau} \right)E_{ab}\left(k^b - n^b\,{\tau} \right)} }{\tau_2^\frac{d}{2}\, \eta^d(\tau)\, \eta^d(\bar{\tau})} \ , \label{narain2}
\eeq
where
\beq
\alpha'E_{ab} \ = \ G_{ab} \ + \ B_{ab} \ . \label{EGB}
\eeq
The inverse of this matrix is
\beq
\alpha'\, E^{-1} \ = \ G' \ + \ B'  \ ,
\eeq
with
\beq
G'\ = (\alpha')^2\ \left( G \ - \ B\,G^{-1}\,B  \right)^{-1} , \qquad B' \, \left(G'\right)^{-1}\ = \left[ B \ G^{-1}\right]^T \ = \ - G^{-1}\,B \ , \label{GG'BB'}
\eeq
and eqs.~\eqref{numerators} are manifestly invariant under this transformation, up to the interchange of $m$ and $n$. Note also that the preceding relations imply that
\beq
G \,B'\ + \  B \, G'\ = \ 0 \ , \qquad G \, G'\ + \ B\,B'\ = \ \alpha'^2\ , \label{TdualB}
\eeq
which will prove convenient in the following. 
These inversions, together with the integer shifts of $B_{ab}$ are part of the $O(d,d,Z)$ T-duality group of these toroidal compactification. The $SL(2,Z)$ group of the torus that we have already encountered in the $d=2$ case is also part of that group.

%Together with $SL(d,Z)$ transformations, the shifts of $B_{ab}$, which are manifest symmetries of %eq.~\eqref{narain1}, and the preceding inversions are part of the $O(d,d;Z)$ duality group for these %toroidal compactifications.

\subsection{\sc Open Strings in a \texorpdfstring{$B_{ab}$} \ \ Background} \label{sec:open_B}

We can now turn to the effect of a generic constant $B_{ab}$ on open--string spectra. For simplicity, we shall continue to focus on the bosonic string, but these considerations
can be extended to the superstring in ten--dimensional noncompact space, or toroidal compactifications thereof, provided the open strings end on lower--dimensional branes, in the absence of an $\Omega$ projection.

To begin with, the coupling between the $B_{ab}$ background and open strings is very similar to the coupling to a constant magnetic field, which we shall explore in the following sections, since letting
\beq
B \ = \ d\,A \ ,
\eeq
with
\beq
A \ = \ \frac{1}{2}\ B_{ab}\,X^a\,d X^b \ ,
\eeq
yields
\beq
- \ \frac{1}{2\pi\alpha'} \ \int_{\Sigma} \ B \ = \ - \ \frac{1}{2\pi\alpha'} \ \left( \int_{\sigma=\pi} A \ - \ \int_{\sigma=0} A \right) \ . \label{BAA}
\eeq
This is indeed reminiscent of the coupling to a constant magnetic field, but with opposite charges at the endpoints of the open string.

The momentum zero mode is still determined by eq.~\eqref{momentum}, but the novelty is the boundary condition
\beq
\left(G_{ab}\,X^{'b} \ + \ B_{ab}\,\dot{X}^b\right) \delta\,X^a \ = \ 0 \ . \label{neumannB}
\eeq
The two options are therefore Dirichlet boundary conditions or modified Neumann boundary conditions. 

In the former case the open strings do not feel the presence of $B_{ab}$, their zero modes have no $\tau$ dependence, so that
\beq
X^a \ = \ x^a \ + \ 2\,n^a \, \sigma \ + \
\sqrt{2\alpha^\prime}
\ \sum_{n \not= 0} \frac{\alpha^a_n}{n} \ e^{- i n \tau} \sin(n
\sigma) \ , \label{Xasin}
\eeq
where
\beq
[\alpha_m^a\,,\,\alpha_n^b] \ = \ m \, \left(G^{-1}\right)^{ab}\, \delta_{m+n,0} \ .
\eeq
It is now convenient return to the moving basis, which we already introduced before eq.~\eqref{T-inv}, letting
\beq
{\alpha}{}^A  \ = \ {\cal E}_a{}^A \, \alpha^a \ ,
\eeq
so that the ${\alpha}{}^A$ continue to satisfy the standard commutation relations
\beq
\left[ {\alpha}_m^A\,,\,{\alpha}_n^B\right] \ = \ m\, \delta^{AB} \, \delta_{m+n,0} \ .
\eeq
The mass formula in the non--compact $26-d$ dimensions can be cast in the form
\beq
\alpha'\, M^2 \ = \  \frac{1}{\alpha'}\, n^T \, G \, n \ + \ N \ - \ 1  \ .
\eeq

With these boundary conditions the annulus amplitude does not involve the $B_{ab}$ background, and consequently the partition function for a D${}_p$-brane stack, with all toroidal Dirichlet directions, can be deduced adapting eq.~\eqref{annulus_In} (so that 9 is replaced by 25, for the bosonic string), and reads
\beq
{{\cal A}} \ = \ N \, \bar{N}\, \int_0^\infty \ \frac{d\tau_2}{\tau_2^2} \ \frac{1}{\tau_2^{12} \ \eta^{24}\left(\frac{i \tau_2}{2}\right)} \ \tau_2^\frac{25-p}{2} \ \sum_{n^a \in \mathbb{Z}} \ e^{-\,\frac{\pi\,\tau_2}{\alpha'}\, n^a \,G_{ab} \,n^b}\ , \label{annulus_In2}
\eeq
where the $N \bar{N}$ factor reflects the presence of a $U(N)$ Chan-Paton group.

If the branes are split into two distinct sets, letting 
\beq
N_1 \ + \ N_2 \ = N \ ,
\eeq
the amplitude becomes
\bea
{{\cal A}} &=& \left(N_1 \, \bar{N}_1 \,+\, N_2 \, \bar{N}_2\right)\, \int_0^\infty \ \frac{d\tau_2}{\tau_2^2} \ \frac{1}{\tau_2^{12} \ \eta^{24}\left(\frac{i \tau_2}{2}\right)} \ \tau_2^\frac{25-p}{2} \ \sum_{n^a \in \mathbb{Z}} \ e^{-\,\pi\,\tau_2 \frac{n^a \,G_{ab} \,n^b}{\alpha'}} \label{annulus_In2_split} \\
&+&\left(N_1 \, \bar{N}_2 \,+\, N_2 \, \bar{N}_1\right)\, \int_0^\infty \ \frac{d\tau_2}{\tau_2^2} \ \frac{1}{\tau_2^{12} \ \eta^{24}\left(\frac{i \tau_2}{2}\right)} \ \tau_2^\frac{25-p}{2} \ \sum_{n^a \in \mathbb{Z}} \ e^{-\,\pi\,\tau_2 \frac{\left(n^a \,+\, \frac{\delta^a}{2\pi}\right) \,G_{ab} \,\left(n^b \,+\, \frac{\delta^b}{2\pi}\right)}{\alpha'}} 
\ , \nonumber
\eea
and the gauge group breaks to $U(N_1) \times U(N_2)$. In the transverse channel, after Poisson summations, the amplitude becomes (here we are using the condition $d+p+1=26$, for definiteness, but in general $d+p+1\leq 26$.)
\beq
{\widetilde{\cal A}} \ = \ 2^{-13} \ \left[\det\left(\frac{G}{\alpha'}\right) \right]^{-\,\frac{1}{2}}\ \int_0^\infty \frac{d\ell}{\eta^{24}(i\ell)} \ \sum_{m^a \in \mathbb{Z}} \ e^{-\,\frac{\pi \alpha'}{2}\, \ell \,m_a \,G^{ab}\,m_b} \left| N_1 \,+\, e^{i m_a \, \delta^a} \, N_2\right|^2 \ ,
\label{Annt_DB}
\eeq
consistently with the propagation of the closed spectrum in the tube with no windings as in eqs.~\eqref{numerators}. Brane displacements translate into phase factors, as we had already seen in Section~\ref{sec:wilson}, but in the absence of an orientifold projection the amplitudes involve absolute squares.

The second option leads to the modified Neumann conditions of eq.~\eqref{neumannB}, which in matrix form read
\beq
G\, X' \ + \ B \dot{X} \ = \ 0  \ ,  \label{mod_Neu}
\eeq
at $\sigma=0,\pi$. Taking momentum quantization into account, the corresponding zero--mode contribution to $X^a$ then reads, in matrix notation,
\beq
x(\tau,\sigma) \ = \ x \ + \ \frac{2}{\alpha'}\left(G - B G^{-1} B \right)^{-1} \left(\tau - B G^{-1} \sigma\right) m \ ,
\eeq
or, alternatively, in terms of the two matrices of eqs.~\eqref{GG'BB'} encoding the $T$-dual data,
\beq
x(\tau,\sigma) \ = \ x \ + \ \frac{2}{\alpha'} \left( G'\tau \ + \ B'\sigma\right) m \ .
\eeq

This second option is in fact related to the previous one by $T$-duality.  With a single internal direction, as we reviewed in Section~\ref{sec:1Dtorus}, $T$-duality links the derivatives of $X$ and the dual coordinate $X'$ as in eq.~\eqref{2d-duality}, and interchanges Neumann and Dirichlet conditions for open strings. Eq.~\eqref{2d-duality} can be recast in the form
\beq
d X \ = \ - \ \star d X' \ , \qquad d X' \ = \ - \ \star d X \ .
\eeq
In matrix notation, a possible extension of this result to the present $d$--dimensional case with $G$ and $B$ backgrounds, which also interchanges the modified Neumann condition~\eqref{mod_Neu} for $X$ with the Dirichlet condition for $X'$, is
\beq
\alpha'\, d\,X' \ = \  - \ G\, \star d X \ + \ B \, dX \ . \label{XtoX'}
\eeq
This reflects the two--dimensional duality relations $\star \,d\tau = -\,d \sigma$, $\star \,d\sigma = -\,d \tau$.
In two dimensions $\star^2=1$, and the dual metric and two-form thus satisfy, for consistency, the conditions in eq.~\eqref{TdualB}. These conditions are equivalent to those in eq.~\eqref{GG'BB'}, as we have seen, so that the inverse transformation is
\beq
\alpha'\, d\,X \ = \  - \ G'\, \star d X' \ + \ B'\, dX' \ . \label{X'toX}
\eeq
Consequently, from eq.~\eqref{Xasin} one can deduce that the complete mode expansion for the $X$ coordinates is
\bea
X &=&  x \ +\ \frac{2}{\alpha'} \left( G'\tau \ + \ B'\sigma\right) m \nonumber \\  &+& \frac{i}{\sqrt{2 \alpha'}} \ \sum_{n \neq 0} e^{-\,i n \tau} \left[  \left(G'+B'\right) e^{-\,i n \sigma} \ + \ \left(G'-B'\right) e^{i n \sigma}\right] \frac{\alpha'_n}{n} \ , \label{X_open_B}
\eea
where
\beq
    [{\alpha'}_m^a\,,\,{\alpha'}_n^b] \ = \ m \, \left(G'^{\,-1}\right)^{\,ab}\, \delta_{m+n,0}  \ . \label{primed_osc}
\eeq
Here $G'$ and $B'$ are the $T$-dual data, which are related to $G$ and $B$ by eqs.~\eqref{GG'BB'}. The $\alpha_n'$ and the ${\alpha}_n$ oscillators are now related according to
\beq
{\alpha}{}^A  \ = \ {\cal E}'_a{}^A \, \alpha'^a \ , 
\eeq
and finally the mass formula in the $26-d$ non--compact dimensions can be cast in the form
\bea
\alpha'\, M^2 &=& \frac{1}{\alpha'}\, m^T G'm \ + \ N \ - \ 1 \nonumber \\
&=& {\alpha'}\, m^T \left(G \ - \ B G^{-1} B\right)^{-1} m \ + \ N \ - \ 1 \ . \label{m2ab}
\eea
Note that the presence of a term linear in $\sigma$ introduces an ambiguity in the definition of the zero modes, and consequently in their commutation relations, since a shift of $\sigma$ affects $x$. In~\cite{schomerus,seibergwitten} the two--point function of the $X^a$ is related to the commutator of the coordinates of the string endpoints, which in our notation would read
\bea
\left.\left[ X^a(\tau)\,,\,X^b(\tau)\right]\right|_{\sigma=\pi} &=& \frac{2 \pi i}{\alpha'}\ \left(B'\right)^{ab} \ , \nonumber \\
\left. \left[ X^a(\tau)\,,\,X^b(\tau)\right]\right|_{\sigma=0} &=& - \ \frac{2 \pi i}{\alpha'}\ \left(B'\right)^{ab} \ .
\eea
This surprising behavior is due to the mixed boundary conditions~\eqref{mod_Neu}, and contradicts the naive expectation that string coordinates should commute among themselves. It can be deduced by focusing on the oscillator contributions to eq.~\eqref{X_open_B}, and demanding that the two--point function be translationally invariant. The preceding relations imply a non--vanishing commutator for the $x^a$, but this can be avoided by recasting eq.~\eqref{X_open_B} into the form
\bea
X &=&  x \ +\ \frac{2}{\alpha'} \left[ G'\tau \ + \ B'\left(\sigma-\frac{\pi}{2}\right)\right] m \nonumber \\  &+& \frac{i}{\sqrt{2 \alpha'}} \ \sum_{n \neq 0} e^{-\,i n \tau} \left[  \left(G'+B'\right) e^{-\,i n \sigma} \ + \ \left(G'-B'\right) e^{i n \sigma}\right] \frac{\alpha'_n}{n} \ , \label{X_open_B2}
\eea
and then one can retain the standard conditions
\beq
\left[ x^a, x^b\right] \ = \ 0 \ , \qquad \left[ x^a, p^b\right] \ = \ i\ \delta^{ab} \ , \qquad \left[ p^a, p^b\right] \ = \ 0 \ .
\eeq

The partition function for $N$ D25 branes with $d$ dimensions compactified on a $d$-torus and in the presence of $G$ and $B$ backgrounds reads
\beq
{{\cal A}} \ = \ N \, \bar{N}\, \int_0^\infty \ \frac{d\tau_2}{\tau_2^2} \ \frac{1}{\tau_2^{12} \ \eta^{24}\left(\frac{i \tau_2}{2}\right)} \ \tau_2^\frac{d}{2} \ \sum_{m_a \in \mathbb{Z}} \ e^{-\,{\pi\,\tau_2\,\alpha'}\, m^T \,\left(G - B G^{-1} B \right)^{-1} \,m}\  \label{annulus_In2N}
\eeq
Adding Wilson lines 
\beq
{\cal A} \ = \ \mathrm{diag} \left[\left(0\right)^{N_1},\left(\frac{A}{2\,\pi}\right)^{N_2}  \right] 
\eeq
in the Cartan subalgebra of $U(N)$ that affect the two string ends as in eq.~\eqref{BAA} splits the Chan--Paton charges into two subsets, and letting 
\beq
N_1 \ + \ N_2 \ = N \ ,
\eeq
the amplitude becomes
\bea
{{\cal A}} \!\!\!&=&\!\!\! \left(N_1 \, \bar{N}_1 \,+\, N_2 \, \bar{N}_2\right)\! \int_0^\infty \ \frac{d\tau_2}{\tau_2^2} \ \frac{1}{\tau_2^{12} \ \eta^{24}\left(\frac{i \tau_2}{2}\right)} \ \tau_2^\frac{d}{2} \!\!\! \sum_{m_a \in \mathbb{Z}} \ e^{-\,{\pi\,\tau_2\,\alpha'}\, m^T \,\left(G - B G^{-1} B \right)^{-1} \,m} \label{annulus_In2_splitN} \\
\!\!\!&+&\!\!\!\left(N_1 \, \bar{N}_2 \,+\, N_2 \, \bar{N}_1\right)\! \int_0^\infty \ \frac{d\tau_2}{\tau_2^2} \ \frac{1}{\tau_2^{12} \ \eta^{24}\left(\frac{i \tau_2}{2}\right)} \ \tau_2^\frac{d}{2} \!\!\! \sum_{m_a \in \mathbb{Z}} \ e^{-\,{\pi\,\tau_2\,\alpha'}\, \left(m + \frac{A}{2\pi} \right)^T \,\left(G - B G^{-1} B \right)^{-1} \,\left(m + \frac{A}{2\pi}\right)} \nonumber
\ , \label{A2branes}
\eea
while the gauge group breaks to $U(N_1) \times U(N_2)$. In the transverse channel, after Poisson summations, this expression becomes
\beq
{\widetilde{\cal A}}  =  2^{-13} \left[\det\left(\frac{G - B G^{-1} B}{\alpha'} \right) \right]^{\frac{1}{2}}\! \int_0^\infty \frac{d\ell}{\eta^{24}(i\ell)} \! \sum_{n^a \in \mathbb{Z}} \ e^{-\,\frac{\pi \alpha'}{2}\, \ell \,n^T  \,\left(G - B G^{-1} B\right)\,n} \left| N_1 \,+\, e^{i n^a \, A_a} \, N_2\right|^2 , \label{Annt_GB}
\eeq
consistently with the propagation of the closed spectrum in the tube with only windings, as in eqs.~\eqref{numerators}. Wilson lines translate into phase factors, as we had already seen in Section~\ref{sec:wilson}, and in the absence of an orientifold projection they build absolute squares. This is also along the lines of what was found in~\cite{bcd}, in a Scherk-Schwarz circle compactification with sectors not affected by the orientifold projection, as we shall see in Section~\ref{sec:toroidal_ss}. 

One can recognize that the amplitudes with modified Neumann conditions in the $(G,B)$ background with Wilson lines of  eqs.~\eqref{annulus_In2_splitN} and \eqref{Annt_GB} coincide those of eqs.~\eqref{annulus_In2_split} and \eqref{Annt_DB} for the Dirichlet case, which are defined in the $T$--dual $(G',B')$ background, where the Wilson lines translate into brane separations.

\subsection{\sc Orientifold Models on Tori} 

The bosonic open--string spectra of Section~\ref{sec:open_B} are invariant under $\Omega P$, where $P$ denotes the parity in the internal directions, and this setting is particularly convenient to address the effect of $B_{ab}$, since with Dirichlet boundary conditions the oscillators do not feel its presence. However, in the superstring one is confronted with a well--known subtlety related to Fermi fields. In fact, parity along a pair of internal directions is equivalent to a $\pi$ rotation, which squares to minus one on them. In fact, in orbifold constructions, applying $P$ to a $T^2$ compactification would yield a $\mathbb{Z}_2$ orbifold with fully broken supersymmetry. This complication can be avoided by combining the operations with $(-1)^{G_L}$, as first pointed out in~\cite{dp}.

Let us begin by asking what conditions can grant that $\Omega\,P$ be an automorphism of the internal lattice~\cite{orientifolds7}. This combination acts on momenta and windings according to
\beq
\Omega\,P \ : \ \left(m_a,n^b\right) \ \rightarrow \ \left(- m_a \,+\,\frac{2}{\alpha'} \, B_{ac} \,n^c, n^b\right) \ , \label{omegaPmn}
\eeq
which is a consistent transformation only if
\beq
\frac{2}{\alpha'} \, B_{ab} \ \in \ \mathbb{Z} \ ,
\eeq
so that a non--trivial $\frac{B_{ab}}{\alpha'}$ should be half-integer quantized.

The Klein bottle amplitude involves contributions from states invariant under $\Omega\,P$, for which $p_L=-p_R$. This condition translates into
\beq
m_a \ = \ \frac{1}{\alpha'} \, B_{ab}\,n^b \ , \label{mBn}
\eeq
and forces $n^a$ to be even numbers in the non--trivial case of half--quantized $B_{ab}$ of maximal rank, on which we focus to begin with. Consequently in this case
\beq
{\cal K} \ = \ \frac{1}{2} \int_0^\infty \frac{d\tau_2}{\tau_2^2}\, \frac{\left(V_8 \, - \, S_8\right) }{\tau_2^{4-\frac{d}{2}} \, \eta^{8}} \sum_{n^a \in \mathbb{Z}} \, e^{-\,\frac{4 \pi \tau_2}{\alpha'} n^T \, G\, n } \ ,
\eeq
where the actual windings are $2 n^a$ and the implicit arguments are, as usual, $2 i \tau_2$.
The corresponding transverse--channel amplitude
\beq
\widetilde{\cal K} \ = \ \frac{2^{5-{d}}}{2} \left[ \det\left(\frac{G}{\alpha'}\right)\right]^{-\frac{1}{2}} \int_0^\infty d\ell \, \frac{\left(V_8 \, - \, S_8\right) }{\eta^{8}}\, \sum_{m^a \in \mathbb{Z}} \, e^{-\,\frac{\pi \alpha'\,\ell}{2} \,m^T \, G^{-1}\, m } \ , \label{tildeK2B}
\eeq
where now the implicit arguments are $i \ell= \frac{i}{2\,\tau_2}$, involves \emph{all} the Kaluza--Klein momenta of the internal torus, while the overall coefficient is reduced accordingly. This contrasts with the standard case without $B_{ab}$, in which the internal momenta would be even, as in Section~\ref{sec:wilson}. 

The annulus amplitude that extends eq.~\eqref{annulus_In2_split} to the case of generic brane sets reads
\bea
{{\cal A}} \!\!\!&=&\!\!\! \sum_{(k,k')=1}^{2r} \frac{N_k  \bar{N}_{k'}}{2} \int_0^\infty \frac{d\tau_2}{\tau_2^2}\, \frac{\left(V_8 \, - \, S_8\right)\left(\frac{i \tau_2}{2}\right) }{\tau_2^{4-\frac{d}{2}} \ \eta^{8}\left(\frac{i \tau_2}{2}\right)} \, \sum_{n^a \in \mathbb{Z}} \, e^{-\,\frac{\pi \tau_2}{\alpha'} \left(n^a \,+\, \frac{\Delta_{k,k'}^a }{2\pi}\right) G_{ab} \left(n^b \,+\, \frac{\Delta_{k,k'}^b }{2\pi}\right)} \label{Atorusd}
, 
\eea
with
\beq
\Delta_{k,k'} \ = \ \delta_k \ - \ \delta_{k'} 
\eeq
the difference between the coordinates for the two brane sets with labels $k$ and $k'$, which generally do not to lie at fixed points. The configurations that are symmetrical under $\Omega P$ involve an even set of branes $(k=1,\ldots,2r)$ with $\delta_{r+k}^a = -\, \delta_k^a$ ($k=1,\ldots,n)$, with additional identifications $N_{r+k}\equiv \bar{N}_k$ and, as usual for complex charges, numerically ${N}_{k}={\bar N}_k$.

A first guess for the corresponding M\"obius amplitude could be
\bea
{{\cal M}} \!\!\!&=&\!\!\! -\,\sum_{k=1}^{2r} \frac{\gamma_{k,n}\,N_k}{2} \! \int_0^\infty \frac{d\tau_2}{\tau_2^2}\, \frac{\left(\hat{V}_8 \, - \, \hat{S}_8\right)}{\tau_2^{4-\frac{d}{2}} \ \hat{\eta}^{8}} \, \sum_{n^a \in \mathbb{Z}} \, e^{-\,\frac{\pi \tau_2}{\alpha'} \left(n + \frac{\delta_{k} }{\pi}\right)^T G \left(n + \frac{\delta_{k} }{\pi}\right)} \label{Mtorusd}
,
\eea
allowing for different signs $\gamma_{k,n}$, since the contributions to $\widetilde{\cal M}$ are geometric means of those to $\widetilde{K}$ and $\widetilde{\cal A}$.  Since these signs determine the (anti)symmetry of open--string states under the interchange of the string ends, they should comply with the fusion rules on the lattice, an important property that we shall return to shortly.

The minimal choice for the $\gamma_{k,n}$ only depends on the parity of the windings, which can be exhibited replacing $n$ by $2 n+\epsilon$, with the components of $\epsilon$ equal to 0 or 1. This choice leads to
\bea
{{\cal M}} \!\!\!&=&\!\!\! -\,\sum_{k=1}^{2r} \sum_{\epsilon^a=(0,1)} \!\! \frac{\gamma_{k,\epsilon}\,N_k}{2} \! \int_0^\infty \frac{d\tau_2}{\tau_2^2}\, \frac{\left(\hat{V}_8 \, - \, \hat{S}_8\right)}{\tau_2^{4-\frac{d}{2}} \ \hat{\eta}^{8}} \, \sum_{n^a \in \mathbb{Z}} \, e^{-\,\frac{\pi \tau_2}{\alpha'} \left(2 n + \epsilon + \frac{\delta_{k} }{\pi}\right)^T G \left(2 n + \epsilon + \frac{\delta_{k} }{\pi}\right)} \label{mobius_epsilon} \ ,
\eea
and allows a Poisson summation to the transverse channel.
The implicit argument of the ``hatted'' characters in the preceding expressions is, as usual, $\frac{1}{2} + \frac{i \tau_2}{2}$. Note that in this fashion, $\Omega$ acts differently on open strings with even and odd lattice windings.

In the transverse channel, the annulus amplitude becomes
\beq
{\widetilde{\cal A}} = \frac{2^{-5}}{2}\, \left[\det\left(\frac{G}{\alpha'}\right) \right]^{-\,\frac{1}{2}} \int_0^\infty d\ell\,  \frac{\left(V_8 - S_8\right)(i \ell)}{\eta^{8}(i\ell)} \, \sum_{m^a \in \mathbb{Z}} e^{-\,\frac{\pi \alpha'}{2}\, \ell \,m_a \,G^{ab}\,m_b} \left| \sum_{k=1}^{2r} N_k\, e^{i m_a \, \delta_k^{a}} \right|^2 ,
\label{Annt_DB_gen}
\eeq
while the M\"obius amplitude~\eqref{mobius_epsilon} becomes
\bea
\widetilde{{\cal M}} \!\!\!&=&\!\!\!  -\,2^{-d} \left[ \det\left(\frac{G}{\alpha'}\right)\right]^{-\,\frac{1}{2}} \sum_{k=1}^{r} \int_0^\infty {d\ell}\, \frac{\left(\hat{V}_8 \, - \, \hat{S}_8\right)}{\hat{\eta}^{8}}  \times \nonumber \\ &&   \sum_{\epsilon^a=(0,1)}\sum_{m^a \in \mathbb{Z}} e^{-\,{\frac{\pi}{2}\, \alpha'}\, \ell \,m_a \,G^{ab}\,m_b} \sum_{k=1}^{2r} \gamma_{k,\epsilon}\, N_k \,e^{i  m_a \left(\pi\,\epsilon^a \,+\,\delta_k^{a}\right)} \, 
,
\eea
where as usual the implicit argument of the ``hatted'' characters is now $\frac{1}{2} + i \ell$. 
For a two-torus, the signs $\gamma_{k,\epsilon}$ will be linked to the propagation between the $D7$ branes and four O7 planes, three of which will be standard O${}7_-$ while one will be O${}7_+$.
Note that the requirement that all branes have the same tension translates into the conditions that $\gamma_{k,\epsilon}=\gamma_\epsilon$, and therefore one is confronted with a set of signs that is independent of $k$. These signs were introduced in~\cite{orientifolds7}, where the construction was apparently based on Neumann conditions. In Section~\ref{sec:geometric} we shall see how to reconcile those early results with the present treatment of the boundaries, which does not allow Neumann conditions, as we have seen.

The tadpole condition takes the form
\beq
2^{5-d} \ + \ 2^{-5} \left(\sum_{k=1}^{2 r} N_k \right)^2 \ - \ 2 \ 2^{-d}\,\sum_{k=1}^{2 r}  N_k \sum_{\epsilon^a=(0,1)} \gamma_{\epsilon} \ = \ 0 \ ,
\eeq
and it has the integer solution
\beq
\sum_{k=1}^{2 r} N_k \ = \ 2^{5-\frac{d}{2}}  
\eeq
only if
\beq
\sum_{\epsilon^a=(0,1)} \gamma_{\epsilon} \ = \ 2^\frac{d}{2} \ , \label{tadpolegamma}
\eeq
with the $\gamma_\epsilon$ that are $\pm 1$. For example, in two dimensions, three $\gamma_\epsilon$ must be positive and one negative, and in general there are $2^{\frac{d}{2}-1}\left(2^\frac{d}{2}+1\right)$ positive and $2^{\frac{d}{2}-1}\left(2^\frac{d}{2}-1\right)$ negative $\gamma$'s. In this case, the gauge group is generically a product of $U(N_k)$ factors, but if some branes lie at fixed points, there are enhancements, as in Section~\ref{sec:wilson}. The novelty here is that, depending on whether the fixed points have positive or negative $\gamma_\epsilon$, the enhancement can give rise to orthogonal or symplectic groups, as noted in~\cite{orientifolds7}. Moreover, the two different types of groups are connected by continuous displacements. The maximal gauge groups are therefore $SO\left(2^{5-\frac{d}{2}}\right)$ or $USp\left(2^{5-\frac{d}{2}}\right)$ in the two cases. Symplectic gauge groups are obtained when the D7 branes sit on top of the $O7_+$ planes, orthogonal ones are obtained when the D7 branes sit on top of the $O7_-$ planes, and finally unitary gauge groups are obtained when the D7 branes are in the bulk, away from the $O7$ planes.

We can now elaborate on the fusion issue that we previously raised, considering for simplicity the two--dimensional case and referring to the bosonic string, while starting, for definiteness, from
\beq
\left(\gamma_{00},\gamma_{01},\gamma_{10},\gamma_{11} \right) \ = \ \left(-1,1,1,1 \right) \ .
\eeq
This choice assigns symmetric Chan--Paton matrices to the $00$ lattice states and antisymmetric ones to the rest, which would seem to conflict with the $\mathbb{Z}_2\times \mathbb{Z}_2$ naive fusion rules reflecting the momentum sums on the lattice. In fact, there is no contradiction, since additional signs are available when combining different amplitudes that contribute to a given pole, once one takes into account the flip property
\beq
A(1,\ldots,n) \ = \ \left(-1\right)^{\sum_{i}^n N_i + n_D} A(n,\ldots,1)
\eeq
here written for the superstring, which reflects the $\Omega$-symmetry of the spectrum. $N_i$ indicates the oscillator level of the $i$-th external state, and the additional $(-1)^{n_D}$ is present if there are $n_D$ Dirichlet coordinates in the amplitude. This example is described in more detail in fig.~\ref{fig:3pointtwist}. Consistent solutions for the $\gamma'$ in higher--dimensional tori can be obtained as direct products of the preceding two--dimensional ones, and the relative signs of the different amplitudes can be assigned consistently.
\begin{figure}[ht]
\centering
\includegraphics[width=100mm]{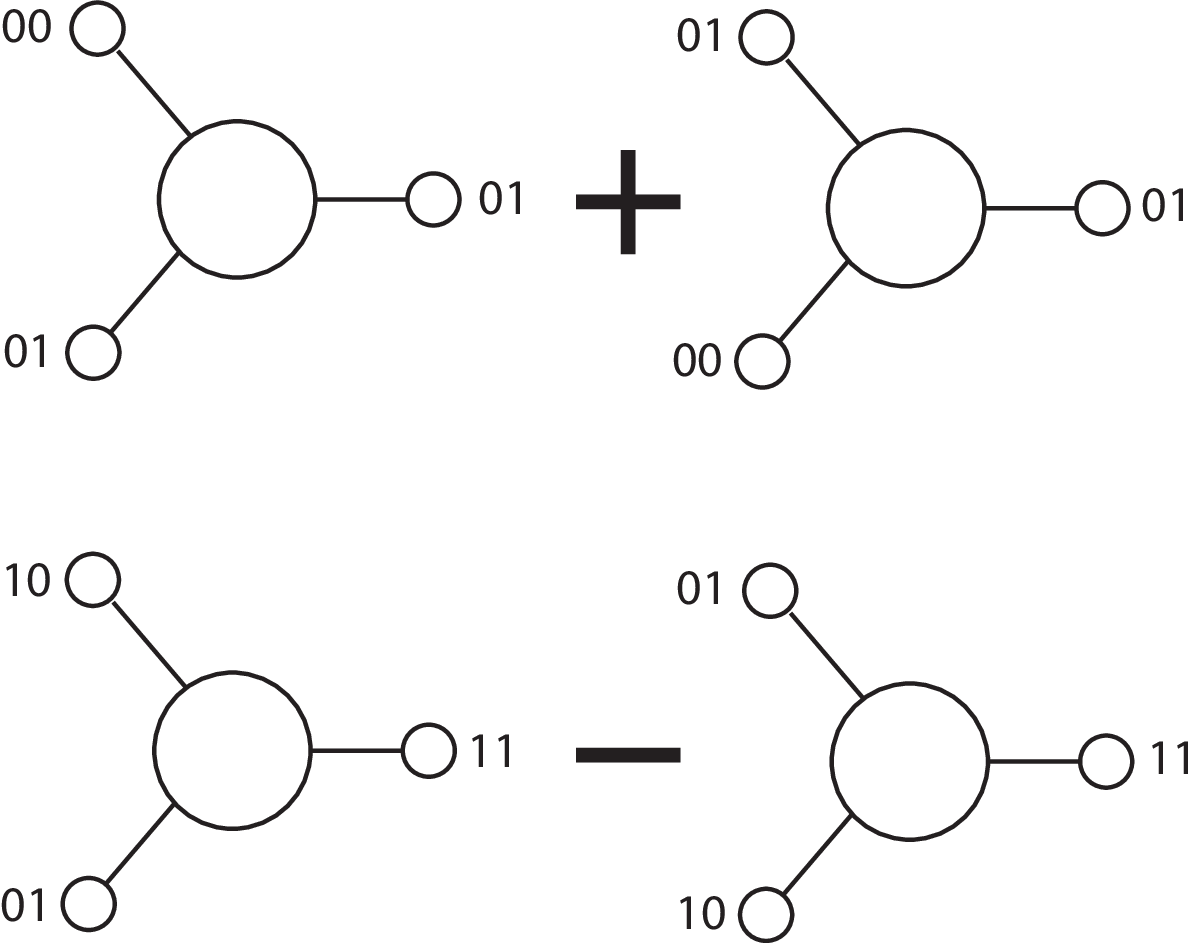}
%\vspace{-5cm}
\caption{\small Pairs of amplitudes related by the ``flip property'', which implies in particular that $A(123)=-A(213)$ for three massless vectors, build the three point functions. Together with momentum compositions on the lattice, their relative signs can conspire to grant the consistency of the Chan--Paton assignments determined by $\left(\gamma_{00},\gamma_{01},\gamma_{10},\gamma_{11} \right) \ = \ \left(-1,1,1,1 \right)$.}
\label{fig:3pointtwist}
\end{figure}

With a generic $B_{ab}$ that is possibly of lower rank, the Klein--bottle amplitude takes the form
\beq
{\cal K} \ = \ \frac{2^{-d}}{2} \int_0^\infty \frac{d\tau_2}{\tau_2^2}\, \frac{\left(V_8 \, - \, S_8\right) }{\tau_2^{4-\frac{d}{2}} \, \eta^{8}} \sum_{\epsilon^a=(0,1)} \sum_{n^a \in \mathbb{Z}} \, e^{-\,\frac{\pi \tau_2}{\alpha'} n^T \, G\, n } \ e^{2\pi i \epsilon^T\frac{B}{\alpha'} n}\ ,
\eeq
where the sum over $\epsilon^a$ enforces the constraint on $n^a$ determined by eq.~\eqref{mBn}, and consequently the transverse--channel Klein--bottle amplitudes is
\beq
\widetilde{\cal K} \ = \ \frac{2^{5-{d}}}{2} \left[ \det\left(\frac{G}{\alpha'}\right)\right]^{-\frac{1}{2}} \!\!\! \int_0^\infty d\ell \, \frac{\left(V_8 \, - \, S_8\right) }{\eta^{8}}\! \sum_{\epsilon^a=(0,1)} \sum_{m^a \in \mathbb{Z}} \, e^{-\,{2\,\pi \alpha'\,\ell} \,\left(2m + \frac{2}{\alpha'}\, B\epsilon\right)^T \, G^{-1}\, \left(2m + \frac{2}{\alpha'}\, B\epsilon \right) } \ . \label{tildeK2Bgen}
\eeq
If $r$ denotes the rank of $B$, $2^{d-r}$ different values of $\epsilon$ contribute to the tadpole condition, so that the Klein--bottle contribution now carriers a factor $\frac{1}{2}\, 2^{5-r}$. The annulus and M\"obius amplitudes retain the same form as before, so that the tadpole condition now becomes
\beq
2^{5-r} \ + \ 2^{-5} \left(\sum_{k=1}^{2 r} N_k \right)^2 \ - \ 2 \ 2^{-d}\,\sum_{k=1}^{2 r}  N_k \sum_{\epsilon^a=(0,1)} \gamma_{\epsilon} \ = \ 0 \ ,
\eeq
and has the integer solution
\beq
\sum_{k=1}^{2 r} N_k \ = \ 2^{5-\frac{r}{2}}  
\eeq
only if
\beq
\sum_{\epsilon^a=(0,1)} \gamma_{\epsilon} \ = \ 2^{d - \frac{r}{2}} \ ,
\eeq
with the $\gamma_\epsilon$ that are $\pm 1$. In this case, there are $2^{\frac{d}{2}-1}\left(2^\frac{d}{2} \pm 2^\frac{d-r}{2}\right)$ orientifolds of the two types.
The $\gamma_\epsilon$ determine the signs of the (identical) tensions and charges of the orientifolds. 
However, as we have seen, these charges and tensions cannot have the same overall signs. Consequently, in the presence of a nontrivial $B_{ab}$ local tadpole cancellation is not possible, in contrast to what we saw for the circle compactification in Section~\ref{sec:wilson}. The arguments in~\cite{pw} then suggest that strong coupling should accompany the large--radius limit. 

\subsection{\sc \texorpdfstring{$T$}\--Dual Formulation}

It is instructive to discuss in detail the $T$-dual version of the construction involving $(X,G,B)$ and Dirichlet boundary conditions for the open sector. To this end, one can start from eq.~\eqref{XtoX'}, which also holds for the closed sector, to deduce from it that
\beq
\alpha'\,\partial_+ X'\ = \ E\, \partial_+ X \ , \qquad \alpha'\,\partial_- X'\ = \  - \ E^T\, \partial_- X \ ,
\eeq
where the matrix $E = G + B$ was introduced in eq.~\eqref{EGB}.
The action of $\Omega\,P$ on 
\beq
X \ = \ X_+(\sigma_+) \ + \ X_-(\sigma_-) 
\eeq
gives
\beq
\Omega\,P\, X \ = \  \ - \ X_+(\sigma_-) \ - \ X_-(\sigma_+) \ .
\eeq
Making use of eq.~\eqref{XtoX'}, one can see that the action induced on $X'$, which we shall denote $\left(\Omega P\right)'$, is
\bea
\left(\Omega P\right)'\, X' &=&  \ - \ \frac{1}{\alpha'}\, E\, X_-(\sigma_+) \ + \ \frac{1}{\alpha'}\,  E^T X_+(\sigma_-) \nonumber \\
&=& \ E\, \left(E^{-1}\right)^T\,X'_-(\sigma_+) \ + \ E^T\,E^{-1}\, X'_+(\sigma_-)  \ . \label{T'}
\eea
Note that this transformation involves two matrices, one of which is the inverse of the other, and consequently squares to one, as it should. Moreover, when $B=0$ it reduces to the standard left--right interchange.
If one applies it to the zero modes, $X$ reduces to
\beq
X \ = \ \left[ \alpha'\,G^{-1}\left(m - \frac{B}{\alpha'}\,n\right) \ + \ n \right]\left(\tau+\sigma\right) \ + \ \left[ \alpha'\,G^{-1}\left(m - \frac{B}{\alpha'}\,n\right) \ - \ n \right]\left(\tau-\sigma\right) \ ,
\eeq
and then eq.~\eqref{T'} implies that
\beq
X'\ = \ \left[ \alpha'\,\left(G'\right)^{-1}\left(m' - \frac{B'}{\alpha'}\,n'\right) \ + \ n' \right]\left(\tau+\sigma\right) \ + \ \left[ \alpha'\,\left(G'\right)^{-1}\left(m' - \frac{B'}{\alpha'}\,n'\right) \ - \ n' \right]\left(\tau-\sigma\right) \ ,
\eeq
with $G'$ and $B'$ related to $G$ and $B$ as in eqs.~\eqref{GG'BB'}, and
\beq
m'\ = \ n \ , \qquad n'\ = \ m \ .
\eeq
Consequently, the action of $\left(\Omega P\right)'$ can be deduced from eq.~\eqref{omegaPmn} and is given by
\beq
\left(n', m' \right) \  \rightarrow \  \left(- n'\ + \ \frac{2}{\alpha'}\,B\,m', m'\right) \ = \  \left[- n'\ + \ 2 \,\alpha'\,\left(B'- G' \left(B'\right)^{-1} G'\right)^{-1}\!\!m', m'\right] \ .
\eeq
This is a symmetry provided $\frac{2}{\alpha'}\,B \in \mathbb{Z}$, a condition that takes the form
\beq
2 \,\alpha'\,\left(B'- G' \left(B'\right)^{-1} G'\right)^{-1} \ \in \ \mathbb{Z}
\eeq
in terms of the ``primed'' variables.

When expressed in terms of the ``primed variables'' the amplitudes read
\bea
{\cal K} \!\!&=&\!\! \sum_{\epsilon^a=(0,1)} \sum_{m_a \in \mathbb{Z}} \, e^{-\,{\pi \tau_2 \alpha'} m^T \, \left[G'- B'\left(G'\right)^{-1}B'\right]^{-1}\!\! m } \ e^{2\pi i\alpha' \epsilon^T\left(B'- G' \left(B'\right)^{-1} G'\right)^{-1} \!\!m}\ ,\nonumber \\
{{\cal A}} \!\!\!&=&\!\!\! \sum_{(k,k')=1}^{2r} \frac{N_k  \bar{N}_{k'}}{2} \, \sum_{m_a \in \mathbb{Z}} \, e^{-\,\pi \tau_2 \alpha' \left(m \,+\, \frac{\Delta_{k,k'} }{2\pi}\right)^T \left[G'- B'\left(G'\right)^{-1}B'\right]^{-1}\!\left(m \,+\, \frac{\Delta_{k,k'} }{2\pi}\right)}\, \nonumber \\
{{\cal M}} \!\!\!&=&\!\!\! -\,\sum_{k=1}^{2r} \sum_{\epsilon^a=(0,1)} \!\! \frac{\gamma_{\epsilon}\,N_k}{2} \!  \sum_{m_a \in \mathbb{Z}} \, e^{-\,\pi \tau_2 \alpha' \left(2 m + \epsilon + \frac{\delta_{k} }{\pi}\right)^T \left[G'- B'\left(G'\right)^{-1}B'\right]^{-1}\! \left(2 m + \epsilon + \frac{\delta_{k} }{\pi}\right)}\ ,
\eea
so that momenta and windings are interchanged. In these expressions, the integral and the factors that are not involved directly,
\beq
\frac{2^{-d}}{2} \int_0^\infty \frac{d\tau_2}{\tau_2^2}\, \frac{\left(V_8 \, - \, S_8\right) }{\tau_2^{4-\frac{d}{2}} \, \eta^{8}} \ ,
\eeq
with the appropriate arguments for the three cases, has been left implicit for brevity. As before, the sum over $\epsilon$ in ${\cal K}$ projects on even values of $\frac{B\,m}{\alpha'}$, the $\delta_k$ are now Wilson lines, and the $\gamma_\epsilon$ determine the (anti)symmetry of the ground states in sectors with identical charges. For general $\delta_k$ the groups are unitary, as in the previous descriptions, and special choices can result in orthogonal or symplectic enhancements. No special points associated to the orientifolds are present in this $T$-dual description.

The two--dimensional case is again an instructive playground. Let us therefore take a closer look at these results, specializing to
\beq
G \ = \ R^2\, 1_2 \ , \qquad \frac{B}{\alpha'} \ = \ \frac{1}{2}\, i\,\sigma_2 \ .
\eeq
In this case
\be
G' \ = \  \frac{4\,R^2}{1 \ + \ \left(\frac{2\,R^2}{\alpha'}\right)^2}\ 1_2 \ , \qquad \ B' \ = \ - \ \frac{2\,i\,\alpha'\,\sigma_2}{1 \ + \ \left(\frac{2 R^2}{\alpha'}\right)^2} \ ,
\eeq
so that the ``dual size'' is at most of order $\sqrt{\alpha'}$, independently of $R$. In the next section, we discuss a different duality link, which connects large and small radii, as is the absence of $B$, while maintaining $B$ invariant. This will make the geometric interpretation more transparent, while also justifying the original treatment in~\cite{orientifolds7}.

\subsection{\sc Geometric Interpretation} \label{sec:geometric}

We can now illustrate an alternative description of the quantized $B_{ab}$ background~\cite{wittenOplus} (see also~\cite{bianchitor,emiliantor}). Considering for definiteness a two-dimensional square torus with $B_{ab} = \frac{\alpha'}{2}$, the torus amplitude can then be cast in the form
\bea
{\cal T} &=& \left[  \Lambda_{m_1,2n_1} \Lambda_{m_2,2n_2} +
\Lambda_{m_1+\frac{1}{2},2n_1} \Lambda_{m_2,2n_2+1} +
\Lambda_{m_1,2n_1+1} \Lambda_{m_2+\frac{1}{2},2n_2} \right. \nonumber \\ &+&
\left. \Lambda_{m_1 +\frac{1}{2} ,2n_1+1} \Lambda_{m_2 + \frac{1}{2},2n_2+1} \right]  \times |V_8-S_8|^2
\ , \label{eq:O+1}
\eea
where we are leaving integrations and bosonic contributions implicit for brevity, and the shorthand notation indicates the lattice sums 
\beq
\Lambda_{m,n} \ = \ \frac{\sum_{m,n} q^{\frac{\alpha'}{4} \left[ \frac{m}{R}\, +\, \frac{n\,
R}{\alpha'} \right]^2} \, \bar{q}^{\frac{\alpha'}{4} \left[
\frac{m}{R} \,-\, \frac{n\, R}{\alpha'} \right]^2}}{\eta \, \bar{\eta}} \ .
\eeq

This torus amplitude is left invariant by $O(2,2;Z)$ transformations, as we saw in Section~\ref{sec:narain}, and there is a special transformation that leaves $B$ invariant but interchanges small and large radii according to
\beq
R_1 \ \to \ \frac{\alpha'}{2 R_1} \ , \qquad R_2 \ \to \ \frac{\alpha'}{2 R_2}
\ . \label{eq:O+2}
\eeq 

Letting
\beq
\tau \ = \ \frac{B}{\alpha'} \ + i\, A \ ,
\eeq
where $A$ denotes the area of the two-torus in units of $\alpha'$ and $B$ the corresponding two-form, it was noted in~\cite{wittenOplus} that the $SL(2,Z)$ transformation 
\beq
\tau \ \rightarrow \ \frac{\tau \ - \ 1}{2\,\tau \ - \ 1}  \label{tauP}
\eeq
preserves the non--trivial value $\frac{B}{\alpha'}=\frac{1}{2}$ while also sending $A$ to $\frac{1}{4 A}$. This transformation therefore connects large and small areas in the presence of $B$, just as the conventional $T$-duality would with $B=0$. This transformation actually already played a role in our discussion of the M\"obius amplitude in Section~\ref{sec:critical_strings}: it is a spacetime version of Pradisi's $P$ transformation, here applied to the K\"ahler cone, which is built by the sequence
\beq
P \ = \ T \,S\, T^2\, S \ ,
\eeq
where, as in Section~\ref{sec:critical_strings}, $T:\tau \to \tau+1$ and $S:\tau \to - \frac{1}{\tau}$.
Therefore, starting with Dirichlet boundary conditions, the presence of two $S$ transformations eliminates the mixed Neumann boundary conditions at the ends. This justifies the consistency of the original amplitudes in~\cite{orientifolds7}~\footnote{The implicit arguments in three contributions are, as usual, $2 i \tau_2$, $i \frac{\tau_2}{2}$ and $\frac{1}{2}+i \frac{\tau_2}{2}$. Note that Pradisi~\cite{pradisi_03} came close to these arguments, elaborating on analogies with the shift orbifolds that we shall discuss in the following section, but without emphasizing the role of $P$ in connection with the boundary conditions.}
\bea
{\cal K} &=& {\frac{1}{2}}\, (V_8 - S_8) \ \sum_m
\frac{q^{\frac{\alpha'}{2} m^T g^{-1} m}} {\eta^d} \, , \nonumber \\
{\cal A}^{(r)} &=& \frac{2^{r-d}}{2} \ N^2 \, (V_8 - S_8 )
\sum_{\epsilon =0,1} \sum_{m}  {q^{{\alpha ' \over 2} ( m +
{1\over \alpha '} B
\epsilon) ^{T} g^{-1} ( m + {1\over \alpha '} B \epsilon )}
\over \eta^d } \, , \nonumber \\
\!\!\!\!\!\!\!\!\!\!{\cal M}^{(r)} &=& - \ \frac{2^{r/2 -d/2}}{2}    N  (\hat V _8 -
\hat S _8 ) 
\! \sum_{\epsilon =0,1} \sum_{m} {q^{{\alpha
' \over 2} (m +  {1\over \alpha '} B
\epsilon )^{T} g^{-1} (m + {1\over \alpha '} B \epsilon )}
\gamma_\epsilon \over \hat\eta^d} \ ,
\eea
here written for a rank--$r$ $B_{ab}$ in a $d$--dimensional torus, which were regarded as relying on Neumann boundary conditions, but in fact relied on Dirichlet conditions on tori with sides redefined as in eqs.~\eqref{eq:O+2}. The $\gamma_\epsilon$ in the M\"obius amplitude grant the proper factorization in the transverse channel, and continuous deformations link, as above, orthogonal and symplectic gauge groups. The $P$--dual description in~\cite{wittenOplus} linked the $\gamma_\epsilon$'s with different signs to the presence of O${}_+$ orientifolds, with positive tension and charge.

Returning to the $T^2$ case, after the transformation in eq.~\eqref{tauP}, the model contains O7 planes and D7 branes, and the orientifold amplitudes that we have discussed become
\bea
&& {\cal K} \ =  \ \frac{1}{2} W_{2n_1} W_{2n_2}(V_8-S_8) \ , \nonumber \\
&& {\cal A} \ = \ \frac{N^2}{2} W_{n_1} W_{n_2}(V_8-S_8) \ , \nonumber \\
&& {\cal M} \ = \ \frac{N}{2} \left[ (-1)^{n_1} W_{n_1} W_{2n_2} - W_{n_1} W_{2n_2+1} \right] (V_8-S_8) \ , \label{eq:O+3}
\eea
where
\beq
W_{n} \ = \ \frac{\sum_{n} q^{\frac{n^2 \,
R^2}{2\,\alpha'}} }{\eta} \ , \ \qquad P_{m} \ = \ \frac{\sum_{m} q^{\frac{\alpha'}{2}\, \frac{m^2}{R^2}} }{\eta} \ .
\eeq
The tree-level, or transverse channel, amplitudes are then
\bea
&& {\tilde {\cal K}} \ =  \ \frac{2^5}{8} \frac{R_1 R_2}{\alpha'} P_{m_1} P_{m_2} (V_8-S_8) \ , \nonumber \\
&& {\tilde {\cal A}} \ =  \ \frac{2^{-5} N^2}{2} \frac{R_1 R_2}{\alpha'} P_{m_1} P_{m_2} (V_8-S_8) \ , \nonumber \\
&& {\tilde {\cal M}} \ =  \ \frac{N}{2} \frac{R_1 R_2}{\alpha'} \Big[  P_{2m_1+1} P_{m_2} - (-1)^{m_2} P_{2m_1} P_{m_2} \Big] (V_8-S_8) \ , \label{eq:O+4}
\eea
where integrations and momentum sums are again implicit and indicated relying on a compact notation.

These amplitudes encode positions and charges of the branes and orientifolds in the internal space: the model contains three O7$_-$ planes and one O7$_+$ plane located at the four fixed points of the orientifold projection $\Omega' \,= \,\Omega \,\Pi_1 \Pi_2 (-1)^{G_L}$, with  $\Pi_i$ being the parity operations in the two internal coordinates. Since the O7$_+$
has opposite charge and tension compared to a O7$_-$, the total orientifold charge is halved with respect to type I superstring, and from this vantage point the rank reduction of the gauge group on D7 branes is directly linked to the presence of the O7$_+$. Moreover, the gauge group is $USp(16)$, since the D7 branes lie on top of each other, and are located at the fixed point where the  O7$_+$ sits. The O$_+$ planes induce an opposite projection on the Chan-Paton factors compared to O$_-$ planes so that, if the D7 branes were located at one of the three fixed points with
O7$_-$ planes, the gauge group would be $SO(16)$. Finally, moving the whole stack of D7 branes into the bulk would have led to a $U(8)$ gauge group, precisely as we saw in the previous section.  From this vantage point, the $\gamma_\epsilon$ coefficients that were introduced above characterize the $\Omega$ projections induced by the two different types of orientifold planes. 

\section{\sc Scherk--Schwarz Circle Compactifications }  \label{sec:toroidal_ss}

The Scherk-Schwarz mechanism~\cite{ss1,ss2} exploits the
presence of compact internal spaces to induce the breaking of supersymmetry in field
theories or superstrings. It relies on symmetry groups ${\cal G}$ of the higher-dimensional
theory that do not commute with supersymmetry, typically R-symmetries or the fermion number $(-1)^F$.
In what follows, it will suffice to focus on the latter option.

In Field Theory, the masses of Bose and Fermi excitations emerging from a circle compactification can be separated by demanding that Bose fields be periodic and Fermi fields anti--periodic. The corresponding Kaluza--Klein mass spectra are then proportional to
\beq
{\cal M}^2_{\mathrm Bose} \ = \ \frac{m^2}{R^2} \ , \qquad {\cal M}^2_{\mathrm Fermi} \ = \ \frac{\left(m+\frac{1}{2}\right)^2}{R^2} \ , \label{ss_ft}
\eeq
where $m \in Z$. The tree--level mass separation indicates that supersymmetry, if initially present, is broken by the compactification. For small values of $R$, the separation can be made arbitrarily large, and conversely the effect disappears in the $R \to \infty$ limit.
		
The Scherk--Schwarz mechanism can be applied in globally supersymmetric models, in supergravity and even in superstrings. The breaking is induced by the different periodicity conditions for bosons and fermions and is therefore explicit in global supersymmetry. However, in compactifications of supergravity it is spontaneous: it can be described via a modification of the superpotential that breaks supersymmetry spontaneously in the ground state, while the internal component of the gravitino plays the role of a goldstino.
Consistent truncations at the effective field theory level are possible if setting massive modes to zero is consistent with their field equations, \emph{i.e.} if there are no contributions in them from the fields that are retained.  This is true if one uses R-symmetries for the boundary conditions, while using $(-1)^F$, although consistent at the string theory level,  does not lead to a consistent truncation in supergravity.  
However, regardless of whether or not a truncation is consistent, there is no sizable Fermi--Bose mass gap in Scherk-Schwarz compactifications, when compared with the Kaluza--Klein scale. All massive Kaluza-Klein modes should be kept in an effective field-theory description, in order to correctly capture the remarkable softness of the breaking granted by this mechanism. We shall return to some of these issues in Section~\ref{sec:KKLT}.

\subsection{\sc Scherk--Schwarz Circle Compactification for Type IIB }

In order to illustrate the novelties that emerge at the string level, let us consider the partition function for the type--IIB string compactified on a circle and subject to a ``shift orbifold'' projection. This construction combines the orbifold projection by $(-1)^{G_L+G_R}$, which we already encountered in our discussion of ten--dimensional models, where $G_{L,R}$ denote the left and right spacetime fermion numbers, with a translation along the circle by half of its length. The resulting modular invariant torus amplitude comprises four terms:
\begin{eqnarray}
{\cal T} \ &=& \ \int_{\cal F} \frac{d^2 \tau}{\tau_2^2} \ \frac{1}{\left( \sqrt{\tau_2}\, \eta \ \bar{\eta}\right)^7}\ \Bigg[ \frac{1}{2} \left| V_8 \, - \, S_8 \right|^2\,
\frac{\sum_{m,n} q^{\frac{\alpha'}{4} \left( \frac{m}{R} + \frac{n
R}{\alpha'} \right)^2} \, \bar{q}^{\frac{\alpha'}{4} \left(
\frac{m}{R} - \frac{n R}{\alpha'} \right)^2}}{\eta \, \bar{\eta}}
\nonumber \\
&+& \frac{1}{2} \left| V_8 \, + \, S_8 \right|^2\,
\frac{\sum_{m,n} \, (-1)^m \, q^{\frac{\alpha'}{4} \left(
\frac{m}{R} + \frac{n R}{\alpha'} \right)^2} \,
\bar{q}^{\frac{\alpha'}{4} \left(
\frac{m}{R} - \frac{n R}{\alpha'} \right)^2}}{\eta \, \bar{\eta}} \nonumber \\
&+& \frac{1}{2} \left| O_8 \, - \, C_8 \right|^2\,
\frac{\sum_{m,n} q^{\frac{\alpha'}{4} \left( \frac{m}{R} +
\frac{(n+1/2) R}{\alpha'} \right)^2} \, \bar{q}^{\frac{\alpha'}{4}
\left( \frac{m}{R} - \frac{(n + 1/2) R}{\alpha'} \right)^2}}{\eta
\, \bar{\eta}} \nonumber \\
&+& \frac{1}{2} \left| O_8 \, + \, C_8 \right|^2\,
\frac{\sum_{m,n} \, (-1)^m \, q^{\frac{\alpha'}{4} \left(
\frac{m}{R} + \frac{(n+1/2) R}{\alpha'} \right)^2} \,
\bar{q}^{\frac{\alpha'}{4} \left( \frac{m}{R} - \frac{(n + 1/2)
R}{\alpha'} \right)^2}}{\eta \, \bar{\eta}}\Bigg]  \ , \label{torus_shift}
\end{eqnarray}
where, as in previous sections,
\beq
q \ = \ e^{2\pi i \tau} \ , \qquad \bar{q} \ = \ e^{- 2\pi i \bar{\tau}} \ .
\eeq

The first two contributions describe the untwisted sector and project the original type--IIB spectrum, while the last two are demanded by modular invariance and describe the twisted sector.
The Kaluza--Klein states with $m$ even or odd are also even or odd under the translation along the circle by half of its length: $y \to y \,+\,\pi\,R$, which is responsible for the presence of the $(-1)^m$ factor in the second line. There is an important novelty: this is the emergence, in the twisted sector, of the $O_8$ character, which starts with a tachyonic scalar but is lifted in mass by the lattice sums.

The presentation in eq.~\eqref{torus_shift} makes the orbifold structure manifest, but the link with the ordinary Scherk--Schwarz mechanism is perhaps less evident. There is an alternative way of presenting the result,
which is obtained collecting separately the contributions of Bose and Fermi modes, while also redefining the circle radius according to $R = 2 R_s$, which leads to
\bea
{\cal T} \ &=& \int_{\cal F} \frac{d^2\,\tau}{\tau_2^2} \ \frac{1}{\left( \sqrt{\tau_2}\, \eta \ \bar{\eta}\right)^7}\ \Bigg[ \left(\left|V_8\right|^2 \ + \ \left|S_8\right|^2 \right) \Lambda_{m,2n} \ - \ \left(V_8\,\overline{S}_8 \ + \ S_8\,\overline{V}_8 \right) \Lambda_{m\,+\,\frac{1}{2},2n} \nonumber \\
&+& \left(\left|O_8\right|^2 \ + \ \left|C_8\right|^2 \right) \Lambda_{m,2n+1} \ - \ \left(O_8\,\overline{C}_8 \ + \ C_8\,\overline{O}_8 \right) \Lambda_{m\,+\,\frac{1}{2},2n+1} \Bigg] \ . \label{torus_ss}
\eea
As in preceding sections, we are resorting to the compact notation
\beq
\Lambda_{m\,+\,\alpha,n\,+\,\beta} \ = \ \frac{\sum_{m,n} q^{\frac{\alpha'}{4} \left[ \frac{m\,+\,\alpha}{R_s}\, +\, \frac{\left(n\,+\,\beta\right)
R_s}{\alpha'} \right]^2} \, \bar{q}^{\frac{\alpha'}{4} \left[
\frac{m\,+\,\alpha}{R_s} \,-\, \frac{\left(n\,+\,\beta\right) R_s}{\alpha'} \right]^2}}{\eta \, \bar{\eta}}
\eeq
for the lattice sums.

Note that in this construction, the mass separation between the Bose and Fermi modes is $\Delta\,m^2 \sim \frac{1}{R^2}$. If one tried to decrease R so as to increase it, for $R \leq \sqrt{\alpha'}$ tachyons would emerge from the twisted $O_8$ contribution. Consequently, in sharp contrast with what happens in Field Theory, the string scale $\frac{1}{\sqrt{\alpha'}}$ is an upper bound on the Bose--Fermi mass separations that can be obtained, in this context, within stable string vacua. Note that the correspondence with the Scherk--Schwarz mechanism of Field Theory is only recovered in the large--radius limit, and when working in terms of $R_s$ the portions of the original lattice corresponding to even and odd windings are associated to different oscillator states. In the small--radius limit all Fermi modes disappear from the spectrum, which approaches that of the tachyonic 0B theory reduced to nine dimensions on a circle of vanishing radius $R$.

\subsection{\sc \texorpdfstring{$\Omega$}\ \ and \texorpdfstring{$\Omega P$}\ \  Scherk--Schwarz Orientifolds }

We can now illustrate how the procedure extends to the open descendants of this type--IIB compactification. To begin with, starting from eq.~\eqref{torus_ss} the standard $\Omega$ projection determines the direct--channel Klein--bottle amplitude
\beq
{\cal K}_{1} \ = \  \frac{1}{2}\int_0^\infty \frac{d\tau_2}{\tau_2^2}\,  \frac{(V_8 - S_8)}{\left(\sqrt{\tau_2}\,\eta\right)^7} \; P_{2m}(R) \ \equiv \ \frac{1}{2} \int_0^\infty \frac{d\tau_2}{\tau_2^2} \, \frac{(V_8 - S_8)}{\left(\sqrt{\tau_2}\,\eta\right)^7} \; P_{m}(R_s) \ , 
\eeq
as can be seen from the diagonal contributions to the torus partition function~\eqref{torus_shift}. In the following, for brevity, we shall omit the integrations, with the corresponding measure and the contributions of the bosonic coordinates, which are identical to those described in Section~\ref{sec:1Dtorus}.

Turning to the open sector, one can build the annulus amplitude by enforcing on the standard circle contribution of eq.~\eqref{AomegaR} a coordinate shift by $\pi R$ accompanied by $(-1)^G$, thus allowing even momenta for bosons and odd ones for fermions. The result is
\beq
{\cal A}_{1} \ = \ 
\frac{N^2}{2}
\left[ V_8\, P_{2m}(R) - S_8 P_{2m+{1} }(R)\right] \,  \ = \ \frac{N^2}{2}
\left[ V_8\, P_{m}(R_s) - S_8 \, P_{m+\frac{1}{2} }(R_s)\right] \ ,
\eeq
and in terms of $R_s$ it realizes the standard Scherk--Schwarz separation of Field Theory. The corresponding M\"obius amplitude is then
\beq
{\cal M}_{1} \ = \  - \ \frac{N}{2}
\left[ \hat V _8 \hat{P}_{2m}(R)\, -\, \hat S _8 \,
\hat{P}_{2m+1}(R) \right]\ = \  - \ \frac{N}{2}
\left[ \hat V _8 \hat{P}_m(R_s) \, -\, \hat S _8\,
\hat{P}_{m+\frac{1}{2}}(R_s)  \right]  \ , \label{eq:ss1-1}
\eeq
where
\beq
P_{m+a}(R) \ = \ \sum_{m} \ \frac{ q^{\alpha' \frac{\left(m+a\right)^2}{2 R^2}}}{
\eta\left(\frac{i \tau_2}{2}\right)} \ , \qquad  \hat{P}_{m+a}(R) \ = \ \sum_{m} \ \frac{ q^{\alpha' \frac{\left(m+a\right)^2}{2 R^2}}}{
\hat{\eta}\left( \frac{i \tau_2}{2} \,+\, \frac{1}{2}\right)} \ . \label{eq:ss1-2}
\eeq

The corresponding vacuum amplitudes, here expressed in terms of $R_s$, are
\bea
\widetilde{\cal K}_{1} &=& \frac{2^5}{2\,\sqrt{\alpha'}}\, R_s \,
(V_8- S_8) W_{2n}(R_s) \ , \nonumber \\
\widetilde{\cal A}_{{\rm 1}} &=& \frac{2^{-5} N^2}{2\,\sqrt{\alpha'}}  R_s
\Bigl[ (V_8  - S_8) W_{2n}(R_s) + (O_8  - C_8) W_{2n+1}(R_s)  \Bigr] \ , \nonumber \\
\widetilde{\cal M}_{{\rm 1}} &=& -  \ \frac{2}{2\,\sqrt{\alpha'}}\, R_s \,  N  \Bigl[
 \, \hat V_8(R_s)\,  -  \, (-1)^n \hat S_8(R_s) \Bigr]  \,
\hat{W}_{2n}  \ , \label{eq:ss1-3}
\eea
where
\beq
W_{n+a}(R) \ = \ \sum_{n} \ \frac{ q^{R^2 \frac{\left(n+a\right)^2}{2 \alpha'}}}{
\eta\left(\frac{i \tau_2}{2}\right)} \ , \qquad  \hat{W}_{n+a}(R) \ = \ \sum_{n} \ \frac{ q^{R^2 \frac{\left(n+a\right)^2}{2 \alpha'}}}{
\hat{\eta}\left( \frac{i \tau_2}{2} \,+\, \frac{1}{2}\right)} \ . \label{eq:ss1-22}
\eeq
and reflect the winding sectors present in eq.~\eqref{torus_ss}.
Note that the D9 branes couple to winding excitations of the "twisted sector" containing the scalar from the NS-NS sector $|O_8|^2$ and the RR field in $|C_8|^2$.  

The preceding partition functions, where $N=32$ on account of the tadpole condition, are tailored to the standard Scherk--Schwarz setup, and afford the standard type--I interpretation in terms of D9 branes and O9${}_-$ planes.  In particular, the annulus and M\"obius contributions display the typical Scherk-Schwarz mass separations of eq.~\eqref{ss_ft} between Bose and Fermi modes. However, as we have seen, the stability of the closed sector sets an upper bound on it, since it demands that $R > \sqrt{\alpha'}$.
One can also add Wilson lines to this construction, breaking the $SO(32)$ gauge group to various subgroups, along the lines of what we saw in Section~\ref{sec:wilson}.

There is an interesting variant of this construction that rests on the orientifold projection $\Omega\,\Pi$, where $\Pi$ is a parity operation in the internal direction, so that only states with vanishing lattice momentum contribute to the Klein--bottle amplitude
\beq
{\cal K}_{2} \ = \  \frac{1}{2} (V_8 - S_8 ) \; W_{2n}(R_s) \ \ + \ \frac{1}{2} (O_8 - C_8 ) \; W_{2n+1}(R_s) \ ,
\eeq
here written in terms of $R_s$. This option corresponds to breaking supersymmetry perpendicularly to the branes, which are D8-branes in this case. The corresponding amplitudes for the open sector are obtained starting from eqs.~\eqref{AomegaPR} and \eqref{MomegaPR}, and read
\bea
 \nonumber \\
{\cal A}_{2} &=& \frac{1}{2} (N_1^2 + N_2^2)\,
(V_8 - S_8 )\;
W_n(R_s) \ + \  N_1 N_2 \, (O_8 - C_8) \;
W_{n+\frac{1}{2}}(R_s) \ , \nonumber \\
{\cal M}_{2} &=& -\, \frac{1}{2}
 (N_1+N_2) \left[ \hat V _8  \,  \hat{W}_{n}(R_s)  \ - \ \hat S_8\, (-1)^n \hat{W}_n(R_s) \right] \ , \label{M_breaking}
\eea
where the lattice sums are defined in eq.~\eqref{eq:ss1-22}, and where the motivation for the presence of two sets of branes will become clear shortly.

The striking feature of this construction is that the massless modes, which involve the $n=0$ terms in the lattice sums, remain supersymmetric. This phenomenon was often dubbed ``brane supersymmetry", but it should be appreciated that the massive spectrum does feel the breaking of supersymmetry. The annulus amplitude clearly shows that $N_1$ D8 branes and $N_2$ $\overline{\mathrm D8}$ antibranes are present, and that the two stacks are mutually separated by a half circle in the internal space. Alternatively, one could derive these amplitudes starting from a different IIB torus amplitude where the shifts, in the untwisted sector, concern windings rather than momenta, with the standard orientifold projection.

The tree-level (or transverse-channel) amplitudes corresponding to eqs.~\eqref{M_breaking} are
\bea
\widetilde{\cal K}_{2} &=& \frac{2^5}{2}\, \frac{\sqrt{\alpha'}}{R_s} \,
\left( V_8\; P_{2m}(R_s) - S_8 \; P_{2m+1}(R_s) \right) \ , \nonumber \\
\widetilde{\cal A}_{{\rm 2}} &=& \frac{2^{-5}}{2}  \, \frac{\sqrt{\alpha'}}{R_s} \,
\Bigl[ \left(N_1+ (-1)^m N_2\right)^2 V_8 \, - \, \left(N_1-(-1^m )N_2\right)^2 \; S_8 \Bigl]
P_m(R_s) \ , \nonumber \\
\widetilde{\cal M}_{{\rm 2}} &=& -  \ \frac{2}{2}\, \frac{\sqrt{\alpha'}}{R_s} \, (N_1+N_2) \Bigl[
 \, \hat V_8\, \hat{P}_{2m}(R_s) \,-  \, \hat S_8 \,
\hat{P}_{2m+1}(R_s)  \Bigr] \ .
\eea
These amplitudes reveal that the model contains an ${\mathrm O}8_-$ plane and $N_1$ D8 branes at the origin $y=0$, and an
$\overline{\mathrm O8}{}_-$ plane and $N_2$ $\overline{\mathrm D8}$ branes at the other fixed point $y=\pi R_s$. The positions are encoded in the projectors $\frac{1}{2}\left(1\pm (-1)^m\right)$ that build $P_{2m}$ and $P_{2m+1}$, as in the $SO(16)\times SO(16)$ spectrum arising from the standard circle compactification of Section~\ref{sec:wilson}. The opposite signs of the orientifold charges are reflected in the presence of only odd momenta in the lattice sum that accompanies $S_8$ in $\widetilde{K}$. As a result, the total orientifold charge vanishes and the massive RR tadpole does not require the introduction of D8 branes. However, if one wants to cancel the tadpoles locally, which is necessary not to incur in strong coupling in the $R_s \to \infty$ limit, one must introduce both the $D8$ branes at the origin and the $\overline{D8}$ branes at $\pi R_s$.  In the partition function, this choice corresponds to setting $N_1=N_2$ and the standard RR tadpole conditions then determine $N_1=N_2=16$, while the structure of the M\"obius amplitude implies that the gauge group is $SO(16) \times SO(16)$. Note that in this case, the NS-NS tadpole condition coincides with the RR one. Consequently, there is no tadpole potential emerging at the (projective) disk level, in contrast to what we saw in Section~\ref{sec:nonsusy_10d} for brane supersymmetry breaking. In this example all tadpoles are canceled locally, and one can argue that, at the one-loop level, supersymmetry breaking effects on the branes are exponentially suppressed, since they are mediated by very heavy modes in the large radius limit. This was explicitly checked in~\cite{ads1,ads2} for the masses of Wilson lines.  Duality arguments indicate~\cite{ads1} that this model realizes the breaking of supersymmetry in the eleventh dimension of the Horava-Witten M-theory scenario~\cite{HW1,HW2}. A field theory description of this model was provided in~\cite{fh}, where the effect of its instability associated to the tachyon mode was discussed in some detail.

There are two additional orientifold projections for Scherk--Schwarz circle compactifications, which we now turn to describe.

\subsection{\sc A Tachyon--free Scherk-Schwarz Orientifold}

The first additional projection is motivated by the analogy with what happens in the 0'B theory. There is an orientifold projection that eliminates the closed--string tachyon for any radius, at the  price of introducing a global NS-NS tadpole \cite{dm-tachyonfree1,dm-tachyonfree2}. This orientifold projection is $\Omega_3 = \Omega \Pi (-1)^{F_L}$, where $(-1)^{F_L}$ is the left world-sheet fermion number. The resulting model contains $O8_{-}$ and $\overline{O8}_{+}$ O-planes, and the RR tadpole cancellation requires the introduction of D8 branes. When all branes lie on top of the $O8_{-}$ plane, the loop--channel vacuum amplitudes read
\bea
{\cal K}_{3} &=& \frac{1}{2} (V_8 - S_8 ) \; W_{2n}(R_s) \ \ - \ \frac{1}{2} (O_8 - C_8 ) \; W_{2n+1}(R_s) \ , \nonumber \\
{\cal A}_{{\rm 3}} &=& \frac{N^2}{2} \,
(V_8 - S_8 )\;
W_n(R_s)  \ , \nonumber \\
{\cal M}_{{\rm 3}} &=& -\, \frac{N}{2} \left[ (-1)^n \hat V _8  \,  \hat{W}_{n}(R_s)  \ - \ \hat S_8\, \hat{W}_n(R_s) \right]  \ . \label{eq: ss3-1}
\eea

These amplitudes reveal that the massless open--string spectrum is supersymmetric, with a gauge group $SO(32)$, while massive states originating from the propagation between the D8 branes and the distant $\overline{O8_{+}}$ plane break supersymmetry.
One can also note that the Klein--bottle amplitude interpolates between those of the type-I superstring and of the 0'B non--supersymmetric string of~\cite{as95,as97}. The corresponding tree-channel amplitudes are
\bea
\tilde{\cal K}_{3} &=& \frac{2^5}{2}\, \frac{\sqrt{\alpha'}}{R_s} \,
\left( V_8\; P_{2m+1}(R_s) - S_8 \; P_{2m}(R_s) \right) \ , \nonumber \\
\tilde{\cal A}_{{\rm 3}} &=& \frac{2^{-5}}{2} \, \frac{\sqrt{\alpha'}}{R_s} \, N^2 \left( V_8- S_8\right)
P_m(R_s) \nonumber \\
\tilde{\cal M}_{{\rm 3}} &=& -  \ \frac{2}{2}\, \frac{\sqrt{\alpha'}}{R_s} \,  N  \Bigl[
 \, \hat V_8\, \hat{P}_{2m+1}(R_s) -  \, \hat S_8 \,
\hat{P}_{2m}(R_s)  \Bigr] \ .  \label{eq: ss3-2}
\eea
The lack of a massless $V_8$ contribution in the Klein bottle amplitude of eq. (\ref{eq: ss3-2}) shows that the (non-BPS) $O8_{-}-\overline{O8}_{+}$ O-plane system has no tension, while the presence of a massless $S_8$ contribution shows that it has the same RR charge as in the type I string. Therefore, adding D8 branes is instrumental in canceling the RR charge, at the price of introducing an uncanceled NS-NS tension and thus a NS-NS tadpole potential. 

If one puts the D8 branes on top of the $\overline{O8}_{+}$ plane, the gauge group becomes
$USp (32)$, with a non-linearly realized supersymmetry in the open sector.  In addition, one can break the gauge symmetry by splitting the D8 branes in various stacks.  

\subsection{\sc  A Scherk--Schwarz Orientifold with Twisted O-Planes} \label{sec:orientifold_twisted}

There is a fourth option for the orientifold projection, which was discussed only recently in~\cite{bcd}. It is implemented by the Klein bottle amplitudes
\bea
{\cal K}_{4} &=& \frac{1}{2} \,(V_8 + S_8) \; (-1)^m P_{m}(R_s) \, , \nonumber \\
{\widetilde{{\cal K}}_{4}} &=& \frac{2^5}{2}\, \frac{R_s}{\sqrt{\alpha'}}\,(O_8 - C_8) \; W_{2n+1}(R_s) \ . \label{ss4-1}
\eea
This last construction is based on the
projection
\beq
\Omega_4 \ = \ \Omega \ (-1)^{G_L} \,\delta \ ,
\eeq
where $(-1)^{G_L}$ is the left spacetime fermion number. Note that $\Omega'$ does not square to one, but
\begin{equation}
(\Omega_4)^2 = g     \ ,
\label{ss4-2}
\end{equation}
where $g$ is the freely-acting orbifold operation used to construct the Scherk-Schwarz circle compactification of the type IIB superstring of eq.~\eqref{torus_shift}, after performing the rescaling from the orbifold basis to the SS basis. Indeed, in the SS basis
\beq
g = (-1)^{G_L}\, \delta^2 \ ,
\eeq
where
$\delta^2 y = y + 2 \pi R_s$ is  a full displacement around the inner circle.
This explains why $\Omega_4$ is not a consistent operation in the type IIB string, but it is consistent in its Scherk-Schwarz deformation.

This new orientifold projection has the peculiar feature of symmetrizing the RR sector, which therefore contains, in its massless spectrum, a zero-form and a {selfdual} four-form rather than the usual two-form. In fact, there are no massless two-forms altogether, since the NS-NS one present in the original type IIB is also removed by ${\cal K}_4$. The O9 plane thus introduced is non-BPS, and has no counterpart in the type I string. It has neither tension nor RR charge, and couples only to massive closed string states coming from the twisted sector of the original freely-acting operation. Since the non-BPS O-plane has no RR charge, the model does not require the addition of open strings, and in the minimal configuration it is fully described by the Klein bottle amplitude and the (halved) torus amplitude of eq.~\eqref{torus_ss}.

Despite the unusual O-plane that does not couple to the standard supergravity fields, the field-theory (KK) spectrum is fully captured by a standard Scherk-Schwarz supergravity reduction, using the symmetry $\Omega (-1)^{G_L}$ of the 10d IIB supergravity, whose action on the forms is inherited from String Theory. Together with the more standard $(-1)^G$, these are the only perturbative symmetries of type IIB supergravity available for a supersymmetry-breaking Scherk-Schwarz reduction to nine dimensions.  The spectrum that  {emerges} from this projection contains a surprising indication. It is well-known that the Type IIB string has an $SL(2,Z)$ strong-weak coupling S-duality symmetry, which acts on the combination
\beq
\tau \ = \  C_0 \ + \ i e^{- \phi} \ ,  \label{ss4-9}
\eeq
where $C_0$ is the axion and $\phi$ the dilaton,
as the fractional linear transformation
\beq
\tau \ \longrightarrow \ \frac{a \tau + b}{c \tau + d} \ ,  \ \label{ss4-10}
\eeq
with $a,b,c,d$ integer numbers such that $a d - b c =1$. Under S-duality, the pair of two-forms $(B_2,C_2)$ transform as a doublet while the selfdual four form $C_4^{+}$ is a singlet. The $SL(2,Z)$ symmetry is broken by the usual $\Omega$ projection of type IIB leading to the type I string, but appears intact in this non-supersymmetric orientifold. The reason is that 
\beq 
S^2 = \Omega \ (-1)^{G_L} \quad , \quad 
S^4 =  (-1)^{G}
\ , \ \label{ss4-811}
\eeq 
so the orientifold projection commutes with the S-transformation.  
If this property were confirmed, this orientifold would also possess a strong--weak coupling self--duality.
Some further arguments in favor of this duality, based on the D-brane spectrum of the model, were presented in~\cite{bcd}.  

\section{\sc Six-Dimensional Orientifolds} \label{sec:6dstrings}

We can now turn to the open descendants or orientifolds of the $T^4/\mathbb{Z}_2$
compactification of the type-IIB superstring. Partial breaking of supersymmetry and chirality are conveniently induced in perturbative string spectra via orbifold compactifications \cite{orbifolds1,orbifolds2}, where the whole perturbative string spectrum can be determined exactly. The supersymmetric orbifold compactifications are singular limits of the smooth Calabi--Yau compactifications~\cite{calabiyau} that we shall describe in Section~\ref{sec:calabi-yau}. Here we focus on the simplest setting that yields a partial breaking of supersymmetry, together with some important variants. For brevity, we shall confine our attention to the six--dimensional case, which suffices to illustrate several important novelties. The interested reader can find more details in the original works~\cite{4d1,4d2,4d3,4d4,4d5,4d6,4d7,abpss,kakudual,4d8,4d9,4d10,ibanez1,ibanez2,ibanez3,ibanez4,bert1,bert2,bert3,bert4}, in the reviews~\cite{orientifolds_rev1,orientifolds_rev2,berlinmadrid_rev,bert_rev,orientifolds_rev3,orientifolds_rev4}, and in the references contained therein. The four--dimensional constructions are somewhat more involved but do not entail new ingredients.

The orbifold idea has actually a more general algebraic rationale, and we have seen examples of this procedure even in ten dimensions, where projections and corresponding completions allow one to link different superstrings to one another. We have also seen how a variant involving internal shifts allows one to extend Scherk-Schwarz compactifications to String Theory. Six--dimensional toroidal orbifolds are highly constrained by the possible emergence of gauge and gravitational anomalies, and yet they allow a wide range of options, so that they are a very instructive playground for the construction of perturbative string spectra.

The six--dimensional models of interest in this section are obtained starting from $T^4$ compactifications and subjecting them to a ${Z}_2$ projection that acts on the two internal complex bosonic coordinates $z_1=x_6+i x_7$ and $z_2=x_8+i x_9$ as a $\pi$ rotation, $R=e^{i\pi\left(J_{67}+J_{89}\right)}$, turning them into $(-z_1,-z_2)$, and squares to one on all modes. It is thus convenient to introduce the
${\rm SO}(4) \times {\rm SO}(4)$ decomposition of the SO(8)
characters
\bea
V_8 &= \ V_4 O_4 + O_4 V_4 \, , \qquad O_8 &= \ O_4 O_4 + V_4 V_4 \, ,
\nonumber \\
S_8 &= \ C_4 C_4 + S_4 S_4 \, , \qquad C_8 &= \ S_4 C_4 + C_4 S_4 \, , \label{orbOVSC}
\eea
where the first SO(4) factor refers to the transverse space-time
directions and the second to the internal
ones. World-sheet supersymmetry demands that
the ${Z}_2$ actions on bosonic and fermionic coordinates
be properly correlated \cite{orbifolds1,orbifolds2,4d1,4d2,4d3,4d4,4d5,4d6,4d7,4d8,4d9,4d10}. This is guaranteed by the link with $\pi$ rotations, and assigns
positive eigenvalues to the internal $O_4$ and $C_4$
and negative ones to the internal $V_4$ and $S_4$, so that the four internal characters respond to the $\mathbb{Z}_2$ as
\beq
\left(O_4,C_4\right) \ \rightarrow \ \left(O_4,C_4\right) \ , \qquad \left(V_4,S_4\right) \ \rightarrow \ - \  \left(V_4,S_4\right) \ .
 \eeq
Combining bosonic and fermionic contributions, one can define the four supersymmetric characters~\cite{orientifolds5,orientifolds6}
\bea
Q_o &= \ V_4 O_4 \ - \ C_4 C_4 \, , \qquad \quad
Q_v &= \ O_4 V_4 \ -  \ S_4 S_4 \, ,
\nonumber \\
Q_s &=\  O_4 C_4 \ - \ S_4 O_4 \, , \qquad \quad
Q_c &=\  V_4 S_4 \ - \ C_4 V_4 \, ,
\eea
which encode six--dimensional $(1,0)$ supermultiplets and are eigenvectors of the ${Z}_2$ generator. For open strings, the low--lying states of $Q_0$, $Q_v$ and $Q_s$ are massless: they describe a six--dimensional vector multiplet (which comprises a vector and a Weyl spinor), a hypermultiplet (which comprises four scalars and a Weyl spinor) and a ``half-hypermultiplet'' (which comprises half of the preceding content, but states of this type will always emerge in even numbers). On the other hand, the low--lying modes of $Q_c$ are massive. In view of eqs.~\eqref{orbOVSC}, $Q_o$ and $Q_s$ are even under the orbifold action, while $Q_v$ and $Q_c$ are odd.

The massless IIB closed--string sectors that we shall shortly encounter are built out of the two possible massless $(2,0)$ multiplets:
\begin{itemize}
\item{\bf the (2,0) gravity multiplet:} graviton, five self--dual two--forms and two right Weyl gravitini;
\item{\bf the (2,0) tensor multiplet:} one antiself--dual two--form, five scalars and two left Weyl spinors.
\end{itemize}

Alternatively, one can describe the six--dimensional Weyl spinors $\psi$ present in these multiplets as $Sp(4)$ quartets of symplectic Majorana spinors $\psi^a$, subject to the constraint
\beq
\psi^a \ = \ \Omega^{ab} \ C \ \bar{\psi}_b^T \ ,
\eeq
with $\Omega$ the $Sp(4)$ symplectic form, where $Sp(4)\sim SO(5)$ is the $R$-symmetry of the $(2,0)$ supersymmetry algebra. In this respect, one can describe the content of the preceding multiplets as follows:
\begin{itemize}
\item{\bf the (2,0) gravity multiplet:} graviton, five self--dual two--forms in the fundamental of $SO(5)$ and a symplectic quartet of right Majorana--Weyl gravitini;
\item{\bf the (2,0) tensor multiplet:} one antiself--dual two--form, scalars in the fundamental of $SO(5)$ and a symplectic quartet of left Majorana--Weyl spinors.
\end{itemize}

The $\Omega$ projection will give rise to $(1,0)$ supersymmetry multiplets of the following types:
\begin{itemize}
\item{\bf the (1,0) gravity multiplet:} graviton, one self--dual two--form and one right Weyl gravitino;
\item{\bf the (1,0) tensor multiplet:} one anti-selfdual two--form, one scalar and one left Weyl spinor;
\item{\bf the (1,0) vector multiplet:} one vector and one right Weyl spinor;
\item{\bf the (1,0) hypermultiplet:} 4 scalars and one left Weyl spinor.
\end{itemize}
Alternatively, one can describe the spinors as $Sp(2)$ Majorana-Weyl doublets, emphasizing the $Sp(2)$ $R$-symmetry of the $(1,0)$ multiplets.

Note that:
\begin{itemize}
\item $|Q_o|^2$ describes the $(2,0)$ gravity multiplet and one $(2,0)$ tensor multiplet;
\item $|Q_v|^2$ describes four $(2,0)$ tensor multiplets;
\item $|Q_s|^2$ describes one $(2,0)$ tensor multiplet.
\end{itemize}

\subsection{\sc Supersymmetric \texorpdfstring{$T^4/\mathbb{Z}_2$} \ \ Orientifolds} \label{sec:susyT4}

The preceding considerations determine the
modular invariant torus amplitude for the $T^4 /{Z}_2$ orbifold compactification of the type--IIB string~\footnote{Many details on the $T^4/\mathbb{Z}_2$ orientifolds, also related to the introduction of a quantized $B_{ab}$, can be found in~\cite{cab}.},
\bea
{\cal T} &=& {\textstyle\frac{1}{2}} \Biggl[ |Q_o + Q_v|^2
\sum_{m,n} \frac{q^{\frac{\alpha'}{4} p_{{\rm L}}^{{\rm T}} G^{-1} p_{{\rm L}}} \bar{q}^{\frac{\alpha'}{4} p_{{\rm R}}^{{\rm T}} G^{-1} p_{{\rm R}}}
}{\eta^4 \bar \eta ^4 }
 + |Q_o - Q_v |^2 \left| \frac{2 \eta}{
\vartheta_2} \right|^4
\nonumber \\
&+& 16 \,|Q_s + Q_c |^2 \left| \frac{\eta}{\vartheta_4}\right|^4
+ 16 \,|Q_s - Q_c |^2 \left| \frac{\eta}{\vartheta_3} \right|^4 \Biggr]\, , \label{torusQ}
\eea
where the left and right momenta are as in eq.~(\ref{plrd}) with
vanishing $B_{ab}$,
and where the overall multiplicity of the twisted contributions reflects the presence of
16 fixed points in the orbifold. The first line in eq.~\eqref{torusQ} describes the result obtained enforcing on the toroidally compactified IIB string the $\mathbb{Z}_2$ projection that rests on the internal parity $\Pi$
\beq
\mathrm{Tr} \left[\frac{1+\Pi}{2} \ q^{L_0 - 1/2}\, q^{\bar{L}_0 - 1/2} \right] \ ,
\eeq
where the GSO projection is left implicit,
while the second line described the corresponding twisted sector demanded by modular invariance, which is confined to the 16 orbifold fixed points.

In view of eq.~\eqref{torusQ}, the massless modes are described by
\beq
{\cal T}_\mathrm{m=0} \ = \ |Q_o|^2 \ + \ |Q_v|^2 \ + \ 16 |Q_s|^2 \ ,
\eeq
and correspond to the $(2,0)$ gravitational multiplet from $\left|Q_o\right|^2$ and 21 $(2,0)$ tensor multiplets, 16 of which originate from the twisted sector. This is the anomaly--free spectrum that was first identified in~\cite{agwitt}.

As usual, the construction of the open descendants begins with the
Klein-bottle amplitude. The standard supersymmetric choice based on $\Omega$,
\beq
{\cal K} \,=\, {\textstyle \frac{1}{4}} \Biggl[ (Q_o + Q_v) \left(
\sum_m \frac{e^{-\,\pi\alpha'\tau_2\, m^{{\rm T}} G^{-1} m}}{\eta^4} +
\sum_n \frac{e^{-\,\frac{\pi \tau_2}{\alpha'} \, n^{{\rm T}} G n}}{\eta^4}
\right) \, + \,  2 \times 16 (Q_s + Q_c) \left(
\frac{\eta}{\vartheta_4} \right)^2 \Biggr] \, ,
\label{kt4s}
\eeq
where the implicit arguments are equal to $2 \,i \tau_2$,
yields a projected $(1,0)$ closed spectrum, obtained from the original $(2,0)$ spectrum by halving the 
fermionic content of the multiplets, symmetrizing their NS-NS sectors and anti-symmetrizing the RR ones. Acting on the GSO projected spectrum, the Klein--bottle amplitude ${\cal K}$ in eq.~\eqref{kt4s} computes 
\beq
\mathrm{Tr} \left[\frac{1}{2}\ \Omega \,\frac{1+\Pi}{2} \ q^{L_0 - 1/2}\, q^{\bar{L}_0 - 1/2} \right] \ ,
\eeq
and the first two terms correspond to the untwisted sector, while the last corresponds to the twisted sector. In fact, the Klein bottle completes the symmetrization of the diagonal terms in the torus amplitude, which are precisely those with zero momentum or zero winding in the lattice, in cosine combinations of the toroidal vertex operators. In the twisted sector, if one symmetrizes the NS-NS contributions and antisymmetrizes the RR ones, the diagonal portions of the last two terms in eq.~\eqref{torusQ} yield identical contributions to ${\cal K}$, which add up to
\beq
2 \times 16 \ \frac{q^\frac{1}{12}\left(Q_s+Q_c\right)}{\prod_n \left(1 - q^{n-1/2}\right)^4}  \ = \  2 \times 16 \left(Q_s+Q_c\right) \left(
\frac{\eta}{\vartheta_4} \right)^2 \ .
\eeq
The projected closed spectrum comprises the ${\cal N}= (1,0)$
gravitational multiplet, a single tensor multiplet and 20 hypermultiplets, 
16 of which originate from the twisted sector. By itself, this projected spectrum would have gravitational anomalies.

The corresponding transverse-channel
amplitude
\bea
\tilde{\cal K} &=& \frac{2^5}{4} \Biggl[ (Q_o + Q_v) \left(
v_4 \sum_n \frac{e^{-\,\frac{2 \pi \ell}{\alpha'} n^{{\rm T}} G n}}{\eta^4}
+ \frac{1}{v_4} \sum_m \frac{e^{\,2\pi\ell\alpha' m^{{\rm T}} C^{-1} m} }{\eta^4}
\right)  \nonumber \\ &+&
2 (Q_o - Q_v)  \left(
\frac{2 \eta}{\vartheta_2} \right)^2 \Biggr] \, , \label{ktilde6}
\eea
where the implicit arguments are $i \ell$ and
\beq
v_4\ = \ \sqrt{\frac{{\rm det}\ G}{(\alpha')^4}} \ , \label{v4def}
\eeq
is proportional to the
internal volume, determines the massless tadpole contributions
\bea
\tilde{\cal K}_0 &=& \frac{2^5}{4} \left[ Q_o \left(
\sqrt{v_4} + \frac{1}{\sqrt{v_4}} \right)^2  \ + \
Q_v \left(
\sqrt{v_4} - \frac{1}{\sqrt{v_4}} \right)^2
 \right] \nonumber \\
 &=& \frac{2^5}{4} \left[ \left(V_4 O_4 - C_4 C_4\right) \left(
\sqrt{v_4} + \frac{1}{\sqrt{v_4}} \right)^2  \ + \
\left(O_4 V_4 - S_4 S_4\right) \left(
\sqrt{v_4} - \frac{1}{\sqrt{v_4}} \right)^2
 \right] \,. \label{kt4st}
\eea
This expression describes the exchange of the closed sector between the orientifolds: the NS contributions reflect the couplings to the six--dimensional dilaton (contained in $Q_o$) and to the internal volume (contained in $Q_v$), while the R ones reflect the coupling to two six-forms, one of which is obtained from the ten--dimensional ten-form, and is described by $\frac{C_4 C_4+S_4 S_4}{\sqrt{v_4}}$, while the other is described by $\frac{C_4 C_4-S_4 S_4}{\sqrt{v_4}}$. One can thus see that the usual
O9$_-$ planes are supplemented
with additional O5$_-$ ones, with standard negative values for
tension and RR charge, since the two contributions associated to $Q_o$ have the same sign. Referring for simplicity to the tensions,
the two NS-NS contributions to $\widetilde{\cal K}_0$ associated to $Q_o$ and $Q_v$
are indeed the derivatives of
\be
\Delta S \ \sim \ - \  \sqrt{v_4} \ T_9\ \int d^6 x \, \sqrt{-g} \ e^{-\varphi_6} \ - \
\frac{1}{\sqrt{v_4}} \, T_5 \ \int d^6 x \, \sqrt{-g} \ e^{-\varphi_6}
\label{oplanecoupls}
\ee
with respect to deviations of the
six-dimensional dilaton $\varphi_6$ and of the internal volume $v_4$
from their background values, defined via
\be
\varphi_6 \to \varphi_6 + \delta \varphi_6 \, , \qquad
\sqrt{v_4} \to \left( 1 + \delta v \right) \, \sqrt{v_4} \, .
\ee
The meaning of (\ref{oplanecoupls})  is perhaps more transparent in terms of
the ten-dimensional dilaton, which is related to $\varphi_6$ by
\be
 v_4 \, e^{-2 \varphi_{10}} \ = \  e^{-2 \varphi_{6}} \, ,
\ee
as demanded by the compactification of the Einstein term in the string
frame. The two terms then become
\be
\Delta S \ \sim \ - \ v_4 \, \, T_9\ \int d^6 x \, \sqrt{-g} \ e^{-\varphi_{10}} \ - \
 \, T_5\ \int d^6 x \, \sqrt{-g} \ e^{-\varphi_{10}} \, , \label{oplanecoupls2}
\ee
and clearly refer to O9 and O5 planes, since the fist originates from the whole internal space while the second is localized, and determine
precisely their relative tensions.

We can now turn to the open sector accompanying the Klein-bottle
amplitude of eq.~(\ref{kt4s}). Here we shall focus on the simplest option, which corresponds to
introducing $D$ D5 branes sitting at a given fixed point and
$N$ D9 branes with no Wilson lines. As we shall see, all their contributions are needed to cancel the RR tadpoles introduced by $\widetilde{\cal K}$. 
This configuration reflects a trace over open--string states for the supersymmetric $\mathbb{Z}_2$ orbifold based on the GSO--projected sectors described by $Q_o$, $Q_v$, $Q_s$ and $Q_c$. 

The annulus amplitude reads
\bea
{\cal A} &=& {\textstyle \frac{1}{4}} \Biggl[ (Q_o + Q_v) \left(
N^2 \sum_m \frac{q^{\frac{\alpha'}{2} m^{{\rm T}} G^{-1} m} }{\eta^4}
\ + \ D^2 \sum_n \frac{q^{\frac{1}{2\alpha '} n^{{\rm T}} G n} }{\eta^4}
\right)
\nonumber \\
&+&
\left(R_N^2 + R_D^2 \right) (Q_o - Q_v) \left( \frac{2\eta}{
\vartheta_2}\right)^2 \label{annz2orb}
\\
&+& 2 N D \, (Q_s + Q_c ) \left( \frac{\eta}{\vartheta_4}\right)^2
\ + \ 2 R_N R_D \, (Q_s - Q_c ) \left( \frac{\eta}{\vartheta_3}\right)^2 
\Biggr] \, , \nonumber
\eea
and we can now interpret the different terms. To begin with, let us recall that $Q_o$ contains a vector multiplet, which includes the spacetime components of the original ten-dimensional vector, while $Q_v$ contains a hypermultiplet, which includes its internal components. The orbifold projection leaves $Q_o$ invariant and flips the sign of $Q_v$, as can be seen from the second line, where $R_N$ and $R_D$ define the corresponding actions on the Chan-Paton charges. The two lattice sums involve internal momenta and windings, and reflect the presence of both Neumann--Neumann (NN) and Dirichlet--Dirichlet (DD) strings in this construction, but still carry an overall factor $\frac{1}{4}$, rather than $\frac{1}{2}$ as in the toroidal case, consistently with the presence of both orientifold and orbifold projections, which halve the number of leftover operators away from the origin of the lattice with respect to the toroidal case. Note also that the terms completing the projections for the NN and DD strings are accompanied by the factor $\left( \frac{2\eta}{
\vartheta_2}\right)^2$, whose numerator compensates the overall 2 present in $\vartheta_2$, as needed to describe a single sector.  The last line in ${\cal A}$ concerns Neumann--Dirichlet (ND) strings, together with the corresponding orbifold action described by the $R_N R_D$ terms.

The transverse-channel amplitude reads
\bea
\tilde {\cal A} &=& \frac{2^{-5}}{4} \Biggl[ (Q_o + Q_v) \left(
N^2 v_4 \sum_n \frac{q^{\frac{1}{4\alpha '} n^{{\rm T}} G n} }{\eta^4}
 + \frac{D^2}{v_4} \sum_m \frac{q^{\frac{\alpha'}{4} m^{{\rm T}} G^{-1} m}
}{\eta^4} \right) \nonumber \\
&+&
2 N D \, (Q_o - Q_v ) \left( \frac{2 \eta}{\vartheta_2}\right)^2
\ + \  16 \left(R_N^2 + R_D^2 \right) (Q_s + Q_c) \left( \frac{\eta}{
\vartheta_4}\right)^2 \nonumber \\
&-& 2 \times 4 R_N R_D \, (Q_s - Q_c )
\left( \frac{\eta}{\vartheta_3}\right)^2
\Biggr] \label{Atilde6} \ ,
\eea
where the factor 16 accompanying $Q_s+Q_c$ reflects the number of fixed points: when transforming ${\cal A}$ it originates from two combined sources, the lack of internal momenta or windings in this sector and the factors of two originally accompanying $\vartheta_2$, which is turned into $\vartheta_4$ by the $S$ modular transformation connecting direct and transverse channels. 
Note that the transverse annulus amplitude thus obtained is consistent, since it propagates states that are present in the closed--string spectrum determined by eq.~\eqref{torusQ}.
One can extract from $\widetilde{\cal A}$ the massless tadpole contributions
\bea
\tilde {\cal A}_0 &=& \frac{2^{-5}}{4} \Biggl\{ Q_o \left(
N \sqrt{v_4} + \frac{D}{\sqrt{v_4}} \right)^2 +
Q_v \left( N \sqrt{v_4} - \frac{D}{\sqrt{v_4}} \right)^2
\nonumber \\
&+&  Q_s \left[ 15 R_N^2 + \left(R_N - 4 R_D \right)^2 \right]
\Biggr\} \ , \label{atilde60}
\eea
where the second line reflects the presence of 15 fixed points without D5 branes and a single one hosting all of them. Displaying the NS and R portions of the characters gives
\bea
\tilde {\cal A}_0 &=& \frac{2^{-5}}{4} \Biggl\{ \left(V_4 O_4 - C_4 C_4\right) \left(
N \sqrt{v_4} + \frac{D}{\sqrt{v_4}} \right)^2 +
\left(O_4 V_4 - S_4 S_4\right)  \left( N \sqrt{v_4} - \frac{D}{\sqrt{v_4}} \right)^2
\nonumber \\
&+&  \left(O_4 C_4 - S_4 O_4\right) \left[ 15 R_N^2 + \left(R_N - 4 R_D \right)^2 \right]
\Biggr\} \ , \label{atilde6}
\eea
and from the coefficient of $V_4 O_4$ (or $O_4 V_4$, as we have seen) one can read that the branes have tensions of the same sign, while from the coefficient of $C_4 C_4$, to which they couple with the same sign (or $S_4 S_4$, to which they couple with opposite signs) one can see that the two types of branes also have RR charges of the same sign, as pertains to BPS D9's and D5's. 

The two transverse amplitudes of eqs.~\eqref{ktilde6} and \eqref{atilde6} determine the transverse--channel M\"obius amplitude
\bea
\tilde{\cal M} &=& -\ \frac{1}{2} \Biggl[
(\hat Q_o + \hat Q_v ) \Biggl(
N v_4 \sum_n \frac{q^{\frac{1}{\alpha'} n^{{\rm T}} G n}}{\hat\eta^4}
+ \frac{D}{v_4} \sum_m
\frac{q^{\alpha ' m^{{\rm T}} G^{-1} m} }{\hat\eta^4}
\Biggr)  \nonumber \\ &+&
\left( N +  D\right)
(\hat Q_o - \hat Q_v ) \left( \frac{2\hat \eta}{\hat \vartheta_2}\right)^2
\Biggr] \, , 
\eea
which is strongly constrained by the structure of the zero modes in eqs.~\eqref{kt4st} and \eqref{atilde6}, and a $P$ transformation finally determines the M\"obius amplitude
\bea
{\cal M} &=& -\ {\frac{1}{4}} \Biggl[ (\hat Q _o + \hat Q _v ) \Biggl(
N \sum_m \frac{q^{\frac{\alpha'}{2} m^{{\rm T}} G^{-1} m} }{\hat\eta^4}
+ D \sum_n \frac{q^{\frac{1}{2\alpha '} n^{{\rm T}} G n} }{
\hat\eta^4} \Biggr) \nonumber \\ &-& \left( N + D \right)
(\hat Q _o - \hat Q _v )
\left( \frac{2\hat\eta}{\hat\vartheta_2} \right)^2 \Biggr] \ .
\nonumber
\eea

Note that there is no contribution from the origin of the lattice involving $Q_o$, so the gauge vectors are not symmetrized by ${\cal M}$. This indicates that the gauge groups are unitary, and the proper parametrization, 
\bea
N &= \ n \ + \ \bar n \, ,
\qquad \quad R_N &= i (n- \bar n) \, ,
\nonumber \\
D &= \ d \ + \ \bar d \, ,
\qquad \quad
R_D &= \ i (d - \bar d ) \, ,
\eea
is tailored to describe representations of $U(n) \times U(d)$.
The resulting massless open spectrum determined by ${\cal A}$ and ${\cal M}$ is captured by
\bea
 \left(n {\bar n} \ + \  d {\bar d}\right) Q_o &+&  \left[\frac{n(n-1)}{2} \ + \ \frac{{\bar n}({\bar n}-1)}{2} \ + \ \frac{d(d-1)}{2} \ + \ \frac{{\bar d}({\bar d}-1)}{2} \right] Q_v \nonumber \\  &+& \left(n {\bar d} \ + \ {\bar n} d \right) Q_s \ .
\eea
There are thus vector multiplets in the adjoint, hypermultiplets from the untwisted sector in the antisymmetric representations and their conjugates and, finally, the contributions involving the half--hypermultiplet in $Q_s$ combine to yield a hypermultiplet in the bi--fundamental. The values of $n$ and $d$ are determined by tadpole cancellation.
The total massless tadpole terms from $\widetilde{\cal K}$, $\widetilde{\cal A}$ and $\widetilde{\cal M}$ are proportional to
\bea
&& Q_o \left[
\left(n+{\bar n}-32\right) \sqrt{v_4} \ + \ \frac{d+{\bar d}-32}{\sqrt{v_4}} \right]^2 \ + \ 
Q_v \left[
\left(n+{\bar n}-32\right) \sqrt{v_4} \ -\  \frac{d+{\bar d}-32}{\sqrt{v_4}} \right]^2
\nonumber \\
&-&  Q_s \left[ 15 \left(n-{\bar n}\right)^2 + \left(n-4 d - {\bar n} + 4 {\bar d} \right)^2 \right], 
\eea 
so that the coefficients in $Q_s$ must vanish identically and give the conditions $n={\bar n}$ and $d={\bar d}$, while the rest gives $n = 16$ and $d =16$. Therefore, the gauge group is $U(16) \times U(16)$, with NN and DD hypermultiplets in the $(120+\overline{120},1)$ and $(1,120+\overline{120})$, together with ND hypermultiplets in the $(16,\overline{16})$.

This spectrum, first derived in \cite{orientifolds6} in a rational CFT setup, was later recovered in \cite{gimpol}. 
It is free of all irreducible gravitational and gauge anomalies, as a
result of tadpole cancellation \cite{pc1,pc2,pc3,pc4,pc5}, and in particular it satisfies the constraint
\beq
n_H \ - \ n_V \ + \ 29 \, n_T \ = \ 273 \ , \label{grav_anomaly_6}
\eeq
linking the number of massless hypermultiplets with those of massless vector and tensor multiplets. Additional
reducible non-Abelian anomalies are eliminated by a conventional
Green-Schwarz mechanism involving a single two-form, whose selfdual and anti-selfdual
parts originate from the gravitational multiplet and from the single
untwisted tensor multiplet present in the model \cite{gs,six}~\footnote{The introduction of a quantized $B_{ab}$ brings along a number of tensor multiplets, whose couplings conspire to grant the cancellation of reducible anomalies, as proposed in~\cite{orientifolds8}. Different formulations of the low--energy supergravity can be related to consistent and covariant anomalies. The detailed construction of the unconventional low--energy supergravity is discussed in~\cite{fabio1,fabio2,fabio3,fabio4,fabio5}.}. Vacuum expectation values for the scalars in the $(120 + \overline{120} ,1)$, which reflect the introduction of Wilson lines, or the scalars in the $(1, 120 + \overline{120})$, which reflect D5-brane displacements along the lines of what we saw in previous sections, can break the two unitary groups into symplectic ones, with a consequent rank reduction, as discussed in detail in~\cite{orientifolds_rev2,bsb2}.

In this orbifold the Klein--bottle projection of eq.~(\ref{kt4s}) is not the only option. There is another choice that affects momentum and winding lattices in a way similar to what in Section~\ref{sec:wilson} eliminated the massless tadpole,
\bea
{\cal K} &=& {\textstyle \frac{1}{4}} \Biggl[ (Q_o + Q_v) \left(
\sum_m \, e^{i \pi {m}^T {\epsilon}_1} \, \frac{q^{\frac{\alpha'}{2} m^{{\rm T}} G^{-1} m}}{\eta^4}
\ + \ \sum_n e^{i \pi {n}^T{\epsilon}_2} \,
\frac{q^{\frac{1}{2\alpha'} n^{{\rm T}} G n}}{\eta^4}
\right)
\nonumber \\
&+& 2 \,\times (8-8) (Q_s + Q_c)
\left(
\frac{\eta}{\vartheta_4} \right)^2 \Biggr] \ .
\label{kt4c}
\eea
The absence of a net twisted--sector contribution is reflected by the replacement of $8+8$ with $8-8$, and can be ascribed to the replacement of an $O5-$ with an $O5+$ at eight of the fixed points.  
Here ${\epsilon}_1$ and ${\epsilon}_2$ are two non--vanishing four--dimensional vectors with components equal to zero or one but otherwise arbitrary. 

All these options are compatible with supersymmetry and do not need an open sector. They result in identical massless anomaly-free ${\cal N}= (1,0)$ {\it closed} spectra~\cite{dp,gepner} that comprise, together with the
gravitational
multiplet, nine tensor multiplets, eight of which originate from the
twisted sector, and twelve hypermultiplets, eight of which originate from the
twisted sector~\cite{poltens}. One can verify that with 9 tensor multiplets and 12 hypermultiplets eq.~\eqref{grav_anomaly_6} is satisfied, so that a generalized Green--Schwarz mechanism as in~\cite{orientifolds8} can eliminate the residual reducible contributions to the gravitational anomaly. 

The new Klein bottle has replaced a hypermultiplet with a tensor multiplet at eight of the fixed points. The corresponding transverse channel amplitude is
\beq
\tilde{\cal K} \ = \  \frac{2^5}{4} (Q_o + Q_v)
\Biggl( v_4 \sum_n \frac{q^{\frac{1}{\alpha'} (n+\frac{\epsilon_1}{2})^{{\rm T}} G
(n+\frac{\epsilon_1}{2})}}{\eta^4} \ + \
\frac{1}{v_4} \sum_m \frac{q^{\alpha' (m+\frac{\epsilon_2}{2})^{{\rm T}}
G^{-1} (m+\frac{\epsilon_2}{2})}}{\eta^4} \, , \label{Ktilde6}
\Biggr)
\eeq
and has only massive contributions, since the vectors $\epsilon_1$ and $\epsilon_2$ are not vanishing. Consequently, no massless tadpoles are
generated, and no open sector is needed, as in the nine--dimensional model discussed in Section~\ref{sec:1Dtorus}. These examples also indicate that 05${}_+$ planes imply the emergence of twisted tensor multiplets, while 05${}_-$ planes imply the emergence of hypermultiplets. 

Several generalizations of these supersymmetric constructions, in six and four dimensions, are discussed in~\cite{orientifolds_rev2} and in many works devoted to potential applications to Particle Physics, including~\cite{ibanez1,ibanez2,ibanez3,ibanez4,bert1,bert2,bert3,bert4,bert_rev,abpss,kakudual}, but we leave out further details to the references. We can now turn to illustrate the differences between this case and a variant where supersymmetry is broken.

\vskip 12pt
\subsection{\sc Brane Supersymmetry Breaking}

The $T^4/\mathbb{Z}_2$ orbifold compactification allows for a third type of consistent Klein--bottle projection, 
\beq
{\cal K} \ = \  {\textstyle \frac{1}{4}} \Biggl[ (Q_o + Q_v) \left(
\sum_m {q^{{\alpha'\over 2} m^{{\rm T}} g^{-1} m} \over \eta^4}
\ + \ \sum_n {q^{{1\over 2\alpha'} n^{{\rm T}} g n} \over \eta^4}
\right) \ - \  2 \,\times 16 (Q_s + Q_c)
\left(
\frac{\eta}{\vartheta_4} \right)^2 \Biggr] \, ,
\label{kt4bsb}
\eeq
where the sign of $\Omega$ is flipped, with respect to eq.~\eqref{kt4s}, in the whole twisted sector. As a result, the projected closed spectrum is still supersymmetric, and contains the $(1,0)$ gravitational multiplet and 17 tensor multiplets, 16 of which originate from twisted contributions, together with four untwisted hypermultiplets. This content is anomalous, as it violates eq.~\eqref{grav_anomaly_6}, so an open sector is needed in this case. However, as we shall see shortly, it is inevitably \emph{non-supersymmetric} \cite{bsb1,bsb2,bsb3,bsb4,cab,au}. In our discussion of ten-dimensional models, we have already encountered a surprising phenomenon of this type: in the USp(32) model, a
projected closed sector with a residual amount of
supersymmetry is tied to an open sector where
supersymmetry is broken at the string scale \cite{sugimoto}.  Whereas in ten dimensions this choice was an alternative to the supersymmetric one resulting in the SO(32) type--I superstring, here a non--linear realization supersymmetry~\cite{dmnonlinear1,dmnonlinear2,dmnonlinear3} is inevitable.
This surprising setup, where supersymmetry appears non--linearly realized in the low--energy effective field theory, is often referred to as ``brane supersymmetry breaking''.
In the ten--dimensional case, the phenomenon was ascribed
to the replacement of the conventional ${\rm O9}_-$ plane with an ${\rm O9}_+$. As a result, RR tadpole cancellation required
D9 antibranes, with a consequent breaking of supersymmetry. In
lower-dimensional models with ${Z}_2$ orbifold projections, the simultaneous
presence of O9 and O5 planes allows for additional possibilities. The first
option, directly related to the ten-dimensional example, would be to reverse
simultaneously tensions and charges of both O9 and O5 planes. This choice,
consistent with the standard Klein-bottle projection, would
not alter the supersymmetric closed spectrum, but
the reversed RR charges would call for the introduction
of antibranes, with the end result that supersymmetry would be
broken in the whole open sector. Models with ${Z}_2$ orbifold projections, however,
allow another option \cite{bsb1,bsb2,bsb3,bsb4}: one can reverse tension
and charge of only one type of orientifold plane, say the O5. This
option lies behind the Klein-bottle projection of eq.~\eqref{kt4bsb} and requires the introduction of D5 antibranes,
where the supersymmetry preserved by the D9 branes
is thus broken at the string scale. The origin
of the breaking is along the lines of what happens in ten dimensions: D5 antibranes and O5$_+$ planes break
two different halves of the original supersymmetry, and therefore
when they are simultaneously present no residual supersymmetry is
left.

Turning the amplitude in eq.~\eqref{kt4bsb} to the vacuum channel, one can identify the tadpole contributions
\bea
\tilde{\cal K}_0 &=& \frac{2^5}{4} \left[ Q_o \left( \sqrt{v_4}  -
\frac{1}{\sqrt{v_4}}\right)^2 \ + \ Q_v \left( \sqrt{v_4}  +
\frac{1}{\sqrt{v_4}}\right)^2 \right] \nonumber \\ &=& 
\frac{2^5}{4} \left[ \left(V_4 O_4 - C_4 C_4\right) \left( \sqrt{v_4}  -
\frac{1}{\sqrt{v_4}}\right)^2 \ + \ \left(O_4 V_4 - S_4 S_4\right) \left( \sqrt{v_4}  +
\frac{1}{\sqrt{v_4}}\right)^2 \right]
\label{bsb3} ,
\eea
whose coefficients are as usual perfect squares. Comparing with the amplitude of eq.~\eqref{kt4st} clearly reveals that
the two types of O-planes are BPS O$9_-$ (with $T_9$ and $Q_9 <0$) and O$5_+$ (with $T_5$ and $Q_5 >0$). 
In order to cancel the RR charges, one is thus forced to introduce an open sector originating from D9 branes and D5 antibranes.
In particular, the untwisted terms in ${\widetilde{\cal A}}_0$ must be of the form
\bea
\tilde{\cal A}_0  &=&  \frac{2^{-5}}{4} \left[ (V_4 O_4 - S_4 S_4) \left( N
\sqrt{v_4}  +
\frac{D}{\sqrt{v_4}}\right)^2 \ + \   (O_4 V_4 - C_4 C_4) \left( N\sqrt{v_4}  -
\frac{D}{\sqrt{v_4}}\right)^2 \right]\nonumber \\
&=& \frac{2^{-5}}{4} \left[ \left(Q_o+ Q_v\right)\left(v_4 N^2 + \frac{D^2}{v_4}\right) \ + \ 2 N D\left(V_4 O_4 - O_4 V_4 - S_4 S_4 + C_4 C_4\right)\right]
\ , \label{bsb5_2}
\eea
which suffices to determine the complete contributions from the internal lattice, and then, after a modular transformation, the complete annulus amplitude is
\bea
{\cal A} &=& {\textstyle \frac{1}{4}}
\Biggl[(Q_o + Q_v) \left( N^2 \sum_m \frac{q^{\frac{\alpha'}{2} m^{{\rm T}} G^{-1} m} }{\eta^4}  + D^2 \sum_n \frac{q^{\frac{1}{2\alpha '} n^{{\rm T}} G n} }{\eta^4} \right) \nonumber \\  &+& (R_N^2 + R_D^2) (Q_o - Q_v) {\left(\frac{2
\eta}{\vartheta_2}\right)}^2 \,+\,
2 N D ( O_4 S_4 - C_4 O_4 + V_4 C_4 -
S_4 V_4) {\left(\frac{\eta}{\vartheta_4}\right)}^2
\nonumber
\\
&+& 2 R_N R_D ( - O_4 S_4 - C_4 O_4 + V_4 C_4 +
S_4 V_4 ){\left(\frac{
\eta}{\vartheta_3}\right)}^2 \Biggr]  \ , \label{bsb_4}
\eea
where the projection terms involving $R_N$ and $R_D$ are determined by the orbifold action on the corresponding terms involving $N$ and $D$. The corresponding $\widetilde{\cal A}$, obtained including the terms involving $R_N$ and $R_D$, is
\bea
\tilde {\cal A} &=& \frac{2^{-5}}{4} \Biggl[ (Q_o + Q_v) \left(
N^2 v_4 \sum_n \frac{q^{\frac{1}{4\alpha '} n^{{\rm T}} G n} }{\eta^4}
 + \frac{D^2}{v_4} \sum_m \frac{q^{\frac{\alpha'}{4} m^{{\rm T}} G^{-1} m}
}{\eta^4} \right) \nonumber \\
&+&
2 N D \, \left(V_4 O_4 - O_4 V_4 + C_4 C_4 -S_4 S_4 \right) \left( \frac{2 \eta}{\vartheta_2}\right)^2
\nonumber \\ &+&  16 \left(R_N^2 + R_D^2 \right) (Q_s + Q_c) \left( \frac{\eta}{
\vartheta_4}\right)^2 \nonumber \\ &-& 2 \times 4 R_N R_D \, (O_4 C_4 + S_4 O_4 - V_4 S_4 - C_4 V_4)
\left( \frac{\eta}{\vartheta_3}\right)^2
\Biggr] \label{Atilde6c} \ ,
\eea
and the complete $\tilde {\cal A}_0$ is then
\bea
\tilde{\cal A}_0 
&=& \frac{2^{-5}}{4} \Big\{ \left(Q_o+ Q_v\right)\left(v_4 N^2 + \frac{D^2}{v_4}\right) \ + \ 2 N D\left(V_4 O_4 - O_4 V_4 - S_4 S_4 + C_4 C_4\right) \nonumber \\
&+& \left[ 15 \,R_N^2 \ + \ \left(R_N - 4 R_D\right)^2 \right] O_4 C_4 \ - \ \left[ 15 \,R_N^2 \ + \ \left(R_N + 4 R_D\right)^2 \right] S_4 O_4 \Big\}
\ . \label{bsb5_21}
\eea
The tadpole conditions for the twisted RR sector can only originate from $\widetilde{\cal A}$ and demand that
\beq
R_N \ = \ 0 \ , \qquad R_D \ = \ 0  \ .
\eeq
If these conditions hold, the twisted NS-NS tadpoles also disappear.
As we saw in eq.~\eqref{bsb3}, these types of contributions are absent in $\widetilde{\cal K}$, and therefore must also be absent in $\widetilde{\cal M}$.

Finally, the contributions to the transverse--channel  M\"obius amplitude from
the origin of the internal lattices
\bea
\tilde{\cal M}_0 &=&  {\frac{2}{4}} \Biggl[ \,-\, \hat{V}_4
\hat{O}_4 \left( \sqrt{v_4}  -
\frac{1}{\sqrt{v_4}}\right) \left( N \sqrt{v_4}  +
\frac{D}{\sqrt{v_4}}\right)  -  \hat{O}_4
\hat{V}_4 \left( \sqrt{v_4}  +
\frac{1}{\sqrt{v_4}}\right) \left( N \sqrt{v_4}  -
\frac{D}{\sqrt{v_4}}\right) \nonumber \\
\!\!\!\!&+&\!\!\! \hat{C}_4 \hat{C}_4 \left(
\sqrt{v_4}  -
\frac{1}{\sqrt{v_4}}\right) \left( N \sqrt{v_4}  -
\frac{D}{\sqrt{v_4}}\right)  -  \hat{S}_4 \hat{S}_4 \left( \sqrt{v_4}  +
\frac{1}{\sqrt{v_4}}\right) \left( N \sqrt{v_4}  +
\frac{D}{\sqrt{v_4}}\right)
\Biggr] , \label{bsb6}
\eea
where the implicit arguments are $i \ell+\frac{1}{2}$,
are obtained combining the different contributions from
${\tilde {\cal K}_0}$ of eq.~\eqref{bsb3} and ${\tilde {\cal A}_0}$ of eq.~\eqref{bsb5_2}. These contributions suffice again to determine the full lattice sums and then, after a $P$ transformation, determine the complete M{\"o}bius amplitude
\bea
{\cal M} &=& -\  {\frac{1}{4}} \Biggl[ N  ( \hat{O}_4
\hat{V}_4  + \hat{V}_4 \hat{O}_4  - \hat{S}_4 \hat{S}_4 - \hat{C}_4
\hat{C}_4 ) \sum_m  \frac{q^{\frac{\alpha'}{2} m^{{\rm T}} G^{-1} m} }{\hat{\eta}^4}  \nonumber \\
&-&  D ( \hat{O}_4
\hat{V}_4  + \hat{V}_4 \hat{O}_4  + \hat{S}_4 \hat{S}_4 + \hat{C}_4
\hat{C}_4 ) \sum_n \frac{q^{\frac{1}{2\alpha '} n^{{\rm T}} G n} }{\hat{\eta}^4}   \nonumber \\
&-& N(
\hat{O}_4 \hat{V}_4 - \hat{V}_4 \hat{O}_4 - \hat{S}_4 \hat{S}_4
+ \hat{C}_4 \hat{C}_4 )\left(
{2{\hat{\eta}}\over{\hat{\vartheta}}_2}\right)^2  \nonumber \\
&+&  D( \hat{O}_4
\hat{V}_4 - \hat{V}_4 \hat{O}_4 + \hat{S}_4 \hat{S}_4
- \hat{C}_4 \hat{C}_4)\left(
{2{\hat{\eta}}\over{\hat{\vartheta}}_2}\right)^2  \Biggr] \ ,
\label{bsb7}
\eea
where the implicit arguments are now $i \frac{\tau_2}{2}+\frac{1}{2}$.

Since the vectors flow in ${\cal M}$, which is signaled by the $V_4 O_4$ terms including contributions from the origin of the lattices, one is led to
introduce real Chan-Paton multiplicities, so that
\bea  N&=\ n_1+ n_2 \ , \qquad D&=\ d_1+ d_2 \ , \nonumber \\
R_N&=\ n_1- n_2 \ , \qquad R_D&=\ d_1- d_2 \ . \label{bsb8}
\eea
The orbifold action on the Chan-Paton charges in~\eqref{bsb_4} implies that the D-branes present in this model have physical couplings to the twisted tensors at the orbifold fixed points, which renders them fractional branes in due right~\footnote{This solution, originally considered in \cite{bsb1,bsb2,bsb3,bsb4}, rests on the simplest option, since the twisted RR tadpoles are canceled by adding equal numbers of D9 branes (D5 antibranes) with positive and negative twisted charges. More general solutions, in which nontrivial twisted charge cancellations occur between D9 and D5 antibranes, were recently discussed in \cite{acdl}.}

The untwisted RR tadpole contributions that result from $\widetilde{\cal K}{}_0$, $\widetilde{\cal A}{}_0$ and $\widetilde{\cal M}{}_0$ are
\beq
\frac{2^{-5}}{4} \left\{ - C_4 C_4 \left[\left(32-N\right) \sqrt{v_4} \ - \ \frac{32-D}{\sqrt{v_4}} \right]^2 \ -\  S_4 S_4 \left[\left(32-N\right) \sqrt{v_4} \ + \ \frac{32-D}{\sqrt{v_4}} \right]^2 \right\}\ ,
\eeq
and cancel if $N=32$ and $D=32$.
The resulting massless open spectrum is summarized in
\bea
{\cal A}_0 + {\cal M}_0 &=& \frac{n_1(n_1-1) + n_2(n_2-1) +
d_1(d_1+1) +  d_2(d_2+1) }{2} \ V_4 O_4 \nonumber \\
 &&- \frac{n_1(n_1-1) + n_2(n_2-1) + d_1(d_1-1) + d_2(d_2-1) }{2} \ C_4
C_4 \nonumber \\ &&+ (n_1 n_2 + d_1 d_2 ) ( O_4 V_4 - S_4 S_4 ) + (
n_1 d_2 + n_2 d_1 ) \ O_4 S_4 \nonumber \\
&& - (n_1 d_1 + n_2 d_2 ) \ C_4 O_4 \,.
\label{bsb9}
\eea

 Taking into account the RR tadpole conditions $N=D=32,R_N=R_D=0$
($n_1=n_2=d_1=d_2=16$) into account determines the gauge group $[ {\rm SO}(16) \times
{\rm SO}(16) ]_9 \times  [ {\rm USp}(16) \times {\rm USp}(16) ]_5$,
where the subscripts refer to the D9 and $\overline{{\rm D} 5}$ branes.
The NN spectrum is supersymmetric and comprises the (1,0)
vector multiplet for the ${\rm SO}(16) \times {\rm SO}(16)$ gauge group and a
hypermultiplet in $(16,16,1,1)$.
On the other hand, the DD spectrum is
not supersymmetric and contains, aside from the gauge vectors of $[
{\rm USp}(16)
\times {\rm USp}(16) ]$, quartets of scalars in the $(1,1,16,16)$,
right-handed Weyl fermions in the $(1,1,120,1)$ and in the $(1,1,1,120)$,
and left-handed Weyl fermions in the
$(1,1,16,16)$. Finally, the ND sector, also not supersymmetric,
comprises doublets of scalars in the $(16,1,1,16)$ and in
the $(1,16,16,1)$, together with additional symplectic
Majorana-Weyl fermions in the $(16,1,16,1)$ and $(1,16,1,16)$.
As we have seen, these Majorana-Weyl fermions
are a peculiar feature of
six-dimensional spacetime, where the fundamental Weyl fermion can be subjected to an additional Majorana condition,  if
this is supplemented by the conjugation in a pseudo-real representation of the gauge group
\cite{romans,witsmall}. In this case, this is indeed possible, since the ND
fermions are valued in the pseudo--real fundamental representation of ${\rm USp}(16)$.

Note that the $\overline{{\rm D} 5}$
spectrum possesses some features that have already emerged in
the discussion of the ten-dimensional USp(32) model of~\cite{sugimoto}, so that all bosonic
and fermionic modes affected by the M\"obius projection are valued in different
representations, while the remaining NN and DD matter in bi-fundamental
representations fills complete hypermultiplets. The novelty here
is the ND sector, where supersymmetry is broken due to the reversed GSO
projection resulting from brane-antibrane exchanges. As in the ten-dimensional
model of \cite{sugimoto}, the open spectrum contains singlet spinors, one combination of which is the Volkov-Akulov goldstino and plays a key role in the counterparts of the low-energy couplings discussed in \cite{dmnonlinear1,dmnonlinear2,dmnonlinear3}.

As is typically the case for non-supersymmetric orientifolds, a dilaton potential,
here localized on the $\overline{{\rm D} 5}$ branes, is generated. This can be
easily deduced from the transverse-channel amplitudes, which
in general encode the one-point functions of bulk fields on branes
and orientifold planes. In this case, the uncanceled tadpoles are determined by
\be
\left[ (N-32){\sqrt v_4}\,+\,{D+32 \over{\sqrt
v_4}}\right]^2 \! V_4O_4 \ + \ \left[ (N-32){\sqrt v_4}\,-\,{D+32 \over{\sqrt
v_4}}\right]^2 \! O_4V_4
 \label{bsb10}
\ee
after setting $N=32$ and $D=32$, and thus originate solely from the D5-O5 system. They are associated with the characters $V_4O_4$ and $O_4V_4$, and thus
with the deviations of the six-dimensional dilaton $\varphi_6$ and
of the internal volume $v_4$ with respect to their background
values. In the string frame, the residual potential reads
\be
V_{\rm eff}\ = \ c\,{e^{-\varphi_6}\over{\sqrt v}}\ = \ c\,e^{-\varphi_{10}} \ , \label{bsb11}
\ee
where we have also expressed the result in terms of
$\varphi_{10}$, the ten-dimensional dilaton, and where
$c$ is a {\it positive}
numerical  constant. The potential (\ref{bsb11}) is in fact localized in the $\overline{{\rm D} 5}$'s and is clearly positive. This can be understood by
noting that
the negative contribution of the O9 plane to the vacuum energy
exactly cancels against that of the D9 branes for $N=32$, and this fixes
the sign of the $\overline{{\rm D} 5}$ and ${\rm O}5_+$ contributions,
both positive, consistent with the interpretation of this mechanism as
supersymmetry breaking on the $\overline{{\rm D} 5}$ branes. The potential (\ref{bsb11}) has the usual
runaway behavior, as expected by general arguments. Potentials of this type can have interesting effects, which will be analyzed in the following sections.

String vacua with brane supersymmetry breaking and a quantized $B_{ab}$, containing different numbers of tensor multiplets, are discussed in~\cite{orientifolds_rev2}, and more examples were recently added in~\cite{acdl}, for all possible six-dimensional orbifolds. These include cases where twisted tadpoles are canceled between D9 and $\overline{{\rm D} 5}$ branes, in configurations that are more rigid, since the anti-brane positions are fixed. In addition to the usual untwisted NS-NS tadpoles, these models also lead to the emergence of twisted NS-NS tadpoles.

As in the higher-dimensional examples, the configuration that we have described can be enriched by adding brane-antibrane pairs \cite{au,bsb1,bsb2,bsb3,bsb4}.
This leads in general to instabilities,
which manifest themselves via the emergence of tachyonic modes, but in some cases one can follow the adjustment of these systems~\cite{abg}.

\subsection{\sc Magnetic Deformations} \label{sec:magnetic_susyb}

Homogeneous internal magnetic fields are interesting deformations
compatible with two-dimen\-sion\-al conformal invariance. They do not modify the closed--string spectrum but have interesting effects on the annulus and M\"obius amplitudes, since they can
modify the boundary conditions at the ends of the string~\cite{ftse,acny}. They also provide an interesting
way to break supersymmetry in four--dimensional open--string vacua
\cite{wito32,bachas,penta}, which was extensively investigated in \cite{berlinmadrid1,berlinmadrid2}. In the following, we shall elaborate upon the link between these developments, the Landau levels of Quantum Mechanics and flux quantization, while still confining our treatment, as in previous sections, to a relatively simple and instructive six--dimensional case.

\subsubsection{\sc Magnetizing the Bosonic String} \label{sec:magnetic_susyb_bose}

Let us begin by considering the bosonic string in the presence of
a uniform magnetic field $F_{ab}$ in an internal two--torus, which can be described by the
Abelian vector potential
\be
A_a \ =\  - \ {\frac{1}{2}} \, F_{ab} X^{b} \ , \label{mag1}
\ee
with a constant $F_{ab}$.
The variational principle for the world-sheet action reads
\bea
S &=& \ - \ \frac{1}{4 \pi \alpha'} \, \int d\tau \int_0^{\pi} d\sigma\ 
\partial_\alpha X^a \, G_{ab}\, \partial^\alpha X^b \nonumber \\
&+&
\left. q_{\rm L} \int d\tau A_a \partial_\tau X^{a} \right|_{\sigma =0}
\ + \
\left. q_{\rm R} \int d\tau A_a \partial_\tau X^{a} \right|_{\sigma =\pi}
\,, \label{mag2}
\eea
where $q_L$ and $q_R$ are the charges of the open--string ends under the Abelian generator corresponding to $A_a$~\footnote{The case with opposite charges will recover the discussion in Section~\ref{sec:open_B} provided $2\pi \alpha' q_{L,R}$ are replaced by $\pm\,1$.}. The preceding expressions are written in the conformal gauge for a strip of width $\pi$, and yield the
wave equations
\be
\left( {\partial^2 \over \partial \tau^2} -
{\partial^2 \over \partial \sigma^2} \right) X^{a} =0 \,, \label{mag3}
\ee
together with the boundary conditions
\bea
\delta\,X^a\left[ {1\over 2\pi\alpha '} \, G_{ab}\, \partial_\sigma X^b \ + \ q_{\rm L} F_{ab} \,
\partial_\tau X^b \right]_{\sigma=0} &=& 0 \ ,
\nonumber \\
\delta\,X^a\left[  {1\over 2\pi\alpha '} \,  G_{ab}\,\partial_\sigma X^b \ - \ q_{\rm R} F_{ab} \,
\partial_\tau X^b \right]_{\sigma=\pi} &=& 0 \ . \label{mag4}
\eea
One thus finds either Dirichlet or modified Neumann conditions, which interpolate between the Neumann and Dirichlet cases. Here we concentrate on the latter choice, which is sensitive to the magnetic field.

To be definite, let us focus on the two--dimensional case, with 
\beq
F_{12} \ = \ H\, R^2 \ , \qquad G_{ab} \ = \ R^2  \, \delta_{ab} \ , 
\eeq
while working with $X^a$ coordinates that are periodic with period $2\pi$. The gauge potential is a linear function of the $X^a$, so that an increment $\delta\,X^a = 2 \pi n^a$ leads to
\beq
\delta\,A_a \ = \ - \ \frac{1}{2}\ F_{ab} \ 2 \pi \, n^b \ .
\eeq
This variation corresponds to  the gauge transformation $\partial_a\,\Lambda$, with
\beq
\Lambda \ = \ \pi \, X^a\, F_{ab}\ n^b \ , 
\eeq
so that the wavefunction of a particle of charge $q$ acquires a phase given by
\beq 
e^{iq\pi F_{ab}x^an^b} \ .
\eeq
Performing now a second translation with $\delta X^a=2\pi  m^a$ leads to phase
\beq 
e^{2 i q\pi^2 m^a F_{ab} n^b} \ . 
\eeq  
One gets a contractible loop on the torus by performing the two translations in the opposite order. Imposing that the two paths have the same effect on wavefunctions leads to the consistency conditions
\beq 
e^{4 i q\pi^2 m^a F_{ab} n^b} = 1 \ . 
\eeq 
The strongest of these comes from the elementary displacements $m^a = (1,0)$ and $n^a = (0,1)$, and therefore for the two-torus
\beq 
e^{4 i q\pi^2 F_{12}} = 1 \ , 
\eeq 
which leads to the quantization condition
\beq 
2 \pi q F_{12} \ = 2 \pi q H R^2 \ = \ \ n \in Z \ . 
\eeq 
If the total charge is $q_L+q_R$ this condition becomes
\beq
 2\pi \left(q_{L} \,+\, q_R\right) H \,R^2 \ \in \ \mathbb{Z} \ , \label{Dirac_q}
\eeq
and must hold for all values of ${q_L}$ and ${q_R}$ in the spectrum. The flux $\Phi= (2 \pi R)^2 H$ is thus quantized in units of $\frac{2\pi}{q_L+q_R}$.

It is now convenient to introduce the complex combinations
\be
Z \ = \ {\textstyle{1\over \sqrt{2}}} (X^1 \ + \ i X^2)\ , \qquad
\bar Z \ = \ {\textstyle{1\over \sqrt{2}}} (X^1 \ -\  i X^2)\ , \label{mag5}
\ee
so that the action becomes
\bea
S &=&  \frac{R^2}{2\pi\alpha '} \left[ - \ \int d \tau \int_{0}^{\pi} d\sigma \,
\partial_\alpha \bar Z \, \partial^\alpha Z \right.\nonumber \\
&-& \left. \frac{i \alpha_L}{2}\, \left. \int d\tau \left( \bar Z \partial_\tau Z \,-\, Z \partial_\tau \bar Z\right)
\right|_{\sigma = 0}
 \,- \,\frac{i \alpha_R}{2}\, \left. \int d\tau \left( \bar Z \partial_\tau Z \,-\, Z \partial_\tau \bar Z\right)
\right|_{\sigma = \pi}\right] \, , \label{mag6}
\eea
with
\beq
\alpha_L \ = \ {2 \pi \alpha'\, q_{\rm L} H} \,, \qquad \alpha_R \ =\  {2 \pi \alpha'\,
q_{\rm R} H} \ .\label{mag8}
\eeq
The introduction of the two charges $q_{L,R}$ will be instrumental to describe the embedding of a magnetized $U(1)$ in the open--string gauge group.

The momentum canonically conjugate to ${\bar Z}$ is
\beq
{\overline \Pi}(\tau,\sigma) \ =\  \frac{R^2}{2 \pi \alpha'}
\Big\{ \partial_\tau  Z (\tau,\sigma) \ +
\ i Z (\tau,\sigma) \left[ \alpha_L \delta(\sigma) +
\alpha_R \delta(\pi - \sigma) \right]
\Big\} \ ,\label{mag12}
\eeq
while the boundary conditions (\ref{mag4}) at the two ends reduce to
\bea
\partial_\sigma Z \ -\ {i} \, \alpha_L \,
\partial_\tau Z \
 \Big|_{\sigma=0}  &=& 0\,,
\nonumber \\
\partial_\sigma Z \ +\ {i}\, \alpha_R \,
\partial_\tau Z \ \Big|_{\sigma=\pi} &=& 0 \ .
\label{mag7}
\eea
The solution of the wave equation is generally of the form
\beq
Z\left(\tau,\sigma\right) \ = \ Z_+(\tau+\sigma) \ + \ Z_-(\tau - \sigma) \ ,
\eeq
and the boundary conditions~\eqref{mag7} become
\bea
\left(1 - i \alpha_L\right) Z'{}_+(\tau) &=& \left(1 + i \alpha_L\right) Z'{}_-(\tau) \ , \nonumber \\
\left(1 + i \alpha_R\right) Z'{}_+(\tau+\pi) &=& \left(1 - i \alpha_R\right) Z'{}_-(\tau - \pi) \ . \label{bc_Z}
\eea
Letting
\beq
e^{2\pi i \zeta_L} \ = \  \frac{1 - i \alpha_L}{1 + i \alpha_L}\ , \qquad e^{2\pi i \zeta_R} \ = \ \frac{1 - i \alpha_R}{1 + i \alpha_R} \ , \qquad  \zeta \ = \ \zeta_L \ + \ \zeta_R \ , \label{zetaLR}
\eeq
so that one can define
\beq
\cos \pi \zeta_L \ = \ \frac{1}{\sqrt{1 \ + \ \alpha_L^2}} \ , \qquad \sin \pi \zeta_L \ = \ - \ \frac{\alpha_L}{\sqrt{1 \ + \ \alpha_L^2}} \ ,
\eeq
one can link $Z'{}_+$ and $Z'{}_-$ according to
\beq
Z'{}_-(\tau) \ = \ e^{2\pi i \zeta_L} \ Z'{}_+(\tau) \ .
\eeq
Using the first of eqs.~\eqref{bc_Z}, one can also obtain the quasi--periodicity condition 
\beq
Z'{}_+(\tau+2\pi) \ = \ e^{2 \pi i \zeta}\ Z'{}_+(\tau)  \ , \label{qper}
\eeq
so that
\beq
Z'{}_+(\tau) \ = \ e^{i \zeta \tau} f(\tau) \label{quasi_periodic}
\eeq
with $f(\tau)$ periodic of period $2 \pi$. According to eq.~\eqref{qper}, the relevant range for $\zeta$ is $0 \leq \zeta < 1$.

We must now distinguish two cases: 
\begin{enumerate}
\item if $\alpha_L+\alpha_R=0$, $\zeta=0$ and $Z'{}_+$ is periodic, and one finally obtains
\beq
Z \ = \ z \ + \ c_0\ \frac{\tau \ + \  i \alpha_L \left(\sigma-\frac{\pi}{2}\right)}{\sqrt{1 \ + \ \alpha_L^2}} \ + \ i\,\sum_{n \neq 0} \frac{c_n}{n} \ e^{- i n \tau} \cos(n \sigma + \pi \zeta_L)   \ . \label{like_B2}
\eeq
This case can be mapped to the setup discussed in Section~\ref{sec:open_B}, in which a constant $B_{ab}$ is present. As in that case, we have shifted $\sigma$ by $- \,\frac{\pi}{2}$, so that the constant mode $z$ and its conjugate $\bar{z}$ commute. The counterpart of eq.~\eqref{like_B2} is eq.~\eqref{X_open_B2}, and the mapping is effected by letting
\beq
\alpha_L \ = \ \frac{B_{12}}{R^2} \ , \qquad c_0 \ = \ \frac{\sqrt{2}\, \alpha'}{R^2} \frac{m_1 + i m_2}{\sqrt{1 \ + \ \alpha_L^2}} \ ,
\eeq
and for the oscillator modes
\beq
c_n \ = \ \frac{\alpha'}{2\,R} \ \frac{\alpha'{}_n^1 \ + \ i\,\alpha'{}_n^2}{1 \ + \ i\,\alpha_L} \ .
\eeq
The commutation relations for the $c_n$'s can be deduced from eq.~\eqref{primed_osc}, and read
\beq
\left[ c_m,c_n\right] \ = \ 0 \ , \qquad \left[ \bar{c}{}_m,\bar{c}{}_n\right] \ = \ 0 \ , \qquad  \left[ {c}_m,\bar{c}{}_n\right] \ = \ \frac{1}{2}\ m \, \delta_{m,n} \ .
\eeq

The different modes are orthogonal with respect to the Klein--Gordon scalar product
\beq
\int _0^\pi d\sigma \bar{\psi}_1 \left[\stackrel{\leftrightarrow }{i \,\partial_\tau} \ - \ 2\left( \alpha_L \, \delta(\sigma)  \ + \ \alpha_R\, \delta(\pi-\sigma)\right) \right]{\psi}_2
\eeq
where we included boundary contributions (for this sector $\alpha_R=-\alpha_L$). The contribution to $M^2$ from this sector can be deduced from what was done about the constant $B_{ab}$ background in eq.~\eqref{m2ab}, and reads
\beq
M^2 \ = \ \frac{1}{R^2} \ \frac{m^T \, m}{1 \ + \ \alpha_L^2} \ + \frac{1}{\alpha'}\left( N \ - \ 1 \right) \ .
\eeq

\item if $\alpha_L+\alpha_R \neq0$, $\zeta$ does not vanish and $Z$ is not periodic. Consequently, $f$ can now be expanded as
\beq
f(\tau) \ = \ \sum_{n \in \mathbb{Z}} c_n \ e^{- i n \tau} \ ,
\eeq
and then
\beq
Z_+(\tau) \ = \  \sum_{n \in \mathbb{Z}} \frac{i}{n - \zeta} \ c_n \ e^{-i \tau(n - \zeta)}  \ , \qquad
Z_-(\tau) \ = \  e^{2\pi i \zeta_L} \, Z_+(\tau) \ ,
\eeq
up to a zero mode, so that
\beq
Z \ = \ z \ + \ i \sum_{n \in \mathbb{Z}} \frac{c_n}{n - \zeta} \ \left[ e^{-i (\tau+\sigma)(n - \zeta)} \ + \ e^{2\pi i \zeta_L} \  e^{-i (\tau-\sigma)(n - \zeta)}\right]  \ ,
\eeq 
Alternatively, after absorbing a phase in the $c_n$'s, this expression becomes
\beq
Z \ = \ z \ + \ i \sum_{n \in \mathbb{Z}} \frac{c_n}{n - \zeta} \ e^{-i \tau(n - \zeta)} \ \cos\left[(n - \zeta)\sigma \ +  \ \pi\,\zeta_L \right]  \ ,
\eeq
\end{enumerate}
and the commutation relations read
\beq
\left[ c_m,c_n\right] \ = \ 0 \ , \qquad \left[ \bar{c}{}_m,\bar{c}{}_n\right] \ = \ 0 \ , \qquad  \left[ {c}_m,\bar{c}{}_n\right] \ = \ \frac{1}{2}\ \left(m- \zeta\right) \, \delta_{m,n} \ , 
\eeq
where now $m$ and $n$ are arbitrary integers. Only a constant is left as zero mode, and~\cite{acny}
\beq
\left[ z\,,\,\bar{z}\right] \ = \ \frac{2\alpha'\pi}{R^2(\alpha_L \ + \ \alpha_R)} \ ,
\eeq
so that these solutions are not continuously connected to those with vanishing total charge. In terms of the Cartesian components $(x,y)$, the preceding commutation relation becomes
\beq
\left[ x\,,\,{y}\right] \ = \ \frac{i\,2\alpha'\, \pi}{R^2(\alpha_L \ + \ \alpha_R)} \ .
\eeq
The analogy with Quantum Mechanics leads one to conclude that the torus, of area $v=(2 \pi)^2$, is effectively divided in cells of area
\beq
a \ = \ \frac{4 \pi^2\alpha'}{R^2(\alpha_L\ + \ \alpha_R)} \ ,
\eeq
and the total number of cells, which counts the degeneracy $k$ of Landau levels, is thus
\beq
k \ = \ 2 \, \frac{R^2}{2\alpha'}\left(\alpha_L \ + \ \alpha_R \right) \ = \ 2\pi R^2 \,H \left(q_L \ + \ q_R \right)\ .  \label{degeneracy}
\eeq
The quantization conditions~\eqref{Dirac_q} guarantees that this quantity is an integer number.

The contribution to $M^2$ from this sector is
\beq
\alpha'\, M^2 \ = \ \frac{1}{24} \ - \ \frac{1}{8}\left({1-2 \zeta}\right)^2 \ + \ \sum_{m=1}^\infty (m-\zeta) a_m^\dagger\,a_m \ + \ \sum_{m=0}^\infty (m+\zeta) b_m^\dagger\,b_m \ , \label{H_zeta}
\eeq
where the $a$'s and $b$'s are two sets of annihilation operators obtained from the $c$'s and ${\bar c}$'s, together with their Hermitian conjugate creation operators.

We now have all the ingredients to compute the partition function of
open strings in the presence of a uniform internal magnetic field. As we shall
see, the annulus amplitude encodes interesting
details on the low-energy interactions. 

Let
us begin by referring to the simpler case of
bosonic strings \cite{acny}, concentrating on a pair of
coordinates, whose contribution
to the annulus amplitude, in the absence of a magnetic field, would be
\be
{\cal A} \ = \ \frac{N^2}{2 \, \tau_2 \,\eta^2} \label{mag16}
\ee
in the case of $N$ D25 branes. 

The case in which the magnetic field only affects some of the open--string ends provides an instructive example. Starting from an orthogonal gauge group, three subsets of Chan--Paton charges can be distinguished, letting
\beq
N = N_0 + M + \bar M, \label{mag_charges}
\eeq
where $N_0$ counts the unaffected ends, while $M$ and $\bar M$
count the equal numbers of ends that are affected oppositely. There is a numerical identification of $M$ and $\bar{M}$, as pertains to the embedding of the magnetized $U(1)$ in the original orthogonal gauge group. This example corresponds to a magnetized $U(1)$ generator $T$, such that
\beq
\mathcal{Q} \ F_{12} \ = \ H \, R^2 \ T \ ,  \label{F12Q}
\eeq
where $T$ is the matrix
\beq
T \ = \ \left(\begin{array}{cc} \!\! \mathbf{0}_{N_0 \times N_0} \ \ \  \ \ \ \  \ \ \ \ \ \ \mathbf{0} \\ \!\! \mathbf{0}\ \ \ \ \ \ \   \mathbf{1}_{M \times M} \ \ \  \ \ \ \mathbf{0}\\ \!\! \ \ \ \ \ \ \ \mathbf{0} \ \ \ \ \ \ \  \mathbf {0} \ \ \ \ \  \ \ \ \ \ \mathbf{- 1}_{{M} \times {M}} \end{array} \right) \ .  \label{F12matrix}
\eeq
Consequently, there are several open--string sectors.
There are neutral strings of two types, those with multiplicity $N_0^2$, for which $q_L=q_R=0$ so that they are not sensitive to the magnetic field, and those with multiplicity
$M \, {\bar M}$, which are sensitive to the magnetic field but have $q_L=- q_R$. There are also charged strings of different types: only one end is charged for those with multiplicities $N_0 \,M$, $N_0\, \bar
M$, while both ends carry identical charges for those with multiplicities $M^2$ and $\bar M^2$. This configuration breaks an orthogonal gauge group to $SO(N_0) \times U(M)$, consistent with the traceless nature of the group generators.

Both the uncharged strings, with overall Chan--Paton factor $N_0^2$, and the
``dipole'' ones, with overall Chan--Paton factor $M\,{\bar M}$, have unshifted oscillators, as we have seen, which give identical contributions to the partition function,
but differ crucially in their zero modes. From eq.~\eqref{like_B2}, one can see that for dipole strings the effective radius is increased by a factor $\sqrt{1+\alpha_L{}^2}$, where $\alpha_L$ is defined in eq.~\eqref{mag8}, so that in the large--radius limit that we are exploring the total contribution of neutral strings to the partition function becomes
\beq
{\cal A}_0 \sim \left\{ {\textstyle{1\over 2}} N_0^2 \ + \ 
 M\bar M \left[1 \,+\, (2 \pi \alpha' q_L H)^2 \right] \right\}
{1 \over \tau_2 \eta^2}\ . \label{mag18}
\eeq

The charged-string contributions differ in two respects: their
modes are shifted and, as a result, their annulus amplitudes involve theta functions with non--vanishing arguments, but no factors of $\tau_2$ accompany them, due to the absence of zero modes.
Altogether, taking eqs.~\eqref{degeneracy} and \eqref{H_zeta} into account, these contributions add up to
\beq
{\cal A}_{\pm} \ \sim \ - \ i N_0 ( M + \bar M ) \, \frac{k_L \ \eta}{
q^{{1\over 2}\zeta_L^2}\vartheta_1 (\zeta_L
\tau | \tau)} \ - \ {\textstyle{1\over 2}}i
(M^2 + \bar M ^2 )\, \frac{2\,k_L \ \eta}
{q^{2\zeta_L^2} \vartheta_1 (2 \zeta_L
\tau | \tau )} \ , \label{mag19}
\eeq
and, as we have seen, Dirac quantization forces $k_L$ to be an integer. These issues are discussed in more detail in~\cite{bbblw}.

Up to overall normalizations, one
can read from (\ref{mag18}) and (\ref{mag19}) the open-string spectrum. However, more interesting results
can be extracted from the
transverse-channel amplitudes. After an $S$ modular transformation, taking into account the contribution of the measure, the vacuum--channel annulus amplitude becomes
\bea
\widetilde{\cal A} &\sim& \left(2 \pi\right)^2 \left[ N_0^2 + 2 M \bar M
(1 + (2 \pi \alpha' q_L H)^2 ) \right] \frac{1}{\eta^2}
\nonumber \\
&-&  2 N_0 ( M + \bar M ) \, \frac{k_L \ \eta}{\vartheta_1 (\zeta_L
| \tau)} \ - \ (M^2 + \bar M ^2 ) \,\frac{2\,k_L \ \eta}{\vartheta_1 (2 \zeta_L
| \tau )} \,, \label{mag20}
\eea
and the low--lying contribution is proportional to
\be
\left[ N_0 + (M+\bar M) \sqrt{ 1 + (2 \pi \alpha'
q_L H)^2} \right]^2 \ .
\ee

In the bosonic string, this would be associated to the tachyon, while in the superstring it concerns the
dilaton. Since these tree-level interactions originate from the
disk, in the latter case one can link these charged contributions to
\be
{\cal S}_{\rm {DBI}}\  \sim \
\int e^{-\varphi} \sqrt{-\ {\rm det} (g_{\mu\nu} + 2 \pi \alpha ' q_L
F_{\mu\nu} )} \,,
\ee
recovering the celebrated result that the low-energy
open-superstring dynamics is governed by the Dirac-Born-Infeld action
\cite{bidirac1,bidirac2,ftse}.

\subsubsection{\sc Magnetizing the Superstring} \label{sec:magnetic_susyb_super}

We can now analyze how the supersymmetric $T^4/\mathbb{Z}_2$ compactification of Section~\ref{sec:susyT4} can be deformed by allowing for uniform internal magnetic fields. In detail, we shall focus on a $[T^2(H_1)\times T^2(H_2)]/{Z}_2$ orbifold,
with a pair of uniform Abelian magnetic
fields $H_1$ and $H_2$ in the two internal tori that are aligned with the same $U(1)$ generator. These magnetic fields will deform the open--string sector and will also affect the NSR fermions. 
The simultaneous presence of two magnetized tori will bring about an interesting
new effect~\cite{aads}.

Let us first describe how the magnetic deformation affects the NSR contributions to the open--string mass spectrum. To this, one can start from the boundary conditions of eqs.~\eqref{mag4} and deduce the corresponding conditions on the fermions by applying two--dimensional supersymmetry transformations. Letting
\beq
E_L \ = \ G \ + \ 2\,\pi\,q_L\,\alpha'\, F \ , \qquad E_R \ = \ G \ + \ 2\,\pi\,q_R\,\alpha'\, F
\eeq
the boundary conditions for Fermi fields are
\bea
\left. \left(E_L\right)_{ab} \tilde{\psi}^b\right|_{\sigma=0} &=& \left.\left(E_L^T \right)_{ab} {\psi}^b\right|_{\sigma=0} \ , \nonumber \\
\left.\left(E_R^T\right)_{ab} \tilde{\psi}^b\right|_{\sigma=\pi} &=&  \pm \ \left.\left(E_R \right)_{ab} {\psi}^b\right|_{\sigma=\pi} \ , \label{bc_psi}
\eea
where the upper (lower) sign refers to the Ramond (Neveu--Schwarz) sector. For the two-torus it is convenient to introduce complex combinations that diagonalize the boundary conditions, along the lines of what we did for bosonic coordinates, letting
\beq
{\psi} \ = \ \frac{\psi^1 \ + \ i\,\psi^2}{\sqrt{2}} \ , \qquad \tilde{\psi} \ = \ \frac{\tilde{\psi}^1 \ + \ i\,\tilde{\psi}^2}{\sqrt{2}} \ ,
\eeq
Eqs.~\eqref{bc_psi} then become
\bea
\left(1 + i \alpha_L\right) \tilde{\psi}(\tau) &=& \left(1 - i \alpha_L\right) \psi(\tau) \ , \nonumber \\
\left(1 - i \alpha_R\right) \tilde{\psi}(\tau + \pi) &=& \pm \ \left(1 + i \alpha_R\right) \psi(\tau - \pi)  \ . \label{bc_ZF}
\eea
Combining them, one can conclude that
\beq
\psi(\tau+2\pi) \ = \ \pm \ e^{2\pi i \zeta}\ \psi(\tau) \ ,
\eeq
so that the Fermi modes have shifted frequencies with respect to the standard case with no magnetic field unless $q_L+q_R=0$. In the presence of a magnetized two torus the four fermionic characters of eqs.~\eqref{fermi10D} are modified, and in the direct--channel annulus amplitude become
\bea
O_8 &\to& O_6 O_2(\zeta \tau) \ + \ V_6 V_2(\zeta \tau) \ , \qquad V_8 \ \to \ V_6 O_2(\zeta \tau) \ + \ O_6 V_2(\zeta \tau) \ , \nonumber \\
S_8 &\to& S_6 S_2(\zeta \tau) \ + \ C_6 C_2(\zeta \tau) \ , \qquad C_8 \ \to \ S_6 C_2(\zeta \tau) \ + \ C_6 S_2(\zeta \tau) \ ,
\eea
where
\bea
O_{2}(x) &=& \frac{1}{2 \eta (\tau)} \left[
\vartheta_3(x|\tau)
+ \vartheta_4(x|\tau)\right] \, ,
\quad 
V_{2}(x)  \ = \ \frac{1}{2 \eta (\tau)} \left[
\vartheta_3(x|\tau)
- \vartheta_4(x|\tau) \right] \, , \label{mag18k} \\
S_{2}(x) &=& \frac{1}{2 \eta(\tau)} \left[
\vartheta_2(x|\tau) - i \vartheta_1(x|\tau)\right] \,
, \quad
C_{2}(x)  \ = \ \frac{1}{2 \eta (\tau)} \left[
\vartheta_2(x|\tau)
+ i \vartheta_1(x|\tau) \right] \ . \nonumber
\eea

In the configuration~\eqref{mag_charges}, the deformed annulus amplitude that includes contributions from a two torus of finite area reads
\bea
{\cal A}_{T^2} &\sim& \left\{ {\textstyle{1\over 2}} N_0^2 \ P \ + \ 
 M\bar M \ \widetilde{P} \right\}
{\left(V_8 \ - \ S_8\right)} \\
&-&  i\,N_0 ( M + \bar M ) \, \frac{k_L \ \eta}{
\vartheta_1 (\zeta_L
\tau | \tau)} \,\left(V_6 O_2(\zeta_L\tau)+O_6 V_2(\zeta_L\tau) - S_6 S_2(\zeta_L\tau) - C_6 C_2(\zeta_L\tau) \right)   \nonumber \\  &-& \frac{i}{2}\
(M^2 + \bar M ^2 )\, \frac{2\,k_L \ \eta}
{\vartheta_1 (2 \zeta_L
\tau | \tau )} \,\left(V_6 O_2(2 \zeta_L\tau)+O_6 V_2(2\zeta_L\tau) - S_6 S_2(2\zeta_L\tau) - C_6 C_2(2\zeta_L\tau) \right)  \ , \nonumber
\eea
where the contributions of the non--compact bosonic coordinates are left implicit and $N_0+2 M=32$. Here $P$ is a shorthand for $P(R)$, 
\beq
{P}(R) \ = \ \frac{1}{\eta^2\left(i \frac{\tau_2}{2}\right)} \ \sum_{m_1,m_2} e^{- \, \frac{\pi \tau_2\alpha'\left(m_1^2+m_2^2\right)}{R^2}} \label{PR}
\eeq
the contribution of a pair of bosonic coordinates on a product of circles of radius $R$, while  in
\beq
\widetilde{P} \ = \  P\left(R \sqrt{1 \ + \ \left(2 \pi \alpha'q_L H \right)^2 } \right) \label{PtildeR}
\eeq
the radius is scaled as pertains to open strings with opposite non--vanishing charges at the ends.

Supersymmetry is broken in the deformed system, and ${\cal A}_{T^2}$ does not vanish. Moreover, there are inevitable instabilities induced by the magnetic moments of the low--lying excitations. This can be seen by tracking their mass deformations, which gives
\beq
\Delta M^2 \ = \  \frac{1}{2 \pi \alpha'} \;
\Bigl[  (2 n + 1) \left|\zeta_{L} \ + \ \zeta_{R}\right| \ + \ 2 \left(\zeta_{L} \ + \ \zeta_{R}\right) \Sigma\Bigr] \, , \label{deltaMgen0}
\eeq
where $\zeta_{L,R}$ can vanish or, when this is not the case, can be identical or opposite, and where the first contribution is from the Landau levels. $\Sigma$ denotes the internal helicities of the modes that were originally massless (0 for scalars, $\pm \frac{1}{2}$ for fermions, $\pm 1$ for vectors), and the problem originates from the internal components of the vectors. Consequently, the actual vacuum is not determined by the present analysis.

The generalization to the deformed supersymmetric $T^4/\mathbb{Z}_2$ compactification obtained via a pair of magnetized internal two tori rests on the four characters
\begin{eqnarray}
Q_o (\zeta_1 ; \zeta_2) &=& V_4 (0) \left[ O_2 (\zeta_1 ) O_2 (\zeta_2 ) +
V_2 (\zeta_1 ) V_2 (\zeta_2 ) \right] \nonumber\\
&-& C_4 (0) \left[ S_2 (\zeta_1 ) C_2 (\zeta_2 ) +
C_2 (\zeta_1 ) S_2 (\zeta_2 ) \right] \, ,
\nonumber
\\
Q_v (\zeta_! ; \zeta_2) &=&  O_4 (0) \left[ V_2 (\zeta_1 ) O_2 (\zeta_2 ) +
O_2 (\zeta_1 ) V_2 (\zeta_2 ) \right] \nonumber \\
&-& S_4 (0) \left[ S_2 (\zeta_1 )
S_2 (\zeta_2 ) + C_2 (\zeta_1 ) C_2 (\zeta_2 ) \right] \, ,
\nonumber
\\
Q_s (\zeta_1 ; \zeta_2 ) &=& O_4 (0) \left[ S_2 (\zeta_1 ) C_2 (\zeta_2 ) +
C_2 (\zeta_1 ) S_2 (\zeta_2 ) \right] \nonumber \\
&-& S_4 (0) \left[  O_2 (\zeta_1 ) O_2 (\zeta_2 ) +
V_2 (\zeta_1 ) V_2 (\zeta_2 ) \right] \, ,
\nonumber
\\
Q_c (\zeta_1 ; \zeta_2)
&=&  V_4 (0) \left[ S_2 (\zeta_1 ) S_2 (\zeta_2 ) +
C_2 (\zeta_1 ) C_2 (\zeta_2 ) \right] \nonumber \\
&-& C_4 (0)
\left[  V_2 (\zeta_1 ) O_2 (\zeta_2 ) +
O_2 (\zeta_1 ) V_2 (\zeta_2 ) \right] \, . \label{mag17k}
\end{eqnarray}

As we have seen in Section~\ref{sec:susyT4}, in the absence of a magnetic field and for coincident branes located at a given fixed point, the open strings of the supersymmetric $T^4/\mathbb{Z}_2$ compactification carry a $U(16) \times U(16)$ gauge group~\cite{orientifolds5,gimpol}. Here we are deforming this system by introducing magnetic fields in the two internal $T^2$ factors, which are only felt by D9 branes. Therefore, D5 contributions are not affected and continue to be described by the complex multiplicities $d$ and $\bar{d}$ of Section~\ref{sec:susyT4}, with $d=16$, due to the D5 tadpole condition. The deformation of the D9 spectrum is induced by a pair of uniform magnetic fields in the two internal $T^2$ such that
\beq
\mathcal{Q} \ F_{12} \ = \ H_1 \, R^2 \ T \ ,  \qquad \mathcal{Q} \ F_{34} \ = \ H_2 \, R^2 \ T \ ,  \label{F12Q_susy}
\eeq
where now $T$ is the matrix
\beq
T \ = \ \left(\begin{array}{cc} \!\!\! \mathbf{0}_{n \times n} \ \ \ \ \ \   \mathbf{0} \\ \ \ \mathbf{0}\ \ \ \ \ \ \  \ \mathbf{1}_{m \times m} \end{array} \right) \ ,  \label{F12matrix2}
\eeq
with $n+m=16$, as the deformation concerns the original $U(16)$ group.

Within the contributions associated with D9 branes, one must thus distinguish two types of complex multiplicities, $(m,\bar{m})$ for the string ends that are charged with respect to the magnetic U(1) and $(n,\bar{n})$ for the remaining uncharged ones, with $n+m=16$ due to the D9 tadpole condition.
As a result, the annulus amplitude involves several types of open strings:
dipole strings, with Chan-Paton multiplicity $m \bar{m}$,
uncharged ones, with multiplicities independent
of $m$ and $\bar{m}$, singly-charged ones, with multiplicities
linear in $m$ or $\bar{m}$, and finally doubly-charged ones, with
multiplicities proportional to $m^2$ or $\bar{m}^2$. 

The
annulus amplitude is then~\footnote{In the preceding expressions, for brevity we have left the subscripts $L_i$ for the two internal tori implicit.}
\begin{eqnarray}
{\cal A}_{T^4/\mathbb{Z}_2} &=& {\textstyle{1\over 4}} \Biggl\{ (Q_o + Q_v)(0;0) \left[
(n+\bar n)^2 P_1 P_2 + (d+\bar d)^2 W_1 W_2
+ 2 m \bar{m} \tilde P_1 \tilde P_2 \right]
\nonumber
\\
&-& 2 (m+\bar m) (n + \bar{n}) (Q_o + Q_v )(\zeta_1 \tau ; \zeta_2 \tau
) {k_1 \eta \over
\vartheta_1 (\zeta_1 \tau)} {k_2 \eta \over \vartheta_1 (\zeta_2 \tau)}
\nonumber
\\
&-& ( m^2 + \bar{m}^2 ) (Q_o + Q_v ) (2 \zeta_1 \tau ; 2 \zeta_2 \tau )
{2 k_1 \eta \over
\vartheta_1 (2 \zeta_1 \tau)} {2 k_2 \eta \over \vartheta_1 (2 \zeta_2 \tau)}
\nonumber
\\
&-& \left[ (n-\bar n)^2 -2 m\bar m + (d-\bar d)^2 \right] (Q_o - Q_v ) (0;0)
\left( {2\eta \over \vartheta_2 (0)}\right)^2
\nonumber
\\
&-& 2 (m-\bar m) (n - \bar{n}) (Q_o - Q_v ) (\zeta_1 \tau ; \zeta_2 \tau)
{2\eta \over \vartheta_2
(\zeta_1 \tau)} {2\eta \over \vartheta_2 (\zeta_2 \tau)}
\nonumber
\\
&-& (m^2 + \bar{m}^2) (Q_o - Q_v ) (2\zeta_1 \tau ; 2\zeta_2 \tau)
{2\eta \over \vartheta_2
(2\zeta_1 \tau)} {2\eta \over \vartheta_2 (2\zeta_2 \tau)}
\nonumber
\\
&+& 2 (n+\bar n ) (d+\bar d) (Q_s + Q_c) (0;0) \left({\eta \over
\vartheta_4 (0)}\right)^2
\nonumber
\\
&+& 2 (m + \bar{m})(d+\bar d)(Q_s + Q_c) (\zeta_1 \tau ; \zeta_2 \tau)
{\eta \over \vartheta_4
(\zeta_1 \tau )} {\eta \over \vartheta_4 (\zeta_2 \tau )}
\nonumber
\\
&-& 2 (n-\bar n) (d - \bar d) (Q_s - Q_c )
(0;0) \left( {\eta \over \vartheta_3 (0)}\right)^2 \label{annsusy}
\\
&-& 2  (m - \bar{m})(d-\bar d) (Q_s - Q_c) (\zeta_1 \tau ; \zeta_2 \tau)
 {\eta \over \vartheta_3
(\zeta_1 \tau )} {\eta \over \vartheta_3 (\zeta_2 \tau )} \Biggr\} \ ,
\nonumber
\end{eqnarray}
where $P_i=P(R_i)$ and $\widetilde{P}_i$ are defined as in Eqs.~\eqref{PR} and \eqref{PtildeR}, while $W_i=W(R_i)$
\beq
W(R) \ = \ \frac{1}{\eta^2\left(i \frac{\tau_2}{2}\right)} \ \sum_{n_1,n_2} e^{- \, \frac{\pi \tau_2 \left(n_1^2+n_2^2\right)R^2 }{\alpha'}} \ .
\eeq
The contributions of the non--compact bosonic coordinates are again implicit.

The orientifold construction includes the Klein-bottle projection that we
already met, in eq.~\eqref{kt4s} of Section~\ref{sec:susyT4}, which we write in the form
\beq
{\cal K} \ = \  {\textstyle{1\over 4}} \Biggl\{ (Q_o + Q_v) (0;0) \left[ P_1 P_2 +
W_1 W_2 \right] \ + \  16\times 2 \,(Q_s + Q_c ) (0;0) \left( {\eta \over
\vartheta_4 (0)} \right)^2 \Biggr\} \, , \label{mag16k}
\eeq
where $P_i$ and $W_i$ denote momentum and winding sums in the two tori, 
and where the six-dimensional $Q$ characters are now endowed with
a pair of arguments, as above. 

The construction is completed by the M\"obius amplitude~\footnote{Here we can correct a typo that, unfortunately, propagated from~\cite{aads}. It concerns the arguments depending on the $\zeta_i$ that enter the direct--channel M\"obius amplitude, which should involve the combinations $\zeta_i \, i \frac{\tau_2}{2}$ rather than $\zeta_i\,\tau$. This subtlety is determined by Pradisi's transformation $P = T S T^2 S$ that connects the direct and transverse M\"obius channels, and the resulting expression for the direct channel is then consistent with the structure of the Landau levels. The transition to the real ``hatted'' basis just makes the leading power of $q$ real, and therefore has no effect on all this.}
\begin{eqnarray}
{\cal M} &=& -{\textstyle{1\over 4}} \Biggl[
(\hat Q_o + \hat Q_v )(0;0) \left[ (n+\bar n) \hat{P}_1 \hat{P}_2 + (d+\bar d) \hat{W}_1
\hat{W}_2 \right]
\nonumber
\\
&-& ( m + \bar{m}) (\hat Q_o + \hat Q_v ) (i \zeta_1 \tau_2 ; i \zeta_2
\tau_2) {2 k_1
\hat\eta \over \hat \vartheta_1 (i \zeta_1\tau_2)} {2 k_2
\hat\eta \over \hat \vartheta_1 (i \zeta_2\tau_2)}
\nonumber
\\
&-& \left( n+ \bar n + d + \bar d \right) (\hat Q_o - \hat Q_v )(0;0) \left(
{2\hat\eta \over \hat \vartheta_2 (0)}\right)^2 \label{mobsusy}
\\
&-& (m + \bar{m}) (\hat Q_o - \hat Q_v ) (i \zeta_1 \tau_2 ; i \zeta_2 \tau_2 )
{2\hat\eta \over \hat\vartheta_2 (i\zeta_1\tau_2)}
{2\hat\eta \over \hat\vartheta_2 (i\zeta_2\tau_2)} \Biggr] \,.
\nonumber
\end{eqnarray}
In the preceding expressions, the arguments $\zeta_i$  and $2\zeta_i$
are associated to strings with one
or two charged ends, and, for brevity, both the imaginary modulus
$\frac{1}{2} i \tau_2$ of ${\cal A}$ and the complex modulus $
\frac{1}{2} + \frac{1}{2} i \tau_2$ of real--valued ``hatted'' quantities are left implicit.
Note that terms with
opposite U(1) charges, and thus with opposite $\zeta_i$ arguments,
have been grouped here and in ${\cal A}$, using the symmetries of
Jacobi theta functions.

We can now comment on how the spectrum of the original supersymmetric model of Section~\ref{sec:susyT4} is modified by the internal magnetic fields. Note that supersymmetry is generically broken and, as expected in these cases, stability is not guaranteed.
In fact, for generic magnetic fields the open spectrum is not supersymmetric
and can develop Nielsen-Olesen instabilities \cite{nole1,nole2,nole3}, in the form of tachyonic modes generated by the magnetic moments of internal
Abelian gauge bosons \cite{bachasmag}\footnote{The fate of this type of tachyonic modes
and the corresponding final state were recently addressed in \cite{bdt}.}. For untwisted string modes one finds the mass correction
\beq
\Delta M^2 \ = \  \frac{1}{2 \pi \alpha'} \; \sum_{i=1,2}
\Bigl[  (2 n_i + 1) \left|\zeta_{i\,L} \ + \ \zeta_{i\,R}\right| \ + \ 2 \left(\zeta_{i\,L} \ + \ \zeta_{i\,R}\right) \Sigma_i\Bigr] \, , \label{deltaMgen}
\eeq
where the $\zeta$'s are the deformation parameters that enter the preceding equations. The first term originates from the Landau levels and
the second from the magnetic moments associated to the internal helicities $\Sigma_i$ in the two tori. For scalar modes in the internal space $\Sigma_{1}=\Sigma_2=0$, for internal vectors $\Sigma_{1}=\pm 1$ and $\Sigma_2=0$ or $\Sigma_{1}=0$ and $\Sigma_2=\pm 1$, while for fermions they are both $\pm \frac{1}{2}$.
As a result, for the internal
components of the vectors, the magnetic moment coupling generally
overrides the zero-point contribution, leading to tachyonic modes,
unless $|H_1|=|H_2|$,
while for spin-$\frac{1}{2}$ modes it can at most compensate for it, giving rise to chiral zero modes, even in toroidal compactifications, in compliance to the Atiyah--Singer index theorem. For
small magnetic fields, eq.~\eqref{deltaMgen} reduces to the standard result in Field Theory,
\beq
\Delta M^2 \ = \  \frac{1}{2 \pi \alpha'} \; \sum_{i=1,2}
\Bigl[  (2 n_i + 1) |2 \pi \alpha ' (q_{\rm L} + q_{\rm R}) H_i| \ + \ 4 \pi\alpha' (q_{\rm L} + q_{\rm R}) \Sigma_i H_i \Bigr] \, .
\eeq

On the other hand, for twisted modes the first contribution
is absent, since ND strings have no Landau levels, but in this case
the low-lying space-time fermions, which originate from the fermionic portion
$S_4 O_4$ of $Q_s$, are scalars in the internal space and
have no magnetic moment couplings. However, their bosonic partners
originating from $O_4 C_4$
are affected by magnetic deformations and have mass shifts
$\Delta M^2 \sim \pm (H_1 - H_2)$, which generally cause tachyonic instabilities.  However,
if $H_1=H_2$ these, and in fact all tachyonic instabilities,
are absent. Actually, with this choice, the low--lying $C_4 C_4$ fermionic sector is also unaffected, so that complete Yang--Mills multiplets emerge from open strings~\footnote{Type II
branes at angles preserving some supersymmetry
were originally considered in \cite{douglas21,douglas22}.  After T-dualities, these
can be related to special choices for the internal
magnetic fields. Type I toroidal models cannot lead to
supersymmetric vacuum configurations, since the resulting RR tadpoles would
require the introduction of antibranes.}. A residual supersymmetry is indeed present for the entire string spectrum, and using Jacobi identities for nonvanishing arguments \cite{jacobi},
one can see that for $\zeta_1=\zeta_2$
both ${\cal A}$ and ${\cal M}$ vanish identically. Still, the
resulting supersymmetric models are rather peculiar,
as can be seen from the deformed tadpole conditions,
to which we now turn.

The untwisted RR tadpoles contain interesting novelties.
For $C_4 S_2 C_2$ one finds
\bea
&& \Big[ n+\bar n + m + \bar{m} - 32 \Big] \, \frac{R_1 R_2}{\alpha'}
\nonumber \\
&+&  \Big[ d+\bar d + \left(2 \pi q_L H_1 R_1^2\right)\left(2 \pi q_L H_2 R_2^2\right) 
- 32\Big] \frac{\alpha'}{R_1 R_2}   \ = \  0 \ , \label{rrutad}
\eea
aside from terms that vanish after identifying the multiplicities
of conjugate representations $(m,\bar{m})$, $(n,\bar{n})$ and $(d,\bar{d})$,
while the additional untwisted RR tadpole conditions
from $Q_o$ and $Q_v$ are compatible with (\ref{rrutad}) and do not lead to additional
constraints. 

Note that we grouped the terms involving the magnetic fields with those arising from the D5 branes. In fact, this expression reflects the familiar Wess-Zumino
coupling of D-branes \cite{dgm1,dgm2,dgm3,dgm4,dgm5,dgm6}, 
\beq
{\cal S}_{{\rm WZ}} \  \sim \ \sum_{p}
\int_{{\cal M}_{p+1}}  \mathrm{Tr}\left[ e^{Q F} \right] \wedge C 
\eeq
where $Q$ is defined in eq.~\eqref{F12Q}, and
\beq
F \ = \ H_1 \ dz_1 \wedge d\bar{z}{}_1 \ + \ H_2 \ dz_2 \wedge d\bar{z}{}_2
\eeq
is a Hermitian two-form containing the magnetic fields in the two tori, with $C$ a sum of forms of different degrees. Only the resulting $(p+1)$-form contribution is retained, and consequently different powers of $F$
couple, in general, to RR forms of different degrees, and in particular \emph{magnetized D9 branes acquire D5 charges}.
In the class of models under scrutiny the term bilinear in the magnetic fields has precisely this effect: it charges the D9 brane with respect to the
six-form potential, and as a result one can replace some of the D5 branes
with their blown-up counterparts thus obtained. 

This process
reverses the familiar relation~\cite{witsmall} between small--size
instantons and D5 branes: a fully blown-up instanton, resulting from
a uniform magnetic field, provides an exact description of
a D5 brane smeared over the internal torus via a magnetized D9 brane.
This can be clearly seen making use of the Dirac
quantization condition~\eqref{Dirac_q},
\be
\pi \,q_L H_i R_i^2 \ = \  k_i \qquad (i=1,2) \ ,
\label{dirac}
\ee
which turns (\ref{rrutad}) into
\begin{eqnarray}
& & m\ + \ \bar m \ + \ n \ + \ \bar n \ = \  32 \ ,
\nonumber
\\
& & 4 k_1 k_2 (m + \bar m )  \ + \ d \ + \ \bar d \ =  \ 32 \ . \label{urrt}
\end{eqnarray}
Thus, if $k_1 k_2 > 0$, the D9 branes acquire the
RR charge of $4|k_1 k_2|$ D5 branes, while if $k_1 k_2 < 0$
they acquire the RR charge of as many $\overline{{\rm D} 5}$
branes. 

The untwisted NS-NS tadpoles exhibit very nicely their relation to the
Born-Infeld term. For instance, the dilaton tadpole
\bea
& & \left[ n+\bar n + (m + \bar m) \sqrt{\left( 1 +  (2 \pi
\alpha' q_L)^2 H_1^2 \right)
\left( 1 + (2 \pi \alpha'q_L) ^2 H_2^2 \right) }  -32 \right] \frac{R_1 R_2}{\alpha'}
\nonumber \\
& & + \ \frac{\alpha'}{R_1 R_2} \left[ d +
\bar d - 32 \right] \label{diltad}
\eea
originates from $V_4 O_2 O_2$, and can be clearly linked to
the $\varphi$-derivative of ${\cal S}_{{\rm DBI}}$, computed for this
background. On the other hand, the volume of the first internal torus
originates from $O_4 V_2 O_2$, and the corresponding tadpole,
\bea
& & \left[ n+\bar n + (m + \bar m) \, {1 - (2\pi\alpha' q_L H_1)^2 \over
\sqrt{ 1 + (2\pi\alpha' q_L H_1)^2 }}\,
\sqrt{ 1 + (2 \pi\alpha' q_L H_2)^2 } -32 \right] \frac{R_1 R_2}{\alpha'} \nonumber \\
& & - \ \frac{\alpha'}{R_1 R_2} \left[ d+\bar
d - 32 \right]  \, , \label{mettad}
\eea
can be linked to the derivative of the Dirac-Born-Infeld action
with respect to
the corresponding breathing mode. A similar result holds for the
volume of the second torus, with the proper interchange
of $H_1$ and $H_2$, and, for the sake of brevity, in these NS-NS tadpoles
we have omitted all terms
that vanish using the constraint $m = \bar m $.

The complete form
of eq.~(\ref{mettad}) is also
rather interesting, since, in contrast to the usual structure
of unoriented string amplitudes, it is {\it not} a perfect square.
This unusual feature can be ascribed to the behavior of the
internal magnetic fields under time
reversal. Indeed, as stressed long ago in \cite{cardy2}, these
transverse-channel amplitudes involve a time-reversal operation ${\cal T}$,
and are thus of the form $\langle {\cal T} (B) | q^{L_0} | B \rangle$.
In the present examples, additional signs are introduced by the
magnetic fields, which are odd under time reversal. Consequently, to derive from factorization the M\"obius amplitudes of these models,
it is crucial to add the two contributions $\langle {\cal T} (B)
| q^{L_0} | C \rangle $ and $\langle {\cal T} (C) | q^{L_0} |
B \rangle $. These two contributions are different and effectively eliminate the additional
terms from the transverse channel.

Both (\ref{mettad}) and the dilaton tadpole (\ref{diltad}) simplify
drastically in the interesting case $H_1=H_2$ where, using the
Dirac quantization conditions (\ref{dirac}), they become
\be
\left[ n+\bar n + m + \bar m -32 \right]  \frac{R_1 R_2}{\alpha'}
\mp \frac{\alpha'}{R_1 R_2} \left[ 4 k_1 k_2 (m
+ \bar m) + d +
\bar d - 32 \right] \ .
\ee
Therefore, they both vanish, as they should, in these supersymmetric
configurations, once the corresponding RR tadpole conditions
(\ref{urrt}) are enforced.

The twisted RR tadpoles
\begin{equation}
15 \left[ {\textstyle{1\over 4}} (m-\bar m + n -\bar n )
\right]^2
+ \left[ {\textstyle{1\over 4}} (m-\bar m + n - \bar n ) - (d-\bar d)
\right]^2 \label{twistedrrmag}
\end{equation}
come from the $S_4 O_2 O_2$ sector,
whose states are scalars in the internal space. As in the undeformed
model of subsection 5.6, these conditions
reflect the distribution of the branes among the sixteen
fixed points. Only one fixed point hosts D5 branes in our
examples, and these tadpole conditions are not affected by the magnetic fields and vanish identically for
unitary gauge groups.
The corresponding twisted NS-NS tadpoles that originate from the
$O_4 S_2 C_2$ and $O_4 C_2 S_2$ sectors
are somewhat more
involved, and after the identification of conjugate multiplicities,
are proportional to
\be
{2 \pi \alpha ' q \, (H_1 - H_2 )  \over \sqrt{ (1 + (2 \pi \alpha' q H_1)^2 )
(1 + (2\pi \alpha' q H_2)^2 ) }} \,.
\ee
They encode new couplings for twisted
NS-NS fields that, as expected,  vanish for $H_1 = H_2$.

More details and some examples can be found in~\cite{orientifolds_rev2}, while the combination with a quantized $B_{ab}$ is discussed in~\cite{fradkin}.

\subsection{\sc The Intersecting  Brane Picture }\label{sec:intersecting_branes}

$T$--duality leads to an interesting alternative picture for these systems, where magnetic deformations leave way to
branes at angles~\cite{douglas21,douglas22}. To be specific, let us consider the type IIB string with a pair of branes wrapped on a two-torus in the 12 plane and a uniform magnetic
flux through one of them, so that
\beq
F_{12} \ = \ H \,R_1\,R_2 \ .
\eeq
We can now examine the open strings stretched between the two branes. One of the ends has charge $q$ while the other is uncharged and, as we have seen, the flux is quantized so that
\beq
q_L\, H \,R_1\,R_2 \ = \ \frac{m}{2\pi} \ , \label{Dirac_rot}
\eeq
with $m \in Z$. Let us perform a $T$-duality along the $X^2$
direction, so that with dimensionless coordinates spanning the $[0,2\pi)$ range as in the previous section, eq.~\eqref{2d-duality} gives
\beq
R_2\, \partial_\tau X^2 \ = \ 
R_2'\, \partial_\sigma X^{'\,2} \ , \qquad R_2\, \partial_\sigma X^2 \ = \ 
R_2'\, \partial_\tau X^{'\,2} \ ,
\eeq 
which link $X^2$ to the dual coordinate $X^{'\,2}$, with 
\beq
R_2'=\frac{\alpha'}{R_2} \ . \label{Tdual_rot}
\eeq
In terms of the dual coordinate $X^{'\,2}$, the boundary conditions~\eqref{mag4} induced by $F_{12}$ become
\beq
\partial_\sigma \left( R_1 \,X^1 \ +\  2 \pi \alpha '
q_L\, H \, {R_2'} \, X^{'\,2} \right) \,=\, 0 \ , \qquad \partial_\tau \left( R_2'\,X^{'\,2} \ - \  2 \pi \alpha '
q_L\, H \, {R}_1 \, X^{1} \right) \, = \,  0 \label{mag4bis1}
\eeq
at $\sigma=0$, and
\beq
\partial_\sigma \,X^1  \,=\, 0 \ , \qquad \partial_\tau \,X^{'\,2}  \, = \,  0  \label{mag4bis2}
\eeq
at $\sigma=\pi$.  After $T$--duality, the branes lose one dimension, and the one where the magnetic field is felt is thus rotated by an angle $\theta$ such that
\beq
\tan \theta \ = \   2 \pi \alpha ' q_L H   \ = \ \frac{m\,R_2'}{R_1} \ , \label{tantheta}
\eeq
where we have taken into account that the actual lengths are $R_1 X^1$ and $R_2'\,X^{'\,2}$. The last expression is obtained after using eqs.~\eqref{Dirac_rot} and \eqref{Tdual_rot}, and describes a brane rotated in such a way that one complete horizontal translation is accompanied by $m$ vertical ones.
This setup thus lends itself to a different view of Dirac quantization in terms of rotations, with an important generalization that was first noted in~\cite{berlinmadrid1,berlinmadrid2}. 
    \begin{figure}[ht]
\centering
%\begin{tabular}{cc}
\includegraphics[width=65mm]{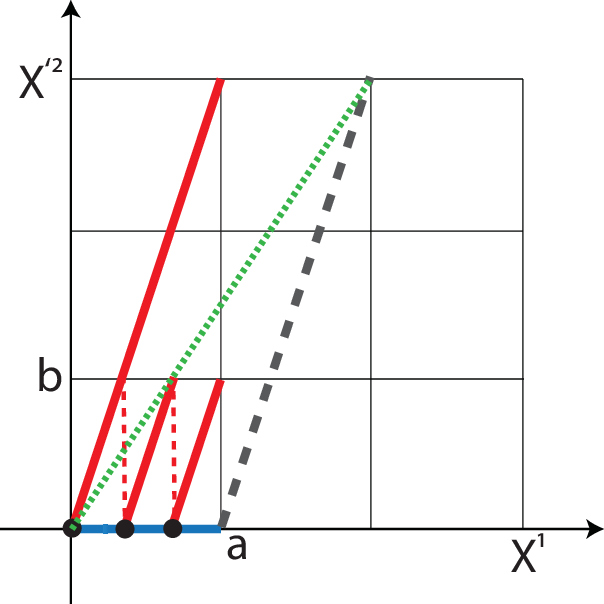} \hskip 1cm
%\end{tabular}
\caption{\small A brane with wrapping numbers $(1,3)$ (red, solid) drawn in the fundamental cell and in its extension, another brane with wrapping numbers $(1,0)$ (blue, solid), their three intersections in the fundamental cell (black dots), which reflect themselves in the multiplicity of chiral fermion families, and a brane with wrapping numbers $(2,3)$ (green, dotted). }
\label{fig:rotated_branes}
\end{figure}

The general rotation angles that grant that a brane lies on internal cycles of finite length obtain if it wraps $m$ times the vertical axis and $n$ times the horizontal one, so that the condition~\eqref{tantheta} generalizes to
\beq
\tan \theta \ = \ \frac{m R'_2}{n R_1} \ . \label{int1}
\eeq
In the original magnetic setup this type of configuration obtains starting from $n$ coincident branes, so that the charge is $n \,q_L$.

A rotated D-brane $A$ of wrapping numbers $(m_A,n_A)$ on a torus wraps the one-cycle
\beq
\Pi_A \ = \  n_A \ |a \rangle \ + \ m_A \ |b \rangle   \ , \label{int2}
\eeq
where $|a \rangle$ denotes the horizontal cycle and $|b \rangle$ the vertical one of the torus, and the associated gauge groups are generally unitary. For a pair of D-branes
of wrapping numbers $(m_A,n_A)$, $(m_B,n_B)$, there are chiral zero modes valued in bi--fundamental representations that are localized at their intersection. The resulting number of generations equals the number of times the two branes intersect in the internal torus, which is given by the intersection number~\cite{berlinmadrid1,berlinmadrid2}
\beq
I_{AB} \ = \q \Pi_A \circ \Pi_B \ = \ n_A m_B \ - \ n_B m_A \ .  \label{int3}
\eeq

If the rotations occur in several tori, the formalism generalizes in a straightforward way. For example, the number of chiral zero modes is generally given by the total number of intersections in the internal space. For a product of tori $T^i$ and factorizable cycles in them, eqs.~\eqref{int2})-\eqref{int3} thus generalize to
\bea
\Pi_A &=& \otimes_i  \left( n_A^{(i)} \ |a_i \rangle \ + \ m_A^{(i)} \ |b_i \rangle \right) \ , \nonumber \\
I_{AB} &=&  \Pi_A \circ \Pi_B \ =\
  \prod_i (n_A^{(i)} m_B^{(i)} \ - \ n_B^{(i)} m_A^{(i)})
 \ ,  \label{int4}
\eea
and the sign of the intersection reflects the chirality.

One can also discuss in this language the number of supersymmetries that are preserved by brane rotations. Indeed, starting from the original ten dimensional supersymmetry generators $Q$, ${\tilde Q}$, say in type IIB, a configuration involving branes rotated by ${\cal R}$ preserves the supercharges that satisfy the conditions
\beq
Q \ = \ {\cal R}\, {\tilde Q} \ , \qquad  Q \ = \ {\tilde Q}  \ , \label{int5}
\eeq
which imply the eigenvalue equation
\beq
Q \ = \ {\cal R}\, {Q} \ . \label{int6}
\eeq
The spinorial charges are here specified by the collection of internal helicities $| \{s_i\} \rangle$, $n$ for $2n$ internal dimensions, which are equal to $\pm \frac{1}{2}$, while the rotation acts as
\beq
{\cal R} | \{s_i\} \rangle = e^{2 \pi i \,\sum_i\, \theta_i s_i}\ | \{s_i\} \rangle \ .   \label{int61}
\eeq
Considering for definiteness compactifications from ten to four dimensions of the type IIA string, which is reached from type IIB after three $T$-dualities, where rotated D6 branes identify lines as above in the three tori, depending on the internal rotation angles $\theta_1$,
$\theta_2$, $\theta_3$, the number of supersymmetries preserved by one rotated brane in four dimensions are
\bea
&& {\rm N=0 \quad supersymmetry \ :\quad if} \quad \pm \theta_1 \pm \theta_2 \pm \theta_3 \not= 0 \ , \nonumber \\
&& {\rm N=1 \quad supersymmetry \ :\quad if} \quad \pm \theta_1 \pm \theta_2 \pm \theta_3 = 0 \ , \nonumber \\
&& {\rm N=2 \quad supersymmetry \ : \quad if} \quad \pm \theta _1 \pm \theta _2 = 0, \theta _3 = 0, \label{int7}
\eea
if there is a solution to the previous conditions for a given choice of signs.

In orientifolds of type II strings with rotated D-branes, the inclusion of O-planes requires for each stack of D-branes to include their images. For example, for $D6_A$ rotated branes, in toroidal compactifications there are also 8 $O6$ planes. Their presence requires to add
branes images $D6_{A'}$  which wrap the cycle
\beq
\Pi_{A'} = \otimes_i  \left( n_A^{(i)} \ |a_i \rangle \ - \ m_A^{(i)} \ |b_i \rangle \right) \ . 
\eeq

\section{\sc Calabi-Yau Compactifications} \label{sec:calabi-yau}
The orbifold compactifications that we discussed in Section~\ref{sec:6dstrings} allow for detailed analyses of the resulting string spectra. The supersymmetric cases actually capture singular limits of compactifications on smooth Ricci--flat Calabi-Yau manifolds~\cite{calabiyau}. These can be regarded as generalizations of tori that connect ten--dimensional strings to four--dimensional Minkowski space, while preserving one quarter of the original supersymmetries. Although the corresponding string spectra are not exactly calculable, and even the metric tensors of Calabi-Yau manifolds are not explicitly known, the low--lying modes can be nicely characterized in the language of Field Theory and reflect the topology of the internal space. Despite the language difference, the main lessons of orbifolds and Calabi--Yau spaces are thus closely related, as we are about to see.

\subsection{\sc Conditions for the Existence of Killing Spinors}

The starting point to describe this class of compactifications is the ten--dimensional $N=1$ supergravity Lagrangian of Section~\ref{sec:sugra1110}, which captures the low--energy limits of both heterotic and type I superstrings. 
The corresponding supersymmetry transformations of eqs.~\eqref{susy_typeI}, which we repeat here for the reader's convenience since they will play a central role in the following, read
\bea{}{}{}{}{}{}{}{}{}{}{}{}{}{}{}{}{}{}
\delta\,{\psi}_M &=& D_M\,\epsilon \ + \ \frac{1}{8}\,\Gamma^{NP}\,  {\cal H}_{MNP}\, \epsilon \ , \nonumber \\
\delta\,\lambda &=& \Gamma^M\,\epsilon\, \partial_M\,\phi
\ + \  \frac{1}{12} \ \Gamma^{MNP} H_{MNP} \, \epsilon\, \ ,\qquad  \delta \chi^a\ = \ - \ \frac{1}{8}\ F_{MN}^a \Gamma^{MN}\epsilon \ .\label{susy_final2cy}
\eea

We shall now focus on compactifications to four dimensions in product manifolds ${\cal M}_4\times \Sigma_6$, so that the ten-dimensional Lorentz symmetry breaks down to
$SO(1,3) \times SO(6)$ and the metric is block diagonal. 
The ten--dimensional supersymmetry parameter $\epsilon$ is a Majorana-Weyl spinor in the ${\bf 16}$ representation of $SO(1,9)$, whose $SO(1,3) \times SO(6)$ decomposition
\beq
{\bf 16} \ \to \ ({\bf 2},{\bf 4}) +  ({\bf \bar 2},{\bf \bar 4}) \ \label{eq:cy2}
\eeq
reveals that it can give rise to at most four Weyl spinors in four dimensions. Consequently, one could obtain at most an ${\cal N}=4$ supersymmetric theory from the compactifications, which is the case if the internal manifold is a torus. 
In general, one can consider ten-dimensional spinors with a nontrivial dependence on internal coordinates, which decompose according to
\beq
\epsilon = u (x) \otimes \zeta (y) \label{eq:cy3} \ .
\eeq
Fermi fields vanish in the vacuum, but their supersymmetry transformations generally do not, since they involve bosonic profiles. The internal parameters $\zeta (y)$ for which 
\beq
\delta_\epsilon \left(\mathrm{Fermi}\right) \ =\ 0 
\eeq
identify the unbroken supersymmetries in four dimensions.

In the simplest setting with a vanishing three-form field strength $H_{MNP}=0$, to which we shall confine our attention here, one can see from eqs.~\eqref{susy_final2cy} that unbroken supersymmetry requires a constant dilaton, while the spinors $u(x)$ and $\zeta(y)$ must satisfy
\beq
D_\mu \, u \  = \ 0 \ , \qquad \nabla_i \, \zeta \ = \ 0  \ .\label{eq:cy4}
\eeq
Taking the product structure of the manifold into account, these conditions imply
\bea
\left[ D_\mu \,,\, D_\nu \right]  u &=& R_{\mu\nu}{}^{ab} \ \gamma_{ab} \, u \ = \ 0 \ , \nonumber \\
\left[ \nabla_i \,,\, \nabla_j \right]  \zeta &=& R_{ij}{}^{pq} \ \gamma_{pq} \, \zeta \ = \ 0 \ .
\eea
Multiplying the first equation by $\gamma^\nu$ and the second by $\gamma^i$ and taking the cyclic identity into account, these equations imply the simpler conditions
\beq
R_{\m\nu}\,\gamma^\nu u \ = \ 0 \ , \qquad R_{ij}\,\gamma^j \zeta \ = \ 0 \ .
\eeq
With maximal symmetry, $R_{\m\nu}=\Lambda \,g_{\mu\nu}$, with $\Lambda$ proportional to the cosmological constant, so that the first equation demands a Minkowski spacetime. In a similar fashion, the second condition can be turned into
\beq
R_{ij} \ R^{ij} \ = \ 0 \ ,
\eeq
which demands the vanishing of $R_{ij}$ if the Euclidean signature of the internal space is taken into account. Summarizing, this setup leads directly to compactifications to Minkowski space with Ricci flat internal manifolds.

As we have seen in eq.~\eqref{eq:cy4}, the internal spinor profile $\zeta(y)$ associated with the unbroken four--dimensional supersymmetries must be covariantly constant in the internal space. The space of solutions for $\zeta$ in eq.~(\ref{eq:cy4}) is related to the holonomy group of the internal manifold. This is the group generated by all rotations that affect the spinor when it is transported along all possible closed paths. For a generic manifold, the holonomy group is $SO(6)$, and there is no covariantly constant spinor, so that all supersymmetries are broken. Conversely, if the internal space is a six-torus, the holonomy group is trivial, and one ten-dimensional spinor gives rise to four covariantly constant spinors in four dimensions, and $\zeta$ is independent of the internal coordinates. In this case, starting with ${\cal N}=1$ supersymmetry in ten dimensions, one ends up with ${\cal N}=4$ supersymmetry in four dimensions. This is true for both heterotic and type I strings, which have one supersymmetry to begin with, while if one starts with IIA and IIB strings, which have ${\cal N}=2$ supersymmetry in ten dimensions, one ends up with
${\cal N}=8$ supersymmetry in four dimensions. Type II compactifications are interesting and instructive, but they are incompatible with the need to recover four--dimensional chiral spectra and interactions, in order to try and connect ten--dimensional strings to the Standard Model.

Since the $\gamma_{pq}$ are rotation generators in the spinorial representation, in order to preserve a fraction of the original supersymmetry one must consider internal spaces with reduced holonomy. In particular, a single invariant spinor in four dimensions is obtained from internal spaces with $SU(3) \subset SO(6)$ holonomy. In fact, an internal spinor valued in the $\mathbf{4}$ of $SO(6)$ decomposes according to
\beq
{\bf 4} \to {\bf 3} + {\bf 1} \label{eq:cy6} \ ,
\eeq
and the presence of the $SU(3)$ singlet grants the preservation of precisely ${\cal N}=1$ supersymmetry in four dimensions. This is the only amount of supersymmetry that is compatible with chirality, as we have stressed.
It is now convenient to use complex $SU(3)$ coordinates for the internal space, rather than the original six real ones. This corresponds to resorting to the decomposition ${\bf 6} \to {\bf 3} + {\bf \bar 3}$, and then an internal vector
$A_m$ can be decomposed as $A_i$, $A_{\bar i}$, with $i=1,2,3$.  

In addition to what was previously done for the gravitino and dilatino fields, one should also set to zero the supersymmetry variation of the ten--dimensional gaugini in eqs.~\eqref{susy_final2cy}. One can show that this enforces on the gauge field strengths the conditions~\cite{stringtheory}
\beq
F_{ij} \ = \ F_{\bar i \bar j} \ = \ 0 \ , \qquad
g^{i \bar j} F_{i \bar j} \ = \ 0 \ , \label{eq:cy7}
\eeq
which can be obtained by carefully analyzing the condition
\beq
\left(\gamma^{ij}\,F_{ij} \ + \ \gamma^{\bar i \bar j}\,F_{\bar i \bar j} \ + \ 2\, \gamma^{i \bar j} \ F_{i \bar j} \right) \zeta \ = \ 0 \ ,
\eeq
while also taking into account that the $\gamma$'s act on the components of $\zeta$ as fermionic creation and annihilation operators. Consequently, one can conclude that the preceding condition only constrains the trace portion of $F_{i \bar j}$, and not all of it, to vanish~\cite{stringtheory}. 

Moreover,
one must satisfy the Bianchi identity for the antisymmetric tensor field
\beq
d H_3 \ = \ \frac{\alpha'}{4} \left[ \ tr (R \wedge R) \ - \ tr
(F \wedge F) \ \right] \ , \label{eq:cy8}
\eeq
taking into account the Yang--Mills Chern--Simons term first introduced in~\cite{I} and its higher--derivative gravitational counterpart introduced by Green and Schwarz~\cite{gs} to grant the cancellation of gauge and gravitational anomalies in ten--dimensional strings.
Embedding the spin connection into the gauge group yields a natural solution to this Bianchi identity for vanishing $H$, since it cancels the two contributions against one another, and to the additional condition on the gauge potential. For example, when applied to the $E_8 \times E_8$ heterotic string this option breaks the $E_8 \times E_8$ gauge group to $E_6 \times SU(3)$, and one can identify the $SU(3)$ factor with the spin connection granting the needed $SU(3)$ holonomy. In this case, the (maximum) resulting gauge group is $E_6 \times E_8$, where $E_6$ can be identified with a grand--unified gauge group and $E_8$ can be associated with a hidden sector. Before this analysis, it was known that $E_6$ is the largest simple unified gauge group that can accommodate the chiral spectrum of the Standard Model~\cite{e6}. Moreover, via gaugino condensation a strongly coupled hidden $E_8$ factor could generate a dynamical scale $\Lambda$ leading to supersymmetry breaking, which could be transmitted to the observable sector via gravitational interactions at a hierarchically lower scale $\Lambda^2/M_P$, along the lines of what we saw in Section~\ref{sec:SUGRA}.
With a constant dilaton and a vanishing antisymmetric field strength background,  the supergravity field equations finally reduce to
\beq
R_{mn} \ = \ 0 \ , \qquad \nabla^m F_{mn} \ = \ 0
\ ,  \label{eq:cy9}
\eeq
so that one recovers, in this fashion, the conditions that the internal manifold must be Ricci--flat, which already emerged from the conditions for $N=1$ four--dimensional supersymmetry. 

\subsection{\sc Brief Review of Differential Topology}

The topology of the internal space plays a crucial role in determining the massless modes of these string compactifications. In order to address these low--lying spectra, let us begin by recalling some basic concepts for real internal spaces $\Sigma$. 

A p-form $\omega_p$ is \emph{closed} if $d \omega_p = 0$ and is \emph{exact} if $\omega_p = d \alpha_{p-1}$. Since $d$ is nilpotent $(d^2 = 0)$, every exact form is also closed, but the converse is not true in general. The de Rham $p$-th cohomology class of the manifold $\Sigma$
\beq 
H^p_{dR} (\Sigma) \ = \ \frac{{\rm closed \ }p{\rm -forms \ on \ \Sigma} }{{\rm exact \ }p{\rm -forms \ on \ \Sigma}} \label{eq:dR1}
\eeq 
is the set of closed $p$-forms $\omega_p$ modulo the equivalence relation 
$\omega_p \sim \omega_p + d \alpha_{p-1}$. 
The Betti numbers, 
\beq
b_p \ = \ {\rm dim} \ H^p_{dR} (\Sigma) \ , 
\eeq
are the dimensions of the different cohomology classes, and the inequivalent classes $[\omega_p]$ capture the cohomology of the space. 
They also determine the Euler number of the manifold according to
\beq 
\chi (\Sigma) \ = \ \sum_{p=0}^d (-1)^p \, b_p \ . \label{eq:dR2}
\eeq 
A p-form is harmonic if $\Delta \,\omega = 0$, where
$\Delta$ is the Laplacian operator for the space $\Sigma$. It can be shown that the number of independent harmonic p-forms on $\Sigma$ equals $b_p$, and moreover $b_p = b_{d-p}$. 

A different but equivalent way of encoding the topology of manifolds rests on its submanifolds $N \subset \Sigma$. A collection of submanifolds $N_p$ of dimension $p$ is called a $p$-chain. An important notion is that of boundary, $\partial N_p$, which is a submanifold of dimension $p-1$. When there is no boundary, the chain is called a $p$-cycle $a_p$, which therefore satisfies $\partial a_p=0$. There is an analogy between the boundary operator and the differential operator $d$ for the forms that define the cohomology, since $\partial (\partial a_p)=0$. Moreover, a $p$-chain that is the boundary of a $p+1$-dimensional submanifold $a_p = \partial b_{p+1}$ is said to be trivial. Any trivial $p$-chain is clearly a $p$-cycle, but the converse is true only for topologically trivial manifolds. One can therefore define homology classes according to  
\beq 
H_p  (\Sigma) \ = \ \frac{  \ p {\rm-cycles \ on \ \Sigma} }{{\rm trivial \ } p{\rm -chains \ on \ \Sigma}} \ , \label{eq:dR01}
\eeq 
and the homology class $[a_p]$ of a $p$-chain can be defined via the equivalence relation $a_p \sim a_p + \partial b_{p+1}$, where $b_{p+1}$ is a $(p+1)$-chain. By definition, trivial cycles lie in the trivial class $[0_p]$. There is a one-to-one correspondence between the cohomology $H^p (\Sigma)$ and the homology $H_{d-p} (\Sigma)$ of a manifold, called Poincar\'e duality. For every $p$-form $\omega_p$, there is a corresponding $d-p$ cycle $N(\omega)$ such that
\beq 
\int_\Sigma \omega_p \wedge \alpha_{d-p} \ = \ \int_{N(\omega)} \alpha_{d-p} \ , \label{eq:dR02}
\eeq 
for arbitrary choices of $\alpha_{d-p}$.
This correspondence generalizes the form constructed out of the projection $\delta (a_p)$ of the volume form into the coordinates normal to the $p$-cycle $a_p$. 

Cohomology and homology have the structure of vector spaces. For a linear combination of chains or cycles $N_p = \sum_k c_k N_p^k$, one defines the corresponding linear combination of integrals 
\beq 
\int_{N_p} \omega_p \ = \ \sum_k c_k \int_{N_p^k} \omega_p  \ . \label{eq:dR03}
\eeq 
If the coefficients $c_k$ are integers, one talks about integral homology $H_p(\Sigma,Z)$, and similarly one can define integral cohomology $H^p(\Sigma,Z)$.   
As for usual integrals, Stokes's  theorem implies that
\beq 
\int_{a_p} d B_{p-1} \ = \  \int_{\partial a_p} B_{p-1}  \ , \label{eq:dR04}
\eeq 
and using this result one can easily prove that the integrals of forms $A$ over cycles $a$ only depend on the classes $[A]$ and $[a]$:
\beq 
\int_{a'} A' \ = \ \int_{a} A \ . \label{eq:dR05}
\eeq 
The Hodge duality between $p$-forms and $d-p$-forms implies that $H^p (\Sigma)$ and $H_p (\Sigma)$ are dual spaces, so that to each $p$-form one can associate a $p$-cycle. It also implies that one can always choose a basis of forms and cycles such that 
\beq 
\int_{a_i} A_j \ = \ \delta_{ij} \ . \label{eq:dR06}
\eeq 
One can also define the intersection number of two integer cycles $a_p \subset H_p (\Sigma,Z)$ and $b_{d-p} \subset H_{d-p} (\Sigma,Z)$, which is a topological invariant in a manifold $\Sigma$, via
\beq 
[a_p] \cdot  [b_{d-p}] \ = \ \int_\Sigma \omega_{a,p} \wedge \omega_{b,d-p} \ , \label{eq:dR07}
\eeq 
where $\omega_{a,p}$ ($\omega_{b,d-p}$) is the dual form to the cycle $a_p$ ($b_{d-p}$). 

As we have seen, the internal spaces for string compactifications with $N=1$ supersymmetry have a complex structure, so their dimension is even, $d=2n$. Consider, therefore, spaces with complex dimensions $n$, letting $\Sigma_{2n}= {\cal M}_n $. On a complex manifold ${\cal M}_n$, one can perform a finer classification of cohomology classes relying on holomorphic and anti-holomorphic quantities, so that $p$-forms on ${\cal M}_n$ can be further decomposed according to 
\beq 
\omega_k (\Sigma_{2n}) \ = \ \sum_{p=0}^n \ \omega_{p,k-p} ({\cal M}_n) \ . \label{eq:dR3}
\eeq 
One can also introduce (anti)holomorphic differential operators
\beq 
\partial\ =\  d z^i \,\partial_i \ , \qquad 
{\overline \partial}\ = \ d {\bar z}^{\bar i} \,
\partial_{\bar i} \   \label{eq:dR4}
\eeq 
such that 
\beq
\partial \ : \  \omega_{p,q} ({\cal M})\  \to \ \omega_{p+1,q} ({\cal M}) \ , \qquad {\overline \partial} \ :  \ \omega_{p,q} ({\cal M}) \ \to \ \omega_{p,q+1} ({\cal M}) \ .  
\eeq
These operators are nilpotent, and therefore one can define the complex  
counterpart of de Rham cohomology, which is called Dolbeault cohomology:
\beq 
H^{p,q}_{\bar \partial} ({\cal M})\ = \ \frac{{\rm {\bar \partial}-closed \ (p,q)-forms \ on \ {\cal M}} }{{\rm {\bar \partial}-exact \ (p,q)-forms \ on \ {\cal M}}} \ . \label{eq:dR5}
\eeq 
The dimensions of these spaces define the Hodge numbers $h^{p,q}= H^{p,q}_{\bar \partial} ({\cal M})$, and the topology of complex manifolds is completely encoded in these data. 

Complex manifolds are naturally endowed with Hermitian metrics $g_{i \bar j}$, and
one can define the {\it K\"ahler form}
\beq
J \ = \ i \ g_{i \bar j} \ dz^i d {\bar z}^{\bar j} \ .
 \label{eq:cy10}
\eeq
The relevant complex spaces for supersymmetric string compactifications are K\"ahler manifolds, for which the K\"ahler form is closed~ \footnote{In the presence of torsion introduced by form fluxes, which we leave out here for brevity, these conditions are violated and the solutions have a more general structure~\cite{nonkahler}.}
\beq
d J \ = \ 0
\ . \label{eq:cy11}
\eeq
This implies that the metric can be written locally in the form
\beq
g_{i \bar j} \ = \ \partial_i \,\partial_{\bar j} \,{\cal K} (z, {\bar z}) \ , \label{eq:cy12}
\eeq
where the real function ${\cal K}$ is the {\it K\"ahler potential}.  In Section~\ref{sec:broken_global} we have described the role played by ${\cal K}$ in field--theory models of low--energy supersymmetry, and in Section~\ref{sec:SUGRA} we have described its role in $N=1$ four--dimensional supergravity. Here we can see its emergence from the internal manifold.

On a compact K\"ahler manifold $\Sigma_{2n}= {\cal M}_n$,  the Betti and Hodge numbers satisfy
\beq 
b_k = b_{2n-k} \quad , \quad b_k = \sum_{p=0}^k 
h^{p,k-p} \quad , \quad h^{p,q} = h^{q,p} \quad , \quad  h^{p,q} = h^{n-p,n-q} \quad , \quad 
h^{p,p} \geq 1  \ . \label{eq:dR6}
\eeq 
These relations imply, in particular, that all non--vanishing Betti numbers of odd order are even, and all
Betti numbers of even order do not vanish.

Defining the first Chern class of a generic complex manifold ${\cal M}$ as
\beq
c_1 \ = \ \frac{1}{2 \pi} \ \int \  R_{i \bar j} \,dz^i \, d {\bar z}^{\bar j} \ , \label{eq:cy17}
\eeq
a K\"ahler manifold with vanishing first Chern class $(c_1=0)$ is called a {\it Calabi--Yau (CY) manifold. Calabi’s conjecture, which was later proved by Yau, states that an N-dimensional K\"ahler manifold with vanishing first Chern class admits a metric of SU(N) holonomy.} Intuitively, since the Christoffel connection has purely holomorphic or purely anti-holomorphic labels, as we saw in eq.~(\ref{bri3}), the rotations on closed loops do not mix the two types. Consequently, one expects the holonomy group to be at most $U(N)= SU(N) \times U(1)$. A vanishing of the first Chern class guarantees that the holonomy in the overall $U(1)$ factor is trivial. Complex three--dimensional Calabi-Yau manifolds (CY 3-folds)  play a key role in string compactifications to four dimensions, since they have $SU(3)$ holonomy and preserve $\frac{1}{4}$ of the maximal supersymmetry. A smaller holonomy group leads to a larger residual amount of supersymmetry in four dimensions. For example, there is a unique compact complex two--dimensional Calabi--Yau space of $SU(2)$ holonomy, called $K3$. When combined with an additional $T^2$, it preserves half of the original supersymmetries, so that $K3 \times T^2$ compactifications of the heterotic string can lead to vacua with $N=2$ supersymmetry. In Section~\ref{sec:toroidal_ss}, we have described in detail type--IIB compactifications on the $T^4/\mathbb{Z}_2$ orbifold limit of $K_3$, the corresponding supersymmetric orientifold and its variant with ``brane supersymmetry breaking''.

The number of parameters entering the choice of a metric on a CY 3-fold depends on the Hodge numbers.
The Hodge numbers on a CY 3-fold are usually displayed in a diamond shape
\bea 
&& \qquad \qquad \qquad h^{0,0} \qquad \qquad \qquad \qquad \qquad \qquad \quad \quad \ 1 \qquad \qquad \nonumber \\
&& \qquad \qquad h^{1,0} \qquad  h^{0,1} \qquad \qquad \qquad \qquad \quad \quad \quad \ 0  \qquad 0 \nonumber \\
&& \qquad h^{2,0} \qquad  h^{1,1} \qquad  h^{0,2} \qquad \qquad \qquad \quad \quad 0  
\qquad \ \ h^{1,1} \qquad 0 \nonumber \\
&& h^{3,0} \qquad h^{2,1} \qquad  h^{1,2}  \qquad h^{0,3} \qquad = \qquad 1  
\qquad h^{2,1} \qquad  h^{2,1} \qquad 1 \nonumber \\
&& \qquad h^{3,1} \qquad  h^{2,2} \qquad  h^{1,3} \qquad \qquad \qquad \quad \quad \ 0  
\qquad \ h^{1,1} \qquad \ 0 \nonumber \\
&& \qquad \qquad \ h^{3,2} \qquad  h^{2,3} \qquad \qquad \qquad \qquad \qquad \quad \ \ 0  \qquad  0 \nonumber \\
&& \qquad \qquad \qquad \ h^{3,3} \qquad \qquad \qquad \qquad \qquad \qquad \quad \quad \ \ 1 \qquad \qquad \qquad \qquad \nonumber \ . \label{eq:cy31}
\eea 
On a Calabi-Yau three-manifold, there is a nowhere vanishing, holomorphic $(3,0)$ form
$\Omega_3$, which is covariantly constant in the Ricci-flat metric
\beq
\Omega_3 = \Omega_{ijk} dz^i dz^j dz^k \ . \label{eq:cy18}
\eeq
The zero modes of ten-dimensional fields on Calabi-Yau spaces are completely determined by $h^{3,0}=1$, and by the two numbers $h^{1,1}$ and $h^{2,1}$, which are not determined a priori. $h^{1,1}$ counts the infinitesimal deformations of the K\"ahler structure, while $h^{2,1}$ counts the infinitesimal deformations of the complex structure. The Euler number $\chi$ of the CY space,
\beq
\chi \ = \ \sum_{p,q=0}^3 (-1)^{p+q} \ h^{p,q} \ = \ 2 \ ( h^{1,1} - h^{2,1} ) \ . \label{eq:cy182}
\eeq
determines the number of chiral zero modes in four dimensions.

\subsection{\sc Massless Modes from Calabi--Yau Manifolds}

Let us briefly discuss the compactification from ten to four dimensions, starting from the gauge sector, taken for definiteness to be $E_8 \times E'_8$. We can concentrate on the first ``observable" $E_8$, since the second factor $E'_8$ is not affected by the geometry of the CY space. Since $E_8$ is broken to $E_6$ when the spin connection is embedded in the gauge group, one can decompose the adjoint of $E_8$ into $SU(3) \times E_6$ representations according to
\beq 
{\bf 248} = ({\bf 8, 1}) + ({\bf 1, 78}) +
({\bf 3, 27}) + ({\bf \bar 3, \overline{27}}) 
\ . \label{eq:cy248}
\eeq
One can also decompose the ten--dimensional vector index $M = (\mu, i,  {\bar i})$ and the adjoint labels of $E_8$ into
$i \bar j$, $i x$, ${\bar i} {\bar x}$ and $a$, where $x$ refers to the fundamental representation ${\bf 27}$ of $E_6$, while $a$ labels the adjoint of $E_6$. Due to identification, the same labels $i, {\bar j}$ are used for the $SU(3)$ factors, regardless of whether they originate from $E_6$ or from the internal manifold.
  The four-dimensional zero modes arising from the gauge sector then comprise a gauge field $A_{\mu}^{a}$ in the adjoint of $E_6$, $h^{(1,1)}$ scalars $A_i^{\bar j, {\bar x}}$ in the $\overline{27}$ of $E_6$ and $h^{(2,1)}$ scalars $A_i^{j, x}$ in the ${27}$ of $E_6$. Including the corresponding fermions, one can recognize the emergence of vector multiplets in the adjoint of the $E_6$ gauge group,  $h^{(1,1)}$ chiral multiplets $\Phi_{\bar x}^A$ valued in the $\overline{27}$ of $E_6$ and $h^{(2,1)}$  chiral multiplets $\chi_{ x}^I$  in the ${27}$ of $E_6$. 
 The compactification of the ten-dimensional Yang-Mills action generates directly a potential for the $h^{(1,1)}$ matter chiral multiplets
\beq 
g_{10}^2 \int d^6 y \ Tr \left( [A_M, A_N] [A^M, A^N] \right)
   \ , \label{eq:Wphi}
\eeq 
which can be associated to the superpotential
 \beq
{\cal W} \ = \  {\cal K}_{ABC} \ d^{\bar x \bar y \bar z} \Phi_{\bar x}^A  \Phi_{\bar y}^B  \Phi_{\bar z}^C \ .
 \eeq
The intersection numbers 
\beq 
{\cal K}_{ABC} = \int \omega^A \wedge \omega^B \wedge \omega^C
 \label{eq:cy251}
\eeq 
count the number of intersections of triples of four-cycles $N_A$ dual to the two-forms $\omega^A$. Three four-cycles in a Calabi-Yau manifold intersect generically at isolated points, and their total number 
\beq
\# (N_A,N_B,N_C) \ =  \ {\cal K}_{ABC}
\eeq
is a topological quantity, which does not depend on moduli fields. 

One can analyze along similar lines the compactification of the bosonic ten-dimensional supergravity fields 
$g_{MN}, B_{MN}, \Phi$. The field components along the four--dimensional noncompact directions $g_{\mu \nu}, B_{\mu \nu}, \Phi$ are accompanied by singlet scalars in the internal space, and thus have one zero mode each. The four--dimensional graviton, together with the gravitino zero mode, builds the $N=1$ gravitational multiplet $(g_{\mu \nu}, \Psi_{\mu})$. On the other hand, $B_{\mu \nu}$ is dual to an axion, and together with the dilaton it builds the universal complex axion-dilaton field
\beq
S \ = \ e^{-\,\phi}  \ + \ i\, a \ , \label{eq:cyS}
\eeq 
which is part of a four--dimensional superfield $S$, together with the fermionic dilatino. Its definition depends on the ten--dimensional string theory one starts with, and in particular the preceding discussion is tailored to the heterotic string, where $B_{\mu\nu}$ originates from the NS sector. On the other hand, in the type--I string the axion is dual to the RR 2-form left invariant by the orientifold projection. 

The fields with mixed indices $g_{\mu i}, B_{\mu i}$ have no zero modes, since there are no $(1,0)$ forms on Calabi-Yau spaces, which do not have global symmetries. The fields 
$g_{i \bar j}, B_{i \bar j}$ generate $h^{1,1}$ scalar zero-modes, while the $g_{ij}$ generate $h^{(2,1)}$ scalar zero modes, since one can write
\beq 
\delta\,g_{i \bar j \bar k} \ =  \ \delta\,g_{il} \,g^{l \bar m} \, \Omega_{\bar m \bar j \bar k} \ , \label{eq:cy012}
\eeq 
where $g^{l \bar m}$ denotes the inverse background metric of the Calabi--Yau manifold.
Together with their complex conjugates, these modes identify complex scalars of four--dimensional chiral multiplets. 
On the other hand, the $B_{ij}$ have no zero modes, since there are no $(2,0)$ forms on a Calabi-Yau space.

\subsection{\sc Special K\"ahler Geometry}

The K\"ahler form can be expanded into a basis of 
$H^{1,1} ({\cal M})$ forms $\omega_A$
\beq
J \ = \ \sum_A v^A \omega^A \ , \label{eq:cy13}
\eeq
where the real parameters $v^A$ are interpreted as K\"ahler moduli fields in four dimensions.
One can also define a $(1,1)$ form that contains the internal components of the antisymmetric tensor,
\beq
B \ = \ B_{i \bar j} \,dz^i \, d {\bar z}^{\bar j} \ ,
\  \label{eq:cy14}
\eeq
which leads to introduce complex K\"ahler moduli fields $T^A$ according to
\beq
J + i B \ = \ \sum_A T^A \, \omega^A \ , \label{eq:cy15}
\eeq
where
\beq
T^A \ = \ v^A \ + \ i \,b^A  \ . \label{eq:cy16}
\eeq

There is a simple relation between the string theory condition~(\ref{eq:cy4})  of having one covariantly constant spinor and the basic forms of the CY 3-fold: one can construct them explicitly as
\beq  
J_{i \bar j} \ = \ - i \ \zeta^{\dagger} \Gamma_i 
\Gamma_{\bar j} \zeta \ , \qquad 
\Omega_{ijk} \ = \ \zeta^T \Gamma_i 
\Gamma_{j}  \Gamma_k \zeta \ , \label{eq:cy32}
\eeq 
where the spinor $\zeta$ was defined in eq.~\eqref{eq:cy3}.

It is useful to define {\it special coordinates} on the moduli space, relying on an orthonormal basis of three-cycles ${A^I, B_J}$, $I,J = 0, \ldots, h^{2,1}$, such that the intersection numbers are
\beq
\# (A^I, B_J) \ = \ \delta_I^J \ , \qquad  \# (A^I, A^J) \ = \ 0 \ , \qquad \# (B_I, B_J) \ = \ 0 \ . \label{eq:cy19}
\eeq
The four-dimensional {\it complex structure} moduli are then defined via
\beq
Z^I \ = \ \int_{A^I} \Omega_3  \ . \label{eq:cy20}
\eeq
The effective field theory of complex structure moduli depends on the holomorphic function ${\cal F}$, such that
\beq
{\cal F}_I \ = \  \frac{\partial {\cal F}}{\partial Z^I} \ = \ \int_{B_I} \Omega_3  \ , \label{eq:cy21}
\eeq
where ${\cal F}$ is a homogeneous holomorphic function of degree two, so that ${\cal F} (\lambda Z) =
\lambda^2 {\cal F} (Z) $. With these definitions, the holomorphic $3$-form can be expanded as
\beq 
\Omega_3 \ =\  Z^I \alpha_I \ - \ {\cal F}_I \beta^I 
\ , \label{eq:cy22}
\eeq 
where $(\alpha_I, \beta^I)$ is a basis of $3$-forms,  and $\alpha_I$ are dual to the $A^I$  $3$-cycles, while $\beta^I$ are dual to the $B_I$. This means more explicitly 
\beq 
\int \Omega_3 \wedge \beta^I \ = \ \int_{A^I } \Omega_3 \ , \qquad \int \Omega_3 \wedge \alpha_I \ = \ \int_{B_I } \Omega_3 \ . \label{eq:cy23}
\eeq 
The data of the four-dimensional effective Lagrangian are thus encoded in the K\"ahler potential
\beq 
{\cal K} \ = \ - \ln {\cal V} \ - \  \ln \left( i \int \Omega_3 \wedge {\overline \Omega_3}  \right) \ , 
\eeq
where
\beq
 {\cal V}  \ = \ \int J \wedge J \wedge J = {\cal K}_{ABC}
\ v^A v^B v^C 
\  \label{eq:cy24}
\eeq
is the volume of the Calabi-Yau space.  
The last term in the K\"ahler potential can also be written in the form
\beq
\ln \left( i \int \Omega_3 \wedge {\overline \Omega_3}  \right)\ = \  \ln \, i  \
\left( {\bar Z}^I \partial_I  {\cal F} -  {Z}^I \partial_I  {\cal \bar F} \right)
\ , \label{eq:cy25}
\eeq
and this structure identifies a special K\"ahler manifold.

In summary, the compactification of the $E_8 \times E_8$ heterotic string on a Calabi--Yau 3-fold yields, at low energies, $N=1$ supergravity with an $E_6$ vector multiplet, together with $h^{(1,1)}$ chiral multiplets in the $\overline{27}$ of $E_6$ and $h^{(2,1)}$ chiral multiplets in the $27$ of $E_6$. The result is generally a chiral spectrum, with an even number of generations determined by the Euler character of the manifold~\eqref{eq:cy182}. Aside from the $N=1$ supergravity multiplet reviewed in Appendix~\ref{app:superfields_local} and the preceding modes, the low--energy spectrum also includes the universal axion--dilaton multiplet and $h^{(1,1)} + h^{(2,1)}$ neutral chiral multiplets. The K\"ahler manifold is of a special type, since the K\"ahler potential, as we have seen, has a holomorphic substructure.

One can extend these considerations to type--II theories, whose Calabi--Yau compactifications lead to $N=2$ supersymmetry in four dimensions. To this end, it suffices to consider the additional bosonic fields present in the two cases:
\bea
\mathrm{Type \ IIA\ } &:& A_M, \ A_{MNP} \ , \nonumber \\
\mathrm{Type \ IIB\ } &:& a, \ A_{MN}, \ A_{MNPQ}^{(+)} \ ,
\eea
where the superscript in the last term is meant to stress that the corresponding field strength is selfdual.

For type IIA, additional zero modes come from
\beq
A_\mu , \ A_{\mu i \bar{j}}, \ A_{ijk} \ = \ \Omega_{ijk} \,c \ , A_{ij\bar{k}}
\eeq
where the last field is complex. $A_\mu$ completes the bosonic content of the $N=2$ gravitational multiplet, the $ A_{\mu i \bar{j}}$, together with $g_{i \bar{j}}$ and $B_{i\bar{j}}$, complete the bosonic content of $h^{1,1}$ $N=2$ vector multiplets, while $\phi$, $B_{\mu\nu}$, $c$ and its conjugate complete the bosonic content of the ``universal'' hypermultiplet. Finally, there are $h^{2,1}$ additional hypermultiplets whose bosonic content originates from the $A_{ij\bar{k}}$, the $g_{ij}$ and their conjugates. In conclusion, there are $h^{2,1}+1$ hypermultiplets and $h^{1,1}$ vector multiplets in the massless spectrum originating from type IIA. In a similar fashion, one can show that there are $h^{2,1}$ vector multiplets and $h^{1,1}+1$ hypermultiplets in the massless spectrum originating from type IIB.

\subsection{\sc K3 and Six--Dimensional Compactifications}

The six--dimensional $T^4/\mathbb{Z}_2$ reduction of type IIB discussed in Section~\ref{sec:6dstrings} has led to the anomaly--free six--dimensional spectrum first noted in~\cite{agwitt}, which comprises the $(2,0)$ gravitational multiplet and 21 tensor multiplets. As we saw in Section~\ref{sec:6dstrings}, the former multiplet contains five self--dual two-forms, while each of the latter multiplets contains an antiself--dual two--form. In this context, the counterpart of the Calabi--Yau spaces is a four--dimensional manifold of $SU(2)$ holonomy called K3~\cite{aspinwall}, with a unique topological structure characterized 
by the independent Hodge numbers $h^{0,0}=1$, $h^{1,0}=0$, $h^{2,0}=1$, $h^{1,1}=20$, 
whose Hodge diamond is
\bea 
&& \qquad \qquad \qquad h^{0,0} \qquad \qquad \qquad \qquad \qquad \qquad \quad \quad \ 1 \qquad \qquad \nonumber \\
&& \qquad \qquad h^{1,0} \qquad  h^{0,1} \qquad \qquad \qquad \qquad \quad \quad \quad \ 0  \qquad 0 \nonumber \\
&& \ \qquad h^{2,0} \qquad  h^{1,1} \qquad  h^{0,2} \qquad  = \qquad \quad \quad \  1  
\qquad \ 20 \qquad 1 \nonumber \\
&& \qquad \qquad \ h^{1,2} \qquad  h^{2,1} \qquad \qquad \qquad \qquad \qquad \quad \ 0  \qquad  0 \nonumber \\
&& \qquad \qquad \qquad \ h^{2,2} \qquad \qquad \qquad \qquad \qquad \qquad \quad \quad \ \ 1 \qquad \qquad \qquad \qquad \nonumber \ . \label{eq:cy31k}
\eea 
Moreover, one can show that out of the 22 two--forms listed in third line of the Hodge diamond, 19 are anti-selfdual and 3 are selfdual. This information suffices to characterize the spectrum emerging from type IIB, starting from its ten--dimensional fields $g_{MN}$, $B_{MN}$, $A_{MN}$, $\phi$, $a$, $A_{MNPQ}^{+}$, simply counting the massless tensor modes.
$B_{MN}$ and $A_{MN}$ give one each, for a total of two selfdual and two anti selfdual tensor modes, since these come with internal scalar wavefunctions, while $A^{+}$ gives 19 anti-selfdual and 3 selfdual two-forms, since  these come with internal tensor wavefunctions. Altogether, one thus finds 5 selfdual two-forms and 21 anti-selfdual ones. The former are part of the gravitational multiplet, whereas each of the latter belongs to a tensor multiplet. The massless spectrum must thus combine the gravitational multiplet and 21 tensor multiplets, as in Section~\ref{sec:6dstrings} and in the original anomaly analysis of~\cite{agwitt}. This is the unique spectrum preserving half of the supersymmetries that emerges from type--IIB in six dimensions, independently of the type of orbifold construction~\cite{dp,gj}. The different constructions differ in the splitting of this spectrum between untwisted and twisted sectors, but the overall result is unique. 

In conclusion, in this section we have reviewed some basic facts about the smooth internal manifolds that connect ten--dimensional superstrings to supersymmetric Minkowski vacua in four and six dimensions. Over the years, a huge literature was devoted to these results and to several important generalizations, also establishing close connections with ongoing research in Mathematics. More details can be found in the recent book~\cite{tomasiello} and references therein.

\section{\sc Branes and Vacua of (Non--)Supersymmetric Strings} \label{sec:SUSY_breaking_com}

The vacua that can be exactly addressed within String Theory, by orbifold and orientifold constructions alike, are a tiny subset of what can be explored with field theory methods. In the supersymmetric case, the former correspond to singular limits of smooth internal manifolds, and in particular of the Calabi--Yau spaces that were explored in Section~\ref{sec:calabi-yau}. The exact limits reveal how strings cope with singular spaces, but the field--theory treatment has the virtue of illuminating the geometric nature of the compactifications. The D-branes that emerged from two--dimensional Conformal Field Theory in the previous sections also afford a geometrical interpretation, as extended objects that break at least part of supersymmetry, if present in the vacuum. They can be addressed with the field theory techniques that we are about to describe, and the development of the AdS/CFT correspondence~\cite{adscft} was also stimulated by a comparison between the different approaches. In the non--supersymmetric case, the field theory approach becomes inevitable, since the emergence of the tadpole potentials that we first met in Section~\ref{sec:10D} drives the vacuum away from Minkowski space, into regions where current string technology is of little help. The formalism that we are about to discuss can serve as a starting point to explore the vacua that we have mentioned and also the branes that they can host, both in supersymmetric cases and in the presence of tadpole potentials.

The cases of interest are captured by the class of metrics
\beq
ds^{\,2}\ = \ e^{2A(r)}\, \gamma_{\mu\nu}(x)\, dx^\mu\,dx^\nu \ + \ e^{2B(r)}\,dr^2\ + \ e^{2C(r)}\, \gamma_{mn}(\xi)\, d \xi^m\,d \xi^n \ , \label{metric_sym}
\eeq
where $\gamma_{\mu\nu}$ is a $(p+1)$--dimensional metric of constant curvature $k$ and $\gamma_{mn}(\xi)$ is a $(D-p-2)$--dimensional metric of constant curvature $k'$. In addition to $\gamma_{\mu\nu}$ and $\gamma_{mn}$, the complete metric thus involves three dynamical functions of a single variable $r$, and there are also symmetric form fluxes supporting it, which will be characterized in Section~\ref{sec:tensor_profiles}. Supersymmetric brane backgrounds can be obtained by solving first--order equations that grant the existence of Killing spinors~\cite{susy_branes}, a procedure that will be illustrated in Section~\ref{sec:BPS_branes}. In the non--supersymmetric case this option does not exist, and one must solve the complete second--order equations of Section~\ref{sec:symmetries}. In this fashion, one can also obtain non--supersymmetric vacua or branes of supersymmetric theories, as in~\cite{mrs24_1} or in the example that will be illustrated in Section~\ref{sec:directsusybcom}. However, characterizing brane profiles in the presence of tadpole potentials requires more general starting points, as in~\cite{mrs24_2}.

\subsection{\sc Effective Action and Equations of Motion} \label{sec:eqs_branes}

In the string frame, the bosonic portions of the low--energy effective field theories that we shall consider in the following include the contributions in
 \bea
 {\cal S} &=&\frac{1}{2\,(\alpha')^\frac{D-2}{2}}\ \int d^{D}x\,\sqrt{-\,G} \Big\{ e^{-2\phi}\Big[\, R\, + \,4(\partial\phi)^2 \Big] \ - \ \frac{\tau_{D-1}}{\alpha'} \, e^{\,\gamma_S\,\phi}
 \nonumber \\ &-& \frac{1}{2\,(p+2)!}\ e^{-2\,\beta_S\,\phi}\, {\cal H}_{p+2}^2 \Big\} \ . \label{eqs1}
 \eea
This prototype action involves two types of fields aside from gravity that play in general a prominent role, the dilaton $\phi$ and a $(p+1)$--form gauge potential ${\cal B}_{p+1}$, of field strength ${\cal H}_{p+2}= d\,{\cal B}_{p+1}$. The values of $p$ and of the two constants $\beta_S$ and $\gamma_S$ for the supersymmetric strings and the three non--tachyonic non--supersymmetric strings are collected in Table~\ref{table:tab_1s}, while the corresponding Einstein--frame parameters are collected in Table~\ref{table:tab_1e}. The ``tadpole term'', proportional to $\tau_{D-1}$, is absent in the supersymmetric case and can describe a noncritical potential if $\gamma_S=-2$ and $D \neq 10$, with $\tau_{D-1} \sim D - 10$, or a tension due to branes and orientifolds if $\gamma_S=-1$ and $D=10$, or a contribution emerging from genus--one amplitudes if $\gamma_S=0$ and $D=10$. In the following, we shall describe the formalism for generic values of $D$, before specializing to the ten--dimensional cases of primary interest here.
 \begin{table}
 \begin{center}
\begin{tabular}{ ||c||c|c|c|| } 
 \hline\hline
  Model & $p$ & $\beta_S$ & $\gamma_S$ \\ [0.5ex] 
  \hline\hline
 $IIA$ & $I,V$;$(0,2,4,6,8)$ & $1,\,-1$;\,$(0,0,0,0,0)$ & - \\ [0.5ex] 
  \hline
 $IIB$ & $I,V$;$(-\,1,1,3,5,7)$ &  $1,\,-1$;\,$(0,0,0,0,0)$ & - \\ [0.5ex] 
  \hline
 $SO(32)$ open & $(1,5)$ &  $(0,0)$ &  - \\ [0.5ex] 
  \hline
 $SO(32)$, $E_8 \times E_8$ het. & $I,V$ &  $1,\,-1$ & - \\ [0.5ex] \hline
    $USp(32)$ & (1,5)  & $(0,0)$ &  $-1$  \\ [0.5ex] 
  \hline
  $U(32)$ & $I,V$;(-1,1,3,5,7)  & $1,\,-1$;\,$(0,0,0,0,0)$ & $-1$ \\ [0.5ex] 
  \hline
 $SO(16) \times SO(16)$ het. & $I,V$ &  $1,-\,1$ & $0$  \\ [0.5ex] 
 \hline\hline
\end{tabular}
 \end{center}
 \vskip 12pt 
 \caption{\small String--frame parameters $\beta_S$ and tadpole--potential parameters $\gamma_S$, whenever present, for the supersymmetric ten--dimensional string models and the three tachyon--free non--supersymmetric strings. Roman numerals refer to NS-NS branes, while entries within parentheses refer to RR ones.}\vskip 12pt
 \label{table:tab_1s}
 \end{table}

 \begin{table}
 \begin{center}
\begin{tabular}{ ||c||c|c|c|| } 
 \hline\hline
  Model & $p$ & $\beta_p$ & $\gamma$ \\ [0.5ex] 
  \hline\hline
 $IIA$ & $I,V$;$(0,2,4,6,8)$ & $\frac{1}{2},-\,\frac{1}{2}$;\,$\left(-\,\frac{3}{4},-\,\frac{1}{4},\frac{1}{4},\frac{3}{4},\frac{5}{4}\right)$ & - \\ [0.5ex] 
  \hline
 $IIB$ & $I,V$;$(-\,1,1,3,5,7)$ &  $\frac{1}{2},-\,\frac{1}{2}$;\,$\left(-\,1,-\,\frac{1}{2},0,\frac{1}{2},1\right)$ & - \\ [0.5ex] 
  \hline
 $SO(32)$ open & $(1,5)$ &  $\left( -\,\frac{1}{2},\frac{1}{2}\right)$ &  - \\ [0.5ex] 
  \hline
 $SO(32)$, $E_8 \times E_8$ het. & $I,V$ &  $\frac{1}{2}$,-\,$\frac{1}{2}$ & - \\ [0.5ex] \hline
    $USp(32)$ & (1,5)  & $\left( -\,\frac{1}{2},\frac{1}{2}\right)$ &  $\frac{3}{2}$  \\ [0.5ex] 
  \hline
  $U(32)$ & $I,V$;(-1,1,3,5,7)  & $\frac{1}{2},-\,\frac{1}{2}$;\,$\left(-\,1,-\,\frac{1}{2},0,\frac{1}{2},1\right)$ & $\frac{3}{2}$ \\ [0.5ex] 
  \hline
 $SO(16) \times SO(16)$ het. & $I,V$ & $\frac{1}{2},-\,\frac{1}{2}$  & $\frac{5}{2}$  \\ [0.5ex] 
 \hline\hline
\end{tabular}
 \end{center}
 \vskip 12pt 
 \caption{\small Einstein--frame parameters $\beta_p$ and tadpole--potential parameters $\gamma_p$, whenever present, for the supersymmetric ten--dimensional string models and the three tachyon--free non--supersymmetric strings. Roman numerals refer to NS-NS branes, while entries within parentheses refer to RR ones.}\vskip 12pt
 \label{table:tab_1e}
 \end{table}
In the Einstein frame, with the corresponding metric $g$ related to $G$ according to
\beq
G_{MN} \ = \ e^\frac{4\phi}{D-2} \ g_{MN} \ , \label{string_vs_einstein}
\eeq
the action of eq.~\eqref{eqs1} becomes
\bea
{\cal S} &=& \frac{1}{2\,\kappa_D^2}\int d^{D}x\sqrt{-g}\left[R\ - \ \frac{4}{D-2}\ (\partial\phi)^2\ - \ \frac{\tau_{D-1}}{\alpha'} \, e^{\,\gamma\,\phi} -  \frac{e^{-2\,\beta_p\,\phi}}{2\,(p+2)!}\, {\cal H}_{p+2}^2 \right] , \label{eqs4}
\eea
with
\beq
\beta_p \,=\, \beta_S \ - \ \frac{D-2(p+2)}{D-2} \ , \quad
\gamma \,=\, \gamma_S \ + \ \frac{2 D}{D-2} \ .
 \label{alphaE}
\eeq

The Einstein--frame field equations, written for a generic potential $V(\phi)$, read
\bea
R_{MN} \,-\, \frac{1}{2}\ g_{MN}\, R &=& \!\!\frac{4}{D-2}\  \pr_M\phi\, \pr_N\phi\, + \, \frac{e^{\,-\,2\,\beta_p\,\phi} }{2(p+1)!}\,  \left({\cal H}_{p+2}^2\right)_{M N}  \nonumber\\
&-& \!\!\frac{1}{2}\,g_{MN}\Big[\frac{4\,(\pr\phi)^2}{D-2}+ \frac{e^{\,-\,2\,\beta_p\,\phi}}{2(p+2)!}\,{\cal H}_{p+2}^2 + V(\phi)\Big] \ ,   \label{eqsbeta}
\\
\frac{8}{D-2} \ \Box\phi &=& \,-\,\frac{\beta_p\, e^{\,-\,2\,\beta_p\,\phi}}{(p+2)!} \ {\cal H}_{p+2}^2 + V^\prime(\phi)  \ , \qquad 
d\Big(e^{\,-\,2\,\beta_p\,\phi}\ {}^{*}\,{\cal H}_{p+2}\Big)  \ = \ 0  \ , \nonumber
\eea
and the tadpole potentials are recovered letting
\beq
V(\phi) \ = \ \frac{\tau_{D-1}}{\alpha'} \ e^{\gamma\,\phi} \ . \label{tadpole_potential}
\eeq
 
 In our conventions the Riemann curvature tensor is defined via
\beq
[ \nabla_M, \nabla_N] V_Q \ = \ {R_{MNQ}}^{P}\, V_P \ ,
\eeq
and therefore
\beq
{R_{MNQ}}^{P} \ = \ \partial_N\,{\Gamma^P}_{MQ} \ - \ \partial_M\,{\Gamma^P}_{NQ}  \ + \ {\Gamma^P}_{NR}\, {\Gamma^R}_{MQ} \ - \ {\Gamma^P}_{MR}\, {\Gamma^R}_{NQ} \ ,
\eeq
and we define the Ricci tensor as
\beq
R_{MQ} \ = \ {R_{MNQ}}^{N} \ .
\eeq
As we have seen in Section~\ref{sec:sugra1110}, in the type IIB string there is a five--form field strength that satisfies the first--order self--duality equation
 \beq{}
{\cal H}_5 \ = \ {}^{*}\,{\cal H}_{5} \ ,
\eeq
and taking it into account will require some slight amendments of the formalism.

\subsection{\sc Isometry Groups and Metric Profiles}
The familiar supersymmetric $p$--branes live in asymptotically flat $D$--dimensional spacetimes, and are special solutions of the low--energy supergravity with isometry groups ${ISO(1,p)}\times {SO(D-p-1)}$. In the following, we shall review basic properties of three important sets of supersymmetric objects of this type, the M2 and M5 branes and the D-branes of String Theory. 

$D$--dimensional spaces with the isometry groups ${G_{k}(p+1)}\times { H_{k'}(D-p-2)}$ of the maximally symmetric Lorentzian and Euclidean manifolds with dimensions equal to $p+1$ and $D-p-2$ provide useful generalizations. The two integers $k$ and $k'$, equal to $\pm 1$ or $0$, determine the corresponding constant curvatures. As is well known, the space--time isometry groups $G_k$ for the three cases $k=1,0,-1$ are $SO(1,p+1)$, $ISO(1,p)$ and $SO(2,p)$. Similarly, the internal isometry groups $H_{k'}$ for $k'=1,0,-1$ are $SO(D-p-1)$, $ISO(D-p-2)$, and $SO(1,D-p-2)$. In special cases, as in~\cite{gm,ms_16}, the radial coordinate combines with one set or the other, and the corresponding symmetry is enhanced, so that different values of $k$ or $k'$ simply reflect different coordinate choices. The spacetimes hosting the ordinary supersymmetric branes are of this type, with $k=0$ and $k'=1$. 

The preceding options are available insofar as $p<D-3$, and lead one to distinguish $p+1$ space--time coordinates $x^\mu$, $(\mu=0,\ldots,p)$, $D-p-2$ internal coordinates $\xi^m$, $(m=1,\ldots,D-p-2)$, and a radial coordinate $r$, invariant under the isometry groups, on which the profiles we are after can depend.  If $p=D-3$ there is a single $\xi$--coordinate and only the choice $k'=0$ is possible, while if $p=D-2$ there are no $\xi$--coordinates altogether. We shall work at times with general values of $D$, but we have in mind primarily critical superstrings, for which $D=10$. The three choices $k=\pm 1,0$ select de Sitter, anti de Sitter and Minkowski space--time manifolds, while $k'=0$ selects a Euclidean internal space (or, more generally, a product of tori, where only continuous internal translations are left), $k'=1$ selects a sphere, while $k'=-1$ selects an internal hyperbolic space.

The redundancy resulting from the introduction of an independent function $B(r)$ has the virtue, as in~\cite{dm_vacuum}, of allowing for a wider class of analytic solutions. In fact, we shall find it convenient to work in the ``harmonic gauge'', whereby
\beq
B \ = \ (p+1)A \ + \ (D-p-2) C \ , \label{harm_gauge}
\eeq
which will simplify the resulting equations.

\subsection{\sc Symmetric Tensor Profiles} \label{sec:tensor_profiles}

We can now characterize symmetric tensor profiles compatible with the isometries of the class of metrics in eq.~\eqref{metric_sym}. To this end, we can begin by considering the closed and invariant volume forms~\footnote{In our conventions $\epsilon_{01\ldots}=+1$ and $\epsilon^{01\ldots}=-1$.}
\beq
{\epsilon_{(p+1)}} \ = \ \sqrt{-\gamma(x)} \ dx^0 \wedge \ldots \wedge dx^{p} \ , \qquad {\widetilde{\epsilon}_{(D-p-2)}}\ = \ \sqrt{\gamma(\xi)} \ d\xi^1 \wedge \ldots \wedge d\xi^{D-p-2} \ ,
\eeq
since combining them with $dr$ one can build profiles of closed $r$--dependent form fields as
\beq
b'_{p+1}(r)\, {\epsilon_{(p+1)}} \, dr \ , \qquad \tilde{b}'_{D-p-2}(r)\, \widetilde{\epsilon}_{(D-p-2)}\, dr \ , \label{clo}
\eeq
where $b$ and $\tilde{b}$ are functions of $r$ and also the closed $r$-independent forms 
\beq 
h_{p+1}\ {\epsilon_{(p+1)}}  \ , \qquad \tilde{h}_{D-p-2}\ {\widetilde{\epsilon}_{(D-p-2)}} \ , \label{clo2}
\eeq 
where $h_{p+1}$ and $\tilde{h}_{D-p-2}$ are two constants. In components
\bea
{\cal H}_{p+2,\ \mu_1 \ldots \mu_{p+1} r} &=& \sqrt{-\gamma(x)}\ \epsilon_{\mu_1 \ldots \mu_{p+1}} \, b'{}_{p+1}(r) \ , \nonumber \\ {\cal H}_{D-p-1,\ i_1 \ldots i_{D-p-2}\,r} &=& \sqrt{\gamma(\xi)}\ \epsilon_{i_1 \ldots i_{D-p-2}} \, \tilde{b}'{}_{D-p-2}(r) \ ,  \label{profile_H}
\eea
and
\bea
{\cal H}_{p+1,\ \mu_1 \ldots \mu_{p+1}} &=& \sqrt{-\gamma(x)}\ \epsilon_{\mu_1 \ldots \mu_{p+1}} \, h_{p+1} \ , \nonumber \\
{\cal H}_{D-p-2,\ i_1 \ldots i_{D-p-2}} &=& \sqrt{\gamma(\xi)}\ \epsilon_{i_1 \ldots i_{D-p-2}} \, \tilde{h}_{D-p-2} \ . \label{profile_h}
\eea
We can now concentrate on the first member of each of the preceding pairs, since the others are related to them by dualities.

The forms in eqs.~\eqref{clo} and \eqref{clo2} have degrees $p+2$, $D-p-1$, $p+1$ and $D-p-2$, and correspond to field strengths of forms of degrees $p+1$, $D-p-2$, $p$ and $D-p-3$. The two cases in eqs.~\eqref{clo2} are special, in that they are $r$--independent, and moreover the corresponding gauge fields are invariant under the isometry groups only up to gauge transformations. They are generalizations of the constant magnetic field strengths discussed in Section~\ref{sec:magnetic_susyb}, and therefore must be treated with care. We shall return to this issue in the following section, after eq.~\eqref{delta_h1}. 

The Bianchi identities are automatically satisfied by the tensor profiles in eqs.~\eqref{clo} and \eqref{clo2}. Moreover, taking into account that
\bea
 {}^{*} \epsilon_{(p+1)}&=& e^{B-(p+1)A+(D-p-2)C} dr \wedge \widetilde{\epsilon}_{(D-p-2)} \ , \nonumber \\
{}^{*} \widetilde{\epsilon}_{(D-p-2)} &=& (-1)^{(p+2)(D-p-2)} \ e^{B+(p+1)A-(D-p-2)C} \epsilon_{(p+1)}\wedge dr \ .
\label{star_clo2}
\eea
one can see that the dynamical equations~\eqref{eqsbeta} are identically satisfied by the $r$--independent forms in eq.~\eqref{clo2}, for arbitrary $r$--dependent scalar profiles $\phi(r)$ and the metric profiles in eq.~\eqref{metric_sym}. On the other hand, the two profiles in eq.~\eqref{clo} satisfy eqs.~\eqref{eqsbeta} if
\bea
b'_{p+1}(r) &=&   H_{p+2}\ e^{\,2\,\beta_p\phi + B +(p+1)A-(D-p-2)C} \, , \nonumber \\  b'_{D-p-2}(r) &=&  \widetilde{H}_{D-p-1}\, e^{\,2\,\beta_{D-p-3}\phi + B -(p+1)A+(D-p-2)C} \, , \label{h12}
\eea
where the overall factors are a pair of constants. 

Summarizing, in form language the four symmetric tensor profiles are described by
\bea
{\cal H}_{p+2} &=& H_{p+2}\ e^{\,2\,\beta_p\phi + B +(p+1)A-(D-p-2)C}\ \sqrt{-\gamma(x)} \ dx^0 \wedge \ldots \wedge dx^p \wedge dr \ , \nonumber \\
{\cal H}_{p+1} &=& h_{p+1}\, \sqrt{-\gamma(x)} \ dx^0 \wedge \ldots \wedge dx^p \ . \label{Hhforms1}
\eea
and by
\bea
{\cal H}_{D-p-1} &=& \widetilde{H}_{D-p-1}\ e^{\,2\,\beta_{D-p-3}\phi + B -(p+1)A+(D-p-2)C} \ dy^1 \wedge \ldots \wedge dy^{D-p-2} \wedge dr \ , \nonumber \\
{\cal H}_{D-p-2} &=& \tilde{h}_{D-p-2}\, \sqrt{\gamma(\xi)} \ dy^1 \wedge \ldots \wedge dy^{D-p-2} \ . \label{Hhforms2}
\eea

For brane profiles, $k'$ would be equal to one, $r$ would be a radial coordinate, and the internal space would be a sphere. The total ``electric'' charge of the first profile in eq.~\eqref{Hhforms1} would be finite and given by
\beq
{q}_e \ = \ H_{p+2}\, \Omega_{D-p-2} \ ,
\eeq
with $\Omega_{D-p-2}$ the volume of a unit internal sphere. The first profile would then be the configuration sourced by an electric $p$-brane, whose counterpart in Maxwell's theory is the Coulomb field of a point charge. On the other hand, with a sphere as internal space, the second profile in eq.~\eqref{Hhforms1}, which also respects the symmetry of the background, would be a uniform field in spacetime, which would result from uncharged open $p$-branes carrying, on their boundaries, opposite charges of one lower dimension, associated to $(p-1)$-(anti)branes. Its counterpart in Maxwell's theory would be a uniform ``electric'' field resulting from a pair of opposite charges $q$ and $-q$ displaced, in opposite directions, to a very large mutual distance $r_0$, in such a way that the ratio $\frac{q}{r_0^2}$ remains finite. The second profile in eq.~\eqref{Hhforms2} would be the configuration sourced by a magnetic $p$-brane, with magnetic charge
\beq
{q}_m \ = \ h_{D-p-2}\, \Omega_{D-p-2} \ ,
\eeq
whose counterpart in the standard Maxwell theory would be the field of a magnetic monopole, while the first profile in eq.~\eqref{Hhforms2} is the dual of the second profile in eq.~\eqref{Hhforms1}. 

There are also some special tensor profiles that are relevant for type--IIB supergravity in ten dimensions. They lead to an interesting class of vacua, which will be dealt with in Section~\ref{sec:directsusybcom}. A proper account of the contribution of these profiles requires a few additional comments, since the corresponding field strength is self--dual. To begin with, one can start from the solution of the self--duality condition, which is proportional to
\beq
{\cal H}_{5} \ \sim \ H_5 \left(e^{B +4\,A\,-\,5\,C}\, {\epsilon}_{(4)}  \wedge dr  \,+\, \widetilde{\epsilon}_{(5)} \right) \ , \label{H5H}
\eeq
since $\beta=0$ in this case. For $k'=1$ this type of profile would be associated to a dyon. In a similar fashion, a second type of profile, proportional to
\beq
{\cal H}_{5} \ = \ {h_{5}}\left(\epsilon_{(5)} \ + \ e^{ - 5 A + B + 4 C} \ dr \wedge \widetilde{\epsilon}_{(4)} \right) \ , \label{H5h}
\eeq
is the counterpart of eq.~\eqref{H5H} for the $h$ field strengths discussed above. 

\subsection{\sc BPS D--Branes} \label{sec:BPS_branes}

In this section, we briefly review how the space-time profiles of D-branes, which are usually presented in the string frame, can be deduced from the supersymmetry transformations of Fermi fields.
The metric profiles for these extended objects are asymptotically flat symmetric solutions that are captured by eq.~\eqref{metric_sym} with $(k,k')=(0,1)$, and are compatible with the existence of Killing spinors. They could be determined by solving the second--order equations that we shall introduce in Section~\ref{sec:symmetries}, but in the BPS limit one can also solve the first--order equations that guarantee the existence of Killing spinors, which in the present setting read
\beq
\delta\,\psi_M \ = \ 0 \ , \qquad \delta\,\lambda \ = \ 0 \ ,
\eeq
The string--frame supersymmetry transformations for the type--II theories can be granted a common presentation, following~\cite{bergshoeff}, which is discussed in Section~\ref{sec:susyIIAB} and provides the starting point for the following considerations.

Note that, when turning to the string frame, the three functions in eq.~\eqref{metric_sym} combine with the dilaton and become
\beq{}{}
A_s\,=\,A\,+\,\frac{\phi}{4} \ , \qquad B_s\,=\,B\,+\,\frac{\phi}{4} \ , \qquad C_s\,=\,C\,+\,\frac{\phi}{4} \ , \ \label{Einststring}
\eeq{}
and consequently the harmonic gauge condition~\eqref{harm_gauge} takes the form
\beq{}{}
B_s \ = \ (p+1)\,A_s \ + \ (8-p)\,C_s \ - \ 2\,\phi \ . \label{harm_S}
\eeq
In the string frame, as recalled in Section~\ref{sec:eqs_branes}, $\beta_S=0$ for RR fields, and taking  the gauge condition into account, one can translate the first of eqs.~\eqref{Hhforms1}, into
\beq
{\cal H}_{p+2} \,=\, H_{p+2}\,e^{B_s+(p+1)A_s-(8-p)C_s}\,dx^0 \wedge \ldots \wedge dx^p \wedge dr \ . \label{form_p}
\eeq
Here all $\gamma$-matrices are flat,
so that adding pairs of dual contributions (in the non--selfdual cases $p \neq 3$) and using
\beq
\frac{\slashed{\cal H}_{p+2}}{(p+2)!} \,=\, H_{p+2}\ e^{\,-\,(8-p)C_s} \ \gamma^{0\ldots p}\,\gamma^r \ ,
\eeq
in the backgrounds~\eqref{metric_sym}
the supersymmetry transformations of Fermi fields in eqs.~\eqref{fermi_transf_gen} finally read
\bea
\delta\,\lambda &=& e^{-B_s}\left(\gamma_r \,\phi' \,\epsilon \ + \ H_{p+2}\ e^{\,-\,\phi\, +\, (p+1)A_s}\ (-1)^p \ \frac{ (3-p)}{4} \ \gamma^{0\ldots p} \,\gamma_r \,{\cal P}_{\frac{p}{2}+1} \epsilon\right) \ , \nonumber \\
\delta\,\psi_r &=& \partial_r \, \epsilon \, + \, \frac{H_{p+2}}{8} \ e^{\,-\,\phi\, +\, (p+1)A_s}\ \gamma^{0\ldots p} \,{\cal P}_{\frac{p}{2}+1} \epsilon \ ,
\label{killing_spinors2} \\
\delta\,\psi_\mu &=& \partial_\mu \, \epsilon \,+\, \frac{1}{2}\ \gamma_\mu \gamma_r\ e^{-pA_s-(8-p)C_s+2\phi}\, A_s'\,\epsilon \, + \, \frac{H_{p+2}}{8} \ e^{\,\phi\, +\, A_s\,-\,(8-p)C_s}\  \gamma^{0\ldots p} \gamma_r\,\gamma_\mu\,{\cal P}_{\frac{p}{2}+1} \epsilon \ ,
\nonumber \\
\delta\,\psi_m &=& \widetilde{D}_m \, \epsilon \,+\, \frac{1}{2}\ \gamma_m \gamma_r\ e^{2\phi-(p+1)A_s-(7-p)C_s}\, C_s'\,\epsilon \, + \, \frac{H_{p+2}}{8} \ e^{\,\phi\,-\,(7-p)C_s}\  \gamma^{0\ldots p} \gamma_r\,\gamma_m\,{\cal P}_{\frac{p}{2}+1} \epsilon \ , \nonumber
\eea
after making use of the gauge condition~\eqref{harm_S}. These results involve the spin connection that is determined by the vielbein postulate
\beq
d\,\theta^A \ + \ \omega^{AB}\, \wedge\, \theta_B \ = \ 0 \ ,
\eeq

whose non--vanishing components read
\beq 
\omega^\mu{}_r \ = \ A_s' \ e^{A_s-B_s} \ dx^\mu \ ,\quad \omega^m{}_r \ = \ C_s' \ e^{C_s-B_s} \ \widetilde{\theta}^m \ , \quad \omega^m{}_n \ = \ \widetilde{\omega}^m{}_n \ .
\eeq
Here ``tilde'' refers to the unit internal sphere, and $\widetilde{\theta}^m$ denotes the corresponding ``moving frame''. The corresponding covariant derivatives satisfy
\beq
\left[\widetilde{D}_m\,,\, \widetilde{D}_n \right] \ = \ \frac{1}{4}\, \widetilde{\Omega}_{mn}{}^{ab}\, \gamma_{ab} \ , \label{sphere1}
\eeq
where
\beq
\widetilde{\Omega}^{ab} \ = \ - \ \widetilde{\theta}^a\,\widetilde{\theta}^b \ . \label{sphere2}
\eeq

For $p \neq 3$, one can conveniently start from the first of eqs.~\eqref{killing_spinors2}, demanding that
\bea
\delta\,\lambda &=& e^{-B}\gamma_r \,\phi' \left(\epsilon \ - \ \frac{H_{p+2}}{\phi'}\ e^{\,-\,\phi\, +\, (p+1)A_s}\ \frac{ (3-p)}{4} \ \gamma^{0\ldots p} \,{\cal P}_{\frac{p}{2}+1} \epsilon\right) \ = \ 0 \ . \label{delta_lambda}
\eea
This equation involves a projection on $\epsilon$ provided
\beq
\phi'\ = \ \pm \ \frac{(3-p)}{4} \ {H_{p+2}}\ e^{\,(p+1)A_s-\phi}\ , \label{cons_phi}
\eeq
since then it reduces to
\beq
\left(1 \ \mp \ \gamma^{0\ldots p} \,{\cal P}_{\frac{p}{2}+1} \right) \epsilon \ = \ 0 \ , \label{susy_proj}
\eeq
which halves the number of supersymmetries with respect to those present in the ten--dimensional Minkowski vacuum.
One can see that in all cases, for both IIA and IIB, 
\beq
\left(\gamma^{0\ldots p} \,{\cal P}_{\frac{p}{2}+1}\right)^2 \ = \ 1 \ ,
\eeq
and therefore the preceding condition is indeed a consistent projection. For $p \neq 3$ the condition $\delta\,\psi_r=0$ can be turned into
\beq
\left(\partial_r \ + \ \frac{1}{2(3-p)}\ \phi' \right) \epsilon \ = \ 0 \ ,
\eeq
whose solution is 
\beq
\epsilon \ = \ e^{\,-\,\frac{\phi}{2(3-p)}} \ \epsilon_0(x,\xi) \ , \label{epsilon_epsilon0}
\eeq
while the condition $\delta\,\psi_\mu=0$ reads
\beq
\partial_\mu \epsilon \ + \ \frac{1}{2}\ \gamma_\mu\,\gamma_r\, e^{-p A_s-(8-p)C_s+2\phi} \left( A_s' \ + \ \frac{\phi'}{(3-p)} \right) \epsilon \ = \ 0 \ .
\eeq
Since the projection anticommutes with $\gamma_\nu\,\gamma_r$, both terms must vanish, and one gets the two conditions
\beq
A_s' \ + \ \frac{\phi'}{(3-p)}  \ = \ 0 \ , \qquad \partial_\mu\, \epsilon \ = \ 0 \ , \label{eqsAphi}
\eeq
so that
\beq
\epsilon \ = \ e^{\,-\,\frac{\phi(r)}{2(3-p)}} \ \epsilon_0(\xi) \ . \label{epsilon_epsilon01}
\eeq
In a similar fashion, from $\delta\,\psi_m=0$ one can deduce that
\beq
\widetilde{D}_m \epsilon \ + \ \frac{1}{2}\ \gamma_m\,\gamma_r\, e^{2\phi-(p+1)A_s-(7-p)C_s}\left( C_s' \ - \ \frac{\phi'}{(3-p)} \right) \epsilon \ = \ 0 \ . \label{deltapsii}
\eeq
Making use of eq.~\eqref{epsilon_epsilon0}, one can now replace $\epsilon$ with $\epsilon_0$, and then all residual $r$--dependence must disappear, so that
\beq
e^{2\phi-(p+1)A_s-(7-p)C_s} \left( C_s' \ - \ \frac{\phi'}{(3-p)} \right)  \ = \ -\ 2\,\sigma \ , \label{eq_Cs}
\eeq
with $\sigma$ a constant. Eq.~\eqref{deltapsii} thus reduces to
\beq
\widetilde{D}_m \epsilon_0 \ = \ \sigma\ \gamma_m\,\gamma_r\,  \epsilon_0 \ , \label{e0xi}
\eeq
which implies that
\beq
\left[ \widetilde{D}_m\,,\,\widetilde{D}_n\right]\epsilon_0 \ = \ - \ 2\,\sigma^2 \, \gamma_{mn}\, \epsilon_0 \ ,
\eeq
and comparing with eq.~\eqref{sphere1} and \eqref{sphere2} one can conclude that for $p<7$
\beq
\sigma^2 \ = \ \frac{k'}{4} \ .
\eeq
This result reduces to the flat solution in~\cite{ms21_1} for $k'=0$, which is the only option for $p \geq 7$, while for the branes we are after $k'=1$, while eq.~\eqref{e0xi} determines the dependence of $\epsilon_0$ on the sphere coordinates $\xi$.
Eq.~\eqref{eqsAphi} is solved by
\beq
A_s  \ = \ - \ \frac{\phi}{(3-p)} \ + \ a_s \ , \label{Aphi}
\eeq
where $a_s$ is a constant, and making use of this result in eq.~\eqref{eq_Cs} reduces it to
\beq
e^{-\,(p-7)\left(A_s + C_s\right)}\,\left(A_s\,+\,C_s\right) ' \ = \ - \ 2\,\sigma\,e^{2a_s(p-3)} \ .
\eeq
Therefore, for $p=7$ $A_s+C_s$ is a constant, while for $p<7$
\beq
e^{(p-7)(A_s+C_s)} \ = \ 2\,\sigma(7-p) e^{2a_s(p-3)}\, r \ + \ c\ , \label{eqAsCs}
\eeq
where $c$ is a constant. Note that flipping the sign of $\sigma$  can be compensated by letting $r \to -\, r$ and by multiplying $\epsilon_0$ by $\gamma_r$, so that we shall assume that $\sigma>0$.

Moreover, combining eqs.~\eqref{cons_phi} and \eqref{Aphi} one can obtain
\beq
e^{\frac{4\phi}{(3-p)}}\,\phi' \ = \ \pm\ \frac{3-p}{4}\ H_{p+2}\ e^{(p+1)a_s} \ ,
\eeq
so that for $p=3$ the dilaton is  constant, while for $p \neq 3$ its string--frame profile reads
\beq
e^\phi \,=\, e^{\phi_s}\ \left[|H_{p+2}|\left(r\,+\,r_1\right)\right]^{\,\frac{3-p}{4}} \ ,
\eeq{}
in the region $r> -\,r_1$, where
\beq{}{}
e^{\phi_s} \ = \ e^{\frac{(p+1)(3-p)}{4}\,a_s} \ .
\eeq{}
Using this relation, eq.~\eqref{Aphi} then gives
\beq
e^{2A_s} \ = \ e^{\frac{(3-p)}{2}\, a_s}  \left[|H_{p+2}|\left(r\,+\,r_1\right)\right]^{\,-\,\frac{1}{2}} \ , 
\eeq
and combining this result with eq.~\eqref{eqAsCs} determines
\beq
e^{2 C_s} \ = \  \, \left[ 2\,\sigma(7-p) e^{2a_s(p-3)}\, r \ + \ c \right]^{- \,\frac{2}{7-p}} \, e^{\frac{(p-3)}{2}\, a_s} \left[|H_{p+2}|\left(r\,+\,r_1\right)\right]^{\,\frac{1}{2}}  \ ,
\eeq
while $B_s$ can be deduced from eq.~\eqref{harm_S}.

The nature of these results takes a more familiar form in isotropic coordinates, which can be reached letting
\beq
\left( \frac{d\rho}{\rho} \right)^2 \ = \ e^{2(B_s-C_s)}\, dr^2 \ = \ e^{4(p-3)a_s} \ e^{2(7-p)\left(A_s\,+\,C_s\right)} \, dr^2\ .
\eeq
Demanding that the Minkowski metric is approached as $\rho \to\infty$ and using the harmonic gauge condition~\eqref{harm_S} and eq.~\eqref{Aphi}, one is led do
\beq
\frac{d\rho}{\rho} \ = \ - \ \frac{e^{2 a_s (p-3)}}{2\,\sigma(7-p) e^{2a_s(p-3)}\, r \ + \ c} \ dr \ .
\eeq
The solution is
\beq
\frac{\rho}{\rho_0} \ = \ \left|\frac{b \,r_2\ + \ c}{b \, r \ + \ c}\right|^\frac{1}{7-p} \ ,
\eeq
where
\beq
b \ = \ \left(7 - p\right) e^{2a_s(p-3)} \ ,
\eeq
and for definiteness we have chosen $2 \sigma=1$. Consequently, the asymptotically flat region is reached as $\rho \to \infty$ or as $r \to - \, \frac{c}{b}$, and choosing
\beq
r_2 \ = \ r_1 \ - \ \frac{2\,c}{b} 
\eeq
leads to
\beq
r \ + \ r_1 \ = \ \left(r_1 \ - \ \frac{c}{b}\right) \left[ 1 \ + \ \left(\frac{\rho_0}{\rho} \right)^{7-p}\right] \ .
\eeq
In terms of the isotropic coordinates thus defined the D-brane solutions take the form
\bea
ds^2 &=& \frac{e^{\frac{(3-p)}{2}\,a_s}}{\left|H_{p+2}\left(r_1 \ - \  \frac{c}{b}\right) \right|^\frac{1}{2}} \ \frac{dx^2}{\left|1 \ + \ \left(\frac{\rho_0}{\rho}\right)^{7-p}  \right|^\frac{1}{2}} \nonumber \\  &+& \frac{\left|H_{p+2}\left(r_1 - \frac{c}{b}\right) \right|^\frac{1}{2} e^{{\frac{(p-3)}{2}\,a_s}}}{\rho_0^2} \ \frac{\left|1 \ + \ \left(\frac{\rho_0}{\rho}\right)^{7-p}  \right|^\frac{1}{2}}{\left|b r_1-c \right|^\frac{2}{7-p}} \left(d \rho^2\ + \ \rho^2\,d\Omega^2\right)  \ , \nonumber \\
e^\phi &=&  e^{\phi_s}\ \left|H_{p+2} \left(r_1 \ - \  \frac{c}{b}\right) \right|^\frac{3-p}{4} \left|1 \ + \ \left(\frac{\rho_0}{\rho}\right)^{7-p}  \right|^\frac{3-p}{4}\ ,  \\
{\cal H}_{p+2} &=& - \ H_{p+2}\ \frac{ e^{2(3-p)a_s}\,{\left|b r_1 - c\right|}}{{\left|H_{p+2}\left(r_1 - \frac{c}{b}\right) \right|^2}\left|1 \ + \ \left(\frac{\rho_0}{\rho}\right)^{7-p} \right|^2} \left(\frac{\rho_0}{\rho}\right)^{7-p} \,dx^0 \wedge \ldots \wedge dx^p \wedge \frac{d\rho}{\rho} \ , \nonumber
\eea
for $p<7$ and $p \neq 3$.

The $x$ coordinates and $\rho$ can be rescaled and, choosing $\rho_0$ in such a way that
\beq
\frac{\left|H_{p+2}\left(r_1 \ - \ \frac{c}{b}\right) \right|^\frac{1}{2} e^{\frac{(p-3)}{2}\, a_s}}{\rho_0^2 \left[b \,r_1 \ - \ c \right]^\frac{2}{7-p}} \ \ = \ 1 \ ,
\eeq
and letting
\beq
e^{\phi_0} \ = \ e^{\phi_s}\ \left|H_{p+2} \left(r_1 \ - \ \frac{c}{b}\right) \right|^\frac{3-p}{4} \ ,
\eeq
the background can be finally presented in the form
\bea
ds^2 &=&  \frac{dx^2}{\left|1 \ + \ \left(\frac{\rho_0}{\rho}\right)^{7-p}  \right|^\frac{1}{2}} \ + \   \left|1 \ + \ \left(\frac{\rho_0}{\rho}\right)^{7-p}  \right|^\frac{1}{2} \left(d \rho^2\ + \ \rho^2\,d\Omega^2\right)  \ , \nonumber \\
e^\phi &=&  e^{\phi_0}\ \left|1 \ + \ \left(\frac{\rho_0}{\rho}\right)^{7-p}  \right|^\frac{3-p}{4}\ ,  \\
{\cal H}_{(p+2)} &=&  - \  \frac{(7-p) \, e^{-\phi_0}\, \rho_0^{7-p}}{\rho^{8-p} \left|1 \ + \ \left(\frac{\rho_0}{\rho}\right)^{7-p} \right|^2}\,dx^0 \wedge \ldots \wedge dx^p \wedge {d\rho} \ ,  \label{profile_branes} \nonumber
\eea
which depends on the two constants $\rho_0$ and $\phi_0$.

Summarizing, we have identified a Killing spinor,
\beq
\epsilon \ = \  e^{\,-\,\frac{\phi(r)}{2(3-p)}} \ \epsilon_0(\xi) \ , 
\eeq
where $\epsilon_0(\xi)$ is subject to the projection~\eqref{susy_proj} and solves eq.~\eqref{e0xi} with $\sigma=\frac{1}{2}$.

The tension can be extracted by noting that D-branes couple to the background via the combination
\beq
- \ {\cal T}_{p} \int  e^{-\phi} \ \sqrt{- \ \gamma} \ d^{p+1} x \ ,
\eeq
where $\gamma$ is the induced metric on their world volume. In the asymptotic region, the deviation from the flat metric can be parametrized as
\beq
ds^2 \ \sim \ \Big[1 \ + V\left(\rho\right)\Big] dx_{p+1}^2 \ + \ d\rho^2 \ + \rho^2 \, d\Omega_{8-p}^2 \ ,
\eeq
and the linearized Einstein equations with a localized source reduce to the Poisson equation for $V$,
\beq
\nabla_{\rho}^2\, V \ = \ \frac{\kappa_{10}^2}{4}\ \left(7-p\right)\, {\cal T}_p \,e^{-\phi_0}\, \delta(\vec{\rho}) \ ,
\eeq
where
\beq
\kappa_{10}^2 \ = \ \left(\alpha'\right)^4 \ e^{2\,\phi_0} .
\eeq

The $\rho$ dependence in eq.~\eqref{profile_branes} characterizes a harmonic function in the space transverse to the brane world volume, since
\beq
- \ \frac{1}{2}\ \nabla^2\,  \left(\frac{\rho_0}{\rho} \right)^{7-p} \ = \  \frac{1}{2}\ \delta\left( \vec{\rho} \right) \, \rho_0^{7-p} \ {(7-p)\, \Omega_{8-p}} \ ,
\eeq
and consequently the string--frame tension is
\beq \label{eq:BPS_tension}
e^{-\,\phi_0} \ {\cal T}_p^{(S)} \ = \ \frac{2\, \rho_0^{7-p}}{\kappa_{10}^2}\ \Omega_{8-p}  \ .
\eeq
The charge can be calculated from the equation for the form field with a localized source
\beq
\frac{1}{2\kappa_{10}^2} \ d \ \star {\cal H}_{p+2} \ =  \ Q_p \ \delta(\vec{\rho}) \ ,
\eeq
or from Gauss's theorem. This leads to
\beq \label{eq:BPS_charge}
Q_p \ = \ e^{-\phi_0} \ \frac{(7-p) \ \rho_0^{7-p} \ \Omega_{8-p}}{2\kappa_{10}^2} \ ,
\eeq
so that charge and tension are proportional for these extended objects in the string frame. 

In the Einstein frame, the exponent of the harmonic function in the spacetime portion of the metric changes, and from $-\,\frac{1}{2}$ it becomes $-\,\frac{(7-p)}{8}$, so that the resulting tension is
\beq \label{eq:BPS_tension_E}
e^{\frac{p-3}{2}\,\phi_0} \ {\cal T}_p^{(E)} \ = \  \frac{(7-p) \ \rho_0^{7-p} \ \Omega_{8-p}}{2\kappa_{10}^2}  \ ,
\eeq
while
\beq \label{eq:BPS_charge_E}
Q_p \ = \ e^{-\, \frac{p-3}{2}\,\phi_0} \ \ \frac{(7-p) \ \rho_0^{7-p} \ \Omega_{8-p}}{2\kappa_{10}^2} \ ,
\eeq
so that the two expressions coincide. The self--dual $p=3$ case will play a role in Section~\ref{sec:directsusybcom}. More details can be found in the reviews in~\cite{susy_branes}.

\subsection{\sc The Second--Order Equations}\label{sec:symmetries}

One can obtain a reduced action principle by inserting in eq.~\eqref{eqs4} the symmetric profiles that we have described.
To begin with, up to an overall factor that we shall leave out, the metric of eq.~\eqref{metric_sym} and a symmetric scalar profile lead to the reduced action principle
\begin{align}
{\cal S} &= \frac{1}{2} \ \int dr \, \Big\{ \, e^{(p+1)A-B+(D-p-2)C}\Big[  p(p+1)\left(A'\right)^2 \ +\ (D-p-2)(D-p-3) \left( C'\right)^2 \nonumber \\ & - \ \frac{4\,(\phi')^2}{D-2} \ + \ 2(p+1)(D-p-2)A'\,C' \Big] \ + \, \frac{k}{\alpha'} \, p\,(p+1)\,e^{(p-1)A+B+(D-p-2)C} \nonumber \\ & - \, {T}_{D-1}\, e^{(p+1)A+B+(D-p-2)C+\gamma \phi}\ + \  \frac{k'}{\alpha'} \, (D-p-2)(D-p-3)\, e^{(p+1)A+B+(D-p-4)C} \Big\} \ ,
\end{align}
where 
\beq
T_{D-1} \ = \ \frac{\tau_{D-1}}{\alpha'} \ ,
\eeq
and $k$ and $k'$, as we have already stated, are the curvatures of the $(p+1)$--dimensional metric $\gamma_{\mu\nu}(x)$ and of the $(D-p-2)$--dimensional metric $\gamma_{mn}(\xi)$. From now on, for brevity, $T_{D-1}$ will be concisely denoted by $T$.

As we have seen, there are two independent options for the inclusion of symmetric tensor fluxes in the class of metric of eq.~\eqref{metric_sym}. The first corresponds to the first profile in eq.~\eqref{clo} for a $(p+1)$--form gauge field, with $b'{}_{p+1}(r)$ given in eq.~\eqref{h12}, and contributes to the dynamical action principle the term
\beq
\Delta\,{\cal S}_{\cal H}^{(1)} \ = \ \frac{1}{4} \ \int dr\ e^{\,-\,2\,\beta_p\,\phi\,-\,(p+1)A \,-\, B \,+\, (D-p-2)C} \left(b'_{p+1}\right)^2 \ . \label{delta_h1}
\eeq
The second independent option corresponds to the first profile in eq.~\eqref{clo2} for a $p$--form gauge field, but is not described in these simple symmetrical terms, as we have stressed. Therefore, we shall only include its contribution to the equations of motions, deducing it from eq.~\eqref{eqsbeta}.
With this proviso, and with different choices of $p$ and $\beta_p$, one can describe in this fashion, as we have anticipated, all symmetric ``electric'' and ``magnetic''  fluxes of interest.
 
One can now combine the different contributions described in Section~\ref{sec:symmetries}, aside from the one related to $h_{p+1}$, and the end result reads
\begin{align}
{\cal S} &= \frac{1}{2} \ \int dr \, \Big\{ \, e^{(p+1)A-B+(D-p-2)C}\Big[  p(p+1)\left(A'\right)^2 \ +\ (D-p-2)(D-p-3) \left( C'\right)^2 \nonumber \\ & - \ \frac{4\,(\phi')^2}{D-2} \ + \ 2(p+1)(D-p-2)A'\,C' \Big] \ + \, \frac{k}{\alpha'} \, p\,(p+1)\,e^{(p-1)A+B+(D-p-2)C} \nonumber \\ & - \, {T} \, e^{(p+1)A+B+(D-p-2)C+\gamma \phi}\ + \  \frac{k'}{\alpha'} \, (D-p-2)(D-p-3)\, e^{(p+1)A+B+(D-p-4)C}\nonumber \\
&+ \, \frac{1}{2}\ e^{\,-\,2\,\beta_p\,\phi\,-\,(p+1)A \,-\, B \, +\, (D-p-2)C} \ \left(b'_{p+1}\right)^2 
 \Big\} \label{reduced_action_part_int} \ .
\end{align}
In the resulting equations of motion, we shall also soon include the contribution related to $h_{p+1}$.  

To begin with, the first tensor profile of Section~\ref{sec:tensor_profiles} satisfies the simple equation
  \beq
 \left(e^{\,-\,2\,\beta_p\,\phi\,-\,(p+1)A \,-\, B \, +\, (D-p-2)C} \ b'_{p+1}\right)^\prime \ = \ 0 \ ,
 \label{Eqb_red1}
\eeq
which is solved by
\beq
b'_{p+1}(r) \ = \   H_{p+2}\ e^{\,2\,\beta_p\phi + B +(p+1)A-(D-p-2)C}\ ,
\eeq
as we had seen more generally in eq.~\eqref{Hhforms1}. It is now convenient to work in the ``harmonic'' gauge 
\beq
 (p+1)A \ - \ B \ + (D-p-2) C \ = \ 0 \ , \label{F_gauge}
\eeq
which reduces the equations of motion for $A$, $C$ and $\phi$ to
 \bea
 A'' \!\!\!&=& - \ \frac{T}{(D-2)} \ e^{2\,B\,+\,\gamma\,\phi}\ +\ \frac{k\,p}{\alpha'}\ e^{2(B-A)} \label{EqA_red}\\
  &+& \!\!\frac{(D-p-3)}{2\,(D-2)} \ e^{2\,B\,+\,2\,\beta_p\,\phi\,-\,2(D-p-2)C } H_{p+2}^2 \,+\, \frac{(D-p-2)}{2\,(D-2)} \ e^{2\,B\,-\,2\,\beta_{p-1}\,\phi\,-\,2 (p+1) A} h_{p+1}^2  \nonumber \  ,  \\
C''\!\!\!&=& - \ \frac{T}{(D-2)} \ e^{2\,B\,+\,\gamma\,\phi}\ +\ \frac{k'(D-p-3)}{\alpha'}\ e^{2(B-C)} \label{EqC_red} \\
   &-& \frac{(p+1)}{2\,(D-2)}\ e^{2\,B\,+\,2\,\beta_p\,\phi\,-\,2(D-p-2)C }  H_{p+2}^2\,-\, \frac{p}{2\,(D-2)}\ e^{2\,B\,-\,2\,\beta_{p-1}\,\phi\,-\,2 (p+1) A} h_{p+1}^2 \ ,  \nonumber  \\
   \phi'' \!\!\!&=& \frac{T\,\gamma\,(D-2)}{8}\ e^{2\,B\,+\,\gamma\,\phi} \label{Eqphi_red} \\ &+& \frac{\beta_p\,(D-2)}{8}\ e^{2\,B\,+\,2\,\beta_p\,\phi\,-\,2(D-p-2)C } H_{p+2}^2 \,+\, \frac{\beta_{p-1}\,(D-2)}{8}\ e^{2\,B\,-\,2\,\beta_{p-1}\,\phi\,-\,2 (p+1) A} h_{p+1}^2\ . \nonumber
  \eea
  Here we have included the contributions related to $h_{p+1}$, and moreover the equation for $B$, which we call ``Hamiltonian constraint'', reads
 \begin{align}
&(p+1)A'[p\,A' \,+\, (D-p-2)C']\,+\, (D-p-2)C'[(D-p-3)C'+(p+1)A'] \nonumber \\
 &- \, \frac{4\,(\phi')^2}{D-2} \, + \, {T} \, e^{\, 2\,B\,+\,\gamma\,\phi}\, - \, \frac{k\,p(p+1)}{\alpha'}\ e^{2(B-A)}\, - \, \frac{k'(D-p-3)(D-p-2)}{\alpha'}\ e^{2(B-C)} \nonumber \\
 & + \, \frac{1}{2}\, e^{\,2\,\beta_p\,\phi\,+\,2\,B\,-\,2\,(D-p-2)\,C} \ H_{p+2}^2 \,-\, \frac{1}{2}\, e^{\,-\,2\,\beta_{p-1}\,\phi\,-\,2(p+1)A\,+\,2\,B} \ h_{p+1}^2  \,= \, 0 \ , \label{EqB_red}
 \end{align}
independently of the gauge choice for the metric.
  
Note that this system has an interesting discrete symmetry: its equations are left invariant by the redefinitions
\bea
&& \left[A,C,\,p,\,k,k'\right] \ \longleftrightarrow \ \left[C,A,\,D-p-3,\,k',k \right] \ , \nonumber \\
&& \left[H_{p+2}^2,\beta_p;h_{p+1}^2,\beta_{p-1} \right] \ \longleftrightarrow \ \left[-h_{p+1}^2,- \beta_{p-1};- H_{p+2}^2,-\beta_p  \right] \label{sym_AC}
\ , \eea
which can be regarded as implementing an ``electric-magnetic'' duality.
The simultaneous presence of $H_{p+2}$ and $h_{p'+1}$ profiles is only relevant in one special case, for type IIA, with one of them of NS-NS type and the other of RR type.  
In the following, we shall focus on solutions with $k=k'=0$. Other types of configurations were considered in~\cite{mrs24_1,mrs24_2}.

As we have anticipated, two special cases, related to the type--IIB string, must be treated separately, since they involve fluxes of the self--dual five--form field strength, for which we refer the reader to eqs.~\eqref{H5H} and \eqref{H5h}.
The complete equations of motion for the first case are
\beq
R_{MN}  \ = \  \frac{1}{24}\ \left({\cal H}_{5}^2\right)_{M N} \ + \ \frac{1}{2}\, \partial_M\,\phi \,\partial_N\,\phi \ ,
\eeq
and their reduced form for the class of metrics of interest in the ``harmonic'' gauge $B=4 A + 5 C$ and for the symmetric $H_5$ profile of eq.~\eqref{H5H} reads
\bea
A'' &=&  \frac{H_5^2}{8}\ e^{8 A} \ , \nonumber \\
C'' &=&  - \ \frac{H_5^2}{8}\ e^{8 A}\ , \nonumber \\
\phi'' &=& 0 \ . \label{eqABC_sdual}
\eea
The corresponding Hamiltonian constraint is
\beq
3\left(A'\right)^2 \ + \ 10\, A'\, C' \ + \ 5 \left(C'\right)^2 \ = \ \frac{1}{8}\, \left(\phi'\right)^2 \ - \ \frac{H_5^2}{16} \ e^{8 A} \ . \label{ham_sdual}
\eeq
The counterpart of these results for the $h_{p+1}$--fluxes corresponds to $p=4$, and in this case
\bea
A'' &=& \frac{h_5^2}{8}\ e^{8 C} \ , \nonumber \\
C'' &=&  - \ \frac{h_5^2}{8}\ e^{8 C}\ , \nonumber \\
\phi'' &=& 0 \ ,  \label{eqABC_sdual_h}
\eea
while the Hamiltonian constraint becomes
\beq
5\left(A'\right)^2 \ + \ 10\, A'\, C' \ + \ 3 \left(C'\right)^2 \ = \ \frac{1}{8}\, \left(\phi'\right)^2 \ + \ \frac{h_5^2}{16} \ e^{8 C} \ . \label{ham_sdual_h}
\eeq

\subsection{\sc 9D Interval Compactifications} \label{sec:dm_solutions}

In this section, we address static backgrounds with $p=8$, reviewing how the tadpole potentials can give rise to one--dimensional compactifications with an internal interval whose finite length is determined by the residual tension (or vacuum energy, in the heterotic case). This class of solutions is the counterpart of the standard Kaluza--Klein scenario, where the size of the internal space, a circle rather than an interval, would be a free parameter.

Insofar as these configurations are concerned, the USp(32) and U(32) orientifold models have identical Lagrangians, since their tadpole potentials emerge at the (projective) disk level and only fields present in both cases have non--trivial profiles. On the other hand, in the $SO(16) \times SO(16)$ model, the tadpole potential emerges from the torus amplitude, and its exponent is different. However, as we shall see, the resulting dynamics has some similarities in all these cases.
In the Einstein frame, the relevant terms of the bosonic Lagrangian are captured, in the three cases, by
\beq
{\cal S} \,=\, \frac{1}{2\,k_{10}^2}\, \int d^{10}x\sqrt{-g}\bigg\{R\ - \ \frac{1}{2}\ (\partial\phi)^2 \, - \ T \, e^{\,\gamma\,\phi} \label{einstein_bsb}
\bigg\} \ ,
\eeq
where $\gamma = \frac{3}{2}$ for the orientifold models, and $\gamma = \frac{5}{2}$
for the SO(16) $\times$ SO(16) heterotic model, as we have seen.

The first classical solutions of these effective Lagrangians were presented in~\cite{dm_vacuum} by two of us, and we shall often refer to them as interval compactifications. 

In the Einstein frame, the solution for the $USp(32)$ and $U(32)$ orientifold models can be presented in the form
\beq
ds^2 \ = \  e^{\,-\,\frac{u}{6}} \, {u}^\frac{1}{18}\, dx^2 \,+\, \frac{2}{3\,T\,u^\frac{3}{2}}\ e^{\,-\,\frac{3}{2}\left(u+\phi_0\right)}\, {du^2} \ , \qquad
e^\phi \ = \  e^{\,u\,+\,{\phi}_0}\, {u}^\frac{1}{3} \ , \quad (0<  u < \infty) \label{9D_solution}
\eeq
where $T$ reflects the overall tension resulting from the branes and orientifolds that are present.
Near $u=0$, the end where the first curvature singularity resides, the limiting behavior of the solution is captured by
\beq
ds^2 \ = \ \xi^\frac{2}{9}\, dx^2 \ + \ d\xi^2 \ , \qquad e^\phi \ \sim \ \xi^\frac{4}{3} \ ,
\eeq
where $\xi=0$ at the endpoint, 
while as $u\to \infty$ the limiting behavior is captured by
\beq
ds^2 \ = \ \left(\xi_m \ - \ \xi\right)^\frac{2}{9}\, dx^2 \ + \ d\xi^2 \ , \qquad e^\phi \ \sim \ \left(\xi_m \ - \ \xi\right)^{\,-\,\frac{4}{3}} \ .
\eeq
$\xi=\xi_m$ at the other endpoint, where the second curvature singularity lies, with $\xi_m$ a finite constant related to the length of the internal interval.
Both limiting behaviors have the interesting feature of corresponding to \emph{exact solutions in the absence of the tadpole potential}, although the string coupling vanishes in the first limit but diverges in the second~\cite{ms21_1,ms21_2}.

For the SO(16)$\times$ SO(16) model, the solution is slightly more complicated, and there are actually three branches of solutions, two of which are described by
\bea
ds^2 &=& e^{\,\mp\, \frac{5\,r}{24}}\ {\left[\Delta\,\sinh\left({r}\right)\right]^{\frac{1}{8}}} \  dx^2
  \ + \ {\rho^2} \left[\Delta\,\sinh\left({r}\right)\right]^{\frac{9}{8}} \ e^{\,\mp\, \frac{15\,r}{8}}\ dr^2 \ , \nonumber \\
e^{\phi} &=& \ \frac{e^{\,\pm\, \frac{3\,r}{4}}}{\left[\Delta\,\sinh\left({r}\right)\right]^{\frac{5}{4}} } \qquad (0 < r < \infty) \ , \label{het_9D_solution}
\eea
where
\beq
\Delta \ = \ \ \rho\,\sqrt{2\,{T}} \ ,
\eeq
$r$ is a dimensionless variable and $\rho$ is an integration constant with the dimension of length.
The third solution can be obtained as the $r \to 0$ limit of the previous ones, and only the upper branch in eqs.~\eqref{het_9D_solution} corresponds to an internal interval of finite length. The three types of solutions approach each other as $r \to 0$, where the string coupling diverges, but now the limiting behavior depends on $T$. At the opposite end, the string coupling vanishes, and the limiting behavior for the first two classes of solutions is independent of $T$~\cite{ms21_1,ms21_2}. Curvature singularities are again present at both ends.

The solutions with $\gamma=\frac{3}{2}$ and $\gamma=\frac{5}{2}$ and an internal interval of finite length have the interesting property of leading to a flat nine--dimensional space--time, just like circle compactification in the absence of a tadpole potential. Moreover, as in that case, in nine dimensions the inherited values for Newton's constant and the gauge coupling are both finite. However, these vacua have several novel features. To begin with, the internal space is an interval, so boundary conditions are to be specified at its endpoints. Moreover, the internal size is not a modulus, as we have stressed, but is determined by the string tension. As we shall see, in contrast to more symmetric AdS $\times $ compactifications~\cite{gm,ms_16,raucci_ads}, where large values of string coupling and curvature can be absent, these vacua are perturbatively stable~\cite{bms,ms23_1}. However, stability is possible in the less symmetric AdS${}_3 \times$ S${}_3 \times$ S${}_3 \times$ S${}_1$ vacua~\cite{sethiO16}.

There is a widespread expectation in the literature that, in a complete theory like String Theory, the ends of the internal interval should host some effective extended objects that emerge dynamically, a conjecture often referred to as ``dynamical cobordism''~\cite{dynamicalcobordism1,dynamicalcobordism2,dynamicalcobordism3,dynamicalcobordism4,dynamicalcobordism5,dynamicalcobordism6,dynamicalcobordism7}. Quantifying this natural conjecture is generally difficult when large-curvature and/or string-loop corrections are present. This is true, in particular, in this setting, but in other vacua where some limiting behavior is protected by supersymmetry, as in the example reviewed in Section~\ref{sec:directsusybcom}, one can identify, at one end where supersymmetry is recovered, effective tensions and charges proportional to background fluxes and to one another, as pertains to BPS extended objects that emerge dynamically~\cite{jphysA}.

\subsection{\sc Interval Compactifications and their Stability} \label{sec:int_comp_stab}

Supersymmetry breaking in String Theory results in strong back-reactions on the vacuum, which typically lead to the emergence of unstable modes. The tadpole potentials of non--tachyonic ten--dimensional strings make the interval compactifications that we just described the most symmetric options at stake.
Since one ends up in Minkowski space, one can ascertain their stability examining the signs of the square masses of bosonic modes. Complete sets of modes in the intervals are needed to describe arbitrary perturbations, and thus one is led to perform a detailed scrutiny of the possible self--adjoint extensions for the corresponding Schr\"odinger operators~\cite{ms23_1}.

A similar question arises in AdS $\times$ S compactifications~\cite{gm,ms_16}, where squared masses are to be compared with Breitenlohner--Freedman bounds~\cite{breitf}. In these cases, the internal excitations are nicely ordered by the spherical symmetry of the internal space, whose size is comparable to the AdS radius, and there is no way to avoid instabilities resulting from mixings of Kaluza--Klein modes~\cite{bms}. On the other hand, the original vacua of~\cite{dm_vacuum} are surprisingly stable. We can now illustrate this result, starting from some simple considerations before addressing the problem in full generality.

Recasting the backgrounds of eqs.~\eqref{9D_solution} and \eqref{het_9D_solution} in terms of a conformal coordinate $z$ for the interval, the equation for scalar perturbations, which affect the conformal--gauge metric and the dilaton perturbation $\varphi$ according to
\bea
ds^2 &=& e^{2\Omega(z)} \left[ (1+ A) \,dx^\mu \, dx_\mu \ + \ (1-7A)\, dz^2 \right] \ , \nonumber \\
\varphi &=& - \ \frac{8}{\phi'}\left(A'\ + \ 7\,A\,\Omega'\right) \ ,  \label{metric_conforma}
\eea
takes the form
\beq{}
A'' + A'\left(24\, \Omega'\ - \ \frac{2}{\phi'} \ e^{2\Omega}\, V_{\phi} \right) \ + \
A\left(m^2 \ - \ \frac{7}{4} \ e^{2\Omega}\, V \ - \ 14 \, e^{2\Omega}\, \Omega'\, \frac{V_\phi}{\phi'} \right) \ = \ 0 \ . \label{A_perturb}
\eeq
In this gauge, the dilaton background $\phi(z)$ is implicitly determined by eqs.~\eqref{9D_solution} and \eqref{het_9D_solution}.
This result is apparently complicated,
but one can make formal statements on the sign of $m^2$ resorting to a very useful trick. This recurs, in various forms, in Physics and Mathematics, and consists in recasting eq.~\eqref{A_perturb} in the form
\beq
\left( {\cal A}\, {\cal A}^\dagger \ + \ b \right) \Psi \ = \ m^2\, \Psi  \ , \label{AAdagpsi}
\eeq
after a suitable redefinition leading from $A$ to a Schr\"odinger field $\Psi$. Nine--dimensional masses thus emerge as eigenvalues of Schr\"odinger--like Hamiltonians built of the first--order operators
\bea
{\cal A} &=& \frac{d}{dz} \ + \ \frac{a(z)}{2} \ , \qquad  {\cal A}^\dagger \ = \  - \ \frac{d}{dz} \ + \ \frac{a(z)}{2}\ ,
\eea
where
\beq
a \ = \  24\,\Omega'\ - \ \frac{2}{\phi'} \ e^{2\Omega}\, V_\phi \ , \qquad
b \ = \  \frac{7}{4}\, e^{2\Omega}\, V\left[1\ + \ 8\,\gamma\, \frac{\Omega'}{\phi'}\right] \ .
\eeq
Note that $b>0$ for the two relevant cases $\gamma=\frac{3}{2}$ and $\gamma=\frac{5}{2}$. For future use, let us also note that for the orientifolds
\beq
z(u) \ = \ \int_0^u d t \ \sqrt{\frac{2}{3\,T}} \ e ^{-\,\frac{2}{3}\,t} \ t^{-\,\frac{7}{9}} \ ,  \label{z_orientifold}
\eeq
so that
\bea
z(u) &\sim& \sqrt{\frac{2}{3\,T}} \ \frac{9}{2}\, u^\frac{2}{9} \ , \nonumber \\
z_m \,-\, z(u) &\sim&  \sqrt{\frac{3}{2\,T}}\ e ^{-\,\frac{2}{3}\,u} \ u^{-\,\frac{7}{9}} \ ,
\eea
in the limiting regions of small and large $u$, and
\beq
z_m \ = \ z(\infty) \ \simeq \  \frac{3.67}{\sqrt{T}}  \   . \label{zm_orientifold}
\eeq
On the other hand, for the $SO(16) \times SO(16)$ model
\beq
z(r) \ = \ \rho \int_0^r dt\  \left(\Delta\,\sinh t\right)^\frac{1}{2} e ^{-\,\frac{5}{6}\,t} \ , 
\eeq
so that
\bea
z(r) &\sim& \frac{2}{3}\ \rho\,\sqrt{\Delta}\,r^\frac{3}{2}  \ , \nonumber \\
z_m \,-\, z(r) &\sim&  3 \  \rho\,\sqrt{\frac{\Delta}{2}} \ e ^{-\,\frac{r}{3}} \ , \label{z_heterotic}
\eea
in the limiting regions of small and large $r$, and
\beq
z_m \ = \ z(\infty) \ \simeq \  1.93 \ \rho\,\sqrt{\Delta}  \  .  \label{zm_heterotic}
\eeq

When applied in this context, the preceding steps yield the definite prediction that $m^2 > 0$, insofar as the boundary conditions grant that ${\cal A}^\dagger$ be the adjoint of ${\cal A}$. When this is the case
the product ${\cal A}^\dagger\, {\cal A}$ is a positive operator, and perturbative stability follows from the positivity of $b$. With a proper choice of self--adjoint boundary conditions, which we shall characterize shortly, one can thus conclude that the vacua of~\cite{dm_vacuum} for the orientifold models and for the $SO(16) \times SO(16)$ heterotic string are~\emph{perturbatively stable}.

\subsubsection{\sc Self--Adjoint Boundary Conditions} \label{sec:selfadjoint}
The non--trivial profiles considered in the previous section for the metric, and also their generalizations including $p+1$-form potentials, are defined on intervals of finite length covered by finite spans $0 \leq z \leq z_m$ of the conformal coordinate. As we are about to see, the corresponding Sch\"odinger potentials have double poles at the ends, due to the asymptotics of the metric.

 Self--adjoint boundary conditions grant the reality of the eigenvalues and the completeness of the resulting modes, which is crucial to address all possible origins of instability. However, the standard setup does not exclude, in general, the presence of negative $m^2$ eigenvalues, which are precisely the signature of perturbative instabilities. The relevance of the sign is precisely where our problem departs from conventional treatments, and, as we shall see, negative eigenvalues can emerge even in cases where, at first sight, one would be inclined to exclude them. The conclusion will be that vacuum stability can only hold, in general, with special choices of self--adjoint boundary conditions, which are essential to grant the positivity of expressions as eq.~\eqref{AAdagpsi}.

For the nine--dimensional vacuum of~\cite{dm_vacuum} of the orientifold models, one finds
 \beq
ds^2 \ \sim \ z^\frac{1}{4}\left(dx^2 \ + \ dz^2 \right) \ , \qquad ds^2 \ \sim \ \left(z_m-z\right)^\frac{1}{4} \left|\log\left(z_m-z\right) \right|^\frac{1}{4}\left(dx^2 \ + \ dz^2 \right) \ , \label{metric_orientifold}
\eeq
with the same leading exponent at both ends. Note that here we also include subleading logarithms, which were not taken into account in~\cite{ms23_1}. The limiting behaviors of the string coupling are
\beq
 e^\phi \ \sim \ z^\frac{3}{2} \ , \qquad
e^\phi  \ \sim \ \left(z_m-z\right)^{-\,\frac{3}{2}} \left|\log\left(z_m-z\right) \right|^{-\,\frac{5}{6}}\ , \label{dilaton_orientifold}
\eeq
where we also included subleading logarithmic terms.
In fact, these results hold for all $\gamma \leq \frac{3}{2}$, since within this range the results
are insensitive to the tadpole potential, as discussed in detail in~\cite{ms21_1,ms21_2}.

For $\gamma>\frac{3}{2}$, the limiting behavior at the strong--coupling end depends on the value of $\gamma$~\cite{ms21_2,ms23_1}. In particular, for the SO(16) $\times$ SO(16) heterotic model, for which $\gamma=\frac{5}{2}$, one finds
 \beq
ds^2 \ \sim \ z^\frac{1}{12}\left(dx^2 \ + \ dz^2 \right) \ , \qquad ds^2 \ \sim \ \left(z_m-z\right)^\frac{1}{4}\left(dx^2 \ + \ dz^2 \right) \ ,
\eeq
so that the exponents are different at the two ends,
while the corresponding behaviors of the string coupling are
\beq
 e^\phi \ \sim \ z^{-\,\frac{5}{6}} \ , \qquad
e^\phi  \ \sim \ \left(z_m-z\right)^{\frac{3}{2}} \ .
\eeq

For all types of perturbations, one is led to Schr\"odinger--like equations with Hamiltonians
\beq
H \ = \ - \ \frac{d^2}{dz^2} \ + \ V(z) \label{ham}
\eeq
that can be cast in the form
\beq
{H} \ = \ b \ + \ {\cal A}\,{\cal A}^\dagger \ ,
\eeq
where
\beq
{\cal A} \ = \ \frac{d}{dz} \ +\ \frac{a}{2} \ , \qquad  {\cal A}^\dagger \ = \ - \ \frac{d}{dz} \ +\ \frac{a}{2} \ ,
\eeq
so that the Schr\"odinger potential is
\beq
V \ = \ b \ + \ \frac{1}{2}\, a'\ + \ \frac{1}{4}\, a^2 \ , \label{V_schrod}
\eeq
where the ``prime'' indicates a derivative with respect to $z$.

This setup actually applies to all types of perturbations. In detail, for tensor perturbations
\beq
a \ = \ 8 \,\Omega' \ , \qquad b \ = \ 0 \ ,
\eeq
where $\Omega$ is defined as in eq.~\eqref{metric_conforma},
while for scalar perturbations~\footnote{There is a subtlety, explained in~\cite{bms}, when using the first type of expressions as $\gamma \to \gamma_c$.}
\bea
a &=& 24\,\Omega' \ - \ \frac{2 \ T\ \gamma}{\phi'} \ e^{2\Omega\,+\,\gamma\,\phi} \ = \  24 \, \Omega' \ - \ \frac{\gamma}{\phi'}\left[\left(\phi'\right)^2 \ - \ \left(12\,\Omega'\right)^2\right] \ , \nonumber \\
b &=& \frac{7}{4}\, T\, e^{2\,\Omega\,+\,\gamma\,\phi}\left(1\ + \ 8 \, \gamma \ \frac{\Omega'}{\phi'}\right) \ = \ \frac{7}{8\,\phi'}\left[\left(\phi'\right)^2 \ - \ \left(12\,\Omega'\right)^2\right]\left( \phi'\ + \ 8\,\gamma\,\Omega'\right)\ . \label{ab}
\eea
Using the preceding results, one can identify the singular limiting behavior of the Schr\"odinger potential
at the two ends, which we shall parametrize in general, for later convenience, as
\beq
V \ \sim \ \frac{\mu^2 \,-\,\frac{1}{4}}{z^2} \ , \qquad V \ \sim \ \frac{1}{\left(z_m \,-\,z\right)^2} \left\{ \tilde{\mu}^2 \,-\,\frac{1}{4} \ + \ \frac{2\,\alpha\,\tilde{\mu}}{\log\left(z_m-z\right)} \ + \ \frac{\alpha\left(\alpha-1\right)}{\left[\log\left(z_m-z\right)\right]^2} \right\} , \label{double_poles}
\eeq
where $\mu$, $\tilde{\mu}$ and $\alpha$ are constant parameters characterizing the asymptotics, since $b$ is subdominant in all cases. These results are obtained relying on
\beq
\frac{a}{2} \ \sim\  \frac{\nu}{z} \ , \qquad \frac{a}{2} \ \sim \ - \ \frac{\widetilde{\nu}}{z_m-z} \ \mp \ \frac{\alpha}{\left(z_m-z\right)\log\left(z_m-z\right)} \label{a_singular} \ ,
\eeq
where
\beq
\nu \ = \ \frac{1}{2} \ \pm \ \mu \ , \qquad \widetilde{\nu} \ = \ \frac{1}{2} \ \pm \ \tilde{\mu} \ . \label{nunutilde}
\eeq
Here $\mu \geq 0$ and $\tilde{\mu} \geq 0$, so that, say, with the upper signs in eqs.~\eqref{a_singular} and \eqref{nunutilde}, $\nu \geq \frac{1}{2}$ and $\widetilde{\nu} \geq \frac{1}{2}$.

In the potential of eq.~\eqref{double_poles} the last contribution, proportional to $\alpha(\alpha-1)$, is negligible for $\tilde{\mu} \neq 0$, but should be retained for $\tilde{\mu}=0$.
The logarithms are only present for the orientifolds and originate from the power in eqs.~\eqref{z_orientifold}.

For the orientifolds of~\cite{as95,sugimoto}, with $\gamma=\frac{3}{2}$, taking into account that
\beq
\Omega_z \ \sim \ \frac{1}{8\,z} \ , \qquad \phi_z \ \sim \ \frac{3}{2\,z}
\eeq
near $z=0$, and
\bea
\Omega_z &\sim& - \ \frac{1}{8}\ \frac{1}{z_m-z} \ - \ \frac{1}{8} \ \frac{1}{\left(z_m-z\right)\,\log\left(z_m-z\right)} \ , \nonumber \\
\phi_z &\sim& \frac{3}{2}\ \frac{1}{z_m-z} \ + \ \frac{5}{6} \ \frac{1}{\left(z_m-z\right)\, \log\left(z_m-z\right)}
\eea
near $z=z_m$, one finds the following results:
\bea
\mathrm{Scalar\ perturbations} &:& \left(\mu,\tilde{\mu},\alpha,\nu,\widetilde{\nu}\right) = \left(1,1,\frac{1}{2},\frac{3}{2},\frac{3}{2} \right) \ , \nonumber \\
\mathrm{Tensor\ perturbations} &:& \left(\mu,\tilde{\mu},\alpha,\nu,\widetilde{\nu}\right) = \left(0,0,\frac{1}{2},\frac{1}{2},\frac{1}{2}\right) \ .
 \eea
 
In the orientifolds there are RR perturbations, for which, in general
\beq
a \ = \ - \ 2\left[\beta_p\,\phi_z \ +\ (p-3) \Omega_z\right] \ .
\eeq
For RR perturbations in the orientifolds, this expression becomes
\beq
a \ = \ - 2(p-3)\left[\frac{1}{4}\,\phi_z \ +\ \Omega_z\right] \ ,
\eeq
and consequently near the origin
\beq
a \sim \ - \ \frac{p-3}{z} \ ,
\eeq
while near the other end
\beq
a \sim \ - \ \frac{p-3}{2\left(z_m \,-\,z\right)} \ - \ \frac{1}{6} \ \frac{p-3}{\left(z_m \,-\,z\right)\,\log\left(z_m \,-\,z\right)} \ .
\eeq
In all these cases $b=0$, and therefore the corresponding potentials for RR perturbations have
\beq
\mu \ = \ \frac{\left|p-2\right|}{2} \ , \quad \tilde{\mu} \ = \ \frac{\left|p-5\right|}{4}  \ , \quad \alpha \ = \ \frac{1}{12}\left|p-3\right| \ , \quad \nu \ = \ \frac{3-p}{2}\ , \quad \widetilde{\nu} \ = \ \frac{p-3}{4} \ .
\eeq
The values $p=1,5$, for which
\bea
p=1 &:&  \left(\mu,\tilde{\mu},\alpha,\nu,\widetilde{\nu}\right) = \left(\frac{1}{2},1,\frac{1}{6},1,-\,\frac{1}{2}\right)\ , \nonumber \\
p=5 &:&  \left(\mu,\tilde{\mu},\alpha,\nu,\widetilde{\nu}\right) = \left(\frac{3}{2},0,\frac{1}{6},-\,1,\frac{1}{2}\right)
\eea
concern both the USp(32)~\cite{sugimoto} and U(32)~\cite{as95,as97} cases, while the values $p=-1,3,7$, for which
\bea
p=-1 &:&  \left(\mu,\tilde{\mu},\alpha,\nu,\widetilde{\nu}\right) = \left(\frac{3}{2},\frac{3}{2},\frac{1}{3},2,-\,{1}\right) \ , \nonumber \\
p=3 &:&   \left(\mu,\tilde{\mu},\alpha,\nu,\widetilde{\nu}\right) = \left(\frac{1}{2},\frac{1}{2},0,0,0\right) \ , \nonumber \\
p=7 &:&   \left(\mu,\tilde{\mu},\alpha,\nu,\widetilde{\nu}\right) = \left(\frac{5}{2},\frac{1}{2},\frac{1}{3},-2,1 \right)
\eea
only concern the latter orientifold. These values correct those listed in~\cite{ms23_1}, and complete the result of~\cite{mrs24_2} with the values of $\alpha$. In addition, there are gauge vector perturbations in the orientifolds, for which $\beta=-\,\frac{1}{4}$, so that
\beq
a \ = \ \frac{1}{2}\,\phi_z \ + \ 6\,\Omega_z \ ,
\eeq
and consequently
\beq
p=0\ : \quad  \left(\mu,\tilde{\mu},\alpha,\nu,\widetilde{\nu}\right) = \left(\frac{1}{4},\frac{1}{2},-\,\frac{1}{6},\frac{3}{4},0 \right) \ .
\eeq

Let us stress that one ought to distinguish gauge vectors from the Kaluza-Klein vector originating from the metric. The latter has a non--normalizable zero mode in all cases of interest, as shown in~\cite{bms}, so that we can concentrate on gauge vectors, which originate from the open sectors of the orientifolds.
These results are summarized in Table~\ref{tab:tab_munu_orientifolds}.
\begin{table}[ht]
\centering
\begin{tabular}{||c || c | c |c | c | c  | c  | c ||}
 \hline
Perturbations & $\mu$ & $\tilde{\mu}$ & $\alpha$ & $\nu$ & $\widetilde{\nu}$ & Case & Sufficient Stability Conditions \\ [0.5ex]
 \hline\hline
Scalar & $1$ &  $1$  & $\frac{1}{2}$ & $\frac{3}{2}$ & $\frac{3}{2}$ & 1 &  always stable \\   [0.3ex] \hline
Vector & $\frac{1}{4}$ &  $\frac{1}{2}$ & $-\,\frac{1}{6}$ & $\frac{3}{4}$ & $0$ & 4 & $C_2=0$ and $C_3 C_4 = 0$\\ [0.3ex] \hline
Tensor & $0$ &  $0$ & $\frac{1}{2}$ & $\frac{1}{2}$ & $\frac{1}{2}$ & 2 & $C_1=0$ and $C_3=0$\\ [0.3ex] \hline
RR $0$ form & $\frac{3}{2}$ &  $\frac{3}{2}$ & $\frac{1}{3}$ & ${2}$ & $-\,{1}$ & 4 & always stable \\  [0.3ex]
RR $2$ form & $\frac{1}{2}$ &  $1$ & $\frac{1}{6}$ & $1$ & $-\,\frac{1}{2}$ & 4 & $C_2=0$ and $C_3 C_4 = 0$ \\  [0.3ex]
RR $4$ form & $\frac{1}{2}$ &  $\frac{1}{2}$ & $0$ & $0$ & $0$ & 5 & $C_1 C_2=0$ and $C_3 C_4=0$ \\  [0.3ex]
RR $6$ form & $\frac{3}{2}$ &  $0$ & $\frac{1}{6}$ & $-\,1$ & $\frac{1}{2}$ & 3 & $C_1 C_2=0$ and $C_3=0$ \\
RR $8$ form & $\frac{5}{2}$ &  $\frac{1}{2}$ & $\frac{1}{3}$ & $-\,2$ & $1$ & 3 & $C_1 C_2=0$ and $C_3=0$ \\
 [0.5ex]
 \hline
\end{tabular}
\caption{Values of $\mu$, $\tilde{\mu}$, $\nu$ and $\widetilde{\nu}$ for the different sectors of the ten--dimensional orientifolds. The different cases refer to the sufficient conditions for positivity discussed in Section~\ref{sec:positivity}.}
\label{tab:tab_munu_orientifolds}
\end{table}

For the $SO(16) \times SO(16)$ heterotic model of~\cite{agmv1,agmv2}, taking into account that
\beq
\Omega_z \ \sim \ \frac{1}{24\,z} \ , \qquad \phi_z \ \sim \ - \ \frac{5}{6\,z}
\eeq
near $z=0$, and
\beq
\Omega_z \ \sim \ - \ \frac{1}{8}\ \frac{1}{z_m-z}  \ , \qquad
\phi_z \ \sim \ - \ \frac{3}{2}\ \frac{1}{z_m-z}
\eeq
near $z=z_m$, so that $\alpha=0$ in all cases,
one finds the following results:
\bea
\mathrm{Scalar\ perturbations} &:& \left(\mu,\tilde{\mu},\nu,\widetilde{\nu}\right) = \left(\frac{2}{3},1,\frac{7}{6},\frac{3}{2}\right) \ , \nonumber \\
\mathrm{Tensor\ perturbations} &:& \left(\mu,\tilde{\mu},\nu,\widetilde{\nu}\right) = \left(\frac{1}{3},0,\frac{1}{6},\frac{1}{2}\right) \ .
 \eea
In addition, there are NS perturbations, for which, in general
\beq
a \ = \ - \ 2\left[\beta_p\,\phi_z \ +\ (p-3) \Omega_z\right] \ ,
\eeq
where
\beq
\beta_p \ = \ \frac{3-p}{4} \ ,
\eeq
with $p=1,5$. Therefore in the first case
\beq
p=1 \ : \quad a \ = \ - \ \phi_z \ +\ 4\, \Omega_z \ , \qquad \left(\mu,\tilde{\mu},\nu,\widetilde{\nu}\right) = \left(0,1,\frac{1}{2},- \frac{1}{2} \right) \ ,
\eeq
while in the second case
\beq
p=5 \ : \quad a \ = \ \phi_z \ - \ 4\, \Omega_z \ , \qquad \left(\mu,\tilde{\mu},\nu,\widetilde{\nu}\right) = \left(1,0,- \frac{1}{2},\frac{1}{2}\right)  \ .
\eeq

Finally, for gauge vector perturbations in the SO(16)$\times$SO(16) string, one finds
\beq
a \ = \ - \ \frac{1}{2}\,\phi_z \ + \ 6\,\Omega_z \ , \qquad \left(\mu,\tilde{\mu},\nu,\widetilde{\nu}\right) = \ \left(\frac{1}{6},\frac{1}{2},\frac{1}{3},0\right)  \ .
\eeq
These results are summarized in Table~\ref{tab:tab_munu_heterotic}.

\begin{table}[ht]
\centering
\begin{tabular}{||c || c | c | c | c | c | c | c ||}
 \hline
Perturbations & $\mu$ & $\tilde{\mu}$ & $\alpha$ & $\nu$ & $\widetilde{\nu}$ & Case & Sufficient Stability Conditions \\ [0.5ex]
 \hline\hline
Scalar & $\frac{2}{3}$ &  $1$  & $0$ & $\frac{7}{6}$ & $\frac{3}{2}$ & 2 & always stable \\ \hline
Vector & $\frac{1}{6}$ &  $\frac{1}{2}$  & $0$ & $\frac{1}{3}$ & $0$ & 5 & $C_1 C_2=0$ and $C_3 C_4=0$ \\ \hline
Tensor & $\frac{1}{3}$ &  $0$ & $0$ & $\frac{1}{6}$ & $\frac{1}{2}$ & 3 & $C_1 C_2=0$ and $C_3=0$\\ \hline
NS $2$ form & $0$ &  $1$ & $0$ & $\frac{1}{2}$ & $- \frac{1}{2}$ & 4 & $C_1=0$ and $C_3 C_4=0$\\
NS $6$ form & $1$ &  $0$ & $0$ & $- \frac{1}{2}$ & $\frac{1}{2}$ & 3 & $C_1C_2=0$ and $C_3=0$ \\
 [1ex]
 \hline
\end{tabular}
\caption{Values of $\mu$, $\tilde{\mu}$, $\alpha$, $\nu$ and $\widetilde{\nu}$ for the different sectors of the ten--dimensional heterotic SO(16) $\times$ SO(16) model. The different cases refer to the sufficient conditions for positivity discussed in Section~\ref{sec:positivity}.}
\label{tab:tab_munu_heterotic}
\end{table}

In addition, as discussed in~\cite{bms}, there are in principle Kaluza--Klein vector modes, but in fact they are absent, since there is only a non-normalizable zero mode of this type in all cases.

\subsubsection{\sc Singular Potentials}

In all preceding cases, and more generally for the vacua in~\cite{dm_vacuum, ms21_1,ms21_2}, the potential $V(z)$ is singular at the ends of the interval, which we shall continue to denote by $z=0$ and $z=z_m$, where it develops double poles, which are accompanied by logarithmic terms for the orientifolds. The limiting behavior can be parametrized as in eq.~\eqref{double_poles}. It depends, in general, on the three real parameters $\mu$, $\tilde{\mu}$ and $\alpha$, whose values we have discussed in detail in the preceding section for the two ten--dimensional orientifolds of~\cite{as95,as97} and~\cite{sugimoto} and for the SO(16)$\times$ SO(16) heterotic model of~\cite{agmv1,agmv2}.

The condition that $H\,\psi$ be in $L^2$ restricts in general the choice of wavefunctions~\cite{math_literature_1,math_literature_2,tables}. Confining our attention to real and non--negative values for $\mu^2$ and $\tilde{\mu}^2$, which characterize all cases of interest for our problem, the allowed asymptotic behaviors at the ends of the $z$ interval are
\bea
\psi &\sim \  \frac{C_1}{\sqrt{2\,\mu}} \ \left(\frac{z}{z_m}\right)^{\frac{1}{2} \ + \ \mu} \ + \  \frac{C_2}{\sqrt{2\,\mu}} \ \left(\frac{z}{z_m}\right)^{\frac{1}{2} \ - \ \mu}  \qquad\qquad\qquad\qquad &\left( 0 \ < \ \mu \ < \ 1 \right)\ ; \nonumber \\
\psi &\sim \  C_1 \ \left(\frac{z}{z_m}\right)^{\frac{1}{2}}\,\log\left(\frac{z}{z_m}\right) \ + \ C_2 \ \left(\frac{z}{z_m}\right)^{\frac{1}{2} } \ \ \qquad\qquad\qquad\qquad &\left( \mu \ = \ 0 \right)\ ; \nonumber \\
\psi &\sim \ \frac{C_1}{\sqrt{2\,{\mu}}} \ \left(\frac{z}{z_m}\right)^{\frac{1}{2}+{\mu}}   \qquad\qquad\qquad \qquad \qquad \qquad \qquad \qquad\qquad &\left( {\mu} \ \geq \ 1\right) \ ;
\label{b.10}
\eea
and
\bea
\psi &\sim&  \frac{C_3}{\sqrt{2\,\tilde{\mu}}} \left(1\,-\,\frac{z}{z_m}\right)^{\frac{1}{2}+ \ \tilde{\mu}} \left|\log\left(1\,-\,\frac{z}{z_m}\right) \right|^\alpha \nonumber \\ &+&  \frac{C_4}{\sqrt{2\,\tilde{\mu}}}\ \left(1\,-\,\frac{z}{z_m}\right)^{\frac{1}{2} \ - \ \tilde{\mu}} \left|\log\left(1\,-\,\frac{z}{z_m}\right) \right|^{-\alpha}  \qquad\qquad\qquad \left( 0 \ < \ \tilde{\mu} \ < \ 1\right) \ ; \nonumber \\
\psi &\sim& \frac{C_3}{\sqrt{1-2\alpha}}\ \left(1\,-\,\frac{z}{z_m}\right)^{\frac{1}{2}}\,\left|\log\left(1\,-\,\frac{z}{z_m}\right) \right|^{1-\alpha} \nonumber \\
&+& \frac{C_4}{\sqrt{1-2\alpha}}\ \left(1\,-\,\frac{z}{z_m}\right)^{\frac{1}{2}} \left|\log\left(1\,-\,\frac{z}{z_m}\right) \right|^{\alpha} \ \ \quad\quad\qquad\qquad \left(\tilde{\mu} \ = \ 0 , {\alpha< \frac{1}{2}}\right) \ ; \nonumber \\
\psi &\sim& \frac{C_3}{\sqrt{2\,\tilde{\mu}}} \ \left(1\,-\,\frac{z}{z_m}\right)^{\frac{1}{2}+\tilde{\mu}}  \left|\log\left(1\,-\,\frac{z}{z_m}\right) \right|^{\alpha}  \ \ \quad\qquad \qquad\qquad \left(\tilde{\mu} \ \geq \ 1 \right)  \ .
\label{b.11}
\eea

The condition granting $H$ to be Hermitian is the vanishing of the boundary contribution
\beq
\Big[ \psi^\star \, \partial_z \, \chi \ - \ \partial_z \,\psi^\star \ \chi \Big]_0^{z_m} \ = \ 0 \ , \label{b.2}
\eeq
for all $\psi$ and $\chi$ wavefunctions belonging to the domain ${\cal D}$ of $H$. This domain identifies, in general, sets of functions $\psi$ subject to proper boundary conditions that are in $L^2$, and such that $H\,\psi$ and $H\,\chi$ are also in $L^2$.

When $\mu \geq 1$ and $\tilde{\mu} \geq 1$ the boundary conditions are fixed at both ends, since only the limiting behaviors associated to $C_1$ and $C_3$ are allowed.
 When $\mu<1$ and $\tilde{\mu} \geq 1$, the possible self--adjoint boundary conditions depend on a single parameter $\theta$, defined as
\beq
\frac{C_2}{C_1} \ = \ \tan\left(\frac{\theta}{2}\right) \ , \label{C21}
\eeq
and in a similar fashion, if ${\mu} \geq 1$ and $\tilde{\mu}<1$, the possible self--adjoint boundary conditions depend on the single parameter
\beq
\frac{C_4}{C_3} \ = \ \tan\left(\frac{\tilde{\theta}}{2}\right) \ . \label{C43}
\eeq
When $0 \leq \mu < 1$ and $0 \leq \tilde{\mu}<1$, defining the two vectors
\beq
\underline{C}(0) \ = \ \left(\begin{array}{c} C_1 \\ C_2  \end{array} \right) \ , \qquad
\underline{C}\left(z_m\right) \ = \  \left(\begin{array}{c} C_4 \\ C_3  \end{array} \right)  \ , \label{mutildezero}
\eeq
the self--adjointness condition becomes
\beq
\underline{C}(z_m) \ = \ {\cal U} \ \underline{C}\left(0\right) \ , \label{P_constraint}
\eeq
where ${\cal U}$ is a generic $U(1,1)$ matrix, so that
\beq
{\cal U}^\dagger\,\sigma_2\ {\cal U} \ = \ \sigma_2 \ .
\eeq
${\cal U}$ can be decomposed as
\beq
{\cal U} \ = \ e^{i\,\beta} \ U \ ,
\eeq
where $\beta$ is a phase and $U$ is a generic $SL(2,R)$ matrix, which can be parametrized as
\beq
U\left(\rho,\theta_1,\theta_2\right) \ = \ \cosh\rho\left(\cos\theta_1 \, \underline{1} \,-\, i\,\sigma_2\,\sin\theta_1\right) \,+\, \sinh\rho\left(\sigma_3\, \cos\theta_2\ + \ \sigma_1\, \sin\theta_2 \right)
\ , \label{global_ads3_1}
\eeq
where $0 \leq \rho< \infty$, $-\,\pi \leq \theta_{1,2} < \pi$.

The equation Schr\"odinger determines, in general, the relation
\beq
\underline{C}\left(z_m\right) \ = \ V \ \underline{C}\left(0\right) \ , \label{C0mV}
\eeq
where $V$ is an $SL(2,R)$ matrix, consistent with our definitions and the constancy of the Wronskian, and
the eigenvalue equation is then, in general~\cite{ms23_1}
\beq
\mathrm{Tr}\left[ {U}^{-1}\ V \right] \ = \ 2\,\cos\beta \ . \label{eigenveq}
\eeq
The large--$\rho$ limit of eq.~\eqref{P_constraint} yields independent boundary conditions at the ends involving the two parameters $\theta_1$ and $\theta_2$, which can be cast in the form
\bea
&& \cos\left(\frac{\theta_1-\theta_2}{2}\right) C_1 \ - \ \sin\left(\frac{\theta_1-\theta_2}{2}\right)C_2  \ = \ 0 \ , \nonumber \\
&&  \sin\left(\frac{\theta_1+\theta_2}{2}\right) C_4 \ - \ \cos\left(\frac{\theta_1+\theta_2}{2}\right)C_3  \ = \ 0    \  , \label{cond_sing1n1}
\eea
when $0 \leq \mu< 1$ and $0 \leq \tilde{\mu} < 1$. We can now elaborate on these different cases for the vacuum solutions of~\cite{dm_vacuum}.

\subsubsection{\sc The Issue of Positivity} \label{sec:positivity}

In Section~\ref{sec:selfadjoint} we have recast the spectral problems for the different sectors of the 9D compactifications of~\cite{dm_vacuum} in terms of Schr\"odinger--like operators of the form
\beq
\left[ {\cal A} \, {\cal A}^\dagger \ + \ b \right] \psi \ = \ m^2 \ \psi \ , \label{b.282}
\eeq
where $b>0$ for scalar perturbations and vanishes in all other cases, and~\cite{bms}
with
\beq
{\cal A} \ = \ \partial_z \ + \ \frac{1}{2}\,a(z) \ , \qquad {\cal A}^\dagger \ = \ - \  \partial_z \ + \ \frac{1}{2}\,a(z) \ . \label{AAdagger}
\eeq

Leaving aside $b$, which gives a positive contribution whenever it is present, we can now examine under what conditions the operator ${\cal A} {\cal A}^\dagger$ gives a positive contribution to $m^2$.
Multiplying eq~\eqref{b.282} by $\psi^\star$ and integrating gives
\beq
m^2 \int_{z_1}^{z_2} dz \ \left| \psi\right|^2 \ = \ \int_{z_1}^{z_2} dz \  \left| {\cal A}^\dagger\ \psi\right|^2 \ + \ \left[ \psi^\star \, {\cal A}^\dagger\,\psi \right]_{z_1}^{z_2} \ , \label{m2_boundary}
\eeq
and one can thus conclude that if
\beq
\lim_{z_1\to 0}\ \lim_{z_2 \to z_m} \ \left[ \psi^\star \, {\cal A}^\dagger\,\psi \right]_{z_1}^{z_2}  \ \equiv \  \left.\psi^\star \, {\cal A}^\dagger\,\psi\right|_{z_m} \ - \ \left.\psi^\star \, {\cal A}^\dagger\,\psi\right|_{0} \ \geq \ 0 \ ,  \label{bc_pos}
\eeq
$m^2 \,\geq\, 0$ and the corresponding sector is perturbatively stable. If this condition does not hold, positivity is not guaranteed, and as we saw in~\cite{ms23_1} self--adjoint boundary conditions can lead to the emergence of tachyonic modes even if the factorization of the Schr\"odinger operator holds, and even in the absence of a potential. Here we would like to summarize some considerations for the singular potentials of interest for the 9D compactifications~\cite{dm_vacuum} of the three non--supersymmetric ten--dimensional strings.

\begin{enumerate}
\item if $\mu \geq 1$ the boundary condition at the left end is fixed and positivity is guaranteed there, and if $\tilde{\mu} \geq 1 $ similar considerations hold at the right end. This is the case, at both ends, for the scalar perturbation in the orientifolds.
    \item If $\nu \geq \frac{1}{2}$ and $\widetilde{\nu} \geq \frac{1}{2}$, which occurs for orientifold and heterotic scalar perturbations, and also for orientifold tensor perturbations, close to the left end of the interval
\beq
{\cal A}^\dagger \ \sim \ - \ \partial_z \ + \ \frac{\mu+ \frac{1}{2}}{z} \ ,
\eeq
and similarly close to the right end
\beq
{\cal A}^\dagger \ \sim \ - \ \partial_z \ - \ \frac{\tilde{\mu}+\frac{1}{2}}{z_m-z} \ - \ \frac{\alpha}{\left(z_m-z\right)\log\left(z_m-z\right)} \ .
\eeq
Analyzing the different cases one can conclude that positivity is guaranteed if the boundary conditions select the less singular behavior at both ends when $\mu<1$ and/or $\tilde{\mu}<1$, and is automatically guaranteed if $\mu$ and $\tilde{\mu}$ are both larger than one. In detail, when $\mu=0$ positivity is guaranteed if $C_1=0$, and when $0<\mu<1$ it is guaranteed if $C_2=0$. Similarly, when $\tilde{\mu}=0$ positivity holds if $C_3=0$, and when $0<\tilde{\mu}<1$ it holds if $C_4=0$. Note that, for these values of $\nu$ and $\widetilde{\nu}$, the solutions of
\beq
{\cal A}^\dagger\,\psi \ = \ 0
\eeq
always verify these conditions, so that the boundary conditions of these zero modes lead to non--tachyonic spectra. These considerations apply to the graviton sector of the orientifolds and to the scalar sector of the heterotic model.
 \item If $\nu < \frac{1}{2}$ and $\widetilde{\nu} \geq \frac{1}{2}$, the sufficient conditions for positivity at the right end is as before, so that when $\tilde{\mu}=0$ it is $C_3=0$, and when $0<\tilde{\mu}<1$ it is $C_4=0$, but at the left end the sufficient condition becomes $C_1 C_2=0$.
  \item If $\nu \geq \frac{1}{2}$ and $\widetilde{\nu} < \frac{1}{2}$, the sufficient conditions at the left end are $C_2=0$ when $0<\mu<1$ and $C_1=0$ for $\mu=0$, while at the right end the sufficient condition is $C_3 C_4=0$.
   \item If $\nu < \frac{1}{2}$ and $\widetilde{\nu} < \frac{1}{2}$, the sufficient conditions at the two ends are $C_1 C_2=C_3 C_4=0$.
\end{enumerate}

\subsubsection{\sc Heterotic  Perturbations} \label{sec:tsv_heterotic}

The values of $\mu$ and $\tilde{\mu}$ for the SO(16)$\times$SO(16) heterotic model of~\cite{agmv1,agmv2} are summarized in Table~\ref{tab:tab_munu_heterotic}. We begin our analysis from this case, which is a bit simpler since $\alpha=0$, and the correspondence with the exactly solvable hypergeometric potentials of Appendix~\ref{app:hypergeometric} is more direct. We work again with $\rho=1$ and $\Delta=1$, so that $z_m \simeq 1.93$.
    \begin{figure}[ht]
\centering
\begin{tabular}{cc}
\includegraphics[width=70mm]{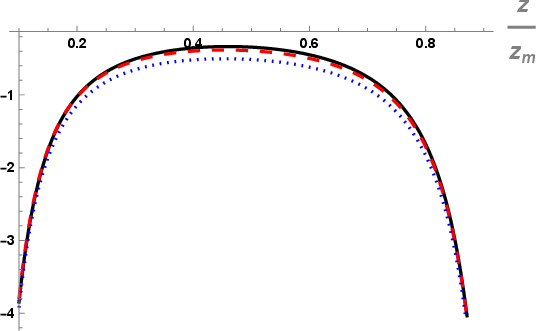} \hskip 0.8cm
\includegraphics[width=70mm]{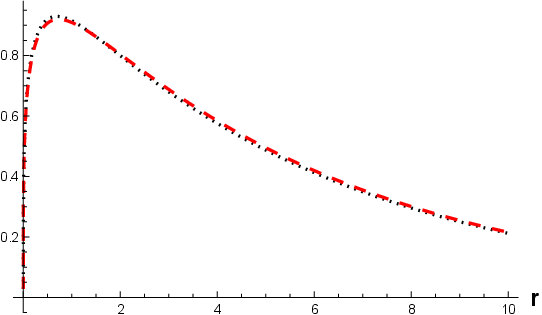}
\end{tabular}
\caption{\small Left panel: the graviton potential for the heterotic model (black,solid), the hypergeometric approximation of eq.~\eqref{pot_hyp_main} with $\xi=0$ (blue, dotted) and with $\xi\simeq 0.21$ (red, dashed). Right panel: the graviton wavefunction (black dotted) and its hypergeometric approximation (red dashed).}
\label{fig:potential_grav_1616}
\end{figure}

\begin{enumerate}
    \item \textbf{The graviton sector}\\
This sector corresponds to case 3, and stable boundary conditions are guaranteed to exist if $C_1 C_2 =0$ and $C_3=0$, but these choices do not guarantee the existence of a zero mode. It is interesting to identify boundary conditions that grant stability together with a zero mode, which would describe a massless graviton in the residual nine--dimensional Minkowski space. In this case one can approximate $V$ by a shifted hypergeometric potential of the form
\beq
V_{\mu,\tilde{\mu}}(z) \ = \ \frac{\pi^2}{4\,z_m^2}\left[ \frac{\mu^2 \ - \ \frac{1}{4}}{\sin^2\left(\frac{\pi\,z}{2\,z_m}\right)} \ + \ \frac{\tilde{\mu}^2 \ - \ \frac{1}{4}}{\cos^2\left(\frac{\pi\,z}{2\,z_m}\right)} \right]  \ + \ \frac{\pi^2}{z_m^2} \xi \ , \label{pot_hyp_main}
\eeq
which has the same leading singularities at both ends if $\mu=\frac{1}{3}$ and $\tilde{\mu}=0$. The shift $\xi$ is determined using the exact zero mode,
\beq
\psi_0 \ = \ e^{4\,\Omega} \ , \label{zero_mode_grav_1616}
\eeq
which has limiting behaviors at the two ends corresponding to
\beq
C_1 \ \simeq \ - \ 0.68\, C_2 \ , \qquad C_3 \ = \ 0 \ .
\eeq
\begin{figure}[ht]
\centering
\begin{tabular}{cc}
\includegraphics[width=65mm]{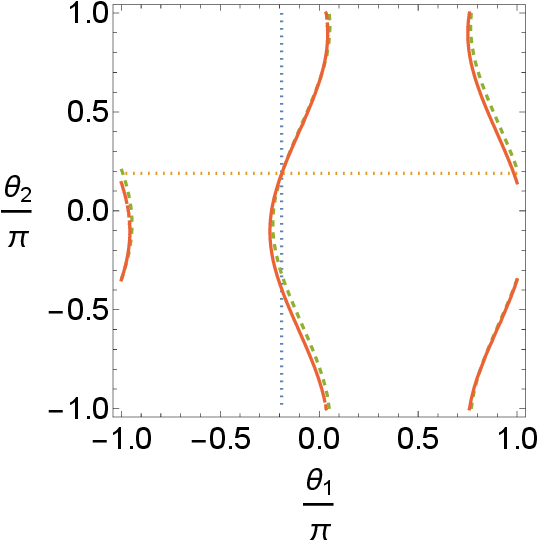} \qquad\qquad &
\includegraphics[width=65mm]{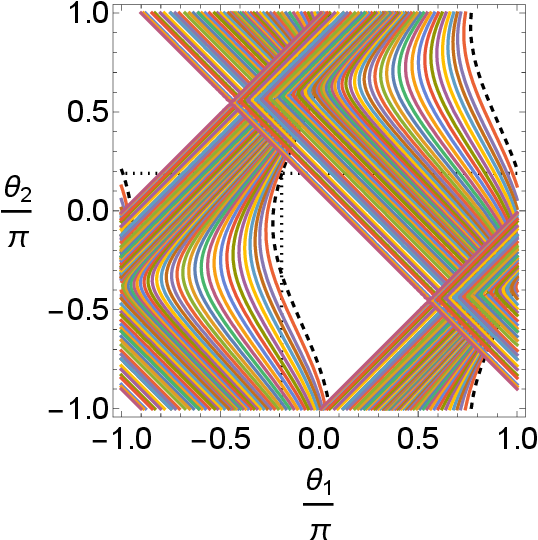} \\
\end{tabular}
\caption{\small Left panel: the boundary conditions leading to a massless mode (dashed line: exact, continuous line: hypergeometric approximation). The zero mode~\eqref{zero_mode_grav_1616} lies at the intersection of the vertical and horizontal dotted lines for $(\theta_1,\theta_2) \simeq (-0.19,0.19)\pi$.  Right panel: the shaded regions identify the boundary conditions leading to instabilities, in the hypergeometric approximation. The dashed lines in the white regions correspond to boundary conditions with zero modes but no instabilities.}
\label{fig:grav_instabilities_1616}
\end{figure}

With these boundary conditions, the eigenvalue equation determined by eqs.~\eqref{xi12def} and \eqref{eigenvmu0_rhoinf} for the hypergeometric setup reduces to
\beq
0.68  \ \simeq \ \left(\frac{\pi}{2}\right)^\frac{2}{3} \frac{\Gamma\left(\frac{2}{3}\right)}{\Gamma\left(\frac{4}{3}\right)} \left|\frac{\Gamma\left(\frac{2}{3} + i \,\xi\right)}{\Gamma\left(\frac{1}{3} + i \,\xi\right)}\right|^2 \ ,
\eeq
whose solution is a tachyonic mode with $\xi\simeq 0.21$.
These boundary conditions determine an unshifted hypergeometric spectrum that starts with a node-free tachyonic mode, which is displayed in the right panel of fig.~\ref{fig:potential_grav_1616}. If this wavefunction is identified with a (faithful) approximation to the actual zero mode for the problem at stake, this condition determines the shift $\xi\simeq 0.21$ to be included in eq.~\eqref{pot_hyp_main}, after which the actual potential and the shifted hypergeometric approximation are almost indistinguishable, as in fig.~\ref{fig:potential_grav_1616}.

One can actually identify all self--adjoint boundary conditions leading to the presence of a zero mode. In general, the self--adjoint boundary conditions are identified by the ratios $\frac{C_1}{C_2}$ and $\frac{C_3}{C_4}$, which can be conveniently labeled by a pair of angles $(\theta_1,\theta_2)$, defined in eqs.~\eqref{cond_sing1n1}, such that
\beq
\frac{C_1}{C_2} \ = \ \tan\left(\frac{\theta_1 \,-\,\theta_2}{2}\right) \ , \qquad \frac{C_3}{C_4} \ = \ \tan\left(\frac{\theta_1 \,+\,\theta_2}{2}\right) \ . \label{C1234}
\eeq
The general zero mode is a linear combination of the zero mode in eq.~\eqref{zero_mode_grav_1616} and the other independent solution of the massless Schr\"odinger problem, which can be obtained from it with the Wronskian method. The corresponding wavefunction has a simple analytic form in terms of $r$, and reads
\beq
\psi(r) \ = \ \left[\left(3\,C_3\,\log\left(\frac{\sqrt{2} {z_m}}{3}\right) \ - \ 3\,C_4 \right) \ + \ C_3\,r\right] e^{- \,\frac{5 r}{12}} \, \left(\sinh r\right)^\frac{1}{4} \ .  \label{zero_mode_1616}
\eeq
By expanding this exact solution near the two endpoints $z=0$ and $z=z_m$, making use of eq.~\eqref{C1234}, one finds that the boundary conditions leading to the exact zero modes satisfy
\beq
\tan \left(\frac{\theta_1+\theta_2}{2} \right)\ = \ \frac{3 \left(\frac{2}{3\, z_m}\right)^{2/3} \tan \left(\frac{\theta_1\,-\,\theta_2}{2} \right)\ + \ 1}{\log \left(\frac{\sqrt{2}\,  z_m}{3}\right) \left(3 \left(\frac{2}{3 \,  z_m}\right)^{2/3}\tan \left(\frac{\theta_1\,-\,\theta_2}{2} \right)\ +\ 1\right)-1} \ .
\eeq
The corresponding locus is the dashed curve in the left panel of fig.~\ref{fig:grav_instabilities_1616}, which can be compared with the nearby continuous curve, which is obtained from the hypergeometric approximation. The results in figs.~\ref{fig:potential_grav_1616} and fig.~\ref{fig:grav_instabilities_1616} manifest its accuracy. No tachyons are present when the zero--mode wavefunction in eq.~\eqref{zero_mode_1616} has no nodes in the physical range $r \geq 0$. One can also see, in this fashion, that at most one tachyon can be present. Only the zero modes corresponding to the dashed curves in the white regions of the right panel of fig.~\ref{fig:grav_instabilities_1616}, which comprise two disconnected components, are not accompanied by tachyonic instabilities.

Summarizing, we have provided some evidence for the accuracy of the hypergeometric approximation, and we have identified two disconnected curves corresponding to boundary conditions leading a massless mode but no tachyonic instabilities. The remaining choices for the boundary conditions in fig.~\ref{fig:grav_instabilities_1616} lead to instabilities (shaded regions) or to purely massive spectra (white regions).
    \begin{figure}[ht]
\centering
%\begin{tabular}{cc}
\includegraphics[width=90mm]{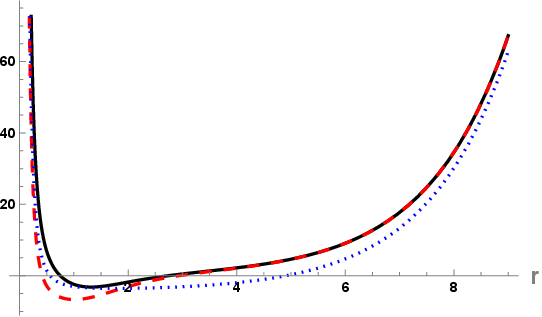} \hskip 1cm
%\end{tabular}
\caption{\small The scalar potential including the shift b(r) (black, solid), without it (red, dashed) and the shifted hypergeometric approximation (blue, dotted).}
\label{fig:potential__1616}
\end{figure}

\item \textbf{The scalar sector}\\
In this case the Hamiltonian is of the form
\beq
H \ = \ {\cal A}\,{\cal A}^\dagger \ + \ b \ \equiv \ H_0 \ + \ b \ ,
\eeq
with $b>0$, as shown in~\cite{bms}. Therefore, the preceding setup does not a priori apply. However,
\beq
b(r) \ = \  \frac{14\ e^{\frac{5 r}{3}}}{\left(\sinh r\right)^3 \left(15 \coth r-9\right)}
\eeq
is subdominant at both ends, with the leading behaviors
\beq
b \ \sim \ z^{-\,\frac{4}{3}} \ , \qquad b \ \sim \ \left(1 \ - \  \frac{z}{z_m}\right)^4 \ .
\eeq
One can therefore resort once more to a shifted hypergeometric approximation for $H_0$, treating $b$ as a perturbation. The starting point, as before, is provided by the exact zero mode wavefunction $\psi_0$, which solves the first--order equation ${\cal A}^\dagger\,\psi_0=0$, and reads
\beq
\psi_0(r) \ = \ N\, \frac{\left(\tanh r\right)^\frac{7}{4} \ \left(1\,-\,\tanh r\right)^\frac{1}{4}}{(1 \,+\,\tanh r ) (5\,-\,3 \tanh r)} \ ,
\eeq
where $N$ is a normalization constant. The limiting behavior of $\psi_0$ at the two ends is
\beq
\psi_0 \ \sim  \ z^\frac{7}{6} \ , \qquad \psi_0 \ \sim \  \ \sim \ \left(1 \ - \  \frac{z}{z_m}\right)^\frac{3}{2} \ , \label{lim_scalar}
\eeq
which grants its normalizability. This is actually the only normalizable zero mode in this case, and is characterized by the boundary condition
\beq
C_2 \ = 0 \ .
\eeq
Eq.~\eqref{eigenvmutildelarger} then implies that the unperturbed spectrum of the unshifted hypergeometric potential is determined by the poles of the $\Gamma$ functions in the denominator, so that
\beq
m^2 \ = \ \frac{\pi^2}{z_m^2}\left(n \ + \ \frac{4}{3}\right)^2 \qquad (n=0,1,\ldots) \ ,
\eeq
with a ground--state wavefunction
\beq
\psi_{0h}(z) \ = \ N_h \ \left[\sin\left( \frac{\pi z}{2\,z_m}\right)\right]^\frac{7}{6} \left[\cos\left( \frac{\pi z}{2\,z_m}\right)\right]^\frac{3}{2}
\eeq
that corresponds to $n=0$, with no nodes and the limiting behavior of eq.~\eqref{lim_scalar}. The shifted potential that approximates $H_0$ is thus given by eq.~\eqref{pot_hyp_main} with $\left(\mu,\tilde{\mu}\right) = \left(\frac{2}{3},1\right)$ and $\xi = - \,\frac{16}{9}$.

Leaving aside momentarily the positive shift $b$, one can use the results in Appendix~\ref{app:hypergeometric} to estimate the dependence of the spectrum of $H_0$ on the boundary conditions. There are indeed boundary conditions leading to the emergence of one tachyonic mode for $H_0$. These present themselves, in the hypergeometric approximation, in the range $- \,0.43< \frac{C_2}{C_1} <0$, which can be deduced solving eq.~\eqref{eigenvmutildelarger} for imaginary values of $m$, and the tachyonic mass becomes arbitrarily large as $\frac{C_2}{C_1} \to 0^-$. Taking into account the positive contribution $b$ makes the massless mode $\psi_0$ massive, while also reducing the instability range. One can thus conclude that, for this sector, there are boundary conditions leading to instabilities or to stable spectra with a massless mode within the range identified above, or purely massive spectra for $\frac{C_2}{C_1} \geq 0$.
\item \textbf{The gauge vector sector}\\
In this case $b=0$, and
\beq
a \ = \ \frac{e^{\frac{5 r}{6}} (\coth r \ -\ 1)}{\sqrt{\sinh r}} \ ,
\eeq
so that the solution to ${\cal A}^\dagger\,\psi_0 = 0$ reads
\beq
\psi_0(r) \ = \ N\, e^{-\,\frac{r}{2}} \, \sqrt{\sinh r} \ ,
\eeq
with $N$ a normalization factor. Its limiting behavior at the two ends is
\beq
\psi_0 \ \sim \ z^\frac{1}{3} \ + \ {\cal O}(z) \ , \qquad \psi_0 \ \sim \ \mathrm{const} \ + \ {\cal O}\left((z_m-z)^6\right)
\eeq
and therefore $C_1=C_3=0$.

Enforcing these conditions in eqs.~\eqref{hyper_les_less} gives the unshifted hypergeometric spectrum
\beq
m^2 \ = \ \frac{\pi^2}{z_m^2} \left(n \ + \ \frac{1}{6}\right)^2 \qquad (n=0,1,\ldots) \ ,
\eeq
so that one recovers a massless mode considering a shifted hypergeometric potential~\eqref{pot_hyp_main} with
$\left(\mu,\tilde{\mu},\xi\right)=\left(\frac{1}{6},\frac{1}{2},- \, \frac{1}{36}\right)$.

However, the shifted hypergeometric potential does not agree with limiting behavior of the actual potential, which vanishes at $z_m$. Here we encounter a limitation of the method, which manifests itself due to lack of singular dominant terms at the right end since $\tilde{\mu}=\frac{1}{2}$. Nevertheless, the sufficient conditions of Section~\ref{sec:positivity} guarantee that, since $\mu$ and $\tilde{\mu}$ are both less than $\frac{1}{2}$, case 5 applies and the boundary conditions of the actual zero mode guarantee positivity. However, we are unable to perform a reliable analysis for generic boundary conditions in this case.

\item \textbf{The NS 6-form sector}\\
    \begin{figure}[ht]
\centering
%\begin{tabular}{cc}
\includegraphics[width=80mm]{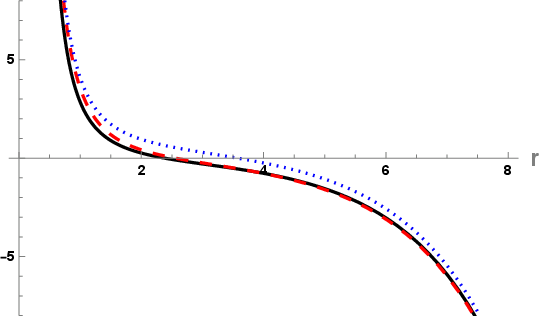}
%\end{tabular}
\caption{\small The six--form potential (black, solid), the shifted hypergeometric approximation (red, dashed) and the unshifted hypergeometric approximation (blue, dotted).}
\label{fig:potential_6form_1616}
\end{figure}
In this case the solution of ${\cal A}^\dagger \psi = 0$ is not normalizable, but a normalizable zero mode can be found by the Wronskian method, and reads
\beq
\psi_0(r) \ = \ \frac{e^{\frac{7 r}{12}} \left(4 r-e^{-4 r}+4 e^{-2 r}-3\right)}{16 \left(\sinh r\right) ^{\frac{3}{4}}} \ .
\eeq
It has the limiting behaviors
\bea
\psi_0 &\sim& \left(\frac{z}{z_m}\right)^\frac{3}{2} \ , \nonumber \\
\psi_0 &\sim& \left(1 - \frac{z}{z_m}\right)^\frac{1}{2} \left[\log\left(1 \,- \,\frac{z}{z_m}\right) \ + \ \frac{1}{4} \ + \ \log\left(\frac{\sqrt{2}\,z_m}{3}\right)\right]
\eea
near the two ends, up to proportionality constants, so that its boundary conditions correspond to $C_2=0$ and
\beq
\frac{C_4}{C_3} \ = \ \frac{1}{4} \ + \ \log\left(\frac{\sqrt{2}\,z_m}{3}\right) \ \simeq 0.15 \ .
\eeq
\begin{figure}[ht]
\centering
\begin{tabular}{ccc}
%\mbox{graphic} & \mbox{table} \\
\includegraphics[width=65mm]{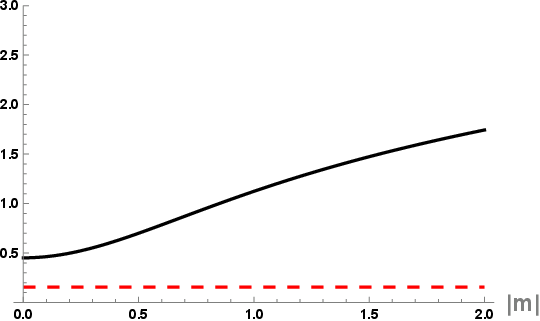} \quad  &
\includegraphics[width=65mm]{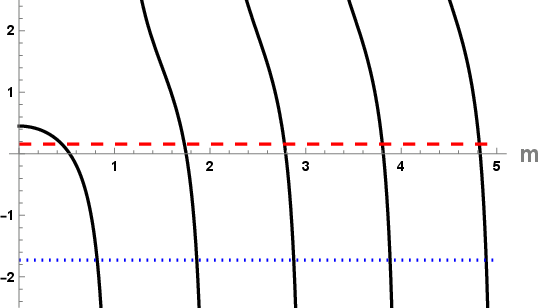}  \\
\end{tabular}
 \caption{\small The hypergeometric eigenvalue equation for the six-form for imaginary values of the mass (left panel) and for real values of the mass (right panel). In both cases the red dashed horizontal line corresponds to $\frac{C_4}{C_3}\simeq 0.15$. The blue dotted line indicates the value of $\frac{C_2}{C_1}$ for the 2-form case, as explained in the text.}
\label{fig:ns6_eigenv}
\end{figure}

One can determine the shift of the hypergeometric potential starting from the ratio of the two equations in~\eqref{eigenvmu0_rhoinf} with $C_2=0$, and the eigenvalue equation then reads
\beq
\frac{C_4}{C_3} \ = \ \frac{1}{2} \left[ \psi(m+1)\  + \ \psi(-m+1) \ -\  2 \psi(1) \ +\  2 \log \frac{\pi}{2} \right] \ ,
\eeq
where $\psi(x)=\frac{\Gamma'(x)}{\Gamma(x)}$ is Euler's digamma function.

The lowest eigenvalue for the given ratio $\frac{C_4}{C_3}$ is real and positive, as can be seen from fig.~\ref{fig:ns6_eigenv}, so that the correspondence with the actual potential, where the solution should be massless, requires a negative shift $\xi \simeq - (0.45)^2$. The resulting spectrum is then positive in this case, for all boundary conditions such that
\beq
\frac{C_4}{C_3} \ \leq \ \frac{1}{4} \ + \ \log\left(\frac{\sqrt{2}\,z_m}{3}\right) \ \simeq 0.15 \ ,
\eeq
but otherwise tachyons are present. Note how the negative shift improves the agreement between the hypergeometric potential and the exact one in fig.~\ref{fig:potential_6form_1616}.

\item \textbf{\sc The NS-NS 2-form sector}\\
This case is similar to the previous one, up to a reflection with respect to the middle of the $z$ interval. The solution of ${\cal A}^\dagger \psi = 0$ is again not normalizable, but a normalizable zero mode can be found by the Wronskian method. It is somewhat involved, but the limiting behaviors near the two ends, which suffice for our considerations, are
\bea
\psi_0 &\sim& \left(\frac{z}{z_m}\right)^\frac{1}{2} \left[\log\left(\frac{z}{z_m}\right) \ + \ \log\left( \frac{3 z_m}{2}\right) \ + \ \frac{3}{2}\ \log\left(\frac{2}{3\sqrt{3}} \right) \ -\  \frac{\sqrt{3}}{4}\ \pi \right] \ , \nonumber \\
\psi_0 &\sim& \left(1 - \frac{z}{z_m}\right)^\frac{3}{2}  \ ,
\eea
up to proportionality constants, so that
\beq
\frac{C_2}{C_1} \ = \  \log\left( \frac{3 z_m}{2}\right) \ + \ \frac{3}{2}\ \log\left(\frac{2}{3\sqrt{3}} \right) \ -\  \frac{\sqrt{3}}{4}\ \pi \ \simeq \ - \ 1.73 \ .
\eeq
This result determines the shift in the hypergeometric approximation, solving the eigenvalue equation
\beq
\frac{C_2}{C_1} \ = \ \frac{1}{2} \left[ \psi(m+1)\  + \ \psi(-m+1) \ -\  2 \psi(1) \ +\  2 \log \frac{\pi}{2} \right] \ ,
\eeq
and the result is $\xi = - (0.82)^2$. No tachyons are present when $\frac{C_2}{C_1} < - 1.73$.
\end{enumerate}

\subsubsection{\sc Orientifold Perturbations} \label{sec:tsv_orientifolds}

The values of $\mu$ and $\tilde{\mu}$ for the two ten--dimensional orientifolds of~\cite{as95,as97} and~\cite{sugimoto} are summarized in Table~\ref{tab:tab_munu_orientifolds}~\footnote{They also correct some results for RR forms and gauge vectors that were given in~\cite{ms23_1}.}. All perturbations lead to formally positive Hamiltonians of the form ${\cal A}{\cal A}^\dagger + b$, with $b$ a positive potential, as was the case for the heterotic model. As explained in Section~\ref{sec:positivity}, in all these cases, one can find sufficient conditions that grant stability, which are summarized in the last column of Table~\ref{tab:tab_munu_orientifolds}. In principle, these grant the existence of stable boundary conditions for these vacua.

More detailed global information on the actual stability regions can be obtained, as in the previous section, using approximate descriptions of the Sch\"odinger potentials. The Legendre approximation has the virtue of being exactly solvable and was considered for the orientifolds in~\cite{ms23_1}. The corresponding results are admittedly less precise than those presented for the heterotic case, since they leave out the logarithmic corrections, which also affect the limiting behavior of the wavefunctions. Logarithmic corrections are actually present in all cases, except for the selfdual five-form, which has no potential altogether. Variational tests comparing the results with and without logarithmic terms showed no significant differences between them, so that in the following we shall set $\alpha=0$, in order to summarize and complete the results presented in~\cite{ms23_1}. The hypergeometric potentials that we have reviewed in Appendix~\ref{app:hypergeometric} provide good approximations for the potentials of all these sectors, which we can now analyze in detail. Whenever $\mu$ and $\tilde{\mu}$ coincide, they reduce to the Legendre potentials discussed in~\cite{ms23_1}.

From Table~\ref{tab:tab_munu_orientifolds} one can also see that the special value $\tilde{\mu}=\frac{1}{2}$ is present for gauge vector perturbations, while $\mu=\frac{1}{2}$ is also present for the RR 2-form. In both cases, as we have seen, the hypergeometric approximation is not accurate, and we can only rely on the sufficient conditions of Section~\ref{sec:positivity}. Our main indications are summarized below.

\begin{figure}[ht]
\centering
\begin{tabular}{ccc}
%\mbox{graphic} & \mbox{table} \\
\includegraphics[width=65mm]{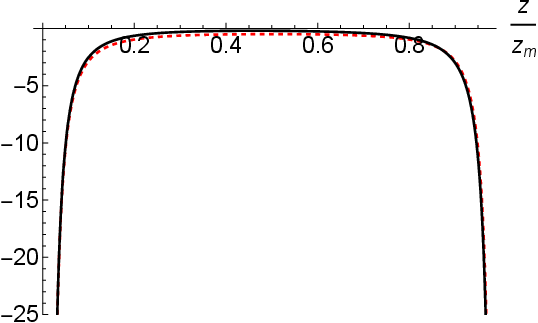} \quad  &
\includegraphics[width=65mm]{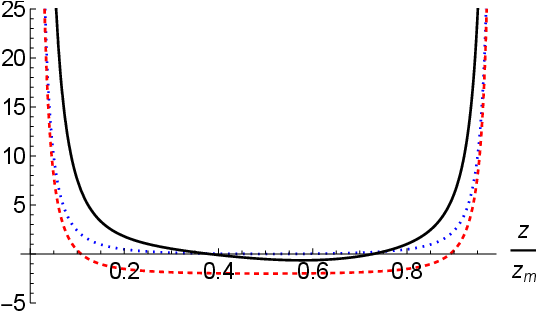}  \\
\end{tabular}
\caption{\small The potentials for tensor perturbations (left panel, solid) and scalar perturbations (right panel, solid) (right panel, solid) for $\gamma=\gamma_c=\frac{3}{2}$, in units of $\frac{1}{z_m^2}$ and as functions of $\frac{z}{z_m}$. The potential of eq.~\eqref{pot_hyp} with $\mu=\tilde{\mu}=0$ is the red dashed curve in the left panel, while the $H_{+-}$ potential for $\mu=\tilde{\mu}=1$ is the red dashed curve in the right panel. Finally, the $H_{-+}$ potential for $\mu=\tilde{\mu}=1$ is the green dotted curve in the right panel.}
\label{fig:exact_vs_sin}
\end{figure}

\begin{enumerate}
\item  \textbf{The gravity sector}\\
    In this case the independent boundary conditions are parametrized by a pair of angles $(\theta_1,\theta_2)$, and fig.~\ref{fig:exact_vs_sin} shows that the hypergeometric (or Legendre) potential provides a close approximation to the actual one.
    \begin{figure}[ht]
\centering
\includegraphics[width=70mm]{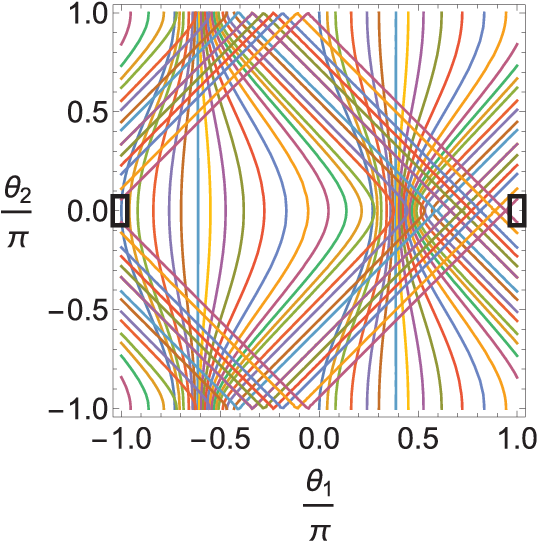}
\caption{\small For $\mu=\tilde{\mu}=0$, there are instabilities for all values of $\theta_1$ and $\theta_2$ away from the special point $(\pi,0)$ in the hypergeometric approximation, and thus in the actual potential at most outside a small region surrounding it (schematically, outside the two half-squares in the figure).}
\label{fig:instabilities_mu0}
\end{figure}

    In the hypergeometric or Legendre approximation, all choices of independent boundary conditions lead to instabilities, aside from the special one corresponding to a wavefunction with no logarithmic singularities at the ends~\cite{ms23_1}, so that the sufficient conditions summarized in Table~\ref{tab:tab_munu_orientifolds} appear also necessary in this case. Variational tests with and without the inclusion of the logarithmic terms exhibited similar very small stability regions around this point, thus providing compatible indications. It is thus safe to conclude that at most a small region around this point is compatible with stability. In particular, the hypergeometric approximation yields the spectrum
    \beq
    M^2 \ = \ \left(\frac{\pi}{z_m}\right)^2 \, n(n+1) \ , \qquad n=0, 1, \ldots \ ,
    \eeq
    which includes a massless graviton as its low--lying mode, whose normalized ground--state wavefunction is
    \beq
\psi_0 \ = \ \sqrt{\frac{\pi}{2\,z_m} \, \sin \left(\frac{\pi\, z}{z_m}\right)} \ .
    \eeq
    The nine--dimensional vacuum of~\cite{dm_vacuum} thus leads to a non--vanishing value for Newton's constant in nine dimensions, with an effective long--range gravitational force.

\item  \textbf{The scalar sector}\\
In this case the
boundary conditions are fixed, and the issue is optimizing the overall shift for the hypergeometric (or Legendre) potential, which is admittedly less precise for this sector, in order to bring it as close as possible to the actual one.

With $\mu=\tilde{\mu}$, the shifts in eqs.~\eqref{V12} and \eqref{m212} only depend on the combination $\epsilon_1-\epsilon_2$ of Appendix~\ref{app:hypergeometric}, and the highest potential obtains for $\epsilon_1=1$, $\epsilon_2=1$. In this case eq.~\eqref{m11} gives
\beq
M^2 \ = \ \left(\frac{\pi}{z_m}\right)^2 \left(2 \ + \ n\right)\left(n+1\right) \ , \qquad n=0,1,2,\ldots \ .
\eeq

Alternatively, one can estimate a shift $a$, starting from $H_{+-} + \frac{\pi^2}{z_m^2}\,a$ and relying on the normalized zero--mode wavefunction of $H_{+-}$,
\beq
\psi_{+-} \ = \ \sqrt{\frac{3\,\pi}{4\,z_m} }\left[ \sin \left(\frac{\pi\, z}{z_m}\right) \right]^\frac{3}{2} \ ,
\eeq
which grants a better convergence at the ends, demanding that the deviation
\beq
\Delta\,M^2 \ = \ \langle \psi_{+-} | \left[ V_{true}(z) \ - \ V_{1,-1}(z) \ - \ \frac{\pi^2}{z_m^2}\, \alpha \right] | \psi_{+-}\rangle  \label{pertas}
\eeq
be as small as possible in absolute value. In fact, in this case the first contribution is negligible with respect to the second, and therefore one can choose
\beq
\alpha \ \simeq \ - \ \frac{z_m^2}{\pi^2}\,\langle \psi_0 | V_{1,-1}(z) | \psi_0\rangle \ = \ \frac{9}{8} \ ,
\eeq
so that the best hypergeometric potential is
\beq
V \ \simeq \ \left(\frac{\pi}{z_m}\right)^2 \left[ \frac{3}{4 \left[\sin\left(\frac{\pi\,z}{z_m}\right)\right]^2} \ - \ \frac{9}{8} \right] \ = \ V_{-1,1} \ - \ \frac{9}{8}\, \left(\frac{\pi}{z_m}\right)^2\ .
\eeq
With $H_{1,-1}$ the spectrum would contain a massless mode, but taking the correction into account this estimate for the scalar spectrum is
   \beq
    M^2 \ \simeq \ \left(\frac{\pi}{z_m}\right)^2 \Big[ \left(n+\frac{3}{2}\right)^2 \,-\, \frac{9}{8}\Big] \ , \qquad n= 0, \ldots \ ,
    \eeq
which is again purely massive. This analysis indicates that the low--lying dilaton mode that emerges in the nine--dimensional effective theory is not a modulus, as in the conventional Kaluza--Klein setting, but is stabilized. The pressing general goal of stabilizing moduli is thus realized in this example.

Note that these considerations are merely adding details on the spectrum to the positivity arguments of Section~\ref{sec:int_comp_stab}. Positivity is indeed guaranteed by the ${\cal A}{\cal A}^\dagger$ form of the Hamiltonian whenever $\mu \geq 1$ and $\tilde{\mu} \geq 1$, so that a single choice of boundary condition is allowed at both ends (see Section 3.2.5 in~\cite{ms23_1}).
\begin{figure}[ht]
\centering
\includegraphics[width=70mm]{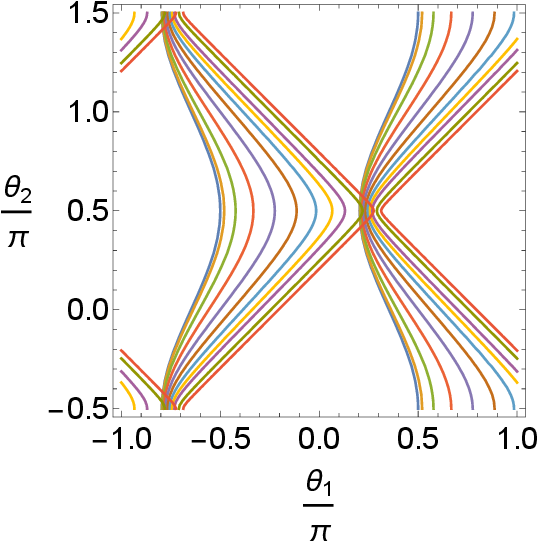}
\caption{\small Curves of constant tachyonic mass for the $p=3$ RR sector. The independent boundary conditions are labeled by a pair of angles $(\theta_1,\theta_2)$,  and the filled regions identify those leading to instabilities.}
\label{fig:tachyon_free}
\end{figure}

\item \textbf{The $p=3$ RR sector}\\  In this case $\mu=\tilde{\mu}=\frac{1}{2}$ and the Schr\"odinger equation has a vanishing potential. The independent boundary conditions are labelled by a pair of angles $(\theta_1,\theta_2)$, and those granting positivity correspond to the white regions in fig.~\ref{fig:tachyon_free}. More details can be found in Section 3.1 in~\cite{ms23_1}).

\item \textbf{The $p=-1$ RR sector}\\ In this case, which only concerns the 0'B theory of~\cite{as95}, $\mu=\tilde{\mu}=\frac{3}{2}$, so that the boundary conditions are fixed at both ends, where normalizable wavefunctions vanish quadratically. Hence, the formal positivity argument relying on the ${\cal A}\,{\cal A}^\dagger$ form of the potential is reliable, and the mass eigenvalues are positive or zero. The equation ${\cal A}^\dagger\,\psi =0$ is solved by
\beq
\psi(z) \ = \ e^{\phi(z) + 4\Omega(z)} \ ,
\eeq
but this wavefunction is not normalizable, due to its behavior at the right end, where is approaches $\left(z_m-z\right)^{-1}$. The same conclusion holds, at the left end, for the other solution obtained by the Wronskian method, so that no normalizable zero mode is present. The spectrum is thus purely massive, so that no instabilities are encountered.
\begin{figure}[ht]
\centering
\begin{tabular}{ccc}
%\mbox{graphic} & \mbox{table} \\
\includegraphics[width=65mm]{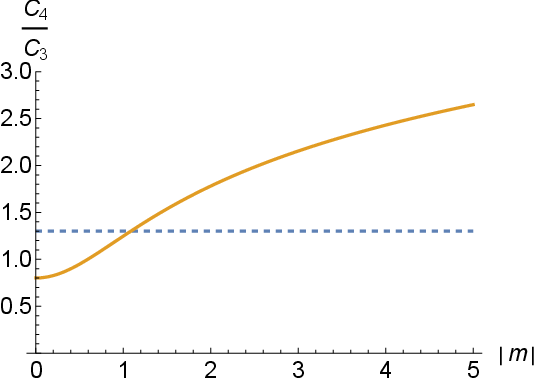} \quad  &
\includegraphics[width=65mm]{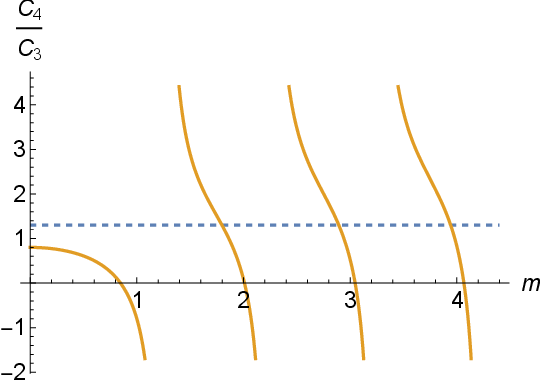}  \\
\end{tabular}
 \caption{\small The hypergeometric eigenvalue equation for the RR six-form for imaginary values of the mass (left panel) and for real values of the mass (right panel). In both cases the red dashed horizontal line corresponds to $\frac{C_4}{C_3}\simeq 1.3$.}
\label{fig:RR6_eigenv}
\end{figure}

\item \textbf{The $p=1$ RR sector} \\ In this case $\left(\mu=\frac{1}{2},\tilde{\mu}={1}\right)$, and the exact zero mode $\psi=e^{\frac{\phi}{2}+2 \Omega}$ is not normalizable, but the independent solution obtained by the Wronskian method is, as discussed in~\cite{mrs24_2}. Our general arguments suffice to conclude that its boundary conditions lead to a stable spectrum including a zero mode, but as we have seen the hypergeometric approximation is not reliable when $\mu=\frac{1}{2}$, so that we are unable to identify the global stability region for this sector.

\item  \textbf{The $p=5$ RR sector} \\ In this case $\left(\mu=\frac{3}{2},\tilde{\mu}=0\right)$, and one has only the freedom of tuning the $\frac{C_4}{C_3}$ ratio. There is a normalizable zero mode, which can be obtained starting from $\psi=e^{-\,\frac{\phi}{2}-2\Omega}$ with the Wronskian method, and reads
\beq
\tilde{\psi}(u) \ = \ u^\frac{4}{9} e^{- \,\frac{u}{3}} \ .
\eeq
Its expansion near the right end is dominated by
\beq
\tilde{\psi}(z) \ \sim \left(1 \ - \ \frac{z}{z_m}\right)^\frac{1}{2} \left[ \log z_m \ + \ \log\left(1 \ - \ \frac{z}{z_m}\right)\right] \ ,
\eeq
which identifies
\beq
\frac{C_4}{C_3} \ = \ \log z_m \ \simeq \ 1.3 \ .
\eeq
This determines the shift $\xi$ in the hypergeometric potential via the eigenvalue equation
\beq
\log z_m \ = \ \frac{1}{2} \left[ \psi\left(\frac{5}{4} + i \xi\right) \ + \ \psi\left(\frac{5}{4} \ - \ i \xi\right) \ - \ 2 \psi(1) \ + \ 2 \log\left(\frac{\pi}{2}\right)\right] \ ,
\eeq
which can be solved graphically and yields
\beq
\xi \simeq \ 1.1 \ ,
\eeq
so that the hypergeometric potential must contain an additional contribution equal to $\frac{1.21 \,\pi^2}{z_m^2}$. The spectrum is tachyon-free for all choices of $\frac{C_4}{C_3}<1.3$ and contains a tachyon in the complementary region.

\item  \textbf{The gauge vector sector} \\  In this case $\left(\mu=\frac{1}{4},\tilde{\mu}=\frac{1}{2}\right)$, so that we cannot rely on the hypergeometric approximation. There a first normalizable zero mode,
$\psi = e^{3 \Omega+ \frac{1}{4}\ \phi}$, and a second normalizable zero mode can be constructed with the Wronskian method. The boundary conditions of the first zero mode are guaranteed to lead to a positive spectrum, and actually there will be a line of zero modes in the $(\theta_1,\theta_2)$ plane: those without a node will correspond to stable boundary conditions.

\end{enumerate}

\subsection{\sc Fermi modes in Internal Intervals} \label{sec:Fermi modes}
We can now discuss the key features of Fermi modes in the presence of an internal interval, a setup that plays a role in the vacuum of~\cite{dm_vacuum} for the Sugimoto model of~\cite{sugimoto} or of the 0'B model of~\cite{as95,as97}. In this case one starts from the Einstein--frame fermionic contributions to the ten--dimensional $N=1$ Lagrangian
\begin{eqnarray}
{\cal L} &=&  \,-\,\frac{e}{2\,k_{10}^2}\ \Big\{ 2\,\bar{\psi}_M\,\Gamma^{MNP}\,D_N\,\psi_P \ - \ 2\,\bar{\lambda}\,\Gamma^M\,D_M\,\lambda \ + \ 4 \bar{\lambda}\,\Gamma^{MN}\,D_M\,\psi_N  \nonumber \\
&-& \frac{1}{2}\, \partial_M\,\phi \ \bar{\lambda} \,\Gamma^{N}\,\Gamma^{M}\,\psi_N  \Big\} \ ,
\label{action_berg2}
\end{eqnarray}
where the gravitino $\psi_M$ and the dilatino $\lambda$ are both subject to the Weyl--Majorana constraints.
This result can be deduced from the corresponding string--frame expression, and for simplicity we have left out the goldstino and the gaugini, which play a role in the setup of~\cite{dmnonlinear1,dmnonlinear2}. We shall actually concentrate on the spin--$\frac{3}{2}$ gravitino modes, which are simpler to discuss, leaving aside the additional spin--$\frac{1}{2}$ modes that arise from the mixing of the two fields, which can be addressed along the lines of~\cite{ms22_1}. We begin by considering a single spin--$\frac{1}{2}$ field, as a first illustration of the role of boundaries.

\subsubsection{\sc Spin--\texorpdfstring{$\frac{1}{2}$}\ \ Modes}

In this section, we would like to elaborate on the behavior of Fermi fields in an interval, in order to highlight analogies and differences with circle compactification, which was the site of Scherk--Schwarz supersymmetry breaking. The main new feature of this case is the need for a non--linear realization of part of the Lorentz symmetry, when the boundary conditions for Fermi field must respect and Majorana and/or Weyl constraint~\cite{ms_20}.

Let us consider a single spin--$\frac{1}{2}$ field $\Psi$ in a background of the form
\beq
ds^2 \ = \ e^{2\,\Omega(z)}\left( dx^2 \ + \  dz^2 \right)
\eeq
that, as we have seen, can capture the vacua of non--supersymmetric strings introduced in~\cite{dm_vacuum}, with the conformal coordinate valued in a finite open interval $\left]0,z_m\right[$ where singularities are present at the ends. Our aim is to address how to define a self--adjoint Dirac operator in the presence of such an internal interval.

To begin with, let us note that the spin connection one--form for this class of metrics has the non--vanishing components
\beq
{\omega}^{\nu z} \ = \ dx^\mu\, \delta_\mu^\nu\, \Omega'(z) \ ,
\eeq
so that the massless Dirac action reads~\footnote{Here we are abiding to the conventional choice of introducing an overall factor of i in front of the action. The difference with the conventions drawn from~\cite{bergshoeff} resides in a different definition of conjugation for Fermi fields: here $(AB)^\dagger = B^\dagger\,A^\dagger$, or alternatively in the introduction of a factor $i$ in the definition of $\bar{\psi}$.}
\beq
{\cal S} \ = \ - \ i\int d^{9}x \ dz \ e^{9\Omega}\ \bar{\Psi} \left[ \gamma^z\left( \partial_z \ + \frac{9}{2}\ \Omega'  \right) \ + \ \gamma^\mu\, \partial_\mu\right] \Psi \ ,
\eeq
where all $\gamma$--matrices have flat indices, and are thus constant. The redefinition
\beq
\Psi \ = \ e^{\,-\,\frac{9}{2}\,\Omega} \, \chi
\eeq
leads to a free Fermi problem, with
\beq
{\cal S} \ = \ - \ i\int d^{9}x \,dz \ \bar{\chi} \big[ \gamma^z\,\partial_z  \ + \ \gamma^\mu\, \partial_\mu\big] \chi \ .
\eeq
The Dirac equation for $\chi$ thus becomes
\beq
\big[ \gamma^z\,\partial_z  \ + \ \gamma^\mu\, \partial_\mu\big] \chi(x,z) \ = \ 0 \  ,
\eeq
provided $\chi$ belongs to the self--adjointness domain of the free Dirac operator $\gamma^0 \,i\, \slash {\!\!\! \partial}$. This condition demands that, for any pair of admissible wavefunctions $\chi$ and $\psi$, the boundary term
\beq
\left. \overline{\psi} \, \gamma^z\,\chi \right|_{z=0}^{z=z_m} \ = \ 0  \label{boundary_term_fermi}
\eeq
vanish.

Linear boundary conditions at $z=(0,z_m)$ granting the vanishing of this term are of the form~\cite{ms_20}
\beq
\psi \ = \ \Lambda\,\psi \ , \qquad \chi \ = \ \Lambda\,\chi \ , \label{Lambdaproj}
\eeq
where without loss of generality one can demand that
\beq
\Lambda \ = \ \Lambda^\dagger \ , \qquad \Lambda^2 \ = \ 1 \ ,
\eeq
provided
\beq
\left\{ \gamma^0\,\gamma^z, \Lambda\right\} \ = \ \left(1- \Lambda\right) X \left(1- \Lambda\right) \ ,
\eeq
where $X$ is an arbitrary Hermitian matrix.

When working with $\Lambda$-projected spinors, this condition is equivalent to
\beq
\left\{ \gamma^0\,\gamma^z, \Lambda\right\} \ = \ 0 \ , \label{Lambda_constraint}
\eeq
and we can now elaborate on how to solve it~\cite{ms_20} while taking into account the Majorana and/or Weyl conditions on the spinors present in the different versions of ten--dimensional (super)gravity of interest for ten--dimensional strings, and in particular for the non--supersymmetric ones of~\cite{as95,as97,agmv1,agmv2,sugimoto}.
The Weyl constraint demands that
\beq
\left[\Lambda, \gamma_{11}\right] \ = \ 0 \ ,
\eeq
while the Majorana constraint
\beq
\chi \ = \ C\, \overline{\chi}^T \ , \qquad \psi \ = \ C\, \overline{\psi}^T \label{majchipsi}
\eeq
demands that $\Lambda\,\chi$ or $\Lambda\,\psi$ obey eq.~\eqref{majchipsi} as $\chi$ and $\psi$.
In the preceding conditions $\gamma_{11}$ is the ten--dimensional chirality matrix and $C$ is the ten--dimensional charge--conjugation matrix, which satisfies the conditions
\beq
C \,\gamma^\mu \, C \ = \ - \ \left(\gamma^\mu\right)^T \ , \qquad C \ = \ - \ C^T \ , \qquad C \ = \ C^\dagger \ , \qquad C^2 \ = \ 1 \ .
\eeq
The compatibility between eq.~\eqref{Lambdaproj} and the Majorana condition demands that 
\beq
\Lambda \,C \, \gamma^{0,T} \ = \ - \ \gamma^0 \,C\,\Lambda^T \ ,  \label{LambdaprojC}
\eeq
so that $\Lambda \,C \, \gamma^0$ must be a symmetric matrix. 
In the absence of a Weyl constraint, $\Lambda=\gamma^z$ is the simplest solution of eq.~\eqref{LambdaprojC}. This choice also applies to a Majorana spinor (the case of interest for type IIA), and clearly respects the residual nine--dimensional Lorentz symmetry.

In type IIB the Weyl constraint requires the introduction of at least another $\gamma$ matrix, so that the simplest choice would be
\beq
\Lambda \ = \ i \sigma_2 \, \gamma^8 \,\gamma^z \ .
\eeq
Here $\sigma_2$ interchanges the two spinors of a pair, and is instrumental to make the whole expression compatible with the Majorana condition. Note that in type IIB there are no solutions preserving the whole leftover Lorentz symmetry. This phenomenon becomes even more evident when a single type of Majorana--Weyl spinor is present, as in the type-I and Sugimoto models, where the simplest solution is~\cite{ms_20}
\beq
\Lambda \ = \ \gamma^6\,\gamma^7\,\gamma^8\,\gamma^z \ , \label{Lambda6789}
\eeq
which is equivalent to a six--dimensional chiral projection when acting on ten--dimensional Weyl spinors.
More generally, any solution obtained from this via nine--dimensional Lorentz transformation would be an equally good choice.
All in all, one is thus confronted with the constraint~\eqref{boundary_term_fermi}, which is invariant under the residual Lorentz symmetry, but whose solutions provide non--linear realizations of it.

It is instructive to see how the projection impinges on a massless ten--dimensional Dirac equation
\beq
\gamma^\mu \, \partial_\mu \, \psi \ + \ \gamma^a \, \partial_a \, \psi \ + \ \gamma^z\,\partial_z \,\psi \ = \ 0 \ ,  \label{Dirac_10}
\eeq
where we split the ten--dimensional label into a portion, $\mu$, such that the corresponding $\gamma$'s commute with $\Lambda$, and others, $a$ and $z$, such that the corresponding $\gamma$'s anticommute with it.

One can define the nine--dimensional mass as
\beq
\left(\partial^\mu\,\partial_\mu \ + \ \partial^a\,\partial_a\right) \psi \ = \ m^2 \, \psi \ , \label{m9}
\eeq
and then, making use of eq.~\eqref{Dirac_10}, one finds that
\beq
\left(\partial_z^2 \ + \ m^2\right) \psi \ = \ 0 \ .
\eeq
Consequently
\beq
\psi \ = \ \psi_1(x^\mu,x^a) \cos mz \ + \ \psi_2(x^\mu,x^a) \ \frac{\sin mz}{m} \ ,\label{psi_dec}
\eeq
where the choice of dividing the second term by $m$ retains as $ m \to 0$ the linear solution, which cannot be dismissed a priori in an interval. Using this decomposition in eq.~\eqref{Dirac_10} leads to the system
\bea
&& \left(\gamma^\mu \, \partial_\mu \ + \ \gamma^a \, \partial_a \right) \psi_1 \ + \ \gamma^z\,\psi_2 \ = \ 0 \ , \nonumber \\
&& \left(\gamma^\mu \, \partial_\mu \ + \ \gamma^a \, \partial_a \right) \psi_2 \ - \ m^2 \,\gamma^z\,\psi_1 \ = \ 0 \ ,
\eea
which is equivalent to
\beq
\psi_2 \ = \ - \gamma^z \left(\gamma^\mu \, \partial_\mu \ + \ \gamma^a \, \partial_a \right) \psi_1 \ ,
\eeq
together with the mass shell condition
\beq
\left(\partial^\mu\,\partial_\mu \ + \ \partial^a\,\partial_a\right) \psi_1 \ = \ m^2 \, \psi_1 \ . \label{m9_1}
\eeq
As a result, the solution can be finally cast in the form
\beq
\psi \ = \ \psi_1\ \cos mz \ - \ \gamma^z \left(\gamma^\mu \, \partial_\mu \ + \ \gamma^a \, \partial_a \right) \psi_1 \ \frac{\sin mz}{m} \ . \label{psizm}
\eeq

Without any loss of generality, we can now demand that
\beq
\psi(0) \ = \ \Lambda\, \psi(0) \ ,
\eeq
to then examine the two options
\beq
\psi\left(z_m\right) \ = \ \pm \ \Lambda\, \psi\left(z_m\right) \ .
\eeq
In both cases, the preceding conditions translate into
\beq
\psi_1(x^\mu,x^a) \ = \ \Lambda\, \psi_1(x^\mu,x^a) \ ,
\eeq
and then
\beq
\Lambda\, \psi(z_m) \ = \ \psi_1\ \cos m z_m \ - \ \gamma^z \left(- \ \gamma^\mu \, \partial_\mu \ + \ \gamma^a \, \partial_a \right) \psi_1 \ \frac{\sin m z_m}{m} \ . \label{lambdapsizm}
\eeq

One must now distinguish two cases:
\begin{enumerate}
    \item[a. ] The choice $\Lambda\, \psi(z_m)=\psi(z_m)$ is the counterpart of a standard circle Kaluza--Klein compactification, and leads to the condition
    \beq
\gamma^\mu \, \partial_\mu \, \psi_1 \  \frac{\sin m z_m}{m} \ = \ 0  \ . \label{constraint_psizm}
    \eeq
    Contrary to the circle case, when the nine--dimensional mass $m$ vanishes this condition implies that 
\beq
\gamma^\mu \, \partial_\mu \, \psi_1  \ = \ 0 \ , \label{diracmu}
\eeq
so that it constrains $\psi_1$ to be a massless fermion from the vantage point of the $x^\mu$ coordinates. 

For general values of $m$, eq.~\eqref{constraint_psizm} has actually two types of solutions. If eq.~\eqref{diracmu} holds, in view of eq.~\eqref{m9_1},
\beq
\partial^a\,\partial_a \ \psi_1 \ = \ m^2 \ \psi_1 \ ,
\eeq
but the only option is then $m=0$, since $\partial^a\,\partial_a$ is a negative operator, so that the nine--dimensional zero mode
\beq
\psi \ = \  \psi_1(x^\mu) \ ,
\eeq
is also bound to have vanishing momenta in the $x^a$ directions. 

This is different from what happens for circle compactification, where a vanishing nine--dimensional mass is compatible with non--vanishing momenta in the $x^a$ directions.
The zero modes in the $x^\mu$ directions are similar,  since in both cases there is no dependence on $x^a$ and $z$. However, even if these zero modes have the same type of coordinate dependence in the two cases, with an internal interval the $\Lambda$ projection halves their number.

The second way of solving~\eqref{constraint_psizm} determines a massive spectrum, with masses that are quantized according to
    \beq
 m \ = \ \frac{k\,\pi}{z_m} \qquad (k=1,2,\ldots) \ ,
    \eeq
and then
\beq
\psi \ = \ \Big[ \cos \left(\frac{k\,\pi\,z}{z_m}\right) \ - \ \frac{z_m}{k\,\pi}\ \sin \left(\frac{k\,\pi\,z}{z_m}\right) \gamma^z \left( \gamma^\mu \, \partial_\mu \ + \ \gamma^a \, \partial_a \right)\Big] \psi_1 \ .
\eeq
Altogether, in nine dimensions there is thus a massive Kaluza--Klein tower of states, together with a massless mode that is independent of the $x^a$. From the $x^\mu$ vantage point, the massive Kaluza--Klein tower is accompanied by the spectrum associated to the $x^a$ directions, which is discrete if the $x^a$ are compact, and continuous otherwise.

The quantitative comparison is with a circle of radius $2 z_m$. This setting has the same spectrum as the interval, but $\psi$ is not projected and moreover all momentum quantum numbers $k$ are independent, so that there is a double degeneracy with respect to the interval for nonzero values of $k$, since with an internal interval modes with $k$ and $-k$ combine, as for bosons. In addition, as we have stressed, $\Lambda$ halves the overall number of Fermi components, and in the Majorana--Weyl case it singles out the $x^a$ coordinates, so that a nine--dimensional zero mode must be also a zero mode for the $x^\mu$.

    \item[b. ] The choice $\Lambda\, \psi(z_m)= -\,\psi(z_m)$ is the counterpart of Scherk--Schwarz circle reduction. In this case, one is led to the condition
    \beq
 \gamma^a \, \partial_a \  \psi_1 \ = \ \gamma^z \ m \ \cot m z_m \ \psi_1 \ ,
    \eeq
    which implies that
    \beq
\partial^a\, \partial_a\,\psi_1 \ = \ - \ m^2\, \cot^2 m z_m \ \psi_1 \ . \label{massa}
    \eeq
    Making use of this result in the mass-shell condition~\eqref{m9} gives
    \beq
    \partial^\mu\, \partial_\mu\,\psi_1 \ = \ \frac{m^2}{\sin^2 m z_m} \ \psi_1 \ , \label{massmu}
    \eeq
    so that, from the vantage point of the $x^\mu$ coordinates, the squared mass is 
    \beq
\frac{m^2}{\sin^2 m z_m} \ .
    \eeq
    $\psi$ can finally be cast in the form
    \beq
\psi \ = \ \psi_1(x^\mu,x^a) \ \frac{\sin m(z_m-z)}{\sin m z_m} \ - \ \gamma^z\ \gamma^\mu\,\partial_\mu\,\psi_1(x^\mu,x^a) \ \frac{\sin m z}{m} \ ,
    \eeq
where the $\Lambda$ projections are manifest. 
\begin{figure}[ht]
\centering
\includegraphics[width=80mm]{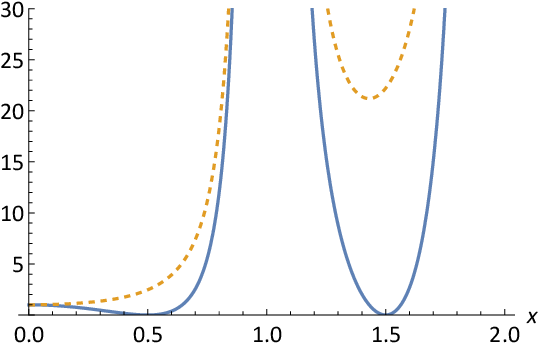}
\caption{\small The first two branches for the functions $(\pi x)^2 \cot^2 \pi x$ (blue, solid) and $\frac{x^2\,\pi^2}{\sin^2 \pi x}$ (orange, dashed), which determine the contributions to the mass from the $x^a$ and $x^\mu$ coordinates.}
\label{fig:Lambdaplusminus}
\end{figure}

Now the possible values of $m$ are determined by eq.~\eqref{massa}: any fixed momenta $p^a$ in the $a$ directions give rise to an unbounded spectrum for $m(p^a p^a, k)$, which depends on an additional discrete label $k$, as can be seen from fig.~\ref{fig:Lambdaplusminus}, which illustrates two of the possible branches of solutions. For example, if the $x^a$ directions correspond to a products of circles of radius $R$, eq.~\eqref{massa} becomes
\beq
\frac{\vec{n} \cdot \vec{n}}{R^2} \ = \ m^2 \ \cot^2 m z_m \ ,
\eeq
and the $m$ spectrum is then genuinely discrete. For large values of $\left|\vec{n}\right|$, the eigenvalues for $m$ approach $\frac{k\,\pi}{z_m}$, with $k=1,2,\ldots$, while for bounded values of $\left|\vec{n}\right|$ and large values of $m$ the eigenvalues approach $\left(k \,+\,\frac{1}{2}\right)\frac{\pi}{z_m}$.
For special choices of $R$ there can be a nine--dimensional zero mode (a solution with $m \to 0$), but it is not a zero mode for the $x^\mu$ directions, since the corresponding mass is $\frac{1}{z_m^2}$, as can be seen from eq.~\eqref{massmu}. Moreover, there is an infinite number of zero modes for the $x^a$ coordinates, which correspond to
\beq
m \ = \ \left(k \ + \ \frac{1}{2}\right)\frac{\pi}{z_m} \qquad (k=0,1,\ldots) \ ,
\eeq
and in these cases
\beq
    \partial^\mu\, \partial_\mu\,\psi_1 \ = \ \left(k \ + \ \frac{1}{2}\right)^2 \frac{\pi^2}{z_m^2} \ \psi_1 \ ,
    \eeq
which is akin to the Scherk--Schwarz spectrum for the circle. 

To reiterate, differently from case a, with this choice of $\Lambda$ projections no zero modes exist in the $x^\mu$ directions, where the effective mass is bounded from below by $\frac{1}{z_m}$.
\end{enumerate}

\subsubsection{\sc Spin--\texorpdfstring{$\frac{3}{2}$} \ \ Modes}

Let us now consider spin--$\frac{3}{2}$ modes, with nonzero components $\psi_\mu$ and $\psi_a$, which we shall collectively denote by $\psi_{\hat{\mu}}$. The conditions
\bea
&& \psi_z \ = \ 0 \ , \nonumber \\
&& \gamma^{\hat{\mu}}\,\psi_{\hat{\mu}} \ \equiv \ \gamma^\mu\,\psi_\mu \ + \ \gamma^a\,\psi_a \ = \ 0 
\eea
grant the removal of spin--$\frac{1}{2}$ modes.
The choice for $\Lambda$ made in eq.~\eqref{Lambda6789} demands that one distinguish two types of indices, as in the previous section. The starting point is provided by the $\mu$, $a$ and $z$ components of the Rarita--Schwinger equation
\bea
 &&\gamma^{\mu\hat{\nu}\hat{\rho}}\,\partial_{\hat{\nu}} \psi_{\hat{\rho}}  \ + \  \gamma_z \left( \partial_z\,+\, \frac{7}{2}\,\Omega_z\right)\psi^{\mu} \ = \  0 \ ,
\nonumber \\
 &&\gamma^{a \hat{\nu}\hat{\rho}}\,\partial_{\hat{\nu}} \psi_{\hat{\rho}}  \ + \  \gamma_z \left( \partial_z\,+\, \frac{7}{2}\,\Omega_z\right)\psi^a \ = \  0 \ ,
\nonumber \\
&&\partial^\mu\,\psi_\mu \ + \ \partial^a\,\psi_a \ = \  0 \ , \label{app:kin32} \nonumber
\eea
which are brought to their flat--space form by the field redefinitions
\beq
\psi_\mu \ = \ e^{\,-\,\frac{7}{2}\,\Omega}\, \chi_\mu \ , \qquad \psi_a \ = \ e^{\,-\,\frac{7}{2}\,\Omega}\, \chi_a  \ .
\eeq
The constraints then make the system equivalent to Dirac equations for $\chi_\mu$ and $\chi_a$ 
\beq
\left(\gamma^{\rho}\,\partial_\rho \ + \ \gamma^b \,\partial_b \ + \ \gamma^z \,\partial_z\right)\chi^{\hat{\mu}} \ \ = \  0 \ ,
\eeq
supplemented by two conditions
\beq
\gamma^\mu\,\chi_\mu \ + \ \gamma^a\,\chi_a \ = \ 0 \ , \qquad
\partial^\rho\,\chi_\rho \ + \ \partial^b\,\chi_b \ = \  0 \ , \label{app:kin322}
\eeq
and by the boundary conditions at the ends.

The Dirac equation demands that
\bea
&&\chi^{\hat{\mu}} \ = \ \chi_1^{\hat{\mu}} \, \cos mz \ - \ \gamma^z \ \gamma^{\hat{\mu}} \, \partial_{\hat{\mu}} \,  \chi_1^{\hat{\mu}} \ \frac{\sin mz}{m} \ , \nonumber \\
&&\Box_9\,\chi_1^{\hat{\mu}} \ = \ m^2 \, \chi_1^{\hat{\mu}} \ , \nonumber \\
&& \gamma^{\hat{\mu}}\,\chi_{1\,\hat{\mu}}  \ = \ 0 \ , \qquad
\partial^{\hat{\rho}}\,\chi_{1\,\hat{\rho}}  \ = \  0 \ .
\label{chizm}
\eea

Without loss of generality, one can demand that
\beq
\chi^{\hat{\mu}}\left(0\right) \ = \ \pm \ \Lambda \, \chi^{\hat{\mu}} \left(0\right)\ ,
\eeq
which translates into
\beq
\chi_1 ^{\hat{\mu}}\ = \ \Lambda\,\chi_1^{\hat{\mu}} \ ,
\eeq
so that the $\gamma$-trace condition in the last of eqs.~\eqref{chizm} implies that
\beq
\gamma^\mu\,\chi_{1\,\mu} \ = \ 0  \ , \qquad \gamma^a\,\chi_{1\,a} \ = \ 0 \ .
\eeq
The decomposition is then different according to whether
\beq
\chi^{\hat{\mu}}\left(z_m\right) \ = \ \pm \ \Lambda \, \chi^{\hat{\mu}} \left(z_m\right)\ ,
\eeq
but follows exactly the lines of the spin--$\frac{1}{2}$ case, so that there are two options.
\begin{enumerate}
\item[a. ] The choice $\Lambda\,\chi^{\hat{\mu}}\left(z_m\right) \ = \ \chi^{\hat{\mu}} \left(z_m\right)$ corresponds to the standard circle compactifications, and
again one finds the condition
    \beq
\gamma^{\rho} \, \partial_{\rho} \, \chi_1^{\hat{\mu}} \  \frac{\sin m z_m}{m} \ = \ 0 \ . \label{constraint_psizm2}
    \eeq
    This admits a zero--mode solution such that
    \beq
\gamma^{\rho}\,\partial_{\rho} \, \chi_1^{\hat{\mu}} \ = \ 0 \ ,
    \eeq
    which does not depend on the $x^a$ coordinates since $\Box_9 \,\chi_1^{\hat{\mu}} \ = \ 0$,
    together with a whole spectrum of massive modes, with masses
    \beq
m_k \ = \ \frac{k\,\pi}{z_m} \quad (k=1,2,\ldots) \ .
    \eeq
\item[b. ] The choice $\Lambda\,\chi_1^{\hat{\mu}}(z_m) = - \,\chi_1^{\hat{\mu}}(z_m)$ corresponds to the Scherk--Schwarz reduction, and one recovers the same types of results obtained for the spin--$\frac{1}{2}$ case, which must be subjected to the additional constraints in the last line of eqs.~\eqref{chizm}.
\end{enumerate}

\section{\sc Supersymmetry Breaking with a Finite \texorpdfstring{$g_s$} \ } \label{sec:directsusybcom}

Different scenarios for supersymmetry breaking in String Theory~\cite{stringtheory} have been explored over the years, but they all entail, in one way or another, strong back reactions on the vacuum. Some of the main options were already discussed in the previous sections. In particular, in Section~\ref{sec:toroidal_ss} we discussed the Scherk--Schwarz mechanism in a one--dimensional internal circle, also with four types of orientifold configurations. All these cases afford exact tree--level descriptions in String Theory, although a back-reaction is induced by the breaking of supersymmetry. Moreover, in Section~\ref{sec:SUSY_breaking_com} we discussed in detail one--dimensional interval compactifications~\cite{dm_vacuum} and we mentioned semi--analytical counterparts of toroidal reductions in the presence of tadpole potentials~\cite{mrs24_1,mrs24_2}. We also presented detailed arguments indicating that the former vacua are perturbatively stable, despite the presence in them of regions with strong coupling and/or large curvatures. 

Although overcoming the large--curvature problem requires higher--derivative corrections, stable vacua with broken supersymmetry and a finite string coupling can be found within the standard two--derivative formulation. From the vantage point of the low--energy effective field theory, one can explore many options, and here we would like to consider one of them, which relies on a class of relatively simple supergravity solutions inspired by~\cite{dm_vacuum}. It presents some peculiar features that, in our opinion, deserve some attention.

There are also non--singular AdS$\times$S solutions~\cite{gm,ms_16,raucci_ads}, where the coupling can be weak everywhere, but these vacua are inevitably unstable~\cite{bms1}. For this reason, we were led to investigate different backgrounds where supersymmetry is broken by warped compactifications supported by self--dual five--form fluxes~\cite{ms21_1,ms21_2,ms22_1,ms23_2} originating from the type--IIB string, so that no tadpole potential is present. The starting point is thus
\bea
ds^2 &=& e^{2 A(r)}\,dx^\mu\,dx^\nu \,+\, e^{2 B(r)}\, dr^2 \ + \ e^{2 C(r)} \, \left(d\,{y}^i\right)^2 \ , \nonumber \\
{{\cal H}_5^{(0)}} &=&
\left(1 \ +  \star\right) B'(r) \ dx^0 \wedge ...\wedge dx^3\wedge dr , \nonumber \\
\phi &=& \phi(r) \ , \label{back_epos_fin}
\eea
where $\star$ denotes the Hodge dual, the $x^\mu$ are the coordinates of a four--dimensional Minkowski space, and the positive values of $r$ parametrize the interior of an internal interval. The five $y^i$ coordinates have a finite range,
\beq{}{}{}{}{}{}{}{}{}{}{}{}{}{}{}{}{}{}{}{}{}{}{}{}{}{}{}{}{}
0 \ \leq \ y^i \ \leq \ 2\,\pi\,R \ ,
\eeq
and parametrize an internal torus, which we take to be the direct product of five circles of radius $R$.

In fact, these backgrounds have some similarities to those of~\cite{dm_vacuum}: they also depend on a single variable and involve an internal interval. However, a special instance of them is characterized by a constant dilaton profile, which we take for simplicity to be
\beq
\phi \ = \ 0 \ ,
\eeq
so that they have the interesting feature of being devoid of strong--coupling regions~\cite{ms21_1,ms22_1}. In the harmonic gauge
\beq
B \ = \ 4 \,A \ + \ 5\,C \label{harmonicgauge}
\eeq
these background profiles have a simple form, and read
    \bea{}{}{}{}{}{}{}{}{}{}{}{}{}
ds^2 &=& \frac{\eta_{\mu\nu}\,dx^\mu\,dx^\nu}{\left[2\left|H\right|\rho\,\sinh\left(\frac{r}{\rho}\right)\right]^\frac{1}{2}} \,+\, \left[2 \left|H\right|\,\rho\,\sinh\left(\frac{r}{\rho}\right)\right]^\frac{1}{2} \left[e^{ \,- \, \frac{\sqrt{10}}{2\rho}\, r} \, dr^2 \ + \ e^{\,- \, \frac{\sqrt{10}}{10\rho}\, r} \, \left(d\,{y}^i\right)^2\right] \ , \nonumber \\
{{\cal H}_5^{(0)}} &=&
H\left\{ \frac{dx^0 \wedge ...\wedge dx^3\wedge dr}{\left[2\left|H\right|\,\rho\,\sinh\left(\frac{r}{\rho}\right)\right]^2} \ + \ dy^1 \wedge ... \wedge dy^5\right\} \ . \label{back_epos_fin2}
\eea
In the $\rho \to \infty$ or $r\to 0$ limit, these solutions take a simpler form, and read
    \bea{}{}{}{}{}{}{}{}{}{}{}{}{}
ds^2 &=& \frac{\eta_{\mu\nu}\,dx^\mu\,dx^\nu}{\left(2\left|H\right| r\right)^\frac{1}{2}} \,+\, \left(2 \left|H\right|r\right)^\frac{1}{2} \left[ dr^2 \ + \  \left(d\,{y}^i\right)^2\right] \ , \nonumber \\
{{\cal H}_5^{(0)}} &=&
H\left\{ \frac{dx^0 \wedge ...\wedge dx^3\wedge dr}{\left(2\left|H\right| r\right)^2} \ + \ dy^1 \wedge ... \wedge dy^5\right\} \ . \label{back_epos_fin3}
\eea

\subsection{\sc Physical Properties of the Background} \label{sec:background2}

These vacua depend on the two constants $H$ and $\rho$, in addition to the toroidal radius $R$~\cite{ms22_1}, but a simpler formulation is possible.
In fact, it is instructive to perform some redefinitions in Eqs.~\eqref{back_epos_fin2}, and for later convenience we thus let
\beq{}{}{}{}{}{}
h \ = \ 2\,H\,\rho \ ,  \label{hH5}
\eeq
while also introducing the new dimensionless variables
\beq{}{}{}{}{}{}
\tilde{r} \,=\, \frac{r}{\rho} \ , \qquad
\tilde{y}^i \,=\, \frac{y^i}{2\pi R}   \ . \label{xry}
\eeq
It is also convenient to introduce the length scale
\beq{}{}{}{}{}{}
\ell \,=\, \rho \ h^\frac{1}{4} \ , \label{ell h}
\eeq
and the five-form flux in the internal torus,
\beq{}{}{}{}{}{}
\Phi \ = \ H \, \left(2\pi R\right)^5  \ = \ \frac{\left(2\pi\,h^\frac{1}{4}\,R\right)^5}{{2}\,\ell}\  \ .
\eeq

In terms of the new variables, the solution becomes~\footnote{We drop ``tilde's'' for brevity, while also warning the reader that we are using the same symbol $r$ for a coordinate that is now dimensionless.}
\bea
ds^2 &=& \frac{\eta_{\mu\nu}\,dx^\mu\,dx^\nu}{\left[h\,\sinh\left(r\right)\right]^\frac{1}{2}} \,+\, \ell^2 \,\left[\sinh\left(r\right)\right]^\frac{1}{2}\, e^{ - \ \frac{\sqrt{10}}{2}\, r} \, dr^2 \nonumber \\ &+& \left({2}\,\Phi\,\ell\right)^\frac{2}{5}\left[\sinh\left({r}\right)\right]^{\,\frac{1}{2}} \  e^{- \ \frac{\sqrt{10}}{10}\, r}   \,d\vec{y}^{\,2}\ , \nonumber \\
\phi &=& 0 \ , \nonumber \\
{{\cal H}_5^{(0)}} &=&
\frac{1}{2\,h}\, \frac{dx^0 \wedge ...\wedge dr}{\left[\sinh \left({r}\right)\right]^2} \ + \ \Phi\, dy^1 \wedge ... \wedge dy^5  \ . \label{4d_PhiEpos}
\eea
With these new coordinates the background no longer depends on $R$, while the ``harmonic'' gauge condition becomes
\beq{}{}{}{}{}{}{}{}{}{}{}
e^B \ = \ \frac{h}{{2}\,\Phi} \ e^{4A+5C} \ .
\eeq
In the $r \to 0$ limit, these profiles become
\bea{}{}{}{}{}{}{}{}{}{}{}{}{}
ds^2 &=& \frac{dx^2}{\sqrt{h\,r}} \ + \ r^\frac{1}{2} \left[\ell^2\,dr^2 \ + \ \left({2}\,\Phi\,\ell\right)^\frac{2}{5}\,d\vec{y}^2\right] \ , \nonumber \\
\phi &=& 0 \ , \nonumber \\
{{\cal H}_5^{(0)}} &=&
\frac{1}{2\,h} \, \frac{dx^0 \wedge ...\wedge dx^3\wedge dr}{r^2} \ + \ \Phi \,dy^1 \wedge ... \wedge dy^5 \ . \label{back_ezero_scale_2}
\eea
Note that, referring to $\Phi$, $\ell$ and $h$, the scale $R$ has completely disappeared, in all cases. Note also that the flux $\Phi$ should be quantized~\cite{witten_flux} according to
\beq
Q_3\,\Phi \ = \ n \ , \label{Dirac}
\eeq
where $N$ is an integer and $Q_3$ is the D3-brane charge~\cite{stringtheory}
\beq{}{}{}{}{}{}
Q_3 \ = \ \sqrt{\pi}\,m_{Pl(10)}^4 \ . \label{q3}
\eeq

As $r \to 0$, the volume of the internal torus shrinks to zero and the scale factor of the spacetime coordinates diverges, while conversely as $r \to \infty$ the volume of the internal torus diverges while the scale factor of the spacetime coordinates shrinks to zero. Both limits are thus delicate within supergravity although, as we shall see in Section~\ref{sec:killingspinors}, this limiting form of the vacuum preserves half of the supersymmetries of ten--dimensional Minkowski space.

Note that the $r=0$ end of the internal interval hosts a true singularity, since there
\beq
R_{MN} \,{R}^{MN} \ \sim \ \frac{1}{\ell^4\,r^5} \ .
\eeq
As a result, sizable $\alpha'$-corrections are expected in String Theory within a region close to it, while the present classical treatment ought to be reliable for
\beq
r \ > \ \left[\frac{\sqrt{\alpha'}}{\ell}\right]^\frac{4}{5} \ . \label{left_bound}
\eeq
Moreover, for finite values of $\rho$ there is again a singularity as $r \to \infty$, as signaled by the divergent behavior of
\beq
R_{MNPQ}\,R^{MNPQ} \ \sim \ \frac{1}{\ell^4}\, e^{\,{r}\left(\frac{ 10\,-\,\sqrt{10}}{\sqrt{10}}\right)} \ . \label{squared_curv}
\eeq
Therefore, one expects small $\alpha'$-corrections to the low--energy effective field theory only for
\beq
r \ < \ \log\left(\frac{\ell}{\sqrt{\alpha'}}\right) \ . \label{cond_zeta}
\eeq
This value should be larger than the bound~\eqref{left_bound} in order for the current treatment to have some intermediate domain of validity, which is guaranteed provided
\beq
\ell \ \gg \ \sqrt{\alpha'} \ .
\eeq

\subsection{\sc Internal Length and Effective Planck Mass}

The length of the $r$-interval,
\beq
L \ = \ \int_0^\infty e^B \, dr \ = \ \ell  \, \int_0^\infty dx \ e^{\,-\,\frac{5\,x}{2\sqrt{10}} } \,\left(\sinh x\right)^\frac{1}{4} \, \simeq \  1.43\, \ell \ , \label{ell1}
\eeq
is finite for finite values of $\rho$ or $\ell$.
The corresponding behavior of the Planck mass is determined by the Einstein term in the background, and thus by the combination $\sqrt{-g} \ g^{\mu\nu}$ integrated over the internal dimensions:
\beq
m_{Pl(4)}^2 \,=\, m_{Pl(10)}^8\,\int\, dr \, d^{\,5} y \, \sqrt{-g} \ e^{-2A}\,=\, \frac{2\,m_{Pl(10)}^8\,\Phi\, \ell}{h} \, \int\, dr \ e^{2(B-A)} \ .
\eeq
The $r$ integral is again finite for finite values of $\rho$ or $\ell$, and in our solutions
\beq{}{}{}{}{}{}{}
m_{Pl(10)}^8 \ = \ \frac{1}{\left(\alpha'\right)^4}\ , \label{mpl84}
\eeq
with $\alpha'$ the Regge slope, since the string coupling $g_s=1$. Consequently, the effective four--dimensional Planck mass,
\bea
m_{Pl(4)}^2 &=&  \frac{4 \,m_{Pl(10)}^8\,\ell^2\,\Phi}{3\,\sqrt{h}} \ , \label{planck4}
\eea
is also finite.

Since our analysis rests on the effective field theory, the results can be reliable in String Theory only if the Kaluza--Klein excitations in the $r$-interval and in the internal torus are much lighter than the string modes. These conditions result in the inequalities
\beq{}{}{}{}{}
\frac{\ell}{\sqrt{\alpha'}}\ \gg \ 1  \ , \qquad \frac{\left({2}\,\Phi\,\ell\right)^\frac{1}{5}}{\sqrt{\alpha'}} \ \gg \ 1 \ , \label{bounds}
\eeq
which also grant that one can ignore winding modes on the internal torus. Once the first inequality holds, the second does not impose stringent conditions, in general, on the flux $\Phi$, and thus on the quantum number $n$.

\subsection{\sc A Probe Brane in the \texorpdfstring{$r$}--Interval}\label{app: probe brane}

The effective Lagrangian for a probe $D3$--brane spanning the four--dimensional Minkowski space, with fixed internal coordinates and an $r$ coordinate that evolves in time, is determined by the induced metric and the coupling to the gauge field $b$ corresponding to the ${\cal H}_5^{(0)}$ field strength
\beq
{\cal H}_5^{(0)} \ = \ \sqrt{\frac{2}{\chi}}\ dx^0 \wedge \ldots \wedge dx^3 \wedge b'(r) \, dr  + \star \ ,
\eeq
where $\star$ denotes the Hodge dual and $b$ is only a function of $r$.

In general, if one starts from the kinetic term
\beq
S\ = \ \frac{1}{2\kappa_{10}^2}\int \frac{\chi}{2} \, {\cal H}_5 \,\wedge \,\star\, {\cal H}_5  \ ,
\eeq
where self--duality is to be imposed at the end and $\chi$ is a real parameter that reflects the choice of normalization, the  background five-form field strength of our self--dual solution becomes
\beq
{{\cal H}_5^{(0)}} \ = \  \sqrt{\frac{2}{\chi}}\
H\left\{ \frac{dx^0 \wedge ...\wedge dx^3\wedge dr}{\left[2\left|H\right|\,\rho\,\sinh\left(\frac{r}{\rho}\right)\right]^2} \ + \ dy^1 \wedge ... \wedge dy^5\right\}\ .
\eeq
One can define the (identical) electric and magnetic charges according to
\beq
d \star {\cal H}_5 \ = \ d \ {\cal H}_5 \ = \  2\kappa_{10}^2  \ Q_3 \  \delta(\vec{\rho}) \ , \label{H5source}
\eeq
where $\delta(\vec{\rho})$ is a six-form localized at a point in the six--dimensional space comprising the internal torus and the interval, so that for any four-form $\Omega$
\beq
\int_{{\cal M}_{10}} \delta(\vec{\rho}) \wedge \Omega \ = \ \int_{{\cal M}_{4}} \Omega \ .
\eeq

The source term in eq.~\eqref{H5source} corresponds to a probe D3 brane coupling~\footnote{Note that the probe brane coupling used in~\cite{ms22_1} differs from eq.~\eqref{d3coupling} by a factor of two, so that tension and charge should rather appear in the combination ${\cal T}_3 - Q_3$. There is actually an additional factor of two, as we have stressed, since in this selfdual case an electric coupling brings along an identical magnetic one. As a result, the complete interaction potential is ${T}_3\,{T}_3'\,-\,2\,\chi\,Q_3\,Q_3'$,  and so the no--force condition is exactly as demanded by eq.~\eqref{BPSchi}.}
\beq
\delta S\ = \ 2\,\chi \,Q_3 \int B_4 \ , \label{d3coupling}
\eeq
where $B_4$ denotes the four-form gauge field and the overall factor of two reflects the simultaneous presence of electric and magnetic interactions for the dyonic probe brane.
The choice $\chi=2$ corresponds to the convention of~\cite{ms21_1}, while $\chi=1$ is the standard convention for the nonself--dual case that we use in this paper, and finally $\chi=\frac{1}{2}$ corresponds to the convention of~\cite{bergshoeff}.
Note that the BPS condition is also convention dependent and becomes in general
\beq
\left|{T}_3 \right| \ = \ \sqrt{2 \, \chi} \ \left| Q_3 \right| \ , \label{BPSchi}
\eeq

and we can recover this result as follows. For a background of the form~\eqref{back_epos_fin}, in the harmonic gauge and in the Einstein frame, the brane action takes the form
\beq
\frac{{\cal S}}{V_3} \ = \ - \ T_3 \int \ dt \, e^{4A(r(t))} \, \sqrt{1 \ - \ e^{2(B-A)(r(t))}\,\dot{r}(t)^2} \ + \ 2\,\sqrt{2\,\chi}\,\,Q_3\, \int b[r(t)] \ dt \ ,
\eeq
where $T_3$, $q_3$ and $V_3$ denote the tension, charge, and volume of the brane. For the solutions with $E>0$ in eqs.~\eqref{back_epos_fin2}
\beq
b'(r) \ =\ \frac{1}{4\,H}\,\frac{1}{\left[\rho\,\sinh\left(\frac{r}{\rho}\right)\right]^2}\ , \label{eqbprime}
\eeq
so that
\beq
b(r) \ = \ - \ \frac{1}{{4\,\rho\,H}} \left[\coth\left(\frac{r}{\rho}\right) \ - \ 1\right] \ .
\eeq
The corresponding results for the solutions with $E=0$ can be obtained from these in the limit $\rho \to \infty$.

The energy conservation condition for the probe is then
\beq
\frac{T_3\ e^{4A(r(t))}}{\sqrt{1 \ - \ e^{2(3A+5C)(r(t))}\,\dot{r}(t)^2}} \ - \ 2\,\sqrt{2\,\chi}\,Q_3\, b \ = \ E \ . \label{en_cons}
\eeq
Close to $r=0$ the limiting behavior of the background, as we have seen, is universal, and in the non--relativistic limit the preceding equation becomes
\beq{}{}{}{}{}{}
\frac{T_3}{2}\, \dot{r}^2 \ + \ \frac{1}{2\,|H|\,r}\left[T_3\ + \ \sqrt{2\,\chi}\,{Q_3}\, \sign(H)\right] \ = E \ , \label{en_cons_0}
\eeq
from which one can identify the potential
\beq{}{}{}{}{}{}
V \ \sim \  \frac{1}{r}\left[T_3 \ + \ \sqrt{2\,\chi}\,{Q_3}\,\sign\left(H\right)\right] \ ,
\eeq
up to a positive overall factor.
This potential describes a gravitational repulsion sized by the $T_3$ term and an ``electric'' interaction that is repulsive for $Q_3 \,H >0$ and attractive for $Q_3\,H <0$. As a result, one can see that the origin behaves as an orientifold.

For finite values of $\rho$, near the right end of the finite interval the energy conservation condition becomes
\beq{}{}{}{}{}{}
\frac{T_3}{2}\,e^{\,-\,\frac{5\, r}{\rho\,\sqrt{10}}} \, \dot{r}^2\,+\,\frac{T_3}{\rho\,|H|} \,e^{\,-\,\frac{r}{\rho}} \,+\,\frac{Q_3\,\sqrt{2\chi}}{\rho\,H}\,e^{\,-\,\frac{2\,r}{\rho}} \ \simeq \ E \ .
\eeq
In order to recover a non--relativistic kinetic term as in eq.~\eqref{en_cons_0},
one can perform the change of variable
\beq
\alpha\,\rho\left(1 \ - \ e^{\,-\,\frac{r}{\alpha\,\rho}}\right) \ = \ u \ ,
\eeq
with
\beq{}{}{}{}{}
\alpha \ = \ \frac{2\sqrt{10}}{5} \ \simeq \ {1.26}  \ ,
\eeq
which implies
\beq{}{}{}{}{}{}{}
e^{\,-\,\frac{r}{\rho}} \ = \ \left(1 \ - \ \frac{u}{\alpha\,\rho}\right)^\alpha \ ,
\eeq
and leads to
\beq{}{}{}{}{}{}
\frac{T_3}{2}\, \dot{u}^2 \, + \, \frac{T_3}{\rho\,|H|} \,\left(1 \ - \ \frac{u}{\alpha\,\rho}\right)^\alpha \,+\,\frac{Q_3\,\sqrt{2\chi}}{\rho\,H}\left(1 \ - \ \frac{u}{\alpha\,\rho}\right)^{2\alpha} \ \simeq \  E \ . \label{en_cons_inf}
\eeq
Thus, one can see that the gravitational force attracts the brane toward the right end, owing to the second term above, while the electric force attracts it there for $Q_3\,H>0$ and repels it for $Q_3\,H<0$, owing to the third term. However, both forces tend to zero as $u$ approaches $\alpha\,\rho$, and are not proportional. One may well wonder about the fate of the electric tensor flux, which seems to wane across the finite interval. In fact, there is no contradiction with the conservation of electric flux, since the solution is precisely the counterpart of a uniform electric field in our metric background and satisfies
\beq
b' \,e^{5C-4A-B} \ = \ \Phi \ ,
\eeq
as can be deduced by taking the dual of the constant internal components along the torus.

One can gain some qualitative insights on the overall motion of the brane, noting that the energy $E$ is bounded from below by the static potential
\beq
V(r) \, = \, T_3e^{4A}\,-\,2Q_3\,\sqrt{2\chi}\,b\, =\,  \frac{1}{2\,|H|\,\rho} \left[ \frac{T_3}{\sinh\left(\frac{r}{\rho}\right)} \, + \, {Q_3\,\sqrt{2\chi}\,\sign(H)} \left(\coth\left(\frac{r}{\rho}\right) \,-\,1\right)\right] \, , \label{pot_rho}
\eeq
and the brane  has turning points where $E=V(r)$. Note that the static potential $V$ contains two contributions, which are singular at $r=0$ and tend to zero as $r\to \infty$. As we have seen, the first contribution, proportional to $T_3$, looks like a gravitational interaction but \emph{repels} the brane from the origin, while the second, proportional to $Q_3$, \emph{attracts} it to the origin if $Q_3\,\sign(H) <0$ and \emph{repels} it if $Q_3\,\sign(H)>0$. Hence, the origin behaves as an orientifold with negative tension and positive or negative charge, depending on the sign of $H$. As we shall see in the next section, half of the original supersymmetry is recovered as $r\to 0$, which also points to the BPS nature of the extended object present there.

\subsection{\sc Supersymmetric Vacua} \label{sec:killingspinors}

We can now prove that the background with constant dilaton profile of eqs.~\eqref{back_epos_fin3}, which is also the $\rho \to \infty$ limit of the background of eqs.~\eqref{back_epos_fin2} and captures its limiting behavior as $r \to 0$, preserves half of the original 32 supercharges of type IIB. In the following, we shall work with $\chi=2$.
We shall see that this is the only option within the class of metrics in eq.~\eqref{back_epos_fin}, together with flat space, where some supersymmetry is present, looking for Killing spinors in IIB backgrounds, within the class of metrics~\eqref{back_epos_fin},
with a generic $r$--dependent dilaton profile and the self--dual tensor field strength
\beq
{\cal H}_5 \ = \ H \left\{ e^{4A+B-5C} \, dx^0 \wedge ...\wedge dx^3 \wedge dr\ + \ dy^1 \wedge ... \wedge dy^5 \right\} \ .
\eeq

The relevant supersymmetry transformations of the ten--dimensional IIB theory in the presence of non--trivial dilaton and five--form backgrounds can be cast in the convenient form~\cite{ms22_1}
\beq{}{}{}{}{}{}{}{}{}{}{}{}{}{}{}{}{}{}
\delta\,{\psi}_M \ = \  D_M\,\epsilon \ + \ \frac{1}{8}\, {\cal H}\!\!\!\!/ \ \Gamma_M\, \, i\,\sigma_2\,\epsilon \ , \qquad
\delta\,\lambda \ = \ \Gamma^M\,\epsilon\,\partial_M\,\phi \ , \label{susy_final}
\eeq
where $\Gamma_M = {e_M}^a \, \gamma_a$ is a curved ten--dimensional $\gamma$-matrix, $\gamma$ is a flat one, $e$ denotes the vielbein and $\epsilon$ is a doublet of ten--dimensional Majorana--Weyl spinors.
The supersymmetry invariance of $\lambda$ demands a constant dilaton profile, and there is thus no essential distinction between Einstein and string frames in supersymmetric vacua of this type.
The remaining Killing--spinor equations reduce to
\bea
\delta\,\psi_r &=& \partial_r \, \epsilon \, + \, \frac{H}{4} \ e^{B-5C}\ \gamma^{0\ldots 3}  \,i\,\sigma_2\, \epsilon \ = \ 0 \ ,
\nonumber \\
\delta\,\psi_\mu &=& \partial_\mu \, \epsilon \,+\, \frac{1}{2}\ \gamma_\mu \gamma_r\ e^{A-B}\, A'\,\epsilon \, + \, \frac{H}{4} \ e^{A\,-\,5\,C}\  \gamma^{0\ldots 3} \gamma_r\,\gamma_\mu \,i\,\sigma_2\,\epsilon  \ = \ 0  \ ,
\nonumber \\
\delta\,\psi_i &=& \partial_i \, \epsilon \,+\, \frac{1}{2}\ \gamma_i \gamma_r\ e^{C-B}\, C'\,\epsilon \, + \, \frac{H}{4} \ e^{\,-\,4\,C}\  \gamma^{0\ldots 3} \gamma_r\,\gamma_i \,i\,\sigma_2\, \epsilon  \ = \ 0  \ ,
\label{killing_spinors3}
\eea
after taking into account the self--dual nature of the tensor field strength and the spinor chirality projections.
One can now decompose $\epsilon$ into eigenstates $\epsilon_\pm$ of the Hermitian matrix
\beq
\Lambda=\gamma^{0\ldots 3}\,i\,\sigma_2 \label{Lambda}
\eeq
corresponding to its eigenvalues $\pm 1$, and it is also convenient to define
\beq
J'(r) \ = \ \frac{H}{4} \ e^{B-5C} \ , \label{eqJ}
\eeq
so that eqs.~\eqref{killing_spinors3} become
\bea
&&\partial_r\,\epsilon_\pm \ \pm \ J'(r) \, \epsilon_\pm \ = \ 0 \ , \nonumber \\
&&\partial_\mu\, \epsilon_\pm  \ + \ \frac{1}{2} \,\gamma_\mu \gamma_r\ e^{A-B}\left( A' \ \mp \ 2\,J'\right) \epsilon_\mp \ = \ 0 \ , \nonumber \\
&&\partial_i\, \epsilon_\pm  \ + \ \frac{1}{2} \,\gamma_i \gamma_r\ e^{C-B}\left( C' \ \mp \ 2\,J'\right) \epsilon_\pm \ = \ 0 \ . \label{killing_J}
\eea
The first of these equations is solved by
\beq
\epsilon_\pm \ = \ e^{\,\mp\,J(r)}\, \epsilon_{0\,\pm}(x,y) \ ,
\eeq
where $\epsilon_{0\,\pm}$ are arbitrary functions of the spacetime coordinates $x$ and the toroidal coordinates $y$, but are independent of $r$.
The remaining equations now reduce to
\bea
&&\partial_\mu\, \epsilon_{0\,\pm}(x,y)  \ + \ \frac{1}{2} \,\gamma_\mu \gamma_r\ e^{A-B\pm 2J}\left( A' \ \mp \ 2\,J'\right) \epsilon_{0\,\mp}(x,y) \ = \ 0 \ , \nonumber \\
&&\partial_i\, \epsilon_{0,\pm}(x,y)  \ + \ \frac{1}{2} \,\gamma_i \gamma_r\ e^{C-B}\left( C' \ \mp \ 2\,J'\right) \epsilon_{0\,\pm}(x,y) \ = \ 0 \ . \label{killeqs}
\eea
For consistency, the $x$--derivative of the first and the $y$--derivative of the second of these equations imply the conditions
\beq
\left( A'\right)^2  \ - \ 4\,\left(J'\right)^2 \ = \ 0 \ , \qquad \left( C'  \ \mp \ 2\,J'\right)^2\epsilon_{0\,\pm} \ = \ 0 \ ,
\eeq
which are solved if
\beq
A'\ = \ 2\,\varepsilon_A\, J' \ , \qquad  C'\ = \ 2\,\varepsilon_C\, J' \ , \qquad \epsilon_{0\,-\,\epsilon_C}(x,y)\  = \ 0 \ ,
\eeq
where $\varepsilon_A$ and $\varepsilon_C$ are signs, so that one of the two $\epsilon_0$'s must vanish. Moreover, the very form of eqs.~\eqref{killing_J} constrains the two signs $\varepsilon_A$ and $\varepsilon_C$ to be opposite, so that the solutions must finally satisfy
\beq
A'\ = \ 2\,\sigma\, J' \ , \qquad  C'\ = \ -\, 2\,\sigma\, J' \ , \qquad \epsilon_{0\,\sigma} = 0 \ , \label{eqsAC}
\eeq
where $\sigma=\pm1$. Eqs.~\eqref{killeqs} then imply that $\epsilon_0$ is a constant spinor.

Combining these results with the definition~\eqref{eqJ} now leads to the differential equation
\beq
A'\ = \ \frac{\sigma\,H}{2} \ e^{B-5C} \ ,  \label{aprimesusy}
\eeq
whose solution in the Harmonic gauge~\eqref{harmonicgauge} reads
\beq
e^{-4\,A} \ = \ - \ 2\,\sigma\,H\,r \ ,
\eeq
up to a shift of $r$. One can work conveniently in the region $r>0$ taking
\beq
\sigma \ = \ - \ \sign(H) \ ,
\eeq
and the solution of eqs.~\eqref{eqsAC} finally reads
\beq
e^{2\,A} \ = \ e^{\,-\,2\left(C-c_s\right)} \ = \ \left[\frac{1}{2\,|H|\,r}\right]^\frac{1}{2}   \ ,
\eeq
where $c_s$ is a constant that can be scaled out of the following expressions.

The end results for the metric and the form field strength accurately recover eqs.~\eqref{back_epos_fin3}. Moreover, as we have stressed, these results also capture the limiting behavior of the solutions in eqs.~\eqref{back_epos_fin2} as $\rho \to \infty$. In this limit, one is thus approaching a supersymmetric background, since the preceding analysis indicates the existence of the Killing spinor
\beq
\epsilon \ = \  \frac{1}{\left(2\,|H|\,r\right)^\frac{1}{8}}\, \epsilon_{0} \ . \label{killingeps}
\eeq
The limiting form of the background thus preserves 16 of the original 32 supersymmetries of ten--dimensional flat space, since $\epsilon_0$ is a constant spinor subject to the condition
\beq
\Lambda\,\epsilon_0 \ = \ \gamma^{0\ldots 3}\,i\,\sigma_2\, \epsilon_0 \ = \ \sign\left(H\right) \epsilon_0 \ , \label{Lambdasusy}
\eeq
which halves the number of its independent components.
The fact that the supersymmetric case is recovered in the $\rho \to \infty$ limit is consistent with a scale of supersymmetry breaking that takes the form
\beq{}{}{}{}{}{}{}{}{}{}{}{}{}{}{}{}{}{}{}{}{}
\mu_S \ \sim \ \frac{1}{\rho^\frac{3}{2} \sqrt{H}}
\eeq
when expressed in terms of $\rho$.

\subsection{\sc Comments on Low--Lying Modes and the Issue of Stability}

We can now briefly address the modes present in the backgrounds of eqs.~\eqref{back_epos_fin2}, or equivalently in eqs.~\eqref{4d_PhiEpos}, which are characterized by finite values for the length scale $\ell$ of the interval and by a constant dilaton profile. From now on, for definiteness, we shall assume that $H>0$.

The fermionic zero modes were determined in~\cite{ms22_1}: for finite values of $\rho$ or $\ell$ they are four Majorana gravitini and 20 Majorana spinors, the massless fermions of $N=4$ supergravity coupled to \emph{five} $N=4$ vector multiplets, despite the breaking of supersymmetry~\footnote{More precisely, as explained in~\cite{ms22_1}, this is the case if Fermi fields are subject to identical ``$\Lambda$ projections'' at the two ends of the $r$ interval. Self--adjoint extensions granting these results we recently discussed in~\cite{mourad_25}. Here we focus on this interesting option, but opposite ``$\Lambda$ projections'' would eliminate the massless Fermi modes.}.

In order to analyze the bosonic modes, in~\cite{ms23_2} we relied on the analogy with the Schr\"odinger problem, as in Section~\ref{sec:selfadjoint}. Even in this context, in fact, the modes emerge from Schr\"odinger--like equations whose potentials have double poles at the two ends in the conformal variable
\beq
z(r) \ = \ z_0 \int_0^r d\xi \ \sinh\,\xi^\frac{1}{2} \ e^{-\,\frac{\sqrt{10}\, \xi}{4}} \ . \label{zr_i}
\eeq
Here
\beq{}{}{}{}{}{}{}{}{}{}{}{}{}{}{}{}{}{}{}{}
z_0  \ = \ \left(2 H \rho^3\right)^\frac{1}{2} \ , \label{z0_i}
\eeq
and the upper bound for $z$ is
\beq
z_m \ \simeq \ 2.24 \ z_0 \ .
\eeq

The field equations lead, in general, to operators that are not manifestly Hermitian, so that the replacement of the independent variable $r$ with the ``conformal'' variable of eq.~\eqref{zr_i}, whose range $0 \leq z \leq z_m$ is finite and proportional to $z_0$, together with redefinitions of the different fields, were instrumental to cast them into standard forms. Remarkably, as in the nine--dimensional vacua discussed in Section~\ref{sec:selfadjoint}, in all cases the resulting potentials develop double poles at the two ends, where they behave as
\beq{}{}{}{}{}
V \ \sim \ \frac{\mu^2\,-\,\frac{1}{4}}{z^2} \ ,
\qquad\quad
V \ \sim \ \frac{\tilde{\mu}^2\,-\,\frac{1}{4}}{\left(z-z_m\right)^2} \ , \label{lim_dil_axion_intro}
\eeq
a behavior that is akin to that found for the SO(16)$\times$SO(16) heterotic model, since there are no logarithmic corrections.
The constants $\mu$ and $\tilde{\mu}$ depend on the mode sector, while the scale dependence is encoded in $z_m$. Moreover, $\tilde{\mu}$ is zero for $h_{\mu\nu}$, $h_{ij}$, dilaton, and axion perturbations, which share the same Schr\"odinger operator, while it is a real number between zero and about 2.3 in all other sectors. On the other hand, the parameter $\mu$ associated to the end at $z=0$ is a rational number, which is either $\frac{1}{3}$ or $\frac{2}{3}$ in all cases. The squared masses are eigenvalues of Hermitian operators and, as we saw in~\cite{ms22_1} for Fermi fields, these steps also determine the normalization conditions, a necessary ingredient to identify the actual physical modes. In most cases, these normalizations can be simply recovered from the four--dimensional kinetic terms determined by ten--dimensional action, but the self--dual tensor field does introduce some complications. Despite its reduced manifest symmetry, the non--standard Henneaux--Teitelboim action of~\cite{henneaux}, when properly adapted, suffices to grant covariant descriptions in the backgrounds of eq.~\eqref{back_epos_fin2}.

The completeness of the modes thus identified is essential for making statements on perturbative stability. It is granted if the Schr\"odinger--like operators are not only Hermitian but also self--adjoint, and this property demands judicious choices of boundary conditions. These are determined by the asymptotics of the wavefunctions at the ends of the interval~\cite{ms23_1}, which reflects, in its turn, the singular behavior~\eqref{lim_dil_axion_intro} of the potentials. Thus, additional sets of parameters emerge, related to the choices of boundary conditions, which impinge on the positivity of Hermitian Schr\"odinger--like operators, and the stability of the resulting mass spectra generally places some constraints on them~\cite{ms23_1}.

Massless modes with special choices of boundary conditions are exactly calculable in most cases, while in a few instances the allowed squared masses emerge as eigenvalues of operators that are manifestly positive, again with suitable boundary conditions. Two sectors with toroidal momentum $\mathbf{k} \neq 0$, the non--singlet vector modes and the non--singlet scalar modes, do not allow exact statements, and approximation methods were necessary to address their stability. However, the variational principle of non--relativistic quantum mechanics can be adapted to the present setting and allows reliable numerical estimates of the lowest eigenvalues and of their dependence on $\mathbf{k}$ and on the boundary conditions. In this fashion, one can identify background--dependent constraints on the boundary conditions that grant stability, which are discussed in detail in the last Appendix of~\cite{ms23_2}.

With these ingredients, one can classify the stable boundary conditions for the different sectors of the spectrum, and it is also possible to tackle, in rather general terms, the possible instabilities of Kaluza--Klein excitations. These represent an insurmountable problem~\cite{bms} for the non--supersymmetric AdS vacua of the ten--dimensional strings of~\cite{agmv1,agmv2,as95,as97,sugimoto}, since the length scales of the internal sphere and of the AdS spacetimes are correlated by the Einstein equations. In the present setting, which includes an internal torus, or more generally with a Ricci--flat internal manifold, there are boundary conditions compatible with a stable spectrum, although the available choices can depend on the background when mixings involving Kaluza--Klein modes occur. For the Kaluza--Klein excitations, the boundary conditions that yield massless modes with $\mathbf{k}=0$ can also give rise to a finite number of tachyons~\cite{ms23_2}. However, this pathology can be avoided if the parameter $\rho$ in eqs.~\eqref{back_epos_fin2} lies one or two orders of magnitude above the radius $R$ of the internal torus. Equivalently, this condition sets on the scale of supersymmetry breaking
\beq{}{}{}{}{}{}{}{}{}{}{}{}{}{}{}{}{}{}{}{}{}{}{}{}{}{}{}{}{}{}
\mu_S \ = \ \frac{1}{\ell\,h^\frac{1}{4}} \ ,
\eeq
 which we identified in~\cite{ms22_1}, the upper bound
\beq{}{}{}{}{}{}{}{}{}{}{}{}{}{}{}{}{}{}{}{}{}{}{}{}{}{}{}{}{}{}
\mu_S \left(\Phi\,h\right)^\frac{1}{4} \ < \ {\cal O}\left(10^{-2}\right) \ . \label{stability_window}
\eeq
Alternatively, one can select boundary conditions granting the absence of tachyons for all values of $\mathbf{k}$, but these typically eliminate the massless modes with $\mathbf{k}=0$.

\begin{table}[ht]
\centering
\begin{tabular}{||c c c ||}
 \hline
$4D\,\mathrm{hel.} \times SO(5)$ & 4D $m=0$ Content & $10D$ origin \\ [0.5ex]
 \hline\hline
 $\left(0,1\right)$ & {1 dilaton} & $\phi$   \\
  $\left(0,1\right)$ & {1 axion} &  $a$  \\
  $\left(\pm 1,0\right)$ & {1 real vector,  1 real scalar}   & $B_{\mu \nu}^{1,2}$,$B_{\mu r}^{1,2}$   \\
    $2\left(\pm 1,5\right)$ & 10 real vectors  & $B_{\mu i}^{1,2}$, $B_{r i}^{1,2}$   \\
    $2\left(0,15\right)$ & 30 real scalars    & $B^{1,2}_{ij}$   \\
$\left(\pm 1,10\right)$ & 10 real vectors    & $B_{\mu\nu i j}$   \\
$\left(\pm\,{2},1\right)$ & { 1  graviton}   & $h_{\mu\nu}$  \\
$\left(0,14\right)$ & 14 real scalars    & $h_{ij}$   \\
$\left(\pm 1,5\right)$ & 5 real vectors    & $h_{\mu i},B_{\mu\nu\rho i}$   \\
$\left(0,5\right)$ & 5 real scalars   & $h_{r i}, \ B_{\mu\nu\rho i}, \ B_{ijkl}$   \\
$\left(0,1\right)$ & 1 real scalar & $b,b^i, b_\mu{}^i,h_{\mu\n}, h_{r r}$
\\[1ex]
 \hline
\end{tabular}
\caption{The maximum numbers of four--dimensional real massless bosonic modes that can arise from the bulk, for generic values of $R$, within the stability window of eq.~\eqref{stability_window}. There are at most 26 vectors and 53 scalars, for a total of 107 massless bosonic degrees of freedom after including the graviton.}
\label{table:tab_B}
\end{table}

Although the preceding results are very encouraging, the analysis presented in~\cite{ms23_2} is incomplete, since the preceding setup does not apply to the Kaluza--Klein excitations of singlet scalars. These modes appear resilient to the approach, since they lead to a three--component Schr\"odinger system where the potential cannot be put in a symmetric form with the techniques used in the other cases. A proper identification of the norm of these perturbations, and of its correspondence with the convenient Henneaux--Teitelboim action~\cite{henneaux}, was thus left for future work, together with a final statement on the stability region allowed by this sector.

The first two columns in Table~\ref{table:tab_B} collect the maximum numbers of massless modes found explicitly in the previous sections. In four dimensions, these correspond to a graviton, 26 real vectors and 53 real scalars, which are a large fraction of the modes that would emerge, from the type--IIB theory, after a toroidal compactification to four dimensions. These numbers are purely indicative, since we are focusing on the quadratic terms, and interactions and/or quantum corrections could lift in mass many of these modes. In addition, their number could be reduced if special choices of boundary conditions were dictated by symmetry requirements of the interacting theory. For example, some of these modes lead to the flow~\cite{ms_20}, across the boundary, of charges that would be conserved in its absence. This was the case for the ten vector modes arising from the four-form gauge field, and for the five vector modes from $h_{\mu i}$ arising from the metric field, and self--adjoint boundary conditions eliminating the flow can make all these modes massive. Table~\ref{table:tab_C} collects the additional massless modes that could be present on the boundary.

\begin{table}[ht]
\centering
\begin{tabular}{||c c c ||}
 \hline
$4D\,\mathrm{hel.} \times SO(5)$ & 4D $m=0$ Content & corresponding gauge parameters \\ [0.5ex]
 \hline\hline
  $\left(\pm 1,0\right)$ & {2 real vectors}   & $\Lambda_{\mu}^{1,2}$ \\
    $\left(0,5\right)$ & {10 real scalars}   & $\Lambda_{i}^{1,2}$ \\
$\left(0,5\right)$ & {5 real scalars}    & $\Lambda_{\mu\nu i}$   \\
$\left(\pm 1,10\right)$ & {10 real vectors}    & $\Lambda_{\mu i j}$   \\
$\left(0,10\right)$ & {10 real scalars}    & $\Lambda_{i j k}$   \\
$\left(\pm\,{1},0\right)$ & { 1 real vector}   & $\xi_{\mu}$  \\
$\left(\pm\,{0},5\right)$ & { 5 real scalars}   & $\xi_{i}$
\\[1ex]
 \hline
\end{tabular}
\caption{The maximum numbers of four--dimensional real massless bosonic modes that can arise from the boundary of the internal interval. There are in principle 56 degrees of freedom of this type, if one concentrates them on one of the two boundaries. The resulting vector equations are gauge invariant, in view of the discussion presented in~\cite{jphysA}.}
\label{table:tab_C}
\end{table}

Summarizing, the bosonic spectrum of the vacua of eqs.~\eqref{back_epos_fin2} presents a number of technical difficulties, reveals some novelties and brings along some surprises. The main novelty is the indication that \emph{stable vacua may be attained in non--supersymmetric compactifications to four--dimensional Minkowski space with a finite string coupling}.
The main surprises are the emergence of additional moduli related to boundary conditions and of corresponding boundary modes, and two technical findings. First, the massless modes of the type--IIB three-forms lead to unfamiliar dynamical vector equation with three derivatives,
\beq
\partial_{[\mu}\,\partial^\rho\,F_{\nu]\,\rho} \ =  \ 0 \ ,
\eeq
which is equivalent to
\beq
\partial^\rho\,F_{\nu\,\rho} \ = \ \partial_\nu\,\sigma \ , \qquad  \Box\,\sigma \ = \ 0 \ , \label{eqsbmn}
\eeq
and thus describes both vector and scalar modes. Moreover, massless modes emerge, for all internal momenta, from the first--order equations of tensor modes. However, this last result is not pathological, as these modes are excitations of tachyonic ground states with different $\mathbf{k}$-dependent boundary conditions. For Fermi fields, enforcing identical $\Lambda$ projections at the ends of the interval removes one-half of the original modes, leaving the massless spectrum of $N=4$ supergravity coupled to five vector multiplets. There is apparently some tension between the massless option and some recent conjectures on the limiting behaviors that should allow for ultraviolet completions~\cite{luest}, but gravitino masses are expected to arise when radiative corrections are taken into account. This issue will be clearly worthy of further investigation, once the full perturbative spectrum and its stability properties are known.

\section{\sc Moduli stabilization, fluxes and the KKLT Setup} \label{sec:KKLT}

As we saw in Sections~\ref{sec:toroidal_ss} and~\ref{sec:calabi-yau}, string compactifications typically give rise to a plethora of massless scalar modes with undetermined vacuum values, called ``moduli fields'', which are related to the topology and geometry of the internal space and to the dilaton--axion system. These moduli fields would generally result in violations of the equivalence principle, so that they should
acquire masses and/or have very weak couplings to visible matter in realistic setups for Particle Physics and Cosmology. The process of giving masses to these fields, called {\it moduli stabilization}, can be driven by perturbative and/or non-perturbative effects, which are accompanied by the partial or total breaking of supersymmetry. Here, we restrict our attention to a few widely explored setups:
\begin{itemize}
    \item \textbf{perturbative Calabi--Yau reductions of M-theory to five dimensions}, which include non--perturbative string effects. Combined with a further Scherk--Schwarz reduction to four dimensions, this setting can stabilize the volume of the internal Calabi--Yau space;
    
    \item \textbf{type--IIB compactifications on warped Calabi-Yau manifolds}. In this setting, the NS-NS and RR 3-form fluxes on appropriate cycles can stabilize, at tree level, complex structure moduli and the dilaton-axion pair.
    The stabilization, although perturbative at the supergravity level, is actually non--perturbative from the string vantage point. K\"ahler moduli remain flat directions after adding these fluxes, and the resulting four-dimensional low-energy action is of no-scale type;
    
    \item \textbf{gaugino condensation}.
    In this scenario, non--perturbative effects in a hidden sector, with a dynamics similar to QCD but arising at much higher energies, generate scalar potentials for moduli fields, stabilizing some of them. In type IIB strings, effects of this type typically originate from D7 branes wrapping four cycles, which host non-Abelian gauge groups in their world volumes. The superpotentials thus generated can stabilize K\"ahler moduli fields. The vacuum energy from the combination of this setting with the previous one is negative, so that the ground state is anti--de Sitter;
    
    \item \textbf{KKLT scenario}. It combines the preceding ingredients with an uplift, an additional positive contribution to the vacuum energy. The uplift can be induced by D3 antibranes or D7 branes with internal magnetic fields, and the resulting four-dimensional theory can have a positive vacuum energy, so that the ground state can be de Sitter.
\end{itemize}

\subsection{\sc Scherk--Schwarz Breaking in M-theory}

The Scherk--Schwarz setup for supersymmetry breaking by compactification, which we described in Section~\ref{sec:toroidal_ss}, can provide a simple setup for stabilizing some moduli fields in the presence of internal Calabi--Yau spaces. The example that we shall focus on is a version of Scherk-Schwarz supersymmetry breaking in M-theory, whose low-energy description is provided by the eleven--dimensional supergravity of Cremmer, Julia and Scherk~\cite{CJS}, leading eventually to four dimensions. 

To this end, let us consider a simple truncation of eleven--dimensional supergravity compactified to five dimensions on a Calabi--Yau manifold with Hodge numbers $(h^{1,1}, h^{2,1})=(1,0)$~\cite{cadavid}, thus somehow retaining only the breathing mode $\sigma$ of the compact space, so that
\beq
g_{i \bar j} \ = \ \delta_{i
\bar j} \, e^{\sigma} \ .
\eeq
In general, a Calabi--Yau with Hodge numbers $(h^{1,1},h^{2,1})$ gives rise, in five dimensions, to $h^{1,1}-1$ vector multiplets and $h^{2,1}$ hypermultiplets.
In our case, in addition to the five--dimensional gravitational multiplet, with bosonic fields $(g_{MN},A_M)$, where the ``graviphoton'' $A_M$ originates from the eleven--dimensional 3 form according to
\beq
A_{M i \bar j} \ =  \ \frac{\sqrt{2}}{6}\,  A_M \,\delta_{i \bar j}  \ ,
\eeq
the low--energy spectrum includes the universal hypermultiplet, with bosonic fields ($\sigma, C_{MNP}, a$), where
\beq
A_{ijk} \ = \ \frac{\sqrt{2}}{6}\ \epsilon_{ijk} \, a \ , \label{Cijka}
\eeq
so that $a$ is a complex scalar, while $C_{MNP}$ is dual to an axion that we shall call $a_1$.
The four scalar fields thus identified parametrize
the ${{SU(2,1)}/ {SU(2) \times U(1)}}$ coset~\cite{ferrara_coset}, and
the bosonic contributions to the five--dimensional supergravity Lagrangian can be identified starting from the eleven--dimensional action of eq.~\eqref{eqd11}, and read~\footnote{There would be additional terms
originating from the modified Bianchi identity in the Horava-Witten theory, which corrects the
Lagrangian (\ref{9.05}). They do not play a role in our discussion, but
can be found, for example, in~\cite{ovrut_stelle}.}
\bea
{\cal S}_5 &=& {1 \over 2\kappa_5^2} \int d^5 x \sqrt{g} \left\{ R
\!-\! {9 \over 2} (\partial_M
\sigma)^2 \!-\! {1 \over 48} e^{6 \sigma} F_{MNPQ} \, F^{MNPQ} -{3 \over 2}
F_{MN} F^{MN} \!-\! 2 \ e^{-6 \sigma} |\partial_M a|^2 \right\} \nonumber \\
&-& {1 \over \kappa_5^2} \int d^5 x \ \epsilon^{MNPQR} \left\{ \frac{i}{\sqrt{2}}\
A_{MNP} \,\partial_Q a \,\partial_R a^{\dagger} + {1 \over 2 \sqrt{2}}
\,A_{M} F_{NP} F_{QR} \right\} \
, \label{9.05}
\eea
where $F_{MN} = \partial_M A_N- \partial_N A_M$ is the Maxwell field strength of $A_M$.
The final compactification to four dimensions rests on the $S^1/\mathbb{Z}_2^{HW}$ orbifold,
where the $\mathbb{Z}_2^{HW}$ orbifold action will be defined shortly. This Lagrangian has an $SU(2)$ R-symmetry that will be used shortly, after some convenient field redefinitions, to implement the Scherk--Schwarz mechanism.

Following~\cite{ferrara_coset}, one can obtain the complete five--dimensional kinetic Lagrangian of the universal hypermultiplet starting from the
K{\"a}hler potential
\beq
\kappa_5^2\, {\cal K} \ = \ - \ln \left(S+S^{\dagger} \ - \ 2\, a^{\dagger} a\right) \ , \label{eq:s1}
\eeq
where $a$ is the complex scalar introduced in eq.~\eqref{Cijka}.
In the result, one then replaces $S$ according to
\beq
S\ = \ \exp (3\sigma ) \ + \ a^{\dagger} a \ + \ i\, a_1  \ .
\eeq
The axion field $a_1$ is related to the four--form field strength by Hodge duality, according to
\beq
\sqrt{2} \,\exp (6 \sigma ) \, F_{MNPQ} \ = \ \epsilon_{MNPQR}
\left[ \partial^R a_1 \ + \  i \, a^\dagger \stackrel{\leftrightarrow}{\partial^R} a\right] \ .
\eeq
The 
$SU(2)_R$ symmetry acts linearly on the redefined hypermultiplet fields
\beq
z_1 \ = \ {1-S \over 1+S} \ , \qquad z_2 \ = \ {2 a \over 1+S} \  \label{eq:s2}
\eeq
and the $\mathbb{Z}_2^{HW}$ projection acts on the hypermultiplet as
\beq
\mathbb{Z}_2^{HW} S \ = \ S \ , \qquad \mathbb{Z}_2^{HW} a \ =\  - \ a \ ,
\eeq
which translates on the $SU(2)_R$ doublet into
\bea
\mathbb{Z}_2^{HW}
\left(
\begin{array}{c}
z_1 \\
z_2
\end{array}
\right)
&=&
\left(
\begin{array}{cc}
1 & 0 \\
0 & -1
\end{array}
\right)
\left(
\begin{array}{c}
z_1 \\
z_2
\end{array}
\right) \ . \label{eq:s3}
\eea
This projection breaks the original $N=2$ supersymmetry that would be inherited after circle compactification to $N=1$.

The Scherk-Schwarz deformation of the entire set of five--dimensional fields is now obtained twisting the periodicity along the circle in such a way that each $SU(2)_R$ doublet verifies
\bea
\left(
\begin{array}{c}
{z}_1(x_5+2 \pi) \\
{z}_2(x_5+2 \pi)
\end{array}
\right)
&=&
\left(
\begin{array}{cc}
\cos \left( 2 \pi  \omega\right) & \sin \left( 2 \pi  \omega\right) \\
-\sin \left( 2 \pi  \omega\right) & \cos \left( 2 \pi  \omega\right)
\end{array}
\right)
\left(
\begin{array}{c}
{z}_1(x_5) \\
{z}_2(x_5)
\end{array}
\right) \ ,\label{eq:s4}
\eea
where $\omega$ is a dimensionless parameter characterizing the Scherk--Schwarz twist.
These periodicity conditions are solved by letting
\bea
\left(
\begin{array}{c}
{z}_1 \\
{z}_2
\end{array}
\right)
&=&
\left(
\begin{array}{cc}
\cos \left( \omega\,x_5 \right) & \sin \left(\omega\,x_5\right) \\
-\sin \left(\omega\,x_5\right) & \cos \left(\omega\,x_5\right)
\end{array}
\right)
\left(
\begin{array}{c}
{\hat z}_1 \\
{\hat z}_2
\end{array}
\right) \ ,\label{eq:s42}
\eea
where ${\hat z}_1$ and ${\hat z}_2$ are periodic.
Note that,
thanks to the structure of eq.~\eqref{eq:s42},
the fields ${\hat z}_i$ have the same $\mathbb{Z}_2^{HW}$ parities as the $z_i$.
The 4D complex superfields of the model are $S$ (without a zero mode
for $a$),
\beq
S\ = \ \exp (3\sigma ) \ + \ i\, a_1  ,
\eeq
and $T$, which is defined as
\beq
T\ = \ M_5 \ \sqrt{g_{55}} \ + \ i \, A_5 \ , \label{TM5}
\eeq
where $M_5^3=\frac{1}{\kappa_5^2}$.
The axion $A_5$ is the
fifth component of the five--dimensional graviphoton, and $\sqrt{g_{55}}=R$ in the vacuum.

In the four--dimensional Einstein frame, which is reached by the rescaling
\beq
g_{\mu \nu}^{(5)}\ = \ (\pi\, Re T)^{-1} g_{\mu \nu}^{(4)} \ , 
\eeq
the four--dimensional scalar potential originates from the
kinetic terms of the ${z}_i$ fields determined by eq.~(\ref{eq:s1}).
After setting ${\hat z}_2=0$, since it lacks zero modes, one finds
\beq
\kappa^4 \ V \ = \ M_5 \ \int dx_5 \,\frac{\sqrt{- g^{(5)}}}{\sqrt{- g^{(4)}}} \ g^{55} \,{\cal K}^{a \bar b} \ \partial_5 {z}_a \,
\partial_5 {\bar z}_{\bar b} \ = \ \frac{4 \omega^2}{\pi\left(T+T^{\dagger}\right)^3} {|1-S|^2 \over \left(S+S^{\dagger}\right)}
\ , \label{eq:s40}
\eeq
where $a,b=1,2$ and ${\cal K}^{a \bar b}$ is the inverse of the
K{\"a}hler metric ${\cal K}_{a \bar b}= \partial_a \partial_{\bar b}
{\cal K}$, and
\beq
\frac{1}{\kappa^2} \ = \ \frac{\pi R}{\kappa_5^2} \ . \label{kappa45}
\eeq
The preceding potential can be related to a superpotential that is generated for $S$. In fact, the four--dimensional
theory is completely described by \cite{dg97}
\beq
\kappa^2\, {\cal K} \ =\  - \ \ln (S + S^{\dagger}) \ - \ 3 \ln (T+T^{\dagger}) \ ,
\qquad \kappa^3\, {\cal W} \ = \  \frac{2 \omega}{\sqrt{\pi}} \left(1 \ + \ S\right) \ . \label{eq:s5}
\eeq

Note that the superpotential originates from a
non-perturbative effect from the heterotic viewpoint, since the real part of $S$ is the inverse of the heterotic string coupling.
The scalar potential has a minimum at $S=1$, which corresponds to a
spontaneously broken supergravity with a vanishing cosmological constant.
The important point about (\ref{eq:s5}), or about any other similar supergravity example, is that \emph{the breaking of supersymmetry {\`a} la Scherk-Schwarz appears to be spontaneous, of the F-type, with a cosmological constant that vanishes at tree level}. The end result is thus a no-scale model of the type discussed in Section~\ref{sec:no_scale_models}~\cite{Cremmer:1983bf}. This is a generic feature of tree-level string compactifications.  However, the no-scale structure is typically broken by quantum corrections, which can also stabilize K\"ahler moduli.

The five--dimensional Dirac gravitino is equivalent to two four--dimensional Majorana gravitinos,
transforming as an $SU(2)_R$ doublet. One of the
gravitini is even under the $\mathbb{Z}_2^{HW}$ projection and has a zero mode (before the
Scherk-Schwarz twisting~\eqref{eq:s42}), while the other is odd and has only massive KK
excitations. 

As we saw in Section~\ref{sec:SUGRA} (see eq.~\eqref{5.14}), the order parameter for supersymmetry breaking is the gravitino mass
\beq
m_{3/2}^{2} \ = \ \kappa^4 e^{{\kappa^2}\, {\cal K}} \,\left|{\cal W}\right|^2 \ = \ \frac{1}{\pi\,\kappa^2} \frac{8\, \omega^2}{
(T+T^{\dagger})^3} \ .  \label{eq:s6}
\eeq
Then, replacing $T+T^\dagger$ by its vacuum value $2 R M_5$ from eq.~\eqref{TM5} and using~\eqref{kappa45}, one can conclude that 
\beq
m_{3/2}^2 = \frac{\omega^2 }{R^2}\ ,
\eeq
consistent with the link between the gravitino mass and its R-symmetry charge, so that it could be deduced directly by the Scherk--Schwarz reduction of its five--dimensional kinetic term. The Goldstone fermion (or goldstino) is the fifth component
of the ($\mathbb{Z}_2^{HW}$ even) five--dimensional gravitino $\Psi_5$, which belongs to the multiplet of the no-scale modulus $T$, whose auxiliary field is responsible for breaking supersymmetry.

The breaking of supersymmetry in M-theory discussed here is linked by duality to the special setting for supersymmetry breaking by compactification in type I strings presented in~\cite{ads1} and reviewed in Section~\ref{sec:toroidal_ss}. Indeed, while supersymmetry breaking in M-theory is a non-perturbative phenomenon from the viewpoint of heterotic strings, since in particular it fixes the dilaton, heterotic-type I duality turns into the perturbative M-theory breaking mechanism in type I strings that we discussed in Section~\ref{sec:toroidal_ss}.

There is a subtlety with Scherk-Schwarz compactifications, the absence of a {\it gap} between the scale of supersymmetry
breaking $\frac{\omega}{R}$ (where $\omega = \frac{1}{2}$ in the example above) and the Kaluza-Klein scale $m_n = \frac{n}{R}$. This implies that the effective theory obtained by truncating to the would-be zero mode is not reliable.  For example, let us consider the contribution to the vacuum energy from a tower containing equal numbers $n_*$ of bosonic and fermionic degrees of freedom, as would pertain to an originally supersymmetric theory in $D+1$ spacetime dimensions, one of which is compact and of radius $R$ while the remaining $D$ dimensions are noncompact, and let us take for definiteness $\omega = \frac{1}{2}$.
The field--theory expression for the vacuum energy is
\beq
{\cal E} \ = \ - \ \frac{1}{2 (4 \pi)^{\frac{D}{2}}} Str
\int_0^{\infty} \frac{dt }{t^{1+\frac{D}{2}}} \ e^{-t M^2}  \ , \label{eq:s7}
\eeq
where $Str$ takes into account the numbers of degrees of freedom and the relative sign for bosons and fermions ($1$ for a real scalar, $-2^{\frac{D}{2}}$ for a Dirac fermion, etc).  Retaining only the would-be zero mode, the vacuum energy can be estimated to be
\beq
{\cal E} \ = \ - \ \frac{n_*}{2 (4 \pi)^{\frac{D}{2}}}
\int_{\frac{1}{\Lambda^2}}^{\infty} \frac{dt }{t^{1+\frac{D}{2}}} \ \left( 1 - e^{-t \left(\frac{1}{2R}\right)^2} \right) \ \sim \ - \ \frac{1}{R^2}\, \Lambda^{D-2} \ , \label{eq:s8}
\eeq
where $\Lambda$ is an UV cutoff. As expected for a standard Quantum Field Theory, the one-loop vacuum energy is ultraviolet divergent.

The result is completely different for the whole Kaluza--Klein tower. In this case the contributions to the vacuum energy are very similar to Matsubara sums for a finite-temperature quantum field theory, which are UV finite.   Indeed, one finds
\bea
{\cal E} &=& - \ \frac{n_*}{2 (4 \pi)^{\frac{D}{2}}}
\int_{0}^{\infty} \frac{dt }{t^{1+\frac{D}{2}}} \ \sum_{n=-\infty}^{\infty}\left[ e^{-t \left(\frac{n}{R}\right)^2} \ - \  e^{-t \left(\frac{n+\frac{1}{2}}{R}\right)^2} \right] \nonumber \\
&=&
- \ \frac{n_*}{2 (4 \pi)^{\frac{D}{2}}}
\int_{0}^{\infty} \frac{dt }{t^{1+\frac{D}{2}}} \ \left[  \theta_3 \left(\frac{i t}{R^2}\right)  \ -
\ \theta_2 \left(\frac{i t}{R^2}\right)  \right]
\ , \label{eq:s9}
\eea
and after a modular transformation, letting $y = R^2 l$, the result becomes
\bea
{\cal E} &=&
- \ \frac{n_*}{R^D (4 \pi)^{\frac{D}{2}}}
\int_{0}^{\infty} dy \ y^{\frac{D-1}{2}} \ \Big[  \theta_3 (i y) -
\theta_4 (i y)  \Big] \nonumber \\
 &=& - \ \frac{n_*}{R^D (4 \pi)^{\frac{D}{2}}}
\int_{0}^{\infty} dy \ y^{\frac{D-1}{2}} \ \sum_{n=-\infty}^{\infty} \left[ 1 - (-1)^n \right]
e^{-\pi n ^2 y}
\ . \label{eq:s10}
\eea
The last expression in eq.~\eqref{eq:s10} is ultraviolet finite and proportional to $- \,\frac{1}{R^D}$, in analogy with finite-temperature computations but in contrast with eq.~(\ref{eq:s8}), which was obtained by truncating the Kaluza--Klein tower.
The upshot of these considerations is that in Scherk-Schwarz reductions the whole tower of Kaluza--Klein states must be included in the effective field theory.

The preceding construction can be extended to more general Calabi-Yau compactifications. 
The moduli fields that arise when reducing from eleven to five dimensions fall into two categories~\cite{cadavid,ovrut_stelle}: {\it K\"ahler moduli} $T_{i \bar j}$, of which $T$ is a simple example, and {\it complex structure moduli}, supplemented by dilaton-axion combination $S$ defined above. The total number of moduli of different types is topological in nature, as we saw in Section~\ref{sec:calabi-yau}, and equals $h^{1,1}$ for K\"ahler moduli and $h^{2,1}+1$ for complex structure moduli and $S$. These moduli fields emerge in Calabi--Yau compactifications of type II strings to four dimensions, which preserve ${\cal N}=2$ supersymmetry.  In more general Calabi--Yau settings, the Scherk-Schwarz mechanism generates a superpotential for complex structure moduli, while K\"ahler moduli remain flat directions in four dimensions.

\subsection{\sc Flux Compactifications in Type--IIB Supergravity} \label{sec:gkp}

Type II string compactifications on a Calabi-Yau manifold lead to ${\cal N}=2$ 
theories in four dimensions. For the RR sector in type IIB, which is our main interest in this section, one must rely on the expansions
\bea 
&& C_0 ({\bf x},y) \ = \ C_0 ({\bf x}) \quad , \quad
C_2 ({\bf x},y) \ = \ C_2 ({\bf x}) + c^A ({\bf x}) \ \omega_A (y) \ , \nonumber \\
&& C_4 ({\bf x},y) \ = \ V_1^I ({\bf x}) \ \alpha_I (y) + C_A ({\bf x}) \ {\tilde \omega}^A (y) \ , \label{eq:IIBcy1}
\eea 
where $\omega_A$ is a basis of $(1,1)$ cycles of the CY space, ${\tilde \omega}^A$ is the dual basis of $(2,2)$ cycles and $\alpha_I$ is a basis of $3$-cycles. The NS-NS $B_2$ has an expansion similar to that of $C_2$.To reduce supersymmetry to ${\cal N}=1$, an orientifold projection can be added with a geometric action $\sigma$ that affects the K\"ahler form and the $(3,0)$ form $\Omega_3$. Three notable options are
\bea 
&& IIB \ {\rm with \ 03/07 \ planes}  \quad : \quad \sigma \ J = J \ , \qquad  \sigma \ \Omega_3 = - \Omega_3 \ , \nonumber \\
&& IIB \ {\rm with \ 05/09 \ planes}  \quad : \quad \sigma \ J = J \ , \qquad  \sigma \ \Omega_3 =  \Omega_3 \ , \nonumber \\
&& IIA \ {\rm with \ 06 \ planes}  \ \ \quad \quad : \quad \sigma \ J = - J \ , \qquad \sigma \ \Omega_3 =  {\overline \Omega}_3 \ . \label{eq:IIBcy2}
\eea 
The first option is the generalization to Calabi-Yau spaces of the IIB orientifold projection $\Omega' = \Omega \ \sigma (-1)^{G_L } $ involving six internal parities. 
The second is related to the type I string compactified on Calabi-Yau manifolds (with orientifold projection $\Omega' = \Omega \ \sigma$), while the third is the appropriate framework to discuss intersecting branes in IIA compactified to four dimensions (with orientifold projection $\Omega' = \Omega \ \sigma (-1)^{F_L }$). In the following, we shall concentrate on the first option. 

In type IIB strings $\sigma$ acts holomorphically, and consequently the cohomology of the CY space splits into even and odd components
$H^{(p,q)} \ = \ H^{(p,q)}_+ \oplus H^{(p,q)}_-$.  
The four--dimensional zero modes are invariant under the orientifold projection $\Omega'$, and are obtained from the axion-dilaton combination $S = e^{- \Phi} - i\,C_0$ and from
\bea 
&& J \ = \ \sum_{\alpha=1}^{h_{1,1}^+} v^{\alpha}  ({\bf x}) \ \omega_{\alpha} \ , \qquad 
B_2  \ = \ \sum_{a=1}^{h_{1,1}^-} b^a ({\bf x}) \ \omega_a \ , \nonumber \\
&& C_2  \ = \ \sum_{a=1}^{h_{1,1}^-} c^a ({\bf x}) \ \omega_a \ , \qquad
C_4  \ = \ \sum_{k=1}^{h_{2,1}^+} V_1^k ({\bf x}) \ \alpha_k  \ + \ \sum_{a=1}^{h_{1,1}^+} C_{\alpha} ({\bf x}) \ {\tilde \omega}^{\alpha}  \ , \nonumber \\
&& Z^I = \int \Omega_3 \wedge \beta^I  \ , 
\label{eq:IIBcy3}
\eea 
where $I = 1,2, \ldots , h_{2,1}^-$, and $\pm$ in general identify Calabi--Yau forms with positive or negative eigenvalues under $\Omega'$. The four-dimensional ${\cal N}=1$ massless spectrum comprises the gravitational multiplet, the axion-dilaton chiral multiplet $\tau$, $h_{1,1}^+$ chiral multiplets
$T_{\alpha}$ (with bosonic component $v^{\alpha}+ i\, C_{\alpha}$), $h_{2,1}^-$ complex structure moduli chiral multiplets $Z^I$, 
$h_{1,1}^-$ chiral multiplets
$G^a$ (with bosonic components $b^a+i\,c^a$) and $h_{2,1}^+$ vector multiplets
$V_{\mu}^k$.

String theory includes several options for internal fluxes originating from various NS-NS and RR forms in the theory. Nonvanishing fluxes along cycles of the internal space generate potentials for moduli fields that can stabilize some of them~s\footnote{For comprehensive reviews on flux compactifications, see e.g. \cite{kachru_douglas-flux}, \cite{grana-flux}, \cite{hebecker}.}. In the case of interest, the relevant fluxes correspond to the two forms $B_2, C_2$ and to the four-form $C_4$, whose field strengths are denoted by $H_3 = dB_2$,
$F_3 = dC_2$, $F_5 = dC_4$. The four-form is peculiar, since it is subject to a selfduality condition
\beq
{\tilde F}_5 \equiv F_5 \ - \ \frac{1}{2} C_2 \wedge H_3 \ + \ \frac{1}{2} B_2 \wedge F_3 \ = \
\star {\tilde F}_5 \ ,
\eeq
as we have seen, which one typically imposes on the field equations.
The fluxes of these fields along the cycles of the internal space defined in Section~\ref{sec:calabi-yau} are quantized according to
\bea
\frac{1}{2 \pi \alpha'} \int_{A_I} F_3 &=&  2 \pi n^I \ , \qquad \frac{1}{2 \pi \alpha'} \int_{B^I} F_3 \ = \  - 2 \pi f_I  \nonumber \\ 
\frac{1}{2 \pi \alpha'} \int_{A_I} H_3 &=&  2 \pi m^I \ , \qquad \frac{1}{2 \pi \alpha'} \int_{B^I} H_3 \ = \  - 2 \pi e_I   \ , 
\label{eq:fluxes4}
\eea
where $n^i,f_ I,m^I,e_I$ are integers.
Gauss's law for the four-form field $C_4$ leads to the RR tadpole condition
\beq
\frac{1}{2 \kappa_{10}^2 T_3} \int H_3 \wedge F_3 \ + \ Q_3^{\rm loc} \ = \ 0  \ , \label{eq:D3tadpole}
\eeq
where $T_3$ is the tension of D3 (anti)branes, and
$T_3 \, Q_3^{\rm loc}$ is the corresponding contribution to the charge, which can also originate from orientifold planes~\footnote{The orientifold projection used in this case is $\Omega' = \Omega \,\Pi_6$, where
$\Pi_6$ is a parity operation in the six internal coordinates.}.

It is customary to define the complex three-form
\beq
G_3 \ = \ F_3 \ - \ i\,S\, H_3 \quad , \quad {\rm where}
\quad S \ = \ e^{- \phi} \ - \ i\, C_0\ .  \label{eq:fluxes3}
\eeq
The compactification preserves $N=1$ supersymmetry in four dimensions if 
\beq
i\, G_3 \ = \ \star \,G_3 \ , \label{eq:fluxes5}
\eeq
 \emph{i.e.} if the three-form fluxes are imaginary anti-selfdual, and if, in addition,  $G_3$ contains no $(0,3)$ part~\cite{gkp}~\footnote{Note that the paper~\cite{drs} was a precursor of this setup.}.  

The ten-dimensional metric and the five-form flux are described by
\bea
d s_{10}^2 &=& H^{\frac{1}{2}} \eta_{\mu \nu} dx^{\mu} dx^{\nu}  \ + \ H^{-\frac{1}{2}} {\tilde g}_{m n} dy^{m} dy^{n} \ , \nonumber \\
 {\tilde F}_5 &=& (1 \ + \ \star) d H \wedge dx^0 \wedge dx^1 \wedge dx^2 \wedge dx^3
\ , \label{eq:gkp1}
\eea
where ${\tilde g}_{m n}$ is the internal metric. These expressions depend on a scalar function of the internal coordinates $H(y)$ that is usually called the warp factor.

In this case the scalar potential can be described in supergravity, and originates from the Gukov-Vafa-Witten superpotential~\cite{GVW}
\bea 
&&  {\cal W}(Z^I,S) \ = \ \frac{1}{(2 \pi)^2 \alpha'} \int (F_3\ - \ i S \, H_3) \wedge \Omega_3 \nonumber \\
&& = (f_I - i S e_I) Z^I + (n^I - i S m^I) {\cal F}_I 
, \label{WGVW}
\eea 
where $H_3$ and $F_3$ are background fluxes, $\Omega_3$ is the holomorphic $(3,0)$ form of the CY space and ${\cal F}_I$ is the derivative of the prepotential defined in eq.~\eqref{eq:cy21}. In deriving this result we have also used eqs.~\eqref{eq:cy20} and eq.~\eqref{eq:cy22}. The condition~(\ref{eq:fluxes5}) allows for $(2,1)$ and $(0,3)$ fluxes. According to the superpotential (\ref{WGVW}), the four-dimensional theory is supersymmetric if the $(0,3)$ flux vanishes, since then ${\cal W}$ and the gravitino mass vanish.

On the other hand, the K\"ahler potential has a no-scale form, and singles out the overall volume included in the K\"ahler modulus $T$ and the axion-dilaton $\tau$. It is given by
\beq
{\cal K} \ =\  - \ 3 \log \left( T + {\overline T} \right) \ -  \ \log \left( S + {\overline S} \right)   \ -\  \log \left( - i \int \Omega_3 \wedge {\overline \Omega_3} \right) \ , \label{eq:gkp2}
\eeq
and the resulting scalar potential is
\beq
V \ = \ e^{\cal K} \ {\cal K}^{i {\bar j}} \ D_i {\cal W}\ 
{\overline D_j {\cal W} }\ , \label{eq:gkp3}
\eeq
where $i, {\bar j}$ refer to the moduli fields, excluding $T$. It is positive definite due to its no-scale structure, so that supersymmetric solutions have zero vacuum energy, and can generally fix the complex--structure moduli $Z^I$ and $S$~\cite{gkp,lmicu}.
If the $(0,3)$ flux does not vanish, a gravitino mass is generated, signaling supersymmetry breaking with zero vacuum energy. The overall modulus $T$ is a flat direction of the four-dimensional scalar potential, in analogy with the original phenomenological no-scale supergravity constructions.

Flux compactifications have several attractive features. First of all, they generate warped internal spaces, which can induce hierarchies along the lines of Randall-Sundrum models~\cite{randall_sundrum}. 
Moreover, one can argue, as in~\cite{nilles}, that the masses induced by fluxes also behave like
\beq
M \ \sim \ \frac{1}{R^3} \ ,
\eeq
where $R$ is a typical internal size, and are thus parametrically smaller than those of Kaluza--Klein excitations within the regime of validity of Field Theory.
Consequently, one can formulate a four-dimensional supergravity for the zero modes of the compactification, taking the effects of fluxes on their masses into account and ignoring massive Kaluza-Klein
states. This is an important difference with respect to the Scherk-Schwarz reduction discussed in the previous section.
Flux compactifications stabilize the same type of (complex structure) moduli as those considered in the supersymmetry breaking in M-theory discussed in the previous section.  This is not a coincidence: the Scherk-Schwarz compactification of M-theory can be viewed as a particular example of flux compactification that relies on different types of fluxes.

So far, we have reviewed mechanisms that can stabilize complex--structure moduli and the dilaton--axion system. We can now turn to K\"ahler moduli.

\subsection{\sc Gaugino Condensation and Stringy Instantons}
We can now address non--perturbative effects that are instrumental to stabilize K\"ahler moduli fields, which were not affected by the preceding mechanisms. Some of these effects rely on string instantons, and thus on Euclidean branes (generically called Ep branes) wrapping cycles in the internal space. For type II strings, these are E3 instantons. Their effects are proportional to the overall tension of these Euclidean branes, $e^{-T_{Ep}\, A_{E_p}}$, where the exponent is proportional to the area $A_{E_p}$  of the internal cycles wrapped by them, which is governed by K\"ahler moduli.

Other types of non-perturbative effects are of field-theoretical origin, and gaugino condensation on D7 branes in type IIB is a notable example.  In this case, the current state-of-the-art technology forces one to rely on holomorphy and other symmetry arguments that apply in the presence of supersymmetry.

Let us briefly discuss the effective Lagrangian approach to gaugino condensation on D7 branes for a pure $SU(N)$ super Yang-Mills theory containing gauge bosons $A_\mu^a$ and gaugini $\lambda^a$ in the adjoint representation of the gauge group. The theory will confine at a scale determined by $\Lambda$, giving rise to a  bilinear gaugino condensate $\langle \lambda^a \lambda^a \rangle$~(this is reviewed, for example, in the last two books in~\cite{susy-books}).  The tree-level gauge coupling is determined by
\beq
\frac{1}{g_{D7}^2} \ = \ \frac{Re \ T}{4 \pi} \ , \label{gc1}
\eeq
where~\cite{gkp,KKLT}
\beq
Re T \ = \ e^{-\,\phi} \ V_4 \ ,
\eeq
where $V_4$ denotes the volume of the four-cycle wrapped by the D7 brane. 

In what follows, $T$ will actually denote the complete chiral superfield.  The SUSY Yang--Mills (SYM) theory has classically an R-symmetry, which acts on the gauge--field superfield strength according to
\beq
W'_{\alpha} \left(e^{- \frac{3 i \alpha}{2}} \theta\right)
\ = \ e^{- \frac{3 i \alpha}{2}} \ W_{\alpha} (\theta)
\ . \label{gc2}
\eeq
At the quantum level, this R-symmetry enters an
$U(1)_R \times SU(N)^2$ anomaly. Using the holomorphy of the superpotential, this suffices to determine
unambiguously the low-energy effective superpotential, below the scale of gaugino condensation.  The low-energy degrees of freedom are bound states captured by the ``glueball" chiral superfield
\beq
U \ = \ \mathrm{Tr} 
\left(W_{\alpha} W^{\alpha} \right) \ , \label{gc3}
\eeq
and the low-energy superpotential is
\cite{vy}
\beq
{\cal W} \ = \ \frac{T}{16 \pi} U \ + \ \frac{b_0}{6} U
\left( \log \frac{U}{\Lambda^3} \ - \ 1 \right)
\ , \label{gc4}
\eeq
where
\beq
b_0 \ = \ \frac{C_2(G)}{16 \pi^2} \ = \ \frac{3 N }{16 \pi^2}
\label{gc5}
\eeq
is the one-loop beta function coefficient of the SYM theory. Note that the R-symmetry is apparently violated by the log-term, but it can be restored by shifting the modulus field under $U(1)_R$ according to
\beq
\delta T \ = \ 8 \pi i \ b_0 \xi \ = \ \frac{3 i N}{2 \pi }
\xi \ . \label{gc6}
\eeq
Minimizing the superpotential with respect to $U$ gives
\beq
\left(\frac{U}{\Lambda^3}\right)^N \ = \ e^{- 2 \pi T} \ ,
\eeq
which determines $N$ solutions for the gaugino condensate superfield
\beq
U_0 \ = \ e^\frac{2\pi i k}{N} \ \Lambda^3 \,e^{- \frac{3 T}{8 \pi b_0}} \ =\ 
e^\frac{2\pi i k}{N} \ \Lambda^3 e^{- \frac{2 \pi T}{N}} \ , \label{gc7}
\eeq
where $k=0,\ldots,N-1$. Consequently, there are $N$ independent vacua, consistently with the Witten index discussed in Section~\ref{sec:wittenindex}.
This superfield equation encodes the condensation scale
\beq
\left| \langle \lambda^a \lambda^a \rangle\right|    \ = \ \Lambda^3 \ e^{- \frac{8 \pi^2 }{N g_{\mathrm{D7}}^2}}  \ . \label{gc07}
\eeq

Integrating out the glueball field, \emph{i.e.} replacing the resulting gaugino condensation scale
(\ref{gc7}) in the effective superpotential (\ref{gc4}), finally yields the low--energy moduli-dependent superpotential
\beq
{\cal W} (U_0) \ = \ - \ e^\frac{2\pi i k}{N}\ \frac{N}{32 \pi^2} \, \Lambda^3 \, e^{- \frac{2 \pi T}{N}}  \ . \label{gc8}
\eeq
Note that after integrating out the glueball superfield, one obtains a superpotential with the correct R-charge $R_W = -3$.

Summarizing, stringy instantons and gaugino condensation on D7 branes can generate non-perturbative superpotentials for K\"ahler moduli fields. When a single K\"ahler modulus $T$ is present, the result is typically of the form
\beq
{\cal W}_{np} \ = \ A \ e^{- \frac{2 \pi T}{N}} \ ,
\ \label{eq:np1}
\eeq
and in general $Re(T)$ is related to an internal area, as above, but $N=1$ for E3 string instantons. In gaugino condensation one therefore talks about ``fractional instantons'', and the factor $N$ is related to the rank of the $SU(N)$ gauge group in pure super-Yang-Mills hidden sectors. Moreover, the factor $A$ typically depends on complex--structure moduli.
Several scenarios proposed in the literature can generate superpotentials of this type. They rely on various configurations, where hidden sectors arise from branes different from those hosting the Standard Model and far away from them.

\subsection{\sc The KKLT Scenario}

Kachru, Kallosh, Linde and Trivedi (KKLT)~\cite{KKLT} proposed a scenario where all moduli in type IIB string theory are stabilized in a de Sitter vacuum, which comprises the following three steps:

\begin{enumerate}
\item stabilize all complex--structure moduli and the axion--dilaton pair by three--form fluxes in a warped compactification, as in Section~\ref{sec:gkp}, which results in a constant superpotential;
\item stabilize K\"ahler moduli by string theory or field theory instanton effects of the types reviewed in the preceding section, and combine their effects with the preceding result;
\item add $\overline{D3}$ antibranes at the tip of the resulting warped throat to uplift the vacuum energy to zero or positive values.
\end{enumerate}

The consistency of this three-step procedure relies on the assumption that the axion--dilaton and the complex structure moduli can be made much heavier than the K\"ahler moduli, whose masses will be eventually generated by non--perturbative effects in the type--IIB theory. The first two steps of the procedure give rise to an anti-de Sitter supersymmetric vacuum with stabilized K\"ahler moduli, while the third step adds a positive, non--supersymmetric contribution, to the vacuum energy. If this is not too large, it does not affect the stabilization of the K\"ahler moduli, while finally generating a positive (and hopefully very small) vacuum energy. The reader will recognize that the source of the uplift in the last step can be traced to the tension unbalance, or NS-NS tadpole, which also lies at the heart of brane supersymmetry breaking~\cite{bsb1,bsb2,bsb3,bsb4}.

We can now address the minimal supergravity description of the KKLT scenario, which relies an on orientifold projection involving six internal parities, so that the relevant branes and orientifolds are D3-O3 and D7-O7. Since the dilaton and complex structure moduli are very heavy, they are integrated out and can be ignored at low energies, except for a possible leftover constant contribution to the superpotential, which is denoted by ${\cal W}_0$. For the simplest possible setting with only one K\"ahler modulus, the effective Lagrangian is then described by
\bea
{\cal K} &=& - \ 3 \log (T + {\bar T}) \ , \qquad {\cal W} \ = \ {\cal W}_0 \ + \ A\, e^{- a\, T} \ , \nonumber \\
V &=& V_{SUGRA} \ + \ V_{\overline{D3}}
\ , \label{eq:kklt1}
\eea
where the parameter $a$ was $\frac{2\pi}{N}$ in the preceding discussion of gaugino condensation, but in general can take a wider range of values.
The second contribution to ${\cal W}$ is of non-perturbative origin, while the uplift $V_{\overline{D3}}$ originates from an anti-D3 brane.
The two contributions to the potential $V$ in eq.~(\ref{eq:kklt1}) are
\beq
V_{SUGRA} \ = \ e^{\cal K} \left({\cal K}^{T \bar T} \,|D_T {\cal W}|^2 \ - \ 3\, |{\cal W}|^2\right) \  , \qquad
V_{\overline{D3}} \ = \ \frac{c}{\left(T \ + \ {\bar T}\right)^n}
\ , \label{eq:kklt2}
\eeq
and the resulting setup is thus $N=1$ supergravity where supersymmetry is non--linearly realized due to the uplift. $V_{\overline{D3}}$ depends on the exponent $n$ and on $c$, a positive parameter that reflects the combined effects of anti--D3 tension and charge. In fact, while the two contributions would cancel for D3 branes, they add up in this case~\footnote{An uplift potential with $n=3$ can be recovered compactifying the tadpole potential of the ten--dimensional Sugimoto orientifold~\cite{sugimoto}, while also taking into account the Weyl rescaling leading to a four--dimensional Einstein frame. The same value of $n$ is obtained if the uplift originates from non-perturbative dynamics.}. The exponent $n$ depends on the details of the compactification: one finds that $n=3$ in the absence of strong warping, which is needed to red--shift $c$ to small enough values. If the antibrane is placed at the tip of a highly warped throat, the exponent becomes $n=2$~\cite{kklmmt}. 

In the absence of uplift, $V_{SUGRA}$ has a supersymmetric anti de Sitter minimum, which can be identified by solving the $F$-term condition
\beq 
D_T {\cal W} \ = \ 0 \ . 
\eeq
This leads to
\beq
{\cal W}_0 \ = \ -\  A \left(\frac{a}{3} \ + \ \frac{1}{T_0 \,+\, {\overline T}_0}\right) e^{-a T_0} 
\ , \label{eq:kkltw_0}
\eeq
but this result points to a possible problem for the scenario. For the consistency of the effective field theory, the internal volume must be large, so that $Re T_0 >> 1$. Consequently, ${\cal W}_0$, which is induced by the $(0,3)$ flux, must be very small. Although it was shown that this is possible in principle, numerical values
compatible with low-energy supersymmetry, so that $m_\mathrm{soft} \sim \mathrm{TeV}$, require extremely small values for $\left|{\cal W}_0\right|$, of order $10^{-13}-10^{-14}$ in Planck units. These are difficult to obtain, although they are not excluded~\cite{macallister}. 
Moderately small values for the gravitino mass are possible for
${\cal W}_0 \ll 1$ provided $c \ll 1$, in appropriate string units, which is possible if the uplift is highly red-shifted. 

The string realization of the KKLT construction of de Sitter vacua \cite{KKLT} rests on warped Calabi-Yau compactifications, so that
\begin{eqnarray}\label{eq:warpedbackground}
d\,s^2 &=&  H^{-1/2} d\, s^2_4 \ + \ H^{1/2} d\,s^2_{6} \ , \nonumber \\
F_5 &=& \left(1\,+\,\ast\right) \mathrm{vol}_4 \wedge d\, H^{-1} \ \equiv \ \ast {\cal F}_5\ + \ {\cal F}_5 \ ,
\end{eqnarray}
in the presence of constant profiles for the dilaton and three-form fluxes \cite{gkp},
where $H$ is the warp factor and $d s^2_{6}$ denotes the unwarped metric of the internal manifold. As argued in \cite{gkp}, and as we said above, this manifold can be considered to comprise a region that hosts a highly warped throat, which is glued to a compact Calabi-Yau space.
In the region of high warping, the local six-dimensional geometry is that of the deformed conifold~\cite{klebanov-strassler}, defined by its embedding into \(\mathbb{C}^4\) as
\begin{equation}
\sum_{a = 1}^4 \omega_a^4 \ = \ t \ .
\end{equation}
The deformation parameter $t$ is the complex structure modulus, whose absolute value determines the size of the 3-sphere at the tip of the cone. Its addition introduces in the low--energy effective field theory a very light field, due to the intense redshift reflecting its very origin from the tip of the cone, and this light field can destabilize other moduli unless special conditions are met~\cite{bdgl}. 
 The setup involves a cutoff \(\Lambda_0\), which reflects the transition between the highly warped region, modeled as a Klebanov-Strassler throat~\cite{klebanov-strassler}, and the (relatively unwarped) rest of the compact Calabi-Yau manifold.

The 3-form fluxes on the 3-cycles are\footnote{The setup only requires one type of flux on each cycle.}
\begin{eqnarray}\label{eq:fluxes2}
\frac{1}{(2\pi)^2 \alpha'} \,F_3 &=& M \alpha \ + \ M^0 \alpha_0 \ - \ M_i \beta^i \ , \\
\frac{1}{(2\pi)^2 \alpha'} \,H_3 &=& - \ K \beta  \ - \ K_0 \beta^0 \ + \ K^i \alpha_i  \ .
\end{eqnarray}
where $\alpha_i, \beta^i$ are Poincar\'e duals to the cycles $B_i, A^i$, and we have singled out the RR flux on the $S^3$ cycle at the tip of the throat, $M$, and its NS-NS partner $K$. These fluxes are responsible for the deformation of the conifold by the parameter $t$, as explained in~\cite{ks}.

On a compact manifold, the Bianchi identity for the five-form flux leads to a tadpole cancellation condition that forces the total D3-charge of the solution to vanish,
\beq \label{tadpolebianchi}
M K \ + \ Q^\mathrm{loc}_3 \ + \ \ldots \ = \  0 \ ,
\eeq
where the contributions that are left implicit originate from other supersymmetry-preserving fluxes and are typically positive.

In terms of the total number of (anti)branes, in the notation of the preceding sections, the tadpole condition reads
\beq \label{tadpoleT6}
M K \ + \ \ldots \ =\ -\ Q^\mathrm{loc}_3 \ =\ 32 \- \ N_{D3} \ + \ N_{\overline{D3}}\ .
\eeq
This sets on the product of fluxes the upper bound
\beq
M K  \ \leq \ | Q^\mathrm{loc}_3| \ . \label{ks9}
\eeq
As we have seen, the Gukov-Vafa-Witten superpotential
\eqref{WGVW} translates into a potential for the axion-dilaton pair and the complex structure moduli that can generically stabilize all of them. The conifold modulus \emph{v.e.v.} is dual to a gaugino condensate in the gauge theory, and its value is of order~\cite{klebanov-strassler}
\beq
t \ \sim \ \Lambda_0^3 \ e^{- \frac{2\pi K}{g_s M}}
\ , \label{ks10}
\eeq
which should lie well below $\Lambda_0$, the UV cutoff that we introduced above.
As we mentioned, the masses of complex structure moduli are typically larger than those of K\"ahler moduli, which are stabilized by non-perturbative corrections. 

The KKLT setup has the clear virtue of providing a complete scenario leading to a de Sitter vacuum, but there are various caveats that were discussed in the literature. One of them, pointed out in~\cite{bdgl}, is related to the wavefunction of the conifold modulus, which is localized at the tip of the throat, so that its mass undergoes a large redshift. Its inclusion in the effective low-energy theory gives rise to an additional constraint,  which is typically in conflict with the tadpole constraint~\eqref{ks9}. The need for large contributions from fluxes to the tadpole needed to stabilize moduli fields was recently promoted to a more general ``tadpole problem" \cite{tadpole_problem}.
For these and other reasons, all three steps of the construction of the KKLT setup, and the assumptions underlying them, have been widely debated over the years, in many papers, including~\cite{sethi_rev,KKLT_disc1,KKLT_disc2,bdgl,KKLT_disc3,lisa-KKLT,KKLT_disc4,KKLT_disc5,KKLT_disc6}. For example, in~\cite{KKLT_disc5} it was argued that the bulk KKLT metric becomes singular because of the need to combine warping and a large internal volume, while in~\cite{KKLT_disc6} holography was used to infer that the first two steps typically lead to small AdS spaces, raising doubts about the validity of the effective field theory approach.

The large--volume scenario (LVS) of~\cite{balasubramanian} (for a review, see~\cite{stringcosmo-rev-quevedo}) is an interesting variant of this setup. Its simplest version relies on Calabi--Yau spaces with some small missing internal portions, so that there are two K\"ahler moduli $T_b$ and $T_s$ and the volume of the space is
\beq
{\cal V} \ = \ \tau_b^\frac{3}{2} \ - \ \tau_s^\frac{3}{2} \ ,
\eeq
with
\beq
\tau_b \ = \ T_b \ + \ \overline{T}_b \ , \qquad \tau_s \ = \ T_s \ + \ \overline{T}_s 
\eeq
and $\tau_s \ll \tau_b$. After the stabilization of the complex--structure moduli and the axion--dilaton system, the K\"ahler potential and the superpotential read
\beq
{\cal K} \ = \ - \ 2 \, \log\left[ {\cal V} \ + \ \frac{\xi}{2} \left(\frac{S \ + \ \overline{S}}{2}\right)^\frac{3}{2} \right]\ , \qquad {\cal W} \ = \ {\cal W}_0 \ +\ A_s\, e^{-\,a_s\,T_s} \ .
\eeq
The extremum of the potential results from a balance between the $\alpha'$ corrections, parametrized by
\beq
\xi \ = \ \frac{\chi(X)\,\zeta(3)}{2\,(2\pi)^3} \ ,
\eeq
with $\chi$ the Euler character of the Calabi--Yau space,
and the non--perturbative effects encoded in the superpotential ${\cal W}$.  The resulting volume,
\beq
\langle {\cal V} \rangle \ = \ \frac{3 \sqrt{\langle \tau_s \rangle} \left| {\cal W}_0\right|}{4 \, a_s\,A_s}\ e^{a_s\langle \tau_s\rangle} \ ,
\eeq
where
\beq
\langle \tau_s\rangle \ = \ \frac{1}{g_s}\ \left(\frac{\xi}{2} \right)^\frac{2}{3} \ ,
\eeq
is exponentially large in units of $\tau_s$, and consequently one is not compelled to demand very low values for ${\cal W}_0$ as in the standard KKLT setup. However, the gravitino mass remains small and one still ends up in AdS, so that the uplift is needed also in this case.

\section{\sc Early Cosmology and Non--Supersymmetric Strings} \label{sec:SUSY_breaking_cosmo}

This last section is devoted to Cosmology. We begin by recalling some basic facts about cosmological models, with emphasis on string--inspired settings for inflation, describing various ingredients that are used to give rise to slow--roll phases. We then elaborate on the simplifications that emerge in cosmological models based on supergravity when supersymmetry is non--linearly realized, and on the difficulties that are generally encountered when trying to embed de Sitter space in String Theory. The final part of this section is devoted to illustrating how the steep tadpole potentials that we met in our discussion of tachyon--free non--supersymmetric ten--dimensional strings of Section~\ref{sec:critical_strings} force a scalar field to emerge from the initial singularity while climbing them, so that the dynamics is generally confined to regions of weak string coupling. We conclude by elaborating on the potential lessons of this ``climbing dynamics'' for the onset of inflation, and on their possible imprints on the low--$\ell$ end of the CMB spectrum. 

\subsection{\sc  FLRW Universe and Acceleration} 

The homogeneity and isotropy of the Universe at large scales are the two basic tenets of Cosmology. A homogeneous and isotropic Universe can be described by a metric of the form
\beq 
ds^2 \ = \ - \  dt^2 \ + \ a^2 (t) 
\left[  dr^2 \ + \ f_k^2(r)\, d \Omega^2 \right] \ , \label{frw1}
\eeq 
where $t$, usually called ``cosmic time'', represents the time felt by observers located at fixed values of the spatial coordinates $(r,\Omega)$, whose mutual distances thus evolve in time as demanded by the ``scale factor'' $a(t)$. There are three classes of Friedmann-Lemaitre-Robertson-Walker (FLRW) solutions of this type, with
\bea
 f_k (r) &=& \sin r \ , \qquad k=1 \qquad {\rm (closed \ universe)} \ , \nonumber \\
 f_k (r) &=&  r \ , \qquad k=0  \qquad {\rm ( flat \ universe)} \ , \nonumber \\
f_k (r) &=& \sinh r \, \qquad k=-1 \qquad  {\rm (open \ universe)}  \ . \label{frw2}
\eea 
They describe a closed, flat or open Universe, and the parameter $k$ is related to the curvature of the spatial slices.  These solutions are determined by the Einstein equations
\beq 
R_{\mu \nu} \ - \ \frac{1}{2} \,g_{\mu \nu} R \ = \  \frac{1}{M_P^2}\, T_{\mu \nu} \ , 
\label{frw3}
\eeq 
written for the class of metrics~\ref{frw1}, using the energy-momentum tensor
\beq 
T_{\nu}{}^{\mu} \ = \ {\rm diag} \  
( - \rho, p , p , p ) \ , \label{frw4}
\eeq 
where $\rho$ and $p$ are density and pressure of the cosmological fluid. The $00$ and $ij$ components of the Einstein equations then reduce to the Friedmann equations
\bea 
H^2 &=& \frac{\rho}{{3}\,M_P^2} \ - \ \frac{k}{a^2} \ , \nonumber \\
\frac{\ddot a}{a} &=& {\dot H} \ +\  H^2 \ = \ 
- \ \frac{1}{{6} \,M_P^2} \left(\rho \ +\  3\, p\right)  
\ , \label{frw5}
\eea 
where 
\beq 
H \ = \ \frac{\dot a}{a} \  \label{frw6}
\eeq 
is the ``Hubble parameter", which is generally a time-dependent quantity on cosmological time scales.   
The conservation of the energy--momentum tensor translates into the condition
\beq 
{\dot \rho} \ + \ 3 H \left(\rho \ + \ p\right) \ = \ 0  \ , \label{frw7}
\eeq 
which is a consequence of eqs.~(\ref{frw5}). 

The second Friedmann equation indicates that the Universe typically decelerates during the expansion and that acceleration can only occur if
\beq 
\rho + 3 p \ < \ 0 \ . \label{frw8}
\eeq 
It is customary to define an equation of state for the fluid
\beq 
p \ = \ w \ \rho \ , \label{frw9}
\eeq 
and then the inequality in eq.~\eqref{frw8} translates into
\beq
\rho \left(1 \ + \ 3\, {w} \right) \ < \ 0 \ . 
\eeq
The energy density $\rho$ is positive for all consistent stable systems, and therefore this condition implies that acceleration can only occur if
\beq 
w  \ < \ - \ \frac{1}{3}  
\ . \label{frw10}
\eeq 
The parameter $w$ is generally time dependent, but it is constant for the simplest and most significant examples in Cosmology:   
\bea 
{\rm radiation-dominated \ Universe} & : &  \qquad w \ = \ \frac{1}{3} \ , \nonumber \\
{\rm matter-dominated \ Universe} & : &  \qquad w \ = \ 0 \ , \nonumber \\
{\rm cosmological \ constant} & : &  \qquad w \ = \ - 1  \ . \label{frw11}
\eea 
Consequently, both radiation and the usual forms of matter (baryonic or cold dark) cannot sustain an accelerated expansion. This is only possible if the energy-momentum tensor includes some exotic contribution like the cosmological constant, or a dynamical (time-dependent) generalization of it that is generally dubbed ``dark-energy''.  

Following standard practice in Cosmology, let us define the critical energy density and the fraction of it contributed by the fluid, as 
\beq 
\rho_c \ = \ 3 M_P^2 H^2 , \qquad 
\Omega \equiv \frac{\rho (t)}{\rho_c (t)} \ . \label{frw12}
\eeq 
One can then recast the first Friedmann equation in~\eqref{frw5} in the form 
\beq 
\Omega (t) \ = \ 1 + \frac{k}{(aH)^2}  \ , \label{frw13}
\eeq
which links the total energy density to the curvature of the Universe as follows:
\bea 
\rho > \rho_c \qquad & \to & \qquad k = + 1 \qquad ({\rm \Omega > 1}) \ , \nonumber \\
\rho = \rho_c \qquad & \to & \qquad k = 0 \qquad \ \ ({\rm \Omega = 1}) \ , \nonumber \\
\rho < \rho_c \qquad & \to & \qquad k = - 1 \qquad ({\rm \Omega < 1}) \ . 
\label{frw14}
\eea 

Current observations indicate that, with good accuracy, we live in a flat Universe, and thus with $k=0$. In this case, and for a cosmological fluid with a constant $w$, integrating the Friedmann equations one finds that, for $w \neq -\,1$,
\beq 
a \ = \ a_0 \left(\frac{t}{t_0}\right)^{\frac{2}{3 (1+w)}} \ , \qquad H \ = \ \frac{2}{3(1+w)} \frac{1}{t} \ , \qquad \rho \ = \ \rho_0 \left(\frac{a}{a_0}\right)^{-3 (1+w)} \ . \label{frw15}
\eeq 
The special case $w=-1$ corresponds to a cosmological constant. The Hubble parameter is then constant in time, and the scale factor is an exponential in terms of the cosmic time $t$:
\beq
a \ = \ a_0 \  e^{H \left(t-t_0\right)} \ .
\eeq

It is convenient to separate the contributions to the energy density of the various components, which evolve differently with the scale factor, letting
\beq 
\Omega_i \ \equiv \ \frac{\rho_i (t)}{\rho (t)} \ ,
\eeq
so that
\beq 
\sum_i \Omega_i \ = \ 1  \ . \label{frw16}
\eeq 
Considering for definiteness the contributions of radiation ($\Omega_r$), matter ($\Omega_m$) and cosmological constant ($\Omega_{\Lambda}$), one can link the current energy density to the present values $H_0$ and $a_0$ of the Hubble parameter and the scale factor $a$, according to
\beq 
H^2 (t) \ = \ \frac{\rho (t)}{3 M_P^2} \ = \ 
H_0^2 \left[ \Omega_m \left(\frac{a_0}{a}\right)^3 \ +\   \Omega_r \left(\frac{a_0}{a}\right)^4
\ + \ \Omega_{\Lambda} \right] \ . \label{frw17}
\eeq 
This decomposition gives a glimpse of the evolution of our Universe: it was dominated by radiation at early times, an epoch that was followed by a matter-dominated era, and only recently it entered an era of accelerated expansion dominated by $\Omega_\Lambda$. 
There is, however, a curious ``coincidence problem", since despite their different scaling properties, the three contributions are comparable today. More precisely, the term $\Omega_\Lambda$, which identifies the ``dark energy'' component, contributes about 70\%, dark matter contributes another 25\%, and finally visible matter contributes about 5\% of the present energy density in the Universe. 

\subsection{\sc  Inflation, and Some String--Inspired Settings} 

In addition to the current stage of mildly accelerated expansion, there is significant evidence for an early stage of very fast, almost exponential, expansion of the Universe. This scenario is usually called inflation~\cite{inflation1,inflation2,inflation3,inflation4,inflation5,inflation6,inflation7,inflation_rev}, and solves various puzzles related to the extreme level of homogeneity and isotropy manifested by the CMB and to the apparent lack of topological defects in our Universe. Even more importantly, it gives rise to small deviations from scale invariance, of quantum origin, that manifest themselves in the Chibisov--Mukhanov tilt~\cite{cm,cm_rev} of the primordial power spectrum of CMB curvature perturbations, which was neatly detected by the {\it Planck} collaboration~\cite{planck_cosmo}. These features would not be guaranteed by the FLRW cosmology. Inflation can greatly dilute inhomogeneities, but should have lasted for a very short time. This accelerated expansion should have been followed by the epoch of reheating, when the energy was converted into the matter that we observe. A cosmological constant could not grant this transition, and one thus needs a mechanism to temporarily simulate its effects. 

The simplest realization of the inflationary scenario postulates the existence of a scalar field called inflaton, here denoted by $\varphi$, with an action  
\beq
{\cal S} \ =\  \, \int d^{4}x\sqrt{-g} \, \Bigl\{\frac{M_P^2}{2} R\ -\  \frac{1}{2}\ (\partial \varphi)^2  \ - \ V (\varphi)  \label{iast1}
\Bigr\} \ .
\eeq
The scalar equation in the FLRW background is then
\beq 
{\ddot \varphi} \ + \ 3 H {\dot \varphi} \ + \ V' \ = \ 0 \ , \label{iast2}
\eeq 
where $V' = \partial_{\varphi} V$, while letting
\beq
\epsilon \ =\  - \ \frac{\dot H}{H^2} \ , \label{iast3}
\eeq 
the second Friedmann equation in~\eqref{frw5} becomes
\beq
\frac{\ddot a}{a} \ = \ H^2 \left(1 \ - \ \epsilon\right) \  \label{iast4}
\eeq 
so that a positive acceleration
requires $\epsilon < 1$. The energy density and pressure stored in the inflaton field are 
\beq 
\rho_{\varphi} \ = \ \frac{1}{2}\, {\dot \varphi}^2 \ +\   V (\varphi) \ , \qquad 
p_{\varphi} \ =\  \frac{1}{2} \, {\dot \varphi}^2 \ -\   V (\varphi) 
\ , \label{iast5}
\eeq 
and using the Friedmann equations~\eqref{frw5} one can obtain the equivalent expression
\beq 
\epsilon \ = \ \frac{{\dot \varphi}^2}{2 M_P^2 H^2} \ . \label{iast6}
\eeq 

The relevant type of evolution for $\varphi$ leading to inflation is therefore the {\it slow-roll} regime, in which the field moves slowly, so that ${\dot \varphi}^2 \ll V$. In this case $\epsilon \ll 1$, and one
also requires that ${\ddot \varphi} \ll 3 H {\dot \varphi}$ and ${\ddot \varphi} \ll V'$, in order to grant a long enough duration to this special dynamics. One can then show that the slow-roll approximation is equivalent to the conditions
\beq 
\epsilon \ \simeq \ \epsilon_V \ \equiv \ \frac{M_P^2}{2} \left( \frac{V'}{V}\right)^2 \ \ll\  1 \ , \qquad 
\eta_V \ \equiv \ M_P^2 \frac{V{''}}{V} \ \ll\  1 \ . \label{iast7}
\eeq 
An important quantity is the number of inflationary $e$-folds, defined as
\beq
N \ = \ \ln \frac{a_e}{a} \ , \label{efold1}
\eeq
where $a_{e}$ and $a$ are the scale factors at the end and at the beginning of inflation. In order to solve the puzzles of early-time Cosmology, a minimum number of about 60 $e$-folds is needed. 
In the slow-roll regime, using the field equations, one can show that 
$N$ can also be conveniently obtained as 
\beq 
N \ = \ \frac{1}{M_P^2} \int_{\varphi}^{\varphi_e} \frac{V}{V'} \ d \varphi 
\ . \label{efold2}
\eeq 

We can now present some string--inspired examples that lead to a slow--roll phase. 
\begin{itemize}
\item \textbf{Volume modulus in a KKLT-like model with antibrane tension}, prior to adding non-perturbative effects.

Assuming that all fields except the real part of the volume modulus $t = Re \ T$ are stabilized by the dynamics, and adding an $\overline{D3}$ brane tension, the effective Lagrangian is described by
\beq 
{\cal L}\  = \ - \ \frac{3}{4 t^2} (\partial t)^2 \ - \ \frac{c}{t^n} 
\ , \label{iast8}
\eeq 
with $c$ a positive constant.
As we have seen in our discussion of the KKLT model, $n=2$ for an antibrane at the tip of a Klebanov-Strassler throat, while $n=3$ in the presence of a weak warping, or ignoring it altogether. This type of effective action is also obtained by dimensionally reducing the Sugimoto model~\cite{sugimoto} to four dimensions and retaining only the volume modulus as a light field in the effective field theory \cite{bsb_cosmology1}. One can obtain a canonically normalized kinetic term via the field redefinition
\beq 
t \ = \ e^{- \sqrt{\frac{2}{3}} \chi} \ ,  \label{iast9} 
\eeq
and then
\beq
\quad 
{\cal L} \ =\  -\ \frac{1}{2 } (\partial \chi)^2 \ - \ c \ e^{n \sqrt{\frac{2}{3}} \chi} \ .
\eeq

\item \textbf{Radius/radion with Scherk-Schwarz supersymmetry breaking potential.} 

In this case the relevant field is an internal radius, with different boundary conditions for bosons and fermions, breaking supersymmetry. At one loop, a vacuum energy is generated similar to the Casimir effect in quantum field theory. The relevant starting Lagrangian is
\beq 
{\cal L} \ =\  - \ \frac{3}{ R^2} \left(\partial R\right)^2 \ - \ \frac{c}{R^4} 
\ . \label{iast10}
\eeq 
The kinetic term is actually as in the preceding example, as can be seen identifying $R^2$ with $t$. 
In this case, one can canonically normalize the kinetic term via the redefinition
\beq 
R \ =\  e^{- \,\frac{\chi}{\sqrt{6}}} \ , 
\eeq
obtaining
\beq
{\cal L} \ =\  -\  \frac{1}{2 } (\partial \chi)^2 \ -\  c \ e^{ \sqrt{\frac{8}{3}} \chi} \ . \label{iast11} 
\eeq
In both cases, and actually in all known examples drawn from String Theory, the scalar potential is an exponential, which can be written more generally as 
\beq 
{\cal L} \ = \ - \ \frac{1}{2 } (\partial \chi)^2 \ -\  V_0 \ e^{ \frac{\lambda\, \chi}{M_P}} \ . \label{iast12} 
\eeq
In order to have a bounded potential, in what follows we shall consider $V_0 > 0$, and moreover one can also assume that $\lambda > 0$, up to a redefinition of $\chi$. 

The cosmological solutions
for the exponential potential are surprisingly rich and will be discussed in detail in Section~\ref{sec:climbing}. For the time being, let us point out that, if $\lambda < \sqrt{6}$, the Friedmann equations have a simple exact solution discovered long ago by Lucchin and Matarrese \cite{lm}, which takes the form
\bea 
\chi (t) & = & - \ \frac{2 M_P}{\lambda} \ln \left( \sqrt{\frac{V_0 \lambda^4}{2 (6-\lambda^2)}} \ \frac{t}{M_P} \right) \ , \nonumber \\
a (t) & \sim & t^{\frac{2}{\lambda^2}}   \label{iast13} 
\eea 
when expressed in terms of the cosmic time $t$.
Positive acceleration in this case is achieved for $\lambda < \sqrt{2}$. The slow-roll parameters in eq. (\ref{iast7}) are computed to be
\beq 
\epsilon_V \ = \ \frac{\lambda^2}{2} \, \qquad \eta_V \ =\  \lambda^2 
\ . \label{iast14} 
\eeq 
Clearly the numerical values of $\lambda$ in the preceding examples of eqs.~(\ref{iast9}) and (\ref{iast11}) are too large to give rise to an acceleration, let alone to attain a slow-roll inflationary regime. This is actually a generic feat for string compactifications: the scalar potentials are typically too steep for both an early phase of slow-roll inflation or the current accelerating phase of the Universe. Still, as we shall see, steep potentials have a surprise in store.   

There are various proposals to overcome these difficulties, which involve D-branes or string-inspired setups. Some of them use non--canonical kinetic terms for scalar fields, D-brane excitations and supergravity. 

\item \textbf{DBI inflation}~\cite{dbi-inflation}. 
This setting relies on the Dirac--Born--Infeld action that can describe D-brane motions in the internal space. The action is typically of the form
\beq 
{\cal L} \ = \ f (\phi) \sqrt{ 1 \ +\ f (\phi)^{-1} (\partial \phi)^2 } \ -\  V(\phi) \ , \label{iast15} 
\eeq 
with $f(\phi) \sim \phi^4$ for D3 positions. The non--linear nature of the dynamics affects the sound speed of curvature perturbations, which becomes
\beq
c_s \ = \ \left[1 \ - \ f (\phi)^{-1} \left(\dot{\phi}\right)^2 \right]^\frac{1}{2} \ ,
\eeq
and its reduction has the effect of enhancing their power spectrum by a factor $\frac{1}{c_s}$ while also depressing the tensor-to-scalar ratio by a factor $c_s$, since tensor perturbations are not sensitive to this effect.
In addition to these features, non--gaussianities are also enhanced by a factor $\frac{1}{c_s^2}$, and this setting embodies, after all, an interesting and surprising lesson: even \emph{fast} D-brane motions in an internal space can sustain inflation.

\item \textbf{$\alpha$-attractors}~\cite{alpha-attractors}.
This setting is motivated by the typical logarithmic K\"ahler potentials of moduli fields in string-inspired supergravity, 
\beq 
{\cal K} \ = \ - \,3 \,\alpha \ln \left(1\,-\,|\Phi|^2\right) \ ,
\eeq
which give rise to the non--canonical kinetic terms
\beq
{\cal L}_{\rm kin} \ =\  - \ \frac{3 \alpha}{\left(1\,-\,|\Phi|^2\right)} \ |\partial \Phi|^2 \ . \label{iast16} 
\eeq
Assuming that the imaginary part of the complex field $\Phi$ be stabilized by the dynamics and adding a scalar potential, the effective action to analyze only depends on its real part $\varphi$ and is
\beq 
{\cal L} \ =\   - \ \frac{3 \alpha}{\left(1\ -\ \varphi^2\right)}\, (\partial \varphi)^2 \ -\  V (\varphi) \ . \label{iast17} 
\eeq 
The kinetic term becomes canonical with the change of variables
\beq 
\varphi \ =\ \tanh \frac{\chi}{\sqrt{6 \alpha}} 
\ , \label{iast18} 
\eeq 
and the action finally takes the form
\beq 
{\cal L} \ =\   -\  \frac{1}{2} (\partial \chi)^2 \ - \ V \left(\tanh \frac{\chi}{\sqrt{6 \alpha}} \right) \ , \label{iast19} 
\eeq
so that when working with the new field variable $\chi$ the potential flattens significantly. For example, a polynomial potential $V(\varphi) \sim \ \lambda \varphi^n$  becomes exponentially flat in terms of the canonically normalized field $\chi$. In this case,  one can link the number of $e$-folds to the slow-roll parameters in eq.~(\ref{iast7}) (\ref{efold2}) according to
\beq 
\epsilon_V \ \simeq \ \frac{3 \alpha}{4 N^2} \ , \qquad \eta_V \ \simeq \ \frac{1}{N}  \ , \label{iast20} 
\eeq 
and they are clearly small for the least number of $e$-folds needed. 
\end{itemize}

One potential problem with this scenario (and actually with many others) is that the candidate inflaton
experiences large, super-Planckian  excursions in field space. There are various arguments pointing out that, along such very long trajectories, additional light degrees of freedom can spoil the validity of the effective field theory. There are several slightly different formulations of this paradigm, which is usually referred to as the distance conjecture \cite{distance-conj} or the trans-Planckian conjecture \cite{transplanckian-conj}. 

\subsection{\sc Nonlinear Supersymmetry and Minimal Models of Inflation} 

Supergravity models of inflation typically contain many scalar fields, in addition to the inflaton.  However, most scalars can become heavy and decouple during the inflationary phase.
When discussing realizations of nonlinear supersymmetry, we have seen that when a scalar becomes heavy and decouples, it can be eliminated from the spectrum using an appropriate superfield constraint, compatibly with supersymmetry, which remains present, albeit in a non--linear phase.
This option was first implemented in cosmological models in~\cite{adfs}, where a Starobinsky model with non--linearly realized supersymmetry was constructed, relying on a nilpotent goldstino superfield. 

In this case the K\"ahler potential ${\cal K}$ and the superpotential ${\cal W}$ read
\beq
{\cal K}\ = \ - \ 3 \,\log \left[T \,+\, \overline T \,-\, 2\, X \,\overline X  \right]  \ , \qquad {\cal W} \ =\ 6\, m\, X\,T\ + \ fX\ + \ W_0 \ , \label{mm1}
\eeq 
and the goldstino superfield $X$ is subject to the quadratic constraint $X^2\ = \ 0$, as in Section~\ref{section:constrainedSuperfields}.  
The parameter $f$ is redundant and can be set to $-3m$ by appropriate rescalings. 
The parameter $m$ is instead physical and controls the scale of inflation, 
while $W_0$ controls the gravitino mass but does not enter the scalar potential. 
The only scalars that participate in the dynamics are the real and imaginary parts of the lowest component of $T$
\beq 
T \ = \  e^{\sqrt{\frac{2}{3} \varphi}} \ + \ i 
\sqrt{\frac{2}{3}} \, a \ , \label{mm2}
\eeq 
where $\varphi$ is the inflaton, while $a$ is an axion that is effectively heavy during inflation, due to the dynamics. In contrast, the goldstino superfield $X$ only contributes Fermi modes. The relevant portion of the Lagrangian is
\beq 
{\cal L} \ = \  - \ \frac{1}{2} \,\left(\partial \varphi\right)^2 \ - \ \frac{1}{2} \,e^{- 2 \sqrt{\frac{2}{3}} \varphi}  (\partial a)^2 \ - \  V \ , 
\eeq
where
\beq
V \ = \  \frac{1}{12} \left( m \ + \ f \, 
e^{- \sqrt{\frac{2}{3}} \varphi} \right)^2 \ + \  \frac{m^2}{18} 
e^{- 2 \sqrt{\frac{2}{3}} \varphi}
a^2 \ . \label{mm3}
\eeq
After a shift of $\varphi$ and ignoring the heavy axion, one recovers the Starobinsky potential of eq.~\eqref{starobinsky}. 

This model approaches a non-supersymmetric Minkowski vacuum at the end of inflation, 
where the description in terms of a nilpotent $X$ breaks down because $F^X$ vanishes. 
In order to rely on the same effective description throughout the process, the vacuum at the end of inflation must be suitably modified.
Versatile models with this property were discussed in \cite{DallAgata:2014qsj}: they all include an inflationary epoch, but supersymmetry remains broken in the $X$ sector when inflation ends, albeit with a vanishing vacuum energy.
The models studied in~\cite{Kallosh:2014hxa,Scalisi:2015qga} improved further on this, by introducing a non-vanishing vacuum energy at the end of inflation that can accommodate the current phase of our Universe.
The idea that KKLT-like de Sitter vacua can be constructed by resorting to constrained superfields was first proposed in~\cite{Ferrara:2014kva}.

The success of inflationary models with nilpotent superfields led to further investigations based on different types of constrained superfields.
At first sight, one could construct even simpler models by imposing constraints that eliminate all additional fields, aside from the inflaton.  
The first efforts used a new chiral superfield subject to a constraint that eliminates all component fields, apart from a single scalar \cite{rocek,nonlinear,brignole,ks,kt}.
Cosmological models of this type were studied, for example, in~\cite{Kahn:2015mla,Ferrara:2015tyn,Carrasco:2015iij}.
In these constructions, the constraint $X^2=0$ is accompanied by an additional constraint on the chiral superfield that contains the inflaton,
\beq 
X \left({\cal A} \ - \ \overline{\cal A}\right) \ = \ 0 \ , \qquad \overline{\cal D}_{\dot \alpha}\, {\cal A} = 0 \ . \label{orthogonal1}
\eeq 
In this fashion, one eliminates all components fields, aside from the real part of the lowest component $a$ of ${\cal A}$.
One can verify this by deducing that eq.~\eqref{orthogonal1} is equivalent to the three constraints
\begin{eqnarray}
&  |X|^2 \left({\cal A} \ - \ \overline{\cal A}\right) \ = \ 0  \quad &{\rm : \ eliminates } \quad \text{Im}{\cal A}| \ , \nonumber \\
& |X|^2 \, {\cal D}_{\alpha} \,{\cal A} \ = \ 0
\quad  &{\rm : \ eliminates \ the \ fermion} \quad {\cal D}_{\alpha} {\cal A} | \ , \nonumber \\
& |X|^2  \,{\cal D}^2 \,{\cal A} \ =  \ 0 \quad &{\rm : \ eliminates \ the \ auxiliary \ field} \quad    {\cal D}^2 {\cal A}| \ .   \label{orthogonal3}
\end{eqnarray}
In global supersymmetry, the component fields thus eliminated become functions of the goldstino and the real scalar $a$, compatibly with the supersymmetry algebra.
In supergravity, the component fields of the supergravity multiplet also enter these functions.

These models leave at first sight much freedom for model building.
Indeed, letting
\beq 
{\cal K} \ = \ X \,\overline X \ - \ \frac14 \left({\cal A} \ - \ \overline {\cal A}\right)^2 \ , \qquad
{\cal W} \ = \ g({\cal A}) \ +\ X \,f({\cal A}) \,
\eeq 
with $\overline{ f(z)} \,=\, f (\overline z)$ and $\overline{ g(z)} \,= \,g (\overline z)$, one
finds the Lagrangian
\begin{eqnarray}
e^{-1} {\cal L}|_{G=0} &=& \frac12\, R
\ + \ \frac{1}{2}\, \epsilon^{klmn} \left(\overline \psi_k \overline \sigma_l {\cal D}_m \psi_n \ - \ \psi_k \sigma_l {\cal D}_m \psi_n\right)
\\[2mm] \nonumber
&-& \frac12 \,\partial^m a \, \partial_m a
\ -\  g(a) \left( \overline \psi_a \overline \sigma^{ab} \overline \psi_b \ + \ \psi_a \sigma^{ab} \psi_b \right)
\ - \ \left(f^2(a) \ - \ 3 \, g^2(a)\right) \, ,
\end{eqnarray}
whose scalar potential can be adjusted to take any form.
There are problematic cosmological implications at the end of inflation~\cite{Hasegawa:2017hgd}, unless the function $g$ is finely tuned.
The constraint of eq.~\eqref{orthogonal1} eliminates the auxiliary field, among other component fields, and
it was shown in~\cite{DallAgata:2016syy} that this can possibly require non-unitary UV physics. This was confirmed in \cite{Dudas:2021njv}, where it was shown that models based on the ``orthogonal'' constraint of eq.~(\ref{orthogonal1}) have potentially superluminal propagation for the gravitino, and in \cite{Bonnefoy:2022rcw}, where this feature was related to positivity constraints of certain operators in the low-energy goldstino-inflaton Lagrangian. However, these problems disappear if only the first two constraints in (\ref{orthogonal3}) are retained, abandoning the last one, which eliminates the auxiliary field.

As we have seen, in addition to the scalar partner of the inflaton and the fermion the constraint of eq.~\eqref{orthogonal1} also removes the auxiliary field from the spectrum, giving rise to the problems of the effective field theory that were highlighted in~\cite{Hasegawa:2017hgd}.
The extra constraint can be avoided following two different proposals discussed in~\cite{Dalianis:2017okk} and in~\cite{Bonnefoy:2022rcw}, which we can now review.

The setup discussed in~\cite{Dalianis:2017okk} aims at constructing models in which the inflaton lives in a chiral superfield that is only subject to the constraint
\beq 
X \,\overline X \left(\Phi \ - \ \overline \Phi\right) \ = \ 0 \ .
\eeq 
In this fashion,  only the imaginary part of the lowest component of $\Phi$ is removed, while the other component fields are not affected. 

For model building purposes one can choose, for example,
\beq 
{\cal K} \ =\  X \,\overline X  \ - \ \frac14 \left(\Phi \ - \ \overline \Phi \right)^2 \ , \qquad   {\cal W} \ = \  f(\Phi) \,  X \ + \ g(\Phi) \,  ,
\eeq 
and then the potential for the canonically normalized real scalar $\phi$ takes the form 
\begin{eqnarray}
\label{XFlagr22}
V \ = \  |f(\phi)|^2 \ +\ 2 \left| g'(\phi) \right|^2 \ - \ 3 \left|g(\phi)\right|^2   \ . 
\end{eqnarray}  
This scalar potential has now the form expected from standard supergravity, 
and in particular the term $2 \left| g'(\phi) \right|^2$ originates from integrating out the auxiliary field $F^\Phi$. 
The bosonic sector in this setup is \emph{minimal}: it contains only gravity and the inflaton.

The inflationary physics of models of this form was studied in \cite{Dalianis:2017okk}, where it was shown that they provide a versatile framework for model building, while post-inflationary physics does {\it not} suffer from the pathologies that we have previously mentioned.
In particular, the gravitino is not overproduced, once a hierarchy between the inflationary scale and the supersymmetry breaking scale due to the nilpotent $X$ superfield is invoked.
The inflationary scale is fixed by the Hubble scale during inflation, but in some models it is
characterized by the inflaton mass $m_\phi$ in the vacuum,
while supersymmetry breaking is characterized by the gravitino mass.
In this case, following \cite{Nilles:2001ry}, one requires that  $m_\phi|_\text{vacuum} \gg m_{3/2}|_\text{vacuum}$
and $\langle F^\Phi \rangle |_\text{vacuum} = 0$ while $\langle F^X \rangle \ne  0$.
However, during the preheating phase supersymmetry breaking is dominated by the inflaton energy density, and the longitudinal component of the gravitino should be identified with the inflatino $\chi_\alpha^\Phi$, assuming $\langle F^\Phi \rangle |_\text{inflation}>>\langle F^X\rangle$,  rather than with the {\it true vacuum goldstino} $G_\alpha$ \cite{Nilles:2001ry}.

In~\cite{Bonnefoy:2022rcw} the fermion partner of the inflaton was also removed from the spectrum, with the combined use of the constraints
\beq 
\label{Im+Ferm}
X \,\overline X \left(\Phi \ - \ \overline \Phi\right) \ =  \ 0 \ , \qquad  X \,\overline X \,{\cal D}_\alpha \Phi \ = \ 0  \,.
\eeq 
In this fashion, the inflaton superfield is brought to a minimal form, so that in the unitary gauge
\beq 
\Phi|_{G=0} \ = \ \phi \ + \ \Theta^2 \,F^\Phi \, ,
\eeq 
while the full solution of the constraints can be found in \cite{Bonnefoy:2022rcw}.
In models with an inflaton superfield of the form~\eqref{Im+Ferm}, the scalar potential has the standard supergravity form, with a {\it minimal spectrum}: a graviton, a massive gravitino and a real inflaton.
For example, one can choose
\beq 
{\cal K} \ = \ -\ \frac14\left(\Phi \ - \ \overline \Phi\right)^2 \ +\  |X|^2 \ , \qquad {\cal W} \ = \ f(\Phi) \,X \ + \ g (\Phi) \ ,
\eeq 
and then the scalar potential takes the standard supergravity form 
\beq
V \ = \  |f(\phi)|^2  \ + \  |g'(\phi)|^2  \ - \  3\, | g(\phi) |^2 \ .
\eeq
In particular, resorting to the choice proposed in \cite{DallAgata:2014qsj}, one obtains
\beq 
f \ = \ \sqrt 3\, g(\phi) \ , 
\eeq 
and
\beq
V \ = \  |g'(\phi)|^2 \ ,
\eeq
which leaves indeed much freedom for model building.
The physical properties of models with the constraints in~\eqref{Im+Ferm} were analyzed in \cite{Bonnefoy:2022rcw}, where it was shown that they {\it do not} suffer from physical inconsistencies.

\subsection{\sc Difficulties with de Sitter vacua in string theory}  

In Section~\ref{sec:KKLT}, we elaborated on some criticisms concerning the KKLT scenario, which is the most popular framework for addressing acceleration and the emergence of de Sitter space in string theory.  Other string and D-brane examples that address the acceleration/dark energy issue are reviewed in \cite{stringcosmo-rev-mcallister} and in~\cite{stringcosmo-rev-quevedo}. Essentially all explicit examples entail a potential breakdown of calculability. Obtaining a constant positive vacuum energy requires two ingredients that are notoriously difficult to attain: stabilization of all moduli fields and supersymmetry breaking to uplift vacuum energy to positive values, together with a very small dark energy, which is certainly the hardest problem in modern physics; see e.g. \cite{rev-cosmo-const,bousso-polchinski}. Moduli stabilization requires several perturbative and non--perturbative contributions to the moduli potential, and calculability requires a careful evaluation of the dominant contributions, while other potential contributions must be subleading and irrelevant in the analysis. This is delicate and difficult to achieve in practice. Supersymmetry breaking, the main subject of our review, is generally plagued by various instabilities and control issues, and these are currently unsolved problems. The difficulty of obtaining a controllable de Sitter vacuum was even elevated to a conjecture, the de Sitter conjecture~\cite{desitter-conj}, which is arguably the most controversial one in the Swampland program. 

An observable positive dark energy, on the other hand, does not necessarily have a unique solution related to the existence of a pure (small and positive) cosmological constant. The vacuum energy could be dynamical with a time-dependent equation of state. 
One popular scenario called quintessence \cite{quintessence} features a very light scalar field that rolls along a very flat scalar potential, with an energy density that is almost constant on cosmological time scales. This is a priori an attractive option, since, as we have seen, in the presence of broken supersymmetry String Theory does generate runaway potentials for moduli fields, which asymptote to zero at infinity.  There are, however, many potential problems that hamper the realization of quintessence in String Theory:
\begin{itemize}
\item In order for the scalar to evolve today, its mass must be at most comparable to the present Hubble scale, which is about $10^{-33}$ eV. If such a field were to couple to the observable sector, even with couplings of gravitational strength, it would give rise to a time dependence of the fundamental couplings, which are highly constrained by cosmological data \cite{uzan} and/or would modify gravitational interactions, with observable consequences in Astrophysics and in gravity experiments.   

\item It is difficult to engineer a model with a very light scalar field whose mass is stable under radiative corrections. However, axion-like fields, with derivative couplings to matter, could potentially fulfill this requirement. 

\item Although runaway potentials are ubiquitous in String Theory, as we already stressed, they are way too steep to implement inflation and, therefore, quintessence, which require much flatter potentials. 
\end{itemize}
\begin{figure}[ht]
\centering
\includegraphics[width=85mm]{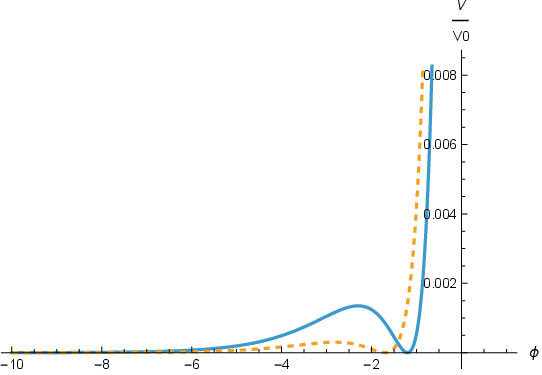}
%\vspace{-5cm}
\caption{\small The ratio $\frac{V}{V_0}$ in eq.~\eqref{des} for $\phi_2=\phi_1-\log 3$, for $\phi_1=-1.2$ (solid line) and for $\phi_1=-1.7$ (dashed line). Note how moving the extrema further into the perturbative region lowers the potential barrier.}
\label{fig:desitter}
\end{figure}

An example can help illustrate the types of difficulties that one encounters when trying to realize perturbative de Sitter vacua in String Theory. The simple potential
\beq
V \ = \ V_0 \left[ e^{\phi_1+\phi_2 + \phi} \ - \ \frac{1}{2} \left( e^{\phi_1} + e^{\phi_2}\right)\, e^{2 \phi} \ + \ \frac{1}{3} \, e^{3 \phi} \right] \ , \label{des}
\eeq
with $V_0$ a positive constant and $\phi$ the dilaton, so that $e^\phi$ is the string coupling, contains contributions from three successive orders of perturbation theory, and 
has extrema at $\phi=\phi_1$ and at $\phi=\phi_2$, which lie within the perturbative region if $\phi_1$ and $\phi_2$ are both negative. The potential is also positive at both extrema, provided
\beq
\left|\phi_1 \ - \ \phi_2 \right| \ < \ \log 3 \ ,
\eeq
and the extremum occurring for a larger value of $\phi$, say $\phi_1$, is a minimum, while the other is a maximum, since
\beq
V''(\phi_1) \ = \ V_0 \ e^{2\phi_1}\left( e^{\phi_1} \ - \ e^{\phi_2} \right) \ , \qquad V''(\phi_2) \ = \ - \ V_0 \ e^{2\phi_2} \left( e^{\phi_1} \ - \ e^{\phi_2} \right) \ .
\eeq
However, the requirement that the extrema lie in the perturbative region demands that two unusual conditions be met: the lowest--order contribution to the potential should be suppressed with respect to the others, and the next one should be similarly suppressed with respect to the third. To the best of our knowledge, no mechanism attaining this for the tension term with respect to torus term in orientifolds, is known. On the other hand, the relative suppression of the first term with respect to the second, without removing it altogether, was achieved in Scherk--Schwarz compactifications of closed strings, demanding that Fermi--Bose degeneracy be present at the massless level, as in~\cite{admavr,kounpart,hervenbnf}.  Moreover, the second term must be negative. Only if these requirements are met can one have a metastable de Sitter vacuum that lies within the perturbative region, and then the higher--order corrections should reasonably play a minor role. Note also that lower values for $\phi_1$ and $\phi_2$ have the effect of lowering the barrier height.

The reader can find more details about these and other problems, together with proposals for potential solutions, in the review~\cite{stringcosmo-rev-quevedo}. For different approaches, see also~\cite{tatar,andriot}.

\subsection{\sc The Climbing Scalar: a Mechanism for the Onset of Inflation ?} \label{sec:climbing}

We can now describe a surprising feature of the some very steep exponential potentials that arise in String Theory.
Cosmological solutions for non--supersymmetric ten--dimensional strings were introduced in~\cite{dm_vacuum,russo}, but a closer look at generic exponential potentials of the type
\beq
V = T \, e^{\gamma \phi} \label{climbing_pot}
\eeq
revealed~\cite{bsb_cosmology1} a sharp change of behavior that occurs precisely when, increasing $\gamma$, one reaches the orientifold value $\gamma_c$ ($\frac{3}{2}$ in ten dimensions, in the Einstein frame, with the standard string normalization of the scalar kinetic term, so that
\beq
{\cal S} \,=\, \frac{1}{2\,k_{D}^2}\, \int d^{D}x\sqrt{-g}\Bigl\{ R\ - \frac{1}{2}\, (\partial\phi)^2  \ - \ V \label{einstein_frame_bsb_d}
\Bigr\} \ ,
\eeq
or overcomes it. In fact, for $\gamma \geq \frac{3}{2}$, the solutions can only describe expanding Universes where the string coupling is \emph{bounded from above} during the whole evolution! 

The critical value depends on the dimension $D$ of spacetime and also on the convention for the kinetic term of the inflation. In the Einstein frame, with string normalization for the scalar kinetic term as in eq.~\eqref{einstein_frame_bsb_d}
\beq
\gamma_c \ = \ \sqrt{\frac{2(D-1)}{D-2}} \ , \label{gc_string}
\eeq
while with canonical normalization
\beq
{\cal S} \,=\, \int d^{D}x\sqrt{-g}\Bigl\{ \frac{1}{2\,k_{D}^2}\, R\ - \frac{1}{2}\, \left(\partial \tilde{\phi}\right)^2  \ - \ V \label{einstein_frame_ft_d}
\Bigr\} \ ,
\eeq
the exponent becomes
\beq
\tilde{\gamma}_c \ = \ 2\,\kappa_D\,\sqrt{\frac{D-1}{D-2}} \ , \label{gc_ft}
\eeq
which becomes $\kappa\,\sqrt{6}$ in four dimensions.

The study of general exponential potentials as in eqs.~\eqref{einstein_frame_bsb_d} or \eqref{einstein_frame_ft_d} in inflationary Cosmology has a long history~\cite{cosmo_power1,cosmo_power2,cosmo_power3,cosmo_power4,cosmo_power5,cosmo_power6,cosmo_power7,cosmo_power8,cosmo_power9,cosmo_power10}. Much of the early activity was stimulated by the work of Lucchin and Matarrese~\cite{lm}, who showed that exponential potentials can support a phase of power--like inflation for low enough values of $\gamma$. We already met their solution, for the case of four dimensions and the action~\eqref{einstein_frame_ft_d}, in terms of cosmic time, in eqs.~\eqref{iast13}. For $\gamma$ below a critical value ($\kappa\,\sqrt{6}$ in four dimensions with canonical normalization), the complete solutions we are about to derive do approach the Lucchin--Matarrese attractor at late times. On the other hand, when $\gamma$ reaches $\gamma_c$ or overcomes it, the Lucchin--Matarrese attractor ceases to exist and the nature of the solutions changes drastically.

The key step that led to the exact solutions in~\cite{dm_vacuum}, and then to their extensions for arbitrary values of $\gamma$ in~\cite{russo}, was to formulate the dynamics in terms of a special choice of parametric time $\tau$, via a convenient gauge choice. We can now trace these steps working in a generic space--time dimension $D$ with the action~\eqref{einstein_frame_bsb_d},
in the class of metrics
\beq
ds^{\,2} \ = \ - \ e^{\,2\,{\cal B}(t)} \, dt^2 \ + \ e^{\,2\,{\cal A}(t)}\
d{\bf x} \,\cdot \, d{\bf x} \ ,
\eeq
where the equations of motion read
\begin{eqnarray}
&&(D-2)(D-1) \dot{\cal A}^{\,2} \ - \ \frac{1}{2}\,\dot{\phi}^{\,2} \ = \  e^{2{\cal B}} \ V \ , \nonumber \\
&& \ddot{\phi} \ + \ \dot{\phi} \left[(D-1) \dot{\cal A} - \dot{\cal B} \right] \ +
\ e^{2\,{\cal B}}\, V' \ = \
\ 0 \ , \label{system_climb}
\end{eqnarray}
where ``dots'' indicate derivatives with respect to $t$.
Combining the two redefinitions
\beq
{\cal A} \ = \ \frac{a}{D-1} \ , \qquad \phi \ = \ \varphi \,\sqrt{\frac{2(D-2)}{D-1}} \ ,
\eeq
so that
\beq
\varphi \ = \ \frac{\gamma_c}{2}\, \phi \ , \label{phivsvarphi}
\eeq
with the gauge choice
\beq
V \, e^{2\,{\cal B}} \ = \ M^2 \ ,
\eeq
which is possible if the potential never vanishes, and with the introduction of the dimensionless parametric time
\beq
\tau \ = \ M\,t\, \sqrt{\frac{D-1}{D-2}} 
\eeq
reduces eqs.~\eqref{system_climb} for an expanding Universe to
\begin{eqnarray}
&&\dot{a}^{\,2} \ - \ \dot{\varphi}^{\,2} \ = \  1 \ , \nonumber \\
&& \ddot{\varphi} \ + \ \dot{\varphi} \sqrt{1\,+\,\dot{\varphi}^{\,2}} \ +
\ \frac{V'}{2\, V} \left(1\,+\,\dot{\varphi}^{\,2} \right) \ = \
\ 0 \ ,  \label{system_climb2}
\end{eqnarray}
where now ``dots'' indicate derivatives with respect to $\tau$, although we do not change notation.
\begin{figure}[ht]
\centering
\includegraphics[width=60mm]{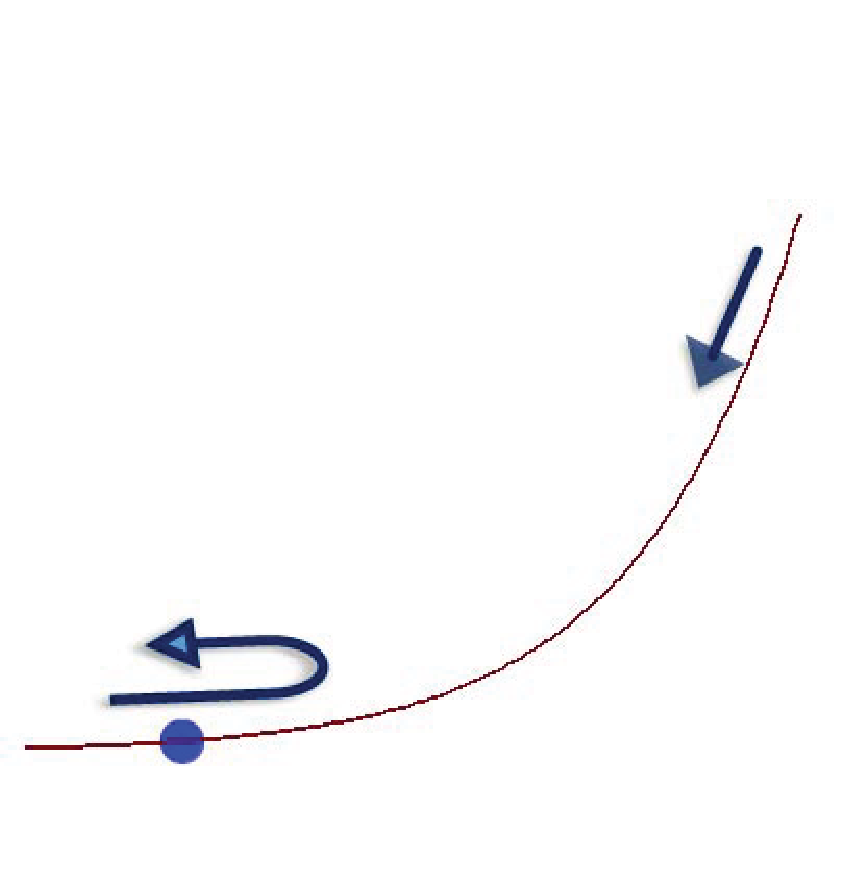}
%\vspace{-5cm}
\caption{\small The two widely distinct scenarios of a climbing and a descending scalar.}
\label{fig:climbing}
\end{figure}

Parametrizing the exponential potential as
\beq
V \ = \ M^2 \, e^{2\,\lambda\,\varphi} \ = \ M^2 \, e^{\lambda\,\gamma_c\,\phi}  \ , \label{Vphivarphi}
\eeq
where the second expression is obtained using eq.~\eqref{phivsvarphi},
leads finally to the scalar equation
\beq
\ddot{\varphi} \ + \ \dot{\varphi} \sqrt{1\,+\,\dot{\varphi}^{\,2}} \ +
\ \lambda \left(1\,+\,\dot{\varphi}^{\,2} \right) \ = \
\ 0 \ .
\eeq
Close to the initial singularity, where $\left|\dot{\varphi}\right|$ is large, this reduces to
\beq
 \ddot{\varphi} \ + \ \dot{\varphi} \left|\dot{\varphi}\right| \ +
\ \lambda \, \dot{\varphi}^{\,2} \ = \ 0 \ , \label{eq_lim}
\eeq
which clearly admits solutions of the type
\beq
\dot{\varphi} \ \sim \ \frac{C}{t} \ .
\eeq
However, substituting this limiting form in eq.~\eqref{eq_lim}, the existence of non--trivial solutions is subject to the constraint
\beq
1 \ = \ {\left|C\right|}\left(1 \ + \ {\lambda}\,\epsilon \right) \ ,
\eeq
where $\epsilon$ denotes the sign of $C$.
The peculiar structure of this algebraic equation lies at the heart of the climbing phenomenon. If $\epsilon = -\,1$, which corresponds to a ``descending scalar'' with $\dot{\varphi}<0$, the equation cannot be solved if $\lambda$ exceeds one, or equivalently if $\gamma$ exceeds the critical value~\eqref{gc_string}.
In ten dimensions this ``critical'' value is precisely $\gamma=\frac{3}{2}$, as can be seen from eqs.~\eqref{Vphivarphi} and \eqref{gc_string}, the value that characterizes the two non-tachyonic orientifolds or, more generally, the (projective)disk contributions to the Polyakov series. The other ten--dimensional non--tachyonic option with broken supersymmetry, the $SO(16) \times SO(16)$ heterotic model, has $\gamma=\frac{5}{2}$, which lies beyond this critical value, and therefore climbing is also the only option in that case.

The Einstein--frame value of $\gamma=\frac{3}{2}$ for the non--tachyonic non--supersymmetric orientifolds in ten dimensions thus separates two vastly different dynamical regimes. At least another scalar mode enters the compactifications to lower dimensions, the one parametrizing the internal volume, which mixes with the dilaton. However, as shown in~\cite{bsb_cosmology3,bsb_cosmology4}, one combination of the two retains a critical potential for all lower dimensions $D<10$. Hence, up to the stabilization of the second scalar, the ``critical'' behavior can persist in lower dimension.

Proceeding along the lines of~\cite{dm_vacuum}, the exact solutions can be built solving identically the first of eqs.~\eqref{system_climb} via the hyperbolic parametrization~\cite{bsb_cosmology1,bsb_cosmology2,bsb_cosmology3,bsb_cosmology4,bsb_cosmology5,bsb_cosmology6,bsb_cosmology7}
\beq
\dot{a} \ = \  \cosh\zeta \ , \qquad \dot{\varphi} \ = \ \sinh\zeta \ ,
\eeq
and then solving by separation of variables the resulting first--order equation for $\zeta$, which reads
\beq
\dot{\zeta} \ +\ \sinh\zeta \ + \ \lambda \ \cosh\zeta \ = \ 0 \ .
\eeq
In fact, it is convenient to let
\beq
y \ = \ e^\zeta \ ,
\eeq
thus turning the preceding equation into
\beq
\frac{dy}{1\,-\,\lambda \,-\, \left(1\,+\,\lambda \right)y^2} \ = \ \frac{1}{2}\ d\tau \ . \label{eq_climbing}
\eeq

For $\lambda< 1$, it is also convenient to let
\beq
\sigma \ = \ \sqrt{\frac{1\,+\,\lambda}{1\,-\,\lambda}} \ ,
\eeq
and then eq.~\eqref{eq_climbing} is solved by
\beq
\log \left|\frac{1\,+\,\sigma\,y}{1\,-\,\sigma\,y} \right| \ = \ \tau\,\sqrt{1\,-\,\lambda^2}  \ , \label{logsigmay}
\eeq
up to a choice of initial time. There are thus two different classes of solutions:
\begin{enumerate}
    \item if $\sigma\, y < 1$ eq.~\eqref{logsigmay} is solved by
    \beq
y \ = \ \frac{1}{\sigma} \ \tanh\left(\frac{\tau}{2}\,\sqrt{1\,-\,{\lambda^2}}\right) \ ,
\eeq
    and then
    \bea
\dot{\varphi} &=& \frac{ \sqrt{\frac{1\,-\,\lambda}{1\,+\,\lambda}}\,
\tanh\left(\frac{\tau}{2}\,\sqrt{1\,-\,\lambda^2}\right) \, - \, \sqrt{\frac{1\,+\,\lambda}{1\,-\,\lambda}}\,
\coth\left(\frac{\tau}{2}\,\sqrt{1\,-\,\lambda^2}\right) }{2}\ , \nonumber \\
\dot{a} &=& \frac{ \sqrt{\frac{1\,-\,\lambda}{1\,+\,\lambda}}\,
\tanh\left(\frac{\tau}{2}\,\sqrt{1\,-\,\lambda^2}\right) \, + \, \sqrt{\frac{1\,+\,\lambda}{1\,-\,\lambda}}\,
\coth\left(\frac{\tau}{2}\,\sqrt{1\,-\,\lambda^2}\right) }{2} \ .
    \eea
    We say that this solution, for which $\dot{\varphi}<0$, describes a \emph{descending scalar};
    \item if $\sigma\, y > 1$ eq.~\eqref{logsigmay} is solved by
      \beq
y \ = \ \frac{1}{\sigma} \ \coth\left(\frac{\tau}{2}\,\sqrt{1\,-\,{\lambda^2}}\right) \ ,
\eeq
and then
    \bea
\dot{\varphi} &=& \frac{ \sqrt{\frac{1\,-\,\lambda}{1\,+\,\lambda}}\,
\coth\left(\frac{\tau}{2}\,\sqrt{1\,-\,\lambda^2}\right) \, - \, \sqrt{\frac{1\,+\,\lambda}{1\,-\,\lambda}}\,
\tanh\left(\frac{\tau}{2}\,\sqrt{1\,-\,\lambda^2}\right) }{2}\ , \nonumber \\
\dot{a} &=& \frac{ \sqrt{\frac{1\,-\,\lambda}{1\,+\,\lambda}}\,
\coth\left(\frac{\tau}{2}\,\sqrt{1\,-\,\lambda^2}\right) \, + \, \sqrt{\frac{1\,+\,\lambda}{1\,-\,\lambda}}\,
\tanh\left(\frac{\tau}{2}\,\sqrt{1\,-\,\lambda^2}\right) }{2} \ .
    \eea
    We say that this solution, for which $\dot{\varphi}$ is initially positive, and after the inversion time $\tau^\star$, defined by
\beq
\tanh\left(\frac{\tau^\star}{2}\,\sqrt{1\,-\,\lambda^2}\right) \ = \ \sqrt{\frac{1\,-\,\lambda}{1\,+\,\lambda}} \ ,
\eeq
it becomes negative, describes a \emph{climbing scalar}. This behavior is particularly interesting, since for this class of solutions $\varphi$ is bounded from above, and the same is true for the string coupling, as we had anticipated.
\end{enumerate}

The limiting behavior for large times
    \beq
\dot{\varphi} \ = \  - \ \frac{\lambda}{\sqrt{1\,-\,\lambda^2}}\ , \qquad
\dot{a} \ = \  \frac{1}{\sqrt{1 \ - \ \lambda^2}} \ ,
    \eeq
is identical for both solutions, and corresponds to the Lucchin--Matarrese attractor~\cite{lm}, which in this coordinate system thus takes the simple form
    \beq
{\varphi} \ = \ \varphi_0\ - \ \frac{\lambda\, \tau}{\sqrt{1\,-\,\lambda^2}}\ , \qquad
{a} \ = \  \frac{\tau}{\sqrt{1 \ - \ \lambda^2}} \ .
    \eeq
    
\begin{itemize}
\item As $\lambda \,\to\, 1^-$, the descending solution becomes singular and thus ceases to exist, while the climbing solution approaches the simple limiting form
\beq
\dot{\varphi} \ = \ \frac{1}{{2}} \left(\frac{1}{\tau} \ - \ \tau \right) \ , \qquad \dot{a} \ = \ \frac{1}{{2}} \left(\frac{1}{\tau} \ + \ \tau \right) \ . \label{sol_orientifold}
\eeq
\item For $\lambda> 1$, the nature of the solutions of eq.~\eqref{eq_climbing} changes, and the hyperbolic functions become trigonometric ones, while the range of $\tau$ becomes finite:
\beq
0 < \tau < \frac{\pi}{\sqrt{\lambda^2 \, - \, 1}} \ , \label{taurange}
\eeq
and for an expanding solution the only allowed option in this range reads
    \bea
\dot{\varphi} &=& \frac{ \sqrt{\frac{\lambda \, - \, 1 }{\lambda\,+\,1}}\,
\cot\left(\frac{\tau}{2}\,\sqrt{\lambda^2 \, - \, 1}\right) \, - \,  \sqrt{\frac{\lambda \, + \, 1 }{\lambda\,-\,1}}\,
\tan\left(\frac{\tau}{2}\,\sqrt{\lambda^2 \, - \, 1}\right) }{{2}}\ , \nonumber \\
\dot{a} &=& \frac{ \sqrt{\frac{\lambda \, - \, 1 }{\lambda\,+\,1}}\,
\cot\left(\frac{\tau}{2}\,\sqrt{\lambda^2 \, - \, 1}\right) \, + \,  \sqrt{\frac{\lambda \, + \, 1 }{\lambda\,-\,1}}\,
\tan\left(\frac{\tau}{2}\,\sqrt{\lambda^2 \, - \, 1}\right) }{{2}} \ .
    \eea
\end{itemize}

Integrating the preceding results yields the metric and dilaton in all preceding cases.
\begin{itemize}
\item For $\lambda < 1$, the integrated \emph{descending} solution is
  \beq
\begin{split} & ds^2 = 
  \,\left|\cosh\left(\frac{\tau}{2}\, \sqrt{1-\lambda^2} \right)\right|^{\frac{2}{(1+\lambda)(D-1)}} \
\!\!\! \left[\sinh\left(\frac{\tau}{2}\, \sqrt{1-\lambda^2} \right)\right]^{\frac{2}{(1-\lambda)(D-1)}}\ d{\bf x} \cdot d{\bf x}
\\ &- \ e^{- \,2 \,\lambda \,\vf_0} \,
\left|\cosh\left(\frac{\tau}{2}\, \sqrt{1-\lambda^2} \right)\right|^{-\,\frac{2\lambda}{1+\lambda}}
\ \!\!\! \left[\sinh\left(\frac{\tau}{2}\, \sqrt{1-\lambda^2} \right)\right]^{\frac{2\lambda}{1-\lambda}}\,
dt^2 \ ,  \\
&  e^{\frac{\gamma_c}{2} \phi} \ = \ e^{\vf_0 }\,
\left[\cosh\left(\frac{\tau}{2}\, \sqrt{1-\lambda^2} \right)\right]^{\frac{1}{1+\lambda}}
\left[\sinh\left(\frac{\tau}{2}\, \sqrt{1-\lambda^2} \right)\right]^{-\, \frac{1}{1-\lambda}} \ ,
\end{split}
    \eeq
\item For $\lambda < 1$, the integrated \emph{climbing} solution is
  \beq
\begin{split} & ds^2 \ = \ 
  \left|\sinh\left(\frac{\tau}{2}\, \sqrt{1-\lambda^2} \right)\right|^{\frac{2}{(1+\lambda)(D-1)}} \
\!\!\! \left[\cosh\left(\frac{\tau}{2}\, \sqrt{1-\lambda^2} \right)\right]^{\frac{2}{(1-\lambda)(D-1)}}\ d{\bf x} \cdot d{\bf x}
\\ &- \ e^{- \,2 \,\lambda \,\vf_0} \,
\left|\sinh\left(\frac{\tau}{2}\, \sqrt{1-\lambda^2} \right)\right|^{-\,\frac{2\lambda}{1+\lambda}}
\ \!\!\! \left[\cosh\left(\frac{\tau}{2}\, \sqrt{1-\lambda^2} \right)\right]^{\frac{2\lambda}{1-\lambda}}\,
dt^2 \ ,  \\
&  e^{\frac{\gamma_c}{2} \phi} \ = \ e^{\vf_0 }\,
\left[\sinh\left(\frac{\tau}{2}\, \sqrt{1-\lambda^2} \right)\right]^{\frac{1}{1+\lambda}}
\left[\cosh\left(\frac{\tau}{2}\, \sqrt{1-\lambda^2} \right)\right]^{-\, \frac{1}{1-\lambda}} \ ,
\end{split}
    \eeq
and in both cases $t>0$.
\item For $\lambda=1$ the integrated \emph{climbing} solution is
\bea  ds^2 &=& |\tau|^{\frac{1}{D-1}} \
e^{\frac{\tau^2}{2(D-1)}}  \ d\,\mathbf{x} \cdot d\,\mathbf{x}
\ - \ e^{-\, 2\, \varphi_{\,0}}\
|\tau|^{\, -\, 1} \ e^{\frac{\tau^2}{2}}\ {d t}^2 \ , \nonumber \\
e^{\frac{\gamma_c}{2} \phi} &=& e^{\varphi_0 }\ |\tau|^{\frac{1}{2}} \ e^{-\frac{\tau^2}{4}}
\, .
\eea
\item Finally, for $\lambda> 1$ the integrated \emph{climbing} solution is
  \beq
\begin{split} & ds^2 \ = \ \left[\sin\left(\frac{\tau}{2}\, \sqrt{\lambda^{\,2}-1} \right)\right]^{\frac{2}{(1+\lambda)(D-1)}} \
\!\!\! \left[\cos\left(\frac{\tau}{2}\, \sqrt{\lambda^{\,2}-1} \right)\right]^{\,-\, \frac{2}{(\lambda-1)(D-1)}}\ d{\bf x} \cdot d{\bf x}
\\ &-\ e^{- \,2 \,\lambda \,\vf_0} \,
\left[\sin\left(\frac{\tau}{2}\, \sqrt{\lambda^{\,2}-1} \right)\right]^{-\,\frac{2\lambda}{1+\lambda}}
\ \!\!\! \left[\cos\left(\frac{\tau}{2}\, \sqrt{\lambda^{\,2}-1} \right)\right]^{\,-\, \frac{2\lambda}{\lambda-1}}\,
dt^2 \ ,  \\
&  e^{\frac{\gamma_c}{2} \phi} \ = \ e^{\vf_0 }\,
\left[\sin\left(\frac{\tau}{2}\, \sqrt{\lambda^{\,2}-1} \right)\right]^{\frac{1}{1+\lambda}}
\left[\cos\left(\frac{\tau}{2}\, \sqrt{\lambda^{\,2}-1} \right)\right]^{\, \frac{1}{\lambda-1}}\,
, \label{supercrit}
\end{split}
    \eeq
where now, as we have seen, $\tau$ is limited to the range~\eqref{taurange}.
\end{itemize}

In particular, for the ten--dimensional orientifolds of~\cite{as95,as97,sugimoto},
\bea  ds^2 &=& |\tau|^{\frac{1}{9}} \
e^{\frac{\tau^2}{18}}  \ d\,\mathbf{x} \cdot d\,\mathbf{x}
\ - \ e^{-\, 2\, \varphi_{\,0}}\
|\tau|^{\, -\, 1} \ e^{\frac{\tau^2}{2}}\ {d t}^2 \ , \nonumber \\
e^{\frac{3}{4} \phi} &=& e^{\varphi_0 }\ |\tau|^{\frac{1}{2}} \ e^{-\frac{\tau^2}{4}}
\ ,
\eea
while for the $SO(16)\times SO(16)$ heterotic string of~\cite{agmv1,agmv2}
  \beq
\begin{split} & ds^2 \ = \ 
  \left[\sin\left(\frac{\tau\, \sqrt{21}}{4} \right)\right]^{\frac{4}{63}} \
\!\!\! \left[\cos\left(\frac{\tau\, \sqrt{21}}{4}\right)\right]^{\,-\, \frac{4}{27}}\ d{\bf x} \cdot d{\bf x}
\\ &- \ e^{- \,2 \,\lambda \,\vf_0} \,
\left[\sin\left(\frac{\tau\, \sqrt{21}}{4} \right)\right]^{-\,\frac{10}{7}}
\ \!\!\! \left[\cos\left(\frac{\tau\, \sqrt{21}}{4} \right)\right]^{\,-\, \frac{10}{3}}\,
dt^2 \ ,  \\
&  e^{\frac{3}{4} \phi} \ = \ e^{\vf_0 }\,
\left[\sin\left(\frac{\tau\, \sqrt{21}}{4}\right)\right]^{\frac{2}{7}}
\left[\cos\left(\frac{\tau\, \sqrt{21}}{4} \right)\right]^{\, \frac{2}{3}}\
. \label{supercrit2}
\end{split}
\eeq

As we have seen, the descending solution disappears for $\lambda \geq 1$ (or, equivalently, for $\gamma \geq \gamma_{c}$), and the only option left for $\phi$ (or for the right combination in $D< 10$~\cite{bsb_cosmology3,bsb_cosmology4}) is to climb the potential up to a turning point, then revert the motion and start a descent. In String Theory $e^\phi$ is the coupling that sizes the loop expansion, so that the climbing behavior is potentially under control in perturbation theory. This is not the whole story, of course, since close to the initial singularity curvature corrections become very large, but one is somehow half of the way in the right direction. Other interesting solutions for the potential of eq.~\eqref{climbing_pot} were discussed in~\cite{cd}, where some puzzles related to this dynamics are also addressed. We shall return to them at the end of this section.

The ``climbing'' picture~\cite{bsb_cosmology1,bsb_cosmology2,bsb_cosmology3,bsb_cosmology4,bsb_cosmology5} has counterparts in a wide class of integrable cosmologies~\cite{bsb_cosmology3}, and suggests very naturally a mechanism to \emph{start inflation}~\cite{inflation1,inflation2,inflation3,inflation4,inflation5,inflation6,inflation7} (for reviews see~\cite{inflation_rev}), provided one identifies a proper combination of dilaton and internal volume mode with the inflaton: a fast inflaton compelled to climb when emerging from the initial singularity would reach a turning point after releasing part of its original energy, stopping there momentarily before descending. However, inflation needs other ingredients, and slow roll can only be achieved in the presence of an additional, flatter contribution to the potential. Let us stress this point: from a top--down perspective, this mechanism can explain how inflation could have been injected, but bottom--up details are needed to complete the picture. 

String theory gives at present no clear top--down indications on the emergence of a portion of the potential capable of sustaining a slow--roll phase, but if an additional flatter contribution were present, a decelerating inflaton could have left, in principle, detectable signatures in the sky. A natural target of these considerations is the observed low--$\ell$ tail of the angular power spectrum of the CMB, which appears to be suppressed with respect to the $\Lambda$ CDM concordance model~\cite{WMAP,PLANCK1,PLANCK2,PLANCK3}. If this effect were not a mere fluctuation, a decelerating inflaton could account for a depression of the primordial power spectrum of scalar perturbations, which a short--enough inflation could have imprinted, almost verbatim, on the first few CMB multipoles~\cite{bsb_cosmology2,bsb_cosmology3}.

Any combination of a (super)critical exponential potential with a milder one would still enforce the climbing phenomenon. In this respect, perhaps a most interesting case is obtained combining the Starobinsky potential~\cite{starobinsky} 
\beq
V(\phi) \ = \ V_0 \left( 1 \ - \ e^{\,-\,\sqrt{\frac{2}{3}}\,\phi}\right)^2 \ + \ T e^{\sqrt{6}\,\phi}
\eeq
with a critical exponential for a canonically normalized scalar field. 
 In this fashion, the standard inflationary picture, with the scalar undergoing a slow-roll phase before falling into the potential well, would be preceded by a pre--inflationary fast-roll as the scalar descends along the sub-critical Starobinsky slope before bouncing against the critical wall, and slow-roll would only be reached after this initial phase. Note that in this setup the two exponentials are non--perturbative with respect to one another.

Even a simple phenomenological combination
\beq{}
V \ = \ T\left( e^{\gamma\,\phi} \ + \ e^{\gamma'\,\phi}\right) \ , \label{pheno_climbing}
\eeq
of a steep exponential with $\gamma \geq \gamma_c$ and a second exponential with an exponent $\gamma'$ small enough to support inflation can grant a slow--roll phase~\cite{bsb_cosmology1,bsb_cosmology2} in the eventual descent, and is simpler to analyze: in all cases the fast--roll injection of inflation would induce a depression in the power spectrum of scalar (and tensor) perturbations~\footnote{The reversal of the scalar motion at the end of its ascent, or at the end of the pre--inflationary phase in the Starobinsky potential, would actually leave another distinct feature in primordial power spectra: a small peak, which seems however beyond reach in present CMB experiments.}.
These scenarios also entail a definite prediction: the tensor--to--scalar ratio would grow, by almost one order of magnitude~\cite{bsb_cosmology2,bsb_cosmology3}, in the transition region between the low-$k$ depression and the usual power--like behavior of~\cite{cm} (for a review, see~\cite{cm_rev}).

The main features of this transition can be captured by the simple formula
\beq{}
P(k) \ = P_0 \ \frac{k^3}{\left[k^2+ \ \Delta^2\right]^{2 \,-\,\frac{n_s}{2}}} \ , \label{PDelta}
\eeq
which generalizes the Chibisov--Mukhanov spectrum~\cite{cm}, and where a new scale $\Delta$ is present.
The region hosting the depression of $P(k)$ terminates around the scale $\Delta$, where the primordial power spectrum begins to converge to the standard form
\beq
P(k) \ = P_0 \ k^{\,3 - 2 \nu} \ . \label{Pcm}
\eeq
As is well known, the spectral index in this key expression was measured to high precision by the {\sc Planck} collaboration~\cite{PLANCK1,PLANCK2,PLANCK3}.
\begin{figure}[ht]
\centering
\includegraphics[width=80mm]{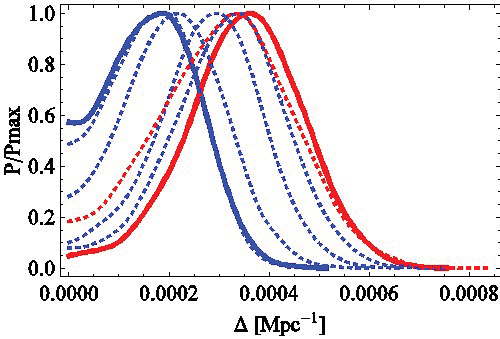}
\caption{\small Posteriors for the parameter $\Delta$ in eq.~\eqref{PDelta} with different masking around the Galactic plane, from~\cite{bsb_cosmology7}. The color coding is as follows:  solid blue for the 94\% mask, thick red for a $+30^\circ$ extension, dotted blue for the intermediate masks with $+6^\circ$, $+12^\circ$, $+18^\circ$ and $+24^\circ$, dotted red for $36^\circ$.}
\label{fig:delta}
\end{figure}

In fact, one can play a more sophisticated game and extend the standard concordance model $\Lambda$ CDM to what might be called $\Lambda$CDM$\Delta$, while also attempting to determine the scale $\Delta$ in Eq.~\eqref{PDelta}. This was done in~\cite{bsb_cosmology6,bsb_cosmology7}, and higher Galactic latitudes 
improve the determination of $\Delta$.
The best detection level rises to about 3 $\sigma$ with a Galactic mask obtained by a $30^\circ$ blind extension of the minimal {\sc Planck} mask (see fig.~\ref{fig:delta}). Notice, however, that one is thus left with about 39\% of the CMB sky, to be compared with the 94\% of it that is allowed by the minimal {\sc Planck} mask. The CMB should be cleaner far away from the Galactic plane, where fewer sources of backgrounds are present, and in this fashion one finds
\beq
\Delta \ = \ \left(0.35 \ \pm \ 0.11\right) \times 10^{-3}\ {\rm Mpc}^{-1} \ ,
\eeq
where the error indicated corresponds to 68$\%$ C.L.. This distance scale, as expected, is of the order of the Cosmic Horizon. In contrast, in the minimal {\sc Planck} mask one attains a lower detection level~\cite{bsb_cosmology6,bsb_cosmology7}, of order $88.5\%$, with $\Delta \ = \ \left(0.17 \ \pm \ 0.09\right) \times 10^{-3}\ {\rm Mpc}^{-1}$.

If the effect were not a mere fluctuation, these findings could have interesting implications for Cosmology. To begin with, the new scale could signal the onset of inflation (or perhaps more prosaically, a local feature of the inflationary potential), but it would then also influence the formation of structures, which should appear less effective on the very large scales affected by $\Delta$.
As we have anticipated, a general feature of these types of setup is an increase of the tensor-to-scalar ratio by about one order of magnitude around the multipoles associated to $\Delta$, while the effective increase in the local value for $n_s$ in the region below $\Delta$ would also suggest a sizable enhancement of non--Gaussian effects. 

Non--gaussian effects were recently analyzed in~\cite{meo_as} for this climbing scalar setup. With some simplifications, and assuming that String Theory resolves the initial singularity turning it into a bounce, as the Mukhanov--Sasaki potential seems to indicate up to the early epoch when the dynamics is overridden by the initial singularity, one can compute these effects analytically. They comprise two contributions to curvature perturbations: the first is associated with the evolution from the bounce onward, and oscillates around Maldacena's~\cite{maldacena} original result, but the second, which arises from the vicinity of the turning point, can be sizable and yet within limits of the {\sc Planck} collaboration~\cite{Planck_ng,forecasts} if the number of $e$-folds lies in the interesting range $62<N< 65$. To this end, one must note that, retracing the history of the Universe, one can translate $\Delta$ into an energy scale at the beginning of inflation, given by
\beq
\Delta_\mathrm{inf} \ \simeq \ 3 \times 10^{14} \ e^{N-60} \ \sqrt{\frac{H_\mathrm{inf}}{\mu_{\mathrm{Pl}}}} \ \mathrm{GeV} \ \simeq \ 2 \times 10^{12} \ \ e^{N-60} \ \mathrm{GeV} \ , \label{DeltaH}
\eeq
where $H_\mathrm{inf} \simeq 10^{14}\ \mathrm{GeV}$, $\mu_{\mathrm{Pl}} \simeq 2.4 \times 10^{18}\ \mathrm{GeV}$ is the reduced Planck energy and $N$ denotes the number of inflationary $e$-folds. 

There is a wide set of models that can be solved exactly~\cite{bsb_cosmology3}, where an early climbing phase injects inflation, along the lines of the preceding arguments. One of them is qualitatively close to eq.~\eqref{pheno_climbing}, since the potential reads
\beq
V(\phi) \ = \ V_0 \left(e^{2\,\gamma\,\varphi} \ + \ e^{\frac{2}{\gamma}\,\varphi} \right) \ ,
\eeq
the main difference being the presence of a steeper exponential wall.
In the same notation used above, the equations of this model follow from the reduced Lagrangian
\beq
{\cal L} \ = \ e^{a\,-\,{\cal B}}\Big[ - \ \dot{a}^2 \ + \ \dot{\varphi}^2 \ - \ e^{2{\cal B}}\,V_0\left( e^{2\,\gamma\,\varphi} \ + \ e^{\frac{2}{\gamma}\,\varphi} \right)\Big] \ , \label{L_integrable}
\eeq
and the Lorentz symmetry of the kinetic terms suggests how to obtain exact solutions in the convenient gauge
\beq
{\cal B} \ = \ a \ .
\eeq
 In fact, two new decoupled variables $\hat{\varphi}$ and $\hat{a}$ can be defined by a suitable ``boost'', and the resulting equations can be turned into the ``energy conservation'' conditions
\bea
\dot{\hat{\varphi}}^{\,2}  &=& V_0 \left[ e^{\frac{2}{\gamma}\,\sqrt{1-\gamma^2}\,\hat{\varphi}_0} \ - \  e^{\frac{2}{\gamma}\,\sqrt{1-\gamma^2}\,\hat{\varphi}} \right] \ , \nonumber \\
\dot{\hat{a}}^{\,2} &=& V_0 \left[ e^{2 \sqrt{1-\gamma^2}\,\hat{a}} \ + \ e^{2 \sqrt{1-\gamma^2}\,\hat{a}_0} \right]\ ,
\eea
and the Hamiltonian constraint, which is obtained by varying ${\cal B}$ in eq.~\eqref{L_integrable}, links the two integration constants according to
\beq
\hat{\varphi}_0 \ = \ \gamma\, \hat{a}_0 \ .
\eeq
The model is thus exactly solvable by quadratures in terms of the ``boosted'' variables, and returning to $a$ and $\varphi$ the solutions read
\bea
e^{a} &=& e^{a_0 \sqrt{1-\gamma^2}}\, \frac{\Big[ \cosh \omega\left(\tau - \tau_0\right)\Big]^\frac{\gamma^2}{1-\gamma^2}}{\Big[ \sinh\left(\omega\,\gamma \,\tau\right) \Big]^\frac{1}{1-\gamma^2}} \ , \nonumber \\
e^{\varphi} &=& \left[\frac{\sinh\left(\omega\,\gamma \,\tau\right) }{\cosh \omega \left(\tau - \tau_0\right)}\right]^\frac{\gamma}{1-\gamma^2}  \ ,
\eea
where $\tau_0$ is an integration constant and
\beq
\omega \ = \ V_0 \, \sqrt{1\,-\,\gamma^2} \ e^{\sqrt{1-\gamma^2}\,\hat{a}_0} \ .
\eeq

Note that the factor dependent on $a_0$ can be eliminated by rescaling the spatial coordinates, so that $\tau_0$ is the only relevant integration constant, which determines the maximum value attained by the string coupling during the evolution.
The initial singularity corresponds to $\tau \to +\infty$, when the scalar starts the climbing phase, while large times correspond to $\tau \to 0$, when the scalar has overcome its turning point and is descending. Although not manifestly, the limiting behaviors of $\phi$ are again as for the single exponential potential with $\gamma$.

The climbing phenomenon actually raises an interesting question: what would happen in potentials like
\beq
V(\phi) \ = \ V_0 \ \cosh\left(\sqrt{6}\,\phi\right) \ ,
\eeq
where an initial descending phase is not allowed on either side? This question was dealt with in detail in~\cite{cd}, and the answer is very interesting. Since the system cannot originate from a descending phase, the scalar field can only undergo wilder and wilder oscillations when approaching the singularity, which point to a chaotic beginning of the dynamics.

Let us conclude this section by mentioning a spring-off of the stability analysis presented in previous ones. An interesting issue is, in general, how perturbations of this type of background would evolve over time. In this case, one can show that the potential instabilities that can build up during the time evolution of the system concern tensor perturbations and can manifest themselves in their growth during the cosmic evolution. Tensor perturbations satisfy in these models the equation
\beq
h_{ij}'' \ + \  \frac{1}{\eta}\  h_{ij}'\ + \ {\bf k}^2\,h_{ij} \ = \ 0 \ ,
\eeq
which is solved by
\begin{eqnarray}
h_{ij} &\sim& A_{ij}\, J_0(k\eta) \ + \ B_{ij}\, Y_0(k \eta) \qquad ( {\bf k} \neq 0) \ , \nonumber \\
h_{ij} &\sim& A_{ij} \ + \ {B_{ij}} \ \log\left(\frac{\eta}{\eta_0}\right) \quad\qquad ( {k} = 0) \ .
\eea
While for $k\neq0$ the Bessel functions decay as $\eta$ grows beyond a certain point, the same is not true for
${k}=0$, where the possible logarithmic growth signals an \emph{instability of isotropy}. This seems an enticing result, if taken at face value: it indicates that, in this setting, the compactification of extra dimensions needed to connect String Theory to Nature might be the result of a mere accident!
\newpage
\section*{\sc Acknowledgments} \label{sec:Acknowledgments}

%%%%%%%%%%%%%%%%%%%%%%%%%%%%%%%%%%%%%%%%%%%%%%%%%%%%%%%
\vskip 12pt

Various portions of the material collected in this review originate from lectures or seminars given by one or more of us, over the years, at various Schools, or in our Institutes. 
We are indebted to the participants and also to many colleagues and/or collaborators with whom we have discussed extensively, over the years, many of the topics reviewed in the preceding sections, or other issues that played a central role in our education. These include S.~Abel, C.~Angelantonj, I.~Antoniadis, C.~Bachas, R.~Barbieri, I.~Basile, C.~Becchi, X. Bekaert, I.~Bena, K.~Benakli, M.~Bianchi, P.~Binetruy, R.~Blumenhagen, Q.~Bonnefoy, G.~Bossard, Ph.~Brax, W.~Buchm\"uller, N.~Cabibbo, P. ~Camara, A.~Campoleoni, G.~Casagrande, C.~Condeescu, T.~Coudarchet, E.~Cremmer, B.~Crosignani, G.~Dall'Agata, G.~D'Appollonio, C.~Deffayet, K.~Dienes, P.~Di Porto, F.~Farakos, P.~Fayet, S.~Ferrara, D.~Fioravanti, D.~Francia, G.~von Gersdorff,  T.~Gherghetta, D.~Ghilencea, L.~Girardello, F.~Gliozzi,  M.~H.~Goroff, M.~Grana, C.~Grojean, A.~Gruppuso, C.~Iazeolla, G.~Jona-Lasinio, E.~Joung, R.~Kallosh, A.~Kehagias, E.~Kiritsis, N.~Kitazawa, C.~Kounnas, S.~Lavignac, A.~Linde, G.~Lo Monaco, A.~Loty, D.~Luest, S.~Lust, J.~Madore, Y.~Mambrini, M.~Meo, N.~Marcus, R.~Minasian, P.~Natoli, M.~Nicolosi, K.~Noui, K.~Olive, E.~Palti, R.~Parentani, G.~Parisi, H.~Partouche, S.~Patil, R.~Petronzio~\cite{petronzio-stanev}, G.~L.~Pimentel, S.~Pokorski, M.~Porrati, G.~Pradisi, P.~Ramond, S.~Raucci, F.~Riccioni, V.~Rubakov, C.~A.~Savoy, J.~H.~Schwarz, H. Sazdjian, W.~Siegel, Y.~S.~Stanev~\cite{petronzio-stanev}, D.~Steer,  P.~Sundell, M.~Taronna, C.~Timirgaziu,  A.~Tomasiello, D.~Toulikas, C.~Vafa, E.~Witten, A.~Zaffaroni and G.~Zahariade. We are especially grateful to C.~Angelantonj and C.~Condeescu, who offered many constructive suggestions and helped us to eliminate several misprints and to streamline some derivations.
The work of E.~D. was supported in part by the IRP UCMN France-USA, while the work of A.~S. was supported in part by INFN (I.S. GSS-Pi) and by Scuola Normale Superiore.

\newpage 

\begin{appendices}

\section{\sc General Conventions} \label{app:conventions}

\subsection{\sc Four Dimensions}

In this review, we use Minkowski metrics of ``mostly plus'' signature in four dimensions, so that
\beq
\eta_{\mu\nu} \ =\  \mathrm{diag} (-1,1,1,1)  \label{mink}
\ .
 \eeq
In four dimensions we also use a Weyl (chiral) basis for gamma matrices, along the lines of the Wess--Bagger book~\cite{susy-books}, so that the \emph{flat} $\gamma$ matrices in four--dimensional Minkowski space are
\bea
&& \gamma^{i} =
\begin{pmatrix}
0 & \sigma^i \\
- \sigma^i & 0
\end{pmatrix} \ , \qquad \gamma^{0} =
\begin{pmatrix}
0 & - 1 \\
-1 & 0
\end{pmatrix} \ , \nonumber \\
&& \gamma^5 \ =\  - \,i\, \gamma^0\gamma^1\gamma^2\gamma^3 = \begin{pmatrix}
-1 & 0 \\
0 & 1
\end{pmatrix} \ ,
\eea
with $\sigma^i$ the three Pauli matrices. As a result, with our signature choice~\eqref{mink}
\beq
\left\{ \gamma_\mu\,,\,\gamma_\nu\right\} \ = \ - \ 2\,\eta_{\mu\nu} \ .
\eeq
Note that these conventions, which are often used in four dimensions, rely on $\gamma$ matrices tailored for the ``mostly minus'' signature.

We often use the compact notation
\beq
 \gamma^{\mu} =
\begin{pmatrix}
0 & \sigma^{\mu} \\
 {\bar \sigma}^{\mu} & 0
\end{pmatrix}  \ , \label{sp200}
\eeq
where 
\beq
\sigma^{\mu} \ = \ \left(- 1, \sigma^i \right) \ , \qquad {\bar \sigma}^{\mu} = (-1, - \sigma^i ) \ . \label{sigmasigmabar}
\eeq
The charge--conjugation matrix,
\beq
\qquad C = \begin{pmatrix}
i \,\sigma^2 & 0 \\
0 & - \,i \,\sigma^2
\end{pmatrix} \ , \label{sp1}
\eeq
satisfies
\beq
C^{-1} \ \gamma_{\mu} \ C \ = \ - \ \gamma_{\mu}^T \ , \qquad C \ = \ - \,C^{-1} \ = \ - \,C^T \ = \ - \, C^{\dagger} \ . \label{sp2}
\eeq

 In the Weyl representation of the Dirac matrices, one can define two--component spinors, starting from a four-component Dirac fermion, according to
\begin{equation}
\Psi = \begin{pmatrix}
\psi_\alpha \\ {\bar \chi}^{\dot \alpha}
\end{pmatrix} \ , \label{sp3}
\end{equation}
where $\alpha, \dot \alpha =1,2$. The two--component spinors $\psi, \chi$ are related to the original four--component spinors according to
\beq
\Psi_L \ = \ \frac{1-\gamma_5}{2}\, \Psi \ = \ \begin{pmatrix}
\psi_\alpha    \\ 0
\end{pmatrix} \ , \qquad \Psi_R \ = \ \frac{1+\gamma_5}{2}\, \Psi \ =  \ \begin{pmatrix}
0 \\  {\bar \chi}^{\dot \alpha}
\end{pmatrix} \ . \label{sp4}
\eeq
In the two--dimensional notation, the charge conjugate spinors read
\beq
\Psi^c{}_L \ = \ \begin{pmatrix}
0 \\ \bar{\chi}^{c\,\dot{\alpha} }
\end{pmatrix} \ , \qquad \Psi^c{}_R \ = \ \begin{pmatrix}
{\psi}^c{}_\alpha \\ 0
\end{pmatrix} \ , \label{sp4c}
\eeq
with
\beq
\psi^{c}_1 \ = \ \left(\bar{\chi}{}^{2}\right)^* \ , \qquad \psi^{c}_2 \ = \ - \ \left(\bar{\chi}{}^{1}\right)^* \ , \qquad \bar{\chi}^{c\,1} \ = \ - \ \left(\psi^* \right)_2 \ , \qquad \bar{\chi}^{c\,2} \ = \ \left(\psi^* \right)_1 \ .
\eeq

Lorentz transformations on vectors, for example on the four-momentum $P^{\mu} = (P^0,{\bf P})$, act according to
\beq
P'^{\mu} \ = \ {\Lambda^{\mu}}{}_{\nu} \, P^{\nu} \ ,
\eeq
and the familiar conditions for the Lorentz transformation matrices
\beq
{\Lambda^{\mu}}{}_{\rho} \ \eta_{\mu \nu} \ {\Lambda^{\nu}}{}_{\sigma} \ = \ \eta_{\rho \sigma}   \label{sp5}
\eeq
grant the invariance of the norm $P^{\mu} P_{\mu} $.
For infinitesimal transformations  $ {\Lambda^{\mu}}{}_{\nu} \simeq {\delta^{\mu}}{}_{\nu} +  {\omega^{\mu}}{}_{\nu}$, and the preceding condition translates into the antisymmetry of the Lorentz parameters, $\omega_{\mu \nu} = - \omega_{\nu \mu} $.

The general theory of four--dimensional Lorentz representations characterizes fields via a pair of angular-momentum quantum numbers $(j_1,j_2)$, since the Lorentz algebra can be split into a pair of $su(2)$ algebras in terms of the non--hermitian combinations
\beq
\mathbf{J} \ \pm \ i \, \mathbf{K} \ ,
\eeq
where $\mathbf{J}$ and $\mathbf{K}$ denote the two vectors that generate rotations and boosts. The actual spin eigenvalues then range from $|j_1-j_2| \leq j \leq j_1+j_2$, and fields characterized by integer (half-odd-integer) values of
$j_1+j_2$ describe bosons (fermions).  The number of degrees of freedom contained in a field characterized by a pair $\left(j_1,j_2\right)$ is $(2j_1+1)(2j_2+1)$.

A  Dirac spinor is a reducible representation corresponding to the combination $\left(\frac{1}{2},0\right) \oplus \left(0,\frac{1}{2}\right)$. A Lorentz transformation acts according to
\beq
\Psi' (x') \ = \ e^{- \,\frac{i}{2} \,\omega_{\mu \nu} J^{\mu \nu}} \ \Psi (x)  \ , \label{sp6}
\eeq
and the Lorentz generators in the spinorial representation have the block--diagonal form
\beq
J^{\mu \nu} \ = \ \frac{i}{4} [  \gamma^{\mu} , \gamma^{\nu} ]  \  = \  \frac{i}{4}
\begin{pmatrix}
\sigma^{\mu} {\bar \sigma}^{\nu} \ - \  \sigma^{\nu} {\bar \sigma}^{\mu} & 0 \\
0  &  {\bar \sigma}^{\mu} \sigma^{\nu} \ -\ {\bar \sigma}^{\nu} \sigma^{\mu}  \end{pmatrix} \ . \label{sp7}
\eeq

It is convenient to define the matrices
\beq
\sigma^{\mu \nu} \ = \  \frac{1}{4} (\sigma^{\mu} {\bar \sigma}^{\nu} \ - \  \sigma^{\nu} {\bar \sigma}^{\mu}) \ , \quad {\bar \sigma}^{\mu \nu} \ = \  \frac{1}{4}  ({\bar \sigma}^{\mu} \sigma^{\nu} \ - \ {\bar \sigma}^{\nu} \sigma^{\mu})
\  , \label{sp8}
\eeq
which are antisymmetric in $\mu,\nu$.
The two component spinor $\psi$ is valued in the $\left(\frac{1}{2},0\right)$ representation of the Lorentz group, while ${\bar \chi}$ is valued in the $(0,\frac{1}{2})$ representation, and
the two--component spinors  transform independently according to
 \beq
 \psi' (x') \ = \ e^{\frac{1}{2} \omega_{\mu \nu} \sigma^{\mu \nu}} \ \psi (x)  \ , \quad    {\bar \chi}' (x') \ = \ e^{\frac{1}{2} \omega_{\mu \nu} {\bar \sigma}^{\mu \nu}} \ {\bar \chi} (x) \ .  \label{sp9}
 \eeq
In terms of the complex $2 \times 2$ matrix of unit determinant $M = e^{\frac{1}{2} \omega_{\mu \nu} \sigma^{\mu \nu}} \in SL(2,C)$, one can therefore write
\beq
\psi'_{\alpha} (x') \ = \ {M_{\alpha}}{}^{\ \beta}  \ \psi_{\beta} (x) \ ,   \quad {\bar \chi}'_{\dot \alpha} (x') \ = \ {(M^*)_{\dot \alpha}}{}^{\ \dot \beta} \ {\bar \chi}_{\dot \beta} (x) \ .  \label{sp10}
\eeq

Alternatively, one can see that the matrix $M$ generates Lorentz transformations starting with a vector $P^{\mu}$ and defining the hermitian matrix $P = P_{\mu} \sigma^{\mu}$, so that $-\,\det P = - P_0^2  {\bf P}^2$ is the Lorentzian norm of the four-vector $P^{\mu}$.
Since $M$ has unit determinant, the transformed matrix $P' = M P M^{\dagger}$ yields a vector with the  same norm, and consequently $P^{\mu}$ and $P'^{\mu}$ are related by Lorentz transformations.

The two-dimensional counterparts of the metric $\eta^{\mu \nu}$, which is invariant under Lorentz transformations, are the antisymmetric tensors
 \beq
 \epsilon_{\alpha \beta} \ =  \  \begin{pmatrix}
0 & -1 \\ 1 & 0
\end{pmatrix} \quad , \quad \epsilon^{\alpha \beta} \ = \  \begin{pmatrix}
0 & 1 \\ -1 & 0
\end{pmatrix} \ .  \label{sp11}
 \eeq
 They satisfy $\epsilon_{\alpha \beta} \ \epsilon^{\beta \gamma} \ = \ {\delta_{\alpha}}{}^{\gamma}$, and their Lorentz invariance is reflected in the identities
\beq
\epsilon_{\alpha \beta} \ = \ {M_{\alpha}}{}^{\ \gamma} \ {M_{\beta}}{}^{\ \delta}\ \epsilon_{\gamma \delta} \quad , \quad \epsilon^{\alpha \beta} \ = \ \epsilon^{\gamma \delta} \ {M_{\ \gamma}}{}^{\alpha} \ {M_{\ \delta}}{}^{\delta}
\ .  \label{sp12}
\eeq
One can verify that
\beq
{\bar \sigma}^{\mu\, \alpha \dot{\alpha}} \ = \ \epsilon^{\alpha\beta}\,\epsilon^{\dot{\alpha}\dot{\beta}}\,\sigma_{\beta\dot{\beta}}^\mu \ .
\eeq
The antisymmetric tensor $\epsilon$ allows one to raise and lower spinor indices, according to
\beq
 \psi^{\alpha} \ = \ \epsilon^{\alpha \beta}\  \psi_{\beta} \ , \quad \psi_{\alpha} \ = \ \epsilon_{\alpha \beta}\  \psi^{\beta} \ .  \label{sp13}
 \eeq

Taking into account the definitions  (\ref{sp13}) and the invariance of the $\epsilon$ tensor under Lorentz transformations  (\ref{sp12}), one can construct Lorentz invariant bilinears out
of a two-component fermion, or equivalently from a fermion of given chirality, as
\beq
\psi \psi  \equiv  \psi^{\alpha} \psi_{\alpha}  \ = \ \epsilon_{\alpha \beta} \psi^{\alpha}\psi^{\beta} \quad , \quad  {\bar \chi}  {\bar \chi}   \equiv  {\bar \chi}_{\dot \alpha} {\bar \chi}^{\dot \alpha}
\ = \ \epsilon^{\dot \alpha \dot \beta} {\bar \chi}_{\dot \alpha} {\bar \chi}_{\dot \beta}  \ .  \label{sp14}
\eeq
Thus, a Dirac mass term for the Dirac spinor (\ref{sp3}) can be expressed as
 \beq
- m {\bar  \Psi}  \Psi \ = \ m (\psi \chi + {\bar \psi}  {\bar \chi})
   \label{sp15}
\eeq
in terms of two-component fields.
For a Majorana fermion  $\psi = \chi$, and one can define a Majorana mass term,
\beq
 \frac{M}{2} ({\bar  \Psi}^c  \Psi \,+\, {\rm c.c.}) \ = \ \frac{M}{2} (\psi \psi \,+\, {\bar \psi}  {\bar \psi})  \ ,  \label{sp16}
\eeq
which involves a single chiral component and its conjugate.
The choice of characterizing Lorentz representations via undotted ($\alpha$) and dotted ($\dot \alpha$) indices is usually referred to as van der Waerden notation.

\subsection{\sc Gravity}

For gravity, we define the Riemann curvature in terms of the Christoffel symbols from the relation
\beq
[ D_M, D_N] V_Q \ = \ {R_{MNQ}}^{P}(\Gamma)\, V_P \ ,
\eeq
with late capital Latin indices $M,N=0,\ldots, D-1$ are curved, so that
\beq
{R_{MNQ}}^{P}(\Gamma) \ = \ \partial_N\,{\Gamma^P}_{MQ} \ - \ \partial_M\,{\Gamma^P}_{NQ}  \ + \ {\Gamma^P}_{NR}\, {\Gamma^R}_{MQ} \ - \ {\Gamma^P}_{MR}\, {\Gamma^R}_{NQ} \ ,
\eeq
where the metric is covariantly constant and
\beq
{\Gamma^M}_{PQ} \ = \ \frac{1}{2}\ g^{MN} \left( \partial_Q\,g_{NP} \ + \ \partial_P\,g_{NQ} \ - \ \partial_N\,g_{PQ} \right) \ .
\eeq
We also define the Ricci tensor as
\beq
R_{MQ} \ = \ {R_{MNQ}}^{N}(\Gamma) \ ,
\eeq
and consequently, in the metric formulation, the Einstein--Hilbert action takes the form
\beq
{\cal S} \ = \ \frac{1}{2\,k^2}\ \int d^D \,x \ \sqrt{-g} \ R(\Gamma) \ .
\eeq
On the other hand, in the vielbein formulation we define the curvature starting from
\beq
[ D_M, D_N] V^A \ = \ R_{MN}{}^{AB}(\omega)\, V_B \ ,
\eeq
where early capital Latin indices $A,B=0,\ldots , D-1$ are flat,
so that, in terms of the spin connection,
\beq
R_{MNA}{}^{B}(\omega) \ = \ \partial_M\,\omega_{NA}{}^{B} \ - \  \partial_N\,\omega_{MA}{}^{B} \ + \ \omega_{MA}{}^{C}\ \omega_{NC}{}^{B} \ - \ \omega_{NA}{}^{C}\ \omega_{MC}{}^{B} \ . \label{Romega}
\eeq
The vielbein is covariantly constant
\beq
\partial_M \ e_N{}^A \ + \ \omega_M{}^{AB}\ e_{N B} \ - \ \Gamma^P{}_{MN}\ e_P{}^A \ = \ 0 \ ,
\eeq
and torsion does not play a role in our considerations. As a result, one can conclude that
\beq
{R_{MNQ}}^{P}(\Gamma) \ = \ R_{MNA}{}^{B}(\omega) \ e_{Q}{}^{A} \ e_B{}^P \ ,
\eeq
so that the Einstein--Hilbert action can also be presented in the form
\beq
{\cal S} \ = \ \frac{1}{2\,k^2}\ \int d^D \,x \ e \ R_{MNA}{}^{B}(\omega) \ e^{MA} \ e^N{}_B \ ,
\eeq
since $e = \sqrt{-g}$.

\subsection{\sc \texorpdfstring{$\gamma$ and $\Gamma$} \ \ Matrices, and Majorana Spinors}
With a slight change of notation, whereby the preceding $
\gamma$'s are multiplied by the imaginary unit, let us now consider the algebra of flat $\gamma$ matrices with the standard signature, so that
\beq
\left\{ \gamma_A\,,\,\gamma_B\right\} \ = \ 2\, \eta_{AB} \ ,
\eeq
where flat indices in generic dimensions $D$ are early Latin letters,
and the spinors satisfy the Majorana condition~\footnote{The Majorana condition can be imposed in this form for $D=2,3,4$ (mod 8), while Majorana condition and Weyl conditions can be imposed simultaneously $D=2$ (mod 8), as discussed in~\cite{GSO}.}
\beq
\psi \ = \ C\, \overline{\psi}{}^{\,T} \ , \label{majorana_c}
\eeq
where
\beq
\overline{\psi} \ = \ \psi^\dagger \, \gamma^0 \ ,
\eeq
$C$ is now Hermitian, and satisfies
\beq
C \, \gamma_A \, C \ =  \ - \ \gamma_A^T \ , \qquad C \ =\ - \ C^T\ , \qquad C\ = \ C^\dagger\ = \ C^{-1} \ .
\eeq
Majorana bilinears have definite symmetry properties, which follow from eq.~\eqref{majorana_c} and the antisymmetry of $C$. For example
\beq
\bar{\psi}\, \chi \ = \ - \ \psi^T \, C \, \chi \ = \ - \ \chi^T \, C \, \psi \ = \ \bar{\chi}\, \psi \ ,
\eeq
taking into account the antisymmetry of $C$ and the anticommuting nature of the two spinor fields. Another example is
\beq
\bar{\psi}\, \gamma_A \, \chi \ = \ - \ \psi^T \, C \, \gamma_A\,\chi \ = \ - \ \chi^T \, \gamma_{A}^{T}\,C\,\chi \ = \ - \ \bar{\chi}\, \gamma_A \, \psi \ ,
\eeq
and so on.

In ten dimensions, the spinors are also simultaneously subject to the Weyl condition
\beq
\psi \ = \ \pm \ \gamma_{11}\,\psi \ ,
\eeq
and in our discussion of eleven--dimensional supergravity we defined the eleven--dimensional Levi--Civita symbol so that
\beq
\epsilon_{0 1 \ldots 11} \ = \ 1 \ .
\eeq

Together with the flat $\gamma$ matrices, it is often convenient to introduce their curved counterparts, which we shall denote by $\Gamma^M$, reserving late Latin letters to generic curved indices. The relation is
\beq
\Gamma_M \ = \ e_M{}^A \ \gamma_A \ ,
\eeq
and consequently
\beq
\left\{ \Gamma_M\,,\,\Gamma_N\right\} \ = \ 2\, g_{MN} \ .
\eeq

%%%%%%%%%%%%%%%%%%%%%%%%%%%%

\section{\sc \texorpdfstring{$N=1$} \ \  Supersymmetry and Superfields}
\label{app:superfields_global}

In this appendix, we collect some properties of four--dimensional $N=1$ superfields that are used in the first portion of the review.

%%%%%%%%%%%%%%%%%%%%%%%%%%%%%%%%%%%%%%%%%%%%%%%%%%%%%%
\subsection{\sc On-Shell and Off-Shell Multiplets: Auxiliary Fields}
\label{app:wz}

In Chapter~\ref{sec:susy_algebras} we saw that all supersymmetric multiplets contain equal numbers of bosonic and fermionic degrees of freedom.
To be specific, let us focus on the case of $N=1$ or minimal supersymmetry and on the simplest multiplets that we constructed there.
The chiral multiplet contains a complex scalar $z$ and a Weyl fermion $\psi$, both of which have two propagating degrees of freedom. As we have seen, however, there is a subtlety: a complex Weyl fermion has four {\it off-shell}
degrees of freedom, which are halved by the Dirac equation to finally end up with the two on--shell ones. When formulating supersymmetric actions and performing quantum calculations with them, it is technically convenient to rely on a
formalism where supersymmetry is manifest off-shell, i.e. without using the field equations.  To this end, one needs to introduce an auxiliary field $F$, an additional complex boson that does not propagate any degrees of freedom on shell. Consequently, the {\it off-shell chiral multiplet} contains four fermionic and four bosonic degrees of freedom, described by
$(z,\psi,F)$.  In the massless $N=1$ vector multiplet, containing a gauge boson $A_m$ and a two-component fermion, the gaugino $\lambda$. On-shell there are two
bosonic and two fermionic degrees of freedom, but off shell a massless gauge boson has three degrees of freedom and a fermion four. Starting with four components, one degree of freedom
can in fact be removed by a gauge transformation $A_m \to A_m + \partial_m \alpha$, and on-shell another degree of freedom becomes unphysical.  As a result, one needs a real bosonic auxiliary field $D$ and the {\it off-shell massless vector multiplet} $(A_m, \lambda, D)$ contains four fermionic and four bosonic degrees of freedom.

Note that the supersymmetry algebra (\ref{repsusy13}) implies that the supercharges have mass dimension $\frac{1}{2}$. It is convenient to define fermionic transformation parameters
$\epsilon, {\bar \epsilon}$  of mass dimension $-\,\frac{1}{2}$, so that for any (bosonic or fermionic) field  $A$, supersymmetry transformations are implemented infinitesimally via
\beq
\delta_{\epsilon} A \ = \  \left(i \epsilon Q \ + \ i \,{\bar \epsilon} \,{\bar Q}\right) A \ , \label{repsusy28}
\eeq
where the supercharges $Q,{\bar Q}$ are realized as differential operators.
This allows one to express the supersymmetry algebra in terms of commutators only. For example, for minimal $N=1$ supersymmetry it takes the form
\bea
&& \left[\epsilon Q  \,,\,{\bar \epsilon} {\bar Q} \right] \ = \ 2 \,\epsilon \,\sigma^{\mu} \,{\bar \epsilon} \ P_{\mu} \ , \quad \left[P_{\mu} \,,\, \epsilon\, Q  \right] \ = \  \left[P_{\mu} \,,\,{\bar \epsilon}\, {\bar Q} \right]  \ = \ 0 \ , \nonumber \\
&&  \left[\epsilon \,Q  \,,\,{\epsilon} \,{Q} \right]  \ = \  \left[{\bar \epsilon}\, {\bar Q}  \,,\,{\bar \epsilon}\, {\bar Q} \right]  \ = \ 0  \ , \label{repsusy29}
\eea
and the first of these translates into
\beq
\left( \delta_{\epsilon_1}  \delta_{\epsilon_2} \ - \  \delta_{\epsilon_2}  \delta_{\epsilon_1} \right) A \ = \  -\ 2\, i \,\left(  \epsilon_1 \sigma^{\mu} {\bar \epsilon}_2 \ - \   \epsilon_2 \sigma^{\mu} {\bar \epsilon}_1  \right)
\partial_{\mu} A   \ . \label{repsusy30}
\eeq

The simplest supersymmetric model in four dimensions describes the interactions of one massive chiral multiplet, and is usually called Wess-Zumino model~\cite{Wess:1974tw1,Wess:1974tw2}.
With only renormalizable couplings its component Lagrangian is
\beq
 {\cal L} \ = \ - \ i\, \psi \sigma^{\mu} \partial_{\mu} \,{\bar \psi} \ - \ |\partial_{\mu} z|^2 \ - \ \frac{m}{2} \left(\psi \psi \,+\, {\bar \psi} {\bar \psi}\right)  \ - \ \lambda \left(z \psi \psi \,+\, z^* {\bar \psi} {\bar \psi}\right)  \ - \ \left|m z \,+\, \lambda z^2\right|^2  . \label{repsusy31}
\eeq
Note that the masses of the Bose and Fermi fields coincide, while the strength of the Yukawa interaction, $\lambda$, is closely related to that of the scalar self-interaction, $\lambda^2$, and to the
trilinear scalar vertex $\lambda m$. These are simple instances of generic features of supersymmetric theories: equal masses of bosons and fermions in the same multiplet and, in general, a reduction of the number of independent interaction parameters. In this case, Yukawa couplings and scalar self-interactions are all determined by a single independent coupling, $\lambda$.
Note also that the Lagrangian  (\ref{repsusy31}) depends only on one holomorphic function
\beq
W (z) \ = \ \frac{m}{2}\, z^2\ + \ \frac{\lambda}{3}\, z^3  \ , \label{superpot1}
\eeq
often referred to as superpotential, since it can be cast in the form
\beq
{\cal L} \ =\  -\ i \,\psi \sigma^{\mu} \partial_{\mu} {\bar \psi} \ - \ |\partial_{\mu} z|^2 \ - \ \frac{1}{2}  \left(\frac{d^2 W}{dz^2} \psi \psi \,+\, {\rm c.c.}\right) \ - \  \left|\frac{d W}{dz}\right|^2  \ . \label{repsusy031}
\eeq
The supersymmetry algebra and dimensional arguments determine the transformation of the scalar $z$, and the standard convention is
\beq
\delta z \ = \ \sqrt{2} \ \epsilon \, \psi  \ . \label{repsusy32}
\eeq
The transformation of the fermion is more complicated, but again dimensional arguments imply that it must involve the derivative of the scalar, and with our conventions it reads
\beq
\delta \psi \ = \ i\, \sqrt{2} \ \sigma^{\mu} \, {\bar \epsilon} \ \partial_{\mu} z \ - \ \sqrt{2} \  {\epsilon} \left(m z^* \,+\, \lambda z^{*2}\right) \ . \label{repsusy33}
\eeq

 In order to verify the supersymmetry algebra (\ref{repsusy30}), it is essential to use the field equation for the Fermi field,
 \beq
 i\,  {\bar \sigma}^{\mu} \,\partial_{\mu} \,{\psi} \ + \ \left(m \,+\, 2\, \lambda z^*\right) {\bar \psi}\ = \ 0 \ .
 \eeq
This is the case since we are now using the transformations of the on-shell chiral multiplet, without including the auxiliary field.
 Note also that the supersymmetry transformations  are non-linear in the fields (due to the interaction sized by $\lambda$), which complicates the construction of the correct supersymmetry transformations of general interacting models. One can introduce the off-shell multiplet, defining the non-dynamical complex auxiliary field $F$ according to
 \beq
 F^* \ =\  -\  \frac{\partial W}{\partial z} \ = \ -\  \left(m z \,+\, \lambda z^{2}\right) \ . \label{repsusy34}
\eeq
 With this definition, one can write the Lagrangian and the supersymmetry transformations as
 \bea
 {\cal L} &=& - \ i \ \psi \,\sigma^{\mu} \partial_{\mu} {\bar \psi} \ -\  |\partial_m z|^2 \ -\  \frac{m}{2} \left(\psi \psi \,+ \,{\bar \psi} {\bar \psi}\right)  \ -\  \lambda \left(z \psi \psi \,+\, z^* {\bar \psi} {\bar \psi}\right)  \  \nonumber \\
 &+& F^* F \ +\  F \left(m \,z \ + \ \lambda \,z^2\right) \ + \ F^* \left(m \,z^* \ + \ \lambda\, z^{*2}\right)  \ ,  \label{repsusy35}
\eea
\bea
 && \delta z \ = \ \sqrt{2} \ \epsilon \,\psi  \ , \nonumber \\
  && \delta \psi \ = \ i \ \sqrt{2} \ \sigma^{\mu} \, {\bar \epsilon} \ \partial_{\mu} z \ + \ \sqrt{2} \  {\epsilon} \  F \ , \nonumber \\
  && \delta F \ =\  i \,\sqrt{2} \ {\bar \epsilon} \,{\bar \sigma}^{\mu} \partial_{\mu} \psi  \ . \label{repsusy035}
 \eea

 Note that in eqs.~(\ref{repsusy35}) the complex bosonic field $F$ has no kinetic term, and therefore, as anticipated, {\it on-shell} it does not propagate any degree of freedom. This is the reason behind the name auxiliary field.  $F$  can be eliminated from the Lagrangian by its algebraic field  equation (\ref{repsusy34}), recovering eq.~(\ref{repsusy31}). There are three significant advantages of the formulation with an
 auxiliary field (\ref{repsusy35}).  The first is that the balance of Bose and Fermi degrees of freedom holds off-shell. The second is that supersymmetry is realized linearly, i.e. as a linear map
 between the components $(z,\psi,F)$ of the off-shell chiral multiplet, and the algebra closes without using the equations of motion. The third will become manifest in the next subsection: the action can be constructed rather simply, and in a systematic way, using the superspace approach, which makes the invariance manifest.

  %%%%%%%%%%%%%%%%%%%%%%%%%%%%%%%%%%%%%%%%%%%%%%%%%%%%%%
\subsection{\sc Superspace and Superfields}

There is a very convenient way of formulating $N=1$ supersymmetric theories, adjoining to the spacetime coordinates $x_m$ additional anti-commuting, fermionic
ones. For the minimal $N=1$ supersymmetry, a point in {\it superspace} is characterized by a triple $(x^{\mu}, \theta, {\bar \theta})$, where $\theta$ is a two-component constant spinor and ${\bar \theta}$
its complex conjugate. A {\it superfield}~\cite{Salam:1974yz}, \cite{Salam:1974jj} can be defined via a power series expansion in  $\theta$, ${\bar \theta}$, which can only contain a finite number of terms since $\theta^3=0$, due to the anticommuting nature of the spinorial coordinates $\theta$ and ${\bar \theta}$. The most general complex superfield that one can define is then
\bea
 F (x^m, \theta, {\bar \theta}) &=&  f(x) \ + \ \theta \,\psi (x)\ + \ {\bar \theta}\, {\bar \chi} (x) \ + \ \theta^2 M(x) \ + \ {\bar \theta}^2 N(x) \ +  \theta \sigma^{\mu}  {\bar \theta} \, A_{\mu} (x) \nonumber \\
&+&  \theta^2 {\bar \theta}\ {\bar \lambda} (x) \ + \ {\bar \theta}^2 \theta  \  {\xi} (x)\  + \ \theta^2 {\bar \theta}^2 D (x)  \ , \label{superspace1}
\eea
where the coefficients in the expansion are spacetime fields. Without special symmetry requirements, off shell the superfield $F$ contains sixteen real bosonic degrees of freedom, $(f,M,N,A_{\mu},D)$, and as many fermionic degrees of freedom,
$(\psi,\chi,\lambda,\xi)$. A superfield of this type, with no conditions imposed on it, is called unconstrained.  A linear combination of superfields $\alpha F_1 + \beta F_2$, or a product of superfields $F_1 F_2$, clearly yields other superfields.

If one wants to represent supersymmetry transformations as in  (\ref{repsusy28})
\beq
\delta_{\epsilon} F (x^{\mu}, \theta, {\bar \theta}) \ = \ \left(i\, \epsilon\, Q \ + \ i \,{\bar \epsilon}\, {\bar Q}\right)  F (x^{\mu}, \theta, {\bar \theta})  \ , \label{superspace2}
\eeq
mapping bosons into fermions and vice-versa within the multiplet, the supercharges  $Q$, ${\bar Q}$ must act as differential operators in superspace.
Since the supersymmetry algebra is an extension of an ordinary Lie algebra including anticommuting parameters, one can also introduce the group element
\beq
G (x , \theta, {\bar \theta}) \ = \ e^{i \left(- \,x^{\mu} P_{\mu} \,+\, \theta Q \,+\, {\bar \theta} {\bar Q}\right)} \ . \label{superspace3}
\eeq
Using the Hausdorff formula $e^A e^B = e^{A+B + \frac{1}{2} [A,B]+ \cdots}$ and realizing that in our case there are no higher order terms since $[A,B]$ commutes with $A$ and $B$, one finds for a product of such
group elements
\beq
G (x_1^{\mu} , \theta_1, {\bar \theta}_1) G (x_2^{\mu} , \theta_2, {\bar \theta}_2) \ = \ G (x_1^{\mu}+x_2^{\mu}+ i \theta_2 \sigma^{\mu}   {\bar \theta}_1 - i \theta_1 \sigma^{\mu}   {\bar \theta}_2  , \theta_2 + \theta_1 ,
 {\bar \theta}_2+{\bar \theta}_1 ) \ . \label{superspace4}
 \eeq
In general, one can implement spacetime translations on a field according to
 \beq
 e^{i a P} \varphi (x) \	e^{-i a P} \ = \  \varphi (x+ a) \ \simeq \  \varphi (x) \ + \ i a^{\mu} P_{\mu} \varphi \ \rightarrow  \ P_{\mu} \ = \ -\ i \,\partial_{\mu} \ ,
  \label{superspace5}
  \eeq
  where we could identify $P_{\mu}$ considering infinitesimal transformations. Similarly, in our case one can define the supercharges as differential operators, by implementing super-translations in the superspace according to
 \bea
e^{i  \epsilon Q +i {\bar \epsilon} {\bar Q}}  \ \Phi (x^{\mu}, \theta^{\alpha},  {\bar \theta^{\dot \alpha}} ) \ e^{- i \epsilon Q - i {\bar \epsilon} {\bar Q}} &=&
 \Phi  \left(x^{\mu}\ + \ i\, \theta \sigma^{\mu}   {\bar \epsilon} \ - \ i \epsilon \sigma^{\mu}   {\bar \theta}  \,, \,\theta \ + \ \epsilon ,
 {\bar \theta}\ + \ {\bar \epsilon} \right)  \nonumber \\
 &\simeq& \ \left(1 \ + \ i\, \epsilon Q \ + \ i\, {\bar \epsilon} {\bar Q}\right)   \ \Phi (x^{\mu}, \theta^{\alpha},  {\bar \theta}^{\dot \alpha} )  \ .
  \label{superspace6}
 \eea
 With infinitesimal parameters one thus finds
 \beq
 Q_{\alpha} \ = \  - \ i \,\frac{\partial}{\partial \theta^{\alpha}} \ - \  \sigma_{\alpha \dot \alpha}^{\mu} \, {\bar \theta}^{\dot \alpha} \partial_{\mu} \quad , \quad
 {\bar Q}_{\dot \alpha} \ = \ i \,\frac{\partial}{\partial {\bar \theta}^{\dot \alpha}} \ + \  {\theta}^{\alpha}  \sigma_{\alpha \dot \alpha}^{\mu}   \partial_{\mu} \ ,
  \label{superspace7}
  \eeq
and by a direct computation one can verify that the anticommutation relations
\beq
  \left\{ Q_{\alpha} ,  {\bar Q}_{\dot \alpha}  \right\} \ =\  - \ 2 \,i \,\sigma_{\alpha \dot \alpha}^{\mu} \, \partial_{\mu} \quad , \quad  \{ Q_{\alpha} ,  {Q}_{\beta}  \}  \ = \   \{ {\bar Q}_{\dot \alpha} ,  {\bar Q}_{\dot \beta}  \}  \ = \  0
  \label{superspace8}
 \eeq
 hold.
Starting from the ($\theta^2 {\bar \theta}^2$) component of  eq.~(\ref{superspace2}) and using eq.~(\ref{superspace7}), one can see that the highest component of a superfield transforms
under supersymmetry into a total derivative. This result is the key to construct supersymmetric actions in superspace.

As we saw, a general superfield has many more degrees of freedom than what we found in the simplest supersymmetric representations constructed in the previous sections. In order to find
appropriate superfields describing the simplest representations, the chiral and vector multiplets, one must reduce the degrees of freedom by imposing constraints compatible with supersymmetry, as we can now review.

%%%%%%%%%%%%%%%%%%%%%%%%%%%%%%%%%%%%%%%%%%%%%%%%%%%%%%%%
\subsection{\sc Chiral Superfields}

A chiral superfield contains as off-shell components the three fields $(z,\psi, F)$, and will be called $\Phi$ in what follows.
Let us define the two differential operators in superspace
\beq
 D_{\alpha} \ = \ \frac{\partial}{\partial \theta^{\alpha}} \ + \ i\, \sigma_{\alpha \dot \alpha}^{\mu} \,{\bar \theta}^{\dot \alpha} \,\partial_{\mu} \ , \qquad
 {\bar D}_{\dot \alpha} \ = \ - \ \frac{\partial}{\partial {\bar \theta}^{\dot \alpha}} \ - \ i \, {\theta}^{\alpha}  \,\sigma_{\alpha \dot \alpha}^{\mu}  \, \partial_{\mu} \ ,
  \label{chirals1}
 \eeq
which satisfy the anticommutation relations
\bea
&&  \{ D_{\alpha} ,  {\bar D}_{\dot \alpha}  \} \ =\  -\ 2 \,i\, \sigma_{\alpha \dot \alpha}^{\mu} \partial_{\mu} \ , \qquad  \{ D_{\alpha} ,  {D}_{\beta}  \}  \ = \  \{ {\bar D}_{\dot \alpha} ,  {\bar D}_{\dot \beta}  \}  \ = \ 0 \ ,
\nonumber \\
&&
  \{ D_{\alpha}  \ {\rm or }  \ {\bar D}_{\dot \alpha} , Q_{\beta} \  {\rm or }  \ {\bar Q}_{\dot \beta} \} \ = \ 0
  \ ,   \label{chirals2}
 \eea
 so that they anticommute with the supersymmetry generators.
The superfield constraint
\beq
{\bar D}_{\dot \alpha}  \Phi \ = \ 0 \ ,   \label{chirals3}
\eeq
is compatible with supersymmetry, due to the anticommutation relations in the last line of (\ref{chirals2}). As we shall see shortly, this constraint reduces the number of
bosonic and fermionic components in the superfield $\Phi$, and leads precisely to the off-shell chiral multiplet. In order to write the solution of eq.~\eqref{chirals3}, it is simpler to redefine momentarily the superspace coordinates, passing from $(x^m, \theta, {\bar \theta})$ to $(y^{\mu}, \theta, {\bar \theta})$, where
\beq
y^{\mu} \ = \  x^{\mu} \ + \ i \,\theta \sigma^{\mu}  {\bar \theta} \ . \label{ytheta}
\eeq
In  the coordinates $(y^m, \theta, {\bar \theta})$, the relevant operators and supercharges become
\bea
D_{\alpha} &=& \frac{\partial}{\partial \theta^{\alpha}} \ + \ 2 \,i \sigma_{\alpha \dot \alpha}^{\mu} \,{\bar \theta}^{\dot \alpha} \frac{\partial}{\partial y^{\mu}} \ , \qquad
  {\bar D}_{\dot \alpha} \ = \ - \ \frac{\partial}{\partial {\bar \theta}^{\dot \alpha}} \ , \nonumber \\
Q_{\alpha} &=&  -\ i\,\frac{\partial}{\partial \theta^{\alpha}} \ , \qquad
   {\bar Q}_{\dot \alpha} \ = \ i \,\frac{\partial}{\partial {\bar \theta}^{\dot \alpha}} \ + \ 2 \,  {\theta}^{\alpha}  \sigma_{\alpha \dot \alpha}^{\mu}  \, \frac{\partial}{\partial y^{\mu}}   \ .
  \label{chirals4}
 \eea
 Since
 \beq
  {\bar D}_{\dot \alpha} y^m \ = \  {\bar D}_{\dot \alpha} \theta \ = \ 0 \ , \label{chirals5}
 \eeq
in the new coordinates  an arbitrary function of $y,\theta$ satisfies the constraint  (\ref{chirals3}).  Its general solution is therefore
 \beq
 \Phi (y,\theta) \ = \ z(y) \ + \ \sqrt{2} \,\theta \psi (y) \ +\  \theta^2 F (y)  \  \label{chirals6}
 \eeq
 or equivalently
\bea
 \Phi (x,\theta, {\bar \theta}) &=& z(x) \ + \ i\, \theta \sigma^{\mu}  {\bar \theta} \partial_{\mu} z(x) \ + \ \frac{1}{4} \, \theta^2  {\bar \theta}^2 \Box z(x) \ + \  \sqrt{2} \,\theta \psi (x)
  \nonumber \\  &-& \frac{i}{\sqrt{2}} \, \theta^2 \partial_{\mu} \psi (x) \sigma^{\mu} {\bar \theta} \ + \ \theta^2  F(x)\  , \label{chirals7}
 \eea
 after expressing $y$ in terms of $x$ and expanding in $\theta$.
In this fashion one recovers the degrees of freedom of the off-shell chiral multiplet $(z,\psi,F)$ introduced in the previous Section, as we had anticipated.

It is often convenient to define the component fields via the relations
\bea
\Phi | \ = \ z \ , \qquad  D_{\alpha} \Phi | \ = \ \sqrt{2} \,\psi_{\alpha}  \ , \qquad   D^2 \Phi | \ = \ - \ 4 F   \ ,  \label{chirals06}
\eea
where the symbol $\Phi |$ means that we take the lowest component of the superfield.

Note also that if $\Phi_i$ are chiral superfields, the products $\Phi_i \Phi_j$ , $\Phi_i  \Phi_j \Phi_k$ are also chiral superfields and the highest components $\left.\Phi_i(y) \Phi_j(y)\right|_{\theta^2}$ , $\left.\Phi_i(y)  \Phi_j(y)\Phi_k(y)\right|_{\theta^2}$
transform under supersymmetry into total derivatives. They can thus be used to construct supersymmetric Lagrangians.
However, as can be explicitly verified,
such terms contain masses and interactions, but not kinetic terms. On the other hand, the non-chiral product  $\left.\Phi_i^{\dagger} \Phi_j\right|_{\theta^2 {\bar \theta}^2} $, being the highest component of a
superfield, transforms under supersymmetry into a total derivative and contains kinetic terms.  The most general renormalizable Lagrangian constructed out of chiral superfields is then
\beq
{\cal L} \ = \ (\Phi_i^{\dagger} \Phi_i)_{\theta^2 {\bar \theta}^2}\ + \ \left[ \left(\lambda_i \Phi \ + \  \frac{1}{2} m_{ij} \Phi_i \Phi_j \ + \ \frac{1}{3} \lambda_{ijk} \Phi_i \Phi_j  \Phi_k \right)_{\theta^2} \ + \ {\rm h.c.} \right]
\  . \label{chirals8}
\eeq
The chiral function
\beq
{\cal W} (\Phi) \ = \ \lambda_i \Phi_i \ + \ \frac{1}{2} m_{ij} \Phi_i \Phi_j \ + \ \frac{1}{3} \lambda_{ijk} \Phi_i \Phi_j  \Phi_k   \label{chirals9}
\eeq
is called {\it superpotential}. The restriction of the superpotential to terms that are at most cubic grants the renormalizability of the resulting Lagrangian.
Expanding the different terms one finds the component Lagrangian
\beq
 {\cal L}\  = \ -\ i\, \psi_i \sigma^{\mu} \partial_{\mu} {\bar \psi}_i \ -\  |\partial_{\mu} z_i|^2 \ +\  |F_i|^2 \ +\
\left[ - \ \frac{1}{2} \frac{\partial^2 {\cal W}}{\partial z_i \partial z_j} \psi_i \psi_j  \ + \ \frac{\partial {\cal W}}{\partial z_i} F_i \ +\  {\rm h.c.}  \right]  \  . \label{chirals10}
\eeq
The auxiliary field can be eliminated via its classical field equation
\beq
F_i \ = \ - \ \frac{\partial  \overline {\cal W}}{\partial z_i^*}   \  , \label{chirals11}
\eeq
leading to the final Lagrangian
\beq
 {\cal L} \ = \  - \ i \psi_i \sigma^{\mu} \partial_{\mu} {\bar \psi}_i \ - \  \left|\partial_{\mu} z_i\right|^2  \ - \
\left[ \frac{1}{2} \frac{\partial^2 {\cal W}}{\partial z_i \partial z_j} \psi_i \psi_j \ +\  {\rm h.c.}  \right]  \ - \  \left|\frac{\partial {\cal W}}{\partial z_i}\right|^2
\  , \label{chirals12}
\eeq
and we see that the potential so defined cannot be negative.

 %%%%%%%%%%%%%%%%%%%%%%%%%%%%%%%%%%%%%%%%%%%%%%%%%%%%%%%%%%
\subsection{\sc Vector Superfields}

Gauge symmetries play a key role for the vector multiplets, which contain gauge fields, and are thus an important step in the construction of supersymmetric extensions of the Standard Model.
Let us begin with Abelian gauge symmetries, taking into account that the usual gauge parameter
$\alpha (x)$ should be extended to a superfield. Moreover, chiral fermions like those in the Standard Model are naturally associated to the chiral multiplets constructed in the previous subsection.

The supersymmetric version of the standard gauge transformations
\beq
\psi_i \to e^{- i q_i \alpha} \psi_i \ ,
\eeq
with $q_i$ the charge of the fermion $\psi_i$, will be of the form
\beq
\Phi_i\  \to \ e^{- i q_i \Lambda}\,  \Phi_i  \  . \label{vectors1}
\eeq
with $\Lambda$ a chiral superfield containing as the real part of its lowest component the gauge parameter $\alpha$. These are often called supergauge transformations.

The superpotential will be gauge invariant if
each monomial in (\ref{chirals9}) is gauge invariant. This implies constraints on the Abelian charges
\beq
\lambda_i \not=0 \ \to \ q_i \ =\  0 \ , \qquad m_{ij} \not=0 \ \to\  q_i + q_j\ = \ 0 \ , \qquad   \lambda_{ijk} \not=0 \ \to\  q_i + q_j + q_k\ = \ 0 \  . \label{vectors2}
\eeq

However, the supersymmetric kinetic terms are not supergauge invariant  since
\beq
\Phi_i^{\dagger} \Phi_i \ \to \ \Phi_i^{\dagger} e^{i q_i (\Lambda - {\bar \Lambda})}\Phi_i \ , \label{local_gauge_chiral}
\nonumber
\eeq
but this is reminiscent of the fact that standard kinetic terms
are not invariant under local gauge transformations. In the non-supersymmetric case, the solution rests on the introduction of covariant derivatives, and its supersymmetric counterpart relies on a real vector superfield $V = V^{\dagger}$, transforming under gauge transformations as
\beq
V \ \to \ V \ + \ i \left(\Lambda \,-\, {\bar \Lambda}\right)  \  , \label{vectors3}
\eeq
 which enters the kinetic terms of the chiral multiplets according to
 \beq
 \Phi_i^{\dagger} \Phi_i \  \to \  \Phi_i^{\dagger} \,e^{q_i V} \Phi_i \  . \label{vectors4}
\eeq
On account of eq.~\eqref{local_gauge_chiral}, the modified kinetic terms are then gauge invariant.

The component expansion of the real superfield $V$ can be parametrized according to
\bea
V (x^{\mu}, \theta, {\bar \theta}) &=& C+ i \theta \chi - i {\bar \theta} {\bar \chi} + \frac{i}{2} \theta^2 (M+i N) -   \frac{i}{2} {\bar \theta}^2 (M-iN)  -  \theta \sigma^{\mu}  {\bar \theta} A_{\mu} \nonumber \\
&+&  i  \theta^2 {\bar \theta} ( {\bar \lambda} + \frac{i}{2} {\bar \sigma}^{\mu} \partial_{\mu} \chi)   - i  {\bar \theta}^2 \theta   ( {\lambda} + \frac{i}{2} {\sigma}^{\mu} \partial_{\mu} {\bar \chi})
 + \frac{1}{2} \theta^2 {\bar \theta}^2 (D + \frac{1}{2} \Box C)    \ . \label{vectors5}
\eea
In components, the gauge transformations  (\ref{vectors3}) translate into
\bea
&& C \to C + i (\alpha- \alpha^*) \quad , \quad \chi \to \chi + \sqrt{2} \psi \ , \nonumber \\
&& M+i N \to  M+i N + 2 F \quad , \quad A_{\mu} \to A_{\mu} + \partial_{\mu}  (\alpha+ \alpha^*) \ , \nonumber \\
 && \lambda \to \lambda \quad , \quad D \to D   \ , \label{vectors6}
 \eea
 where
 \beq
 \Lambda (y,\theta) \ = \ \alpha(y) \ + \ \sqrt{2}\,\theta\, \psi(y) \ + \ \theta^2 F(y) \ ,
 \eeq
 and the variable $y$ was defined in eq.~\eqref{ytheta}
There is thus a gauge choice, called {\it Wess-Zumino gauge} in the literature, which eliminates altogether $C,\chi$ and $M+iN$, thus reducing the real superfield $V$ to
\beq
V (x^{\mu}, \theta, {\bar \theta}) \ = \  - \  \theta \sigma^{\mu}  {\bar \theta} A_{\mu}  \ + \ \ i\,  \theta^2 {\bar \theta} \, {\bar \lambda} \ - \ i \, {\bar \theta}^2 \theta\,  {\lambda}
\  +  \ \frac{1}{2} \,\theta^2 {\bar \theta}^2 D    \ . \label{vectors7}
\eeq
The degrees of freedom in the Wess-Zumino gauge are thus $(A_{\mu},\lambda,D)$, which are precisely the fields of the off-shell massless vector multiplet that we discussed in  Section \ref{sec:susy_algebras}.
Note that in the Wess-Zumino gauge there are simplifications, since
\beq
V^2 \ = \ - \ \frac{1}{2} \,\theta^2 \, {\bar \theta}^2 \,A_{\mu} \, A^{\mu} \ , \qquad V^3=0  \ , \label{vectors8}
\eeq
and the renormalizability with standard couplings becomes manifest.

The supersymmetry transformations can be worked out relying on the standard formula
\beq
\delta V \ = \ i \left(\epsilon Q \ +\ {\bar \epsilon} {\bar Q}\right) V \ ,
\eeq
and in the Wess-Zumino gauge one finds
\bea
\delta A_{\mu} &=& i \left({\bar \epsilon}\, {\bar \sigma}_{\mu} \lambda \ - \  {\bar \lambda} \,{\bar \sigma}_{\mu} \epsilon \right) \ , \nonumber \\
\delta \lambda &=& i\, \epsilon D \ + \ \sigma^{\mu \nu} \epsilon \ F_{\mu \nu} \ , \nonumber \\
\delta D &=& - \ \epsilon \,\sigma^{\mu} \partial_{\mu} {\bar \lambda} \ - \  \partial_{\mu} {\lambda}\, \sigma^{\mu} {\bar \epsilon} \ . \label{vectors9}
\eea
There is actually a subtlety: while the Wess-Zumino (WZ) gauge is compatible with the usual gauge symmetry, the solution (\ref{vectors7}) in the WZ gauge is not preserved by supersymmetry transformations.
Eqs.~(\ref{vectors9}) actually combine standard supersymmetry transformations with a particular gauge transformation
\bea
\delta V_{WZ} &=& i (\epsilon Q +  {\bar \epsilon} {\bar Q}) V_{WZ} \ + \ i (\Lambda - {\bar \Lambda})  \ , {\rm where }\nonumber \\
\Lambda &=& - \  i \theta \sigma^{\mu} {\bar \epsilon} A_{\mu} \- \ \theta^2 ({\bar \epsilon} {\bar \lambda})
  \label{vectors09}
\eea
that, altogether, preserve the Wess--Zumino gauge.

As was the case for the chiral multiplet, supersymmetry transformations are linear in the fields in the off-shell vector multiplet with the auxiliary field $D$ while. However, as we shall see
shortly, they become nonlinear on-shell, after eliminating the auxiliary field via its field equation.
The compensating gauge transformation  (\ref{vectors09}) in the Wess-Zumino gauge also affects the supersymmetry transformations of the chiral multiplets in  (\ref{repsusy035}), which become
\beq
 \delta \Phi^i \ = \   i \left(\epsilon Q \,+\,  {\bar \epsilon} {\bar Q}\right)  \Phi^i  \ -\  i\, q_i\, \Lambda \,\Phi^i  \ ,  \label{wz2}
 \eeq
 with $\Lambda$ given in  (\ref{vectors09}). The compensating gauge transformation gives rise to gauge covariant expressions, so that now
  \bea
 \delta z^i &=& \sqrt{2} \,\epsilon \psi^i  \ , \nonumber \\
  \delta \psi^i &=& i\, \sqrt{2}\, \sigma^{\mu} {\bar \epsilon} \ {\cal D}_{\mu} z^i \ + \ \sqrt{2} \, {\epsilon} \  F^i \ , \nonumber \\
  \delta F^i &=& i\, \sqrt{2} \,{\bar \epsilon}\, {\bar \sigma}^{\mu} {\cal D}_{\mu} \psi^i \ + \ i\, q^i  \left({\bar \epsilon} {\bar \lambda}\right) z^i  \ , \label{wz3}
 \eea
 where
 \beq
 {\cal D}_{\mu} z^i  \ = \ \partial_{\mu} z^i \ - \ \frac{i}{2} \, q_i \,A_{\mu} \,z^i  \ , \qquad  {\cal D}_{\mu} \psi^i  \ =  \ \partial_{\mu} \psi^i \ - \ \frac{i}{2} \, q_i \,A_{\mu} \,\psi^i \ . \label{wz4}
 \eeq

The next step is the construction of the counterpart of the electromagnetic field strength $F_{\mu \nu} = \partial_{\mu} A_{\nu} \,-\, \partial_{\nu} A_{\mu}$. The corresponding superfield should be
gauge invariant,  should contain $F_{\mu \nu}$ in one of its components, and should lead to the Maxwell kinetic term in the Lagrangian. The correct choice turns out to be a spinorial superfield,
\beq
W_{\alpha} \ = \  - \ \frac{1}{4} \, {\bar D}^2 D_{\alpha} V \ , \label{vectors10}
\eeq
whose gauge invariance under eq.~\eqref{vectors3} can be simply verified, since the combination ${\bar D}^2 D_{\alpha}$ annihilates both the chiral superfield $\Lambda$ and its conjugate $\bar{\Lambda}$.  Note that $W_{\alpha}$ is a chiral superfield, but not
a generic one, since it satisfies the additional constraint
\beq
D^{\alpha} W_{\alpha} \ = \ {\bar D}_{\dot \alpha} {\bar W}^{\dot \alpha} \ . \label{vectors11}
\eeq
The constraint is a Bianchi identity, which reveals that $W_{\alpha}$ is not an arbitrary chiral superfield but rather a special one, since the auxiliary field $D$ is real.
In the Wess-Zumino gauge and in the $(y,\theta,{\bar \theta})$ superspace coordinate system,
\beq
W_{\alpha} \ =\  - \,i \lambda_{\alpha} (y) \ + \ \left[ \delta_{\alpha}^{\beta} D \ -\  i\, (\sigma^{\mu \nu})_{\alpha}^{\ \beta} F_{\mu \nu} (y)   \right] \theta_{\beta}
\ + \ \theta^2 \sigma_{\alpha \dot \alpha}^{\mu} \partial_{\mu} {\bar \lambda}^{\dot \alpha} (y)  \ . \label{vectors12}
\eeq
 Since $W_{\alpha}$ is a chiral superfield, the same argument used for the superpotential implies that $(W^{\alpha} W_{\alpha})_{\theta^2}$ transforms under supersymmetry into a total derivative and therefore it is a good candidate for a supersymmetric Maxwell Lagrangian. A short calculation leads to
 \beq
 (W^{\alpha} W_{\alpha})_{\theta^2} \ = \  -\ 2\, i\, \lambda \,\sigma^{\mu} \partial_{\mu} {\bar \lambda} \ -\  \frac{1}{2} \, F_{\mu \nu}^2 \ + \ D^2 \ + \ \frac{i}{4} \,\epsilon_{\mu \nu \rho \sigma} F^{\mu \nu} F^{\rho \sigma}   \ , \label{vectors120}
  \eeq
 and the Lagrangian is therefore
\beq
\frac{1}{4} \left( W^{\alpha} W_{\alpha}|_{\theta^2} \ + \ {\bar W}_{\dot \alpha}  {\bar W}^{\dot \alpha}|_{\bar \theta^2} \right)
\  = \  - \ i \, \lambda \, \sigma^{\mu} \partial_{\mu} {\bar \lambda} \ - \  \frac{1}{4} \,F_{\mu \nu}^2 \ + \ \frac{1}{2}\, D^2   \ . \label{vectors13}
  \eeq

The Lagrangian for an arbitrary number of chiral multiplets coupled to a massless $U(1)$ vector multiplet is then
\beq
{\cal L} \ = \  \left(  \Phi_i^{\dagger} e^{q_i V} \Phi_i  \right)_{\theta^2  {\bar \theta^2} } \ +  \
\left[ \left( \frac{1}{4}  W^{\alpha} W_{\alpha} + \lambda_i \Phi_i + \frac{1}{2} m_{ij} \Phi_i \Phi_j + \frac{1}{3} \lambda_{ijk} \Phi_i \Phi_j  \Phi_k   \right)_{\theta^2} \ + \ {\rm h.c.} \right]     \, . \label{vectors14}
\eeq
As before, gauge invariance of the superpotential requires the conditions (\ref{vectors2}).
For a $U(1)$ gauge theory, another  possible gauge invariant term exists, the Fayet-Iliopoulos term:
\beq
{\cal L}_{\rm FI} \ = \ 2\, \xi \left(V\right)_{\theta^2 {\bar \theta}^2} \ = \ \xi \,D  \ , \label{vectors014}
\eeq
  which transforms into a total derivative, in view of~(\ref{vectors9}) .

We turn now to the construction of non-Abelian supersymmetric gauge theories. Similarly to the replacement of gauge fields by matrices, we construct matrix valued vector superfields
$V = V_a T^a$ and similarly for the gauge parameters $\Lambda = \Lambda_a T^a$. Chiral matter superfields $\Phi$, containing components $\Phi_i$ transform in various representations
of the gauge group. For example, if they are in the fundamental representation, they transform like
\beq
\Phi \to e^{-i \Lambda} \Phi \quad , \quad \Phi^{\dagger} \to \Phi^{\dagger} e^{i \Lambda^{\dagger}} \ . \label{superym1a} 
\eeq 
A kinetic term $ \Phi^{\dagger} e^{2gV} \Phi$ is invariant if
\beq 
e^{2gV} \to e^{- i \Lambda^{\dagger}} e^{2gV} e^{i \Lambda} \ . \label{superym2a}
\eeq 
By extending the usual adjoint representation transformation of the Yang-Mills field strength, one has to define a chiral superfield containing $F_{mn}^a$ transforming as
\beq 
W_{\alpha} \to  e^{- i \Lambda} W_{\alpha}  e^{i \Lambda} \ , \label{superym3a}
\eeq 
such that $tr (W^{\alpha} W_{\alpha}) $ is gauge invariant. The correct definition turns out to be
\beq 
2g W_{\alpha} = - \frac{1}{4} {\bar D}^2 ( e^{-2gV} D_{\alpha}  e^{2gV})  \  . \label{superym4a}
\eeq 
In the Wess-Zumino gauge the vector superfield is
\beq 
V^a (x^m, \theta, {\bar \theta}) =  -  \theta \sigma^{\mu}  {\bar \theta} A_{\mu}^a  + \ i  \theta^2 {\bar \theta}  {\bar \lambda^a} - i  {\bar \theta}^2 \theta  {\lambda^a} 
 + \frac{1}{2} \theta^2 {\bar \theta}^2 D^a    \ . \label{superym5a}
\eeq 
and the chiral superfield field strength reduces to
\beq 
W_{\alpha} = - \frac{1}{4} {\bar D}^2  D_{\alpha}  V +  \frac{g}{4} {\bar D}^2  [V, D_{\alpha}  V]   \ . \label{superym6a}
\eeq 
Its component expansion in superspace coordinates $(y,\theta,{\bar \theta})$ turns out to be
\beq 
W_{\alpha}^a = - i \lambda_{\alpha}^a + \left[ \delta_{\alpha}^{\beta} D^a - i (\sigma^{\mu \nu})_{\alpha}^{\ \beta} F_{\mu \nu}^a   \right] \theta_{\beta} 
+ \theta^2 \sigma_{\alpha \dot \alpha}^{\mu} {\cal D}_{\mu} {\bar \lambda}^{\dot \alpha,a}   \ , \label{superym7a}
\eeq 
where 
\beq 
 F_{\mu \nu}^a = \partial_{\mu} A_{\nu}^a - \partial_{\nu} A_{\mu}^a - g f^{abc}  A_{\mu}^b A_{\nu}^c \quad , \quad {\cal D}_{\mu} {\lambda}^{a} =  {\partial}_{\mu} {\lambda}^{a} - g f^{abc} A_{\mu}^b  {\lambda}^{c}
\  \label{superym8a}  
\eeq 
and where $f_{abc}$ are the structure constants of the non--Abelian gauge group. The most general Lagrangian describing the interaction of a matter chiral multiplet $\Phi^i = (z^i, \psi^i, F^i)$ with the 
Yang-Mills supermultiplet is
\beq 
{\cal L} = \int d^4 \theta \ \Phi^{\dagger} e^{2gV} \Phi + \left( \int d^2 \theta   \left [\frac{1}{4} tr (W^{\alpha} W_{\alpha}) + {\cal W}(\Phi)    \right] + {\rm h.c.}   \right)  \ . \label{superym9a}
\eeq 
If there are $U(1)$ factors, one can also add Fayet-Iliopoulos terms $2 \xi_a V^a$. 
In components, one finds
\bea 
&& {\cal L} = - | {\cal D}_{\mu}  z^i|^2  - i \psi^i  \sigma^{\mu}  {\cal D}_{\mu} {\bar \psi}_i   + |F^i|^2 - \frac{1}{4} (F_{\mu \nu}^{a})^2  - i \lambda^a \sigma^{\mu}  {\cal D}_{\mu} {\bar \lambda}^a   + \frac{1}{2} D^a D^a  \nonumber \\
&& - i \sqrt{2} g   {\bar \lambda^a} {\bar \psi}_i (T^a)_j^i z^j   + i \sqrt{2} g z_j^* (T^a)_i^j  {\psi}^i   {\lambda^a}    -
\frac{1}{2} \frac{\partial^2 {\cal W}}{\partial z^i \partial z^j} \psi^i \psi^j - \frac{1}{2} \frac{\partial^2 \overline{\cal W}}{\partial z_i^* \partial z_j^*} {\bar \psi}_i {\bar \psi}_j \nonumber \\  
&&  + F^i \frac{\partial {\cal W}}{\partial z^i} +    {\bar F}_i  \frac{\partial \overline{\cal W}}{\partial z_i^*}  + g \sum_a  z_j^* (T^a)_i^j  z^i D^a +   \xi_a D^a   \  ,  \label{superym010a}
\eea
where summation over repeated indices are implicit and 
\beq 
{\cal D}_{\mu}  z^i = \partial_{\mu} z^i + i g A_{\mu}^a (T^a)_j^i  z^j \quad , \quad  {\cal D}_{\mu}  \psi^i = \partial_{\mu} \psi^i + i g A_{\mu}^a (T^a)_j^i  \psi^j  \  .  \label{superym11a}
\eeq 
The auxiliary fields are determined algebraically by their classical field equations 
 \beq 
 F^i = - \frac{\partial \overline{\cal W}}{\partial z_i^*} \quad , \quad D^a = - g  z_j^* (T^a)_i^j  z^i - \xi^a \ , \label{superym13a}
\eeq 
where as we mentioned above the Fayet-Iliopoulos terms $\xi^a$ only exist for Abelian generators.
After eliminating the auxiliary fields $F^i, D^a$, one finds
\bea 
&& {\cal L} = - | {\cal D}_{\mu}  z^i|^2  - i \psi_i \sigma^{\mu}  {\cal D}_{\mu} {\bar \psi}_i   - \frac{1}{4} (F_{\mu \nu}^{a})^2  - i \lambda^a \sigma^{\mu}  {\cal D}_{\mu} {\bar \lambda}^a     \nonumber \\
&& - i \sqrt{2} g   {\bar \lambda^a} {\bar \psi}_i (T^a)_j^i z^j   + i \sqrt{2} g z_j^* (T^a)_i^j  {\psi}^i   {\lambda^a}    -
\frac{1}{2} \frac{\partial^2 {\cal W}}{\partial z^i \partial z^j} \psi^i \psi^j - \frac{1}{2} \frac{\partial^2 \overline{\cal W}}{\partial z_i^* \partial z_j^*} {\bar \psi}_i {\bar \psi}_j \nonumber \\  
&& - \left|\frac{\partial {\cal W}}{\partial z^i}\right|^2  
- \frac{g^2}{2} \sum_a [ z_j^* (T^a)_i^j  z^i ]^2     \  .  \label{superym10a}
\eea 
The supersymmetry transformations in the Wess-Zumino gauge are
\bea 
&& \delta z^i = \sqrt{2} \epsilon \psi^i  \ , \nonumber \\
 && \delta \psi^i = i \sqrt{2} \sigma^{\mu} {\bar \epsilon} \ {\cal D}_{\mu} z^i + \sqrt{2}  {\epsilon} \  F^i \ , \nonumber \\
 && \delta F^i = i \sqrt{2} {\bar \epsilon} {\bar \sigma}^{\mu} {\cal D}_{\mu} \psi^i + 2 i g   (T^a)_j^i  z^j {\bar \epsilon}  {\bar \lambda^a} \ , \nonumber \\
 && \delta A_{\mu}^a = i ({\bar \epsilon} {\bar \sigma}_{\mu} \lambda^a -  {\bar \lambda^a} {\bar \sigma}_{\mu} \epsilon ) \ , \nonumber \\
&& \delta \lambda^a = i \epsilon D^a + \sigma^{\mu \nu} \epsilon F_{\mu \nu}^a \ , \nonumber \\
&& \delta D^a = - \epsilon \sigma^{\mu} {\cal D}_{\mu} {\bar \lambda^a} -  {\cal D}_{\mu} {\lambda^a} \sigma^{\mu} {\bar \epsilon} \ . \label{superym12a}
\eea 
Note that in this case the auxiliary fields $F^i$ and $D^a$ do not transform anymore into total derivatives: for $D^a$ this happens only in the non-Abelian case. 
This implies in particular that Fayet-Iliopoulos terms are not gauge invariant in the non-Abelian case. 
The most general renormalizable superpotential is again of the form (\ref{chirals9}), where only gauge invariant operators should be considered, as in the Abelian case of eq.~\eqref{vectors2}.  The scalar potential, defined by the last line in (\ref{superym10}), can be written
in terms of the auxiliary fields  (\ref{superym13a}) as 
\beq 
V (z^i, z_j^*) = \sum_i |F^i|^2 + \frac{1}{2} \sum_a D^a D^a \ .  \label{superym14a}
\eeq 
Note that, in analogy with the Abelian example, gauge couplings also govern some Yukawa couplings of charged fermions and quartic interactions for charged scalars. 
As a final remark, note that, when searching for supersymmetric vacua, there is no need to solve classical field equations for scalar fields. If there is a solution of
\beq 
F^i = D^a =0 \ ,  \label{superym15a}
\eeq  
which are equations for the scalar fields, it is automatically an extremum of the scalar potential. Since the corresponding vacuum energy is zero, whereas any solution
breaking supersymmetry has positive energy, such a solution is also a stable minimum. Supersymmetric vacua are then easier to find than in arbitrary non-supersymmetry theories. Using the explicit form of the auxiliary fields (\ref{superym13a}), it is relatively easy to find supersymmetric minima if they exist or to prove that supersymmetry is broken if eqs.~\eqref{superym15} have no solution. 

A supersymmetric Lagrangian can have two types of internal symmetries:

\begin{itemize}

\item Ordinary {\it internal symmetries} $U$, which commute with supersymmetry, $[Q,U] =0$. In this case, all components of a supermultiplet have the same charge and the invariance of the action demands that
the superpotential be invariant

\bea
(z'_i, \psi'_i) &=& e^{-i q_i \alpha} \left(z_i, \psi_i\right) \qquad \longleftrightarrow \qquad \Phi'_i (x,\theta,{\bar \theta}) \ = \ e^{-i q_i \alpha} \,\Phi_i (x,\theta,{\bar \theta}) \ , \nonumber \\
{\cal W}'(\Phi'_i ) &=& {\cal W} (\Phi_i ) \ . \label{vectors15a}
\eea

\item Internal {\it R-symmetries}, which do not commute with supersymmetry, $[Q,R] \not=0$. In this case,  the different component fields in a supermultiplet have different charges, the superspace coordinates transform, and the invariance of the action demands that the superpotential have a definite charge. The different transformations read
\bea
\theta' &=& e^{\frac{3 i \beta}{2}} \,\theta \ , \qquad {\bar \theta}' \ = \ e^{\frac{-3 i \beta}{2}} \,{\bar \theta} \ , \qquad W'(\Phi'_i ) \ = \ e^{3 i \beta} \,W (\Phi_i ) \ , \nonumber \\
z'_i &=& e^{i r_i \beta} z_i \ , \qquad  \psi'_i \ = \ e^{i (r_i - 3/2) \beta} \psi_i \qquad \leftrightarrow \qquad \Phi'_i (x,\theta',{\bar \theta'}) \ = \ e^{i r_i \beta} \Phi_i (x,\theta,{\bar \theta}) \ , \nonumber \\
A'_{\mu} &=& A_{\mu} \ , \qquad  \lambda' \ = \ e^{ \frac{3i \beta}{2}}  \lambda  \qquad \leftrightarrow \qquad  W'_{\alpha} (x,\theta',{\bar \theta'}) \ = \ e^{ \frac{3i \beta}{2} } \, W_{\alpha}  (x,\theta,{\bar \theta})  \ . \label{vectors16a}
\eea
In the off-shell formulation, the auxiliary fields transform as
\beq
 F'_i \ = \ e^{i (r_i - 3) \beta}\, F_i \ , \qquad D'\ = \ D  \ . \label{vectors016}
  \eeq

The supercharges also transform under R-symmetries, which implies the commutation relations
\bea
Q' &=&  e^{-\frac{3 i \beta}{2}} \,Q \ , \qquad [R,Q] \ = \ - \ \frac{3}{2} \,Q \ , \nonumber \\
{\bar Q'} &=&  e^{\frac{3 i \beta}{2}} \,{\bar Q} \ , \qquad [R,{\bar Q}] \ = \ \frac{3}{2}\, {\bar Q}   \ , \label{vectors0016}
\eea
and the superpotential transforms in such a way as to make the Lagrangian invariant. This condition demands that each term contained in a renormalizable superpotential of the type
\beq
{\cal W} \ = \ \lambda_i\,\Phi^i \ + \ \frac{1}{2}\, \lambda_{ij}\,\Phi^i\,\Phi^j \ + \ \frac{1}{3}\, \lambda_{ijk}\, \Phi^i\,\Phi^j\,\Phi^k 
\eeq
have an R-charge equal to three, so that
\beq
\lambda_i \, \neq \, 0 \, \to \, r_i \, = \, 3 \, , \quad m_{ij} \, \neq \, 0 \, \to \, r_i \,+\, r_j \, = \, 3 \, , \quad   \lambda_{ijk} \, \neq \, 0  \, \to \, r_i \,+\, r_j \,+\, r_k \, = \, 3 \  . \label{vectors17a}
\eeq

\end{itemize}

\subsection{\sc Some Exact Results in Global Supersymmetry}
\label{app:exact}

Supersymmetric theories have very special ultraviolet properties. In particular, there are no quadratically divergent contributions to scalar masses, since the superpotential is not renormalized in perturbation theory.  This result can be understood, noting that counterterms are local expressions involving full superspace integrals~\cite{superspace_review}. The arguments of~\cite{seiberg-nonren} and \cite{intriligator-seiberg-review} can provide additional indications to the same effect. In order to illustrate them, let us consider the superpotential of a theory of chiral multiplets $\Phi^i $ and couplings $g^a$,  so that $W (\Phi^i, g^a)$, regarding the couplings $g^a$ as non-dynamical background fields. Moreover, one should also assume that
\begin{itemize}   
\item the theory is invariant under the larger symmetry group obtained if transformation rules are assigned to the background fields $g^a$;
\item  the superpotential is holomorphic in the couplings/background fields, and in particular it does not depend on the conjugate couplings ${\overline g}{}^a$;   
\item  the superpotential can be analyzed at weak coupling.
\end{itemize}

Consider, as an example, the Wess-Zumino model with tree-level superpotential
\beq  
W_0 \ = \ \frac{m}{2} \, \Phi^2 \ + \  \frac{\lambda}{3} \, \Phi^3   \ ,  \label{exact1}
\eeq   
and regard the couplings $m,\lambda$ as background fields. The model has two relevant $U(1)$ symmetries, under which $\Phi$, $m$ and $\lambda$ transform with the charges collected in Table~\ref{tab:u1 charges}.
\begin{table}[ht]
\centering
\begin{tabular}{||c || c || c ||}
 \hline
 & $U(1)$ & $U(1)_R$ \\ [0.5ex]
 \hline\hline
$\Phi$ & $1$ &  $1$ \\ \hline
$m$ & $-\,2$ &  $1$ \\ \hline
$\lambda$ & $-\,3$ &  $0$ \\ \hline
\end{tabular}
\caption{$U(1)$ and $U(1)_R$ charges for the Wess--Zumino model of eq.~\eqref{exact1}.}
\label{tab:u1 charges}
\end{table}

The two $U(1)$ symmetries and holomorphy restrict the exact superpotential to the form
\beq 
W_{\rm exact} \ = \  \frac{m}{2} \, \Phi^2 \ f \left(z \,=\, \frac{\lambda \Phi}{m}\right)   \ .  \label{exact2}
\eeq  
The comparison with weak coupling can be obtained taking the limit $\lambda \to 0$, $m \to 0$, with fixed ratio. $W_{\rm exact}$ then approaches $W_0$,  and therefore one can conclude that
\beq
f (z) \ = \  1 \ + \ \frac{2}{3}\, z \ .
\eeq
However, $z$ is arbitrary in the preceding argument, and therefore this perturbative value of $f(z)$ is exact to all orders in perturbation theory.  However, it is well known that non-perturbative effects can modify the superpotential. 

When quantum corrections are taken into account, the general Lagrangian describing the interaction of a matter chiral multiplet $\Phi^i $ with the 
Yang-Mills supermultiplet takes the form
\beq 
{\cal L}_W = \int d^4 \theta \ Z  \ \Phi^{\dagger} e^{2gV} \Phi \ + \ \left( \int d^2 \theta   \left [\frac{1}{4 g_W^2} tr (W^{\alpha} W_{\alpha}) \ + \ W(\Phi, g^a)    \right] \ + \ {\rm h.c.}   \right)  \ . \label{exact3}
\eeq 
An effective action defined by integrating out heavy degrees of freedom from a UV scale $M_0$ down to a scale $\mu$ is usually called a Wilsonian action. In this case, the Wilsonian gauge coupling $g_W$ is also holomorphic. It is also subject to a non-renormalization theorem, which states that it is only renormalized at one loop. This means that the one-loop running of the gauge coupling
\beq  
\frac{1}{g_W^2 (\mu)} = \frac{1}{g_W^2 (M_0)} \ + \  \frac{T (R) \ - \ 3 T (G)}{8 \pi^2} \ln \frac{M_0}{\mu}    \  , \label{exact4}
\eeq  
does not receive higher-order corrections and is thus exact in perturbation theory. 
In the preceding formula, $G$ refers to the adjoint representation, while in general
\beq
Tr_R T^A T^B \ = \  \delta^{AB} T (R) \ ,
\eeq
and $T(R)$ defines the Dynkin index in the group representation $R$.
However, the effective action obtained by integrating over all momenta is different.  In particular, the physical gauge coupling is non-holomorphic and its beta function receives contributions from all orders
of perturbation theory. Its energy dependence is governed by the equation \cite{shifman-vainshtein-beta}
\beq  
\frac{1}{g^2 (\mu)} \ = \ \frac{1}{g_0^2 }  \ + \  \frac{T (R)}{8 \pi^2} \ln \left( \frac{M_0}{ Z \mu}  \right)  \ - \ \frac{3 T (G)}{8 \pi^2}   \ln \left( \frac{M_0}{  \mu}  \left[  \frac{g^2 (\mu)}{g_0^2} \right]^{\frac{1}{3}} \right)  \  , \label{exact5}
\eeq  
where the additional factors of $Z$ and $g^2 (\mu)$ in  eq.~(\ref{exact5}) can be understood by rescaling of the matter fields and the gauge fields, due to renormalization effects.  Taking the derivative of the physical coupling with respect to $\ln \mu$ in eq. (\ref{exact5}) 
defines the beta function. One thus obtains the all-loop exact beta function \cite{shifman-vainshtein-beta}
\beq  
\beta (g) = - \frac{g^3}{16 \pi^2}  \frac{3 T (G) - T (R) (1- \gamma_R)}{1 - T (G) \frac{g^2}{8 \pi ^2}}   \  , \label{exact6}
\eeq 
where 
\beq  
\gamma_R = - \frac{\partial \ln Z}{\partial \ln \mu}  \   \label{exact7}
\eeq 
is the anomalous dimension of the chiral field $\Phi$, which transforms in the representation $R$ of the gauge group. 

\section{\sc Four--Dimensional \texorpdfstring{$N=1$} \ \ Local Supersymmetry}
\label{app:superfields_local}
We can now review some properties of the gravity multiplet and of four--dimensional $N=1$ supergravity, which is based on it.

\subsection{\sc The Rarita--Schwinger Field}
\label{app:grav_mult}

In supergravity, the graviton is part of a multiplet that also includes one or more (Majorana) gravitini $\Psi_{M}$. In a four--dimensional supersymmetric vacuum, both the graviton and a Majorana gravitino have two degrees of freedom on-shell, but the balance is again lost off shell.

Before providing some details on $N=1$ supergravity in four dimensions, it is perhaps useful to recall some properties of the Rarita--Schwinger action for the gravitino in flat space. Let us begin by considering the case of a massless Dirac gravitino in $D$--dimensional Minkowski space, which is described by
\beq
{\cal S} \ = \ \,-\, i\, \int \, d^{\,D} x \  {\overline \Psi}_M\, \Gamma^{MNP} \, \partial_N\, \Psi_P  \ ,
\eeq
in a generic number $D$ of dimensions. Here $\Gamma^{MNP}$ is totally antisymmetric in $(M,N,P)$, and if $M \neq N \neq P$
\beq
\Gamma^{MNP}  \ = \ \Gamma^{M} \, \Gamma^{N} \, \Gamma^{P} \ .
\eeq
In four dimensions, the gravitino field can be subject to a Majorana or Weyl constraint, while in ten dimensions, where they are independent, it can be even subject to both.

The Lagrangian is invariant under 
\beq
\delta\,\Psi_M \ = \ \partial_M\,\epsilon \ , 
\eeq
where $\epsilon$ is a local supersymmetry parameter, which is a spinor field subject to the same projections as the gravitino field $\Psi_M$. 
The Rarita--Schwinger equation of motion reads
\beq
\Gamma^{MNP} \, \partial_N\, \Psi_P  \ = \ 0 \ , \label{eq_RS}
\eeq
and its $\gamma$ trace is
\beq
\Gamma^{MN} \, \partial_M\, \Psi_N \ = \ 0 \ . \label{divergence}
\eeq
The counterpart of the Lorentz gauge of Electrodynamics for this case is the condition
\beq
\Gamma^M\, \Psi_M \ = \ 0 \ , \label{b91}
\eeq
which can be reached by solving a Dirac equation, 
and when combined with eq.~\eqref{divergence} yields the second constraint
\beq
\partial^{\,M}\, \Psi_M \ = \ 0 \ . \label{b92}
\eeq
Making use of these results in the equation of motion \eqref{eq_RS} then reduces it to the Dirac equation
\beq
\Gamma \cdot \partial \ \Psi_M  \ = \ 0 \ , 
\eeq
which implies the massless Klein--Gordon equation
\beq
\Box\,\Psi_M \ = \ 0 \ .
\eeq

Selecting a light--like momentum, one can repeat almost verbatim the steps that associate $D-2$ polarizations to a massless vector.  In the present case, taking into account the presence of a residual on--shell gauge transformation, one can eliminate $\Psi^\pm$ leaving only $(D-2)$ transverse $\Psi_i$. However, only $(D-3)$ of them are independent, in view of eq.~\eqref{b91}, and finally the Dirac equation halves, on shell, the remaining components. These arguments thus show that a massless Dirac gravitino has ${color{red} {(D-3)}\, {2^{[D/2]}}}$ degrees of freedom, a number that can be further reduced by Majorana and/or Weyl projections, when these are possible.

The massive Rarita--Schwinger action is
\beq
{\cal S} \ = \ \int \, d^{\,D} x \left[ - \ i\,  {\overline \Psi}_M \, \Gamma^{MNP} \, \partial_N\, \Psi_P  \ - \ i\, m \ {\overline \Psi}_M \, \Gamma^{MN} \, \Psi_N \right]\ ,
\eeq
and now the equation of motion is
\beq
\Gamma^{MNP} \, \partial_N\, \Psi_P  \ + \ m\, \Gamma^{MN} \, \Psi_N \ = \ 0 \ .
\eeq
Its divergence now yields~\eqref{divergence}, and combining it with its $\Gamma$-trace one obtains eqs.~\eqref{b91} and ~\eqref{b92}. Taking these results into account, one is finally led to the massive Dirac equation
\beq
\left( \Gamma \cdot \partial \ + \ m \right)\Psi_M  \ = \ 0 \ , 
\eeq
which implies the massive Klein--Gordon equation. Taking a rest--frame momentum $P^M=(m,0,\ldots,0)$, one can now conclude that $\Psi^0=0$, so that the field contains $\frac{(D-2)}{2}\, {2^{[D/2]}}$ degrees of freedom, in view of eq.~\eqref{b91}. Note that these degrees of freedom are as many as those of a massless gravitino and a massless fermion in $D+1$ dimensions, in the spirit of what we saw for the Kaluza--Klein theory in the main body of the review.

\subsection{\sc Pure \texorpdfstring{$N=1$} \ \ Supergravity in Four Dimensions}

We can now briefly discuss $N=1$ Supergravity in four dimensions~\cite{sugra1,sugra2,sugrarev}, and the first step to this end involves a combination of the Einstein--Hilbert term in the vielbein formalism and a covariant completion of the Rarita--Schwinger action for the gravitino field $\Psi_\mu$, here written in Majorana notation, in terms of a yet unspecified spin connection $\omega$,
\beq
{\cal S}\left[e,\omega,\Psi\right] \ = \ \frac{1}{2\, \kappa^2}\ \int d^4 x \ e \ \left(\  e^\mu{}_a \, e^\nu{}_b \, R_{\mu\nu}{}^{ab} \ - \ {i}\ \overline{\Psi}_\mu\,\Gamma^{\mu\nu\rho}\,D_\nu(\omega)\,\Psi_\rho \,  \right) \ , \label{sugra}
\eeq
where $R_{\mu\nu}{}^{ab}$ is given in terms of the spin connection $\omega_\mu{}^{ab}$ by eq.~\eqref{Romega} and the $\Gamma^{\mu\nu\rho}$ are curved, as explained in Appendix~\ref{app:conventions}, and are related to the flat $\gamma$ matrices according to
\beq
\Gamma^\mu \ = \ e^\mu{}_a \ \gamma^a \ .
\eeq

Following the original works, we exclude torsion terms in the covariant derivative entering the Rarita--Schwinger action contained in eq.~\eqref{sugra},
so that taking the antisymmetry in $(\mu,\rho)$ of eq.~\eqref{sugra} into account, it suffices to work with
\beq
D_\mu(\omega)\,\Psi_\nu \ = \ \partial_\mu\,\Psi_\nu \ + \ \frac{1}{4}\ {\omega_\mu}^{ab}\, \gamma_{ab}\ \Psi_\nu  \ .
\eeq

It is now convenient to eliminate from ${\cal S}$ the determinant of the vielbein, using the identities
\bea
&& e \left(e^\mu{}_a \, e^\nu{}_b \ - \ e^\mu{}_b \, e^\nu{}_a\right) \ = \ - \ \frac{1}{2}\ \epsilon^{\mu\nu\rho\sigma}\, \epsilon_{abcd} \ e_\rho{}^c \,e_\sigma{}^d  \ ,\label{levicivita} \\
&& e\, \Gamma^{\mu\nu\rho} \ = \ - \ i\  \epsilon^{\mu\nu\rho\sigma}\,\gamma_5\,\Gamma_\sigma \ ,
\label{epsilongamma}
\eea
where $\epsilon^{\mu\nu\rho\sigma}$ is the completely antisymmetric tensor in four dimensions with values $\pm 1$,
turning the action into
\beq
{\cal S} \ = \  \frac{1}{2\, \kappa^2}\ \epsilon^{\mu\nu\rho\sigma} \int d^4 x \left(-\,\frac{1}{4}\   \epsilon_{abcd} \ e_\rho{}^c \,e_\sigma{}^d \, {R_{\mu\nu}}^{ab} \  -\  \overline{\Psi}_\mu\,\gamma_5\,\Gamma_\sigma\,D_\nu\,\Psi_\rho \right) \ . \label{sugra_2}
\eeq

We now \emph{define} $\omega$ as the solution of its field equation
\beq
\epsilon^{\mu\nu\rho\sigma}\left[ \epsilon_{abcd}\,e_\rho{}^c \left( D_\mu\,e_\sigma{}^d \ - \ D_\sigma\,e_\mu{}^d \right) \ - \ \frac{1}{2} \ \overline{\Psi}_\mu\,\gamma_5\,\gamma_c\,\gamma_{ab}\,\Psi_\rho\ e_\sigma{}^c\right] \ = \ 0 \
, \label{torsion_1}
\eeq
where
\beq
D_\mu\,e_\nu{}^a \ = \ \partial_\mu\,e_\nu{}^a \ + \ \omega_\mu{}^{ab}\,e_{\nu\, b} \ .
\eeq
The Majorana flip symmetries of Fermi bilinears reviewed in Appendix~ \ref{app:conventions} imply that the only contribution to the Fermi bilinear originates from the three--$\gamma$ term, and from eq.~\eqref{epsilongamma} one can deduce that
\beq
\gamma_5\,\gamma_{abc} \ = \ - \ i \, \epsilon_{abcd}\,\gamma^d \ .
\eeq
In this fashion, eq.~\eqref{torsion_1} implies the ``gravitino torsion equation''
\beq
D_\mu\, e_\nu{}^a \ - \ D_\nu\,e_\mu{}^a \ = \ \frac{i}{2} \ {\overline \Psi}_\mu\,\gamma^a\,\Psi_\nu \ , \label{torsion_2}
\eeq
which reflects indeed the presence of a torsion contribution in $\omega$ that is determined by the gravitino field $\psi_\mu$. This is directly implied by the ``vielbein postulate''
\beq
D_\mu\, e_\nu{}^a \ - \ \Gamma^\rho{}_{\mu\nu}\, e_\rho{}^a \ = \ 0 \ ,
\eeq
from which one can deduce that
\beq
D_\mu\, e_\nu{}^a \ - \ D_\nu\, e_\mu{}^a \ = \ \left(\Gamma^\rho{}_{\mu\nu} \ - \ \Gamma^\rho{}_{\nu\mu}\right)\, e_\rho{}^a \ .
\eeq

We can now sketch how one can prove that the action \eqref{sugra} is invariant under the supersymmetry transformation
\beq
\delta\, e_\mu{}^a \ = \ \frac{i}{2}\ \overline{\epsilon}\, \gamma^a \, \Psi_\mu \ , \qquad \delta\,\Psi_\mu \ = \  D_\mu(\omega) \,\epsilon \ ,
\eeq
up to the torsion equation \eqref{torsion_2} and up to total derivatives. To this end, let us begin by examining the variation of the Einstein--Hilbert term,
\beq
\delta\,{\cal S}_{EH} \ = \ - \ \frac{i}{8\, \kappa^2 } \int d^4 x \,  \epsilon^{\mu\nu\rho\sigma} \epsilon_{abcd} \ {\overline \epsilon}\,\gamma^c\,\Psi_\rho \ e_\sigma{}^d \, {R_{\mu\nu}}^{ab}\ , \label{deltaseh}
\eeq
which is simpler and should be canceled by the variation of the Rarita--Schwinger action. 

There is no need to vary $\omega$, both here and in the Rarita--Schwinger action, since it is a non--dynamical field defined to satisfy its field equation, and therefore its variation vanishes identically. Still, the variation of the Rarita--Schwinger action comprises three types of contribution:
\begin{enumerate}
\item the variations of ${\overline{\Psi}}_\mu$ and $\Psi_\sigma$;
\item the variation of the vielbein contained in $\Gamma_\sigma$;
\item a torsion term generated by a partial integration, after variation ${\overline \Psi}_\mu$.
\end{enumerate}
\noindent
The two terms arising from the variations of the gravitino fields yield
\beq
\delta_1\,{\cal S}_{RS} \ = \ \frac{1}{16\,\kappa^2}\ \int d^4 x \, \epsilon^{\mu\nu\rho\sigma} \ e_\nu{}^c\, \overline{\epsilon}\,\gamma_5\left\{\gamma_c,\gamma_{ab} \right\}\Psi_\sigma\ {R_{\mu\rho}}^{ab}(\omega) \ ,
\eeq
where the Riemann curvature tensor has emerged from commutators of covariant derivatives.
The $\gamma$--matrix anticommutator now yields $2\,\gamma_{abc}$, and then, using eq.~\eqref{epsilongamma}, one can conclude that
\beq
\delta_1\,{\cal S}_{RS} \ = \ - \ \frac{i}{8\,\kappa^2}\ \int d^4 x \, \epsilon^{\mu\nu\rho\sigma} \ \epsilon_{abcd}\ \overline{\epsilon}\,\gamma^d \, \Psi_\sigma\ e_\nu{}^c \ {R_{\mu\rho}}^{ab}(\omega) \ ,
\eeq
an expression that cancels exactly against $\delta\,{\cal S}_{EH}$ in eq.~\eqref{deltaseh}, after some relabeling.
However, the preceding result was obtained integrating by parts the variation of ${\overline{\Psi}}_\mu$, which yields the additional contribution
\beq
\delta_2\,{\cal S}_{RS} \ = \ \frac{1}{4\,\kappa^2} \ \int d^4 x \,\epsilon^{\mu\nu\rho\sigma} \left( D_\mu\, e_\nu{}^a \ - \ D_\nu\,e_\mu{}^a \right)\overline{\epsilon}\,\gamma_5\,\gamma_a\,D_\rho\,\Psi_\sigma \ ,
\eeq
or
\beq
\delta_2\,{\cal S}_{RS} \ = \ \frac{i}{8\,\kappa^2} \ \int d^4 x \,\epsilon^{\mu\nu\rho\sigma} \ {\overline \Psi}_\mu\,\gamma^a\,\Psi_\nu \ \, \overline{\epsilon}\,\gamma_5\,\gamma_a\,D_\rho\,\Psi_\sigma \ ,
\eeq
after making use of the torsion equation \eqref{torsion_2}.
Finally, the variation of the vielbein in eq.~\eqref{sugra_2} yields
\beq
\delta_3\,{\cal S}_{RS} \ = \ -\ \frac{i}{4\,\kappa^2} \ \int d^4 x \,\epsilon^{\mu\nu\rho\sigma} \ {\overline \epsilon}\,\gamma^a\,\Psi_\nu \ \, \overline{\Psi}_\mu\,\gamma_5\,\gamma_a\,D_\rho\,\Psi_\sigma \ ,
\eeq
which cancels against $\delta_2\,{\cal S}_{RS}$ after a Fierz rearrangement. All these identities follow from the completeness of antisymmetric products of $\gamma$ matrices, and are contained in
\beq
\lambda\,\bar{\psi} \ = \ - \ \frac{1}{4}\ \bar{\psi}\,\lambda \ - \ \frac{1}{4}\ \gamma_5 \bar{\psi}\,\gamma_5\, \lambda  \ - \ \frac{1}{4}\ \gamma^\mu \bar{\psi}\,\gamma_\mu\,\lambda \ + \ \frac{1}{4}\ \gamma^\mu\,\gamma_5 \,\bar{\psi}\,\gamma_\mu\,\gamma_5\, \lambda \ - \ \frac{1}{8}\,\gamma^{\mu\nu}\,\bar{\psi}\,\gamma_{\mu\nu}\,\lambda \ ,
\eeq
where the coefficients can be computed via traces, taking the anticommuting nature of the spinors into account.

$N=1$ supergravity admits an important deformation~\cite{townsend}, which comprises the addition of a gravitino mass term and a correlated \emph{negative} cosmological constant that are encoded in
\beq
\Delta\,{\cal L} \ = \ \frac{1}{2\,\kappa^2} \ \int d^4 x \, e\left( \frac{i}{L} \ \bar{\psi}_\mu\,\Gamma^{\mu\nu}\,\psi_\nu \ + \ \frac{6}{L^2}\right) \ .
\eeq
The supersymmetry transformations are also deformed, and become
\beq
\delta\, e_\mu{}^a \ = \ \frac{i}{2}\ \overline{\epsilon}\, \gamma^a \, \Psi_\mu \ , \qquad \delta\,\Psi_\mu \ = \  D_\mu(\omega) \,\epsilon \ + \ \frac{1}{L}\ \Gamma_\mu\,\epsilon \ .
\eeq
The mass term thus added is precisely needed to describe a gauge invariant spin--$\frac{3}{2}$ in an $AdS_4$ background.

The supersymmetry algebra only closes on shell for $N=1$ supergravity. In the following section, we shall see how auxiliary fields can be introduced to arrive at an off--shell formulation of the gravity multiplet.

\subsection{\sc \texorpdfstring{$N=1$} \ \ Supergravity in Four Dimensions in Superspace}

There are various off-shell extensions of four--dimensional supergravity, and here we briefly describe the ``old-minimal" one, which contains two auxiliary fields, a complex scalar $u$ and a vector $A_\mu$. $A_\mu$ is not a gauge field in Poincar\'e supergravity, but a non--propagating auxiliary four-vector, which therefore has four off--shell (and zero on--shell) degrees of freedom.  Taking the auxiliary fields and the gauge symmetries into account, the off--shell bosonic and fermionic degrees of freedom are then balanced, since $g_{\mu \nu}$ contributes $10-4=6$ of them, while $u$ and $A_m$ contribute 2 and 4, for a total of $12$, to be compared with $4 \times 4-4 = 12$ degrees of freedom from the gravitino.

The off-shell gravity multiplet is therefore
\beq
(g_{\mu \nu} , \Psi_{\mu},u, A_{\mu}) \ . \label{eq:gravm1}
\eeq
In superspace, the gravity multiplet is encoded in the compensator chiral field $S_0$, which defines a chiral curvature superfield
\beq
{\cal R} \ = \ \frac{1}{S_0} \Sigma ({\overline S_0})
\ , \label{eq:gravm2}
\eeq
where $\Sigma ({\overline S_0})$ denotes the chiral projector in supergravity. The components of the chiral curvature superfield are
\beq
{\cal R} \ = \ \left( \ {\bar u} \ , \ \gamma^{\mu \nu} {\cal D}_{\mu} \Psi_{\nu} \ , \  -\frac{1}{2} R - \frac{1}{3} A_{\mu} A^{\mu} + i {\cal D}^{\mu} A_{\mu} - \frac{1}{3} |u|^2 \ \right) \ , \label{eq:gravm3}
\eeq
where ${\cal D}_{\mu}$ denotes the proper covariant derivatives acting on the gravitino and on the four-vector auxiliary field.
With these ingredients, the pure supergravity Lagrangian can be written in superspace as
\beq
{\cal L}_{SUGRA} \ = \  - \left[S_0 {\overline S_0}\right]_D \ + \
\left[W_0 S_0^3\right]_F \ , \label{eq:gravm4}
\eeq
where $W_0$ is a constant that can be regarded as a constant superpotential, but whose physical interpretation will become more transparent shortly.

A useful supergravity identity converts a D-density containing the compensator field into a F-density according to
\beq
\left[ \left( g(\phi_i) + \overline{g(\phi_i)} \right) S_0 {\overline S_0} \right]_D =
\left[ g(\phi_i)  {\cal R}  S_0^2 \right]_F
\ , \label{eq:gravm5}
\eeq
where $g(\phi_i)$ is an arbitrary holomorphic function of chiral superfields $\phi_i$. Using eq.~(\ref{eq:gravm5}) one can also present the supergravity Lagrangian of eq.~(\ref{eq:gravm4}) as a pure F-density
\beq
{\cal L}_{SUGRA} =
\left[ \ - \frac{1}{2} {\cal R}  S_0^2  \ + \  W_0 S_0^3 \ \right]_F \ . \label{eq:gravm6}
\eeq
In components, one obtains
\bea
 {\cal S}_{SUGRA} &=& \int d^4 x \ \det e \ \left\{  - \ \frac{1}{2} R \ - \  \frac{i}{2} {\bar \Psi}_{\mu}  \gamma^{\mu \nu \rho} D_{\nu} \Psi_{\rho} \ + \
\frac{1}{3} A_{\mu } A^{\mu} \ - \  \frac{1}{3} |u|^2 \right.  \nonumber \\
&+& \left. \left[ W_0  \left(u \,+ \,\frac{i}{2}\, {\bar \Psi}_{\mu}
\gamma^{\mu \nu} \Psi_{\nu} \right)  \ + \  {\rm h.c.}  \right]  \right\} \ . \label{eq:gravm7}
\eea

From the action eq. (\ref{eq:gravm7}) it is clear that $u$ and $A_{\mu}$ are non-propagating auxiliary fields, which can be readily eliminated via their field equations. In this simplest pure-supergravity case one obtains
\beq
A_{\mu} \  = \  0 \ , \qquad u \ = \ 3 \, \overline{ W}{}_0
\  \label{eq:gravm8}
\eeq
and finally the on-shell supergravity action becomes
\beq
 {\cal S}_{SUGRA} = \int d^4 x \ \det e \ \left\{  - \frac{1}{2} R - \frac{i}{2} {\bar \Psi}_{\mu}  \gamma^{\mu \nu \rho} D_{\nu} \Psi_{\rho} +  \left( \frac{i}{2} W_0  {\bar \Psi}_{\mu}
\gamma^{\mu \nu} \Psi_{\nu} + {\rm h.c.}  \right) + 3 |W_0|^2  \right\} \ . \label{eq:gravm9}
\eeq
This describes a theory with a gravitino mass parameter $m_{3/2} = W_0$ and negative vacuum energy/cosmological constant  $\Lambda = - 3 m_{3/2}^2$ in Planck units. Supersymmetry grants that, despite the mass parameter, the gravitino propagates only two degrees of freedom, which match those of the graviton.  The negative contribution to the scalar potential presents itself, more generally, in all $N=1$ supergravity theories coupled to matter, where in general the gravitino has a field--dependent mass $m_{3/2} = W e^{K/2}$. This negative contribution, instrumental to allow for supersymmetry breaking with vanishing (or very small) cosmological constant, originates from the auxiliary field $u$ of the gravity multiplet, whose contribution to the scalar potential is opposite in sign compared to those of the $F^i$ auxiliary fields of chiral multiplets and the $D^a$ auxiliary fields of vector multiplets.
%%%%%%%%%%%%%%%%%%%%%%%%%%%%%%%%%%%%

\section{\sc Open Descendants of the Bosonic String} \label{app:bosonic_orientifold}
This appendix is devoted to illustrating in detail how to build unoriented closed and open strings, starting from oriented closed strings only with spectra that are symmetric under the interchange of left and right modes. This procedure is referred to as orientifolding or building open descendants. It rests on extending the sum over closed--oriented Riemann surfaces of arbitrary genera by also allowing, in them, crosscaps and/or boundaries. The resulting spectra are determined by combining the torus amplitude with additional contributions from the Klein-bottle, the annulus, and the M\"obius strip. These additional contributions to the vacuum energy encode, in their integrands, the key constraints underlying the construction. New infrared divergences appear in general, which can signal the emergence of potentials in the low--energy effective theory or even of anomalies in the conservation of gauge and gravitational currents.

The bosonic string is the simplest example whose partition functions
\beq
{\cal T}\  = \ \int_{\cal F} \ \frac{d^2 \tau}{\tau_2^2} \ \frac{1}{\tau_2{}^{12}} \ \mathrm{Tr}\left[ q^{N-1} \ \bar{q}{}^{\bar{N}-1} \right] \ = \  \int_{\cal F} \ \frac{d^2 \tau}{\tau_2^2} \
\frac{1}{\left(\sqrt{\tau_2}\left|\eta(\tau)\right|^2\right)^{24}} \label{bosonictorus}
\ ,
\eeq
is symmetric under the interchange of left and right modes. We already saw in eq.~\eqref{bosonictorusn}, and we repeat here for the reader's convenience.
 The spectrum can be projected with the operator $\Omega$ that swaps the two sets of modes. 
\begin{figure}[ht]
\centering
%\begin{figure}
\begin{tabular}{ccc}
%\mbox{graphic} & \mbox{table} \\
\includegraphics[width=28mm]{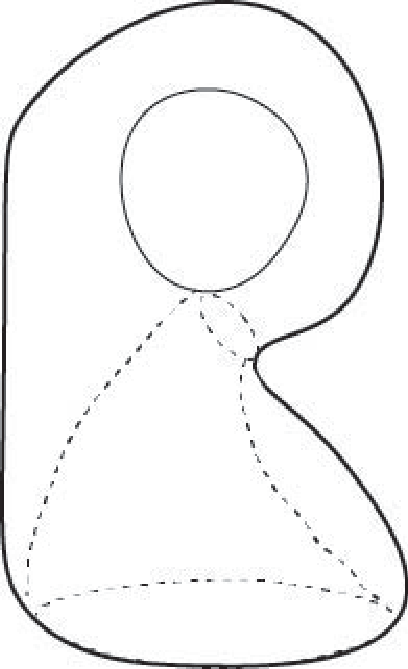} & \hskip 1cm \includegraphics[width=38mm]{Klein.eps} & \hskip 1cm \includegraphics[width=60mm]{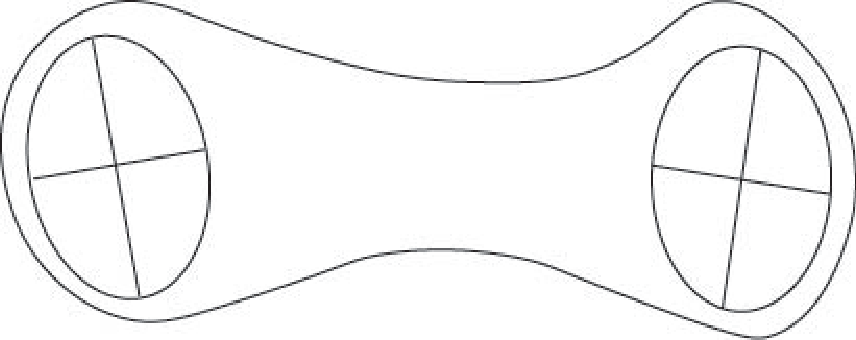}  \\
\end{tabular}
\caption{\small When embedded in three--dimensional Euclidean space, the Klein--bottle has the peculiar self-intersecting shape in the left panel. In its complex plane representation, given in the middle panel, one can distinguish the fundamental rectangle for the ``direct channel'' (the lower rectangle, with vertices at 1 and $i \tau_2$), the doubly covering torus (the union of the two rectangles, with vertices at 1 and $2 i \tau_2$) and the fundamental polygon for the ``transverse channel'' (the shaded rectangle). Finally, the right panel illustrates the transverse--channel representation as a tube terminating at two crosscaps.}
\label{fig:Klein_bottle}
\end{figure}
This projection replaces the preceding expression with
\beq
 \frac{1}{2}\, {\cal T} \ + \ {\cal K} \ ,
\eeq
where
\beq
{\cal K} \ = \ {\textstyle\frac{1}{2}} \ \int_{{\cal F}_{\cal K}} \ \frac{d^2
\tau}{\tau_2^2} \
\frac{1}{\tau_2^{12}}\ {\rm Tr} \left[ q^{N - 1} \
\bar{q}^{\bar{N} - 1} \ \Omega \right] \ ,
\eeq
and the integration domain ${\cal F}_{\cal K}$ will be identified shortly. The inner trace is, in fact,
\beq
\sum_{{\rm L},{\rm R}} \langle {\rm L},{\rm R}| \ q^{N - 1} \
\bar{q}^{\bar{N} - 1} \ \Omega \ |{\rm L},{\rm R}\rangle \ = \ \sum_{{\rm L},{\rm R}} \langle {\rm L},{\rm R}| \ q^{N - 1} \
\bar{q}^{\bar{N} - 1} \ | {\rm R},{\rm L} \rangle \ ,
\eeq
since $\Omega | {\rm L},{\rm R} \rangle  =  | {\rm
R},{\rm L} \rangle $. Consequently, the integrand reduces to
\beq
\sum_{\rm L}
\langle {\rm L},{\rm L}| \ (q \bar{q} )^{N - 1} | {\rm L},{\rm L} \label{D5}
\rangle
\eeq
and the restriction to the diagonal subset $| {\rm L},{\rm L}
\rangle $  has effectively identified $N$ and
$\bar{N}$. As a result, the amplitude depends on $q \bar{q}=e^{-\,4\pi\tau_2}$ and $2 i \tau_2$, as we are about to explain, is the modulus of the doubly covering torus of the Klein bottle, while the integration over $\tau_1$ is trivialized. However, the leftover integration domain is the whole $\tau_2$ half--line, since the different Klein bottles thus defined are inequivalent.
\begin{figure}[ht]
\centering
%\begin{figure}
\begin{tabular}{ccc}
%\mbox{graphic} & \mbox{table} \\
\includegraphics[width=28mm]{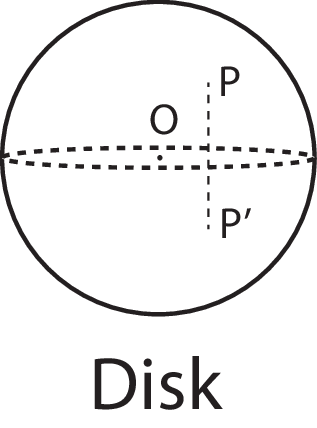} &
{} \hskip 5cm & \includegraphics[width=28mm]{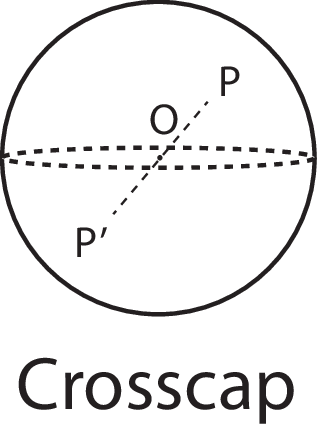}  \\
\end{tabular}
\caption{\small Left panel: the disk is an open and orientable surface obtained from a sphere by identifying points of opposite latitude. Right panel: the crosscap, or real projective plane, is a closed non--orientable surface obtained from a sphere identifying antipodal points.}
\label{fig:boundary_crosscap}
\end{figure}

When embedded in three--dimensional Euclidean space, the Klein bottle has self--intersections, as can be seen in the left panel of fig.~\ref{fig:Klein_bottle}. 
This non--orientable surface can be defined by subjecting the complex $z$ plane to the (anti)holomorphic identifications
\beq
z \ \sim \ z \ + \ 1 \, \qquad z \ \sim \ - \ \bar{z} \ + \ i \,\tau_2  \ ,
\eeq
the second of which implies that
\beq
\qquad z \ \sim \ {z} \ + \ 2\, i \,\tau_2 \ ,
\eeq
so that it is doubly covered, indeed, by a torus with $\tau = 2 \,i\,\tau_2$, consistently with the result in eq.~\eqref{D5}. In
conclusion, after performing the trace, for the bosonic string one finds
\beq
{\cal K} = {\textstyle\frac{1}{2}} \, \int_0^\infty \ \frac{d
\tau_2}{\tau_2^{14}}
\ \frac{1}{\eta^{24}(2 i \tau_2)} \ , \label{klein1d}
\eeq
and the ultraviolet problem at $\tau_2=0$ is not eliminated, in sharp contrast to what we saw for the torus amplitude. As we shall see shortly, this divergence can be eliminated by introducing open strings, whose contributions will rest on the other two surfaces of vanishing Euler character, the annulus and the M\"obius strip.

It is instructive to compare the $q$-expansions of the integrands of
${\cal T}$ and ${\cal K}$, retaining in the former only contributions
with equal powers of $q$ and $\bar{q}$, which correspond to on-shell
physical states satisfying the level-matching condition.
 In
addition to the powers of $\tau_2$, the relevant portions of the integrands of ${\cal T}$ and ${\cal K}$ are then
\bea
{\cal T} &\rightarrow& \left (q \bar{q})^{-1} + (24)^2 +
\ldots\right) \, , \nonumber \\
{\cal K} &\rightarrow& {\textstyle\frac{1}{2}} \left(
(q \bar{q})^{-1} + (24) + \ldots\right) \, ,
\eea
which indicate that the right counting of states in the projected spectrum
is indeed attained by halving the torus amplitude ${\cal T}$
and adding to it the Klein-bottle amplitude ${\cal K}$, as defined in eq.~\eqref{klein1d}.

Following \cite{gsop,cp5}, let us now use as an integration variable
the modulus $t = 2
\tau_2$ of the doubly--covering torus of the Klein bottle, so that the amplitude becomes
\beq
{\cal K} \ = \ \frac{2^{13}}{2} \ \int_0^\infty \ \frac{d t}{t^{14}}
\ \frac{1}{\eta^{24}(it)}  \,. \label{klein2d}
\eeq
This step is important, since the Klein bottle allows two distinct natural choices of
``time'', and with one of them, as we shall see, this contribution is similar to those of the other two surfaces of the same genus, the annulus and the M\"obius strip, which will soon emerge in this discussion. The vertical time, $\tau_2$, enters the operatorial
definition of the trace, as we have seen, and defines the amplitude of the {\it direct channel} or
{\it loop channel}. On the other hand, when referring to the horizontal
time $\ell=\frac{1}{t}$,
the Klein bottle becomes a tube terminating at two ``cross-caps'', and defines
the {\it transverse channel} or {\it tree--level} closed--string amplitude. 

The disk and the crosscap (or real projective place) are two surfaces that are doubly covered by spheres, as shown in fig.~\ref{fig:boundary_crosscap}. The former is obtained by identifying points of opposite latitude, and the equator is then a set of fixed points, the boundary of the disk. The latter is obtained by identifying antipodal points, so that it is a closed non--orientable surface where opposite points on the equator are also identified.

Since we focus on the integrand, it is convenient to use  a different name, $\tilde{\cal K}$, for the transverse--channel amplitude, which can be obtained from eq. (\ref{klein2d}) by the $S$ modular
transformation in eq.~\eqref{STeta}. In this case
\beq
\tilde{\cal K} \ = \ \frac{2^{13}}{2} \ \int_0^\infty \ d \ell
\ \frac{1}{\eta^{24}(i\ell)}  \label{klein1t} \ ,
\eeq
and, in this fashion, the ultraviolet divergence at $\tau_2=0$ has become an infrared one at $\ell=\infty$. Leaving aside the tachyon contribution, there is a massless exchange at zero momentum, proportional to 
\beq
\frac{2^{13}}{2} \ \int_0^\infty \ d \ell
  \label{klein1tadpole} \ ,
\eeq
which can regarded as a zero--mass limit of the expression
\beq
\frac{2^{13}}{2} \ \int_0^\infty \ d \ell \ e^{ -\,\alpha'\, m^2 \ell} \ = \ \frac{2^{13}}{2} \ \frac{1}{\alpha'\, m^2}  \ . 
  \label{klein1tadpole2}
\eeq
This contribution can be associated to an on--shell zero--momentum string amplitude for the dilaton on the projective plane. 
In fact, in the large--$\ell$ limit, the vacuum--channel amplitude~\eqref{klein1t} degenerates to a tube terminating at two crosscaps, and therefore, up to an overall factor, the square root of the residue in eq.~\eqref{tadpole2} does define a zero--momentum on--shell one--point amplitude. The divergence can be regulated by adding to the low--energy effective theory the ``tadpole potential''
\beq
V \ \sim \ \pm \ 2^{13}\ \int \, d^{26} x \ \sqrt{-g} \ e^{- \phi} \, ,
\eeq
whose overall sign is still undetermined.
The exponential of the dilaton present in this string--frame expression reflects its emergence from the crosscap or projective plane that, as illustrated in fig.~\ref{fig:boundary_crosscap}, is a genus--$\frac{1}{2}$ surface, with Euler character $\chi=1$, since its double is a sphere that has $\chi=2$. In fact, the genus expansion for unoriented closed strings includes surfaces with arbitrary numbers of handles and zero, one, or two crosscaps. The structure of the genus expansion is again captured by eq.~\eqref{genus_expansion}, but now the
Euler characters of the surfaces involved are
\beq
\chi \ = \ 2 \ - \ 2 \, h \ - \ b\ - \ c \ , \label{chibc}
\eeq
where $b$ and $c$ denote the numbers of boundaries and crosscaps. The reason why one stops at two crosscaps is due to an equivalence between three crosscaps and the combination of one handle and one crosscap, which is illustrated, for example, in~\cite{orientifolds_rev2}.

\begin{figure}[ht]
\centering
%\begin{figure}
\begin{tabular}{cc}
%\mbox{graphic} & \mbox{table} \\
\includegraphics[width=40mm]{Annulus.eps} & \hskip 1cm \includegraphics[width=60mm]{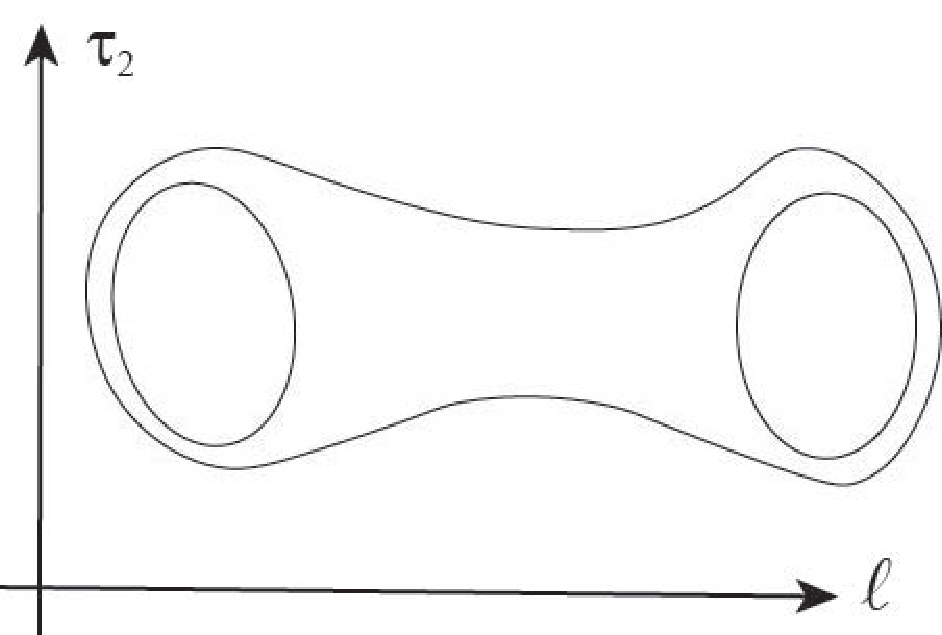}   \\
\end{tabular}
\caption{\small In the complex--plane representation of the annulus, displayed in the left panel, the wiggly lines denote the boundaries. One can distinguish there the fundamental rectangle for the ``direct channel'' (the left rectangle, with vertices at 1 and $i \tau_2$) and the doubly covering torus (the union of the two rectangles, with vertices at 2 and $i \tau_2$), where the second rectangle is obtained by a reflection across the right boundary from the original one. The right panel illustrates the familiar embedding of the annulus in three--dimensional Euclidean space, which also highlights its transverse--channel representation, parametrized by $\ell$, as a tube terminating at a pair of disks.}
\label{fig:Annulus}
\end{figure}
The divergence that we just identified does not signal an inconsistency but an important modification of the vacuum, which cannot be a flat space because of the tadpole potential. The tadpole potential can be eliminated by the introduction of open strings, whose loop amplitude depends, in general, on contributions from the annulus and the M\"obius strip. 

In describing the annulus amplitudes, let us also associate a multiplicity ${\cal N}$ with each end of the
string, in order to account for the internal
Chan-Paton symmetry. In the resulting expression, after the momentum integral discussed in Section~\ref{sec:critical_strings}, one is left with
\beq
{\cal A}\  = \ \frac{{\cal N}^2}{2} \ \int_0^\infty \ \frac{d
\tau_2}{\tau_2^{14}} \ {\rm tr} \left[ e^{-
\,\pi\,\tau_2\,(N - 1)} \right],
\eeq
where the exponent is rescaled by a factor four with respect to what we saw for the Klein--bottle amplitude,
consistent with the different Regge slope of the open spectrum that we identified in Section~\ref{sec:critical_strings}. Computing the trace, one finds
\beq
{\cal A} \ =\  \frac{{\cal N}^2}{2} \ \int_0^\infty \ \frac{d
\tau_2}{\tau_2^{14}} \ \frac{1}{\eta^{24}\left( \frac{1}{2}i \tau_2
\right)} \, , \label{ann1d}
\eeq
and even this amplitude depends on the
modulus, now ${1\over 2} i \tau_2$, of the corresponding doubly-covering torus, as can be seen from fig.~\ref{fig:Annulus}. The annulus can be defined, in fact, subjecting the complex plane
to the identifications
\beq
z \ \sim  \ - \ \bar{z} \ \sim \ z \ + \ i \, \tau_2 \ \sim \ z \ + \ 2 \ .
\eeq

The first terms in the expansion of the integrand in powers of $\sqrt{q}= e^{\,-\,\pi\,\tau_2}$
give
\beq
{\cal A} \rightarrow \frac{{\cal N}^2}{2} \left(
(\sqrt{q})^{-1} + (24) + \ldots \right)  \label{annexp}
\eeq
and, as for the Klein bottle, it is convenient to move to the modulus
of the double cover, now $t= \frac{\tau_2}{2}$, as
integration variable, obtaining
\beq {\cal A} \ =  \ \frac{{\cal N}^2 \ 2^{-13}}{2} \ \int_0^\infty \ \frac{d
t}{t^{14}} \ \frac{1}{\eta^{24}(i t)} \,. \label{ann2d}
\eeq
The alternative integration variable, $\ell=\frac{1}{t}$, displays the
annulus as a tube terminating at two boundaries and defines the
transverse-channel amplitude. We denote by $\tilde{\cal A}$ the corresponding expression,
\beq
\tilde{\cal A} = \frac{{\cal N}^2 \ 2^{-13}}{2} \ \int_0^\infty \ d \ell
\ \frac{1}{\eta^{24}(i\ell)}  \, , \label{ann1t}
\eeq
which can be obtained from eq.~(\ref{ann2d}) by the modular transformation $S$ of eq.~\eqref{STeta}. In this transverse tree--level amplitude,
the multiplicity $\tilde{\cal N}$ of the Chan-Paton charge spaces associated to the
ends of the open string determines the reflection coefficients for
the closed spectrum in front of the two boundaries.
\begin{figure}[ht]
\centering
%\begin{figure}
\begin{tabular}{ccc}
%\mbox{graphic} & \mbox{table} \\
\includegraphics[width=28mm]{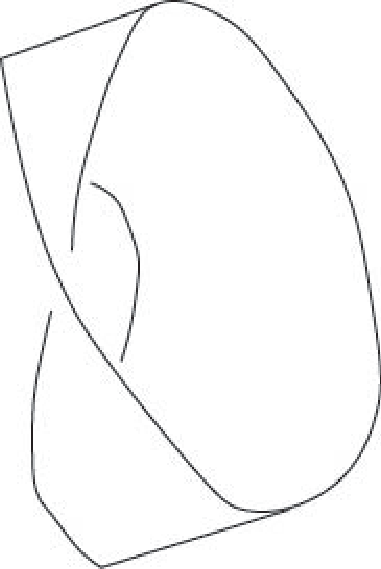} & \hskip 1cm \includegraphics[width=42mm]{Moebius.eps} & \hskip 1cm \includegraphics[width=40mm]{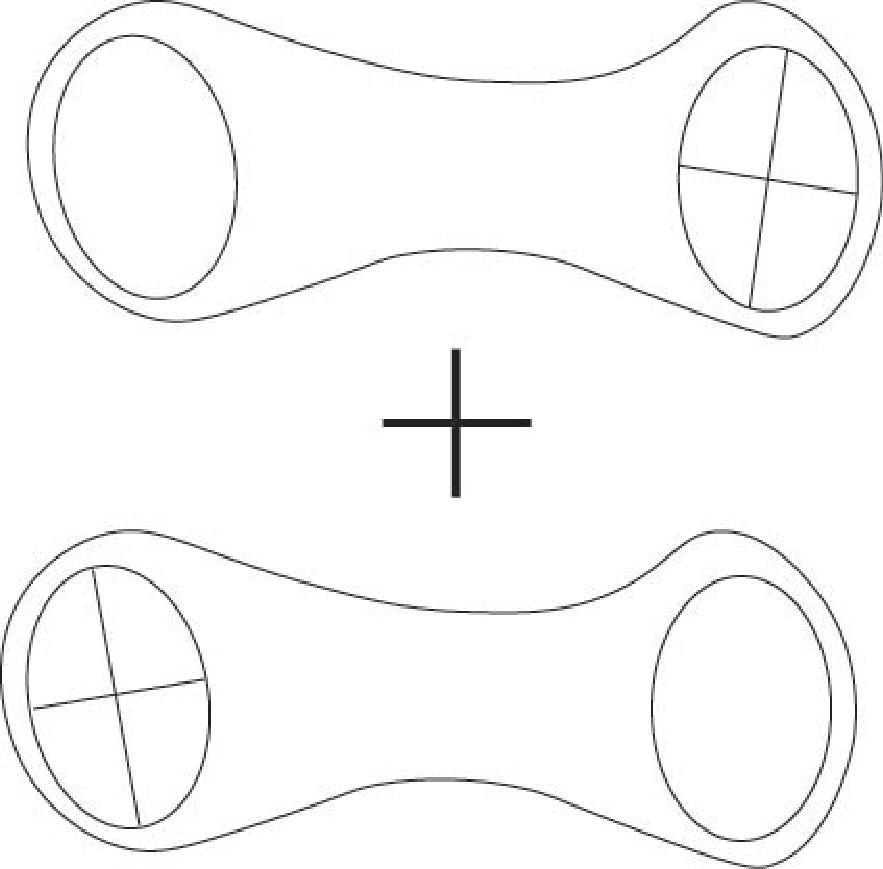}  \\
\end{tabular}
\caption{\small The M\"obius strip is an unoriented surface that, when embedded in three--dimensional Euclidean space, has the peculiar shape in the left panel. In its complex plane representation, displayed in the middle panel, one can distinguish the fundamental rectangle for the ``direct--channel'' (the left rectangle, with vertices at 1 and $i \tau_2$), the skew doubly covering torus (with vertices at 2 and $1+i \tau_2$) and the fundamental polygon for the ``transverse--channel'' representation (the shaded rectangle). Finally, the right panel illustrates its transverse--channel representation as a tube terminating at a crosscap and a disk.}
\label{fig:Moebius}
\end{figure}

Proceeding as for the Klein--bottle amplitude, one can identify a zero--momentum infrared divergence proportional to
\beq
\frac{2^{-13}}{2} \ {\cal N}^{\,2} \ \int_0^\infty \ d \ell   \ , \label{ann1tad}
\eeq
which reflects the presence on a non--vanishing one--point disk amplitude at zero momentum for the dilaton, and can be regulated adding to the string--frame effective action the tadpole potential
\beq
V \ \sim \ 2^{-\,13}\ {\cal N}^{\,2} \,  \int \, d^{26} x \ \sqrt{-g} \ e^{- \phi} \, .
\eeq
The exponential factor reflects the emergence of this additional contribution from another genus--$\frac{1}{2}$ surface, the disk in this case.

According to eq.~\eqref{chibc}, the M\"obius strip, with $b=c=1$, is the last surface of vanishing Euler character. It can be defined subjecting the complex plane to the identifications
\beq
z \ \sim  \ - \ \bar{z} \ \sim \ z \ + \ 2 \ \sim \  z \ + \ 1 \ + \ i\,\tau_2 \ .
\eeq
and contributes to the vacuum amplitude for unoriented open strings. It presents some additional subtleties, since the discussion of the other two
amplitudes suggests that the corresponding integrand should depend on the modulus of
the doubly-covering torus, 
\beq
\tau \ = \ \frac{1}{2} \ + \ i \ \frac{\tau_2}{2} \ ,
\eeq
which is not purely imaginary in this case. Its real part,
equal to
$\frac{1}{2}$, introduces relative signs for oscillator
modes at different mass levels, which
reflect the behavior of open-string states under an orientation flip.

Returning to the open bosonic string, the M\"obius amplitude takes the form
\beq
{\cal M}\  = \ \epsilon\, \frac{{\cal N}}{2} \ \int_0^\infty \ \frac{d
\tau_2}{\tau_2^{14}} \ {\rm tr} \left[ \Omega\, q^{\frac{1}{2}(N - 1)} \right]
\eeq
where the overall factor ${\cal N}$ is associated to its single boundary, and with Neumann boundary conditions
\beq
\Omega \, X(\sigma, \tau) \Omega^{-1} \ =\  X(\pi - \sigma, \tau) \ ,
\eeq
which introduces relative signs between the different  oscillator  contributions, consistent with the shift by $\frac{1}{2}$ of the arguments in ${\cal M}$. In this way, one obtains
\beq
{\cal M} = \frac{\epsilon\,{\cal N}}{2} \int_0^\infty \ \frac{d \tau_2}{\tau_2^{14}} \
\frac{1}{\hat{\eta}^{24}({\textstyle{1\over 2}}i \tau_2 +
{\textstyle{1\over 2}} )} \, ,
\eeq
where $\epsilon$ is again an overall sign that completes the definition of $\Omega$.
Although the integrand is
obviously real for both ${\cal K}$ and ${\cal A}$, which depend on an
imaginary modulus, the same is not true for the M\"obius amplitude
${\cal M}$, where
$\tau_1=\frac{1}{2}$. We have thus defined
\beq
\hat{\eta} \left(\frac{1}{2} + i\,\frac{\tau_2}{2}\right) \ = \ e^{-\,\frac{\pi \tau_2}{24}} \prod_{n=1}^\infty \left[ 1 \ - \ (-1)^n \ e^{-\,\frac{n\, \pi \tau_2}{24}}\right] \ ,
\eeq
removing from $\eta$ an overall phase factor, so as to define a real function. The expansion in powers of $\sqrt{q}$ gives
\beq
{\cal M} \rightarrow \frac{\epsilon \, {\cal N}}{2} \left(
(\sqrt{q})^{-1} - (24) + \ldots \right) \ , \label{mobexp}
\eeq
so that comparing with the annulus  contribution~\eqref{ann1d} one can see that the choice $\epsilon=+1$ corresponds to a total of $N(N-1)/2$
massless vectors, and thus to an orthogonal $O(N)$ gauge group, while
(for even $N$) the choice $\epsilon=-1$ corresponds to a symplectic gauge group~\footnote{One could describe a unitary group $U(N)$ leaving out the M\"obius strip and replacing $\frac{{\cal N}^{\,2}}{2}$ with ${\cal N}\,\overline{\cal N}$, with ${\cal N}$ ($\overline{\cal N}$) the (identical) dimensions of its fundamental and conjugate fundamental representations. This option will play a role in Section~\ref{sec:critical_strings} and the following.}

In this case, the transition to the transverse channel requires to connect 
\beq
\frac{1}{2} \ + \ i \ \frac{1}{2 \,t} \qquad \mathrm{and} \qquad \frac{1}{2} \ + \ i \ \frac{t}{2} \ ,
\eeq
via a special sequence and $T$ and $S$ transformations,
\beq
P \ = \ T^\frac{1}{2} \ S \ T^2 \ S \ T^\frac{1}{2}  \ , \label{P-trans}
\eeq
which was originally identified by G.~Pradisi and is usually called a $P$ transformation. One can then show that
\beq
\hat{\eta}\left( \frac{i}{2 t} + \frac{1}{2} \right) \ = \
\sqrt{t} \ \hat{\eta}\left( \frac{i t}{2} + \frac{1}{2} \right)
\ ,
\eeq
and therefore
\beq
\tilde{\cal M} \ = \ \frac{\epsilon \, {\cal N}}{2} \int_0^\infty \ d t \
\frac{1}{\hat{\eta}^{24}({\textstyle{1\over 2}} i t + {\textstyle{1\over 2}}
)}
\eeq
or, in terms of $\ell = \frac{t}{2}$,
\beq
\tilde{\cal M} \ = \ 2 \ \frac{\epsilon \, {\cal N}}{2} \int_0^\infty \ d \ell \
\frac{1}{\hat{\eta}^{24}(i \ell + {\textstyle{1\over 2}} )} \ . \label{mob1t}
\eeq

The additional factor of two introduced by the last redefinition is
very important, since it reflects the combinatorics of the vacuum
channel: $\tilde{\cal M}$ may be associated
to a tube with one boundary and one crosscap at the two
ends, and the non--symmetric combination needs precisely a combinatoric factor of two compared to
$\tilde{\cal K}$ and $\tilde{\cal A}$, while the sign $\epsilon$ is
a relative phase between crosscap and boundary reflection coefficients that is a priori a free parameter. 
The Chan-Paton multiplicity ${\cal N}$ corresponds, in this case, to
the reflection coefficient for the closed string in front
of the single boundary present in the transverse channel.
The geometric prescription of referring the different amplitudes in the tree--level channel to the double covers identifies a regularization of the overall infrared divergence.

We can now address the ultraviolet behavior of the four amplitudes of
vanishing Euler character that emerges in the limit of small vertical time. As we have seen, the torus
${\cal T}$
is formally protected by modular invariance, which
excludes the ultraviolet region from its
integration domain. On the other hand,  the integration domains for the other three surfaces touch the real axis, and thus
introduce corresponding ultraviolet divergences. In order to take a closer look, it is
convenient
to turn the three additional amplitudes to the transverse channel, where they acquire very similar forms. The divergences then appear in the infrared, or
large $\ell$, limit of eqs.~(\ref{klein1t}), (\ref{ann1t}) and (\ref{mob1t}) and, as we have seen, include on--shell massless exchanges that identify corrections to the low--energy effective theory. In
general, a
state of mass $M$ gives a contribution proportional to
\beq
\int_0^\infty \ d \ell \ e^{- M^2 \ \ell} = \frac{1}{M^2} \, ,
\label{limtransv}
\eeq
and therefore there is
no way to regulate individual massless exchanges. However, once the massless
contribution is eliminated by the tadpole contribution, the vertical-time ultraviolet region
inherits a natural cutoff of the order of the string scale, along the lines of what happens for oriented closed strings due to modular invariance.
\begin{figure}[ht]
\centering

\includegraphics[width=70mm]{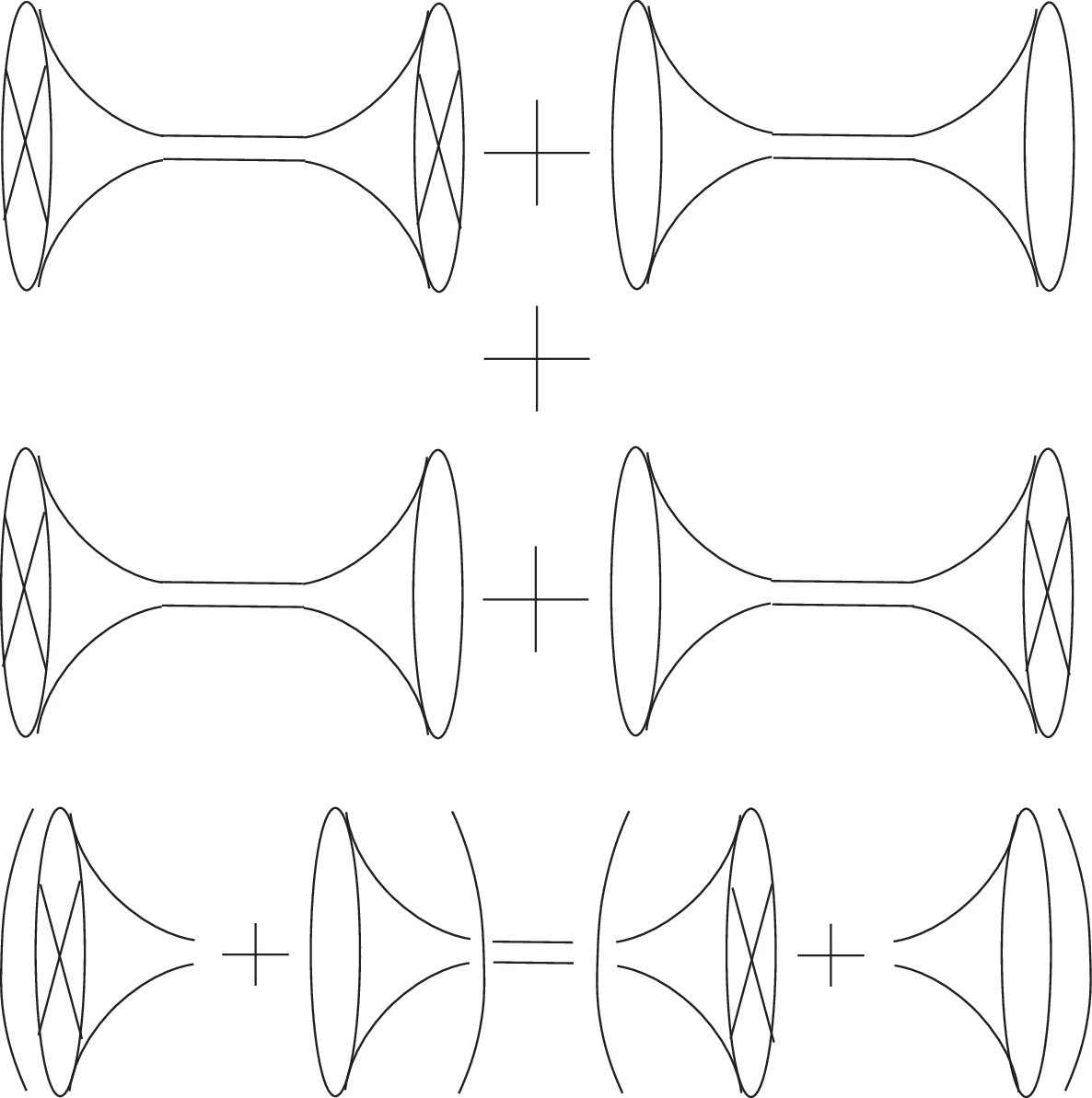}
 \caption{\small The tadpole condition: the contributions of Klein bottle, annulus and M\"obius strip to the massless exchange combine in such a way the overall coefficients, in general for different sectors, are perfect squares.}
\label{fig:tadpole}
\end{figure}

Putting together the different contributions that we have already identified finally gives
\beq
\tilde{\cal K} \ + \ \tilde{\cal A} \ + \ \tilde{\cal M} \sim { \textstyle
\frac{1}{2}} \
\left( 2^{13} \ + \ 2^{-13} \ {\cal N}^2 \ - \ 2 \, \epsilon \, {\cal N} \right) \  = \ \frac{2^{-13}}{2} \
\left({\cal N} \  - \ \epsilon \, 2^{13}\right)^2 \ ,
\eeq
which vanishes for the special choice ${\cal N}= 2^{13} = 8192$ and $\epsilon=+1$, or if you will for the $SO(8192)$ gauge group~\cite{cp5,douglas86,weinberg86,bs88}. In the geometrical picture of~\cite{Dbranes}, the contribution proportional to ${\cal N}$ reflects the D-brane tension, which is positive as it should be for dynamical objects, while the relative sign with the other contribution identifies the orientifold tension, which in this model is $2^{13}$ times the tension of a single brane, and is negative for $\epsilon=1$ and positive for $\epsilon=-1$. More general choices
of ${\cal N}$ and $\epsilon$ lead in this case to a final form for the dilaton ``tadpole potential'',
\beq
V \ \sim \ \left({\cal N} \  - \ \epsilon \, 2^{13}\right) \int \, d^{26} x \ \sqrt{-g} \ e^{- \phi} \, ,
\eeq
which is fully determined by general covariance and by
the Euler characters of disk and crosscap. In more
complicated cases, as we saw in Section~\ref{sec:sugra1110}, one can similarly
dispose of some {\it inconsistent} contributions,  eliminating the corresponding tadpoles that signal the presence of irreducible
anomalies in gauge and gravitational currents~\cite{pc1,pc2,pc3,pc4,pc5}.
In these cases, tadpole
cancellations make open sectors inevitable.

In Section~\ref{sec:critical_strings} and the following we described more complicated models whose GSO -- projected spectra involve various sectors. In these cases, it is convenient to introduce a basis of real ``hatted" characters, generalizing what we saw for the bosonic string. The hatted characters are defined as
\beq
\hat{\chi}_i\left(i \tau_2 +{\textstyle{1\over 2}}\right) \ =
\ q^{h_i - c/24} \sum_k (-1)^k \, d_{(i) k} \,q^k \, ,
\label{chihats}
\eeq
where $h_i$ is the conformal weight of the primary and $q=e^{-2 \pi \tau_2}$, which differ from $\chi_i(i \tau_2 +
\frac{1}{2})$ by the overall
phases
$e^{- i \pi(h_i -c/24)}$. 
For a generic conformal field theory, using the constraints
\beq
S^2 = (S T)^3 = {\cal C} \, ,
\eeq
where ${\cal C}$ is the conjugation matrix,
one can show that
\beq
P^2 = {\cal C} \, ,
\eeq
so that $P$ shares with $S$ the important property of squaring to ${\cal C}$.

\section{\sc Modular Functions and Characters} \label{app:so2n}

In discussing string partition functions, we used extensively the Jacobi theta functions and the SO(2n) level--one characters, defined according to
\begin{eqnarray}
O_{2n} &=& \frac{\theta^n\left[\substack{0\\0}\right]\left(0|\tau\right)\, + \ \theta^n\left[\substack{0\\{1/2}}\right]\left(0|\tau\right)}{2\,\eta^n(\tau)}\ ,  \quad S_{2n} \ = \ \frac{\theta^n\left[\substack{{1/2}\\0}\right]\left(0|\tau\right)\, + \ i^{-n}\, \theta^n\left[\substack{{1/2}\\{1/2}}\right]\left(0|\tau\right)}{2\,\eta^n(\tau)} \, , \nonumber\\
V_{2n} &=& \frac{\theta^n\left[\substack{0\\0}\right]\left(0|\tau\right)\, - \ \theta^n\left[\substack{0\\{1/2}}\right]\left(0|\tau\right)}{2\,\eta^n(\tau)}\ ,  \quad C_{2n} \ = \ \frac{\theta^n\left[\substack{{1/2}\\0}\right]\left(0|\tau\right)\,- \, i^{-n}\, \theta^n\left[\substack{{1/2}\\{1/2}}\right]\left(0|\tau\right)}{2\,\eta^n(\tau)} \, , \label{E1fermi}
\eea
where
\bea
\eta(\tau) &=& q^\frac{1}{24}\ \prod_{n=1}^\infty (1 \ - \ q^n ) \, , \qquad q \,=\, e^{2\pi i \tau}\,,\nonumber \\  \vartheta\left[\substack{\alpha\\\beta}\right]\left(z|\tau\right) &=& \sum_{n \in Z} \ q^{\,\frac{1}{2}\, (n+\alpha)^2} \, e^{i 2 \pi(n+\alpha)(z+\beta)} \ , \\
\vartheta \left[\substack{\alpha\\\beta}\right] (z|\tau) &=&
e^{2 i \pi \alpha (z+\beta)} \ q^{\alpha^2/2} \prod_{n=1}^\infty \
( 1 - q^n) (1 + q^{n + \alpha - 1/2} e^{2 i \pi (z+\beta)} )
(1 + q^{n - \alpha - 1/2} e^{-2 i \pi (z+\beta)} ) \nonumber
\, . \label{characters}
\end{eqnarray}
The SO(2n) level--one characters thus rest on the Jacobi $\vartheta$--functions with characteristics~(see, for example, \cite{jacobi}) and on the Dedekind $\eta$ function, which is also needed to encode the contributions of bosonic oscillators.

In detail, the four $\vartheta$ functions with half--integer characteristics, which play a key role in the partition functions of ten--dimensional strings, admit the series and product decompositions
\bea
\vartheta_3(z|\tau) \!\!&\equiv&\!\! \vartheta\left[\substack{0\\ 0}\right]\left(z|\tau\right)  =  \sum_{n \in Z} \ q^{\,\frac{n^2}{2}} \, e^{i\, 2 \pi\,n\,z} =  \prod_{n=1}^{\infty} (1-q^n)  (1 + q^{n-\frac{1}{2}}e^{2 \pi i z} ) (1 + q^{n-\frac{1}{2}}e^{-2 \pi i z} ) \,, \nonumber \\
\vartheta_4(z|\tau) \!\!&\equiv&\!\! \vartheta\left[\substack{0\\ \frac{1}{2}}\right]\left(z|\tau\right)  =  \sum_{n \in Z} \ q^{\,\frac{n^2}{2}} \, e^{i \,2 \pi n\left(z-\frac{1}{2}\right)}  = \prod_{n=1}^{\infty} (1-q^n)  (1 - q^{n-\frac{1}{2}}e^{2 \pi i z} ) (1 - q^{n-\frac{1}{2}}e^{-2 \pi i z} ) \,, \nonumber \\
\vartheta_2(z|\tau) \!\!&\equiv&\!\! \vartheta\left[\substack{\frac{1}{2} \\ 0}\right]\left(z|\tau\right) = \sum_{n \in Z} \ q^{\,\frac{1}{2}\, \left(n+ \frac{1}{2}\right)^2} \, e^{i \,2 \pi \left(n+ \frac{1}{2}\right)z} \nonumber \\ &=& 2 \cos (\pi z) \ q^{\frac{1}{8}} \prod_{n=1}^{\infty} (1-q^n)  (1 + q^{n} e^{2 \pi i z} ) (1 + q^{n} e^{-2 \pi i z} ) \ , \nonumber \\
\vartheta_1(z|\tau) \!\!&\equiv&\!\!\vartheta\left[\substack{\frac{1}{2} \\ \frac{1}{2}}\right]\left(z|\tau\right) = \sum_{n \in Z} \ q^{\,\frac{1}{2} \left(n+ \frac{1}{2}\right)^2} \, e^{i \,2 \pi \left(n+ \frac{1}{2}\right)\left(z + \frac{1}{2}\right)} \nonumber \\
&=& \ - \  2 \sin (\pi z) \ q^{\frac{1}{8}} \prod_{n=1}^{\infty} (1-q^n)  (1 - q^{n} e^{2 \pi i z} ) (1 - q^{n} e^{-2 \pi i z} ) \ . \label{explicit-theta}
\eea
We have also seen how the argument $z$ reflects the presence of internal magnetic fields, and thus vanishes for the ten--dimensional models.

The torus amplitude can be defined working on the complex plane with the two identifications $z \sim z+1$ and $z \sim z+\tau$. The corresponding modular transformations act on $\tau$ via the fractional linear transformations
\beq
\tau \ \to \ \frac{a\, \tau \ + \ b}{d\, \tau \ + \ d} \ , \qquad ad \,-\, bc \,=\,1 \ , \label{sl2z}
\eeq
and can be built out of two generators
\beq
T: \tau \to \tau + 1 \ , \qquad S : \tau \rightarrow \,- \ \frac{1}{\tau}\ .
\eeq
$S$ and $T$ act on the four SO(2n) level--one characters via the two matrices
\beq
 S \ = \ \frac{1}{2} \left( \begin{array}{rrrr} 1 & 1 & 1 & 1 \\
1 & 1 & -1 & -1 \\
1 & -1 & i^{-n} & -i^{-n} \\
1 & -1 & -i^{-n} & i^{-n}  \end{array} \right)\, ,  \quad T \ = \ e^{\,-\,\frac{i n \pi}{12}} \ \left( \begin{array}{rrrr} 1 & 0 & 0 & 0 \\
0 & -1 & 0 & 0 \\
0 & 0 & e^{\frac{i n \pi}{4}} & 0 \\
0 & 0 & 0 & e^{\frac{i n \pi}{4}}  \end{array} \right)\,  .  \label{modular}
\eeq
and on the Dedekind function $\eta(\tau)$ as
\beq
\eta\left(\,-\,\frac{1}{\tau} \right) \ = \ (-i\,\tau)^\frac{1}{2} \ \eta (\tau) \ , \qquad  \eta(\tau+1) \ = \ e^{\,\frac{i \pi}{12}}\ \eta(\tau) \ .
\eeq
The preceding $S$ and $T$ matrices are determined by the modular transformations of $\eta$ and of the $\vartheta$ functions,
\bea
\vartheta \left[\substack{\alpha\\\beta}\right] (z|\tau+1)
&=& e^{-i \pi \alpha
(\alpha -1)} \vartheta \left[\substack{\alpha\\\beta+\alpha - 1/2}\right] 
(z|\tau) \ , \nonumber \\
\vartheta \left[\substack{\alpha\\\beta}\right]
\left(\frac{z}{\tau}\right|\left.-\,\frac{1}{\tau}\right) &=&
(-i \tau)^{1/2} \ e^{2 i \pi \alpha \beta + i \pi z^2/\tau} \   \vartheta \left[\substack{- \beta\\ \alpha}\right] (z|\tau)  \ .
\eea

On the other hand, the modular parameters $\frac{1}{2}\ + \ i \,\frac{\tau_2}{2}$ and  $\frac{1}{2}\ + \ i \,\frac{1}{2\,\tau_2}$ are connected by Pradisi's $P$ sequence,
\beq
T\,S\,T^2\,S \ .
\eeq
This acts on the real basis of SO(2n) ``hatted'' characters via the matrix
\beq
 P \ = \  \left( \begin{array}{rrrr} \cos\left(\frac{n\pi}{4}\right) & \sin\left(\frac{n\pi}{4}\right) & 0 & 0 \\
\sin\left(\frac{n\pi}{4}\right) & - \ \cos\left(\frac{n\pi}{4}\right) & 0 & 0 \\
0 & 0 & e^{-\frac{i n \pi}{4}}\,\cos\left(\frac{n\pi}{4}\right) & i\ e^{-\frac{i n \pi}{4}}\,\sin\left(\frac{n\pi}{4}\right) \\
0 & 0 & i\ e^{-\frac{i n \pi}{4}}\,\sin\left(\frac{n\pi}{4}\right) & e^{-\frac{i n \pi}{4}}\,\cos\left(\frac{n\pi}{4}\right) \end{array} \right)\, ,
\eeq
which can be defined as
\beq
P \ = \ T^\frac{1}{2}\,S\,T^2\,S\,T^\frac{1}{2} \ ,
\eeq
where $T^\frac{1}{2}$ is the diagonal matrix
\beq
T^\frac{1}{2} \ = \ e^{\,-\,\frac{i n \pi}{24}} \ \left( \begin{array}{rrrr} 1 & 0 & 0 & 0 \\
0 & -1 & 0 & 0 \\
0 & 0 & e^{\frac{i n \pi}{8}} & 0 \\
0 & 0 & 0 & e^{\frac{i n \pi}{8}}  \end{array} \right) \ .
\eeq

In general, given a character
\beq
\chi\left(\tau\right) \ = \ q^{h\,-\,\frac{c}{24}}\ \sum_n \ d_n\ q^n \ ,
\eeq
where $h$ and $c$ denote the weight of the primary and the central charge of the conformal field theory, the corresponding ``hatted'' character is defined as
\beq
\widehat{\chi}\left(\tau\ \,+\,\frac{1}{2}\right) \ = \ q^{h\,-\,\frac{c}{24}}\ \sum_n \ d_n\ (-1)^n \ q^n \ ,
\eeq
so that the overall phase is removed, and for the Dedekind function all this leads to
\beq
\widehat{\eta}\left(\frac{i}{2\,\tau_2} \,+\,\frac{1}{2} \right) \ = \ \tau_2^\frac{1}{2}\ \widehat{\eta}\left(\frac{i \tau_2}{2} \,+\,\frac{1}{2} \right) \ .
\eeq

It is important to appreciate that the definitions in eq.~\eqref{characters} rest on a key step, whereby  contributions that are \emph{physically different} but \emph{numerically identical} are distinguished. In two--dimensional Conformal Field Theory~\cite{bpz} this step is usually called ``resolution of ambiguities''. In the present examples the ``odd spin structure'' contribution $\theta\left[\substack{{1/2}\\{1/2}}\right]\left(0|\tau\right)$ vanishes, consistently with the fact that it involves a counterpart of the trace of the four--dimensional chirality matrix $\gamma_5$~\footnote{The ambiguity is lifted in the presence of internal magnetic fields, as we saw in Section~\ref{sec:magnetic_susyb}.}. In this fashion, one can consistently distinguish the $S_{2n}$ and $C_{2n}$ sectors, which describe Fermi modes of different chiralities, and this choice has the additional virtue of bringing the matrix $S$ into the symmetric and unitary form of eq.~\eqref{modular}. In more complicated examples of two--dimensional Conformal Field Theory, this procedure can actually reveal the presence of different sectors, which here is evident for physical reasons.

In the ten--dimensional Minkowski background, supersymmetry demands that equal numbers of Bose and Fermi excitations be present at every mass level. One can verify when this condition holds enforcing on partition functions Jacobi's \emph{aequatio}, which takes the form
\beq
V_8 \ = \ S_8 \ = \ C_8   \ ,  \label{aequatio}
\eeq
in the notation relying on the $SO(2n)$ level--one characters, or~(see \cite{jacobi})
\beq
\vartheta\left[\substack{0\\0}\right]\left(0|\tau\right)^4 \ -\  \vartheta\left[\substack{0\\ \frac{1}{2}}\right]\left(0|\tau\right)^4  \ - \ \vartheta\left[\substack{\frac{1}{2} \\ 0}\right]\left(0|\tau\right)^4 \ \equiv \ \vartheta_3(0|\tau)^4 \ - \ \vartheta_4(0|\tau)^4 \ - \ \vartheta_2(0|\tau)^4 \ = \ 0
\eeq
in terms of $\vartheta$ functions.

\section{\sc Exactly Solvable Hypergeometric Potentials} \label{app:hypergeometric}

This Appendix is devoted to a class of Schr\"odinger problems defined on the interval $]0,z_m[$, where the potential $V(z)$ exhibits the singular behaviors
\beq
V \ \sim \ \frac{\mu^2 - \frac{1}{4}
}{z^2}
\eeq
and
\beq
V \ \sim \ \frac{\tilde{\mu}^2 - \frac{1}{4}
}{\left(z_m \,-\,z\right)^2} 
\eeq
near the two ends at $z=0$ and $z=z_m$. As we saw in Sections~\ref{sec:SUSY_breaking_com} and ~\ref{sec:directsusybcom}, this type of singular behavior recurs in string compactifications with broken supersymmetry. In particular, we shall determine how the spectra depend on the choice of self--adjoint boundary conditions, which led us to identify stability regions in those cases in the main body of this review.

The potentials of interest are
\beq
V_{\mu,\tilde{\mu}}(z) \ = \ \frac{\pi^2}{4\,z_m^2}\left[ \frac{\mu^2 \ - \ \frac{1}{4}}{\sin^2\left(\frac{\pi\,z}{2\,z_m}\right)} \ + \ \frac{\tilde{\mu}^2 \ - \ \frac{1}{4}}{\cos^2\left(\frac{\pi\,z}{2\,z_m}\right)} \right]  \ , \label{pot_hyp}
\eeq
and can be obtained starting from the hypergeometric equation and performing a change of independent variable to confine its range to the region $0 < z< z_m$, together with a redefinition of the function to finally reach the Schr\"odinger form
\beq
- \ \Psi''(z) \ + \ V_{\mu,\tilde{\mu}}(z)\,\Psi(z) \ = \ \frac{\pi^2\,m^2}{z_m^2}\, \Psi(z) \ . \label{schrod_hyper}
\eeq
For $\mu=\tilde{\mu}$, the hypergeometric potentials reduce to
\beq
V_\mu(z) \ = \ \frac{\pi^2}{z_m^2}\ \frac{\mu^2 \ - \ \frac{1}{4}}{\sin^2\left(\frac{\pi\,z}{z_m}\right)} \ ,
\eeq
which are related to Legendre functions, and were discussed in detail in~\cite{ms23_1}.

For $\mu \neq 0$, the general solution of eq.~\eqref{schrod_hyper} reads
\beq
\Psi(z) = \frac{A\ w_1(z) \,+\, B\ w_2(z)}{u(z)^{\mu-\frac{1}{2}} \ v(z)^{-\tilde{\mu}-\frac{1}{2}} } \ , \label{psihyp}
\eeq
where
\bea
w_1(z) &=& {}_2F_1\left[a,b\,;c\,;u^2(z)\right]  \ , \nonumber \\
w_2(z) &=&  \ u(z)^{2(1-c)}\ {}_2F_1\left[a-c+1,b-c+1\,;2-c\,;u^2(z)\right] \ .
\eea
The ${}_2F_1$ are hypergeometric functions (for details on them, see~\cite{tables}), and moreover
\bea
u(z) &=& \sin\left(\frac{\pi\,z}{2\,z_m}\right) \ , \qquad v(z) \ = \ \cos\left(\frac{\pi\,z}{2\,z_m}\right) \ , \nonumber \\
a &=& \frac{\tilde{\mu} \ - \ \mu \ + \ 1}{2} \ + \ m \ , \quad b \ = \  \frac{\tilde{\mu} \ - \ \mu \ + \ 1}{2} \ - \ m \ , \quad c \ = \ 1 \ - \ \mu \ ,
\eea
where, without loss of generality, $\mu$ and $\tilde{\mu}$ are not negative.

The two functions
\bea
w_3(z) &=& {}_2F_1\left[a,b\,;a+b-c+1\,;v^2(z)\right] \ , \nonumber \\
w_4(z) &=&  v(z)^{2(c-a-b)}\ {}_2F_1\left[c-a,c-b\,;c-a-b+1\,;v^2(z)\right]  \label{hypers12}
\eea
provide an alternative basis of solutions, and are related to previous pair according to~\cite{tables}
\bea
w_1(z) &=& \frac{\Gamma\left(c\right)\,\Gamma\left(c-a-b\right)}{\Gamma\left(c-a\right)\,\Gamma\left(c-b\right)}\ w_3(z) \ + \ \frac{\Gamma\left(c\right)\,\Gamma\left(a+b-c\right)}{\Gamma\left(a\right)\,\Gamma\left(b\right)} \ w_4(z) \ , \nonumber \\
w_2(z) &=& \frac{\Gamma\left(2-c\right)\,\Gamma\left(c-a-b\right)}{\Gamma\left(1-a\right)\,\Gamma\left(1-b\right)} \ w_3(z) \ + \ \frac{\Gamma\left(2-c\right)\,\Gamma\left(a+b-c\right)}{\Gamma\left(a-c+1\right)\,\Gamma\left(b-c+1\right)} \ w_4(z) \ . \label{connections_hyper}
\eea

One can also introduce first--order operators ${\cal A}_{\epsilon_1,\epsilon_2}$ and ${\cal A}_{\epsilon_1,\epsilon_2}^\dagger$, where
\beq
{\cal A}_{\epsilon_1,\epsilon_2} \ = \ \partial_z \ + \ \frac{\pi}{4\,z_m}\left(2\,\epsilon_1\,\mu\,+\,1\right) \ \cot\left(\frac{\pi\,z}{2\,z_m}\right) \ +\ \frac{\pi}{4\,z_m}\left(2\,\epsilon_2\,\tilde{\mu}\,-\,1\right) \ \tan\left(\frac{\pi\,z}{2\,z_m}\right) \ .
\eeq
which depend on the signs $\epsilon_1$ and $\epsilon_2$, and satisfy
\beq
{\cal A}_{\epsilon_1,\epsilon_2}\,{\cal A}_{\epsilon_1,\epsilon_2}^\dagger \ = \ - \ \partial_z^2 \ + \ V_{\epsilon_1,\epsilon_2,\mu,\tilde{\mu}}(z) \ ,
\eeq
where
\beq
V_{\epsilon_1,\epsilon_2,\mu,\tilde{\mu}}(z) \ = \  V_{\mu,\tilde{\mu}}(z) \ - \ \frac{\pi^2}{4\,z_m^2}\Big(1\ + \ \epsilon_1\,\mu \ - \ \epsilon_2\,\tilde{\mu}\Big)^2 \ . \label{V12}
\eeq
The different Hamiltonians
\beq
H_{\epsilon_1,\epsilon_2,\mu,\tilde{\mu}} \ = \ - \ \partial_z^2 \ + \ V_{\epsilon_1,\epsilon_2,\mu,\tilde{\mu}}(z)
\eeq
have the same eigenvectors as $H$ with shifted eigenvalues, so that
\beq
m^2_{\epsilon_1,\epsilon_2} \ = \ m^2 \ - \ \frac{1}{4} \ \Big(1\ + \ \epsilon_1\,\mu \ - \ \epsilon_2\,\tilde{\mu}\Big)^2 \ . \label{m212}
\eeq
The solutions of
\beq{\cal A}_{\epsilon_1,\epsilon_2}^\dagger\ \Psi_{\epsilon_1,\epsilon_2} \ = \ 0 \ ,
\eeq
are
\beq
\Psi_{\epsilon_1,\epsilon_2}(z) \ = \ C \left[\sin\left(\frac{\pi\,z}{2\,z_m}\right)\right]^{\frac{1}{2} \,+\, \epsilon_1\,\mu}\ \left[\cos\left(\frac{\pi\,z}{2\,z_m}\right)\right]^{\frac{1}{2} \,-\, \epsilon_2\,\tilde{\mu}}  \label{psi12} \ .
\eeq
When they are normalizable, they are zero modes of $H_{\epsilon_1,\epsilon_2,\mu,\tilde{\mu}}$.

In order to discuss the possible self--adjoint boundary conditions for $V(z)$ in the different sectors of the spectrum, one must distinguish different ranges for $\mu$ and $\tilde{\mu}$.
\begin{itemize}
\item If $\mu \geq 1$ and $\tilde{\mu} \geq 1 $, the $L^2$ condition at the origin implies that $A=0$ in eq.~\eqref{psihyp}, and the limiting behavior at the other end of the interval is determined by eqs.\eqref{connections_hyper}. The corresponding $L^2$ condition demands that the coefficient of $w_4(z)$ vanish, so that $a-c+1$ or $b-c+1$ must be negative integers, which determines the stable spectrum
\beq
m^2 \ = \ \left(\frac{\mu \, + \, \tilde{\mu} \, + \, 1}{2} \ + \ n\right)^2 \ , \qquad n=0,1,\ldots \ . \label{m2hyper}
\eeq
Consequently
\beq
m^2_{\epsilon_1,\epsilon_2} \ = \ \frac{1}{4}\left[\left(1+\epsilon_1\right)\mu \, + \, \left(1-\epsilon_2\right)\tilde{\mu} \ + \ 2\left(n+1\right)\right]\left[\left(1-\epsilon_1\right)\mu \, + \, \left(1+\epsilon_2\right)\tilde{\mu} \ + \ 2\,n\right] \ . \label{m11}
\eeq
In this case, among the zero mode wavefunctions~\eqref{psi12}, only $\Psi_{+-}$ is normalizable, and the corresponding zero--mass eigenvalue is recovered for $n=0$.
\item If $0 < \mu < 1$ and $\tilde{\mu} \geq 1$, both solutions in eq.~\eqref{psihyp} are normalizable, but one must again demand that the resulting contribution proportional to $w_4(z)$ vanish near the right end of the interval. In this case, the allowed self--adjoint boundary conditions are related to the ratio of the two coefficients $A$ and $B$ and, according to eq.~\eqref{C21}, they can be parametrized via an angle $\alpha$, so that
\beq
\frac{A}{B} \ = \ \tan\left(\frac{\alpha}{2}\right) \left(\frac{\pi}{2}\right)^{\,2\,\mu} \  .
\eeq
The resulting eigenvalue equation reads
\beq
\tan\left(\frac{\alpha_1}{2}\right) \ \equiv \  \frac{C_2}{C_1} \ = \ - \ \left(\frac{\pi}{2}\right)^{\,-\,2\,\mu} \ \frac{\Gamma\left(1+\mu\right)\Gamma\left(\frac{\tilde{\mu}-\mu+1}{2} \,+\, m\right) \Gamma\left(\frac{\tilde{\mu}-\mu+1}{2} \,-\, m\right)}{\Gamma\left(1-\mu\right)\Gamma\left(\frac{\tilde{\mu}+\mu+1}{2} \,+\, m\right) \Gamma\left(\frac{\tilde{\mu}+\mu+1}{2} \,-\, m\right)} \ , \label{eigenvmutildelarger}
\eeq
where the two coefficients $C_1$ and $C_2$ were defined in eq.~\eqref{b.10},
and can be solved graphically, for both real values of $m$, which correspond to stable modes, and for imaginary ones, which correspond to tachyonic modes.

\item If ${\mu} \geq 1$ and $0 < \tilde{\mu} < 1$ , the eigenvalue equation concerns the coefficients at the right end, and reads
\beq
\tan\left(\frac{\alpha_2}{2}\right) \ \equiv \  \frac{C_4}{C_3} \ = \ - \ \left(\frac{\pi}{2}\right)^{\,-\,2\,\tilde{\mu}} \ \frac{\Gamma\left(1+\tilde{\mu}\right)\Gamma\left(\frac{{\mu}-\tilde{\mu}+1}{2} \,+\, m\right) \Gamma\left(\frac{{\mu}-\tilde{\mu}+1}{2} \,-\, m\right)}{\Gamma\left(1-\tilde{\mu}\right)\Gamma\left(\frac{{\mu}+\tilde{\mu}+1}{2} \,+\, m\right) \Gamma\left(\frac{{\mu}+\tilde{\mu}+1}{2} \,-\, m\right)} \ , \label{eigenvmutildelarger_right}
\eeq
where the two coefficients $C_3$ and $C_4$ were defined in eq.~\eqref{b.11}.

\item If $0< \mu < 1$ and $0<\tilde{\mu} <1$, one is free to use arbitrary combinations of the independent solutions at the two ends of the interval, and the self--adjoint boundary conditions relate them by a $U(1,1)$ matrix, according to eq.~\eqref{P_constraint}. Taking eqs.~\eqref{b.10}, \eqref{b.11} and \eqref{psihyp} into account, one can first conclude that
\beq
C_1 \ = \ B \, \sqrt{2\mu} \left(\frac{\pi}{2}\right)^{\frac{1}{2}+\mu} \ , \qquad C_2 \ = \ A \, \sqrt{2\mu} \left(\frac{\pi}{2}\right)^{\frac{1}{2}-\mu} \ , \label{C12AB}
\eeq
and then eqs.~\eqref{connections_hyper} determine $C_4$ and $C_3$ as
\bea
\frac{\left(\frac{\pi}{2}\right)^{\tilde{\mu}}\sqrt{\frac{\mu}{\tilde{\mu}}} \, C_4}{\Gamma\left(a+b-c\right)} \!\!&=&\!\!C_1\, \frac{\Gamma\left(2-c\right)\, \left(\frac{\pi}{2}\right)^{-\mu}}{\Gamma\left(a-c+1\right)\,\Gamma\left(b-c+1\right)} \ + \  C_2\,\frac{\Gamma\left(c\right)\, \left(\frac{\pi}{2}\right)^{\mu}}{\Gamma\left(a\right)\,\Gamma\left(b\right)} \ , \nonumber \\
\frac{\left(\frac{\pi}{2}\right)^{-\tilde{\mu}}\sqrt{\frac{\mu}{\tilde{\mu}}} \, C_3}{\Gamma\left(c-a-b\right)} \!\!&=&\!\!C_1\, \frac{\Gamma\left(2-c\right)\, \left(\frac{\pi}{2}\right)^{-\mu}}{\Gamma\left(1-a\right)\,\Gamma\left(1-b\right)} \ + \  C_2\,\frac{\Gamma\left(c\right)\, \left(\frac{\pi}{2}\right)^{\mu}}{\Gamma\left(c-a\right)\,\Gamma\left(c-b\right)} \ . \label{hyper_les_less}
\eea
One can verify that the two pairs $(C_1,C_2)$ and $(C_4,C_3)$ are related by an $SL(2,R)$ transformation $V$, as in eq.~\eqref{C0mV}. The boundary conditions can now be parametrized via eq.~\eqref{global_ads3_1} and
an additional phase $\beta$, and the eigenvalue equation~\eqref{eigenveq}
reads
\bea
&& \xi\left(-\,\mu,\tilde{\mu},m\right) \left(\cos\theta_1 \cosh\rho-\cos\theta_2 \sinh\rho\right)
 \nonumber \\
 &&+ \ \xi\left(\mu,-\,\tilde{\mu},m\right)\left(\cos\theta_1\cosh\rho+\cos\theta_2 \sinh\rho\right) \nonumber \\
&&- \ \xi\left(\mu,\tilde{\mu},m\right) \left(\sin\theta_1 \cosh\rho + \sin\theta_2 \sinh\rho\right) \nonumber \\
&&+ \ \xi\left(-\,\mu,-\,\tilde{\mu},m\right) \left(\sin\theta_1 \cosh\rho-\sin\theta_2 \sinh\rho\right) \ = \ 2 \cos\beta \, ,
\eea
where
\beq
\xi\left(\mu,\tilde{\mu},m\right) \ = \  \frac{\left(\frac{\pi }{2}\right)^{\mu -\tilde{\mu} } \sqrt{\left|\frac{\tilde{\mu} }{\mu }\right|} \ \Gamma (1-\mu ) \Gamma (\tilde{\mu} ) }{\Gamma \left[\frac{1}{2} (-\mu +\tilde{\mu} +1)-m\right] \Gamma \left[m+\frac{1}{2} (-\mu +\tilde{\mu} +1)\right]} \ .
\eeq
In the $\rho \to \infty$ limit, which translates into independent boundary conditions at the ends of the interval, this expression reduces to
\bea
&& \xi\left(-\,\mu,\tilde{\mu},m\right) \left(\cos\theta_1 -\cos\theta_2 \right) \ + \ \xi\left(\mu,-\,\tilde{\mu},m\right)\left(\cos\theta_1+\cos\theta_2 \right) \nonumber \\
&&- \ \xi\left(\mu,\tilde{\mu},m\right) \left(\sin\theta_1 + \sin\theta_2 \right) \ + \ \xi\left(-\,\mu,-\,\tilde{\mu},m\right) \left(\sin\theta_1 -\sin\theta_2 \right) \ = \ 0 \, . \label{eigenvfinmumutilde}
\eea
\item If $0< \mu < 1$ and $\tilde{\mu} = 0$, both solutions in eq.~\eqref{psihyp} are normalizable, and the asymptotic behavior at the left end defines again the two coefficients $C_1$ and $C_2$ according to eq.~\eqref{C12AB}, while $C_3$ and $C_4$ are defined according to
\beq
\psi \ \sim \ \sqrt{1 - \ \frac{z}{z_m}}\left[C_4 \ + \  C_3\, \log\left(1 - \ \frac{z}{z_m}\right)\right] \ .
\eeq
There is a small subtlety, since the self--adjoint boundary conditions relate in this case the two vectors $\left(C_1,C_2\right)$ and $\left(C_4,C_3\right)$ by a $U(1,1)$ matrix, which led us to the definitions in eq.~\eqref{mutildezero}.

One can obtain the connection formulas as a limit, as $\tilde{\mu} \to 0$, of the preceding expressions in eqs.~\eqref{hypers12} and \eqref{connections_hyper}. Consequently, the behavior in the vicinity of the right end of the interval is now
\bea
w_1(z) &\sim& \xi_1(\mu,m) \ + \ \xi_2(\mu,m) \log\left(1\,-\,\frac{z}{z_m}\right) \ , \nonumber \\
w_2(z) &\sim& \xi_1(-\mu,m) \ + \ \xi_2(-\mu,m) \log\left(1\,-\,\frac{z}{z_m}\right) \ , \label{xi12}
\eea
where
\bea
\xi_1(\mu,m) \!\!\!&=&\!\!\! - \  \frac{\left(\frac{\pi}{2}\right)^\mu \Gamma(1-\mu) \left[\psi\left(-m-\frac{\mu }{2}+\frac{1}{2}\right)\,+\,\psi\left(m-\frac{\mu }{2}+\frac{1}{2}\right) \ - \ 2\,\psi\left(1\right)+ 2\log\left(\frac{\pi}{2}\right)\right]}{\sqrt{2\left|\mu\right|}\,\Gamma \left(-m-\frac{\mu }{2}+\frac{1}{2}\right) \Gamma \left(m-\frac{\mu }{2}+\frac{1}{2}\right) } , \nonumber \\
\xi_2(\mu,m) \!\!\!&=&\!\!\! - \ \frac{ 2\,\left(\frac{\pi}{2}\right)^\mu \Gamma(1-\mu)}{\sqrt{2\left|\mu\right|}\,\Gamma \left(-m-\frac{\mu }{2}+\frac{1}{2}\right) \Gamma \left(m-\frac{\mu }{2}+\frac{1}{2}\right) } \ , \label{xi12def}
\eea
with $\psi(1)=- \gamma$ and $\gamma\sim 0.577$ the Euler--Mascheroni constant. The linear relations among the coefficients are now
\bea
C_4  &=& C_2 \  \xi_1(\mu,m) \ + \ C_1 \ \xi_1(-\,\mu,m) \ , \nonumber \\
C_3  &=& C_2 \  \xi_2(\mu,m) \ + \ C_1 \ \xi_2(-\,\mu,m)   \ , \label{eigenvmu0_rhoinf}
\eea
and define an $SL(2,R)$ transformation $V$. The resulting eigenvalue equation is
\bea
&&\left(\cosh\rho\,\cos\theta_1 \,-\, \sinh\rho\,\cos\theta_2\right)  \xi_1(-\,\mu,m)\,+\,\left(\cosh\rho\,\sin\theta_1 \,-\, \sinh\rho\,\sin\theta_2\right)  \xi_2(-\,\mu,m) \nonumber \\
&&+\left(\cosh\rho\,\cos\theta_1 \,+\, \sinh\rho\,\cos\theta_2\right)  \xi_2(\mu,m)\,-\,\left(\cosh\rho\,\sin\theta_1 \,+\, \sinh\rho\,\sin\theta_2\right)  \xi_1(\mu,m) \nonumber \\
&&-\ 2 \,\cos \beta \ = \ 0\ . \label{muzerorhofinite}
\eea
In the $\rho \to \infty$ limit, which translates into independent boundary conditions at the ends of the interval, this expression reduces to
\bea
&&\left(\cos\theta_1 \,-\, \cos\theta_2\right)  \xi_1(-\,\mu,m)\,+\,\left(\sin\theta_1 \,-\, \sin\theta_2\right)  \xi_2(-\,\mu,m) \nonumber \\
&&+ \left(\cos\theta_1 \,+\, \cos\theta_2\right)  \xi_2(\mu,m)\,-\,\left(\sin\theta_1 \,+\, \sin\theta_2\right)  \xi_1(\mu,m)  \ = \ 0\ . \label{spec_mu0}
\eea
\item If ${\mu} =0$ and $0< \tilde{\mu} < 1$, taking into account the symmetry of the potential under
$z \to z_m - z$ combined with the interchange of $\mu$ and $\tilde{\mu}$, together with the interchange between $C_3$ and $C_4$ in eq.~\eqref{mutildezero}, one can conclude that the eigenvalue equations can be deduced from those of the previous case by interchanging in them $\mu$ and $\tilde{\mu}$ and letting $\theta_1 \to - \theta_1$ and $\theta_2 \to \pi - \theta_2$. Therefore, eq.~\eqref{muzerorhofinite} becomes
\bea
&&\left(\cosh\rho\,\cos\theta_1 \,+\, \sinh\rho\,\cos\theta_2\right)  \xi_1(-\,\tilde{\mu},m)\,-\,\left(\cosh\rho\,\sin\theta_1 \,+\, \sinh\rho\,\sin\theta_2\right)  \xi_2(-\,\tilde{\mu},m) \nonumber \\
&&+\left(\cosh\rho\,\cos\theta_1 \,-\, \sinh\rho\,\cos\theta_2\right)  \xi_2(\tilde{\mu},m)\,+\,\left(\cosh\rho\,\sin\theta_1 \,-\, \sinh\rho\,\sin\theta_2\right)  \xi_1(\tilde{\mu},m) \nonumber \\
&&-\ 2 \,\cos \beta \ = \ 0\ , \label{muzerorhofinite2}
\eea
while eq.~\eqref{spec_mu0} becomes
\bea
&&\left(\cos\theta_1 \,+\, \cos\theta_2\right)  \xi_1(-\,\tilde{\mu},m)\,-\,\left(\sin\theta_1 \,+\, \sin\theta_2\right)  \xi_2(-\,\tilde{\mu},m) \nonumber \\
&&+ \left(\cos\theta_1 \,-\, \cos\theta_2\right)  \xi_2(\tilde{\mu},m)\,+\,\left(\sin\theta_1 \,-\, \sin\theta_2\right)  \xi_1(\tilde{\mu},m)  \ = \ 0\ . \label{spec_mu02}
\eea
\item If ${\mu} =0$ and $\tilde{\mu} =0$, one can avoid the need for further singular limits and directly rely on the result of~\cite{ms23_1}, so that the eigenvalue equation reads
\bea
&&\left(\cos\theta_1\,+\,\cos\theta_2\right)\left[\sin\pi\nu \left(\frac{\pi}{4} \,-\,\frac{1}{\pi}\, \sigma^2(\nu)\right) \ - \ \sigma(\nu)\,\cos\pi\nu \right] \label{mu0nuerhoinf} \\
&&+\left(\cos\theta_1\,-\,\cos\theta_2\right) \frac{\sin \pi\nu}{\pi} \ + \ \sin\theta_1\left[ \frac{2}{\pi}\, \sin\pi\nu\, \sigma(\nu) \ + \ \cos\pi\nu \right] \ = \ 0 \ , \nonumber
\eea
where
\beq
\sigma(\nu) \ = \ \log\left(\frac{\pi}{2}\right) \ - \ \psi(1) \ + \ \psi\left(\nu + 1\right)  \ ,
\label{sigma}
\eeq
with
\beq
\nu \ = \ - \  \frac{1}{2} \ + \ \sqrt{m^2 \ + \ \frac{1}{4}} \label{eingevmu0} \ ,
\eeq
in the most relevant $\rho \to \infty$ limit. This case was analyzed in detail in~\cite{ms23_1}, and concerns the graviton mode in the vacuum of~\cite{dm_vacuum}. There is only one stable boundary condition, which corresponds to $\left(\theta_1,\theta_2\right)=\left(\pi,0\right)$ and leads to a spectrum starting with a massless graviton mode in nine dimensions.

The remaining cases can be deduced from these by specializing them, removing some coefficients that are not allowed when $\mu>1$ or $\tilde{\mu}>1$, or by a reflection about the middle point of the interval.
\end{itemize}

\end{appendices}


\newpage
\begin{thebibliography}{999}
\bibitem{coleman-mandula}
S.~R.~Coleman and J.~Mandula,
``All Possible Symmetries of the S Matrix,''
Phys. Rev. \textbf{159} (1967), 1251.
%1395 citations counted in INSPIRE as of 18 Sep 2020

%\cite{Haag:1974qh}
\bibitem{Haag:1974qh}
R.~Haag, J.~T.~Lopuszanski and M.~Sohnius,
``All Possible Generators of Supersymmetries of the s Matrix,''
Nucl. Phys. B \textbf{88} (1975), 257.
%1424 citations counted in INSPIRE as of 18 Sep 2020

\bibitem{stringtheory}
M.~B.~Green, J.~H.~Schwarz and E.~Witten, ``Superstring Theory'', 2 vols., Cambridge Univ. Press (1987); \\
J.~Polchinski, ``String theory'', 2 vols. Cambridge, UK: Cambridge Univ. Press (1998);  \\
C.~V.~Johnson, ``D-branes,'' Cambridge Univ. Press (2003); \\
B.~Zwiebach, ``A first course in string theory,'' Cambridge Univ. Press (2004); \\
K.~Becker, M.~Becker and J.~H.~Schwarz,
``String theory and M-theory: A modern introduction'' Cambridge, UK: Cambridge Univ.
Press (2007); \\
E.~Kiritsis, ``String theory in a nutshell,'' Princeton Univ. Press (2007);\\
P.~West, ``Introduction to strings and branes,'' Cambridge Univ. Press (2012).
  %%CITATION = INSPIRE-1190041;%%


%\cite{Golfand:1971iw}
\bibitem{Golfand:1971iw}
Y.~A.~Golfand and E.~P.~Likhtman,
``Extension of the Algebra of Poincare Group Generators and Violation of p Invariance,''
JETP Lett. \textbf{13} (1971), 323.
%1779 citations counted in INSPIRE as of 29 Jan 2021

\bibitem{NSR1}
   A.~Neveu and J.~H.~Schwarz,
  ``Factorizable dual model of pions,''
  Nucl.\ Phys.\ B {\bf 31} (1971) 86.
 % doi:10.1016/0550-3213(71)90448-2
  %%CITATION = doi:10.1016/0550-3213(71)90448-2;%%

  \bibitem{NSR2}
  P.~Ramond,
  ``Dual Theory for Free Fermions,''
  Phys.\ Rev.\ D {\bf 3} (1971) 2415.
 % doi:10.1103/PhysRevD.3.2415
  %%CITATION = doi:10.1103/PhysRevD.3.2415;%%

%\cite{Gervais:1971ji}
\bibitem{Gervais:1971ji}
J.~L.~Gervais and B.~Sakita,
``Field Theory Interpretation of Supergauges in Dual Models,''
Nucl. Phys. B \textbf{34} (1971), 632.
%doi:10.1016/0550-3213(71)90351-8
%503 citations counted in INSPIRE as of 29 Jan 2021

%\cite{Volkov:1973ix}
\bibitem{Volkov:1973ix}
D.~V.~Volkov and V.~P.~Akulov,
``Is the Neutrino a Goldstone Particle?,''
Phys. Lett. B \textbf{46} (1973), 109.
%1878 citations counted in INSPIRE as of 18 Sep 2020

\bibitem{goldstone}
J.~Goldstone,
``Field Theories with Superconductor Solutions,''
Nuovo Cim. \textbf{19} (1961), 154-164
%doi:10.1007/BF02812722
%2715 citations counted in INSPIRE as of 22 Aug 2025

%\cite{Wess:1974tw}
\bibitem{Wess:1974tw1}
J.~Wess and B.~Zumino,
``Supergauge Transformations in Four-Dimensions,''
Nucl. Phys. B \textbf{70} (1974), 39.
%3236 citations counted in INSPIRE as of 18 Sep 2020

\bibitem{Wess:1974tw2}
J.~Wess and B.~Zumino,
``A Lagrangian Model Invariant Under Supergauge Transformations,''
Phys. Lett. B \textbf{49} (1974), 52.
%1664 citations counted in INSPIRE as of 18 Sep 2020

\bibitem{Ferrara:1974pu}
S.~Ferrara and B.~Zumino,
``Supergauge Invariant Yang-Mills Theories,''
Nucl. Phys. B \textbf{79} (1974), 413.
%doi:10.1016/0550-3213(74)90559-8
%769 citations counted in INSPIRE as of 05 Jun 2023

%\cite{Salam:1974yz}
\bibitem{Salam:1974yz}
A.~Salam and J.~A.~Strathdee,
``Supergauge Transformations,''
Nucl. Phys. B \textbf{76} (1974), 477.
%doi:10.1016/0550-3213(74)90537-9
%748 citations counted in INSPIRE as of 29 Jan 2021

%\cite{fayet}
\bibitem{fayet1}
  P.~Fayet,
``Spontaneously Broken Supersymmetric Theories of Weak, Electromagnetic and Strong Interactions,''
Phys. Lett. B \textbf{69} (1977), 489.
%1336 citations counted in INSPIRE as of 04 Oct 2020

\bibitem{fayet2}
 P.~Fayet,``Supergauge Invariant Extension of the Higgs Mechanism and a Model for the electron and Its Neutrino,''
Nucl. Phys. B \textbf{90} (1975), 104.
%1119 citations counted in INSPIRE as of 04 Oct 2020

\bibitem{fayet3}
 P.~Fayet,``Supersymmetry and Weak, Electromagnetic and Strong Interactions,''
Phys. Lett. B \textbf{64} (1976), 159.
%711 citations counted in INSPIRE as of 04 Oct 2020

\bibitem{susy-books}
P.~Fayet and S.~Ferrara,
``Supersymmetry,''
Phys. Rept. \textbf{32} (1977), 249;
%1211 citations counted in INSPIRE as of 04 Oct 2020
M.F.~Sohnius, "Introducing supersymmetry", Phys.Rept. \textbf{128} (1985), 39;\\
J.~Wess and J.~Bagger,
``Supersymmetry and supergravity,''
%424 citations counted in INSPIRE as of 04 Oct 2020
Princeton Univ. Press, 1992; 
S.~P.~Martin,
``A Supersymmetry primer,''
Adv. Ser. Direct. High Energy Phys. \textbf{21} (2010), 1
% doi:10.1142/9789812839657\_0001
[arXiv:hep-ph/9709356 [hep-ph]];
A.~Bilal,
``Introduction to supersymmetry,''
[arXiv:hep-th/0101055 [hep-th]];
%118 citations counted in INSPIRE as of 02 Dec 2020
M.~Drees, R.~Godbole and P.~Roy,
``Theory and phenomenology of sparticles: An account of four-dimensional N=1 supersymmetry in high energy physics,'' World Scientific (2004);
P.~Binetruy,
``Supersymmetry: Theory, experiment and cosmology,''
Oxford, UK: Oxford Univ. Press,  2006;
        %50 citations counted in INSPIRE as of 04 Oct 2020
M.~Dine, "Supersymmetry and String Theory : Beyond the Standard Model," Cambridge Univ. Press (2007);
%104 citations counted in INSPIRE as of 16 Nov 2020
P.G.O.~Freund, "Introduction to Supersymmetry", Cambridge Univ. Press (2012);
S.~Weinberg, "The quantum theory of fields. Vol. 3: Supersymmetry", Cambridge Univ. Press (2013).

\bibitem{sugra1}
D.~Z.~Freedman, P.~van Nieuwenhuizen and S.~Ferrara,
``Progress Toward a Theory of Supergravity,''
Phys. Rev. D \textbf{13} (1976), 3214.
%1299 citations counted in INSPIRE as of 06 Oct 2020

\bibitem{sugra2}
S.~Deser and B.~Zumino,
``Consistent Supergravity,''
Phys. Lett. B \textbf{62} (1976), 335.
%1268 citations counted in INSPIRE as of 06 Oct 2020

\bibitem{sugrarev}
P.~Van Nieuwenhuizen,
``Supergravity,''
Phys. Rept. \textbf{68} (1981), 189;
%doi:10.1016/0370-1573(81)90157-5
%1423 citations counted in INSPIRE as of 26 Jan 2024
H.~P.~Nilles,
``Supersymmetry, Supergravity and Particle Physics,''
Phys. Rept. \textbf{110} (1984), 1;
%doi:10.1016/0370-1573(84)90008-5
P.~Binetruy, G.~Girardi and R.~Grimm,
``Supergravity couplings: A Geometric formulation,''
Phys. Rept. \textbf{343} (2001), 255,
[arXiv:hep-th/0005225 [hep-th]];
 D.~Z.~Freedman and A.~Van Proeyen,
  ``Supergravity,''
  Cambridge Univ. Press (2012);
%154 citations counted in INSPIRE as of 06 Oct 2020
  G.~Dall\textquoteright{}Agata and M.~Zagermann,
``Supergravity: From First Principles to Modern Applications,''
Lect. Notes Phys. \textbf{991} (2021), 1
2021.
For an elementary overview of some of the main developments, see also:
  S.~Ferrara and A.~Sagnotti,
  ``Supergravity at 40: Reflections and Perspectives,''
  Riv.\ Nuovo Cim.\  {\bf 40} (2017) no.6,  1
   [J.\ Phys.\ Conf.\ Ser.\  {\bf 873} (2017) no.1,  012014]
  %doi:10.1088/1742-6596/873/1/012014, 10.1393/ncr/i2017-10136-6
  [arXiv:1702.00743 [hep-th]].
  %%CITATION = doi:10.1088/1742-6596/873/1/012014, 10.1393/ncr/i2017-10136-6;%%

\bibitem{noglobalgrav}
S.~W.~Hawking,
``Particle Creation by Black Holes,''
Commun. Math. Phys. \textbf{43} (1975), 199
[erratum: Commun. Math. Phys. \textbf{46} (1976), 206].
%doi:10.1007/BF02345020
%12526 citations counted in INSPIRE as of 21 Aug 2025
  
\bibitem{GSO}
F.~Gliozzi, J.~Scherk and D.~I.~Olive,
  ``Supersymmetry, Supergravity Theories and the Dual Spinor Model,''
  Nucl.\ Phys.\ B {\bf 122} (1977) 253.
  %doi:10.1016/0550-3213(77)90206-1
  %%CITATION = doi:10.1016/0550-3213(77)90206-1;%%

  \bibitem{gs82}
  M.~B.~Green and J.~H.~Schwarz,
``Supersymmetrical String Theories,''
Phys. Lett. B \textbf{109} (1982), 444.
%doi:10.1016/0370-2693(82)91110-8
%505 citations counted in INSPIRE as of 05 Mar 2025

\bibitem{IIB}
M.~B.~Green and J.~H.~Schwarz,
``Extended Supergravity in Ten-Dimensions,''
Phys. Lett. B \textbf{122} (1983), 143;
%doi:10.1016/0370-2693(83)90781-5
%156 citations counted in INSPIRE as of 20 Jan 2025
J.~H.~Schwarz and P.~C.~West,
%``Symmetries and Transformations of Chiral N=2 D=10 Supergravity,''
Phys. Lett. B \textbf{126} (1983), 301;
%doi:10.1016/0370-2693(83)90168-5
J.~H.~Schwarz,
``Covariant Field Equations of Chiral N=2 D=10 Supergravity,''
Nucl. Phys. B \textbf{226} (1983), 269;
%doi:10.1016/0550-3213(83)90192-X
%839 citations counted in INSPIRE as of 20 Jan 2025
P.~S.~Howe and P.~C.~West,
``The Complete N=2, D=10 Supergravity,''
Nucl. Phys. B \textbf{238} (1984), 181.
%doi:10.1016/0550-3213(84)90472-3

\bibitem{gs}
M.~B.~Green and J.~H.~Schwarz,
``Anomaly Cancellation in Supersymmetric D=10 Gauge Theory and Superstring Theory,''
Phys. Lett. B \textbf{149} (1984), 117.
%doi:10.1016/0370-2693(84)91565-X
%2861 citations counted in INSPIRE as of 01 Jun

\bibitem{heterotic1}
 D.~J.~Gross, J.~A.~Harvey, E.~J.~Martinec and R.~Rohm,
  ``The Heterotic String,''
  Phys.\ Rev.\ Lett.\  {\bf 54} (1985) 502.
  %doi:10.1103/PhysRevLett.54.502
  %%CITATION = doi:10.1103/PhysRevLett.54.502;%%

  \bibitem{heterotic2}
   D.~J.~Gross, J.~A.~Harvey, E.~J.~Martinec and R.~Rohm,
 ``Heterotic String Theory. 1. The Free Heterotic String,''
  Nucl.\ Phys.\ B {\bf 256} (1985) 253.
  %doi:10.1016/0550-3213(85)90394-3
  %%CITATION = doi:10.1016/0550-3213(85)90394-3;%%

  \bibitem{heterotic3}
   D.~J.~Gross, J.~A.~Harvey, E.~J.~Martinec and R.~Rohm,
  ``Heterotic String Theory. 2. The Interacting Heterotic String,''
  Nucl.\ Phys.\ B {\bf 267} (1986) 75.
  %doi:10.1016/0550-3213(86)90146-X
  %%CITATION = doi:10.1016/0550-3213(86)90146-X;%%
  
\bibitem{witten}
E.~Witten,
``String theory dynamics in various dimensions,''
Nucl. Phys. B \textbf{443} (1995), 85
%doi:10.1016/0550-3213(95)00158-O
[arXiv:hep-th/9503124 [hep-th]].

\bibitem{recent_books_SM}
  C.~Itzykson and J.B.~Zuber, ''Quantum Field Theory,'' McGraw-Hill (1980);\\
  P. Ramond, ``Field Theory: a Modern Primer,'' Routledge (1997);\\
  S.~Weinberg, ``The Quantum Theory of Fields, 3 vols.,'' Cambridge Univ. Press (2013);\\
  M.~D.~Schwartz, ``Quantum Field Theory and the Standard Model,'' Cambridge Univ. Press (2014);\\
  J~Iliopoulos and T.~N.~Tomaras, ``Elementary Particle Physics,'' Oxford Univ. Press (2021).

%\citee{fayetilio}
\bibitem{fayetilio}
P.~Fayet and J.~Iliopoulos,
``Spontaneously Broken Supergauge Symmetries and Goldstone Spinors,''
Phys. Lett. B \textbf{51} (1974), 461.
%916 citations counted in INSPIRE as of 09 Oct 2020

 %\cite{O'R}
\bibitem{O'R}
L.~O'Raifeartaigh,
``Spontaneous Symmetry Breaking for Chiral Scalar Superfields,''
Nucl. Phys. B \textbf{96} (1975), 331.
%734 citations counted in INSPIRE as of 09 Oct 2020


%\cite{fgp}
\bibitem{fgp}
S.~Ferrara, L.~Girardello and F.~Palumbo,
``A General Mass Formula in Broken Supersymmetry,''
Phys. Rev. D \textbf{20} (1979), 403.
%doi:10.1103/PhysRevD.20.403
%316 citations counted in INSPIRE as of 02 Aug 2023


%\cite{girardello-grisaru}
\bibitem{girardello-grisaru}
L.~Girardello and M.~T.~Grisaru,
``Soft Breaking of Supersymmetry,''
Nucl. Phys. B \textbf{194} (1982), 65.
%959 citations counted in INSPIRE as of 11 Oct 2020


\bibitem{gravity-mediation1}
R.~Barbieri, S.~Ferrara and C.~A.~Savoy,
``Gauge Models with Spontaneously Broken Local Supersymmetry,''
Phys. Lett. B \textbf{119} (1982), 343.
%1808 citations counted in INSPIRE as of 02 Nov 2020

\bibitem{gravity-mediation2}
A.~H.~Chamseddine, R.~L.~Arnowitt and P.~Nath,
``Locally Supersymmetric Grand Unification,''
Phys. Rev. Lett. \textbf{49} (1982), 970.
%doi:10.1103/PhysRevLett.49.970
%2098 citations counted in INSPIRE as of 02 Nov 2020

\bibitem{hlw}
L.~J.~Hall, J.~D.~Lykken and S.~Weinberg,
``Supergravity as the Messenger of Supersymmetry Breaking,''
Phys. Rev. D \textbf{27} (1983), 2359.
%doi:10.1103/PhysRevD.27.2359
%1652 citations counted in INSPIRE as of 16 Apr 2025

 \bibitem{SM1}
 S.~L.~Glashow,
``Partial Symmetries of Weak Interactions,''
Nucl. Phys. \textbf{22}, 579 (1961).
%doi:10.1016/0029-5582(61)90469-2
%9457 citations counted in INSPIRE as of 06 Jun 2023

 \bibitem{SM2}
S.~Weinberg,
``A Model of Leptons,''
Phys. Rev. Lett. \textbf{19}, 1264 (1967).
%doi:10.1103/PhysRevLett.19.1264
%13995 citations counted in INSPIRE as of 06 Jun 2023

 \bibitem{SM3}
A.~Salam,
``Weak and Electromagnetic Interactions,''
Conf. Proc. C \textbf{680519}, 367 (1968).
%doi:10.1142/9789812795915\_0034
%5820 citations counted in INSPIRE as of 06 Jun 2023

%\cite{Dimopoulos:1981zb}
\bibitem{dg}
S.~Dimopoulos and H.~Georgi,
``Softly Broken Supersymmetry and SU(5),''
Nucl. Phys. B \textbf{193}, 150 (1981).
%doi:10.1016/0550-3213(81)90522-8
%3145 citations counted in INSPIRE as of 06 Jun 2023


\bibitem{Tduality1}
V.~P.~Nair, A.~D.~Shapere, A.~Strominger and F.~Wilczek,
``Compactification of the Twisted Heterotic String,''
Nucl. Phys. B \textbf{287} (1987), 402.
%doi:10.1016/0550-3213(87)90112-X
%306 citations counted in INSPIRE as of 02 Jun 2021

\bibitem{Tduality2}
B.~Sathiapalan,
``Duality in Statistical Mechanics and String Theory,''
Phys. Rev. Lett. \textbf{58} (1987), 1597.
%doi:10.1103/PhysRevLett.58.1597
%115 citations counted in INSPIRE as of 02 Jun 2021

\bibitem{Tduality3}
J.~Dai, R.~G.~Leigh and J.~Polchinski,
``New Connections Between String Theories,''
Mod. Phys. Lett. A \textbf{4} (1989), 2073.
%doi:10.1142/S0217732389002331
%946 citations counted in INSPIRE as of 16 Nov 2023

\bibitem{Tduality4}
P.~H.~Ginsparg,
``Comment on Toroidal Compactification of Heterotic Superstrings,''
Phys. Rev. D \textbf{35} (1987), 648.
%doi:10.1103/PhysRevD.35.648
%230 citations counted in INSPIRE as of 16 Nov 2023

\bibitem{Tduality_rev}
A.~Giveon, M.~Porrati and E.~Rabinovici,
``Target space duality in string theory,''
Phys. Rept. \textbf{244} (1994), 77
%doi:10.1016/0370-1573(94)90070-1
[arXiv:hep-th/9401139 [hep-th]].
%1064 citations counted in INSPIRE as of 03 Aug 2023


\bibitem{Dbranes}
J.~Polchinski,
``Dirichlet Branes and Ramond-Ramond charges,''
Phys. Rev. Lett. \textbf{75} (1995), 4724
% doi:10.1103/PhysRevLett.75.4724
[arXiv:hep-th/9510017 [hep-th]].

\bibitem{orientifolds1}
A.~Sagnotti, ``Open Strings And Their Symmetry Groups,'' in Cargese '87, ``Non-Perturbative Quantum Field
Theory'', eds. G. Mack et al (Pergamon Press, 1988), p. 521,
arXiv:hep-th/0208020.
%%CITATION = HEP-TH 0208020;%%

\bibitem{orientifolds2}
G.~Pradisi and A.~Sagnotti,
``Open String Orbifolds,''
Phys.\ Lett.\ {\bf B 216} (1989) 59.
%%CITATION = PHLTA,B216,59;%%

\bibitem{orientifolds3}
P.~Horava,
``Strings On World Sheet Orbifolds,''
Nucl.\ Phys.\ {\bf B 327} (1989) 461.
%%CITATION = NUPHA,B327,461;%%

\bibitem{orientifolds4}
P.~Horava, ``Background Duality Of Open String Models,''
Phys.\ Lett.\ {\bf B 231} (1989) 251.
%%CITATION = PHLTA,B231,251;%%

\bibitem{orientifolds5}
M.~Bianchi and A.~Sagnotti,
``On The Systematics Of Open String Theories,''
Phys.\ Lett.\ {\bf B 247} (1990) 517.
%%CITATION = PHLTA,B247,517;%%

\bibitem{orientifolds6}
M.~Bianchi and A.~Sagnotti,
``Twist Symmetry And Open String Wilson Lines,''
Nucl.\ Phys.\ {\bf B 361} (1991) 519.
%%CITATION = NUPHA,B361,519;%%

\bibitem{orientifolds7}
M.~Bianchi, G.~Pradisi and A.~Sagnotti,
``Toroidal compactification and symmetry breaking in open string theories,''
Nucl.\ Phys.\ {\bf B 376} (1992) 365.
%%CITATION = NUPHA,B376,365;%%

\bibitem{orientifolds8}
A.~Sagnotti,
 ``A Note on the Green-Schwarz mechanism in open string theories,''
 Phys.\ Lett.\  {\bf B 294} (1992) 196
 [arXiv:hep-th/9210127].
 %%CITATION = PHLTA,B294,196;%%

\bibitem{orientifolds_rev1}
E.~Dudas,
``Theory and phenomenology of type I strings and M-theory,''
Class.\ Quant.\ Grav.\  {\bf 17} (2000) R41 [arXiv:hep-ph/0006190].
%%CITATION = HEP-PH 0006190;%%

\bibitem{orientifolds_rev2}
C.~Angelantonj and A.~Sagnotti,
``Open strings,''
Phys.\ Rept.\  {\bf 371} (2002) 1 [Erratum-ibid.\  {\bf 376} (2003)
339] [arXiv:hep-th/0204089].
%%CITATION = HEP-TH 0204089;%%

\bibitem{orientifolds_rev3}
C.~Angelantonj and I.~Florakis,
``A Lightning Introduction to String Theory,''
%doi:10.1007/978-981-19-3079-9\_53-1
[arXiv:2406.09508 [hep-th]].

\bibitem{orientifolds_rev4}
G.~Leone and S.~Raucci,
``Aspects of strings without spacetime supersymmetry,''
[arXiv:2509.24703 [hep-th]].
%1 citations counted in INSPIRE as of 15 Oct 2025

\bibitem{CJS}
E.~Cremmer, B.~Julia and J.~Scherk,
``Supergravity Theory in Eleven Dimensions,''
Phys. Lett. B \textbf{76} (1978), 409.
%doi:10.1016/0370-2693(78)90894-8
%1714 citations counted in INSPIRE as of 01 Jun 2021


\bibitem{agmv1}
L.~J.~Dixon and J.~A.~Harvey,
``String Theories in Ten-Dimensions Without Space-Time Supersymmetry,''
Nucl. Phys. B \textbf{274} (1986), 93.
%doi:10.1016/0550-3213(86)90619-X
%390 citations counted in INSPIRE as of 01 Jun 2021

\bibitem{agmv2}
L.~Alvarez-Gaume, P.~H.~Ginsparg, G.~W.~Moore and C.~Vafa,
``An O(16) x O(16) Heterotic String,''
Phys. Lett. B \textbf{171} (1986), 155.
%doi:10.1016/0370-2693(86)91524-8
%313 citations counted in INSPIRE as of 01 Jun 2021

\bibitem{as95}
A.~Sagnotti,
``Some properties of open string theories,''
[arXiv:hep-th/9509080 [hep-th]].
%179 citations counted in INSPIRE as of 31 May 2021

\bibitem{as97}
A.~Sagnotti,
``Surprises in open string perturbation theory,''
Nucl. Phys. B Proc. Suppl. \textbf{56} (1997), 332
%doi:10.1016/S0920-5632(97)00344-7
[arXiv:hep-th/9702093 [hep-th]].
%157 citations counted in INSPIRE as of 31 May 2021

\bibitem{sugimoto}
S.~Sugimoto,
``Anomaly cancellations in type I D9-D9-bar system and the USp(32)  string
%theory,''
Prog.\ Theor.\ Phys.\  {\bf 102} (1999) 685 [arXiv:hep-th/9905159].
%%CITATION = HEP-TH 9905159;%%

\bibitem{dmnonlinear1}
E.~Dudas and J.~Mourad,
  ``Consistent gravitino couplings in nonsupersymmetric strings,''
  Phys.\ Lett.\ B {\bf 514} (2001) 173
  [hep-th/0012071].
  %%CITATION = doi:10.1016/S0370-2693(01)00777-8;%%

  \bibitem{dmnonlinear2}
  G.~Pradisi and F.~Riccioni,
  ``Geometric couplings and brane supersymmetry breaking,''
  Nucl.\ Phys.\ B {\bf 615} (2001) 33
  [hep-th/0107090].
  %%CITATION = doi:10.1016/S0550-3213(01)00441-2;%%

  \bibitem{dmnonlinear3}
  N.~Kitazawa,
  ``Brane SUSY Breaking and the Gravitino Mass,''
  JHEP {\bf 1804} (2018) 081
  %doi:10.1007/JHEP04(2018)081
  [arXiv:1802.03088 [hep-th]].

  %\cite{Scherk:1979zr}
\bibitem{ss1}
J.~Scherk and J.~H.~Schwarz,
``How to Get Masses from Extra Dimensions,''
Nucl. Phys. B \textbf{153} (1979), 61.
%doi:10.1016/0550-3213(79)90592-3;
%1265 citations counted in INSPIRE as of 16 Nov 2023

\bibitem{ss2}
E.~Cremmer, J.~Scherk and J.~H.~Schwarz,
``Spontaneously Broken N=8 Supergravity,''
Phys. Lett. B \textbf{84} (1979), 83.
%doi:10.1016/0370-2693(79)90654-3
%284 citations counted in INSPIRE as of 16 Nov 2023

\bibitem{ss_closed1}
R.~Rohm,
``Spontaneous Supersymmetry Breaking in Supersymmetric String Theories,''
Nucl. Phys. B \textbf{237} (1984), 553.
%doi:10.1016/0550-3213(84)90007-5
%386 citations counted in INSPIRE as of 08 Jul 2021

\bibitem{ss_closed2}
C.~Kounnas and M.~Porrati,
``Spontaneous Supersymmetry Breaking in String Theory,''
Nucl. Phys. B \textbf{310} (1988), 355.
%doi:10.1016/0550-3213(88)90153-8
%265 citations counted in INSPIRE as of 08 Jul 2021

\bibitem{ss_closed3}
S.~Ferrara, C.~Kounnas, M.~Porrati and F.~Zwirner,
``Superstrings with Spontaneously Broken Supersymmetry and their Effective Theories,''
Nucl. Phys. B \textbf{318} (1989), 75.
%doi:10.1016/0550-3213(89)90048-5
%296 citations counted in INSPIRE as of 08 Jul 2021

\bibitem{ss_closed4}
C.~Kounnas and B.~Rostand,
``Coordinate Dependent Compactifications and Discrete Symmetries,''
Nucl. Phys. B \textbf{341} (1990), 641.
%doi:10.1016/0550-3213(90)90543-M
%225 citations counted in INSPIRE as of 08 Jul 2021

\bibitem{ss_closed5}
I.~Antoniadis and C.~Kounnas,
``Superstring phase transition at high temperature,''
Phys. Lett. B \textbf{261} (1991), 369.
%doi:10.1016/0370-2693(91)90442-S
%157 citations counted in INSPIRE as of 08 Jul 2021

\bibitem{ss_closed6}
E.~Kiritsis and C.~Kounnas,
``Perturbative and nonperturbative partial supersymmetry breaking: N=4 ---\ensuremath{>} N=2 ---\ensuremath{>} N=1,''
Nucl. Phys. B \textbf{503} (1997), 117
%doi:10.1016/S0550-3213(97)00430-6
[arXiv:hep-th/9703059 [hep-th]].
%192 citations counted in INSPIRE as of 08 Jul 2021

\bibitem{ads1}
I.~Antoniadis, E.~Dudas and A.~Sagnotti,
``Supersymmetry breaking, open strings and M theory,''
Nucl. Phys. B \textbf{544} (1999), 469
%doi:10.1016/S0550-3213(98)00806-2
[arXiv:hep-th/9807011 [hep-th]].
%159 citations counted in INSPIRE as of 20 Nov 2023

\bibitem{ads2}
I.~Antoniadis, G.~D'Appollonio, E.~Dudas and A.~Sagnotti,
``Partial breaking of supersymmetry, open strings and M theory,''
Nucl. Phys. B \textbf{553} (1999), 133
%doi:10.1016/S0550-3213(99)00232-1
[arXiv:hep-th/9812118 [hep-th]].


%\cite{Dudas:2000sn}
\bibitem{dm-tachyonfree1}
E.~Dudas and J.~Mourad,
``D-branes in nontachyonic 0B orientifolds,''
Nucl. Phys. B \textbf{598} (2001), 189
%doi:10.1016/S0550-3213(00)00781-1
[arXiv:hep-th/0010179 [hep-th]].
%33 citations counted in INSPIRE as of 04 Mar 2024

\bibitem{bcd}
G.~Bossard, G.~Casagrande and E.~Dudas,
``Twisted orientifold planes and S-duality without supersymmetry,''
JHEP \textbf{02} (2025), 062
%doi:10.1007/JHEP02(2025)062
[arXiv:2411.00955 [hep-th]].
%2 citations counted in INSPIRE as of 25 Aug 2025

%\cite{Fabinger:2000jd}
\bibitem{fh}
M.~Fabinger and P.~Horava,
``Casimir effect between world branes in heterotic M theory,''
Nucl. Phys. B \textbf{580} (2000), 243
%doi:10.1016/S0550-3213(00)00255-8
[arXiv:hep-th/0002073 [hep-th]].
%164 citations counted in INSPIRE as of 07 Dec 2023

\bibitem{bsb1}
I.~Antoniadis, E.~Dudas and A.~Sagnotti,
``Brane supersymmetry breaking,''
Phys. Lett. B \textbf{464} (1999), 38
%doi:10.1016/S0370-2693(99)01023-0
[arXiv:hep-th/9908023 [hep-th]].

\bibitem{bsb2}
C.~Angelantonj,
``Comments on open-string orbifolds with a non-vanishing B(ab),''
Nucl.\ Phys.\ {\bf B 566} (2000) 126 [arXiv:hep-th/9908064].
%%CITATION = HEP-TH 9908064;%%

\bibitem{bsb3}
G.~Aldazabal and A.~M.~Uranga,
``Tachyon-free non-supersymmetric type IIB orientifolds via  brane-antibrane
%systems,''
JHEP {\bf 9910} (1999) 024 [arXiv:hep-th/9908072].
%%CITATION = HEP-TH 9908072;%%

\bibitem{bsb4}
C.~Angelantonj, I.~Antoniadis, G.~D'Appollonio, E.~Dudas and A.~Sagnotti,
``Type I vacua with brane supersymmetry breaking,''
Nucl. Phys. B \textbf{572} (2000), 36
%doi:10.1016/S0550-3213(00)00052-3
[arXiv:hep-th/9911081 [hep-th]].

\bibitem{ftse} E.~S.~Fradkin and A.~A.~Tseytlin,
``Nonlinear electrodynamics from quantized strings'',
Phys.\ Lett.\ B163 (1985) 123.
%%CITATION = PHLTA,B163,123;%%

\bibitem{acny} A.~Abouelsaood, C.~G.~Callan, C.~R.~Nappi and S.~A.~Yost,
``Open strings in background gauge fields'',
Nucl.\ Phys.\ B280 (1987) 599.
%%CITATION = NUPHA,B280,599;%%

\bibitem{bachas}  C.~Bachas,
 ``A Way to break supersymmetry,''
  hep-th/9503030.

  \bibitem{aads}
   C.~Angelantonj, I.~Antoniadis, E.~Dudas and A.~Sagnotti,
 ``Type I strings on magnetized orbifolds and brane transmutation,''
 Phys.\ Lett.\ B {\bf 489} (2000) 223
%doi:10.1016/S0370-2693(00)00907-2
 [hep-th/0007090].

\bibitem{berlinmadrid1} R.~Blumenhagen, L.~G\"orlich, B.~K\"ors and D.~L\"ust,
``Asymmetric orbifolds, noncommutative geometry and type I string vacua'',
Nucl.\ Phys.\ B582 (2000) 44
[arXiv:hep-th/0003024].
%%CITATION = HEP-TH 0003024;%%

\bibitem{berlinmadrid2}
 R.~Blumenhagen, L.~G\"orlich, B.~K\"ors and D.~L\"ust,
 ``Noncommutative compactifications of type I strings on tori with
magnetic background flux'',
JHEP 0010 (2000) 006
[arXiv:hep-th/0007024].
%%CITATION = HEP-TH 0007024;%%

\bibitem{berlinmadrid3}
R.~Blumenhagen, B.~K\"ors and D.~L\"ust,
``Type I strings with F- and B-flux'',
JHEP 0102 (2001) 030
[arXiv:hep-th/0012156].
%%CITATION = HEP-TH 0012156;%%


 \bibitem{berkoozdl}
 M.~Berkooz, M.~R.~Douglas and R.~G.~Leigh,
  ``Branes intersecting at angles,''
  Nucl.\ Phys.\ B {\bf 480} (1996), 265
%doi:10.1016/S0550-3213(96)00452-X
  [hep-th/9606139].

\bibitem{berlinmadrid4}
L.~E.~Ibanez, F.~Marchesano and R.~Rabadan,
``Getting just the standard model at intersecting branes,''
JHEP \textbf{11} (2001), 002
%doi:10.1088/1126-6708/2001/11/002
[arXiv:hep-th/0105155 [hep-th]].

\bibitem{berlinmadrid5}
R.~Rabadan,
``Branes at angles, torons, stability and supersymmetry,''
Nucl.\ Phys.\ B620 (2002), 152
[arXiv:hep-th/0107036].
%%CITATION = HEP-TH 0107036;%%

\bibitem{berlinmadrid6}
S.~F\"orste, G.~Honecker and R.~Schreyer,
``Supersymmetric $\ {Z}_N \times \ {Z}_M$ orientifolds
in $4D$ with D-branes at angles'',
Nucl.\ Phys.\ B593 (2001), 127
[arXiv:hep-th/0008250].
%%CITATION = HEP-TH 0008250;%%

\bibitem{berlinmadrid7}
S.~F\"orste, G.~Honecker and R.~Schreyer,
``Orientifolds with branes at angles,''
JHEP 0106 (2001) 004
[arXiv:hep-th/0105208].
%%CITATION = HEP-TH 0105208;%%

\bibitem{berlinmadrid8}
G.~Honecker,
``Intersecting brane world models from D8-branes on (T**2 x T**4/Z(3))/Omega R(1) type IIA orientifolds,''
JHEP \textbf{01} (2002), 025
%doi:10.1088/1126-6708/2002/01/025
[arXiv:hep-th/0201037 [hep-th]].

\bibitem{berlinmadrid9}
R.~Blumenhagen, B.~Kors and D.~Lust,
``Moduli stabilization for intersecting brane worlds in type 0-prime string theory,''
Phys. Lett. B \textbf{532} (2002), 141
%doi:10.1016/S0370-2693(02)01504-6
[arXiv:hep-th/0202024 [hep-th]].
%%CITATION = HEP-TH 0202024;%%

\bibitem{berlinmadrid10}
L.~F.~Alday and G.~Aldazabal,
``In quest of just the standard model on D-branes at a singularity,''
JHEP \textbf{05} (2002), 022
%doi:10.1088/1126-6708/2002/05/022
[arXiv:hep-th/0203129 [hep-th]].
%%CITATION = HEP-TH 0203129;%%

\bibitem{berlinmadrid11}
D.~Cremades, L.~E.~Ibanez and F.~Marchesano,
``Intersecting brane models of particle physics and the Higgs mechanism,''
JHEP \textbf{07} (2002), 022
%doi:10.1088/1126-6708/2002/07/022
[arXiv:hep-th/0203160 [hep-th]].
%%CITATION = HEP-TH 0203160;%%


\bibitem{berlinmadrid_rev}
R.~Blumenhagen, B.~Kors, D.~Lust and S.~Stieberger,
``Four-dimensional String Compactifications with D-Branes, Orientifolds and Fluxes,''
Phys. Rept. \textbf{445} (2007), 1
%doi:10.1016/j.physrep.2007.04.003
[arXiv:hep-th/0610327 [hep-th]].
%898 citations counted in INSPIRE as of 27 Nov 2023

\bibitem{bsb_rev}
J.~Mourad and A.~Sagnotti,
``An Update on Brane Supersymmetry Breaking,''
[arXiv:1711.11494 [hep-th]].
%59 citations counted in INSPIRE as of 25 Aug 2025

\bibitem{4d1} I.~Antoniadis, C.~P.~Bachas and C.~Kounnas,
``Four-Dimensional Superstrings,''
Nucl. Phys. B \textbf{289} (1987), 87.
%doi:10.1016/0550-3213(87)90372-5
%910 citations counted in INSPIRE as of 23 Aug 2025

\bibitem{4d2}
I.~Antoniadis and C.~Bachas,
``$4D$ fermionic superstrings with arbitrary twists'',
Nucl.\ Phys.\ B298 (1988), 586.
%%CITATION = NUPHA,B298,586;%%

\bibitem{4d3}
H.~Kawai, D.~C.~Lewellen and S.~H.~Tye,
``Construction of four-dimensional fermionic string models'',
Phys.\ Rev.\ Lett.\  57 (1986), 1832
[Erratum-ibid.\  58 (1986), 429].
%%CITATION = PRLTA,57,1832;%%

\bibitem{4d4}
H.~Kawai, D.~C.~Lewellen and S.~H.~Tye,
``Classification of closed fermionic string models'',
Phys.\ Rev.\ D34 (1986), 3794.
%%CITATION = PHRVA,D34,3794;%%

\bibitem{4d5}
H.~Kawai, D.~C.~Lewellen and S.~H.~Tye,
``Construction of fermionic string models in four-dimensions'',
Nucl.\ Phys.\ B288 (1987), 1.
%%CITATION = NUPHA,B288,1;%%

\bibitem{4d6}
I.~Antoniadis, J.~R.~Ellis, J.~S.~Hagelin and D.~V.~Nanopoulos,
``The flipped ${\rm SU}(5) \times {\rm U}(1)$ string model revamped'',
Phys.\ Lett.\ B231 (1989), 65.
%%CITATION = PHLTA,B231,65;%%


\bibitem{4d7}
A.~E.~Faraggi,
``Construction of realistic standard-like models in the free
fermionic superstring formulation'',
Nucl.\ Phys.\ B387 (1992), 239
[arXiv:hep-th/9208024].
%%CITATION = HEP-TH 9208024;%%

\bibitem{4d8}
J.~L.~Lopez, D.~V.~Nanopoulos and K.~j.~Yuan,
``The search for a realistic flipped SU(5) string model'',
Nucl.\ Phys.\ B399 (1993), 654
[arXiv:hep-th/9203025].
%%CITATION = HEP-TH 9203025;%%

\bibitem{4d9}
A.~E.~Faraggi,
``Custodial non-Abelian gauge symmetries in realistic
superstring derived models'',
Phys.\ Lett.\ B339 (1994), 223
[arXiv:hep-ph/9408333].
%%CITATION = HEP-PH 9408333;%%

\bibitem{4d10}
S.~Chaudhuri, G.~Hockney and J.~Lykken,
``Three generations in the fermionic construction'',
Nucl.\ Phys.\ B469 (1996), 357
[arXiv:hep-th/9510241].
%%CITATION = HEP-TH 9510241;%%

\bibitem{abpss}
C.~Angelantonj, M.~Bianchi, G.~Pradisi, A.~Sagnotti and Y.~S.~Stanev,
``Chiral asymmetry in four-dimensional open string vacua,''
Phys. Lett. B \textbf{385} (1996), 96
%doi:10.1016/0370-2693(96)00869-6
[arXiv:hep-th/9606169 [hep-th]].
%271 citations counted in INSPIRE as of 15 Apr 2024

\bibitem{kakudual} Z.~Kakushadze,
``Aspects of $N = 1$ type I-heterotic duality in four dimensions'',
Nucl.\ Phys.\ B512 (1998), 221
[arXiv:hep-th/9704059].
%%CITATION = HEP-TH 9704059;%%


\bibitem{ibanez1}
G.~Aldazabal, A.~Font, L.~E.~Ibanez and G.~Violero,
``D = 4, N=1, type IIB orientifolds,''
Nucl. Phys. B \textbf{536} (1998), 29
%doi:10.1016/S0550-3213(98)00666-X
[arXiv:hep-th/9804026 [hep-th]].

\bibitem{ibanez2}
G.~Aldazabal, L.~E.~Ibanez, F.~Quevedo and A.~M.~Uranga,
``D-branes at singularities: A Bottom up approach to the string embedding of the standard model,''
JHEP \textbf{08} (2000) 002
%doi:10.1088/1126-6708/2000/08/002
[arXiv:hep-th/0005067 [hep-th]].

\bibitem{ibanez3}
L.~E.~Ibanez, F.~Marchesano and R.~Rabadan,
``Getting just the standard model at intersecting branes,''
JHEP \textbf{11} (2001), 002
%doi:10.1088/1126-6708/2001/11/002
[arXiv:hep-th/0105155 [hep-th]].

\bibitem{ibanez4}
G.~Aldazabal, S.~Franco, L.~E.~Ibanez, R.~Rabadan and A.~M.~Uranga,
``Intersecting brane worlds,''
JHEP \textbf{02} (2001), 047
%doi:10.1088/1126-6708/2001/02/047
[arXiv:hep-ph/0011132 [hep-ph]].

\bibitem{bert1}
T.~P.~T.~Dijkstra, L.~R.~Huiszoon and A.~N.~Schellekens,
``Chiral supersymmetric standard model spectra from orientifolds of Gepner models,''
Phys. Lett. B \textbf{609} (2005), 408
%doi:10.1016/j.physletb.2004.04.094
[arXiv:hep-th/0403196 [hep-th]].

\bibitem{bert2}
T.~P.~T.~Dijkstra, L.~R.~Huiszoon and A.~N.~Schellekens,
``Supersymmetric standard model spectra from RCFT orientifolds,''
Nucl. Phys. B \textbf{710} (2005), 3
%doi:10.1016/j.nuclphysb.2004.12.032
[arXiv:hep-th/0411129 [hep-th]].

\bibitem{bert3}
P.~Anastasopoulos, T.~P.~T.~Dijkstra, E.~Kiritsis and A.~N.~Schellekens,
``Orientifolds, hypercharge embeddings and the Standard Model,''
Nucl. Phys. B \textbf{759} (2006), 83
%doi:10.1016/j.nuclphysb.2006.10.013
[arXiv:hep-th/0605226 [hep-th]].

\bibitem{bert4}
L.~E.~Ibanez, A.~N.~Schellekens and A.~M.~Uranga,
``Discrete Gauge Symmetries in Discrete MSSM-like Orientifolds,''
Nucl. Phys. B \textbf{865} (2012), 509
% doi:10.1016/j.nuclphysb.2012.08.008
[arXiv:1205.5364 [hep-th]].

\bibitem{bert_rev}
A.~N.~Schellekens,
``Life at the Interface of Particle Physics and String Theory,''
Rev. Mod. Phys. \textbf{85} (2013) no.4, 1491
% doi:10.1103/RevModPhys.85.1491
[arXiv:1306.5083 [hep-ph]].
%80 citations counted in INSPIRE as of 11 Jan 2025

%\cite{Candelas:1985en}
\bibitem{calabiyau}
P.~Candelas, G.~T.~Horowitz, A.~Strominger and E.~Witten,
``Vacuum configurations for superstrings,''
Nucl. Phys. B \textbf{258} (1985), 46.
%doi:10.1016/0550-3213(85)90602-9
%3037 citations counted in INSPIRE as of 05 Jan 2024

\bibitem{tomasiello}
A.~Tomasiello,
``Geometry of String Theory Compactifications,''
Cambridge University Press (2022).
%ISBN 978-1-108-63574-5, 978-1-108-47373-6
%doi:10.1017/9781108635745
%25 citations counted in INSPIRE as of 25 Jul 2025

\bibitem{aspinwall}
P.~S.~Aspinwall,
``K3 surfaces and string duality,''
[arXiv:hep-th/9611137 [hep-th]];
%440 citations counted in INSPIRE as of 22 Aug 2025
A.~Dabholkar,
``Lectures on orientifolds and duality,''
[arXiv:hep-th/9804208 [hep-th]].
%130 citations counted in INSPIRE as of 22 Aug 2025

\bibitem{dm_vacuum}
E.~Dudas and J.~Mourad,
 ``Brane solutions in strings with broken supersymmetry and dilaton
 tadpoles,''
 Phys.\ Lett.\  {\bf B 486} (2000) 172
 [arXiv:hep-th/0004165].
 %%CITATION = PHLTA,B486,172;%%

\bibitem{gm}
S.~S.~Gubser and I.~Mitra,
``Some interesting violations of the Breitenlohner-Freedman bound,''
JHEP \textbf{07} (2002), 044
%doi:10.1088/1126-6708/2002/07/044
[arXiv:hep-th/0108239 [hep-th]].
%18 citations counted in INSPIRE as of 01 Jun 2021

\bibitem{ms_16}
J.~Mourad and A.~Sagnotti,
  ``$AdS$ Vacua from Dilaton Tadpoles and Form Fluxes,''
  Phys.\ Lett.\ B {\bf 768} (2017), 92
  %doi:10.1016/j.physletb.2017.02.053
  [arXiv:1612.08566 [hep-th]].
  %%CITATION = doi:10.1016/j.physletb.2017.02.053;%%
 
\bibitem{bms1}
I.~Basile, J.~Mourad and A.~Sagnotti,
``On Classical Stability with Broken Supersymmetry,''
JHEP \textbf{01} (2019), 174
%doi:10.1007/JHEP01(2019)174
[arXiv:1811.11448 [hep-th]].
%15 citations counted in INSPIRE as of 01 Jun 2021

\bibitem{bms2}
R.~Antonelli, I.~Basile and A.~Bombini,
``AdS Vacuum Bubbles, Holography and Dual RG Flows,''
Class. Quant. Grav. \textbf{36} (2019) no.4, 045004
%doi:10.1088/1361-6382/aafef9
[arXiv:1806.02289 [hep-th]].
%10 citations counted in INSPIRE as of 02 Jun 2021

\bibitem{bms3}
R.~Antonelli and I.~Basile,
``Brane annihilation in non-supersymmetric strings,''
JHEP \textbf{11} (2019), 021
%doi:10.1007/JHEP11(2019)021
[arXiv:1908.04352 [hep-th]].
%7 citations counted in INSPIRE as of 02 Jun 2021

\bibitem{bms4}
I.~Basile,
``On String Vacua without Supersymmetry: brane dynamics, bubbles and holography,''
Ph.D. Thesis, [arXiv:2010.00628 [hep-th]].
%1 citations counted in INSPIRE as of 02 Jun 2021

  \bibitem{ms23_1}
  J.~Mourad and A.~Sagnotti,
``Non-supersymmetric vacua and self-adjoint extensions,''
JHEP \textbf{08} (2023), 041
%doi:10.1007/JHEP08(2023)041
[arXiv:2305.09587 [hep-th]].

\bibitem{KKLT}
S.~Kachru, R.~Kallosh, A.~D.~Linde and S.~P.~Trivedi,
  ``De Sitter vacua in string theory,''
  Phys.\ Rev.\ D {\bf 68} (2003) 046005
  %doi:10.1103/PhysRevD.68.046005
  [hep-th/0301240].
  %%CITATION = doi:10.1103/PhysRevD.68.046005;%%


\bibitem{inflation1}
 A.~A.~Starobinsky,
  ``A New Type of Isotropic Cosmological Models Without Singularity,''
  Phys.\ Lett.\ {\bf B 91} (1980) 99.
 %%CITATION = PHLTA,B91,99;%%

 \bibitem{inflation2}
  D.~Kazanas,
 ``Dynamics of the Universe and Spontaneous Symmetry Breaking,''
  Astrophys.\ J.\  {\bf 241} (1980) L59.
 %%CITATION = ASJOA,241,L59;%%

 \bibitem{inflation3}
  K.~Sato,
``Cosmological Baryon Number Domain Structure and the First Order Phase Transition of a Vacuum,''
  Phys.\ Lett.\  {\bf B 99} (1981) 66.
 %%CITATION = PHLTA,B99,66;%%

 \bibitem{inflation4}
 A.~H.~Guth,
 ``The Inflationary Universe: A Possible Solution to the Horizon and Flatness Problems,''
 Phys.\ Rev.\ {\bf D 23} (1981) 347.
  %%CITATION = PHRVA,D23,347;%%

  \bibitem{inflation5}
 A.~D.~Linde,
 ``A New Inflationary Universe Scenario: A Possible Solution of the Horizon, Flatness, Homogeneity, Isotropy and Primordial Monopole Problems,''
  Phys.\ Lett.\ {\bf B 108} (1982) 389.
  %%CITATION = PHLTA,B108,389;%%

  \bibitem{inflation6}
   A.~Albrecht and P.~J.~Steinhardt,
``Cosmology for Grand Unified Theories with Radiatively Induced Symmetry Breaking,''
  Phys.\ Rev.\ Lett.\  {\bf 48} (1982) 1220.

  \bibitem{inflation7}
   A.~D.~Linde,
``Chaotic Inflation,''
  Phys.\ Lett.\ {\bf B 129} (1983) 177.
 %%CITATION = PHLTA,B129,177;%%

 \bibitem{inflation_rev}
 N.~Bartolo, E.~Komatsu, S.~Matarrese and A.~Riotto,
``Non-Gaussianity from inflation: Theory and observations,''
  Phys.\ Rept.\  {\bf 402} (2004) 103
  [astro-ph/0406398].\\
  %%CITATION = ASTRO-PH/0406398;%%
V.~Mukhanov,
  ``Physical foundations of cosmology,''
  Cambridge Univ. Press (2005); \\
S.~Weinberg, ``Cosmology,''
 Oxford Univ. Press (2008); \\
D.~H.~Lyth and A.~R.~Liddle,
 ``The primordial density perturbation: Cosmology, inflation and the origin of structure,''
  Cambridge Univ. Press (2009); \\
  D.~S.~Gorbunov and V.~A.~Rubakov,
  ``Introduction to the theory of the early universe: Cosmological perturbations and inflationary theory,''
  World Scientific (2011);\\
  %doi:10.1142/7874
  %%CITATION = doi:10.1142/7874;%%
  %15 citations counted in INSPIRE as of 04 Aug 2017
 J.~Martin, C.~Ringeval and V.~Vennin,
  ``Encyclopaedia Inflationaris,''
  Phys.\ Dark Univ.\  {\bf 5-6} (2014) 75
  [arXiv:1303.3787 [astro-ph.CO]];\\
  %%CITATION = doi:10.1016/j.dark.2014.01.003;%%
    D.~Baumann, ``Cosmology,''
 Cambridge Univ. Press (2022).

\bibitem{WMAP}
D.~N.~Spergel {\it et al.}  [WMAP Collaboration],
  ``First year Wilkinson Microwave Anisotropy Probe (WMAP) observations: Determination of cosmological parameters,''
  Astrophys.\ J.\ Suppl.\  {\bf 148}, 175 (2003)
  [astro-ph/0302209].
  %%CITATION = ASTRO-PH/0302209;%%
  %6820 citations counted in INSPIRE as of 12 Apr 2013

\bibitem{PLANCK1}
N.~Aghanim \textit{et al.} [Planck],
``Planck 2018 results. VI. Cosmological parameters,''
Astron. Astrophys. \textbf{641} (2020), A6
[erratum: Astron. Astrophys. \textbf{652} (2021), C4]
%doi:10.1051/0004-6361/201833910
[arXiv:1807.06209 [astro-ph.CO]].
%13414 citations counted in INSPIRE as of 17 Apr 2024

\bibitem{PLANCK2}
Y.~Akrami \textit{et al.} [Planck],
``Planck 2018 results. VII. Isotropy and Statistics of the CMB,''
Astron. Astrophys. \textbf{641} (2020), A7
%doi:10.1051/0004-6361/201935201
[arXiv:1906.02552 [astro-ph.CO]].
%275 citations counted in INSPIRE as of 17 Apr 2024

\bibitem{PLANCK3}
N.~Aghanim \textit{et al.} [Planck],
``Planck 2018 results. V. CMB power spectra and likelihoods,''
Astron. Astrophys. \textbf{641} (2020), A5
%doi:10.1051/0004-6361/201936386
[arXiv:1907.12875 [astro-ph.CO]].
%893 citations counted in INSPIRE as of 17 Apr 2024

\bibitem{bsb_cosmology1}
E.~Dudas, N.~Kitazawa and A.~Sagnotti,
``On Climbing Scalars in String Theory,''
Phys. Lett. B \textbf{694} (2011), 80
%doi:10.1016/j.physletb.2010.09.040
[arXiv:1009.0874 [hep-th]].

\bibitem{bsb_cosmology2}
E.~Dudas, N.~Kitazawa, S.~P.~Patil and A.~Sagnotti,
``CMB Imprints of a Pre-Inflationary Climbing Phase,''
JCAP \textbf{05} (2012), 012
%doi:10.1088/1475-7516/2012/05/012
[arXiv:1202.6630 [hep-th]].

\bibitem{bsb_cosmology3}
P.~Fr\'e, A.~Sagnotti and A.~S.~Sorin,
``Integrable Scalar Cosmologies I. Foundations and links with String Theory,''
Nucl. Phys. B \textbf{877} (2013), 1028
%doi:10.1016/j.nuclphysb.2013.10.015
[arXiv:1307.1910 [hep-th]].

\bibitem{bsb_cosmology4}
A.~Sagnotti,
``Brane SUSY breaking and inflation: implications for scalar fields and CMB distortion,''
%doi:10.1134/S1547477114070395
[arXiv:1303.6685 [hep-th]].

\bibitem{bsb_cosmology5}
N.~Kitazawa and A.~Sagnotti,
``Pre-inflationary clues from String Theory?,''
JCAP \textbf{04} (2014), 017
%doi:10.1088/1475-7516/2014/04/017
[arXiv:1402.1418 [hep-th]].

\bibitem{bsb_cosmology6}
A.~Gruppuso and A.~Sagnotti,
%``Observational Hints of a Pre--Inflationary Scale?,''
Int. J. Mod. Phys. D \textbf{24} (2015) no.12, 1544008
%doi:10.1142/S0218271815440083
[arXiv:1506.08093 [astro-ph.CO]];
%31 citations counted in INSPIRE as of 11 Sep 2025
A.~Gruppuso, N.~Kitazawa, N.~Mandolesi, P.~Natoli and A.~Sagnotti,
``Pre-Inflationary Relics in the CMB?,''
Phys. Dark Univ. \textbf{11} (2016), 68
%doi:10.1016/j.dark.2015.12.001
[arXiv:1508.00411 [astro-ph.CO]].

\bibitem{bsb_cosmology7}
A.~Gruppuso, N.~Kitazawa, M.~Lattanzi, N.~Mandolesi, P.~Natoli and A.~Sagnotti,
``The Evens and Odds of CMB Anomalies,''
Phys. Dark Univ. \textbf{20} (2018), 49
%doi:10.1016/j.dark.2018.03.002
[arXiv:1712.03288 [astro-ph.CO]].

\bibitem{swampland}
C.~Vafa,
``The String landscape and the swampland,'' [arXiv:hep-th/0509212 [hep-th]]; H.~Ooguri and C.~Vafa,
``On the Geometry of the String Landscape and the Swampland,''
Nucl. Phys. B \textbf{766} (2007), 21, 
[arXiv:hep-th/0605264 [hep-th]]. For reviews, see \emph{e.g.} E.~Palti,
``The Swampland: Introduction and Review,''
Fortsch. Phys. \textbf{67} (2019) no.6, 1900037, 
[arXiv:1903.06239 [hep-th]];  
M.~van Beest, J.~Calder{\'o}n-Infante, D.~Mirfendereski and I.~Valenzuela,
``Lectures on the Swampland Program in String Compactifications,''
Phys. Rept. \textbf{989} (2022), 1, [arXiv:2102.01111 [hep-th]]; 
M.~Gra{\~n}a and A.~Herr{\'a}ez,
``The Swampland Conjectures: A Bridge from Quantum Gravity to Particle Physics,''
Universe \textbf{7} (2021) no.8, 273, 
[arXiv:2107.00087 [hep-th]]; 
N.~B.~Agmon, A.~Bedroya, M.~J.~Kang and C.~Vafa,
``Lectures on the string landscape and the Swampland,''
[arXiv:2212.06187 [hep-th]].

  

\bibitem{gurseyradicati}
F.~Gursey and L.~A.~Radicati,
``Spin and unitary spin independence of strong interactions,''
Phys. Rev. Lett. \textbf{13} (1964), 173.
%doi:10.1103/PhysRevLett.13.173
%452 citations counted in INSPIRE as of 04 Jul 2023

\bibitem{eightfold}
M.~Gell-Mann,
``The Eightfold Way: A Theory of strong interaction symmetry'';
%doi:10.2172/4008239
Y.~Ne'eman,
``Derivation of strong interactions from a gauge invariance,''
Nucl. Phys. \textbf{26} (1961), 222.
%doi:10.1016/0029-5582(61)90134-1

\bibitem{FSZ1}
A.~Salam and J.~A.~Strathdee,
``Supergauge Transformations,''
Nucl. Phys. B \textbf{76} (1974), 477.
% doi:10.1016/0550-3213(74)90537-9
%775 citations counted in INSPIRE as of 31 Jul 2023

\bibitem{FSZ2}
M. Gell-Mann, talk given at the 1977 Washington meeting of the American
Physical Society.

\bibitem{FSZ3}
S.~Ferrara, C.~A.~Savoy and B.~Zumino,
``General Massive Multiplets in Extended Supersymmetry,''
Phys. Lett. B \textbf{100} (1981), 393.
% doi:10.1016/0370-2693(81)90144-1
%148 citations counted in INSPIRE as of 31 Jul 2023


\bibitem{N8SUGRA}
E.~Cremmer and B.~Julia,
``The SO(8) Supergravity,''
Nucl. Phys. B \textbf{159} (1979), 141;
%doi:10.1016/0550-3213(79)90331-6
B.~de Wit and H.~Nicolai,
``N=8 Supergravity,''
Nucl. Phys. B \textbf{208} (1982), 323.
%doi:10.1016/0550-3213(82)90120-1

\bibitem{BPS1}
E.~B.~Bogomolny,
``Stability of Classical Solutions,''
Sov. J. Nucl. Phys. \textbf{24} (1976), 449
PRINT-76-0543 (LANDAU-INST.).
%1934 citations counted in INSPIRE as of 31 Jul 2023

\bibitem{BPS2}
M.~K.~Prasad and C.~M.~Sommerfield,
``An Exact Classical Solution for the 't Hooft Monopole and the Julia-Zee Dyon,''
Phys. Rev. Lett. \textbf{35} (1975), 760.
% doi:10.1103/PhysRevLett.35.760.
%1645 citations counted in INSPIRE as of 31 Jul 2023

%\cite{Salam:1974jj}
\bibitem{Salam:1974jj}
A.~Salam and J.~A.~Strathdee,
``On Superfields and Fermi-Bose Symmetry,''
Phys. Rev. D \textbf{11} (1975), 1521.
%419 citations counted in INSPIRE as of 24 Sep 2020

\bibitem{superspace_review}
S.~J.~Gates, M.~T.~Grisaru, M.~Rocek and W.~Siegel,
``Superspace Or One Thousand and One Lessons in Supersymmetry,''
Front. Phys. \textbf{58} (1983), 1
[arXiv:hep-th/0108200 [hep-th]].
%828 citations counted in INSPIRE as of 24 Sep 2020

\bibitem{fayet-goldstino}
P.~Fayet,
``Mixing Between Gravitational and Weak Interactions Through the Massive Gravitino,''
Phys. Lett. B \textbf{70} (1977), 461; 
%doi:10.1016/0370-2693(77)90414-2
P.~Fayet,
``Scattering Cross-Sections of the Photino and the Goldstino (Gravitino) on Matter,''
Phys. Lett. B \textbf{86} (1979), 272; 
%doi:10.1016/0370-2693(79)90836-0
R.~Casalbuoni, S.~De Curtis, D.~Dominici, F.~Feruglio and R.~Gatto,
``A Gravitino - Goldstino High-Energy Equivalence Theorem,''
Phys. Lett. B \textbf{215} (1988), 313.
%doi:10.1016/0370-2693(88)91439-6. 

 %\cite{ISS}
\bibitem{ISS}
K.~A.~Intriligator, N.~Seiberg and D.~Shih,
``Supersymmetry breaking, R-symmetry breaking and metastable vacua,''
JHEP \textbf{07} (2007), 017
[arXiv:hep-th/0703281 [hep-th]].
%185 citations counted in INSPIRE as of 09 Oct 2020

\bibitem{fayetor}
P.~Fayet,
``Spontaneous Supersymmetry Breaking Without Gauge Invariance,''
Phys. Lett. B \textbf{58} (1975), 67.
%doi:10.1016/0370-2693(75)90730-3
%148 citations counted in INSPIRE as of 19 Nov 2025

%\cite{gauge-mediation}
\bibitem{gauge-mediation}
G.~F.~Giudice and R.~Rattazzi,
``Theories with gauge mediated supersymmetry breaking,''
Phys. Rept. \textbf{322} (1999), 419
%doi:10.1016/S0370-1573(99)00042-3
[arXiv:hep-ph/9801271 [hep-ph]].
%1656 citations counted in INSPIRE as of 02 Nov 2020

\bibitem{gim}
  S.~L.~Glashow, J.~Iliopoulos and L.~Maiani,
``Weak Interactions with Lepton-Hadron Symmetry,''
  Phys.\ Rev.\ D {\bf 2} (1970) 1285.
  %%CITATION = PHRVA,D2,1285;%%
  
%\cite{ Zumino:1979e}
\bibitem{Zumino:1979et}
B.~Zumino,
``Supersymmetry and Kahler Manifolds,''
Phys. Lett. B \textbf{87} (1979), 203.
%doi:10.1016/0370-2693(79)90964-X
%669 citations counted in INSPIRE as of 04 Dec 2020

\bibitem{dsb}
E.~Witten,
``Dynamical Breaking of Supersymmetry,''
Nucl. Phys. B \textbf{188} (1981), 513
%doi:10.1016/0550-3213(81)90006-7
%3586 citations counted in INSPIRE as of 25 Aug 2025

\bibitem{windex}
E.~Witten,
``Constraints on Supersymmetry Breaking,''
Nucl. Phys. B \textbf{202} (1982), 253
%doi:10.1016/0550-3213(82)90071-2
%2010 citations counted in INSPIRE as of 25 Aug 2025

  %\cite{bd}
\bibitem{bd}
P.~Binetruy and E.~Dudas,
``Gaugino condensation and the anomalous U(1),''
Phys. Lett. B \textbf{389} (1996), 503
%doi:10.1016/S0370-2693(96)01305-6
[arXiv:hep-th/9607172 [hep-th]].
%218 citations counted in INSPIRE as of 16 Nov 2020

%\cite{split}
\bibitem{split1}
N.~Arkani-Hamed and S.~Dimopoulos,
``Supersymmetric unification without low energy supersymmetry and signatures for fine-tuning at the LHC,''
JHEP \textbf{06} (2005), 073
%doi:10.1088/1126-6708/2005/06/073
[arXiv:hep-th/0405159 [hep-th]].
%1178 citations counted in INSPIRE as of 16 Nov 2020

\bibitem{split2}
G.~F.~Giudice and A.~Romanino,
``Split supersymmetry,''
Nucl. Phys. B \textbf{699} (2004), 65
[erratum: Nucl. Phys. B \textbf{706} (2005), 487-487]
%doi:10.1016/j.nuclphysb.2004.08.001
[arXiv:hep-ph/0406088 [hep-ph]].
%941 citations counted in INSPIRE as of 16 Nov 2020


\bibitem{iliopoulos25}
J.~Iliopoulos,
``From Many Models to ONE THEORY,''
[arXiv:2501.10233 [physics.hist-ph]].
%0 citations counted in INSPIRE as of 01 Feb 2025

\bibitem{QED} J.~Schwinger,
``Selected Papers on Quantum Electrodynamics,''
Dover Publications, 1958.
%34 citations counted in INSPIRE as of 03 Aug 2023

\bibitem{fermi}
E.~Fermi,
``An attempt of a theory of beta radiation. 1.,''
Z. Phys. \textbf{88} (1934), 161.
%%doi:10.1007/BF01351864
%1600 citations counted in INSPIRE as of 03 Aug 2023


\bibitem{cabibbo}
 N.~Cabibbo,
``Unitary Symmetry and Leptonic Decays,''
  Phys.\ Rev.\ Lett.\  {\bf 10} (1963) 531.
  %%CITATION = PRLTA,10,531;%%

\bibitem{glashow}
 S.~L.~Glashow,
 ``Partial Symmetries of Weak Interactions,''
  Nucl.\ Phys.\  {\bf 22} (1961) 579.
  %%CITATION = NUPHA,22,579;%%

 \bibitem{weinberg}
 S.~Weinberg,
 ``A Model of Leptons,''
  Phys.\ Rev.\ Lett.\  {\bf 19} (1967) 1264.
  %%CITATION = PRLTA,19,1264;%%

  \bibitem{salam}
 A. Salam (1968), N. Svartholm. ed. Elementary Particle Physics: Relativistic Groups and Analyticity,
 Eighth Nobel Symposium, Stockholm: Almquvist and Wiksell. pp. 367.

 \bibitem{beh1}
 F.~Englert and R.~Brout,
 ``Broken Symmetry and the Mass of Gauge Vector Mesons,''
Phys. Rev. Lett. \textbf{13} (1964), 321.
%doi:10.1103/PhysRevLett.13.321
%6893 citations counted in INSPIRE as of 03 Aug 2023

 \bibitem{beh2}
P.~W.~Higgs,
``Broken Symmetries and the Masses of Gauge Bosons,''
Phys. Rev. Lett. \textbf{13} (1964), 508.
%doi:10.1103/PhysRevLett.13.508

 \bibitem{beh3}
G.~S.~Guralnik, C.~R.~Hagen and T.~W.~B.~Kibble,
``Global Conservation Laws and Massless Particles,''
Phys. Rev. Lett. \textbf{13} (1964), 585.
% doi:10.1103/PhysRevLett.13.585

\bibitem{gurhagkib}
G.~S.~Guralnik, C.~R.~Hagen and T.~W.~B.~Kibble,
``Global Conservation Laws and Massless Particles,''
Phys. Rev. Lett. \textbf{13} (1964), 585.
%doi:10.1103/PhysRevLett.13.585
%5289 citations counted in INSPIRE as of 30 Jan 2025

\bibitem{quarks1}
M.~Gell-Mann,
``A Schematic Model of Baryons and Mesons,''
Phys. Lett. \textbf{8} (1964), 214.
%doi:10.1016/S0031-9163(64)92001-3
%3990 citations counted in INSPIRE as of 03 Aug 2023

\bibitem{quarks2}
G.~Zweig,
``An SU(3) model for strong interaction symmetry and its breaking. Version 1,''
CERN-TH-401.
%792 citations counted in INSPIRE as of 03 Aug 2023

\bibitem{TV1}
 G.~'t Hooft,
 ``Renormalizable lagrangians for massive Yang-Mills fields,''
  Nucl.\ Phys.\  B {\bf 35} (1971) 167.
  %%CITATION = NUPHA,B35,167;%%

  \bibitem{TV2}
 G.~'t Hooft and M.~J.~G.~Veltman,
 ``Regularization And Renormalization Of Gauge Fields,''
  Nucl.\ Phys.\  B {\bf 44} (1972) 189.
  %%CITATION = NUPHA,B44,189;%%

  \bibitem{TV3}
   G.~'t Hooft and M.~J.~G.~Veltman,
 ``Combinatorics of gauge fields,''
  Nucl.\ Phys.\  B {\bf 50} (1972) 318.
  %%CITATION = NUPHA,B50,318;%%

  \bibitem{TV4}
B.~W.~Lee and J.~Zinn-Justin,
``Spontaneously Broken Gauge Symmetries Part 1: Preliminaries,''
Phys. Rev. D \textbf{5} (1972), 3121;
%doi:10.1103/PhysRevD.5.3121
B.~W.~Lee and J.~Zinn-Justin,
``Spontaneously Broken Gauge Symmetries Part 2: Perturbation Theory and Renormalization,''
Phys. Rev. D \textbf{5} (1972), 3137
[erratum: Phys. Rev. D \textbf{8} (1973), 4654].
%doi:10.1103/PhysRevD.5.3137

  \bibitem{ren_rev}
J.~C.~Collins,
  ``Renormalization. An Introduction To Renormalization, The Renormalization Group, And The Operator Product Expansion,''  Cambridge, Uk: Univ. Pr. ( 1984).

  \bibitem{BIM}
 C.~Bouchiat, J.~Iliopoulos and P.~Meyer,
  ``An Anomaly Free Version of Weinberg's Model,''
  Phys.\ Lett.\ B {\bf 38} (1972) 519.
  %%CITATION = PHLTA,B38,519;%%

\bibitem{GJACK}
D.~J.~Gross and R.~Jackiw,
  ``Effect of anomalies on quasirenormalizable theories,''
  Phys.\ Rev.\ D {\bf 6} (1972) 477.
  %%CITATION = PHRVA,D6,477;%%

  \bibitem{ABJ1}
  S.~L.~Adler,
``Axial vector vertex in spinor electrodynamics,''
Phys. Rev. \textbf{177} (1969), 2426.
%doi:10.1103/PhysRev.177.2426
%4635 citations counted in INSPIRE as of 23 Nov 2023

  \bibitem{ABJ2}
J.~S.~Bell and R.~Jackiw,
``A PCAC puzzle: $\pi^0 \to \gamma \gamma$ in the $\sigma$ model,''
Nuovo Cim. A \textbf{60} (1969), 47.
%doi:10.1007/BF02823296
%3835 citations counted in INSPIRE as of 23 Nov 2023

\bibitem{ABJ3}
S.~L.~Adler and W.~A.~Bardeen,
``Absence of higher order corrections in the anomalous axial vector divergence equation,''
Phys. Rev. \textbf{182} (1969), 1517.
%doi:10.1103/PhysRev.182.1517

\bibitem{wzcc}
J.~Wess and B.~Zumino,
``Consequences of anomalous Ward identities,''
Phys. Lett. B \textbf{37} (1971), 95
%doi:10.1016/0370-2693(71)90582-X
%3150 citations counted in INSPIRE as of 23 May 2025

  \bibitem{ckm}
 M.~Kobayashi and T.~Maskawa,
 ``CP Violation in the Renormalizable Theory of Weak Interaction,''
  Prog.\ Theor.\ Phys.\  {\bf 49} (1973) 652.
  %%CITATION = PTPKA,49,652;%%

  \bibitem{croninfitch}
J.~H.~Christenson, J.~W.~Cronin, V.~L.~Fitch and R.~Turlay,
``Evidence for the $2\pi$ Decay of the $K_2^0$ Meson,''
Phys. Rev. Lett. \textbf{13} (1964), 138.
%doi:10.1103/PhysRevLett.13.138

\bibitem{gw}
   D.~J.~Gross and F.~Wilczek,
 ``Ultraviolet Behavior of Nonabelian Gauge Theories,''
  Phys.\ Rev.\ Lett.\  {\bf 30} (1973) 1343.
  %%CITATION = PRLTA,30,1343;%%

  \bibitem{politzer}
   H.~D.~Politzer,
 ``Reliable Perturbative Results for Strong Interactions?,''
  Phys.\ Rev.\ Lett.\  {\bf 30} (1973) 1346.
  %%CITATION = PRLTA,30,1346;%%

\bibitem{QCD1}
H.~Fritzsch, M.~Gell-Mann and H.~Leutwyler,
``Advantages of the Color Octet Gluon Picture,''
Phys. Lett. B \textbf{47} (1973), 365.
%doi:10.1016/0370-2693(73)90625-4
%2454 citations counted in INSPIRE as of 03 Aug 2023

\bibitem{QCD2}
S.~Weinberg,
``Nonabelian Gauge Theories of the Strong Interactions,''
Phys. Rev. Lett. \textbf{31} (1973), 494.
%doi:10.1103/PhysRevLett.31.494
%1066 citations counted in INSPIRE as of 03 Aug 2023


\bibitem{gargamelle}
F.~J.~Hasert \textit{et al.} [Gargamelle Neutrino],
``Observation of Neutrino Like Interactions Without Muon Or Electron in the Gargamelle Neutrino Experiment,''
Phys. Lett. B \textbf{46} (1973), 138.
%doi:10.1016/0370-2693(73)90499-1
%1892 citations counted in INSPIRE as of 03 Aug 2023

\bibitem{ivb1}
G.~Arnison \textit{et al.} [UA1],
``Experimental Observation of Isolated Large Transverse Energy Electrons with Associated Missing Energy at $\sqrt{s}= 540$ GeV,''
Phys. Lett. B \textbf{122} (1983), 103.
%doi:10.1016/0370-2693(83)91177-2
%3210 citations counted in INSPIRE as of 04 Aug 2023

\bibitem{ivb2}
G.~Arnison \textit{et al.} [UA1],
``Experimental Observation of Lepton Pairs of Invariant Mass Around 95-GeV/c**2 at the CERN SPS Collider,''
Phys. Lett. B \textbf{126} (1983), 398.
%doi:10.1016/0370-2693(83)90188-0
%2759 citations counted in INSPIRE as of 04 Aug 2023

\bibitem{ivb3}
M.~Banner \textit{et al.} [UA2],
``Observation of Single Isolated Electrons of High Transverse Momentum in Events with Missing Transverse Energy at the CERN anti-p p Collider,''
Phys. Lett. B \textbf{122} (1983), 476.

\bibitem{ivb4}
P.~Bagnaia \textit{et al.} [UA2],
``Evidence for $Z^{0} \to e^+ e^-$ at the CERN $\bar{p} p$ Collider,''
Phys. Lett. B \textbf{129} (1983), 130.
%doi:10.1016/0370-2693(83)90744-X
%2288 citations counted in INSPIRE as of 04 Aug 2023
%doi:10.1016/0370-2693(83)91605-2
%2453 citations counted in INSPIRE as of 04 Aug 2023

%\cite{ATLAS:2012yve}
\bibitem{higgs1}
G.~Aad \textit{et al.} [ATLAS],
``Observation of a new particle in the search for the Standard Model Higgs boson with the ATLAS detector at the LHC,''
Phys. Lett. B \textbf{716} (2012), 1
%doi:10.1016/j.physletb.2012.08.020
[arXiv:1207.7214 [hep-ex]].
%14441 citations counted in INSPIRE as of 04 Aug 2023

\bibitem{higgs2}
S.~Chatrchyan \textit{et al.} [CMS],
``Observation of a New Boson at a Mass of 125 GeV with the CMS Experiment at the LHC,''
Phys. Lett. B \textbf{716} (2012), 30
%doi:10.1016/j.physletb.2012.08.021
[arXiv:1207.7235 [hep-ex]].
%14013 citations counted in INSPIRE as of 04 Aug 2023

\bibitem{strong_CP}
G.~'t Hooft,
``Symmetry Breaking Through Bell-Jackiw Anomalies,''
Phys. Rev. Lett. \textbf{37} (1976), 8.
%doi:10.1103/PhysRevLett.37.8
%4073 citations counted in INSPIRE as of 04 Aug 2023

\bibitem{PQ1}
 R.~D.~Peccei and H.~R.~Quinn,
  ``CP Conservation in the Presence of Instantons,''
  Phys.\ Rev.\ Lett.\  {\bf 38} (1977), 1440.
  %%CITATION = PRLTA,38,1440;%%

  \bibitem{PQ2}
R.~D.~Peccei and H.~R.~Quinn,
``Constraints Imposed by CP Conservation in the Presence of Instantons,''
Phys. Rev. D \textbf{16} (1977), 1791.
%doi:10.1103/PhysRevD.16.1791
%3552 citations counted in INSPIRE as of 06 Dec 2022

\bibitem{wolfenstein}
L.~Wolfenstein,
``Parametrization of the Kobayashi-Maskawa Matrix,''
Phys. Rev. Lett. \textbf{51} (1983), 1945.
%doi:10.1103/PhysRevLett.51.1945
%3385 citations counted in INSPIRE as of 04 Aug 2023

\bibitem{GlashWei}
S.~L.~Glashow and S.~Weinberg,
%``Natural Conservation Laws for Neutral Currents,''
Phys. Rev. D \textbf{15} (1977), 1958.
%doi:10.1103/PhysRevD.15.1958
%2180 citations counted in INSPIRE as of 02 Apr 2025

\bibitem{GaiL}
 M.~K.~Gaillard and B.~W.~Lee,
 ``Rare Decay Modes of the K-Mesons in Gauge Theories,''
  Phys.\ Rev.\ D {\bf 10} (1974), 897.
  %%CITATION = PHRVA,D10,897;%%

  
\bibitem{custodial}
 P.~Sikivie, L.~Susskind, M.~B.~Voloshin and V.~I.~Zakharov,
  ``Isospin Breaking in Technicolor Models,''
  Nucl.\ Phys.\ B {\bf 173} (1980), 189.
  %%CITATION = NUPHA,B173,189;%%

\bibitem{thooft_anom}
G.~'t Hooft,
``Naturalness, chiral symmetry, and spontaneous chiral symmetry breaking,''
NATO Sci. Ser. B \textbf{59} (1980), 135.
%doi:10.1007/978-1-4684-7571-5\_9
%1166 citations counted in INSPIRE as of 23 May 2025
  
\bibitem{zwyz}
B.~Zumino, Y.~S.~Wu and A.~Zee,
``Chiral Anomalies, Higher Dimensions, and Differential Geometry,''
Nucl. Phys. B \textbf{239} (1984), 477.
%doi:10.1016/0550-3213(84)90259-1
%451 citations counted in INSPIRE as of 23 May 2025

\bibitem{msz}
J.~Manes, R.~Stora and B.~Zumino,
``Algebraic Study of Chiral Anomalies,''
Commun. Math. Phys. \textbf{102} (1985), 157.
%doi:10.1007/BF01208825
%197 citations counted in INSPIRE as of 21 Aug 2025

%\cite{Anastasopoulos:2006cz}
\bibitem{abdk}
J.~Preskill,
``Gauge anomalies in an effective field theory,''
Annals Phys. \textbf{210} (1991), 323; 
%doi:10.1016/0003-4916(91)90046-B
P.~Anastasopoulos, M.~Bianchi, E.~Dudas and E.~Kiritsis,
``Anomalies, anomalous U(1)'s and generalized Chern-Simons terms,''
JHEP \textbf{11} (2006), 057
%doi:10.1088/1126-6708/2006/11/057
[arXiv:hep-th/0605225 [hep-th]].

\bibitem{gaisei}
D.~Gaiotto, A.~Kapustin, N.~Seiberg and B.~Willett,
``Generalized Global Symmetries,''
JHEP \textbf{02} (2015), 172
%doi:10.1007/JHEP02(2015)172
[arXiv:1412.5148 [hep-th]].
%1514 citations counted in INSPIRE as of 04 Sep 2025

\bibitem{rev_anom_unif}
I.~Garcia Etxebarria,
``Branes and Non-Invertible Symmetries,''
Fortsch. Phys. \textbf{70} (2022) no.11, 2200154
%doi:10.1002/prop.202200154
[arXiv:2208.07508 [hep-th]];
%125 citations counted in INSPIRE as of 21 Nov 2025
S.~H.~Shao,
``What's Done Cannot Be Undone: TASI Lectures on Non-Invertible Symmetries,''
[arXiv:2308.00747 [hep-th]];
%338 citations counted in INSPIRE as of 21 Nov 2025
S.~Schafer-Nameki,
``ICTP lectures on (non-)invertible generalized symmetries,''
Phys. Rept. \textbf{1063} (2024), 1
%doi:10.1016/j.physrep.2024.01.007
[arXiv:2305.18296 [hep-th]];
%344 citations counted in INSPIRE as of 21 Nov 2025
L.~Bhardwaj, L.~E.~Bottini, L.~Fraser-Taliente, L.~Gladden, D.~S.~W.~Gould, A.~Platschorre and H.~Tillim,
``Lectures on generalized symmetries,''
Phys. Rept. \textbf{1051} (2024), 1
%doi:10.1016/j.physrep.2023.11.002
[arXiv:2307.07547 [hep-th]];
%252 citations counted in INSPIRE as of 21 Nov 2025
D.~Costa, C.~Cordova, M.~Del Zotto, D.~Freed, J.~G\"odicke, A.~Hofer, D.~Jordan, D.~Morgante, R.~Moscrop and K.~Ohmori, \emph{et al.}
``Simons Lectures on Categorical Symmetries,''
[arXiv:2411.09082 [math-ph]].
%36 citations counted in INSPIRE as of 21 Nov 2025

  \bibitem{baluni}
  V.~Baluni,
``CP Violating Effects in QCD,''
Phys. Rev. D \textbf{19} (1979), 2227.
%doi:10.1103/PhysRevD.19.2227
%713 citations counted in INSPIRE as of 04 Feb 2025

\bibitem{crewther}
R.~J.~Crewther, P.~Di Vecchia, G.~Veneziano and E.~Witten,
``Chiral Estimate of the Electric Dipole Moment of the Neutron in Quantum Chromodynamics,''
Phys. Lett. B \textbf{88} (1979), 123
[erratum: Phys. Lett. B \textbf{91} (1980), 487].
%doi:10.1016/0370-2693(79)90128-X

\bibitem{flag}
Y.~Aoki \textit{et al.} [Flavour Lattice Averaging Group (FLAG)],
``FLAG Review 2024,''
[arXiv:2411.04268 [hep-lat]].

\bibitem{nelson-barr}
A.~E.~Nelson,
``Naturally Weak CP Violation,''
Phys. Lett. B \textbf{136} (1984), 387;
%doi:10.1016/0370-2693(84)92025-2
S.~M.~Barr,
``Solving the Strong CP Problem Without the Peccei-Quinn Symmetry,''
Phys. Rev. Lett. \textbf{53} (1984), 329.
%doi:10.1103/PhysRevLett.53.329


\bibitem{PQ}
 R.~D.~Peccei and H.~R.~Quinn,
  ``CP Conservation in the Presence of Instantons,''
  Phys.\ Rev.\ Lett.\  {\bf 38} (1977) 1440;
  %%CITATION = PRLTA,38,1440;%%
R.~D.~Peccei and H.~R.~Quinn,
``Constraints Imposed by CP Conservation in the Presence of Instantons,''
Phys. Rev. D \textbf{16} (1977), 1791.
%doi:10.1103/PhysRevD.16.1791
%3552 citations counted in INSPIRE as of 06 Dec 2022

\bibitem{axion}
S.~Weinberg,
  ``A New Light Boson?,''
  Phys.\ Rev.\ Lett.\  {\bf 40} (1978) 223;
  %%CITATION = PRLTA,40,223;%%
F.~Wilczek,
``Problem of Strong  $P$  and  $T$  Invariance in the Presence of Instantons,''
Phys. Rev. Lett. \textbf{40} (1978), 279.
%doi:10.1103/PhysRevLett.40.279
%4629 citations counted in INSPIRE as of 06 Dec 2022

\bibitem{rubakov}
V.~A.~Kuzmin, V.~A.~Rubakov and M.~E.~Shaposhnikov,
``On the Anomalous Electroweak Baryon Number Nonconservation in the Early Universe,''
Phys. Lett. B \textbf{155} (1985), 36
%doi:10.1016/0370-2693(85)91028-7
%3589 citations counted in INSPIRE as of 16 Apr 2025


\bibitem{vw}
C.~Vafa and E.~Witten,
``Restrictions on Symmetry Breaking in Vector-Like Gauge Theories,''
Nucl. Phys. B \textbf{234} (1984), 173.
%doi:10.1016/0550-3213(84)90230-X
%691 citations counted in INSPIRE as of 06 Dec 2022

\bibitem{ksvz}
J.~E.~Kim,
``Weak Interaction Singlet and Strong CP Invariance,''
Phys. Rev. Lett. \textbf{43} (1979), 103;
%doi:10.1103/PhysRevLett.43.103;
%2702 citations counted in INSPIRE as of 06 Dec 2022
M.~A.~Shifman, A.~I.~Vainshtein and V.~I.~Zakharov,
``Can Confinement Ensure Natural CP Invariance of Strong Interactions?,''
Nucl. Phys. B \textbf{166} (1980), 493.
%doi:10.1016/0550-3213(80)90209-6
%2392 citations counted in INSPIRE as of 06 Dec 2022


\bibitem{dvv}
P.~Di Vecchia and G.~Veneziano,
``Chiral Dynamics in the Large n Limit,''
Nucl. Phys. B \textbf{171} (1980), 253;
%doi:10.1016/0550-3213(80)90370-3
%725 citations counted in INSPIRE as of 08 Dec 2022
H.~Georgi, D.~B.~Kaplan and L.~Randall,
``Manifesting the Invisible Axion at Low-energies,''
Phys. Lett. B \textbf{169} (1986), 73.
%doi:10.1016/0370-2693(86)90688-X
%262 citations counted in INSPIRE as of 08 Dec 2022

\bibitem{sikivie}
P.~Sikivie,
``Experimental Tests of the Invisible Axion,''
Phys. Rev. Lett. \textbf{51} (1983), 1415
[erratum: Phys. Rev. Lett. \textbf{52} (1984), 695].
% doi:10.1103/PhysRevLett.51.1415

\bibitem{primakoff}
H.~Primakoff,
``Photoproduction of neutral mesons in nuclear electric fields and the mean life of the neutral meson,''
Phys. Rev. \textbf{81} (1951), 899.
%doi:10.1103/PhysRev.81.899

\bibitem{tremaine}
L.~Hui, J.~P.~Ostriker, S.~Tremaine and E.~Witten,
``Ultralight scalars as cosmological dark matter,''
Phys. Rev. D \textbf{95} (2017) no.4, 043541
%doi:10.1103/PhysRevD.95.043541
[arXiv:1610.08297 [astro-ph.CO]].

%\cite{svrcek-witten}
\bibitem{svrcek-witten}
See, \emph{e.g.}, P.~Svrcek and E.~Witten,
``Axions In String Theory,''
JHEP \textbf{06} (2006), 051
%doi:10.1088/1126-6708/2006/06/051
[arXiv:hep-th/0605206 [hep-th]];
J.~E.~Kim and G.~Carosi,
``Axions and the Strong CP Problem,''
Rev. Mod. Phys. \textbf{82} (2010), 557
[erratum: Rev. Mod. Phys. \textbf{91} (2019) no.4, 049902]
%doi:10.1103/RevModPhys.82.557
[arXiv:0807.3125 [hep-ph]].


\bibitem{gs85}
M.~H.~Goroff and A.~Sagnotti,
``Quantum Gravity at Two Loops,''
Phys. Lett. B \textbf{160} (1985), 81.
%doi:10.1016/0370-2693(85)91470-4
%434 citations counted in INSPIRE as of 10 May 2024

\bibitem{gs86}
M.~H.~Goroff and A.~Sagnotti,
``The Ultraviolet Behavior of Einstein Gravity,''
Nucl. Phys. B \textbf{266} (1986), 709.
%doi:10.1016/0550-3213(86)90193-8
%835 citations counted in INSPIRE as of 10 May 2024

\bibitem{wk}
K.~G.~Wilson and J.~B.~Kogut,
``The Renormalization group and the epsilon expansion,''
Phys. Rept. \textbf{12} (1974), 75.
%doi:10.1016/0370-1573(74)90023-4
%3366 citations counted in INSPIRE as of 03 Feb 2025


\bibitem{statistical}
For reviews, see:~G.~Parisi,
``Statistical field Theory,'' CRC Press (1998);
%20 citations counted in INSPIRE as of 03 Feb 2025
C.~Itzykson and J.~M.~Drouffe,
``Statistical field Theory Vol. 1: from Brownian Motion to Renormalization and Lattice Gauge Theory,''
Cambridge Univ. Press (1989);
%doi:10.1017/CBO9780511622779
%25 citations counted in INSPIRE as of 03 Feb 2025
C.~Itzykson and J.~M.~Drouffe,
``Statistical field Theory. Vol. 2: Strong Coupling, Monte Carlo Methods, Conformal Field Theory, and Random Systems,''
Cambridge Univ. Press (1989).
%doi:10.1017/CBO9780511622786
%7 citations counted in INSPIRE as of 03 Feb 2025

\bibitem{RamMoh}
R.~N.~Mohapatra, S.~Antusch, K.~S.~Babu, G.~Barenboim, M.~C.~Chen, S.~Davidson, A.~de Gouv\^ea, P.~de Holanda, B.~Dutta and Y.~Grossman, \textit{et al.}
``Theory of neutrinos: A White paper,''
Rept. Prog. Phys. \textbf{70} (2007), 1757
%doi:10.1088/0034-4885/70/11/R02
[arXiv:hep-ph/0510213 [hep-ph]].

\bibitem{GMRS}
P.~Minkowski,
``$\mu \to e\gamma$ at a Rate of One Out of $10^{9}$ Muon Decays?,''
Phys. Lett. B \textbf{67} (1977), 421;
%doi:10.1016/0370-2693(77)90435-X 
M.~Gell-Mann, P.~Ramond and R.~Slansky,
``Complex Spinors and Unified Theories,''
Conf. Proc. C \textbf{790927} (1979), 315 [arXiv:1306.4669 [hep-th]];
T.~Yanagida,
``Horizontal gauge symmetry and masses of neutrinos,''
Conf. Proc. C \textbf{7902131} (1979), 95-99 KEK-79-18-95;
R.~N.~Mohapatra and G.~Senjanovic,
``Neutrino Mass and Spontaneous Parity Nonconservation,''
Phys. Rev. Lett. \textbf{44} (1980), 912.
%doi:10.1103/PhysRevLett.44.912

\bibitem{su51}
  H.~Georgi and S.~L.~Glashow,
  ``Unity of All Elementary Particle Forces,''
  Phys.\ Rev.\ Lett.\  {\bf 32} (1974) 438.
  %%CITATION = PRLTA,32,438;%%
  
\bibitem{so10}
H.~Fritzsch and P.~Minkowski,
``Unified Interactions of Leptons and Hadrons,''
Annals Phys. \textbf{93} (1975), 193.
%doi:10.1016/0003-4916(75)90211-0
%2240 citations counted in INSPIRE as of 15 Jan 2025

\bibitem{e6}
F.~Gursey, P.~Ramond and P.~Sikivie,
``A Universal Gauge Theory Model Based on E6,''
Phys. Lett. B \textbf{60} (1976), 177.
% doi:10.1016/0370-2693(76)90417-2

\bibitem{susyunif}
S.~Dimopoulos, S.~Raby and F.~Wilczek,
``Supersymmetry and the Scale of Unification,''
Phys. Rev. D \textbf{24} (1981), 1681;
%doi:10.1103/PhysRevD.24.1681
J.~R.~Ellis, S.~Kelley and D.~V.~Nanopoulos,
``Probing the desert using gauge coupling unification,''
Phys. Lett. B \textbf{260} (1991), 131;
%doi:10.1016/0370-2693(91)90980-5
U.~Amaldi, W.~de Boer and H.~Furstenau,
``Comparison of grand unified theories with electroweak and strong coupling constants measured at LEP,''
Phys. Lett. B \textbf{260} (1991), 447.
%doi:10.1016/0370-2693(91)91641-8


  \bibitem{su52}
 H.~Georgi, H.~R.~Quinn and S.~Weinberg,
  ``Hierarchy of Interactions in Unified Gauge Theories,''
  Phys.\ Rev.\ Lett.\  {\bf 33} (1974) 451.
  %%CITATION = PRLTA,33,451;%%

\bibitem{DDG}
K.~R.~Dienes, E.~Dudas and T.~Gherghetta,
``Extra space-time dimensions and unification,''
Phys. Lett. B \textbf{436} (1998), 55
%doi:10.1016/S0370-2693(98)00977-0
[arXiv:hep-ph/9803466 [hep-ph]];
K.~R.~Dienes, E.~Dudas and T.~Gherghetta,
%``Grand unification at intermediate mass scales through extra dimensions,''
Nucl. Phys. B \textbf{537} (1999), 47
%doi:10.1016/S0550-3213(98)00669-5
[arXiv:hep-ph/9806292 [hep-ph]];
%975 citations counted in INSPIRE as of 03 Feb 2025
K.~R.~Dienes, E.~Dudas and T.~Gherghetta,
``Neutrino oscillations without neutrino masses or heavy mass scales: A Higher dimensional seesaw mechanism,''
Nucl. Phys. B \textbf{557} (1999), 25
%doi:10.1016/S0550-3213(99)00377-6
[arXiv:hep-ph/9811428 [hep-ph]].
%523 citations counted in INSPIRE as of 03 Feb 2025

\bibitem{hierarchy}
E.~Gildener,
``Gauge Symmetry Hierarchies,''
Phys. Rev. D \textbf{14} (1976), 1667.
%doi:10.1103/PhysRevD.14.1667

%\cite{nfw
\bibitem{nfw}
See,  \emph{e.g.}, J.~F.~Navarro, C.~S.~Frenk and S.~D.~M.~White,
``The Structure of cold dark matter halos,''
Astrophys. J. \textbf{462} (1996), 563
%doi:10.1086/177173
[arXiv:astro-ph/9508025 [astro-ph]];
D.~Clowe, M.~Bradac, A.~H.~Gonzalez, M.~Markevitch, S.~W.~Randall, C.~Jones and D.~Zaritsky,
``A direct empirical proof of the existence of dark matter,''
Astrophys. J. Lett. \textbf{648} (2006), L109
%doi:10.1086/508162
[arXiv:astro-ph/0608407 [astro-ph]].

%\cite{bertone}
\bibitem{bertone}
See e.g. G.~Bertone, D.~Hooper and J.~Silk,
``Particle dark matter: Evidence, candidates and constraints,''
Phys. Rept. \textbf{405} (2005), 279
%doi:10.1016/j.physrep.2004.08.031
[arXiv:hep-ph/0404175 [hep-ph]];
J.~L.~Feng,
``Dark Matter Candidates from Particle Physics and Methods of Detection,''
Ann. Rev. Astron. Astrophys. \textbf{48} (2010), 495
%doi:10.1146/annurev-astro-082708-101659
[arXiv:1003.0904 [astro-ph.CO]];
G.~Bertone and D.~Hooper,
``History of dark matter,''
Rev. Mod. Phys. \textbf{90} (2018) no.4, 045002
%doi:10.1103/RevModPhys.90.045002
[arXiv:1605.04909 [astro-ph.CO]]; 
Y.~Mambrini,
``Particles in the Dark Universe. A Student\textquoteright{}s Guide to Particle Physics and Cosmology,''
Springer (2021).
%ISBN 978-3-030-78138-5, 978-3-030-78139-2. 
%doi:10.1007/978-3-030-78139-2


\bibitem{sugra_div_recent}
Z.~Bern, J.~J.~Carrasco, W.~M.~Chen, A.~Edison, H.~Johansson, J.~Parra-Martinez, R.~Roiban and M.~Zeng,
``Ultraviolet Properties of $\mathcal N = 8$ Supergravity at Five Loops,''
Phys. Rev. D \textbf{98} (2018) no.8, 086021
%doi:10.1103/PhysRevD.98.086021
[arXiv:1804.09311 [hep-th]].
%169 citations counted in INSPIRE as of 10 May 2024

\bibitem{hierarchy1}
S.~Weinberg,
  ``Implications of Dynamical Symmetry Breaking,''
  Phys.\ Rev.\ D {\bf 13} (1976) 974.

  \bibitem{hierarchy2}
S.~Weinberg,``Implications of Dynamical Symmetry Breaking: An Addendum,''
  Phys.\ Rev.\ D {\bf 19} (1979) 1277.
  %%CITATION = PHRVA,D19,1277;%%;

  \bibitem{hierarchy3}
 E.~Gildener,
  ``Gauge Symmetry Hierarchies,''
  Phys.\ Rev.\ D {\bf 14} (1976) 1667.
  %%CITATION = PHRVA,D14,1667;%%;

  \bibitem{hierarchy4}
   L.~Susskind,
  ``Dynamics of Spontaneous Symmetry Breaking in the Weinberg-Salam Theory,''
  Phys.\ Rev.\ D {\bf 20} (1979) 2619.
  %%CITATION = PHRVA,D20,2619;%%;\\

  \bibitem{hierarchy5}
 G. 't Hooft, in "Recent developments in gauge theories", Proceedings of the NATO Advanced Summer Institute, Cargese
1979, (Plenum Press, 1980).


\bibitem{susyunif1}
 S.~Dimopoulos, S.~Raby and F.~Wilczek,
  ``Supersymmetry and the Scale of Unification,''
  Phys.\ Rev.\ D {\bf 24} (1981) 1681.
  %%CITATION = PHRVA,D24,1681;%%

  \bibitem{susyunif2}
  U.~Amaldi, W.~de Boer and H.~Furstenau,
``Comparison of grand unified theories with electroweak and strong coupling constants measured at LEP,''
Phys. Lett. B \textbf{260}, 447 (1991).

\bibitem{susyunif3}
J.~R.~Ellis, S.~Kelley and D.~V.~Nanopoulos,
``Probing the desert using gauge coupling unification,''
Phys. Lett. B \textbf{260}, 131 (1991).
%doi:10.1016/0370-2693(91)90980-5
%1126 citations counted in INSPIRE as of 06 Jun 2023
%doi:10.1016/0370-2693(91)91641-8
%2010 citations counted in INSPIRE as of 06 Jun 2023;

\bibitem{keith}
K.~R.~Dienes,
  ``String theory and the path to unification: A Review of recent developments,''
  Phys.\ Rept.\  {\bf 287} (1997) 447
  [hep-th/9602045].
  %%CITATION = HEP-TH/9602045;%%


%\cite{farrar-fayet}
\bibitem{farrar-fayet1}
G.~R.~Farrar and P.~Fayet,
``Phenomenology of the Production, Decay, and Detection of New Hadronic States Associated with Supersymmetry,''
Phys. Lett. B \textbf{76} (1978), 575.
%1720 citations counted in INSPIRE as of 20 Oct 2020

\bibitem{farrar-fayet2}
J.~R.~Ellis, J.~S.~Hagelin, D.~V.~Nanopoulos, K.~A.~Olive and M.~Srednicki,
``Supersymmetric Relics from the Big Bang,''
Nucl. Phys. B \textbf{238} (1984), 453.
%doi:10.1016/0550-3213(84)90461-9
%2071 citations counted in INSPIRE as of 06 Jun 2023

\bibitem{dark_matter}
G.~Jungman, M.~Kamionkowski and K.~Griest,
``Supersymmetric dark matter,''
Phys. Rept. \textbf{267} (1996), 195
%doi:10.1016/0370-1573(95)00058-5
[arXiv:hep-ph/9506380 [hep-ph]].
%4741 citations counted in INSPIRE as of 06 Jun 2023


\bibitem{weinberg_cc}
S.~Weinberg,
``The Cosmological Constant Problem,''
Rev. Mod. Phys. \textbf{61} (1989), 1;
%doi:10.1103/RevModPhys.61.1
S.~M.~Carroll,
``The Cosmological constant,''
Living Rev. Rel. \textbf{4} (2001), 1
%doi:10.12942/lrr-2001-1
[arXiv:astro-ph/0004075 [astro-ph]].

%\cite{susy-FNCN}
\bibitem{susy-FCNC}
See e.g. 
F.~Gabbiani, E.~Gabrielli, A.~Masiero and L.~Silvestrini,
``A Complete analysis of FCNC and CP constraints in general SUSY extensions of the standard model,''
Nucl. Phys. B \textbf{477} (1996), 321
%doi:10.1016/0550-3213(96)00390-2
[arXiv:hep-ph/9604387 [hep-ph]];
M.~Misiak, S.~Pokorski and J.~Rosiek,
``Supersymmetry and FCNC effects,''
Adv. Ser. Direct. High Energy Phys. \textbf{15} (1998), 795
%doi:10.1142/9789812812667\_0012
[arXiv:hep-ph/9703442 [hep-ph]].


%\cite{Barbier:2004ez}
\bibitem{r-parity-review}
R.~Barbier, C.~Berat, M.~Besancon, M.~Chemtob, A.~Deandrea, E.~Dudas, P.~Fayet, S.~Lavignac, G.~Moreau and E.~Perez, \textit{et al.}
``R-parity violating supersymmetry,''
Phys. Rept. \textbf{420} (2005), 1
[arXiv:hep-ph/0406039 [hep-ph]].
%1155 citations counted in INSPIRE as of 24 Oct 2020

%\cite{Goldberg:1983nd}
\bibitem{lsp}
H.~Goldberg,
``Constraint on the Photino Mass from Cosmology,''
Phys. Rev. Lett. \textbf{50} (1983), 1419
[erratum: Phys. Rev. Lett. \textbf{103} (2009), 099905];
%1449 citations counted in INSPIRE as of 24 Oct 2020
J.~R.~Ellis, J.~S.~Hagelin, D.~V.~Nanopoulos, K.~A.~Olive and M.~Srednicki,
``Supersymmetric Relics from the Big Bang,''
Nucl. Phys. B \textbf{238} (1984), 453-476. 
%doi:10.1016/0550-3213(84)90461-9


%\cite{instability-sm}
\bibitem{sm-instability}
D.~Buttazzo, G.~Degrassi, P.~P.~Giardino, G.~F.~Giudice, F.~Sala, A.~Salvio and A.~Strumia,
``Investigating the near-criticality of the Higgs boson,''
JHEP \textbf{12} (2013), 089
%doi:10.1007/JHEP12(2013)089
[arXiv:1307.3536 [hep-ph]].


%\cite{ATLAS-SUSY}
\bibitem{ATLAS-SUSY}
G.~Aad \textit{et al.} [ATLAS],
``The quest to discover supersymmetry at the ATLAS experiment,''
[arXiv:2403.02455 [hep-ex]].

%\cite{direct-detection-dm}
\bibitem{direct-detection-dm}
M.~Schumann,
``Direct Detection of WIMP Dark Matter: Concepts and Status,''
J. Phys. G \textbf{46} (2019) no.10, 103003
%doi:10.1088/1361-6471/ab2ea5
[arXiv:1903.03026 [astro-ph.CO]]; 
J.~Billard, M.~Boulay, S.~Cebri\'an, L.~Covi, G.~Fiorillo, A.~Green, J.~Kopp, B.~Majorovits, K.~Palladino and F.~Petricca, \textit{et al.}
%``Direct detection of dark matter\textemdash{}APPEC committee report*,''
Rept. Prog. Phys. \textbf{85} (2022) no.5, 056201
%doi:10.1088/1361-6633/ac5754
[arXiv:2104.07634 [hep-ex]]. 

%\cite{R-parity-searches}
\bibitem{R-parity-searches}
G.~Aad \textit{et al.} [ATLAS],
``Search for R-parity-violating supersymmetry in a final state containing leptons and many jets with the ATLAS experiment using $\sqrt{s} = 13 { TeV}$ proton\textendash{}proton collision data,''
Eur. Phys. J. C \textbf{81} (2021) no.11, 1023
%doi:10.1140/epjc/s10052-021-09761-x
[arXiv:2106.09609 [hep-ex]].

\bibitem{atlas}
G.~Aad \textit{et al.} [ATLAS],
``Observation of a new particle in the search for the Standard Model Higgs boson with the ATLAS detector at the LHC,''
Phys. Lett. B \textbf{716} (2012), 1
% doi:10.1016/j.physletb.2012.08.020
[arXiv:1207.7214 [hep-ex]].

\bibitem{cms}
S.~Chatrchyan \textit{et al.} [CMS],
``Observation of a New Boson at a Mass of 125 GeV with the CMS Experiment at the LHC,''
Phys. Lett. B \textbf{716} (2012), 30
% doi:10.1016/j.physletb.2012.08.021
[arXiv:1207.7235 [hep-ex]].

%\cite{B-anomalies}
\bibitem{B-anomalies}
Y.~Li and C.~D.~L\"u,
``Recent Anomalies in B Physics,''
Sci. Bull. \textbf{63} (2018), 267
%doi:10.1016/j.scib.2018.02.003
[arXiv:1808.02990 [hep-ph]];
A.~Crivellin and B.~Mellado,
``Anomalies in particle physics and their implications for physics beyond the standard model,''
Nature Rev. Phys. \textbf{6} (2024) no.5, 294
%doi:10.1038/s42254-024-00703-6
[arXiv:2309.03870 [hep-ph]].

\bibitem{velo-zwanziger}
G.~Velo and D.~Zwanziger,
``Propagation and quantization of Rarita-Schwinger waves in an external electromagnetic potential,''
Phys. Rev. \textbf{186} (1969), 1337;
%doi:10.1103/PhysRev.186.1337
G.~Velo and D.~Zwanziger,
``Noncausality and other defects of interaction Lagrangians for particles with spin one and higher,''
Phys. Rev. \textbf{188} (1969), 2218.
%doi:10.1103/PhysRev.188.2218
%313 citations counted in INSPIRE as of 07 Feb 2025



\bibitem{cfgvp}
 E.~Cremmer, S.~Ferrara, L.~Girardello and A.~Van Proeyen,
  ``Yang-Mills Theories with Local Supersymmetry: Lagrangian, Transformation Laws and SuperHiggs Effect,''
  Nucl.\ Phys.\ B {\bf 212} (1983), 413.
  %doi:10.1016/0550-3213(83)90679-X
  %%CITATION = doi:10.1016/0550-3213(83)90679-X;%%

\bibitem{deser-zumino}
S.~Deser and B.~Zumino,
``Broken Supersymmetry and Supergravity,''
Phys. Rev. Lett. \textbf{38} (1977), 1433.
%doi:10.1103/PhysRevLett.38.1433

%\cite{freedman}
\bibitem{freedman}
D.~Z.~Freedman,
``Supergravity with Axial Gauge Invariance,''
Phys. Rev. D \textbf{15} (1977), 1173.
%doi:10.1103/PhysRevD.15.1173
%125 citations counted in INSPIRE as of 19 Sep 2023

\bibitem{cftvp}
N.~Cribiori, F.~Farakos, M.~Tournoy and A.~van Proeyen,
%``Fayet-Iliopoulos terms in supergravity without gauged R-symmetry,''
JHEP \textbf{04} (2018), 032
% doi:10.1007/JHEP04(2018)032
[arXiv:1712.08601 [hep-th]].
%84 citations counted in INSPIRE as of 10 Jun 2025

%\cite{Dine:1987xk}
\bibitem{Dine:1987xk}
M.~Dine, N.~Seiberg and E.~Witten,
``Fayet-Iliopoulos Terms in String Theory,''
Nucl. Phys. B \textbf{289} (1987), 589.
%doi:10.1016/0550-3213(87)90395-6
%725 citations counted in INSPIRE as of 21 Sep 2023

\bibitem{giudice-masiero}
G.~F.~Giudice and A.~Masiero,
``A Natural Solution to the mu Problem in Supergravity Theories,''
Phys. Lett. B \textbf{206} (1988), 480.
%doi:10.1016/0370-2693(88)91613-9
%1062 citations counted in INSPIRE as of 16 Jan 2025

%\cite{sw}
\bibitem{sw}
S.~K.~Soni and H.~A.~Weldon,
``Analysis of the Supersymmetry Breaking Induced by N=1 Supergravity Theories,''
Phys. Lett. B \textbf{126} (1983), 215.
%doi:10.1016/0370-2693(83)90593-2
%388 citations counted in INSPIRE as of 21 Sep 2023

%\cite{kl}
\bibitem{kl}
V.~S.~Kaplunovsky and J.~Louis,
``Model independent analysis of soft terms in effective supergravity and in string theory,''
Phys. Lett. B \textbf{306} (1993), 269
%doi:10.1016/0370-2693(93)90078-V
[arXiv:hep-th/9303040 [hep-th]].
%737 citations counted in INSPIRE as of 21 Sep 2023

%\cite{fkz}
\bibitem{fkz}
S.~Ferrara, C.~Kounnas and F.~Zwirner,
``Mass formulae and natural hierarchy in string effective supergravities,''
Nucl. Phys. B \textbf{429} (1994), 589
[erratum: Nucl. Phys. B \textbf{433} (1995), 255]
%doi:10.1016/0550-3213(94)00494-Y
[arXiv:hep-th/9405188 [hep-th]].
%221 citations counted in INSPIRE as of 21 Sep 2023

%\cite{bim}
\bibitem{bim}
A.~Brignole, L.~E.~Ibanez and C.~Munoz,
``Towards a theory of soft terms for the supersymmetric Standard Model,''
Nucl. Phys. B \textbf{422} (1994), 125
[erratum: Nucl. Phys. B \textbf{436} (1995), 747]
%doi:10.1016/0550-3213(94)00068-9
[arXiv:hep-ph/9308271 [hep-ph]].
%710 citations counted in INSPIRE as of 21 Sep 2023

%\cite{sudhir}
\bibitem{sudhir}
E.~Dudas and S.~K.~Vempati,
``Large D-terms, hierarchical soft spectra and moduli stabilisation,''
Nucl. Phys. B \textbf{727} (2005), 139
%doi:10.1016/j.nuclphysb.2005.08.034
[arXiv:hep-th/0506172 [hep-th]].
%79 citations counted in INSPIRE as of 21 Sep 2023

%\cite{Cremmer:1983bf}
\bibitem{Cremmer:1983bf}
E.~Cremmer, S.~Ferrara, C.~Kounnas and D.~V.~Nanopoulos,
``Naturally Vanishing Cosmological Constant in N=1 Supergravity,''
Phys. Lett. B \textbf{133} (1983), 61.
%doi:10.1016/0370-2693(83)90106-5
%869 citations counted in INSPIRE as of 21 Sep 2023

\bibitem{Ivanov:1978}
E.~A.~Ivanov and A.~A.~Kapustnikov,
``Relation Between Linear and Nonlinear Realizations of Supersymmetry,''
JINR-E2-10765;
E.~A.~Ivanov and A.~A.~Kapustnikov,
``General Relationship Between Linear and Nonlinear Realizations of Supersymmetry,''
J. Phys. A \textbf{11} (1978), 2375;
%doi:10.1088/0305-4470/11/12/005
%260 citations counted in INSPIRE as of 05 Oct 2023
E.~A.~Ivanov and A.~A.~Kapustnikov,
``THE NONLINEAR REALIZATION STRUCTURE OF MODELS WITH SPONTANEOUSLY BROKEN SUPERSYMMETRY,''
J. Phys. G \textbf{8} (1982), 16.
%doi:10.1088/0305-4616/8/2/004
%113 citations counted in INSPIRE as of 11 Nov 2025
For a more recent review, see:
E.~A.~Ivanov,
``Gauge Fields, Nonlinear Realizations, Supersymmetry,''
Phys. Part. Nucl. \textbf{47} (2016) no.4, 508
%doi:10.1134/S1063779616040080
[arXiv:1604.01379 [hep-th]].

\bibitem{Pashnev:1974}
A.~I.~Pashnev,
``Nonlinear realisation for symmetry group with spinor parameters,''
Teor. Mat. Fiz. \textbf{20} (1974), 141.
%doi:10.1007/BF01038765

\bibitem{Clark:1996aw}
T.~E.~Clark and S.~T.~Love,
``Goldstino couplings to matter,''
Phys. Rev. D \textbf{54} (1996), 5723
%doi:10.1103/PhysRevD.54.5723
[arXiv:hep-ph/9608243 [hep-ph]].

\bibitem{Clark:2000rv}
T.~E.~Clark and S.~T.~Love,
``The Akulov-Volkov Lagrangian, symmetry currents and spontaneously broken extended supersymmetry,''
Phys. Rev. D \textbf{63} (2001), 065012
%doi:10.1103/PhysRevD.63.065012
[arXiv:hep-th/0007225 [hep-th]].

\bibitem{Kapustnikov:1981de}
A.~A.~Kapustnikov,
``Non--linear Realization of Einsteinian Supergravity,''
Theor. Math. Phys. \textbf{47} (1981), 406;
%doi:10.1007/BF0108639
E.~A.~Ivanov and A.~A.~Kapustnikov,
``On a Model Independent Description of Spontaneously Broken $N=1$ Supergravity in Superspace,''
Phys. Lett. B \textbf{143} (1984), 379;
%doi:10.1016/0370-2693(84)91486-2
E.~A.~Ivanov and A.~A.~Kapustnikov,
``Geometry of Spontaneously Broken Local $N=1$ Supersymmetry in Superspace,''
Nucl. Phys. B \textbf{333} (1990), 439.
%doi:10.1016/0550-3213(90)90046-G

\bibitem{Samuel:1982uh}
S.~Samuel and J.~Wess,
``A Superfield Formulation of the Nonlinear Realization of Supersymmetry and Its Coupling to Supergravity,''
Nucl. Phys. B \textbf{221} (1983), 153.
% doi:10.1016/0550-3213(83)90622-3

   \bibitem{rocek}
  M.~Rocek,
  ``Linearizing the Volkov-Akulov Model,''
  Phys.\ Rev.\ Lett.\  {\bf 41} (1978) 451;
 % doi:10.1103/PhysRevLett.41.451
  %%CITATION = doi:10.1103/PhysRevLett.41.451;%%
E.~A.~Ivanov and A.~A.~Kapustnikov, in~\cite{Ivanov:1978};
U.~Lindstrom and M.~Rocek,
  ``Constrained Local Superfields,''
  Phys.\ Rev.\ D {\bf 19} (1979), 2300.
  % doi:10.1103/PhysRevD.19.2300
  %%CITATION = doi:10.1103/PhysRevD.19.2300;%%
  %123 citations counted in INSPIRE as of 15 Jun 2017

\bibitem{nonlinear}
R.~Casalbuoni, S.~De Curtis, D.~Dominici, F.~Feruglio and R.~Gatto,
 ``Nonlinear Realization of Supersymmetry Algebra From Supersymmetric Constraint,''
  Phys.\ Lett.\ B {\bf 220} (1989), 569.
%  doi:10.1016/0370-2693(89)90788-0
  %%CITATION = doi:10.1016/0370-2693(89)90788-0;%%

\bibitem{brignole}
  A.~Brignole, F.~Feruglio and F.~Zwirner,
  ``On the effective interactions of a light gravitino with matter fermions,''
  JHEP {\bf 9711} (1997), 001
  %doi:10.1088/1126-6708/1997/11/001
  [hep-th/9709111].
  %%CITATION = doi:10.1088/1126-6708/1997/11/001;%%

\bibitem{ks}
 Z.~Komargodski and N.~Seiberg,
 ``From Linear SUSY to Constrained Superfields,''
  JHEP {\bf 0909} (2009), 066
 %  doi:10.1088/1126-6708/2009/09/066
  [arXiv:0907.2441 [hep-th]].
  %%CITATION = doi:10.1088/1126-6708/2009/09/066;%%

\bibitem{kt}
  S.~M.~Kuzenko and S.~J.~Tyler,
 ``Relating the Komargodski-Seiberg and Akulov-Volkov actions: Exact nonlinear field redefinition,''
  Phys.\ Lett.\ B {\bf 698} (2011), 319
  % doi:10.1016/j.physletb.2011.03.020
  [arXiv:1009.3298 [hep-th]].
  %%CITATION = doi:10.1016/j.physletb.2011.03.020;%%

\bibitem{sorokin1}
I. Bandos, L. Martucci, D.Sorokin and M. Tonin,
``Brane induced supersymmetry breaking and de Sitter supergravity,''
JHEP 02 (2016), 080
%doi:10.1007/JHEP02(2016)080
[arXiv:1511.03024 [hep-th]].

\bibitem{sorokin2}
I. Bandos, M. Heller, S.M. Kuzenko, L. Martucci and D. Sorokin,
``The Goldstino brane, the constrained superfields and matter in N=1 supergravity,''
JHEP 11 (2016), 109
%doi:10.1007/JHEP11(2016)109
[arXiv:1608.05908 [hep-th]].

%\cite{DallAgata:2016syy}
\bibitem{DallAgata:2016syy}
G.~Dall'Agata, E.~Dudas and F.~Farakos,
``On the origin of constrained superfields,''
JHEP \textbf{05} (2016), 041
%doi:10.1007/JHEP05(2016)041
[arXiv:1603.03416 [hep-th]].
%75 citations counted in INSPIRE as of 12 Oct 2023

%\cite{DallAgata:2015zxp}
\bibitem{DallAgata:2015zxp}
G.~Dall'Agata and F.~Farakos,
``Constrained superfields in Supergravity,''
JHEP \textbf{02} (2016), 101
%doi:10.1007/JHEP02(2016)101
[arXiv:1512.02158 [hep-th]].
%55 citations counted in INSPIRE as of 19 Oct 2023


%\cite{Antoniadis:2010hs}
\bibitem{Antoniadis:2010hs}
I.~Antoniadis, E.~Dudas, D.~M.~Ghilencea and P.~Tziveloglou,
``Non-linear MSSM,''
Nucl. Phys. B \textbf{841} (2010), 157
%doi:10.1016/j.nuclphysb.2010.08.002
[arXiv:1006.1662 [hep-ph]].

\bibitem{Antoniadis:2012hs}
%71 citations counted in INSPIRE as of 05 Oct 2023
I.~Antoniadis, E.~Dudas, D.~M.~Ghilencea and P.~Tziveloglou,
``Nonlinear supersymmetry and goldstino couplings to the MSSM,''
Theor. Math. Phys. \textbf{170} (2012), 26.
%doi:10.1007/s11232-012-0004-y
%10 citations counted in INSPIRE as of 05 Oct 2023

%\cite{Bellazzini:2017neg}
\bibitem{Bellazzini:2017neg}
B.~Bellazzini, A.~Mariotti, D.~Redigolo, F.~Sala and J.~Serra,
``R-axion at colliders,''
Phys. Rev. Lett. \textbf{119} (2017) no.14, 141804
%doi:10.1103/PhysRevLett.119.141804
[arXiv:1702.02152 [hep-ph]].
%57 citations counted in INSPIRE as of 12 Oct 2023

\bibitem{adfs}
I.~Antoniadis, E.~Dudas, S.~Ferrara and A.~Sagnotti,
``The Volkov\textendash{}Akulov\textendash{}Starobinsky supergravity,''
Phys. Lett. B \textbf{733} (2014), 32
%doi:10.1016/j.physletb.2014.04.015
[arXiv:1403.3269 [hep-th]].

  \bibitem{fkl1}
 S.~Ferrara, R.~Kallosh and A.~Linde,
  ``Cosmology with Nilpotent Superfields,''
  JHEP {\bf 1410} (2014) 143
  doi:10.1007/JHEP10(2014)143
  [arXiv:1408.4096 [hep-th]].
  %%CITATION = doi:10.1007/JHEP10(2014)143;%%

    \bibitem{fkl2}
  R.~Kallosh and A.~Linde,
  ``Inflation and Uplifting with Nilpotent Superfields,''
  JCAP {\bf 1501} (2015) 025
  %doi:10.1088/1475-7516/2015/01/025
  [arXiv:1408.5950 [hep-th]].
  %%CITATION = doi:10.1088/1475-7516/2015/01/025;%%

    \bibitem{fkl3}
  G.~Dall'Agata and F.~Zwirner,
  ``On sgoldstino-less supergravity models of inflation,''
  JHEP {\bf 1412} (2014) 172
  %doi:10.1007/JHEP12(2014)172
  [arXiv:1411.2605 [hep-th]].
  %%CITATION = doi:10.1007/JHEP12(2014)172;%%

\bibitem{starobinsky}
 A.~A.~Starobinsky,
  ``A New Type of Isotropic Cosmological Models Without Singularity,''
  Phys.\ Lett.\  {\bf 91B} (1980) 99.
 % doi:10.1016/0370-2693(80)90670-X
  %%CITATION = doi:10.1016/0370-2693(80)90670-X;%%


 \bibitem{follow_ups1}
 L.~Alvarez-Gaume, A.~Kehagias, C.~Kounnas, D.~Lust and A.~Riotto,
  ``Aspects of Quadratic Gravity,''
  Fortsch.\ Phys.\  {\bf 64} (2016) no.2-3,  176
  %doi:10.1002/prop.201500100
  %%CITATION = doi:10.1002/prop.201500100;%%
 [arXiv:1505.07657 [hep-th]].

  \bibitem{follow_ups2}
  S.~Ferrara, A.~Kehagias and M.~Porrati,
  ``${\cal R}^2$ Supergravity,''
  JHEP {\bf 1508} (2015) 001
  %doi:10.1007/JHEP08(2015)001
  [arXiv:1506.01566 [hep-th]].
  %%CITATION = doi:10.1007/JHEP08(2015)001;%%

   \bibitem{follow_ups3}
E.~Dudas, S.~Ferrara, A.~Kehagias and A.~Sagnotti,
  ``Properties of Nilpotent Supergravity,''
  JHEP {\bf 1509} (2015) 217
  %doi:10.1007/JHEP09(2015)217
  [arXiv:1507.07842 [hep-th]].
  %%CITATION = doi:10.1007/JHEP09(2015)217;%%

   \bibitem{follow_ups4}
     F.~Hasegawa and Y.~Yamada,
  ``Component action of nilpotent multiplet coupled to matter in 4 dimensional $ \mathcal{N}=1 $ supergravity,''
  JHEP {\bf 1510} (2015) 106
  %doi:10.1007/JHEP10(2015)106
  [arXiv:1507.08619 [hep-th]].

   \bibitem{follow_ups5}
E.~A.~Bergshoeff, D.~Z.~Freedman, R.~Kallosh and A.~Van Proeyen,
  ``Pure de Sitter Supergravity,''
  Phys.\ Rev.\ D {\bf 92} (2015) no.8,  085040
   Erratum: [Phys.\ Rev.\ D {\bf 93} (2016) no.6,  069901]
  %doi:10.1103/PhysRevD.93.069901, 10.1103/PhysRevD.92.085040
  [arXiv:1507.08264 [hep-th]].
  %%CITATION = doi:10.1103/PhysRevD.93.069901, 10.1103/PhysRevD.92.085040;%%

   \bibitem{follow_ups6}
  S.~Ferrara, M.~Porrati and A.~Sagnotti,
  ``Scale invariant Volkov–Akulov supergravity,''
  Phys.\ Lett.\ B {\bf 749} (2015) 589
  %doi:10.1016/j.physletb.2015.08.066
  [arXiv:1508.02939 [hep-th]].
  %%CITATION = doi:10.1016/j.physletb.2015.08.066;%%

   \bibitem{follow_ups7}
I.~Antoniadis and C.~Markou,
  ``The coupling of Non-linear Supersymmetry to Supergravity,''
  Eur.\ Phys.\ J.\ C {\bf 75} (2015) no.12,  582
 % doi:10.1140/epjc/s10052-015-3783-0
  [arXiv:1508.06767 [hep-th]].
  %%CITATION = doi:10.1140/epjc/s10052-015-3783-0;%%

   \bibitem{follow_ups8}
  N.~Cribiori, G.~Dall'Agata, F.~Farakos and M.~Porrati,
  ``Minimal Constrained Supergravity,''
  Phys.\ Lett.\ B {\bf 764} (2017) 228
  %doi:10.1016/j.physletb.2016.11.040
  [arXiv:1611.01490 [hep-th]].
  %%CITATION = doi:10.1016/j.physletb.2016.11.040;%%

  %\cite{Bonnefoy:2022rcw}
\bibitem{Bonnefoy:2022rcw}
Q.~Bonnefoy, G.~Casagrande and E.~Dudas,
``Causality constraints on nonlinear supersymmetry,''
JHEP \textbf{11} (2022), 113
%doi:10.1007/JHEP11(2022)113
[arXiv:2206.13451 [hep-th]].
%0 citations counted in INSPIRE as of 19 Oct 2023

\bibitem{casagrande}|
G.~Casagrande,
``Linear and nonlinear supersymmetry in field and string theory,''
Ph.D. Thesis, [arXiv:2506.20396 [hep-th]].
%0 citations counted in INSPIRE as of 26 Sep 2025

\bibitem{nonlinear_reviews}
A.~Sagnotti and S.~Ferrara,
  ``Supersymmetry and Inflation,''
  PoS PLANCK {\bf 2015} (2015) 113
  [arXiv:1509.01500 [hep-th]];
  %%CITATION = ARXIV:1509.01500;%%
  S.~Ferrara, A.~Kehagias and A.~Sagnotti,
  ``Cosmology and Supergravity,''
  Int.\ J.\ Mod.\ Phys.\ A {\bf 31} (2016) no.25,  1630044
  %doi:10.1142/S0217751X16300441
  [arXiv:1605.04791 [hep-th]];
  %%CITATION = doi:10.1142/S0217751X16300441;%%
  I.~Antoniadis, E.~Dudas, F.~Farakos and A.~Sagnotti,
``Non-Linear Supergravity and Inflationary Cosmology,''
[arXiv:2409.14943 [hep-th]].
%6 citations counted in INSPIRE as of 11 Jun 2025

\bibitem{apt}
I.~Antoniadis, H.~Partouche and T.~R.~Taylor,
  ``Spontaneous breaking of N=2 global supersymmetry,''
  Phys.\ Lett.\ B {\bf 372} (1996), 83
  %doi:10.1016/0370-2693(96)00028-7
  [hep-th/9512006];
  %%CITATION = doi:10.1016/0370-2693(96)00028-7;%%
E.~A.~Ivanov and B.~M.~Zupnik,
``Modified N=2 supersymmetry and Fayet-Iliopoulos terms,''
Phys. Atom. Nucl. \textbf{62} (1999), 1043
[arXiv:hep-th/9710236 [hep-th]].
%63 citations counted in INSPIRE as of 11 Nov 2025

\bibitem{bg}
J.~Bagger and A.~Galperin,
  ``A New Goldstone multiplet for partially broken supersymmetry,''
  Phys.\ Rev.\ D {\bf 55} (1997) 1091
  %doi:10.1103/PhysRevD.55.1091
  [hep-th/9608177].
  %%CITATION = doi:10.1103/PhysRevD.55.1091;%%


  \bibitem{hughes_polchinski1}
J.~Hughes and J.~Polchinski,
  ``Partially Broken Global Supersymmetry and the Superstring,''
  Nucl.\ Phys.\ B {\bf 278} (1986) 147.
 % doi:10.1016/0550-3213(86)90111-2
  %%CITATION = doi:10.1016/0550-3213(86)90111-2;%%

   \bibitem{hughes_polchinski2}
J.~Hughes, J.~Liu and J.~Polchinski,
  ``Supermembranes,''
  Phys.\ Lett.\ B {\bf 180} (1986), 370.
  %doi:10.1016/0370-2693(86)91204-9
  %%CITATION = doi:10.1016/0370-2693(86)91204-9;%%

\bibitem{cgfp1}
S.~Cecotti, L.~Girardello and M.~Porrati,
  ``An Exceptional $N=2$ Supergravity With Flat Potential and Partial Superhiggs,''
  Phys.\ Lett.\  {\bf 168B} (1986), 83.
  %doi:10.1016/0370-2693(86)91465-6
  %%CITATION = doi:10.1016/0370-2693(86)91465-6;%%

  \bibitem{cgfp2}
  S.~Ferrara, L.~Girardello and M.~Porrati,
  ``Minimal Higgs branch for the breaking of half of the supersymmetries in N=2 supergravity,''
  Phys.\ Lett.\ B {\bf 366} (1996), 155
  %doi:10.1016/0370-2693(95)01378-4
  [hep-th/9510074].
  %%CITATION = doi:10.1016/0370-2693(95)01378-4;%%

  \bibitem{cgfp3}
S.~Ferrara, L.~Girardello and M.~Porrati,
  ``Spontaneous breaking of N=2 to N=1 in rigid and local supersymmetric theories,''
  Phys.\ Lett.\ B {\bf 376} (1996), 275
  %doi:10.1016/0370-2693(96)00229-8
  [hep-th/9512180].
  %%CITATION = doi:10.1016/0370-2693(96)00229-8;%%

  \bibitem{cecottiferrara}
  S.~Cecotti and S.~Ferrara,
``Supersymmetric Born-Infeld Lagrangians,''
Phys. Lett. B \textbf{187} (1987), 335.
%doi:10.1016/0370-2693(87)91105-1
%170 citations counted in INSPIRE as of 11 Jun 2025

  \bibitem{fps1}
S.~Ferrara, M.~Porrati and A.~Sagnotti,
  ``N = 2 Born-Infeld attractors,''
  JHEP {\bf 1412} (2014), 065
  %doi:10.1007/JHEP12(2014)065
  [arXiv:1411.4954 [hep-th]].
  %%CITATION = doi:10.1007/JHEP12(2014)065;%%

    \bibitem{fps2}
  S.~Ferrara, M.~Porrati, A.~Sagnotti, R.~Stora and A.~Yeranyan,
  ``Generalized Born--Infeld Actions and Projective Cubic Curves,''
  Fortsch.\ Phys.\  {\bf 63} (2015), 189
  %doi:10.1002/prop.201400087
  [arXiv:1412.3337 [hep-th]].
  %%CITATION = doi:10.1002/prop.201400087;%%

\bibitem{tseytlin_review}
A.~A.~Tseytlin,
  ``Born-Infeld action, supersymmetry and string theory,''
  In *Shifman, M.A. (ed.): The many faces of the superworld* 417-452
  %doi:10.1142/9789812793850_0025
  [hep-th/9908105].
  %%CITATION = doi:10.1142/9789812793850_0025;%%

\bibitem{dfs}
E.~Dudas, S.~Ferrara and A.~Sagnotti,
  ``A superfield constraint for $ \mathcal{N} $ = 2 → $ \mathcal{N} $ = 0 breaking,''
  JHEP {\bf 1708} (2017) 109
%  doi:10.1007/JHEP08(2017)109
  [arXiv:1707.03414 [hep-th]].
  %%CITATION = doi:10.1007/JHEP08(2017)109;%%

  \bibitem{recent201}
N.~Cribiori, G.~Dall'Agata and F.~Farakos,
  ``Interactions of N Goldstini in Superspace,''
  Phys.\ Rev.\ D {\bf 94} (2016) no.6,  065019
 % doi:10.1103/PhysRevD.94.065019
  [arXiv:1607.01277 [hep-th]].
  %%CITATION = doi:10.1103/PhysRevD.94.065019;%%

    \bibitem{recent202}
   S.~M.~Kuzenko and G.~Tartaglino-Mazzucchelli,
  ``New nilpotent ${\cal N}= 2$ superfields,''
  arXiv:1707.07390 [hep-th].
  %%CITATION = ARXIV:1707.07390;%%
  %1 citations counted in INSPIRE as of 26 Nov 2017

%\cite{witten-partial}
\bibitem{witten-partial}
E.~Witten,
``Dynamical Breaking of Supersymmetry,''
Nucl. Phys. B \textbf{188} (1981), 513.
%doi:10.1016/0550-3213(81)90006-7
%3517 citations counted in INSPIRE as of 07 Aug 2024

\bibitem{I}
A.~H.~Chamseddine,
``Interacting Supergravity in Ten-Dimensions: The Role of the Six-Index Gauge Field,''
Phys. Rev. D \textbf{24} (1981), 3065;
%doi:10.1103/PhysRevD.24.3065
E.~Bergshoeff, M.~de Roo, B.~de Wit and P.~van Nieuwenhuizen,
``Ten-Dimensional Maxwell-Einstein Supergravity, Its Currents, and the Issue of Its Auxiliary Fields,''
Nucl. Phys. B \textbf{195} (1982), 97;
%doi:10.1016/0550-3213(82)90050-5
%592 citations counted in INSPIRE as of 20 Jan 2025
G.~F.~Chapline and N.~S.~Manton,
``Unification of Yang-Mills Theory and Supergravity in Ten-Dimensions,''
Phys. Lett. B \textbf{120} (1983), 105.
%doi:10.1016/0370-2693(83)90633-0

\bibitem{BSS}
L.~Brink, J.~H.~Schwarz and J.~Scherk,
``Supersymmetric Yang-Mills Theories,''
Nucl. Phys. B \textbf{121} (1977), 77.
%doi:10.1016/0550-3213(77)90328-5

\bibitem{fayetn4}
P.~Fayet,
``Spontaneous Generation of Massive Multiplets and Central Charges in Extended Supersymmetric Theories,''
Nucl. Phys. B \textbf{149} (1979), 137.
%doi:10.1016/0550-3213(79)90162-7
%193 citations counted in INSPIRE as of 19 Nov 2025

\bibitem{nahm}
W.~Nahm,
``Supersymmetries and Their Representations,''
Nucl. Phys. B \textbf{135} (1978), 149.
%doi:10.1201/9781482268737-2
%918 citations counted in INSPIRE as of 20 Jan 2025

\bibitem{add}
N.~Arkani-Hamed, S.~Dimopoulos and G.~R.~Dvali,
``The Hierarchy problem and new dimensions at a millimeter,''
Phys. Lett. B \textbf{429} (1998), 263
%doi:10.1016/S0370-2693(98)00466-3
[arXiv:hep-ph/9803315 [hep-ph]];  
``Phenomenology, astrophysics and cosmology of theories with submillimeter dimensions and TeV scale quantum gravity,''
Phys. Rev. D \textbf{59} (1999), 086004
%doi:10.1103/PhysRevD.59.086004
[arXiv:hep-ph/9807344 [hep-ph]].

\bibitem{dark}
M.~Montero, C.~Vafa and I.~Valenzuela,
``The dark dimension and the Swampland,''
JHEP \textbf{02} (2023), 022
%doi:10.1007/JHEP02(2023)022
[arXiv:2205.12293 [hep-th]].

\bibitem{DJG}
D.~J.~Gross,
``Oscar Klein and gauge theory,''
[arXiv:hep-th/9411233 [hep-th]].

\bibitem{agwitt}
L.~Alvarez-Gaume and E.~Witten,
``Gravitational Anomalies,''
Nucl. Phys. B \textbf{234} (1984), 269.
%doi:10.1016/0550-3213(84)90066-X
  
\bibitem{cp1}
J.~E.~Paton and H.~M.~Chan,
  ``Generalized Veneziano model with isospin,''
  Nucl.\ Phys.\ B {\bf 10} (1969) 516.
 % doi:10.1016/0550-3213(69)90038-8
  %%CITATION = doi:10.1016/0550-3213(69)90038-8;%%

  \bibitem{cp2}
  J.~H.~Schwarz,
  ``Superstring Theory,''
  Phys.\ Rept.\  {\bf 89} (1982) 223.
  %doi:10.1016/0370-1573(82)90087-4
  %%CITATION = doi:10.1016/0370-1573(82)90087-4;%%

  \bibitem{cp3}
  J.~H.~Schwarz,
``Gauge Groups for Type I Superstrings,''
CALT-68-906-REV.

\bibitem{cp4}
N.~Marcus and A.~Sagnotti,
  ``Tree Level Constraints on Gauge Groups for Type I Superstrings,''
  Phys.\ Lett.\  {\bf 119B} (1982) 97.
 % doi:10.1016/0370-2693(82)90253-2
  %%CITATION = doi:10.1016/0370-2693(82)90253-2;%%

  \bibitem{cp5}
  N.~Marcus and A.~Sagnotti,
``Group Theory from Quarks at the Ends of Strings,''
  Phys.\ Lett.\ B {\bf 188} (1987) 58.
  %doi:10.1016/0370-2693(87)90705-2
  %%CITATION = doi:10.1016/0370-2693(87)90705-2;%%

\bibitem{karapet}
O.~Evnin and K.~Mkrtchyan,
``Three approaches to chiral form interactions,''
Differ. Geom. Appl. \textbf{89} (2023), 102016
[arXiv:2207.01767 [hep-th]].
  
\bibitem{hull-townsend1}
C.~M.~Hull and P.~K.~Townsend,
``Unity of superstring dualities,''
Nucl. Phys. B \textbf{438} (1995), 109
%doi:10.1016/0550-3213(94)00559-W
[arXiv:hep-th/9410167 [hep-th]].
%1759 citations counted in INSPIRE as of 25 Nov 2023

\bibitem{hull-townsend2}
P.~K.~Townsend,
``D-branes from M-branes,''
Phys. Lett. B \textbf{373} (1996), 68
%doi:10.1016/0370-2693(96)00104-9
[arXiv:hep-th/9512062 [hep-th]].


\bibitem{bergshoeff}
E.~Bergshoeff, R.~Kallosh, T.~Ortin, D.~Roest and A.~Van Proeyen,
  ``New formulations of D = 10 supersymmetry and D8 - O8 domain walls,''
  Class.\ Quant.\ Grav.\  {\bf 18} (2001), 3359
  %doi:10.1088/0264-9381/18/17/303
  [hep-th/0103233].
  
  \bibitem{romansm}
  L.~J.~Romans,
``Massive N=2a Supergravity in Ten-Dimensions,''
Phys. Lett. B \textbf{169} (1986), 374.
%doi:10.1016/0370-2693(86)90375-8
%616 citations counted in INSPIRE as of 05 Aug 2025

\bibitem{Veneziano68}
G.~Veneziano,
``Construction of a crossing - symmetric, Regge behaved amplitude for linearly rising trajectories,''
Nuovo Cim. A \textbf{57} (1968), 190.
%doi:10.1007/BF02824451
%1662 citations counted in INSPIRE as of 20 Oct 2025

\bibitem{Virasoro}
M.~A.~Virasoro,
``Alternative constructions of crossing-symmetric amplitudes with regge behavior,''
Phys. Rev. \textbf{177} (1969), 2309.
%doi:10.1103/PhysRev.177.2309
%311 citations counted in INSPIRE as of 20 Oct 2025

\bibitem{Shapiro}
J.~A.~Shapiro,
``Electrostatic analog for the virasoro model,''
Phys. Lett. B \textbf{33} (1970), 361.
%doi:10.1016/0370-2693(70)90255-8
%181 citations counted in INSPIRE as of 20 Oct 2025

\bibitem{chrysoula}
C.~Markou and E.~Skvortsov,
``An excursion into the string spectrum,''
JHEP \textbf{12} (2023), 055
%doi:10.1007/JHEP12(2023)055
[arXiv:2309.15988 [hep-th]];
T.~Basile and C.~Markou,
``On the deep superstring spectrum,''
JHEP \textbf{07} (2024), 184
%doi:10.1007/JHEP07(2024)184
[arXiv:2405.18467 [hep-th]]. For a review, see: C.~Markou,
``On the deep string spectrum,''
[arXiv:2505.00177 [hep-th]].

\bibitem{bucciotti}
B.~Bucciotti, F.~Figueroa and G.~L.~Pimentel,
``Unraveling the Spectrum of the Open String,''
[arXiv:2511.07524 [hep-th]].
%0 citations counted in INSPIRE as of 24 Nov 2025

\bibitem{BDVH1}
L.~Brink, P.~Di Vecchia and P.~S.~Howe,
``A Locally Supersymmetric and Reparametrization Invariant Action for the Spinning String,''
Phys. Lett. B \textbf{65} (1976), 471.
%doi:10.1016/0370-2693(76)90445-7

\bibitem{BDVH2}
S.~Deser and B.~Zumino,
``A Complete Action for the Spinning String,''
Phys. Lett. B \textbf{65} (1976), 369.
%doi:10.1016/0370-2693(76)90245-8

\bibitem{shapmod}J.~A.~Shapiro,
``Loop graph in the dual tube model'',
Phys.\ Rev.\ D5 (1972) 1945.
%%CITATION = PHRVA,D5,1945;%%


\bibitem{jacobi} See, for instance:
E.T. Whittaker and G.N. Watson, ``A course of modern analysis,''
 Cambridge Univ. Press (1927);
{\it Higher Transcendental Functions}, 3 vols, A. Erde'lyi ed., McGraw-Hill (1953);
J.D. Fay, ``Theta functions on Riemann surfaces'',
Lecture notes in Mathematics 352, Springer-Verlag (1973); D. Mumford,
``Tata lectures on theta'', Birkhauser (1983).

  \bibitem{verlinde}
E.~P.~Verlinde,
  ``Fusion Rules and Modular Transformations in 2D Conformal Field Theory,''
  Nucl.\ Phys.\ B {\bf 300} (1988) 360.
  %doi:10.1016/0550-3213(88)90603-7
  %%CITATION = doi:10.1016/0550-3213(88)90603-7;%%

  \bibitem{schellekens_spacetime}
  W.~Lerche, A.~N.~Schellekens and N.~P.~Warner,
``Ghost Triality and Superstring Partition Functions,''
Phys. Lett. B \textbf{214} (1988), 41.
%doi:10.1016/0370-2693(88)90448-0
%18 citations counted in INSPIRE as of 17 May 2024

  \bibitem{schellekens_spacetime_rev}
  W.~Lerche, A.~N.~Schellekens and N.~P.~Warner,
``Lattices and Strings,''
Phys. Rept. \textbf{177} (1989), 1.
%doi:10.1016/0370-1573(89)90077-X
%209 citations counted in INSPIRE as of 17 May 2024

%\cite{Dixon:1985jw}
\bibitem{orbifolds1}
L.~J.~Dixon, J.~A.~Harvey, C.~Vafa and E.~Witten,
``Strings on Orbifolds,''
Nucl. Phys. B \textbf{261} (1985), 678.
%doi:10.1016/0550-3213(85)90593-0
%1806 citations counted in INSPIRE as of 05 Jan 2024

\bibitem{orbifolds2}
L.~J.~Dixon, J.~A.~Harvey, C.~Vafa and E.~Witten,
``Strings on Orbifolds. 2.,''
Nucl. Phys. B \textbf{274} (1986), 285.
%doi:10.1016/0550-3213(86)90287-7

  \bibitem{clasha}
 L.~Clavelli and J.~A.~Shapiro,
``Pomeron factorization in general dual models,''
Nucl. Phys. B \textbf{57} (1973), 490.
%doi:10.1016/0550-3213(73)90113-2


\bibitem{cremm_gerv}
E.~Cremmer and J.~L.~Gervais,
``Combining and Splitting Relativistic Strings,''
Nucl. Phys. B \textbf{76} (1974), 209.
%doi:10.1016/0550-3213(74)90383-6

\bibitem{ext_superstring}
C.~G.~Callan, Jr., C.~Lovelace, C.~R.~Nappi and S.~A.~Yost,
``Adding Holes and Crosscaps to the Superstring,''
Nucl. Phys. B \textbf{293} (1987), 83.
%doi:10.1016/0550-3213(87)90065-4


\bibitem{pc1} J.~Polchinski and Y.~Cai,
``Consistency of open superstring theories'',
Nucl.\ Phys.\ B296 (1988), 91.
%%CITATION = NUPHA,B296,91;%%


%\cite{Polchinski:1995df}
\bibitem{pw}
J.~Polchinski and E.~Witten,
``Evidence for heterotic - type I string duality,''
Nucl. Phys. B \textbf{460} (1996), 525
%doi:10.1016/0550-3213(95)00614-1
[arXiv:hep-th/9510169 [hep-th]].
%732 citations counted in INSPIRE as of 16 Nov 2023

\bibitem{HW1}
P.~Horava and E.~Witten,
``Heterotic and type I string dynamics from eleven-dimensions,''
Nucl. Phys. B \textbf{460} (1996), 506
%doi:10.1016/0550-3213(95)00621-4
[arXiv:hep-th/9510209 [hep-th]].
%2477 citations counted in INSPIRE as of 01 Jun 2021

\bibitem{HW2}
P.~Horava and E.~Witten,
``Eleven-dimensional supergravity on a manifold with boundary,''
Nucl. Phys. B \textbf{475} (1996), 94
%doi:10.1016/0550-3213(96)00308-2
[arXiv:hep-th/9603142 [hep-th]].
%1834 citations counted in INSPIRE as of 01 Jun 2021

  \bibitem{dp} A.~Dabholkar and J.~Park,
``A note on orientifolds and F-theory'',
Phys.\ Lett.\ B394 (1997), 302
[arXiv:hep-th/9607041].
%%CITATION = HEP-TH 9607041;%%

\bibitem{gepner}
C.~Angelantonj, M.~Bianchi, G.~Pradisi, A.~Sagnotti and Y.~S.~Stanev,
``Comments on Gepner models and type I vacua in string theory'',
Phys.\ Lett.\ B387 (1996), 743
[arXiv:hep-th/9607229].
%%CITATION = HEP-TH 9607229;%%

\bibitem{BST}
E.~Bergshoeff, E.~Sezgin and P.~K.~Townsend,
``Supermembranes and Eleven-Dimensional Supergravity,''
Phys. Lett. B \textbf{189} (1987), 75.
%doi:10.1016/0370-2693(87)91272-X
%832 citations counted in INSPIRE as of 01 Jun 2021


\bibitem{wittenOplus}
E.~Witten,
  ``Toroidal compactification without vector structure,''
  JHEP {\bf 9802} (1998), 006
 % doi:10.1088/1126-6708/1998/02/006
  [hep-th/9712028].
  %%CITATION = doi:10.1088/1126-6708/1998/02/006;%%

\bibitem{sen1}
A.~Sen,
``Stable non-BPS states in string theory'',
JHEP 9806 (1998), 007
[arXiv:hep-th/9803194].
%%CITATION = HEP-TH 9803194;%%

\bibitem{sen2}
A.~Sen,
``Stable non-BPS bound states of BPS D-branes'',
9808 (1998), 010
[arXiv:hep-th/9805019].
%%CITATION = HEP-TH 9805019;%%

\bibitem{sen3}
A.~Sen,
``Tachyon condensation on the brane antibrane system'',
9808 (1998), 012
[arXiv:hep-th/9805170].
%%CITATION = HEP-TH 9805170;%%

\bibitem{sen4}
O.~Bergman and M.~R.~Gaberdiel,
``Stable non-BPS D-particles'',
Phys.\ Lett.\ B441 (1998), 133
[arXiv:hep-th/9806155].
%%CITATION = HEP-TH 9806155;%%

\bibitem{sen5}
A. Sen,
``SO(32) spinors of type I and other solitons on brane-antibrane pairs'',
JHEP 9809 (1998), 023
[arXiv:hep-th/9808141].
%%CITATION = HEP-TH 9808141;%%

\bibitem{sen6}
A.~Sen,
``Type I D-particle and its interactions'',
JHEP 9810 (1998), 021
[arXiv:hep-th/9809111].
%%CITATION = HEP-TH 9809111;%%

\bibitem{sen7}
A.~Sen,
``BPS D-branes on non-supersymmetric cycles'',
JHEP 9812 (1998), 021
[arXiv:hep-th/9812031].
%%CITATION = HEP-TH 9812031;%%

\bibitem{dms}
E.~Dudas, J.~Mourad and A.~Sagnotti,
``Charged and uncharged D-branes in various string theories,''
Nucl. Phys. B \textbf{620} (2002), 109
%doi:10.1016/S0550-3213(01)00552-1
[arXiv:hep-th/0107081 [hep-th]].

\bibitem{fayet-large}
P.~Fayet,
``Supersymmetric Grand Unification in a Six-dimensional Space-time,''
Phys. Lett. B \textbf{159} (1985), 121; 
%doi:10.1016/0370-2693(85)90869-X
``Six-dimensional Supersymmetric {QED}, R Invariance and $N=2$ Supersymmetry Breaking by Dimensional Reduction,''
Nucl. Phys. B \textbf{263} (1986), 649; 
%doi:10.1016/0550-3213(86)90280-4; 
I.~Antoniadis,
``A Possible new dimension at a few TeV,''
Phys. Lett. B \textbf{246} (1990), 377. 
%doi:10.1016/0370-2693(90)90617-F

\bibitem{aadd}
I.~Antoniadis, N.~Arkani-Hamed, S.~Dimopoulos and G.~R.~Dvali,
``New dimensions at a millimeter to a Fermi and superstrings at a TeV,''
Phys. Lett. B \textbf{436} (1998), 257
%doi:10.1016/S0370-2693(98)00860-0
[arXiv:hep-ph/9804398 [hep-ph]].

  \bibitem{sw0}
  N.~Seiberg and E.~Witten,
``Spin Structures in String Theory,''
Nucl. Phys. B \textbf{276} (1986), 272.
%doi:10.1016/0550-3213(86)90297-X
%450 citations counted in INSPIRE as of 16 Apr 2024


\bibitem{opentachyon1}
A.~Sen and B.~Zwiebach,
``Tachyon condensation in string field theory,''
JHEP \textbf{03} (2000), 002
%doi:10.1088/1126-6708/2000/03/002
[arXiv:hep-th/9912249 [hep-th]].
%393 citations counted in INSPIRE as of 02 Jun 2021

\bibitem{opentachyon2}
N.~Berkovits, A.~Sen and B.~Zwiebach,
``Tachyon condensation in superstring field theory,''
Nucl. Phys. B \textbf{587} (2000), 147
%doi:10.1016/S0550-3213(00)00501-0
[arXiv:hep-th/0002211 [hep-th]].
%239 citations counted in INSPIRE as of 02 Jun 2021


  \bibitem{erice}
  A.~Sagnotti,
``Anomaly cancellations and open string theories,''
[arXiv:hep-th/9302099 [hep-th]].

  \bibitem{armoni1}
A.~Armoni and B.~Kol,
  ``Nonsupersymmetric large N gauge theories from type 0 brane configurations,''
  JHEP {\bf 9907} (1999), 011
  %doi:10.1088/1126-6708/1999/07/011
  [hep-th/9906081].
  %%CITATION = doi:10.1088/1126-6708/1999/07/011;%%

    \bibitem{armoni2}
 C.~Angelantonj and A.~Armoni,
  ``Nontachyonic type 0B orientifolds, nonsupersymmetric gauge theories and cosmological RG flow,''
  Nucl.\ Phys.\ B {\bf 578} (2000), 239
  %doi:10.1016/S0550-3213(00)00136-X
  [hep-th/9912257].
  %%CITATION = doi:10.1016/S0550-3213(00)00136-X;%%

    \bibitem{armoni3}
 A.~Armoni, M.~Shifman and G.~Veneziano,
  ``Exact results in non-supersymmetric large N orientifold field theories,''
  Nucl.\ Phys.\ B {\bf 667} (2003), 170
  %doi:10.1016/S0550-3213(03)00538-8
  [hep-th/0302163].
  %%CITATION = doi:10.1016/S0550-3213(03)00538-8;%%

    \bibitem{armoni4}
   A.~Armoni, M.~Shifman and G.~Veneziano,
``SUSY relics in one flavor QCD from a new 1/N expansion,''
  Phys.\ Rev.\ Lett.\  {\bf 91} (2003), 191601
 % doi:10.1103/PhysRevLett.91.191601
  [hep-th/0307097].
  %%CITATION = doi:10.1103/PhysRevLett.91.191601;%%

    \bibitem{armoni5}
   A.~Armoni, M.~Shifman and G.~Veneziano,
  ``From superYang-Mills theory to QCD: Planar equivalence and its implications,''
  In *Shifman, M. (ed.) et al.: From fields to strings, vol. 1* 353
  %$doi:10.1142/9789812775344_0013$
  [hep-th/0403071].
  %%CITATION = doi:10.1142/9789812775344_0013;%%

    \bibitem{armoni6}
A.~Armoni and V.~Niarchos,
  ``Phases of QCD$_3$ from Non-SUSY Seiberg Duality and Brane Dynamics,''
  arXiv:1711.04832 [hep-th].
  %%CITATION = ARXIV:1711.04832;%%


\bibitem{pss1}
G.~Pradisi, A.~Sagnotti and Y.~S.~Stanev,
  ``Planar duality in SU(2) WZW models,''
  Phys.\ Lett.\ B {\bf 354} (1995), 279
  %doi:10.1016/0370-2693(95)00532-P
  [hep-th/9503207].
  %%CITATION = doi:10.1016/0370-2693(95)00532-P;%%

  \bibitem{pss2}
  G.~Pradisi, A.~Sagnotti and Y.~S.~Stanev,
  ``The Open descendants of nondiagonal SU(2) WZW models,''
  Phys.\ Lett.\ B {\bf 356} (1995), 230
  %doi:10.1016/0370-2693(95)00840-H
  [hep-th/9506014].
  %%CITATION = doi:10.1016/0370-2693(95)00840-H;%%

  \bibitem{pss3}
   G.~Pradisi, A.~Sagnotti and Y.~S.~Stanev,
     ``Completeness conditions for boundary operators in 2-D conformal field theory,''
  Phys.\ Lett.\ B {\bf 381} (1996), 97
  %doi:10.1016/0370-2693(96)00578-3
  [hep-th/9603097].
  %%CITATION = doi:10.1016/0370-2693(96)00578-3;%%

\bibitem{crosscap_constraint}
D.~Fioravanti, G.~Pradisi and A.~Sagnotti,
  ``Sewing constraints and nonorientable open strings,''
  Phys.\ Lett.\ B {\bf 321} (1994), 349
  %doi:10.1016/0370-2693(94)90255-0
  [hep-th/9311183].
  %%CITATION = doi:10.1016/0370-2693(94)90255-0;%%

\bibitem{lewellen}
D.~C.~Lewellen,
  ``Sewing constraints for conformal field theories on surfaces with boundaries,''
  Nucl.\ Phys.\ B {\bf 372} (1992), 654.
 % doi:10.1016/0550-3213(92)90370-Q
  %%CITATION = doi:10.1016/0550-3213(92)90370-Q;%%

\bibitem{schellekens1}
L.~R.~Huiszoon, A.~N.~Schellekens and N.~Sousa,
  ``Klein bottles and simple currents,''
  Phys.\ Lett.\ B {\bf 470} (1999), 95
  %doi:10.1016/S0370-2693(99)01241-1
  [hep-th/9909114].
  %%CITATION = doi:10.1016/S0370-2693(99)01241-1;%%

  \bibitem{schellekens2}
J.~Fuchs, L.~R.~Huiszoon, A.~N.~Schellekens, C.~Schweigert and J.~Walcher,
  ``Boundaries, crosscaps and simple currents,''
  Phys.\ Lett.\ B {\bf 495} (2000), 427
 % doi:10.1016/S0370-2693(00)01271-5
  [hep-th/0007174].
  %%CITATION = doi:10.1016/S0370-2693(00)01271-5;%%

\bibitem{WZW}
D.~Gepner and E.~Witten,
  ``String Theory on Group Manifolds,''
  Nucl.\ Phys.\ B {\bf 278} (1986), 493.
  %doi:10.1016/0550-3213(86)90051-9
  %%CITATION = doi:10.1016/0550-3213(86)90051-9;%%
  
\bibitem{branecft1}
E.~Dudas and J.~Mourad,
  ``D-branes in nontachyonic 0B orientifolds,''
  Nucl.\ Phys.\ B {\bf 598} (2001), 189
  %doi:10.1016/S0550-3213(00)00781-1
  [hep-th/0010179].
  %%CITATION = doi:10.1016/S0550-3213(00)00781-1;%%

  
\bibitem{ang_98}
C.~Angelantonj,
``Nontachyonic open descendants of the 0B string theory,''
Phys. Lett. B \textbf{444} (1998), 309
%doi:10.1016/S0370-2693(98)01430-0
[arXiv:hep-th/9810214 [hep-th]].
%97 citations counted in INSPIRE as of 28 Aug 2025

\bibitem{4dnonsusyheterotic}
See e.g. K.~R.~Dienes,
``Statistics on the heterotic landscape: Gauge groups and cosmological constants of four-dimensional heterotic strings,''
Phys. Rev. D \textbf{73} (2006), 106010, 
[arXiv:hep-th/0602286 [hep-th]].

\bibitem{admavr}
S.~Abel, K.~R.~Dienes and E.~Mavroudi,
``Towards a nonsupersymmetric string phenomenology,''
Phys. Rev. D \textbf{91} (2015) no.12, 126014, 
[arXiv:1502.03087 [hep-th]].

\bibitem{recent_v}
Z.~K.~Baykara, H.~C.~Tarazi and C.~Vafa,
``New Non-Supersymmetric Tachyon-Free Strings,''
[arXiv:2406.00185 [hep-th]];
C.~Angelantonj, I.~Florakis, G.~Leone and D.~Perugini,
``Non-supersymmetric non-tachyonic heterotic vacua with reduced rank in various dimensions,''
JHEP \textbf{10} (2024), 216, 
[arXiv:2407.09597 [hep-th]];
L.~A.~Detraux, A.~R.~D.~Avalos, A.~E.~Faraggi and B.~Percival,
``Vacuum energy of nonsupersymmetric S{\textasciitilde} heterotic string models,''
Phys. Rev. D \textbf{110} (2024) no.8, 086006, 
[arXiv:2407.19980 [hep-th]]; 
E.~Basaad, L.~A.~Detraux, A.~R.~D.~Avalos, A.~E.~Faraggi and B.~Percival,
``Vacuum energy in non-supersymmetric quasi-realistic heterotic-string vacua with fixed moduli,''
Eur. Phys. J. C \textbf{85} (2025) no.2, 209, 
[arXiv:2408.03928 [hep-th]].

  
\bibitem{vacuum_redefinitions1}
W.~Fischler and L.~Susskind,
 ``Dilaton Tadpoles, String Condensates And Scale Invariance,''
 Phys.\ Lett.\  {\bf B 171} (1986), 383.
 %$doi:10.1016/0370-2693(86)91425-5$,
 %%CITATION = PHLTA,B171,383;%%

 \bibitem{vacuum_redefinitions2}
 W.~Fischler and L.~Susskind,
``Dilaton Tadpoles, String Condensates And Scale Invariance. 2,''
 Phys.\ Lett.\  {\bf B 173} (1986) 262.
 %$doi:10.1016/0370-2693(86)90514-9$;
 %%CITATION = PHLTA,B173,262;%%

 \bibitem{vacuum_redefinitions3}
E.~Dudas, M.~Nicolosi, G.~Pradisi and A.~Sagnotti,
 ``On tadpoles and vacuum redefinitions in string theory,''
 Nucl.\ Phys.\  {\bf B 708} (2005), 3
 %$doi:10.1016/j.nuclphysb.2004.11.028$
 [arXiv:hep-th/0410101].
 %%CITATION = NUPHA,B708,3;%%

 \bibitem{vacuum_redefinitions4}
N.~Kitazawa,
 ``Tadpole Resummations in String Theory,''
 Phys.\ Lett.\  {\bf B 660} (2008), 415
 [arXiv:0801.1702 [hep-th]].
% $doi:10.1016/j.physletb.2008.01.028$;
 %%CITATION = PHLTA,B660,415;%%

   \bibitem{vacuum_redefinitions5}
 R.~Pius, A.~Rudra and A.~Sen,
  ``String Perturbation Theory Around Dynamically Shifted Vacuum,''
  JHEP {\bf 1410} (2014), 70
  %$doi:10.1007/JHEP10(2014)070$
  [arXiv:1404.6254 [hep-th]].
  %%CITATION = doi:10.1007/JHEP10(2014)070;%%

 
\bibitem{mrs24_1}
J.~Mourad, S.~Raucci and A.~Sagnotti,
``Brane-like solutions and other non-supersymmetric vacua,''
JHEP \textbf{10} (2024), 054
%doi:10.1007/JHEP10(2024)054
[arXiv:2406.14926 [hep-th]].

\bibitem{mrs24_2}
J.~Mourad, S.~Raucci and A.~Sagnotti,
``Brane profiles of non-supersymmetric strings,''
JHEP \textbf{09} (2024), 019
%doi:10.1007/JHEP09(2024)019
[arXiv:2406.16327 [hep-th]].

\bibitem{kut_sei} D.~Kutasov and N.~Seiberg, ``Number of degrees of freedom, density of states and tachyons in string theory and CFT,'' Nucl. Phys. B \textbf{358} (1991), 600. 
% doi:10.1016/0550-3213(91)90426-X %241 citations counted in INSPIRE as of 01 Sep 2025

\bibitem{mis_1} K.~R.~Dienes, ``Modular invariance, finiteness, and misaligned supersymmetry: New constraints on the numbers of physical string states,'' Nucl. Phys. B \textbf{429} (1994), 533 
% doi:10.1016/0550-3213(94)90153-8 
[arXiv:hep-th/9402006 [hep-th]]. 

\bibitem{mis_2} K.~R.~Dienes, M.~Moshe and R.~C.~Myers, ``String theory, misaligned supersymmetry, and the supertrace constraints,'' Phys. Rev. Lett. \textbf{74} (1995), 4767
% doi:10.1103/PhysRevLett.74.4767 
[arXiv:hep-th/9503055 [hep-th]].

\bibitem{carlo1} C.~Angelantonj, M.~Cardella, S.~Elitzur and E.~Rabinovici, ``Vacuum stability, string density of states and the Riemann zeta function,'' JHEP \textbf{02} (2011), 024 
%doi:10.1007/JHEP02(2011)024 
arXiv:1012.5091 [hep-th]].

\bibitem{carlo2} N.~Cribiori, S.~Parameswaran, F.~Tonioni and T.~Wrase, ``Misaligned Supersymmetry and Open Strings,'' JHEP \textbf{04} (2021), 099 
%doi:10.1007/JHEP04(2021)099 
[arXiv:2012.04677 [hep-th]].

\bibitem{carlo3} C.~Angelantonj, I.~Florakis and G.~Leone, ``Tachyons and misaligned supersymmetry in closed string vacua,'' JHEP \textbf{06} (2023), 174 
% doi:10.1007/JHEP06(2023)174 
[arXiv:2301.13702 [hep-th]]. 

\bibitem{carlo4} G.~Leone, ``Tachyons and Misaligned Supersymmetry in orientifold vacua,'' JHEP \textbf{11} (2023), 066 %doi:10.1007/JHEP11(2023)066 
[arXiv:2308.09757 [hep-th]]. 

\bibitem{harvey}
J.~A.~Harvey,
``String duality and nonsupersymmetric strings,''
Phys. Rev. D \textbf{59} (1999), 026002
%doi:10.1103/PhysRevD.59.026002
[arXiv:hep-th/9807213 [hep-th]].

\bibitem{kks} S.~Kachru, J.~Kumar and E.~Silverstein, ``Vacuum energy cancellation in a nonsupersymmetric string,'' Phys. Rev. D \textbf{59} (1999), 106004 
%doi:10.1103/PhysRevD.59.106004 
[arXiv:hep-th/9807076 [hep-th]]. 

\bibitem{carlo5} C.~Angelantonj, I.~Antoniadis and K.~Forger, ``Nonsupersymmetric type I strings with zero vacuum energy,'' Nucl. Phys. B \textbf{555} (1999), 116 
% doi:10.1016/S0550-3213(99)00344-2 
[arXiv:hep-th/9904092 [hep-th]]. 

\bibitem{carlo6} C.~Angelantonj and M.~Cardella, 
``Vanishing perturbative vacuum energy in nonsupersymmetric orientifolds,'' Phys. Lett. B \textbf{595} (2004), 505 
% doi:10.1016/j.physletb.2004.06.058 
[arXiv:hep-th/0403107 [hep-th]]. 

%\cite{Dienes:1994np}
\bibitem{dienes}
K.~R.~Dienes,
``Modular invariance, finiteness, and misaligned supersymmetry: New constraints on the numbers of physical string states,''
Nucl. Phys. B \textbf{429} (1994), 533
%doi:10.1016/0550-3213(94)90153-8
[arXiv:hep-th/9402006 [hep-th]].
%119 citations counted in INSPIRE as of 13 Nov 2023

%\cite{Angelantonj:2023egh}
\bibitem{carlogiorgio1}
N.~Cribiori, S.~Parameswaran, F.~Tonioni and T.~Wrase,
``Modular invariance, misalignment and finiteness in non-supersymmetric strings,''
JHEP \textbf{01} (2022), 127
%doi:10.1007/JHEP01(2022)127
[arXiv:2110.11973 [hep-th]].
%12 citations counted in INSPIRE as of 13 Nov 2023

\bibitem{carlogiorgio2}
C.~Angelantonj, I.~Florakis and G.~Leone,
``Tachyons and misaligned supersymmetry in closed string vacua,''
JHEP \textbf{06} (2023), 174
%doi:10.1007/JHEP06(2023)174
[arXiv:2301.13702 [hep-th]].
%3 citations counted in INSPIRE as of 13 Nov 2023

\bibitem{carlogiorgio3}
G.~Leone,
``Tachyons and Misaligned Supersymmetry in orientifold vacua,''
JHEP \textbf{11} (2023), 066
%doi:10.1007/JHEP11(2023)066
[arXiv:2308.09757 [hep-th]].
%0 citations counted in INSPIRE as of 16 Apr 2024

\bibitem{narain1} K.~S.~Narain,
``New Heterotic String Theories in Uncompactified Dimensions \ensuremath{<} 10,''
Phys. Lett. B \textbf{169} (1986), 41.
%doi:10.1016/0370-2693(86)90682-9
%1075 citations counted in INSPIRE as of 16 Apr 2024

\bibitem{narain2}
K.~S.~Narain, M.~H.~Sarmadi and E.~Witten,
``A Note on Toroidal Compactification of Heterotic String Theory,''
Nucl. Phys. B \textbf{279} (1987), 369.
%doi:10.1016/0550-3213(87)90001-0
%838 citations counted in INSPIRE as of 16 Apr 2024

\bibitem{T-duality}
A.~Giveon, M.~Porrati and E.~Rabinovici,
``Target space duality in string theory,''
Phys. Rept. \textbf{244} (1994), 77
%doi:10.1016/0370-1573(94)90070-1
[arXiv:hep-th/9401139 [hep-th]].
%1117 citations counted in INSPIRE as of 17 Apr 2025

\bibitem{schomerus}
V.~Schomerus,
``D-branes and deformation quantization,''
JHEP \textbf{06} (1999), 030
%doi:10.1088/1126-6708/1999/06/030
[arXiv:hep-th/9903205 [hep-th]].
%612 citations counted in INSPIRE as of 03 Jul 2025

\bibitem{seibergwitten}
N.~Seiberg and E.~Witten,
``String theory and noncommutative geometry,''
JHEP \textbf{09} (1999), 032
%doi:10.1088/1126-6708/1999/09/032
[arXiv:hep-th/9908142 [hep-th]].
%4805 citations counted in INSPIRE as of 03 Jul 2025

\bibitem{bianchitor} M.~Bianchi,
``A note on toroidal compactifications of the type I superstring and
other superstring vacuum configurations with 16 supercharges'',
Nucl.\ Phys.\ B528 (1998), 73
[arXiv:hep-th/9711201].
%%CITATION = HEP-TH 9711201;%%

\bibitem{emiliantor}
T.~Coudarchet, E.~Dudas and H.~Partouche,
``Geometry of orientifold vacua and supersymmetry breaking,''
JHEP \textbf{07} (2021), 104
%doi:10.1007/JHEP07(2021)104
[arXiv:2105.06913 [hep-th]].
%16 citations counted in INSPIRE as of 14 Aug 2025

\bibitem{pradisi_03}
G.~Pradisi,
``Magnetic fluxes, NS NS B field and shifts in four-dimensional orientifolds,''
[arXiv:hep-th/0310154 [hep-th]].
%13 citations counted in INSPIRE as of 02 Nov 2025

\bibitem{dm-tachyonfree2}
E.~Dudas, J.~Mourad and C.~Timirgaziu,
``Time and space dependent backgrounds from nonsupersymmetric strings,''
Nucl. Phys. B \textbf{660} (2003), 3
%doi:10.1016/S0550-3213(03)00248-7
[arXiv:hep-th/0209176 [hep-th]].
%34 citations counted in INSPIRE as of 04 Mar 2024

\bibitem{cab} C.~Angelantonj,
``Comments on open-string orbifolds with a non-vanishing $B_{ab}$'',
Nucl.\ Phys.\ B566 (2000), 126
[arXiv:hep-th/9908064].
%%CITATION = HEP-TH 9908064;%%

\bibitem{gimpol}
E.~G.~Gimon and J.~Polchinski,
``Consistency conditions for orientifolds and D-manifolds,''
Phys. Rev. D \textbf{54} (1996), 1667
%doi:10.1103/PhysRevD.54.1667
[arXiv:hep-th/9601038 [hep-th]].
%652 citations counted in INSPIRE as of 02 Jan 2024


\bibitem{pc2}
G.~Aldazabal, D.~Badagnani, L.~E.~Ib\'a\~nez and A.~M.~Uranga,
``Tadpole versus anomaly cancellation in $D = 4, 6$
compact IIB  orientifolds'',
JHEP 9906 (1999) 031
[arXiv:hep-th/9904071].
%%CITATION = HEP-TH 9904071;%%

\bibitem{pc3}
C.~A.~Scrucca and M.~Serone,
``Anomaly cancellation in K3 orientifolds,''
Nucl.\ Phys.\ B564 (2000) 555
[arXiv:hep-th/9907112].
%%CITATION = HEP-TH 9907112;%%

\bibitem{pc4}
C.~A.~Scrucca and M.~Serone,
``Gauge and gravitational anomalies in $D = 4$ $N = 1$ orientifolds,''
JHEP 9912 (1999) 024
[arXiv:hep-th/9912108].
%%CITATION = HEP-TH 9912108;%%

\bibitem{pc5}
M.~Bianchi and J.~F.~Morales,
``Anomalies and tadpoles'',
JHEP 0003 (2000) 030
[arXiv:hep-th/0002149].
%%CITATION = HEP-TH 0002149;%%


\bibitem{six} M.~Berkooz, R.~G.~Leigh, J.~Polchinski, J.~H.~Schwarz,
N.~Seiberg and E.~Witten,
``Anomalies, dualities, and topology of $D=6$ $N=1$ superstring vacua'',
Nucl.\ Phys.\ B475 (1996), 115
[arXiv:hep-th/9605184].
%%CITATION = HEP-TH 9605184;%%


\bibitem{fabio1}
S.~Ferrara, R.~Minasian and A.~Sagnotti,
``Low-energy analysis of M and F theories on Calabi-Yau threefolds'',
Nucl.\ Phys.\ B474 (1996), 323
[arXiv:hep-th/9604097].
%%CITATION = HEP-TH 9604097;%%

\bibitem{fabio2}
S.~Ferrara, F.~Riccioni and A.~Sagnotti,
``Tensor and vector multiplets in six-dimensional supergravity'',
Nucl.\ Phys.\ B519 (1998), 115
[arXiv:hep-th/9711059].
%%CITATION = HEP-TH 9711059;%%

\bibitem{fabio3}
F.~Riccioni and A.~Sagnotti,
``Consistent and covariant anomalies in six-dimensional supergravity'',
Phys.\ Lett.\ B436 (1998), 298
[arXiv:hep-th/9806129].
%%CITATION = HEP-TH 9806129;%%

\bibitem{fabio4}
F.~Riccioni,
``Abelian vector multiplets in six-dimensional supergravity'',
Phys.\ Lett.\ B474 (2000), 79
[arXiv:hep-th/9910246].
%%CITATION = HEP-TH 9910246;%%

\bibitem{fabio5}
F.~Riccioni,
``All couplings of minimal six-dimensional supergravity'',
Nucl.\ Phys.\ B605 (2001), 245
[arXiv:hep-th/0101074].
%%CITATION = HEP-TH 0101074;%%


\bibitem{poltens} J.~Polchinski,
``Tensors from K3 orientifolds'',
Phys.\ Rev.\ D55 (1997), 6423
[arXiv:hep-th/9606165].
%%CITATION = HEP-TH 9606165;%%

\bibitem{au} G.~Aldazabal and A.~M.~Uranga,
``Tachyon-free non-supersymmetric type IIB orientifolds
via  brane-antibrane systems'',
JHEP 9910 (1999) 024
[arXiv:hep-th/9908072].
%%CITATION = HEP-TH 9908072;%%

\bibitem{acdl}
C.~Angelantonj, C.~Condeescu, E.~Dudas and G.~Leone,
``Rigid vacua with Brane Supersymmetry Breaking,''
JHEP \textbf{04} (2024), 103
%doi:10.1007/JHEP04(2024)103
[arXiv:2403.02392 [hep-th]].

\bibitem{romans} L.~J.~Romans,
``Selfduality for interacting fields: covariant field equations
for six-dimensional chiral supergravities'',
Nucl.\ Phys.\ B276 (1986), 71.
%%CITATION = NUPHA,B276,71;%%

\bibitem{witsmall} E.~Witten,
``Small instantons in string theory'',
Nucl.\ Phys.\ B460 (1996), 541
[arXiv:hep-th/9511030].
%%CITATION = HEP-TH 9511030;%%

\bibitem{fradkin}
C.~Angelantonj and A.~Sagnotti,
``Type-I vacua and brane transmutation,''
[arXiv:hep-th/0010279 [hep-th]].

\bibitem{abg}
C.~Angelantonj, R.~Blumenhagen and M.~R.~Gaberdiel,
``Asymmetric orientifolds, brane supersymmetry breaking and nonBPS branes,''
Nucl. Phys. B \textbf{589} (2000), 545
%doi:10.1016/S0550-3213(00)00518-6
[arXiv:hep-th/0006033 [hep-th]].
%57 citations counted in INSPIRE as of 25 Aug 2025

\bibitem{wito32} E.~Witten,
``Some properties of O(32) superstrings'',
Phys.\ Lett.\ B149 (1984), 351.
%%CITATION = PHLTA,B149,351;%%

\bibitem{penta} M.~Bianchi and Y.~S.~Stanev,
``Open strings on the Neveu-Schwarz pentabrane'',
Nucl.\ Phys.\ B523 (1998), 193
[arXiv:hep-th/9711069].
%%CITATION = HEP-TH 9711069;%%

\bibitem{bbblw}
C.~Bachas, M.~Bianchi, R.~Blumenhagen, D.~Lust and T.~Weigand,
``Comments on Orientifolds without Vector Structure,''
JHEP \textbf{08} (2008), 016
%doi:10.1088/1126-6708/2008/08/016
[arXiv:0805.3696 [hep-th]].

\bibitem{bidirac1}
M.~Born and L.~Infeld,
``Foundations of the new field theory'',
Proc.\ Roy.\ Soc.\ Lond.\ A144 (1934), 425.
%%CITATION = PRSLA,A144,425;%%

\bibitem{bidirac2}
P.~A.~Dirac,
``An extensible model of the electron'',
Proc.\ Roy.\ Soc.\ Lond.\ A268 (1962), 57.
%%CITATION = PRSLA,A268,57;%%


\bibitem{nole1} N.~K.~Nielsen and P.~Olesen,
``An unstable Yang-Mills field mode'',
Nucl.\ Phys.\ B144 (1978), 376.
%%CITATION = NUPHA,B144,376;%%

\bibitem{nole2}
J.~Ambjorn, N.~K.~Nielsen and P.~Olesen,
``A hidden Higgs Lagrangian in QCD'',
Nucl.\ Phys.\ B152 (1979), 75.
%%CITATION = NUPHA,B152,75;%%

\bibitem{nole3}
H.~B.~Nielsen and M.~Ninomiya,
``A bound on bag constant and Nielsen-Olesen unstable mode in QCD'',
Nucl.\ Phys.\ B156 (1979), 1.
%%CITATION = NUPHA,B156,1;%%

\bibitem{bachasmag} C.~Bachas,
``A way to break supersymmetry'',
arXiv:hep-th/9503030.
%%CITATION = HEP-TH 9503030;%%


\bibitem{bdt}
W.~Buchmuller, E.~Dudas and Y.~Tatsuta,
``Tachyon condensation in magnetic compactifications,''
JHEP \textbf{03} (2021), 070
%doi:10.1007/JHEP03(2021)070
[arXiv:2010.10891 [hep-th]].
%2 citations counted in INSPIRE as of 07 Feb 2024


\bibitem{dgm1} M.~B.~Green, J.~A.~Harvey and G.~Moore,
``I-brane inflow and anomalous couplings on D-branes'',
Class.\ Quant.\ Grav.\ 14 (1997), 47
[arXiv:hep-th/9605033].
%%CITATION = HEP-TH 9605033;%%

\bibitem{dgm2}
J.~Mourad,
``Anomalies of the SO(32) five-brane and their cancellation'',
Nucl.\ Phys.\ B512 (1998), 199
[arXiv:hep-th/9709012].
%%CITATION = HEP-TH 9709012;%%

\bibitem{dgm3}
J.~F.~Morales, C.~A.~Scrucca and M.~Serone,
``Anomalous couplings for D-branes and O-planes'',
Nucl.\ Phys.\ B552 (1999), 291
[arXiv:hep-th/9812071].
%%CITATION = HEP-TH 9812071;%%

\bibitem{dgm4}
B.~J.~Stefanski,
``Gravitational couplings of D-branes and O-planes'',
Nucl.\ Phys.\ B548 (1999), 275
[arXiv:hep-th/9812088].
%%CITATION = HEP-TH 9812088;%%.

\bibitem{dgm5}
R.~C.~Myers,
``Dielectric-branes,''
JHEP 9912 (1999) 022;
[arXiv:hep-th/9910053].
%%CITATION = HEP-TH 9910053;%%

\bibitem{dgm6}
S.~F.~Hassan and R.~Minasian,
``D-brane couplings, RR fields and Clifford multiplication,''
arXiv:hep-th/0008149.
%%CITATION = HEP-TH 0008149;%%


\bibitem{douglas21} M.~Berkooz, M.~R.~Douglas and R.~G.~Leigh,
``Branes intersecting at angles'',
Nucl.\ Phys.\ B480 (1996), 265
[arXiv:hep-th/9606139].
%%CITATION = HEP-TH 9606139;%%


\bibitem{douglas22}
V.~Balasubramanian and R.~G.~Leigh,
``D-branes, moduli and supersymmetry'',
Phys.\ Rev.\ D55 (1997), 6415
[hep-th/9611165].
%%CITATION = HEP-TH 9611165;%%

\bibitem{cardy2} J.~L.~Cardy,
``Boundary conditions, fusion rules and the Verlinde formula'',
Nucl.\ Phys.\ B324 (1989), 581.
%%CITATION = NUPHA,B324,581;%

\bibitem{nonkahler}
C.~M.~Hull,
``Superstring Compactifications with Torsion and Space-Time Supersymmetry,'' Print-86-0251 (Cambridge);
%3 citations counted in INSPIRE as of 24 Nov 2025
A.~Strominger,
``Superstrings with Torsion,''
Nucl. Phys. B \textbf{274} (1986), 253;
%doi:10.1016/0550-3213(86)90286-5
J.~P.~Gauntlett, D.~Martelli, S.~Pakis and D.~Waldram,
``G structures and wrapped NS5-branes,''
Commun. Math. Phys. \textbf{247} (2004), 421
%doi:10.1007/s00220-004-1066-y
[arXiv:hep-th/0205050 [hep-th]];
%344 citations counted in INSPIRE as of 21 Nov 2025
S.~Gurrieri, J.~Louis, A.~Micu and D.~Waldram,
``Mirror symmetry in generalized Calabi-Yau compactifications,''
Nucl. Phys. B \textbf{654} (2003), 61
%doi:10.1016/S0550-3213(03)00045-2
[arXiv:hep-th/0211102 [hep-th]];
%351 citations counted in INSPIRE as of 21 Nov 2025
G.~Lopes Cardoso, G.~Curio, G.~Dall'Agata, D.~Lust, P.~Manousselis and G.~Zoupanos,
``NonKahler string backgrounds and their five torsion classes,''
Nucl. Phys. B \textbf{652} (2003), 5
%doi:10.1016/S0550-3213(03)00049-X
[arXiv:hep-th/0211118 [hep-th]];
%353 citations counted in INSPIRE as of 21 Nov 2025
C.~Jeschek and F.~Witt,
``Generalised G(2) - structures and type IIb superstrings,''
JHEP \textbf{03} (2005), 053
%doi:10.1088/1126-6708/2005/03/053
[arXiv:hep-th/0412280 [hep-th]];
M.~Grana, R.~Minasian, M.~Petrini and A.~Tomasiello,
``Generalized structures of N=1 vacua,''
JHEP \textbf{11} (2005), 020
%doi:10.1088/1126-6708/2005/11/020
[arXiv:hep-th/0505212 [hep-th]].

\bibitem{gj}
E.~G.~Gimon and C.~V.~Johnson,
``K3 orientifolds,''
Nucl. Phys. B \textbf{477} (1996), 715
%doi:10.1016/0550-3213(96)00356-2
[arXiv:hep-th/9604129 [hep-th]].
%216 citations counted in INSPIRE as of 25 Aug 2025

\bibitem{adscft}
J.~M.~Maldacena,
``The Large $N$ limit of superconformal field theories and supergravity,''
Adv. Theor. Math. Phys. \textbf{2} (1998), 231
%doi:10.4310/ATMP.1998.v2.n2.a1
[arXiv:hep-th/9711200 [hep-th]];
%21064 citations counted in INSPIRE as of 04 Aug 2025
S.~S.~Gubser, I.~R.~Klebanov and A.~M.~Polyakov,
``Gauge theory correlators from noncritical string theory,''
Phys. Lett. B \textbf{428} (1998), 105
%doi:10.1016/S0370-2693(98)00377-3
[arXiv:hep-th/9802109 [hep-th]];
%11025 citations counted in INSPIRE as of 04 Aug 2025
E.~Witten,
``Anti de Sitter space and holography,''
Adv. Theor. Math. Phys. \textbf{2} (1998), 253
%doi:10.4310/ATMP.1998.v2.n2.a2
[arXiv:hep-th/9802150 [hep-th]].
%13408 citations counted in INSPIRE as of 04 Aug 2025
For a review, see: O.~Aharony, S.~S.~Gubser, J.~M.~Maldacena, H.~Ooguri and Y.~Oz,
``Large N field theories, string theory and gravity,''
Phys. Rept. \textbf{323} (2000), 183
%doi:10.1016/S0370-1573(99)00083-6
[arXiv:hep-th/9905111 [hep-th]].
%6005 citations counted in INSPIRE as of 04 Aug 2025

\bibitem{susy_branes}
C.~G.~Callan, Jr., J.~A.~Harvey and A.~Strominger,
``World sheet approach to heterotic instantons and solitons,''
Nucl. Phys. B \textbf{359} (1991), 611;
%doi:10.1016/0550-3213(91)90074-8
%553 citations counted in INSPIRE as of 25 Sep 2023
C.~G.~Callan, Jr., J.~A.~Harvey and A.~Strominger,
``Worldbrane actions for string solitons,''
Nucl. Phys. B \textbf{367} (1991), 60;
%doi:10.1016/0550-3213(91)90041-U
G.~T.~Horowitz and A.~Strominger,
``Black strings and P-branes,''
Nucl. Phys. B \textbf{360} (1991), 197.
% doi:10.1016/0550-3213(91)90440-9
%1210 citations counted in INSPIRE as of 25 Sep 2023
For reviews, see:
C.~G.~Callan, Jr., J.~A.~Harvey and A.~Strominger,
``Supersymmetric string solitons,''
[arXiv:hep-th/9112030 [hep-th]];
M.~J.~Duff, R.~R.~Khuri and J.~X.~Lu,
``String solitons,''
Phys. Rept. \textbf{259} (1995), 213
%doi:10.1016/0370-1573(95)00002-X
[arXiv:hep-th/9412184 [hep-th]].

  \bibitem{ms21_1}
J.~Mourad and A.~Sagnotti,
``On warped string vacuum profiles and cosmologies. Part I. Supersymmetric strings,''
JHEP \textbf{12} (2021), 137
%doi:10.1007/JHEP12(2021)137
[arXiv:2109.06852 [hep-th]].

\bibitem{ms21_2}
J.~Mourad and A.~Sagnotti,
``On warped string vacuum profiles and cosmologies. Part II. Non-supersymmetric strings,''
JHEP \textbf{12} (2021), 138
%doi:10.1007/JHEP12(2021)138
[arXiv:2109.12328 [hep-th]].

  \bibitem{raucci_ads}
  S.~Raucci,
``On new vacua of non-supersymmetric strings,''
Phys. Lett. B \textbf{837} (2023), 137663
%doi:10.1016/j.physletb.2022.137663
[arXiv:2209.06537 [hep-th]].

  \bibitem{bms}
  I.~Basile, J.~Mourad and A.~Sagnotti,
``On Classical Stability with Broken Supersymmetry,''
JHEP \textbf{01} (2019), 174
%doi:10.1007/JHEP01(2019)174
[arXiv:1811.11448 [hep-th]].
%52 citations counted in INSPIRE as of 27 Nov 2023

\bibitem{sethiO16}
Z.~K.~Baykara, D.~Robbins and S.~Sethi,
``Non-supersymmetric AdS from string theory,''
SciPost Phys. \textbf{15} (2023) no.6, 224
%doi:10.21468/SciPostPhys.15.6.224
[arXiv:2212.02557 [hep-th]].

  \bibitem{dynamicalcobordism2}
M.~Montero and C.~Vafa,
``Cobordism Conjecture, Anomalies, and the String Lamppost Principle,''
JHEP \textbf{01} (2021), 063
%doi:10.1007/JHEP01(2021)063
[arXiv:2008.11729 [hep-th]].

  \bibitem{dynamicalcobordism3}
G.~Buratti, M.~Delgado and A.~M.~Uranga,
``Dynamical tadpoles, stringy cobordism, and the SM from spontaneous compactification,''
JHEP \textbf{06} (2021), 170
%doi:10.1007/JHEP06(2021)170
[arXiv:2104.02091 [hep-th]].

  \bibitem{dynamicalcobordism4}
G.~Buratti, J.~Calder\'on-Infante, M.~Delgado and A.~M.~Uranga,
``Dynamical Cobordism and Swampland Distance Conjectures,''
JHEP \textbf{10} (2021), 037
%doi:10.1007/JHEP10(2021)037
[arXiv:2107.09098 [hep-th]].

\bibitem{dynamicalcobordism5}
R.~Blumenhagen, N.~Cribiori, C.~Kneissl and A.~Makridou,
``Dynamical cobordism of a domain wall and its companion defect 7-brane,''
JHEP \textbf{08} (2022), 204
%doi:10.1007/JHEP08(2022)204
[arXiv:2205.09782 [hep-th]].

  \bibitem{dynamicalcobordism6}
S.~Raucci,
``On codimension-one vacua and string theory,''
Nucl. Phys. B \textbf{985} (2022), 116002
%doi:10.1016/j.nuclphysb.2022.116002
[arXiv:2206.06399 [hep-th]].

  \bibitem{dynamicalcobordism7}
R.~Blumenhagen, C.~Kneissl and C.~Wang,
``Dynamical Cobordism Conjecture: solutions for end-of-the-world branes,''
JHEP \textbf{05} (2023), 123
%doi:10.1007/JHEP05(2023)123
[arXiv:2303.03423 [hep-th]].

  \bibitem{dynamicalcobordism1}
J.~McNamara and C.~Vafa,
``Cobordism Classes and the Swampland,''
[arXiv:1909.10355 [hep-th]].

\bibitem{jphysA}
J.~Mourad and A.~Sagnotti,
``Effective orientifolds from broken supersymmetry,''
J. Phys. A \textbf{57} (2024) no.3, 035401
%doi:10.1088/1751-8121/ad16f8
[arXiv:2309.05268 [hep-th]].


%\cite{Breitenlohner:1982jf}
\bibitem{breitf}
P.~Breitenlohner and D.~Z.~Freedman,
``Stability in Gauged Extended Supergravity,''
Annals Phys. \textbf{144} (1982), 249.
%doi:10.1016/0003-4916(82)90116-6
%1716 citations counted in INSPIRE as of 27 Feb 2024

\bibitem{math_literature_1}
K.~M.~Case, Phys. Rev.  \textbf{80} (1950) 797;
M.~A.~Naimark, ``Linear Differential Operators'', vols.~I and II, Ge.C. Harrap \& C (1968); N.~I.~Akhiezer and I.~Glazman, ``Theory of Linear Operators in Hilbert Space", parts I and II, Dover Publications (1993); C.~Lanczos, ``Linear Differential Operators, ''SIAM (1996).

\bibitem{math_literature_2}  D.~M.~Gitman, I.~V.~Tyutin and B.~L.~Voronov, ``Self--Adjoint Extensions in Quantum Mechanics,'' Birkhauser-Springer (2010); B.~C.~Hall, ``Quantum Theory for Mathematicians,'' Springer (2013).

\bibitem{tables}
NIST Handbook of Mathematical Functions, eds.~F.W. J.~Olver et al, Cambridge Univ. Press (2015).

\bibitem{ms22_1}
J.~Mourad and A.~Sagnotti,
``A 4D IIB flux vacuum and supersymmetry breaking. Part I. Fermionic spectrum,''
JHEP \textbf{08} (2022), 301
%doi:10.1007/JHEP08(2022)301
[arXiv:2206.03340 [hep-th]].

\bibitem{ms_20}
J.~Mourad and A.~Sagnotti,
``On boundaries, charges and Fermi fields,''
Phys. Lett. B \textbf{804} (2020), 135368
%doi:10.1016/j.physletb.2020.135368
[arXiv:2002.05372 [hep-th]].
%8 citations counted in INSPIRE as of 24 Jul 2024

  \bibitem{vanishing_nonsusy1}
S.~Kachru, J.~Kumar and E.~Silverstein,
  ``Vacuum energy cancellation in a nonsupersymmetric string,''
  Phys.\ Rev.\ D {\bf 59} (1999) 106004
  %doi:10.1103/PhysRevD.59.106004
  [hep-th/9807076].
  %%CITATION = doi:10.1103/PhysRevD.59.106004;%%

\bibitem{ms23_2}
J.~Mourad and A.~Sagnotti,
``A 4D IIB flux vacuum and supersymmetry breaking. Part II. Bosonic spectrum and stability,''
JHEP \textbf{11} (2023), 061
%doi:10.1007/JHEP11(2023)061
[arXiv:2309.04026 [hep-th]].

\bibitem{witten_flux}
E.~Witten,
``On flux quantization in M theory and the effective action,''
J. Geom. Phys. \textbf{22} (1997), 1
%doi:10.1016/S0393-0440(96)00042-3
[arXiv:hep-th/9609122 [hep-th]].
%424 citations counted in INSPIRE as of 09 Jun 2021


\bibitem{mourad_25} J.~Mourad, talk at ``Gravity, Strings and Supersymmetry Breaking'' (Pisa, Scuola Normale, April 3-5 2025), https://indico.sns.it/event/79/.

  \bibitem{henneaux}
M.~Henneaux and C.~Teitelboim,
``Dynamics of Chiral (Selfdual) $P$ Forms,''
Phys. Lett. \textbf{B 206} (1988), 650.
%doi:10.1016/0370-2693(88)90712-5
%224 citations counted in INSPIRE as of 28 May 2022

\bibitem{luest} {
N.~Cribiori, D.~Lust and M.~Scalisi,
``The gravitino and the swampland,''
JHEP \textbf{06} (2021), 071
%doi:10.1007/JHEP06(2021)071
[arXiv:2104.08288 [hep-th]];
A.~Castellano, A.~Font, A.~Herraez and L.~E.~Ib\'a\~nez,
``A gravitino distance conjecture,''
JHEP \textbf{08} (2021), 092
%doi:10.1007/JHEP08(2021)092
[arXiv:2104.10181 [hep-th]].}

\bibitem{cadavid}
A.~C.~Cadavid, A.~Ceresole, R.~D'Auria and S.~Ferrara,
``Eleven-dimensional supergravity compactified on Calabi-Yau threefolds,''
Phys. Lett. B \textbf{357} (1995), 76
%doi:10.1016/0370-2693(95)00891-N
[arXiv:hep-th/9506144 [hep-th]].
%268 citations counted in INSPIRE as of 30 Jul 2025

%\cite{Ferrara:1989ik}
\bibitem{ferrara_coset}
S.~Ferrara and S.~Sabharwal,
``Quaternionic Manifolds for Type II Superstring Vacua of Calabi-Yau Spaces,''
Nucl. Phys. B \textbf{332} (1990), 317.
%doi:10.1016/0550-3213(90)90097-W
%348 citations counted in INSPIRE as of 17 Apr 2024


%\cite{Lukas:1998tt}
\bibitem{ovrut_stelle}
A.~Lukas, B.~A.~Ovrut, K.~S.~Stelle and D.~Waldram,
``Heterotic M theory in five-dimensions,''
Nucl. Phys. B \textbf{552} (1999), 246
%doi:10.1016/S0550-3213(99)00196-0
[arXiv:hep-th/9806051 [hep-th]].
%433 citations counted in INSPIRE as of 17 Apr 2024

%\cite{Dudas:1997jn}
\bibitem{dg97}
E.~Dudas and C.~Grojean,
``Four-dimensional M theory and supersymmetry breaking,''
Nucl. Phys. B \textbf{507} (1997), 553
%doi:10.1016/S0550-3213(97)00590-7
[arXiv:hep-th/9704177 [hep-th]].
%154 citations counted in INSPIRE as of 04 Mar 2024

%\cite{Douglas:2006es}
\bibitem{kachru_douglas-flux}
M.~R.~Douglas and S.~Kachru,
``Flux compactification,''
Rev. Mod. Phys. \textbf{79} (2007), 733
%doi:10.1103/RevModPhys.79.733
[arXiv:hep-th/0610102 [hep-th]].
%972 citations counted in INSPIRE as of 17 Apr 2024

%\cite{Grana:2005jc}
\bibitem{grana-flux}
M.~Grana,
``Flux compactifications in string theory: A Comprehensive review,''
Phys. Rept. \textbf{423} (2006), 91
%doi:10.1016/j.physrep.2005.10.008
[arXiv:hep-th/0509003 [hep-th]].
%954 citations counted in INSPIRE as of 17 Apr 2024

%\cite{hebecker}
\bibitem{hebecker}
A.~Hebecker,
``Naturalness, String Landscape and Multiverse: A Modern Introduction with Exercises,''
Lect. Notes Phys. \textbf{979} (2021), 1-313
2021,
ISBN 978-3-030-65150-3, 978-3-030-65151-0
%doi:10.1007/978-3-030-65151-0
[arXiv:2008.10625 [hep-th]].

\bibitem{gkp}
S.~B.~Giddings, S.~Kachru and J.~Polchinski,
``Hierarchies from fluxes in string compactifications,''
Phys. Rev. D \textbf{66} (2002), 106006
%doi:10.1103/PhysRevD.66.106006
[arXiv:hep-th/0105097 [hep-th]].
%2080 citations counted in INSPIRE as of 04 Mar 2024

\bibitem{drs}
K.~Dasgupta, G.~Rajesh and S.~Sethi,
``M theory, orientifolds and G - flux,''
JHEP \textbf{08} (1999), 023
%doi:10.1088/1126-6708/1999/08/023
[arXiv:hep-th/9908088 [hep-th]].
%874 citations counted in INSPIRE as of 11 Nov 2025

\bibitem{GVW}
S.~Gukov, C.~Vafa and E.~Witten,
``CFT's from Calabi-Yau four folds,''
Nucl. Phys. B \textbf{584} (2000), 69
[erratum: Nucl. Phys. B \textbf{608} (2001), 477]
%doi:10.1016/S0550-3213(00)00373-4
[arXiv:hep-th/9906070 [hep-th]].
%1190 citations counted in INSPIRE as of 04 Mar 2024


\bibitem{lmicu}
J.~Louis and A.~Micu,
``Heterotic string theory with background fluxes,''
Nucl. Phys. B \textbf{626} (2002), 26
%doi:10.1016/S0550-3213(02)00040-8
[arXiv:hep-th/0110187 [hep-th]].
%54 citations counted in INSPIRE as of 30 Jul 2025


\bibitem{randall_sundrum}
L.~Randall and R.~Sundrum,
``Out of this world supersymmetry breaking,''
Nucl. Phys. B \textbf{557} (1999), 79
%doi:10.1016/S0550-3213(99)00359-4
[arXiv:hep-th/9810155 [hep-th]];
%2010 citations counted in INSPIRE as of 25 Jul 2025
L.~Randall and R.~Sundrum,
``A Large mass hierarchy from a small extra dimension,''
Phys. Rev. Lett. \textbf{83} (1999), 3370
%doi:10.1103/PhysRevLett.83.3370
[arXiv:hep-ph/9905221 [hep-ph]];
%10395 citations counted in INSPIRE as of 25 Jul 2025
L.~Randall and R.~Sundrum,
``An Alternative to compactification,''
Phys. Rev. Lett. \textbf{83} (1999), 4690
%doi:10.1103/PhysRevLett.83.4690
[arXiv:hep-th/9906064 [hep-th]].
%7843 citations counted in INSPIRE as of 25 Jul 2025
%\cite{Dudas:1997cd}

\bibitem{nilles}
K.~Choi, A.~Falkowski, H.~P.~Nilles and M.~Olechowski,
``Soft supersymmetry breaking in KKLT flux compactification,''
Nucl. Phys. B \textbf{718} (2005), 113
%doi:10.1016/j.nuclphysb.2005.04.032
[arXiv:hep-th/0503216 [hep-th]].
%509 citations counted in INSPIRE as of 31 Jul 2025

%\cite{Veneziano:1982ah}
\bibitem{vy}
G.~Veneziano and S.~Yankielowicz,
``An Effective Lagrangian for the Pure N=1 Supersymmetric Yang-Mills Theory,''
Phys. Lett. B \textbf{113} (1982), 231.
%doi:10.1016/0370-2693(82)90828-0
%786 citations counted in INSPIRE as of 17 Apr 2024


\bibitem{kklmmt}
S.~Kachru, R.~Kallosh, A.~D.~Linde, J.~M.~Maldacena, L.~P.~McAllister and S.~P.~Trivedi,
``Towards inflation in string theory,''
JCAP \textbf{10} (2003), 013
%doi:10.1088/1475-7516/2003/10/013
[arXiv:hep-th/0308055 [hep-th]].
%1336 citations counted in INSPIRE as of 31 Jul 2025

\bibitem{macallister}
M.~Demirtas, M.~Kim, L.~McAllister, J.~Moritz and A.~Rios-Tascon,
``Small cosmological constants in string theory,''
JHEP \textbf{12} (2021), 136
%doi:10.1007/JHEP12(2021)136
[arXiv:2107.09064 [hep-th]];
%95 citations counted in INSPIRE as of 31 Jul 2025
M.~Demirtas, M.~Kim, L.~McAllister, J.~Moritz and A.~Rios-Tascon,
``Exponentially Small Cosmological Constant in String Theory,''
Phys. Rev. Lett. \textbf{128} (2022) no.1, 011602
%doi:10.1103/PhysRevLett.128.011602
[arXiv:2107.09065 [hep-th]].

\bibitem{sethi_rev}
S.~Sethi,
``Supersymmetry Breaking by Fluxes,''
JHEP \textbf{10} (2018), 022
%doi:10.1007/JHEP10(2018)022
[arXiv:1709.03554 [hep-th]].
%182 citations counted in INSPIRE as of 11 Nov 2025

%\cite{Bena:2018fqc}
\bibitem{bdgl}
I.~Bena, E.~Dudas, M.~Gra\~na and S.~L\"ust,
``Uplifting Runaways,''
Fortsch. Phys. \textbf{67} (2019) no.1-2, 1800100
%doi:10.1002/prop.201800100
[arXiv:1809.06861 [hep-th]].
%136 citations counted in INSPIRE as of 06 Mar 2024


\bibitem{klebanov-strassler}
I.~R.~Klebanov and M.~J.~Strassler,
``Supergravity and a confining gauge theory: Duality cascades and chi SB resolution of naked singularities,''
JHEP \textbf{08} (2000), 052
%doi:10.1088/1126-6708/2000/08/052
[arXiv:hep-th/0007191 [hep-th]].
%1999 citations counted in INSPIRE as of 31 Jul 2025

%\cite{tadpole_problem}
\bibitem{tadpole_problem}
I.~Bena, J.~Blab\"ack, M.~Gra\~na and S.~L\"ust,
``The tadpole problem,''
JHEP \textbf{11} (2021), 223
%doi:10.1007/JHEP11(2021)223
[arXiv:2010.10519 [hep-th]].
%89 citations counted in INSPIRE as of 17 Apr 2024


%\cite{Bena:2009xk}
\bibitem{KKLT_disc1}
I.~Bena, M.~Grana and N.~Halmagyi,
``On the Existence of Meta-stable Vacua in Klebanov-Strassler,''
JHEP \textbf{09} (2010), 087
%doi:10.1007/JHEP09(2010)087
[arXiv:0912.3519 [hep-th]].
%211 citations counted in INSPIRE as of 17 Apr 2024

%\cite{Bena:2011wh}
\bibitem{KKLT_disc2}
I.~Bena, G.~Giecold, M.~Grana, N.~Halmagyi and S.~Massai,
``The backreaction of anti-D3 branes on the Klebanov-Strassler geometry,''
JHEP \textbf{06} (2013), 060
%doi:10.1007/JHEP06(2013)060
[arXiv:1106.6165 [hep-th]].
%96 citations counted in INSPIRE as of 17 Apr 2024

%\cite{Blumenhagen:2019qcg}
\bibitem{KKLT_disc3}
R.~Blumenhagen, D.~Kl\"awer and L.~Schlechter,
``Swampland Variations on a Theme by KKLT,''
JHEP \textbf{05} (2019), 152
%doi:10.1007/JHEP05(2019)152
[arXiv:1902.07724 [hep-th]].
%89 citations counted in INSPIRE as of 17 Apr 2024

%\cite{Randall:2019ent}
\bibitem{lisa-KKLT}
L.~Randall,
``The Boundaries of KKLT,''
Fortsch. Phys. \textbf{68} (2020) no.3-4, 1900105
%doi:10.1002/prop.201900105
[arXiv:1912.06693 [hep-th]].
%43 citations counted in INSPIRE as of 17 Apr 2024

%\cite{Dudas:2019pls}
\bibitem{KKLT_disc4}
E.~Dudas and S.~L\"ust,
``An update on moduli stabilization with antibrane uplift,''
JHEP \textbf{03} (2021), 107
%doi:10.1007/JHEP03(2021)107
[arXiv:1912.09948 [hep-th]].
%42 citations counted in INSPIRE as of 17 Apr 2024

%\cite{Gao:2020xqh}
\bibitem{KKLT_disc5}
X.~Gao, A.~Hebecker and D.~Junghans,
``Control issues of KKLT,''
Fortsch. Phys. \textbf{68} (2020), 2000089
%doi:10.1002/prop.202000089
[arXiv:2009.03914 [hep-th]].
%73 citations counted in INSPIRE as of 17 Apr 2024

%\cite{Lust:2022lfc}
\bibitem{KKLT_disc6}
S.~L\"ust, C.~Vafa, M.~Wiesner and K.~Xu,
``Holography and the KKLT scenario,''
JHEP \textbf{10} (2022), 188
%doi:10.1007/JHEP10(2022)188
[arXiv:2204.07171 [hep-th]].
%38 citations counted in INSPIRE as of 17 Apr 2024

\bibitem{balasubramanian}
V.~Balasubramanian, P.~Berglund, J.~P.~Conlon and F.~Quevedo,
``Systematics of moduli stabilisation in Calabi-Yau flux compactifications,''
JHEP \textbf{03} (2005), 007
%doi:10.1088/1126-6708/2005/03/007
[arXiv:hep-th/0502058 [hep-th]].
%1189 citations counted in INSPIRE as of 31 Oct 2025

\bibitem{stringcosmo-rev-quevedo}
M.~Cicoli, J.~P.~Conlon, A.~Maharana, S.~Parameswaran, F.~Quevedo and I.~Zavala,
``String cosmology: From the early universe to today,''
Phys. Rept. \textbf{1059} (2024), 1
%doi:10.1016/j.physrep.2024.01.002
[arXiv:2303.04819 [hep-th]].

\bibitem{cm}
V.~F.~Mukhanov and G.~V.~Chibisov,
``Quantum Fluctuations and a Nonsingular Universe,''
JETP Lett. \textbf{33} (1981), 532.
%1682 citations counted in INSPIRE as of 03 Jun 2021

\bibitem{cm_rev}
V.~F.~Mukhanov, H.~A.~Feldman and R.~H.~Brandenberger,
``Theory of cosmological perturbations. Part 1. Classical perturbations. Part 2. Quantum theory of perturbations. Part 3. Extensions,''
Phys. Rept. \textbf{215} (1992), 203.
%doi:10.1016/0370-1573(92)90044-Z
%3462 citations counted in INSPIRE as of 27 Nov 2023

\bibitem{planck_cosmo}
N.~Aghanim \textit{et al.} [Planck],
``Planck 2018 results. VI. Cosmological parameters,''
Astron. Astrophys. \textbf{641} (2020), A6
[erratum: Astron. Astrophys. \textbf{652} (2021), C4]
%doi:10.1051/0004-6361/201833910
[arXiv:1807.06209 [astro-ph.CO]].
%19165 citations counted in INSPIRE as of 15 Oct 2025

 \bibitem{lm}
 F.~Lucchin and S.~Matarrese,
``Power Law Inflation,''
Phys. Rev. D \textbf{32} (1985), 1316.
% doi:10.1103/PhysRevD.32.1316
%944 citations counted in INSPIRE as of 15 Feb 2024

\bibitem{dbi-inflation}
E.~Silverstein and D.~Tong,
``Scalar speed limits and cosmology: Acceleration from D-cceleration,''
Phys. Rev. D \textbf{70} (2004), 103505.
%doi:10.1103/PhysRevD.70.103505
[arXiv:hep-th/0310221 [hep-th]]; 
M.~Alishahiha, E.~Silverstein and D.~Tong,
``DBI in the sky,''
Phys. Rev. D \textbf{70} (2004), 123505
%doi:10.1103/PhysRevD.70.123505
[arXiv:hep-th/0404084 [hep-th]].


\bibitem{alpha-attractors}
R.~Kallosh, A.~Linde and D.~Roest,
``Superconformal Inflationary $\alpha$-Attractors,''
JHEP \textbf{11} (2013), 198
%doi:10.1007/JHEP11(2013)198
[arXiv:1311.0472 [hep-th]];
M.~Galante, R.~Kallosh, A.~Linde and D.~Roest,
``Unity of Cosmological Inflation Attractors,''
Phys. Rev. Lett. \textbf{114} (2015) no.14, 141302
%doi:10.1103/PhysRevLett.114.141302
[arXiv:1412.3797 [hep-th]].

\bibitem{distance-conj}
H.~Ooguri, E.~Palti, G.~Shiu and C.~Vafa,
``Distance and de Sitter Conjectures on the Swampland,''
Phys. Lett. B \textbf{788} (2019), 180
%doi:10.1016/j.physletb.2018.11.018
[arXiv:1810.05506 [hep-th]].

\bibitem{transplanckian-conj}
A.~Bedroya and C.~Vafa,
``Trans-Planckian Censorship and the Swampland,''
JHEP \textbf{09} (2020), 123
%doi:10.1007/JHEP09(2020)123
[arXiv:1909.11063 [hep-th]].

\bibitem{DallAgata:2014qsj}
G.~Dall'Agata and F.~Zwirner,
``On sgoldstino-less supergravity models of inflation,''
JHEP \textbf{12} (2014), 172
%doi:10.1007/JHEP12(2014)172
[arXiv:1411.2605 [hep-th]].
%128 citations counted in INSPIRE as of 16 Oct 2025

\bibitem{Kallosh:2014hxa}
R.~Kallosh, A.~Linde and M.~Scalisi,
``Inflation, de Sitter Landscape and Super-Higgs effect,''
JHEP \textbf{03} (2015), 111
%doi:10.1007/JHEP03(2015)111
[arXiv:1411.5671 [hep-th]].
%65 citations counted in INSPIRE as of 16 Oct 2025

\bibitem{Scalisi:2015qga}
M.~Scalisi,
``Cosmological $\alpha$-attractors and de Sitter landscape,''
JHEP \textbf{12} (2015), 134
%doi:10.1007/JHEP12(2015)134
[arXiv:1506.01368 [hep-th]].

\bibitem{Ferrara:2014kva}
S.~Ferrara, R.~Kallosh and A.~Linde,
``Cosmology with Nilpotent Superfields,''
JHEP \textbf{10} (2014), 143
%doi:10.1007/JHEP10(2014)143
[arXiv:1408.4096 [hep-th]].

\bibitem{Kahn:2015mla}
Y.~Kahn, D.~A.~Roberts and J.~Thaler,
``The goldstone and goldstino of supersymmetric inflation,''
JHEP \textbf{10} (2015), 001
%doi:10.1007/JHEP10(2015)001
[arXiv:1504.05958 [hep-th]].

\bibitem{Ferrara:2015tyn}
S.~Ferrara, R.~Kallosh and J.~Thaler,
``Cosmology with orthogonal nilpotent superfields,''
Phys. Rev. D \textbf{93} (2016) no.4, 043516
%doi:10.1103/PhysRevD.93.043516
[arXiv:1512.00545 [hep-th]].
%72 citations counted in INSPIRE as of 16 Oct 2025

\bibitem{Carrasco:2015iij}
J.~J.~M.~Carrasco, R.~Kallosh and A.~Linde,
``Minimal supergravity inflation,''
Phys. Rev. D \textbf{93} (2016) no.6, 061301
%doi:10.1103/PhysRevD.93.061301
[arXiv:1512.00546 [hep-th]].

\bibitem{Hasegawa:2017hgd}
F.~Hasegawa, K.~Mukaida, K.~Nakayama, T.~Terada and Y.~Yamada,
``Gravitino Problem in Minimal Supergravity Inflation,''
Phys. Lett. B \textbf{767} (2017), 392
%doi:10.1016/j.physletb.2017.02.030
[arXiv:1701.03106 [hep-ph]].

\bibitem{Dudas:2021njv}
E.~Dudas, M.~A.~G.~Garcia, Y.~Mambrini, K.~A.~Olive, M.~Peloso and S.~Verner,
``Slow and Safe Gravitinos,''
Phys. Rev. D \textbf{103} (2021), 123519
%doi:10.1103/PhysRevD.103.123519
[arXiv:2104.03749 [hep-th]].

\bibitem{Dalianis:2017okk}
I.~Dalianis and F.~Farakos,
``Constrained superfields from inflation to reheating,''
Phys. Lett. B \textbf{773} (2017), 610
%doi:10.1016/j.physletb.2017.09.020
[arXiv:1705.06717 [hep-th]].

\bibitem{Nilles:2001ry}
H.~P.~Nilles, M.~Peloso and L.~Sorbo,
``Nonthermal production of gravitinos and inflatinos,''
Phys. Rev. Lett. \textbf{87} (2001), 051302
%doi:10.1103/PhysRevLett.87.051302
[arXiv:hep-ph/0102264 [hep-ph]].

\bibitem{stringcosmo-rev-mcallister}
L.~McAllister and E.~Silverstein,
``String Cosmology: A Review,''
Gen. Rel. Grav. \textbf{40} (2008), 565
%doi:10.1007/s10714-007-0556-6
[arXiv:0710.2951 [hep-th]].

\bibitem{rev-cosmo-const}
S.~Weinberg,
``The Cosmological Constant Problem,''
Rev. Mod. Phys. \textbf{61} (1989), 1. 
%doi:10.1103/RevModPhys.61.1

\bibitem{bousso-polchinski}
R.~Bousso and J.~Polchinski,
``Quantization of four form fluxes and dynamical neutralization of the cosmological constant,''
JHEP \textbf{06} (2000), 006
%doi:10.1088/1126-6708/2000/06/006
[arXiv:hep-th/0004134 [hep-th]].

\bibitem{desitter-conj} 
G.~Obied, H.~Ooguri, L.~Spodyneiko and C.~Vafa,
``De Sitter Space and the Swampland,''
[arXiv:1806.08362 [hep-th]];
P.~Agrawal, G.~Obied, P.~J.~Steinhardt and C.~Vafa,
``On the Cosmological Implications of the String Swampland,''
Phys. Lett. B \textbf{784} (2018), 271
%doi:10.1016/j.physletb.2018.07.040
[arXiv:1806.09718 [hep-th]].

\bibitem{quintessence} 
B.~Ratra and P.~J.~E.~Peebles,
``Cosmological Consequences of a Rolling Homogeneous Scalar Field,''
Phys. Rev. D \textbf{37} (1988), 3406
%doi:10.1103/PhysRevD.37.3406; 
R.~R.~Caldwell, R.~Dave and P.~J.~Steinhardt,
``Cosmological imprint of an energy component with general equation of state,''
Phys. Rev. Lett. \textbf{80} (1998), 1582
%doi:10.1103/PhysRevLett.80.1582
[arXiv:astro-ph/9708069 [astro-ph]];
for a review, see e.g. 
P.~J.~E.~Peebles and B.~Ratra,
``The Cosmological Constant and Dark Energy,''
Rev. Mod. Phys. \textbf{75} (2003), 559
%doi:10.1103/RevModPhys.75.559
[arXiv:astro-ph/0207347 [astro-ph]].

\bibitem{uzan}
J.~P.~Uzan,
``The Fundamental Constants and Their Variation: Observational Status and Theoretical Motivations,''
Rev. Mod. Phys. \textbf{75} (2003), 403
%doi:10.1103/RevModPhys.75.403
[arXiv:hep-ph/0205340 [hep-ph]].

\bibitem{kounpart}
C.~Kounnas and H.~Partouche,
``Super no-scale models in string theory,''
Nucl. Phys. B \textbf{913} (2016), 593
%doi:10.1016/j.nuclphysb.2016.10.001
[arXiv:1607.01767 [hep-th]].
%63 citations counted in INSPIRE as of 29 Oct 2025

\bibitem{hervenbnf}
S.~Abel, E.~Dudas, D.~Lewis and H.~Partouche,
``Stability and vacuum energy in open string models with broken supersymmetry,''
JHEP \textbf{10} (2019), 226
%doi:10.1007/JHEP10(2019)226
[arXiv:1812.09714 [hep-th]];
S.~Abel, T.~Coudarchet and H.~Partouche,
``On the stability of open-string orbifold models with broken supersymmetry,''
Nucl. Phys. B \textbf{957} (2020), 115100
[erratum: Nucl. Phys. B \textbf{1004} (2024), 116548]
%doi:10.1016/j.nuclphysb.2020.115100
[arXiv:2003.02545 [hep-th]].
%19 citations counted in INSPIRE as of 29 Oct 2025

\bibitem{tatar}
S.~Brahma, K.~Dasgupta and R.~Tatar,
``Four-dimensional de Sitter space is a Glauber-Sudarshan state in string theory,''
JHEP \textbf{07} (2021), 114
%doi:10.1007/JHEP07(2021)114
[arXiv:2007.00786 [hep-th]].
%41 citations counted in INSPIRE as of 11 Nov 2025

\bibitem{andriot}
D.~Andriot,
``New constraints on classical de Sitter: flirting with the swampland,''
Fortsch. Phys. \textbf{67} (2019) no.1-2, 1800103
%doi:10.1002/prop.201800103
[arXiv:1807.09698 [hep-th]];
D.~Andriot, P.~Marconnet and T.~Wrase,
``New de Sitter solutions of 10d type IIB supergravity,''
JHEP \textbf{08} (2020), 076
%doi:10.1007/JHEP08(2020)076
[arXiv:2005.12930 [hep-th]].

\bibitem{russo}
J.~G.~Russo,
  ``Exact solution of scalar-tensor cosmology with exponential potentials  and
  transient acceleration,''
  Phys.\ Lett.\  {\bf B 600} (2004), 185
  [arXiv:hep-th/0403010].
  %%CITATION = PHLTA,B600,185;%%

    \bibitem{cosmo_power1}
J.~J.~Halliwell,
  ``Scalar Fields In Cosmology With An Exponential Potential,''
  Phys.\ Lett.\  {\bf B 185} (1987) 341.
  %%CITATION = PHLTA,B185,341;%%

    \bibitem{cosmo_power2}
L.~F.~Abbott and M.~B.~Wise,
 ``Constraints On Generalized Inflationary Cosmologies,''
 Nucl.\ Phys.\  {\bf B 244} (1984) 541.
 %%CITATION = NUPHA,B244,541~;%%

   \bibitem{cosmo_power3}
D.~H.~Lyth and E.~D.~Stewart,
 ``The Curvature perturbation in power law (e.g. extended)
 %inflation,''
 Phys.\ Lett.\  {\bf B 274} (1992) 168.
 %%CITATION = PHLTA,B274,168~;%%

   \bibitem{cosmo_power4}
I.~P.~C.~Heard and D.~Wands,
  ``Cosmology with positive and negative exponential potentials,''
  Class.\ Quant.\ Grav.\  {\bf 19} (2002), 5435
  [arXiv:gr-qc/0206085].
  %%CITATION = CQGRD,19,5435;%%

    \bibitem{cosmo_power5}
N.~Ohta,
  ``Accelerating Cosmologies from S-Branes,''
  Phys.\ Rev.\ Lett.\  {\bf 91} (2003), 061303
  [arXiv:hep-th/0303238].
  %%CITATION = PRLTA,91,061303;%%

    \bibitem{cosmo_power6}
S.~Roy,
  ``Accelerating cosmologies from M/string theory compactifications,''
  Phys.\ Lett.\  {\bf B 567} (2003), 322
  [arXiv:hep-th/0304084].
  %%CITATION = PHLTA,B567,322;%%

    \bibitem{cosmo_power7}
P.~K.~Townsend and M.~N.~R.~Wohlfarth,
  ``Accelerating cosmologies from compactification,''
  Phys.\ Rev.\ Lett.\  {\bf 91} (2003), 061302
  [arXiv:hep-th/0303097].
  %%CITATION = PRLTA,91,061302;%%

    \bibitem{cosmo_power8}
R.~Emparan and J.~Garriga,
  ``A note on accelerating cosmologies from compactifications and S-branes,''
  JHEP {\bf 0305} (2003), 028
  [arXiv:hep-th/0304124].
  %%CITATION = JHEPA,0305,028;%%

    \bibitem{cosmo_power9}
    E.~Bergshoeff, A.~Collinucci, U.~Gran, M.~Nielsen and D.~Roest,
  ``Transient quintessence from group manifold reductions or how all roads lead to Rome,''
  Class.\ Quant.\ Grav.\  {\bf 21} (2004), 1947
  [hep-th/0312102].
  %%CITATION = HEP-TH/0312102;%%

    \bibitem{cosmo_power10}
A.~A.~Andrianov, F.~Cannata and A.~Y.~Kamenshchik,
  ``General solution of scalar field cosmology with a (piecewise) exponential potential,''
  JCAP {\bf 1110} (2011), 004
  [arXiv:1105.4515 [gr-qc]].

  \bibitem{meo_as}
  M.~Meo and A.~Sagnotti,
``On Pre-Inflationary non Gaussianities,''
[arXiv:2510.01360 [hep-th]], to appear in JHEP; M.~Meo, Master Thesis,
%``Primordial Non-Gaussianity from a String-Inspired Cosmology,''
[arXiv:2510.23377 [hep-th]].

  \bibitem{maldacena}
  J.~M.~Maldacena,
``Non-Gaussian features of primordial fluctuations in single field inflationary models,''
JHEP \textbf{05} (2003), 013
%doi:10.1088/1126-6708/2003/05/013
[arXiv:astro-ph/0210603 [astro-ph]].
%3079 citations counted in INSPIRE as of 16 Oct 2025

\bibitem{Planck_ng}
P.A.R.~Ade \textit{et al.} [Planck],
%``Planck 2013 Results. XXIV. Constraints on primordial non-Gaussianity,''
Astron. Astrophys. \textbf{571} (2014), A24
%doi:10.1051/0004-6361/201321554
[arXiv:1303.5084 [astro-ph.CO]];
%840 citations counted in INSPIRE as of 15 Sep 2025
P.~A.~R.~Ade \textit{et al.} [Planck],
%``Planck 2015 results. XVII. Constraints on primordial non-Gaussianity,''
Astron. Astrophys. \textbf{594} (2016), A17
%doi:10.1051/0004-6361/201525836
[arXiv:1502.01592 [astro-ph.CO]].
%862 citations counted in INSPIRE as of 15 Sep 2025
Y.~Akrami \textit{et al.} [Planck],
%``Planck 2018 results. IX. Constraints on primordial non-Gaussianity,''
Astron. Astrophys. \textbf{641} (2020), A9
%doi:10.1051/0004-6361/201935891
[arXiv:1905.05697 [astro-ph.CO]].
%904 citations counted in INSPIRE as of 15 Sep 2025

\bibitem{forecasts}
M.~Abitbol \textit{et al.} [Simons Observatory],
%``The Simons Observatory: science goals and forecasts for the enhanced Large Aperture Telescope,''
JCAP \textbf{08} (2025), 034
%doi:10.1088/1475-7516/2025/08/034
[arXiv:2503.00636 [astro-ph.IM]];
%38 citations counted in INSPIRE as of 15 Sep 2025
W.~Sohn and J.~Fergusson,
%``CMB-S4 forecast on the primordial non-Gaussianity parameter of feature models,''
Phys. Rev. D \textbf{100} (2019) no.6, 063536
%doi:10.1103/PhysRevD.100.063536
[arXiv:1902.01142 [astro-ph.CO]].
%15 citations counted in INSPIRE as of 15 Sep 2025

\bibitem{cd}
 C.~Condeescu and E.~Dudas,
  ``Kasner solutions, climbing scalars and big-bang singularity,''
  JCAP {\bf 1308} (2013), 013
 % doi:10.1088/1475-7516/2013/08/013
  [arXiv:1306.0911 [hep-th]].
  %%CITATION = doi:10.1088/1475-7516/2013/08/013;%%

  %\cite{seiberg-nonren}
\bibitem{seiberg-nonren}
N.~Seiberg,
``Naturalness versus supersymmetric nonrenormalization theorems,''
Phys. Lett. B \textbf{318} (1993), 469
%doi:10.1016/0370-2693(93)91541-T
[arXiv:hep-ph/9309335 [hep-ph]].
%548 citations counted in INSPIRE as of 24 Jan 2025
  
  %\cite{intriligator-seiberg-review}
\bibitem{intriligator-seiberg-review}
K.~A.~Intriligator and N.~Seiberg,
``Lectures on supersymmetric gauge theories and electric-magnetic duality,''
Nucl. Phys. B Proc. Suppl. \textbf{45BC} (1996), 1
%doi:10.1016/0920-5632(95)00626-5
[arXiv:hep-th/9509066 [hep-th]].
%847 citations counted in INSPIRE as of 24 Jan 2025

  %\cite{shifman-vainshtein-beta}
\bibitem{shifman-vainshtein-beta}
M.~A.~Shifman and A.~I.~Vainshtein,
``Solution of the Anomaly Puzzle in SUSY Gauge Theories and the Wilson Operator Expansion,''
Nucl. Phys. B \textbf{277} (1986), 456.
%doi:10.1016/0550-3213(86)90451-7
%622 citations counted in INSPIRE as of 24 Jan 2025


\bibitem{townsend}
D.~Z.~Freedman and A.~K.~Das,
``Gauge Internal Symmetry in Extended Supergravity,''
Nucl. Phys. B \textbf{120} (1977), 221;
%doi:10.1016/0550-3213(77)90041-4
%298 citations counted in INSPIRE as of 07 Feb 2025
P.~K.~Townsend,
``Cosmological Constant in Supergravity,''
Phys. Rev. D \textbf{15} (1977), 2802;
%doi:10.1103/PhysRevD.15.2802
%180 citations counted in INSPIRE as of 07 Feb 2025
S.~W.~MacDowell and F.~Mansouri,
``Unified Geometric Theory of Gravity and Supergravity,''
Phys. Rev. Lett. \textbf{38} (1977), 739
[erratum: Phys. Rev. Lett. \textbf{38} (1977), 1376].
%doi:10.1103/PhysRevLett.38.739

\bibitem{gsop}
M.B.~Green and J.H.~Schwarz,
``Infinity Cancellations in SO(32) Superstring Theory,''
Phys. Lett. B \textbf{151} (1985), 21.
%doi:10.1016/0370-2693(85)90816-0

\bibitem{douglas86}
M.R.~Douglas and B.~Grinstein,
``Dilaton Tadpole for the Open Bosonic String,''
Phys. Lett. B \textbf{183} (1987), 52
[erratum: Phys. Lett. B \textbf{187} (1987), 442].
%doi:10.1016/0370-2693(87)91416-X
%96 citations counted in INSPIRE as of 03 Apr 2024

\bibitem{weinberg86}
S.~Weinberg,
``Cancellation of One Loop Divergences in SO(8192) String Theory,''
Phys. Lett. B \textbf{187} (1987), 278.
%doi:10.1016/0370-2693(87)91096-3
%72 citations counted in INSPIRE as of 03 Apr 2024

\bibitem{bs88}
M.~Bianchi and A.~Sagnotti,
``The Partition Function of the SO(8192) Bosonic String,''
Phys. Lett. B \textbf{211} (1988), 407.
%doi:10.1016/0370-2693(88)91884-9
%75 citations counted in INSPIRE as of 03 Apr 2024

\bibitem{bpz}
A.~A.~Belavin, A.~M.~Polyakov and A.~B.~Zamolodchikov,
``Infinite Conformal Symmetry in Two-Dimensional Quantum Field Theory,''
Nucl. Phys. B \textbf{241} (1984), 333.
%doi:10.1016/0550-3213(84)90052-X
%4223 citations counted in INSPIRE as of 17 Apr 2024

\bibitem{petronzio-stanev}
A.~Sagnotti,
``Roberto Petronzio and the {\textquotedblleft}Tor Vergata{\textquotedblright} Theory Group,''
Nuovo Cim. C \textbf{40} (2017) no.4, 156.
%doi:10.1393/ncc/i2017-17156-8
%0 citations counted in INSPIRE as of 19 Nov 2025

%%%%%%%%%%%%%%%%%%%%%%%%%%%%%%%%%%%%%%%%%%%%%%%%%%%%%%%%%%%%%%%%%%%%%%%%%%%%%%%%%%%

\end{thebibliography}
\end{document}